%% file: feynman_book.tex
\documentclass[12pt]{book}
\usepackage{a4wide}

\usepackage{latexsym}
\usepackage{amsmath}
\usepackage{amsfonts}
\usepackage{amssymb}
\usepackage{amscd}
\usepackage{amsthm}
\usepackage{shuffle}
\usepackage{slashed}
\usepackage{stmaryrd}
\usepackage{graphicx}
\usepackage{float}
\usepackage{axodraw}
\usepackage{color}
\usepackage{tcolorbox}
\tcbuselibrary{breakable}
\usepackage{graphbox}
\usepackage{colortbl}
\usepackage{multirow}
\usepackage{algpseudocode}
\usepackage{url}

\usepackage{pslatex}
\usepackage[latin1,utf8]{inputenc}
\usepackage[OT2,T1]{fontenc}

\usepackage{tikz}
\usetikzlibrary{hobby}
\usetikzlibrary{cd}

\usepackage{cite}

\usepackage{makeidx}

\usepackage[symbols,nogroupskip,nonumberlist,nomain]{glossaries-extra}

\usepackage{bm}

\newcommand{\bq}{\begin{eqnarray}}
\newcommand{\eq}{\end{eqnarray}}
\newcommand{\bs}{\begin{small}}
\newcommand{\es}{\end{small}}
\newcommand{\slashoperator}[2]{|_{#2} #1}
\newcommand{\hypergeometric}[2]{\vphantom{F}_{#1}F_{#2}}

\newcommand{\mathcalI}{\mbox{\fontencoding{OMS}\fontfamily{cmr}\fontseries{m}\fontshape{it}\fontsize{12pt}{12pt}\selectfont I}}

\newcommand{\keepstyleunderbrace}[1]{
  \mathop{
    \mathchoice
    {\underbrace{\displaystyle#1}}
    {\underbrace{\textstyle#1}}
    {\underbrace{\scriptstyle#1}}
    {\underbrace{\scriptscriptstyle#1}}
  }\limits
}

\newcommand{\eps}{\varepsilon}
\newcommand{\Eulerconstant}{\gamma_{\mathrm{E}}}
\newcommand{\loopnumber}{l}
\newcommand{\nedges}{n}
\newcommand{\nexternal}{n_{\mathrm{ext}}}
\newcommand{\nexternalindependent}{e}
\newcommand{\ninternal}{n_{\mathrm{int}}}
\newcommand{\nvertices}{r}
\newcommand{\ninternalvertices}{r_{\mathrm{int}}}
\newcommand{\nverticesquiver}{r}
\newcommand{\NB}{N_B}
\newcommand{\NV}{N_V}
\newcommand{\Nmaster}{N_{\mathrm{master}}}
\newcommand{\Ncohom}{N_{\mathrm{cohom}}}
\newcommand{\NL}{N_L}
\newcommand{\Dint}{D_{\mathrm{int}}}

\newcommand{\Obj}{\mathrm{Obj}}
\newcommand{\Hom}{\mathrm{Hom}}

\newcommand{\lideal}{\langle}
\newcommand{\rideal}{\rangle}

\newcommand{\genuszero}[2]{
 \draw (#1+1.0,#2+1.0) circle (1.0);
}

\newcommand{\genusone}[2]{
 \draw (#1+2.0,#2+1.0) ellipse (2.0 and 1.0);
 \draw (#1+1.5,#2+1.15) to [curve through={(#1+1.7,#2+1.0)(#1+2.0,#2+0.85)(#1+2.3,#2+1.0)}] (#1+2.5,#2+1.15);
 \draw (#1+1.7,#2+1.0) to [curve through={(#1+2.0,#2+1.15)}] (#1+2.3,#2+1.0);
}

\glsxtrnewsymbol[description={speed of light},sort={c}]{speedoflight}{\ensuremath{c}}
\glsxtrnewsymbol[description={reduced Planck constant},sort={hbar}]{reducedPlanckconstant}{\ensuremath{\hbar}}
\glsxtrnewsymbol[description={dimensional regularisation parameter},sort={epsilon}]{eps}{\ensuremath{\eps}}
\glsxtrnewsymbol[description={Euler-Mascheroni constant},sort={gammaE}]{gammaE}{\ensuremath{\Eulerconstant}}
\glsxtrnewsymbol[description={loop number},sort={l}]{loopnumber}{\ensuremath{\loopnumber}}
\glsxtrnewsymbol[description={number of edges},sort={n}]{numberofedges}{\ensuremath{\nedges}}
\glsxtrnewsymbol[description={number of external edges},sort={next}]{numberofexternaledges}{\ensuremath{\nexternal}}
\glsxtrnewsymbol[description={number of internal edges},sort={nint}]{numberofinternaledges}{\ensuremath{\ninternal}}
\glsxtrnewsymbol[description={number of vertices},sort={r}]{numberofvertices}{\ensuremath{\nvertices}}
\glsxtrnewsymbol[description={number of internal vertices},sort={rint}]{numberofinternalvertices}{\ensuremath{\ninternalvertices}}
\glsxtrnewsymbol[description={dimension of space-time},sort={D}]{spacetimedimension}{\ensuremath{D}}
\glsxtrnewsymbol[description={(integer) expansion point within dimensional regularisation},sort={Dint}]{integerspacetimedimension}{\ensuremath{\Dint}}
\glsxtrnewsymbol[description={zeta value},sort={zetan}]{zetan}{\ensuremath{\zeta_n}}
\glsxtrnewsymbol[description={number of kinematic variables},sort={NB}]{numberofkinematicvariables}{\ensuremath{\NB}}
\glsxtrnewsymbol[description={number of master integrals},sort={Nmaster}]{numberofmasterintegrals}{\ensuremath{\Nmaster}}
\glsxtrnewsymbol[description={dimension of the twisted cohomology group},sort={Ncohom}]{dimensionoftwistedcohomology}{\ensuremath{\Ncohom}}
\glsxtrnewsymbol[description={number of letters},sort={NL}]{numberofletters}{\ensuremath{\NL}}
\glsxtrnewsymbol[description={number of Baikov variables},sort={NV}]{numberofBaikovvariables}{\ensuremath{\NV}}
\glsxtrnewsymbol[description={first Symanzik polynomial},sort={U}]{firstgraphpolynomial}{\ensuremath{{\mathcal U}}}
\glsxtrnewsymbol[description={second Symanzik polynomial},sort={F}]{secondgraphpolynomial}{\ensuremath{{\mathcal F}}}
\glsxtrnewsymbol[description={Lee-Pomeransky polynomial},sort={G}]{LeePomeranskypolynomial}{\ensuremath{{\mathcal G}}}
\glsxtrnewsymbol[description={Baikov polynomial},sort={B}]{Baikovpolynomial}{\ensuremath{{\mathcal B}}}
\glsxtrnewsymbol[description={Kirchhoff polynomial},sort={K}]{Kirchhoffpolynomial}{\ensuremath{{\mathcal K}}}
\glsxtrnewsymbol[description={Schwinger parameter},sort={alphaj}]{Schwingerparameter}{\ensuremath{\alpha_j}}
\glsxtrnewsymbol[description={Feynman parameter},sort={aj}]{Feynmanparameter}{\ensuremath{a_j}}
\glsxtrnewsymbol[description={Lee-Pomeransky variable},sort={uj}]{LeePomeranskyvariable}{\ensuremath{u_j}}
\glsxtrnewsymbol[description={Baikov variable},sort={zj}]{Baikovvariable}{\ensuremath{z_j}}
\glsxtrnewsymbol[description={objects of a category},sort={Obj}]{Objectscategory}{\ensuremath{\Obj}}
\glsxtrnewsymbol[description={morphisms of a category},sort={Hom}]{Morphismscategory}{\ensuremath{\Hom}}
\glsxtrnewsymbol[description={category of finite-dimensional ${\mathbb F}$-vector spaces},sort={VectF}]{CategoryVectF}{\ensuremath{\mathrm{\bf Vect}_{{\mathbb F}}}}
\glsxtrnewsymbol[description={category of sets},sort={Set}]{CategorySet}{\ensuremath{\mathrm{\bf Set}}}
\glsxtrnewsymbol[description={category of groups},sort={Grp}]{CategoryGrp}{\ensuremath{\mathrm{\bf Grp}}}
\glsxtrnewsymbol[description={category of finitely generated $R$-modules},sort={ModR}]{CategoryModR}{\ensuremath{\mathrm{\bf Mod}_R}}
\glsxtrnewsymbol[description={category of finitely generated projective $R$-modules},sort={ProjR}]{CategoryProjR}{\ensuremath{\mathrm{\bf Proj}_R}}
\glsxtrnewsymbol[description={category of algebraic varieties defined over ${\mathbb Q}$},sort={VarQ}]{CategoryVarQ}{\ensuremath{\mathrm{\bf Var}_{\mathbb Q}}}
\glsxtrnewsymbol[description={category of smooth projective varieties over ${\mathbb Q}$},sort={SmProjQ}]{CategorySmProjQ}{\ensuremath{\mathrm{\bf SmProj}_{\mathbb Q}}}
\glsxtrnewsymbol[description={category of mixed motives},sort={MixMot}]{CategoryMixMot}{\ensuremath{\mathrm{\bf MixMot}}}
\glsxtrnewsymbol[description={category of pure motives},sort={PureMot}]{CategoryPureMot}{\ensuremath{\mathrm{\bf PureMot}}}
\glsxtrnewsymbol[description={category of mixed Hodge structures},sort={MHS}]{CategoryMHS}{\ensuremath{\mathrm{\bf MHS}}}
\glsxtrnewsymbol[description={category of (pure) Hodge structures},sort={HS}]{CategoryHS}{\ensuremath{\mathrm{\bf HS}}}

\newcounter{exercise}
\theoremstyle{plain}
\newtheorem{theoremcounter}{}[]

\newtheorem{algorithmcounter}{}[]
\newtheorem{definitioncounter}{}[]

\newtheorem{theorem}[theoremcounter]{Theorem}
\newtheorem{corollary}[theoremcounter]{Corollary}
\newtheorem{proposition}[theoremcounter]{Proposition}
\newtheorem{lemma}[theoremcounter]{Lemma}
\newtheorem{myalgorithm}[algorithmcounter]{Algorithm}

\newtheorem*{digression}{Digression}

\newtheorem{definition}[definitioncounter]{Definition}

\makeglossaries

\makeindex

\begin{document}

\thispagestyle{empty}

\begin{flushright}
  MITP/22-001
\end{flushright}

\vspace*{1cm}

\begin{center}
 {\Huge {\bf Feynman Integrals}}
 \\
 \vspace{2cm}
 {\Large Stefan Weinzierl} 
 \\
 \vspace{1cm}
\end{center}

\vspace{2cm}

\begin{center}
\includegraphics[scale=1.0]{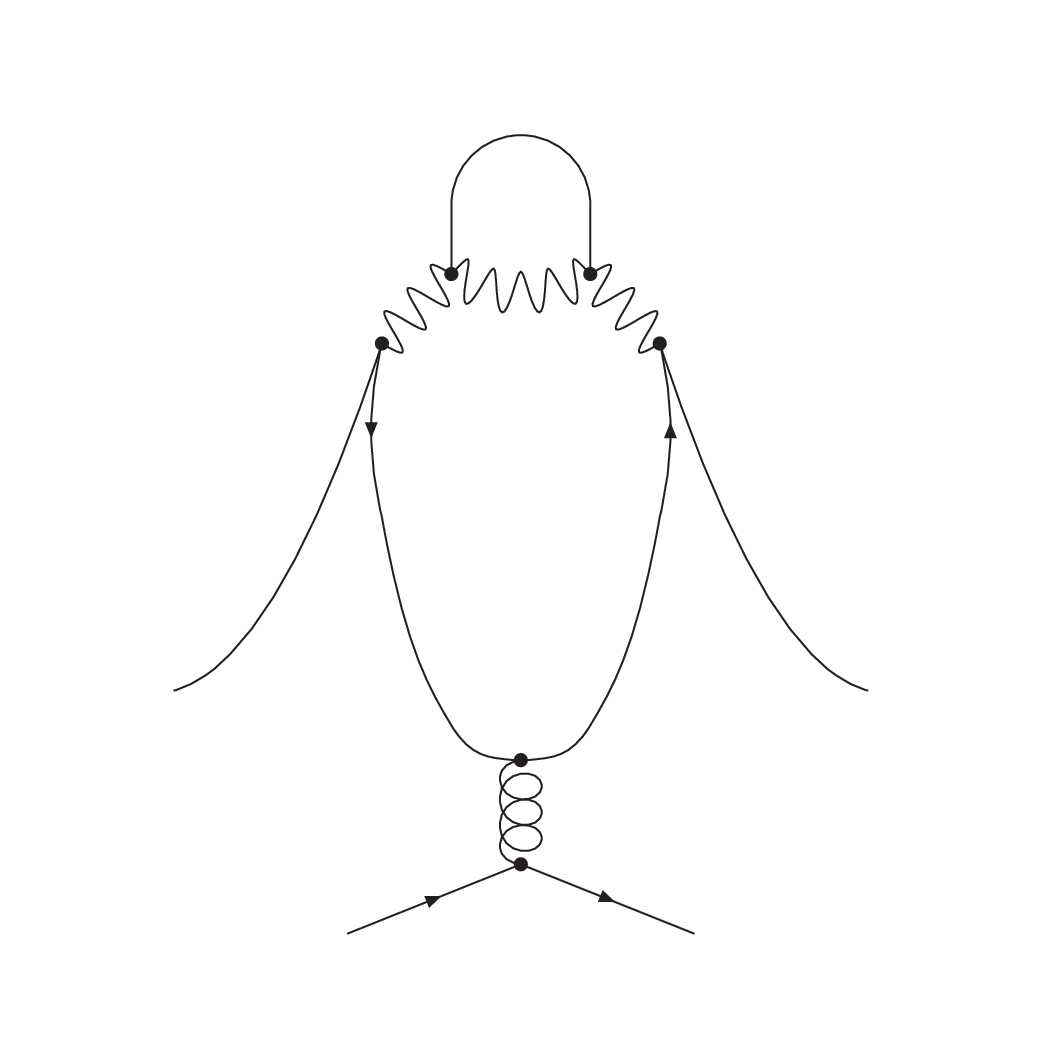}
\end{center}

\vspace*{\fill}

\newpage

\vspace*{\fill}
This is a preprint of the following work: Stefan Weinzierl, Feynman Integrals, 2022,
Springer reproduced with permission of Springer. 
The final authenticated version is available online at:
https://doi.org/10.1007/978-3-030-99558-4

\newpage

\tableofcontents


\input introduction.tex

\input basics.tex

\input graph_polynomials.tex

\input quantum_field_theory.tex

\input oneloop.tex

\input iterated_integrals.tex

\input transformations.tex

\input multiple_polylogs.tex

\input nested_sums.tex

\input sector_decomposition.tex

\input hopf.tex

\input cluster.tex

\input elliptics.tex

\input motives.tex

\input numerics.tex

\input final_project.tex

\begin{appendix}
\input spinor.tex

\input oneloopintegrals.tex

\input transcendental.tex

\input lie_algebra.tex

\input dirichlet.tex

\input moduli_space.tex

\input algebraic_geometry.tex

\input algorithms.tex

\input finite_fields.tex

\input solutions.tex
\end{appendix}

\newpage
\addcontentsline{toc}{chapter}{References}
{\small
\bibliography{/home/stefanw/notes/biblio}
\bibliographystyle{/home/stefanw/latex-style/JHEP}
}

\newpage
\printglossary[type=symbols,style=long,title={List of symbols}]

\newpage
\addcontentsline{toc}{chapter}{Index}
\printindex

\end{document}

%% file: introduction.tex
\newpage
\chapter{Introduction}
\label{chapter_intro}

Feynman integrals are indispensable for precision calculations in quantum field theory.
They occur as soon as one goes beyond the leading order in perturbative quantum field theory.
Feynman integrals are also fascinating from a mathematical point of view.
They can be used to teach and illustrate a large part of modern mathematics 
with concrete non-trivial examples.

In recent years there were some exiting developments in the field of Feynman integrals,
enlarging significantly our knowledge and understanding of Feynman integrals.
Although there are some excellent older books on the subject \cite{Hwa,Nakanishi,Todorov,Smirnov:2004ym,Smirnov:2006ry,Smirnov:2012gma}, 
a modern introduction to the theory of Feynman integrals, 
which includes the recent developments,
will be helpful.

This book is intended for two types of readers:
First, there is the physicist interested in precision calculations in quantum field theory,
where she/he encounters Feynman integrals.
Her/his primary motivation is to be able to calculate these integrals.
For this audience the book provides current state-of-the-art techniques, covering all aspects
of the computations, from the starting definition of a Feynman integral to the final step of getting a number
out.

Secondly, there is the mathematician, interested in the mathematical aspects underlying Feynman integrals.
These are rich and the book provides wherever possible the connection to mathematics.

Of course, these two topics are not independent but interwoven, which makes the theory of Feynman integrals
so enthralling. 
The book is written in this spirit, showing the deep connection between physics and mathematics
in the field of Feynman integrals.

This book is intended for students at the master level in physics or mathematics.
I tried to keep the essential requirements to a minimum.
As minimum requirements I assume that all readers are familiar with special relativity on the physics side and
the theory of complex functions (i.e. Cauchy's residue theorem) and differential forms on the mathematics side.
Students of physics or mathematics with an interest in mathematical physics should have covered these topics 
during their bachelor studies.
Of course, a knowledge of quantum field theory or algebraic geometry is extremely helpful for the topic of this book.
However, as most readers might have followed one of these courses, but not both, I arranged the material covered
in this book in such a way that no prior knowledge of quantum field theory nor algebraic geometry is assumed.
The relevant topics are introduced as they are needed.
In this way the book can complement a course in quantum field theory or algebraic geometry. 
Of course, the book cannot substitute a course in quantum field theory nor 
a course in algebraic geometry
and readers are encouraged to familiarise themselves with quantum field theory and algebraic geometry
beyond the topics required and introduced in this book.

There are always some readers, who are impatient: Reading chapters~\ref{chapter_basics}, \ref{chapter_iterated_integrals} and \ref{chapter_transformations}
should bring them to the point that they can perform state-of-the-art Feynman integral calculations 
with the method of differential equations.
One might also be lucky that a particular Feynman integral is already known in the literature.
The database Loopedia \cite{Bogner:2017xhp} is a good place to check out first.

For all others, who would like to take the recommended long and scenic route, we start in chapter~\ref{chapter_basics}
with introducing the central objects of this book: Feynman integrals.
We do this by requiring only a basic knowledge of special relativity, avoiding quantum field theory as a prerequisite.
The chapter also introduces the most popular integral representations for Feynman integrals.

The Feynman parameter representation and the Schwinger parameter representation involve two graph
polynomials. These graph polynomials have many interesting properties and we discuss them in detail
in chapter~\ref{chapter_graph_polynomials}.

As we deliberately did not build upon quantum field theory in chapter~\ref{chapter_basics}, we should
nevertheless discuss how Feynman integrals arise in quantum field theory. This is done in chapter~\ref{chapter_qft}.

In many applications within perturbation theory the next-to-leading order correction requires
one-loop integrals. The one-loop integrals are therefore of particular importance.
We devote a special chapter to them (chapter~\ref{chapter_one_loop}).

For all other Feynman integrals the most commonly method to compute these integrals 
(at the time of writing this book) is the method of differential equations.
We introduce this technique in chapter~\ref{chapter_iterated_integrals}.
The most important result of this chapter is the fact, that the computation of Feynman integrals can be reduced to the problem
of finding a suitable transformation for the associated differential equation.
Methods to find such a transformation are discussed in chapter~\ref{chapter_transformations}.

In chapter~\ref{chapter_multiple_polylogarithms} we discuss an important class of functions, which appear in Feynman integral computations:
These are the multiple polylogarithms.

Apart from an iterated integral representation 
(which we use extensively in chapters~\ref{chapter_iterated_integrals} and \ref{chapter_transformations})
Feynman integrals may also be represented as nested sums. 
We discuss this aspect in chapter~\ref{chapter_nested_sums}.
In this chapter we also show the relation of Feynman integrals to 
Gelfand-Kapranov-Zelevinsky hypergeometric systems.

Chapter~\ref{chapter_sector_decomposition} is devoted to sector decomposition.
On the one hand sector decomposition (or in a more mathematical language: blow-ups) allow us to device
an algorithm, which computes numerically the coefficients of the Laurent expansion in the dimensional regularisation
parameter of a Feynman integral.
On the other hand (and on the more formal side), we may use this algorithm to prove that these coefficients
are numerical periods for rational input parameters.

In chapter~\ref{chapter_multiple_polylogarithms} we discussed
the algebraic properties of the multiple polylogarithms.
Chapter~\ref{chapter_hopf} continues this theme and explores the coalgebra side: Coproducts, Hopf algebras,
coactions, symbols and single-valued projections are discussed in this chapter.

With the methods of chapters~\ref{chapter_iterated_integrals} and \ref{chapter_transformations}
we may transform the differential equation of a Feynman integral, which evaluates to multiple polylogarithms, to
a dlog-form. 
We may ask if there is any relation between the arguments of the dlog's and the original kinematic variables.
This will lead us to cluster algebras, which we introduce in chapter~\ref{chapter_cluster}.

In chapter~\ref{chapter_elliptics} we discuss integrals, which do not evaluate to multiple polylogarithms.
We focus on the next-more-complicated case: These are Feynman integrals related to an elliptic curve.
We introduce elliptic curves, elliptic functions, modular transformations, modular forms and the moduli space
of a genus one curve with marked points.

Chapter~\ref{chapter_motives} is the most mathematical chapter of this book: We introduce motives
and mixed Hodge structures and their relation to Feynman integrals.
This chapter continues a thread, which started on the one hand in chapter~\ref{chapter_sector_decomposition}
with (numerical) periods and on the other hand in chapter~\ref{chapter_hopf} with coactions.
In chapter~\ref{chapter_motives} we bring these concepts together. We will see that each
coefficient of the Laurent expansion in the dimensional regularisation parameter of a Feynman integral
corresponds to a motivic period.

At the end of the day physics is about numbers: We would like to get for specified input parameters
(i.e. for specified kinematic variables) a number for a Feynman integral 
(more precisely for the coefficients of the Laurent expansion in the dimensional regularisation parameter).
In chapter~\ref{chapter_numerics} we discuss numerical evaluation routines.
These methods can be used to obtain numerical values to a high numerical precision (up to a few hundred or thousand digits).
This in turn opens the possibility to use the heuristic PSLQ algorithm to simplify analytic expressions.
The main application is the simplification of boundary constants.
In this chapter we also introduce the PSLQ algorithm.

In the last chapter of this book (chapter~\ref{chapter_final_project}) we carry out a full project:
We show in detail how the two-loop penguin diagram on the title page is computed from the starting Feynman diagram
to the final numerical result.
The purpose of this chapter is to show how the methods and algorithms introduced in this book 
are used in practice.

This book is supplemented with several appendices:

In appendix~\ref{appendix_spinors} we review spinors in four space-time dimensions.
These are useful for the methods discussed in chapter~\ref{chapter_one_loop}.
Chapter~\ref{chapter_one_loop} is devoted to one-loop integrals.
There are only a finite number of one-loop integrals which we need to know.
We list all relevant one-loop integrals for massless theories in appendix~\ref{appendix_one_loop_integrals}.

Appendix~\ref{appendix_trancendental} is a supplement to chapter~\ref{chapter_nested_sums}:
We summarise the definitions and main properties of a few transcendental functions:
Hypergeometric functions,
Appell functions,
Lauricella functions and
Horn functions are reviewed in this appendix.

Appendix~\ref{appendix_lie_algebra} is devoted to Lie groups and Lie algebras.
Of course Lie groups and Lie algebras are omnipresent in particle physics.
In the context of Feynman integrals it is useful to know the classification of simple Lie algebras and their
relation to Dynkin diagrams (which we will need in chapter~\ref{chapter_cluster}).
The appendix gives a concise discussion of the classification of simple Lie algebras.

Appendices~\ref{appendix_dirichlet}
and \ref{appendix_moduli_space} supplement chapter~\ref{chapter_elliptics}:
Appendix~\ref{appendix_dirichlet} introduces Dirichlet characters, while appendix~\ref{appendix_moduli_space}
discusses the moduli space ${\mathcal M}_{g,n}$ of a smooth algebraic curve of genus $g$ with $n$ marked points.

Appendix~\ref{appendix_algebraic_geometry} is a concise introduction to the main concepts
of algebraic geometry. We give the definitions of sheaves and schemes.
These are avoided (as much as possible) in the main text of the book, but one is confronted with these terms as soon
as one consults the mathematical literature.

Appendices~\ref{appendix_algorithms} and \ref{appendix_finite_fields} are supplements to chapters~\ref{chapter_iterated_integrals} and \ref{chapter_transformations}.
Appendix~\ref{appendix_algorithms} reviews standard algorithms in polynomial rings 
for computing a Gr\"obner basis, a Nullstellensatz certificate and an annihilator.
These are used in chapters~\ref{chapter_iterated_integrals} and \ref{chapter_transformations}.
Appendix~\ref{appendix_finite_fields} introduces finite field methods, which can be used to speed-up
the integration-by-parts reduction discussed in chapter~\ref{chapter_iterated_integrals:integration_by_parts}.

There are many exercises included in the main text of this book. The solutions to the exercises
are given in appendix~\ref{appendix_solutions}.

This book grew out of lectures on Feynman integrals given at the Johannes Gutenberg University Mainz
in the summer term 2021 (covering the first half of the book) and of lectures given at the 
Higgs Centre School of Theoretical Physics 2021 (covering some of the more advanced topics).
I am grateful to the students attending these lectures for their feedback 
and to Luigi del Debbio, Einan Gardi and Roman Zwicky
for organising the Higgs Centre School.
My particular thanks go to
Alexander Aycock,
Christian Bogner,
Ina H\"onemann,
Philipp Kreer,
Sascha Kromin,
Hildegard M\"uller,
Farroukh Peykar Negar Khiabani, 
Robert Runkel,
Juan Pablo Vesga 
and
Xing Wang
for valuable suggestions on the manuscript.
I am also grateful to John Gracey for helpful comments.

In writing this book I recycled some existing material.
In particular, chapter~\ref{chapter_graph_polynomials} is a revision of a review article \cite{Bogner:2010kv}
on graph polynomials written together with Christian Bogner.
Other chapters have their origin in shorter contributions to summer schools and conference proceedings 
\cite{Weinzierl:2002cg,Weinzierl:2003jx,Weinzierl:2003ub,Weinzierl:2006qs,Weinzierl:2010ps,Weinzierl:2013yn,Weinzierl:2015nda,Weinzierl:2020kyq}.

I would like to thank my collaborators 
Luise Adams,
Marco Besier,
Isabella Bierenbaum,
Christian Bogner,
Ekta Chaubey,
Andre van Hameren,
Philipp Kreer,
Dirk Kreimer,
Sven Moch,
Stefan M\"uller-Stach,
Armin Schweitzer,
Duco van Straten,
Kirsten Tempest,
Peter Uwer,
Jens Vollinga,
Moritz Walden,
Pascal Wasser
and
Raphael Zayadeh,
for their shared work and research interest related to various topics in relation with Feynman integrals.

Finally, I would like to thank a few colleagues, from whom I learned through discussions, conversations or lectures
about different aspects of Feynman integrals:
My thanks go to
David Broadhurst,
Johannes Br\"odel,
Francis Brown,
Lance Dixon,
Claude Duhr,
Johannes Henn,
David Kosower,
Erik Panzer
and
Lorenzo Tancredi.
I also would like to thank Jacob Bourjaily and Henriette Elvang for useful information on book projects.

The figures in this book have been produced with the help of the programs {\tt Axodraw} \cite{Vermaseren:1994je}, {\tt TikZ} \cite{tikz} and {\tt ROOT} \cite{Brun:1996}.

%% file: basics.tex
\newpage
\chapter{Basics}
\label{chapter_basics}

In this chapter we introduce the central object of this book: Feynman integrals.
We first review special relativity in section~\ref{chapter_basics:special_relativity}
and the basic concepts of graphs in section~\ref{chapter_basics:graphs}.
With these preparations and the Feynman rules for a scalar theory we define
Feynman integrals in section~\ref{chapter_basics:feynman_rules}.
Before we embark on calculating the first Feynman integrals we need to introduce
two fundamental concepts: Wick rotation and regularisation.
We do this in section~\ref{chapter_basics:fundamental_concepts}.
We conclude this chapter 
with a section containing
an overview of various integral representations
for Feynman integrals 
(section~\ref{chapter_basics:representations_of_Feynman_integrals}).
This includes the Feynman parameter representation,
the Schwinger parameter representation,
the Baikov representation,
the Lee-Pomeransky representation
and the Mellin-Barnes representation.

\section{Special relativity}
\label{chapter_basics:special_relativity}

Let us denote by 
$\gls{spacetimedimension}$
the number of space-time dimensions.
In our real world $D$ equals $4$ (one time dimension and three spatial
dimensions), but it is extremely helpful to keep this number arbitrary.
We will always assume that space-time consists of one time dimension
and $(D-1)$ spatial dimensions.

The momentum of a particle is a $D$-dimensional vector, whose first
component gives the energy $E$ (divided by the speed of light $c$)
and the remaining $(D-1)$ components
give the components of the spatial momentum, which we label with superscripts:
\bq
 p & = & \left( \frac{E}{c}, p^1, \dots, p^{D-1} \right).
\eq
It is common practice in high-energy physics to work in 
\index{natural units}
natural units, where
\bq
 \gls{speedoflight}
 \; = \;
 \gls{reducedPlanckconstant}
 \; = \;
 1.
\eq
We use this convention throughout this book from now on.
Let us set $p^0=E$.
We then have
\bq
 p & = & \left( p^0, p^1, \dots, p^{D-1} \right).
\eq
We write
\bq
 p^\mu & \mbox{with} & 0 \; \le \; \mu \; \le \; D-1
\eq
for a component of $p$. 
The index $\mu$ is called a 
\index{Lorentz index}
{\bf Lorentz index}.
 
We denote by $g_{\mu\nu}$ the components of the metric tensor. 
The indices $\mu$ and $\nu$ take integer values between $0$ and $(D-1)$.
We are primarily concerned with the 
\index{Minkowski metric}
{\bf Minkowski metric}.
Our convention for the Minkowski metric is
\bq
\label{chapter_basics:def_Minkowski_metric}
 g_{\mu\nu}
 & = &
 \left\{
  \begin{array}{rl}
   1, & \mu=\nu=0, \\
   -1, & \mu=\nu \in \{1,\dots,D-1\}, \\
   0, & \mbox{otherwise}.
  \end{array}
 \right.
\eq
This convention is the standard convention in high-energy physics phenomenology.
Some authors (mostly in the field of formal high-energy physics theory) use a convention, where the roles of
$(+1)$ and $(-1)$ are interchanged.
Working consistently within one or the other convention does not change physics.
The transition from one convention to the other is rather easy and given by a minus sign.
In this book we use the convention as given by eq.~(\ref{chapter_basics:def_Minkowski_metric}).

The 
\index{Minkowski scalar product}
{\bf Minkowski scalar product} of two momentum vectors $p_a$ and $p_b$ is
\bq
 p_a \cdot p_b
 & = &
 \sum\limits_{\mu=0}^{D-1}
 \sum\limits_{\nu=0}^{D-1}
 p_a^\mu \; g_{\mu\nu} \; p_b^\nu.
\eq
Einstein's summation convention is the convention to drop the summation symbol for any Lorentz index, which occurs twice,
once as an upper index and once as a lower index.
The summation is then implicitly assumed.
With Einstein's summation convention we may write
\bq
 p_a \cdot p_b
 & = &
 p_a^\mu \; g_{\mu\nu} \; p_b^\nu.
\eq
The Minkowski scalar product $p_a \cdot p_b$ is an example of a 
\index{Lorentz invariant}
{\bf Lorentz invariant}: 
A Lorentz invariant is a quantity, whose value is not changed under Lorentz transformations.

With the Minkowski scalar product at hand, we may in particular take the scalar product of a momentum vector with itself.
Let us write this out explicitly:
\bq
 p^2 \; = \; p \cdot p
 \; = \;
 \left(p^0\right)^2 - \left(p^1\right)^2 - \left(p^2\right)^2 - \dots - \left(p^{D-1}\right)^2.
\eq
Please note that on the left-hand side $p^2$ denotes the Minkowski scalar product of $p$ with itself,
while on the right-hand side $p^2$ (appearing in the term $(p^2)^2$) denotes the third component (the second spatial component) of $p$.
As the meaning should be clear from the context, we follow common practice and do not disambiguate the notation.

Apart from the momentum, we also associate a mass $m$ to a particle.
We represent a particle propagating in space-time by a line:
\bq
\label{chapter_basics:basic_line}
 \begin{picture}(100,20)(0,0)
 \Line(20,10)(80,10)
\end{picture} 
\eq
If we would like to indicate the direction of the momentum flow, 
we optionally put an arrow:
\bq
 \begin{picture}(100,20)(0,0)
 \ArrowLine(20,8)(80,8)
 \Text(50,12)[b]{\footnotesize $p$}
\end{picture} 
\eq
The line in eq.~(\ref{chapter_basics:basic_line}) is our first building block
for a Feynman graph.
A Feynman graph is a graphical notation for a mathematical formula and a line 
for a particle with momentum $p$ and mass $m$ stands
for
\bq
\label{chapter_basics:scalar_propagator}
 \begin{picture}(90,20)(0,5)
 \ArrowLine(20,8)(80,8)
 \Text(50,12)[b]{\footnotesize $p, m$}
\end{picture} 
 & = &
 \frac{1}{-p^2+m^2}.
\eq
On the right-hand side the Minkowski scalar product of $p$ with itself appears.
Note that the right-hand side is independent of the orientation of $p$.

If one follows standard conventions in quantum field theory the propagator of a scalar particle
with momentum $p$ and mass $m$ is given by
\bq
 \frac{i}{p^2-m^2}.
\eq
This differs by a factor $(-i)$ from eq.~(\ref{chapter_basics:scalar_propagator}).
This is just a prefactor and easily adjusted.
Throughout this book we use the convention that the mathematical expression corresponding to a line
is given by eq.~(\ref{chapter_basics:scalar_propagator}).

\section{Graphs}
\label{chapter_basics:graphs}

Let us now turn to graphs.
An {\bf unoriented graph} consists of edges and vertices, where an edge connects two vertices.
A graph may be connected or disconnected. We will mainly consider connected graphs.
An {\bf oriented graph} is a graph, where for every edge an orientation is chosen.
An oriented graph is also called a 
\index{quiver}
{\bf quiver}.
As any edge connects two vertices, say $v_a$ and $v_b$, an orientation is equivalent to declaring one of the two vertices the 
\index{source}
source for this edge (for example $v_a$) and the other vertex the 
\index{sink}
sink for this edge (for example $v_b$).
An orientated edge is usually drawn with an arrow line:
\bq
 \begin{picture}(90,20)(0,5)
 \Vertex(20,8){2}
 \Vertex(80,8){2}
 \ArrowLine(20,8)(80,8)
 \Text(20,12)[b]{\footnotesize $\mathrm{source}$}
 \Text(80,12)[b]{\footnotesize $\mathrm{sink}$}
\end{picture} 
\eq
The 
\index{valency}
{\bf valency of a vertex} is the number of edges attached to it.
Vertices of valency $0$, $1$ and $2$ are special.
A vertex of valency $0$ is necessarily disconnected from the rest of graph and therefore not relevant for connected graphs.
A vertex of valency $1$ has exactly one edge attached to it. 
This edge is called an 
\index{edge, external}
{\bf external edge}. 
All other edges are called 
\index{edge, internal}
{\bf internal edges}.
In the physics community it is common practice not to draw a vertex of valency 1, but just the external edge.
A vertex of valency $2$ is also called a 
\index{dot}
{\bf dot}.
In physics the use of the word ``vertex'' sometimes implies a vertex of valency $3$ or greater.
This derives from the fact that in a particle picture a vertex of valency $3$ or greater corresponds
to a genuine interaction among particles.

\begin{figure}
\begin{center}
\includegraphics[scale=1.0]{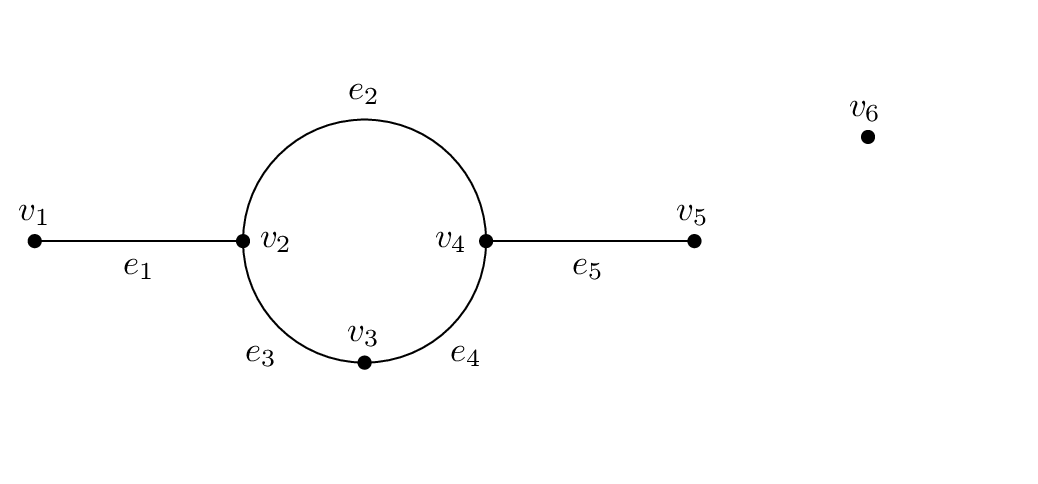}
\end{center}
\caption{
A (disconnected) graph with five edges and six vertices.
The vertex $v_6$ is disconnected from the rest of the graph.
}
\label{chapter_basics:fig_intro_graph}
\end{figure}

As an example for a graph let us look at fig.~\ref{chapter_basics:fig_intro_graph}.
Fig.~\ref{chapter_basics:fig_intro_graph} shows a disconnected graph with five edges and six vertices.
There, vertex $v_6$ has valency $0$ and is disconnected from the rest of the graph.
Vertices $v_1$ and $v_5$ have valency $1$. The edges attached to them ($e_1$ and $e_5$) are the external edges of the graph.
Vertex $v_3$ has valency $2$ and is an example of a dot.
Vertices $v_2$ and $v_4$ have valency $3$ and are genuine interaction vertices.
The internal edges of the graph are $e_2$, $e_3$ and $e_4$.

A Feynman graph is a graph with additional information.
In the basic version we associate to each edge an orientation (thus our graph becomes an oriented graph),
a $D$-dimensional vector $p$ (the momentum) and a number $m$ (the mass).
The physics picture is that an oriented edge represents the propagation of a particle with momentum $p$
and mass $m$.
The momentum flow is in the direction of the orientation of the edge.
Note that the choice of the orientation of an edge does not matter if a change in the orientation is accompanied
by reversing the momentum $p \rightarrow -p$.
Consider again an edge $e$, which connects the vertices $v_a$ and $v_b$.
We have
\bq
 \begin{picture}(90,20)(0,5)
 \Vertex(20,8){2}
 \Vertex(80,8){2}
 \ArrowLine(20,8)(80,8)
 \Text(50,12)[b]{\footnotesize $p, m$}
 \Text(20,12)[b]{\footnotesize $v_a$}
 \Text(80,12)[b]{\footnotesize $v_b$}
\end{picture} 
 & = &
 \begin{picture}(90,20)(0,5)
 \Vertex(20,8){2}
 \Vertex(80,8){2}
 \ArrowLine(80,8)(20,8)
 \Text(50,12)[b]{\footnotesize $-p, m$}
 \Text(20,12)[b]{\footnotesize $v_a$}
 \Text(80,12)[b]{\footnotesize $v_b$}
\end{picture} 
\eq
Let us now consider a graph $G$ with 
$\gls{numberofedges}$
edges and 
$\gls{numberofvertices}$
vertices. Assume that the graph has $k$ connected components.
The 
\index{loop number}
{\bf loop number} 
$\gls{loopnumber}$
is defined by
\bq
\label{chapter_basics:def_loop_number}
 \loopnumber & = & \nedges-\nvertices+k.
\eq
If the graph is connected we have $\loopnumber=\nedges-\nvertices+1$.
The loop number $\loopnumber$ is also called the 
\index{Betti number, first}
{\bf first Betti number} of the graph or the 
\index{cyclomatic number}
{\bf cyclomatic number}.
In the physics context it has the following interpretation:
If we fix all momenta of the external lines and if we impose momentum conservation at each vertex, then the loop number is equal to the number of
independent momentum vectors not constrained by momentum conservation.

A connected graph of loop number $0$ is called a 
\index{tree}
{\bf tree}.
A graph of loop number $0$, connected or not, is called a 
\index{forest}
{\bf forest}.
If the forest has $k$ connected components, it is called a $k$-forest.
A tree is a $1$-forest.
Feynman graphs which are trees pose no conceptual problem.
Our focus in this book is on connected Feynman graphs, which are not trees, e.g. Feynman graphs with loop number $\loopnumber > 0$.

Unless stated otherwise we consider in this book {\bf from now on only connected graphs}.

\begin{figure}
\begin{center}
\includegraphics[scale=1.0]{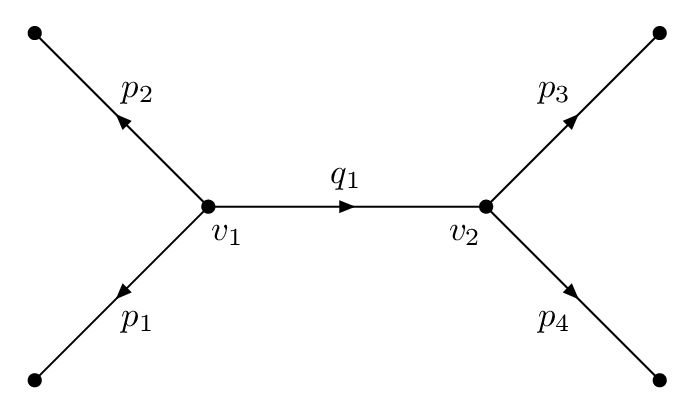}
\end{center}
\caption{
A tree graph with five edges and six vertices. There are four external edges (labelled with momenta $p_1,\dots,p_4$)
and one internal edge (labelled with momentum $q_1$).
The orientation of the external edges is chosen such that all external momenta are outgoing.
}
\label{chapter_basics:fig_tree_graph}
\end{figure}

Let us fix some notation:
Consider a graph $G$ with 
$\gls{numberofexternaledges}$
external edges, 
$\gls{numberofinternaledges}$
internal edges and loop number $\loopnumber$.
As we now always assume that the graph is connected, 
we know from eq.~(\ref{chapter_basics:def_loop_number}) that the graph $G$ must
have
\bq
 \nvertices & = & \nexternal + \ninternal + 1 - \loopnumber
\eq
vertices.
Out of these $\nvertices$ vertices exactly $\nexternal$ vertices are vertices of valency $1$, this leaves
\bq
 \gls{numberofinternalvertices}
 & = &
 \ninternal + 1 - \loopnumber
\eq
vertices of valency $>1$.
For each edge we choose an orientation.
We label the edges such that $e_1,\dots,e_{\ninternal}$ are the internal edges
and $e_{\ninternal+1},\dots,e_{\ninternal+\nexternal}$ are the external edges.
For any edge $e_j$ (internal or external) we denote the momentum flowing 
through this edge (with respect to the chosen orientation) by $q_j$.
For external momenta we use a second notation: We label the momentum flowing through
the external edge $e_{\ninternal+j}$ by $p_j$.
Thus we have
\bq
 p_j & = & q_{\ninternal+j}.
\eq
We will soon see that this redundant notation is useful and simplifies the notation in some formulae.
Consider a vertex $v_a$. We denote by 
\bq
 E^{\mathrm{source}}(v_a) & : &  \mbox{set of edges, which have vertex $v_a$ as source},
 \nonumber \\
 E^{\mathrm{sink}}(v_a) & : & \mbox{set of edges, which have vertex $v_a$ as sink}.
\eq
At each vertex $v_a$ of valency $>1$ we impose momentum conservation:
\bq
 \sum\limits_{e_j \in E^{\mathrm{source}}(v_a)} q_j 
 & = &
 \sum\limits_{e_j \in E^{\mathrm{sink}}(v_a)} q_j.
\eq
Furthermore, we denote by
\bq
 E^{\mathrm{in}} & : &  \mbox{set of edges, which have a vertex of valency $1$ as source},
 \nonumber \\
 E^{\mathrm{out}} & : &  \mbox{set of edges, which have a vertex of valency $1$ as sink}.
\eq
The edges in $E^{\mathrm{in}}$ and $E^{\mathrm{out}}$ are necessarily external edges.
\\
\\
\bs
{\it \refstepcounter{exercise}
{\bf Exercise \theexercise}: 
Consider a connected graph $G$ with the notation as above.
Show that momentum conservation at each vertex of valency $>1$ implies momentum conservation
of the external momenta:
\bq
 \sum\limits_{e_j \in E^{\mathrm{in}}} q_j 
 & = &
 \sum\limits_{e_j \in E^{\mathrm{out}}} q_j.
\eq
If we choose an orientation such that all external edges have a vertex of valency $1$ as sink 
(e.g. $E^{\mathrm{in}} = \emptyset$) this translates to
\bq
 \sum\limits_{j=1}^{\nexternal} p_j & = & 0.
\eq
}
\es
\\
Let us now investigate how many independent momenta we have.
Our graph has $\nedges=\nexternal+\ninternal$ edges and thus we start from $\nedges$ momenta $q_i$ ($1\le i \le \nedges$).
Clearly, in a space of dimension $D$ there can only be $D$ linear independent momenta.
This is not the effect we want to study here. Therefore we assume that the dimension $D$ of space-time
is large enough ($D \ge \nexternal -1 + l$ will be sufficient).
We have seen in the exercise above that the external momenta satisfy momentum conservation.
Thus they are not independent and there is at least one linear relation among them.
We will assume that the external momenta are generic, e.g. there are besides momentum conservation
no further linear relations among the external momenta.
Thus we have $(\nexternal-1)$ linear independent external momenta.
(A non-generic or special configuration of external momenta is for example given by 
the four momenta $p_1$, $p_2$, $p_3=2p_1$, $p_4=-3p_1-p_2$.)
Each vertex of valency $>1$ gives us through momentum conservation at this vertex a relation among the
$\nedges$ momenta $q_j$.
Assuming that the external momenta are known quantities,
this leaves
\bq
 \nedges - \left( \nexternal - 1 \right) - \ninternalvertices
 & = & l
\eq
momenta undetermined.
We label these momenta by $k_1,\dots,k_\loopnumber$ and call them the {\bf independent loop momenta}.
We may then express any other momentum $q_j$ as linear combination of 
the $\loopnumber$ independent loop momenta and the $(\nexternal - 1)$ independent external momenta
with coefficients $\{-1,0,1\}$:
\bq
 q_j & = &
 \sum\limits_{r=1}^\loopnumber \lambda_{jr} k_r
 +
 \sum\limits_{r=1}^{\nexternal - 1} \sigma_{jr} p_r,
 \;\;\;\;\;\;
 \lambda_{jr}, \sigma_{jr} \in \{-1,0,1\}.
\eq
Let us look at some examples: We start with a tree graph, shown in fig.~\ref{chapter_basics:fig_tree_graph}.
This graph has five edges ($\nedges=5$) and six vertices ($\nvertices=6$).
The graph is connected, hence $k=1$.
Eq.~(\ref{chapter_basics:def_loop_number}) gives then the loop number as
\bq
 \loopnumber & = & 5 - 6 + 1 \; = \; 0,
\eq
confirming that it is a tree graph.
Two vertices have valency $>1$, in fig.~\ref{chapter_basics:fig_tree_graph} these two vertices are labelled $v_1$ and $v_2$.
Momentum conservation at these two vertices yields
\begin{alignat}{3}
 v_1 & : & \;\;\; && p_1 + p_2 + q_1 & = 0,
 \nonumber \\
 v_2 & : & \;\;\; && p_3 + p_4 & = q_1.
\end{alignat}
From the first equation we have $q_1=-p_1-p_2$ and combining the two equations we obtain momentum
conservation for the external momenta:
\bq
 p_1 + p_2 + p_3 + p_4 & = & 0.
\eq
In particular, $p_4=-p_1-p_2-p_3$ and all momenta can be expressed as a linear combination
of $p_1,p_2,p_3$ with coefficients $\{-1,0,1\}$.
Since we considered a tree graph there are no independent loop momenta in this example.

\begin{figure}
\begin{center}
\includegraphics[scale=1.0]{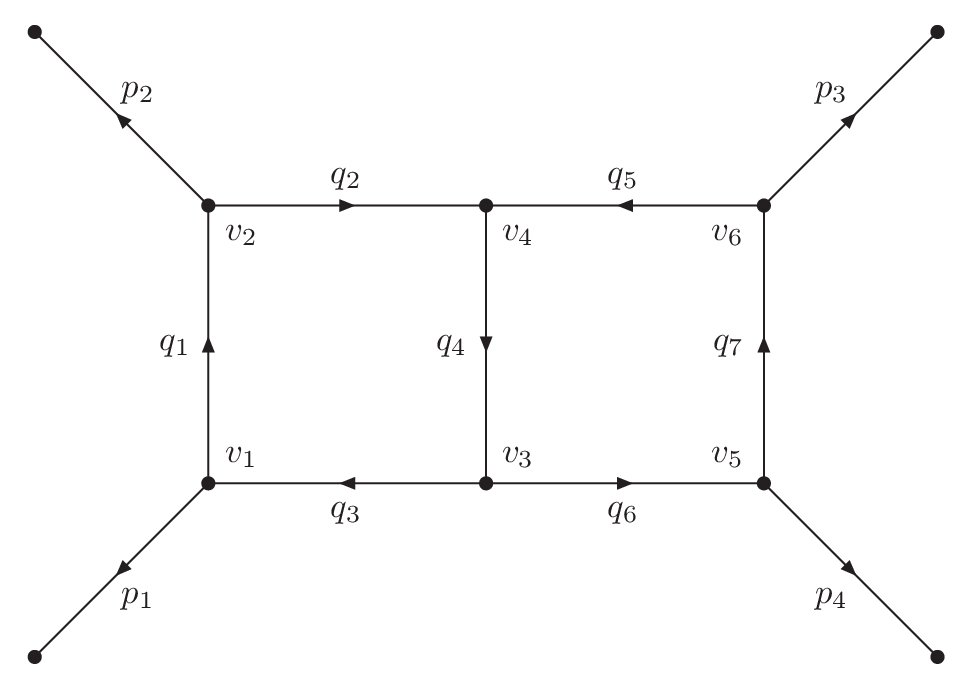}
\end{center}
\caption{
A two-loop graph with eleven edges and ten vertices. 
There are four external edges (labelled with momenta $p_1,\dots,p_4$)
and seven internal edges (labelled with momenta $q_1,\dots,q_7$).
Six vertices have valency $>1$. These are labelled by $v_1,\dots,v_6$.
The orientation of the external edges is chosen such that all external momenta are outgoing.
}
\label{chapter_basics:fig_doublebox}
\end{figure}

In the next example we look at a loop graph. The graph is shown in fig.~\ref{chapter_basics:fig_doublebox}.
This graph has eleven edges ($\nedges=11$) and ten vertices ($\nvertices=10$).
Again, the graph is connected, hence $k=1$.
The loop number is therefore
\bq
 \loopnumber & = & 11 - 10 + 1 \; = \; 2.
\eq
Six vertices have valency $>1$, in fig.~\ref{chapter_basics:fig_doublebox} these vertices are labelled by $v_1,\dots,v_6$.
Momentum conservation at these vertices gives
\begin{alignat}{3}
\label{chapter_basics:example_two_loop_momentum_conservation}
 v_1 & : & \;\;\; && p_1 + q_1 & = q_3,
 \nonumber \\
 v_2 & : & \;\;\; && p_2 + q_2 & = q_1,
 \nonumber \\
 v_3 & : & \;\;\; && q_3 + q_6 & = q_4,
 \nonumber \\
 v_4 & : & \;\;\; && q_4 & = q_2 + q_5,
 \nonumber \\
 v_5 & : & \;\;\; && p_4 + q_7 & = q_6,
 \nonumber \\
 v_6 & : & \;\;\; && p_3 + q_5 & = q_7.
\end{alignat}
Let us take $p_1,p_2,p_3$ as the independent external momenta and
\bq
 k_1 \; = \; q_3,
 & &
 k_2 \; = \; q_6
\eq
as the independent loop momenta.
All other momenta may then be expressed as a linear combination of $k_1, k_2, p_1, p_2, p_3$ with coefficients
$\{-1,0,1\}$.
This is nothing else than solving the linear system in eq.~(\ref{chapter_basics:example_two_loop_momentum_conservation})
for the momenta $q_1,q_2,q_4,q_5,q_7,p_4$.
Explicitly we have
\bq
 q_1 & = & k_1 - p_1,
 \nonumber \\
 q_2 & = & k_1 - p_1 - p_2,
 \nonumber \\
 q_4 & = & k_1 + k_2,
 \nonumber \\
 q_5 & = & k_2 + p_1 + p_2,
 \nonumber \\
 q_7 & = & k_2 + p_1 + p_2 + p_3
\eq
and $p_4=-p_1-p_2-p_3$.
\\
\\
\bs
{\it \refstepcounter{exercise}
{\bf Exercise \theexercise}: 
We stepped from a tree example immediately to a two-loop example.
As an exercise consider the one-loop graph shown in fig.~\ref{chapter_basics:fig_oneloopbox}.
\begin{figure}
\begin{center}
\includegraphics[scale=1.0]{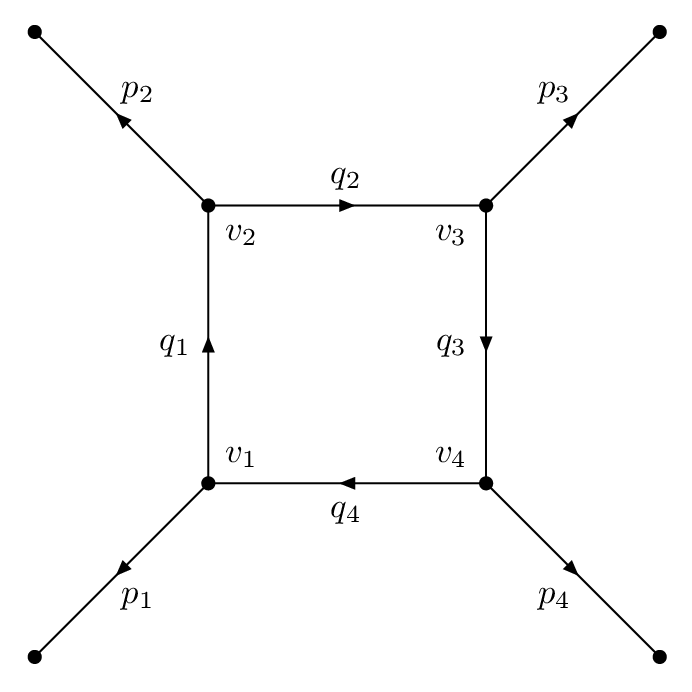}
\end{center}
\caption{
A one-loop graph with eight edges and eight vertices. 
There are four external edges (labelled with momenta $p_1,\dots,p_4$)
and four internal edges (labelled with momenta $q_1,\dots,q_4$).
Four vertices have valency $>1$. These are labelled by $v_1,\dots,v_4$.
The orientation of the external edges is chosen such that all external momenta are outgoing.
}
\label{chapter_basics:fig_oneloopbox}
\end{figure}
Write down the equations expressing momentum conservation at each vertex of valency $>1$.
Use $p_1,p_2,p_3$ as independent external momenta and $k_1=q_4$ as the independent loop momentum.
Express all other momenta as linear combinations of these.
}
\es
\\
\\
Let us now return to a general graph with $\nexternal$ external momenta (satisfying momentum conservation)
and $\loopnumber$ independent loop momenta.
We may consider the external momenta as input data, but what shall we do with the $\loopnumber$ independent loop momenta?
As they are independent of the external momenta there is no reason to prefer a particular configuration over any other configuration.
Quantum field theory instructs us to integrate over the independent loop momenta.
Thus we include for every independent loop momentum $k_r$ ($1 \le r \le \loopnumber$) a $D$-dimensional integration
\bq
 \int \frac{d^Dk_r}{i \pi^{\frac{D}{2}}}.
\eq
$i$ denotes the imaginary unit.
This is the measure which we will use in this book. It is normalised conveniently.
If one follows standard conventions in quantum field theory, the measure is given by
\bq
 \int \frac{d^Dk_r}{\left(2\pi\right)^D}.
\eq
The difference is a simple prefactor and one easily converts from one convention to the other convention.

We should also specify the integration contour. 
Our naive expectation is that we integrate each of the $D$ components of $k_r$ along the real axis 
from $-\infty$ to $+\infty$.
However, we have to be more careful.
An internal edge with momentum $k_r$ and mass $m$ contributes a factor
\bq
 \begin{picture}(90,20)(0,5)
 \ArrowLine(20,8)(80,8)
 \Text(50,12)[b]{\footnotesize $k_r, m$}
\end{picture} 
 & = &
 \frac{1}{-k_r^2+m^2}
\eq
to the integrand and there will be poles on the real axis.
For example, for the $k_r^0$-integration we will have poles at
\bq
 k_r^0 & = & \pm \sqrt{\sum\limits_{i=1}^{D-1} \left(k_r^i\right)^2 + m^2}.
\eq
($k_r^i$ denotes the $i$-th component of the $D$-dimensional vector $k_r$.)
We can go around these poles in the complex plane, but for two poles there a four possibilities:
For each poles we may either escape above the real axis or below the real axis into the complex plane.
In order to avoid to write down the square root more often, let us define
\bq
 E_{\vec{k}_r}
 & = &
 \sqrt{\sum\limits_{i=1}^{D-1} \left(k_r^i\right)^2 + m^2}.
\eq
Quantum field theory and causality in particular dictates us that the correct integration contour is the following:
\begin{center}
\begin{picture}(440,100)(0,0)
\Line(200,10)(200,90)
\Line(20,50)(380,50)
\ArrowLine(20,50)(90,50)
\CArc(100,50)(10,180,0)
\Line(110,50)(290,50)
\CArc(300,50)(10,0,180)
\ArrowLine(310,50)(390,50)
\Text(400,50)[l]{$\mbox{Re}(k_r^0)$}
\Text(205,90)[l]{$\mbox{Im}(k_r^0)$}
\Text(100,55)[b]{$-E_{\vec{k}_r}$}
\Text(300,45)[t]{$E_{\vec{k}_r}$}
\Line(100,47)(100,53)
\Line(300,47)(300,53)
\end{picture}
\end{center}
Thus we escape for the pole at $k_r^0=-E_{\vec{k}_r}$ into the complex lower half-plane and
for the pole at $k_r^0=E_{\vec{k}_r}$ into the complex upper half-plane.
Alternatively, we may keep the contour along the real axis and move the pole at $k_r^0=-E_{\vec{k}_r}$
an infinitesimal amount above the real axis and the pole as $k_r^0=E_{\vec{k}_r}$
an infinitesimal amount below the real axis.
This can be done by adding an infinitesimal small imaginary part to the Feynman rule for an internal edge:
\bq
 \begin{picture}(90,20)(0,5)
 \ArrowLine(20,8)(80,8)
 \Text(50,12)[b]{\footnotesize $q, m$}
\end{picture} 
 & = &
 \frac{1}{-q^2+m^2-i\delta}
\eq
with $\delta$ an infinitesimal small positive number.
This is 
\index{Feynman's $i\delta$-prescription}
{\bf Feynman's $i\delta$-prescription}.
In this book we usually don't write the $i\delta$ explicitly and only include it where it is needed.
This is common practice in the field.

We started defining the basic version of a Feynman graph as a graph, where we associate to every edge a momentum and a mass.
Let us add some additional information.
The motivation is as follows:
Consider a vertex of valency $2$ inside a Feynman graph.
By momentum conservation, the momenta flowing through the two edges attached to this vertex 
are (with an appropriate choice of orientation of the two edges) identical.
Let us assume that the masses associated with these two edges are identical as well.
The two edges would then contribute two identical factors
\bq
 \begin{picture}(150,20)(0,5)
 \ArrowLine(20,8)(80,8)
 \Vertex(80,8){2}
 \ArrowLine(80,8)(140,8)
 \Text(50,12)[b]{\footnotesize $q, m$}
 \Text(110,12)[b]{\footnotesize $q, m$}
\end{picture} 
 & = &
 \frac{1}{\left(-q^2+m^2\right)^2}.
\eq
We get the same effect if we associate to each edge in addition to the momentum and the mass a further number $\nu$, 
corresponding to the power to which the propagator occurs: 
\bq
 \begin{picture}(90,20)(0,5)
 \ArrowLine(20,8)(80,8)
 \Text(50,12)[b]{\footnotesize $q, m, \nu$}
\end{picture} 
 & = &
 \frac{1}{\left(-q^2+m^2\right)^\nu}
\eq
This convention will prove useful in later chapters of the book.

\section{Feynman rules}
\label{chapter_basics:feynman_rules}
\index{Feynman rules}

Feynman rules translate a Feynman graph into a mathematical formula.
In the previous paragraph we already discussed the essential ingredients.
Let's wrap them up and finalise them.

We consider a (connected) Feynman graph $G$ with $\nexternal$ external edges, $\ninternal$ internal edges and $\loopnumber$ loops.
This graph has
\bq
 \nvertices
 & = &
 \nexternal + \ninternal + 1 - \loopnumber
\eq
vertices.
To each external edge we associate an external momentum.
We label the external momenta by $p_1,\dots,p_{\nexternal}$.
To each internal edge $e_j$ we associate a triple $(q_j,m_j,\nu_j)$, 
where $q_j$ is the momentum flowing through this edge, $m_j$ the mass and $\nu_j$ the power to which the propagator occurs.
Momentum conservation at each vertex of valency $>1$ allows us to express any $q_j$ as a linear combination of $(\nexternal-1)$
linear independent external momenta and $\loopnumber$ independent loop momenta.
We denote the latter by $k_1,\dots,k_{\loopnumber}$.
Thus
\bq
 q_j & = &
 \sum\limits_{r=1}^\loopnumber \lambda_{jr} k_r
 +
 \sum\limits_{r=1}^{\nexternal - 1} \sigma_{jr} p_r,
 \;\;\;\;\;\;
 \lambda_{jr}, \sigma_{jr} \in \{-1,0,1\}.
\eq
The Feynman integral $I$ corresponding to this Feynman graph is obtained as follows:
\begin{description}
\item{1.} For each internal edge $e_j$ include a factor
\bq
\label{chapter_basics:summary_scalar_propagator}
 \frac{1}{\left(-q_j^2+m_j^2\right)^{\nu_j}}.
\eq
\item{2.} For each independent loop momentum $k_r$ include an integration
\bq
\label{chapter_basics:summary_integral_measure}
 \int \frac{d^Dk_r}{i \pi^{\frac{D}{2}}}.
\eq
\item{3.} Multiply by the prefactor
\bq
\label{chapter_basics:summary_prefactor}
 e^{\loopnumber \eps \Eulerconstant} \left(\mu^2\right)^{\nu-\frac{\loopnumber D}{2}}
 & \mbox{where} &
 \eps \; = \; \frac{\left[D\right]-D}{2},
 \;\;\;
 \nu \; = \; \sum\limits_{j=1}^{\ninternal} \nu_j.
\eq
\end{description}
The prefactor in rule 3 requires some explanation.
First of all, it is just a prefactor. All complications in computing Feynman integrals come from the integrations
in rule 2.
The sole purpose of the prefactor is to make the final result as simple as possible.
We will see examples later on.

Let us discuss the ingredients of the prefactor.
$\gls{eps}$
is called the dimensional regularisation parameter.
For a start we may define $\eps$ as follows: If $D$ denotes the dimension of space-time, 
we define 
$\left[D\right]$ 
to be the closest integer
to $D$. 
The dimensional regularisation parameter $\eps$ is then defined to be
\bq
\label{chapter_basics:def_eps}
 \eps & = & \frac{\left[D\right]-D}{2}.
\eq
This may sound weird at first sight. 
Up to now we implicitly assumed that the dimension $D$ of space-time is a positive integer and the closest integer to an
integer is the integer itself:
If $D$ is an integer we have $\left[D\right]=D$ and therefore $\eps=0$.
Later on in this book we will work with the assumption that $D$ is an arbitrary complex number, not necessarily an integer.
This is called dimensional regularisation.
Very often we will write
\bq
\label{chapter_basics:four_minus_two_eps}
 D & = & 4 - 2 \eps,
\eq
where we assume that $\left|\eps\right|$ is a small number.
In this case
\bq
 \left[ D \right] & = & 4
\eq
and eq.~(\ref{chapter_basics:four_minus_two_eps}) agrees with the definition in eq.~(\ref{chapter_basics:def_eps}).
Rearranging eq.~(\ref{chapter_basics:def_eps}) gives
\bq
\label{chapter_basics:def_Ddim}
 D & = & \left[ D \right] - 2 \eps,
\eq
and once we specify $\Dint=[D]$ as the integer dimension of the physical space-time we are interested in,
we may give up the restriction $|\mathrm{Re}(\eps)|<1/2$ and consider $\eps$ as an arbitrary complex 
parameter.
In other words, as we will be using perturbation theory in $\eps$, the quantity 
$\gls{integerspacetimedimension}$
denotes
the (integer) expansion point in the complex $D$-plane and $\eps$ the expansion parameter.
We will also denote the integer dimension of the physical space-time by $\Dint$.
Eq.~(\ref{chapter_basics:def_Ddim}) reads then
\bq
\label{chapter_basics:def_Ddim_v2}
 D & = & \Dint - 2 \eps.
\eq

The symbol 
$\gls{gammaE}$
denotes {\bf Euler's constant} 
(also called the 
\index{Euler-Mascheroni constant}
Euler-Mascheroni constant). 
This constant is defined by
\bq
 \Eulerconstant
 & = & 
 \lim\limits_{n \rightarrow \infty} \left( -\ln n + \sum\limits_{j=1}^n \frac{1}{j} \right).
\eq
The numerical value is
\bq
 \Eulerconstant
 & = & 
 0.57721566490153286\dots
\eq
We will later see that without the prefactor $e^{l \eps \Eulerconstant}$ Euler's constant
will appear in the final result for a Feynman integral.
The dependence of the result on $\Eulerconstant$ is rather simple (later we will see explicitly that
it is given by $e^{-l \eps \Eulerconstant}$) and it is convenient to factor this out.
Thus the prefactor $e^{l \eps \Eulerconstant}$ removes this factor and with this prefactor included
Euler's constant will not clutter the final result of a Feynman integral.

In physics we like to work with dimensionless quantities.
The momentum squared $p^2$ is not dimensionless, it has mass dimension $2$ (recall that we work
in natural units where $c=1$).
We have $\ninternal$ internal edges, each bringing a factor $1/(-q_j^2+m_j^2)^{\nu_j}$.
The total mass dimension of these factors is
\bq
 \dim\left( \prod\limits_{j=1}^{\ninternal} \frac{1}{\left(-q_j^2+m_j^2\right)^{\nu_j}} \right)
 & = &
 - 2 \nu,
 \;\;\;\;\;\;\;\;\;
 \nu \; = \; \sum\limits_{j=1}^{\ninternal} \nu_j.
\eq
The integral measure has mass dimension
\bq
 \dim\left( \prod\limits_{r=1}^{\loopnumber} \frac{d^Dk_r}{i \pi^{\frac{D}{2}}} \right)
 & = &
 \loopnumber D.
\eq
In order to enforce that our Feynman integral is dimensionless, we introduce 
an arbitrary parameter $\mu$ with mass dimension
\bq
 \dim\left( \mu \right)
 & = &
 1
\eq
and multiply by $(\mu^2)^{v-\loopnumber D/2}$.

\begin{tcolorbox}
In summary, the Feynman integral 
corresponding to a Feynman graph $G$ with $\nexternal$ external edges, $\ninternal$ internal edges and $\loopnumber$ loops
is given in $D$ space-time dimensions by
\bq
\label{chapter_basics:def_Feynman_integral}
 I
 & = &
 e^{\loopnumber \eps \Eulerconstant} \left(\mu^2\right)^{\nu-\frac{\loopnumber D}{2}}
 \int \prod\limits_{r=1}^{\loopnumber} \frac{d^Dk_r}{i \pi^{\frac{D}{2}}} 
 \prod\limits_{j=1}^{\ninternal} \frac{1}{\left(-q_j^2+m_j^2\right)^{\nu_j}},
\eq
where each internal edge $e_j$ of the graph is associated with a triple $(q_j,m_j,\nu_j)$,
specifying the momentum $q_j$ flowing through this edge,
the mass $m_j$ and the power $\nu_j$ to which the propagator occurs.
The external momenta are labelled by $p_1,\dots,p_{\nexternal}$.
Furthermore
\begin{align}
 \eps & = \frac{\Dint-D}{2},
 &
 \nu & = \sum\limits_{j=1}^{\ninternal} \nu_j,
 &
 q_j & = 
 \sum\limits_{r=1}^\loopnumber \lambda_{jr} k_r
 +
 \sum\limits_{r=1}^{\nexternal - 1} \sigma_{jr} p_r.
\end{align}
The coefficients $\lambda_{jr}$ and $\sigma_{jr}$ can be obtained from momentum conservation
at each vertex of valency $>1$.
The integration contour is given by Feynman's $i\delta$-prescription.
\end{tcolorbox}

Eq.~(\ref{chapter_basics:def_Feynman_integral}) defines the central object of this book.
We would like to compute integrals of this type.
We will learn about methods how to approach this task and the underlying mathematics in the sequel
of this book.

We already mentioned that the Feynman integral defined in eq.~(\ref{chapter_basics:def_Feynman_integral})
differs by a prefactor from standard conventions in quantum field theory.
Let us summarise the (small) differences:
If we come from quantum field theory, each edge corresponds to a single power of a propagator.
Thus $\nu_j=1$.
A scalar propagator is given by
\bq
 \frac{i}{q_j^2-m_j^2},
\eq
the integral measure is given by
\bq
 \frac{d^Dk_r}{\left(2\pi\right)^D}
\eq
and the prefactor $e^{\loopnumber \eps \Eulerconstant} (\mu^2)^{\nu-\loopnumber D/2}$ is replaced by one.
If one later chooses the {\bf modified minimal subtraction scheme} for renormalisation, 
Euler's constant is removed in the same way as the factor $e^{\loopnumber \eps \Eulerconstant}$ removes
$\Eulerconstant$ from the final result.
The $\eps$-dependent part of $(\mu^2)^{\nu-\loopnumber D/2}$ is $(\mu^2)^{\loopnumber \eps}$.
The latter is introduced to keep the action dimensionless (in natural units $\hbar=1$).
The parameter $\mu$ is known as the {\bf renormalisation scale}.

\section{Fundamental concepts}
\label{chapter_basics:fundamental_concepts}

\subsection{Wick rotation}

Minkowski space comes with the Minkowski metric, given by $g_{\mu\nu}=\mathrm{diag}(1,-1,-1,-1,\dots)$.
It will be simpler to work with the standard Euclidean metric $d_{\mu\nu}^{\mathrm{eucl}}=\mathrm{diag}(1,1,1,1,\dots)$.
The Wick rotation \cite{Wick:1954eu} allows us effectively to go from the Minkowski metric to the standard Euclidean metric.

We explain the basic idea for the one-loop case, where the integrand depends on the integration variables only through $k^2$.
The simplest example is given by the one-loop tadpole integral
\bq
 T_1
 & = &
 e^{\eps \Eulerconstant} \left(\mu^2\right)^{1-\frac{D}{2}}
 \int \frac{d^Dk}{i \pi^{\frac{D}{2}}} 
 \frac{1}{\left(-k^2+m^2\right)}.
\eq
Remember, that $k^2$ written out in components in $D$-dimensional Minkowski space reads
\bq
 k^2 = \left(k^0\right)^2 - \left(k^1\right)^2 - \left(k^2\right)^2 - \left(k^3\right)^2 - \dots - \left(k^{D-1}\right)^2.
\eq
Furthermore, when integrating over $k^0$, we encounter poles which are avoided by Feynman's $i\delta$-prescription
\bq
 \frac{1}{-k^2+m^2-i\delta}.
\eq
\begin{figure}
\begin{center}
\includegraphics[scale=1.0]{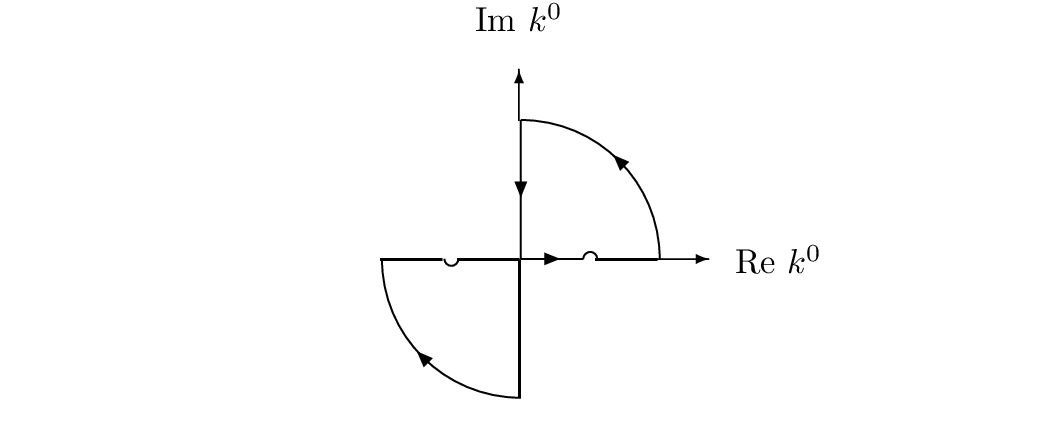}
\end{center}
\caption{
Integration contour for the Wick rotation. The little circles along the real axis
exclude the poles.
}
\label{chapter_basics:fig_Wick_rotation}
\end{figure}
In the complex $k^0$-plane we consider the integration contour shown in fig.~\ref{chapter_basics:fig_Wick_rotation}.
Since the contour does not enclose any poles, the integral along the complete contour is zero:
\bq
\oint dk^{0} f(k^0) & = & 0.
\eq
If the quarter-circles at infinity give a vanishing contribution
(it can be shown that this is the case)
we obtain
\bq
\int\limits_{-\infty}^{\infty} dk^{0} f(k^0) 
 & = & - \int\limits_{i \infty}^{-i \infty} dk^{0} f(k^0).
\eq
We now make the following change of variables:
\bq
\label{chapter_basics:Euclidean_loop_momentum}
 k^{0} & = & i K^{0}, \nonumber \\
 k^j   & = & K^j, \;\;\;\;\;\mbox{for}\; 1 \le j \le D-1.
\eq
As a consequence we have
\bq
k^{2} & = & - K^{2}, \nonumber \\
d^{D}k & = & i d^{D}K,
\eq
where $K^2$ is now given with Euclidean signature:
\bq
 K^2 & = & \left(K^0\right)^2 + \left(K^1\right)^2 + \left(K^2\right)^2 + \left(K^3\right)^2 + \dots + \left(K^{D-1}\right)^2.
\eq
In this book we use lower case letters for vectors in Minkowski space and upper case letters
for vectors in Euclidean space.
Combining the exchange of the integration contour with the change of variables
we obtain for the integration of a function $f(k^2)$ in $D$ dimensions
\bq
 \int \frac{d^Dk}{i \pi^{\frac{D}{2}}} f(-k^2)
 & = & 
 \int \frac{d^DK}{\pi^{\frac{D}{2}}} f(K^2),
\eq
whenever there are no poles inside the contour of fig.~\ref{chapter_basics:fig_Wick_rotation} and the arcs at infinity give a 
vanishing contribution.
The integral on the right-hand side is now in $D$-dimensional Euclidean space.
This equation justifies our convention 
to introduce in the definition of the Feynman integral a factor $i$ in the denominator of the measure
and a minus sign for each propagator.
These conventions are just such that after Wick rotation we have simple formulae.

\subsection{Dimensional regularisation}
\label{chapter_basics:dimensional_regularisation}

Before we start with an actual calculation of a Feynman integral, we should mention one 
complication: Loop integrals are often divergent!
Let us first look at the simple example of a one-loop tadpole integral with a double propagator and vanishing internal mass in four space-time dimensions:
\bq
\label{chapter_basics:example_divergent}
 T_2
 & = &
 \int \frac{d^4k}{i \pi^{2}} \frac{1}{(-k^2)^2}
 \; = \; 
 \int \frac{d^4K}{\pi^{2}} \frac{1}{(K^2)^2}
 \; = \;
 \int\limits_0^\infty \frac{dK^2}{K^2} 
 \; = \;
 \int\limits_0^\infty \frac{dx}{x}. 
\eq
Here, we first performed a Wick rotation to Euclidean space.
We then used spherical coordinates in four dimensions and integrated over the angles.
We are left with the radial integration, where we used as variable the norm squared.
This integral diverges at 
\begin{itemize}
\item 
$x\rightarrow \infty$, which is called an 
\index{divergence, ultraviolet}
{\bf ultraviolet} (UV) divergence and at 
\item 
$x\rightarrow 0$, which is called an 
\index{divergence, infrared}
{\bf infrared} (IR) divergence.
\end{itemize}   
Any quantity, which is given by a divergent integral, is of course an ill-defined quantity.
Therefore the first step is to make these integrals well-defined by introducing a regulator.
There are several possibilities how this can be done.
One possibility is cut-off regularisation with an ultraviolet regulator $\Lambda$ and an infrared regulator $\lambda$:
\bq
 \int\limits_0^\infty \frac{dx}{x}
 & \rightarrow &
 \int\limits_{\lambda}^\Lambda \frac{dx}{x} 
 \; = \;
 \ln \Lambda - \ln \lambda.
\eq
For infrared divergences mass regularisation can be used:
\bq
\int \frac{d^4K}{\pi^2} \frac{1}{(K^2)^2} 
 & \rightarrow &
\int \frac{d^4K}{\pi^2} \frac{1}{(K^2+m^2)^2} 
 \; = \;
 \int\limits_0^\infty dK^2 \frac{K^2}{(K^2+m^2)^2}.
\eq
This cures the problem at $K^2\rightarrow 0$, but not at $K^2\rightarrow \infty$.
A third possibility is lattice regularisation, where in position space space-time is approximated by
a finite lattice with finite lattice spacing.
Ultraviolet divergences are regulated by the finite lattice spacing, infrared divergences are regulated by the finite lattice.
However, within perturbative quantum field theory
the method of {\bf dimensional regularisation} 
\cite{'tHooft:1972fi,Bollini:1972ui,Cicuta:1972jf}
has almost become a standard, as the calculations in this regularisation
scheme turn out to be the simplest.
Within dimensional regularisation one replaces the four-dimensional integral over the loop momentum by a
$D$-dimensional integral, where $D$ is now an additional parameter, which can be a non-integer or
even a complex number.
We consider the result of the integration as a function of $D$ and we are interested in the behaviour of this 
function as $D$ approaches $4$.

At first sight the concept of dimensional regularisation may sound strange, as it is hard to imagine a space of non-integer dimension,
but the concept is closely related 
to the following situation:
Consider a function $f(z)$, which is defined for any positive integer $n \in {\mathbb N}$ by
\bq
 f\left(n\right) & = & n!
 \;\; = \;\; 1 \cdot 2 \cdot 3 \cdot \dots \cdot n.
\eq
We then would like to define $f(z)$ for any value $z \in \mathbb{C}$ (except for a countable set of isolated points, where $f(z)$ is allowed to have poles).
Of course, the answer is well known in mathematics and given by Euler's gamma function
\bq
 f\left(z\right) & = & \Gamma\left(z+1\right).
\eq
We will soon see that dimensional regularisation is based on the analytic properties of Euler's gamma function.

Let us start with general properties of dimensional regularisation:
The $D$-dimensional integration still fulfils the standard laws for integration,
like linearity, translation invariance and scaling behaviour
\cite{Wilson:1972cf,Collins}.
If $f$ and $g$ are two functions, and if $a$ and $b$ are two constants, 
linearity states that
\bq
 \int \frac{d^DK}{\pi^{\frac{D}{2}}} \left[ a f\left(K\right) + b g\left(K\right) \right] 
 & = & 
 a \int \frac{d^DK}{\pi^{\frac{D}{2}}} f\left(K\right) + b \int \frac{d^DK}{\pi^{\frac{D}{2}}} g\left(K\right).
\eq
Translation invariance requires that 
\bq
\label{chapter_basics:translation_invariance}
 \int \frac{d^DK}{\pi^{\frac{D}{2}}} f\left(K+P\right) 
 & = & 
 \int \frac{d^DK}{\pi^{\frac{D}{2}}} f\left(K\right)
\eq
for any vector $P$.
The scaling law states that
\bq
\label{chapter_basics:scaling_integral_measure}
 \int \frac{d^DK}{\pi^{\frac{D}{2}}} f\left(\lambda K\right) 
 & = & 
 \lambda^{-D} \int \frac{d^DK}{\pi^{\frac{D}{2}}} f\left(K\right).
\eq
The integral measure is normalised such that it agrees with the result for the integration of a Gaussian
function for all integer values $D$:
\bq
\label{chapter_basics:normalisation_D_int}
 \int \frac{d^DK}{\pi^{\frac{D}{2}}} \exp \left( - K^2 \right) 
 & = &
 1.
\eq
In eq.~(\ref{chapter_basics:def_Ddim_v2}) we introduced $\Dint$ as an integer, giving the dimension of space-time we 
are interested in. We further introduced the dimensional regularisation parameter $\eps$ such that
\bq
 D - \Dint & = & - 2 \eps.
\eq
Let us take $\Dint$ and $D-\Dint=-2\eps$ as two independent quantities.
We will assume that we can always decompose any vector into 
a $\Dint$-dimensional part and
a $(D-\Dint)$-dimensional part
\bq
\label{chapter_basics:eq_spit_vectors}
 K_{(D)} & = & K_{(\Dint)} + K_{(D-\Dint)},
\eq
and that the $\Dint$-dimensional and $(D-\Dint)$-dimensional subspaces are orthogonal to each other:
\bq
\label{chapter_basics:eq_orthogonality_vectors}
 K_{(\Dint)} \cdot K_{(D-\Dint)} & = & 0.
\eq
If we substitute $k^0=iK^0$ and $k^j = K^j$ for $1 \le j \le (D-1)$ this also implies
the Minkowski version
\bq
 k_{(\Dint)} \cdot k_{(D-\Dint)} & = & 0.
\eq
If $D$ is an integer greater than $\Dint$, eq.~(\ref{chapter_basics:eq_spit_vectors}) and eq.~(\ref{chapter_basics:eq_orthogonality_vectors})
are obvious. 
We postulate that these relations are true
for any value of $D$. One can think of the underlying vector space as a space 
of sufficiently high dimension (possibly infinite), where 
the integral measure mimics the one in $D$ dimensions.
\begin{digression} {\bf Constructing vector spaces associated to dimensional regularisation}
\\
Let us digress 
and discuss how eq.~(\ref{chapter_basics:eq_spit_vectors}) and eq.~(\ref{chapter_basics:eq_orthogonality_vectors}) can actually be realised. 
This will give us some insight how to interpret a space-time of dimension $D=3.99$.

We start with a simpler but related question: Suppose we know the natural numbers ${\mathbb N}$ together addition 
and multiplication. Subtraction and division are not yet known to us.
How do we construct the integer numbers ${\mathbb Z}$, the rational numbers ${\mathbb Q}$, the real numbers ${\mathbb R}$
and the complex numbers ${\mathbb C}$?

The way to do it is well-known to mathematicians: Let's consider the first step, constructing the integer numbers ${\mathbb Z}$
from the natural numbers ${\mathbb N}$.
Consider pairs $(a,b)$ with $a,b \in {\mathbb N}$ together with an equivalence relation.
Two pairs $(a_1,b_1)$ and $(a_2,b_2)$ are equivalent if there is
a $n \in {\mathbb N}$ such that
\bq 
 a_1 + b_2 + n = a_2 + b_1 + n.
\eq
We denote the equivalence classes by $[(a,b)]$.
One defines an addition on the set of equivalence classes by
\bq
\label{chapter_basics:addition_equivalence_classes}
 \left[\left(a_1,b_1\right)\right] + \left[\left(a_2,b_2\right)\right] & = & \left[\left(a_1+a_2,b_1+b_2\right)\right].
\eq
One can show that this definition does not depend on the representatives of the equivalence classes.
The set of equivalence classes together with the addition defined by eq.~(\ref{chapter_basics:addition_equivalence_classes}) forms a group, isomorphic to ${\mathbb Z}$.
The neutral element is the equivalence class $[(1,1)]$,
the inverse of $[(a,b)]$ is $[(b,a)]$.
Note that we never used a minus sign in this construction.

The construction of the rational numbers ${\mathbb Q}$ from the integer numbers ${\mathbb Z}$ proceeds in a similar way:
One considers pairs $(p,q)$ with $p,q \in {\mathbb Z}$ together with an equivalence relation.
Two pairs $(p_1,q_1)$ and $(p_2,q_2)$ are equivalent if there is
a $n \in {\mathbb Z}$ such that
\bq 
 p_1 \cdot q_2 \cdot n = p_2 \cdot q_1 \cdot n.
\eq
Let us denote the equivalence classes by $[(p,q)]$. 
One defines a multiplication on the set of equivalence classes by
\bq
 \left[\left(p_1,q_1\right)\right] \cdot \left[\left(p_2,q_2\right)\right] & = & \left[\left(p_1 \cdot p_2,q_1 \cdot q_2\right)\right].
\eq
Again, one can show this does not depend on the representatives and that the set of equivalence classes together
with the multiplication defines a group isomorphic to ${\mathbb Q}$.
The neutral element is $[(1,1)]$,
the inverse of $[(p,q)]$ is $[(q,p)]$.

This construction is quite general.
We started from the natural numbers ${\mathbb N}$, an Abelian semi-group with respect to addition, 
and constructed the integer numbers ${\mathbb Z}$, an Abelian group with respect to addition.
We then used the integer numbers ${\mathbb Z}$, which form an Abelian semi-group with respect to multiplication
and constructed the rational numbers ${\mathbb Q}$, an Abelian group with respect to multiplication.
The mathematical framework, which associates to each Abelian semi-group an Abelian group, is the domain
of K-theory. 
If $A$ is an Abelian semi-group with composition $\circ$,
the 
\index{Grothendieck group}
{\bf Grothendieck group} $K(A)$ of $A$ 
is constructed in the same way as in the examples above:
We consider pairs $(a,b)$ with $a,b \in A$ together with the equivalence relation
\bq
 \left(a_1,b_1\right) \sim \left(a_2,b_2\right)
 & \Leftrightarrow &
 \exists \; p \in A
 \; : \;
 a_1 \circ b_2 \circ p \; = \; a_2 \circ b_1 \circ p.
\eq
Then by definition $K(A) = A \times A / \sim$. 
The Grothendieck group $K(A)$ is an Abelian group.
Elements of $K(A)$ are denoted $\left[(a,b)\right]$.  

Let us now consider a set of vector space ${\mathcal V} = \{ V_1, V_2, V_3, \dots \}$,
where the vector space $V_j$ has dimension $\dim V_j = j$.
On ${\mathcal V}$  we have two operations, the direct sum and the tensor product, such that
\bq
 V_i \oplus V_j \in {\mathcal V} 
 & \mbox{and} & 
 V_i \otimes V_j \in {\mathcal V}
\eq
are again elements of ${\mathcal V}$. 
It is easy to see that with respect to each of these operations ${\mathcal V}$ is an Abelian semi-group.
The dimensions of the resulting vector spaces are:
\bq
 \dim\left( V_i \oplus V_j \right) = i + j, 
 & &
 \dim\left( V_i \otimes V_j \right) = i \cdot j.
\eq
We may therefore proceed as above and construct the Grothendieck $K$-groups. 
As an example we consider the $K$-group with respect to the direct sum:
One considers pairs $(V_a,V_b)$ with $V_a, V_b \in {\mathcal V}$ together with an equivalence relation.
Two pairs $(V_{a_1},V_{b_1})$ and $(V_{a_2},V_{b_2})$ are equivalent if there is
a $V_n \in {\mathcal V}$ such that
\bq 
 V_{a_1} \oplus V_{b_2} \oplus V_n = V_{a_2} \oplus V_{b_1} \oplus V_n,
\eq
where the equal sign refers to an isomorphism between vector spaces.
The equivalence classes are denoted by $[(V_a,V_b)]$.
On the set of equivalence classes one defines the operation  $\oplus$ by
\bq
 \left[\left(V_{a_1},V_{b_1}\right)\right] \oplus \left[\left(V_{a_2},V_{b_2}\right)\right] 
 & = & 
 \left[\left(V_{a_1} \oplus V_{a_2}, V_{b_1} \oplus V_{b_2}\right)\right].
\eq
Again, one can show that this is independent of the chosen representative.
For example we have
\bq
 \left[\left(V_{5},V_{1}\right)\right]
 & = &
 \left[\left(V_{42},V_{38}\right)\right].
\eq
Since we are considering equivalence classes of pairs of vector spaces, the dimensions 
of the vector spaces representing an equivalence class have no particular meaning.
However, the rank defined by
\bq
 \mathrm{rank}\; [(V_i,V_j)] & = & \dim V_i - \dim V_j
\eq
is an integer number and independent of the representative.
The rank of an equivalence class corresponds to our variable $D$, and we managed to
construct equivalence classes, where the rank is a negative integer.
We then repeat the argumentation where we replace the direct sum with the tensor product.
This gives us equivalence classes, where the rank is a rational number.

It remains to define the integration on these equivalence classes.
It is no problem to define the integration on a representative of an equivalence class.
The subtle point is that we have to show that this definition is independent of the chosen
representative.
Integration is a linear functional on a space of functions, and the space of functions
we are interested is rather special.
The functions we want to integrate depend only
on a few components 
\bq
 k^0, k^1, \dots, k^{(D-1)}
\eq
explicitly and on the rest of the components only through the combination
\bq
 \left(k^0\right)^2 + \left(k^1\right)^2 + \dots + \left( k^{d-1} \right)^2,
\eq
where $d$ is the sum of dimensions of the vector spaces making up the representative.
In this case it is possible to define an integration, which is independent of the chosen
representative and has the desired properties \cite{Weinzierl:1999xb}.

Thus we managed to give a meaning to integration in spaces of dimension $D \in {\mathbb Q}$,
where $D$ corresponds to the rank of the equivalence class.
In the last step one extends the integration to $D \in {\mathbb R}$ and $D \in {\mathbb C}$ 
in the same way as the real and complex numbers are constructed:
Each real number is the limit of a sequence of rational numbers and each complex number
can be represented by a pair of real numbers.
\end{digression}
In practice we will always arrange things such that every function we integrate over $D$ dimensions
is rotational invariant, e.g. is a function of $k^2$.
In this case the integration over the $(D-1)$ angles is trivial and can be expressed in a closed form
as a function of $D$.
Let us assume that we have an integral, original in four space-time dimensions, which has a UV-divergence, but no IR-divergences. 
Let us further assume that this integral would diverge logarithmically, if we would use
a cut-off regularisation instead of dimensional regularisation, e.g. the integral behaves for large $x$ in
four space-time dimensions as
\bq
 \mbox{logarithmically divergent}: & &
 \int\limits_1^\Lambda \frac{dx}{x} \; = \; \ln \Lambda.
\eq 
It turns out that this integral will be convergent if the real part of $D$ is smaller than $4$.
Therefore we may compute this integral under the assumption that $\mbox{Re}(D)<4$ and we will obtain as
a result a function of $D$. This function can be analytically continued to the whole complex plane.
We are mainly interested in what happens close to the point $D=4$. For an ultraviolet divergent one-loop
integral we will find that the analytically continued result will exhibit a pole at $D=4$.
It should be mentioned that there are also integrals which 
in the ultraviolet region diverge stronger.
An integral diverges for large $x$ linearly, respectively quadratically if it behaves as a function of the cut-off as
\bq
 \mbox{linearly divergent}: & & \int\limits_0^\Lambda dx \; = \; \Lambda,
 \nonumber \\
 \mbox{quadratically divergent}: & & 2 \int\limits_0^\Lambda x \; dx \; = \; \Lambda^2.
\eq
For example, if the integral diverges quadratically for $D=4$
we can repeat the argumentation above with the replacement $\mbox{Re}(D)<2$.

The terminology also applies to infrared divergences.
An integral is said to be logarithmically divergent at $x=0$ if it behaves as
\bq
 \mbox{logarithmically divergent}: & &
 \int\limits_{\lambda}^1 \frac{dx}{x} \; = \; - \ln \lambda.
\eq 
The integrand is said to have a simple pole at $x=0$.
Integrands with a double pole correspond to linearly divergent integrals,
integrands with a pole of order three to quadratically divergent integrals.
This is most easily seen by the substitution $\lambda=1/\Lambda'$, for example
\bq
 \int\limits_{\lambda}^\infty \frac{dx}{x^2} \; = \; \frac{1}{\lambda}
 \; = \; \Lambda'.
\eq
Let us now consider 
a logarithmic IR-divergent integral, which has no UV-divergence. This integral
will be convergent if $\mbox{Re}(D)>4$. Again, we can compute the integral in this domain and
continue the result to $D=4$. 
Logarithmic IR-divergent one-loop integral may have a divergence in the radial integration as well as in the angular integration.
The former is called a 
\index{soft divergence}
{\bf soft divergence}, the latter a 
\index{collinear divergence}
{\bf collinear divergence}.
Each divergence will give a pole at $D=4$, and if both divergences are present this will lead in total to a double pole at $D=4$.

We will use dimensional regularisation to regulate both the ultraviolet and infrared divergences.
The attentive reader may ask how this goes together, as we argued above that UV-divergences require
$\mbox{Re}(D)<4$ or even $\mbox{Re}(D)<2$, whereas IR-divergences are regulated by $\mbox{Re}(D)>4$.
Suppose for the moment that we use dimensional regularisation just for the UV-divergences
and that we use a second regulator for the IR-divergences.
For the IR-divergences we could keep all external momenta off-shell, or introduce small masses for all massless
particles or even raise the original propagators to some power $\nu$.
The exact implementation of this regulator is not important, as long as the IR-divergences are screened by this
procedure. We then perform the loop integration in the domain where the integral is UV-convergent.
We obtain a result, which we can analytically continue to the whole complex $D$-plane, in particular
to $\mbox{Re}(D)>4$. There we can remove the additional regulator and the IR-divergences are now regulated
by dimensional regularisation. Then the infrared divergences
will also show up as poles at $D=4$.

In summary, within dimensional regularisation the initial divergences show up as poles in the complex $D$-plane
at the point $D=4$.
What shall we do with these poles? The answer has to come from physics and we distinguish again the case of
UV-divergences and IR-divergences.
The UV-divergences are removed through renormalisation.
On the level of Feynman diagrams we can associate to any UV-divergent part a counter-term, which has exactly
the same pole term at $D=4$, but with an opposite sign, such that in the sum the two pole terms cancel.

As far as infrared-divergences are concerned we first note that any detector has a finite resolution.
Therefore two particles which are sufficiently close to each other in phase space will be detected as one
particle.
The Kinoshita-Lee-Nauenberg theorem 
\cite{Kinoshita:1962ur,Lee:1964is}
guarantees that all infrared divergences cancel, when summed over all
degenerate physical states.
To make this cancellation happen in practice requires usually quite some work, as the different contributions live
on phase spaces of different dimensions.
We will discuss the cancellation of ultraviolet and infrared divergences in chapter~\ref{chapter_qft}.

A Feynman integral $I$ in $D$ space-time dimensions will therefore have a Laurent expansion
around $D=\Dint$:
\bq
\label{chapter_basics:Laurent_expansion}
 I & = &
 \sum\limits_{j=j_{\min}}^\infty
 \eps^j \; I^{(j)},
 \;\;\;\;\;\;\;\;\;
 \eps \; = \; \frac{\Dint-D}{2},
\eq
where $I^{(j)}$ denotes the coefficient of $\eps^j$.
For precision calculations we are interested in the first few terms $I^{(j_{\min})}$, $I^{(j_{\min}+1)}$, $\dots$ of this Laurent series.
The exact number of required terms depends on the order of perturbation theory we are calculating.
Let us stress that we would like to get the $I^{(j)}$'s, not necessarily $I$ itself.
There are situations where a closed form expression for $I$ is readily
obtained, but the Laurent expansion in $\eps$ is not immediate.

Let us now start to get our hands dirty. We compute the first Feynman integrals.
For the moment we focus on one-loop integrals, which only depend on the loop momentum squared
(and no scalar product $k \cdot p$ with some external momentum).
At first sight it seems that there aren't too many Feynman integrals of this type.
The one-loop tadpole integral already exhausts these specifications:
\bq
 T_\nu
 & = &
 e^{\eps \Eulerconstant} \left(\mu^2\right)^{\nu-\frac{D}{2}}
 \int \frac{d^Dk}{i \pi^{\frac{D}{2}}} 
 \frac{1}{\left(-k^2+m^2\right)^\nu}.
\eq
However, we will later see that we can always arrange the integrand of any Feynman integral in such a way that it only depends
on $k^2$. Thus this covers an important case and doing this loop-by-loop allows us to perform
all loop momenta integrations.
However, there is no free lunch: Re-organising the integrand such that it depends only on $k^2$
introduces additional integrations (typically over Schwinger or Feynman parameters) and we merely shifted
the complications from the loop momentum integration to the Schwinger or Feynman parameter integration.

After Wick rotation we have 
\bq
 T_\nu
 & = &
 e^{\eps \Eulerconstant} \left(\mu^2\right)^{\nu-\frac{D}{2}}
 \int \frac{d^DK}{\pi^{\frac{D}{2}}} 
 \frac{1}{\left(K^2+m^2\right)^\nu}.
\eq
As the integrand only depends on $K^2$, it is natural to introduce spherical coordinates. 
In $D$ dimensions they are given by
\bq
 K^{0} & = & K \cos \theta_{1}, \nonumber \\
 K^{1} & = & K \sin \theta_{1} \cos \theta_{2}, \nonumber \\
 & ... & \nonumber \\
 K^{D-2} & = & K \sin \theta_{1} ... \sin \theta_{D-2} \cos \theta_{D-1}, \nonumber \\
 K^{D-1} & = & K \sin \theta_{1} ... \sin \theta_{D-2} \sin \theta_{D-1}.
\eq
In $D$ dimensions we have one radial variable $K$, $(D-2)$ polar angles $\theta_j$ (with $1 \le j \le D-2$)
and one azimuthal angle $\theta_{D-1}$.
The measure becomes
\bq
d^{D}K & = & K^{D-1} dK \; d\Omega_{D},
 \;\;\;\;\;\;
 d\Omega_{D} = \prod\limits_{i=1}^{D-1} \sin^{D-1-i} \theta_{i} \; d\theta_{i}.
\eq
Integration over the angles yields
\bq
\label{chapter_basics:angular_integration}
 \int d\Omega_{D} & = & \int\limits_{0}^{\pi} d\theta_{1} \sin^{D-2} \theta_{1}
 ... \int\limits_{0}^{\pi} d\theta_{D-2} \sin \theta_{D-2} 
 \int\limits_{0}^{2 \pi} d\theta_{D-1} 
 = \frac{2 \pi^{\frac{D}{2}}}{\Gamma\left( \frac{D}{2} \right)}.
\eq
$\Gamma(z)$ denotes Euler's gamma function.
\begin{digression} {\bf Euler's gamma function and Euler's beta function}
\\
It is now the appropriate place to introduce two special functions, Euler's gamma function and Euler's beta function, 
which are 
used within dimensional regularisation to continue the results from integer $D$ towards non-integer values.
The 
\index{Euler's gamma function}
{\bf gamma function} is defined for $\mbox{Re}(z) > 0$ by
\bq
\Gamma(z) & = & \int\limits_{0}^{\infty} e^{-t} t^{z-1} dt.
\eq
It fulfils the functional equation
\bq
\Gamma(z+1) & = & z \; \Gamma(z).
\eq
For positive integers $n$ it takes the values
\bq
\Gamma(n+1) & = & n! \; = \; 1 \cdot 2 \cdot 3 \cdot ... \cdot n.
\eq
The gamma function $\Gamma(z)$ has simple poles located on the negative real axis at $z=0,-1,-2,\dots$.
Quite often we will need the expansion around these poles.
This can be obtained from the expansion around $z=1$ and the functional equation.
The expansion around $z=1$ reads
\bq
\Gamma(1+\eps)  & = & 
  \exp \left( - \Eulerconstant \eps + \sum\limits_{n=2}^\infty \frac{(-1)^n}{n} \zeta_n \eps^n \right),
\eq
where
$\Eulerconstant$ is Euler's constant
and 
$\gls{zetan}$
is given by
\bq
 \zeta_n & = & \sum\limits_{j=1}^\infty \frac{1}{j^n}.
\eq
$\zeta_n$ is called a 
\index{zeta value}
{\bf zeta value}.
As an example we obtain for the Laurent expansion around $z=0$
\bq
\Gamma(\varepsilon) = \frac{1}{\varepsilon} - \Eulerconstant + O(\varepsilon).
\eq 
It will be useful to know the residues of $\Gamma(z+a)$ and $\Gamma(-z+a)$
at the poles
\bq
 \mathrm{res}\left( \Gamma(z+a), z=-a-n \right) & = & \frac{(-1)^n}{n!},
 \;\;\;\;\;\;\;\;\; n \in {\mathbb N}_0,
 \nonumber \\
 \mathrm{res}\left( \Gamma(-z+a), z=a+n \right) & = & -\frac{(-1)^n}{n!}.
\eq
For integers $n$ we have the reflection identity
\bq
\label{chapter_basics:Gamma_function_reflection_identity}
\frac{\Gamma(z-n)}{\Gamma(z)} & = & \left(-1 \right)^n \frac{\Gamma(1-z)}{\Gamma(1-z+n)}.
\eq
Furthermore
\bq
 \Gamma(z) \Gamma(1-z) & = & \frac{\pi}{\sin \pi z},
\eq
from which we may deduce the value at $z=1/2$:
\bq
 \Gamma\left( \frac{1}{2} \right) & = & \sqrt{\pi}.
\eq
There is a duplication formula for the gamma function:
\bq
\prod\limits_{j=0}^{k-1} \Gamma\left(z + \frac{j}{k} \right)
 & = & 
 \left( 2 \pi \right)^{\frac{k-1}{2}} \; k^{\frac{1}{2}-kz} \; \Gamma\left( k z \right),
 \;\;\;\;\;\;\;\;\; 
 k \in {\mathbb N}.
\eq
In particular we have for $k=2$
\bq
 \Gamma\left(z\right) \Gamma\left( z + \frac{1}{2} \right)
 & = & 
 2^{1-2z} \; \sqrt{\pi} \; \Gamma\left(2 z\right).
\eq
\index{Euler's beta function}
{\bf Euler's beta function} 
is defined for $\mbox{Re}(z_1) > 0$ and $\mbox{Re}(z_2) > 0$ by
\bq
 B\left(z_1,z_2\right) & = & \int\limits_{0}^{1} t^{z_1-1} (1-t)^{z_2-1} dt,
\eq
or equivalently by
\bq
 B\left(z_1,z_2\right) & = & \int\limits_{0}^{\infty} \frac{t^{z_1-1}}{(1+t)^{z_1+z_2}} dt.
\eq
The beta function can be expressed in terms of gamma functions:
\bq
 B\left(z_1,z_2\right) & = & \frac{ \Gamma\left(z_1\right) \Gamma\left(z_2\right)}{ \Gamma\left(z_1+z_2\right)}.
\eq
\end{digression}
Note that the integration on the left-hand side
of eq.~(\ref{chapter_basics:angular_integration}) is defined for any natural number $D$, 
whereas the result on the right-hand side is an analytic function of $D$, 
which can be continued to any complex value.
Performing the angular integrations for our tadpole integral we obtain
\bq
\label{chapter_basics:tadpole_after_angular_integration}
 T_\nu
 & = &
 \frac{e^{\eps \Eulerconstant} \left(\mu^2\right)^{\nu-\frac{D}{2}}}{\Gamma\left( \frac{D}{2} \right)}
 \int\limits_0^\infty dK^2
 \frac{\left(K^2\right)^{\frac{D}{2}-1}}{\left(K^2+m^2\right)^\nu}.
\eq
Please note that in eq.~(\ref{chapter_basics:tadpole_after_angular_integration}) we may now allow 
non-integer values for $D$ without any problems.
Let us proceed and let us substitute $t=K^2/m^2$. We obtain
\bq
 T_\nu
 & = &
 \frac{e^{\eps \Eulerconstant}}{\Gamma\left( \frac{D}{2} \right)}
 \left( \frac{m^2}{\mu^2} \right)^{\frac{D}{2}-\nu}
 \int\limits_0^\infty dt
 \frac{t^{\frac{D}{2}-1}}{\left(t + 1\right)^\nu}.
\eq
The remaining integral is just Euler's beta function
\bq
 \int\limits_0^\infty dt
 \frac{t^{\frac{D}{2}-1}}{\left(1 + t\right)^\nu}
 & = &
 \frac{\Gamma\left(\frac{D}{2}\right)\Gamma\left(\nu-\frac{D}{2}\right)}{\Gamma\left(\nu\right)}.
\eq
Thus we computed our first Feynman integral (recall $D=\Dint-2\eps$):
\begin{tcolorbox}
{\bf Tadpole integral}:
\bq
\label{chapter_basics:result_tadpole}
 T_\nu\left(D,\frac{m^2}{\mu^2}\right)
 & = &
 \frac{e^{\eps \Eulerconstant} \Gamma\left(\nu-\frac{D}{2}\right)}{\Gamma\left(\nu\right)}
 \left( \frac{m^2}{\mu^2} \right)^{\frac{D}{2}-\nu}.
\eq
\end{tcolorbox}
We are interested in the Laurent expansion of Feynman integrals.
With $D=4-2\eps$, $\nu=1$ and $L=\ln(m^2/\mu^2)$ we obtain
\bq
\label{chapter_basics:result_tadpole_T1_4D}
 T_1\left(4-2\eps\right)
 & = &
 \frac{m^2}{\mu^2}
 e^{\eps \Eulerconstant} \Gamma\left(-1+\eps\right)
 e^{-\eps L}
 \nonumber \\
 & = &
 \frac{m^2}{\mu^2}
 \left[ 
 - \frac{1}{\eps} + \left( L - 1 \right) + \left( - \frac{1}{2} L^2 - \frac{1}{2} \zeta_2 + L - 1\right) \eps
 \right]
 + {\mathcal O}\left(\eps^2\right).
\eq
The pole at $\eps=0$ originates from the ultraviolet divergence of tadpole integral with $\nu=1$ in four space-time dimensions.

The result in eq.~(\ref{chapter_basics:result_tadpole}) is valid for any $D$, so we may as well expand it around two space-time dimensions.
Just for fun, let's do it:
\bq
\label{chapter_basics:result_tadpole_T1_2D_uniform_weight}
 \eps \; T_1\left(2-2\eps\right)
 & = &
 e^{\eps \Eulerconstant} \Gamma\left(1+\eps\right)
 e^{-\eps L}
 \nonumber \\
 & = &
 1 - L \eps + \left( \frac{1}{2} L^2 + \frac{1}{2} \zeta_2 \right) \eps^2
 + {\mathcal O}\left(\eps^3\right).
\eq
Let us define a 
\index{weight}
{\bf weight}. 
We declare that $L=\ln(m^2/\mu^2)$ has weight one and that $L^n$ and $\zeta_n$ have weight $n$.
A rational number has weight $0$.
The weight of a product is the sum of the weights of its factors.
We then spot a difference between $\eps \; T_1(2-2\eps)$ and $\eps \; (\mu^2/m^2) T_1(4-2\eps)$.
In $\eps \; T_1(2-2\eps)$ we see that the $j$-th term in the $\eps$-expansion only involves terms of weight $j$
(we have only given the first three terms of the $\eps$-expansion, but this statement holds to any order),
whereas in $\eps \; (\mu^2/m^2) T_1(4-2\eps)$ the $j$-th term in the $\eps$-expansion involves terms of weight $j$ and terms of lower weight.
In this sense, $\eps \; T_1(2-2\eps)$ has a simpler $\eps$-expansion.
We call $\eps \; T_1(2-2\eps)$ to be of 
\index{uniform weight}
{\bf uniform weight}.
We will discuss this issue in more detail in chapter~\ref{chapter_iterated_integrals}.

From eq.~(\ref{chapter_basics:result_tadpole}) we also deduce
\bq
\label{chapter_basics:tadpole_dimensional_shift}
 T_\nu\left(2-2\eps\right)
 & = &
 \nu T_{\nu+1}\left(4-2\eps\right).
\eq
This is an example of a {\bf dimensional shift relation}, relating integrals in $D=2-2\eps$ and $D=4-2\eps$
space-time dimensions.
Also dimensional shift relations will be discussed 
in more detail in chapter~\ref{chapter_iterated_integrals}.
\\
\\
\bs
{\it \refstepcounter{exercise}
{\bf Exercise \theexercise}: 
Prove
\bq
 T_\nu\left(D\right)
 & = &
 \nu T_{\nu+1}\left(D+2\right).
\eq
}
\es
\\
\\
Let us set temporarily
\bq
 J_1
 & = &
 \eps \; T_1\left(2-2\eps\right)
 \; = \;
 e^{\eps \Eulerconstant} \Gamma\left(1+\eps\right)
 e^{-\eps L}.
\eq
It is not too difficult to show that 
\bq
\label{chapter_basics:tadpole_reduction}
 T_\nu\left(D\right)
 & = &
 \frac{\Gamma\left(\nu-\frac{\Dint}{2}+\eps\right)}{\Gamma\left(\nu\right)\Gamma\left(1+\eps\right)}
 \left( \frac{m^2}{\mu^2} \right)^{\left(\frac{\Dint}{2}-\nu\right)}
 J_1.
\eq
For $\nu \in {\mathbb N}$ and $\Dint$ even, the prefactor is always
a rational function in $\eps$ and $m^2$, for example for $D=4-2\eps$ and $\nu=1$
we have
\bq
 \frac{\Gamma\left(\nu-\frac{\Dint}{2}+\eps\right)}{\Gamma\left(\nu\right)\Gamma\left(1+\eps\right)}
 \left( \frac{m^2}{\mu^2} \right)^{\left(\frac{\Dint}{2}-\nu\right)}
 & = &
 - \frac{1}{\eps\left(1-\eps\right)}
 \frac{m^2}{\mu^2}.
\eq
Eq.~(\ref{chapter_basics:tadpole_reduction}) expresses any integral
$T_\nu(D)$ as a coefficient times $J_1$.
For the tadpole integrals this is a trivial statement, as the coefficient is just the ratio $T_\nu(D)/J_1$.
Later on we will see, that this generalises as follows:
We may express any member of a family of Feynman integrals as
a linear combination of Feynman integrals from a finite set.
The Feynman integrals from this finite set are called
\index{master integrals}
{\bf master integrals}.
A master integral, which is of uniform weight, is called a
\index{canonical master integral}
{\bf canonical master integral}.
Thus $J_1$ is a canonical master integral.

Let us close this section with some results on related integrals.
The first result is a generalisation of the tadpole integrals and is helpful whenever we iteratively integrate out
loop momenta (after having arranged that the integrand only depends on $k^2$).
We consider
\bq
\label{chapter_basics:def_master_one_loop}
 \tilde{T}
 & = &
 e^{\eps \Eulerconstant} \left(\mu^2\right)^{\nu-\frac{D}{2}-a}
 \int \frac{d^Dk}{i \pi^{\frac{D}{2}}} 
\frac{\left(-k^2\right)^a}{\left( -U k^2 + V \right)^\nu},
\eq
where $U$, $V$ and $a$ do not depend on $k$.
For $a=0$, $U=1$ and $V=m^2$ we recover the tadpole integral.
Following the same steps as for the tadpole integral one finds
\bq
\label{chapter_basics:master_one_loop_v1}
 \tilde{T}
 & = &
 e^{\eps \Eulerconstant} \left(\mu^2\right)^{\nu-\frac{D}{2}-a}
 \frac{\Gamma\left(\frac{D}{2}+a\right)}{\Gamma\left(\frac{D}{2}\right)}
 \frac{\Gamma\left(\nu-\frac{D}{2}-a\right)}{\Gamma\left(\nu\right)} 
 \frac{U^{-\frac{D}{2}-a}}{V^{\nu-\frac{D}{2}-a}}
\eq
and the
\begin{tcolorbox}
{\bf one-loop master formula}:
\bq
\label{chapter_basics:master_one_loop}
 \int \frac{d^Dk}{i \pi^{\frac{D}{2}}} 
\frac{\left(-k^2\right)^a}{\left( -U k^2 + V \right)^\nu}
 & = &
 \frac{\Gamma\left(\frac{D}{2}+a\right)}{\Gamma\left(\frac{D}{2}\right)}
 \frac{\Gamma\left(\nu-\frac{D}{2}-a\right)}{\Gamma\left(\nu\right)} 
 \frac{U^{-\frac{D}{2}-a}}{V^{\nu-\frac{D}{2}-a}}.
\eq
\end{tcolorbox}
In the definition of $\tilde{T}$ we allowed additional powers $(-k^2)^a$ of the loop momentum in the numerator.
Note that the dependency of the result on $a$, apart from a factor 
$\Gamma(D/2+a)/\Gamma(D/2)$, occurs only in the combination 
$D/2+a$.
Therefore adding additional powers $(-k^2)^a$ to the numerator is almost equivalent to consider the integral
without this factor in dimensions $(D+2a)$.
\\
\\
\bs
{\it \refstepcounter{exercise}
{\bf Exercise \theexercise}: 
Derive eq.~(\ref{chapter_basics:master_one_loop_v1}).
}
\es
\\
\\
There is one more generalisation:
Sometimes it is convenient to decompose $k^2$ into a $\Dint$-dimensional piece and a remainder:
\bq
 k_{(D)}^2 & = & k_{(\Dint)}^2 + k_{(-2\eps)}^2.
\eq
If $D$ is an integer greater than $\Dint$ we have
\bq
 k_{(\Dint)}^2 & = & \left(k^0\right)^2 - \left(k^1\right)^2 - ... - \left(k^{\Dint-1}\right)^2, \nonumber \\
 k_{(-2\eps)}^2 & = & -\left(k^{\Dint}\right)^2 - ... - \left(k^{D-1}\right)^2.
\eq
We also need loop integrals where additional powers of $(-k_{(-2\eps)}^2)$ appear in the numerator.
These are related to integrals in higher dimensions as follows:
\begin{tcolorbox}
{\bf $\eps$-components in the numerator}:
\bq
\label{chapter_basics:k2_eps_numerator}
 \int \frac{d^Dk}{i \pi^{\frac{D}{2}}} 
 \left(-k_{(-2\eps)}^2\right)^r 
 f\left(k_{(\Dint)},k_{(-2\eps)}^2\right) 
 & = &
 \frac{\Gamma(r-\eps)}{\Gamma(-\eps)}
 \int \frac{d^{D+2r}k}{i \pi^{\frac{D+2r}{2}}} 
 f\left(k_{(\Dint)},k_{(-2\eps)}^2\right).
\hspace*{8mm}
\eq
\end{tcolorbox}
Here, $f(k_{(\Dint)},k_{(-2\eps)}^2)$ is a function which depends on $k^{\Dint}$, $k^{\Dint+1}$, ..., $k^{D-1}$ only
through $k_{(-2\eps)}^2$.
The dependency on $k^0$, $k^1$, ..., $k^{\Dint-1}$ is not constrained.
\\
\\
\bs
{\it \refstepcounter{exercise}
{\bf Exercise \theexercise}: 
Derive eq.~(\ref{chapter_basics:k2_eps_numerator}).
\\
\\
Hint: Split the $D$-dimensional integration into a $\Dint$-dimensional part and a
$(-2\eps)$-dimensional part.
Eq.~(\ref{chapter_basics:k2_eps_numerator}) can be derived by just considering the $(-2\eps)$-dimensional part.
}
\es
\\
\\
Finally it is worth noting that 
\bq 
\label{chapter_basics:basic_eq_negative_dimensions}
 \int \frac{d^Dk}{i \pi^{\frac{D}{2}}} 
 \left( - k^2 \right)^a & = & 
 \left\{
 \begin{array}{ll}
 \left(-1\right)^{\frac{D}{2}} \Gamma\left(1-\frac{D}{2}\right), & \mbox{if}\; \frac{D}{2}+a=0, \\
 0, & \mbox{otherwise}.
 \end{array}
 \right.
\eq
\\
\\
\bs
{\it \refstepcounter{exercise}
{\bf Exercise \theexercise}: 
Derive eq.~(\ref{chapter_basics:basic_eq_negative_dimensions}).
\\
\\
Hint: Consider the mass dimension of the integral to prove the statement for $D/2+a\neq0$
and the normalisation of the integral measure in eq.~(\ref{chapter_basics:normalisation_D_int}) to prove the statement for $D/2+a = 0$.
}
\es

\section{Representations of Feynman integrals}
\label{chapter_basics:representations_of_Feynman_integrals}

There are several integral representations for Feynman integrals.
We introduce the various integral representations in this section.

As before, we consider a Feynman graph $G$
with $\nexternal$ external edges, $\ninternal$ internal edges and $\loopnumber$ loops.
To each external edge we associate an external momentum,
labelled by $p_1,\dots,p_{\nexternal}$.
To each internal edge $e_j$ we associate a triple $(q_j,m_j,\nu_j)$, 
where $q_j$ is the momentum flowing through this edge, 
$m_j$ the mass and $\nu_j$ the power to which the propagator occurs.
Momentum conservation at each vertex of valency $>1$ allows us to express any $q_j$ 
as a linear combination of $(\nexternal-1)$
linear independent external momenta and $\loopnumber$ independent loop momenta.
We denote the latter by $k_1,\dots,k_{\loopnumber}$.

\subsection{The momentum representation of Feynman integrals}
\label{chapter_basics:momentum_representation_of_Feynman_integrals}

The momentum representation of Feynman integrals is the one we started with in eq.~(\ref{chapter_basics:def_Feynman_integral}):
\bq
\label{chapter_basics:def_Feynman_integral_v2}
 I
 & = &
 e^{\loopnumber \eps \Eulerconstant} \left(\mu^2\right)^{\nu-\frac{\loopnumber D}{2}}
 \int \prod\limits_{r=1}^{\loopnumber} \frac{d^Dk_r}{i \pi^{\frac{D}{2}}} 
 \prod\limits_{j=1}^{\ninternal} \frac{1}{\left(-q_j^2+m_j^2\right)^{\nu_j}},
\eq
where
\begin{align}
 \eps & = \frac{\Dint-D}{2},
 &
 \nu & = \sum\limits_{j=1}^{\ninternal} \nu_j,
 &
 q_j & = 
 \sum\limits_{r=1}^\loopnumber \lambda_{jr} k_r
 +
 \sum\limits_{r=1}^{\nexternal - 1} \sigma_{jr} p_r.
\end{align}
The coefficients $\lambda_{jr}$ and $\sigma_{jr}$ can be obtained from momentum conservation
at each vertex of valency $>1$.
The integration contour is given by Feynman's $i\delta$-prescription.

Let us discuss the variables the Feynman integral depends on.
First of all, the Feynman integrals depends on the dimension of space-time $D \in {\mathbb C}$
and through the prefactor $e^{\loopnumber \eps \Eulerconstant}$ on $\Dint$. 
Secondly, the Feynman integral depends also on the $\ninternal$-tuple $(\nu_1,\dots,\nu_{\ninternal})$.
In principle we may allow $\nu_j \in {\mathbb C}$, but very often we will limit us to the case
$\nu_j \in {\mathbb Z}$.
Thirdly, the Feynman integral depends on kinematic variables.
The Feynman integral in eq.~(\ref{chapter_basics:def_Feynman_integral_v2}) is a scalar integral, thus the dependence
on the $(\nexternal-1)$  linear independent external momenta is only through the Lorentz invariants
\bq
 p_i \cdot p_j.
\eq
The Feynman integral in eq.~(\ref{chapter_basics:def_Feynman_integral_v2}) is dimensionless, therefore the dependence
on the Lorentz invariants, the internal masses and the scale $\mu$ is only through the dimensionless ratios
\bq
\label{chapter_basics:example_kinematic_variables}
 \frac{- p_i \cdot p_j}{\mu^2},
 & &
 \frac{m_i^2}{\mu^2}.
\eq
We call these the 
\index{kinematic variables}
{\bf kinematic variables}.
We  will denote the kinematic variables by $x_1, x_2, \dots$.
Let us count how many kinematic variables there can be.
We may have 
$\nexternal (\nexternal-1)/2$ kinematic variables of the type
\bq
\label{chapter_basics:kinematic_variables_from_Lorentz_invariants}
 \frac{- p_i \cdot p_j}{\mu^2},
 & & 1 \le i \le j \le \left(\nexternal-1\right),
\eq
and $\ninternal$ kinematic variables of the type
\bq
 \frac{m_i^2}{\mu^2}.
\eq
However, if we rescale all kinematic variables by a factor $\lambda$ we have
\bq
\label{chapter_basics:kinematic_variables_scaling_relation}
 I\left( \lambda x_1, \lambda x_2, \dots \right)
 & = &
 \lambda^{\frac{\loopnumber D}{2}-\nu}
 I\left( x_1, x_2, \dots \right).
\eq
This is most easily seen by substituting $\mu^2 \rightarrow \mu^2/\lambda$ in eq.~(\ref{chapter_basics:def_Feynman_integral_v2}).
Thus, we may set one kinematic variable to $1$ and recover the full dependence on all kinematic
variables from the scaling relation in eq.~(\ref{chapter_basics:kinematic_variables_scaling_relation}).

In total we may have up to
\bq
 \frac{\nexternal\left(\nexternal-1\right)}{2} + \ninternal -1
\eq
kinematic variables. 
In typical applications some of them may be zero (for example some internal masses might be zero)
or identical (for example some internal masses might be identical).
We denote the number of independent kinematic variables by 
$\gls{numberofkinematicvariables}$
and the 
independent kinematic variables by $x_1,\dots,x_{\NB}$:
\begin{tcolorbox}
{\bf Notation}:
\begin{center}
\begin{tabular}{ll}
 {\bf number of independent kinematic variables}: & $\NB$ \\
 {\bf independent kinematic variables}: & $x_1,x_2,\dots,x_{\NB}$ \\
\end{tabular}
\end{center}
\end{tcolorbox}
The $x_j$'s are dimensionless quantities of the form as in eq.~(\ref{chapter_basics:example_kinematic_variables}).
\\
\\
\bs
{\it \refstepcounter{exercise}
{\bf Exercise \theexercise}: 
Consider again the one-loop box graph shown in fig.\ref{chapter_basics:fig_oneloopbox}.
Assume first that all internal masses are non-zero and pairwise distinct and that the external momenta
are as generic as possible.
How many kinematic variables are there?
\\
\\
Secondly, assume that all internal masses are zero and that the external momenta satisfy
$p_1^2=p_2^2=p_3^2=p_4^2=0$.
How many kinematic variables are there now?
}
\es
\\
\\
We allow the $x_j$'s to be complex numbers.
We will often encounter the situation, where a kinematic variable is given by a real number
plus an infinitesimal small imaginary part.
The infinitesimal small imaginary part is inherited from 
Feynman's $i\delta$-prescription.
The special case, where all kinematic variables are real and non-negative is called
the 
\index{Euclidean region}
{\bf Euclidean region}.
We have defined the kinematic variables involving Lorentz invariants of the external momenta
with a minus sign as in eq.~(\ref{chapter_basics:kinematic_variables_from_Lorentz_invariants}).
We may define {\bf Euclidean external momenta} in the same way as we defined the Euclidean loop momentum
in eq.~(\ref{chapter_basics:Euclidean_loop_momentum}):
\bq
\label{chapter_basics:Euclidean_external_momentum}
 p^{0} & = & i P^{0}, \nonumber \\
 p^j   & = & P^j, \;\;\;\;\;\mbox{for}\; 1 \le j \le D-1.
\eq
Then
\bq
 \frac{-p_i \cdot p_j}{\mu^2}
 & = &
 \frac{P_i \cdot P_j}{\mu^2},
\eq
where the scalar product on the right-hand side is calculated with Euclidean signature 
\bq
 (+,+,+,+, \dots).
\eq
Thus, if everything is expressed in Euclidean variables, no minus sign appears.
\\
\\
Let us add one word on our notation:
We denote the Feynman integral in eq.~(\ref{chapter_basics:def_Feynman_integral_v2})
by $I$. If we want to emphasise that this integral corresponds to the graph $G$ we write
\bq
 I_G,
\eq
if we want to give the dependence on all variables we write
\bq
\label{chapter_basics:full_notation_Feynman_integral}
 I_{\nu_1 \dots \nu_{\ninternal}}\left(D,x_1,\dots,x_{\NB},\Dint \right).
\eq
In situations, where the dependence on some specific variables is relevant, while the dependence
on the other variables is not, we may write the former and suppress the latter.
In particular we will almost always suppress the dependence on $\Dint$ and write 
\bq
 I_{\nu_1 \dots \nu_{\ninternal}}\left(D,x_1,\dots,x_{\NB}\right).
\eq
Examples of even shorter notations are $I_{\nu_1 \dots \nu_{\ninternal}}$ or $I(D)$.
Thus we will use the notation which is most appropriate within a given context.

\subsection{The Schwinger parameter representation}
\label{chapter_basics:subsection:Schwinger_parameter}

The Schwinger parameter representation is the first representation, where we trade the $(\loopnumber D)$ momentum integrations
for some auxiliary integrations.
In the case of the Schwinger parameter representation we will treat the momentum integration for an integration over
$\ninternal$ Schwinger parameters.
We start from the following identity, also called {\bf Schwinger's trick}:
Let $A>0$ and $\mathrm{Re}(\nu)>0$.
We have
\bq
\label{chapter_basics:Schwinger_trick}
 \frac{1}{A^\nu} & = &
 \frac{1}{\Gamma\left(\nu\right)} \int\limits_0^\infty d\alpha \; \alpha^{\nu-1} \; e^{-\alpha A}.
\eq
Eq.~(\ref{chapter_basics:Schwinger_trick}) follows immediately from the definition of Euler's gamma function.
We apply eq.~(\ref{chapter_basics:Schwinger_trick}) for
$A=Q^2+m^2=-q^2+m^2$ (the quantity $A$ is positive after Wick rotation and for external Euclidean kinematics):
\bq
\label{chapter_basics:Schwinger_trick_propagator}
 \frac{1}{(-q_j^2+m_j^2)^{\nu_j}}
 & = &
 \frac{1}{\Gamma(\nu_j)}
 \int\limits_{\alpha_j \ge 0}  d\alpha_j \;
 \alpha_j^{\nu_j-1}
 \exp\left(-\alpha_j (-q_j^2+m_j^2)\right)
\eq
The variable 
$\gls{Schwingerparameter}$
is called a
\index{Schwinger parameter}
{\bf Schwinger parameter}.
Doing this for all internal edges gives
\bq
\label{chapter_basics:loop_momenta_and_Schwinger_integral}
I  & = &
 \frac{e^{\loopnumber \eps \Eulerconstant} \left(\mu^2\right)^{\nu-\frac{\loopnumber D}{2}}}{\prod\limits_{j=1}^{\ninternal}\Gamma(\nu_j)}
 \int\limits_{\alpha_j \ge 0}  d^{\ninternal}\alpha \;
 \left( \prod\limits_{j=1}^{\ninternal} \alpha_j^{\nu_j-1} \right)
 \int \prod\limits_{r=1}^{\loopnumber} \frac{d^Dk_r}{i\pi^{\frac{D}{2}}}\;
 \exp\left(-\sum\limits_{j=1}^{\ninternal} \alpha_j \left(-q_j^2+m_j^2\right) \right).
 \nonumber \\
\eq
Using
\bq
 q_j & = &
 \sum\limits_{r=1}^\loopnumber \lambda_{jr} k_r
 +
 \sum\limits_{r=1}^{\nexternal - 1} \sigma_{jr} p_r
\eq
we express the argument of the exponential function as
\bq
\label{chapter_basics:eq_poly_calc_1}
 \sum\limits_{j=1}^{\ninternal} \alpha_{j} (-q_j^2+m_j^2)
 & = & 
 - \sum\limits_{r=1}^{\loopnumber} \sum\limits_{s=1}^{\loopnumber} k_r M_{rs} k_s + \sum\limits_{r=1}^{\loopnumber} 2 k_r \cdot v_r + J.
\eq
where $M$ is a $l \times l$ matrix with scalar entries, $v$ is a $l$-vector
with $D$-dimensional momentum vectors as entries and $J$ is scalar.
Let us define 
\bq
\label{chapter_basics:eq_poly_calc_2}
 {\mathcal U} = \mbox{det}(M),
 & &
 {\mathcal F} = \mbox{det}(M) \left( J + v^T M^{-1} v \right)/\mu^2.
\eq
The functions ${\mathcal U}$ and ${\mathcal F}$ are called 
\index{graph polynomials}
{\bf graph polynomials} 
(they are polynomials in the $\alpha_j$'s) 
and are discussed in detail in chapter~\ref{chapter_graph_polynomials}.
The polynomials ${\mathcal U}$ and ${\mathcal F}$ have the following properties:
\begin{itemize}
\item They are homogeneous in the Schwinger parameters, 
${\mathcal U}$ is of degree $l$, ${\mathcal F}$ is of degree $l+1$.
\item ${\mathcal U}$ is linear in each Schwinger parameter. If all internal masses are zero, then
also ${\mathcal F}$ is linear in each Schwinger parameter.
\item In expanded form each monomial of ${\mathcal U}$ has coefficient $+1$.
\end{itemize}
We call 
$\gls{firstgraphpolynomial}$
the 
\index{Symanzik polynomial}
{\bf first Symanzik polynomial} and 
$\gls{secondgraphpolynomial}$
the {\bf second Symanzik polynomial}.

From linear algebra we know that for 
a real symmetric positive definite $(n \times n)$-matrix $A$ we have
\bq
 \int\limits_{-\infty}^\infty
 dy_1 ... dy_n  \; \exp\left(- \vec{y}^T A \vec{y} + 2 \vec{w}^T \vec{y} + c \right) 
 & = & 
 \pi^{n/2} 
 \left( \det A \right)^{-\frac{1}{2}}
 \exp \left( \vec{w}^T A^{-1} \vec{w} + c \right),
\eq
and due to eq.~(\ref{chapter_basics:normalisation_D_int}) this extends to dimensional regularisation.
We may therefore perform the loop momentum integration and obtain the
\begin{tcolorbox}
{\bf Schwinger parameter representation}:
\bq
\label{chapter_basics:Schwinger_parameter_representation}
I  & = &
 \frac{e^{\loopnumber \eps \Eulerconstant}}{\prod\limits_{j=1}^{\ninternal}\Gamma(\nu_j)}
 \;
 \int\limits_{\alpha_j \ge 0}  d^{\ninternal}\alpha \;
 \left( \prod\limits_{j=1}^{\ninternal} \alpha_j^{\nu_j-1} \right)
 \left[ {\mathcal U}\left(\alpha\right) \right]^{-\frac{D}{2}}
 \exp\left( - \frac{{\mathcal F}\left(\alpha\right)}{{\mathcal U}\left(\alpha\right)}\right).
\eq
\end{tcolorbox}
Thus we went from a $(\loopnumber \cdot D)$-fold momentum integration to a
$\ninternal$-fold Schwinger parameter integration.
Note that the number of space-time dimensions $D$ enters only the exponent of ${\mathcal U}^{-D/2}$
(and the prefactor $e^{\loopnumber \eps \gamma}$ through $\eps=(\Dint-D)/2$).

We now encountered for the first time the sum on the left-hand side of eq.~(\ref{chapter_basics:eq_poly_calc_1}) and the two graph polynomials ${\mathcal U}$ and ${\mathcal F}$.
As these ingredients will occur quite frequently throughout this book, it is worth
working them out in a non-trivial example.
We consider the two-loop graph shown in fig.~\ref{chapter_basics:fig_doublebox}
for the case
\bq
\label{chapter_basics:specification_double_box}
 & & p_1^2 = 0, \;\;\; p_2^2 = 0, \;\;\; p_3^2 = 0, \;\;\; p_4^2 = 0,
 \nonumber \\
 & & m_1 = m_2 = m_3 = m_4 = m_5 = m_6 = m_7 = 0.
\eq
We define
\bq
 s = \left(p_1+p_2\right)^2=\left(p_3+p_4\right)^2,
 & &
 t = \left(p_2+p_3\right)^2=\left(p_1+p_4\right)^2.
\eq
We have
\bq
 \sum\limits_{j=1}^7 \alpha_j \left(-q_j^2\right) & = &
 - \left(\alpha_1+\alpha_2+\alpha_3+\alpha_4\right) k_1^2 - 2 \alpha_4 k_1 \cdot k_2 - \left( \alpha_4+\alpha_5+\alpha_6+\alpha_7\right) k_2^2
 \\
 & &
 + 2 \left[ \alpha_1 p_1 + \alpha_2 \left( p_1 + p_2 \right) \right] \cdot k_1
 + 2 \left[ \alpha_5 \left( p_3 + p_4 \right) + \alpha_7 p_4 \right] \cdot k_2
 - \left( \alpha_2 + \alpha_5 \right) s.
 \nonumber
\eq
In comparing with eq.~(\ref{chapter_basics:eq_poly_calc_1})
we find
\bq
 M & = & \left( \begin{array}{cc}
 \alpha_1+\alpha_2+\alpha_3+\alpha_4 & \alpha_4 \\
 \alpha_4 & \alpha_4+\alpha_5+\alpha_6+\alpha_7 \\
 \end{array} \right),
 \nonumber \\
 v & = & \left( \begin{array}{c}
          \alpha_1 p_1 + \alpha_2 \left( p_1 + p_2 \right) \\
          \alpha_5 \left( p_3 + p_4 \right) + \alpha_7 p_4 \\
          \end{array} \right),
 \nonumber \\
 J & = & \left( \alpha_2 + \alpha_5 \right) \left(-s\right).
\eq
Plugging this into eq.~(\ref{chapter_basics:eq_poly_calc_2})
we obtain the graph polynomials as
\bq
\label{chapter_basics:result_U_and_F_double_box}
{\mathcal U} & = & \left( \alpha_1+\alpha_2+\alpha_3 \right) \left( \alpha_5+\alpha_6+\alpha_7 \right) + \alpha_4 \left( \alpha_1+\alpha_2+\alpha_3+\alpha_5+\alpha_6+\alpha_7 \right),
 \nonumber \\
{\mathcal F} & = & \left[ \alpha_2 \alpha_3 \left( \alpha_4+\alpha_5+\alpha_6+\alpha_7 \right)
                        + \alpha_5 \alpha_6 \left( \alpha_1+\alpha_2+\alpha_3+\alpha_4 \right)
                        + \alpha_2 \alpha_4 \alpha_6 + \alpha_3 \alpha_4 \alpha_5 \right] \left( \frac{-s}{\mu^2} \right)
 \nonumber \\
 & &
      + \alpha_1 \alpha_4 \alpha_7 \left( \frac{-t}{\mu^2} \right).
\eq
We see in this example that ${\mathcal U}$ is of degree $2$ and ${\mathcal F}$ is of degree $3$ in the 
Schwinger parameters.
Each polynomial is linear in each Schwinger parameter. Furthermore, when we write ${\mathcal U}$
in expanded form
\bq 
 {\mathcal U} & = &
  \alpha_1 \alpha_5 + \alpha_1 \alpha_6 + \alpha_1 \alpha_7
+ \alpha_2 \alpha_5 + \alpha_2 \alpha_6 + \alpha_2 \alpha_7
+ \alpha_3 \alpha_5 + \alpha_3 \alpha_6 + \alpha_3 \alpha_7
 \nonumber \\
 & &
+ \alpha_1 \alpha_4 + \alpha_2 \alpha_4 + \alpha_3 \alpha_4 + \alpha_4 \alpha_5 + \alpha_4 \alpha_6 + \alpha_4 \alpha_7,
\eq
each term has coefficient $+1$.
\\
\\
\bs
{\it \refstepcounter{exercise}
{\bf Exercise \theexercise}: 
Determine with the method above the graph polynomials ${\mathcal U}$ and ${\mathcal F}$ for
\begin{figure}
\begin{center}
\includegraphics[scale=1.0]{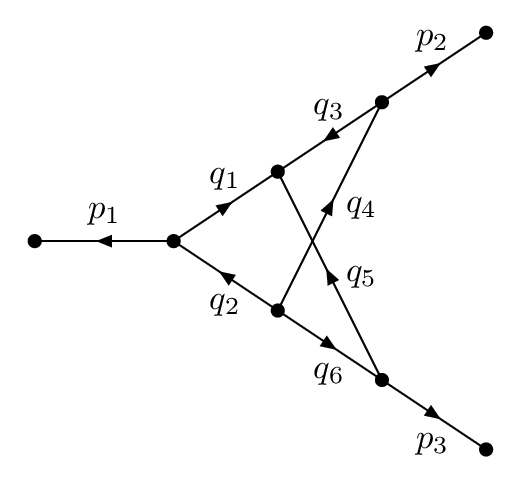}
\end{center}
\caption{
A two-loop non-planar vertex graph.
}
\label{chapter_basics:fig_nonplanar_vertex}
\end{figure}
the graph shown in fig.~\ref{chapter_basics:fig_nonplanar_vertex} for the case where
all internal masses are zero.
}
\es

\subsection{The Feynman parameter representation}

Probably the most popular parameter representation is the Feynman parameter representation.
It is effectively a $(\ninternal-1)$-fold integral representation.
We obtain the Feynman parameter representation from the Schwinger parameter representation as follows:
We first note that the sum of the Schwinger parameters is non-negative:
\bq
 \sum\limits_{j=1}^n \alpha_j  & \ge & 0.
\eq
We then insert a $1$ in the form of
\bq
 1 & = & 
 \int\limits_{-\infty}^\infty dt \; \delta\left(t - \sum\limits_{j=1}^n \alpha_j \right)
 \;\; = \;\;
 \int\limits_{0}^\infty dt \; \delta\left(t - \sum\limits_{j=1}^n \alpha_j \right),
\eq
where in the last step we used the fact that the sum of the Schwinger parameters is non-negative.
$\delta(x)$ denotes Dirac's delta distribution.
Changing variables according to $a_j = \alpha_j/t$ gives us the identity
\bq
 \int\limits_{\alpha_j \ge 0} d^n\alpha \; f\left(\alpha_1,\dots,\alpha_n\right) 
 & = & 
 \int\limits_{a_j \ge 0} d^na \; \delta\left(1-\sum\limits_{j=1}^n a_j \right) \; 
 \int\limits_{0}^\infty dt \; t^{n-1} \; f\left(t a_1,\dots,t a_n\right).
\eq
We apply this identity to the Schwinger parameter representation and use the fact
that ${\mathcal U}$ and ${\mathcal F}$ are homogeneous of degree $\loopnumber$ and $(\loopnumber+1)$, respectively.
We obtain
\bq
 I 
 & = &
 \frac{e^{\loopnumber \eps \Eulerconstant}}{\prod\limits_{j=1}^{\ninternal}\Gamma(\nu_j)}
 \;
 \int\limits_{a_j \ge 0} d^{\ninternal}a \; \delta\left(1-\sum\limits_{j=1}^{\ninternal} a_j \right) \; 
 \left( \prod\limits_{j=1}^{\ninternal} a_j^{\nu_j-1} \right)
 \left[ {\mathcal U}\left(a\right) \right]^{-\frac{D}{2}}
 \\
 & & 
 \times
 \int\limits_{0}^\infty dt \; t^{\nu-\frac{\loopnumber D}{2}-1}
 \exp\left( - \frac{{\mathcal F}\left(a\right)}{{\mathcal U}\left(a\right)} t \right)
 \nonumber \\
 & = &
 \frac{e^{\loopnumber \eps \Eulerconstant}}{\prod\limits_{j=1}^{\ninternal}\Gamma(\nu_j)}
 \;
 \int\limits_{a_j \ge 0} d^{\ninternal}a \; \delta\left(1-\sum\limits_{j=1}^{\ninternal} a_j \right) \; 
 \left( \prod\limits_{j=1}^{\ninternal} a_j^{\nu_j-1} \right)
 \frac{\left[ {\mathcal U}\left(a\right) \right]^{\nu-\frac{\left(\loopnumber+1\right) D}{2}}}{\left[ {\mathcal F}\left(a\right) \right]^{\nu-\frac{\loopnumber D}{2}}}
 \int\limits_{0}^\infty dt \; t^{\nu-\frac{\loopnumber D}{2}-1}
 e^{-t}.
 \nonumber
\eq
In the step towards the last line we substituted $t \rightarrow t {\mathcal U}(a) / {\mathcal F}(a)$.
The final integral over $t$ gives $\Gamma(\nu-\loopnumber D/2)$.
We thus arrive at the
\begin{tcolorbox}
{\bf Feynman parameter representation}:
\bq
\label{chapter_basics:Feynman_parameter_representation}
 I
 & = &
 \frac{e^{\loopnumber \eps \Eulerconstant}\Gamma\left(\nu-\frac{\loopnumber D}{2}\right)}{\prod\limits_{j=1}^{\ninternal}\Gamma(\nu_j)}
 \int\limits_{a_j \ge 0} d^{\ninternal}a \; \delta\left(1-\sum\limits_{j=1}^{\ninternal} a_j \right) \; 
 \left( \prod\limits_{j=1}^{\ninternal} a_j^{\nu_j-1} \right)
 \frac{\left[ {\mathcal U}\left(a\right) \right]^{\nu-\frac{\left(\loopnumber+1\right) D}{2}}}{\left[ {\mathcal F}\left(a\right) \right]^{\nu-\frac{\loopnumber D}{2}}}.
 \hspace*{12mm}
\eq
\end{tcolorbox}
The polynomials ${\mathcal U}$ and ${\mathcal F}$ are as before, with $\alpha_j$ substituted by $a_j$.
The variable 
$\gls{Feynmanparameter}$
is called a
\index{Feynman parameter}
{\bf Feynman parameter}.

We have derived the Feynman parameter representation from the Schwinger parameter representation.
We may go directly from the momentum representation to the Feynman parameter representation
with the help of {\bf Feynman's trick}: For $A_j>0$ and $\mathrm{Re}(\nu_j)>0$ we have
\bq
\label{chapter_basics:Feynman_trick}
 \prod\limits_{j=1}^{n} \frac{1}{A_{j}^{\nu_{j}}} 
 & = &
 \frac{\Gamma(\nu)}{\prod\limits_{j=1}^{n} \Gamma(\nu_{j})} \;\;
 \int\limits_{a_j \ge 0} d^na \; \delta(1-\sum\limits_{j=1}^{n} a_{j})
 \left( \prod\limits_{j=1}^{n} a_{j}^{\nu_{j}-1} \right)
 \frac{1}{\left( \sum\limits_{j=1}^{n} a_{j} A_{j} \right)^{\nu}},
 \nonumber \\
 & &\nu = \sum\limits_{j=1}^{n} \nu_{j}. 
\eq
\bs
{\it \refstepcounter{exercise}
{\bf Exercise \theexercise}: 
Prove eq.~(\ref{chapter_basics:Feynman_trick}).
}
\es
\\
\\
We use this formula with $A_j=-q_j^2+m_j^2$.
We may then use translational invariance (i.e. eq.~(\ref{chapter_basics:translation_invariance}))
for each $D$-dimensional momentum integral
and shift each loop
momentum $k_r$ to complete the square, such that the integrand depends only on $k_r^2$.
Then all $D$-dimensional momentum integrals can be performed with the help of eq.~(\ref{chapter_basics:master_one_loop})
and we recover eq.~(\ref{chapter_basics:Feynman_parameter_representation}).

Let us look at an example.
\begin{figure}
\begin{center}
\includegraphics[scale=1.0]{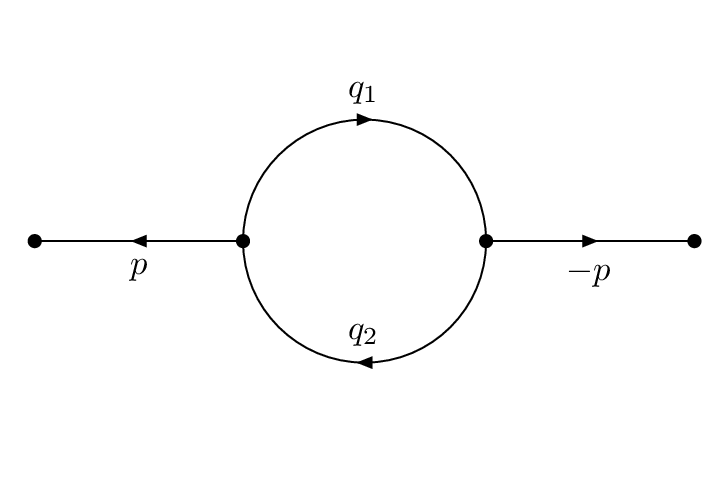}
\end{center}
\caption{
The one-loop bubble diagram.
}
\label{chapter_basics:fig_bubble}
\end{figure}
We consider the one-loop bubble diagram shown in fig.~\ref{chapter_basics:fig_bubble}.
The momentum representation is
\bq
 I_{\nu_1\nu_2}
 & = &
 e^{\eps \Eulerconstant} \left(\mu^2\right)^{\nu-\frac{D}{2}}
 \int \frac{d^Dk}{i \pi^{\frac{D}{2}}} 
 \frac{1}{\left(-q_1^2+m_1^2\right)^{\nu_1} \left(-q_2^2+m_2^2\right)^{\nu_2}},
\eq
with
\bq
 q_1 \; = \; k-p,
 & & 
 q_2 \; = \; k.
\eq
Feynman parametrisation gives us
\bq
 I_{\nu_1\nu_2}
 & = &
 \frac{e^{\eps \Eulerconstant} \Gamma\left(\nu\right)}{\Gamma\left(\nu_1\right)\Gamma\left(\nu_2\right)}
 \left(\mu^2\right)^{\nu-\frac{D}{2}}
 \int d^2a \; \delta\left(1-a_1-a_2\right) a_1^{\nu_1-1} a_2^{\nu_2-1}
 \nonumber \\
 & &
 \times
 \int \frac{d^Dk}{i \pi^{\frac{D}{2}}} 
 \frac{1}{\left[ a_1 \left(-q_1^2+m_1^2\right) + a_2 \left(-q_2^2+m_2^2\right)\right]^{\nu}}.
\eq
Completing the square we obtain
\bq
\lefteqn{
 a_1 \left(-q_1^2+m_1^2\right) + a_2 \left(-q_2^2+m_2^2\right)
 = } & & \nonumber \\
 & &
 - \left(a_1+a_2\right) \left(k- \frac{a_1}{a_1+a_2} p \right)^2
 + \frac{a_1a_2}{a_1+a_2} \left(-p^2\right)
 + a_1 m_1^2 + a_2 m_2^2.
\eq
Let us set
\bq
 {\mathcal U} & = & a_1+a_2,
 \nonumber \\
 {\mathcal F} & = & 
 a_1a_2 \left(\frac{-p^2}{\mu^2}\right)
 + \left(a_1+a_2\right) \left[ a_1 \left(\frac{m_1^2}{\mu^2}\right) + a_2 \left(\frac{m_2^2}{\mu^2}\right) \right].
\eq
These are the two graph polynomials.
With the substitution $k \rightarrow k+ a_1/(a_1+a_2) p$ we obtain
\bq
 I_{\nu_1\nu_2}
 & = &
 \frac{e^{\eps \Eulerconstant} \Gamma\left(\nu\right)}{\Gamma\left(\nu_1\right)\Gamma\left(\nu_2\right)}
 \left(\mu^2\right)^{\nu-\frac{D}{2}}
 \int d^2a \; \delta\left(1-a_1-a_2\right) a_1^{\nu_1-1} a_2^{\nu_2-1}
 \int \frac{d^Dk}{i \pi^{\frac{D}{2}}} 
 \frac{1}{\left[ - {\mathcal U} k^2 + \frac{\mathcal F}{\mathcal U} \mu^2 \right]^{\nu}}.
 \nonumber \\
\eq
This is now in the form of eq.~(\ref{chapter_basics:master_one_loop}) 
and using the one-loop master formula yields
\bq
 I_{\nu_1\nu_2}
 & = &
 \frac{e^{\eps \Eulerconstant} \Gamma\left(\nu-\frac{D}{2}\right)}{\Gamma\left(\nu_1\right)\Gamma\left(\nu_2\right)}
 \int d^2a \; \delta\left(1-a_1-a_2\right) a_1^{\nu_1-1} a_2^{\nu_2-1}
 \frac{{\mathcal U}^{\nu-D}}{{\mathcal F}^{\nu-\frac{D}{2}}}.
\eq
This is again the Feynman parameter representation of eq.~(\ref{chapter_basics:Feynman_parameter_representation}), which we recovered ``by foot''.
Of course, it is just sufficient to determine the two graph polynomials and use
eq.~(\ref{chapter_basics:Feynman_parameter_representation}) directly.
We have seen one method to determine the two graph polynomials
in section~\ref{chapter_basics:subsection:Schwinger_parameter}.
We will learn more (efficient) methods to determine the graph polynomials
in chapter~\ref{chapter_graph_polynomials}.

Let us now specialise to the case, where the internal masses vanish: $m_1=m_2=0$. 
In this case the second graph polynomial simplifies  to
\bq
 {\mathcal F} & = & 
 a_1a_2 \left(\frac{-p^2}{\mu^2}\right)
\eq
and the one-loop bubble integral to
\bq
 I_{\nu_1\nu_2}
 & = &
 \frac{e^{\eps \Eulerconstant} \Gamma\left(\nu-\frac{D}{2}\right)}{\Gamma\left(\nu_1\right)\Gamma\left(\nu_2\right)}
 \left(\frac{-p^2}{\mu^2}\right)^{\frac{D}{2}-\nu}
 \int\limits_0^1 da \; a^{\frac{D}{2}-\nu_2-1} \left(1-a\right)^{\frac{D}{2}-\nu_1-1}.
\eq
The integral over $a$ is just Euler's beta function and we obtain
\bq
\label{chapter_basics:result_bubble}
 I_{\nu_1\nu_2}
 & = &
 e^{\eps \Eulerconstant} 
 \frac{\Gamma\left(\nu-\frac{D}{2}\right)\Gamma\left(\frac{D}{2}-\nu_1\right)\Gamma\left(\frac{D}{2}-\nu_2\right)}{\Gamma\left(\nu_1\right)\Gamma\left(\nu_2\right)\Gamma\left(D-\nu\right)}
 \left(\frac{-p^2}{\mu^2}\right)^{\frac{D}{2}-\nu}.
\eq
If we further specialise to $\nu_1=\nu_2=1$, $D=4-2\eps$ and set $L=\ln(-p^2/\mu^2)$ we find
\bq
\label{chapter_basics:result_bubble_B11_4D}
 I_{11}
 & = &
 \frac{1}{\eps}
 +
 2 - L
 + {\mathcal O}\left(\eps\right).
\eq
Thus we calculated our second Feynman integral, the massless one-loop two-point function.
\\
\\
\bs
{\it \refstepcounter{exercise}
{\bf Exercise \theexercise}: 
Calculate with the help of the Feynman parameter representation 
\begin{figure}
\begin{center}
\includegraphics[scale=1.0]{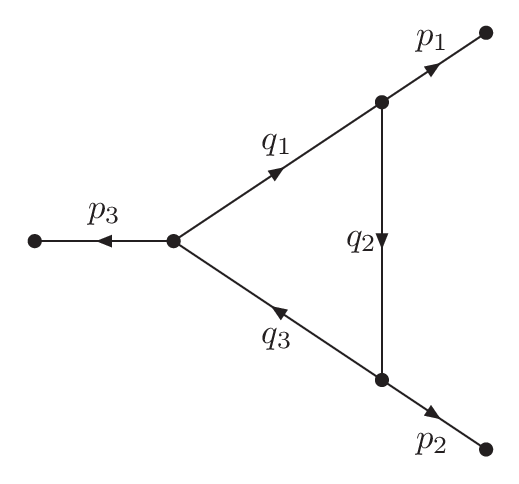}
\end{center}
\caption{
The one-loop triangle diagram.
}
\label{chapter_basics:fig_onelooptriangle}
\end{figure}
the 
one-loop triangle integral 
\bq
 I_{\nu_1\nu_2\nu_3}
 & = &
 e^{\eps \Eulerconstant} \left(\mu^2\right)^{\nu-\frac{D}{2}}
 \int \frac{d^Dk}{i \pi^{\frac{D}{2}}} 
 \frac{1}{\left(-q_1^2\right)^{\nu_1} \left(-q_2^2\right)^{\nu_2} \left(-q_3^2\right)^{\nu_3}},
\eq
shown in fig.~\ref{chapter_basics:fig_onelooptriangle}
for the case where all internal masses are zero ($m_1=m_2=m_3=0$) and 
for the kinematic configuration $p_1^2=p_2^2=0$, $p_3^2 \neq 0$.
}
\es
\\
\\
The Feynman parameter representation of eq.~(\ref{chapter_basics:Feynman_parameter_representation})
treats every internal edge equal. This is called the 
\index{democratic approach, Feynman parameter}
{\bf democratic approach}.
However, this is not the only possibility.
We may also use a 
\index{hierarchical approach, Feynman parameter}
{\bf hierarchical approach}.
The following exercise shows, that there are situations where this is useful.
\\
\\
\bs
{\it \refstepcounter{exercise}
\label{chapter_basics:exercise_oneloopbox_Feynman_parameter_trick}
{\bf Exercise \theexercise}: 
Consider again the one-loop box graph in fig.~\ref{chapter_basics:fig_oneloopbox},
this time for the kinematic configuration
\bq
 p_2^2 \; = \; p_4^2 \; = \; 0,
 & &
 m_1 \; = \; m_2 \; = \; m_3 \; = \; m_4 \; = \; 0.
\eq
Write down the Feynman parameter representation as in eq.~(\ref{chapter_basics:Feynman_parameter_representation}).
Obtain a second integral representation by first combining propagators $1$ and $2$ with a
pair of Feynman parameters, then combining propagators $3$ and $4$ with a second pair of
Feynman parameters and finally the two results with a third pair of Feynman parameters.
}
\es

\subsubsection{Projective integrals}

The Feynman parameter representation is actual a projective integral.
It pays off to rewrite the Feynman parameter representation in terms of differential
forms on projective space.
Below we will state the Cheng-Wu theorem, which follows directly
from the fact the Feynman parameter representation is a projective integral.

We start with introducing the essential facts about projective space.
\begin{digression} 
\index{projective space}
{\bf Projective space}
\\
Let ${\mathbb F}$ be a field. 
Relevant to us are the cases where the field ${\mathbb F}$ is either the field
of real numbers ${\mathbb R}$ or the field of complex numbers ${\mathbb C}$.

The projective space $\mathrm{P}^n\left({\mathbb F}\right)$ is the set of lines 
through the origin in ${\mathbb F}^{n+1}$.
Equivalently, it is the set of points in ${\mathbb F}^{n+1} \backslash \{0\}$ modulo
the equivalence relation
\bq
\label{chapter_basics:projective_space_equivalence_relation}
 \left( x_0, x_1, ..., x_n \right) \sim \left( y_0, y_1, ..., y_n \right)
 & \Leftrightarrow &
 \exists \; \lambda \neq 0 : 
 \left( x_0, x_1, ..., x_n \right) = \left( \lambda y_0, \lambda y_1, ..., \lambda y_n \right).
\hspace*{8mm}
\eq
Points in $\mathrm{P}^n\left({\mathbb F}\right)$ will be denoted by
\bq
\label{chapter_basics:homogeneous_coordinates_projective_space}
 \left[ z_0 : z_1 : ... : z_n \right].
\eq
The coordinates in eq.~(\ref{chapter_basics:homogeneous_coordinates_projective_space}) are called 
\index{homogeneous coordinates in projective space}
{\bf homogeneous coordinates}.
Affine coordinate patches are defined as follows: 
We consider the open subsets
\bq
 U_j & = &
 \left\{ \; \left[ z_0 : z_1 : ... : z_n \right] \; | \; z_j \neq 0 \; \right\},
 \;\;\;\;\;\;
 0 \le j \le n.
\eq
We have the homeomorphisms
\bq
 \varphi_j & : & U_j \rightarrow {\mathbb F}^n,
 \nonumber \\
 & &
 \left[ z_0 : z_1 : ... : z_n \right] \rightarrow \left( \frac{z_0}{z_j}, ..., \frac{z_{j-1}}{z_j}, \frac{z_{j+1}}{z_j}, ..., \frac{z_n}{z_j} \right).
\eq
The inverse mapping is given by
\bq
 \varphi_j^{-1} & : & {\mathbb F}^n \rightarrow U_j,
 \nonumber \\
 & &
 \left( z_0, ..., z_{j-1}, z_{j+1}, ..., z_n \right) \rightarrow
 \left[ z_0 : ... : z_{j-1} : 1 : z_{j+1} : ... : z_n \right].
\eq
The pair $(U_j,\varphi_j)$ defines a chart for $\mathrm{P}^n\left({\mathbb F}\right)$,
and the collection of all $(U_j,\varphi_j)$ for $0 \le j \le n$ provides an atlas
for $\mathrm{P}^n\left({\mathbb F}\right)$.

For ${\mathbb F}={\mathbb C}$ or ${\mathbb F}={\mathbb R}$ we will also use the notation
\bq
 {\mathbb C} {\mathbb P}^n \; = \; \mathrm{P}^n\left({\mathbb C}\right),
 & &
 {\mathbb R} {\mathbb P}^n \; = \; \mathrm{P}^n\left({\mathbb R}\right),
\eq
and we will speak about the complex projective space and the real projective space, respectively.

The 
\index{positive real projective space}
{\bf positive real projective space} ${\mathbb R} {\mathbb P}^n_{>0}$
is the set of all points of ${\mathbb R} {\mathbb P}^n$, which
can be represented by
\bq
 \left[ x_0 : x_1 : ... : x_n \right]
 & \mbox{with} &
 x_j \; > \; 0,
 \;\;\;\;\;\; 
 0 \le j \le n.
\eq
Thus we have $[1:2:3] \in {\mathbb R} {\mathbb P}^2_{>0}$ 
and $[(-4):(-5):(-6)] \in {\mathbb R} {\mathbb P}^2_{>0}$
(we may choose $\lambda=-1$ in eq.~(\ref{chapter_basics:projective_space_equivalence_relation})),
but $[7:(-8):9] \notin {\mathbb R} {\mathbb P}^2_{>0}$.

The 
{\bf non-negative real projective space} ${\mathbb R} {\mathbb P}^n_{\ge 0}$
is the set of all points of ${\mathbb R} {\mathbb P}^n$, which
can be represented by
\bq
 \left[ x_0 : x_1 : ... : x_n \right]
 & \mbox{with} &
 x_j \; \ge \; 0,
 \;\;\;\;\;\; 
 0 \le j \le n.
\eq
In the literature the notation is sometimes used in a sloppy way, e.g.
the word positive real projective space is also used where the non-negative real projective space
is meant.
However, the symbols ${\mathbb R} {\mathbb P}^n_{> 0}$ and ${\mathbb R} {\mathbb P}^n_{\ge 0}$ clearly indicate what is meant.
\end{digression}
We return to the Feynman parameter representation.
We denote the integrand of the Feynman parameter representation by
\bq
\label{chapter_basics:def_f}
 f\left(a\right) & = &
 \frac{e^{\loopnumber \eps \Eulerconstant}\Gamma\left(\nu-\frac{\loopnumber D}{2}\right)}{\prod\limits_{j=1}^{\ninternal}\Gamma(\nu_j)}
 \left( \prod\limits_{j=1}^{\ninternal} a_j^{\nu_j-1} \right)
 \frac{\left[ {\mathcal U}\left(a\right) \right]^{\nu-\frac{\left(\loopnumber+1\right) D}{2}}}{\left[ {\mathcal F}\left(a\right) \right]^{\nu-\frac{\loopnumber D}{2}}}.
\eq
If we rescale all Feynman parameters by $\lambda$ we have
\bq
 f\left(\lambda a_1, \dots, \lambda a_{\ninternal}\right)
 & = &
 \lambda^{-\ninternal}
 f\left(a_1, \dots, a_{\ninternal}\right).
\eq
This follows easily from the homogeneity of ${\mathcal U}$ and ${\mathcal F}$ in the Feynman parameters.
We also have
\bq
\label{chapter_basics:dgl_integrand}
 \left( \ninternal + \sum\limits_{j=1}^{\ninternal} a_j \frac{\partial}{\partial a_j} \right) f\left(a\right) & = & 0.
\eq
\bs
{\it \refstepcounter{exercise}
{\bf Exercise \theexercise}: 
Prove eq.~(\ref{chapter_basics:dgl_integrand}).
}
\es
\\
\\
Let us further introduce the differential $(\ninternal-1)$-form
\bq
\label{chapter_basics:def_omega}
 \omega & = & \sum\limits_{j=1}^{\ninternal} (-1)^{{\ninternal}-j}
  \; a_j \; da_1 \wedge ... \wedge \widehat{da_j} \wedge ... \wedge da_{\ninternal},
\eq
where the hat indicates that the corresponding term is omitted.
Let us denote by $\Delta$ the $(\ninternal-1)$-dimensional 
\index{standard simplex}
standard simplex:
\bq
 \Delta & = & \left\{ \left(a_1,\dots,a_{\ninternal}\right) \in \left. {\mathbb R}^{\ninternal} \right| \sum\limits_{j=1}^{\ninternal} a_j = 1, a_j \ge 0 \right\}.
\eq
With these definitions we may write the Feynman parameter representation as
\bq
\label{chapter_basics:Feynman_representation_differential_forms}
 I
 & = &
 \int\limits_{\Delta} f \omega.
\eq
\bs
{\it \refstepcounter{exercise}
{\bf Exercise \theexercise}: 
Show explicitly that eq.~(\ref{chapter_basics:Feynman_representation_differential_forms})
is equivalent to eq.~(\ref{chapter_basics:Feynman_parameter_representation}).
}
\es
\\
\\
The differential $(\ninternal-1)$-form $\omega$ has the property
that when we integrate it along a line through the origin, the result
vanishes.
In general, integrating a $(\ninternal-1)$-form along a curve
gives a $(\ninternal-2)$-form. 
If the curve is defined by the vector field $X$, 
integrating along an infinitesimal interval gives 
a $(\ninternal-2)$-form proportional to
the 
\index{interior product}
{\bf interior product} $\iota_X \omega$.
A line through the origin is defined by the vector field
\bq
 X & = &
 \lambda_1 e_1 + \dots + \lambda_{\ninternal} e_{\ninternal},
\eq
where $\lambda_1, \dots, \lambda_{\ninternal}$ are constants and 
$(\lambda_1, \dots, \lambda_{\ninternal}) \neq (0, \dots, 0)$.
$e_1, \dots, e_{\ninternal}$ are basis vectors of the tangent space.
The statement above, that the integration of $\omega$ along a line through
the origin vanishes, is equivalent to
\bq
\label{chapter_basics:interior_product}
 \iota_X \omega & = & 0.
\eq
\bs
{\it \refstepcounter{exercise}
{\bf Exercise \theexercise}: 
Prove eq.~(\ref{chapter_basics:interior_product}).
}
\es
\\
\\
From eq.~(\ref{chapter_basics:dgl_integrand}) it follows that 
$f \omega$ is closed:
\bq
\label{chapter_basics:f_omega_closed}
d \left( f \omega \right) & = & 0.
\eq
\bs
{\it \refstepcounter{exercise}
{\bf Exercise \theexercise}: 
Prove eq.~(\ref{chapter_basics:f_omega_closed}).
}
\es
\\
\\
Let us temporarily assume that $f \omega$ is non-singular for all points $a \in \Delta$.
(This assumption is not as innocent as it may seem.
The case, where $f \omega$ is singular for some points $a \in \Delta$
will stalk us in the sequel of the book.
For the moment, we are certainly safe in the Euclidean region, if all internal propagators have a non-zero mass
(this avoids infrared singularities)
and $D<0$ (this avoids ultraviolet singularities)). 
We now have all ingredients to claim that the Feynman parameter representation is a projective
integral over the non-negative real projective space ${\mathbb R} {\mathbb P}^{\ninternal-1}_{\ge 0}$:
\begin{tcolorbox}
{\bf Projective Feynman parameter integral representation}:
\bq
 I 
 & = &
 \int\limits_{{\mathbb R} {\mathbb P}^{\ninternal-1}_{\ge 0}} f \omega,
\eq
where $f$ is defined in eq.~(\ref{chapter_basics:def_f}),
$\omega$ is defined in eq.~(\ref{chapter_basics:def_omega})
and ${\mathbb R} {\mathbb P}^{\ninternal-1}_{\ge 0}$ denotes the non-negative real projective space 
of dimension $(\ninternal-1)$.
\end{tcolorbox}
In particular we may integrate over any hyper-surface
covering the solid angle $a_j \ge 0$.

The proof is based on Stoke's theorem's: Since $d(f\omega)$ is closed, 
integration over any $\ninternal$-dimensional
domain $\Sigma$ in ${\mathbb R}^{\ninternal}$ gives zero.
Thus
\bq
 \int\limits_{\Sigma} d\left( f \omega\right)
 & = &
 \int\limits_{\partial \Sigma} f \omega.
\eq
We may choose $\Sigma$ as a domain bounded by the standard simplex $\Delta$, 
the hyper-surface $\tilde{\Delta}$ we are interested in and covering the solid angle $a_j \ge 0$,
and additional $(\ninternal-1)$-dimensional domains in the coordinate sub-spaces $a_j=0$.
\begin{figure}
\begin{center}
\includegraphics[scale=1.0]{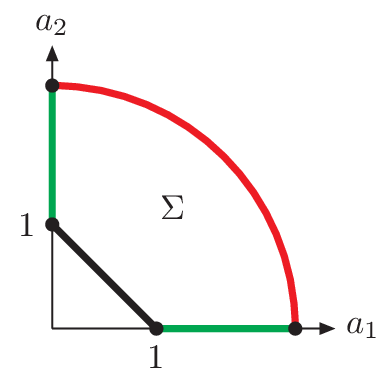}
 \hspace*{15mm}
\includegraphics[scale=1.0]{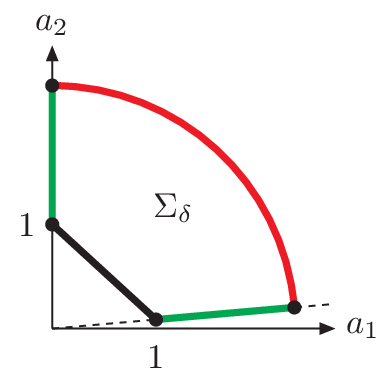}
\end{center}
\caption{
The left picture shows an example for $\ninternal=2$. The domain $\Sigma$ is bounded by the one-dimensional 
standard simplex $\Delta$ (shown in black), a hyper-surface $\tilde{\Delta}$, covering the 
solid angle $a_1, a_2 \ge 0$ (shown in red) and two one-dimensional domains along
the coordinate sub-space $a_1=0$ and $a_2=0$ (shown in green).
Integration over the domains along the coordinate sub-spaces vanishes, and therefore the integration 
over $\Delta$ gives the same result as the integration over $\tilde{\Delta}$.
If the integrand has an integrable singularity for $[a_1:a_2]=[1:0]$ we first consider the domain $\Sigma_\delta$,
where we avoid the $a_1$-axis by an angle $\delta$.
The domains shown in green are still along the radial direction.
In the end we take the limit $\delta \rightarrow 0$.
}
\label{chapter_basics:fig_stokes_theorem}
\end{figure}
An example for $\ninternal=2$ is shown in fig.~\ref{chapter_basics:fig_stokes_theorem}.
Due to eq.~(\ref{chapter_basics:interior_product}) the integration over the 
$(\ninternal-1)$-dimensional domains in the coordinate sub-spaces $a_j=0$ vanishes.
and the result follows.

Let us now relax the assumption, that $f \omega$ is regular for all points $a \in \Delta$.
We may allow integrable singularities on the boundary $\partial \Delta$.
The Feynman integral exists as an improper integral.
In the proof above we replace the domain $\Sigma$ by $\Sigma_\delta$, which avoids all singular points
by an angle $\delta$, such that the domains connecting $\Delta$ and $\tilde{\Delta}$ (these domains are shown in green
in fig.~\ref{chapter_basics:fig_stokes_theorem}) 
have always a direction along a line through the origin.
For $\Sigma_\delta$ we may use the proof as above, and in particular the integration over the domains
connecting $\Delta$ and $\tilde{\Delta}$ give a vanishing contribution.
Taking then the limit $\delta\rightarrow 0$ shows
that the two improper integrals over $\Delta$ and $\tilde{\Delta}$ are equal.

A consequence of the fact, that we may choose any hyper-surface covering the solid angle $a_j \ge 0$
is the 
\index{Cheng-Wu theorem}
{\bf Cheng-Wu theorem} \cite{Cheng:1987ga}.
To state this theorem, let $S$ be a non-empty subset of $\{1,\dots,\ninternal\}$.
\begin{theorem} 
\label{chapter_basics:theorem_1}
(Cheng-Wu theorem):
We may replace the argument 
\bq
 1-\sum\limits_{j=1}^{\ninternal} a_j 
\eq
of the delta distribution in eq.~(\ref{chapter_basics:Feynman_parameter_representation})
by
\bq
 1-\sum\limits_{j \in S} a_j.
\eq
The Feynman integral is then given by 
\bq
\label{chapter_basics:Cheng_Wu_representation}
 I
 & = &
 \frac{e^{\loopnumber \eps \Eulerconstant}\Gamma\left(\nu-\frac{\loopnumber D}{2}\right)}{\prod\limits_{j=1}^{\ninternal}\Gamma(\nu_j)}
 \int\limits_{a_j \ge 0} d^{\ninternal}a \; \delta\left(1-\sum\limits_{j \in S} a_j \right) \; 
 \left( \prod\limits_{j=1}^{\ninternal} a_j^{\nu_j-1} \right)
 \frac{\left[ {\mathcal U}\left(a\right) \right]^{\nu-\frac{\left(\loopnumber+1\right) D}{2}}}{\left[ {\mathcal F}\left(a\right) \right]^{\nu-\frac{\loopnumber D}{2}}}.
 \hspace*{12mm}
\eq
In particular one may choose $S=\{j_0\}$, which sets $a_{j_0}$ to one.
The integration is then over all other Feynman parameters from zero to infinity.
\end{theorem}
\begin{proof}
The proof uses again Stoke's theorem, where the $\ninternal$-dimensional domain $\Sigma$ is now
bounded by the standard simplex $\Delta$, the hyper-plane defined by
\bq
 \sum\limits_{j \in S} a_j & = & 1,
\eq
coordinate sub-spaces $a_j=0$ for $j \in S$ and hyper-surfaces at infinity.
The hyper-surfaces at infinity can be taken as $a_j=\Lambda$ for $j \notin S$
with $\Lambda \rightarrow \infty$.
\begin{figure}
\begin{center}
\includegraphics[scale=1.0]{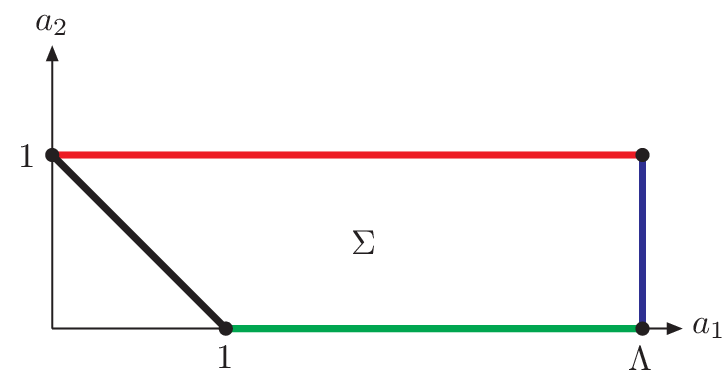}
\end{center}
\caption{
Illustration for the Cheng-Wu theorem for the case $\ninternal=2$.
The domain $\Sigma$ is bounded by the one-dimensional 
standard simplex $\Delta$ (shown in black), the hyper-plane $a_2=1$ (shown in red),
a one-dimensional domain along
the coordinate sub-space $a_2=0$ (shown in green)
and a domain at infinity (shown in blue).
}
\label{chapter_basics:fig_cheng_wu}
\end{figure}
An example for $\ninternal=2$ is shown in fig.~\ref{chapter_basics:fig_cheng_wu}.
The integration over the $(\ninternal-1)$-dimensional domains in the coordinate sub-spaces
gives a vanishing contribution for the same reasons as before.
The integration over the hyper-planes at infinity corresponds to an infinitesimal small solid angle and gives therefore
a vanishing contribution.
Hence, the claim follows.
\end{proof}
\bs
{\it \refstepcounter{exercise}
{\bf Exercise \theexercise}: 
An alternative proof of the Cheng-Wu theorem: Prove the Cheng-Wu theorem directly from the
Schwinger parameter representation by inserting
\bq
 1 & = & 
 \int\limits_{-\infty}^\infty dt \; \delta\left(t - \sum\limits_{j \in S} \alpha_j \right)
 \;\; = \;\;
 \int\limits_{0}^\infty dt \; \delta\left(t - \sum\limits_{j \in S} \alpha_j \right),
\eq
where in the last step we used again the fact that the sum of the Schwinger parameters is non-negative.
}
\es
\\
\\
Let us give an example for the application of the Cheng-Wu theorem:
We consider again the one-loop bubble integral with vanishing internal masses.
We choose
\bq
 S & = & \left\{ 2 \right\}.
\eq
Setting $x=-p^2/\mu^2$,
the Feynman integral is then given by
\bq
 I_{\nu_1\nu_2}
 & = &
 \frac{e^{\eps \Eulerconstant} \Gamma\left(\nu-\frac{D}{2}\right)}{\Gamma\left(\nu_1\right)\Gamma\left(\nu_2\right)}
 \int\limits_{a_j \ge 0} d^2a \; \delta\left(1-a_2\right) a_1^{\nu_1-1} a_2^{\nu_2-1}
 \frac{{\mathcal U}^{\nu-D}}{{\mathcal F}^{\nu-\frac{D}{2}}},
 \nonumber \\
 & & 
 {\mathcal U} \; = \; a_1+a_2,
 \;\;\;\;\;\;\;\;\;
 {\mathcal F} \; = \; a_1 a_2 x.
\eq
The integration over $a_2$ is trivial due to the delta distribution, leaving us with
\bq
 I_{\nu_1\nu_2}
 & = &
 \frac{e^{\eps \Eulerconstant} \Gamma\left(\nu-\frac{D}{2}\right)}{\Gamma\left(\nu_1\right)\Gamma\left(\nu_2\right)}
 \int\limits_0^\infty da_1 \; a_1^{\nu_1-1} 
 \frac{\left(a_1+1\right)^{\nu-D}}{\left(a_1 x\right)^{\nu-\frac{D}{2}}}
 \nonumber \\
 & = &
 \frac{e^{\eps \Eulerconstant} \Gamma\left(\nu-\frac{D}{2}\right)}{\Gamma\left(\nu_1\right)\Gamma\left(\nu_2\right)}
 x^{\frac{D}{2}-\nu}
 \int\limits_0^\infty da_1 \; \frac{a_1^{\frac{D}{2}-\nu_2-1}}{\left(1+a_1\right)^{D-\nu}}.
\eq
The integral over $a_1$ gives Euler's beta function $B(D/2-\nu_2,D/2-\nu_1)$ and we recover
the result of eq.~(\ref{chapter_basics:result_bubble}):
\bq
 I_{\nu_1\nu_2}
 & = &
 e^{\eps \Eulerconstant} 
 \frac{\Gamma\left(\nu-\frac{D}{2}\right)\Gamma\left(\frac{D}{2}-\nu_1\right)\Gamma\left(\frac{D}{2}-\nu_2\right)}{\Gamma\left(\nu_1\right)\Gamma\left(\nu_2\right)\Gamma\left(D-\nu\right)}
 x^{\frac{D}{2}-\nu}.
\eq

\subsection{The Lee-Pomeransky representation}

The Lee-Pomeransky representation is as the Schwinger parameter representation a $\ninternal$-fold integral representation.
It has the advantage that only one polynomial 
$\gls{LeePomeranskypolynomial}$
enters, given by the sum of the two graph polynomials 
${\mathcal G} = {\mathcal U} + {\mathcal F}$.
The Lee-Pomeransky representation reads \cite{Lee:2013hzt}:
\begin{tcolorbox}
{\bf Lee-Pomeransky representation}:
\bq
I  & = &
 \frac{e^{\loopnumber \eps \Eulerconstant}\Gamma\left(\frac{D}{2}\right)}{\Gamma\left(\frac{\left(\loopnumber+1\right)D}{2}-\nu\right)\prod\limits_{j=1}^{\ninternal}\Gamma(\nu_j)}
 \;
 \int\limits_{u_j \ge 0}  d^{\ninternal}u \;
 \left( \prod\limits_{j=1}^{\ninternal} u_j^{\nu_j-1} \right)
 \left[{\mathcal G}\left(u\right)\right]^{-\frac{D}{2}},
 \nonumber \\
 & &  \mbox{with} \;\;\;
 {\mathcal G}\left(u\right) = {\mathcal U}\left(u\right) + {\mathcal F}\left(u\right).
\eq
\end{tcolorbox}
In order to derive the Lee-Pomeransky representation it is simplest to work backwards:
We start with the Lee-Pomeransky representation and show that the Lee-Pomeransky representation is equivalent
to the Feynman parameter representation.
In order to do this, we use the same trick as before and we insert a one in the form of
\bq
 1 & = & \int\limits_{0}^\infty dt \; \delta\left(t - \sum\limits_{j=1}^{\ninternal} u_j \right)
\eq
into the Lee-Pomeransky representation.
We then change variables according to 
$a_j = u_j/t$ and exploit again that ${\mathcal U}$ and ${\mathcal F}$ are homogeneous of degree $l$ and $(l+1)$, respectively.
We arrive at
\bq
I  & = &
 \frac{e^{\loopnumber \eps \Eulerconstant}\Gamma\left(\frac{D}{2}\right)}{\Gamma\left(\frac{\left(\loopnumber+1\right))D}{2}-\nu\right)\prod\limits_{j=1}^{\ninternal}\Gamma(\nu_j)}
 \;
 \int\limits_{a_j \ge 0} d^{\ninternal}a \; \delta\left(1-\sum\limits_{j=1}^{\ninternal} a_j \right) \; 
 \left( \prod\limits_{j=1}^{\ninternal} a_j^{\nu_j-1} \right)
 \nonumber \\
 & &
 \times
 \int\limits_0^\infty dt \; t^{\nu-\frac{\loopnumber D}{2}-1} \; 
 \left[ {\mathcal U}\left(a\right) + {\mathcal F}\left(a\right) t \right]^{-\frac{D}{2}}.
\eq
We then substitute
$t \rightarrow t {\mathcal U}(a) / {\mathcal F}(a)$.
This gives us
\bq
 I
 & = &
 \frac{e^{\loopnumber \eps \Eulerconstant}\Gamma\left(\frac{D}{2}\right)}{\Gamma\left(\frac{\left(\loopnumber+1\right))D}{2}-\nu\right)\prod\limits_{j=1}^{\ninternal}\Gamma(\nu_j)}
 \;
 \int\limits_{a_j \ge 0} d^{\ninternal}a \; \delta\left(1-\sum\limits_{j=1}^{\ninternal} a_j \right) \; 
 \left( \prod\limits_{j=1}^{\ninternal} a_j^{\nu_j-1} \right)
 \frac{\left[{\mathcal U}\left(a\right)\right]^{\nu-\frac{\left(\loopnumber+1\right) D}{2}}}{\left[{\mathcal F}\left(a\right)\right]^{\nu-\frac{\loopnumber D}{2}}}
 \nonumber \\
 & &
 \times
 \int\limits_0^\infty dt \; t^{\nu-\frac{\loopnumber D}{2}-1} \; 
 \left( 1 + t \right)^{-\frac{D}{2}}.
\eq
We recognise the integral over $t$ as the second integral representation of Euler's beta function:
\bq
 \int\limits_0^\infty dt \; t^{\nu-\frac{\loopnumber D}{2}-1} \; \left( 1 + t \right)^{-\frac{D}{2}}
 & = &
 \frac{\Gamma\left(\nu-\frac{\loopnumber D}{2}\right) \Gamma\left(\frac{\left(\loopnumber+1\right) D}{2} -\nu\right)}{\Gamma\left(\frac{D}{2}\right)}.
\eq
With this result we recover the Feynman parameter representation.
We call the variables 
$\gls{LeePomeranskyvariable}$
\index{Lee-Pomeransky variables}
{\bf Lee-Pomeransky variables}.

Let us also consider for the Lee-Pomeransky representation an example.
Again, we choose the one-loop two-point integral with vanishing internal masses.
The Lee-Pomeransky polynomial is in this case
\bq
 {\mathcal G}
 & = &
 u_1+u_2+u_1 u_2 x,
\eq
where we set $x=-p^2/\mu^2$.
The Lee-Pomeransky representation is then given by
\bq
I_{\nu_1\nu_2}  & = &
 \frac{e^{\eps \Eulerconstant}\Gamma\left(\frac{D}{2}\right)}{\Gamma\left(D-\nu\right)\Gamma(\nu_1)\Gamma(\nu_2)}
 \;
 \int\limits_0^\infty du_1 
 \int\limits_0^\infty du_2 
 \;
 u_1^{\nu_1-1} u_2^{\nu_2-1}
 \left[u_1+u_2+u_1 u_2 x\right]^{-\frac{D}{2}}.
\eq

\subsection{The Baikov representation}
\label{chapter_basics:sect_Baikov_representation}

Up to now we always considered arbitrary Feynman graphs
with $\nexternal$ external edges, $\ninternal$ internal edges and $\loopnumber$ loops.
In particular we never assumed that there is an additional relation between
$\nexternal$, $\ninternal$ and $\loopnumber$.

The Baikov representation \cite{Baikov:1996iu} applies to a subset of Feynman graphs, 
where the number of internal edges $\ninternal$
equals the number of independent scalar products involving the loop momenta.

Let $p_1, p_2, ..., p_{\nexternal}$ denote the external momenta
and denote by
\bq
 \nexternalindependent & = &
 \dim \left\langle p_1, p_2, ..., p_{\nexternal} \right\rangle
\eq
the dimension of the span of the external momenta.
For generic external momenta and $D \ge \nexternal-1$ we have $\nexternalindependent=\nexternal-1$.
Lorentz invariants involving the loop momenta are of the form
\bq
 -k_i^2, & & 1 \; \le \; i \; \le \; \loopnumber,
 \nonumber \\
 -k_i \cdot k_j, & & 1 \; \le \; i \; < \; j \; \le \; \loopnumber,
 \nonumber \\
 -k_i \cdot p_j, & & 1 \; \le \; i \; \le \; \loopnumber, \;\;\;\;\;\; 1 \; \le \; j \; \le \; \nexternalindependent.
\eq
In total we have
\bq
 \NV & = &
 \frac{1}{2} \loopnumber \left(\loopnumber+1\right) + \nexternalindependent \loopnumber
\eq
linear independent scalar products involving the loop momenta.
We denote this number by
$\gls{numberofBaikovvariables}$
and the linear independent scalar products involving the loop momenta
by
\bq
 \sigma & = &
 \left( \sigma_1, ..., \sigma_{\NV} \right)
 \; = \; 
 \left( -k_1 \cdot k_1, -k_1 \cdot k_2, ..., -k_{\loopnumber} \cdot k_{\loopnumber}, -k_1 \cdot p_1, ..., -k_{\loopnumber} \cdot p_{\nexternalindependent} \right).
\eq
A Feynman graph $G$ has a Baikov representation if
\bq
\label{chapter_basics:baikov_condition1}
 \NV & = & \ninternal
\eq
and if we may express any internal inverse propagator as a linear combination
of the linear independent scalar products involving the loop momenta and terms independent of the loop momenta.
The second condition says, that there is an invertible $\NV \times \NV$-dimensional matrix $C$ 
and a loop-momentum independent $\NV$-dimensional vector $f$
such that
\bq
\label{chapter_basics:baikov_condition2}
 -q_s^2 + m_s^2 & = & C_{st} \sigma_t + f_s
\eq
for all $1 \le s \le \ninternal$.
At first sight it might seem that the Baikov representation 
applies due to the conditions in eq.~(\ref{chapter_basics:baikov_condition1}) and eq.~(\ref{chapter_basics:baikov_condition2}) 
only to a very special subset 
of Feynman graphs.
However, we will soon see that for a given graph $G$, which not necessarily satisfies
the conditions eq.~(\ref{chapter_basics:baikov_condition1}) and eq.~(\ref{chapter_basics:baikov_condition2}),
we can always find a graph $\tilde{G}$ which does, 
and obtain the induced Baikov representation of the graph $G$ from the Baikov representation of 
the graph $\tilde{G}$.

In order to arrive at the Baikov representation 
we change the integration variables to 
the 
\index{Baikov variables}
{\bf Baikov variables}
$\gls{Baikovvariable}$:
\bq
 z_j & = & -q_j^2+m_j^2.
\eq
The Baikov variables are nothing else than the inverse propagators.
From eq.~(\ref{chapter_basics:baikov_condition2}) we have the inverse relation
\bq
\label{chapter_basics:sigma_to_z}
 \sigma_t
 & = &
 \left( C^{-1} \right)_{t s} \left( z_s - f_s \right).
\eq
The Baikov representation of $I$ is given by
\bq
 I & = &
 e^{\loopnumber \eps \Eulerconstant}
 \left(\mu^2\right)^{\nu-\frac{\loopnumber D}{2}}
 \frac{\pi^{-\frac{1}{2}\left(\NV-\loopnumber\right)}}{\prod\limits_{j=1}^{\loopnumber} \Gamma\left(\frac{D-\nexternalindependent+1-j}{2}\right)}
 \frac{ \left[ \det G\left(p_1,...,p_{\nexternalindependent}\right)\right]^{\frac{-D+{\nexternalindependent}+1}{2}} }{ \det C }
 \nonumber \\
 & &
 \times
 \int\limits_{\mathcal C} d^{\NV}z \;
  \left[ \det G\left(k_1,...,k_{\loopnumber},p_1,...,p_{\nexternalindependent}\right)\right]^{\frac{D-\loopnumber-{\nexternalindependent}-1}{2}}
 \prod\limits_{s=1}^{\NV} z_s^{-\nu_s},
\eq
where the Gram determinants are defined by
\bq
 \det G\left(q_1,...,q_n\right) & = &
 \det\left(-q_i \cdot q_j \right)
 \;\; = \;\;
 \det\left( Q_i \cdot Q_j \right),
\eq
e.g.
\bq
 \det G\left(q_1,q_2\right)
 & = &
 \left| \begin{array}{cc}
 -q_1^2 & -q_1 \cdot q_2 \\
 -q_1 \cdot q_2 & - q_2^2 \\
 \end{array} \right|
 \; = \;
 q_1^2 q_2^2 - \left(q_1 \cdot q_2 \right)^2.
\eq
Note that exchanging two vectors leaves the Gram determinant invariant
\bq
 \det G\left(\dots,q_i,\dots,q_j,\dots\right)
 & = &
 \det G\left(\dots,q_j,\dots,q_i,\dots\right),
\eq
as this corresponds to the exchange of two rows and two columns.

The determinant 
$\det G(k_1,...,k_l,p_1,...,p_e)$ expressed in the variables $z_s$'s through eq.~(\ref{chapter_basics:sigma_to_z}) is called the 
\index{Baikov polynomial}
{\bf Baikov polynomial}:
\bq
\label{chapter_basics:def_Baikov_polynomial}
 \gls{Baikovpolynomial}\left(z_1,...,z_{\NV}\right)
 & = & \det G\left(k_1,...,k_{\loopnumber},p_1,...,p_{\nexternalindependent}\right).
\eq
The domain of integration ${\mathcal C}$ is given by 
\bq
\label{chapter_basics:Baikov_domain_1}
 {\mathcal C}
 & = &
 {\mathcal C}_1 \cap {\mathcal C}_2 \cap \dots \cap {\mathcal C}_l
\eq
with
\bq
\label{chapter_basics:Baikov_domain_2}
 {\mathcal C}_j
 & = &
 \left\{
 \frac{\det G\left(k_j,k_{j+1},...,k_l,p_1,...,p_e\right)}{\det G\left(k_{j+1},...,k_l,p_1,...,p_e\right)} \ge 0
 \right\}.
\eq
Putting everything together we arrive at the
\begin{tcolorbox}
{\bf Baikov representation} (for a graph $G$ satisfying eq.~(\ref{chapter_basics:baikov_condition1}) and eq.~(\ref{chapter_basics:baikov_condition2})):
\bq
\label{chapter_basics:Baikov_representation}
 I & = &
 \frac{e^{\loopnumber \eps \Eulerconstant} \left(\mu^2\right)^{\nu-\frac{\loopnumber D}{2}} \left[ \det G\left(p_1,...,p_{\nexternalindependent}\right)\right]^{\frac{-D+{\nexternalindependent}+1}{2}}}{\pi^{\frac{1}{2}\left(\NV-\loopnumber\right)} \left( \det C \right) \prod\limits_{j=1}^{\loopnumber} \Gamma\left(\frac{D-\nexternalindependent+1-j}{2}\right)}
 \int\limits_{\mathcal C} d^{\NV}z \;
 \left[{\mathcal B}\left(z\right)\right]^{\frac{D-\loopnumber-{\nexternalindependent}-1}{2}}
 \prod\limits_{s=1}^{\NV} z_s^{-\nu_s}.
 \;\;\;\;\;\;
\eq
\end{tcolorbox}
The Baikov representation is very useful if we would like to calculate 
cuts of Feynman integrals, i.e. the residue when one or several propagators go on-shell.

Before we go into the details of the derivation of the Baikov representation, let us first
discuss how to get around the restrictions imposed by
eq.~(\ref{chapter_basics:baikov_condition1}) and eq.~(\ref{chapter_basics:baikov_condition2}).
The most common situation is that the number of internal propagators $\ninternal$ is smaller
than the number $\NV$ of linear independent scalar products involving the loop momenta:
\bq
 \ninternal & < & \NV.
\eq
This does not occur at one-loop, but it frequently occurs beyond one-loop.
As an example let us look at the two-loop double box graph in fig.~\ref{chapter_basics:fig_doublebox}.
In this example we have seven propagators ($\ninternal=7$), 
but nine linear independent scalar products involving the loop momenta:
\bq
 -k_1^2, 
 \;
 -k_2^2, 
 \;
 -k_1 \cdot k_2,
 \;
 -k_1 \cdot p_1,
 \;
 -k_1 \cdot p_2,
 \;
 -k_1 \cdot p_3,
 \;
 -k_2 \cdot p_1,
 \;
 -k_2 \cdot p_2,
 \;
 -k_2 \cdot p_3.
\eq
We may express seven scalar products
\bq
 -k_1^2, 
 \;
 -k_2^2, 
 \;
 -k_1 \cdot k_2,
 \;
 -k_1 \cdot p_1,
 \;
 -k_1 \cdot p_2,
 \;
 -k_2 \cdot \left(p_1+p_2\right),
 \;
 -k_2 \cdot p_3
\eq
in terms of inverse propagators,
but not
\bq
\label{chapter_basic:irreducible_scalar_product}
 -k_1 \cdot p_3,
 & &
 -k_2 \cdot p_1.
\eq
The scalar products in eq.~(\ref{chapter_basic:irreducible_scalar_product})
are called
\index{irreducible scalar product}
{\bf irreducible scalar products}.
The solution is to consider the original graph $G$ as a subgraph of a larger graph $\tilde{G}$.
\begin{figure}
\begin{center}
\includegraphics[scale=1.0]{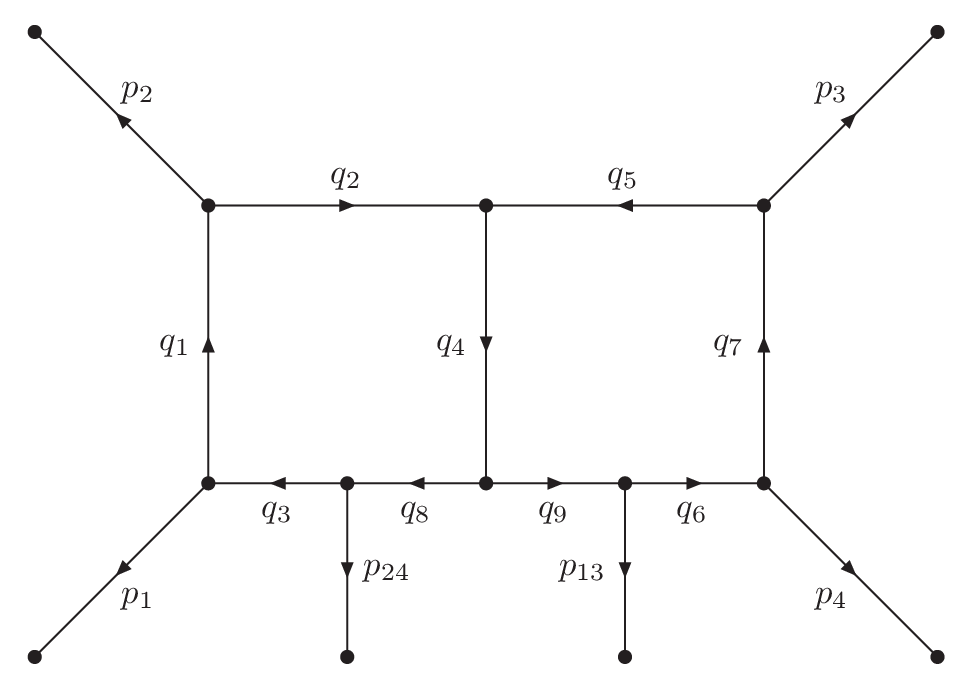}
\end{center}
\caption{
An auxiliary graph $\tilde{G}$ with nine internal propagators.
For this graph we have as many internal propagators
as linear independent scalar products involving the loop momenta.
We use the notation $p_{ij}=p_i+p_j$.
}
\label{chapter_basics:fig_doublebox_auxiliary}
\end{figure}
An example of an auxiliary graph $\tilde{G}$ for the double box graph $G$
is shown in fig.~\ref{chapter_basics:fig_doublebox_auxiliary}.
This graph has nine internal edges and six external edges.
However, the dimension of the span of the external momenta is still 
\bq
 \nexternalindependent & = &
 \dim \left\langle p_1, p_2, p_3, p_4, p_1+p_3, p_2+p_4 \right\rangle
 \; = \; 3,
\eq
and hence the number of linear independent scalar products involving the loop momenta
remains $\NV=9$.
For the graph $\tilde{G}$ we may express any 
internal inverse propagator as a linear combination
of the linear independent scalar products involving the loop momenta and terms independent of the loop momenta.
The Baikov representation of the graph $G$ is then given by the Baikov representation of the graph $\tilde{G}$
with $\nu_8=\nu_9=0$.
Setting $\nu_8=\nu_9=0$ ensures that the two additional propagators corresponding to the edges $e_8$ and
$e_9$ are absent.

It is always possible to find a graph $\tilde{G}$.
To see this, let us first introduce for a graph $G$ the  
associated 
\index{chain graph}
{\bf chain graph} $G^{\mathrm{chain}}$ as follows:
We group the internal propagators into chains.
Two propagators belong to the same chain, if their momenta differ only by a linear combination 
of the external momenta.
Obviously, each internal line can only belong to one chain. 
To each graph $G$ we associate a new graph $G^{\mathrm{chain}}$ called the chain graph by deleting all external lines
and by choosing one propagator for each chain as a representative.
\begin{figure}
\begin{center}
\includegraphics[scale=1.0]{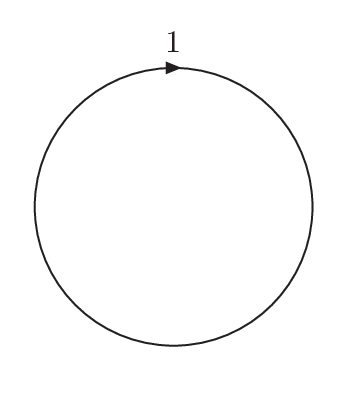}
\hspace*{20mm}
\includegraphics[scale=1.0]{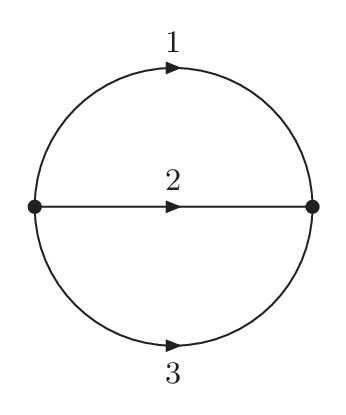}
\hspace*{20mm}
\includegraphics[scale=1.0]{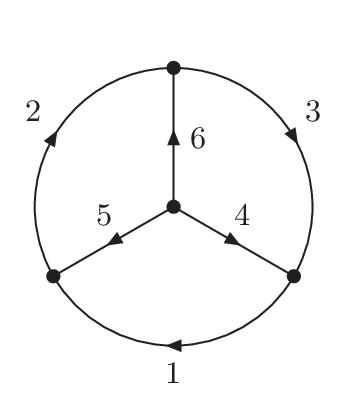}
\end{center}
\caption{
The basic chain graphs up to three loops. Up to this loop order, all other chain graphs are subgraphs of these
three graphs.
}
\label{chapter_basics:chain_graphs}
\end{figure}
Up to three loops, all chain graphs are (sub-) graphs of the three chain graphs shown in fig.~\ref{chapter_basics:chain_graphs}.
A non-trivial example is given by the three-loop ladder graph shown in the left figure of fig.~\ref{chapter_basics:example_chain_graphs}.
\begin{figure}
\begin{center}
\includegraphics[scale=1.0]{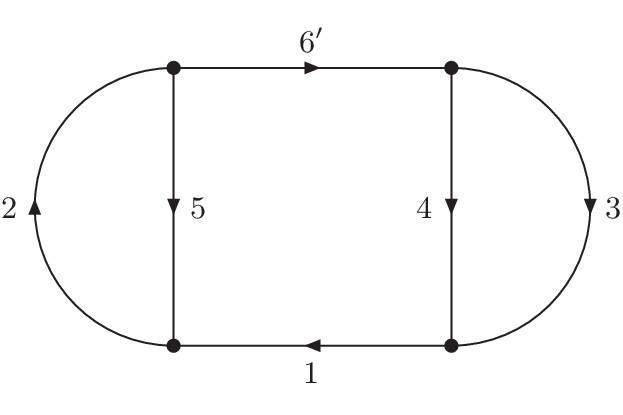}
\hspace*{20mm}
\includegraphics[scale=1.0]{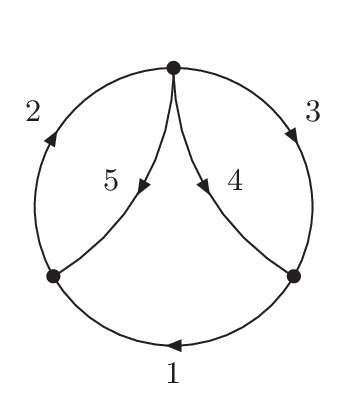}
\end{center}
\caption{
The left figure shows a three-loop graph. Propagators $1$ and $6'$ belong to the same chain. 
The underlying chain graph is the five-propagator graph shown in the right figure.
This chain graph is a subgraph of the three-loop chain graph of fig.~\ref{chapter_basics:chain_graphs}.
}
\label{chapter_basics:example_chain_graphs}
\end{figure}
In this example we note that propagators $1$ and $6'$ belong to the same chain, as the same loop momentum is flowing
through both propagators.
Hence, the associated chain graph is the one shown in the right figure of fig.~\ref{chapter_basics:example_chain_graphs}.
This chain graph is a subgraph of the three-loop Mercedes-Benz graph shown
in the right figure of fig.~\ref{chapter_basics:chain_graphs}.

With the chain graph $G^{\mathrm{chain}}$ at hand, we re-insert
external edges to the chains, 
such that all linear independent scalar products involving the loop momenta
can be expressed in terms of inverse propagators and terms independent of the loop momenta.
In practice one starts with the original propagators and then adds additional propagators 
such that the irreducible scalar products may be expressed in terms of the inverse propagators.
In a final step one adjusts the external momenta appropriately.

\begin{figure}
\begin{center}
\includegraphics[scale=1.0]{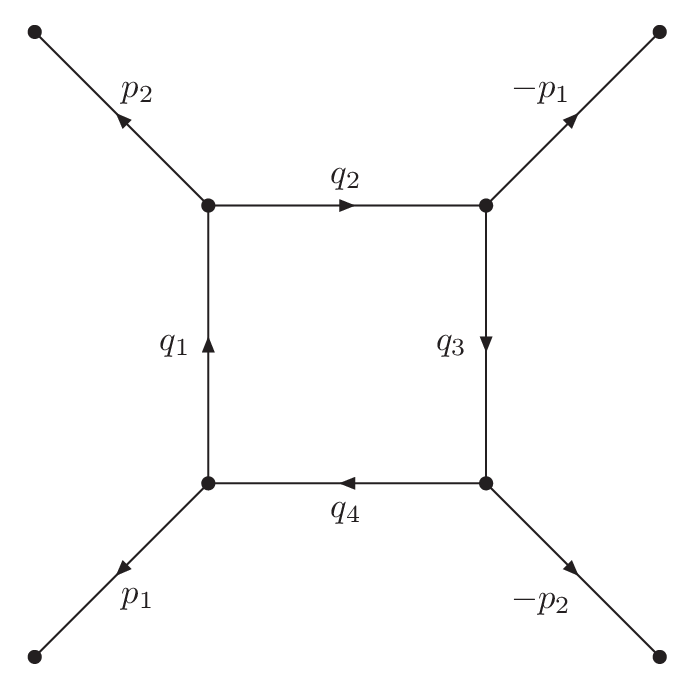}
\end{center}
\caption{
A one-loop box graph with external momenta $p_1, p_2, -p_1, -p_2$.
For this graph we have 
$\nexternalindependent = \dim \langle p_1, p_2, -p_1, -p_2 \rangle = 2$.
}
\label{chapter_basics:fig_forward_oneloopbox}
\end{figure}

Having discussed the case $\ninternal < \NV$, let us now turn to the other case:
\bq
 \ninternal & > & \NV.
\eq
In practice, this case is rather special. We discuss this case for completeness.
An example is shown in fig.~\ref{chapter_basics:fig_forward_oneloopbox}.
This graph has four internal propagators, but with
\bq
 \nexternalindependent & = & \dim \left\langle p_1, p_2, -p_1, -p_2 \right\rangle \; = \; 2
\eq
we only have three
linear independent scalar products involving the loop momenta:
\bq
 -k^2,
 \;\;\;\;
 -k \cdot p_1
 \;\;\;\;
 -k \cdot p_2.
\eq
At the same time, the four inverse propagators of the graph in fig.~\ref{chapter_basics:fig_forward_oneloopbox} are not independent.
We have
\bq
\lefteqn{
 -q_3^2 + m_3^2
 = } & &
 \\
 & &
 - \left[ -q_1^2 + m_1^2 \right]
 + \left[ -q_2^2 + m_2^2 \right]
 + \left[ -q_4^2 + m_4^2 \right]
 + 2 p_1 \cdot p_2 + m_1^2 - m_2^2 + m_3^2 - m_4^2.
 \nonumber
\eq
We may therefore consider as graph $\tilde{G}$ a one-loop triangle graph with internal edges $\{e_1,e_2,e_4\}$.
With
\bq
 z_1 \; = \; -q_1^2 + m_1^2,
 \;\;\;
 z_2 \; = \; -q_2^2 + m_2^2,
 \;\;\;
 z_4 \; = \; -q_4^2 + m_4^2
\eq
we obtain the induced Baikov representation of the graph $G$ from the Baikov representation of the
graph $\tilde{G}$ by replacing
\bq
 \frac{1}{\left(-q_3^2 + m_3^2\right)^{\nu_3}}
 & = &
 \frac{1}{\left(-z_1+z_2+z_4+f\right)^{\nu_3}},
 \;\;\;\;\;\;
 f \; = \; 2 p_1 \cdot p_2 + m_1^2 - m_2^2 + m_3^2 - m_4^2.
 \hspace*{4mm}
\eq
\begin{digression} {\bf Details on the derivation of the Baikov representation}
\\
The basic idea to derive the Baikov representation is to split the loop momentum
variables into a set, on which the integrands depends non-trivially and a set, on which
the integrand depends trivially.
For the former set we perform a change of variables to the Baikov variables, 
while for the latter set we integrate over these variables.

Let us start with the one-loop case.
We decompose the loop momentum
\bq
 k & = & k_\parallel + k_\bot
\eq
into a component living in the parallel space, defined by
\bq
 k_\parallel & \in & \langle p_1, p_2, ..., p_{\nexternal} \rangle
\eq
and a component living in the complement, called the orthogonal space.
The dependence of the integrand on $k_\bot$ is only through
\bq
 k^2 & = & k_{\parallel}^2 + k_\bot^2.
\eq
For the measure we have
\bq
 d^Dk & = & d^{\nexternalindependent}k_\parallel \; d^{D-\nexternalindependent}k_\bot.
\eq
The relation with the Euclidean measure is
\bq
 d^Dk \; = \; i d^DK,
 \;\;\;\;\;\;
 d^{\nexternalindependent}k_\parallel \; = \; i d^{\nexternalindependent}K_\parallel,
 \;\;\;\;\;\;
 d^{D-\nexternalindependent}k_\bot \; = \; d^{D-\nexternalindependent}K_\bot.
\eq
The energy component is part of the parallel space.
From the standard formulae
\bq
 d^DK \; = \; \left(K^2\right)^{\frac{D-2}{2}} dK^2 \; \frac{1}{2} d\Omega_D
 & \mbox{and} &
 \int d\Omega_D \; = \; \frac{2 \pi^{\frac{D}{2}}}{\Gamma\left(\frac{D}{2}\right)}
\eq
we obtain
\bq
 d^{D-\nexternalindependent}K_\bot
 & = &
 \frac{\pi^{\frac{D-\nexternalindependent}{2}}}{\Gamma\left(\frac{D-\nexternalindependent}{2}\right)}
 \left(K_\bot^2\right)^{\frac{D-\nexternalindependent-2}{2}} dK_\bot^2.
\eq
This allows us to perform all angular integrations in the orthogonal space.

The linear independent scalar products involving the loop momenta are
\bq
 \sigma
 & = & 
 \left( -k^2, -k \cdot p_1, \dots -k \cdot p_{\nexternalindependent} \right)
 \; = \;
 \left( K^2, K \cdot P_1, \dots K \cdot P_{\nexternalindependent} \right).
\eq
We have
\bq
 J & = &
 \frac{\partial\left(\sigma_2,...,\sigma_{\nexternalindependent+1}\right)}{\partial\left(K^0,...,K^{\nexternalindependent-1}\right)} 
 \;\;
 = \;\;
 \left(
 \begin{array}{cccc}
  P_1^0 & P_1^1 & ... & P_1^{\nexternalindependent-1} \\
  P_2^0 & P_2^1 & ... & P_2^{\nexternalindependent-1} \\
  & & ... & \\
  P_{\nexternalindependent}^0 & P_{\nexternalindependent}^1 & ... & P_{\nexternalindependent}^{\nexternalindependent-1} \\
 \end{array} \right).
\eq
We have $J J^T = (P_i \cdot P_j)$ and therefore
\bq
 \det J & = & 
 \sqrt{\det G^{\mathrm{eucl}}\left(P_1,\dots,P_{\nexternalindependent}\right)}
 \; = \; \sqrt{\det G\left(p_1,\dots,p_{\nexternalindependent}\right)}.
\eq
Here we defined the Euclidean Gram determinant by
\bq
 \det G^{\mathrm{eucl}}\left(Q_1,\dots,Q_n\right)
 & = &
 \det(Q_i \cdot Q_j)
 \; = \;
 \det(- q_i \cdot q_j)
 \; = \;
 \det G\left(q_1,\dots,q_n\right).
\eq
In addition, we may trade the integration over $dK_\bot^2$ for an integration over 
$\sigma_1=-k^2=K^2$ with Jacobian
\bq
 \frac{\partial \sigma_1}{\partial K_\bot^2}
 & = &
 \frac{\partial \left(K_\parallel^2+K_\bot^2\right)}{\partial K_\bot^2}
 \; = \;
 1.
\eq
Our final change of variables is to change from the variables
$\sigma=(\sigma_1,\dots,\sigma_{\nexternalindependent+1})$
to the Baikov variables
$(z_1,\dots,z_{\NV})$, where $\NV=\nexternalindependent+1$.
From eq.~(\ref{chapter_basics:baikov_condition2})
\bq
\label{}
 z_s & = & C_{st} \sigma_t + f_s
\eq
we obtain the Jacobian
\bq
 \frac{\partial z_s}{\partial \sigma_t} 
 & = & 
 \det C.
\eq
It remains to express $-k_\bot^2=K_\bot^2$ in terms of the Baikov variables.
This can be done by first noting that
\bq
\label{chapter_basics:k_orthogonal_squared}
 -k_\bot^2 
 \; = \;
 K_\bot^2 
 \; = \;
 \frac{\det G^{\mathrm{eucl}}\left(K,P_1,...,P_{\nexternalindependent}\right)}{\det G^{\mathrm{eucl}}\left(P_1,...,P_{\nexternalindependent}\right)}
 \; = \;
 \frac{\det G\left(k,p_1,...,p_{\nexternalindependent}\right)}{\det G\left(p_1,...,p_{\nexternalindependent}\right)},
\eq
and then replacing the scalar products involving the loop momenta 
with the Baikov variables with the help of eq.~(\ref{chapter_basics:sigma_to_z}).
\\
\\
\bs
{\it \refstepcounter{exercise}
{\bf Exercise \theexercise}: 
Prove eq.~(\ref{chapter_basics:k_orthogonal_squared}).
}
\es
\\
\\
Putting everything together we obtain for the measure
\bq
\label{chapter_basics:one_loop_Baikov_measure}
 \frac{d^Dk}{i \pi^{\frac{D}{2}}} 
 & = &
 \frac{1}{\pi^{\frac{\nexternalindependent}{2}}\left(\det C\right)\Gamma\left(\frac{D-\nexternalindependent}{2}\right)}
 \left[\det G\left(p_1,...,p_e\right) \right]^{\frac{-D+\nexternalindependent+1}{2}}
 \left[\det G\left(k,p_1,...,p_{\nexternalindependent}\right)\right]^{\frac{D-\nexternalindependent-2}{2}}
 d^{\NV}z.
 \;\;\;\;\;\;\;\;\;
\eq
The domain of integration follows from the requirement
\bq
 -k_\bot^2
 \; = \;
 K_\bot^2
 \; = \;
 \frac{\det G\left(k,p_1,...,p_{\nexternalindependent}\right)}{\det G\left(p_1,...,p_{\nexternalindependent}\right)}
 \; \ge \;
 0.
\eq
For a multi-loop integral ($\loopnumber > 1$) we may apply the argument above $\loopnumber$-times, starting with loop momentum $k_1$ and ending with loop momentum $k_{\loopnumber}$,
keeping in mind that 
the parallel space for loop momentum $k_j$ is spanned by the external momenta and all loop momenta
not yet integrated out:
\bq
 \mbox{parallel space for $k_j$}:
 & &
 \left\langle k_{j+1}, \dots, k_{\loopnumber}, p_1, p_2, \dots, p_{\nexternal} \right\rangle.
\eq
Doing so, one derives eq.~(\ref{chapter_basics:Baikov_representation}).
\end{digression}
Let us now look at a simple example for the Baikov representation.
We consider the one-loop tadpole integral
\bq
 T_\nu\left(D,x\right)
 & = &
 e^{\eps \Eulerconstant} \left(\mu^2\right)^{\nu-\frac{D}{2}}
 \int \frac{d^Dk}{i \pi^{\frac{D}{2}}} 
 \frac{1}{\left(-k^2+m^2\right)^\nu},
 \;\;\;\;\;\;
 x \; = \; \frac{m^2}{\mu^2},
\eq
which we already computed in eq.~(\ref{chapter_basics:result_tadpole}).
This integral does not depend on any external momenta,
therefore
\bq
 \nexternalindependent \; = \; 0,
 & \mbox{and} &
 \NV \; = \; 1.
\eq
There is one Baikov variable $z_1=-k^2+m^2$.
The Gram determinant is given by
\bq
 \det G\left(k\right)
 \; = \;
 -k^2
 \; = \;
 z_1 - m^2.
\eq
Thus, we obtain the Baikov polynomial as
\bq
 {\mathcal B}\left(z_1\right)
 & = &
 z_1 - m^2.
\eq
The requirement $\det G(k) \ge 0$ defines the integration region $z_1 \in [m^2,\infty[$.
We arrive at the Baikov representation of the tadpole integral:
\bq
\label{chapter_basics:tadpole_Baikov}
 T_{\nu}\left( D, x \right)
 & = &
 \frac{e^{\Eulerconstant \eps} \left(\mu^2\right)^{\nu-\frac{D}{2}}}{\Gamma\left(\frac{D}{2}\right)}
 \int\limits_{\mathcal C} dz_1
 \left[ {\mathcal B}\left(z_1\right) \right]^{\frac{D}{2}-1}
 \frac{1}{z_1^\nu},
\eq
where the contour is given by ${\mathcal C} = [m^2,\infty[$.
\\
\\
\bs
{\it \refstepcounter{exercise}
{\bf Exercise \theexercise}: 
Perform the integration in eq.~(\ref{chapter_basics:tadpole_Baikov}).
}
\es
\\
\\
We close this section by introducing a variant of the Baikov representation.
In the Baikov representation in eq.~(\ref{chapter_basics:Baikov_representation})
we treated any scalar product of a loop momentum with any other momentum on equal footing.
This is referred to as the 
\index{democratic approach, Baikov representation}
{\bf democratic approach}.
This approach allows us to express any irreducible scalar product in terms of the Baikov variables.
This is useful, when irreducible scalar products appear in the numerator of the loop momentum
representation of the Feynman integral and we would like to convert the Feynman integral
to the Baikov representation.
Up to now we didn't discuss this case. 
We will treat this case in chapter~\ref{chapter_qft}.
The price to pay for being able to express any irreducible scalar product in terms of Baikov
variables is the number of Baikov variables:
\bq
 \NV & = &
 \frac{1}{2} \loopnumber \left(\loopnumber+1\right) + \nexternalindependent \loopnumber.
\eq
However, if we are only interested in scalar Feynman integrals (without any 
irreducible scalar product in the numerator)
-- and we will see in chapter~\ref{chapter_qft} that it is enough to focus on these
integrals --
we might have an additional interest in keeping the number of integration variables
as low as possible.
There is a variant of the Baikov representation, known as the 
\index{loop-by-loop approach, Baikov representation}
{\bf loop-by-loop approach} \cite{Frellesvig:2017aai},
which achieves this.
This is best explained by an example.
\begin{figure}
\begin{center}
\includegraphics[scale=1.0]{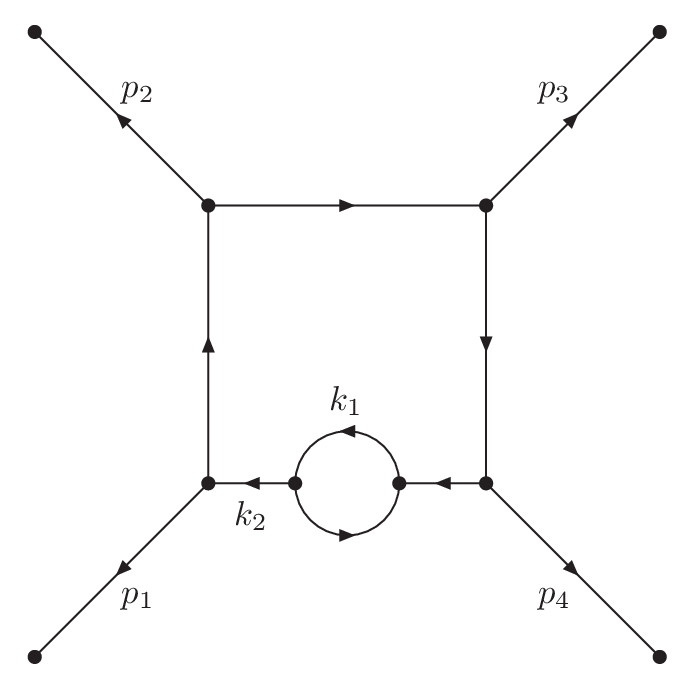}
\end{center}
\caption{
A two-loop graph, where a one-loop bubble is inserted into a one-loop box. 
}
\label{chapter_basics:fig_bubble_insertion}
\end{figure}
Consider the two-loop graph shown in fig.~\ref{chapter_basics:fig_bubble_insertion}.
With
\bq
 \nexternalindependent & = &
 \dim \left\langle p_1, p_2, p_3, p_4 \right\rangle
 \; = \; 3
\eq
we have
\bq
 \NV & = & 9
\eq
independent scalar products involving the loop momenta:
\bq
 -k_1^2, 
 \; 
 -k_1 \cdot k_2, 
 \;
 -k_2^2,
 \;
 -k_1 \cdot p_1, 
 \;
 -k_1 \cdot p_2, 
 \;
 -k_1 \cdot p_3, 
 \;
 -k_2 \cdot p_1, 
 \;
 -k_2 \cdot p_2, 
 \;
 -k_2 \cdot p_3.
\eq
However, the external momenta relative to the loop with $k_1$ are just $k_2$ (and $-k_2$)
and not the full set $k_2,p_1,p_2,p_3$.
Thus we may decompose $k_1$ in
\bq
 k_1 & = & k_{1,\parallel} + k_{1,\bot},
\eq
where $k_{1,\parallel}$ is spanned by
\bq
 \mbox{loop-by-loop parallel space for $k_1$}:
 & &
 \left\langle k_{2} \right\rangle
\eq
instead of
\bq
 \mbox{democratic parallel space for $k_1$}:
 & &
 \left\langle k_{2}, p_1, p_2, p_3 \right\rangle
\eq
We then use eq.~(\ref{chapter_basics:one_loop_Baikov_measure}) for the integration over $k_1$
with the loop-by-loop parallel space for $k_1$.
This introduces only two Baikov variables.
For the integration over $k_2$ the parallel space is spanned by
\bq
 \mbox{parallel space for $k_2$}:
 & &
 \left\langle p_1, p_2, p_3 \right\rangle,
\eq
giving four additional Baikov variables.
We therefore obtain the loop-by-loop Baikov representation with only six integration variables.
\\
\\
\bs
{\it \refstepcounter{exercise}
{\bf Exercise \theexercise}: 
Derive the Baikov representation of the graph shown in fig.~\ref{chapter_basics:fig_sunrise}
\begin{figure}
\begin{center}
\includegraphics[scale=1.0]{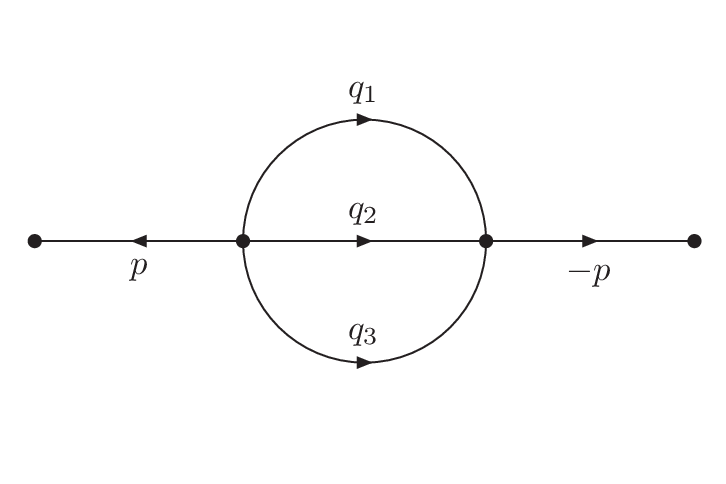}
\end{center}
\caption{
The two-loop sunrise diagram (also known as sunset diagram or London transport diagram).
}
\label{chapter_basics:fig_sunrise}
\end{figure}
within the democratic approach and within the loop-by-loop approach. 
Assume that all internal masses are non-zero and equal.
}
\es

\subsection{The Mellin-Barnes representation}
\label{chapter_basics:Mellin_Barnes}

Up to now we presented for all representations of Feynman integrals
(momentum representation, Schwinger parameter representation, Feynman parameter representation,
Lee-Pomeransky representation and Baikov representation)
a closed integral formula for the Feynman integral.
We won't do this for the Mellin-Barnes representation.
The Mellin-Barnes representation \cite{Mellin:1910,Barnes:1910} can be applied to any representation discussed so far
and allows to trivialise the integrations of the representation we start with at the
expense of introducing new integrations over Mellin-Barnes variables.
We explain the Mellin-Barnes representation by starting from the Feynman parameter representation.

The Feynman parameter representation in eq.~(\ref{chapter_basics:Feynman_parameter_representation}) depends on two
graph polynomials ${\mathcal U}$ and ${\mathcal F}$.
Assume for the moment that the two graph polynomials 
${\mathcal U}$ and ${\mathcal F}$ are absent from the Feynman parameter integral.
In this case we have an integral of the form
\bq
\label{chapter_basics:multi_beta_fct}
 \int\limits_{a_j \ge 0} d^{n}a \; \delta\left(1-\sum\limits_{j=1}^{n} a_j \right) \; 
 \left( \prod\limits_{j=1}^{n} a_j^{\nu_j-1} \right)
 & = & 
 \frac{\prod\limits_{j=1}^{n}\Gamma(\nu_j)}{\Gamma(\nu_1+...+\nu_{n})}.
\eq
The Mellin-Barnes transformation allows us to reduce the Feynman parameter integration to this case.
The Mellin-Barnes transformation reads
\bq
\label{chapter_basics:multi_mellin_barnes}
\lefteqn{
\left(A_1 + A_2 + ... + A_n \right)^{-c} 
 = 
 \frac{1}{\Gamma(c)} \frac{1}{\left(2\pi i\right)^{n-1}} 
 \int\limits_{-i\infty}^{i\infty} d\sigma_1 ... \int\limits_{-i\infty}^{i\infty} d\sigma_{n-1}
 } & & \\
 & & 
 \times 
 \Gamma(-\sigma_1) ... \Gamma(-\sigma_{n-1}) \Gamma(\sigma_1+...+\sigma_{n-1}+c)
 \; 
 A_1^{\sigma_1} ...  A_{n-1}^{\sigma_{n-1}} A_n^{-\sigma_1-...-\sigma_{n-1}-c}.
 \nonumber 
\eq
Each contour is such that the poles of $\Gamma(-\sigma)$ are to the right and the poles
of $\Gamma(\sigma+c)$ are to the left.
This transformation can be used to convert the sum of monomials of the polynomials ${\mathcal U}$ and ${\mathcal F}$ into
a product, such that all Feynman parameter integrals are of the form of eq.~(\ref{chapter_basics:multi_beta_fct}).
As this transformation converts a sum into a product it is 
the ``inverse'' of Feynman parametrisation, which converts a product into a sum.

Eq.~(\ref{chapter_basics:multi_mellin_barnes}) is derived from the theory of Mellin transformations:
Let $h(x)$ be a function which is bounded by a power law for $x\rightarrow 0$ and $x \rightarrow \infty$,
e.g.
\bq
\left| h(x) \right| \le K x^{-c_0} & & \mbox{for}\;\; x \rightarrow 0,
 \nonumber \\
\left| h(x) \right| \le K' x^{c_1} & & \mbox{for}\;\; x \rightarrow \infty.
\eq
Then the Mellin transform is defined for
$c_0 < \mbox{Re}\; \sigma < c_1$
by
\bq
h_{\cal M}(\sigma) & = &
 \int\limits_{0}^\infty dx \; h(x) \; x^{\sigma-1}.
\eq
The inverse Mellin transform is given by
\bq
\label{chapter_basics:inversemellin}
h(x) & = & \frac{1}{2\pi i} \int\limits_{\gamma-i\infty}^{\gamma+i\infty}
 d\sigma \; h_{\cal M}(\sigma) \; x^{-\sigma}.
\eq
The integration contour is parallel to the imaginary axis and $c_0 < \mbox{Re}\; \gamma < c_1$.
As an example for the Mellin transform we consider the function 
\bq
h(x) & = & \frac{x^c}{(1+x)^c}
\eq
with Mellin transform $h_{\cal M}(\sigma)=\Gamma(-\sigma) \Gamma(\sigma+c) / \Gamma(c)$.
For $\mbox{Re}(-c) < \mbox{Re} \; \gamma < 0$ we have
\bq
\label{chapter_basics:baseMellin}
\frac{x^c}{(1+x)^c}
 & = & 
\frac{1}{2\pi i} \int\limits_{\gamma-i\infty}^{\gamma+i\infty}
 d\sigma \; \frac{\Gamma(-\sigma) \Gamma(\sigma+c)}{\Gamma(c)} \; x^{-\sigma}.
\eq
From eq. (\ref{chapter_basics:baseMellin}) one obtains with $x=B/A$ the Mellin-Barnes formula
\bq
\label{chapter_basics:simple_mellin_barnes}
\left(A+B\right)^{-c}
 & = & 
\frac{1}{2\pi i} \int\limits_{\gamma-i\infty}^{\gamma+i\infty}
 d\sigma \; \frac{\Gamma(-\sigma) \Gamma(\sigma+c)}{\Gamma(c)} \; A^\sigma B^{-\sigma-c}.
\eq
Eq.~(\ref{chapter_basics:multi_mellin_barnes}) is then obtained by repeated use of eq.~(\ref{chapter_basics:simple_mellin_barnes}).

With the help of eq.~(\ref{chapter_basics:multi_beta_fct}) and eq.~(\ref{chapter_basics:multi_mellin_barnes})
we may exchange a Feynman parameter integral against a multiple contour integral.
A typical single contour integral is of the form
\bq
\label{chapter_basics:MellinBarnesInt}
I
 & = & 
\frac{1}{2\pi i} \int\limits_{\gamma-i\infty}^{\gamma+i\infty}
 d\sigma \; 
 \frac{\Gamma(\sigma+a_1) ... \Gamma(\sigma+a_m)}
      {\Gamma(\sigma+c_1) ... \Gamma(\sigma+c_p)}
 \frac{\Gamma(-\sigma+b_1) ... \Gamma(-\sigma+b_n)}
      {\Gamma(-\sigma+d_1) ... \Gamma(-\sigma+d_q)} 
 \; x^{-\sigma}.
\eq
If $\;\mbox{max}\left( \mbox{Re}(-a_1), ..., \mbox{Re}(-a_m) \right) < \mbox{min}\left( \mbox{Re}(b_1), ..., \mbox{Re}(b_n) \right)$ the contour can be chosen
as a straight line parallel to the imaginary axis with
\bq
\mbox{max}\left( \mbox{Re}(-a_1), ..., \mbox{Re}(-a_m) \right) 
 \;\;\; < \;\;\; \mbox{Re} \; \gamma \;\;\; < \;\;\;
\mbox{min}\left( \mbox{Re}(b_1), ..., \mbox{Re}(b_n) \right),
\eq
otherwise the contour is indented, such that the residues of
$\Gamma(\sigma+a_1)$, ..., $\Gamma(\sigma+a_m)$ are to the left of the contour,
whereas the residues of 
$\Gamma(-\sigma+b_1)$,  ..., $\Gamma(-\sigma+b_n)$ are to the right of the contour,
\begin{figure}
\begin{center}
\includegraphics[scale=1.0]{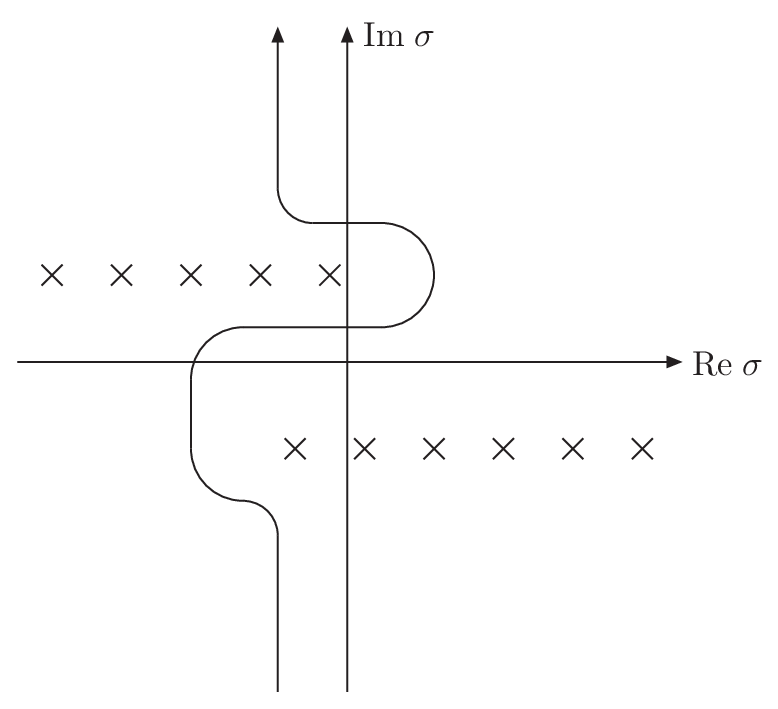}
\end{center}
\caption{
The integration contour for Mellin-Barnes integrals:
Residues of
$\Gamma(\sigma+a_1)$, ..., $\Gamma(\sigma+a_m)$ are to the left of the contour,
residues of 
$\Gamma(-\sigma+b_1)$,  ..., $\Gamma(-\sigma+b_n)$ are to the right of the contour.
}
\label{chapter_basics:fig_Mellin_Barnes}
\end{figure}
as shown in fig.~\ref{chapter_basics:fig_Mellin_Barnes}.

The integral eq. (\ref{chapter_basics:MellinBarnesInt}) is most conveniently evaluated with 
the help of the residue theorem by closing the contour to the left or to the right.
To sum up all residues which lie inside the contour
it is useful to know the residues of the gamma function:
\bq
 \mathrm{res} \; \left( \Gamma(\sigma+a), \sigma=-a-n \right) = \frac{(-1)^n}{n!}, 
 & &
 \mathrm{res} \; \left( \Gamma(-\sigma+a), \sigma=a+n \right) = -\frac{(-1)^n}{n!}. 
\eq
Remember that in the case where we close the contour to the right, there is an extra minus sign from 
the negative winding number.

In general there are multiple contour integrals, and as a consequence one obtains multiple sums.
In particular simple cases the contour integrals can be performed in closed form with
the help of two lemmas of Barnes.
\index{Barnes' lemmata}
{\bf Barnes' first lemma} states that
\bq
\frac{1}{2\pi i} \int\limits_{-i\infty}^{i\infty} d\sigma \;
\Gamma(a+\sigma) \Gamma(b+\sigma) \Gamma(c-\sigma) \Gamma(d-\sigma) 
 =  
\frac{\Gamma(a+c) \Gamma(a+d) \Gamma(b+c) \Gamma(b+d)}{\Gamma(a+b+c+d)},
\eq
if none of the poles of $\Gamma(a+\sigma) \Gamma(b+\sigma)$ coincides with the
ones from $\Gamma(c-\sigma) \Gamma(d-\sigma)$.
{\bf Barnes' second lemma} reads
\bq
\lefteqn{
\frac{1}{2\pi i} \int\limits_{-i\infty}^{i\infty} d\sigma \;
\frac{\Gamma(a+\sigma) \Gamma(b+\sigma) \Gamma(c+\sigma) \Gamma(d-\sigma) \Gamma(e-\sigma)}
{\Gamma(a+b+c+d+e+\sigma)} } & & \nonumber \\
& = & 
\frac{\Gamma(a+d) \Gamma(b+d) \Gamma(c+d) 
      \Gamma(a+e) \Gamma(b+e) \Gamma(c+e)}
{\Gamma(a+b+d+e) \Gamma(a+c+d+e) \Gamma(b+c+d+e)}.
\eq

%% file: graph_polynomials.tex
\newpage
\chapter{Graph polynomials}
\label{chapter_graph_polynomials}

The Schwinger parameter representation and the
Feynman parameter representation involve 
two graph polynomials ${\mathcal U}$ and ${\mathcal F}$,
the Lee-Pomeransky representation involves the sum of these
two polynomials ${\mathcal G} = {\mathcal U} + {\mathcal F}$.
These polynomials encode the essential information of the integrands.
In this chapter we study these polynomials in more detail.
Previously, we defined the polynomials ${\mathcal U}$ and ${\mathcal F}$
in eq.~(\ref{chapter_basics:eq_poly_calc_2}).
In this chapter we will also learn alternative methods to compute
these polynomials.


\section{Spanning trees and spanning forests}
\label{chapter_graph_polynomials:sect_spanning_trees}

In this section we delve deeper into concepts of graph theory.
We define spanning trees and spanning forests. These concepts lead to a second method for
the computation of the graph polynomials.
We consider a connected graph $G$ with $\nedges$ edges and $\nvertices$ vertices.
We label the edges by $\{e_1,\dots,e_{\nedges}\}$ and the vertices by $\{v_1,\dots,v_{\nvertices}\}$.
Vertices of valency $1$ are called external vertices, all other vertices are internal vertices.
There is exactly one edge connected to an external vertex. Such an edge is called an external edge.
Edges, which are not external edges are called internal edges.
We label the edges of the graph $G$ such that the first $\ninternal$ edges are the internal edges
and the remaining $\nexternal$ edges are the external edges:
\bq
 \mbox{internal edges} & : & \{ e_1, e_2, \dots, e_{\ninternal} \},
 \nonumber  \\
 \mbox{external edges} & : & \{ e_{\ninternal+1}, e_{\ninternal+2}, \dots, e_{\ninternal+\nexternal} \}.
\eq
We have $\nedges=\ninternal+\nexternal$.
In a similar way 
we label the vertices such that the first $\ninternalvertices$ vertices are the internal vertices
and the remaining $\nexternal$ vertices are the external vertices 
(there are exactly $\nexternal$ external vertices):
\bq
 \mbox{internal vertices} & : & \{ v_1, v_2, \dots, v_{\ninternalvertices} \},
 \nonumber  \\
 \mbox{external vertices} & : & \{ v_{\ninternalvertices+1}, v_{\ninternalvertices+2}, \dots, v_{\ninternalvertices+\nexternal} \}.
\eq
The distinction between internal edges and external edges (and internal vertices and external vertices)
is necessary for the application towards Feynman integrals: The Feynman rules for the internal and external
objects differ (and the variables of the graph polynomials are in one-to-one correspondence with the internal
edges of the graph).
However, pure mathematicians might prefer to work just with a graph (consisting of edges and vertices), 
without any particular distinction between internal and external vertices and edges.
In order to reconcile the two approaches, let us 
introduce the
\index{internal graph} 
{\bf internal graph} $G_{\mathrm{int}}$ associated to $G$
as the sub-graph of $G$ obtained by deleting the external vertices and the external edges from $G$.
\begin{figure}
\begin{center}
\includegraphics[scale=1.0]{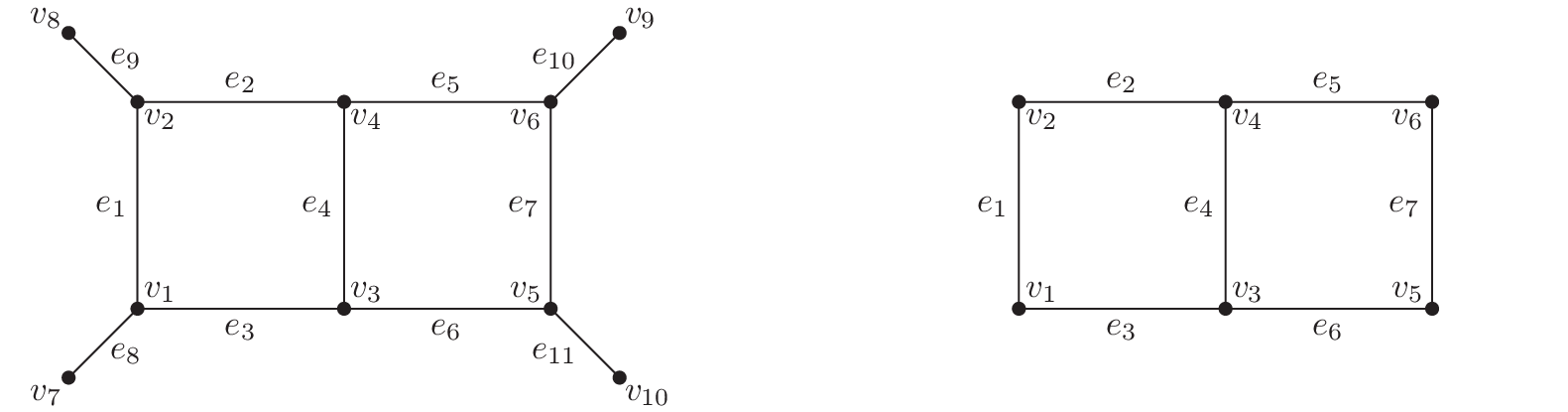}
\end{center}
\caption{\label{chapter_graph_polynomials:fig_internal_graph} 
An example for a Feynman graph $G$ (left) and the associated internal graph $G_{\mathrm{int}}$ (right).
The latter is obtained by deleting all external vertices and all external edges.
}
\end{figure}    
The internal graph $G_{\mathrm{int}}$ has $\ninternal$ edges $\{ e_1, e_2, \dots, e_{\ninternal} \}$
and $\ninternalvertices$ vertices $\{ v_1, v_2, \dots, v_{\ninternalvertices} \}$.
An example is shown in fig.~\ref{chapter_graph_polynomials:fig_internal_graph}.

As before we denote by $\loopnumber$ the first Betti number of the graph (or in physics jargon: the number of loops).
We have the relation
\bq
 \loopnumber & = & \nedges - \nvertices + 1
 \; = \;
 \ninternal - \ninternalvertices + 1.
\eq
If we would allow for disconnected graphs, the corresponding formula for the first Betti number would be
$\nedges-\nvertices+k$, where $k$ is the number of connected components.
\begin{definition} (spanning tree):
\label{chapter_graph_polynomials:definition_spanning_tree}
A 
\index{spanning tree}
{\bf spanning tree} 
for the graph $G$ is a sub-graph $T$ of $G$
satisfying the following requirements:
\begin{itemize}
\item $T$ contains all the vertices of $G$,
\item the first Betti number of $T$ is zero,
\item $T$ is connected.
\end{itemize}
\end{definition}
If $T$ is a spanning tree for $G$, then it can be obtained from $G$ by deleting $\loopnumber$ edges.
In general a given graph $G$ has several spanning trees. 
We will later obtain a formula which counts the number
of spanning trees for a given graph $G$.
\begin{definition} (spanning forest):
\label{chapter_graph_polynomials:definition_spanning_forest}
A 
\index{spanning forest}
{\bf spanning forest} 
for the graph $G$ is a sub-graph $F$ of $G$ satisfying just the first two requirements:
\begin{itemize}
\item $F$ contains all the vertices of $G$,
\item the first Betti number of $F$ is zero.
\end{itemize}
\end{definition}
It is not required that a spanning forest is connected.
If $F$ has $k$ connected components, we say that $F$ is a $k$-forest.
A spanning tree is a spanning $1$-forest.
If $F$ is a spanning $k$-forest for $G$, then it can be obtained from $G$ by deleting $\loopnumber+k-1$ edges.

For the application towards Feynman graphs we need a refinement of this definition:
\begin{definition} (spanning forest with respect to an edge set):
\label{chapter_graph_polynomials:definition_spanning_forest_edge_set}
A 
spanning forest
for the graph $G$ with respect to an edge set $E$ 
is a sub-graph $F$ of $G$ satisfying:
\begin{itemize}
\item $F$ contains all the vertices of $G$,
\item the first Betti number of $F$ is zero.
\item $F$ contains all edges $\{e_1,\dots,e_{\nedges}\}\backslash E$.
\end{itemize}
\end{definition}
The third requirement states that we may only delete edges from the set $E$, but not from the
complement $\{e_1,\dots,e_{\nedges}\}\backslash E$.
For $E=\{e_1,\dots,e_{\nedges}\}$ the two definitions agree:
A spanning forest with respect to the edge set $\{e_1,\dots,e_{\nedges}\}$ is a spanning forest in the sense
of definition~\ref{chapter_graph_polynomials:definition_spanning_forest}.
The typical application towards Feynman graphs is the case, where $E$ is the set of internal edges
\bq
 E & = & \left\{ e_1, e_2, \dots, e_{\ninternal} \right\}.
\eq
In this case the third condition ensures that
no external edges are deleted.
As this is the most common case,
we will from now on always assume that a $k$-forest of a Feynman graph $G$
is a $k$-forest of $G$ with respect to the internal edges, unless stated otherwise.

\begin{figure}
\begin{center}
\includegraphics[scale=1.0]{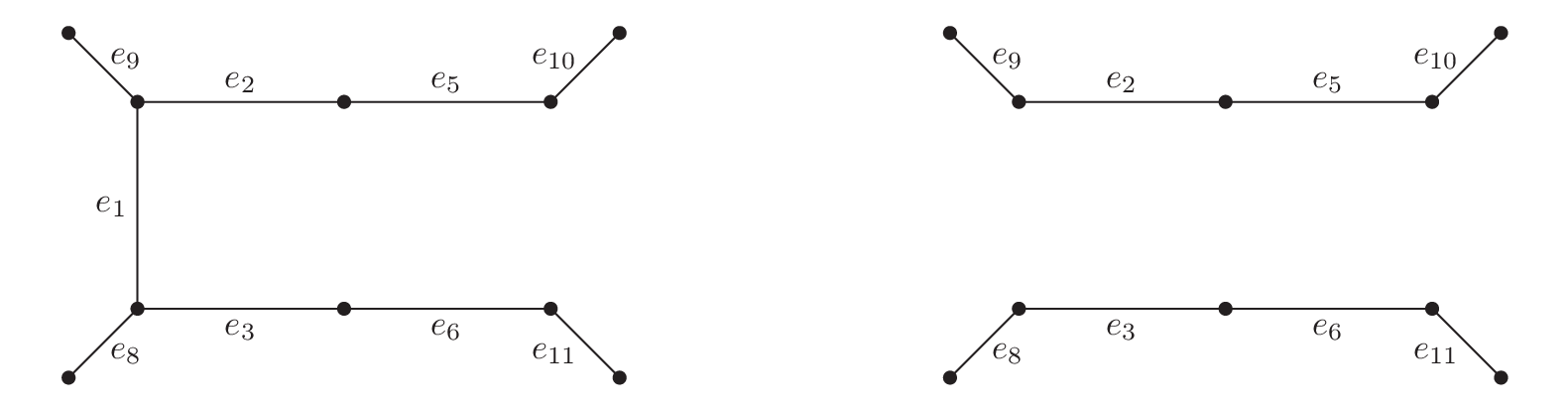}
\end{center}
\caption{\label{chapter_graph_polynomials:fig2} The left picture shows a spanning tree for the graph of fig.~\ref{chapter_basics:fig_doublebox},
the right picture shows a spanning $2$-forest for the same graph.
The spanning tree is obtained by deleting edges $4$ and $7$, the spanning $2$-forest is obtained
by deleting edges $1$, $4$ and $7$.
}
\end{figure}

Fig.~\ref{chapter_graph_polynomials:fig2} shows an example 
for a spanning tree and a spanning $2$-forest for the graph of fig.~\ref{chapter_basics:fig_doublebox}.

We denote by ${\mathcal T}$ the set of spanning forests of $G$ (by default with respect to the internal edges,
unless stated otherwise)
and by ${\mathcal T}_k$ the set of spanning $k$-forests of $G$ (again by default 
with respect to the internal edges unless stated otherwise).
Obviously, we can write ${\mathcal T}$ as the disjoint union
\bq
 {\mathcal T} & = & \bigcup\limits_{k=1}^{\ninternalvertices} {\mathcal T}_k.
\eq
${\mathcal T}_1$ is the set of spanning trees.
For an element of ${\mathcal T}_k$ we write
\bq
 \left( T_1, T_2, \dots, T_k \right) & \in & {\mathcal T}_k.
\eq
The $T_i$ are the connected components of the $k$-forest. They are necessarily trees.
We denote by $P_{T_i}$ the set of external momenta attached to $T_i$.
For the example of the $2$-forest in the right picture of fig.~\ref{chapter_graph_polynomials:fig2} we have
(compare with the momentum labelling in fig.~\ref{chapter_basics:fig_doublebox})
\bq
 P_{T_1} = \{ p_2, p_3 \},
 & &
 P_{T_2} = \{ p_1, p_4 \}.
\eq
The spanning trees and the spanning $2$-forests (with respect to the internal edges)
of a graph $G$ are closely related to the 
graph polynomials ${\mathcal U}$ and ${\mathcal F}$ of the graph:
\begin{tcolorbox}
{\bf Graph polynomials from spanning trees and the spanning $2$-forests}:
\bq
\label{chapter_graph_polynomials:eq_poly_calc_3}	
 {\mathcal U}\left(a\right)
 & = & 
 \sum\limits_{T\in {\mathcal T}_1} \;
     \prod\limits_{e_i\notin T} a_i,
 \\
 {\mathcal F}\left(a\right)
 & = & 
 \sum\limits_{(T_1,T_2)\in {\mathcal T}_2} \;
     \left( \prod\limits_{e_i\notin (T_1,T_2)} a_i \right) 
     \left( \sum\limits_{p_j\in P_{T_1}} \sum\limits_{p_k\in P_{T_2}} \frac{p_j \cdot p_k}{\mu^2} \right)
 \; + \; {\mathcal U}\left(a\right) \sum\limits_{i=1}^{\ninternal} a_i \frac{m_i^2}{\mu^2}.
 \nonumber
\eq
The sum is over all spanning trees for ${\mathcal U}$, and over all spanning $2$-forests
(with respect to the internal edges)
in the first term of the formula for ${\mathcal F}$.
\end{tcolorbox}
Eq.~(\ref{chapter_graph_polynomials:eq_poly_calc_3}) provides a second method for the computation of the graph polynomials
${\mathcal U}$ and ${\mathcal F}$.
Let us first look at the formula for ${\mathcal U}$. For each spanning tree $T$ we take the edges $e_i$,
which have been removed from the graph $G$ to obtain $T$. The product of the
corresponding parameters $a_i$ gives a monomial.
The first formula says, that ${\mathcal U}$ is the sum of all the monomials obtained from all spanning trees.
The formula for ${\mathcal F}$ has two parts: One part is related to the external momenta and the other part
involves the masses.
The latter is rather simple and we write
\bq
\label{chapter_graph_polynomials:def_F_0}
 {\mathcal F}\left(a\right) & = & 
 {\mathcal F}_0\left(a\right) + {\mathcal U}\left(a\right) \sum\limits_{i=1}^{\ninternal} a_i \frac{m_i^2}{\mu^2}.
\eq
We focus on the polynomial ${\mathcal F}_0$. 
Here the $2$-forests with respect to the internal edges are relevant. For each $2$-forest $(T_1,T_2)$
we consider again the edges $e_i$,
which have been removed from the graph $G$ to obtain $(T_1,T_2)$. 
The product of the corresponding parameters $a_i$ defines again a monomial,
which in addition is multiplied by a quantity which depends on the external momenta.
We define the square of the sum of momenta through the deleted lines of $(T_1,T_2)$ by
\bq
 s_{(T_1,T_2)} & = & \left( \sum\limits_{e_j\notin (T_1,T_2)} q_j \right)^2.
\eq
Here we assumed for simplicity that the orientation of the momenta of the deleted internal lines are chosen
such that all deleted momenta flow from $T_1$ to $T_2$ (or alternatively that all deleted momenta flow from $T_2$
to $T_1$, but not mixed).
From momentum conservation it follows that
the sum of the momenta flowing through the deleted edges out of $T_1$
is equal to the negative of the sum of the external momenta of $T_1$.
With the same reasoning 
the sum of the momenta flowing through the deleted edges into $T_2$
is equal to the sum of the external momenta of $T_2$.
Therefore we can equally write
\bq
 s_{(T_1,T_2)} & = &
   - \left( \sum\limits_{p_i\in P_{T_1}} p_i \right) \cdot \left( \sum\limits_{p_j\in P_{T_2}} p_j \right)
\eq
and ${\mathcal F}_0$ is given by
\bq
 {\mathcal F}_0\left(a\right)
 & = & 
 \sum\limits_{(T_1,T_2)\in {\mathcal T}_2} \;
     \left( \prod\limits_{e_i\notin (T_1,T_2)} a_i \right)
     \left( \frac{-s_{(T_1,T_2)}}{\mu^2} \right).
\eq
Since we have to remove $\loopnumber$ edges from $G$ to obtain a spanning tree and $(\loopnumber+1)$ edges to obtain
a spanning $2$-forest, it follows that ${\mathcal U}(a)$ and ${\mathcal F}(a)$ are homogeneous in the 
parameters $a$ of degree $\loopnumber$ and $(\loopnumber+1)$, respectively.
From the fact, that an internal edge can be removed at most once, it follows that ${\mathcal U}$ and
${\mathcal F}_0$ are linear in each parameter $a_j$.
Finally it is obvious from eq.~(\ref{chapter_graph_polynomials:eq_poly_calc_3}) that each monomial in the expanded form of ${\mathcal U}$
has coefficient $+1$.

\begin{figure}
\begin{center}
\includegraphics[scale=1.0]{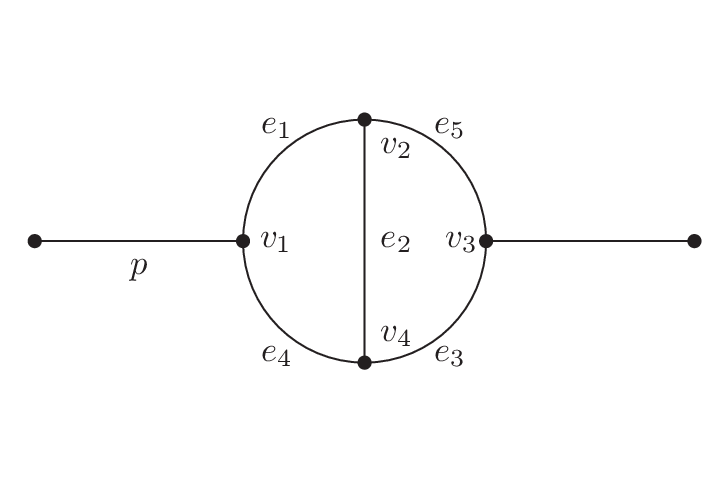}
\end{center}
\caption{\label{chapter_graph_polynomials:fig3} A two-loop two-point graph.}
\end{figure}

\begin{figure}
\begin{center}
\includegraphics[scale=1.0]{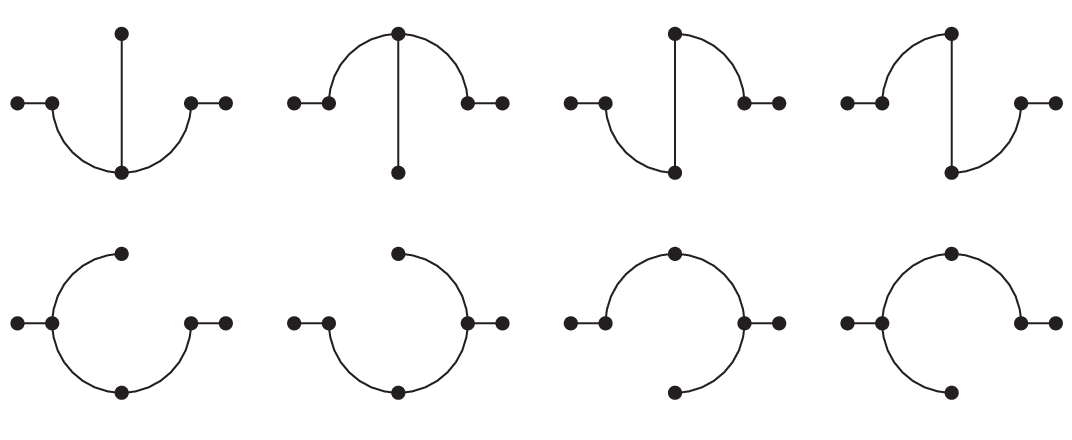}
\end{center}
\caption{\label{chapter_graph_polynomials:fig4} The set of spanning trees for the two-loop two-point graph of fig.~\ref{chapter_graph_polynomials:fig3}.}
\end{figure}

Let us look at an example. Fig.~\ref{chapter_graph_polynomials:fig3} shows the graph of a two-loop two-point integral.
We take again all internal masses to be zero.
The set of all spanning trees for this graph is shown in fig.~\ref{chapter_graph_polynomials:fig4}.
There are eight spanning trees.
\begin{figure}
\begin{center}
\includegraphics[scale=1.0]{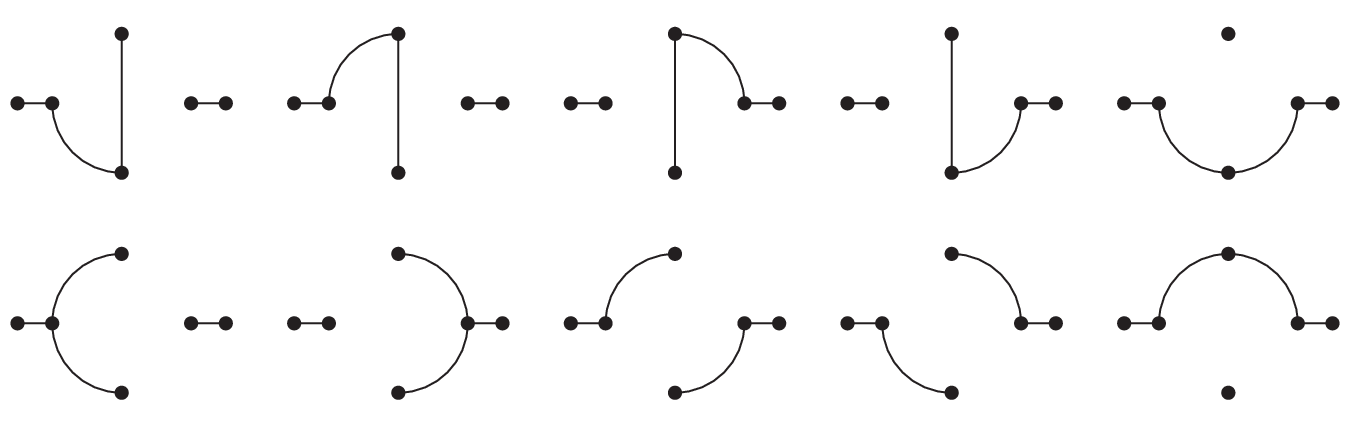}
\end{center}
\caption{\label{chapter_graph_polynomials:fig5} The set of spanning $2$-forests for the two-loop two-point graph of fig.~\ref{chapter_graph_polynomials:fig3}.}
\end{figure}    
Fig.~\ref{chapter_graph_polynomials:fig5} shows the set of spanning $2$-forests with respect to the internal edges for this graph. There are ten spanning $2$-forests.
The last example in each row of fig.~\ref{chapter_graph_polynomials:fig5} does not contribute to the graph polynomial ${\mathcal F}$, since
the momentum sum flowing through all deleted edges is zero. Therefore we have in this case $s_{(T_1,T_2)}=0$.
In all other cases we have $s_{(T_1,T_2)}=p^2$. We arrive therefore at the graph polynomials
\bq
\label{chapter_graph_polynomials:graph_polynomials_ex_tbubble}
 {\mathcal U} & = & (a_1+a_4)(a_3+a_5) + (a_1+a_3+a_4+a_5)a_2,
 \nonumber \\
 {\mathcal F} & = & 
      \left[ (a_1+a_5)(a_3+a_4)a_2
      +a_1a_4(a_3+a_5)
      +a_3a_5(a_1+a_4) \right]
      \left( \frac{-p^2}{\mu^2} \right).
\eq
\\
\\
\bs
{\it \refstepcounter{exercise}
{\bf Exercise \theexercise}: 
Re-compute the first graph polynomial ${\mathcal U}$
for the graph shown in fig.~\ref{chapter_basics:fig_nonplanar_vertex} 
from the set of spanning trees.
}
\es


\section{The matrix-tree theorem}
\label{chapter_graph_polynomials:sect_matrix_tree_theorem}

In this section we introduce the Laplacian of a graph. 
The Laplacian is a matrix constructed from the topology of the graph.
The determinant of a minor of this matrix where the $i$-th row and column have been deleted
gives us the Kirchhoff polynomial of the graph, which in turn
upon a simple substitution leads to the first Symanzik polynomial.
We then show how this construction generalises for the second Symanzik polynomial.
This provides a third method for the computation of the two graph polynomials.
This method is very well suited for computer algebra systems, as it involves just the computation
of a determinant of a matrix. The matrix is easily constructed from the data defining the graph.

We begin with the 
\index{Kirchhoff polynomial}
{\bf Kirchhoff polynomial} of a graph $G$.
We associate a parameter $a_j$ to any edge $e_j$ (internal or external). 
The Kirchhoff polynomial is defined by
\bq
\label{chapter_graph_polynomials:def_Kirchhoff_polynomial}
 \gls{Kirchhoffpolynomial}\left(a_1,\dots,a_{\nedges}\right)
 & = & 
 \sum\limits_{T\in {\mathcal T}_1} \;
     \prod\limits_{e_j \in T} a_j.
\eq
In physics we associate parameters $a_j$ only to internal edges.
We set
\bq
 {\mathcal K}_{\mathrm{int}}\left(G\right)
 & = & 
 {\mathcal K}\left(G_{\mathrm{int}}\right)
 \; = \; 
 \sum\limits_{T\in {\mathcal T}_1} \;
     \prod\limits_{e_j \in \left(T \cap E\right)} a_j,
\eq
where $E$ is the set of internal edges.
The definition is very similar to the expression for the first Symanzik polynomial in eq.~(\ref{chapter_graph_polynomials:eq_poly_calc_3}).
Again we have a sum over all spanning trees, but this time we take for each spanning tree the monomial of the Feynman 
parameters corresponding to the edges which have not been removed.
The Kirchhoff polynomial is therefore homogeneous of degree $(\ninternal-\loopnumber)$ 
in the parameters $a$.
There is a simple relation between the Kirchhoff polynomial ${\mathcal K}_{\mathrm{int}}$ 
and the first Symanzik polynomial ${\mathcal U}$:
\bq
\label{chapter_graph_polynomials:convert_U_K}
 {\mathcal U}(a_1,\dots,a_{\ninternal}) 
 & = &
 a_1 \dots a_{\ninternal} \; {\mathcal K}_{\mathrm{int}}\left(\frac{1}{a_1},\dots,\frac{1}{a_{\ninternal}}\right),
 \nonumber \\
 {\mathcal K}_{\mathrm{int}}(a_1,\dots,a_{\ninternal}) 
 & = &
 a_1 \dots a_{\ninternal} \; {\mathcal U}\left(\frac{1}{a_1},\dots,\frac{1}{a_{\ninternal}}\right).
\eq
These equations are immediately evident from the fact that ${\mathcal U}$ and ${\mathcal K}_{\mathrm{int}}$ are
homogeneous polynomials which are linear in each variable together with
the fact that a monomial corresponding to a 
specific spanning tree in one polynomial contains exactly those Feynman parameters which are not 
in the corresponding monomial in the other polynomial.

We now define the 
\index{Laplacian of a graph}
{\bf Laplacian of a graph $G$}.
\begin{definition} (Laplacian of a graph):
\label{chapter_graph_polynomials:definition_Laplacian}
Let $G$ be a graph with $\nedges$ edges and $\nvertices$ vertices.
To each edge $e_j$ one associates a parameter $a_j$.
The Laplacian of the graph $G$ is a 
$\nvertices \times \nvertices$-matrix $L$, 
whose entries are given by 
\bq
 L_{ij} & = & \left\{ \begin{array}{rl} 
                      \sum a_k & \mbox{if $i =j$ and edge $e_k$ is attached to $v_i$ and is not a self-loop,} \\
                      - \sum a_k & \mbox{if $i \neq j$ and edge $e_k$ connects $v_i$ and $v_j$.}\\
                      \end{array} \right.
\eq
The graph may contain multiple edges and self-loops.
\end{definition}
We speak 
\begin{figure}
\begin{center}
\includegraphics[scale=1.0]{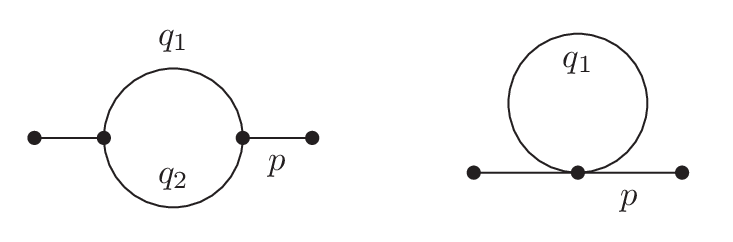}
\end{center}
\caption{\label{chapter_graph_polynomials:fig6} The left picture shows a graph with a double edge, the right picture shows a graph with a self-loop.}
\end{figure}    
of a 
\index{multiple edge}
{\bf multiple edge}, if two vertices are connected by more than one edge.
We speak of a 
\index{self-loop}
{\bf self-loop} if an edge starts and ends at the same vertex. In the physics literature a self-loop is
known as a tadpole. Fig.~\ref{chapter_graph_polynomials:fig6} shows a simple example for a double edge and a self-loop.
If the vertices $v_i$ and $v_j$ are connected by two edges $e_{k_1}$ and $e_{k_2}$, then the Laplacian depends only on the
sum $a_{k_1}+a_{k_2}$.
If an edge $e_k$ is a self-loop attached to a vertex $v_i$, then it does not contribute to the Laplacian.

For the application towards Feynman graphs we need a refinement of this definition.
(This can already be anticipated from the fact, that we associate parameters $a_j$ to all internal edges,
but not to the external edges.)
\begin{definition} (Laplacian of a graph with respect to internal vertices and edges):
\label{chapter_graph_polynomials:definition_Laplacian_internal}
Let $G$ be a graph with $\ninternal$ internal edges 
and $\ninternalvertices$ internal vertices.
To each internal edge $e_j$ one associates a parameter $a_j$.
Denote by $G_{\mathrm{int}}$ the internal graph of $G$.
The Laplacian of the graph $G$ with respect to internal vertices and edges
is the (ordinary) Laplacian of the graph $G_{\mathrm{int}}$:
\bq
 L_{\mathrm{int}}\left(G\right)
 & = &
 L\left(G_{\mathrm{int}}\right).
\eq
Phrased differently, 
the Laplacian of the graph $G$ with respect to internal vertices and edges
is a 
$\ninternalvertices \times \ninternalvertices$-matrix $L_{\mathrm{int}}$, 
whose entries are given by
\bq
\label{chapter_graph_polynomials:definition_Laplacian_internal_eq}
 \left(L_{\mathrm{int}}\right)_{ij} & = & \left\{ \begin{array}{rl} 
                      \sum a_k & \mbox{if $i =j$ and edge $e_k$ is attached to $v_i$ and is not a self-loop,} \\
                      - \sum a_k & \mbox{if $i \neq j$ and edge $e_k$ connects $v_i$ and $v_j$.}\\
                      \end{array} \right.
\eq
In eq.~(\ref{chapter_graph_polynomials:definition_Laplacian_internal_eq}) only internal vertices and edges are considered.
\end{definition}
Unless stated otherwise, we will from now on always assume that the Laplacian of a Feynman graph $G$
refers to the Laplacian of the graph $G$ with respect to internal vertices and edges.

Let us consider an example: The Laplacian of the two-loop two-point graph of fig.~\ref{chapter_graph_polynomials:fig3} is given by
\bq
\label{chapter_graph_polynomials:example_laplacian}
 L_{\mathrm{int}} & = & \left( \begin{array}{cccc}
                a_1+a_4 & -a_1 & 0 & -a_4 \\ 
                -a_1 & a_1+a_2+a_5 & -a_5 & -a_2 \\ 
                0 & -a_5 & a_3+a_5 & -a_3 \\ 
                -a_4 & -a_2 & -a_3 & a_2+a_3+a_4 \\ 
                \end{array} \right).
\eq
In the sequel we will need minors of the matrices $L$ and $L_{\mathrm{int}}$. It is convenient to introduce the
following notation:
For a $\nvertices \times \nvertices$ matrix $A$ we denote by $A[i_1,\dots,i_k;j_1,\dots,j_k]$ the
$(\nvertices-k)\times(\nvertices-k)$ matrix, which is obtained from $A$ by deleting the rows $i_1$, \dots, $i_k$ and the 
columns $j_1$, \dots, $j_k$.
For $A[i_1,\dots,i_k;i_1,\dots,i_k]$ we will simply write $A[i_1,\dots,i_k]$.

Let $v_i$ be an arbitrary vertex of $G$. The 
\index{matrix-tree theorem}
{\bf matrix-tree theorem} 
states \cite{Tutte:1984}
\bq
\label{chapter_graph_polynomials:matrix_tree_theorem}
 {\mathcal K} & = & \det \; L[i],
\eq
i.e. the Kirchhoff polynomial is given by the determinant of the minor of the Laplacian, where
the $i$-th row and column have been removed.
One can choose for $i$ any number between $1$ and $\nvertices$.
For an arbitrary internal vertex $v_i$ of $G$ we have 
\bq
\label{chapter_graph_polynomials:matrix_tree_theorem_v2}
 {\mathcal K}_{\mathrm{int}} & = & \det \; L_{\mathrm{int}}[i].
\eq
Choosing for example $i=4$ in eq.~(\ref{chapter_graph_polynomials:example_laplacian}) one finds
for the Kirchhoff polynomial of the two-loop two-point graph of fig.~\ref{chapter_graph_polynomials:fig3}
\bq
 {\mathcal K}_{\mathrm{int}} & = &
         \left| \begin{array}{ccc}
                a_1+a_4 & -a_1 & 0 \\ 
                -a_1 & a_1+a_2+a_5 & -a_5 \\ 
                0 & -a_5 & a_3+a_5 \\ 
                \end{array} \right|
 \nonumber \\
 & = &
  a_1 a_5 ( a_3 + a_4 ) + ( a_1 + a_5 ) a_3 a_4 
+ \left( a_1 a_5 + a_1 a_3 + a_4 a_5 + a_3 a_5 \right) a_2.
\eq
Using eq.~(\ref{chapter_graph_polynomials:convert_U_K}) one recovers the first Symanzik polynomial of this graph as
given in eq.~(\ref{chapter_graph_polynomials:graph_polynomials_ex_tbubble}).
\\
\\
\bs
{\it \refstepcounter{exercise}
{\bf Exercise \theexercise}: 
Re-compute the first graph polynomial ${\mathcal U}$
for the graph shown in fig.~\ref{chapter_basics:fig_nonplanar_vertex} 
from the Laplacian of the graph.
}
\es
\\
\\
The matrix-tree theorem allows to determine the number of spanning trees of a given graph
$G$. Setting $a_1=\dots=a_n=1$, each monomial in ${\mathcal K}$, ${\mathcal K}_{\mathrm{int}}$ and ${\mathcal U}$ reduces
to $1$. There is exactly one monomial for each spanning tree, therefore one obtains
\bq
 \left| {\mathcal T}_1 \right| & = &
 {\mathcal K}(1,\dots,1)
 \; = \;
 {\mathcal K}_{\mathrm{int}}(1,\dots,1)
 \; = \;
 {\mathcal U}(1,\dots,1).
\eq

\begin{figure}
\begin{center}
\includegraphics[scale=1.0]{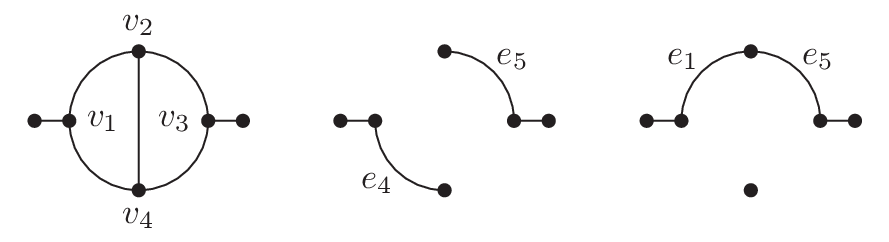}
\end{center}
\caption{\label{chapter_graph_polynomials:fig7} The left picture shows the labelling of the vertices for the two-loop two-point
function. 
The middle and the right picture show the two $2$-forests contributing to ${\mathcal T}_2^{I,J}$
with $I=(2,4)$ and $J=(3,4)$.}
\end{figure}

The matrix-tree theorem as in eq.~(\ref{chapter_graph_polynomials:matrix_tree_theorem}) relates the determinant of 
the minor of the Laplacian, where the
$i$-th row and the $i$-th column have been deleted to a sum over the spanning trees of the graph.
There are two generalisations we can think of:
\begin{enumerate}
\item We delete more than one row and column.
\item We delete different rows and columns, i.e. we delete row $i$ and column $j$ with $i \neq j$.
\end{enumerate}
The all-minors matrix-tree theorem relates the determinant of the corresponding minor to a 
specific sum over spanning forests \cite{Chaiken:1982,Chen:1982,Moon:1994}.
We first state the version for the Laplacian $L$ and then specialise to $L_{\mathrm{int}}$.
To state this theorem we need some notation:
We consider a graph with $\nvertices$ vertices (internal and external).
Let $I=(i_1,\dots,i_k)$ with $1\le i_1 < \dots < i_k \le \nvertices$ denote the rows, which we delete from the
Laplacian $L$, and let $J=(j_1,\dots,j_k)$ with $1\le j_1 < \dots < j_k \le \nvertices$ denote the columns to be deleted
from the Laplacian $L$.
We set $|I|=i_1+\dots+i_k$ and $|J|=j_1+\dots+j_k$.
We denote by ${\mathcal T}_k^{I,J}$ the spanning $k$-forests (in sense of definition~\ref{chapter_graph_polynomials:definition_spanning_forest}), such that 
each tree of an element of ${\mathcal T}_k^{I,J}$ contains exactly one vertex $v_{i_\alpha}$ and
exactly one vertex $v_{j_\beta}$.
The set ${\mathcal T}_k^{I,J}$ is a sub-set of all spanning $k$-forests.
We now consider an element $F$ of ${\mathcal T}_k^{I,J}$. 
Since the element $F$ is a $k$-forest, it consists
therefore of
$k$ trees and we can write it as
\bq 
 F = \left( T_1, \dots, T_k \right) & \in & {\mathcal T}_k^{I,J}.
\eq
We can label the trees such that $v_{i_1} \in T_1$, \dots, $v_{i_k} \in T_k$.
By assumption, each tree $T_\alpha$ contains also exactly one vertex from the set
$\{v_{j_1},\dots,v_{j_k}\}$, although not necessarily in the order $v_{j_\alpha} \in T_\alpha$.
In general it will be in a different order, which we can specify by a permutation
$\pi_F \in S_k$:
\bq
 v_{j_\alpha} \in T_{\pi_F(\alpha)}.
\eq
The 
\index{all-minors matrix-tree theorem}
{\bf all-minors matrix-tree theorem}
reads then
\bq
\label{chapter_graph_polynomials:all_minors_tree_theorem}
 \det \; L[I,J] & = & (-1)^{|I|+|J|} \sum\limits_{F\in {\mathcal T}_k^{I,J}}
 \mbox{sign}(\pi_F) \prod\limits_{e_j \in F} a_j.
\eq
Let us now specialise to $L_{\mathrm{int}}$.
We consider a graph with $\ninternalvertices$ internal vertices.
As before we denote the deleted rows by $I=(i_1,\dots,i_k)$, now with $1\le i_1 < \dots < i_k \le \ninternalvertices$.
The deleted columns are denoted by $J=(j_1,\dots,j_k)$ with $1\le j_1 < \dots < j_k \le \ninternalvertices$.
As before we set $|I|=i_1+\dots+i_k$ and $|J|=j_1+\dots+j_k$.
The set ${\mathcal T}_k^{I,J}$ now denotes the spanning $k$-forests in sense of definition~\ref{chapter_graph_polynomials:definition_spanning_forest_edge_set}, 
such that 
each tree of an element of ${\mathcal T}_k^{I,J}$ contains exactly one vertex $v_{i_\alpha}$ and
exactly one vertex $v_{j_\beta}$.
As before we consider
\bq 
 F = \left( T_1, \dots, T_k \right) & \in & {\mathcal T}_k^{I,J}
\eq
and define 
$\pi_F \in S_k$ by $v_{j_\alpha} \in T_{\pi_F(\alpha)}$.
The all-minors matrix-tree theorem for the internal graph reads
\bq
\label{chapter_graph_polynomials:all_minors_tree_theorem_internal}
 \det \; L_{\mathrm{int}}[I,J] & = & (-1)^{|I|+|J|} \sum\limits_{F\in {\mathcal T}_k^{I,J}}
 \mbox{sign}(\pi_F) \prod\limits_{e_j \in F} a_j.
\eq
In the special case $I=J$ this reduces to
\bq
 \det \; L_{\mathrm{int}}[I] & = & \sum\limits_{F\in {\mathcal T}_k^{I,I}} \;
 \prod\limits_{e_j \in F} a_j.
\eq
If we specialise further to $I=J=(i)$, the sum equals the sum over all spanning trees
(since each spanning $1$-forest of ${\mathcal T}_1^{(i),(i)}$ necessarily contains the vertex $v_i$).
We recover the classical matrix-tree theorem:
\bq
 \det \; L_{\mathrm{int}}[i] & = & \sum\limits_{T\in {\mathcal T}_1} \;
 \prod\limits_{e_j \in T} a_j.
\eq
Let us illustrate the all-minors matrix-tree theorem with an example. We consider again the two-loop two-point graph with the 
labelling of the vertices as shown in fig.~\ref{chapter_graph_polynomials:fig7}.
Taking as an example
\bq
 I = (2,4)
 \;\;\mbox{and}\;\;
 J = (3,4)
\eq
we find for the determinant of $L_{\mathrm{int}}[I;J]$:
\bq
\label{chapter_graph_polynomials:example_all_minors}
 \det \; L_{\mathrm{int}}[2,4;3,4] & = & 
 \left| \begin{array}{cc}
 a_1+a_4 & -a_1 \\
 0 & -a_5 \\
 \end{array} \right|
 = -a_1 a_5 - a_4 a_5.
\eq
On the other hand there are exactly two $2$-forests, such that in each $2$-forest the vertices
$v_2$ and $v_3$ are contained in one tree, while the vertex $v_4$ is contained in the other tree.
These two $2$-forests are shown in fig.~\ref{chapter_graph_polynomials:fig7}.
The monomials corresponding to these two $2$-trees are $a_1 a_5$ and $a_4 a_5$, respectively.
The permutation $\pi_F$ is in both cases the identity and with $|I|=6$, $|J|=7$ we have
an overall minus sign
\bq
 (-1)^{|I|+|J|} & = & -1.
\eq
Therefore, the right hand side of eq.~(\ref{chapter_graph_polynomials:all_minors_tree_theorem}) equals $-a_1a_5 - a_4 a_5$,
showing the agreement with the result of eq.~(\ref{chapter_graph_polynomials:example_all_minors}).

Eq.~(\ref{chapter_graph_polynomials:matrix_tree_theorem}) together with eq.~(\ref{chapter_graph_polynomials:convert_U_K}) allows to determine 
the first Symanzik polynomial ${\mathcal U}$ from the Laplacian of the graph.

We may ask if also the polynomial ${\mathcal F}_0$ can be obtained in a similar way.
We consider again a graph $G$ with $\ninternal$ internal edges $(e_1,\dots,e_{\ninternal})$, 
$\ninternalvertices$ internal vertices $(v_1,\dots,v_{\ninternalvertices})$, 
$\nexternal$ external edges $(e_{\ninternal+1},\dots,e_{\ninternal+\nexternal})$
and $\nexternal$ external vertices $(v_{\ninternalvertices+1},\dots,e_{\ninternalvertices+\nexternal})$.
As before we associate the parameters $a_i$ to the edges $e_i$ ($1\le i \le \ninternal$) 
and new parameters $b_j$ to the edges $e_{\ninternal+j}$ ($1\le j \le \nexternal$).
The Laplacian of $G$ is a $(\ninternalvertices+\nexternal) \times (\ninternalvertices+\nexternal)$ matrix.
We are now considering the Laplacian as in definition~\ref{chapter_graph_polynomials:definition_Laplacian}, not the Laplacian
with respect to internal vertices and edges.

Let us consider the polynomial
\bq
\label{chapter_graph_polynomials:def_W}
 {\mathcal W}(a_1,\dots,a_{\ninternal},b_1,\dots,b_{\nexternal}) & = & \det \; L\left(G\right)[\ninternalvertices+1,\dots,\ninternalvertices+\nexternal].
\eq
${\mathcal W}$ is a polynomial of degree $\ninternalvertices=\ninternal-\loopnumber+1$ in the variables $a_i$ and $b_j$.
We can expand ${\mathcal W}$ in polynomials homogeneous in the variables $b_j$:
\bq
\label{chapter_graph_polynomials:def_expansion}
 {\mathcal W} & = & {\mathcal W}^{(0)} + {\mathcal W}^{(1)} + {\mathcal W}^{(2)} + \dots + {\mathcal W}^{(m)},
\eq
where ${\mathcal W}^{(k)}$ is homogeneous of degree $k$ in the variables $b_j$.
We further write
\bq
\label{chapter_graph_polynomials:def_W_coefficients}
 {\mathcal W}^{(k)} & = &
 \sum\limits_{(j_1,\dots,j_k)} {\mathcal W}^{(k)}_{(j_1,\dots,j_k)}(a_1,\dots,a_{\ninternal}) \; b_{j_1} \dots b_{j_k}.
\eq
The sum is over all indices with $1\le j_1<\dots<j_j \le \nexternal$.
The ${\mathcal W}^{(k)}_{(j_1,\dots,j_k)}$ are homogeneous polynomials of degree $\ninternalvertices-k$ in the variables $a_i$.
For ${\mathcal W}^{(0)}$ and ${\mathcal W}^{(1)}$ one finds
\bq
\label{chapter_graph_polynomials:laplacian_W0_W1}
 {\mathcal W}^{(0)} = 0,
 & &
 {\mathcal W}^{(1)} = {\mathcal K}_{\mathrm{int}}\left(a_1,\dots,a_{\ninternal}\right) \; \sum\limits_{j=1}^{\nexternal} b_j,
\eq
therefore
\bq
\label{chapter_graph_polynomials:laplacian_U}
 {\mathcal U} & = & a_1 \dots a_{\ninternal} \; {\mathcal W}^{(1)}_{(j)}\left(\frac{1}{a_1},\dots,\frac{1}{a_{\ninternal}}\right),
\eq
for any $j\in\{1,\dots,\nexternal\}$.
${\mathcal F}_0$ is related to ${\mathcal W}^{(2)}$:
\bq
\label{chapter_graph_polynomials:laplacian_F_0}
 {\mathcal F}_0 & = & 
   a_1 \dots a_{\ninternal} \sum\limits_{(j,k)} 
   \left( \frac{p_j \cdot p_k}{\mu^2} \right)
   \cdot 
   {\mathcal W}^{(2)}_{(j,k)}\left(\frac{1}{a_1},\dots,\frac{1}{a_{\ninternal}}\right).
\eq
The proof of eqs.~(\ref{chapter_graph_polynomials:laplacian_W0_W1})-(\ref{chapter_graph_polynomials:laplacian_F_0}) follows from the
all-minors matrix-tree theorem. 
The all-minors matrix-tree theorem states
\bq
{\mathcal W}\left(a_1,\dots,a_{\ninternal},b_1,\dots,b_{\nexternal}\right) & = &
 \sum\limits_{F\in {\mathcal T}_{\nexternal}^{I,I}(G)} \;\; \prod\limits_{e_j \in F} c_j,
\eq
with $I=(\ninternalvertices+1,\dots,\ninternalvertices+\nexternal)$ and $c_j=a_j$ if $e_j$ is an internal edge or $c_j=b_{j-\ninternal}$ if $e_j$ is an external edge.
The sum is over all $\nexternal$-forests of $G$ (in the sense of definition~\ref{chapter_graph_polynomials:definition_spanning_forest}), 
such that each tree in an $\nexternal$-forest contains
exactly one of the external vertices $v_{\ninternalvertices+1}$, \dots, $v_{\ninternalvertices+\nexternal}$.
Each $\nexternal$-forest has $\nexternal$ connected components. 
The polynomial ${\mathcal W}^{(0)}$ by definition does not contain
any variable $b_j$. ${\mathcal W}^{(0)}$ would therefore correspond to forests where
all edges connecting the external vertices $e_{\ninternalvertices+1}$, \dots, $e_{\ninternalvertices+\nexternal}$ have been cut.
The external vertices appear therefore as isolated vertices in the forest.
For $\loopnumber>0$, such a forest must necessarily have more than $\nexternal$ connected components. 
This is a contradiction 
with the requirement of having exactly $\nexternal$ connected components and
therefore ${\mathcal W}^{(0)}=0$.
Next, we consider ${\mathcal W}^{(1)}$. Each term is linear in the variables $b_j$. 
Therefore $(\nexternal-1)$ vertices of the external vertices $v_{\ninternalvertices+1}$, \dots, $v_{\ninternalvertices+\nexternal}$ appear as isolated vertices
in the $\nexternal$-forest. The remaining added vertex is connected to a spanning tree of $G_{\mathrm{int}}$.
Summing over all possibilities one sees that ${\mathcal W}^{(1)}$ is given by the product of
$(b_1+\dots+b_{\nexternal})$ with the Kirchhoff polynomial of $G_{\mathrm{int}}$.
Finally we consider ${\mathcal W}^{(2)}$.
Here, $(\nexternal-2)$ of the added vertices appear as isolated vertices. The remaining two are connected to
a spanning $2$-forest of the graph $G_{\mathrm{int}}$, one to each tree of the $2$-forest.
Summing over all possibilities one obtains eq.~(\ref{chapter_graph_polynomials:laplacian_F_0}).

Let us summarise the results on the Laplacian:
\begin{tcolorbox}
{\bf Graph polynomials from the Laplacian of the graph}:
\bq
\label{chapter_graph_polynomials:laplacian_U_and_F_0}
 {\mathcal U} & = & a_1 \dots a_{\ninternal} \; {\mathcal W}^{(1)}_{(j)}\left(\frac{1}{a_1},\dots,\frac{1}{a_{\ninternal}}\right),
 \;\;\;\;\;\;
 \mbox{for any $j\in\{1,\dots,\nexternal\}$},
 \nonumber \\
 {\mathcal F}_0 & = & 
   a_1 \dots a_{\ninternal} \sum\limits_{(j,k)} 
   \left( \frac{p_j \cdot p_k}{\mu^2} \right)
   \cdot 
   {\mathcal W}^{(2)}_{(j,k)}\left(\frac{1}{a_1},\dots,\frac{1}{a_{\ninternal}}\right).
\eq
The quantities ${\mathcal W}^{(1)}_{(j)}$ and ${\mathcal W}^{(2)}_{(j,k)}$ are obtained from the Laplacian of the
graph by eqs.~(\ref{chapter_graph_polynomials:def_W})-(\ref{chapter_graph_polynomials:def_W_coefficients}).
The graph polynomial ${\mathcal F}$ is obtained from ${\mathcal F}_0$ and ${\mathcal U}$ by
eq.~(\ref{chapter_graph_polynomials:def_F_0}).
\end{tcolorbox}
Eq.~(\ref{chapter_graph_polynomials:laplacian_U_and_F_0}) together with eq.~(\ref{chapter_graph_polynomials:def_F_0})
allow the computation of the first and second Symanzik polynomial
from the Laplacian of the graph $G$.
This provides a third method for the computation of the graph polynomials ${\mathcal U}$ and ${\mathcal F}$.

As an example we consider the double-box graph of fig.~\ref{chapter_basics:fig_doublebox}.
\begin{figure}
\begin{center}
\includegraphics[scale=1.0]{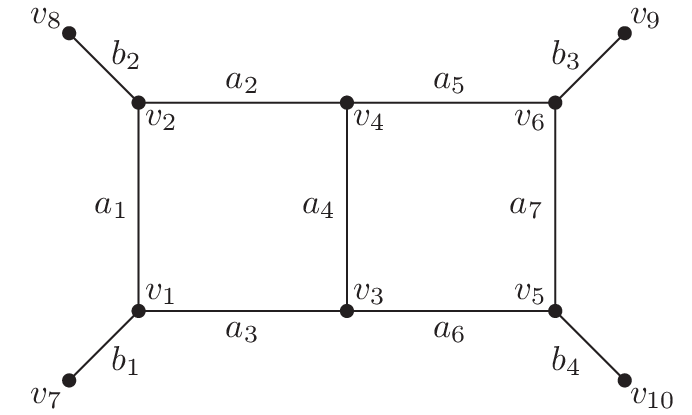}
\end{center}
\caption{\label{chapter_graph_polynomials:fig8} The labelling of the vertices and the Feynman parameters for the ``double box''-graph.}
\end{figure}    
Fig.~\ref{chapter_graph_polynomials:fig8} shows the labelling of the vertices and the Feynman parameters for the graph $G$.
The Laplacian of $G$ (in the sense of definition~\ref{chapter_graph_polynomials:definition_spanning_forest})
is a $10 \times 10$-matrix. We are interested in the minor, where 
-- with the labelling of fig.~\ref{chapter_graph_polynomials:fig8} -- we
delete the rows and columns $7$, $8$, $9$ and $10$.
The determinant of this minor reads
\bq
\lefteqn{
{\mathcal W} = \det \; L[7,8,9,10]
 = } & & 
 \nonumber \\
 & = &
 \left| \begin{array}{cccccc}
 a_1+a_3+b_1 & -a_1 & -a_3 & 0 & 0 & 0 \\
 -a_1 & a_1+a_2+b_2 & 0 & -a_2 & 0 & 0 \\
 -a_3 & 0 & a_3+a_4+a_6 & -a_4 & -a_6 & 0 \\
 0 & -a_2 & -a_4 & a_2+a_4+a_5 & 0 & -a_5 \\
 0 & 0 & -a_6 & 0 & a_6+a_7+b_4 & -a_7 \\
 0 & 0 & 0 & -a_5 & -a_7 & a_5+a_7+b_3 \\
 \end{array} \right|
 \nonumber \\
 & = &
 {\mathcal W}^{(1)} + {\mathcal W}^{(2)} + {\mathcal W}^{(3)}+ {\mathcal W}^{(4)}. 
\eq
For the polynomials ${\mathcal W}^{(1)}$ and ${\mathcal W}^{(2)}$ one finds 
\bq
\label{chapter_graph_polynomials:W1_W2_double_box}
\lefteqn{
{\mathcal W}^{(1)} = 
 \left( b_1 + b_2 + b_3 + b_4 \right) 
} & & 
 \nonumber \\
 & &
 \left(
          a_1 a_2 a_3 a_5 a_6 
        + a_1 a_2 a_3 a_5 a_7 
        + a_1 a_2 a_3 a_6 a_7 
        + a_1 a_2 a_4 a_5 a_6 
        + a_1 a_2 a_4 a_5 a_7 
 \right. \nonumber \\
 & & \left.
        + a_1 a_2 a_4 a_6 a_7 
        + a_1 a_2 a_5 a_6 a_7 
        + a_1 a_3 a_4 a_5 a_6 
        + a_1 a_3 a_4 a_5 a_7  
        + a_1 a_3 a_4 a_6 a_7 
 \right. \nonumber \\
 & & \left.
        + a_1 a_3 a_5 a_6 a_7 
        + a_2 a_3 a_4 a_5 a_6 
        + a_2 a_3 a_4 a_5 a_7  
        + a_2 a_3 a_4 a_6 a_7 
        + a_2 a_3 a_5 a_6 a_7  
 \right),
 \nonumber \\
\lefteqn{
{\mathcal W}^{(2)} =
          (b_1 + b_4) (b_2 + b_3) a_2 a_3 a_5 a_6
 } & & \nonumber \\
 & &  
     + (b_1 + b_2) (b_4 + b_3) 
      \left( 
           a_1 a_2 a_3 a_7 
         + a_1 a_2 a_4 a_7 
         + a_1 a_2 a_6 a_7 
         + a_1 a_3 a_4 a_7 
         + a_1 a_3 a_5 a_7 
         + a_1 a_4 a_5 a_6  
 \right. \nonumber \\
 & & \left.
         + a_1 a_4 a_5 a_7 
         + a_1 a_4 a_6 a_7 
         + a_1 a_5 a_6 a_7 
         + a_2 a_3 a_4 a_7 
      \right)
 \nonumber \\
 & &
         + b_1 (b_2 + b_3 + b_4) a_2
           \left( 
           a_3 a_5 a_7 
         + a_4 a_5 a_6 
         + a_4 a_5 a_7 
         + a_4 a_6 a_7 
         + a_5 a_6 a_7 
           \right)
 \nonumber \\
 & &
         + b_2 (b_1 + b_3 + b_4) a_3 
           \left(
           a_2 a_6 a_7  
         + a_4 a_5 a_6  
         + a_4 a_5 a_7 
         + a_4 a_6 a_7 
         + a_5 a_6 a_7 
           \right)
 \nonumber \\
 & &
         + b_3 (b_1 + b_2 + b_4) a_6
           \left(
           a_1 a_2 a_3  
         + a_1 a_2 a_4  
         + a_1 a_3 a_4  
         + a_1 a_3 a_5 
         + a_2 a_3 a_4 
           \right)
 \nonumber \\
 & &
         + b_4 (b_1 + b_2 + b_3) a_5 
           \left(
           a_1 a_2 a_3  
         + a_1 a_2 a_4 
         + a_1 a_2 a_6
         + a_1 a_3 a_4 
         + a_2 a_3 a_4  
          \right).
\eq
With the help of eq.~(\ref{chapter_graph_polynomials:laplacian_U}) and eq.~(\ref{chapter_graph_polynomials:laplacian_F_0})
and using the kinematic specifications of eq.~(\ref{chapter_basics:specification_double_box}) we recover 
${\mathcal U}$ and ${\mathcal F}$ of eq.~(\ref{chapter_basics:result_U_and_F_double_box}).

We would like to make a few remarks: The polynomial ${\mathcal W}$ is obtained from the determinant
of the matrix $L=L\left(G\right)[\ninternalvertices+1,\dots,\ninternalvertices+\nexternal]$. 
This matrix was constructed from the Laplacian of the graph $G$, taking external vertices and external edges into account.
Then one deletes the rows and columns corresponding to the external vertices.
There are two alternative ways to arrive at the same matrix $L$:

The first alternative consists in merging the external vertices
$v_{\ninternalvertices+1}, v_{\ninternalvertices+2}, \dots, v_{\ninternalvertices+\nexternal}$
into a single new vertex $v_\infty$, 
which connects to all external lines. This defines a new graph $\hat{G}$, which by construction no longer
has any external lines. 
As before we associate variables $b_1$, \dots, $b_{\nexternal}$ to the edges connected to $v_\infty$.
\begin{figure}
\begin{center}
\includegraphics[scale=1.0]{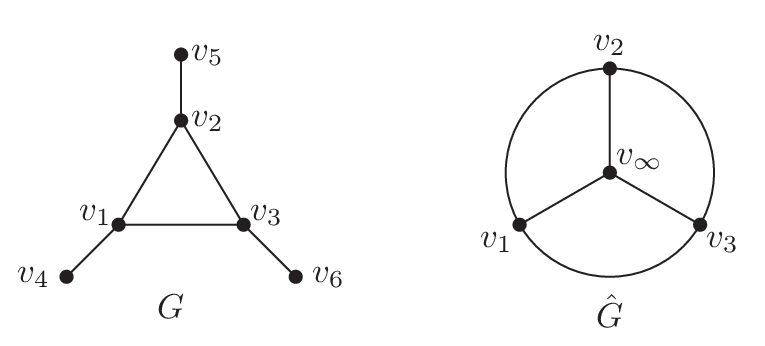}
\end{center}
\caption{\label{chapter_graph_polynomials:fig11} The left picture shows a one-loop graph with three external edges.
The right picture shows the graph $\hat{G}$ associated to $G$, where all external vertices have been joined
in one additional vertex $v_\infty$.}
\end{figure}    
Fig.~\ref{chapter_graph_polynomials:fig11} shows an example for the graph $\hat{G}$ associated to a
one-loop graph with three external legs.
The Laplacian of $\hat{G}$ is a $(\ninternalvertices+1)\times(\ninternalvertices+1)$-matrix.
It is easy to see that
\bq
 L & = & L\left(\hat{G}\right)[\ninternalvertices+1].
\eq
From eq.~(\ref{chapter_graph_polynomials:matrix_tree_theorem}) we see that the polynomial ${\mathcal W}$ is nothing else than
the Kirchhoff polynomial of the graph $\hat{G}$:
\bq
\label{chapter_graph_polynomials:relation_W_K}
 {\mathcal W}(G) & = & {\mathcal K}\left( \hat{G} \right) 
 = \det L\left(\hat{G}\right)[j],
\eq
where $j$ is any number between $1$ and $\ninternalvertices+1$.

For the second alternative one starts from the Laplacian with respect to internal vertices and edges
of the original graph $G$. Let $\pi$ be a 
permutation of $(1,\dots,\ninternalvertices)$. We consider the diagonal matrix
$\mbox{diag}\left(b_{\pi(1)},\dots,b_{\pi(\ninternalvertices)}\right)$.
We can choose the permutation $\pi$ such that
\bq
 \pi(i) & = & j, \;\;\;\mbox{if the external edge $e_{\ninternal+j}$ is attached to vertex $v_i$.}
\eq
We then have
\bq
 L & = & L_{\mathrm{int}}\left( G \right) 
       + \left.\mbox{diag}\left(b_{\pi(1)},\dots,b_{\pi(\ninternalvertices)}\right) \right|_{b_{\nexternal+1}=\dots=b_{\ninternalvertices}=0}.
\eq


\section{Deletion and contraction properties}
\label{chapter_graph_polynomials:sect_deletion_and_contraction}

In this section we study two operations on a graph: the deletion of an edge and the contraction of an edge.
This leads to a recursive algorithm for the calculation of the graph polynomials ${\mathcal U}$ and ${\mathcal F}$.
In addition we discuss the multivariate Tutte polynomial and Dodgson's identity.

In graph theory an edge is called a 
\index{bridge}
{\bf bridge}, 
if the deletion of the edge increases
the number of connected components. 
In the physics literature the term ``one-particle-reducible'' is used for a connected graph 
containing at least one bridge as an internal edge.
The contrary is called ``one-particle-irreducible'', i.e. a connected graph containing no internal bridges.
All external edges are necessarily bridges.
\begin{figure}
\begin{center}
\includegraphics[scale=1.0]{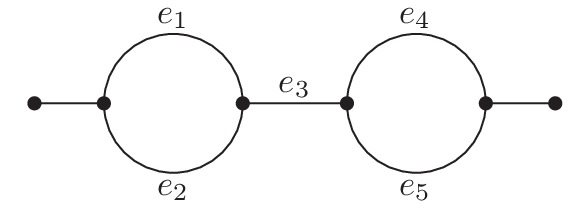}
\end{center}
\caption{\label{chapter_graph_polynomials:fig9} 
A one-particle-reducible graph: The edge $e_3$ is called a bridge. Deleting $e_3$ results in two
connected components.
}
\end{figure}    
Fig.~\ref{chapter_graph_polynomials:fig9} shows an example. The edge $e_3$ is a bridge, while the edges $e_1$, $e_2$, $e_4$ and
$e_5$ are not bridges.
Note that all edges of a tree graph are bridges.
An edge which is neither a bridge nor a self-loop is called a 
\index{edge, regular}
{\bf regular edge}.
All regular edges are internal edges.
For a graph $G$ and a regular edge $e$ we define
\bq
G/e & & \mbox{to be the graph obtained from $G$ by contracting the regular edge $e$,} \nonumber \\
G- e & & \mbox{to be the graph obtained from $G$ by deleting the regular edge $e$.}
\eq
\begin{figure}
\begin{center}
\includegraphics[scale=1.0]{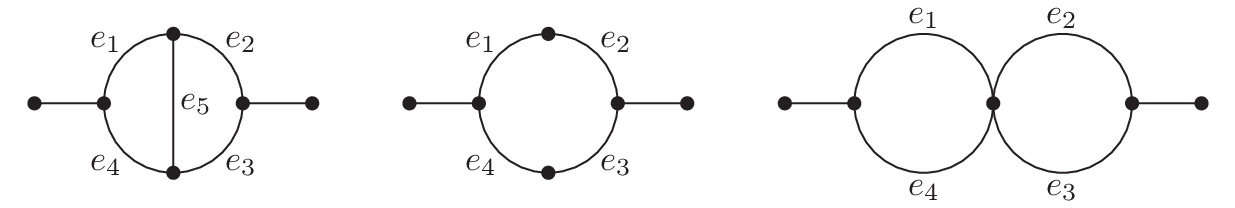}
\end{center}
\caption{\label{chapter_graph_polynomials:fig10} The left picture shows the graph $G$ of the two-loop two-point function.
The middle picture shows the graph $G-e_5$, where edge $e_5$ has been deleted.
The right picture shows the graph $G/e_5$, where the two vertices connected to $e_5$ have been joined and the edge
$e_5$ has been removed.
}
\end{figure}    
Fig.~\ref{chapter_graph_polynomials:fig10} shows an example. 
If the graph $G$ has loop number $l$ it follows that $G-e$ has loop number $(l-1)$, while
$G/e$ has loop number $l$.
This follows easily from the formula $l=n-r+1$ for a connected graph: $G-e$ has one edge less, but the
same number of vertices, while $G/e$ has one edge and one vertex less.

Let us now study the behaviour of the Laplacian under these operations. 
Under deletion the Laplacian behaves as
\bq
 L_{\mathrm{int}}\left(G- e_k\right) & = & \left. L_{\mathrm{int}}(G) \right|_{a_k=0},
\eq
i.e. the Laplacian of the graph $G-e_k$ is obtained from the Laplacian of the graph $G$ by setting the variable
$a_k$ to zero.
The behaviour of the Laplacian under contraction is slightly more complicated:
As before we consider a graph with $\ninternalvertices$ internal vertices.
Assume that edge $e_k$ connects the vertices $v_a$ and $v_{\ninternalvertices}$.
The Laplacian $L_{\mathrm{int}}(G/ e_k)$ is then a $(\ninternalvertices-1) \times (\ninternalvertices-1)$-matrix with entries
\bq
 L_{\mathrm{int}}\left(G/ e_k\right)_{ij}
 & = & \left\{ \begin{array}{ll}
   L_{\mathrm{int}}(G)_{aa}+L_{\mathrm{int}}(G)_{\ninternalvertices\ninternalvertices}+L_{\mathrm{int}}(G)_{a\ninternalvertices}+L_{\mathrm{int}}(G)_{\ninternalvertices a}, & \mbox{if $i=j=a$,} \\
   L_{\mathrm{int}}(G)_{aj}+L_{\mathrm{int}}(G)_{\ninternalvertices j},      & \mbox{if $i=a$, $j\neq a$,} \\
   L_{\mathrm{int}}(G)_{ia}+L_{\mathrm{int}}(G)_{i\ninternalvertices},      & \mbox{if $j=a$, $i\neq a$,} \\
   L_{\mathrm{int}}(G)_{ij},                & \mbox{otherwise.} \\
  \end{array} \right.
 \nonumber \\
\eq
Therefore the Laplacian of $L_{\mathrm{int}}(G/ e_k)$ is identical to the minor $L_{\mathrm{int}}(G)[\ninternalvertices]$ except for the row and column $a$.
The choice that the edge $e_k$ is connected to the last internal vertex $v_{\ninternalvertices}$ was merely made to keep the notation simple.
If the edge connects the vertices $v_a$ and $v_b$ with $a<b$ one deletes from $L_{\mathrm{int}}(G)$ row and column $b$ and modifies
row and column $a$ analogously to the formula above with $b$ substituted for $\ninternalvertices$.
In particular we have \cite{Godsil:2001}
\bq
 L_{\mathrm{int}}\left(G/ e_k\right)[a] & = & L_{\mathrm{int}}\left(G\right)[a,b].
\eq
The deletion/contraction operations can be used for a recursive definition of the graph polynomials.
For any regular edge $e_k$ we have
\bq
\label{chapter_graph_polynomials:recursion_U_F0}
 {\mathcal U}(G) & = & {\mathcal U}(G/e_k) + a_k {\mathcal U}(G-e_k), 
 \nonumber \\
 {\mathcal F}_0(G) & = & {\mathcal F}_0(G/e_k) + a_k {\mathcal F}_0(G-e_k).
\eq
The recursion terminates when all edges are either bridges or self-loops.
This is then a graph, which can be obtained from a tree graph by attaching self-loops to some
vertices. 
These graphs are called 
\index{terminal form}
{\bf terminal forms}.
If a terminal form has $\ninternalvertices$ internal vertices and $\loopnumber$ (self-) loops, 
then there are $(\ninternalvertices-1)$ ``tree-like'' propagators, where the momenta flowing through these propagators
are linear combinations of the external momenta $p_i$ alone and independent of the independent loop momenta $k_j$.
The momenta of the remaining $\loopnumber$ propagators are on the other hand independent of the external momenta 
and can be taken as the independent loop momenta $k_j$, $j=1,\dots,\loopnumber$.
Let us agree that we label the $(\ninternalvertices-1)$ ``tree-like'' internal edges from $1$ to $\ninternalvertices-1$, and the remaining $\loopnumber$
internal edges from $\ninternalvertices$ to $\ninternal$ (with $\ninternal=\ninternalvertices+\loopnumber-1$).
We further denote the momentum squared flowing through edge $j$ by $q_j^2$.
For a terminal form we have
\bq
\label{chapter_graph_polynomials:terminus_U_F0}
 {\mathcal U} = a_{\ninternalvertices} \dots a_{\ninternal},
 & &
 {\mathcal F}_0 = a_{\ninternalvertices} \dots a_{\ninternal} \sum\limits_{j=1}^{\ninternalvertices-1} a_j \left( \frac{-q_j^2}{\mu^2} \right).
\eq
In the special case that the terminal form is a tree graph, this reduces to
\bq
 {\mathcal U} = 1,
 & &
 {\mathcal F}_0 = \sum\limits_{j=1}^{\ninternalvertices-1} a_j \left( \frac{-q_j^2}{\mu^2} \right).
\eq
The Kirchhoff polynomial has for any regular edge the recursion relation
\bq
\label{chapter_graph_polynomials:recursion_K}
 {\mathcal K}(G) & = & a_k {\mathcal K}(G/e_k) + {\mathcal K}(G-e_k),
 \nonumber \\
 {\mathcal K}_{\mathrm{int}}(G) & = & a_k {\mathcal K}_{\mathrm{int}}(G/e_k) + {\mathcal K}_{\mathrm{int}}(G-e_k),
\eq
Note that the factor $a_k$ appears here in combination with the contracted graph $G/e_k$.
The recursion ends again on terminal forms. For these graphs we have with the 
conventions as above
\bq
\label{terminus_K}
 {\mathcal K}_{\mathrm{int}} = a_1 \dots a_{\left(\ninternalvertices-1\right)},
\eq
and a similar formula holds for the terminal forms of ${\mathcal K}$.
The recursion relations eq.~(\ref{chapter_graph_polynomials:recursion_U_F0}) and eq.~(\ref{chapter_graph_polynomials:recursion_K}) are proven with 
the help of the formulae, which express the polynomials ${\mathcal U}$, ${\mathcal K}$ and ${\mathcal K}_{\mathrm{int}}$ in terms of
spanning trees. For ${\mathcal F}_0$ one uses the corresponding formula, which expresses this polynomial
in terms of spanning $2$-forests.
As an example consider the polynomial ${\mathcal U}$ and the set of all spanning trees.
This set can be divided into two sub-sets: the first sub-set is given by the spanning trees, which contain the
edge $e_k$, while the second subset is given by those which do not.
The spanning trees in the first sub-set are in one-to-one correspondence with the spanning trees of $G/e_k$,
the relation is given by contracting and decontracting the edge $e_k$.
The second subset is identical to the set of spanning trees of $G-e_k$. The graphs $G$ and $G-e_k$ differ by the edge
$e_k$ which has been deleted, therefore the explicit factor $a_k$ in front of ${\mathcal U}(G-e_k)$.

We summarise the results on the deletion and contraction properties:
\begin{tcolorbox}
{\bf Graph polynomials from recursion}:
For any regular edge $e_k$ we have
\bq
\label{chapter_graph_polynomials:recursion_U_F0_summary}
 {\mathcal U}(G) & = & {\mathcal U}(G/e_k) + a_k {\mathcal U}(G-e_k), 
 \nonumber \\
 {\mathcal F}_0(G) & = & {\mathcal F}_0(G/e_k) + a_k {\mathcal F}_0(G-e_k).
\eq
The recursion terminates when all edges are either bridges or self-loops, in which case 
the graph polynomials are given
by
\bq
\label{chapter_graph_polynomials:terminus_U_F0_summary}
 {\mathcal U} = a_{\ninternalvertices} \dots a_{\ninternal},
 & &
 {\mathcal F}_0 = a_{\ninternalvertices} \dots a_{\ninternal} \sum\limits_{j=1}^{\ninternalvertices-1} a_j \left( \frac{-q_j^2}{\mu^2} \right),
\eq
where we associate the parameters $a_1, \dots, a_{\ninternalvertices-1}$ to the internal bridges and
the parameters $a_{\ninternalvertices}, \dots, a_{\ninternal}$ to the self-loops.
The graph polynomial ${\mathcal F}$ is obtained from ${\mathcal F}_0$ and ${\mathcal U}$ by
eq.~(\ref{chapter_graph_polynomials:def_F_0}).
\end{tcolorbox}
Eq.~(\ref{chapter_graph_polynomials:recursion_U_F0_summary}) and eq.~(\ref{chapter_graph_polynomials:terminus_U_F0_summary}) together with eq.~(\ref{chapter_graph_polynomials:def_F_0})
provide a fourth method for the computation of the graph polynomials ${\mathcal U}$ and ${\mathcal F}$.
\\
\\
\bs
{\it \refstepcounter{exercise}
{\bf Exercise \theexercise}: 
Consider a massless theory.
Show that in this case the 
Lee-Pomeransky polynomial ${\mathcal G}$ satisfies for any regular edge $e_k$ the recursion
\bq
 {\mathcal G}(G) & = & {\mathcal G}(G/e_k) + a_k {\mathcal G}(G-e_k).
\eq
}
\es
\\
\\
We now look at a generalisation of the Kirchhoff polynomial satisfying a recursion relation similar
to eq.~(\ref{chapter_graph_polynomials:recursion_K}).
For a graph $G$ -- not necessarily connected -- we denote by 
${\mathcal S}$ the set of all 
\index{spanning sub-graph}
{\bf spanning sub-graphs} of $G$, i.e. sub-graphs $H$
of $G$, which contain all vertices of $G$. It is not required that a spanning sub-graph is a forest or a tree.
We denote by $k(H)$ the number of connected components of $H$.
As before we associate to each edge $e_i$ a variable $a_i$. We will need one further formal variable $q$. 
We recall that the loop number of a graph $G$ with $\nedges$ internal edges and $\nvertices$ vertices is given by
\bq
 \loopnumber & = & \nedges - \nvertices + k,
\eq
where $k$ is the number of connected components of $G$.
We can extend the definition of the deletion and contraction properties to edges which are not regular.
It is straightforward to define the operation of deleting a bridge or a self-loop (just delete the edge).
It is also straightforward to define the operation of contracting a bridge (just contract the bridge).
Only the operation of contracting a self-loop needs a dedicated definition:
If the edge $e$ is a self loop, we define the contracted graph $G/e$ to be identical to $G-e$.
The multivariate Tutte polynomial is defined by \cite{Sokal:2005aa}
\bq
\label{chapter_graph_polynomials:def_Tutte}
{\mathcal Z}\left(q,a_1,\dots,a_{\nedges}\right)
 & = &
 \sum\limits_{H \in {\mathcal S}} q^{k(H)}
 \prod\limits_{e_i \in H} a_i.
\eq
It is a polynomial in $q$ and $a_1$, \dots, $a_{\nedges}$.
The multivariate Tutte polynomial generalises the standard
Tutte polynomial \cite{Tutte:1947,Tutte:1954,Tutte:1967,Ellis-Monaghan:2008aa,Ellis-Monaghan:2008ab,Krajewski:2008fa},
which is a polynomial in two variables.
For the multivariate Tutte polynomial we have the recursion relation
\bq
\label{chapter_graph_polynomials:recursion_Z}
 {\mathcal Z}(G) & = & a_k {\mathcal Z}(G/e_k) + {\mathcal Z}(G-e_k),
\eq
where $e_k$ is any edge, not necessarily regular.
The terminal forms are graphs which consists solely of vertices without any edges.
For a graph with $\nvertices$ vertices and no edges one has
\bq
 {\mathcal Z} & = & q^{\nvertices}.
\eq
The multivariate Tutte polynomial starts as a polynomial in $q$ with $q^k$ if $G$ is a graph with $k$
connected components.
If the graph $G$ is not connected we write $G=(G_1,\dots,G_k)$, where $G_1$ to $G_k$ are the connected
components. For a disconnected graph the multivariate Tutte polynomial factorises:
\bq
 {\mathcal Z}\left(G\right)
 & = &
 {\mathcal Z}\left(G_1\right) \dots {\mathcal Z}\left(G_k\right).
\eq
Some examples for the multivariate Tutte polynomial are
\bq
{\mathcal Z}\left(
\begin{picture}(20,10)(0,2)
 \Vertex(2,5){2}
 \Vertex(18,5){2}
 \Line(2,5)(18,5)
\end{picture}
\right) 
 & = & q a + q^2,
\nonumber \\
{\mathcal Z}\left(
\begin{picture}(20,10)(0,3)
 \Vertex(10,2){2}
 \CArc(10,8)(6,0,360)
\end{picture}
\right) 
 & = & q \left( a + 1 \right),
\nonumber \\
{\mathcal Z}\left(
\begin{picture}(20,10)(0,3)
 \Vertex(4,8){2}
 \Vertex(16,8){2}
 \CArc(10,8)(6,0,360)
\end{picture}
\right) 
 & = & q \left( a_1 a_2 + a_1 + a_2 \right) + q^2.
\eq
If $G$ is a connected graph we recover the Kirchhoff polynomial ${\mathcal K}(G)$ from the Tutte
polynomial ${\mathcal Z}(G)$ by first taking the coefficient of the linear term in $q$ and then retaining only
those terms with the lowest degree of homogeneity in the variables $a_i$.
Expressed in a formula we have
\bq
 {\mathcal K}\left(a_1,\dots,a_{\nedges}\right)
 & = & 
 \lim\limits_{\lambda \rightarrow 0} \; \lim\limits_{q\rightarrow 0} \;
 \lambda^{1-\nvertices} q^{-1} {\mathcal Z}\left(q,\lambda a_1,\dots,\lambda a_{\nedges}\right).
\eq
To prove this formula one first notices that the definition in eq.~(\ref{chapter_graph_polynomials:def_Tutte})
of the multivariate Tutte polynomial is equivalent to
\bq
{\mathcal Z}\left(q,a_1,\dots,a_{\nedges}\right)
 & = &
 q^{\nvertices}
 \sum\limits_{H \in {\mathcal S}} q^{\loopnumber(H)}
 \prod\limits_{e_i \in H} \frac{a_i}{q}.
\eq
One then obtains
\bq
 \lambda^{1-\nvertices} q^{-1} {\mathcal Z}\left(q,\lambda a_1,\dots,\lambda a_{\nedges}\right)
 & = &
 \sum\limits_{H \in {\mathcal S}} q^{k(H)-1} \lambda^{\loopnumber(H)-k(H)+1}
 \prod\limits_{e_i \in H} a_i.
\eq
The limits $q\rightarrow 0$ and $\lambda \rightarrow 0$ select $k(H)=1$ and $\loopnumber(H)=0$, hence
the sum over the spanning sub-graphs reduces to a sum over spanning trees and one recovers
the Kirchhoff polynomial.

At the end of this section we want to discuss 
Dodgson's identity\footnote{Dodgson's even more famous literary work contains the
novel ``Alice in wonderland'' which he wrote using the pseudonym
Lewis Carroll.}.
Dodgson's identity states that for any $n \times n$ matrix $A$ and integers $i$, $j$ with $1\le i,j \le n$ and $i \neq j$ 
one has \cite{Dodgson:1866,Zeilberger:1997}
\bq
\label{chapter_graph_polynomials:dodgson_identity}
 \det\left( A \right) \det\left( A[i,j] \right)
 & = & 
 \det\left( A[i] \right) \det\left( A[j] \right)
 -
 \det\left( A[i;j] \right) \det\left( A[j;i] \right).
\eq
We remind the reader that $A[i,j]$ denotes a $(n-2)\times(n-2)$ matrix obtained from $A$ by deleting
the rows and columns $i$ and $j$.
On the other hand $A[i;j]$ denotes a $(n-1)\times(n-1)$ matrix which is obtained from $A$ by deleting
the $i$-th row and the $j$-th column.
The identity in eq.~(\ref{chapter_graph_polynomials:dodgson_identity}) 
has an interesting application towards graph polynomials: 
Let $e_a$ and $e_b$ be two regular edges of a graph $G$, which share one common vertex.
Assume that the edge $e_a$ connects the vertices $v_i$ and $v_k$, while the edge $e_b$ connects the vertices
$v_j$ and $v_k$. The condition $i \neq j$ ensures that after contraction of one edge the other edge
is still regular.
(If we would allow $i=j$ we have a multiple edge 
and the contraction of one edge leads to a self-loop for the other edge.)
For the Kirchhoff polynomial of the graph $G-e_a-e_b$ we have
\bq
 {\mathcal K}\left( G-e_a-e_b \right) & = & \det L\left( G-e_a-e_b\right)[k].
\eq
Let us now consider the Kirchhoff polynomials of the graphs $G/e_a-e_b$ and $G/e_b-e_a$. One finds
\bq
 {\mathcal K}\left( G/e_a-e_b \right) & = & \det L\left( G-e_a-e_b\right)[i,k],
 \nonumber \\
 {\mathcal K}\left( G/e_b-e_a \right) & = & \det L\left( G-e_a-e_b\right)[j,k].
\eq
Here we made use of the fact that the operations of contraction and deletion commute
(i.e. $G/e_a-e_b=(G-e_b)/e_a$) as well as of the fact that the variable $a_a$ occurs in the Laplacian
of $G$ only in rows and columns $i$ and $k$, therefore 
$L\left( G-e_b\right)[i,k] = L\left( G-e_a-e_b\right)[i,k]$.
Finally we consider the Kirchhoff polynomial of the graph $G/e_a/e_b$, for which one finds
\bq
 {\mathcal K}\left( G/e_a/e_b \right) & = & \det L\left( G-e_a-e_b\right)[i,j,k].
\eq
The Laplacian of any graph is a symmetric matrix, therefore
\bq
 \det L( G-e_a-e_b )[i,k;j,k] & = & \det L( G-e_a-e_b )[j,k;i,k].
\eq
We can now apply Dodgson's identity to the matrix $L( G-e_a-e_b )[k]$.
Using the fact that $L\left( G-e_a-e_b\right)[i,k;j,k]=L\left( G \right)[i,k;j,k]$ one finds \cite{Stembridge:1998}
\bq
\label{chapter_graph_polynomials:Dodgson_K}
 {\mathcal K}\left( G/e_a- e_b \right) {\mathcal K}\left( G/e_b- e_a \right)
 -
 {\mathcal K}\left( G- e_a-e_b \right) {\mathcal K}\left( G/e_a/e_b \right)
 = 
 \left( \det L\left( G \right)[i,k;j,k] \right)^2.
\eq
The version for the internal graph reads
\bq
\label{chapter_graph_polynomials:Dodgson_K_v2}
\lefteqn{
 {\mathcal K}_{\mathrm{int}}\left( G/e_a- e_b \right) {\mathcal K}_{\mathrm{int}}\left( G/e_b- e_a \right)
 -
 {\mathcal K}_{\mathrm{int}}\left( G- e_a-e_b \right) {\mathcal K}_{\mathrm{int}}\left( G/e_a/e_b \right)
 = } & & \\
 & &
\hspace*{100mm}
 \left( \det L_{\mathrm{int}}\left( G \right)[i,k;j,k] \right)^2.
 \nonumber 
\eq
This equation shows that the expression on the left-hand side factorises into a square.
The expression on the right-hand side can be re-expressed using the all-minors matrix-tree theorem
as a sum over $2$-forests, such that the vertex $v_k$ is contained in one tree of the 
$2$-forest, while the vertices $v_i$ and $v_j$ are both contained in the other tree.

Expressed in terms of the first Symanzik polynomial we have
\bq
\label{chapter_graph_polynomials:factorisation_U_U}
 {\mathcal U}\left( G/e_a- e_b \right) {\mathcal U}\left( G/e_b- e_a \right)
 -
 {\mathcal U}\left( G- e_a-e_b \right) {\mathcal U}\left( G/e_a/e_b \right)
 & = &
 \left( \frac{\Delta_1}{a_a a_b} \right)^2.
\eq
The expression $\Delta_1$ is given by
\bq
\label{chapter_graph_polynomials:def_Delta_1}
\Delta_1 & = &
\sum\limits_{F\in {\mathcal T}_2^{(i,k),(j,k)}} \; \prod\limits_{e_t \notin F} a_t.
\eq
The sum is over all $2$-forests $F=(T_1,T_2)$ of $G$ such that $v_i, v_j \in T_1$ and $v_k \in T_2$.
Note that each term of $\Delta_1$ contains $a_a$ and $a_b$.
The factorisation property of eq.~(\ref{chapter_graph_polynomials:factorisation_U_U}) plays a crucial role
in the algorithms of refs.~\cite{Brown:2008,Brown:2009a,Brown:2009b}.

A factorisation formula similar to eq.~(\ref{chapter_graph_polynomials:factorisation_U_U}) can be derived for an expression containing
both the first Symanzik polynomial ${\mathcal U}$ and the polynomial ${\mathcal F}_0$.
As before we assume that $e_a$ and $e_b$ are two regular edges of a graph $G$, which share one common vertex.
The derivation uses the results of sect.~\ref{chapter_graph_polynomials:sect_matrix_tree_theorem}
and starts from eq.~(\ref{chapter_graph_polynomials:Dodgson_K}) for the graph $\hat{G}$ associated to $G$. 
Eq.~(\ref{chapter_graph_polynomials:relation_W_K}) relates then the Kirchhoff 
polynomial of $\hat{G}$ to the ${\mathcal W}$-polynomial of $G$.
The ${\mathcal W}$-polynomial is then expanded in powers of $b$. The lowest order terms reproduce eq.~(\ref{chapter_graph_polynomials:factorisation_U_U}).
The next order yields
\bq
 {\mathcal U}\left( G/e_a- e_b \right) {\mathcal F}_0\left( G/e_b- e_a \right)
 -
 {\mathcal U}\left( G- e_a-e_b \right) {\mathcal F}_0\left( G/e_a/e_b \right)
&&
\\
 +
 {\mathcal F}_0\left( G/e_a- e_b \right) {\mathcal U}\left( G/e_b- e_a \right)
 -
 {\mathcal F}_0\left( G- e_a-e_b \right) {\mathcal U}\left( G/e_a/e_b \right)
 & = & 
 2 \left( \frac{\Delta_1}{a_a a_b} \right) \left( \frac{\Delta_2}{a_a a_b} \right).
 \nonumber 
\eq
The quantity $\Delta_2$ appearing on the right-hand side is obtained from the
all-minors matrix-tree theorem. We can express this quantity in terms of spanning three-forests of $G$ as follows:
Let us denote by ${\mathcal T}_3^{((i,j),\bullet,k)}$ the set of spanning 
three-forests $(T_1,T_2,T_3)$ of $G$ such that
$v_i, v_j \in T_1$ and $v_k \in T_3$.
Similar we denote by ${\mathcal T}_3^{(i,j,k)}$ the set of spanning 
three-forests $(T_1,T_2,T_3)$ of $G$ such that
$v_i \in T_1$, $v_j \in T_2$ and $v_k \in T_3$.
Then
\bq
\Delta_2 & = &
\sum\limits_{(T_1,T_2,T_3)\in {\mathcal T}_3^{(i,j,k)}} \; 
 \sum\limits_{v_c \in T_1, v_d \in T_2} \; \left( \frac{p_c \cdot p_d}{\mu^2} \right)
 \prod\limits_{e_t \notin (T_1,T_2,T_3)} \; a_t
 \nonumber \\
 & & 
-\sum\limits_{(T_1,T_2,T_3)\in {\mathcal T}_3^{((i,j),\bullet,k)}} \; 
 \sum\limits_{v_c, v_d \in T_2} \; \left( \frac{p_c \cdot p_d}{\mu^2} \right)
 \prod\limits_{e_t \notin (T_1,T_2,T_3)} \; a_t.
\eq
In this formula we used the convention that the momentum $p_j$ equals zero if no external leg is 
attached to vertex $v_j$.
Expanding the ${\mathcal W}$-polynomial to order $b^4$ we have terms of order $b^2$ squared as well as terms
which are products of order $b$ with order $b^3$. We are interested in an expression which arises 
from terms of order $b^2$ squared alone. In this case we obtain a factorisation formula only for special
kinematic configurations.
If for all external momenta one has
\bq
\label{chapter_graph_polynomials:condition_external_momenta}
 \left( p_{i_1} \cdot p_{i_2} \right) \cdot \left( p_{i_3} \cdot p_{i_4} \right)
 & = & 
 \left( p_{i_1} \cdot p_{i_3} \right) \cdot \left( p_{i_2} \cdot p_{i_4} \right),
 \;\;\;\;\;\;
 i_1, i_2, i_3, i_4 \in \{1,\dots,m\}
\eq
then
\bq
 {\mathcal F}_0\left( G/e_a- e_b \right) {\mathcal F}_0\left( G/e_b- e_a \right)
 -
 {\mathcal F}_0\left( G- e_a-e_b \right) {\mathcal F}_0\left( G/e_a/e_b \right)
 & = &
 \left( \frac{\Delta_2}{a_a a_b} \right)^2.
\eq
Eq.~(\ref{chapter_graph_polynomials:condition_external_momenta}) is satisfied for example if all external momenta are collinear.
A second example is given by a three-point function. In the kinematic configuration where
\bq
 \left( p_1^2 \right)^2 + \left( p_2^2 \right)^2 + \left( p_3^2 \right)^2
 - 2 p_1^2 p_2^2 - 2 p_1^2 p_3^2 - 2 p_2^2 p_3^2 & = & 0,
\eq
eq.~(\ref{chapter_graph_polynomials:condition_external_momenta}) is satisfied. 


\section{Duality}
\label{chapter_graph_polynomials:sect_duality}

We have seen that the Kirchhoff polynomial ${\mathcal K}_{\mathrm{int}}$ and the first Symanzik polynomial ${\mathcal U}$ 
of a graph $G$ with $\ninternal$ internal edges are related by the equations~(\ref{chapter_graph_polynomials:convert_U_K}):
\bq
 {\mathcal U}\left(a_1,\dots,a_{\ninternal}\right) 
 & = & 
 a_1 \dots a_{\ninternal} \; {\mathcal K}_{\mathrm{int}}\left(\frac{1}{a_1},\dots,\frac{1}{a_{\ninternal}}\right),
 \nonumber \\
 {\mathcal K}_{\mathrm{int}}\left(a_1,\dots,a_{\ninternal}\right) 
 & = & 
 a_1 \dots a_{\ninternal} \; {\mathcal U}\left(\frac{1}{a_1},\dots,\frac{1}{a_{\ninternal}}\right).
 \nonumber
\eq
Let $G$ be a graph with $\nedges$ edges (internal and external).
In this section we will ask if one can find a graph $G^\ast$ with $\nedges$ edges such that 
${\mathcal K}(G^\ast)= {\mathcal U}(G)$ and 
${\mathcal K}(G)= {\mathcal U}(G^\ast)$.
Such a graph $G^\ast$ will be called a dual graph of $G$.
In this section we will show that for a planar graph one can always construct a dual graph.
The dual graph of $G$ need not be unique, there might be several topologically distinct graphs $G^\ast$ fulfilling the above mentioned
relation.
In other words two topologically distinct graphs $G_1$ and $G_2$ both of them with $\nedges$ edges can have the same Kirchhoff polynomial.

In this section we associate parameters $a_j$ to all edges (internal and external).
This is no restriction, as the following exercise shows:
\\
\\
\bs
{\it \refstepcounter{exercise}
{\bf Exercise \theexercise}: 
Let $G$ be a graph with $\ninternal$ edges and $\nexternal$ edges and set $\nedges=\ninternal+\nexternal$.
Label the edges as 
\bq
 \mbox{internal edges} & : & \{ e_1, e_2, \dots, e_{\ninternal} \},
 \nonumber  \\
 \mbox{external edges} & : & \{ e_{\ninternal+1}, e_{\ninternal+2}, \dots, e_{\ninternal+\nexternal} \}.
\eq
Let $G_{\mathrm{int}}$ be the internal graph of $G$.
Define ${\mathcal U}$, ${\mathcal K}$ and ${\mathcal K}_{\mathrm{int}}$ as before.
Define $\tilde{{\mathcal U}}$ by
\bq
 \tilde{{\mathcal U}}\left(a_1,\dots,a_{\nedges}\right) 
 & = & 
 a_1 \dots a_{\nedges} \; {\mathcal K}\left(\frac{1}{a_1},\dots,\frac{1}{a_{\nedges}}\right).
\eq
Show
\bq
 \tilde{{\mathcal U}}\left(a_1,\dots,a_{\nedges}\right) 
 & = & 
 {\mathcal U}\left(a_1,\dots,a_{\ninternal}\right),
 \nonumber \\
 {\mathcal K}\left(a_1,\dots,a_{\nedges}\right) 
 & = & 
 a_{\ninternal+1} \dots a_{\nedges}
 {\mathcal K}\left(a_1,\dots,a_{\ninternal}\right).
\eq
}
\es
\\
\\
The exercise also shows that $\tilde{{\mathcal U}}={\mathcal U}$ and in particular $\tilde{{\mathcal U}}$
is independent of $a_{\ninternal+1}$, $\dots,$ $a_{\nedges}$, so we can simply use
${\mathcal U}$ to denote both polynomials.

A graph is called planar if it can be embedded in a plane without
crossings of edges.
\begin{figure}
\begin{center}
\includegraphics[scale=1.0]{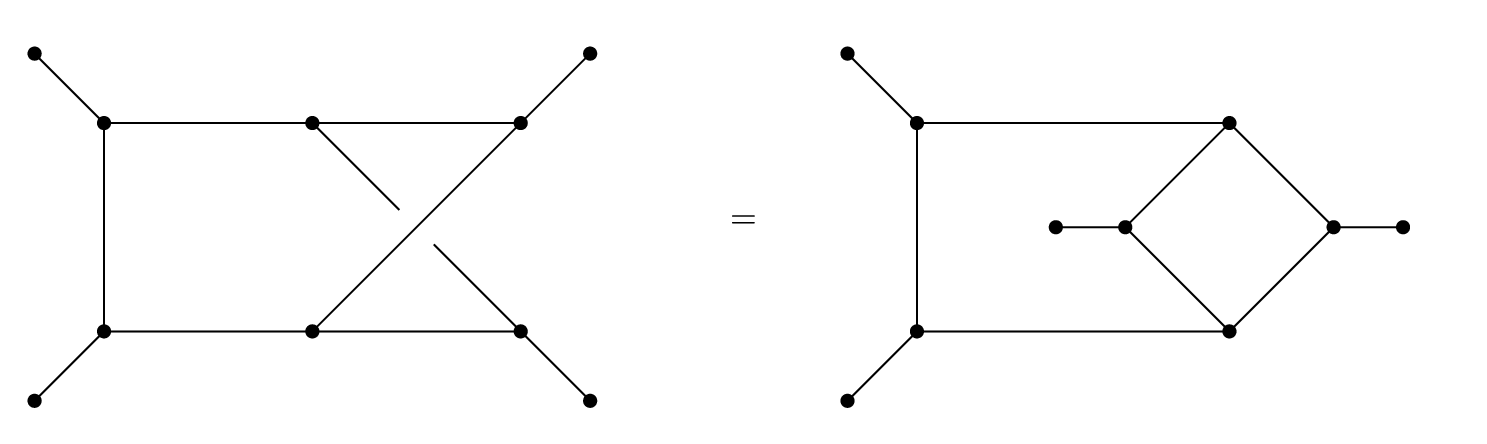}
\end{center}
\caption{\label{chapter_graph_polynomials:fig12} The ``crossed double-box''-graph can be drawn as a planar graph.}
\end{figure}    
We would like to note that the ``crossed double-box''-graph shown in fig.~\ref{chapter_graph_polynomials:fig12} is a planar graph. 
The right picture of fig.~\ref{chapter_graph_polynomials:fig12} shows how this graph can be drawn in the plane without any crossing of edges.

\begin{figure}
\begin{center}
\includegraphics[scale=1.0]{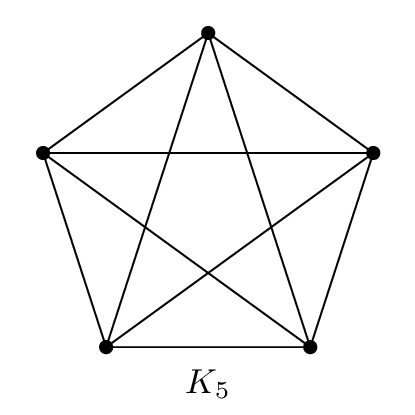}
\hspace*{15mm}
\includegraphics[scale=1.0]{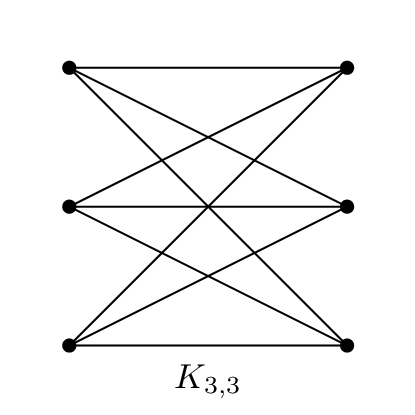}
\end{center}
\caption{\label{chapter_graph_polynomials:fig13} The 'smallest' non-planar graphs.
}
\end{figure}

\begin{figure}
\begin{center}
\includegraphics[scale=1.0]{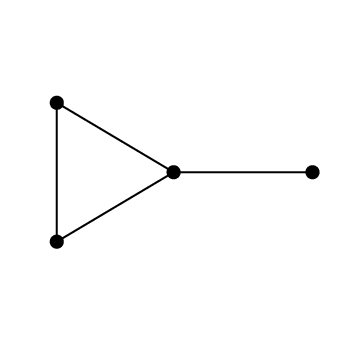}
\hspace*{15mm}
\includegraphics[scale=1.0]{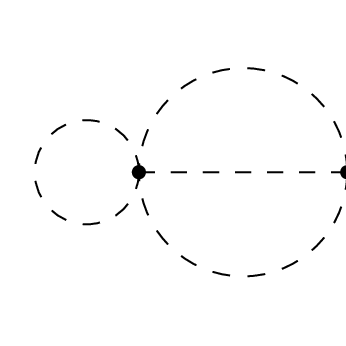}
\hspace*{15mm}
\includegraphics[scale=1.0]{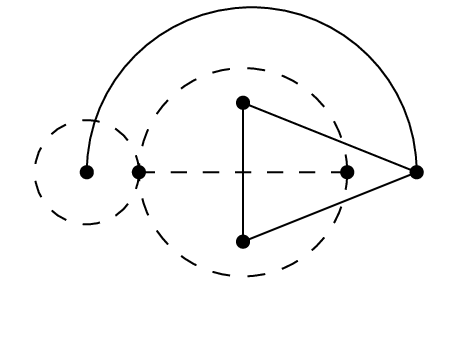}
\end{center}
\caption{\label{chapter_graph_polynomials:fig14} The first two pictures show a graph $G$ and its dual graph $G^\ast$.
The right picture shows the construction of $G^\ast$ from $G$ (or vice versa).
}
\end{figure}

Fig.~\ref{chapter_graph_polynomials:fig13} shows two examples of non-planar graphs. 
The first graph is the complete graph with five vertices $K_5$. The second example is denoted $K_{3,\,3}$.
A theorem states that a graph $G$ is planar
if and only if none of the graphs obtained from $G$ by a (possibly
empty) sequence of contractions of edges contains $K_{5}$ or $K_{3,\,3}$
as a sub-graph \cite{Kuratowski:1930,Wagner:1937,Diestel:2005}. 
\\
\\
\bs
{\it \refstepcounter{exercise}
{\bf Exercise \theexercise}: 
Determine the number of loops for $K_{5}$ and $K_{3,\,3}$.
}
\es
\\
\\
Each planar graph $G$ has a dual graph $G^{\star}$ which
can be obtained as follows:
\begin{itemize}
\item Draw the graph $G$ in a plane, such that no edges intersect. In this
way, the graph divides the plane into open subsets, called faces.
\item Draw a vertex inside each face. These are the vertices of $G^{\star}$.
\item For each edge $e_{i}$ of $G$ draw a new edge $e_i^\ast$ connecting the two
vertices of the faces, which are separated by $e_{i}$. The new edges $e_i^\ast$
are the edges of $G^{\star}$.
\end{itemize}
An example for this construction is shown in fig.~\ref{chapter_graph_polynomials:fig14}.
We note from the construction of the dual graph, that for each external
edge in $G$ there is a self-loop in $G^{\star}$ and that for
each self-loop in $G$ there is an external edge in $G^{\star}$.

\begin{figure}
\begin{center}
\includegraphics[scale=1.0]{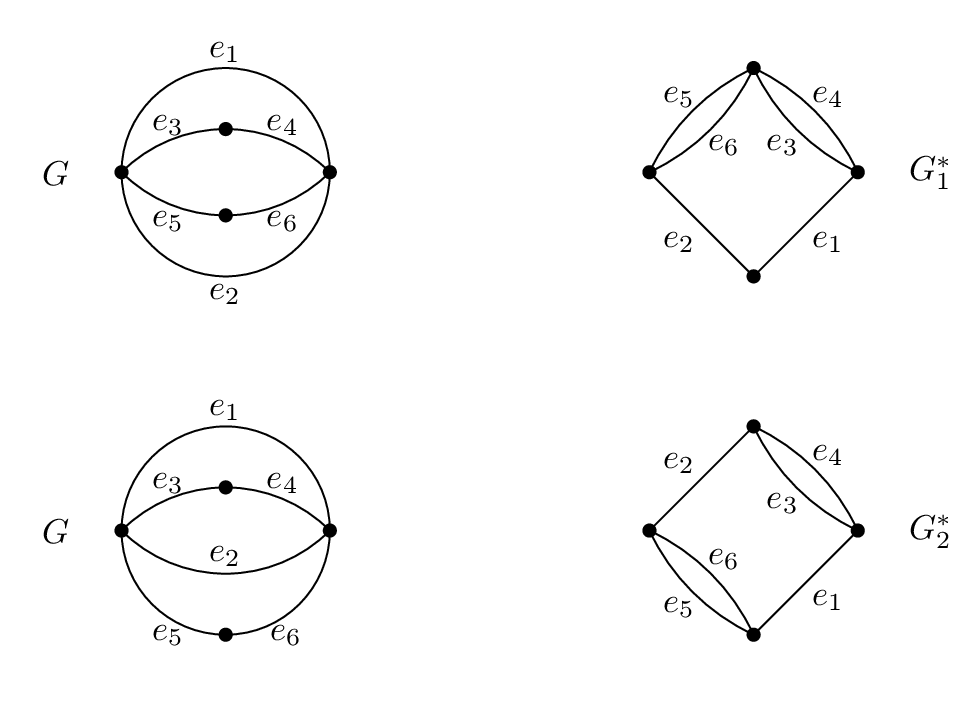}
\end{center}
\caption{\label{chapter_graph_polynomials:fig15} An example showing that different embeddings of a planar graph $G$ into the plane
yield different dual graphs $G^\ast_1$ and $G^\ast_2$.}
\end{figure}

If we now associate the variable $a_i$ to the edge $e_i$ of $G$ as well as to the edge $e_i^\ast$
of $G^\ast$ we have
\bq
{\mathcal K}(G^\ast)= {\mathcal U}(G),
&& 
{\mathcal K}(G)= {\mathcal U}(G^\ast).
\eq
It is important to note that the above construction of the dual graph
$G^{\star}$ depends on the way, how $G$ is drawn in the plane.
A given graph $G$ can have several topologically distinct dual graphs. 
These dual graphs have the same Kirchhoff polynomial. 
An example is shown in fig.~\ref{chapter_graph_polynomials:fig15}.
For this example one finds
\bq
& &
 {\mathcal K}(G) = {\mathcal U}(G^\ast_1) = {\mathcal U}(G^\ast_2)
 = 
 (a_1+a_2)(a_3+a_4)(a_5+a_6) + a_3 a_4 (a_5+a_6) + (a_3+a_4) a_5 a_6,
 \nonumber \\
& &
 {\mathcal U}(G) = {\mathcal K}(G^\ast_1) = {\mathcal K}(G^\ast_2)
 = 
 a_1 a_2 ( a_3 + a_4 + a_5 + a_6 ) + ( a_1 + a_2 ) ( a_3 + a_4 ) ( a_5 + a_6 ).
\eq


\section{Matroids}
\label{chapter_graph_polynomials:sect_matroids}

In this section we introduce the basic terminology of matroid theory.
We are in particular interested in cycle matroids.
A cycle matroid can be constructed from
a graph and it contains the information which is needed for
the construction of the Kirchhoff polynomial 
and therefore as well for the construction of the first Symanzik polynomial $\mathcal{U}$.
In terms of matroids we want to discuss the fact, that two different
graphs can have the same Kirchhoff polynomial. 
We have already encountered an example in fig.~\ref{chapter_graph_polynomials:fig15}.
We review a theorem on matroids which determines the classes of graphs whose Kirchhoff
polynomials are the same. For a detailed introduction to matroid theory
we refer to \cite{Oxley,Oxley:2003aa}.

\begin{figure}
\begin{center}
\includegraphics[scale=1.0]{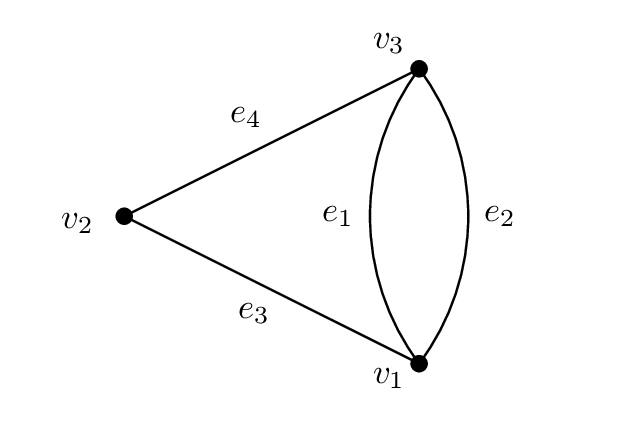}
\end{center}
\caption{\label{chapter_graph_polynomials:fig17} A graph $G$.}
\end{figure}

We introduce cycle matroids by an
example and consider the graph $G$ of fig.~\ref{chapter_graph_polynomials:fig17}.
The graph $G$ has three vertices $V=\left\{ v_{1},\, v_{2},\, v_{3}\right\}$
and four edges $E=\left\{ e_{1},\, e_{2},\, e_{3},\, e_{4}\right\}$.
The graph has five spanning trees given by the sets of
edges $\left\{ e_{1},\, e_{3}\right\}$, $\left\{ e_{1},\, e_{4}\right\}$,
$\left\{ e_{2},\, e_{3}\right\}$, $\left\{ e_{2},\, e_{4}\right\}$,
$\left\{ e_{3},\, e_{4}\right\}$, respectively. We obtain the Kirchhoff
polynomial ${\mathcal K}=a_{1}a_{3}+a_{1}a_{4}+a_{2}a_{3}+a_{2}a_{4}+a_{3}a_{4}$. 
The 
\index{incidence matrix}
{\bf unoriented incidence matrix} of a graph $G$ with $\nvertices$ vertices and $\nedges$ edges is 
a $\nvertices \times \nedges$-matrix $B_{\mathrm{incidence}}=(b_{ij})$, defined by 
\bq
b_{ij} & = & \left\{ \begin{array}{l}
1,\textrm{\ensuremath{\;} if }e_{j}\textrm{ is incident to }v_{i}\textrm{ and }e_{j}\textrm{ is not a self-loop},\\
0,\textrm{\ensuremath{\;} else.}\end{array}\right.
\eq
There is also the definition of the
{\bf oriented incidence matrix} of an oriented graph $G$ with $\nvertices$ vertices and $\nedges$ edges.
This is again $\nvertices \times \nedges$-matrix $B_{\mathrm{oriented}\;\mathrm{incidence}}=(b_{ij})$, whose entries are
\bq
b_{ij} & = & \left\{ \begin{array}{l}
+1,\textrm{\ensuremath{\;} if }v_{i}\textrm{ is the source of }e_{j}\textrm{ and }e_{j}\textrm{ is not a self-loop},\\
-1,\textrm{\ensuremath{\;} if }v_{i}\textrm{ is the sink of }e_{j}\textrm{ and }e_{j}\textrm{ is not a self-loop},\\
0,\textrm{\ensuremath{\;} else.}\end{array}\right.
\eq
The entries in each column of $B_{\mathrm{oriented}\;\mathrm{incidence}}$
sum up to zero, as every edge has exactly one source and one sink.
A self-loop corresponds to a zero column in $B_{\mathrm{oriented}\;\mathrm{incidence}}$
and $B_{\mathrm{incidence}}$.
Given $B_{\mathrm{oriented}\;\mathrm{incidence}}$ for an oriented graph $G$, we obtain
the unoriented incidence matrix $B_{\mathrm{incidence}}$ for $G$ as
\bq
 B_{\mathrm{incidence}}
 & = &
 B_{\mathrm{oriented}\;\mathrm{incidence}} \mod 2.
\eq

For the graph $G$ of fig.~\ref{chapter_graph_polynomials:fig17} the unoriented incidence matrix reads
\[
\begin{array}{cccc}
e_{1} & e_{2} & e_{3} & e_{4}
\end{array}
\]
\[
\left(\begin{array}{cccc}
1\, & \,1\, & \,1\, & 0\\
0 & 0 & 1 & 1\\
1 & 1 & 0 & 1
\end{array}\right),
\]
where we indicated that each column vector corresponds to one edge
of the graph. Let us focus on the set of these four column vectors.
We want to consider all subsets of these vectors which are linearly
independent over $\mathbb{Z}_2$. Obviously the set of all four vectors
and the set of any three of the given vectors are linearly dependent
over $\mathbb{Z}_2.$ Furthermore the first two columns corresponding
to $\left\{ e_{1},\, e_{2}\right\} $ are equal and therefore linearly
dependent. Hence the linearly independent subsets are all sets with
only one vector and all sets with two vectors, except for the just
mentioned one consisting of the first and second column. 
For each set of linearly independent vectors let us now write the set of the
corresponding edges. The set of all these sets shall be denoted ${\mathcalI}$.
We obtain
\bq
\label{chapter_graph_polynomials:example_independent_sets}
{\mathcalI} 
 & = & 
 \left\{ \emptyset, \, \left\{ e_{1}\right\}, \, \left\{ e_{2}\right\}, \,
         \left\{ e_{3}\right\}, \, \left\{ e_{4}\right\},\right. \nonumber \\
 &  & \left.\left\{ e_{1}, \, e_{3}\right\}, \, \left\{ e_{1}, \, e_{4}\right\}, \,
      \left\{ e_{2}, \, e_{3}\right\}, \, \left\{ e_{2}, \, e_{4}\right\}, \,
      \left\{ e_{3}, \, e_{4}\right\} \right\}.
\eq
The empty set is said to be independent and is included here
by definition. Let us make the important observation, that the sets
in ${\mathcalI}$ which have two elements, i.e. the maximal number
of elements, are exactly the sets of edges of the spanning trees of
the graph given above. 

The pair $(E, \, \mathcalI)$ consisting of the set of edges $E$ 
and the set of linearly independent sets $\mathcalI$ is an example of a matroid.
\index{matroid}
{\bf Matroids} are defined as ordered pairs $(E,\,\mathcalI)$
where $E$ is a finite set, the ground set, and where $\mathcalI$
is a collection of subsets of $E$, called the independent sets,
fulfilling the following conditions:
\begin{enumerate}
\item $\emptyset\in\mathcalI.$
\item If $I\in\mathcalI$ and $I'\subseteq I,$ then $I'\in\mathcalI.$
\item If $I_{1}$ and $I_{2}$ are in $\mathcalI$ and $\left|I_{1}\right|<\left|I_{2}\right|,$
then there is an element $e$ of $I_{2}-I_{1}$ such that $I_{1}\cup e\in\mathcalI.$
\end{enumerate}
All subsets of $E$ which do not belong to $\mathcalI$ are called
dependent. 
The definition goes back to Whitney who wanted 
to describe the properties of linearly independent sets in an abstract way. 
In a similar way as a topology on a space
is given by the distinction between open and closed sets, a matroid
is given by deciding, which of the subsets of a ground set $E$ shall
be called independent and which dependent. 
A matroid can be defined on any kind of ground set, but if we choose $E$ to
be a set of vectors, we can see that the conditions for the independent
sets match with the well-known linear independence of vectors. 

Let us go through the three conditions. The first condition simply says
that the empty set shall be called independent. The second condition states that a
subset of an independent set is again an independent set. 
This is fulfilled for sets of linearly independent vectors as we already have seen in
the above example. 
The third condition is called the independence
augmentation axiom and may be clarified by an example. Consider the sets
\[
I_{1}=\left\{ \left(\begin{array}{c}
1\\
0\\
0\\
0\end{array}\right),\,\left(\begin{array}{c}
0\\
1\\
0\\
0\end{array}\right)\right\} ,\quad I_{2}=\left\{ \left(\begin{array}{c}
1\\
0\\
0\\
0\end{array}\right),\,\left(\begin{array}{c}
0\\
1\\
1\\
1\end{array}\right),\,\left(\begin{array}{c}
1\\
0\\
0\\
1\end{array}\right)\right\}.
\]
Both sets are sets of linearly independent vectors.
The set $I_{2}$ has one element more
than $I_{1}$. $I_{2}-I_{1}$ is the set of vectors in $I_{2}$ which
do not belong to $I_{1}$. The set $I_{2}-I_{1}$ contains for example
$e=(1,\,0,\,0,\,1)^{T}$ and if we include this vector in $I_{1}$
then we obtain again a linearly independent set. The third condition states 
that such an $e$ can be found for any two independent sets with different
numbers of elements. 

The most important origins of examples of matroids are linear algebra
and graph theory. The 
\index{cycle matroid}
{\bf cycle matroid} 
(or 
\index{polygon matroid}
{\bf polygon matroid})
of a graph $G$ is the matroid whose ground set $E$ is given by the
edges of $G$ and whose independent sets $\mathcalI$ are given
by the linearly independent subsets over ${\mathbb Z}_2$ of column vectors in the incidence
matrix of $G$. We can convince ourselves, that $\mathcalI$
fulfils the conditions laid out above.

Let us consider the 
{\bf bases} or 
\index{maximal independent set of a matroid}
{\bf maximal independent sets}
of a matroid $(E,\,\mathcalI).$ These are the sets in $\mathcalI$
which are not proper subsets of any other sets in $\mathcalI$.
The set of these bases of a matroid shall be denoted $\mathcal{B}$
and it can be defined by the following conditions:
\begin{enumerate}
\item $\mathcal{B}$ is non-empty.
\item If $B_{1}$ and $B_{2}$ are members of $\mathcal{B}$ and $x\in B_{1}-B_{2},$
then there is an element $y$ of $B_{2}-B_{1}$ such that $(B_{1}-\{x\})\cup y\in\mathcal{B}.$
\end{enumerate}
One can show, that all sets in $\mathcal{B}$ have the same number
of elements. Furthermore $\mathcalI$ is uniquely determined by
$\mathcal{B}$: it contains the empty set and all subsets of members
of $\mathcal{B}$. 

Let $M=(E,\,\mathcalI)$ be the cycle matroid of a connected
graph $G$ and let $\mathcal{B}(M)$ be the set of bases. Then
one can show that $\mathcal{B}(M)$ consists of the sets of edges
of the spanning trees in $G$. In other words, $T$ is a spanning
tree of $G$ if and only if its set of edges are a basis in $\mathcal{B}(M)$.
We can therefore relate the Kirchhoff polynomial to the bases 
of the cycle matroid:
\bq
\label{chapter_graph_polynomials:eq:basis_generating}
\mathcal{K} & = & \sum_{B_{j}\in\mathcal{B}(M)}\; \prod_{e_{i}\in B_{j}}a_{i}.
\eq
The Kirchhoff polynomial of $G$ is called a 
basis generating polynomial of the matroid $M$ associated to $G$. 
The Kirchhoff polynomial allows us to read off the set of bases $\mathcal{B}(M)$.
Therefore two graphs without any self-loops have the same Kirchhoff polynomial if and
only if they have the same cycle matroid associated to them. 

Let us cure a certain ambiguity which is still left when we consider
cycle matroids and Kirchhoff polynomials and which comes from the
freedom in giving names to the edges of the graph. In the graph of
fig.~\ref{chapter_graph_polynomials:fig17} we could obviously choose different
names for the edges and their edge variables, for example the
edge $e_{2}$ could instead be named $e_{3}$ and vice versa. As a
consequence we would obtain a different cycle matroid where compared
to the above one, $e_{3}$ takes the place of $e_{2}$ and vice versa.
Similarly, we would obtain a different Kirchhoff polynomial, where
the variables $a_{2}$ and $a_{3}$ are exchanged. Of course we are
not interested in any such different cases which simply result from
a change of variable names. Therefore it makes sense to consider classes
of isomorphic matroids and Kirchhoff polynomials. 

Let $M_{1}$ and $M_{2}$ be two matroids and let $E(M_{1})$ and
$E(M_{2})$ be their ground sets, respectively. The matroids $M_{1}$ and $M_{2}$
are called isomorphic if there is a bijection $\psi$ from
$E(M_{1})$ to $E(M_{2})$ such that $\psi(I)$ is an independent
set in $M_{2}$ if and only if $I$ is an independent set in $M_{1}$.
The mentioned interchange of $e_{2}$ and $e_{3}$ in the above example
would be such a bijection: $\psi(e_{2})=e_{3}$, $\psi(e_{3})=e_{2}$,
$\psi(e_{i})=e_{i}$ for $i=1,\,4.$ The new matroid obtained this
way is isomorphic to the above one and its independent sets are given by
interchanging $e_{2}$ and $e_{3}$ in $\mathcalI$ of eq.~(\ref{chapter_graph_polynomials:example_independent_sets}).
In the same sense we want to say that two Kirchhoff polynomials are
isomorphic if they are equal up to bijections on their sets of variables. 

\begin{figure}
\begin{center}
\includegraphics[scale=1.0]{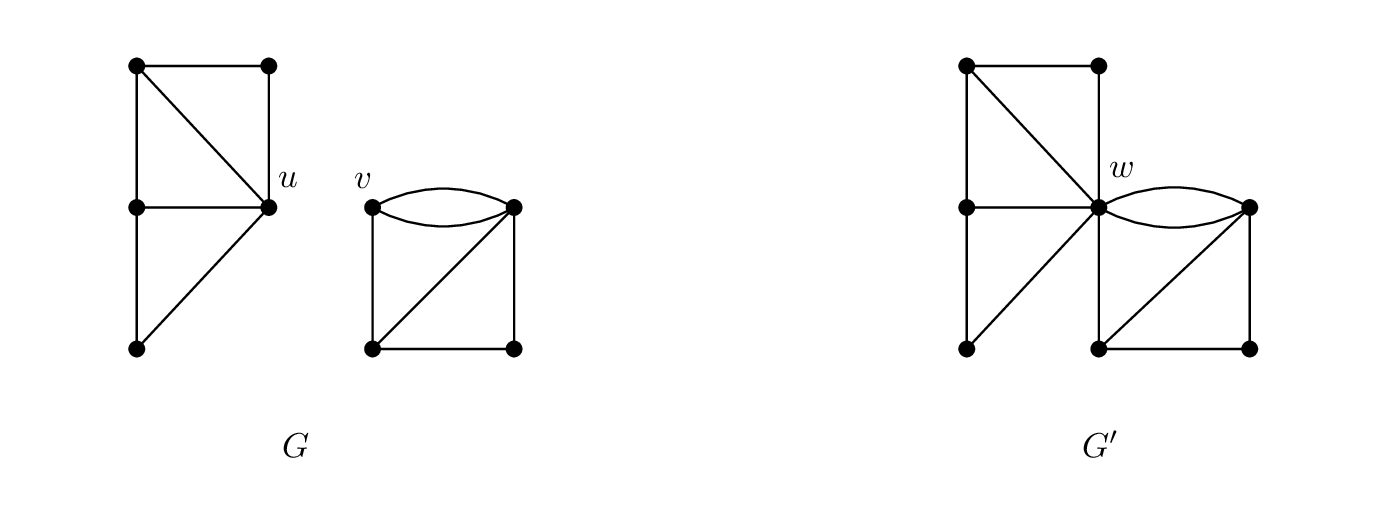}
\end{center}
\caption{Vertex identification and cleaving\label{chapter_graph_polynomials:fig:cleaving identification}}
\end{figure}

\begin{figure}
\begin{center}
\includegraphics[scale=1.0]{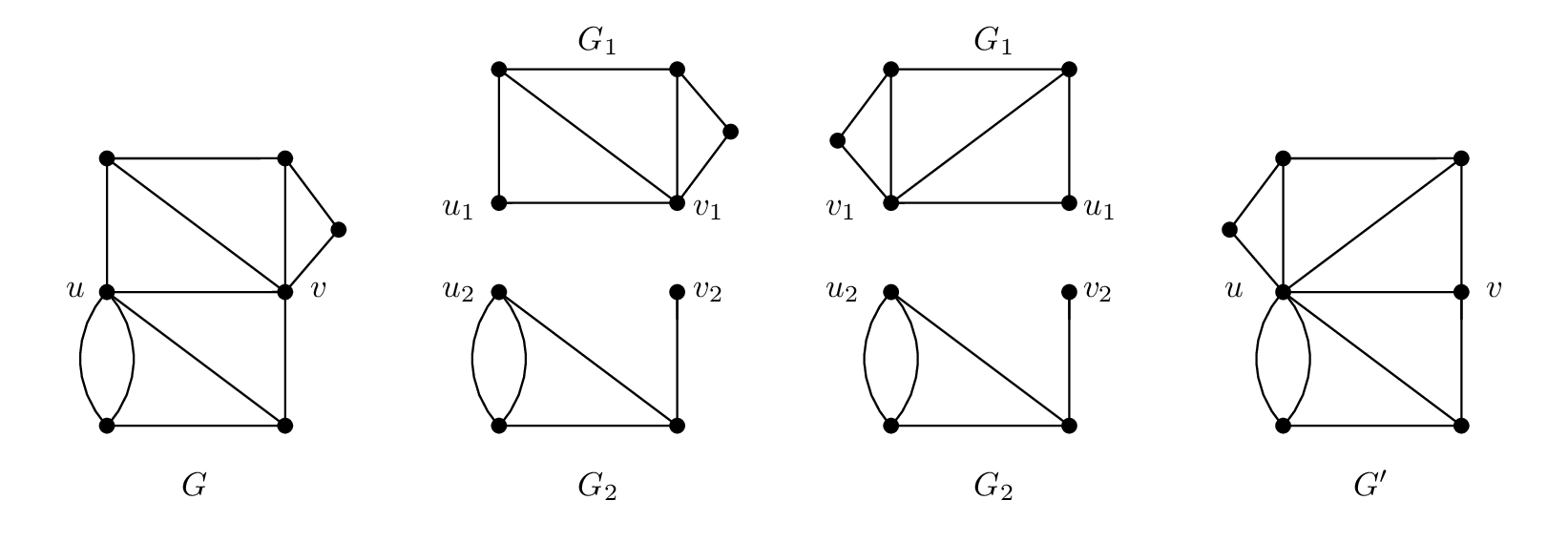}
\end{center}
\caption{Twisting about $u$ and $v$\label{chapter_graph_polynomials:fig:twisting}}
\end{figure}

Now let us come to the question when the Kirchhoff polynomials of
two different graphs are isomorphic, which means that after an appropriate
change of variable names they are equal. From the above discussion
and eq.~(\ref{chapter_graph_polynomials:eq:basis_generating}) it is now clear, that 
a sufficient condition is that the cycle matroids of the graphs are isomorphic.
The question when two graphs have isomorphic cycle matroids was answered
in the following theorem of Whitney \cite{Whitney:1933aa} (also see \cite{Oxley:1986aa})
which was one of the foundational results of matroid theory: 
\begin{tcolorbox}
{\bf Isomorphic cycle matroids}:

Let $G$ and $H$ be graphs having no isolated vertices. Then the
cycle matroids $M(G)$ and $M(H)$ are isomorphic if and only if
$G$ is obtained from $H$ after a sequence of the following three
transformations:
\begin{enumerate}
\item {\bf Vertex identification}: Let $u$ and $v$ be vertices of distinct
components of a graph $G.$ Then a new graph $G'$ is obtained from
the identification of $u$ and $v$ as a new vertex $w$ in $G'$
(see the transition from $G$ to $G'$ in fig.~\ref{chapter_graph_polynomials:fig:cleaving identification}).
\item {\bf Vertex cleaving}: Vertex cleaving is the reverse operation of
vertex identification, such that from cleaving at vertex $w$ in $G'$
we obtain $u$ and $v$ in distinct components of $G$ (see the transition
from $G'$ to $G$ in fig.~\ref{chapter_graph_polynomials:fig:cleaving identification}).
\item {\bf Twisting}: Let $G$ be a graph which is obtained from two disjoint
graphs $G_{1}$ and $G_{2}$ by identifying the vertices $u_{1}$ of $G_{1}$
and $u_{2}$ of $G_{2}$ as a vertex $u$ of $G$ and by identifying
the vertices $v_{1}$ of $G_{1}$ and $v_{2}$ of $G_{2}$ as a vertex $v$
of $G$. Then the graph $G'$ is called the twisting of $G$ about
$\{u,\, v\}$ if it is obtained from $G_{1}$ and $G_{2}$ by identifying
instead $u_{1}$ with $v_{2}$ and $v_{1}$ with $u_{2}$ (see fig.~\ref{chapter_graph_polynomials:fig:twisting}).
\end{enumerate}
\end{tcolorbox}
Proofs can be obtained from \cite{Oxley:1986aa,Whitney:1933aa,Truemper:1980aa,Oxley}. 
Whitney's theorem does not exclude self-loops.
If self-loops are allowed, isomorphic cycle matroids are a sufficient condition
for isomorphic Kirchhoff polynomials, but not a necessary condition, as the next exercise shows:
\\
\\
\bs
{\it \refstepcounter{exercise}
{\bf Exercise \theexercise}: 
\begin{figure}
\begin{center}
\includegraphics[scale=1.0]{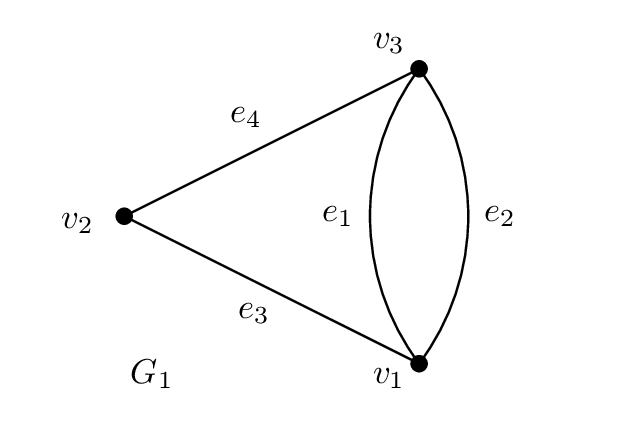}
\includegraphics[scale=1.0]{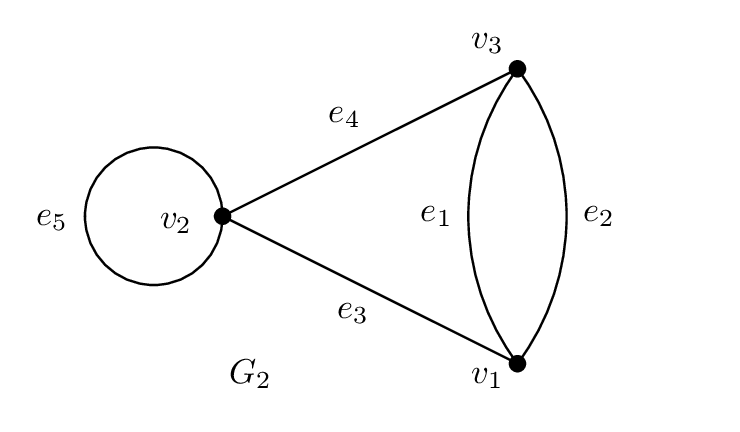}
\end{center}
\caption{\label{chapter_graph_polynomials:fig1735} Two graphs $G_1$ and $G_2$, which differ by the self-loop
formed by $e_5$.}
\end{figure}
Consider the two graphs $G_1$ and $G_2$ shown in fig.~\ref{chapter_graph_polynomials:fig1735}, which differ by a self-loop.
For each of the two graphs, give the Kirchhoff polynomial $\mathcal{K}$ and the first graph polynomial $\mathcal{U}$.
Show that the cycle matroids are not isomorphic.
}
\es
\\
\\
As a consequence of Whitney's theorem, the Kirchhoff polynomials $\mathcal{K}(G)$
and $\mathcal{K}(H)$ of the connected graphs $G$ and $H$, both without any self-loops, are isomorphic
if and only if $G$ is obtained from $H$ by a sequence of the above
three transformations. For the transformations of vertex identification
and vertex cleaving this is obvious from a well-known observation:
If two distinct components $G_{1}$ and $G_{2}$ are obtained from
$G$ after vertex cleaving, then $\mathcal{K}(G)$ is the product
of $\mathcal{K}(G_1)$ and $\mathcal{K}(G_2)$. Therefore any
other graph $G'$ obtained from $G_{1}$ and $G_{2}$ after vertex
identification has the same Kirchhoff polynomial 
$\mathcal{K}(G')=\mathcal{K}(G_1) \cdot \mathcal{K}(G_2)=\mathcal{K}(G)$.
The non-trivial part of the statement on the Kirchhoff polynomials
of $G$ and $H$ is the relevance of the operation of twisting. In
the initial example of fig.~\ref{chapter_graph_polynomials:fig15} the two graphs $G_1^\ast$ and $G_2^\ast$ can be obtained
from each other by twisting.

Let us now discuss the implications for the two graph polynomials $\mathcal{U}$ and $\mathcal{F}_0$.
Consider two connected graphs $G$ and $H$, possibly with external edges and self-loops.
Suppose that $G$ can be obtained from $H$ 
by sequence of vertex identifications, vertex cleavings and twisting. 
Then the first graph polynomials $\mathcal{U}(G)$ and $\mathcal{U}(H)$ are isomorphic.
Denote by $\hat{G}$ the graph obtained from $G$ by merging all external vertices
into a single new vertex $v_\infty$, which connects to all external lines,
as discussed at the end of section~\ref{chapter_graph_polynomials:sect_matrix_tree_theorem}.
Similarly, denote by $\hat{H}$ the graph obtained from $H$ through the same operation.
If $\hat{G}$ can be obtained from $\hat{H}$ 
by sequence of vertex identifications, vertex cleavings and twisting, then
the graph polynomials $\mathcal{F}_0(G)$ and $\mathcal{F}_0(H)$ are isomorphic.
This follows directly from eq.~(\ref{chapter_graph_polynomials:relation_W_K}).
\\
\\
\bs
{\it \refstepcounter{exercise}
{\bf Exercise \theexercise}: 
\begin{figure}
\begin{center}
\includegraphics[scale=1.0]{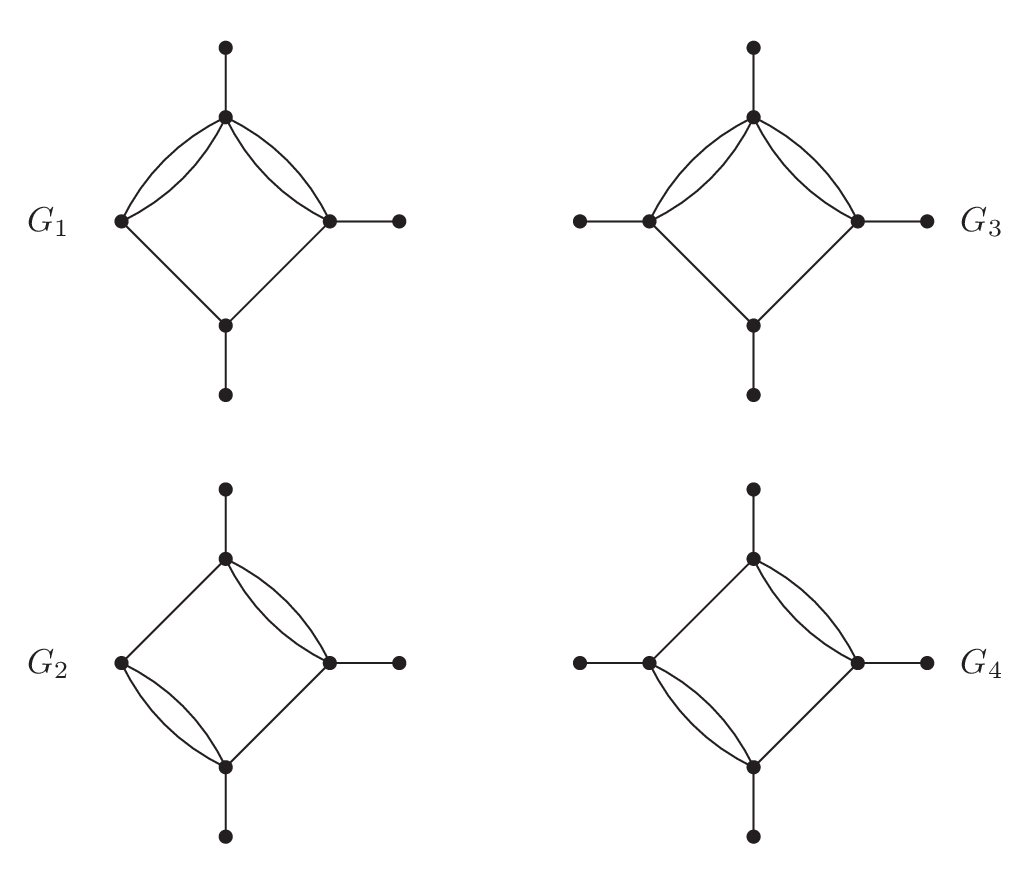}
\end{center}
\caption{\label{chapter_graph_polynomials:fig_twisting} The graphs $G_1$ and $G_2$ have three external edges,
the graphs $G_3$ and $G_4$ have four external edges.}
\end{figure}
Consider first the two graphs $G_1$ and $G_2$ shown in fig.~\ref{chapter_graph_polynomials:fig_twisting},
both with three external legs.
Assume that all internal masses vanish.
Show that
\bq
 \mathcal{U}\left(G_1\right) \; = \; \mathcal{U}\left(G_2\right),
 & &
 \mathcal{F}\left(G_1\right) \; = \; \mathcal{F}\left(G_2\right).
\eq
Consider then the graphs $G_3$ and $G_4$ with four external legs.
Show that
\bq
 \mathcal{U}\left(G_3\right) & = & \mathcal{U}\left(G_4\right),
\eq
but
\bq
 \mathcal{F}\left(G_3\right) & \neq & \mathcal{F}\left(G_4\right).
\eq
}
\es
\\
\\
There is an alternative definition of a matroid: Instead of specifying the set ${\mathcalI}$
of independent sets a matroid can be defined by a 
\index{rank function for a matroid}
{\bf rank function}, 
which associates a non-negative
integer to every sub-set of the ground set.
The rank function has to satisfy for all $S,S' \subseteq E$ the following three conditions:
\begin{enumerate}
\item $\mbox{rk}(S) \le \left|S\right|$.
\item $S' \subset S$ implies $\mbox{rk}(S') \le \mbox{rk}(S)$.
\item $\mbox{rk}(S\cup S') + \mbox{rk}(S\cap S') \le \mbox{rk}(S) + \mbox{rk}(S')$.
\end{enumerate}
The independent sets are exactly those for which $\mbox{rk}(S)=|S|$ holds.
For the cycle matroid of a graph $G$ we can associate to a subset $S$ of $E$ the spanning sub-graph $H$
of $G$ obtained by taking all the vertices of $G$, but just the edges which are in $S$.
In this case the rank of $S$ equals the number of vertices of $H$ minus the number of connected components
of $H$.
The multivariate Tutte polynomial for a matroid is defined by
\bq
\label{chapter_graph_polynomials:def_Tutte_matroid}
\tilde{\mathcal Z}\left(q,a_1,\dots,a_{\nedges}\right)
 & = &
 \sum\limits_{S \subseteq E} q^{-\mbox{rk}(S)}
 \prod\limits_{e_i \in S} a_i.
\eq
It is a polynomial in $1/q$ and $a_1$, \dots, $a_{\nedges}$.
Since a matroid can be defined by giving the rank for each subset $S$ of $E$, it is clear that
the multivariate Tutte polynomial encodes all information of a matroid.
For the cycle matroid of a graph $G$ with $\nvertices$ vertices the multivariate Tutte polynomial $\tilde{\mathcal Z}$
of the matroid is related to the multivariate Tutte polynomial ${\mathcal Z}$ of the graph by
\bq
\tilde{\mathcal Z}\left(q,a_1,\dots,a_{\nedges}\right)
 & = &
 q^{-\nvertices} {\mathcal Z}\left(q,a_1,\dots,a_{\nedges}\right).
\eq
For a matroid there are as well the notions of duality, deletion and contraction.
Let us start with the definition of the dual of a matroid. 
We consider a matroid $M$ with the ground set $E$ and the
set of bases $\mathcal{B}(M)=\{B_{1},\,,B_{2},\,\dots,\, B_{n}\}$.
The dual matroid $M^{\star}$ of $M$ is the matroid with ground set $E$ and whose set of
bases is given by $\mathcal{B}(M^{\star})=\{E-B_{1},\, E-B_{2},\,\dots,\, E-B_{n}\}.$
It should be noted that in contrast to graphs the dual matroid 
can be constructed for any arbitrary matroid.

Deletion and contraction for matroids are defined as follows:
Let us consider a matroid $M=(E,\,\mathcalI)$ with ground
set $E$ and $\mathcalI$ being the set of independent sets. 
Let us divide the ground set $E$ into two disjoint sets $X$ and $Y$:
\bq
 E = X \cup Y, & & X \cap Y = \emptyset.
\eq
We denote by ${\mathcalI}_Y$ the elements of ${\mathcalI}$, which are sub-sets of $Y$.
The matroid $M-X$ is then defined as the matroid with ground set $E-X=Y$ and whose set of independent sets
is given by ${\mathcalI}_Y$.
We say that the matroid $M-X$ is obtained from the matroid $M$ by deleting $X$.
The contracted matroid $M/X$ is defined as follows:
\bq
 M/X & = & \left( M^\ast - X \right)^\ast,
\eq
i.e. one deletes from the dual $M^\ast$ the set $X$ and takes then the dual.
The contracted matroid $M/X$ has the ground set $E-X$.
With these definitions of deletion and contraction we can now state the recursion relation for
the multivariate Tutte polynomial of a matroid:
We have to distinguish two cases. 
If the one-element set $\{e\}$ is of rank zero (corresponding to a self-loop in a graph) we have
\bq
\tilde{\mathcal Z}\left(M\right) 
 & = & \tilde{\mathcal Z}\left(M-\{e\}\right) + a_{e} \tilde{\mathcal Z}\left(M/\{e\}\right).
\eq
Otherwise we have 
\bq
\label{chapter_graph_polynomials:recursion_matroid_non_zero_rank}
\tilde{\mathcal Z}\left(M\right)
 & = & 
 \tilde{\mathcal Z}\left(M-\{e\}\right) + \frac{a_{e}}{q} \tilde{\mathcal Z}\left(M/\{e\}\right).
\eq
The recursion terminates for a matroid with an empty ground set, in this case we have $\tilde{\mathcal Z}=1$.
The fact that one has a different recursion relation for the case where the one-element set
$\{e\}$ is of rank zero is easily understood from the definition of $\tilde{\mathcal Z}$ and the
relation to graphs: 
For a cycle matroid $\tilde{\mathcal Z}$ differs from ${\mathcal Z}$ by the extra factor $q^{-\nvertices}$, where
$\nvertices$ is the number of vertices of the graph.
If $e$ is a self-loop of $G$, the contracted graph $G/e$ equals $G-e$ and in particular it has the same
number of vertices as $G$.
In all other cases the contracted graph $G/e$ has one vertex less than $G$, which is reflected by the 
factor $1/q$ in eq.~(\ref{chapter_graph_polynomials:recursion_matroid_non_zero_rank}).

%% file: quantum_field_theory.tex
\newpage
\chapter{Quantum field theory}
\label{chapter_qft}

We introduced Feynman integrals in chapter~\ref{chapter_basics},
building only on the knowledge of special relativity and graphs.
We did not discuss how Feynman integrals arise in perturbative quantum field theory.
This is of course the main application for Feynman integrals.
In this chapter we fill this gap and give a brief outline in section~\ref{chapter_qft:sect:action}, 
how Feynman integrals
arise in the perturbative expansion for scattering amplitudes in quantum field theory.
This is also covered in depth in many books on quantum field theory and readers not yet familiar
with quantum field theory are invited to consult one of these textbooks \cite{Peskin,Boehm,Srednicki:2007qs,Schwartz}.

We have seen quite early on with the example of eq.~(\ref{chapter_basics:example_divergent})
that Feynman integrals in four space-time dimensions are often divergent.
In order to have a well-defined expression, we introduced dimensional regularisation.
This regulates all divergences.
The original divergences show up as poles in the dimensional regularisation parameter $\eps$.
While this procedure allows us to work with well-defined expressions, it does not tell us anything
what we shall do with these poles.
The answer comes again from quantum field theory. 
In the end we would like to have finite results, where we can take the limit $\eps \rightarrow 0$.
The way divergences cancel is explained in section~\ref{chapter_qft:sect:finite}.

The definition of a Feynman integral in eq.~(\ref{chapter_basics:def_Feynman_integral}) corresponds
to a ``scalar'' integral.
From the Feynman rules for most quantum field theories (like Yang-Mills theory, QED, QCD or more generally
any quantum field theory, which is not a scalar theory) we get Feynman integrals, which are
``tensor'' integrals.
In section~\ref{chapter_qft:sect:tensor_reduction} we show that tensor integrals can be expressed in terms of scalar integrals.
It is therefore sufficient to focus our attention on scalar integrals.

Quantum field theories with spin $1/2$-fermions involve the Dirac matrices and the weak interactions
involve $\gamma_5$.
As we use dimensional regularisation as our regularisation scheme, the Dirac algebra has to be continued
from four space-time dimensions to $D$ space-time dimensions.
For the most part of the Dirac algebra this is straightforward, but the treatment of $\gamma_5$ is
a little bit more subtle.
We discuss this in section~\ref{chapter_qft:sect:dirac_algebra}.

\section{Basics of perturbative quantum field theory}
\label{chapter_qft:sect:action}

Elementary particle physics is described by quantum field theory. To begin with let us start with a single field
$\phi(x)$. Important concepts in quantum field theory are the Lagrangian, the action and the generating functional.
If $\phi(x)$ is a scalar field, a typical Lagrangian is
\bq
\label{chapter_qft:Lagranigan_phi4}
{\mathcal L} & = & 
 \frac{1}{2} \left( \partial_\mu \phi(x) \right) \left( \partial^\mu \phi(x) \right) - \frac{1}{2} m^2 \phi(x)^2
 + \frac{1}{4} \lambda \phi(x)^4.
\eq
The quantity $m$ is interpreted as the mass of the particle described by the field $\phi(x)$, the quantity $\lambda$
describes the strength of the interactions among the particles.
Integrating the Lagrangian over Minkowski space yields the action:
\bq
 S\left[\phi \right] & = & \int d^Dx \; {\mathcal L}\left(\phi \right).
\eq
The action is a functional of the field $\phi$. In order to arrive at the generating functional we introduce
an auxiliary field $J(x)$, called the source field, and integrate over all field configurations $\phi(x)$:
\bq
 Z[J] & = & {\mathcal N} \int {\mathcal D} \phi \; e^{i \left( S[\phi] + \int d^Dx J(x) \phi(x) \right)}.
\eq
The integral over all field configurations is an infinite-dimensional integral. It is called a path integral.
The prefactor ${\mathcal N}$ is chosen such that $Z[0]=1$.
The $\nexternal$-point Green function is given by
\bq
 \langle 0| T( \phi(x_{1}) ... \phi(x_{\nexternal})) |0 \rangle & = &
   \frac{\int {\mathcal D} \phi \; \phi(x_{1}) ... \phi(x_{\nexternal}) e^{i S(\phi)}}
        {\int {\mathcal D} \phi \; e^{i S(\phi)}}.
\eq
With the help of functional derivatives this can be expressed as
\bq
 \langle 0| T( \phi(x_{1}) ... \phi(x_{\nexternal})) |0 \rangle & = &
   \left. \left(-i \right)^{\nexternal}
   \frac{\delta^{\nexternal} Z[J]}{\delta J(x_1) ... \delta J(x_{\nexternal})} \right|_{J=0}.
\eq
We are in particular interested in connected Green functions. These are obtained from a functional $W[J]$, which is
related to $Z[J]$ by
\bq
Z[J] & = & e^{i W[J]}.
\eq
The connected Green functions are then given by
\bq
\label{chapter_qft:functional_derivative}
 G_{\nexternal}\left(x_1,...,x_{\nexternal}\right) 
 & = & 
 \left( -i \right)^{{\nexternal}-1} 
 \left. \frac{\delta^{\nexternal} W[J]}{\delta J(x_1) ... \delta J(x_{\nexternal}) } \right|_{J=0}.
\eq
It is convenient to go from position space to momentum space by a Fourier transformation.
We define the Green functions in momentum space by
\bq
\label{chapter_qft:fourier_trafo}
\lefteqn{
 G_{\nexternal}\left(x_1,...,x_{\nexternal}\right) 
 = } 
 & & \\
 & & 
 \int \frac{d^Dp_1}{(2\pi)^D} ... \frac{d^Dp_{\nexternal}}{(2\pi)^D}
 e^{-i \sum p_j x_j} \left(2 \pi \right)^D \delta^D\left(p_1+...+p_{\nexternal}\right) 
 \tilde{G}_{\nexternal}\left(p_1,...,p_{\nexternal}\right).
 \nonumber
\eq
Note that the Fourier transform $\tilde{G}_{\nexternal}$ is defined by explicitly factoring out the 
$\delta$-function $\delta(p_1+...+p_{\nexternal})$ and a factor $(2 \pi )^D$.
We denote the two-point function in momentum space by $\tilde{G}_2(p)$.
In this case we have to specify only one momentum, since the momentum flowing into the Green function on one side
has to be equal to the momentum flowing
out of the Green function on the other side due to the presence of the $\delta$-function in eq.~(\ref{chapter_qft:fourier_trafo}) .
We now are in a position to define the scattering amplitude: In momentum space the 
\index{scattering amplitude}
{\bf scattering amplitude} 
with $\nexternal$ external particles is
given by the connected $\nexternal$-point Green function multiplied by the inverse two-point function for each external
particle:
\bq
 i {\mathcal A}_{\nexternal}\left(p_1,...,p_{\nexternal}\right)
 & = &
 \tilde{G}_2\left(p_1\right)^{-1}
 ...
 \tilde{G}_2\left(p_{\nexternal}\right)^{-1}
 \tilde{G}_{\nexternal}\left(p_1,...,p_{\nexternal}\right).
\eq
The multiplication with the inverse two-point function for each external
particle amputates the external propagators.
This is the reason, why we distinguish in a graph external and internal edges.

The scattering amplitude enters directly the calculation of a physical observable.
Let us first consider the scattering process of two incoming elementary spinless
particles with no further internal degrees of freedom (like colour) and momenta $p_a'$ and $p_b'$ 
and $(\nexternal-2)$ outgoing particles with momenta $p_1$ to $p_{\nexternal-2}$.
Let us further assume that we are interested in an observable $O\left(p_1,...,p_{\nexternal-2}\right)$ 
which depends on the momenta of the outgoing particles.
In general the observable depends on the experimental set-up 
and 
can be an arbitrary complicated function of the momenta. 
In the simplest case this function is just a constant equal to one,
corresponding to the situation where we count every event with $(\nexternal-2)$ particles in the final state.
In more realistic situations one takes for example 
into account that it is not possible to detect particles close to the beam pipe. 
The function $O$ would then be zero if all final state particles are in this region of phase space.
Furthermore any experiment has a finite resolution.
Therefore it will not be possible to detect particles which are very soft
nor will it be possible to distinguish particles which are very close in angle.
We will therefore sum over the number of final state particles.
In order to obtain finite results within perturbation theory we have to require that 
in the case where one or more particles become unresolved the value of the observable $O$
has a continuous limit agreeing with the value of the observable for a configuration
where the unresolved particles have been merged into ``hard'' (or resolved) pseudo-particles.
Observables having this property are called 
\index{infrared-safe observable}
{\bf infrared-safe observables}. 
The expectation value for an infrared-safe observable $O$ is given by
\bq
\label{chapter_qft:observable_master_electron_positron}
\langle O \rangle
 & = &
 \frac{1}{2 (p_a'+p_b')^2}
 \sum\limits_{\nexternal}
             \int d\phi_{\nexternal-2}
             O\left(p_1,...,p_{\nexternal-2}\right)
             \left| {\mathcal A}_{\nexternal} \right|^2,
\eq
where $1/2/(p_a'+p_b')^2$ is a normalisation factor taking into account the incoming flux.
The phase space measure is given by
\bq
\label{chapter_qft:def_phasespacemeasure}
d\phi_n & = &
 \frac{1}{n! }
 \prod\limits_{i=1}^n \frac{d^{D-1}p_i}{(2 \pi)^{3} 2 E_i} 
 \left(2 \pi \right)^D \delta^D\left(p_a'+p_b'-\sum\limits_{i=1}^n p_i\right).
\;\;\;\;\;\;\;
\eq
The quantity
$E_i$ is the energy of particle $i$, given by
\bq
 E_i 
 \;\; = \;\;
 \sqrt{\vec{p}_i^2 + m_i^2}.
\eq
We see that the expectation value of $O$ is given by the phase space integral over the observable, weighted
by the norm squared of the scattering amplitude.
As the integrand can be a rather complicated function, the phase space integral is usually performed numerically by
Monte Carlo integration.

Let us now look towards a more realistic theory relevant to LHC physics.
LHC is the abbreviation for the Large Hadron Collider at CERN, Geneva.
As an example for a more realistic theory we consider 
quantum chromodynamics
(QCD) consisting of quarks and gluons. 
Quarks and gluons are collectively called partons.
There are a few modifications to eq.~(\ref{chapter_qft:observable_master_electron_positron}). The master formula reads now
\bq
\label{chapter_qft:observable_master_hadron_hadron}
\lefteqn{
\langle O \rangle = 
 \sum\limits_{f_a,f_b}
 \int dx_a f_{f_a}(x_a) \int dx_b f_{f_b}(x_b) 
} & & 
 \\
 & & 
             \frac{1}{2 \hat{s} n_s(a) n_s(b) n_c(a) n_c(b)}
 \sum\limits_{\nexternal}
             \int d\phi_{\nexternal-2}
             O\left(p_1,...,p_{\nexternal-2}\right)
             \sum\limits_{\mathrm{spins,colour}} 
             \left| {\mathcal A}_{\nexternal} \right|^2.
 \nonumber
\eq
The partons have internal degrees of freedom, given by the spin and the colour of the partons.
In squaring the amplitude we sum over these degrees of freedom. 
For the particles in the initial state we would like to average over these degrees of freedom.
This is done by dividing by the factors $n_s(i)$ and $n_c(i)$, giving the number of spin degrees of
freedom ($2$ for quarks and gluons)
and the number of colour degrees of freedom ($3$ for quarks, $8$ for gluons).
The second modification is due to the fact that the particles brought into collision are not partons, but 
composite particles like protons. At high energies the elementary 
constituents of the protons interact and we have to include
a function $f_{f_a}(x_a)$ giving us the probability of finding a parton of flavour $f_a$ 
with momentum fraction $x_a$ of the original
proton momentum inside the proton.
If the momenta of the incoming protons are $P_a'$ and $P_b'$, then the momenta
of the two incoming partons are given by
\bq
 p_a' 
 \;\; = \;\;
 x_a P_a',
 & &
 p_b' 
 \;\; = \;\;
 x_b P_b'.
\eq
$\hat{s}$ is the centre-of-mass energy squared of the two partons entering the hard interaction.
Neglecting particle masses we have
\bq
 \hat{s}
 \;\; = \;\;
 \left(p_a' + p_b'\right)^2
 \;\; = \;\;
 x_a x_b \left(P_a' + P_b'\right)^2.
\eq
In addition there is a small change in eq.~(\ref{chapter_qft:def_phasespacemeasure}). The quantity $(n!)$ is replaced by
$(\prod n_j!)$, where $n_j$ is the number of times a parton of type $j$ occurs in the final state.

It is very convenient to calculate the amplitude with the convention that all particles are out-going.
To this aim we set
\bq
 p_{\nexternal-1} \;\; = \;\; - p_a',
 & &
 p_{\nexternal} \;\; = \;\; - p_b'
\eq
and calculate the amplitude for the momentum configuration
\bq
 \left\{ p_1, ..., p_{\nexternal-2}, p_{\nexternal-1}, p_{\nexternal} \right\}.
\eq
Momentum conservation reads
\bq
 p_1 + ... + p_{\nexternal-2} + p_{\nexternal-1} + p_{\nexternal} & = & 0.
\eq
Note that the momenta $p_{\nexternal-1}$ and $p_{\nexternal}$ have negative energy components.

We have seen through eq.~(\ref{chapter_qft:observable_master_electron_positron}) and eq.~(\ref{chapter_qft:observable_master_hadron_hadron})
that the scattering amplitudes ${\mathcal A}_{\nexternal}$ with $\nexternal$ external particles
enter the theory predictions for an observable $O$.
Thus we need to compute the scattering amplitudes.
Unfortunately, it is usually not possible to calculate the scattering amplitudes exactly.
However, we may calculate scattering amplitudes within perturbation theory, if all couplings
describing the strengths of interactions among the particles are small.
Let us assume for simplicity that there is only one coupling, which we denote by $g$.
We expand the scattering amplitude in powers of $g$:
\bq
\label{chapter_qft:basic_perturbative_expansion}
 {\mathcal A}_{\nexternal} 
 & = & 
 {\mathcal A}_{\nexternal}^{(0)} + {\mathcal A}_{\nexternal}^{(1)} + {\mathcal A}_{\nexternal}^{(2)} + {\mathcal A}_{\nexternal}^{(3)} + ...,
\eq
where ${\mathcal A}_{\nexternal}^{(\loopnumber)}$ contains $({\nexternal}-2+2\loopnumber)$ factors of $g$.
Eq.~(\ref{chapter_qft:basic_perturbative_expansion}) gives the perturbative expansion of the
scattering amplitude.
In this expansion, ${\mathcal A}_{\nexternal}^{(\loopnumber)}$ is an amplitude with ${\nexternal}$ external particles and $\loopnumber$ loops.
The recipe for the computation of ${\mathcal A}_{\nexternal}^{(\loopnumber)}$ based on Feynman diagrams
is as follows:
\begin{tcolorbox}
\begin{myalgorithm}
\label{chapter_qft:algo_amplitude}
Calculation of scattering amplitudes from Feynman diagrams.
\begin{enumerate}
\item Draw all Feynman diagrams for the given number of external particles $\nexternal$ and the given number of loops $\loopnumber$. 
\item Translate each graph into a mathematical formula with the help of the Feynman rules.
\item The quantity $i {\mathcal A}_{\nexternal}^{(\loopnumber)}$ is then given as the sum of all these terms.
\end{enumerate}
\end{myalgorithm}
\end{tcolorbox}
Tree-level amplitudes are amplitudes with no loops and are denoted by ${\mathcal A}_{\nexternal}^{(0)}$.
They give the leading contribution to the full amplitude.
The computation of tree-level amplitudes involves only basic mathematical operations:
Addition, multiplication, contraction of indices, etc..
The above algorithm allows therefore in principle for any $\nexternal$ the computation of the corresponding 
tree-level amplitude.
(In practice, the number of contributing Feynman diagrams is a limiting factor for tree-level amplitudes
with a large number of external particles. For a review of methods to tackle this problem see \cite{Weinzierl:2016bus}.)
The situation is different for loop amplitudes ${\mathcal A}_{\nexternal}^{(\loopnumber)}$ (with $\loopnumber\ge 1$).
Here, the Feynman rules involve an integration over each internal momentum not constrained
by momentum conservation.
That's where Feynman integrals enter the game.

In the algorithm above a Feynman diagram is translated into a mathematical expression with the help of the Feynman rules.
The starting point for a physical theory (or a model) of particle physics is usually the Lagrangian.
In eq.~(\ref{chapter_qft:Lagranigan_phi4}) we specified a scalar $\phi^4$-theory by giving the Lagrangian.
For a more realistic theory let's look at quantum chromodynamics (QCD).
For QCD the Lagrange density reads in Lorenz gauge:
\bq
{\mathcal L}_{\mathrm{QCD}} 
 & = & 
 - \frac{1}{4} F^{a}_{\mu\nu}(x) F^{a \mu\nu}(x)
 - \frac{1}{2 \xi} ( \partial^{\mu} A^{a}_{\mu}(x) )^{2} 
 - \bar{c}^a(x) \partial^\mu D^{ab}_\mu c^b(x),
 \nonumber \\
 & &
 +
 \sum\limits_{\mathrm{quarks } \; q} 
   \bar{\psi}_{q}(x) \left(  i \gamma^{\, \mu} D_{\mu} - m_q \right) \psi_{q}(x), 
\eq
with
\bq
F^{a}_{\mu\nu}(x) & = & \partial_{\mu} A^{a}_{\nu}(x) - \partial_{\nu} A^{a}_{\mu}(x)
 + g f^{abc} A^{b}_{\mu}(x) A^{c}_{\nu}.
\eq
The gluon field is denoted by $A_\mu^a(x)$, the Faddeev-Popov ghost fields are denoted by $c^a(x)$ and 
the quark fields are denoted by $\psi_q(x)$.
The sum is over all quark flavours. The masses of the quarks are denoted by $m_q$.
The quark fields also carry a colour index $j$ and a Dirac index $\alpha$, which we didn't denote explicitly.
The variable $g$ gives the strength of the strong coupling.
QCD is a $SU(3)$-gauge theory. 
Indices referring to the fundamental representation of $SU(3)$ are chosen from the middle of the alphabet $i,j,k,\dots$
and range from $1$ to $3$,
while indices referring to the adjoint representation of $SU(3)$ are chosen from the beginning of the alphabet $a,b,c,\dots$
and range from $1$ to $8$.
The generators of the group $SU(3)$ are denoted by $T^a$ and satisfy
\bq
 \left[ T^a, T^b \right] & = & i f^{abc} T^c.
\eq
The standard normalisation is
\bq
 \mathrm{Tr} \left( T^a T^b \right) & = & T_R \delta^{ab} = \frac{1}{2} \delta^{ab}.
\eq
For later use we set
\bq
 N_c \; = \; 3,
 \;\;\;\;\;\;
 T_R \; = \; \frac{1}{2},
 \;\;\;\;\;\;
 C_A \; = \; N_c \; = \; 3,
 \;\;\;\;\;\;
 C_F \; = \; \frac{N_c^2-1}{2N_c} \; = \; \frac{4}{3}.
\eq
The quantity $F^a_{\mu\nu}$ is called the field strength, 
the quantity $D_\mu=D_{\mu,j k}$ denotes the covariant derivative in the fundamental representation of $SU(3)$,
$D^{ab}_\mu$ denotes the covariant derivative in the adjoint representation of $SU(3)$:
\bq
 D_{\mu,jk} & = & \delta_{jk} \partial_\mu - i g T^a_{jk} A^a_\mu,
 \nonumber \\
  D^{ab}_\mu  & = & \delta^{ab} \partial_\mu - g f^{abc} A^c_\mu.
\eq
The variable $\xi$ is called the gauge-fixing parameter. 
Gauge-invariant quantities like scattering amplitudes are
independent of this parameter.

In order to derive the Feynman rules from the Lagrangian one proceeds as follows:
We first order the terms in the Lagrangian according to the number of fields they involve.
From the terms bilinear in the fields one obtains the propagators, while the terms with three or more
fields give rise to vertices.
Note that a ``normal'' Lagrangian does not contain terms with just one or zero fields.
Furthermore we always assume within perturbation theory that all fields fall off rapidly enough at infinity.
Therefore we can use partial integration and ignore boundary terms.
Using partial integration we may re-write the Lagrangian of $\phi^4$-theory of eq.~(\ref{chapter_qft:Lagranigan_phi4}) as
\bq
\label{chapter_qft:Lagranigan_phi4_expanded}
{\mathcal L} & = & 
 \frac{1}{2} \phi(x) \left[ -\Box - m^2 \right] \phi(x)
 + \frac{1}{4} \lambda \phi(x)^4,
\eq
where we denoted the d'Alembert operator by $\Box=\partial_\mu \partial^\mu$.
As a second example we consider the gluonic part of the QCD Lagrange density:
\bq
\label{chapter_qft:Lagrangian_QCD_expanded}
{\mathcal L}_{\mathrm{QCD}} & = & 
 \frac{1}{2} A^{a}_{\mu}(x) \left[ \partial_\rho \partial^\rho g^{\mu\nu} \delta^{ab}
                  - \left( 1 - \frac{1}{\xi} \right) \partial^\mu \partial^\nu \delta^{ab} \right] A^{b}_{\nu}(x)
 \nonumber \\
 & &
 - g f^{abc} \left( \partial_\mu A^{a}_{\nu}(x) \right) A^{b \mu}(x) A^{c \nu}(x)
 - \frac{1}{4} g^2 f^{eab} f^{ecd} A^{a}_{\mu}(x) A^{b}_{\nu}(x) A^{c \mu}(x) A^{d \nu}(x) 
 \nonumber \\
 & &
 + {\mathcal L}_{\mathrm{quarks}}
 + {\mathcal L}_{\mathrm{FP}}.
\eq
Let us first consider the terms bilinear in the fields, giving rise to the propagators.
A generic term for real boson fields $\phi_i$ has the form
\bq
{\mathcal L}_{\mathrm{bilinear}}(x) 
 & = &
 \frac{1}{2} \phi_i(x) P_{ij}(x) \phi_j(x),
\eq
where $P$ is a real symmetric operator that may contain derivatives and must have an inverse. 
Define the inverse of $P$ by
\bq
 \sum\limits_j P_{ij}(x) P_{jk}^{-1}(x-y) & = & \delta_{ik} \delta^D(x-y),
\eq
and its Fourier transform by
\bq
 P_{ij}^{-1}(x) & = & \int \frac{d^D q}{(2 \pi)^D} e^{-i q \cdot x} \tilde{P}_{ij}^{-1}(q).
\eq
Then the 
\index{propagator, Feynman rule}
{\bf propagator} is given by
\bq
 \Delta_F(q)_{ij} 
 & = & 
 i \tilde{P}_{ij}^{-1}(q).
\eq
Let's see how this works out for the scalar propagator in $\phi^4$-theory and for the gluon propagator in QCD.
We start with $\phi^4$-theory:
From eq.~(\ref{chapter_qft:Lagranigan_phi4_expanded}) we deduce
\bq
 P(x) & = & -\Box - m^2.
\eq
It is not too difficult to show that
\bq
 \tilde{P}^{-1}(q) & = & \frac{1}{q^2-m^2}.
\eq
The propagator of a scalar particle is therefore given by
\bq
\label{chapter_qft:scalar_propagator}
 \Delta_F(k) 
 & = & 
 \begin{picture}(100,20)(0,5)
 \Line(20,10)(70,10)
\end{picture} 
 \;\; = \;\;
 \frac{i}{q^2-m^2}
\eq
and drawn as a line.

Let us now look into a more involved example. We consider the gluon propagator in QCD.
The first line of eq.~(\ref{chapter_qft:Lagrangian_QCD_expanded}) gives the terms bilinear in the gluon fields.
This defines an operator
\bq
 P^{\mu\nu\;ab}(x) & = & \partial_\rho \partial^\rho g^{\mu\nu} \delta^{ab}
                  - \left( 1 - \frac{1}{\xi} \right) \partial^\mu \partial^\nu \delta^{ab}.
\eq
For the propagator we are interested in the inverse of this operator
\bq
\label{chapter_qft:gluon_prop_inverse_x_space}
 P^{\mu\sigma\;ac}(x) \left( P^{-1} \right)_{\sigma\nu}^{cb}(x-y) & = & g^\mu_{\;\;\nu} \delta^{ab} \delta^4(x-y).
\eq
Working in momentum space we are more specifically interested in the Fourier transform of the inverse
of this operator:
\bq
\label{chapter_qft:gluon_prop_Fourier_trafo}
 \left( P^{-1} \right)_{\mu\nu}^{ab}(x) & = & 
  \int \frac{d^D q}{(2 \pi)^D} e^{-i q \cdot x} \left( \tilde{P}^{-1} \right)_{\mu\nu}^{ab}(q).
\eq
The Feynman rule for the propagator is then given by $(\tilde{P}^{-1})_{\mu\nu}^{ab}(q)$ times the imaginary unit.
For the gluon propagator one finds the Feynman rule
\bq
\label{chapter_qft:gluon_propagator}
 \begin{picture}(100,20)(0,5)
 \Gluon(20,10)(70,10){-5}{5}
 \Text(15,12)[r]{\footnotesize $\mu, a$}
 \Text(75,12)[l]{\footnotesize $\nu, b$}
\end{picture} 
 & = & 
  \frac{i}{q^2} \left( - g_{\mu\nu} + \left( 1 -\xi \right) \frac{q_\mu q_\nu}{q^2} \right) \delta^{ab}.
\eq
\bs
{\it \refstepcounter{exercise}
{\bf Exercise \theexercise}: 
Derive eq.~(\ref{chapter_qft:gluon_propagator}) from eq.~(\ref{chapter_qft:gluon_prop_inverse_x_space}) 
and eq.~(\ref{chapter_qft:gluon_prop_Fourier_trafo}).
\\
\\
Hint: It is simpler to work directly in momentum space, using the Fourier representation of
$\delta^D(x-y)$.
}
\es
\\
\\
Let us now consider a generic interaction term with $n\ge 3$ fields.
We may write this term as
\bq
 {\mathcal L}_{\mathrm{int}}(x) & = & 
 O_{i_1 ... i_n}\left(\partial_1,...,\partial_n \right)
 \phi_{i_1}(x) ... \phi_{i_n}(x),
\eq
with the notation that $\partial_j$ acts only on the $j$-th field $\phi_{i_j}(x)$.
For each field we have the Fourier transform
\bq
 \phi_i(x) = 
 \int \frac{d^Dq}{(2\pi)^D} \; e^{-i q x} \; \tilde{\phi}_i(q),
 & &
 \tilde{\phi}_i(q) =
 \int d^Dx \; e^{i q x} \; \phi_i(x),
\eq
where $q$ denotes an in-coming momentum.
We thus have
\bq
 {\mathcal L}_{\mathrm{int}}(x) & = & 
 \int
 \frac{d^Dq_1}{(2\pi)^D} ... \frac{d^Dq_n}{(2\pi)^D}
 e^{-i \left(q_1+...+q_n\right) x}
 O_{i_1 ... i_n}\left(-i q_1,...,-i q_n \right)
 \tilde{\phi}_i\left(q_1\right) ... \tilde{\phi}_i\left(q_n\right).
\eq
Changing to outgoing momenta we replace $q_j$ by $-q_j$.
The 
\index{vertex, Feynman rule}
{\bf vertex} is then given by
\bq
 V & = & 
 i \sum\limits_{\mathrm{permutations}} 
   (-1)^{P_F} O_{i_1 ... i_n}\left(i q_1,...,i q_n \right),
\eq
where the momenta are taken to flow outward.
The summation is over all permutations of indices and momenta of identical particles. 
In the case of identical fermions there is in addition a minus sign for every odd permutation 
of the fermions, indicated by $(-1)^{P_F}$.

Let us also work out some examples here.
We start again with scalar $\phi^4$-theory.
There is only one interaction term, containing four field $\phi(x)$:
\bq
 {\mathcal L}_{\mathrm{int}}(x) & = & 
 \frac{\lambda}{4!}
 \phi(x) \phi(x) \phi(x) \phi(x).
\eq
Thus $O = \lambda/4!$ and the Feynman rule for the vertex is given by
\bq
\begin{picture}(100,30)(0,50)
\Vertex(50,50){2}
\Line(50,50)(71,71)
\Line(50,50)(71,29)
\Line(50,50)(29,29)
\Line(50,50)(29,71)
\end{picture}
 & = &
 i \lambda.
 \\
 \nonumber 
\eq
The factor $1/4!$ is cancelled by summing over the $4!$ permutations of the four identical particles.

It is instructive to consider also a more involved example.
We derive the Feynman rule for the three-gluon vertex in QCD.
The relevant term in the Lagrangian is the first term in the second line of
eq.~(\ref{chapter_qft:Lagrangian_QCD_expanded}):
\bq
 {\mathcal L}_{ggg} & = &
 - g f^{abc} \left( \partial_\mu A^{a}_{\nu}(x) \right) A^{b \mu}(x) A^{c \nu}(x).
\eq
This term contains three gluon fields and will give rise to the three-gluon vertex.
We may rewrite this term as
\bq
 {\mathcal L}_{ggg} 
 & = &
 - g f^{abc} g^{\mu\rho}  \partial^\nu_1 \; A^a_\mu(x) A^b_\nu(x) A^c_\rho(x).
\eq
Thus
\bq
 O^{a b c, \mu \nu \rho}\left(\partial_1,\partial_2,\partial_3\right)
 \;\; = \;\;
 - g f^{abc} g^{\mu\rho}  \partial^\nu_1,
 & &
 O^{a b c, \mu \nu \rho}\left(i q_1, i q_2, i q_3 \right)
 \;\; = \;\;
 - g f^{abc} g^{\mu\rho}  i q^\nu_1.
 \;\;\;\;
\eq
The Feynman rule for the vertex is given by the sum over all permutations of identical particles of the
function $O^{a b c, \mu \nu \rho}(i q_1, i q_2, i q_3 )$ multiplied by the imaginary unit $i$. 
For the case at hand, we have three identical gluons and we have to sum over $3!=6$ permutations.
One finds
\bq
 V_{ggg} & = &
 i
 \sum\limits_{\mathrm{permutations}}
 \left( - g f^{abc} g^{\mu\rho}  i q_1^\nu \right)
  \nonumber \\
 & = &
 - g f^{abc}
 \left[
          g^{\mu\nu} \left( q_1^\rho - q_2^\rho \right)
        + g^{\nu\rho} \left( q_2^\mu - q_3^\mu \right)
        + g^{\rho\mu} \left( q_3^\nu - q_1^\nu \right)
 \right].
\eq
Note that we have momentum conservation at each vertex, for the three-gluon vertex this implies
\bq
 q_1 + q_2 + q_3 & = & 0.
\eq
In a similar way one obtains the Feynman rules for the four-gluon vertex, the ghost-antighost-gluon
vertex and the quark-antiquark-gluon vertex.

Let us summarise the Feynman rules for the propagators and the vertices of QCD:
The gluon propagator (in Feynman gauge, corresponding to $\xi=1$), 
the ghost propagator and the quark propagator are given by
\bq
 \begin{picture}(100,20)(0,5)
 \Gluon(20,10)(70,10){-5}{5}
 \Text(15,12)[r]{\footnotesize $\mu, a$}
 \Text(75,12)[l]{\footnotesize $\nu, b$}
\end{picture} 
& = &
 \frac{-ig^{\mu\nu}}{q^2} \delta^{ab},
 \nonumber \\
\begin{picture}(100,20)(0,5)
 \DashArrowLine(70,10)(20,10){3}
 \Text(15,10)[rb]{\footnotesize $a$}
 \Text(75,10)[lb]{\footnotesize $b$}
\end{picture} 
 & = &
 \frac{i}{q^2} \delta^{ab},
 \nonumber \\
\begin{picture}(100,20)(0,5)
 \ArrowLine(70,10)(20,10)
 \Text(15,10)[rb]{\footnotesize $j$}
 \Text(75,10)[lb]{\footnotesize $k$}
\end{picture} 
 & = &
 i \frac{{\slashed q}+m}{q^2-m^2} \delta_{jk}.
\eq
Here we used the notation ${\slashed q}= q_\mu \gamma^{\, \mu}$.
The Feynman rules for the vertices of QCD are
\bq
\label{chapter_qft:Feynman_rules_vertices}
\begin{picture}(100,35)(0,55)
\Vertex(50,50){2}
\Gluon(50,50)(50,80){3}{4}
\Gluon(50,50)(76,35){3}{4}
\Gluon(50,50)(24,35){3}{4}
\LongArrow(56,70)(56,80)
\LongArrow(67,47)(76,42)
\LongArrow(33,47)(24,42)
\Text(60,80)[lt]{$q_{1}$,$\mu$,$a$}
\Text(78,35)[lc]{$q_{2}$,$\nu$,$b$}
\Text(22,35)[rc]{$q_{3}$,$\rho$,$c$}
\end{picture}
 & = &
 g \left( i f^{abc} \right)
 i
 \left[
          g^{\mu\nu} \left( q_1^\rho - q_2^\rho \right)
        + g^{\nu\rho} \left( q_2^\mu - q_3^\mu \right)
        + g^{\rho\mu} \left( q_3^\nu - q_1^\nu \right)
   \right],
 \nonumber \\
\begin{picture}(100,75)(0,50)
\Vertex(50,50){2}
\Gluon(50,50)(71,71){3}{4}
\Gluon(50,50)(71,29){3}{4}
\Gluon(50,50)(29,29){3}{4}
\Gluon(50,50)(29,71){3}{4}
\Text(72,72)[lb]{$\mu$,$a$}
\Text(72,28)[lt]{$\nu$,$b$}
\Text(28,28)[rt]{$\rho$,$c$}
\Text(28,72)[rb]{$\sigma$,$d$}
\end{picture}
 & = &
   i g^2 \left[
                \left( i f^{abe} \right) \left( i f^{ecd} \right) \left( g^{\mu\rho} g^{\nu\sigma} - g^{\nu\rho} g^{\mu\sigma} \right)
 \right. \nonumber \\
 & & \left.
              + \left( i f^{bce} \right) \left( i f^{ead} \right) \left( g^{\nu\mu} g^{\rho\sigma} - g^{\rho\mu} g^{\nu\sigma} \right)
 \right. \nonumber \\
 & & \left.
              + \left( i f^{cae} \right) \left( i f^{ebd} \right) \left( g^{\rho\nu} g^{\mu\sigma} - g^{\mu\nu} g^{\rho\sigma} \right)
  \right],
 \nonumber \\
\begin{picture}(100,35)(0,55)
\Vertex(50,50){2}
\Gluon(50,50)(80,50){3}{4}
\DashArrowLine(50,50)(29,71){3}
\DashArrowLine(29,29)(50,50){3}
\LongArrow(36,59)(29,66)
\Text(28,71)[rb]{$q_1$,$a$}
\Text(82,50)[lc]{$\mu$,$b$}
\Text(28,29)[rt]{$c$}
\end{picture}
 & = &
 i g \left( i f^{abc} \right) q_1^{\mu}.
 \nonumber \\
 \nonumber \\
 \nonumber \\
\begin{picture}(100,35)(0,55)
\Vertex(50,50){2}
\Gluon(50,50)(80,50){3}{4}
\ArrowLine(50,50)(29,71)
\ArrowLine(29,29)(50,50)
\Text(82,50)[lc]{$\mu$,$a$}
\Text(28,33)[rt]{$l$}
\Text(28,67)[rb]{$j$}
\end{picture}
 & = &
i g \gamma^{\, \mu} T^a_{jl},
 \\
 \nonumber \\ 
 \nonumber
\eq
When translating a Feynman diagram to a mathematical expression, the Feynman rules distinguish
between internal edges and external edges. 
Whereas an internal edge is translated to the mathematical formula for the corresponding propagator, 
an external edge translates to a factor describing the spin polarisation of the corresponding particle.
Thus, there is a 
\index{polarisation vector}
{\bf polarisation vector} $\eps^\mu(q)$ for each external spin-$1$ boson and a 
\index{spinor}
{\bf spinor} 
$\bar{u}(q)$, $u(q)$, $\bar{v}(q)$ or $v(q)$ for each external spin-$1/2$ fermion.
For spin-$0$ bosons there is no non-trivial spin polarisation to be described, hence 
an external edge corresponding to a spin-$0$ boson translates to the trivial factor $1$.

In addition, there are a few additional Feynman rules:
\begin{itemize}
\item There is an integration
\bq
 \int \frac{d^Dq}{(2\pi)^D}
\eq
for each internal momentum not constrained by momentum conservation.
Such an integration is called a ``loop integration'' 
and the number of independent loop integrations in a diagram
is called the loop number of the diagram.

\item A factor $(-1)$ for each closed fermion loop.

\item Each diagram is multiplied by a factor $1/S$, where $S$ is the order of the permutation group
of the internal lines and vertices leaving the diagram unchanged when the external lines are fixed.
\end{itemize}
With the methods outlined above we may obtain the Feynman rules for any theory specified by a Lagrangian.
As examples we considered a scalar $\phi^4$-theory and QCD.
The list of Feynman rules for the various propagators and interaction vertices of the full Standard Model
of particle physics is rather long and not reproduced here.
The Feynman rules for the Standard Model
comprise apart from the Feynman rules for QCD discussed above also
the Feynman rules for the electro-weak sector and the Higgs sector.
These rules can be found in many textbooks of quantum field theory, for example \cite{Boehm}.
However, we would like to show one particular Feynman rule: The Feynman rule
for the coupling of a $Z$-boson to a fermion-antifermion pair reads
\bq
\begin{picture}(100,35)(0,55)
\Vertex(50,50){2}
\Photon(50,50)(80,50){3}{4}
\ArrowLine(50,50)(29,71)
\ArrowLine(29,29)(50,50)
\Text(82,50)[lc]{$\mu$}
\end{picture}
 & = &
 \frac{i e}{2 \sin \theta_W \cos \theta_W} \gamma^{\, \mu} \left( v_f - a_f \gamma_5 \right), 
 \\
 \nonumber \\ 
 \nonumber
\eq
where $e$ denotes the elementary electric charge (i.e. the magnitude of the electric charge of the electron),
$\theta_W$ denotes the Weinberg angle and the quantities $v_f$ and $a_f$ are given by
\bq
 v_f \;\; = \;\; I_3 - 2 Q \sin^2 \theta_W, 
 & &
 a_f \;\; = \;\; I_3.
\eq
Here $Q$ denotes the electric charge of the fermion in units of $e$ and
$I_3$ equals $1/2$ for up-type fermions and $-1/2$ for down-type fermions.
We picked this specific Feynman rule for the following reason: The Feynman rule involves
the Dirac matrix  $\gamma_5$. In four space-time dimensions $\gamma_5$ is defined by
\bq
 \gamma_5 & = & i \gamma^0 \gamma^1 \gamma^2 \gamma^3.
\eq
$\gamma_5$ is an inherently four-dimensional object.
Therefore, the treatment of $\gamma_5$ within dimensional regularisation requires some care
and is discussed in section~\ref{chapter_qft:sect:dirac_algebra}.
\\
\\
\bs
{\it \refstepcounter{exercise}
{\bf Exercise \theexercise}: 
Compute the four-gluon amplitude ${\mathcal A}_4^{(0)}$ from the four diagrams shown in fig.~\ref{chapter_qft:fig1}.
\begin{figure}
\begin{center}
\includegraphics[scale=0.8]{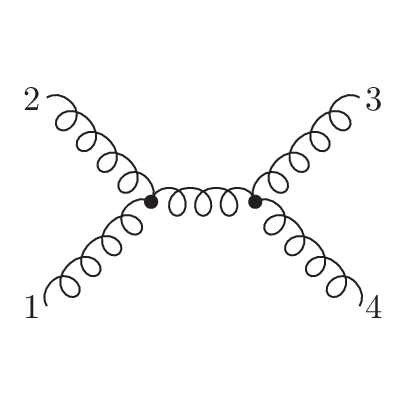}
\hspace*{8mm}
\includegraphics[scale=0.8]{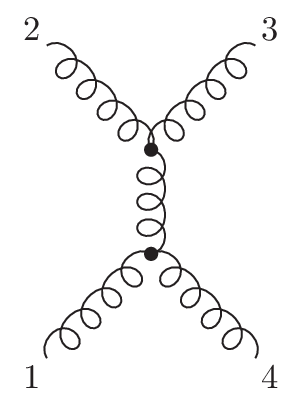}
\hspace*{8mm}
\includegraphics[scale=0.8]{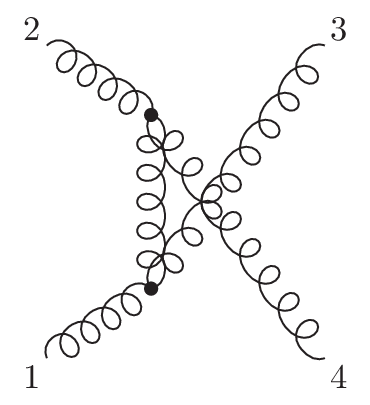}
\hspace*{8mm}
\includegraphics[scale=0.8]{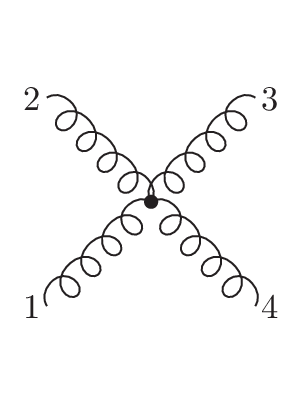}
\caption{\label{chapter_qft:fig1} The four Feynman diagrams contributing to the tree-level four-gluon amplitude ${\mathcal A}_4^{(0)}$.}
\end{center}
\end{figure}
Assume that all momenta are outgoing.
The result will involve scalar products $2 p_i \cdot p_j$, $2 p_i \cdot \eps_j$ and $2 \eps_i \cdot \eps_j$.
For a $2 \rightarrow 2$ process (more precisely for a $0 \rightarrow 4$ process, since we take all momenta to be outgoing), the Mandelstam
variables are defined by
\bq
 s \;\; = \;\; \left( p_1 + p_2 \right)^2,
 \;\;\;
 t \;\; = \;\; \left( p_2 + p_3 \right)^2,
 \;\;\;
 u \;\; = \;\; \left( p_1 + p_3 \right)^2.
\eq
The four momenta $p_1$, $p_2$, $p_3$ and $p_4$ are on-shell, $p_i^2=0$ for $i=1,...,4$, and satisfy
momentum conservation. Derive the Mandelstam relation
\bq 
 s + t + u & = & 0.
\eq
This relation allows to eliminate in the result for ${\mathcal A}_4^{(0)}$ one variable, say $u$.
Furthermore the polarisation vector of gluon $j$ is orthogonal to the momentum of gluon $j$, i.e.
we have the relation $2 p_j \cdot \eps_j = 0$.
Combined with momentum conservation we may eliminate several scalar products
$2 p_i \cdot \eps_j$, such that for a given $j$ we only have 
$2 p_{j-1} \cdot \eps_j$ and $2 p_{j+1} \cdot \eps_j$, where the indices $(j-1)$ and $(j+1)$ are understood modulo $4$.
You might want to use a computer algebra system to carry out the calculations.
The open-source computer algebra systems {\tt FORM} \cite{Vermaseren:2000nd} and {\tt GiNaC} \cite{Bauer:2000cp}
have their roots in particle physics and were originally invented for calculations of this type.
}
\es
\\
\\
Let us summarise:
\begin{tcolorbox}
\index{Feynman rules}
{\bf Feynman rules}:
\begin{itemize}
\item For each internal edge include a propagator.
The propagator is derived from the terms in the Lagrangian bilinear in the fields.
\item For each external edge include a factor, describing the spin polarisation of the particle.
\item For each internal vertex include a vertex factor.
The vertex factor for a vertex of valency $n$ is derived from the terms of the Lagrangian
containing exactly the $n$ fields meeting at this vertex.
\item For each internal momentum not constrained by momentum conservation integrate with measure
\bq
\label{chapter_qft:integral_measure}
 \int \frac{d^Dq}{(2\pi)^D}
\eq
\item Include a factor $(-1)$ for each closed fermion loop.
\item Include a factor $1/S$, where $S$ is the order of the permutation group
of the internal lines and vertices leaving the diagram unchanged when the external lines are fixed.
\end{itemize}
\end{tcolorbox}
Let us now look at an example how to translate a Feynman diagram with a loop into a mathematical
expression.
Fig. \ref{chapter_qft:fig2} shows a Feynman diagram contributing to the one-loop correction
for the process $e^+ e^- \rightarrow q g \bar{q}$.
\begin{figure}
\begin{center}
\includegraphics[scale=1.0]{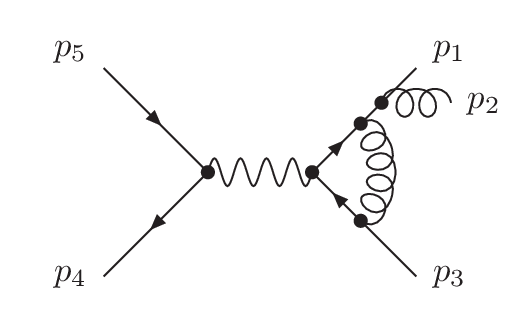}
\end{center}
\caption{\label{chapter_qft:fig2} A one-loop Feynman diagram contributing to the process
$e^+ e^- \rightarrow q g \bar{q}$.}
\end{figure}    
At high energies we can ignore the masses of the electron and the light quarks.
From the Feynman rules one obtains for this diagram:
\bq
\label{chapter_qft:feynmanrules}
- e^2 g^3 
  C_F T^a_{jl} 
  \bar{v}(p_4) \gamma^\mu u(p_5)
  \frac{1}{p_{123}^2}
  \int \frac{d^{D}k_1}{(2\pi)^{D}}
  \frac{1}{k_2^2}
  \bar{u}(p_1) {\slashed \eps}(p_2) \frac{{\slashed p}_{12}}{p_{12}^2}
  \gamma_\nu \frac{{\slashed k}_1}{k_1^2}
  \gamma_\mu \frac{{\slashed k}_3}{k_3^2}
  \gamma^\nu
  v(p_3).
\eq
Here, $p_{12}=p_1+p_2$, $p_{123}=p_1+p_2+p_3$, $k_2=k_1-p_{12}$, $k_3=k_2-p_3$.
Further ${\slashed \eps}(p_2) = \gamma_\tau \eps^\tau(p_2)$, where $\eps^\tau(p_2)$ is the
polarisation vector of the outgoing gluon.
All external momenta are assumed to be
massless: $p_i^2=0$ for $i=1, \dots, 5$.
We can reorganise this formula into a part, which depends on the loop integration and a part, which does not.
The loop integral to be calculated reads:
\bq
\label{chapter_qft:loop_int_example_1}
  \int \frac{d^D k_1}{(2\pi)^{D}}
  \frac{k_1^\rho k_3^\sigma}{k_1^2 k_2^2 k_3^2},
\eq
while the remaining factor, which is independent of the loop integration is given by
\bq
\label{chapter_qft:loop_int_example_remainder}
- e^2 g^3 
  C_F T^a_{jl}
  \bar{v}(p_4) \gamma^\mu u(p_5)
  \frac{1}{p_{123}^2 p_{12}^2}
  \bar{u}(p_1) {\slashed \eps}(p_2) {\slashed p}_{12}
  \gamma_\nu \gamma_\rho
  \gamma_\mu \gamma_\sigma
  \gamma^\nu
  v(p_3).
\eq
The loop integral in eq.~(\ref{chapter_qft:loop_int_example_1}) contains in the denominator three propagator factors
and in the numerator two factors of the loop momentum.
We call a loop integral, in which the loop momentum occurs also in the numerator a 
\index{tensor integral}
``{\bf tensor integral}''.
A loop integral, in which the numerator is independent of the loop momentum is called a 
\index{scalar integral}
``{\bf scalar integral}''.
The scalar integral associated to eq.~(\ref{chapter_qft:loop_int_example_1}) reads
\bq
\label{chapter_qft:loop_int_example_1a}
  \int \frac{d^D k_1}{(2\pi)^{D}}
  \frac{1}{k_1^2 k_2^2 k_3^2}.
\eq
It is always possible to reduce tensor integrals to scalar integrals 
and we will discuss a method to achieve that in section~\ref{chapter_qft:sect:tensor_reduction}.

\section{How to obtain finite results}
\label{chapter_qft:sect:finite}

We have already seen in eq.~(\ref{chapter_basics:result_tadpole_T1_4D}) that the result of a regularised Feynman integral may contain
poles in the regularisation parameter $\eps$.
These poles reflect the original ultraviolet and infrared singularities of the unregularised integral.
What shall we do with these poles? The answer has to come from physics and we distinguish again the case of
UV-divergences and IR-divergences.
The UV-divergences are removed through 
\index{renormalisation}
{\bf renormalisation}.
Ultraviolet divergences are absorbed into a redefinition of the parameters.
As an example we consider the renormalisation of the coupling in QCD:
\bq
 \underbrace{g}_{\mathrm{divergent}} 
 & = & 
 \underbrace{Z_g}_{\mathrm{divergent}} \underbrace{g_r}_{\mathrm{finite}}.
\eq
The renormalisation constant $Z_g$ absorbs the divergent part. However $Z_g$ is not unique: One may always shift
a finite piece from $g_r$ to $Z_g$ or vice versa.
Different choices for $Z_g$ correspond to different 
\index{renormalisation scheme}
{\bf renormalisation schemes}.
Two different renormalisation schemes are always connected by a finite renormalisation.
Note that different renormalisation schemes give numerically different answers.
Therefore one always has to specify the renormalisation scheme.
Some popular renormalisation schemes are the 
\index{on-shell scheme}
{\bf on-shell scheme}, 
where the renormalisation constants are defined by conditions at a scale where the particles are on-shell.
A second widely used scheme is 
\index{modified minimal subtraction}
{\bf modified minimal subtraction} ($\overline{\mathrm{MS}}$-scheme).
In this scheme one always absorbs the combination
\bq
 \Delta & = & \frac{1}{\eps} - \Eulerconstant + \ln 4 \pi
\eq
into the renormalisation constants.
One proceeds similar with all other quantities appearing in the original Lagrangian. For example:
\bq
 A_\mu^a = \sqrt{Z_3} A^a_{\mu,r}, 
 \;\;\;
 \psi_q = \sqrt{Z_2} \psi_{q,r},
 \;\;\;
g = Z_g g_r,
 \;\;\;
m = Z_m m_r,
 \;\;\;
\xi = Z_\xi \xi_r.
\eq
The fact that square roots appear for the field renormalisation is just convention.
Let us look a little bit closer into the coupling renormalisation within dimensional regularisation 
and the $\overline{\mathrm{MS}}$-renormalisation scheme.
Within dimensional regularisation the renormalised coupling $g_r$ is a dimensionfull quantity.
We define a dimensionless quantity $g_R$ by
\bq
\label{chapter_qft:dimensionless_g}
g_r & = & g_R \mu^\varepsilon,
\eq
where $\mu$ is an arbitrary mass scale, called the 
\index{renormalisation scale}
{\bf renormalisation scale}.
From a one-loop calculation one obtains
\bq
\label{chapter_qft:Z_g}
Z_g & = & 
 1 -\frac{1}{2} \beta_0 \frac{g_R^2}{(4 \pi)^2} \Delta + {\mathcal O}(g_R^4),
 \;\;\;\;\;\;
\beta_0 = \frac{11}{3} N_c - \frac{2}{3} N_f.
\eq
$N_c$ is the number of colours and $N_f$ the number of light quarks.
The quantity $g_R$ will depend on the arbitrary scale $\mu$. To derive this dependence one first notes that
the unrenormalised coupling constant $g$ is of course independent of $\mu$:
\bq
\frac{d}{d\mu} g & = & 0
\eq
Substituting $g=Z_g \mu^\eps g_R$ into this equation one obtains
\bq
 \mu \frac{d}{d\mu} g_R
 & = &
 - \eps g_R
 - \left( Z_g^{-1} \mu \frac{d}{d\mu} Z_g \right) g_R.
\eq
From eq.~(\ref{chapter_qft:Z_g}) one obtains
\bq
 Z_g^{-1} \mu \frac{d}{d\mu} Z_g 
 & = & 
 \beta_0 \frac{g_R^2}{(4 \pi)^2} \left( \eps \Delta \right)
 + {\mathcal O}(g_R^4).
\eq
Instead of $g_R$ one often uses the quantity $\alpha_s = g_R^2/(4 \pi)$,
Going to $D=4$ (in this limit we have $\eps \Delta = 1 +{\mathcal O}(\eps)$) one arrives at 
\bq
\mu^2 \frac{d}{d \mu^2} \frac{\alpha_s}{4\pi} 
 & = & 
- \beta_0 \left( \frac{\alpha_s}{4\pi} \right)^2
 + {\mathcal O}\left(\left( \frac{\alpha_s}{4\pi} \right)^3\right)
 + {\mathcal O}\left(\eps\right).
\eq
This differential equation gives the dependence of $\alpha_s$ on the renormalisation scale $\mu$.
At leading order the solution is given by
\bq
\frac{\alpha_s(\mu)}{4 \pi} & = & \frac{1}{\beta_0 \ln \left( \frac{\mu^2}{\Lambda^2} \right)},
\eq
where $\Lambda$ is an integration constant. The quantity $\Lambda$ is called the QCD scale parameter.
For QCD $\beta_0$ is positive and $\alpha_s(\mu)$ decreases with larger $\mu$. This property is called
asymptotic freedom: The coupling becomes smaller at high energies.
In QED $\beta_0$ has the opposite sign and the fine-structure constant $\alpha(\mu)$ increases with larger $\mu$.
The electromagnetic coupling becomes weaker when we go to smaller energies.

Let us now look at the infrared divergences:
We first note that any detector has a finite resolution.
Therefore two particles which are sufficiently close to each other in phase space will be detected as one
particle.
Now let us look again at eqs.~(\ref{chapter_qft:observable_master_electron_positron}) and (\ref{chapter_qft:basic_perturbative_expansion}).
The next-to-leading order term will receive contributions from the interference term of the one-loop amplitude
${\mathcal A}^{(1)}_{\nexternal}$ with the leading-order amplitude ${\mathcal A}^{(0)}_{\nexternal}$, both with $(\nexternal-2)$ final state particles. 
This contribution is of order $g^{2\nexternal-2}$. Of the same order is the square of the leading-order amplitude
${\mathcal A}^{(0)}_{\nexternal+1}$ with $(\nexternal-1)$ final state particles. This contribution we have to take into account whenever
our detector resolves only $(\nexternal-2)$ final-state particles.
It turns out that the phase space integration over the regions where one or more particles become unresolved
is also divergent, and, when performed in $D$ dimensions, leads to poles with the opposite sign as the one
encountered in the loop amplitudes. Therefore the sum of the two contributions is finite.
The {\bf Kinoshita-Lee-Nauenberg theorem} 
\cite{Kinoshita:1962ur,Lee:1964is}
guarantees that all infrared divergences cancel, when summed over all
degenerate physical states.
As an example we consider the NLO corrections to $\gamma^\ast \rightarrow 2 \; \mbox{jets}$, where we treat 
the quarks as massless.
The interference term of the one-loop amplitude with the Born amplitude is given for one flavour by
\bq
2 \; \mbox{Re} \; \left.{\mathcal A}^{(0)}_3\right.^{\ast} {\mathcal A}^{(1)}_3
 & = & 
 \frac{\alpha_s}{\pi} C_F \left( - \frac{1}{\eps^2} - \frac{3}{2\eps} - 4 + \frac{7}{12} \pi^2 \right)
 S_\eps \left| {\mathcal A}^{(0)}_3 \right|^2
 + {\mathcal O}\left(\eps\right).
\eq
$S_\eps=(4\pi)^\eps e^{-\eps\Eulerconstant}$ is the typical phase-space volume factor in $D=4-2\eps$ dimensions.
For simplicity we have set the renormalisation scale $\mu$ equal to the 
centre-of-mass energy squared $s$.
The square of the Born amplitude is given by
\bq
 \left| {\mathcal A}^{(0)}_3 \right|^2
 & = & 16 \pi N_c \alpha \left(1-\eps\right) s.
\eq
This is independent of the final state momenta and the integration over the
phase space can be written as
\bq
 \int d\phi_2 \;
\left( 2 \; \mbox{Re} \; \left.{\mathcal A}^{(0)}_3\right.^{\ast} {\mathcal A}^{(1)}_3 \right)
 = 
 \frac{\alpha_s}{\pi} C_F \left( - \frac{1}{\eps^2} - \frac{3}{2\eps} - 4 + \frac{7}{12} \pi^2 \right)
 S_\eps \int d\phi_2 \; \left| {\mathcal A}^{(0)}_3 \right|^2
 + {\mathcal O}\left(\eps\right).
 \;\;\;\;\;\;
\eq
The real corrections are given by the leading order matrix element for 
$\gamma^\ast \rightarrow q g \bar{q}$ and read
\bq
 \left| {\mathcal A}^{(0)}_4 \right|^2 & = & 128 \pi^2 \alpha \alpha_s C_F N_c ( 1 - \varepsilon)  \left[
         \frac{2}{x_1 x_2} 
         - \frac{2}{x_1}
         - \frac{2}{x_2} 
         + (1-\varepsilon) \frac{x_2}{x_1}
         + (1-\varepsilon) \frac{x_1}{x_2} 
         - 2 \varepsilon 
        \right],
 \;\;\;\;\;\;
\eq
where $x_1=s_{12}/s_{123}$, $x_2=s_{23}/s_{123}$ and $s_{123}=s$ is again the centre-of-mass energy squared.
The quantities $s_{ij}$ and $s_{ijk}$ are defined by
\bq
 s_{ij} \; = \; \left(p_i+p_j\right)^2,
 & &
 s_{ijk} \; = \; \left(p_i+p_j+p_k\right)^2.
\eq
For massless particles these quantities are equal to
\bq
 s_{ij} \; = \; 2 p_i \cdot p_j,
 & &
 s_{ijk} \; = \; 2 p_i \cdot p_j + 2 p_i \cdot p_k + 2 p_j \cdot p_k.
\eq
For this particular simple example we can write the three-particle phase space in $D$ dimensions as
\bq
 d\phi_3 & = & d\phi_2 d\phi_{\mathrm{unres}},
 \nonumber \\
d\phi_{\mathrm{unres}} & = & \frac{\left(4\pi\right)^{\eps-2}}{\Gamma\left(1-\eps\right)}
 s_{123}^{1-\eps} d^3x \delta(1-x_1-x_2-x_3) \left( x_1 x_2 x_3 \right)^{-\eps}.
\eq
Integration over the phase space $\phi_{\mathrm{unres}}$ yields
\bq
 \int d\phi_3 \;
 \left| {\mathcal A}^{(0)}_4 \right|^2 
 & = &
 \frac{\alpha_s}{\pi} C_F \left( \frac{1}{\eps^2} + \frac{3}{2\eps} + \frac{19}{4} - \frac{7}{12} \pi^2 \right)
 S_\eps \int d\phi_2 \; \left| {\mathcal A}^{(0)}_3 \right|^2
 + {\mathcal O}\left(\eps\right).
\eq
We see that in the sum the poles cancel and we obtain the finite result
\bq
 \int d\phi_2 \;
\left( 2 \; \mbox{Re} \; \left.{\mathcal A}^{(0)}_3\right.^{\ast} {\mathcal A}^{(1)}_3 \right)
 +
 \int d\phi_3 \;
 \left| {\mathcal A}^{(0)}_4 \right|^2 
 & = &
 \frac{3}{4} C_F \frac{\alpha_s}{\pi} 
 \int d\phi_2 \; \left| {\mathcal A}^{(0)}_3 \right|^2
 + {\mathcal O}\left(\eps\right).
\eq
In this example we have seen the cancellation of the infrared (soft and collinear) singularities 
between the virtual and the real corrections according to the Kinoshita-Lee-Nauenberg theorem.
In this example we integrated over the phase space of all final state particles.
In practise one is often interested in differential distributions.
In these cases the cancellation is technically more complicated, as the different contributions live
on phase spaces of different dimensions and one integrates only over restricted regions of phase space.
Methods to overcome this obstacle are known under the name ``phase-space slicing'' and ``subtraction method''
\cite{Giele:1992vf,Giele:1993dj,Keller:1998tf,Frixione:1996ms,Catani:1997vz,Dittmaier:1999mb,Phaf:2001gc,Catani:2002hc}.

The Kinoshita-Lee-Nauenberg theorem is related to the finite experimental resolution in detecting
final state particles.
In addition we have to discuss initial state particles.
Let us go back to eq.~(\ref{chapter_qft:observable_master_hadron_hadron}). The differential cross section
we can write schematically
\bq
 d \sigma_{H_1 H_2} 
 & = & 
 \sum\limits_{a,b} 
 \int dx_a f_{H_1 \rightarrow a}(x_a)
 \int dx_b f_{H_2 \rightarrow b}(x_b) 
 d \sigma_{ab}(x_a,x_b),
\eq
where $f_{H \rightarrow a}(x)$ is the parton distribution function, giving us
the probability to find a parton of type $a$ in a hadron of type $H$ carrying a fraction
$x$ to $x+dx$ of the hadron's momentum.
$d \sigma_{ab}(x_a,x_b)$ is the differential cross section for the scattering of partons $a$ and $b$.
Now let us look at the parton distribution function $f_{a \rightarrow b}$ of a parton inside another parton.
At leading order this function is trivially given by $\delta_{ab} \delta(1-x)$, but already at the next order
a parton can radiate off another parton and thus loose some of its momentum and/or convert to another flavour.
One finds in $D$ dimensions
\bq
f_{a \rightarrow b}(x,\eps) & = & \delta_{ab} \delta(1-x)
- \frac{1}{\eps} \frac{\alpha_s}{4 \pi} P^{0}_{a \rightarrow b}(x) 
+ O(\alpha_s^2),
\eq
where $P^{0}_{a \rightarrow b}$ is the lowest order Altarelli-Parisi
splitting function.
To calculate a cross section $d \sigma_{H_1 H_2}$ at NLO involving parton densities one first 
calculates the cross section $d \hat{\sigma}_{a b}$ where the hadrons $H_1$ and $H_2$ are replaced
by partons $a$ and $b$ to NLO:
\bq
d \hat{\sigma}_{a b} & = & d \hat{\sigma}^0_{a b} + \frac{\alpha_s}{4 \pi} d \hat{\sigma}^1_{a b}
+ O(\alpha_s^2)
\eq
The hard scattering part $d \sigma_{a b}$ is then obtained by inserting the perturbative
expansions for $d \hat{\sigma}_{a b}$ and $f_{a \rightarrow b}$ into 
the factorisation formula.
\bq
d \hat{\sigma}^0_{a b} + \frac{\alpha_s}{4 \pi} d \hat{\sigma}^1_{a b} & = &
d \sigma^0_{a b} + \frac{\alpha_s}{4 \pi} d \sigma^1_{a b} 
- \frac{1}{\eps} \frac{\alpha_s}{4 \pi} \sum\limits_c \int dx_1 P^0_{a \rightarrow c} d \sigma^0_{c b}
- \frac{1}{\eps} \frac{\alpha_s}{4 \pi} \sum\limits_d \int dx_2 P^0_{b \rightarrow d} d \sigma^0_{a d}.
 \nonumber
\eq
One therefore obtains for the LO- and the NLO-terms of the hard scattering part
\bq
d \sigma^0_{a b} & = & d \hat{\sigma}^0_{a b} \nonumber \\
d \sigma^1_{a b} & = & d \hat{\sigma}^1_{a b}
+ \frac{1}{\eps} \sum\limits_c \int dx_1 P^0_{a \rightarrow c} d \hat{\sigma}^0_{c b}
+ \frac{1}{\eps} \sum\limits_d \int dx_2 P^0_{b \rightarrow d} d \hat{\sigma}^0_{a d}.
\eq
The last two terms remove the collinear initial state singularities in $d \hat{\sigma}^1_{a b}$.

\section{Tensor reduction}
\label{chapter_qft:sect:tensor_reduction}

In section~\ref{chapter_qft:sect:action} we listed the Feynman rules for a theory like QCD
and worked out one example in eq.~(\ref{chapter_qft:feynmanrules})
how a one-loop Feynman diagram translates into a mathematical formula
involving a Feynman integral.
The attentive reader will have noticed, that the Feynman integral in eq.~(\ref{chapter_qft:loop_int_example_1})
does not fit directly the definition of Feynman integrals in eq.~(\ref{chapter_basics:def_Feynman_integral}).
There are two issues here:
One issue is rather trivial and concerns prefactors:
If we follow standard conventions within quantum field theory, the propagator of a scalar particle is given by
(see eq.~(\ref{chapter_qft:scalar_propagator}))
\bq
 \frac{i}{q^2-m^2},
\eq
and the integral measure for every internal momentum not constrained by momentum conservation is given by
(see eq.~(\ref{chapter_qft:integral_measure}))
\bq
 \int \frac{d^Dq}{(2\pi)^D}.
\eq
On the other hand, we used in chapter~\ref{chapter_basics} the convention, that an edge corresponds
to (see eq.~(\ref{chapter_basics:summary_scalar_propagator}))
\bq
 \frac{1}{-q^2+m^2},
\eq
and the integral measure is given by
(see eq.~(\ref{chapter_basics:summary_integral_measure}))
\bq
 \int \frac{d^Dq}{i \pi^{\frac{D}{2}}}.
\eq
In addition, we included in chapter~\ref{chapter_basics} a factor
(see eq.~(\ref{chapter_basics:summary_prefactor}))
\bq
 e^{\loopnumber \eps \Eulerconstant} \left(\mu^2\right)^{\nu-\frac{\loopnumber D}{2}}
\eq
for each Feynman integral.
This issue is rather trivial and only concerns prefactors: If we are able to compute a Feynman integral
within one convention for prefactors, we also are able to compute this Feynman integral within any other
convention for prefactors.
The only point to remember is, that we should not forget to adjust the prefactors appropriately in the final result.

The second issue is more serious:
The integral in eq.~(\ref{chapter_qft:loop_int_example_1}) is a tensor integral, while in chapter~\ref{chapter_basics}
we only discussed scalar integrals.
Thus we have to show that any tensor integral can always be reduced to a linear combination
of scalar integrals.
This can be done if we allow in the 
linear combination of scalar integrals shifted space-time dimensions $(D \rightarrow D+2$)
and raised propagators ($\nu_j \rightarrow \nu_j+1$) \cite{Tarasov:1996br,Tarasov:1997kx}.
We denote a 
\index{tensor integral}
{\bf tensor integral} 
by giving the numerator in square brackets:
\bq
\label{chapter_qft:def_tensor_integral}
 I_{\nu_1 \dots \nu_{\ninternal}}\left(D,x_1,\dots,x_{\NB}\right)\left[k_{i_1}^{\mu_1} \dots k_{i_t}^{\mu_t}\right]
 & = &
 e^{\loopnumber \eps \Eulerconstant} \left(\mu^2\right)^{\nu-\frac{\loopnumber D}{2}}
 \int \prod\limits_{r=1}^{\loopnumber} \frac{d^Dk_r}{i \pi^{\frac{D}{2}}} 
 \frac{k_{i_1}^{\mu_1} \dots k_{i_t}^{\mu_t}}{\prod\limits_{j=1}^{\ninternal} \left(-q_j^2+m_j^2\right)^{\nu_j}},
 \nonumber \\
 & & 
 k_{i_1}, \dots, k_{i_t} \; \in \; \left\{ k_1, \dots, k_{\loopnumber} \right\}.
\eq
In this notation, a 
\index{scalar integral}
{\bf scalar integral} is an integral, where the numerator equals one:
\bq
\label{chapter_qft:def_scalar_integral}
 I_{\nu_1 \dots \nu_{\ninternal}}\left(D,x_1,\dots,x_{\NB}\right)\left[1\right]
 & = &
 I_{\nu_1 \dots \nu_{\ninternal}}\left(D,x_1,\dots,x_{\NB}\right)
 \nonumber \\
 & = &
 e^{\loopnumber \eps \Eulerconstant} \left(\mu^2\right)^{\nu-\frac{\loopnumber D}{2}}
 \int \prod\limits_{r=1}^{\loopnumber} \frac{d^Dk_r}{i \pi^{\frac{D}{2}}} 
 \frac{1}{\prod\limits_{j=1}^{\ninternal} \left(-q_j^2+m_j^2\right)^{\nu_j}}.
\eq
Let us introduce two operators ${\bf D}^+$ and ${\bf D}^-$, which raise, respectively lower, the
number of space-time dimensions by two units:
\bq
 {\bf D}^\pm
 I_{\nu_1 \dots \nu_{\ninternal}}\left(D,x_1,\dots,x_{\NB}\right)
 & = &
 I_{\nu_1 \dots \nu_{\ninternal}}\left(D \pm 2,x_1,\dots,x_{\NB}\right).
\eq
The operators ${\bf D}^\pm$ are called 
\index{dimensional-shift operator}
{\bf dimensional-shift operators}.
In addition, we introduce 
\index{raising operator}
{\bf raising operators} ${\bf j}^+$ (with $j \in \{1,\dots,\ninternal\}$), which raise
the power of the propagator $j$ by one unit:
\bq
\label{chapter_qft:def_raising_operator}
 {\bf j}^+
 I_{\nu_1 \dots \nu_j \dots \nu_{\ninternal}}\left(D,x_1,\dots,x_{\NB}\right)
 & = &
 \nu_j \cdot
 I_{\nu_1 \dots \left(\nu_j+1\right) \dots \nu_{\ninternal}}\left(D,x_1,\dots,x_{\NB}\right).
\eq
Note that we defined ${\bf j}^+$ such that it raises the index $\nu_j \rightarrow \nu_j+1$ and multiplies
the integral with a factor $\nu_j$.
With this definition we have for example
\bq
 \left( {\bf j}^+ \right)^2
 I_{\nu_1 \dots \nu_j \dots \nu_{\ninternal}}\left(D,x_1,\dots,x_{\NB}\right)
 & = &
 \nu_j \left( \nu_j+1\right) \cdot
 I_{\nu_1 \dots \left(\nu_j+2\right) \dots \nu_{\ninternal}}\left(D,x_1,\dots,x_{\NB}\right).
 \;\;\;
\eq
Let us now study how the operators ${\bf D}^+$ and ${\bf j}^+$ act on the integrand of the Schwinger parameter
representation of a scalar Feynman integral. We will see that they act in a rather simple way.
Let us write for the graph polynomials in the Schwinger parameter
representation ${\mathcal U}={\mathcal U}(\alpha)$ and ${\mathcal F}={\mathcal F}(\alpha)$.
From 
\bq
I_{\nu_1 \dots \nu_{\ninternal}}\left(D\right)  & = &
 \frac{e^{\loopnumber \eps \Eulerconstant}}{\prod\limits_{k=1}^{\ninternal}\Gamma(\nu_k)}
 \;
 \int\limits_{\alpha_k \ge 0}  d^{\ninternal}\alpha \;
 \left( \prod\limits_{k=1}^{\ninternal} \alpha_k^{\nu_k-1} \right)
 \frac{1}{{\mathcal U}^{\frac{D}{2}}}
 e^{- \frac{{\mathcal F}}{{\mathcal U}}}
\eq
we find
\bq
\label{chapter_qft:shift_relations}
 {\bf D}^+ I_{\nu_1 \dots \nu_{\ninternal}}\left(D\right)  & = &
 \frac{e^{\loopnumber \eps \Eulerconstant}}{\prod\limits_{k=1}^{\ninternal}\Gamma(\nu_k)}
 \;
 \int\limits_{\alpha_k \ge 0}  d^{\ninternal}\alpha \;
 \left( \prod\limits_{k=1}^{\ninternal} \alpha_k^{\nu_k-1} \right)
 \frac{1}{{\mathcal U} \cdot {\mathcal U}^{\frac{D}{2}}}
 e^{- \frac{{\mathcal F}}{{\mathcal U}}},
 \nonumber \\
 {\bf j}^+ I_{\nu_1 \dots \nu_j \dots \nu_{\ninternal}}\left(D\right)  & = &
 \frac{e^{\loopnumber \eps \Eulerconstant}}{\prod\limits_{k=1}^{\ninternal}\Gamma(\nu_k)}
 \;
 \int\limits_{\alpha_k \ge 0}  d^{\ninternal}\alpha \;
 \left( \prod\limits_{k=1}^{\ninternal} \alpha_k^{\nu_k-1} \right)
 \frac{\alpha_j}{{\mathcal U}^{\frac{D}{2}}}
 e^{- \frac{{\mathcal F}}{{\mathcal U}}}.
\eq
We see that an additional factor of the first graph polynomial ${\mathcal U}$ in the denominator corresponds
to a shift $D \rightarrow D+2$.
An additional factor of the Schwinger parameter $\alpha_j$ in the numerator corresponds to the application
of ${\bf j}^+$ (i.e. a multiplication of the integral by $\nu_j$ and a shift $\nu_j \rightarrow \nu_j+1$).
\\
\\
\bs
{\it \refstepcounter{exercise}
{\bf Exercise \theexercise}: 
Let $n \in {\mathbb N}$. Show that the action of $({\bf j}^+)^n$ on the integrand  
of the Schwinger parameter representation is given by
\bq
\label{chapter_qft:shift_relations_jplus_power_n}
 \left( {\bf j}^+ \right)^n I_{\nu_1 \dots \nu_j \dots \nu_{\ninternal}}\left(D\right)  & = &
 \frac{e^{\loopnumber \eps \Eulerconstant}}{\prod\limits_{k=1}^{\ninternal}\Gamma(\nu_k)}
 \;
 \int\limits_{\alpha_k \ge 0}  d^{\ninternal}\alpha \;
 \left( \prod\limits_{k=1}^{\ninternal} \alpha_k^{\nu_k-1} \right)
 \frac{\alpha_j^n}{{\mathcal U}^{\frac{D}{2}}}
 e^{- \frac{{\mathcal F}}{{\mathcal U}}}.
\eq
}
\es
\\
\\
Applying ${\bf j}^+$ to a Feynman integral with $\nu_j=0$ gives zero, due to explicit prefactor $\nu_j$
in eq.~(\ref{chapter_qft:def_raising_operator}):
\bq
 {\bf j}^+
 I_{\nu_1 \dots \nu_{j-1} 0 \nu_{j+1} \dots \nu_{\ninternal}}
 & = &
 0 \cdot
 I_{\nu_1 \dots \nu_{j-1} 1 \nu_{j+1} \dots \nu_{\ninternal}}
 \; = \; 0.
\eq

The algorithm for reducing tensor integrals to scalar integrals proceeds as follows:
We start from a tensor integral in the momentum representation.
For each propagator we introduce a Schwinger parameter as in eq.~(\ref{chapter_basics:Schwinger_trick_propagator}).
We then obtain an integral over the loop momenta and the Schwinger parameters similar to eq.~(\ref{chapter_basics:loop_momenta_and_Schwinger_integral}), but with the additional tensor structure in the integrand.
The argument of the exponential function is as in the scalar case the quadric
(see eq.~(\ref{chapter_basics:eq_poly_calc_1}))
\bq
 \sum\limits_{j=1}^{\ninternal} \alpha_{j} (-q_j^2+m_j^2)
 & = & 
 - \sum\limits_{r=1}^{\loopnumber} \sum\limits_{s=1}^{\loopnumber} k_r M_{rs} k_s + \sum\limits_{r=1}^{\loopnumber} 2 k_r \cdot v_r + J.
\eq
By a suitable change of the independent loop momenta variables $k_r \rightarrow k_r'$ we bring this 
quadric to the form
\bq
 \sum\limits_{j=1}^{\ninternal} \alpha_{j} (-q_j^2+m_j^2)
 & = & 
 - \sum\limits_{r=1}^{\loopnumber} \lambda_r {k_r'}^2 + J'.
\eq
This decouples the $\loopnumber$ momentum integrations and we can treat each momentum integration
separately.
We have to consider integrals of the form
\bq
 \int \frac{d^{D}k}{i\pi^{D /2}} k^{\mu_1} \dots k^{\mu_t} f(k^2),
\eq
where $f(k^2)=e^{\lambda k^2}$.
Integrals with an odd power of the loop momentum in the numerator vanish by symmetry:
\bq
\label{chapter_qft:odd_power_loop_momenta}
 \int \frac{d^{D}k}{i\pi^{D /2}} k^{\mu_1} \dots k^{\mu_{2t-1}} f(k^2)
 & = &
 0,
 \;\;\;\;\;\; t \; \in {\mathbb N}.
\eq
Integrals with an even power of the loop momentum must be proportional to a 
symmetric tensor build from the metric tensor due to Lorentz symmetry.
For the simplest cases we have
\bq
\label{chapter_qft:symmetric_integration}
\int \frac{d^{D }k}{i\pi^{D /2}} k^\mu k^\nu f(k^2) & = & 
 - \frac{1}{D } g^{\mu\nu} \int \frac{d^{D }k}{i\pi^{D /2}} (-k^2) f(k^2), \\
\int \frac{d^{D }k}{i\pi^{D /2}} k^\mu k^\nu k^\rho k^\sigma f(k^2) & = & 
 \frac{1}{D (D +2)} 
  \left( g^{\mu\nu} g^{\rho\sigma} + g^{\mu\rho} g^{\nu\sigma} + g^{\mu\sigma} g^{\nu\rho} \right) 
  \int \frac{d^{D }k}{i\pi^{D /2}} (-k^2)^2 f(k^2). \nonumber
\eq
The generalisation to arbitrary higher tensor structures is obvious.
\\
\\
\bs
{\it \refstepcounter{exercise}
{\bf Exercise \theexercise}: 
Work out the corresponding formula for
\bq
\int \frac{d^{D }k}{i\pi^{D /2}} k^{\mu_1} k^{\mu_2} k^{\mu_3} k^{\mu_4} k^{\mu_5} k^{\mu_6} f(k^2).
\eq
}
\es
\\
\\
Each loop momentum integral is now of the form
\bq
\label{chapter_qft:final_momentum_integration}
 \int \frac{d^Dk}{i\pi^{D /2}} \left(-k^2\right)^a e^{\lambda k^2} & = & 
 \int \frac{d^DK}{\pi^{D /2}} \left(K^2\right)^a e^{-\lambda K^2}
 \;\; = \;\;
 \frac{\Gamma\left(\frac{D}{2}+a\right)}{\Gamma\left(\frac{D}{2}\right)}
 \frac{1}{\lambda^{\frac{D}{2}+a}}.
\eq
This leaves us with the Schwinger parameter integrals.
The change of variables $k_r \rightarrow k_r'$ and the integration in eq.~(\ref{chapter_qft:final_momentum_integration})
may introduce additional powers of the Schwinger parameters in the numerator and additional powers
of the first graph polynomial ${\mathcal U}$ in the denominator.
With the help of eq.~(\ref{chapter_qft:shift_relations})
we may write these integrals as scalar integrals with raised powers of the propagators and shifted space-time
dimensions.
This completes the algorithm for the tensor reduction.

Let us consider an example.
We consider the two-loop double-box graph shown in fig.~\ref{chapter_basics:fig_doublebox}
for the case
\bq
 & & p_1^2 = 0, \;\;\; p_2^2 = 0, \;\;\; p_3^2 = 0, \;\;\; p_4^2 = 0,
 \nonumber \\
 & & m_1 = m_2 = m_3 = m_4 = m_5 = m_6 = m_7 = 0.
\eq
This example is a continuation of the example discussed in section~\ref{chapter_basics:subsection:Schwinger_parameter}.
Suppose we would like to reduce the tensor integral
\bq
 I_{1111111}\left(D\right)\left[ k_1^\mu k_2^\nu \right]
 & = &
 e^{2\eps \Eulerconstant} \left(\mu^2\right)^{7-D}
 \int 
 \frac{d^Dk_1}{i \pi^{\frac{D}{2}}} 
 \frac{d^Dk_2}{i \pi^{\frac{D}{2}}} 
 \frac{k_1^\mu k_2^\nu}{\prod\limits_{j=1}^{7} \left(-q_j^2\right)}
\eq
to scalar integrals.
Introducing Schwinger parameters we have
\bq
 I_{1111111}\left(D\right)\left[ k_1^\mu k_2^\nu \right]
 & = &
 e^{2\eps \Eulerconstant} \left(\mu^2\right)^{7-D}
 \int\limits_{\alpha_j \ge 0}  d^7\alpha \;
 \int 
 \frac{d^Dk_1}{i \pi^{\frac{D}{2}}} 
 \frac{d^Dk_2}{i \pi^{\frac{D}{2}}} 
 k_1^\mu k_2^\nu
 e^{-\sum\limits_{j=1}^{7} \alpha_j \left(-q_j^2\right)}.
\eq
We may write the argument of the exponential function
as
\bq
 \sum\limits_{j=1}^7 \alpha_j \left(-q_j^2\right) 
 & = &
 - \alpha_{1234} \left(k_1 - \frac{1}{\alpha_{1234}} \left( \alpha_{12} p_1 + \alpha_2 p_2 - \alpha_4 k_2 \right) \right)^2 
 \nonumber \\
 & &
 - \frac{{\mathcal U}}{\alpha_{1234}} 
    \left( k_2 - \frac{1}{{\mathcal U}} \left( - \alpha_{12} \alpha_4 p_1 - \alpha_2 \alpha_4 p_2 + \alpha_{1234} \alpha_5 p_3 + \alpha_{1234} \alpha_{57} p_4 \right) \right)^2
 \nonumber \\
 & &
 + \mu^2 \frac{{\mathcal F}}{{\mathcal U}}.
\eq
Here we used the notation $\alpha_{i_1 i_2 \dots i_n}=\alpha_{i_1} + \alpha_{i_2} + \dots + \alpha_{i_n}$.
We substitute
\bq
 k_1' & = & k_1 - \frac{1}{\alpha_{1234}} \left( \alpha_{12} p_1 + \alpha_2 p_2 - \alpha_4 k_2 \right)
\eq
followed by
\bq
 k_2' & = & 
 k_2 - \frac{1}{{\mathcal U}} \left( - \alpha_{12} \alpha_4 p_1 - \alpha_2 \alpha_4 p_2 + \alpha_{1234} \alpha_5 p_3 + \alpha_{1234} \alpha_{57} p_4 \right).
\eq
This gives
\bq
 k_1
 & = &
 k_1' - \frac{\alpha_4}{\alpha_{1234}} k_2'
 + \frac{1}{{\mathcal U}} \left[ \alpha_{4567} \left( \alpha_{12} p_1 + \alpha_2 p_2 \right) - \alpha_4 \left(\alpha_5 p_3+\alpha_{57} p_4\right)\right],
 \nonumber \\
 k_2 & = & 
 k_2' + \frac{1}{{\mathcal U}} \left[ - \alpha_4 \left( \alpha_{12} p_1 + \alpha_2 p_2 \right) + \alpha_{1234} \left( \alpha_5 p_3 + \alpha_{57} p_4 \right) \right].
\eq
The Jacobian of the transformation $(k_1,k_2) \rightarrow (k_1',k_2')$ is one.
With the help of eq.~(\ref{chapter_qft:odd_power_loop_momenta}) 
we obtain
\bq
\lefteqn{
 I_{1111111}\left(D\right)\left[ k_1^\mu k_2^\nu \right]
 = } & & \\
 & &
 e^{2\eps \Eulerconstant} \left(\mu^2\right)^{7-D}
 \int\limits_{\alpha_j \ge 0}  d^7\alpha \;
 \int 
 \frac{d^Dk_1'}{i \pi^{\frac{D}{2}}} 
 \frac{d^Dk_2'}{i \pi^{\frac{D}{2}}} 
 e^{\alpha_{1234} {k_1'}^2 +\frac{{\mathcal U}}{\alpha_{1234}} {k_2'}^2 - \mu^2 \frac{{\mathcal F}}{{\mathcal U}}}
 \left\{ - \frac{\alpha_4}{\alpha_{1234}} {k_2'}^\mu {k_2'}^\nu 
 \right. 
 \nonumber \\
 & & \left.
 + \frac{1}{{\mathcal U}^2} 
    \left[ \alpha_{4567} \left( \alpha_{12} p_1^\mu + \alpha_2 p_2^\mu \right) 
           - \alpha_4 \left(\alpha_5 p_3^\mu+\alpha_{57} p_4^\mu\right)\right]
    \left[ - \alpha_4 \left( \alpha_{12} p_1^\nu + \alpha_2 p_2^\nu \right) 
 \right. \right. 
 \nonumber \\
 & & \left. \left.
 + \alpha_{1234} \left( \alpha_5 p_3^\nu + \alpha_{57} p_4^\nu \right) \right] 
 \right\}.
 \nonumber
\eq
Let us consider two terms in more detail (the others are similar).
We first consider the term proportional to ${k_2'}^\mu {k_2'}^\nu$:
\bq
\lefteqn{
 \tilde{I}_1
 \; = \;
 - e^{2\eps \Eulerconstant} \left(\mu^2\right)^{7-D}
 \int\limits_{\alpha_j \ge 0}  d^7\alpha \;
 \int 
 \frac{d^Dk_1'}{i \pi^{\frac{D}{2}}} 
 \frac{d^Dk_2'}{i \pi^{\frac{D}{2}}} 
 \frac{\alpha_4}{\alpha_{1234}} 
 {k_2'}^\mu {k_2'}^\nu 
 e^{\alpha_{1234} {k_1'}^2 +\frac{{\mathcal U}}{\alpha_{1234}} {k_2'}^2 - \mu^2 \frac{{\mathcal F}}{{\mathcal U}}}
 } & & \nonumber \\
 & = &
 \frac{g^{\mu\nu}}{D}
 e^{2\eps \Eulerconstant} \left(\mu^2\right)^{7-D}
 \int\limits_{\alpha_j \ge 0}  d^7\alpha \;
 \frac{\alpha_4}{\alpha_{1234}} 
 \int 
 \frac{d^Dk_1'}{i \pi^{\frac{D}{2}}} 
 \frac{d^Dk_2'}{i \pi^{\frac{D}{2}}} 
 \left(-{k_2'}^2\right)
 e^{\alpha_{1234} {k_1'}^2 +\frac{{\mathcal U}}{\alpha_{1234}} {k_2'}^2 - \mu^2 \frac{{\mathcal F}}{{\mathcal U}}}.
 \nonumber
\eq
With
\bq
 \int
 \frac{d^Dk_1'}{i \pi^{\frac{D}{2}}} 
 e^{\alpha_{1234} {k_1'}^2}
 & = &
 \frac{1}{\alpha_{1234}^{\frac{D}{2}}},
 \nonumber \\
 \int
 \frac{d^Dk_2'}{i \pi^{\frac{D}{2}}} 
 \left(-{k_2'}^2\right)
 e^{\frac{{\mathcal U}}{\alpha_{1234}} {k_2'}^2}
 & = &
 \frac{\Gamma\left(\frac{D}{2}+1\right)}{\Gamma\left(\frac{D}{2}\right)}
 \left(\frac{\alpha_{1234}}{{\mathcal U}} \right)^{\frac{D}{2}+1}
\eq
we obtain
\bq
 \tilde{I}_1
 & = &
 \frac{g^{\mu\nu}}{2}
 e^{2\eps \Eulerconstant}
 \int\limits_{\alpha_j \ge 0}  d^7\alpha \;
 \frac{\alpha_4}{{\mathcal U}^{\frac{D}{2}+1}}
 e^{- \frac{{\mathcal F}}{{\mathcal U}}}
 \; = \;
 \frac{1}{2} g^{\mu\nu} \; I_{1112111}\left(D+2\right).
\eq
Note that the powers of $\alpha_{1234}$ have cancelled out.
As second term we consider
\bq
 \tilde{I}_2
 & =&
 -
 e^{2\eps \Eulerconstant} \left(\mu^2\right)^{7-D}
 \int\limits_{\alpha_j \ge 0}  d^7\alpha \;
 \int 
 \frac{d^Dk_1'}{i \pi^{\frac{D}{2}}} 
 \frac{d^Dk_2'}{i \pi^{\frac{D}{2}}} 
 \frac{\left( \alpha_{4} \alpha_{1} p_1^\mu \right) \left( \alpha_4 \alpha_{1} p_1^\nu \right)}{{\mathcal U}^2} 
 e^{\alpha_{1234} {k_1'}^2 +\frac{{\mathcal U}}{\alpha_{1234}} {k_2'}^2 - \mu^2 \frac{{\mathcal F}}{{\mathcal U}}}
 \nonumber \\
 & = &
 -
 p_1^\mu p_1^\nu
 e^{2\eps \Eulerconstant} \left(\mu^2\right)^{7-D}
 \int\limits_{\alpha_j \ge 0}  d^7\alpha \;
 \frac{\alpha_{1}^2 \alpha_{4}^2}{{\mathcal U}^2} 
 \int 
 \frac{d^Dk_1'}{i \pi^{\frac{D}{2}}} 
 \frac{d^Dk_2'}{i \pi^{\frac{D}{2}}} 
 e^{\alpha_{1234} {k_1'}^2 +\frac{{\mathcal U}}{\alpha_{1234}} {k_2'}^2 - \mu^2 \frac{{\mathcal F}}{{\mathcal U}}}
 \nonumber \\
 & = &
 -
 p_1^\mu p_1^\nu
 e^{2\eps \Eulerconstant}
 \int\limits_{\alpha_j \ge 0}  d^7\alpha \;
 \frac{\alpha_{1}^2 \alpha_{4}^2}{{\mathcal U}^{\frac{D}{2}+2}}
 e^{- \frac{{\mathcal F}}{{\mathcal U}}}
 \; = \;
 -
 4 p_1^\mu p_1^\nu
 I_{3113111}\left(D+4\right).
\eq
All other terms are similar to the last one.
Thus we are able to express the tensor integral 
$I_{1111111}(D)[ k_1^\mu k_2^\nu ]$ in terms of scalar integrals.

\section{Dimensional regularisation and spins}
\label{chapter_qft:sect:dirac_algebra}

We introduced dimensional regularisation to regulate divergent Feynman integrals.
In the calculation of amplitudes the tensor integrals are multiplied by loop momenta independent prefactors.
In theories with particles with spin, these prefactors include the polarisation factors for the external particles
(i.e. spinors for spin-$1/2$ fermions and polarisation vectors for spin-$1$ bosons).
Within dimensional regularisation we have to specify how to continue these polarisation factors 
from four space-time dimensions to $D$ space-time dimensions.

There are several schemes on the market which treat this issue differently.
To discuss these schemes it is best to look how they treat the momenta and the polarisation factors
of observed and unobserved particles.
Unobserved particles are particles circulating inside loops or emitted particles not resolved within
a given detector resolution.
The most commonly used schemes are the 
\index{conventional dimensional regularisation scheme}
{\bf conventional dimensional regularisation scheme} (CDR) \cite{Collins}, 
where all momenta and all polarisation factors are taken to be in $D$ dimensions
(the momenta of the observed particles can be taken to lie in a four-dimensional sub-space of the $D$-dimensional space) and
the 
\index{'t Hooft-Veltman scheme}
{\bf 't Hooft-Veltman scheme} (HV) \cite{'tHooft:1972fi,Breitenlohner:1977hr}, 
where the momenta and the polarisation factors of the unobserved particles are $D$-dimensional,
whereas the momenta and the polarisation factors of the observed particles are four-dimensional.

Let us also mention two further schemes, dimensional reduction and the four-dimensional helicity scheme.
These two schemes introduce an additional space of dimension $D_s$.
In the modern formulation of 
\index{dimensional reduction}
{\bf dimensional reduction} (DRED) \cite{Siegel:1979wq,Siegel:1980qs,Gnendiger:2017pys}
all momenta are taken to be in $D$ dimensions, whereas
all polarisation factors are taken to be in $D_s$ dimensions.
As above, the momenta of the observed particles can be taken to lie in a four-dimensional sub-space of the $D$-dimensional space.
In the 
\index{four-dimensional helicity scheme}
{\bf four-dimensional helicity scheme} (FDH) \cite{Bern:1992aq,Weinzierl:1999xb,Bern:2002zk}
the momenta of the unobserved particles are $D$-dimensional,
the momenta of the observed particles are four-dimensional,
the polarisation factors of the unobserved particles are $D_s$-dimensional
and
the polarisation factors of the observed particles are four-dimensional.
Let us summarise:
\begin{tcolorbox}
\index{dimensional regularisation schemes}
{\bf Dimensional regularisation schemes for particles with spin}:
\begin{center}
\begin{tabular}{l|cccc}
 & momenta & momenta & polarisation & polarisation \\
 & observed & unobserved & observed & unobserved \\
 \hline 
 CDR & $D$ & $D$ & $D$ & $D$ \\
 HV & $4$ & $D$ & $4$ & $D$ \\
 DRED & $D$ & $D$ & $D_s$ & $D_s$ \\
 FDH & $4$ & $D$ & $4$ & $D_s$ \\
\end{tabular}
\end{center}
\end{tcolorbox}
One assumes that the four-dimensional space can be embedded into the $D$-dimensional space,
and that the $D$-dimensional space can be embedded into the $D_s$-dimensional space.
Thus, there is a projection from the $D_s$-dimensional space to the $D$-dimensional space, 
which forgets the $(D_s-D)$-dimensional components.
Likewise, there is a projection from the $D$-dimensional space to the four-dimensional space,
which forgets the $(D-4)$-dimensional components.
This implies for example the algebraic rules
\bq
 g^{(4)}_{\mu\rho} g^{(D) \; \rho}_{\;\;\;\;\;\;\;\;\;\;\nu} \; = \; g^{(4)}_{\mu \nu},
 \;\;\;\;\;\;
 g^{(D)}_{\mu\rho} g^{(D_s) \; \rho}_{\;\;\;\;\;\;\;\;\;\;\;\;\nu} \; = \; g^{(D)}_{\mu \nu},
 \;\;\;\;\;\;
 g^{(4)}_{\mu\rho} g^{(D_s) \; \rho}_{\;\;\;\;\;\;\;\;\;\;\;\;\nu} \; = \; g^{(4)}_{\mu \nu}.
\eq
As mentioned in section~\ref{chapter_basics:dimensional_regularisation}
we may realise spaces of non-integer dimensions as equivalence classes of tuples of vector space.
In this construction, the quantities $D_s$, $D$ and $4$ corresponds to the rank of the tuples of vector spaces.

In dimensional reduction and in the four-dimensional helicity scheme the space of dimension $D_s$
is only used for the polarisations.
The final result will be an analytic function of $D_s$.
As the polarisations only enter in the numerator, the limit $D_s \rightarrow 4$ will not lead to
additional poles.
Thus, we may take the limit $D_s \rightarrow 4$ at the end of the calculation without any problems.
This also explains the name ``four-dimensional helicity scheme''.

It is possible to relate results obtained in one scheme to another scheme, using simple
and universal transition formulae \cite{Kunszt:1994sd,Signer:PhD,Catani:1997pk}.

\subsection{The Dirac algebra within dimensional regularisation}

In four space-time dimensions the Dirac matrices are $4 \times 4$-matrices satisfying
the anti-commu\-tation relation
\bq
\label{chapter_qft:Dirac_anticommutation}
 \left\{ \gamma^\mu_{(4)} , \gamma^\nu_{(4)} \right\} & = & 2 g^{\mu\nu}_{(4)} \cdot {\bf 1},
\eq
where${\bf 1}$ denotes the unit matrix in spinor space.
The hermitian properties are
\bq
\label{chapter_qft:Dirac_hermitian}
 \left( \gamma^0_{(4)} \right)^\dagger \; = \; \gamma^0_{(4)}, 
 \;\;\;\;\;\;
 \left( \gamma^i_{(4)} \right)^\dagger \; = \; - \gamma^i_{(4)}, \;\;\; 1 \le i \le 3.
\eq
The matrix $\gamma_5$ is defined by
\bq
 \gamma_5 & = & i \gamma_{(4)}^0 \gamma_{(4)}^1 \gamma_{(4)}^2 \gamma_{(4)}^3
  \;\; = \;\;
 \frac{i}{24} \eps_{\mu\nu\rho\sigma} \gamma_{(4)}^\mu \gamma_{(4)}^\nu \gamma_{(4)}^\rho \gamma_{(4)}^\sigma,
\eq
where $\eps_{\mu\nu\rho\sigma}$ denotes the totally anti-symmetric tensor with 
$\eps_{0 1 2 3}=1$.
The matrix $\gamma_5$ satisfies
\bq
 \left\{ \gamma_{(4)}^\mu, \gamma_5 \right\} \; = \; 0,
 & &
 \gamma_5^2 \; = \; {\bf 1}
\eq
and
\bq
 \gamma_5^\dagger & = & \gamma_5.
\eq
In evaluating traces of Dirac matrices we have the following rules:
\begin{enumerate}
\item Traces of an even number of Dirac matrices are evaluated with the rules
\bq
\label{chapter_qft:Dirac_trace_rule_1}
 \mathrm{Tr}\left(\gamma_{(4)}^\mu \gamma_{(4)}^\nu\right) 
 & = & 
 4 g_{(4)}^{\mu \nu},
 \nonumber \\
 \mathrm{Tr}\left(\gamma_{(4)}^{\mu_1} \gamma_{(4)}^{\mu_2} ... \gamma_{(4)}^{\mu_{2 n}}\right)
 & = & 
 \sum\limits_{j=2}^{2n} \left(-1\right)^j
 g_{(4)}^{\mu_1 \mu_j} \mathrm{Tr}\left(\gamma_{(4)}^{\mu_2} ... \gamma_{(4)}^{\mu_{j-1}} \gamma_{(4)}^{\mu_{j+1}} ... \gamma_{(4)}^{\mu_{2 n}}\right).
\eq
\item Traces of an odd number of Dirac matrices vanish:
\bq
\label{chapter_qft:Dirac_trace_rule_2}
 \mathrm{Tr}\left(\gamma_{(4)}^{\mu_1} \gamma_{(4)}^{\mu_2} ... \gamma_{(4)}^{\mu_{2 n-1}}\right) 
 & = & 0
\eq
\item For traces involving $\gamma_5$ we have
\bq
\label{chapter_qft:Dirac_trace_rule_3}
 \mathrm{Tr}\left(\gamma_5\right) 
 & = & 0,
 \nonumber \\
 \mathrm{Tr}\left(\gamma_{(4)}^{\mu} \gamma_{(4)}^{\nu} \gamma_5\right) 
 & = & 0,
 \nonumber \\
 \mathrm{Tr}\left(\gamma_{(4)}^{\mu} \gamma_{(4)}^{\nu} \gamma_{(4)}^{\rho} \gamma_{(4)}^{\sigma} \gamma_5\right) 
 & = & 4 i \eps^{\mu \nu \rho \sigma}.
\eq
\end{enumerate}
In particular, we have a non-zero value for 
$\mathrm{Tr}(\gamma_{(4)}^{\mu} \gamma_{(4)}^{\nu} \gamma_{(4)}^{\rho} \gamma_{(4)}^{\sigma} \gamma_5)$.
\\
\\
\bs
{\it \refstepcounter{exercise}
{\bf Exercise \theexercise}: 
Prove eqs.~(\ref{chapter_qft:Dirac_trace_rule_1})-(\ref{chapter_qft:Dirac_trace_rule_3}).
}
\es
\\
\\
Let us now consider the Dirac algebra in $D$ dimensions.
The generalisation of eq.~(\ref{chapter_qft:Dirac_anticommutation}) reads
\bq
\label{chapter_qft:Dirac_anticommutation_D_dim}
 \left\{ \gamma^\mu_{(D)} , \gamma^\nu_{(D)} \right\} & = & 2 g^{\mu\nu}_{(D)} \cdot {\bf 1}.
\eq
This is unproblematic. ${\bf 1}$ denotes again the unit matrix in spinor space.
If $D$ is a positive even integer, the standard representation of the Dirac matrices is
given by matrices of size $2^{\frac{D}{2}} \times 2^{\frac{D}{2}}$.
It is common practice to use for simplicity the convention that the trace of the unit matrix in spinor space equals
\bq
\label{chapter_qft:Dirac_normalisation}
 \mathrm{Tr}\left( {\bf 1} \right)
 & = & 4,
\eq
and not the more natural choice from a mathematical point of view
\bq
 \mathrm{Tr}\left( {\bf 1} \right)
 & = & 2^{\frac{D}{2}}.
\eq
As a trace of Dirac matrices is always associated with a closed fermion loop, we may convert easily between these
two conventions.
We will follow standard conventions and use the convention of eq.~(\ref{chapter_qft:Dirac_normalisation}) from now on.

However, there is no way of continuing the definition of $\gamma_5$ to $D$ dimensions,
maintaining the cyclicity of the trace and the relations
\bq
\label{chapter_qft:gamma_5_conditions}
 \left\{ \gamma_{(D)}^\mu, \gamma_5 \right\} & = & 0,
 \;\;\;\;\;\;\;\;\;\;\;\;\;\;\;\;\;\;\;\;
 0 \; \le \; \mu \; \le \; D-1,
 \nonumber \\
 \mathrm{Tr}\left(\gamma_{(D)}^{\mu} \gamma_{(D)}^{\nu} \gamma_{(D)}^{\rho} \gamma_{(D)}^{\sigma} \gamma_5\right) 
 & = & 4 i \eps^{\mu \nu \rho \sigma},
 \;\;\;\;\;\;\;\;\;
 \mu, \nu, \rho, \sigma \; \in \; \{0,1,2,3\}.
\eq
In order to see this, consider
\bq
\label{chapter_qft:gamma_5_contradiction}
 g^{(D)}_{\alpha\beta} \eps_{\mu\nu\rho\sigma} 
 \mathrm{Tr}\left(\gamma_{(D)}^\alpha \gamma_{(D)}^\mu \gamma_{(D)}^\nu \gamma_{(D)}^\rho \gamma_{(D)}^\sigma \gamma_{(D)}^\beta \gamma_5 \right).
\eq
We first note that from eq.~(\ref{chapter_qft:Dirac_anticommutation_D_dim}) we have
\bq
\label{chapter_qft:Dirac_contraction}
 \gamma^\mu_{(D)} \gamma_\mu^{(D)}
 & = &
 D \cdot {\bf 1}.
\eq
We then evaluate the expression in eq.~(\ref{chapter_qft:gamma_5_contradiction}) in two ways:
We first use the cyclicity of the trace, anti-commute $\gamma_5$ with $\gamma_{(D)}^\alpha$ and use
eq.~(\ref{chapter_qft:Dirac_contraction}):
\bq
\label{chapter_qft:Dirac_contraction_1}
 g^{(D)}_{\alpha\beta} \eps_{\mu\nu\rho\sigma} 
 \mathrm{Tr}\left(\gamma_{(D)}^\alpha \gamma_{(D)}^\mu \gamma_{(D)}^\nu \gamma_{(D)}^\rho \gamma_{(D)}^\sigma \gamma_{(D)}^\beta \gamma_5 \right)
 & = &
 g^{(D)}_{\alpha\beta} \eps_{\mu\nu\rho\sigma} 
 \mathrm{Tr}\left(\gamma_{(D)}^\mu \gamma_{(D)}^\nu \gamma_{(D)}^\rho \gamma_{(D)}^\sigma \gamma_{(D)}^\beta \gamma_5 \gamma_{(D)}^\alpha \right)
 \nonumber \\
 & = &
 - g^{(D)}_{\alpha\beta} \eps_{\mu\nu\rho\sigma} 
 \mathrm{Tr}\left(\gamma_{(D)}^\mu \gamma_{(D)}^\nu \gamma_{(D)}^\rho \gamma_{(D)}^\sigma \gamma_{(D)}^\beta \gamma_{(D)}^\alpha \gamma_5 \right)
 \nonumber \\
 & = &
 - D \eps_{\mu\nu\rho\sigma} 
 \mathrm{Tr}\left(\gamma_{(D)}^\mu \gamma_{(D)}^\nu \gamma_{(D)}^\rho \gamma_{(D)}^\sigma \gamma_5 \right).
\eq
On the other hand, we may anti-commute $\gamma_{(D)}^\alpha$ through the string $\gamma_{(D)}^\mu \gamma_{(D)}^\nu \gamma_{(D)}^\rho \gamma_{(D)}^\sigma$, followed
by the application of eq.~(\ref{chapter_qft:Dirac_contraction}):
\bq
\label{chapter_qft:Dirac_contraction_2}
\lefteqn{
 g^{(D)}_{\alpha\beta} \eps_{\mu\nu\rho\sigma} 
 \mathrm{Tr}\left(\gamma_{(D)}^\alpha \gamma_{(D)}^\mu \gamma_{(D)}^\nu \gamma_{(D)}^\rho \gamma_{(D)}^\sigma \gamma_{(D)}^\beta \gamma_5 \right)
 } & &
 \nonumber \\
 & = &
 2 \eps_{\mu\nu\rho\sigma} \left[ 
 \mathrm{Tr}\left(\gamma_{(D)}^\nu \gamma_{(D)}^\rho \gamma_{(D)}^\sigma \gamma_{(D)}^\mu \gamma_5 \right)
 - \mathrm{Tr}\left(\gamma_{(D)}^\mu \gamma_{(D)}^\rho \gamma_{(D)}^\sigma \gamma_{(D)}^\nu \gamma_5 \right)
 + \mathrm{Tr}\left(\gamma_{(D)}^\mu \gamma_{(D)}^\nu \gamma_{(D)}^\sigma \gamma_{(D)}^\rho \gamma_5 \right)
 \right. \nonumber \\
 & & \left.
 - \mathrm{Tr}\left(\gamma_{(D)}^\mu \gamma_{(D)}^\nu \gamma_{(D)}^\rho \gamma_{(D)}^\sigma \gamma_5 \right)
 \right]
 + D \eps_{\mu\nu\rho\sigma} \mathrm{Tr}\left(\gamma_{(D)}^\mu \gamma_{(D)}^\nu \gamma_{(D)}^\rho \gamma_{(D)}^\sigma \gamma_5 \right)
 \nonumber \\
 & = &
 \left(D-8\right) \eps_{\mu\nu\rho\sigma} \mathrm{Tr}\left(\gamma_{(D)}^\mu \gamma_{(D)}^\nu \gamma_{(D)}^\rho \gamma_{(D)}^\sigma \gamma_5 \right).
\eq
Combining eq.~(\ref{chapter_qft:Dirac_contraction_1}) and eq.~(\ref{chapter_qft:Dirac_contraction_2}) we arrive at
\bq
 2 \left(D-4\right) \eps_{\mu\nu\rho\sigma} \mathrm{Tr}\left(\gamma_{(D)}^\mu \gamma_{(D)}^\nu \gamma_{(D)}^\rho \gamma_{(D)}^\sigma \gamma_5 \right)
 & = & 0.
\eq
At $D=4$ this equation permits the usual non-zero trace of $\gamma_5$ with four
other Dirac matrices. However, for $D \neq 4$ we conclude
that the trace equals zero, and there is no smooth limit $D \rightarrow 4$ which 
reproduces the non-zero trace at $D=4$.

Thus we cannot have simultaneously the cyclicity of the trace, 
the anti-commutation relation as in eq.~(\ref{chapter_qft:gamma_5_conditions}) and 
a non-zero value for the trace as in eq.~(\ref{chapter_qft:gamma_5_conditions}).
On physical grounds we insist on keeping the non-zero value for the trace.
It enters the theoretical description of certain decays of particles. 
If the trace would be zero, the predicted decay rate would be zero as well, in contradiction with experiment.
We also would like to maintain the cyclicity of the trace.
Therefore we have to give up the simple anti-commutation relation of $\gamma_5$ in $D$ dimensions.

The 't Hooft-Veltman prescription \cite{'tHooft:1972fi}
defines $\gamma_5$ as the product of the first four Dirac matrices in $D$ dimensions:
\bq
\label{chapter_qft:def_gamma_5_HV}
 \gamma_5 & = & i \gamma_{(D)}^0 \gamma_{(D)}^1 \gamma_{(D)}^2 \gamma_{(D)}^3.
\eq
With this definition, $\gamma_5$ anti-commutes with the first four Dirac matrices (as in four space-time dimensions),
but commutes with the remaining ones:
\bq
\label{chapter_qft:Dirac_anticommutation_gamma_5_HV}
\begin{array}{rclcl}
 \left\{ \gamma_{(D)}^\mu, \gamma_5 \right\} & = & 0, & & \mbox{if} \;\; \mu \; \in \; \{0,1,2,3\},
 \\
 \left[ \gamma_{(D)}^\mu, \gamma_5 \right] & = & 0, & & \mbox{otherwise}.
\end{array}
\eq
\bs
{\it \refstepcounter{exercise}
{\bf Exercise \theexercise}: 
Show that with the definitions and conventions as above the rules for the traces of Dirac matrices carry over 
to $D$ dimensions.
In detail, show:
\begin{enumerate}
\item Traces of an even number of Dirac matrices are evaluated with the rules
\bq
\label{chapter_qft:Dirac_trace_rule_1_D_dim}
 \mathrm{Tr}\left(\gamma_{(D)}^\mu \gamma_{(D)}^\nu\right) 
 & = & 
 4 g_{(D)}^{\mu \nu},
 \nonumber \\
 \mathrm{Tr}\left(\gamma_{(D)}^{\mu_1} \gamma_{(D)}^{\mu_2} ... \gamma_{(D)}^{\mu_{2 n}}\right)
 & = & 
 \sum\limits_{j=2}^{2n} \left(-1\right)^j
 g_{(D)}^{\mu_1 \mu_j} \mathrm{Tr}\left(\gamma_{(D)}^{\mu_2} ... \gamma_{(D)}^{\mu_{j-1}} \gamma_{(D)}^{\mu_{j+1}} ... \gamma_{(D)}^{\mu_{2 n}}\right).
\eq
\item Traces of an odd number of Dirac matrices vanish:
\bq
\label{chapter_qft:Dirac_trace_rule_2_D_dim}
 \mathrm{Tr}\left(\gamma_{(D)}^{\mu_1} \gamma_{(D)}^{\mu_2} ... \gamma_{(D)}^{\mu_{2 n-1}}\right) 
 & = & 0
\eq
\item For traces involving $\gamma_5$ we have
\bq
\label{chapter_qft:Dirac_trace_rule_3_D_dim}
 \mathrm{Tr}\left(\gamma_5\right) 
 & = & 0,
 \nonumber \\
 \mathrm{Tr}\left(\gamma_{(D)}^{\mu} \gamma_{(D)}^{\nu} \gamma_5\right) 
 & = & 0,
 \nonumber \\
 \mathrm{Tr}\left(\gamma_{(D)}^{\mu} \gamma_{(D)}^{\nu} \gamma_{(D)}^{\rho} \gamma_{(D)}^{\sigma} \gamma_5\right) 
 & = & 
 \left\{ \begin{array}{ll}
 4 i \eps^{\mu \nu \rho \sigma}, & \mu,\nu,\rho,\sigma \in \{0,1,2,3\}, \\
 0, & \mbox{otherwise}. \\
 \end{array} \right.
\eq
\end{enumerate}
}
\es
\noindent
Let us now look at the implications of the 't Hooft-Veltman prescription for $\gamma_5$.
We first show that with this definition of $\gamma_5$ we correctly obtain the triangle anomaly
within dimensional regularisation.
On the other hand, the 't Hooft-Veltman prescription for $\gamma_5$
treats the first four indices differently from the remaining indices, as can be seen in the 
anti-commutation / commutation relations in eq.~(\ref{chapter_qft:Dirac_anticommutation_gamma_5_HV}).
As such, the regularisation scheme breaks a symmetry of the original unregularised theory.
This is unavoidable and not a problem.
However, we have to include finite renormalisations, which restore Ward identities reflecting the original symmetry.
We discuss this in the context of the non-singlet axial vector current.

From now on we always take the Dirac algebra in $D$ dimensions and we drop the subscript $(D)$.

\subsubsection{The singlet axial-vector current and the triangle anomaly}

\begin{figure}
\begin{center}
\includegraphics[scale=1.0]{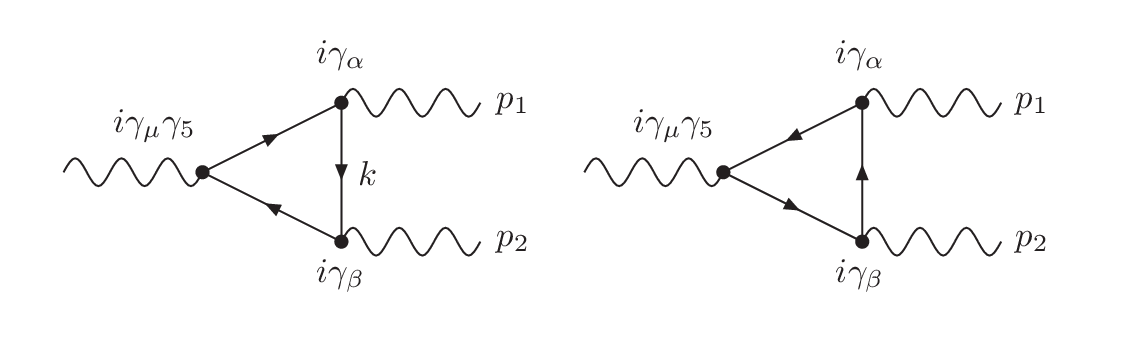}
\caption{\label{chapter_qft:singlet} The triangle graphs for the anomaly}
\end{center}
\end{figure}

The triangle anomaly for one axial-vector coupling and two vector couplings originates from the two
diagrams shown in fig.~\ref{chapter_qft:singlet}.
For massless fermions we obtain for the sum of the two graphs (ignoring coupling factors)
\bq
\label{chapter_qft:starting_point}
 A_{\alpha\beta\mu} 
 & = & 
 \int \frac{d^Dk}{(2 \pi)^D} \frac{N_{\alpha\beta\mu}}{q_0^2 q_1^2 q_2^2},
\eq
where
\bq
 N_{\alpha\beta\mu} 
 & = & 
 \mathrm{Tr}\left({\slashed q}_1 \gamma_\beta {\slashed q}_0 \gamma_\alpha {\slashed q}_2 \gamma_\mu \gamma_5\right)
 - 
 \mathrm{Tr}\left({\slashed q}_2 \gamma_\alpha {\slashed q}_0 \gamma_\beta {\slashed q}_1 \gamma_\mu \gamma_5\right)
\eq
and $q_0=k$, $q_1 = k - p_2$ and $q_2 = k + p_1$.
It is convenient to calculate the graphs for the kinematic
configuration where $p_1^2$, $p_2^2$ and $(p_1+p_2)^2$ are non-zero. 
In that case there will be no infrared divergences, which are not relevant to the discussion of the anomaly.
Contracting $A_{\alpha\beta\mu}$ with $(p_1+p_2)^\mu$ gives the anomaly:
\bq
 A^{AVV} & = & \left( p_1 + p_2 \right)^\mu A_{\alpha\beta\mu}.
\eq
Let us now calculate the anomaly.
In the first trace of $(p_1+p_2)^\mu N_{\alpha\beta\mu}$ we use
\bq
 \left( {\slashed p}_1 + {\slashed p}_2 \right) \gamma_5 
 = \left( {\slashed q}_2 - {\slashed q}_1 \right) \gamma_5
 = {\slashed q}_2 \gamma_5 + \gamma_5 {\slashed q}_1 - 2 {\slashed k}_{(-2\eps)} \gamma_5.
\eq
For the second trace we use
\bq
 \left( {\slashed p}_1 + {\slashed p}_2 \right) \gamma_5 
 = \left( {\slashed q}_2 - {\slashed q}_1 \right) \gamma_5
 = - {\slashed q}_1 \gamma_5 - \gamma_5 {\slashed q}_2 + 2 {\slashed k}_{(-2\eps)} \gamma_5.
\eq
The terms ${\slashed q}_1 {\slashed q}_1$ and ${\slashed q}_2 {\slashed q}_2$ inside the traces cancel propagators and the
resulting tensor bubble integrals can be shown to vanish after integration.
Therefore the only relevant term is
\bq
 - 2 \left( 
 \mathrm{Tr}\; {\slashed q}_1 \gamma_\beta {\slashed q}_0 \gamma_\alpha {\slashed q}_2 {\slashed k}_{(-2\eps)} \gamma_5 
 + 
 \mathrm{Tr}\; {\slashed q}_2 \gamma_\alpha {\slashed q}_0 \gamma_\beta {\slashed q}_1 {\slashed k}_{(-2\eps)} \gamma_5
 \right).
\eq
The traces evaluate to
\bq
\label{chapter_qft:trace_anomaly}
 \mathrm{Tr}\; {\slashed q}_1 \gamma_\beta {\slashed q}_0 \gamma_\alpha {\slashed q}_2 {\slashed k}_{(-2\eps)} \gamma_5 
 & = & 
 k_{(-2\eps)}^2 \cdot 4 i \eps_{\alpha\lambda\beta\kappa} p_1^\lambda p_2^\kappa + \dots,
 \nonumber \\
 \mathrm{Tr}\; {\slashed q}_2 \gamma_\alpha {\slashed q}_0 \gamma_\beta {\slashed q}_1 {\slashed k}_{(-2\eps)} \gamma_5
 & = & 
 k_{(-2\eps)}^2 \cdot 4 i \eps_{\alpha\lambda\beta\kappa} p_1^\lambda p_2^\kappa + \dots,
\eq
where the dots stand for terms, which vanish after integration.
\\
\\
\bs
{\it \refstepcounter{exercise}
{\bf Exercise \theexercise}: 
Derive eq.~(\ref{chapter_qft:trace_anomaly}).
}
\es
\\
\\
We then obtain for the anomaly
\bq
 A^{AVV} 
 & = & 
 16 \eps_{\alpha\lambda\beta\kappa} p_1^\lambda p_2^\kappa 
 \int \frac{d^Dk}{(2 \pi)^D i} \frac{k_{(-2\eps)}^2}{k_0^2 k_1^2 k_2^2} 
 \; = \;
 \frac{1}{(4 \pi)^2} 8 \eps_{\alpha\beta\lambda\kappa} p_1^\lambda p_2^\kappa
 + {\mathcal O}\left(\eps\right),
\eq
which is the well-known result for the anomaly in the 't Hooft-Veltman scheme \cite{'tHooft:1972fi}. 
\\
\\
\bs
{\it \refstepcounter{exercise}
{\bf Exercise \theexercise}: 
Show
\bq
 \int \frac{d^Dk}{(2 \pi)^D i} \frac{k_{(-2\eps)}^2}{k_0^2 k_1^2 k_2^2} 
 & = & 
 - \frac{1}{2} \frac{1}{(4 \pi)^2}
 + {\mathcal O}\left(\eps\right).
\eq
}
\es

\subsubsection{The non-singlet axial-vector current}

The Ward identity for the non-singlet axial-vector current for massless fermions reads
\bq
\label{chapter_qft:ward}
( p_{1} - p_{2} )^{\mu} \Gamma_{\mu 5} & = & {S}_{F}^{-1}(p_{1}) \gamma_{5} + \gamma_{5} {S}_{F}^{-1}(p_{2}),
\eq
where $i S_F(p)$ denotes the full fermion propagator and $i \Gamma_{\mu 5}$ denotes the full 
$i \gamma_\mu \gamma_5$-vertex. We are now going to check the Ward identity at one-loop level.
\begin{figure}
\begin{center}
\includegraphics[scale=1.0]{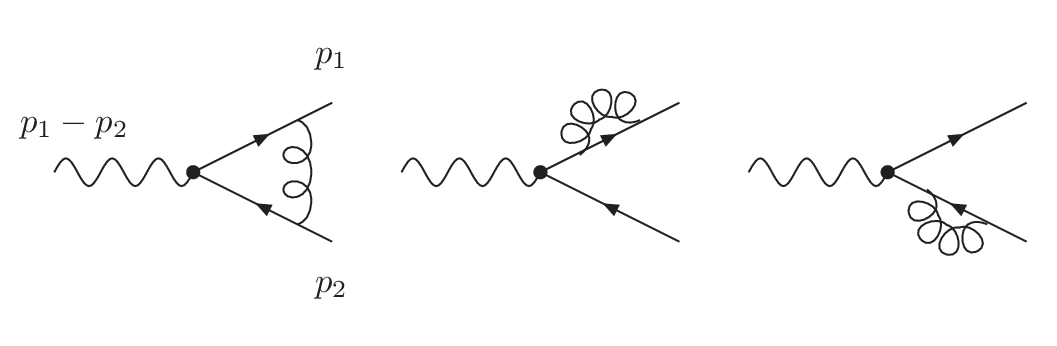}
\caption{\label{chapter_qft:nonsinglet} Feynman graphs for the non-singlet axial-vector Ward identity}
\end{center}
\end{figure}
The relevant diagrams are shown in fig.~\ref{chapter_qft:nonsinglet}.
The momentum $p_1$ is flowing outwards, whereas we take the momentum $p_2$ to be directed inwards.
The one-loop contribution from the right-hand-side of eq.~(\ref{chapter_qft:ward}) reads:
\bq
 - \int \frac{d^Dk}{(2 \pi)^D i} \frac{\gamma_\nu {\slashed q}_2 \gamma^\nu}{q_0^2 q_2^2} \gamma_5
 - \gamma_5 \int \frac{d^Dk}{(2 \pi)^D i} \frac{\gamma_\nu {\slashed q}_1 \gamma^\nu}{q_0^2 q_1^2} \gamma_5,
\eq
where we used the notation $q_0 = k$, $q_1 = k+p_2$ and $q_2 = k + p_1$.
The contribution from the three-point diagram reads:
\bq
 \int \frac{d^Dk}{(2 \pi)^D i} 
 \frac{\gamma_\nu {\slashed q}_2 \gamma_\mu \gamma_5 {\slashed q}_1 \gamma^\nu}{q_0^2 q_1^2 q_2^2}
\eq
Contracting with $(p_1-p_2)^\mu$ and rewriting $p_1-p_2=q_2-q_1$ we obtain
\bq
 \gamma_\nu {\slashed q}_2 \left( {\slashed p}_1 - {\slashed p}_2 \right) \gamma_5 {\slashed q}_1 \gamma^\nu
 & = &
 \frac{1}{2} 
 \left[ 
  \gamma_\nu {\slashed q}_2 \left( {\slashed p}_1 - {\slashed p}_2 \right) \gamma_5 {\slashed q}_1 \gamma^\nu
  -
  \gamma_\nu {\slashed q}_2 \gamma_5 \left( {\slashed p}_1 - {\slashed p}_2 \right) {\slashed q}_1 \gamma^\nu
 \right]
 \nonumber \\
 & = &
 \frac{1}{2} \left[ 
 \gamma_\nu {\slashed q}_2 \left( {\slashed q}_2 - {\slashed q}_1 \right) \gamma_5 {\slashed q}_1 \gamma^\nu
 -
 \gamma_\nu {\slashed q}_2 \gamma_5 \left( {\slashed q}_2 - {\slashed q}_1 \right) {\slashed q}_1 \gamma^\nu 
 \right]
 \nonumber \\
 & = &
 - q_1^2 \gamma_\nu {\slashed q}_2 \gamma^\nu \gamma_5 
 - q_2^2 \gamma_5 \gamma_\nu {\slashed q}_1 \gamma^\nu 
 \\
 & & 
 + 4 \eps q_1^2 {\slashed q}_2 \gamma_5 
 + 4 \eps q_2^2 \gamma_5 {\slashed q}_1 
 - \left( k_{(-2\eps)} \right)^2  \gamma_\nu \left( \gamma_5 {\slashed q}_1 + {\slashed q}_2 \gamma_5 \right) \gamma^\nu.
 \nonumber
\eq
The two terms on the second-to-last line correspond exactly to the right-hand-side of eq.(\ref{chapter_qft:ward}). 
However, the terms on the last line
spoil the Ward identity. These terms give the contribution
\bq
 - 4 \frac{1}{(4 \pi)^2} \left( {\slashed p}_1 - {\slashed p}_2 \right) \gamma_5
 + {\mathcal O}\left(\eps\right).
\eq
In order to restore the Ward identity we have to perform a finite renormalisation on the 
non-singlet axial-vector current
\bq
 \Gamma_{\mu 5}^r & = & Z_{5}^{ns} \Gamma_{\mu 5}^0.
\eq
For QCD the finite renormalisation $Z_{5}^{ns}$ is given
in the 't Hooft-Veltman scheme
(including a factor $g^2 C_F$, where 
$ C_F = \mbox{Tr} T^a T^a$ is the fundamental Casimir of the gauge group)
by
\bq
Z_{5}^{ns} & = & 1 -4 \frac{\alpha_s}{4 \pi} C_F,
\eq
with $\alpha_s = g^2 /(4 \pi)$.

%% file: oneloop.tex
\newpage
\chapter{One-loop integrals}
\label{chapter_one_loop}

The simplest, but most important loop integrals are the one-loop integrals.
Within perturbative quantum field theory they enter the first quantum corrections.
Contributions of this order are called {\bf next-to-leading order (NLO)} contributions.
The {\bf leading-order (LO)} contribution is (usually) the tree-level approximation or Born approximation.
As the perturbative expansion is an expansion in a small coupling, the next-to-leading order corrections
are expected to give numerically the dominant corrections to the tree-level approximation.

The one-loop Feynman integrals are very well understood.
They are simpler than the full class of all Feynman integrals.
First of all, at one-loop there are no irreducible scalar products in the numerator.
As a consequence, there is an alternative algorithm for the reduction of tensor integrals to scalar
integrals. This algorithm is known as Passarino-Veltman reduction and discussed in section~\ref{chapter_one_loop:sect_Passarino_Veltman}.
The Passarino-Veltman method does not shift the space-time dimension, nor does it raise the powers of the propagators.

In relativistic quantum field theory (and by using dimensional regularisation) we are usually interested in 
results for loop integrals in $D=4-2\eps$ space-time dimensions.
A second important result states, that we may reduce any scalar one-loop integral to scalar one-loop integrals
with no more than $5$ external legs.
Furthermore, we are usually only interested in the Laurent expansion up to and including the ${\mathcal O}(\eps^0)$-term.
In this case, only scalar integrals with no more than $4$ external legs are relevant.
This is discussed in section~\ref{chapter_one_loop:sect_reduction_higher_point}.
The number of one-loop integrals is therefore finite, and they can be calculated (and have been calculated) once and for all.
In appendix~\ref{appendix_one_loop_integrals} we provide the full list of scalar one-loop integrals for massless theories 
and give references, where results
for one-loop integrals with internal masses can be found.

As mentioned above, for NLO calculations we are usually only interested 
in the Laurent expansion up to and including the ${\mathcal O}(\eps^0)$-term.
We may ask, what transcendental functions appear in these terms.
The answer at one-loop is amazingly simple: 
Up to and including the ${\mathcal O}(\eps^0)$-term
there are just two transcendental functions. These are the logarithm
and the dilogarithm
\bq
\label{chapter_one_loop:def_logarithm}
 \mathrm{Li}_1\left(x\right) 
 \;\; = \;\;
 - \ln\left(1-x\right)
 \;\; = \;\;
 \sum\limits_{n=1}^\infty \frac{x^n}{n},
 & \hspace*{15mm} & 
 \mathrm{Li}_2\left(x\right) 
 \;\; = \;\;
 \sum\limits_{n=1}^\infty \frac{x^n}{n^2}.
\eq
In section~\ref{chapter_one_loop:sect_dilogarithm} we will study a Feynman integral, which leads to a dilogarithm.
We will also discuss the properties of the dilogarithm.

The Passarino-Veltman method mentioned above is conceptually simple and historically important,
but it also has a short-coming: 
Expressing a tensor integral as a linear combination of scalar integrals, the coefficients of this linear combination
may contain Gram determinants in the denominator.
This can lead to numerical instabilities in certain regions of phase-space.
We discuss a more efficient method in section~\ref{chapter_one_loop:sect_spinor_techniques}.

At the end of the day our real interest are loop amplitudes, i.e. the sum of all relevant Feynman integrals.
Of course, if we know how to compute all relevant Feynman integrals, we may sum up the individual results and
obtain the loop amplitude.
However, as the number of Feynman diagrams growths, this approach becomes inefficient
and methods which directly deal with loop amplitudes are preferred.
For one-loop amplitudes we have efficient methods which bypass individual Feynman diagrams.
These methods exploit the fact that we know all relevant one-loop integrals, therefore only the coefficients in front of
these integrals need to be determined.
These methods are discussed in section~\ref{chapter_one_loop:sect_amplitude_methods}.

\section{Passarino-Veltman reduction}
\label{chapter_one_loop:sect_Passarino_Veltman}

The Passarino-Veltman reduction method \cite{Passarino:1979jh}
reduces one-loop tensor integrals to scalar integrals
and offers for one-loop integrals an alternative to the general method discussed in section~\ref{chapter_qft:sect:tensor_reduction}.
The Passarino-Veltman reduction method exploits the fact, that at one-loop, any scalar product involving the loop momentum and the external
momenta can be expressed as a linear combination of inverse propagators and a loop-momentum independent term.
This is not true for a general Feynman integral beyond one-loop, as there might be irreducible scalar products.
This is exactly the same issue as in our discussion of the Baikov representation in section.~\ref{chapter_basics:sect_Baikov_representation}.
In other words, the Passarino-Veltman reduction method exploits the fact that any one-loop graph $G$ has a Baikov representation,
which implies that we may express
any scalar product involving the loop momentum and the external
momenta as a linear combination of inverse propagators and a loop-momentum independent term.
In particular, there is no need to consider a larger graph $\tilde{G}$.

Let us first introduce the Passarino-Veltman notation for one-loop tensor integrals.
For scalar integrals with one, two or three external legs we write
\bq
A_{0}(m) 
 & = &
 e^{\eps \Eulerconstant} \mu^{2\eps}
 \int \frac{d^{D}k}{i \pi^{D/2}}
        \frac{1}{(-k^{2}+m^{2})},
 \\
B_{0}(p,m_{1},m_{2}) 
 & = & 
 e^{\eps \Eulerconstant} \mu^{2\eps}
 \int \frac{d^{D}k}{i \pi^{D/2}}
        \frac{1}{(-k^{2}+m_{1}^{2})
                 (-(k-p)^{2}+m_{2}^{2})},
 \nonumber \\
C_{0}(p_{1},p_{2},m_{1},m_{2},m_{3})
 & = & 
 \nonumber \\
 & &
 \hspace*{-4cm}
 e^{\eps \Eulerconstant} \mu^{2\eps}
 \int \frac{d^{D}k}{i \pi^{D/2}}
        \frac{1}{(-k^{2}+m_{1}^{2})
                 (-(k-p_{1})^{2}+m_{2}^{2})
                 (-(k-p_{1}-p_{2})^{2}+m_{3}^{2})},
 \nonumber
\eq
with an obvious generalisation towards more external legs.
Four-point functions are denoted with the letter $D$, five-point functions are denoted with the letter $E$, etc..
For tensor integrals we use the notation $X^{\mu_1 \mu_2 \dots \mu_r}$, with $X \in \{A,B,C,\dots\}$ and
the superscripts ${\mu_1 \mu_2 \dots \mu_r}$ indicate that the numerator is 
\bq
 k^{\mu_1} k^{\mu_2} \dots k^{\mu_r}.
\eq
To give an example:
\bq
B^{\mu_1}(p,m_{1},m_{2}) 
 & = & 
 e^{\eps \Eulerconstant} \mu^{2\eps}
 \int \frac{d^{D}k}{i \pi^{D/2}}
        \frac{k^{\mu_1}}{(-k^{2}+m_{1}^{2})
                 (-(k-p)^{2}+m_{2}^{2})},
 \nonumber \\
B^{\mu_1\mu_2}(p,m_{1},m_{2}) 
 & = & 
 e^{\eps \Eulerconstant} \mu^{2\eps}
 \int \frac{d^{D}k}{i \pi^{D/2}}
        \frac{k^{\mu_1}k^{\mu_2}}{(-k^{2}+m_{1}^{2})
                 (-(k-p)^{2}+m_{2}^{2})}.
\eq
The reduction technique according to Passarino and Veltman 
uses the fact that due to Lorentz symmetry the result can only depend on tensor structures which 
can be build from the external momenta $p_j^{\mu}$ and the metric tensor $g^{\mu\nu}$.
We therefore write the tensor integrals 
in the most general form
in terms of form factors times external momenta and/or the metric tensor. For example 
\bq 
\label{chapter_one_loop:passarino}
B^{\mu_1} & = & p^{\mu_1} B_{1}, \nonumber \\
B^{\mu_1 \mu_2} & = & p^{\mu_1} p^{\mu_2} B_{21} + g^{\mu_1 \mu_2} B_{22}, \nonumber \\
 & & \nonumber \\
C^{\mu_1} & = & p^{\mu_1}_{1} C_{11} + p^{\mu_1}_{2} C_{12}, \nonumber \\
C^{\mu_1 \mu_2} & = & p^{\mu_1}_{1} p^{\mu_2}_{1} C_{21} + p^{\mu_1}_{2} p^{\mu_2}_{2} C_{22}
 + \left( p_{1}^{\mu_1} p_{2}^{\mu_2} + p_{1}^{\mu_2} p_{2}^{\mu_1} \right) C_{23} + g^{\mu_1 \mu_2} C_{24}.
\eq
One then solves for the form factors $B_1$, $B_{21}$, $B_{22}$, $C_{11}$, etc. by
first contracting both sides with the external momenta and the metric tensor $g_{\mu \nu}$.
On the left-hand side the resulting scalar products between the loop momentum $k^{\mu}$ and the external
momenta are rewritten in terms of inverse propagators, as for example
\bq
2 p \cdot k & = & \left[ - (k-p)^2 + m_2^2 \right] - \left[ - k^2 + m_1^2 \right] + \left( p^2 + m_1^2 - m_2^2 \right).
\eq
The first two terms of the right-hand side above cancel propagators, whereas the last term does not involve the 
loop momentum any more.
The remaining step is to solve for the form-factors by inverting the matrix which one obtains on the 
right-hand side of
equation (\ref{chapter_one_loop:passarino}).
\\
\\
As an example we consider the two-point function:
Contraction with 
$p_{\mu_1}$ or $p_{\mu_1} p_{\mu_2}$ and $g_{\mu_1 \mu_2}$
yields
\bq
p^{2} B_{1} = -\frac{1}{2} 
 \left( 
 \left( m_{2}^{2} - m_{1}^{2} - p^{2} \right) B_{0} - A_{0}(m_{1}) + A_{0}(m_{2}) \right), \nonumber \\
 & & \nonumber \\
\left( \begin{array}{cc}
 p^{2} & 1 \\
 p^{2} & D \\
\end{array} \right)
\left( \begin{array}{c}
B_{21} \\ B_{22} \end{array} \right) 
=
\left( \begin{array}{c}
 -\frac{1}{2}(m_{2}^{2}-m_{1}^{2}-p^{2}) B_{1} - \frac{1}{2} A_{0}(m_{2}) \\
 m_{1}^{2} B_{0} - A_{0}(m_{2}) \\
\end{array} \right).
\eq
Solving for the form factors we obtain
\bq
B_{1} & = & -\frac{1}{2 p^{2}} 
 \left( \left( m_{2}^{2} - m_{1}^{2} - p^{2} \right) B_{0} - A_{0}(m_{1}) + A_{0}(m_{2}) \right), \nonumber \\
B_{21} & = & 
 \frac{1}{(D-1) p^{2}} 
 \left( 
 -\frac{D}{2} (m_{2}^{2} - m_{1}^{2} - p^{2}) B_{1} - m_{1}^{2} B_{0} - \frac{D-2}{2} A_{0}(m_{2}) 
 \right), \nonumber \\
B_{22} & = & 
 \frac{1}{2(D-1)} \left( ( m_{2}^{2} - m_{1}^{2} - p^{2} ) B_{1} + 2 m_{1}^{2} B_{0} - A_{0}(m_{2})  
 \right).
\eq
Due to the matrix inversion in the last step determinants usually appear in the denominator of the final expression.
For a three-point function we would encounter the Gram determinant 
\bq
 \det G\left(p_1,p_2\right) & = & 
\left|
\begin{array}{cc}
-p_1^2 & -p_1\cdot p_2 \\
-p_1 \cdot p_2 & -p_2^2 \\
\end{array} \right|.
\eq
One drawback of this algorithm is closely related to these determinants: In a phase space
region where $p_1$ becomes collinear to $p_2$, the Gram determinant will tend to zero, and the 
form factors will take large values, with possible large cancellations among them. This makes
it difficult to set up a stable numerical program for automated evaluation of tensor loop
integrals.
Methods to overcome this obstacle are discussed in section~\ref{chapter_one_loop:sect_spinor_techniques}.
\\
\\
\bs
{\it \refstepcounter{exercise}
{\bf Exercise \theexercise}: 
Reduce the tensor integral
\bq
 A^{\mu_1 \mu_2 \mu_3 \mu_4}(m) 
 & = &
 e^{\eps \Eulerconstant} \mu^{2\eps}
 \int \frac{d^{D}k}{i \pi^{D/2}}
        \frac{k^{\mu_1} k^{\mu_2} k^{\mu_3} k^{\mu_4}}{(-k^{2}+m^{2})}
\eq
to $A_0(m)$.
}
\es
\\
\\
\bs
{\it \refstepcounter{exercise}
{\bf Exercise \theexercise}: 
Reduce 
\bq
 g_{\mu_1 \mu_2} g_{\mu_3 \mu_4} C^{\mu_1 \mu_2 \mu_3 \mu_4}(p_{1},p_{2},0,0,0)
 & = & 
 - g_{\mu_1 \mu_2} g_{\mu_3 \mu_4} 
 e^{\eps \Eulerconstant} \mu^{2\eps}
 \int \frac{d^{D}k}{i \pi^{D/2}}
        \frac{k^{\mu_1} k^{\mu_2} k^{\mu_3} k^{\mu_4}}{k^{2}
                 (k-p_{1})^{2}
                 (k-p_{1}-p_{2})^{2}}
 \;\;\;
\eq
to scalar integrals.
}
\es
\\
\\
The Passarino-Veltman algorithm is based on the observation, that for one-loop
integrals a scalar product
of the loop momentum with an external momentum can be expressed
as a combination of inverse propagators and a loop-momentum independent term.
This property does no longer hold if one goes to two or more loops.
If we consider fig.~\ref{chapter_basics:fig_doublebox}
and eq.~(\ref{chapter_basic:irreducible_scalar_product}), we see for example in the double-box Feynman integral
that the scalar product
\bq
 -k_2 \cdot p_1
\eq
cannot be expressed in terms of inverse propagators and a loop-momentum independent term.
In order to be able to express any scalar product involving the loop momenta and the external momenta
as a linear combination of inverse propagators and a loop-momentum independent term
we must introduce a larger graph $\tilde{G}$.
We may still use the Passarino-Veltman ansatz based on Lorentz symmetry, that a tensor integral is written
in terms of form factors times external momenta and/or the metric tensor. 
If one now tries to solve for the form factors by contracting with external momenta and the metric tensor,
not all scalar products cancel propagators or give loop-momentum independent terms.
Inverse propagators related to the propagators in $\tilde{G}$, which are not in the original graph $G$,
may remain in the numerator.
These inverse propagators in the numerator are called irreducible scalar products.
In the notation of eq.~(\ref{chapter_basics:full_notation_Feynman_integral})
\bq
 I_{\nu_1 \dots \nu_{\ninternal}}\left(D,x_1,\dots,x_{\NB}\right)
\eq
for a Feynman integral associated to the graph $\tilde{G}$ these irreducible scalar products correspond
to negative integer values for some $\nu_j$'s.

\section{Reduction of higher point integrals}
\label{chapter_one_loop:sect_reduction_higher_point}

In this section we discuss the reduction of scalar one-loop integrals with $\nexternal$ external legs
to a set of scalar one-, two-, three-, four- and five-point functions.
By an appropriate choice of the basis integrals for the five-point functions it can be arranged, that
the five-point functions only contribute at order ${\mathcal O}(\eps)$ and are thus not relevant 
to NLO calculations, where we only need the $\eps$-expansion of scalar one-loop integrals up to and including
${\mathcal O}(\eps^0)$.
Thus, we may reduce any one-loop Feynman integral to
scalar one-, two-, three- and four-point functions plus terms of order ${\mathcal O}(\eps)$ and beyond, which are
not relevant to NLO calculations.
This is a finite set of one-loop Feynman integrals, which can (and has been) calculated once and for all.

The reason a scalar one-loop $\nexternal$-point function can be reduced to scalar one-loop Feynman integrals
with no more than five external legs is the following:
We assume all external momenta to lie in a four-dimensional space.
Thus, even for $\nexternal>5$, the external momenta span maximally a space of dimension $4$.
The scalar one-loop $n$-point functions with $\nexternal \ge 5$ are always ultraviolet-finite, but they may have infrared-divergences.
Let us first assume that there are no IR-divergences. Then the integral is finite and can be performed 
in four dimensions. In a space of four dimensions we can have no more than four linearly independent vectors,
therefore it comes to no surprise that in a one-loop integral with five or more propagators, one propagator
can be expressed through the remaining ones. This is the basic idea for the reduction of 
the higher point scalar integrals.
For infrared-finite integrals this fact has been known for a long time \cite{Melrose:1965kb,vanNeerven:1984vr}.
For infrared-divergent integrals we have to use a regulator.
With slight modifications the basic idea above can be generalised to dimensional regularisation \cite{Bern:1994kr,Binoth:1999sp,Fleischer:1999hq,Denner:2002ii,Duplancic:2003tv,Binoth:2005ff}.
Within dimensional regularisation, the external momenta and the loop momentum span maximally a space of dimension $5$.

Let us now look at the details. 
We discuss the method for massless one-loop integrals.
For one-loop integrals we have $\nexternal=\ninternal$.
The first step is to set up an appropriate notation. In this section we denote
a scalar one-loop integral with $\nexternal$ legs (and massless propagators) by
\bq
\label{chapter_one_loop:basic_one_loop_scalar_int}
I_{\nexternal}\left(D\right) & = &
 e^{\eps \Eulerconstant} \mu^{2\eps} 
  \int \frac{d^Dk}{i \pi^{\frac{D}{2}}}
  \frac{1}{(-k^2) (-(k-p_1)^2) ... (-(k-p_1-...p_{\nexternal-1})^2)}.
\eq
In terms of our previous notation
\bq
 I_{\nexternal}\left(D\right) 
 & = &
 \left(\mu^2\right)^{\frac{\Dint}{2}-\nexternal}
 I_{\keepstyleunderbrace{1 1 \dots 1}_{\nexternal}}\left(D\right).
\eq
We denote by $I_{\nexternal-1}^{(i)}(D)$ the scalar one-loop integral, where the $i$'th propagator has been
removed:
\bq
 I_{\nexternal-1}^{(i)}\left(D\right) 
 & = &
 \left(\mu^2\right)^{\frac{\Dint}{2}-\left(\nexternal-1\right)}
 I_{\keepstyleunderbrace{1 \dots 1}_{i-1} 0 \keepstyleunderbrace{1 \dots 1}_{\nexternal-i}}\left(D\right).
\eq
Let us also introduce the sums of the external momenta:
\bq
 p^{\mathrm{sum}}_i & = & \sum\limits_{j=1}^i p_j,
 \;\;\;\;\;\;\;\;\;
 1 \; \le \; i \; \le \; \nexternal - 1.
\eq
We associate two matrices $S$ and $G$ to the integral in eq.~(\ref{chapter_one_loop:basic_one_loop_scalar_int}).
The entries of the $\nexternal \times \nexternal$ kinematic matrix $S$ are given by
\bq
 S_{ij} & = & \left( p^{\mathrm{sum}}_i - p^{\mathrm{sum}}_j \right)^2,
\eq
and the entries of the $(\nexternal-1) \times (\nexternal-1)$ Gram matrix are defined by
\bq
G_{ij} & = & - p^{\mathrm{sum}}_i \cdot p^{\mathrm{sum}}_j.
\eq
For the reduction one distinguishes three different cases: 
Scalar pentagons (i.e. scalar five-point functions),
scalar hexagons (scalar six-point functions) and scalar integrals with more than
six propagators.

Let us start with the pentagon.
A five-point function in $D=4-2\eps$ dimensions can be expressed as a sum of four-point functions, where
one propagator is removed, plus a five-point function in $6-2\eps$ dimensions \cite{Bern:1994kr}.
Since the $(6-2\eps)$-dimensional pentagon is finite and comes with an extra factor of $\eps$ in front, it does
not contribute at ${\mathcal O}(\eps^0)$. In detail we have:
\begin{tcolorbox}
{\bf Reduction of the massless five-point integral}:
\bq
\label{chapter_one_loop:scalarfivepoint}
I_5\left(4-2\eps\right) 
 & = & 
 -2\eps B I_5\left(6-2\eps\right)
 - \sum\limits_{i=1}^5 b_i I_4^{(i)}\left(4-2\eps\right)
 \nonumber \\
 & = &
 - \sum\limits_{i=1}^5 b_i I_4^{(i)}\left(4-2\eps\right)
 + {\mathcal O}\left(\eps\right),
\eq
where the coefficients $B$ and $b_i$ are obtained from the kinematic matrix $S_{ij}$ as follows: 
\bq
b_i = \sum\limits_{j=1}^5 \left( S^{-1} \right)_{ij},
 & &
B = \sum\limits_{i=1}^5 b_i.
\eq
\end{tcolorbox}
In eq.~(\ref{chapter_one_loop:scalarfivepoint}) $I_5(6-2\eps)$ denotes the $(6-2\eps)$-dimensional pentagon and
$I_4^{(i)}(4-2\eps)$ denotes the four-point function, 
which is obtained from the pentagon by removing propagator $i$.
The proof of eq.~(\ref{chapter_one_loop:scalarfivepoint}) 
(as the proofs of eq.~(\ref{chapter_one_loop:scalarsixpoint}) and eq.~(\ref{chapter_one_loop:scalarnpoint}) below)
uses integration-by-parts identities, which 
will be introduced in chapter~\ref{chapter_iterated_integrals}.

The six-point function can be expressed as a sum of five-point functions \cite{Binoth:1999sp}
without any correction of ${\mathcal O}(\eps)$:
\begin{tcolorbox}
{\bf Reduction of the massless six-point integral}:
\bq
\label{chapter_one_loop:scalarsixpoint}
I_6\left(4-2\eps\right) & = & - \sum\limits_{i=1}^6 b_i I_5^{(i)}\left(4-2\eps\right).
\eq
The coefficients $b_i$ are again related to the kinematic matrix $S_{ij}$: 
\bq
b_i & = & \sum\limits_{j=1}^6 \left( S^{-1} \right)_{ij}.
\eq
\end{tcolorbox}
For the seven-point function and beyond we can again express the $\nexternal$-point function as a sum over
$(\nexternal-1)$-point functions \cite{Duplancic:2003tv}: 
\begin{tcolorbox}
{\bf Reduction of the massless $\nexternal$-point integral ($\nexternal \ge 7$)}:
\bq
\label{chapter_one_loop:scalarnpoint}
 I_{\nexternal}\left(4-2\eps\right) 
 & = & 
 - \sum\limits_{i=1}^{\nexternal} r_i I_{\nexternal-1}^{(i)}\left(4-2\eps\right),
\eq
where the coefficients $r_i$ are defined below in eq.~(\ref{chapter_one_loop:scalarnpoint_coefficients}).
\end{tcolorbox}
In contrast to eq. (\ref{chapter_one_loop:scalarsixpoint}), the decomposition in eq. (\ref{chapter_one_loop:scalarnpoint}) is no longer unique.
A possible set of coefficients $r_i$ can be
obtained from the singular value decomposition of the Gram matrix 
\bq
G_{ij} & = & \sum\limits_{k=1}^4 U_{ik} w_k \left(V^\dagger\right)_{kj}.
\eq
as follows \cite{Giele:2004iy}
\bq
\label{chapter_one_loop:scalarnpoint_coefficients}
  r_i = \frac{V_{i 5}}{W_5}, \;\;\; 1 \le i \le \nexternal-1,
  \;\;\; \;\;\; 
  r_{\nexternal} = - \sum\limits_{j=1}^{\nexternal-1} r_j,
 \;\;\; \;\;\; 
 W_5 = - \sum\limits_{j=1}^{\nexternal-1} G_{j j} V_{j 5}.
\eq
\begin{digression} 
\index{singular value decomposition}
{\bf Singular value decomposition}
\\
Let $M$ be a complex $m \times n$-matrix of rank $r$.
The singular value decomposition of $M$ is a decomposition of the form
\bq
 M & = & U \Sigma V^\dagger,
\eq
where $U$ is a unitary $m \times m$-matrix, $V$ is a unitary $n \times n$-matrix, $V^\dagger$ denotes 
the Hermitian transpose of $V$ and $\Sigma$ is a real $m \times n$-matrix of the form
\bq
 \Sigma
 & = &
 \left( 
  \begin{array}{ccccc|ccccc}
   w_1 & & & &      & & & \vdots & & \\
   & & \ddots & &    & \dots & & 0 & & \dots \\
   & & & & w_r      & & & \vdots & & \\
   \hline
   & & \vdots & &      & & & \vdots & & \\
   \dots & & 0 & & \dots   & \dots & & 0 & & \dots \\
   & & \vdots & &      & & & \vdots & & \\
  \end{array}
 \right),
\eq
with $w_k > 0$. The diagonal entries $w_k$ are called the singular values of $M$.
If $M$ is real, the matrices $U$ and $V$ can be chosen as orthogonal matrices.
\end{digression}

\section{The basic one-loop integrals}
\label{chapter_one_loop:sect_dilogarithm}

With the results of the previous two sections, we may reduce any one-loop Feynman integral
to scalar one-, two-, three- and four-point functions plus terms of order ${\mathcal O}(\eps)$ and beyond, 
which are
not relevant to NLO calculations.
This is a finite set of one-loop Feynman integrals, which can (and has been) calculated once and for all.
We have already calculated the massive one-loop one-point function in eq.~(\ref{chapter_basics:result_tadpole_T1_4D})
(the massless one-loop one-point function is zero, see eq.~(\ref{chapter_basics:basic_eq_negative_dimensions}))
and the massless one-loop two-point function in eq.~(\ref{chapter_basics:result_bubble_B11_4D}).
In the results of eq.~(\ref{chapter_basics:result_tadpole_T1_4D}) and eq.~(\ref{chapter_basics:result_bubble_B11_4D})
we already saw the appearance of a logarithm ($\ln(m^2/\mu^2)$ for the tadpole and $\ln(-p^2/\mu^2)$ for the massless bubble).
It turns out that up to order ${\mathcal O}(\eps)$ there is just another transcendental function, which we have to know:
This is Euler's dilogarithm.
As an example for the appearance of the dilogarithm 
let us discuss the one-loop three-point function
with no internal masses and the kinematic configuration 
\bq
 p_1^2 \; \neq \; 0, 
 \;\;\;\;\;\;
 p_2^2 \; \neq \; 0, 
 \;\;\;\;\;\;
 p_3^2 \; = \; (p_1+p_2)^2 \; \neq \; 0.
\eq
We consider the integral
\bq
 I_3
 & = & 
 e^{\eps \Eulerconstant} \mu^{2\eps}
 \int \frac{d^{D}k}{i \pi^{D/2}}
        \frac{1}{(-k^{2})
                 (-(k-p_{1})^{2})
                 (-(k-p_{1}-p_{2})^{2})}.
\eq
The integral is finite and can be evaluated in four dimensions.
In the Feynman parameter representation the Feynman integral is given by
\bq 
 I_3
 & = &
 \int\limits_0^1 da_1 \int\limits_0^{a_1} da_2
 \frac{1}{-a_1^2 p_3^2 - a_2^2 p_2^2 +a_1a_2(p_1^2-p_2^2-p_3^2) - a_1 p_3^2 + a_2 (p_3^2-p_1^2)}
 + {\mathcal O}(\eps).
 \nonumber \\
\eq
We follow here closely the original work of 't Hooft and Veltman \cite{'tHooft:1979xw}.
We make the change of variables $a_2' = a_2 - \alpha a_1$ and choose $\alpha$ as a root of the equation
\bq
 - \alpha^2 p_2^2 + \alpha \left( p_1^2 - p_2^2 - p_3^2 \right) - p_3^2 & = & 0.
\eq
With this choice we eliminate the quadratic term in $a_1$.
We then perform the $a_1$-integration and we end up with three integrals of the form
\bq
 \int\limits_0^1 \frac{dt}{t-t_0} \left[ \ln\left(at^2+bt+c\right) - \ln\left(at_0^2+bt_0+c\right) \right].
\eq
Factorising the arguments of the logarithms, these integrals are reduced to the type
\bq
 R & =  & \int\limits_0^1 \frac{dt}{t-t_0} \left[ \ln\left(t-t_1\right) - \ln\left(t_0-t_1\right) \right].
\eq
This integral is expressed in terms of a new function, the dilogarithm, as follows:
\bq
 R & = & 
 \mathrm{Li}_2\left( \frac{t_0}{t_1-t_0} \right)
 -
 \mathrm{Li}_2\left( \frac{t_0-1}{t_1-t_0} \right),
\eq
provided $-t_1$ and $1/(t_0-t_1)$ have imaginary part of opposite sign, otherwise additional logarithms occur.

\begin{digression}
\index{dilogarithm}
{\bf The dilogarithm}
\\
The dilogarithm is defined by
\bq
\mathrm{Li}_{2}(x) & = &- \int\limits_{0}^{1} dt \frac{ \ln(1 - x t)}{t}
= - \int\limits_{0}^{x} dt \frac{\ln(1-t)}{t}.
\eq
If we take the main branch of the logarithm with a cut along the negative real
axis, then the dilogarithm has a cut along the positive real
axis, starting at the point $ x=1 $.
For $\left|x\right| \le 1$ the dilogarithm has the power series expansion
\bq
\mathrm{Li}_{2}(x) & = & \sum\limits_{n=1}^{\infty} \frac{x^{n}}{n^{2}}.
\eq
Some important numerical values are
\bq
\label{chapter_one_loop_Li2_numerical_values}
\mathrm{Li}_{2}(0) = 0, 
\;\;\;\;\;
\mathrm{Li}_{2}(1) = \frac{\pi^{2}}{6},
\;\;\;\;\;
\mathrm{Li}_{2}(-1) = -\frac{\pi^{2}}{12},
\;\;\;\;\;
\mathrm{Li}_{2}\left(\frac{1}{2}\right) 
= \frac{\pi^{2}}{12} - \frac{1}{2} \left( \ln 2 \right)^{2}.
\eq
The dilogarithm with argument $x$ can be related to the dilogarithms with argument $(1-x)$ or $1/x$:
\bq
\label{chapter_one_loop:dilog_identities}
\mathrm{Li}_{2}(x) & = & - \mathrm{Li}_{2}(1-x) + \frac{1}{6} \pi^{2} - \ln(x) \ln(1-x),
 \nonumber \\
\mathrm{Li}_{2}(x) & = & - \mathrm{Li}_{2}\left(\frac{1}{x}\right) - \frac{1}{6} \pi^{2}
 - \frac{1}{2} \left( \ln(-x) \right)^{2}.
\eq
Another important relation is the five-term relation:
\bq
\label{chapter_one_loop:dilog_five_term_relation}
\mathrm{Li}_{2}(xy) 
 & = & 
  \mathrm{Li}_{2}(x) + \mathrm{Li}_{2}(y)
 +\mathrm{Li}_{2}\left(\frac{xy-x}{1-x}\right)
 +\mathrm{Li}_{2}\left(\frac{xy-y}{1-y}\right)
 +\frac{1}{2} \ln^{2}\left(\frac{1-x}{1-y}\right).
\nonumber \\
\eq
\end{digression}
In appendix~\ref{appendix_one_loop_integrals}
we provide a list of all basic one-loop integrals for massless theories up to order ${\mathcal O}(\eps)$
and references for the basic integrals with internal masses.

\section{Spinor techniques}
\label{chapter_one_loop:sect_spinor_techniques}

The reduction methods for one-loop tensor integrals discussed in section~\ref{chapter_one_loop:sect_Passarino_Veltman} 
(and in section~\ref{chapter_qft:sect:tensor_reduction})
are rather general
and independent of the tensor structure into which the tensor integral is contracted.
By taking into account information from this external tensor structure, more efficient reduction
algorithms can be derived
\cite{Pittau:1997ez,Pittau:1998mv,Weinzierl:1998we,delAguila:2004nf,Pittau:2004bc,vanHameren:2005ed}.
These algorithms significantly soften the problem with Gram determinants inherent in the Passarino-Veltman tensor reduction
method.
We will discuss as an example a method for one-loop integrals with massless propagators.
The method is most conveniently explained within the FDH-scheme of dimensional regularisation.
A generic one-loop tensor integral of rank $r$ is denoted by
\bq
\label{chapter_one_loop:basic_one_loop_tensor_int}
I_{\nexternal}^{\mu_1 \dots \mu_r}\left(D\right) & = &
 e^{\eps \Eulerconstant} \mu^{2\eps} 
  \int \frac{d^Dk}{i \pi^{\frac{D}{2}}}
  \frac{k^{\mu_1} \dots k^{\mu_r}}{(-k^2) (-(k-p_1)^2) ... (-(k-p_1-...p_{\nexternal-1})^2)}.
\eq
Let us assume that this integral is contracted into $J_{\mu_1 \dots \mu_r}$, e.g.
we are considering
\bq
 J_{\mu_1 \dots \mu_r} I_{\nexternal}^{\mu_1 \dots \mu_r}\left(D\right).
\eq
The tensor $J_{\mu_1 \dots \mu_r}$ does not depend on the loop momentum $k$.
In the FDH-scheme we can assume without loss of generality that the tensor structure
$J_{\mu_1 ... \mu_r}$
is given by
\bq
 J_{\mu_1 ... \mu_r} & = &
  \left\langle a_1 - \left| \gamma_{\mu_1} \right| b_1 - \right\rangle
  ...
  \left\langle a_r - \left| \gamma_{\mu_r} \right| b_r - \right\rangle,
\eq
where $\langle a_i - |$ and $| b_j - \rangle$ are Weyl spinors of definite helicity.
Spinors are reviewed in appendix~\ref{appendix_spinors}.
Therefore we consider tensor integrals of the form
\bq
\label{chapter_one_loop:basictensorintegral1}
I_{\nexternal}^r & = &
 e^{\eps \Eulerconstant} \mu^{2\eps} 
  \left\langle a_1 - \left| \gamma_{\mu_1} \right| b_1 - \right\rangle
  ...
  \left\langle a_r - \left| \gamma_{\mu_r} \right| b_r - \right\rangle
 \nonumber \\
 & &
  \int \frac{d^Dk}{i \pi^{\frac{D}{2}}}
  \frac{k^{\mu_1}_{(4)} ... k^{\mu_r}_{(4)}}{(-k^2) (-(k-p_1)^2) ... (-(k-p_1-...p_{\nexternal-1})^2)},
\eq
where $k^{\mu}_{(4)}$ denotes the projection of the $D$ dimensional vector $k^\mu$ onto
the four-dimensional subspace.
The quantity $\left\langle a - \left| \gamma_{\mu} \right| b - \right\rangle$ is a vector in a complex
vector-space of dimension $4$ and can therefore be expressed as a linear combination of
four basis vectors.

The first step for the construction of the reduction algorithm based on spinor methods
is to associate to each $n$-point loop integral a pair of two {\bf light-like} momenta $l_1$ and
$l_2$, which are linear combinations of two external momenta $p_i$ and $p_j$ of the loop
integral under consideration \cite{delAguila:2004nf}.
Obviously, this construction only makes sense for three-point integrals and beyond, 
as for two-point integrals there is only one independent external momentum.
This is not a limitation, tensor two-point functions can be reduced with the Passarino-Veltman technique.
The only Gram determinant occurring in this process is the determinant of the $1 \times 1$-matrix $G=-p^2$,
where $p$ denotes the external momentum of the two-point function. This is harmless.

For three-point functions and beyond
we write
\bq
\label{chapter_one_loop_def_l_1_l_2}
l_1 = \frac{1}{1-\alpha_1 \alpha_2} \left( p_i - \alpha_1 p_j \right), 
& &
l_2 = \frac{1}{1-\alpha_1 \alpha_2} \left( -\alpha_2 p_i + p_j \right),
\eq
where $\alpha_1$ and $\alpha_2$ are two constants, which can be determined from $p_i$ and $p_j$.
\\
\\
\bs
{\it \refstepcounter{exercise}
{\bf Exercise \theexercise}: 
Determine the constants $\alpha_1$ and $\alpha_2$ in eq.~(\ref{chapter_one_loop_def_l_1_l_2}) from the requirement that $l_1$ and $l_2$ are light-like, 
i.e. $l_1^2=l_2^2=0$.
Distinguish the cases
\begin{description}
\item{(i)} $p_i$ and $p_j$ are light-like.
\item{(ii)} $p_i$ is light-like, $p_j$ is not.
\item{(iii)} both $p_i$ and $p_j$ are not light-like.
\end{description}
}
\es
\noindent
In the second step we use $l_1$ and $l_2$ to write 
$\left\langle a - \left| \gamma_{\mu} \right| b - \right\rangle$
as a linear combination of the four basis vectors 
\bq
 \left\langle l_1 - \left| \gamma_\mu \right| l_1 - \right\rangle,
 \;\;\;
 \left\langle l_2 - \left| \gamma_\mu \right| l_2 - \right\rangle,
 \;\;\;
 \left\langle l_1 - \left| \gamma_\mu \right| l_2 - \right\rangle,
 \;\;\;
 \left\langle l_2 - \left| \gamma_\mu \right| l_1 - \right\rangle.
\eq
The contraction of $k^\mu_{(4)}$ with the first or second basis vector leads to
\bq
\label{chapter_one_loop:l1orl2scalprod}
 \left\langle l_1 - \left| \gamma_\mu \right| l_1 - \right\rangle k^\mu_{(4)} 
 & = & 2 k_l l_1 
 \; = \;
 \frac{1}{1-\alpha_1\alpha_2} \left( 2p_i k - \alpha_1 2p_j k \right),
 \nonumber \\
 \left\langle l_2 - \left| \gamma_\mu \right| l_2 - \right\rangle k^\mu_{(4)}
 & = & 2 k_l l_2 
 \; = \; 
 \frac{1}{1-\alpha_1\alpha_2} \left( -\alpha_2 2p_i k + 2p_j k \right),
 \nonumber \\
\eq
and therefore reduces immediately the rank of the tensor integral.
Repeating this procedure we end up with integrals, where the numerator is given by
products of
\bq
\label{chapter_one_loop:standard_form_tensor_struc}
 \left\langle l_1 - \left| \slashed{k}_{(4)} \right| l_2 - \right\rangle
 & \mbox{and} & 
 \left\langle l_2 - \left| \slashed{k}_{(4)} \right| l_1 - \right\rangle,
\eq
plus additional reduced integrals. 
Therefore the tensor integral is now in a standard form.
In the next step one reduces any product of factors as in eq.~(\ref{chapter_one_loop:standard_form_tensor_struc}).
For example, if in the tensor structure both spinor types appear, we can use
\bq
\label{chapter_one_loop:mixed_product}
\left\langle l_1- | \slashed{k}^{(4)} | l_2- \right\rangle \left\langle l_2- | \slashed{k}^{(4)} | l_1- \right\rangle
 & = & 
  \left( 2l_1 k \right) \left( 2 l_2 k \right) 
  - \left( 2 l_1 l_2 \right) \left(k^{(4)}\right)^2
\eq
and
\bq
 \left(k^{(4)}\right)^2
 & = &
 \left(k^{(D)}\right)^2 - \left(k^{(-2\eps)}\right)^2.
\eq
After repeated use of eq.~(\ref{chapter_one_loop:mixed_product}) we end up with a tensor integral, where only one spinor type appears.
These tensor integrals can be reduced with formulae, which depend on the number of external legs.
These formulae are not reproduced here, but can be found in the literature \cite{vanHameren:2005ed}.
After completion of this step, all tensor integrals are reduced to rank $1$ integrals.
Finally, the rank $1$ integrals are reduced to scalar integrals.
The relevant formulae for the last step depend again on the number of external legs and can be found in the literature \cite{vanHameren:2005ed}.

Therefore the only non-zero higher-dimensional integrals which occur in the Feynman gauge result from
the two-point function with a single power of $k_{(-2\eps)}^2$ in the numerator ($n=2$ and $s=1$),
the three-point function with a single power of $k_{(-2\eps)}^2$ in the numerator ($n=3$ and $s=1$)
and the four-point function with two powers of $k_{(-2\eps)}^2$ in the numerator ($n=4$ and $s=2$).
With $q_j=k-p_j^{\mathrm{sum}}$ we find for these cases:
\bq
 e^{\eps \Eulerconstant} \mu^{2\eps} \int \frac{d^Dk}{i \pi^{\frac{D}{2}}}
  \frac{\left(-k_{(-2\eps)}^2\right)}{\left(-q_1^2\right) \left(-q_2^2\right)}
 & = &
 - \frac{p^2}{6}  + {\cal O}(\eps),
 \nonumber \\
 e^{\eps \Eulerconstant} \mu^{2\eps} \int \frac{d^{D}k}{i \pi^{\frac{D}{2}}}
  \frac{\left(-k_{(-2\eps)}^2\right)}{\left(-q_1^2\right) \left(-q_2^2\right) \left(-q_3^2\right)}
 & = &
 - \frac{1}{2} + {\cal O}(\eps),
 \nonumber \\
 e^{\eps \Eulerconstant} \mu^{2\eps} \int \frac{d^{D}k}{i \pi^{\frac{D}{2}}}
  \frac{\left(-k_{(-2\eps)}^2\right)^2}{\left(-q_1^2\right) \left(-q_2^2\right) \left(-q_3^2\right) \left(-q_4^2\right)}
 & = &
 - \frac{1}{6} + {\cal O}(\eps).
\eq
Contributions of this type are called
\index{rational terms}
{\bf rational terms} 
(as they do not involve logarithms or dilogarithms at order ${\mathcal O}(\eps^0)$).
\\
\\
\bs
{\it \refstepcounter{exercise}
{\bf Exercise \theexercise}: 
The method above does not apply to a tensor two-point function, as there is only one linear independent external
momentum.
However, the tensor two-point functions is easily reduced with standard methods to the scalar two-point function.
In this exercise you are asked to work this out for the massless tensor two-point function.
The most general massless tensor two-point function is given by
\bq
I_2^{\mu_1 ... \mu_r,s} & = &
 e^{\eps \Eulerconstant} \mu^{2\eps} \int \frac{d^Dk}{i \pi^{\frac{D}{2}}}
  \left(-k_{(-2\eps)}^2\right)^s \frac{k^{\mu_1} ... k^{\mu_r}}{k^2 (k-p)^2}.
\eq
Reduce this tensor integral to a scalar integral.
}
\es

\section{Amplitude methods}
\label{chapter_one_loop:sect_amplitude_methods}

It is the scattering amplitude ${\mathcal A}_{\nexternal}$ which enters 
the formula eq.~(\ref{chapter_qft:observable_master_electron_positron})
for the expectation value of an observable.
By algorithm~\ref{chapter_qft:algo_amplitude} the scattering amplitude is computed through the sum of all relevant
Feynman diagrams.
However, the number of Feynman diagrams growths rapidly with the number of external particles
and this approach can become inefficient in practice.
Fortunately, we have for one-loop amplitudes methods which bypass individual Feynman diagrams.

From section~\ref{chapter_one_loop:sect_Passarino_Veltman} and section~\ref{chapter_one_loop:sect_reduction_higher_point}
we know that we can reduce any one-loop tensor integral 
to a set of scalar one-, two-, three-, four- and five-point functions.
By an appropriate choice of the basis integrals for the five-point functions it can be arranged, that
the five-point functions only contribute at order ${\mathcal O}(\eps)$ and are thus not relevant 
to NLO calculations, where we only need the $\eps$-expansion of scalar one-loop integrals up to and including
${\mathcal O}(\eps^0)$.
Thus, we may reduce any one-loop Feynman integral to
scalar one-, two-, three- and four-point functions plus terms of order ${\mathcal O}(\eps)$ and beyond, which are
not relevant to NLO calculations.
These scalar integrals are known (see section~\ref{chapter_one_loop:sect_dilogarithm}), therefore only the coefficients
of these integrals need to be determined.

In this section it is convenient to use the notation
\bq
\label{chapter_one_loop:amplitude_integral_notation}
I^{(i_1 \dots i_n)}_{n} & = &
 e^{\eps \Eulerconstant} \mu^{2\eps} 
  \int \frac{d^Dk}{i \pi^{\frac{D}{2}}}
  \frac{1}{\left(-q_{i_1}^2\right) \dots \left(-q_{i_n}^2\right)},
\eq
where the superscript $(i_1 \dots i_n)$ indicate the propagators present in the one-loop Feynman integral.
With this notation we may write for a one-loop amplitude in a massless theory
\bq
 {\mathcal A}_{\nexternal}^{(1)}
 & = &
 \sum\limits_{i_1<i_2<i_3<i_4} c_{i_1 i_2 i_3 i_4} I^{(i_1 i_2 i_3 i_4)}_{4}
 + \sum\limits_{i_1<i_2<i_3} c_{i_1 i_2 i_3} I^{(i_1 i_2 i_3)}_{3}
 + \sum\limits_{i_1<i_2} c_{i_1 i_2} I^{(i_1 i_2)}_{2}
 + {\mathcal O}\left(\eps\right).
\eq
$I^{(i_1 i_2)}_2$, $I^{(i_1 i_2 i_3)}_3$ and $I^{(i_1 i_2 i_3 i_4)}_4$ are the scalar bubble, triangle and  box integral functions.
In a massive theory we would have in addition also scalar one-point functions.
In a massless theory these functions are zero within dimensional regularisation.
Note that there are no integral functions with more than four internal propagators.
These higher-point functions can always be reduced to the set above, as we have seen
in section~\ref{chapter_one_loop:sect_reduction_higher_point}.
The coefficients $c_{i_1 i_2}$, $c_{i_1 i_2 i_3}$ and $c_{i_1 i_2 i_3 i_4}$
depend on the external momenta and the dimensional regularisation parameter $\eps$.
All poles in the dimensional regularisation parameter $\eps$ are contained in the scalar integral functions.
The coefficients $c_{i_1 i_2}$, $c_{i_1 i_2 i_3}$ and $c_{i_1 i_2 i_3 i_4}$ have a Taylor expansion in $\eps$.
We write
\bq
\label{chapter_one_loop:coefficient_decomposition}
 c_{i_1 \dots i_n}
 & = &
 c^{(0)}_{i_1 \dots i_n}
 + {\mathcal O}\left(\eps\right),
\eq
where $c^{(0)}_{i_1 \dots i_n}$ is the coefficient in four space-time dimensions.
We therefore have
\bq
\label{chapter_one_loop:one_loop_amplitude}
 {\mathcal A}_{\nexternal}^{(1)}
 & = &
 \sum\limits_{i_1<i_2<i_3<i_4} c^{(0)}_{i_1 i_2 i_3 i_4} I^{(i_1 i_2 i_3 i_4)}_{4}
 + \sum\limits_{i_1<i_2<i_3} c^{(0)}_{i_1 i_2 i_3} I^{(i_1 i_2 i_3)}_{3}
 + \sum\limits_{i_1<i_2} c^{(0)}_{i_1 i_2} I^{(i_1 i_2)}_{2}
 + R
 + {\mathcal O}\left(\eps\right),
 \nonumber \\
\eq
where the correction term $R$ contains all terms up to order $\eps^0$ originating from the ${\mathcal O}(\eps)$-terms
in eq.~(\ref{chapter_one_loop:coefficient_decomposition}) hitting a pole in $\eps$ from the scalar integral functions.
It can be shown that $R$ is of order $\eps^0$ and does not contain any logarithms.
$R$ is called the 
\index{rational terms}
{\bf rational term}.
The set of all occurring integral functions
\bq
 \{ I_2^{(i_1 i_2)}, I_3^{(i_1 i_2 i_3)}, I_4^{(i_1 i_2 i_3 i_4)} \}
\eq
is rather easily obtained from pinching in all possible ways internal propagators in all occurring diagrams.
We can assume that we know this set in advance.
Furthermore all integral functions in this set are known (see section~\ref{chapter_one_loop:sect_dilogarithm}).
To compute the amplitude requires therefore only the determination of the coefficients
$c^{(0)}_{i_1 i_2}$, $c^{(0)}_{i_1 i_2 i_3}$, $c^{(0)}_{i_1 i_2 i_3 i_4}$ and of the rational term $R$.

Below we discuss methods how this information can be obtained.
Readers only interested in the most practical method may directly jump to section~\ref{chapter_one_loop:sect_OPP}, where we discuss the Ossola-Papadopoulos-Pittau (OPP) method.
We follow the historical path and discuss first the unitarity-based method, followed by a discussion
of generalised unitarity before finally arriving at the OPP-method.
We do this, because some of the ideas like generalised cuts will reappear in the next chapter.

\subsection{The unitarity-based method}
\label{chapter_one_loop:sect_unitarity}

Loop amplitudes have branch cuts.
We denote the discontinuity across a branch cut in a particular channel $s$ by
\bq
 \mathrm{Disc}_s \; {\mathcal A}^{(\loopnumber)}_{\nexternal}
 & = &
 {\mathcal A}^{(\loopnumber)}_{\nexternal}\left(s+i \delta\right)
 -
 {\mathcal A}^{(\loopnumber)}_{\nexternal}\left(s-i \delta\right),
\eq
where $\delta>0$ denotes an infinitesimal quantity.
Please note that the discontinuity across a branch cut is a well-defined quantity,
while the imaginary part of a loop amplitude depends on several phase conventions, like the ones used in the expressions for
the polarisation factors for the external particles.
For one-loop amplitudes, the discontinuity across a branch cut
stems from the imaginary parts of the logarithm and the dilogarithm in certain regions of phase space.
We have for example
\bq
 \mathrm{Im} \; \ln \left( \frac{-s-i \delta}{-t-i \delta} \right) 
 & = & 
 - \pi \left[ \theta(s) -\theta(t) \right], 
 \nonumber \\
 \mathrm{Im} \; \mathrm{Li}_2 \left( 1 - \frac{(-s-i \delta)}{(-t-i \delta)} \right)
 & = & 
 - \ln \left( 1 - \frac{s}{t} \right)
        \mathrm{Im} \; \ln \left( \frac{-s-i \delta}{-t-i \delta} \right).
\eq
The unitarity-based method \cite{Bern:1994zx,Bern:1995cg} exploits the fact, that the discontinuities
of the basic integral functions are characteristic: Knowing the discontinuity we may uniquely reconstruct the integral
function and the coefficient accompanying it.
In general there will be discontinuities corresponding to different
channels (e.g. to the different possibilities to cut a one-loop
diagram into two parts).
The discontinuity in one channel of a one-loop amplitude 
can be obtained via unitarity from a phase space integral over
two tree-level amplitudes.
Let us consider a process with $\nexternal$ particles, which we label from $1$ to $\nexternal$.
Divide the set $\{1,\dots,\nexternal\}$ into two disjoint sets $I$ and $J$
\bq
 I \cup J \; = \; \{1,\dots,\nexternal\},
 & &
 I \cap J \; = \; \emptyset,
\eq
and consider the channel
\bq
 s & = & 
 \left( \sum\limits_{i \in I} p_i \right)^2
 \; = \;
 \left( \sum\limits_{j \in J} p_j \right)^2.
\eq
From the unitarity of the $S$-matrix (hence the name unitarity-based method)
\bq
 S^\dagger S & = & 1
\eq
and the Cutkosky rules \cite{Cutkosky:1960sp} one arrives at
\bq
\label{chapter_one_loop:cutconstr}
 \mathrm{Disc}_s \; {\mathcal A}_{\nexternal}^{(1)} 
 & = & 
 \sum\limits_{\lambda_1,\lambda_2}
 \int \frac{d^D k}{(2 \pi)^D i} 
 \left( 2 \pi i \right) \delta_+\left(q_1^2\right)
 \left( 2 \pi i \right) \delta_+\left(q_2^2\right)
 {\mathcal A}_{|I|+2}^{(0)} {\mathcal A}_{|J|+2}^{(0)}.
\eq
${\mathcal A}_{|I|+2}^{(0)}$ and ${\mathcal A}_{|J|+2}^{(0)}$
are tree-level amplitudes appearing on the left and right side
of the cut in a given channel, as shown in the first picture of fig.~\ref{chapter_one_loop:fig_cuts}.
The momenta crossing the cut are $q_1$ and $q_2$. We set $q_1=k$ (setting $q_2=k$ would equally be possible).
The sub-script ``$+$'' of $\delta_+(q^2)$ selects the solution of $q^2=0$ with positive energy. 
The factors of $i$ follow from our convention, that $i {\mathcal A}_{\nexternal}^{(1)}$ equals the sum of all relevant Feynman diagrams. 
The amplitude ${\mathcal A}_{|I|+2}^{(0)}$ has $|I|+2$ external particles, the (outgoing) momenta of these particles 
are
\bq
 q_1, q_2, p_i, & & i \in I.
\eq
The amplitude ${\mathcal A}_{|J|+2}^{(0)}$ has $|J|+2$ external particles, the (outgoing) momenta of these particles 
are
\bq
 -q_1, -q_2, p_j, & & j \in J.
\eq
The sum over $\lambda_1$ and $\lambda_2$ in eq.~(\ref{chapter_one_loop:cutconstr})
is over the spins of the two particles crossing the cut.
Note that in ${\mathcal A}_{|I|+2}^{(0)}$ and ${\mathcal A}_{|J|+2}^{(0)}$ all external particles are on-shell, 
also the ones with momenta $q_1$ and $q_2$ (respectively $(-q_1)$ and $(-q_2)$).

Let us denote by ${\mathcal A}_{|I|+2}^{(0),\mathrm{off}}$ and ${\mathcal A}_{|J|+2}^{(0),\mathrm{off}}$
off-shell continuations with respect to $q_1$ and $q_2$ of ${\mathcal A}_{|I|+2}^{(0)}$ and ${\mathcal A}_{|J|+2}^{(0)}$, respectively. An off-shell continuation is neither unique nor gauge-invariant.
Two off-shell continuations of ${\mathcal A}_{|I|+2}^{(0)}$ (or ${\mathcal A}_{|J|+2}^{(0)}$) 
may differ by terms proportional to $q_1^2$ or $q_2^2$.
However, neither the non-uniqueness nor the gauge-dependence matter for the subsequent argument.
Lifting eq. (\ref{chapter_one_loop:cutconstr}) one obtains
\bq
 {\mathcal A}_{\nexternal}^{(1)} 
 & = & 
 \sum\limits_{\lambda_1,\lambda_2}
 \int \frac{d^D k}{(2 \pi)^D i} 
 \frac{1}{q_1^2}
 \frac{1}{q_2^2} 
 {\mathcal A}_{|I|+2}^{(0),\mathrm{off}} {\mathcal A}_{|J|+2}^{(0),\mathrm{off}}
 + \;\mbox{cut free pieces},
\eq
where ``cut free pieces'' denote contributions which do not develop an imaginary
part in this particular channel.
By evaluating the cut, one determines the coefficients of the integral
functions, which have a discontinuity in this channel.
Iterating over all possible cuts, one finds all coefficients.
One advantage of a cut-based calculation is that one starts with tree amplitudes on both sides of the cut, which are already sums
of Feynman diagrams. 
Therefore cancellations and simplifications, which usually occur between
various diagrams, can already be performed before we start the calculation
of the loop amplitude.
The rational part $R$ can be obtained by calculating higher order terms in $\eps$ within the cut-based method.
At one-loop order an arbitrary scale $\mu^{2\eps}$ is introduced in order to keep the coupling
dimensionless. In a massless theory the factor $\mu^{2\eps}$ is always accompanied
by some kinematical invariant $s^{-\eps}$ for dimensional reasons.
If we write symbolically
\bq
 {\mathcal A}_{\nexternal}^{(1)}
 & = & 
 \frac{c_2}{\eps^2} \left( \frac{s_2}{\mu^2} \right)^{-\eps} 
 + \frac{c_1}{\eps} \left( \frac{s_1}{\mu^2} \right)^{-\eps}
 + c_0 \left( \frac{s_0}{\mu^2} \right)^{-\eps} 
 + \eps \tilde{R}
 + {\mathcal O}\left(\eps^2\right),
\eq
where $\tilde{R}$ is independent of $\eps$ and free of discontinuities,
the cut-free pieces $c_0 (s_0/\mu^2)^{-\eps}$ can be detected at order $\eps$:
\bq
 c_0 \left( \frac{s_0}{\mu^2} \right)^{-\eps} 
 & = & 
 c_0 - \eps c_0 \ln \left(\frac{s_0}{\mu^2}\right) + O(\eps^2).
\eq

\subsection{Generalised unitarity}
\label{chapter_one_loop:sect_generalised_unitarity}

The unitarity-based method allows us to bypass the set of all one-loop diagrams.
We sew two tree amplitudes together (for which very often compact expressions are known) and perform
a traditional tensor reduction on the resulting integrand.

However, we may push things even further.
Apart from the two-particle cut discussed in the previous section, one can also consider triple or quadruple cuts
\begin{figure}
\begin{center}
\includegraphics[scale=0.8]{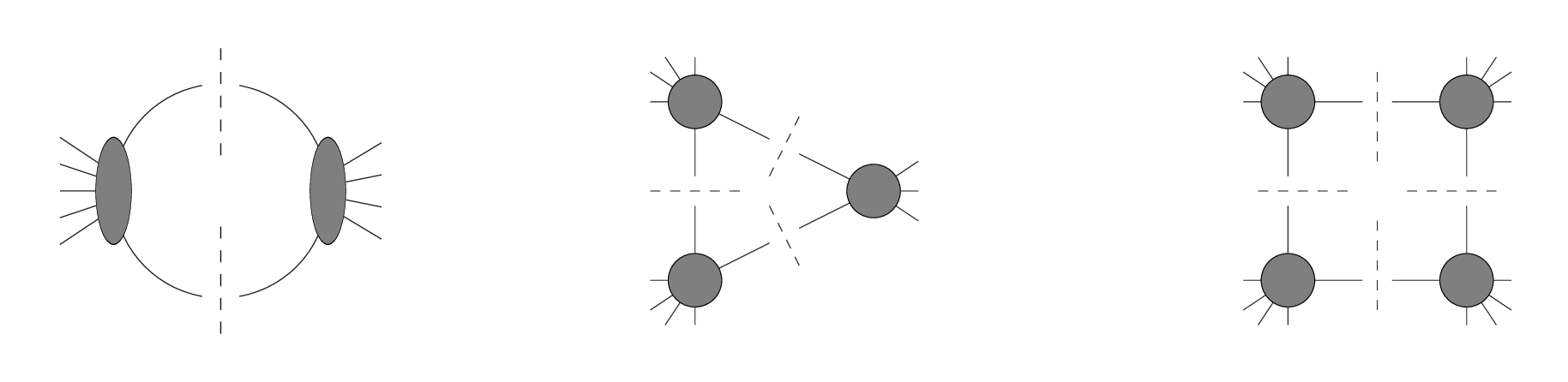}
\end{center}
\caption{Double, triple and quadruple cuts.}
\label{chapter_one_loop:fig_cuts}
\end{figure}
as shown in fig.~\ref{chapter_one_loop:fig_cuts}.
These more general cuts motivate the name ``generalised unitarity''.
A particular nice result follows from quadruple cuts \cite{Britto:2004nc}: 
Let us consider the coefficient $c^{(0)}_{i_1 i_2 i_3 i_4}$ of the scalar box integral functions $I^{(i_1 i_2 i_3 i_4)}_{4}$
in eq.~(\ref{chapter_one_loop:one_loop_amplitude}).
The quadruple cut is defined by the equations
\bq
 q_{i_1}^2 = q_{i_2}^2 = q_{i_3}^2 = q_{i_4}^2 = 0,
\eq
where $q_j = k - p_j^{\mathrm{sum}}$.
These equations have in four space-time dimensions two solutions for $k$, which we denote by $k^+$ and $k^-$.
We also set $q_j^\pm = k^\pm - p_j^{\mathrm{sum}}$.
\\
\\
\bs
{\it \refstepcounter{exercise}
{\bf Exercise \theexercise}: 
Consider a one-loop four-point function with external momenta $p_1, p_2, p_3, p_4$ and 
$p_1^2=p_2^2=p_3^2=p_4^2=0$. The external momenta satisfy momentum conservation $p_1+p_2+p_3+p_4=0$.
For $j \in \{1,2,3,4\}$ set $q_j=k-p_j^{\mathrm{sum}}$.
Solve the equations for the quadruple cut
\bq
 q_{1}^2 = q_{2}^2 = q_{3}^2 = q_{4}^2 = 0.
\eq
Hint: Start from an ansatz
\bq
 k_\mu & = & c \left\langle a- | \gamma_\mu | b- \right\rangle,
\eq
with $c \in {\mathbb C}$ and $a, b$ light-like.
}
\es
\\
\\
The coefficient of the box integral function
is proportional to a product of four tree amplitudes, summed over the spins of the particles crossing the cuts and
averaged over the two solutions of the on-shell conditions.
Let us say that the quadruple cut divides the labels $\{1,\dots,\nexternal\}$ into four disjoint sets $I_1$, $I_2$, $I_3$ and $I_4$,
corresponding to the four corners in the right picture of fig.~\ref{chapter_one_loop:fig_cuts}.
Assume further that $q_{i_1}$, $q_{i_2}$, $q_{i_3}$ and $q_{i_4}$ are labelled such that the external momenta of the four
tree amplitudes are
\bq
 {\mathcal A}_{|I_1|+2}^{(0)}\left(q_{i_1},\dots,-q_{i_4}\right),
 \;\;\;
 {\mathcal A}_{|I_2|+2}^{(0)}\left(q_{i_2},\dots,-q_{i_1}\right),
 \;\;\;
 {\mathcal A}_{|I_3|+2}^{(0)}\left(q_{i_3},\dots,-q_{i_2}\right),
 \;\;\;
 {\mathcal A}_{|I_4|+2}^{(0)}\left(q_{i_4},\dots,-q_{i_3}\right).
 \nonumber
\eq
Then
\bq
\label{chapter_one_loop:box_coefficient}
 c^{(0)}_{i_1 i_2 i_3 i_4}
 & = &
 \frac{1}{2}
 \frac{1}{\left(4\pi\right)^2}
 \sum\limits_{\lambda_1, \lambda_2, \lambda_3, \lambda_4}
 \sum\limits_{\sigma=\pm}
 {\mathcal A}_{|I_1|+2}^{(0)}\left(q_{i_1}^\sigma,\dots,-q_{i_4}^\sigma\right)
 {\mathcal A}_{|I_2|+2}^{(0)}\left(q_{i_2}^\sigma,\dots,-q_{i_1}^\sigma\right)
 \nonumber \\
 & &
 \times
 {\mathcal A}_{|I_3|+2}^{(0)}\left(q_{i_3}^\sigma,\dots,-q_{i_2}^\sigma\right)
 {\mathcal A}_{|I_4|+2}^{(0)}\left(q_{i_4}^\sigma,\dots,-q_{i_3}^\sigma\right).
\eq
The factor $1/(4\pi)^2$ comes from our convention for the integral measure in $d^Dk/\pi^{D/2}$ in eq.~(\ref{chapter_one_loop:amplitude_integral_notation}) instead of $d^Dk/(2\pi)^D$.
The sum over $\lambda_1, \dots, \lambda_4$ is over the spins of the particles crossing the cuts.
\\
\\
\bs
{\it \refstepcounter{exercise}
{\bf Exercise \theexercise}: 
Consider the one-loop eight-point amplitude in massless $\phi^4$ theory.
Verify eq.~(\ref{chapter_one_loop:box_coefficient}) for the box coefficient.
}
\es
\\
\\
Having determined all box coefficients with quadrupole cuts, one may move on to the triangle coefficients by considering triple
cuts.
The triple cut receives contributions from box integrals and triangle integrals.
As we already have determined the coefficient of all box integrals, we may subtract out the box contributions and uniquely identify
the triangle coefficients.
This can then be repeated for the bubble coefficients: Having all the coefficients of the box and triangle integrals at hand,
we consider double cuts. These cuts receive contributions from box integrals, triangle integrals and bubble integrals.
Subtracting out the contributions from the box integrals and triangle integrals, one may extract the coefficients of the bubble 
integrals.

\subsection{The OPP method}
\label{chapter_one_loop:sect_OPP}

We now discuss the method of Ossola, Papadopoulos and Pittau \cite{Ossola:2006us}.
Let us start with a preliminary remark:
It is always possible to decompose a one-loop amplitude ${\mathcal A}_{\nexternal}^{(1)}$
into cyclic-ordered primitive amplitudes $A_{\nexternal}^{(1)}$:
\bq
 {\mathcal A}_{\nexternal}^{(1)}
 & = & 
 \sum\limits_{\sigma \in S_{\nexternal}/{\mathbb Z}_{\nexternal}}
 A_{\nexternal}^{(1)}\left(\sigma\right).
\eq
For non-gauge amplitudes this is a trivial statement, for gauge amplitudes the non-trivial part of this decomposition
is the fact that the primitive amplitudes $A_{\nexternal}^{(1)}$ are themselves gauge-invariant.
The cyclic order of the external legs is specified by the permutation $\sigma \in S_{\nexternal}/{\mathbb Z}_{\nexternal}$.
Without loss of generalisation we will consider in the following the cyclic order
$\sigma = (1,2,\dots,\nexternal)$.
Working with cyclic-ordered primitive amplitudes has the advantage that only $\nexternal$ different loop propagators may appear.
For simplicity let us discuss -- as before -- massless theories.
We may write
\bq
 A_{\nexternal}^{(1)}
 & = & 
 e^{\eps \Eulerconstant} \mu^{2\eps} 
  \int \frac{d^Dk}{i \pi^{\frac{D}{2}}}
  \frac{P\left(k\right)}{\prod\limits_{j=1}^{\nexternal} \left(-q_j^2\right)},
\eq
with $q_j = k - p_j^{\mathrm{sum}}$ and $p_j^{\mathrm{sum}}=\sum_{i=1}^j p_i$.
The numerator $P(k)$ is a polynomial in the loop momentum $k$.
The degree of this polynomial is bounded.
For example, we have in gauge theories in a renormalisable gauge
\bq
 \deg P\left(k\right) & \le & \nexternal.
\eq
Furthermore, $P(k)$ can be computed easily by a tree-like calculation. 
Let us split the numerator into a four-dimensional part and a remainder
\bq
\label{chapter_one_loop:OPP_numerator_splitting}
 P\left(k\right)
 & = &
 P\left(k^{(4)}\right) + \tilde{P}\left(k^{(4)},k^{(-2\eps)}\right),
\eq
The OPP method consists in writing
\bq
\label{chapter_one_loop:OPP_ansatz}
 P\left(k^{(4)}\right)
 & = &
 \sum\limits_{i_1<i_2<i_3<i_4} \left[ c^{(0)}_{i_1 i_2 i_3 i_4} + \tilde{c}_{i_1 i_2 i_3 i_4}\left(k^{(4)}\right) \right] 
  \prod\limits_{i \notin \{i_1,i_2,i_3,i_4\}} \left[-\left(q_i^{(4)}\right)^2\right]
 \nonumber \\
 & &
 +
 \sum\limits_{i_1<i_2<i_3} \left[ c^{(0)}_{i_1 i_2 i_3} + \tilde{c}_{i_1 i_2 i_3}\left(k^{(4)}\right) \right] 
  \prod\limits_{i \notin \{i_1,i_2,i_3\}} \left[-\left(q_i^{(4)}\right)^2\right]
 \nonumber \\
 & &
 +
 \sum\limits_{i_1<i_2} \left[ c^{(0)}_{i_1 i_2} + \tilde{c}_{i_1 i_2}\left(k^{(4)}\right) \right] 
  \prod\limits_{i \notin \{i_1,i_2\}} \left[-\left(q_i^{(4)}\right)^2\right].
\eq
The terms $\tilde{c}_{i_1 i_2 i_3 i_4}$, $\tilde{c}_{i_1 i_2 i_3}$ and $\tilde{c}_{i_1 i_2}$ have the property, that they vanish
after integration.
Their dependence on $k^{(4)}$ is known up to some yet to be determined parameters, on which these terms depend linearly.
To give an example, let's consider $\tilde{c}^{(0)}_{i_1 i_2 i_3 i_4}$. We denote by $p_1', p_2', p_3', p_4'$ the external momenta of the box function
with propagators $(-q_{i_1})^2$, $(-q_{i_2})^2$, $(-q_{i_3})^2$ and $(-q_{i_4})^2$.
Then 
\bq
 \tilde{c}_{i_1 i_2 i_3 i_4}\left(k^{(4)}\right)
 & = &
 \tilde{C}_{i_1 i_2 i_3 i_4}
 \mathrm{Tr}\left(\slashed{q}_{i_4} \slashed{p}_1' \slashed{p}_2' \slashed{p}_3' \gamma_5 \right)
 \; = \;
 4 i \tilde{C}_{i_1 i_2 i_3 i_4} \eps_{\mu\nu\rho\sigma} q_{i_4}^\mu p_1'{}^\nu p_2'{}^\rho p_3'{}^\sigma.
\eq
It is clear that a rank one box integral with this numerator will vanish after integration: We may choose $k'=q_{i_4}$.
From Passarino-Veltman reduction we know that $k'{}^\mu$ will become proportional to either $p_1'{}^\mu$, $p_2'{}^\mu$ or $p_3'{}^\mu$
after integration. But this vanishes when contracted into the antisymmetric tensor.
$\tilde{C}_{i_1 i_2 i_3 i_4}$ is the yet to be determined parameter.

These parameters and the constants $c^{(0)}_{i_1 i_2 i_3 i_4}$, $c^{(0)}_{i_1 i_2 i_3}$ and $c^{(0)}_{i_1 i_2}$
can be determined by evaluating the left-hand side and the right-hand side of eq.~(\ref{chapter_one_loop:OPP_ansatz}) 
for various values of $k^{(4)}$.
Solving this linear system yields the coefficients $c^{(0)}_{i_1 i_2 i_3 i_4}$, $c^{(0)}_{i_1 i_2 i_3}$ and $c^{(0)}_{i_1 i_2}$
of the scalar integral functions.
It remains to extract the rational term $R$.
There are two sources contributing to $R$ and we write \cite{Ossola:2008xq}
\bq
 R & = & R_1 + R_2.
\eq
First of all, the factors $(-(q_i^{(4)})^2)$ in eq.~(\ref{chapter_one_loop:OPP_ansatz}) do not cancel exactly
the denominators $(-q_i^2)$.
There is a mismatch
\bq
 -\left(q_i^{(4)}\right)^2
 & = &
 -q_i^2
 +
 \left(q_i^{(-2\eps)}\right)^2
 \; = \;
 -q_i^2
 +
 \left(k^{(-2\eps)}\right)^2.
\eq
The terms proportional to $(k^{(-2\eps)})^2$ make up the rational term $R_1$.

Secondly, we split in eq.~(\ref{chapter_one_loop:OPP_numerator_splitting}) the numerator $P(k)$ into
a four-dimensional part and a remainder.
The remainder $\tilde{P}(k^{(4)},k^{(-2\eps)})$ makes up the rational term $R_2$.

%% file: iterated_integrals.tex
\newpage
\chapter{Iterated integrals}
\label{chapter_iterated_integrals}

In this chapter we introduce modern methods to tackle Feynman integrals.
The main tool will be the method of differential equations.
This builds on integration-by-parts identities and dimensional shift relations, which we discuss first.
If the system of differential equations can be brought into a particular simple form 
(the $\eps$-form which is introduced in section~\ref{chapter_iterated_integrals:section:eps_form}),
a solution in terms of iterated integrals is immediate.
The methods of this chapter reduce the problem of computing Feynman integrals to the problem
of finding an appropriate transformation for the system of differential equations.
Algorithms to construct an appropriate transformation are discussed in chapter~\ref{chapter_transformations}.
We will see that integration-by-parts identities allow us to express any Feynman integral as a linear
combination of basis integrals, which we call master integrals.
The master integrals span a vector space and viewing the Feynman integrals as functions of the kinematic variables
gives us a vector bundle.
We discuss fibre bundles in section~\ref{chapter_iterated_integrals:fibre_bundles}.

Sections~\ref{chapter_iterated_integrals:cuts} and \ref{chapter_iterated_integrals:singularities}
are devoted to cuts of Feynman integrals and singularities of Feynman integrals, respectively.

As we may express any Feynman integral as a linear
combination of master integrals, we may ask if this vector space is equipped with an inner product.
An inner product would allow us to compute the coefficient in front of each master integral directly,
bypassing the need to solve a linear system of integration-by-parts identities. 
This leads us to twisted cohomology, which we introduce in section~\ref{chapter_iterated_integrals:twisted_cohomology}.

\section{Integration-by-parts}
\label{chapter_iterated_integrals:integration_by_parts}

In this section we study more closely the family of Feynman integrals
\bq
 I_{\nu_1 \dots \nu_{\ninternal}}\left(D,x_1,\dots,x_{\NB}\right)
 & = &
 e^{\loopnumber \eps \Eulerconstant} \left(\mu^2\right)^{\nu-\frac{\loopnumber D}{2}}
 \int \prod\limits_{r=1}^{\loopnumber} \frac{d^Dk_r}{i \pi^{\frac{D}{2}}} 
 \prod\limits_{j=1}^{\ninternal} \frac{1}{\left(-q_j^2+m_j^2\right)^{\nu_j}}.
\eq
We are in particular interested in relations between members of this family, 
which differ by the values of the indices $(\nu_1, \dots, \nu_{\ninternal})$.
Integration-by-parts identities provide these relations \cite{Tkachov:1981wb,Chetyrkin:1981qh}.
Integration-by-parts identities are based on the fact that within dimensional regularisation the integral
of a total derivative vanishes
\bq
\label{chapter_iterated_integrals:basic_ibp_relation}
 \int 
 \frac{d^Dk}{i \pi^{\frac{D}{2}}}
 \;\;
 \frac{\partial}{\partial k^\mu} \; \left[ q^\mu \cdot f\left(k\right) \right]
 & = & 0,
\eq
i.e. there are no boundary terms. The vector $q$ can be any linear combination of the external momenta and the loop momentum $k$.
Eq.~(\ref{chapter_iterated_integrals:basic_ibp_relation}) is derived as follows:
Let us first assume that $q$ is a linear combination of the external momenta.
Integrals within dimensional regularisation are translation invariant 
(see eq.~(\ref{chapter_basics:translation_invariance}))
\bq
 \int \frac{d^Dk}{i \pi^{\frac{D}{2}}} f\left(k\right) 
 & = & 
 \int \frac{d^Dk}{i \pi^{\frac{D}{2}}} f\left(k+\lambda q\right).
\eq
The right-hand side has to be independent of $\lambda$. This implies in particular that the ${\mathcal O}(\lambda)$-term
has to vanish.
From
\bq
 \left. \left[ \frac{d}{d\lambda} f\left(k+\lambda q\right) \right] \right|_{\lambda=0}
 & = &
 q^\mu \frac{\partial}{\partial k^\mu} f\left(k\right)
 \; = \;
 \frac{\partial}{\partial k^\mu} \left[ q^\mu \cdot f\left(k\right) \right]
\eq
eq.~(\ref{chapter_iterated_integrals:basic_ibp_relation}) follows.

Eq.~(\ref{chapter_iterated_integrals:basic_ibp_relation}) also holds for $q=k$.
This is the task of the next exercise:
\\
\\
\bs
{\it \refstepcounter{exercise}
{\bf Exercise \theexercise}: 
Show that eq.~(\ref{chapter_iterated_integrals:basic_ibp_relation}) holds for $q=k$.
\\
Hint: Consider the scaling relation eq.~(\ref{chapter_basics:scaling_integral_measure}).
}
\es
\\
\\
Let us formulate the integration-by-parts identities for $\loopnumber$-loop integrals:
\begin{tcolorbox}
{\bf Integration-by-parts identities}:
\\
Within dimensional regularisation we have for any loop momentum $k_i$ ($1 \le i \le \loopnumber$)
and any vector $q_{\mathrm{IBP}} \in \{p_1,...,p_{N_{\mathrm{ext}}},k_1,...,k_{\loopnumber}\}$
\bq
\label{chapter_iterated_integrals:ibp}
 e^{\loopnumber \eps \Eulerconstant} \left(\mu^2\right)^{\nu-\frac{\loopnumber D}{2}}
 \int 
 \prod\limits_{r=1}^{\loopnumber} \frac{d^Dk_r}{i \pi^{\frac{D}{2}}}
 \;\;
 \frac{\partial}{\partial k_i^\mu} \; q_{\mathrm{IBP}}^\mu
 \;\;
 \prod\limits_{j=1}^{\ninternal} \frac{1}{\left(-q_j^2+m_j^2\right)^{\nu_j}}
 & = & 0.
\eq
Working out the derivatives leads to relations among integrals
with different sets of indices $(\nu_1, \dots, \nu_{\ninternal})$.
\end{tcolorbox}
Let's see how this works in an example:
We consider the one-loop two-point function with an equal internal mass:
\bq
\label{chapter_iterated_integrals:example_equal_mass_bubble}
 I_{\nu_1 \nu_2}\left(D,x\right)
 & = &
 e^{\eps \Eulerconstant} \left(m^2\right)^{\nu_{12}-\frac{D}{2}}
 \int \frac{d^Dk}{i \pi^{\frac{D}{2}}} 
 \frac{1}{\left(-q_1^2+m^2\right)^{\nu_1} \left(-q_2^2+m^2\right)^{\nu_2}}.
\eq
with $q_1=k-p$, $q_2=k$, $\nu_{12}=\nu_1+\nu_2$ and $x=-p^2/m^2$.
We have set $\mu^2=m^2$.
As vector $q_{\mathrm{IBP}}$ we may take $q_{\mathrm{IBP}} \in \{p,k\}$.
Let us start with $q_{\mathrm{IBP}}=p$. We obtain
\bq
 0
 & = &
 e^{\eps \Eulerconstant} \left(m^2\right)^{\nu_{12}-\frac{D}{2}}
 p^\mu
 \int \frac{d^Dk}{i \pi^{\frac{D}{2}}} 
 \frac{\partial}{\partial k^\mu} 
 \frac{1}{\left(-q_1^2+m^2\right)^{\nu_1} \left(-q_2^2+m^2\right)^{\nu_2}}
 \nonumber \\
 & = &
 e^{\eps \Eulerconstant} \left(m^2\right)^{\nu_{12}-\frac{D}{2}}
 \int \frac{d^Dk}{i \pi^{\frac{D}{2}}} 
 \left[ 
 \frac{\nu_1 \left(q_2^2-q_1^2-p^2\right)}{\left(-q_1^2+m^2\right)^{\nu_1+1} \left(-q_2^2+m^2\right)^{\nu_2}}
 \right. \nonumber \\
 & & \left.
 \hspace*{40mm}
 +
 \frac{\nu_2 \left(q_2^2-q_1^2+p^2\right)}{\left(-q_1^2+m^2\right)^{\nu_1} \left(-q_2^2+m^2\right)^{\nu_2+1}}
 \right]
 \nonumber \\
 & = &
 \nu_1 \left[ I_{\nu_1 \nu_2} - I_{(\nu_1+1) (\nu_2-1)} + x I_{(\nu_1+1) \nu_2} \right]
 + \nu_2 \left[ I_{(\nu_1-1) (\nu_2+1)} - I_{\nu_1 \nu_2} - x I_{\nu_1 (\nu_2+1)} \right].
\eq
In deriving this result we used
\bq
\label{chapter_iterated_integrals:ibp_scalar_products}
 2 p \cdot q_1 & = & q_2^2 - q_1^2 - p^2, 
 \nonumber \\
 2 p \cdot q_2 & = & q_2^2 - q_1^2 + p^2.
\eq
Thus we obtain a relation between integrals with different values of $(\nu_1,\nu_2)$:
\bq
\label{chapter_iterated_integrals:ibp_eq1}
 \left( \nu_1 - \nu_2 \right) I_{\nu_1 \nu_2} 
 - \nu_1 I_{(\nu_1+1) (\nu_2-1)} 
 + \nu_2 I_{(\nu_1-1) (\nu_2+1)} 
 + \nu_1 x I_{(\nu_1+1) \nu_2}
 - \nu_2 x I_{\nu_1 (\nu_2+1)}
 = 0. 
\eq
\bs
{\it \refstepcounter{exercise}
{\bf Exercise \theexercise}: 
Repeat the derivation with $q_{\mathrm{IBP}}=k$ and show
\bq
\label{chapter_iterated_integrals:ibp_eq2}
 \left( D - \nu_1 - 2 \nu_2 \right) I_{\nu_1 \nu_2}
 - \nu_1 I_{(\nu_1+1) (\nu_2-1)} + \nu_1 \left(2+x\right) I_{(\nu_1+1) \nu_2} 
 + 2 \nu_2 I_{\nu_1 (\nu_2+1)} 
 & = & 0.
\eq
}
\es
\\
\\
Instead of eq.~(\ref{chapter_iterated_integrals:ibp_eq1}) and eq.~(\ref{chapter_iterated_integrals:ibp_eq2})
we may consider two independent linear combinations, where either the integral
$I_{(\nu_1+1) \nu_2}$ or the integral $I_{\nu_1 (\nu_2+1)}$ is absent:
\bq
\label{chapter_iterated_integrals:ibp_equations}
\lefteqn{
 \nu_1 x \left(4+x\right) I_{(\nu_1+1) \nu_2}
 = } & &
 \nonumber \\
 & &
 \left[ 2 \left( -\nu_1+\nu_2\right) + \left( \nu_1 + 2\nu_2 - D \right) x \right] I_{\nu_1 \nu_2}
 + \nu_1 \left(2+x\right) I_{(\nu_1+1)(\nu_2-1)}
 - 2 \nu_2 I_{(\nu_1-1)(\nu_2+1)},
 \nonumber \\
\lefteqn{
 \nu_2 x \left(4+x\right) I_{\nu_1 (\nu_2+1)}
 = } & &
 \nonumber \\
 & &
 \left[ 2 \left( \nu_1-\nu_2\right) + \left( 2 \nu_1 + \nu_2 -D \right) x \right] I_{\nu_1 \nu_2}
 - 2 \nu_1 I_{(\nu_1+1)(\nu_2-1)}
 + \nu_2 \left(2+x\right) I_{(\nu_1-1)(\nu_2+1)}.
\eq
For $\nu_1>0$ and $\nu_2>0$ we may use either the first or the second equation to reduce the sum $\nu_1+\nu_2$:
In both equations, the sum of the indices equals $\nu_1+\nu_2+1$ on the left-hand side, while
on the right-hand side the sum of the indices 
equals for all terms $\nu_1+\nu_2$.
\begin{figure}
\begin{center}
\includegraphics[scale=1.0]{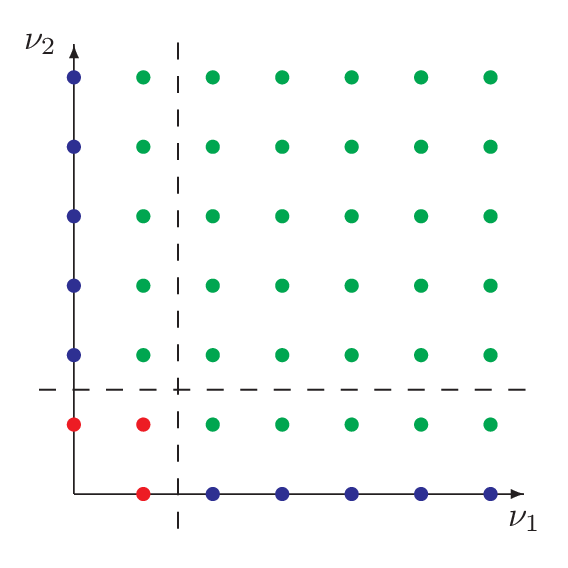}
\end{center}
\caption{
Integration-by-parts reduction for the one-loop two-point function:
For all integrals to the right of the vertical dashed line and indicated by a green dot, we may use
the first equation of eq.~(\ref{chapter_iterated_integrals:ibp_equations}) to lower $\nu_{12}$,
for all integrals above the horizontal dashed line and indicated by a green dot, we may use
the second equation of eq.~(\ref{chapter_iterated_integrals:ibp_equations}) to lower $\nu_{12}$.
Integrals represented by a blue dot are reduced with eq.~(\ref{chapter_iterated_integrals:ibp_equations_tadpole}).
The integrals with a red dot cannot be reduced to simpler integrals.
}
\label{chapter_iterated_integrals:ibp_reduction}
\end{figure}
If either $\nu_1=0$ (and $\nu_2>0$) or $\nu_2=0$ (and $\nu_1>0$)
we have a simpler integral: The integral reduces to a tadpole integral.
As the two internal masses are equal, we have
\bq
 I_{\nu 0} & = & I_{0 \nu}.
\eq
\bs
{\it \refstepcounter{exercise}
{\bf Exercise \theexercise}: 
Derive the integration-by-parts identity for the integral
\bq
 I_{0 \nu_2}
 & = &
 e^{\eps \Eulerconstant} \left(m^2\right)^{\nu_{2}-\frac{D}{2}}
 \int \frac{d^Dk}{i \pi^{\frac{D}{2}}} 
 \frac{1}{\left(-k^2+m^2\right)^{\nu_2}}.
\eq
Verify the identity with the explicit result from eq.~(\ref{chapter_basics:result_tadpole}).
}
\es
\\
\\
In the previous exercise you were supposed to derive the identity
\bq
\label{chapter_iterated_integrals:ibp_equations_tadpole}
 \nu_2 I_{0 (\nu_2+1)}
 & = &
 \left(\nu_2-\frac{D}{2} \right) I_{0 \nu_2}.
\eq
This identity can be used to reduce for $\nu_2>0$ the integral $I_{0 (\nu_2+1)}$.
The situation is summarised in fig.~\ref{chapter_iterated_integrals:ibp_reduction}.
We may reduce with integration-by-parts identities 
any integral $I_{\nu_1 \nu_2}$ with $\nu_1\ge 0$, $\nu_2 \ge 0$ and $\nu_1+\nu_2>0$
to a linear combination of $I_{1 1}$, $I_{1 0}$ and $I_{0 1}$. 
In the equal mass case we have the symmetry $I_{0 1}=I_{1 0}$, and therefore any integral
$I_{\nu_1 \nu_2}$ with $\nu_1\ge 0$, $\nu_2 \ge 0$ and $\nu_1+\nu_2>0$
can be reduced to a linear combination of $I_{1 1}$ and $I_{1 0}$.
We call $I_{1 1}$ and $I_{1 0}$ 
\index{master integrals}
{\bf master integrals}.

Let us now return to the general case.
We consider a graph $G$ which has a Baikov representation.
This ensures that we may express any scalar product involving a loop momentum
as a linear combination of inverse propagators and loop momentum independent terms.
In our example above we needed this property in eq.~(\ref{chapter_iterated_integrals:ibp_scalar_products}).
We consider integrals
\bq
 I_{\nu_1 \dots \nu_{\ninternal}},
 & & 
 \nu_j \in {\mathbb Z}.
\eq
We call integrals, where all indices satisfy $\nu_j > 0$,
integrals of the {\bf top topology} (or of the {\bf top sector}).
Integrals, where one or more indices satisfy $\nu_j < 1$, belong to a sub-topology (or belong to a sub-sector).
\begin{figure}
\begin{center}
\includegraphics[scale=1.0]{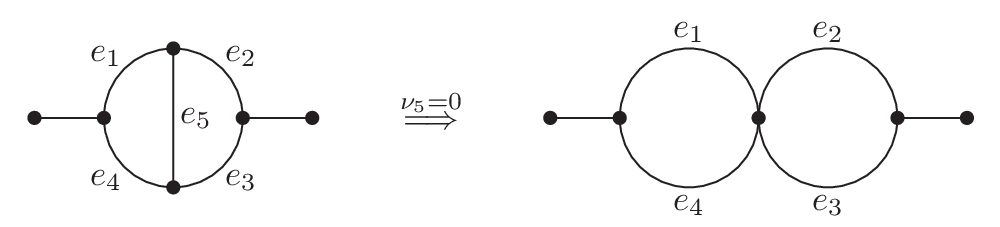}
\end{center}
\caption{
Integrals, where one or more indices satisfy $\nu_j < 1$, belong to a sub-topology.
The figure shows as an example the two-loop two-point integral: 
The case $\nu_5=0$ corresponds to the sub-topology obtained by pinching the edge $e_5$. 
Note that also the case $\nu_5<0$ corresponds to the sub-topology shown in the right picture, in this case
with an irreducible scalar product in the numerator. 
}
\label{chapter_iterated_integrals:pinching}
\end{figure}
This is illustrated in fig.~\ref{chapter_iterated_integrals:pinching}.

For a given set of indices $(\nu_1, \dots, \nu_{\ninternal})$ we define
\bq
\label{chapter_iterated_integrals:def_notation_sectors}
 N_{\mathrm{prop}}
 \; = \;
 \sum\limits_{j=1}^{\ninternal} \Theta\left(\nu_j-\frac{1}{2}\right),
 & &
 N_{\mathrm{id}}
 \; = \;
 \sum\limits_{j=1}^{\ninternal} 2^{j-1} \Theta\left(\nu_j-\frac{1}{2}\right),
 \nonumber \\
 r
 \; = \;
 \sum\limits_{j=1}^{\ninternal} \nu_j \Theta\left(\nu_j-\frac{1}{2}\right),
 & &
 s
 \; = \;
 \sum\limits_{j=1}^{\ninternal} \left|\nu_j\right| \Theta\left(-\nu_j+\frac{1}{2}\right).
\eq
$\Theta(x)$ denotes the Heaviside step function.
(Adding/subtracting the constant $1/2$ avoids ambiguities in the definition of $\Theta(0)$.)
$N_{\mathrm{prop}}$ counts the number of propagators having positive indices.
$N_{\mathrm{id}}$ assigns a sector identity to the integral (a number between $0$ and $2^{\ninternal}-1$).
The variable $r$ counts the sum of the powers of the propagators having positive indices, 
the variable $s$ does the same thing for the propagators having negative indices.
With the help of these variables we may now define a criteria which allows us to compare two integrals
and to decide which integral is considered to be simpler.
One possibility is the tuple
\bq
\label{chapter_iterated_integrals:isp_basis}
 \left( N_{\mathrm{prop}}, N_{\mathrm{id}}, r, s, \dots \right),
\eq
together with the lexicographical order.
Thus integrals with a smaller number of propagators $N_{\mathrm{prop}}$ are considered simpler.
Within the group of integrals with the same number of propagators, 
integrals with a smaller sector identity are considered simpler.
Within one sector, one first selects integrals with a smaller value of $r$ as simpler, and in the case
of an equal value of $r$, one uses as a secondary criteria the variable $s$.
The dots stand for further variables, which are used if two non-identical integrals agree in the first four
variables.

A second possibility is the tuple
\bq
\label{chapter_iterated_integrals:dot_basis}
 \left( N_{\mathrm{prop}}, N_{\mathrm{id}}, s, r, \dots \right),
\eq
again with the lexicographical order.

We may now write down all possible integration-by-parts identities according to eq.~(\ref{chapter_iterated_integrals:ibp}).
This is a system of linear equations for the Feynman integrals $I_{\nu_1 \dots \nu_{\ninternal}}$.
With the help of an ordering criteria as in eq.~(\ref{chapter_iterated_integrals:isp_basis}) or in eq.~(\ref{chapter_iterated_integrals:dot_basis})
we may eliminate the more complicated integrals in favour of the simpler ones.
This procedure is known as the 
\index{Laporta algorithm}
{\bf Laporta algorithm} \cite{Laporta:2001dd}.
At the end of the day we are able to express most of the integrals in terms of a few remaining integrals.
The remaining integrals are called 
\index{master integrals}
{\bf master integrals}.
The set of master integrals depends on the chosen ordering criteria.
It should be noted that for the ordering criteria given in eq.~(\ref{chapter_iterated_integrals:isp_basis}) and eq.~(\ref{chapter_iterated_integrals:dot_basis})
the set of master integrals will also depend on the way we label the internal edges.
This dependence enters through the sector identity $N_{\mathrm{id}}$, which depends on the labelling of the internal edges.
The choice of eq.~(\ref{chapter_iterated_integrals:isp_basis})
will lead to a basis of master integrals with irreducible scalar products (i.e. with some negative indices), 
avoiding positive indices larger than one.
Such a basis is called an {\bf ISP-basis}.
On the other hand, the choice of eq.~(\ref{chapter_iterated_integrals:dot_basis})
will avoid irreducible scalar products (i.e. negative indices) at the expense of allowing positive
indices larger than one.
Such a basis is called a {\bf dot-basis}.

As in the example discussed above we may supplement the integration-by-parts identities
with symmetry relations.
(In the example of the one-loop two-point function with equal internal masses we had the symmetry
$I_{\nu 0} = I_{0 \nu}$.)
We denote the number of master integrals 
obtained by taking
integration-by-parts identities and symmetries into account by
$\gls{numberofmasterintegrals}$.
If we are only interested in the number of unreduced integrals 
obtained from
integration-by-parts identities alone, we denote this number by 
$\gls{dimensionoftwistedcohomology}$.
In physics we are mainly interested in the master integrals, which remain after
applying integration-by-parts identities and symmetries.
However, in section~\ref{chapter_iterated_integrals:twisted_cohomology} we analyse in more detail
the effects of the integration-by-parts identities alone. In this context, $\Ncohom$ is the relevant quantity.

We denote the indices of the master integrals by
\bq
 {\bm{\nu}}_1 & = & \left( \nu_{1 1}, \dots, \nu_{1 \ninternal} \right),
 \nonumber \\
 {\bm{\nu}}_2 & = & \left( \nu_{2 1}, \dots, \nu_{2 \ninternal} \right),
 \nonumber \\
 & & \dots
 \nonumber \\
 {\bm{\nu}}_{\Nmaster} & = & \left( \nu_{\Nmaster 1}, \dots, \nu_{\Nmaster \ninternal} \right).
\eq
We define a $\Nmaster$-dimensional vector $\vec{I}$ by
\bq
 \vec{I}
 & = &
 \left( I_{{\bm{\nu}}_1}, I_{{\bm{\nu}}_2}, \dots, I_{{\bm{\nu}}_{\Nmaster}} \right)^T.
\eq
For the specific example discussed in eq.~(\ref{chapter_iterated_integrals:example_equal_mass_bubble}) and below
we have
\bq
 \vec{I}
 & = &
 \left( \begin{array}{c} I_{1 0} \\ I_{1 1} \\ \end{array} \right)
\eq
Integration-by-parts identities and symmetries allow us to express a generic integral 
$I_{\nu_1 \dots \nu_{\ninternal}}$ as a linear combination of the master integrals.
Only the latter need to be computed.
There are public available computer programs, which perform the task of reducing
Feynman integrals to master integrals.
These programs are {\tt Fire} \cite{Smirnov:2008iw,Smirnov:2019qkx},
{\tt Reduze} \cite{Studerus:2009ye,vonManteuffel:2012np} and
{\tt Kira} \cite{Maierhoefer:2017hyi,Klappert:2020nbg}.
The following exercise will help you to get acquainted with these programs:
\\
\\
\bs
{\it \refstepcounter{exercise}
\label{chapter_iterated_integrals:exercise_doublebox_ibp}
{\bf Exercise \theexercise}: 
Consider the double-box graph $G$ shown in fig.~\ref{chapter_basics:fig_doublebox} and the auxiliary graph $\tilde{G}$ with nine
propagators shown in fig.~\ref{chapter_basics:fig_doublebox_auxiliary}.
This exercise is about the family of Feynman integrals
\bq 
 I_{\nu_1 \nu_2 \nu_3 \nu_4 \nu_5 \nu_6 \nu_7 \nu_8 \nu_9}
\eq 
with $\nu_8, \nu_9 \le 0$. 
Use the notation of the momenta as in fig.~\ref{chapter_basics:fig_doublebox_auxiliary}.
Assume that all external momenta are light-like ($p_1^2=p_2^2=p_3^2=p_4^2=0$) and that all internal propagators are massless. 
Use one of the public available computer programs {\tt Kira}, {\tt Reduze} or {\tt Fire} to reduce the Feynman integral
\bq 
 I_{1 1 1 1 1 1 1 (-1) (-1)}
\eq 
to master integrals.
For the choice of master integrals you may use the default ordering criteria of the chosen computer program. 
}
\es
\\
\\
We note that integration-by-parts reduction is based only on linear algebra with
rational functions in the kinematic variables $x$ and the dimension of space-time $D$.
However, the simplification of the rational functions (i.e. cancelling common factors in the numerator
and in the denominator) is actually a performance bottle-neck.
For this reason, many of the programs mentioned above employ
finite field methods to improve the performance. 
Finite field methods are reviewed in appendix~\ref{appendix_finite_fields}.

\section{Dimensional shift relations}
\label{chapter_iterated_integrals:sect:dimensional_shift_relations}

Given a basis of master integrals $\vec{I}$ in $D$ space-time dimensions
and a basis of master integrals $\vec{I}'$ in $(D+2)$ space-time dimensions, 
we may express any element of the basis $\vec{I}$ (i.e. any component of $\vec{I}$)
as a linear combination of the elements of $\vec{I}'$ and vice versa.
In dimensional regularisation we have $\dim \vec{I} = \dim \vec{I}'$.
The relation between Feynman integrals in $D$ and $(D+2)$ space-time dimensions
(or $(D-2)$ space-time dimensions) is known as the {\bf dimensional shift relations} \cite{Tarasov:1996br,Tarasov:1997kx}.
In this section we will derive these relations.

In section~\ref{chapter_qft:sect:tensor_reduction} we introduced the dimensional-shift operators ${\bf D}^\pm$,
which act on a Feynman integral as
\bq
 {\bf D}^\pm
 I_{\nu_1 \dots \nu_{\ninternal}}\left(D,x\right)
 & = &
 I_{\nu_1 \dots \nu_{\ninternal}}\left(D \pm 2,x\right)
\eq
and the raising operators ${\bf j}^+$ (with $j \in \{1,\dots,\ninternal\}$),
which act on a Feynman integral as
\bq
 {\bf j}^+
 I_{\nu_1 \dots \nu_j \dots \nu_{\ninternal}}\left(D,x\right)
 & = &
 \nu_j \cdot
 I_{\nu_1 \dots \left(\nu_j+1\right) \dots \nu_{\ninternal}}\left(D,x\right).
\eq
We start from the Schwinger parameter representation of the Feynman integral
\bq
\label{chapter_iterated_integrals:shift_relations_starting_point}
I_{\nu_1 \dots \nu_{\ninternal}}\left(D\right)  & = &
 \frac{e^{\loopnumber \eps \Eulerconstant}}{\prod\limits_{k=1}^{\ninternal}\Gamma(\nu_k)}
 \;
 \int\limits_{\alpha_k \ge 0}  d^{\ninternal}\alpha \;
 \left( \prod\limits_{k=1}^{\ninternal} \alpha_k^{\nu_k-1} \right)
 \frac{1}{{\mathcal U}^{\frac{D}{2}}}
 e^{- \frac{{\mathcal F}}{{\mathcal U}}}.
\eq
In eq.~(\ref{chapter_qft:shift_relations}) we have already seen that the operators
${\bf D}^+$ and ${\bf j}^+$ act on the Schwinger parameter representation as
\bq
\label{chapter_iterated_integrals:shift_relations}
 {\bf D}^+ I_{\nu_1 \dots \nu_{\ninternal}}\left(D\right)  & = &
 \frac{e^{\loopnumber \eps \Eulerconstant}}{\prod\limits_{k=1}^{\ninternal}\Gamma(\nu_k)}
 \;
 \int\limits_{\alpha_k \ge 0}  d^{\ninternal}\alpha \;
 \left( \prod\limits_{k=1}^{\ninternal} \alpha_k^{\nu_k-1} \right)
 \frac{1}{{\mathcal U} \cdot {\mathcal U}^{\frac{D}{2}}}
 e^{- \frac{{\mathcal F}}{{\mathcal U}}},
 \nonumber \\
 {\bf j}^+ I_{\nu_1 \dots \nu_j \dots \nu_{\ninternal}}\left(D\right)  & = &
 \frac{e^{\loopnumber \eps \Eulerconstant}}{\prod\limits_{k=1}^{\ninternal}\Gamma(\nu_k)}
 \;
 \int\limits_{\alpha_k \ge 0}  d^{\ninternal}\alpha \;
 \left( \prod\limits_{k=1}^{\ninternal} \alpha_k^{\nu_k-1} \right)
 \frac{\alpha_j}{{\mathcal U}^{\frac{D}{2}}}
 e^{- \frac{{\mathcal F}}{{\mathcal U}}}.
\eq
In order to derive the dimensional shift relations we use eq.~(\ref{chapter_iterated_integrals:shift_relations_starting_point}) and expand the fraction with ${\mathcal U}$:
\bq
I_{\nu_1 \dots \nu_{\ninternal}}\left(D\right)  & = &
 \frac{e^{\loopnumber \eps \Eulerconstant}}{\prod\limits_{k=1}^{\ninternal}\Gamma(\nu_k)}
 \;
 \int\limits_{\alpha_k \ge 0}  d^{\ninternal}\alpha \;
 \left( \prod\limits_{k=1}^{\ninternal} \alpha_k^{\nu_k-1} \right)
 \frac{{\mathcal U}}{{\mathcal U} \cdot {\mathcal U}^{\frac{D}{2}}}
 e^{- \frac{{\mathcal F}}{{\mathcal U}}}.
\eq
The additional factor of ${\mathcal U}$ in the denominator shifts the space-time dimension of the Feynman integral
by two units according to the first formula of eq.~(\ref{chapter_iterated_integrals:shift_relations}).
The additional factor of ${\mathcal U}$ in the numerator is treated as follows:
Recall that the graph polynomial ${\mathcal U}(\alpha_1,\dots,\alpha_{\ninternal})$ is a homogeneous polynomial of degree
$\loopnumber$ in the Schwinger parameters $\alpha$.
Furthermore, the graph polynomial ${\mathcal U}(\alpha_1,\dots,\alpha_{\ninternal})$ is linear
in each Schwinger parameter $\alpha_j$.
We may interpret each occurrence of a Schwinger parameter $\alpha_j$ in the numerator as the result of 
applying the raising operator ${\bf j}^+$ to the Feynman integral, according to the second formula in
eq.~(\ref{chapter_iterated_integrals:shift_relations}).
Thus we have
\bq
\label{chapter_iterated_integrals:shift_relations_1}
 I_{\nu_1 \dots \nu_{\ninternal}}\left(D\right) 
 & = &
 {\mathcal U}\left( {\bf 1}^+,\dots,{\bf \ninternal}^+ \right)
 {\bf D}^+ 
 I_{\nu_1 \dots \nu_{\ninternal}}\left(D\right).
\eq
The action of ${\mathcal U}({\bf 1}^+,\dots,{\bf \ninternal}^+)$ on 
$I_{\nu_1 \dots \nu_{\ninternal}}(D)$ is defined in the obvious way: 
If ${\mathcal U}(\alpha_1,\alpha_2,\alpha_3)= \alpha_1\alpha_2+\alpha_2\alpha_3+\alpha_1\alpha_3$
we have
\bq
 {\mathcal U}\left( {\bf 1}^+, {\bf 2}^+, {\bf 3}^+ \right)
 I_{1 1 1}\left(D\right) 
 & = &
 \left( {\bf 1}^+ {\bf 2}^+ + {\bf 2}^+ {\bf 3}^++ {\bf 1}^+ {\bf 3}^+ \right) I_{1 1 1}\left(D\right)
 \nonumber \\
 & = &
 I_{2 2 1}\left(D\right) 
 +
 I_{1 2 2}\left(D\right) 
 +
 I_{2 1 2}\left(D\right).
\eq
It is also clear that the operators ${\bf i}^+$ and ${\bf j}^+$ commute
\bq
 \left[ {\bf i}^+, {\bf j}^+ \right] & = & 0
\eq
and that the operators ${\bf j}^+$ commute with ${\bf D}^+$:
\bq
 \left[ {\bf j}^+, {\bf D}^{\pm} \right] & = & 0.
\eq
We may bring in eq.~(\ref{chapter_iterated_integrals:shift_relations_1})
the dimensional shift operator to the other side and obtain
\bq
\label{chapter_iterated_integrals:shift_relations_2}
 {\bf D}^- 
 I_{\nu_1 \dots \nu_{\ninternal}}\left(D\right) 
 & = &
 {\mathcal U}\left( {\bf 1}^+,\dots,{\bf \ninternal}^+ \right)
 I_{\nu_1 \dots \nu_{\ninternal}}\left(D\right)
\eq
or
\bq
\label{chapter_iterated_integrals:shift_relations_3}
 I_{\nu_1 \dots \nu_{\ninternal}}\left(D-2\right) 
 & = &
 {\mathcal U}\left( {\bf 1}^+,\dots,{\bf \ninternal}^+ \right)
 I_{\nu_1 \dots \nu_{\ninternal}}\left(D\right).
\eq
Let's go back to eq.~(\ref{chapter_iterated_integrals:shift_relations_1}), which we may also write as
\bq
\label{chapter_iterated_integrals:shift_relations_4}
 I_{\nu_1 \dots \nu_{\ninternal}}\left(D\right) 
 & = &
 {\mathcal U}\left( {\bf 1}^+,\dots,{\bf \ninternal}^+ \right)
 I_{\nu_1 \dots \nu_{\ninternal}}\left(D+2\right).
\eq
The right-hand side consists of integrals with raised propagators in $(D+2)$ space-time dimensions.
Let $\vec{I}=( I_{{\bm{\nu}}_1}, \dots, I_{{\bm{\nu}}_{\Nmaster}} )^T$ be a basis in $D$ space-time dimensions
and $\vec{I}'=( I_{{\bm{\nu}}_1}', \dots, I_{{\bm{\nu}}_{\Nmaster}}' )^T$ be a basis in $(D+2)$ space-time dimensions.
On the left-hand side we may consider all master integrals from the basis $\vec{I}$ in $D$ dimension.
For each of these Feynman integrals we may use on the right-hand side integration-by-parts identities 
and express all integrals as linear combinations of the
master integrals $\vec{I}'$ in $(D+2)$ dimensions.
This allows us to express any master integral of the basis $\vec{I}$ in $D$ dimensions 
as a linear combination of the master integrals $\vec{I}'$ in $(D+2)$ dimensions.
Thus we find a $(\Nmaster \times \Nmaster)$-matrix $S$
\bq
\label{chapter_iterated_integrals:shift_relations_D_to_D_plus_2}
 \vec{I} & = & S \; \vec{I}'.
\eq
Within dimensional regularisation the matrix $S$ is invertible.
Inverting this matrix allows us to express any master integral in $(D+2)$ dimensions
as a linear combination of master integrals in $D$ dimensions:
\bq
\label{chapter_iterated_integrals:shift_relations_D_plus_2_to_D}
 \vec{I}' & = & S^{-1} \; \vec{I}.
\eq
\bs
{\it \refstepcounter{exercise}
\label{chapter_iterated_integrals:exercise_dimensional_shift}
{\bf Exercise \theexercise}: 
Consider the example of the one-loop two-point function with equal internal masses, discussed 
below eq.~(\ref{chapter_iterated_integrals:example_equal_mass_bubble}).
Let 
\bq
 \vec{I} & = & \left( \begin{array}{c} I_{10}\left(D,x\right) \\ I_{11}\left(D,x\right) \\ \end{array} \right)
\eq
be a basis in $D$ space-time dimensions and 
\bq
 \vec{I}' & = & \left( \begin{array}{c} I_{10}\left(D+2,x\right) \\ I_{11}\left(D+2,x\right) \\ \end{array} \right)
\eq
be a basis in $(D+2)$ space-time dimensions.
Work out the $2 \times 2$-matrices $S$ and $S^{-1}$.
}
\es

\section{Differential equations}
\label{chapter_iterated_integrals:differential_equations}

We now introduce one of the most important methods to compute Feynman integrals:
The method of differential equations \cite{Kotikov:1990kg,Kotikov:1991pm,Remiddi:1997ny,Gehrmann:1999as}.
The idea is the following: Instead of calculating the Feynman integral directly, we first
derive a differential equation of the Feynman integral under consideration with respect to the kinematic
variables.
In a second step we solve this differential equation and obtain in this way the answer for the 
sought-after Feynman integral.
To be more precise, we study a system of differential equations, namely the system
of differential equations for a basis of master integrals.
This has the advantage that we have to consider only first-order differential equations.

Solving a differential equation requires boundary values.
As boundary value we may use the master integrals, where one of the kinematic variables 
has a special value, for example zero or equal to another kinematic variable.
The Feynman integrals for this special kinematic configuration are simpler, as they depend
on one kinematic variable less.
We may assume that they are already known.
If not, we first solve this simpler problem first.

The power of the method of differential equations lies in the following facts:
We will soon see that it is always possible to derive the system of differential equations
for a basis of master integrals.
There are no principle obstacles to do this, we only might be limited by the available computing
resources and the fact that the used algorithms are not particular efficient.
We will also see that if the system of differential equations is in a particular nice form
(the $\eps$-form), it is always possible to solve the system of differential equations
in terms of iterated integrals.
Here we assume -- as remarked above -- that the boundary values are known.
Thus the only missing piece is the transformation of an original system of differential
equations to the nice $\eps$-form.
In the cases where we know how to do this, this is achieved by a redefinition of the master integrals
and/or a variable transformation of the kinematic variables.
We call a redefinition of the master integrals a fibre transformation
and a transformation of the kinematic variables a base transformation.
We discuss these transformations in sections~\ref{chapter_iterated_integrals:fibre_transformation} 
and \ref{chapter_iterated_integrals:base_transformation}, respectively.
Let us stress that this reduces the task of computing a Feynman integral to finding a suitable
fibre transformation and/or base transformation for the associated system of differential equations.

\subsection{Deriving the system of differential equation}
\label{chapter_iterated_integrals:deriving_the_dgl}

Let's start to derive the system of differential equations.
We consider a basis of master integrals
\bq
 \vec{I}
 & = & 
 \left( I_{{\bm{\nu}}_1}, I_{{\bm{\nu}}_2}, \dots, I_{{\bm{\nu}}_{\Nmaster}} \right)^T,
\eq
depending on $\NB$ kinematic variables $x_1, x_2, \dots, x_{\NB}$.
Let's recall from section~\ref{chapter_basics:momentum_representation_of_Feynman_integrals}
that we start with $\NB+1$ variables of the form
\bq
 \frac{- p_i \cdot p_j}{\mu^2},
 & &
 \frac{m_i^2}{\mu^2}.
\eq
We denote these variables by $x_1, x_2, \dots, x_{\NB}, x_{\NB+1}$.
Due to the scaling relation eq.~(\ref{chapter_basics:kinematic_variables_scaling_relation})
we may set one of these variables to one, say $x_{\NB+1}=1$.
Having done so, we usually view the Feynman integrals as functions of $x_1, x_2, \dots, x_{\NB}$ (and $D$).
For the moment, let's keep the dependence on all variables
$x_1, x_2, \dots, x_{\NB}, x_{\NB+1}$, without setting one kinematic variable to one.
The second Symanzik polynomial ${\mathcal F}$ is linear in the kinematic variables $x_j$.
We may write
\bq
 {\mathcal F}\left(\alpha,x\right)
 & = &
 \sum\limits_{j=1}^{\NB+1}
 {\mathcal F}_{x_j}'\left(\alpha\right)
 \cdot
 x_j,
\eq
where ${\mathcal F}_{x_j}'$ denotes the coefficient of $x_j$. As ${\mathcal F}$ is linear in $x_j$, we have
\bq
 {\mathcal F}_{x_j}'\left(\alpha\right)
 & = &
 \frac{\partial}{\partial x_j} {\mathcal F}\left(\alpha,x\right).
\eq
In order to derive the differential equation we start again from the Schwinger parameter representation
\bq
I_{\nu_1 \dots \nu_{\ninternal}}  & = &
 \frac{e^{\loopnumber \eps \Eulerconstant}}{\prod\limits_{k=1}^{\ninternal}\Gamma(\nu_k)}
 \;
 \int\limits_{\alpha_k \ge 0}  d^{\ninternal}\alpha \;
 \left( \prod\limits_{k=1}^{\ninternal} \alpha_k^{\nu_k-1} \right)
 \frac{1}{\left[{\mathcal U}\left(\alpha\right)\right]^{\frac{D}{2}}}
 e^{- \frac{{\mathcal F}\left(\alpha,x\right)}{{\mathcal U}\left(\alpha\right)}}.
\eq
The only dependence on the kinematic variables is through the second Symanzik polynomial 
${\mathcal F}(\alpha,x)$.
We therefore find
\bq
 \frac{\partial}{\partial x_j} 
 I_{\nu_1 \dots \nu_{\ninternal}}
 & = &
 -
 \frac{e^{\loopnumber \eps \Eulerconstant}}{\prod\limits_{k=1}^{\ninternal}\Gamma(\nu_k)}
 \;
 \int\limits_{\alpha_k \ge 0}  d^{\ninternal}\alpha \;
 \left( \prod\limits_{k=1}^{\ninternal} \alpha_k^{\nu_k-1} \right)
 \frac{{\mathcal F}_{x_j}'\left(\alpha\right)}{{\mathcal U}\left(\alpha\right) \cdot \left[{\mathcal U}\left(\alpha\right)\right]^{\frac{D}{2}}}
 e^{- \frac{{\mathcal F}\left(\alpha,x\right)}{{\mathcal U}\left(\alpha\right)}}
\eq
for $x_j \in \{x_1, \dots, x_{\NB+1}\}$.
The additional factor of ${\mathcal U}(\alpha)$ in the denominator 
is again equivalent to shifting the space-time dimension of the Feynman
integral by two units.
The additional factor of ${\mathcal F}_{x_j}'(\alpha)$ in the numerator is a polynomial in the Schwinger
parameters, equivalent to the action of the polynomial 
${\mathcal F}_{x_j}'({\bf 1}^+,\dots,{\bf \ninternal}^+)$ in the raising operators on the Feynman integral.
Thus
\bq
\label{chapter_iterated_integrals:partial_derivative}
 \frac{\partial}{\partial x_j} 
 I_{\nu_1 \dots \nu_{\ninternal}}
 & = &
 -
 {\mathcal F}_{x_j}'\left( {\bf 1}^+,\dots,{\bf \ninternal}^+ \right)
 {\bf D}^+ 
 I_{\nu_1 \dots \nu_{\ninternal}}
\eq
for $x_j \in \{x_1, \dots, x_{\NB+1}\}$.

If the kinematic variable $x_j$ corresponds to an internal mass, there is a slightly simpler formula,
which follows directly from the momentum representation.
Let us assume
\bq
 x_j & = & \frac{m_j^2}{\mu^2}.
\eq
Let's further assume that $m_j$ denotes the mass of the $j$-th internal edge, that
this mass is distinct from all other internal masses and that the kinematic configuration is defined
without any reference to this mass (i.e. we exclude on-shell conditions like $p^2=m_j^2$).
In other words, $x_j$ enters only as the mass of the the $j$-th internal propagator.
From
\bq
 \frac{\partial}{\partial x_j} \frac{1}{\left(-q_j^2+m_j^2\right)^{\nu_j}}
 & = &
 - \nu_j \frac{\mu^2}{\left(-q_j^2+m_j^2\right)^{\nu_j+1}}
\eq
we obtain in this case
\bq
\label{chapter_iterated_integrals:partial_derivative_internal_mass}
 \frac{\partial}{\partial x_j} 
 I_{\nu_1 \dots \nu_{\ninternal}}
 & = &
 - {\bf j}^+ I_{\nu_1 \dots \nu_{\ninternal}}.
\eq
We may relax the condition that the mass of the $j$-th internal propagator has to be distinct from all
other internal masses.
Let $S_{x_j}$ be the subset of $\{1,\dots,\ninternal\}$ containing all indices of internal edges whose internal
mass equals $m_j$.
We keep the condition, that 
the kinematic configuration is defined
without any reference to $m_j^2$.
Then
\bq
\label{chapter_iterated_integrals:partial_derivative_internal_mass_generalised}
 \frac{\partial}{\partial x_j} 
 I_{\nu_1 \dots \nu_{\ninternal}}
 & = &
 - \sum\limits_{j \in S_{x_j}} {\bf j}^+ I_{\nu_1 \dots \nu_{\ninternal}},
\eq
which follows directly from the product rule for differentiation.
\\
\\
\bs
{\it \refstepcounter{exercise}
{\bf Exercise \theexercise}: 
Show that
\bq
\label{chapter_iterated_integrals:differential_scaling_relation}
 \sum\limits_{j=1}^{\NB+1}
 x_j \frac{\partial}{\partial x_j} 
 I_{\nu_1 \dots \nu_{\ninternal}}
 & = &
 \left( \frac{\loopnumber D}{2} - \nu \right)
 \cdot
 I_{\nu_1 \dots \nu_{\ninternal}}.
\eq
Hint: Consider the Feynman parameter representation.
}
\es
\\
\\
From eq.~(\ref{chapter_iterated_integrals:differential_scaling_relation})
we may extract the derivative with respect to $x_{\NB+1}$, provided we know all derivatives with respect to
$x_1, \dots, x_{\NB}$.

From now on we consider again the case, where we set one kinematic variable to one (i.e. $x_{\NB+1}=1$).
We view the Feynman integral $I_{\nu_1 \dots \nu_{\ninternal}}(D,x_1,\dots,x_{\NB})$ as a function 
of $D$ and the $\NB$ kinematic variables $x_1, \dots, x_{\NB}$.
The derivatives with respect to the kinematic variables are given by eq.~(\ref{chapter_iterated_integrals:partial_derivative}).
The expression on the right-hand side of eq.~(\ref{chapter_iterated_integrals:partial_derivative})
is a linear combination of Feynman integrals in $(D+2)$ space-time dimensions.
Using integration-by-parts identities we may reduce this expression to a linear combination
of master integrals in $(D+2)$ space-time dimensions.
Using the dimensional shift relations discussed in section~\ref{chapter_iterated_integrals:sect:dimensional_shift_relations} we may express each master integral in $(D+2)$ space-time dimensions
as a linear combination of master integrals in $D$ space-time dimensions.
Combining these two operations we may express the right-hand side of 
eq.~(\ref{chapter_iterated_integrals:partial_derivative}) as 
a linear combination of master integrals in $D$ space-time dimensions.

Let us now specialise to the basis 
\bq
 \vec{I}
 & = & 
 \left( I_{{\bm{\nu}}_1}, I_{{\bm{\nu}}_2}, \dots, I_{{\bm{\nu}}_{\Nmaster}} \right)^T,
\eq
For each $I_{{\bm{\nu}}_i} \in \{I_{{\bm{\nu}}_1}, \dots, I_{{\bm{\nu}}_{\Nmaster}} \}$ we therefore have
\begin{align}
 \frac{\partial}{\partial x_j} 
 I_{{\bm{\nu}}_i}
 & = 
 - \sum\limits_{k=1}^{\Nmaster} A_{x_j, i k} \; I_{{\bm{\nu}}_k},
 &&
 1 \; \le \; i \; \le \; \Nmaster,
 &&
 1 \; \le \; j \; \le \; \NB,
\end{align}
where the coefficients $A_{x_j, i k}$ are rational functions of $D$ and $x_1, \dots, x_{\NB}$.
The fact that the coefficients $A_{x_j, i k}$ are rational functions follows from the fact
that integration-by-parts identities and dimensional shift relations
involve only rational functions.

Let us make a few definitions:
We use the standard notation for the total differential with respect to the kinematic variables $x_1, \dots, x_{\NB}$
($D$ is treated as a parameter)
\bq
 d I_{{\bm{\nu}}_i}
 & = &
 \sum\limits_{j=1}^{\NB}
 \left( \frac{\partial I_{{\bm{\nu}}_i} }{\partial x_j} \right)
 dx_j.
\eq
We denote by $A_{x_j}$ the $(\Nmaster \times \Nmaster)$-matrix with entries $A_{x_j, i k}$.
We also define a matrix-valued one-form $A$ by
\bq
 A 
 & = &
 \sum\limits_{j=1}^{\NB}
 A_{x_j}
 dx_j.
\eq
We may then write the system of differential equations compactly as
\bq
 \left( d + A \right) \vec{I}
 & = & 0.
\eq
This is the sought-after system of first-order differential equations for the master integrals
$I_{{\bm{\nu}}_1}$, $\dots$, $I_{{\bm{\nu}}_{\Nmaster}}$.
This system is integrable, which puts a constraint on $A$.
The integrability condition reads
\bq
\label{chapter_iterated_integrals:integrability_condition}
 d A + A \wedge A & = & 0.
\eq
Let us summarise:
\begin{tcolorbox}
{\bf System of differential equations}:
\\
The vector of master integrals
$\vec{I} =  ( I_{{\bm{\nu}}_1}, I_{{\bm{\nu}}_2}, \dots, I_{{\bm{\nu}}_{\Nmaster}} )^T$ 
satisfies the differential equation
\bq
\label{chapter_iterated_integrals:differential_equation_master_integrals}
 \left( d + A \right) \vec{I}
 & = & 0,
\eq
where $d$ denotes the total differential with respect to the kinematic variables $x_1, \dots, x_{\NB}$
and $A$ is a matrix-valued one-form, which satisfies the integrability condition
\bq
 d A + A \wedge A & = & 0.
\eq
If we write
\bq
 A 
 & = &
 \sum\limits_{j=1}^{\NB}
 A_{x_j}
 dx_j,
\eq
then the $A_{x_j}$'s are $(\Nmaster \times \Nmaster)$-matrices with entries, which are rational functions 
in $x_1, \dots, x_{\NB}$ and $D$.
The matrix-valued one-form $A$ is computable with the help eq.~(\ref{chapter_iterated_integrals:partial_derivative}), 
integration-by-parts identities and
dimensional shift relations.
\end{tcolorbox}

\subsubsection{Example 1}

Let's look at a few examples.
As our first example we 
consider the one-loop two-point function with equal internal masses, discussed 
below eq.~(\ref{chapter_iterated_integrals:example_equal_mass_bubble}):
\bq
 I_{\nu_1 \nu_2}\left(D,x\right)
 & = &
 e^{\eps \Eulerconstant} \left(m^2\right)^{\nu_{12}-\frac{D}{2}}
 \int \frac{d^Dk}{i \pi^{\frac{D}{2}}} 
 \frac{1}{\left(-q_1^2+m^2\right)^{\nu_1} \left(-q_2^2+m^2\right)^{\nu_2}},
\eq
with $x=-p^2/m^2$.
As a basis of master integrals we choose
\bq
 \vec{I} & = & \left( \begin{array}{c} I_{10} \\ I_{11} \\ \end{array} \right).
\eq
The graph polynomials read
\bq
 {\mathcal U} \; = \; \alpha_1 + \alpha_2,
 & &
 {\mathcal F} \; = \; \alpha_1 \alpha_2 x +\left(\alpha_1+\alpha_2\right)^2.
\eq
From eq.~(\ref{chapter_iterated_integrals:partial_derivative}) we have
\bq
\label{chapter_iterated_integrals:example_partial_derivative_1}
 \frac{\partial}{\partial x} I_{\nu_1 \nu_2}\left(D,x\right)
 & = &
 - {\bf 1}^+ {\bf 2}^+ {\bf D}^+ I_{\nu_1 \nu_2}\left(D,x\right)
 \; = \;
 - \nu_1 \nu_2 I_{\left(\nu_1+1\right) \left(\nu_2+1\right)}\left(D+2,x\right).
\eq
Using integration-by-parts identities and
dimensional shift relations we obtain
\bq
\label{chapter_iterated_integrals:example_partial_derivative_2}
 \frac{\partial}{\partial x} I_{1 0}\left(D,x\right)
 & = & 
 0,
 \nonumber \\
 \frac{\partial}{\partial x} I_{1 1}\left(D,x\right)
 & = &
 - \frac{D-2}{x\left(4+x\right)} I_{1 0}\left(D,x\right)
 - \frac{4+\left(4-D\right)x}{2 x\left(4+x\right)} I_{1 1}\left(D,x\right).
\eq
Hence
\bq
\label{chapter_iterated_integrals:example_A_oneloop_twopoint}
 A & = &
 \left(\begin{array}{cc}
 0 & 0 \\
 \frac{D-2}{x\left(4+x\right)} & \frac{4+\left(4-D\right)x}{2 x\left(4+x\right)} \\
 \end{array} \right) dx.
\eq
\bs
{\it \refstepcounter{exercise}
{\bf Exercise \theexercise}: 
The steps from eq.~(\ref{chapter_iterated_integrals:example_partial_derivative_1}) 
to eq.~(\ref{chapter_iterated_integrals:example_partial_derivative_2}) can still be carried out by hand. Fill in the missing details.
\\
Hint: Use eq.~(\ref{chapter_iterated_integrals:ibp_equations}), eq.~(\ref{chapter_iterated_integrals:ibp_equations_tadpole}) 
and the result from exercise~\ref{chapter_iterated_integrals:exercise_dimensional_shift}.
}
\es

\subsubsection{Example 2}

As our second example we consider the one-loop four-point function
shown in fig.~\ref{chapter_basics:fig_oneloopbox}
\bq
\label{chapter_iterated_integrals:example_box}
 I_{\nu_1 \nu_2 \nu_3 \nu_4}\left(D,x_1,x_2\right)
 = 
 e^{\eps \Eulerconstant} \left(-p_4^2\right)^{\nu_{1234}-\frac{D}{2}}
 \int \frac{d^Dk}{i \pi^{\frac{D}{2}}} 
 \frac{1}{\left(-q_1^2\right)^{\nu_1} \left(-q_2^2\right)^{\nu_2} \left(-q_3^2\right)^{\nu_3} \left(-q_4^2\right)^{\nu_4}},
\eq
with $q_1=k-p_1$, $q_2=k-p_1-p_2$, $q_3=k-p_1-p_2-p_3$, $q_4=k$ and vanishing internal masses.
We consider the kinematic configuration $p_1^2=p_2^2=p_3^2=0$ but $p_4^2 \neq 0$.
The integral depends on two kinematic variables, which we take as
\bq
\label{chapter_iterated_integrals:example_box_def_variables}
 x_1 \; = \; \frac{2 p_1 \cdot p_2}{p_4^2},
 & &
 x_2 \; = \; \frac{2 p_2 \cdot p_3}{p_4^2}.
\eq
The graph polynomials are given by
\bq
 {\mathcal U} \; = \; \alpha_1 + \alpha_2 + \alpha_3 + \alpha_4,
 & &
 {\mathcal F} \; = \; 
 \alpha_2 \alpha_4 x_1 
 + \alpha_1 \alpha_3 x_2
 + \alpha_3 \alpha_4.
\eq
There are four master integral and we choose as basis
\bq
\label{chapter_iterated_integrals:master_integrals_box_I}
 \vec{I} 
 & = & 
 \left( \begin{array}{c} 
  I_{0011} \\ 
  I_{0101} \\ 
  I_{1010} \\ 
  I_{1111} \\
 \end{array} \right).
\eq
The computations to determine the differential equation are best done with the help 
of a computer algebra program and we only quote the result here.
One finds
\bq
\label{chapter_iterated_integrals:A_box}
 A 
 & = &
 A_{x_1}
 dx_1
 +
 A_{x_2}
 dx_2,
 \\
 A_{x_1}
 & = &
 \left( \begin{array}{cccc} 
 0 & 0 & 0 & 0 \\
 0 & \frac{D-4}{2 x_1} & 0 & 0 \\
 0 & 0 & 0 & 0 \\
  \frac{2\left(D-3\right)}{x_1\left(1-x_1\right)\left(1-x_1-x_2\right)} 
  & - \frac{2\left(D-3\right)}{x_1\left(1-x_1\right)\left(1-x_1-x_2\right)} 
  & - \frac{2\left(D-3\right)}{x_1 x_2 \left(1-x_1-x_2\right)} 
  & - \frac{2x_1 + \left(D-6\right)\left(1-x_2\right)}{2 x_1 \left(1-x_1-x_2\right)} \\
 \end{array} \right),
 \nonumber \\
 A_{x_2}
 & = &
 \left( \begin{array}{cccc} 
 0 & 0 & 0 & 0 \\
 0 & 0 & 0 & 0 \\
 0 & 0 & - \frac{D-4}{2 x_2} & 0 \\
  \frac{2\left(D-3\right)}{x_2\left(1-x_2\right)\left(1-x_1-x_2\right)} 
  & - \frac{2\left(D-3\right)}{x_1 x_2 \left(1-x_1-x_2\right)} 
  & - \frac{2\left(D-3\right)}{x_2 \left(1-x_2\right)\left(1-x_1-x_2\right)} 
  & - \frac{2x_2 + \left(D-6\right)\left(1-x_1\right)}{2 x_2 \left(1-x_1-x_2\right)} \\
 \end{array} \right).
 \nonumber 
\eq
\bs
{\it \refstepcounter{exercise}
{\bf Exercise \theexercise}: 
This example depends on two kinematic variables $x_1$ and $x_2$, hence the integrability condition is non-trivial.
Check explicitly the integrability condition
\bq
 d A + A \wedge A & = & 0.
\eq
}
\es
In order to derive the differential equation, we first make a choice for the master integrals.
In this example the choice of master integrals is given by eq.~(\ref{chapter_iterated_integrals:master_integrals_box_I}).
However, there is no particular reason for this specific choice 
(except that it is the default choice of the computer program {\tt Kira}).
In general, we may choose a set of four linear independent linear combinations of these four master integrals with
coefficients being functions of $D$ and $x$.
In the simplest case the coefficients are rational functions of $D$ and $x$.
However, we will soon also consider the case, where the coefficients are algebraic functions of $x$ (e.g. expressions with square
roots). The dependence on $D$ of the coefficients will usually remain rational.
In chapter~\ref{chapter_elliptics} we will also consider the case, where the coefficients are transcendental functions of $x$.

Let us now explore this freedom. Suppose we start from the basis
\bq
\label{chapter_iterated_integrals:base_change_box}
 \vec{I}'
 & = & 
 \left( \begin{array}{c} 
  - \frac{1}{2} \left(D-3\right) \left(D-4\right) I_{0011} \\ 
  - \frac{1}{2} \left(D-3\right) \left(D-4\right) I_{0101} \\ 
  - \frac{1}{2} \left(D-3\right) \left(D-4\right) I_{1010} \\ 
  \frac{1}{8} \left(D-4\right)^2 x_1 x_2 I_{1111} \\
 \end{array} \right).
\eq
We repeat the calculation with this basis of master integrals.
Again we find a differential equation
\bq
\label{chapter_iterated_integrals:example_box_dgl_eps_form}
 \left( d + A' \right) \vec{I}' & = & 0,
 \;\;\;\;\;\;
 A' \; = \; A_{x_1}' dx_1 + A_{x_2}' dx_2,
\eq
where the matrices $A_{x_1}'$ and $A_{x_2}'$ are now given by
\bq
 A_{x_1}'
 & = &
 \frac{4-D}{2}
 \left( \begin{array}{cccc} 
 0 & 0 & 0 & 0 \\
 0 & \frac{1}{x_1} & 0 & 0 \\
 0 & 0 & 0 & 0 \\
  \frac{1}{x_1-1} - \frac{1}{x_1+x_2-1}
  &
  - \frac{1}{x_1-1} + \frac{1}{x_1+x_2-1}
  &
  \frac{1}{x_1+x_2-1}
  &
  \frac{1}{x_1} - \frac{1}{x_1+x_2-1}
 \end{array} \right),
 \nonumber \\
 A_{x_2}'
 & = &
 \frac{4-D}{2}
 \left( \begin{array}{cccc} 
 0 & 0 & 0 & 0 \\
 0 & 0 & 0 & 0 \\
 0 & 0 & \frac{1}{x_2} & 0 \\
  \frac{1}{x_2-1} - \frac{1}{x_1+x_2-1}
  &
  \frac{1}{x_1+x_2-1}
  &
  - \frac{1}{x_2-1} + \frac{1}{x_1+x_2-1}
  &
  \frac{1}{x_2} - \frac{1}{x_1+x_2-1}
 \end{array} \right).
 \nonumber \\
\eq
We observe that the only dependence on $D$ is now through the prefactor $(4-D)/2$. 
Within dimensional regularisation we usually set $D=4-2\eps$. Then
\bq
 \frac{4-D}{2} & = & \eps.
\eq
Let us further introduce five one-forms
\begin{align}
\label{chapter_iterated_integrals:dlog_form}
 \omega_1
 & =
 d \ln\left(x_1\right)
 \; = \;
 \frac{dx_1}{x_1},
 &
 \omega_2
 & =
 d \ln\left(x_1-1\right)
 \; = \;
 \frac{dx_1}{x_1-1},
 \nonumber \\
 \omega_3
 & = 
 d \ln\left(x_2\right)
 \; = \;
 \frac{dx_2}{x_2},
 &
 \omega_4
 & = 
 d \ln\left(x_2-1\right)
 \; = \;
 \frac{dx_2}{x_2-1},
 \nonumber \\
 \omega_5
 & = 
 d \ln\left(x_1+x_2-1\right)
 \; = \;
 \frac{dx_1+dx_2}{x_1+x_2-1}.
 & &
\end{align}
Differential one-forms as in eq.~(\ref{chapter_iterated_integrals:dlog_form}) are called 
\index{dlog-form}
{\bf dlog-forms}.
We may then write $A'$ as
\bq
\label{chapter_iterated_integrals:Aprime_box}
 A'
 & = &
 \eps 
 \left( \begin{array}{rrrr} 
 0 & 0 & 0 & 0 \\
 0 & 1 & 0 & 0 \\
 0 & 0 & 0 & 0 \\
 0 & 0 & 0 & 1 \\
 \end{array} \right)
 \omega_1
 +
 \eps 
 \left( \begin{array}{rrrr} 
 0 & 0 & 0 & 0 \\
 0 & 0 & 0 & 0 \\
 0 & 0 & 0 & 0 \\
 1 & -1 & 0 & 0 \\
 \end{array} \right)
 \omega_2
 +
 \eps 
 \left( \begin{array}{rrrr} 
 0 & 0 & 0 & 0 \\
 0 & 0 & 0 & 0 \\
 0 & 0 & 1 & 0 \\
 0 & 0 & 0 & 1 \\
 \end{array} \right)
 \omega_3
 \nonumber \\
 & &
 +
 \eps 
 \left( \begin{array}{rrrr} 
 0 & 0 & 0 & 0 \\
 0 & 0 & 0 & 0 \\
 0 & 0 & 0 & 0 \\
 1 & 0 & -1 & 0 \\
 \end{array} \right)
 \omega_4
 +
 \eps 
 \left( \begin{array}{rrrr} 
 0 & 0 & 0 & 0 \\
 0 & 0 & 0 & 0 \\
 0 & 0 & 0 & 0 \\
 -1 & 1 & 1 & -1 \\
 \end{array} \right)
 \omega_5.
\eq
In the basis $\vec{I}'$ the differential equation has a particular simple form, which we will
discuss in more detail in section~\ref{chapter_iterated_integrals:section:eps_form}.
In particular there is a systematic (and easy) way to solve this differential equation
order-by-order in the dimensional regularisation parameter.
We discuss this in section~\ref{chapter_iterated_integrals:section:solution_eps_form}.

We call the one-forms $\omega_1$, $\omega_2$, $\omega_3$, $\omega_4$ and $\omega_5$ 
\index{letter}
{\bf letters} and the set of all independent letters the 
\index{alphabet}
{\bf alphabet}.
In this example the alphabet contains five letters.
We denote the number of letters in the alphabet by 
$\gls{numberofletters}$.

\subsubsection{Example 3}

As third example 
let us consider the double-box integral discussed in exercise~\ref{chapter_iterated_integrals:exercise_doublebox_ibp}.
We use the same notation and consider 
\bq 
 I_{\nu_1 \nu_2 \nu_3 \nu_4 \nu_5 \nu_6 \nu_7 \nu_8 \nu_9}
\eq 
with $\nu_8, \nu_9 \le 0$. 
We assume that all external momenta are light-like ($p_1^2=p_2^2=p_3^2=p_4^2=0$) and that all internal propagators are massless. 
As usual we define the Mandelstam variables by
\bq
 s \; = \; \left(p_1+p_2\right)^2,
 & &
 t \; = \; \left(p_2+p_3\right)^2.
\eq
We set $\mu^2=t$ and $x=s/t$.
There are eight master integral and we choose as basis
\bq
\label{chapter_iterated_integrals:precanonical_masters_double_box}
 \vec{I} 
 & = & 
 \left( \begin{array}{l} 
  I_{001110000} \\ 
  I_{100100100} \\
  I_{011011000} \\ 
  I_{100111000} \\ 
  I_{111100100} \\
  I_{101110100} \\
  I_{111111100} \\
  I_{1111111\left(-1\right)0} \\
 \end{array} \right).
\eq
The calculations to determine the differential equation are again best carried out with the help of a computer program
and one finds $A=A_x dx$ with
{\footnotesize
\bq
\label{chapter_iterated_integrals:A_x_precanonical_double_box}
\lefteqn{ 
 A_x
 = } & &
 \nonumber \\
 & &
 \hspace*{-5mm}
 \left( \begin{array}{cccccccc}
 -\frac{D-3}{x} & 0 & 0 & 0 & 0 & 0 & 0 & 0 \\
 0 & 0 & 0 & 0 & 0 & 0 & 0 & 0 \\
 0 & 0 & -\frac{D-4}{x} & 0 & 0 & 0 & 0 & 0 \\
 0 & 0 & 0 & -\frac{D-4}{x} & 0 & 0 & 0 & 0 \\
 0 & \frac{\left(3D-8\right)\left(3D-10\right)}{2\left(D-4\right)x\left(1+x\right)} & 0 & \frac{3D-10}{2 x \left(1+x\right)}  & \frac{2x-\left(D-6\right)}{2 x \left(1+x\right)} & 0 & 0 & 0 \\
 - \frac{\left(D-3\right)\left(3D-8\right)\left(3D-10\right)}{\left(D-4\right)^2x^2\left(1+x\right)} & \frac{\left(D-3\right)\left(3D-8\right)\left(3D-10\right)}{\left(D-4\right)^2x\left(1+x\right)} & 0 & 0 & 0 & \frac{x-\left(D-4\right)}{x\left(1+x\right)} & 0 & 0 \\
 \frac{3 \left(D-3\right)\left(3D-8\right)\left(3D-10\right)}{\left(D-4\right)^2x^3\left(1+x\right)} & \frac{3 \left(D-3\right)\left(3D-8\right)\left(3D-10\right)}{\left(D-4\right)^2x^2\left(1+x\right)} & 0 & \frac{3 \left(D-3\right)\left(3D-10\right)}{\left(D-4\right)x^2\left(1+x\right)} & \frac{6\left(D-3\right)}{x\left(1+x\right)} & -\frac{3\left(D-4\right)}{x^2} & \frac{2}{x} & \frac{D-4}{x \left(1+x\right)} \\
 A_{x,81}
 & 
 A_{x,82}
 & 
 A_{x,83}
 & 
 A_{x,84}
 & 
 A_{x,85}
 & 
 A_{x,86}
 & 
 A_{x,87}
 & 
 A_{x,88}
 \\
 \end{array} \right).
 \nonumber
\eq
}
The entries in the eighth row are a little bit longer and we list them separately below:
\bq
 A_{x,81}
 & = &
 \frac{3 \left(D-3\right)\left(3D-8\right)\left(3D-10\right)\left[\left(3D-14\right)x+2\left(D-5\right)\right]}{2\left(D-4\right)^3x^3\left(1+x\right)},
 \nonumber \\
 A_{x,82}
 & = &
 \frac{3 \left(D-3\right)\left(3D-8\right)\left(3D-10\right)\left[\left(3D-14\right)x+2\left(2D-9\right)\right]}{2\left(D-4\right)^3x^2\left(1+x\right)},
 \nonumber \\
 A_{x,83}
 & = &
 \frac{4\left(D-3\right)^2}{\left(D-4\right)x^3},
 \nonumber \\
 A_{x,84}
 & = &
 \frac{3 \left(D-3\right)\left(3D-10\right)\left[\left(3D-14\right)x+2\left(2D-9\right)\right]}{2\left(D-4\right)^2x^2\left(1+x\right)}, 
 \nonumber \\
 A_{x,85}
 & = &
 \frac{3\left(D-3\right)\left(3D-14\right)}{[\left(D-4\right)x}, 
 \nonumber \\
 A_{x,86}
 & = &
 -\frac{3\left[ \left(3D-14\right) x + 2 \left(2D-9\right)\right]}{2 x^2},
 \nonumber \\
 A_{x,87}
 & = &
 -\frac{D-4}{x},
 \nonumber \\
 A_{x,88}
 & = &
 -\frac{\left(3D-16\right)x+4\left(D-5\right)}{2 x \left(1+x\right)}.
\eq
We will always order the master integrals such that master integrals which can be obtained through
pinching (and possibly symmetry relations) from other master integrals appear before their parent integrals.
The matrices $A_{x_j}$ have then always a lower block triangular structure, induced by the sub-sectors.
In eq.~(\ref{chapter_iterated_integrals:def_notation_sectors}) 
we introduced the sector identification number $N_{\mathrm{id}}$. 
In the example of the double box integral we have master integrals from seven sectors:
\begin{alignat}{3}
 N_{\mathrm{id}} & = 28, & \hspace*{5mm} & & & I_{001110000}, 
 \nonumber \\ 
 N_{\mathrm{id}} & = 73, & \hspace*{5mm} & & & I_{100100100},
 \nonumber \\ 
 N_{\mathrm{id}} & = 54, & \hspace*{5mm} & & & I_{011011000},
 \nonumber \\ 
 N_{\mathrm{id}} & = 57, & \hspace*{5mm} & & & I_{100111000},
 \nonumber \\ 
 N_{\mathrm{id}} & = 79, & \hspace*{5mm} & & & I_{111100100},
 \nonumber \\ 
 N_{\mathrm{id}} & = 93, & \hspace*{5mm} & & & I_{101110100},
 \nonumber \\ 
 N_{\mathrm{id}} & = 127, & \hspace*{5mm} & & & I_{111111100}, I_{1111111\left(-1\right)0}.
\end{alignat}
The first six sectors have one master integral per sector, 
while the seventh sector (sector $127$) has two master integrals.

\subsubsection{Example 4}

As fourth and final example 
let us consider the two-loop sunrise integral with equal internal masses, shown in fig.~\ref{chapter_basics:fig_sunrise}:
\bq
 I_{\nu_1 \nu_2 \nu_3}\left(D,x\right)
 & = &
 e^{2 \eps \Eulerconstant} \left(m^2\right)^{\nu_{123}-D}
 \int \frac{d^Dk_1}{i \pi^{\frac{D}{2}}} \frac{d^Dk_2}{i \pi^{\frac{D}{2}}} 
 \frac{1}{\left(-q_1^2+m^2\right)^{\nu_1} \left(-q_2^2+m^2\right)^{\nu_2} \left(-q_3^2+m^2\right)^{\nu_3}},
 \nonumber \\
\eq
with $x=-p^2/m^2$ and $q_1=k_1$, $q_2=k_2-k_1$, $q_3=-k_2-p$.
We have set $\mu^2=m^2$.
There are three master integrals and we choose the basis
\bq
 \vec{I} & = & \left( \begin{array}{c} I_{110} \\ I_{111} \\ I_{211} \\ \end{array} \right).
\eq
We obtain the differential equation
\bq
 \left( d + A \right ) \vec{I} & = & 0
\eq
with
\bq
\label{chapter_iterated_integrals:dgl_equal_mass_sunrise}
 A
 & = &
 \left( \begin{array}{ccc}
 0 & 0 & 0 \\
 0 & -\left(D-3\right) & -3 \\
 0 & \frac{1}{6}\left(D-3\right)\left(3D-8\right) & \frac{1}{2} \left(3D-8\right) \\
 \end{array} \right) \frac{dx}{x}
 \nonumber \\
 & &
 +
 \left( \begin{array}{ccc}
 0 & 0 & 0 \\
 0 & 0 & 0 \\
 -\frac{\left(D-2\right)^2}{16} & - \frac{1}{8}\left(D-3\right)\left(3D-8\right) & -\left(D-3\right) \\
 \end{array} \right) \frac{dx}{x+1}
 \nonumber \\
 & &
 +
 \left( \begin{array}{ccc}
 0 & 0 & 0 \\
 0 & 0 & 0 \\
 \frac{\left(D-2\right)^2}{16} & - \frac{1}{24}\left(D-3\right)\left(3D-8\right) & -\left(D-3\right) \\
 \end{array} \right) \frac{dx}{x+9}.
\eq
The two-loop sunrise integral with equal non-vanishing internal masses is a Feynman integral which cannot be expressed
in terms of multiple polylogarithms.
We will discuss this integral in more detail in chapter~\ref{chapter_elliptics}.

\subsection{The $\eps$-form of the system of differential equations}
\label{chapter_iterated_integrals:section:eps_form}

In the previous section we have seen that we can always systematically obtain the differential equation 
for a set of master integrals $\vec{I}$:
\bq
 \left( d + A \right) \vec{I} & = & 0.
\eq
$A$ is a matrix-valued one-form, which we write as
\bq
 A 
 & = &
 \sum\limits_{j=1}^{\NB}
 A_{x_j}
 dx_j.
\eq
The $(\Nmaster \times \Nmaster)$-matrices $A_{x_j}$ can be computed
with the help eq.~(\ref{chapter_iterated_integrals:partial_derivative}), 
integration-by-parts identities and
dimensional shift relations.
In general, the entries of $A_{x_j}$ are rational functions 
of $x_1, \dots, x_{\NB}$ and $D$.

It will be convenient to exchange the $D$-dependence for a dependence on the dimensional regularisation parameter $\eps$.
We fix an even integer $\Dint$, giving us the dimension of space-time we are interested in and set
$D = \Dint -2 \eps$.
Our usual interest is $\Dint=4$, hence
\bq
 D \; = \; 4 - 2 \eps,
 & &
 \eps \; = \; \frac{4-D}{2}.
\eq
The entries of $A_{x_j}$ are then rational functions of $x$ and $\eps$.
In eq.~(\ref{chapter_iterated_integrals:Aprime_box}) we have already seen an example where the dependence on $\eps$ is rather simple:
In eq.~(\ref{chapter_iterated_integrals:Aprime_box}) the only dependence on $\eps$ is through a prefactor $\eps$.
Furthermore, the dependence on the kinematic variables $x_1$ and $x_2$ has also particular features:
The one-forms $\omega_1, \omega_2, \dots, \omega_5$ have only simple poles.
Thirdly, the entries of the $(4 \times 4)$-matrices multiplying $\eps \cdot \omega_j$ are integer numbers. 

We say that the set of master integrals $\vec{I}$ satisfies a differential equation in $\eps$-form \cite{Henn:2013pwa}
if the following conditions are met:
\begin{tcolorbox}
{\bf $\eps$-form of the differential equation}:
\\
The differential equation
\bq
 \left( d + A \right) \vec{I} & = & 0.
\eq
is in $\eps$-form, if $A$ is of the form
\bq
\label{chapter_iterated_integrals:eps_form}
 A
 & = &
 \eps \sum\limits_{j=1}^{\NL} \; C_j \; \omega_j,
\eq
where
\begin{enumerate}
\item $C_j$ is a $(\Nmaster \times \Nmaster)$-matrix, whose entries are algebraic numbers,
\item the only dependence on $\eps$ is given by the explicit prefactor,
\item the only singularities of the differential one-forms $\omega_j$ are simple poles,
\item the non-zero boundary constants have uniform weight zero.
\end{enumerate}
We call the $\omega_j$'s {\bf letters} and $\NL$ the {\bf number of letters}.
\end{tcolorbox}
The first requirement (the entries of the matrices $C_j$ are algebraic numbers)
forbids transcendental numbers like $\pi^2$ as entries.

The fourth requirement is a condition on the boundary constants. 
This is best explained by an example.
Consider the one-loop tadpole integral, given by eq.~(\ref{chapter_basics:result_tadpole}).
For $\mu^2=m^2$ this integral does not depend on any kinematic variable
and can be considered in the framework of differential equations as a pure boundary constant.
We have
\bq
\label{chapter_iterated_integrals:example_non_uniform_weight}
 I & = &
 T_1\left(4-2\eps\right)
 \; = \;
 e^{\eps \Eulerconstant} \Gamma\left(-1+\eps\right)
 \nonumber \\
 & = &
 - \frac{1}{\eps} 
 - 1 
 - \left( 1 - \frac{1}{2} \zeta_2 \right) \eps
 + \left( \frac{1}{3} \zeta_3 - \frac{1}{2} \zeta_2 - 1 \right) \eps^2
 + {\mathcal O}\left(\eps^3\right). 
\eq
We assign rational numbers and more generally any algebraic expression in the kinematic variables
the weight zero, $\pi$ the weight $1$, zeta values $\zeta_n$ the weight $n$ and the dimensional regularisation parameter
$\eps$ the weight $(-1)$.
The weight of a product is the sum of the weights of its factors.
We say that an expression, given as a sum of terms, is of 
\index{uniform weight}
{\bf uniform weight} 
if every term has the same weight.
With these assignments the expression in eq.~(\ref{chapter_iterated_integrals:example_non_uniform_weight})
is not of uniform weight.
However, if we consider instead of $I$ the integral $I'$, defined by
\bq
 I'
 & = &
 \eps T_2\left(4-2\eps\right)
 \; = \;
 \eps T_1\left(2-2\eps\right)
 \; = \;
 e^{\eps \Eulerconstant} \Gamma\left(1+\eps\right)
 \nonumber \\
 & = &
 1
 + \frac{1}{2} \zeta_2 \eps^2
 - \frac{1}{3} \zeta_3 \eps^3
 + \frac{9}{16} \zeta_4 \eps^4
 - \left( \frac{1}{5} \zeta_5 + \frac{1}{6} \zeta_2 \zeta_3 \right) \eps^5
 + {\mathcal O}\left(\eps^6\right),
\eq
we see that $I'$ is of uniform weight zero.
A uniform weight zero implies that the coefficient of the $\eps^j$-term in the $\eps$-expansion has weight $j$.

The differential equation with $A'$ given by eq.~(\ref{chapter_iterated_integrals:Aprime_box}) is in $\eps$-form.

\begin{digression} {\bf Poles and residues of differential forms}
\\
Let us digress and discuss poles and residues of differential forms.
Let $X$ be a complex manifold of dimension $n$
and $\omega$ a differential $k$-form.
$\omega$ is 
\index{closed form}
{\bf closed},
if
\bq
 d \omega & = & 0.
\eq
$\omega$ is 
\index{exact form}
{\bf exact},
if there is a $(k-1)$-form $\eta$ such that
\bq
 \omega & = & d \eta.
\eq
The 
\index{de Rham cohomology}
$k$-th {\bf de Rham cohomology group} 
$H^k_{\mathrm{dR}}(X)$ is the set of equivalence classes of closed $k$-forms modulo exact $k$-forms.
The group law is the addition of $k$-forms.

We are in particular interested in holomorphic and meromorphic $k$-forms on $X$:
\bq
\label{chapter_iterated_integrals:meromorphic_k_form}
 \omega
 & = &
 \sum\limits_I
 \omega_I\left(x\right) dx_I,
 \;\;\;\;\;\;
 dx_I \; = \; dx_{i_1} \wedge dx_{i_2} \wedge \dots \wedge dx_{i_k}.
\eq
$\omega$ is holomorphic if all the $\omega_I(x)$ are holomorphic functions,
$\omega$ is meromorphic if all the $\omega_I(x)$ are meromorphic functions.
Note that $\omega$ in eq.~(\ref{chapter_iterated_integrals:meromorphic_k_form}) does not contain any antiholomorphic differentials $d\bar{x}_j$.

Let $Y$ be a complex codimension one submanifold, defined locally by an equation
\bq
 Y & = & \left\{ x\in X | f\left(x\right) = 0 \right\},
\eq
where 
$f$ is meromorphic and $df \neq 0$ on $Y$.
(Don't worry about the poles of $f$, $Y$ is defined by the zeros of $f$ and 
as we are only interested in local properties,
we are essentially saying that $f$ is holomorphic in a neighbourhood of $f(x)=0$.)
The $k$-form $\omega$ has a pole of order $r$ on the manifold $Y$, if $r$ is the smallest integer such that
$f^r \cdot \omega$ is holomorphic in a neighbourhood of $Y$.
Let us further assume that $\omega$ is closed.
We may write $\omega$ as
\bq
\label{chapter_iterated_integrals:decomposition_pole_r}
 \omega
 & = &
 \frac{df}{f^r} \wedge \psi + \theta,
\eq
where the $(k-1)$-form $\psi$ is holomorphic in a neighbourhood of $Y$, and the $k$-form $\theta$ 
has at most a pole of order $(r-1)$ on $Y$.
We may reduce poles of order $r>1$ to poles of order $1$ and exact forms due to the identity
\bq
 \frac{df}{f^r} \wedge \psi + \theta
 & = &
 d \left(-\frac{\psi}{\left(r-1\right)f^{r-1}} \right)
 + \frac{d\psi}{\left(r-1\right) f^{r-1}} + \theta.
\eq
Thus every form $\omega$ is equivalent up to an exact form to a form $\omega_1$ with at most a simple pole
on $Y$.
For a form $\omega_1$ with at most a simple pole on $Y$
\bq
 \omega_1 & = &
 \frac{df}{f} \wedge \psi_1 + \theta_1,
\eq
we define the 
\index{Leray residue}
{\bf Leray residue} \cite{Leray:1959} 
of $\omega_1$ along $Y$ by
\bq
 \mathrm{Res}_Y\left(\omega_1\right)
 & = &
 \left. \psi_1 \right|_Y,
\eq
where $\psi_1|_Y$ denotes the restriction of $\psi_1$ to $Y$.
If $\omega_1$ is equivalent to $\omega$ up to an exact form we set
\bq
 \mathrm{Res}_Y\left(\omega\right)
 & = &
 \mathrm{Res}_Y\left(\omega_1\right).
\eq
Since we assumed $\omega$ to be closed and having a pole (of order $r$) along $Y$,
the $k$-form $\omega$
defines a class $[\omega]\in H^k_{\mathrm{dR}}(X\backslash Y)$.
It can be shown that $\mathrm{Res}_Y(\omega)$ is independent of the chosen representative and the
decomposition in eq.~(\ref{chapter_iterated_integrals:decomposition_pole_r}).
Furthermore, $\mathrm{Res}_Y(\omega)$ is again closed.
Therefore the Leray residue defines a map
\bq
 \mathrm{Res}_Y & : & H^k_{\mathrm{dR}}\left(X\backslash Y\right) \rightarrow H^{k-1}_{\mathrm{dR}}\left(Y\right).
\eq
Multivariate (Leray) residues are defined as follows: 
Suppose we have two codimension one sub-varieties $Y_1$ and $Y_2$ defined by
$f_1=0$ and $f_2=0$, respectively.
Again we may reduce higher poles to simple poles modulo exact forms.
Let us therefore consider
\bq
 \omega & = &
 \frac{df_1}{f_1} \wedge \frac{df_2}{f_2} \wedge \psi_{12}
 + \frac{df_1}{f_1} \wedge \psi_1
 + \frac{df_2}{f_2} \wedge \psi_2
 + \theta,
\eq
where $\psi_{12}$ is regular on $Y_1 \cap Y_2$, $\psi_j$ is regular on $Y_j$ and $\theta$ is regular 
on $Y_1 \cup Y_2$.
One sets
\bq
 \mathrm{Res}_{Y_1,Y_2}\left(\omega\right)
 & = &
 \left. \psi_{12} \right|_{Y_1 \cap Y_2}.
\eq
Note that the residue is anti-symmetric with respect to the order of the hypersurfaces:
\bq
 \mathrm{Res}_{Y_2,Y_1}\left(\omega\right)
 & = &
 -
 \mathrm{Res}_{Y_1,Y_2}\left(\omega\right).
\eq
Multivariate residues for several codimension one sub-varieties $Y_1$, ..., $Y_m$ are defined analogously.
\\
\\
\bs
{\it \refstepcounter{exercise}
{\bf Exercise \theexercise}: 
Let $X={\mathbb C}^2$ and 
\bq
 Y & = & \left\{ x \in X | x_1+x_2 \; = \; 0 \right\}.
\eq
Compute
\bq
 \mathrm{Res}_Y\left( \frac{x_1 x_2^2 dx_1 \wedge dx_2}{x_1+x_2} \right).
\eq
}
\es
\\
\\
There is a second definition of the Leray residue: 
Let us first introduce the 
\index{tubular neighbourhood}
{\bf tubular neighbourhood} 
of $Y$:
These are all points in $X$, with distance to $Y$ less or equal to a small quantity $\delta$.
Let's denote the boundary of this tubular neighbourhood by $\delta Y$.
As $X$ has real dimension $2n$, and $Y$ has real dimension $(2n-2)$, $\delta Y$ has real dimension $(2n-1)$.
$\delta Y$ consists of all points in $X$, which are a distance $\delta$ away from $Y$.
By construction, $\delta Y$ does not intersect $Y$.
Locally, we may choose $n$ complex coordinates $(y_1,\dots,y_{n-1},z)$ on $X$ such that $(y_1,\dots,y_{n-1},0) \in Y$.
Then for each point $(y_1,\dots,y_{n-1},0) \in Y$ the points $(y_1,\dots,y_{n-1},z) \in \delta Y$ are the ones with
\bq
 \left| z \right | & = & \delta.
\eq
This is a circle in the complex $z$-plane with radius $\delta$.
Mathematically we say that the boundary $\delta Y$ of the tubular neighbourhood fibres over $Y$ with fibre $S^1$.
We may then integrate for each $y \in Y$ the differential $k$-form $\omega$ over $S^1$, yielding a $(k-1)$-form.
This gives the second definition of the Leray residue:
\bq
 \mathrm{Res}_Y\left(\omega\right)
 & = &
 \frac{1}{2\pi i} \int\limits_{S^1} \left. \omega \right|_{\delta Y}.
\eq
The orientation of $S^1$ is induced by the complex structure on $X$.
Multivariate residues are then defined iteratively.

The two definitions of Leray's residue generalise two expressions for the ordinary residue known from complex analysis:
Let $f(z)$ be meromorphic function of one complex variable with a pole of order $r$ at $z_0$.
We may give the residue of $f(z)$ at $z_0$ either as
\bq
\label{chapter_iterated_integrals:example_residue_1}
 \mathrm{res}\left(f,z_0\right)
 & = &
 \frac{1}{\left(r-1\right)!} 
 \left. \left( \frac{d^{r-1}}{dz^{r-1}} \left[ \left(z-z_0\right)^r f\left(z\right) \right] \right) \right|_{z=z_0}
\eq
or as
\bq
\label{chapter_iterated_integrals:example_residue_2}
 \mathrm{res}\left(f,z_0\right)
 & = &
 \frac{1}{2\pi i}
 \int\limits_{\gamma} f\left(z\right) dz,
\eq
where $\gamma$ denotes a small circle around $z_0$, oriented counter-clockwise.
Eq.~(\ref{chapter_iterated_integrals:example_residue_1}) corresponds to the first definition,
eq.~(\ref{chapter_iterated_integrals:example_residue_2}) corresponds to the second definition.

A special case of the Leray residue is the {\bf Grothendieck residue},
where we consider the $n$-fold residue of a $n$-form.
Let $X$ be a complex manifold of dimension $n$ as above and 
consider $n$ meromorphic functions $f_1, f_2, \dots, f_n$,
defining 
\bq
 Y_j & = & \left\{ x\in X | f_j\left(x\right) = 0 \right\},
 \;\;\;\;\;\;
 1 \; \le \; j \; \le \; n.
\eq
Assume further that the system of equations
\bq
\label{chapter_iterated_integrals:system_Grothendieck_residue}
 f_1\left(x\right) 
 \; = \; 
 f_2\left(x\right) 
 \; = \; 
 \dots
 f_n\left(x\right) 
 \; = \; 
 0
\eq
has as solutions a finite number of isolated points $x^{(j)}=(x_1^{(j)}, ..., x_n^{(j)})$,
where $j$ labels the individual solutions.
Let us further consider a function $g(x)$, regular at the solutions $x^{(j)}$.
We now consider a $n$-form of the form
\bq
\label{chapter_iterated_integrals:def_n_form_omega}
 \omega
 & = &
 \frac{g}{f_1 f_2 \dots f_n} dx_1 \wedge dx_2 \wedge \dots \wedge dx_n.
\eq
We first continue to consider the situation locally.
Let $x^{(j)}$ be one of the solutions of eq.~(\ref{chapter_iterated_integrals:system_Grothendieck_residue}).
We define the 
\index{local residue}
{\bf local residue} or 
\index{Grothendieck residue}
{\bf Grothendieck residue} \cite{Griffiths:book}
of $\omega$ with respect to $Y_1$, ..., $Y_n$ at $x^{(j)}$ by
\bq
\label{chapter_iterated_integrals:local_residue_1}
 \mathrm{Res}_{Y_1,...,Y_n}\left( \omega, x^{(j)} \right)
 & = &
\frac{1}{\left(2\pi i\right)^{n}}
 \oint\limits_{\Gamma_\delta}
 \frac{g\left(x\right) \; dx_1 \wedge ... \wedge dx_n}{f_1\left(x\right) ... f_n\left(x\right)}.
\eq
The integration in eq.~(\ref{chapter_iterated_integrals:local_residue_1}) is around a small $n$-torus
\bq
 \Gamma_\delta & = &
 \left\{
   \; \left( x_1, ..., x_n \right) \in X \; | \; \left| f_i\left(x\right)\right| = \delta \;
 \right\},
\eq
encircling $x^{(j)}$ with orientation
\bq
 d \arg f_1 \wedge d \arg f_2 \wedge ... \wedge d \arg f_n & \ge & 0.
\eq
In order to evaluate a local residue it is advantageous to perform a change of variables
\bq
 x_i' & = & f_i\left(x\right),
 \;\;\;\;\;\; i=1,...,n.
\eq
Let us denote the Jacobian of this transformation by
\bq
 J\left(x\right) & = &
 \frac{1}{\det\left( \frac{\partial\left(f_1,...,f_n\right)}{\partial\left(x_1,...,x_n\right)} \right)}.
\eq
The local residue at $x^{(j)}$ is then given by
\bq
 \mathrm{Res}_{Y_1,...,Y_n}\left( \omega, x^{(j)} \right)
 & = &
\frac{1}{\left(2\pi i\right)^{n}}
 \oint\limits_{\Gamma_\delta}
 \frac{g\left(x\right) \; dx_1 \wedge ... \wedge dx_n}{f_1\left(x\right) ... f_n\left(x\right)}
 \;\; = \;\; 
 J\left(x^{(j)}\right) g\left(x^{(j)}\right).
\eq
This also shows that the local residue at $x^{(j)}$ agrees with the Leray residue at $x^{(j)}$.

The
\index{global residue}
{\bf global residue}
of $\omega$ with respect to $f_1$, ..., $f_n$ is defined as
\bq
\label{chapter_iterated_integrals:global_residue_1}
 \mathrm{Res}_{Y_1,...,Y_n}\left( \omega \right)
 & = &
 \sum\limits_{\mathrm{solutions} \; j}
 \mathrm{Res}_{Y_1,...,Y_n}\left( \omega, x^{(j)} \right),
\eq
where the sum is over all solutions $x^{(j)}$ of  eq.~(\ref{chapter_iterated_integrals:system_Grothendieck_residue}).

Eq.~(\ref{chapter_iterated_integrals:global_residue_1}) defines the global residue of a $n$-form $\omega$.
As this $n$-form is given by the meromorphic function $g$ and the $n$ meromorphic functions $f_1, \dots, f_n$
as in eq.~(\ref{chapter_iterated_integrals:def_n_form_omega}), it will be convenient to define
the 
global residue of the function $g$ by
\bq
\label{chapter_iterated_integrals:global_residue_2}
 \mathrm{res}_{Y_1,...,Y_n}\left( g \right)
 & = &
 \mathrm{Res}_{Y_1,...,Y_n}\left( \omega \right).
\eq
\end{digression}

\subsection{Solution in terms of iterated integrals}
\label{chapter_iterated_integrals:section:solution_eps_form}

In this section we solve a differential equation, which is in $\eps$-form.
We show that this can be done systematically.

We fix an even integer $\Dint$ and set $D = \Dint -2 \eps$.
The main application will be $\Dint=4$, hence
\bq
 D \; = \; 4 - 2 \eps,
 & &
 \eps \; = \; \frac{4-D}{2}.
\eq
The Feynman integrals $I_{\nu_1 \dots \nu_{\ninternal}}$ are then functions of $\eps$ and $x$.
We are interested in the Laurent expansion in $\eps$:
\bq
 I_{\nu_1 \dots \nu_{\ninternal}}\left(\eps,x\right)
 & = &
 \sum\limits_{j=j_{\min}}^\infty
 I_{\nu_1 \dots \nu_{\ninternal}}^{(j)}\left(x\right) \cdot \eps^j.
\eq
We would like to determine the coefficients $I_{\nu_1 \dots \nu_{\ninternal}}^{(j)}(x)$.
In applications towards perturbation theory we usually need only the first few terms of this Laurent expansion.
The method discussed here is systematic and allows us to obtain as many terms of the Laurent expansion as desired.

Let $\vec{I}$ be a vector of $\Nmaster$ master integrals.
\bq
 \vec{I}
 & = &
 \left( I_{{\bm{\nu}}_1}, I_{{\bm{\nu}}_2}, \dots, I_{{\bm{\nu}}_{\Nmaster}} \right)^T.
\eq
We make the following assumptions:
\begin{enumerate}
\item The differential equation for $\vec{I}$ is in $\eps$-form:
\bq
\label{chapter_iterated_integrals:dgl_eps_form_master_integrals}
 \left( d + A \right) \vec{I} \; = \; 0,
 & &
 A
 \; = \;
 \eps \sum\limits_{j=1}^{\NL} \; C_j \; \omega_j.
\eq
\item All master integrals have a Taylor expansion in $\eps$:
\bq
\label{chapter_iterated_integrals:Taylor_expansion_master_integral}
 I_{{\bm{\nu}}_i}\left(\eps,x\right)
 & = &
 \sum\limits_{j=0}^\infty
 I_{{\bm{\nu}}_i}^{(j)}\left(x\right) \cdot \eps^j.
\eq
\item We know suitable boundary values for all master integrals.
\end{enumerate}
Assumption 2 may seem at first sight rather restrictive, as it forbids any pole terms in $\eps$.
However, it isn't. It only means that we multiplied a Feynman integral with a sufficient high power
of $\eps$, such that the Laurent expansion starts with the $\eps^0$-term or later.
We have already seen an example in eq.~(\ref{chapter_iterated_integrals:base_change_box}):
The first three master integrals $I_{0011}$, $I_{0101}$ and $I_{1010}$ are one-loop two-point function 
with vanishing internal masses.
They have been calculated in eq.~(\ref{chapter_basics:result_bubble}).
For example, the Laurent expansion for $I_{0101}$ starts as
\bq
 I_{0101}
 & = &
 \frac{1}{\eps}
 + \left( 2 - L \right)
 + \left( \frac{1}{2} L^2 - 2L - \frac{\pi^2}{12} + 4 \right) \eps
 \nonumber \\
 & &
 + \left( -\frac{1}{6} L^3 + L^2 - 4 L - \frac{7}{3} \zeta_3 - \frac{\pi^2}{6} + \frac{\pi^2}{12} L + 8 \right) \eps^2
 + {\mathcal O}\left(\eps^3\right),
\eq
with $L=\ln x_1$.
This integral has a pole in $\eps$.
However, in going to the $\eps$-form for the differential equation we defined a new master integral 
$I_{0101}'= \eps \left(1-2\eps\right) I_{0101}$.
The Laurent expansion for $I_{0101}'$ starts as 
\bq
 I_{0101}'
 & = &
 1
 - L \eps
 + \left( \frac{1}{2} L^2 - \frac{\pi^2}{12} \right) \eps^2
 + \left( -\frac{1}{6} L^3 - \frac{7}{3} \zeta_3 + \frac{\pi^2}{12} L \right) \eps^3
 + {\mathcal O}\left(\eps^4\right).
\eq
This expansion starts at $\eps^0$.

In order to present the solution of a differential equation in $\eps$-form, we introduce 
{\bf iterated integrals}.
Let us start with the general definition of an iterated integrals \cite{Chen}:
Let $X$ be a $n$-dimensional (complex) manifold and
\bq
 \gamma & : & \left[a,b\right] \rightarrow X
\eq
a path with start point $x_a=\gamma(a)$ and end point $x_b=\gamma(b)$.
Suppose further that $\omega_1, \dots, \omega_r$ are differential $1$-forms on $X$.
Let us write
\bq
 f_j\left(\lambda\right) d\lambda & = & \gamma^\ast \omega_j
\eq
for the pull-backs to the interval $[a,b]$.
If 
\bq
 \omega_j \; = \; \sum\limits_{k=1}^n \omega_{j k}\left(x\right) dx_k,
 & &
 \gamma\left(\lambda\right) \; = \; 
 \left( \begin{array}{c}
  \gamma_1\left(\lambda\right) \\
  \gamma_2\left(\lambda\right) \\
  \vdots \\
  \gamma_n\left(\lambda\right) \\
 \end{array} \right),
\eq
the pull-back $\gamma^\ast \omega_j$ is given by
\bq
 \gamma^\ast \omega_j
 & = &
 \sum\limits_{k=1}^n \omega_{j k}\left(\gamma\left(\lambda\right)\right) \frac{d\gamma_k\left(\lambda\right)}{d\lambda} d\lambda,
\eq
hence
\bq
 f_j\left(\lambda\right) 
 & = &
 \sum\limits_{k=1}^n 
 \omega_{j k}\left(\gamma\left(\lambda\right)\right) \frac{d\gamma_k\left(\lambda\right)}{d\lambda}.
\eq
\bs
{\it \refstepcounter{exercise}
{\bf Exercise \theexercise}: 
Let 
\bq
 \omega & = & 3 dx_1 + (5+x_1) dx_2 + x_3 dx_3
\eq
and
\bq
 \gamma & : & [0,1] \rightarrow {\mathbb C}^3,
 \;\;\;\;\;\;
 \gamma\left(\lambda\right)
 = \left( \begin{array}{c}
 \lambda \\ \lambda^2 \\ 1 + \lambda \\ \end{array} \right).
\eq
Compute
\bq
 \int\limits_\gamma \omega.
\eq
}
\es
\\
\\
For $\lambda \in [a,b]$ the $r$-fold iterated integral 
of $\omega_1, \dots, \omega_r$ along the path $\gamma$ is defined
by
\bq
\label{chapter_iterated_integrals:def_iterated_integral}
 I_{\gamma}\left(\omega_1,\dots,\omega_r;\lambda\right)
 & = &
 \int\limits_a^{\lambda} d\lambda_1 f_1\left(\lambda_1\right)
 \int\limits_a^{\lambda_1} d\lambda_2 f_2\left(\lambda_2\right)
 \dots
 \int\limits_a^{\lambda_{r-1}} d\lambda_r f_r\left(\lambda_r\right).
\eq
We call $r$ the 
\index{depth}
{\bf depth} of the iterated integral.
We define the $0$-fold iterated integral to be
\bq
 I_{\gamma}\left(;\lambda\right)
 & = &
 1.
\eq
We then have the recursive structure
\bq
\label{chapter_iterated_integrals:def_iterated_integral_recursive}
 I_{\gamma}\left(\omega_1,\omega_2,\dots,\omega_r;\lambda\right)
 & = &
 \int\limits_a^{\lambda} d\lambda_1 f_1\left(\lambda_1\right)
 I_{\gamma}\left(\omega_2,\dots,\omega_r;\lambda_1\right).
\eq
\begin{digression} {\bf Basic properties of iterated integrals}
\\
Let $\gamma_1 : [a,b] \rightarrow X$ and $\gamma_2 : [a,b] \rightarrow X$ be two paths with 
$\gamma_1(b)=\gamma_2(a)$.
In this case we may form a new path by gluing the endpoint of $\gamma_1$ to the starting point of $\gamma_2$.
The combined path starts at $\gamma_1(a)$ and ends at $\gamma_2(b)$.
In detail 
we define the path $\gamma_2 \circ \gamma_1 : [a,b] \rightarrow X$ to be given by
\bq
 \left( \gamma_2 \circ \gamma_1 \right)\left(\lambda\right)
 & = & 
 \left\{ \begin{array}{lll}
  \gamma_1\left(2\lambda-a\right) & \mbox{for} & a \le \lambda \le \frac{1}{2} \left(a+b\right),
  \nonumber \\
  \gamma_2\left(2\lambda-b\right) & \mbox{for} & \frac{1}{2} \left(a+b\right) \le \lambda \le b.
  \nonumber \\
 \end{array} \right.
\eq
For the iterated integral along the path $\gamma_2 \circ \gamma_1$ we have
\bq
\label{chapter_iterated_integrals:combined_path}
 I_{\gamma_2 \circ \gamma_1}\left(\omega_1,\dots,\omega_r;\lambda\right)
 & = &
 \sum\limits_{j=0}^r
 I_{\gamma_2}\left(\omega_1,\dots,\omega_j;\lambda\right)
 I_{\gamma_1}\left(\omega_{j+1},\dots,\omega_r;\lambda\right).
\eq
For a path $\gamma : [a,b] \rightarrow X$ we denote by $\gamma^{-1} : [a,b] \rightarrow X$ the reverse path
given by
\bq
 \gamma^{-1}\left(\lambda\right) & = & \gamma\left(a+b-\lambda\right).
\eq
For the iterated integral along the path $\gamma^{-1}$ we have
\bq
\label{chapter_iterated_integrals:reversed_path}
 I_{\gamma^{-1}}\left(\omega_1,\dots,\omega_r;b\right)
 & = &
 \left(-1\right)^r
 I_{\gamma}\left(\omega_r,\dots,\omega_1;b\right).
\eq
\bs
{\it \refstepcounter{exercise}
{\bf Exercise \theexercise}: 
Prove eq.~(\ref{chapter_iterated_integrals:combined_path}) for the case $r=2$, i.e. show
\bq
 I_{\gamma_2 \circ \gamma_1}\left(\omega_1,\omega_2;\lambda\right)
 & = &
 I_{\gamma_1}\left(\omega_1,\omega_2;\lambda\right)
 +
 I_{\gamma_2}\left(\omega_1;\lambda\right)
 I_{\gamma_1}\left(\omega_2;\lambda\right)
 +
 I_{\gamma_2}\left(\omega_1,\omega_2;\lambda\right).
\eq
}
\es
\\
\\
\bs
{\it \refstepcounter{exercise}
{\bf Exercise \theexercise}: 
Prove eq.~(\ref{chapter_iterated_integrals:reversed_path}).
}
\es
\\
\\
Let us discuss the path (in-) dependence of iterated integrals.
We consider two paths $\gamma_1 : [a,b] \rightarrow X$ and $\gamma_2 : [a,b] \rightarrow X$ with
the same starting point and the same end point
\bq
 \gamma_1(a) \; = \; \gamma_2(a) \; = \; x_a,
 & &
 \gamma_1(b) \; = \; \gamma_2(b) \; = \; x_b,
\eq
see 
\begin{figure}
\begin{center}
\includegraphics[scale=1.0]{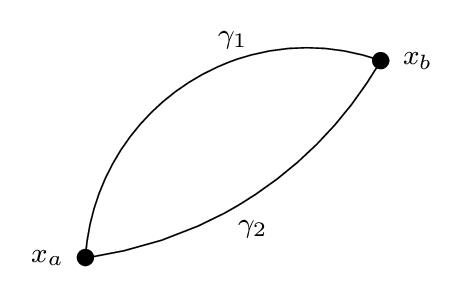}
\end{center}
\caption{
Two paths $\gamma_1$ and $\gamma_2$ with the same starting point $x_a$ and the same end point $x_b$.
}
\label{chapter_iterated_integrals:path_homotopy}
\end{figure}
fig.~\ref{chapter_iterated_integrals:path_homotopy}.
We call two paths $\gamma_1$ and $\gamma_2$ 
\index{homotopic}
{\bf homotopic} if there is a continuous map
$\phi : [a,b] \times [0,1] \rightarrow X$ such that
\bq
 \phi(\lambda,0) \; = \; \gamma_1(\lambda),
 & &
 \phi(\lambda,1) \; = \;  \gamma_2(\lambda)
\eq 
and $\phi(a,\kappa)=x_a$, $\phi(b,\kappa)=x_b$ for all $\kappa\in[0,1]$.
This defines an equivalence relation between paths with the same starting point and the same end point.
We would like to investigate the question under which conditions iterated integrals depend only on the 
equivalence class of homotopic paths and not on a specific representative within this equivalence class.
We call such iterated integrals
\index{homotopy functional}
{\bf homotopy functionals}.
In general, an individual iterated integral is not a homotopy functional.
This is easily seen as follows:
Consider
\bq
 \omega_1 \; = \; dx, 
 & &
 \omega_2 \; = \; dy
\eq
and a family of paths ($\kappa>0$)
\bq
 \gamma_\kappa & : & \left[0,1\right] \rightarrow {\mathbb R}^2,
 \nonumber \\
 & & \gamma_\kappa\left(\lambda\right) \; = \; \left(\lambda,\lambda^\kappa\right).
\eq
$\gamma_\kappa$ defines a family of homotopic paths with starting point $\gamma_\kappa(0)=(0,0)$ and 
end point $\gamma_\kappa(1)=(1,1)$.
Members of this family are indexed by $\kappa$.
Let us consider
\bq
 I_{\gamma_\kappa}\left(\omega_1,\omega_2;1\right)
 & = &
 \int\limits_0^1 d\lambda_1
 \int\limits_0^{\lambda_1} \kappa \lambda_2^{\kappa-1} d\lambda_2
 \; = \;
 \frac{1}{\kappa+1}.
\eq
The result depends on $\kappa$ and $I_{\gamma_\kappa}(\omega_1,\omega_2)$ is therefore not a homotopy functional.

Let's consider an iterated integral of depth $r$
\bq
 I_\gamma\left(\omega_1,\omega_2,\dots,\omega_r;\lambda\right).
\eq
There is a one-to-one correspondence between 
ordered sequences of differential one-forms $\omega_1$, $\omega_2$, $\dots$,$\omega_r$
and elements in the tensor algebra $(\Omega(X))^{\otimes r}$ (where $\Omega(X)$ denotes the space of differential forms on $X$)
of the form
\bq
 \omega_1 \otimes \omega_2 \otimes \dots \otimes \omega_r.
\eq
It is customary to denote the latter as
\bq
 \left[ \omega_1 | \omega_2 | \dots | \omega_r \right]
 & = &
 \omega_1 \otimes \omega_2 \otimes \dots \otimes \omega_r.
\eq
This is called the 
\index{bar construction}
{\bf bar construction}.
In the tensor algebra we define
\bq
 d \left[ \omega_1 | \omega_2 | \dots | \omega_r \right]
 & = &
 \sum\limits_{j=1}^r \left[ \omega_1 | \dots | \omega_{j-1} | d\omega_j | \omega_{j+1} | \dots | \omega_r \right]
 \nonumber \\
 & &
 +
 \sum\limits_{j=1}^{r-1} \left[ \omega_1 | \dots | \omega_{j-1} | \omega_j \wedge \omega_{j+1} | \omega_{j+2} | \dots | \omega_r \right].
\eq
If all our $\omega$'s are closed, this reduces to
\bq
 d \left[ \omega_1 | \omega_2 | \dots | \omega_r \right]
 & = &
 \sum\limits_{j=1}^{r-1} \left[ \omega_1 | \dots | \omega_{j-1} | \omega_j \wedge \omega_{j+1} | \omega_{j+2} | \dots | \omega_r \right].
\eq
Let us now consider a linear combination of iterated integrals of depth $\le r$ with constant coefficients:
\bq
 I
 & = &
 \sum\limits_{j=1}^r \sum\limits_{i_1,\dots,i_j}
 c_{i_1 \dots i_j} I_\gamma\left(\omega_{i_1},\dots,\omega_{i_j};\lambda\right)
\eq
and the corresponding element in the tensor algebra
\bq
 B
 & = &
 \sum\limits_{j=1}^r \sum\limits_{i_1,\dots,i_j}
 c_{i_1 \dots i_j} \left[\omega_{i_1}|\dots|\omega_{i_j}\right].
\eq
$I$ is a homotopy functional if and only if \cite{Chen}
\bq
\label{chapter_iterated_integrals:def_integrable_word}
 dB & = & 0.
\eq
This is the sought-after criteria when a linear combination of iterated integrals is path independent.
We call any $B$ satisfying eq.~(\ref{chapter_iterated_integrals:def_integrable_word}) an 
\index{integrable word}
{\bf integrable word}.
\end{digression}

\noindent
We are interested in differential one-form $\omega_j$, which have simple poles.
It may happen that some $\omega$ has a simple pole at the starting point of the integration path $x=x_a$.
In the special case where
\bq
 \gamma^\ast \omega & = & \frac{d\lambda}{\lambda},
\eq
e.g. the pull-back has just a simple pole and no regular part,
we define the iterated integral by
\bq
\label{chapter_iterated_integrals:def_iterated_integral_trailing_zeros}
 I_{\gamma}\left(\omega,...,\omega;\lambda\right)
 & = &
 \frac{1}{r!} \ln^r\left(\lambda\right).
\eq
In the case, where $\omega$ has a simple pole at $x=x_a$ and a regular part, we may always decompose
$\omega$ as
\bq
 \omega & = & L + \omega_{\mathrm{reg}},
\eq
with $\gamma^\ast L = d\lambda/\lambda$ and $\omega_{\mathrm{reg}}$ having no pole at $x=x_a$.
In general we define the iterated integral by eq.~(\ref{chapter_iterated_integrals:def_iterated_integral_recursive}) 
and eq.~(\ref{chapter_iterated_integrals:def_iterated_integral_trailing_zeros}).
We say that the iterated integral $I_{\gamma}(\omega_1,\omega_2,...,\omega_r;\lambda)$ has a
\index{trailing zero}
{\bf trailing zero},
if the last differential one-form $\omega_r$ has a simple pole at $x=x_a$.

Let us discuss a specific class of iterated integrals:
We take $X={\mathbb C}$ with coordinate $x$ and
\bq
\label{chapter_iterated_integrals:def_omega_mpl}
 \omega^{\mathrm{mpl}}\left(z_j\right) & = & \frac{dx}{x-z_j}.
\eq
In this example we treat $z_j$ as a (fixed) parameter.
Let $\gamma : [0,\lambda] \rightarrow {\mathbb C}$ be the line segment from zero along the positive real axis 
to $y \in {\mathbb R}_+$, e.g. $\gamma(0)=0$ and $\gamma(\lambda)=y$.
Let us assume that none of the $z_j$'s lie on the path $\gamma$.
We introduce a special notation for iterated integrals build from differential one-forms 
as in eq.~(\ref{chapter_iterated_integrals:def_omega_mpl}).
We set
\bq
 G\left(z_1,\dots,z_r;y\right)
 & = &
 I_\gamma\left( \omega^{\mathrm{mpl}}\left(z_1\right), \dots, \omega^{\mathrm{mpl}}\left(z_r\right); \lambda \right).
\eq
The functions $G\left(z_1,\dots,z_r;y\right)$ are called
{\bf multiple polylogarithms}.
We discuss these functions in detail in chapter~\ref{chapter_multiple_polylogarithms}.
The general definition of iterated integrals translates in the case of multiple polylogarithms to
\bq
 G(\underbrace{0,\dots,0}_{r-\mathrm{times}};y)
 & = & 
 \frac{1}{r!} \ln^r\left(y\right),
 \nonumber \\
 G\left(z_1,z_2\dots,z_r;y\right)
 & = &
 \int\limits_0^y
 \frac{dy_1}{y_1-z_1}
 G\left(z_2\dots,z_r;y_1\right).
\eq
Let us now return to the differential equation.
According to eq.~(\ref{chapter_iterated_integrals:Taylor_expansion_master_integral}), 
each master integral has a Taylor expansion in $\eps$.
It is convenient to write
\bq
\label{chapter_iterated_integrals:Taylor_expansion_vector_master_integral}
 \vec{I}\left(\eps,x\right)
 & = & 
 \sum\limits_{j=0}^\infty \vec{I}^{(j)}\left(x\right) \cdot \eps^j.
\eq
We plug eq.~(\ref{chapter_iterated_integrals:Taylor_expansion_vector_master_integral}) 
into the differential equation~(\ref{chapter_iterated_integrals:dgl_eps_form_master_integrals}) 
\bq
 \left( d + \eps \sum\limits_{k=1}^{\NL} \; C_k \; \omega_k \right) 
 \left( \sum\limits_{j=0}^\infty \vec{I}^{(j)}\left(x\right) \cdot \eps^j \right)
 & = & 0,
\eq
and compare term-by-term in the $\eps$-expansion.
We obtain
\bq
\label{chapter_iterated_integrals:dgl_order_by_order}
 d \vec{I}^{(0)}\left(x\right) & = & 0,
 \nonumber \\
 d \vec{I}^{(j)}\left(x\right) & = & - \sum\limits_{k=1}^{\NL} \omega_k \; C_k \; \vec{I}^{(j-1)}\left(x\right),
 \;\;\;\;\;\; j \; \ge \; 1.
\eq
This system can easily be solved:
The first equation of eq.~(\ref{chapter_iterated_integrals:dgl_order_by_order}) states that 
$\vec{I}^{(0)}(x)$ is a constant, which is determined by the boundary condition.
Knowing $\vec{I}^{(j-1)}(x)$ we obtain $\vec{I}^{(j)}(x)$ by integration.
The integration constant is again fixed by the boundary condition.
The integration can be done in the class of iterated integrals. Each integration increases the depth of the iterated
integrals by one.
At order $j$ we obtain iterated integrals of depth $\le j$.
The integrability condition eq.~(\ref{chapter_iterated_integrals:integrability_condition}) ensures
that the result is a homotopy functional.
Thus each
\bq
 I_{{\bm{\nu}}_i}^{(j)},
 & &
 1 \le i \le \Nmaster
\eq
is a path-independent linear combination of iterated integrals.
Recall that $I_{{\bm{\nu}}_i}^{(j)}$ denotes the $\eps^j$-term of the $i$-th master integral. 
The individual iterated integrals appearing in $I_{{\bm{\nu}}_i}^{(j)}$ are in general not path independent.

Let's look at a simple example.
We consider a system with $\Nmaster=1$ and $\NB=1$, 
e.g. one function $I(\eps,x)$ depending on $\eps$ and one variable $x$.
Let us assume that the function $I(\eps,x)$ satisfies the differential equation
\bq
 \left( d + A \right) I \; = \; 0,
 & &
 A \; = \; - \eps \; \frac{dx}{x-1}
\eq
with the boundary condition $I(\eps,0)=1$.
Then
\bq
 I(x) & = & 
 1
 + \eps G\left(1;x\right)
 + \eps^2 G\left(1,1;x\right)
 + \eps^3 G\left(1,1,1;x\right)
 + ...
\eq

Let us also discuss a non-trivial example.
We consider the one-loop four point function 
in eq.~(\ref{chapter_iterated_integrals:example_box}) 
with the basis of master integrals as in eq.~(\ref{chapter_iterated_integrals:base_change_box}).
In this basis the differential equation is in $\eps$-form and given by eq.~(\ref{chapter_iterated_integrals:example_box_dgl_eps_form}).
In this example the Feynman integrals depend on two variables $x_1$ and $x_2$.
Let assume that we would like to integrate the differential equation from the start point $x_a=(0,0)$ 
to the point of interest $x_b=(x_1,x_2)$.
We can do this by first integrating from $(0,0)$ to $(x_1,0)$ along the $x_1$-direction, and then from $(x_1,0)$ to
$(x_1,x_2)$ along the $x_2$ direction.
Doing so, we find
\bq
 I_1'
 & = &
 B_1^{(0)} + B_1^{(1)} \eps + B_1^{(2)} \eps^2 
 + {\mathcal O}\left(\eps^3\right),
 \\
 I_2'
 & = &
 B_2^{(0)} 
 + \left[ B_2^{(1)} - B_2^{(0)} G\left(0;x_1\right) \right] \eps 
 + \left[ B_2^{(2)} - B_2^{(1)} G\left(0;x_1\right) + B_2^{(0)} G\left(0,0;x_1\right) \right] \eps^2 
 + {\mathcal O}\left(\eps^3\right),
 \nonumber \\
 I_3'
 & = &
 B_3^{(0)} 
 + \left[ B_3^{(1)} - B_3^{(0)} G\left(0;x_2\right) \right] \eps 
 + \left[ B_3^{(2)} - B_3^{(1)} G\left(0;x_2\right) + B_3^{(0)} G\left(0,0;x_2\right) \right] \eps^2 
 + {\mathcal O}\left(\eps^3\right),
 \nonumber \\
 I_4'
 & = &
 B_4^{(0)} 
 + \left[ B_4^{(1)} 
          - B_4^{(0)} G\left(0;x_1\right) - B_4^{(0)} G\left(0;x_2\right) 
          + \left( B_4^{(0)} - B_3^{(0)} \right) G\left(1;x_1\right)
 \right. \nonumber \\
 & & \left.
          + \left( B_3^{(0)} - B_1^{(0)} \right) G\left(1;x_2\right)
          + \left( B_1^{(0)} - B_2^{(0)} - B_3^{(0)} + B_4^{(0)} \right) G\left(1-x_1;x_2\right)
   \right] \eps 
 + {\mathcal O}\left(\eps^2\right).
 \nonumber 
\eq
where the $B_i^{(j)}$ are integration constants.
The ${\mathcal O}\left(\eps^2\right)$-term of $I_4'$ is already rather long, therefore we stopped
for $I_4'$ at order ${\mathcal O}\left(\eps^1\right)$.
In order to fix the integration constants $B_i^{(j)}$ we have to know the integrals at one specific
kinematic point.
This does not have to be the starting point $x_a=(0,0)$ of the integration.
For the case at hand it is convenient to choose the point $x_{\mathrm{boundary}}=(1,1)$.
The first three master integrals $I_1'$, $I_2'$ and $I_3'$ are given at this kinematic point by
(see eq.~(\ref{chapter_basics:result_bubble}))
\bq
 I_1' \; = \; I_2' \; = \; I_3'
 & = &
 e^{\eps \Eulerconstant} 
 \frac{\Gamma\left(1+\eps\right)\Gamma\left(1-\eps\right)^2}{\Gamma\left(1-2\eps\right)}
 \; = \;
 1
 - \frac{\pi^2}{12} \eps^2
 + {\mathcal O}\left(\eps^3\right),
\eq
yielding
\bq
 B_1^{(0)} \; = \; B_2^{(0)} \; = \; B_3^{(0)} & = & 1,
 \nonumber \\
 B_1^{(1)} \; = \; B_2^{(1)} \; = \; B_3^{(1)} & = & 0,
 \nonumber \\
 B_1^{(2)} \; = \; B_2^{(2)} \; = \; B_3^{(2)} & = & - \frac{1}{2} \zeta_2,
\eq
with $\zeta_2=\pi^2/6$.
\\
\\
\bs
{\it \refstepcounter{exercise}
{\bf Exercise \theexercise}: 
Show that $I_4'$ is given at the kinematic point $(x_1,x_2)=(1,1)$ by
\bq
\label{chapter_iterated_integrals:boundary_box}
 I_4'
 & = &
 e^{\eps \Eulerconstant} 
 \frac{\Gamma\left(1+\eps\right)\Gamma\left(1-\eps\right)^2}{\Gamma\left(1-2\eps\right)}
 \left( 1 - \sum\limits_{k=2}^\infty \zeta_k \eps^k \right).
\eq
Hint: Use the trick from exercise~\ref{chapter_basics:exercise_oneloopbox_Feynman_parameter_trick} 
and the Mellin-Barnes technique.
}
\es
\\
\\
From eq.~(\ref{chapter_iterated_integrals:boundary_box}) we deduce
\begin{align}
 B_4^{(0)} & = 1,
 &
 B_4^{(1)} & = 0,
 &
 B_4^{(2)} & = \frac{1}{2} \zeta_2,
\end{align}
and obtain as final result
\bq
\label{chapter_iterated_integrals:result_box_v1}
 I_1'
 & = &
 1 - \frac{1}{2} \zeta_2 \eps^2 
 + {\mathcal O}\left(\eps^3\right),
 \\
 I_2'
 & = &
 1 
 - G\left(0;x_1\right) \eps 
 + \left[ G\left(0,0;x_1\right) - \frac{1}{2} \zeta_2 \right] \eps^2 
 + {\mathcal O}\left(\eps^3\right),
 \nonumber \\
 I_3'
 & = &
 1 
 - G\left(0;x_2\right) \eps 
 + \left[ G\left(0,0;x_2\right) - \frac{1}{2} \zeta_2 \right] \eps^2 
 + {\mathcal O}\left(\eps^3\right),
 \nonumber \\
 I_4'
 & = &
 1 
 - \left[ G\left(0;x_1\right) + G\left(0;x_2\right) \right] \eps
 + \left[ 
          G\left(0,0;x_1\right) + G\left(0,0;x_2\right)
        - G\left(1,0;x_1\right) - G\left(1,0;x_2\right)
 \right. \nonumber \\
 & & \left.
        + G\left(0;x_1\right) G\left(0;x_2\right)
        + \frac{1}{2} \zeta_2
  \right] \eps^2 
 + {\mathcal O}\left(\eps^3\right).
 \nonumber 
\eq
Up to order ${\mathcal O}(\eps^2)$ we may express alternatively the result in terms of logarithms and dilogarithms:
\bq
\label{chapter_iterated_integrals:result_box_v2}
 I_1'
 & = &
 1 - \frac{1}{2} \zeta_2 \eps^2 
 + {\mathcal O}\left(\eps^3\right),
 \\
 I_2'
 & = &
 1 
 - \ln\left(x_1\right) \eps 
 + \frac{1}{2} \left[ \ln^2\left(x_1\right) - \zeta_2 \right] \eps^2 
 + {\mathcal O}\left(\eps^3\right),
 \nonumber \\
 I_3'
 & = &
 1 
 - \ln\left(x_2\right) \eps 
 + \frac{1}{2} \left[ \ln^2\left(x_1\right) - \zeta_2 \right] \eps^2 
 + {\mathcal O}\left(\eps^3\right),
 \nonumber \\
 I_4'
 & = &
 1 
 - \left[ \ln\left(x_1\right) + \ln\left(x_2\right) \right] \eps
 + \left[ 
          - \mathrm{Li}_2\left(x_1\right)
          - \mathrm{Li}_2\left(x_2\right)
          + \frac{1}{2} \ln^2\left(x_1\right) 
          + \frac{1}{2} \ln^2\left(x_2\right) 
          + \ln\left(x_1\right) \ln\left(x_2\right) 
 \right. \nonumber \\
 & & \left.
          - \ln\left(x_1\right) \ln\left(1-x_1\right)
          - \ln\left(x_2\right) \ln\left(1-x_2\right)
        + \frac{1}{2} \zeta_2
  \right] \eps^2 
 + {\mathcal O}\left(\eps^3\right).
 \nonumber 
\eq
\bs
{\it \refstepcounter{exercise}
{\bf Exercise \theexercise}: 
Show the equivalence of the ${\mathcal O}(\eps^2)$-term of $I_4'$ 
between eq.~(\ref{chapter_iterated_integrals:result_box_v1}) and eq.~(\ref{chapter_iterated_integrals:result_box_v2}).
}
\es
\\
\\
In practical applications
our main interest is $I_4=I_{1111}$ and not so much $I_4'= 1/2 \cdot \eps^2 x_1 x_2 I_{1111}$.
Inverting this relation gives us
\bq
 I_{1111} & = & \frac{2}{\eps^2} \frac{1}{x_1 x_2} I_4'.
\eq
We have calculated $I_4'$ as a function of $\eps$, $x_1 = s/p_4^2$ and $x_2 = t/p_4^2$.
From the scaling relation of eq.~(\ref{chapter_basics:kinematic_variables_scaling_relation})
we may reinstate the full dependence on $\eps$, $s$, $t$, $p_4^2$ and $\mu^2$:
\bq
 I_{1111}\left(\eps,s,t,p_4^2,\mu^2\right)
 & = &
 \frac{2}{\eps^2} 
 \left( \frac{-p_4^2}{\mu^2} \right)^{-2-\eps}
 \left. \frac{1}{x_1 x_2} I_4'\left(\eps,x_1,x_2\right) \right|_{x_1=s/p_4^2, x_2=t/p_4^2}.
\eq
One may check that this agrees with eq.~(\ref{appendix_one_loop_integrals:one_mass_box}) given in 
appendix~\ref{appendix_one_loop_integrals}.

Let us make a few remarks:
Firstly, instead of integrating from $(0,0)$ to $(x_1,0)$ and from there to $(x_1,x_2)$,
we could have used alternatively a path from $(0,0)$ to $(0,x_2)$ and from there to $(x_1,x_2)$.
This gives the same result.
In general, we are free to choose any path we like, as long as we don't cross any branch cuts.
In the present example there are no problems with branch cuts for real values of $x_1$ and $x_2$ as long as
\bq
 x_1 \; > \; 0,
 \;\;\;\;\;\;
 x_2 \; > \; 0,
 \;\;\;\;\;\;
 x_1 + x_2 \; < \; 1.
\eq
Branch cuts originate from the singularities in the differential equation.
In the present example the singularities are determined by $\omega_1$ - $\omega_5$ 
(defined in eq.~(\ref{chapter_iterated_integrals:dlog_form}))
and located at
\bq
 x_1 \; = \; 0,
 \;\;\;\;\;\;
 x_1 \; = \; 1,
 \;\;\;\;\;\;
 x_2 \; = \; 0,
 \;\;\;\;\;\;
 x_2 \; = \; 1,
 \;\;\;\;\;\;
 x_1 + x_2 \; = \; 1.
\eq
We may integrate the differential equation beyond a singularity.
In this case Feynman's $i\delta$-prescription tells us which branch we have to choose.

As a second remark let us notice that the multiple polylogarithms in eq.~(\ref{chapter_iterated_integrals:boundary_box})
have trailing zeros
and that the coefficients of the $\eps$-expansion of $I_2'$, $I_3'$ and $I_4'$ diverge
as a power of a logarithm in the limit $x_1 \rightarrow 0$ or $x_2 \rightarrow 0$.
This may happen and there is nothing wrong with it.
This was the reason why we didn't use $(x_1,x_2)=(0,0)$ as the kinematic point to fix the integration
constants.
Let's see what happens, if we set $x_1=x_2=0$ from the start and calculate these integrals
within dimensional regularisation.
We denote these integrals, where we set $x_1=x_2=0$ from the start by
$\tilde{I}_2'$, $\tilde{I}_3'$ and $\tilde{I}_4'$.
The first two are scaleless integrals and hence equal to zero within dimensional regularisation
(see eq.~(\ref{chapter_basics:basic_eq_negative_dimensions})).
Also $\tilde{I}_4'$ is zero due to the explicit prefactor $x_1 x_2$.
Thus
\bq
\label{chapter_iterated_integrals:result_boundary_0_0}
 \tilde{I}_2' & = & 0,
 \nonumber \\
 \tilde{I}_3' & = & 0,
 \nonumber \\
 \tilde{I}_4' & = & 0.
\eq
This seems puzzling, as eq.~(\ref{chapter_iterated_integrals:result_boundary_0_0})
is quite different from eq.~(\ref{chapter_iterated_integrals:result_box_v1}).
To analyse the situation, let's consider $I_2'$.
For $x_1>0$ and $\eps<0$ the integral $I_2'$ is unambiguously given by
\bq
\label{chapter_iterated_integrals:one_loop_two_point_unambiguous}
 I_2'
 & = &
 e^{\eps \Eulerconstant} 
 \frac{\Gamma\left(1+\eps\right)\Gamma\left(1-\eps\right)^2}{\Gamma\left(1-2\eps\right)}
 x_1^{-\eps}.
\eq
If we first expand in $\eps$ and then take the limit $x_1 \rightarrow 0$, we find
that the coefficients of the $\eps$-expansion are given
by eq.~(\ref{chapter_iterated_integrals:result_box_v1})
and that they 
diverge as a power of a logarithm in this limit.
If on the other hand we first take the limit $x_1 \rightarrow 0$
and then expand in $\eps$, we recover eq.~(\ref{chapter_iterated_integrals:result_boundary_0_0}).
(In our particular example, we already obtain zero after taking the limit $x_1 \rightarrow 0$,
so for the expansion in $\eps$ there is nothing left to expand.)
In other words,
the expansion in $\eps$ does not commute with the limit $x_1 \rightarrow 0$.
In a nutshell, we have
\bq
 \lim\limits_{x \rightarrow 0+} \lim\limits_{\eps \rightarrow 0-} x^{-\eps}
 & = &
 \lim\limits_{x \rightarrow 0+} 1
 \; = \; 1,
 \nonumber \\
 \lim\limits_{\eps \rightarrow 0-} \lim\limits_{x \rightarrow 0+} x^{-\eps}
 & = &
 \lim\limits_{\eps \rightarrow 0-} 0
 \; = \; 0.
\eq

\section{Fibre bundles}
\label{chapter_iterated_integrals:fibre_bundles}

We have seen that with the help of integration-by-parts identities we may express any Feynman integral
$I_{\nu_1 \dots \nu_{\ninternal}}$ as a linear combination of $\Nmaster$ master integrals
$I_{{\bm{\nu}}_1}, I_{{\bm{\nu}}_2}, \dots, I_{{\bm{\nu}}_{\Nmaster}}$.
We may view the master integrals as a basis of a $\Nmaster$-dimensional vector space.
The master integrals depend on the dimensional regularisation parameter $\eps$ and $\NB$ kinematic variables
$x=(x_1, \dots x_{\NB})$.
Let us now focus on the dependence on $x$. For the moment we treat $\eps$ as a (fixed) parameter. 
For every value of $x$ we have a separate vector space spanned by $I_{{\bm{\nu}}_1}(x), \dots, I_{{\bm{\nu}}_{\Nmaster}}(x)$
The master integrals satisfy a differential equation 
\bq
 \left( d + A \right) \vec{I}\left(x\right) & = & 0,
\eq
where $\vec{I}(x)=(I_{{\bm{\nu}}_1}(x), \dots, I_{{\bm{\nu}}_{\Nmaster}}(x))^T$.

In mathematical terms we are looking at a vector bundle
of rank $\Nmaster$ over a base space parametrised by the coordinates $x$.
The vector bundle is equipped with a flat connection defined locally by the matrix-valued one-form $A$.
In section~\ref{chapter_iterated_integrals:def_fibre_bundle} we introduce these terms.

We also know by now that the computation of Feynman integrals reduces to the task of finding appropriate transformations,
which bring the differential equation for the master integrals into a $\eps$-form.
The mathematical reformulation gives us a clear picture what type of transformation we should consider:
These are transformations, which correspond to a basis change in the fibre or
transformations, which correspond to a coordinate transformation on the base manifold.
These are discussed in section~\ref{chapter_iterated_integrals:fibre_transformation} and
section~\ref{chapter_iterated_integrals:base_transformation}, respectively.

\subsection{Mathematical background}
\label{chapter_iterated_integrals:def_fibre_bundle}

Let us introduce the required terminology for fibre bundles.
We give a concise summary with the essential definitions.
For a more detailed introduction into this topic for a readership with a physics background
we refer to the books by Nakahara \cite{Nakahara} and Isham \cite{Isham:1999qu}.
\\
\\
We start with the definition of a manifold. Let $M$ be a topological space.
The basics of topology are summarised in appendix~\ref{appendix_algebraic_geometry:sect_topology}.
\\
\\
An 
\index{open chart}
{\bf open chart} on $M$ is a pair $(U,\varphi)$, where $U$ is an open subset of $M$ and $\varphi$ is a homeomorphism 
of $U$ onto an open subset of ${\mathbb R}^n$.
\\
\\
A 
\index{differentiable manifold}
{\bf differentiable manifold} of dimension $n$ is a Hausdorff space with a collection of open charts
$(U_\alpha, \varphi_\alpha)_{\alpha\in A}$ such that
\begin{description}
\item{M1:}
\bq
 M & = & \bigcup\limits_{\alpha\in A} U_\alpha.
\eq
\item{M2:} For each pair $\alpha,\beta \in A$ the mapping $\varphi_\beta \circ \varphi_\alpha^{-1}$ is an 
infinitely differentiable mapping of 
$\varphi_\alpha\left(U_\alpha \cap U_\beta \right)$
onto
$\varphi_\beta\left(U_\alpha \cap U_\beta \right)$.
\end{description}
A differentiable manifold is also often denoted as a $C^\infty$ manifold. As we will only be concerned with differentiable
manifolds, we will often omit the word ``differentiable'' and just speak about manifolds.
\\
\\
The collection of open charts $(U_\alpha, \varphi_\alpha)_{\alpha\in A}$ is called an {\bf atlas}.
\\
\\
If $p \in U_\alpha$ and
\bq
 \varphi_\alpha(p) & = & \left( x_1(p), ..., x_n(p) \right),
\eq
the set $U_\alpha$ is called the 
\index{coordinate neighbourhood}
{\bf coordinate neighbourhood} of $p$ and the numbers $x_i(p)$ are called
the 
\index{local coordinates}
{\bf local coordinates} of $p$.
\\
\\
Note that in each coordinate neighbourhood
$M$ looks like an open subset of ${\mathbb R}^n$. But note that we do not
require that $M$ be ${\mathbb R}^n$ globally.
\\
\\
The definition of a 
\index{complex manifold}
{\bf complex manifold} requires only small modifications:
\begin{itemize}
\item An open chart of a complex manifold $M$ is a pair $(U,\varphi)$, where $U$ is an open subset of $M$ 
and $\varphi$ is a homeomorphism of $U$ onto an open subset of ${\mathbb C}^n$ (which we may also view as an
open subset of ${\mathbb R}^{2n}$, so the real dimension of any complex manifold is even).
\item Axiom (M2) in the definition above is modified as follows:
We require that 
for each pair $\alpha,\beta \in A$ the mapping 
$\varphi_\beta \circ \varphi_\alpha^{-1} : \varphi_\alpha\left(U_\alpha \cap U_\beta \right) \rightarrow \varphi_\beta\left(U_\alpha \cap U_\beta \right)$ is holomorphic.
\end{itemize}
Let us now turn to fibre bundles.
A 
\index{fibre bundle}
{\bf differentiable fibre bundle} $(E,M,F,\pi,G)$ (or simply fibre bundle for short) 
consists of the following elements:
\begin{itemize}
\item A differentiable manifold $E$ called the 
\index{total space}
{\bf total space}.
\item A differentiable manifold $M$ called the 
\index{base space}
{\bf base space}.
\item A differentiable manifold $F$ called the 
\index{fibre}
{\bf fibre}.
\item A surjection $\pi : E \rightarrow M$ called the 
\index{projection}
{\bf projection}. The inverse image
$\pi^{-1}(p) = F_p$ is called the fibre at $p$.
\item A Lie group $G$ called the 
\index{structure group}
{\bf structure group}, which acts on $F$ from the left.
\end{itemize}
We require that
\begin{description}
\item{F1:}
there is a set of open coverings $\{U_i\}$ of $M$ with diffeomorphisms 
$\phi_i : U_i \times F \rightarrow \pi^{-1}(U_i)$ such that $\pi \phi_i(p,f) = p$. The map
$\phi_i$ is called the 
\index{local trivialisation}
{\bf local trivialisation}, since $\phi_i^{-1}$ maps $\pi^{-1}(U_i)$ onto the
direct product $U_i \times F$,
\item{F2:}
if we write $\phi_i(p,f) = \phi_{i,p}(f)$, the map $\phi_{i,p} : F \rightarrow F_p$ is a
diffeomorphism. On $U_i \cap U_j \neq \emptyset$ we require that
$t_{i j}(p) = \phi_{i,p}^{-1} \phi_{j,p} : F \rightarrow F$ be an element of $G$, satisfying the
consistency conditions $t_{i i} = \mbox{id}, t_{i j} = t_{j i}^{-1}, t_{i j} t_{j k} = t_{i k}$.
The $\{ t_{i j} \}$ are called the 
\index{transition function}
{\bf transition functions}.
\end{description}
A 
\index{section, local}
{\bf local section} of a fibre bundle $E \stackrel{\pi}{\rightarrow} M$ over $U \subset M$
is a smooth map $\sigma: U \rightarrow E$, which satisfies $\pi \sigma = \mbox{id}_M$.
A 
\index{section, global}
{\bf global section} of a fibre bundle $E \stackrel{\pi}{\rightarrow} M$ 
is a smooth map $\sigma: M \rightarrow E$, which satisfies $\pi \sigma = \mbox{id}_M$.
The space of local sections and global sections is denoted by $\Gamma(U,E)$ and $\Gamma(M,E)$, respectively.
\\
\\
Two important special cases of fibre bundles are vector bundles and principal bundles:
\\
\\
A {\bf principal bundle} is a fibre bundle, whose fibre is identical with the structure group $G$. 
A principal bundle is also often called a $G$-bundle over $M$ and denoted $P(M,G)$.
$P$ denotes the total space.
\\
\\
A 
\index{vector bundle}
{\bf vector bundle} is a fibre bundle, whose fibre is a vector space.
The dimension $r$ of the fibre $F$ is called the 
\index{rank of a vector bundle}
{\bf rank} of the vector bundle.
A vector bundle of rank $1$ is called a 
\index{line bundle}
{\bf line bundle}.
A 
\index{complex vector bundle}
{\bf complex vector bundle} of rank $r$ is a vector space, where the fibre is ${\mathbb C}^r$.
Examples of vector bundles are the tangent bundle $TM$ and the cotangent bundle $T^\ast M$.
For the 
\index{tangent bundle}
{\bf tangent bundle} the fibre at the point $p \in M$ is the vector space of all tangent vectors
to $M$ at $p$.
Similar, the fibre of the 
\index{cotangent bundle}
{\bf cotangent bundle} at the point $p \in M$ is the vector space of all cotangent vectors
at $p$.
\\
\\
A 
\index{frame}
{\bf frame} of a rank $r$ vector bundle over $U \subset M$ is 
an ordered set of local sections $\sigma_j: U \rightarrow E$ with $1 \le j \le r$, such that
$\sigma_1(p), \sigma_2(p), \dots, \sigma_r(p)$ is a basis of $F_p$ for any $p \in U$.
\\
\\
Let $E$ and $M$ be complex manifolds and $\pi : E \rightarrow M$ be a holomorphic surjection. 
$E$ is a 
\index{holomorphic vector bundle}
{\bf holomorphic vector bundle} of rank $r$ if
the typical fibre is ${\mathbb C}^r$, the structure group is $G=\mathrm{GL}_r({\mathbb C})$ and
\begin{description}
\item{H1:} the local trivialisation $\phi_i : U_i \times {\mathbb C}^r \rightarrow \pi^{-1}(U_i)$ is a biholomorphism,
\item{H2:} the transition functions $t_{ij} : U_i \cap U_j \rightarrow G = \mathrm{GL}_r({\mathbb C})$ are holomorphic maps.
\end{description}
Let us now turn to connections. We start from a principal bundle $P(M,G)$.
Let $u$ be a point in the total space $P$ and let $G_p$ be the fibre at $p=\pi(u)$.
The 
\index{vertical subspace}
{\bf vertical subspace} $V_u P$ is a subspace of the tangent space $T_u P$, which is tangent to $G_p$ at $u$.
The vertical subspace $V_u P$ is isomorphic as a vector space 
to the Lie algebra $\mathfrak{g}$ of $G$.
There is a right action of an element $g \in G$ on a point $u \in P$.
Within a local trivialisation a point $u$ in the total space is given by $u=(x,g')$ and the right action
by $g$ is given by $u g = (x, g' g)$. The right action by $g$ maps the point $u$ to another point $u g$
on the same fibre.
\\
\\
A 
\index{connection}
{\bf connection} on $P(M,G)$ is a unique separation of the tangent space $T_u P$ into the vertical 
subspace $V_u P$ and a 
\index{horizontal subspace}
{\bf horizontal subspace} $H_u P$ such that 
\begin{description}
\item{C1:} $T_u P = H_u P \oplus V_u P$
\item{C2:} a smooth vector field $X$ on $P(M,G)$ is separated into smooth vector fields $X^H \in H_u P$ and
$X^V \in V_u P$ as $ X = X^H + X^V$.
\item{C3:} Let $g \in G$. 
The horizontal subspaces $H_u P$ and $H_{ug} P$ on the same fibre are related by a linear
map $R_{g \ast}$ induced by the right action of $g$ : $H_{u g} P = R_{g \ast} H_u P$.
Accordingly a subspace $H_u P$ at $u$ generates all the horizontal subspaces on the same fibre.
\end{description}
A 
\index{connection one-form}
{\bf connection one-form} $\omega \in \mathfrak{g} \otimes T^{\ast} P$, which takes values in the
Lie algebra $\mathfrak{g}$ of $G$, is a projection of $T_u P$ onto the vertical component
$V_u P \cong \mathfrak{g}$, compatible with axiom (C3) above.
In detail we require that for $X \in T_uP$
\begin{description}
\item{CF1:} $\omega(X) \in \mathfrak{g}$,
\item{CF2:} $\omega_{u g}(R_{g \ast} X) = g^{-1} \omega_u(X) g$,
where $R_{g \ast} X$ denotes the push-forward of the tangent vector $X$ at the point $u$ to the point $u g$
by the right action of $g$.
\end{description}
The horizontal subspace $H_u P$ is defined to be the kernel of $\omega$.
Condition (CF2) may be stated equivalently as $R_g^{\ast} \omega = \mathrm{Ad}_{g^{-1}} \omega$, with
$R^{\ast}_g \omega_{u g}(X) = \omega_{u g}(R_{g \ast} X)$ and $\mathrm{Ad}_{g^{-1}} \omega(X) = g^{-1} \omega_u(X) g$
we recover (CF2).

Let $U \subset M$ be an open subset of $M$ and $s : U \rightarrow P$ a local section.
We denote by
$A$ the pull-back of $\omega$ to $M$:
\bq
 A & = & s^\ast \omega.
\eq
Let us now consider two sections $\sigma_1$ and $\sigma_2$, with associated pull-backs $A_1$ and $A_2$.
We can always relate the two sections $\sigma_1$ and $\sigma_2$ by
\bq
\label{chapter_iterated_integrals:relation_sections}
 \sigma_1(x) & = & \sigma_2(x) g(x),
\eq
where $g(x)$ is a $x$-dependent element of the Lie group $G$.
Then we obtain for the pull-backs $A_1$ and $A_2$ of the connection one-form the relation
\bq
\label{chapter_iterated_integrals:relation_pull_backs_orig}
 A_2 & = & g A_1 g^{-1} + g d g^{-1}.
\eq
In the sequel of the book we will use the notation $g(x)=U(x)$.
Eq.~(\ref{chapter_iterated_integrals:relation_pull_backs}) reads then
\bq
\label{chapter_iterated_integrals:relation_pull_backs}
 A_2 & = & U A_1 U^{-1} + U d U^{-1}.
\eq
Readers familiar with gauge theories in physics will recognise that 
eq.~(\ref{chapter_iterated_integrals:relation_pull_backs}) is nothing else than a gauge transformation.
\\
\\
\bs
{\it \refstepcounter{exercise}
{\bf Exercise \theexercise}: 
Derive eq.~(\ref{chapter_iterated_integrals:relation_pull_backs}) from eq.~(\ref{chapter_iterated_integrals:relation_sections}).
\\
\\
Hint: Recall that the action of the pull-back $A_2$ on a tangent vector is defined as the 
action of the original form $\omega$ on the push-forward of the tangent vector.
Recall further that a tangent vector at a point $x$ can be given as a tangent vector to a curve
through $x$. It is sufficient to show that the actions of $A_2$ and $U A_1 U^{-1} + U d U^{-1}$
on an arbitrary tangent vector give the same result.
In order to prove the claim you will need in addition 
the defining relations for the connection one-form $\omega$,
given in (CF1) and (CF2).
}
\es
\\
\\
Given a principal bundle $P(M,G)$ and a $r$-dimensional vector space $F$, 
on which $G$ acts on the left through a $r$-dimensional representation $\rho$,
we may always construct the associated vector bundle $(E,M,F,\pi,G)$.
The total space $E$ is given by $(P \times F)/ G$, where points $(u,v)$ and $(u,\rho(g^{-1})v)$
are identified. 
The projection $\pi$ in the associated vector bundle $E$
is given by $\pi(u,v)=\pi_P(u)$, where $\pi_P$ denotes the
projection of the principal bundle $P$.
The transition functions of $E$ are given by $\rho(t_{ij})$, where $t_{ij}$ denote the transition functions
of $P$.

Conversely, a rank $r$ vector bundle with fibre
${\mathbb R}^r$ or ${\mathbb C}^r$ induces a principal bundle with structure group
$\mathrm{GL}_r({\mathbb R})$ or $\mathrm{GL}_r({\mathbb C})$, respectively.
The transition functions of the induced principal bundle are the ones of the vector bundle,
and this defines the principal bundle.
(In general, the minimal information required to define a fibre bundle are $M$, $F$, $G$, 
a set of open coverings $\{U_i\}$ and the transition functions $t_{ij}$.)

We continue to work in a local chart $(U,\varphi)$.
The local connection one-form $A$ defines the 
\index{covariant derivative}
{\bf covariant derivative}
\bq
 D_A & = & d + A.
\eq
If $A$ is a $\mathfrak{g}$-valued $p$-form and 
$B$ is a $\mathfrak{g}$-valued $q$-form, the commutator of the two is defined by
\bq
 \left[ A, B \right] & = & A \wedge B - (-1)^{p q} B \wedge A,
\eq
the factor $(-1)^{p q}$ takes into account that we have to permute $p$ differentials $dx_i$ from $A$ past $q$ differentials
$dx_j$ from $B$.
If $A$ and $B$ are both one-forms we have
\bq
 \left[ A, B \right] & = & A \wedge B + B \wedge A
 =
 \left[ A_i, B_j \right] dx_i \wedge dx_j.
\eq
In particular
\bq
 A \wedge A & = &
 \frac{1}{2} \left[ A, A \right] = 
 \frac{1}{2} \left[ A_i, A_j \right] dx_i \wedge dx_j.
\eq
We define the 
\index{curvature two-form}
{\bf curvature two-form} of the fibre bundle by
\bq
 F & = & 
   D_A A = dA + A \wedge A
 =
  dA + \frac{1}{2} \left[ A, A \right].
\eq
The fibre bundle is 
\index{flatness}
{\bf flat} (or pure gauge) if
\bq
 F & = & 0.
\eq

\subsection{Fibre bundles in physics}
\label{chapter_iterated_integrals:fibre_bundles_in_physics}

\subsubsection{Gauge theories}

Let us now see where fibre bundles occur in physics.
Our first example is a gauge theory with a real scalar particle in the fundamental representation of a gauge group $G$.
We denote the generators of the gauge group by $T^a$ and the coupling by $g$.
Let us further denote by $r$ the dimension of the fundamental representation of $G$.
The Lagrange density reads
\bq
 {\mathcal L} 
 & = &
 \frac{1}{2} \left(D_{\mu,jk} \phi_k(x) \right)^\dagger \left(D^\mu_{\;\;jl} \phi_l(x) \right) 
 - \frac{1}{4} F^{a}_{\mu\nu}(x) F^{a \mu\nu}(x)
\eq
where the covariant derivative $D_{\mu,jk}$ and the field strength $F^{a}_{\mu\nu}$ are given by
\bq
\label{chapter_iterated_integrals:field_strength}
 D_{\mu,jk} & = & \delta_{jk} \partial_\mu - i g T^a_{jk} A^a_\mu(x),
 \nonumber \\
 F^{a}_{\mu\nu}(x) & = & \partial_{\mu} A^{a}_{\nu}(x) - \partial_{\nu} A^{a}_{\mu}(x)
 + g f^{abc} A^{b}_{\mu}(x) A^{c}_{\nu}.
\eq
There are two fibre bundles here: a principal bundle and a vector bundle.
In both cases the base space $M$ is given by flat Minkowski space.
For simplicity let's assume that we are not interested in topologically non-trivial configurations (like instantons), 
therefore a trivial fibre bundle and a global section are fine for us.
The discussion below can easily be extended to the general case by introducing an atlas of coordinate patches
and local sections glued together in the appropriate way.
The gauge field $A^a_\mu$ defines a local connection one-form by
\bq
A & = & A_\mu \; dx^\mu = \frac{g}{i} T^a A^a_\mu \; dx^\mu,
 \;\;\;\;\;\;
 A_\mu = \frac{g}{i} T^a A^a_\mu.
\eq
Let's see what the curvature of the connection is: We have
\bq
 d A 
 & = & d \left( A_\nu dx^\nu \right) = \partial_\mu A_\nu dx^\mu \wedge dx^\nu 
 =
 \frac{1}{2} \left( \partial_\mu A_\nu - \partial_\nu A_\mu \right) dx^\mu \wedge dx^\nu,
 \nonumber \\
 A \wedge A 
 & = & 
 A_\mu dx^\mu \wedge A_\nu dx^\nu = \frac{1}{2} \left( A_\mu A_\nu - A_\nu A_\mu \right) dx^\mu \wedge dx^\nu 
 =
 \frac{1}{2} \left[ A_\mu, A_\nu \right] dx^\mu \wedge dx^\nu,
\eq
and therefore
\bq
 F 
 & = & 
 \frac{1}{2} \left( \partial_\mu A_\nu - \partial_\nu A_\mu + \left[ A_\mu, A_\nu \right] \right) dx^\mu \wedge dx^\nu
 =
 \frac{1}{2} F_{\mu\nu} dx^\mu \wedge dx^\nu.
\eq
With the notation
\bq
 F_{\mu\nu} & = & \frac{g}{i} T^a F^a_{\mu\nu}
\eq
we have
\bq
F & = & 
  \frac{1}{2} F_{\mu\nu} dx^\mu \wedge dx^\nu
  =
  \frac{1}{2} \frac{g}{i} T^a F^a_{\mu\nu} dx^\mu \wedge dx^\nu,
\eq
and $F^a_{\mu\nu}$ is given by eq.~(\ref{chapter_iterated_integrals:field_strength}).
Thus we see that a gauge field is equivalent to the pull-back of the connection one-form of a principal bundle.

The real scalar field $\phi_k(x)$ (with $1 \le k \le r$)
we may view as a section in a vector bundle with fibre ${\mathbb R}^r$.

A more realistic example is QCD, where the real scalar field is replaced by one or more spinor fields
(depending on the number of quarks).

\subsubsection{Feynman integrals}

Our main interest are Feynman integrals.
In section~\ref{chapter_basics:momentum_representation_of_Feynman_integrals}
we discussed the variables on which a Feynman integral $I_{\nu_1 \dots \nu_{\ninternal}}$ depends.
These were the dimension of space-time $D$ (or alternatively the dimensional regularisation parameter $\eps$)
and $\NB+1$ dimensionless kinematic variables $x_1, \dots, x_{\NB+1}$ of the form
\bq
 \frac{-p_i \cdot p_j}{\mu^2}
 & \mbox{or} &
 \frac{m_i^2}{\mu^2}.
\eq
Due to the scaling relation in eq.~(\ref{chapter_basics:kinematic_variables_scaling_relation})
we may set without loss of information one kinematic variable to one, leaving $\NB$ non-trivial kinematic variables.
That's the way we usually approach the problem from the physics side: We prefer dimensionless quantities and 
the arbitrary scale $\mu$ can be related to the renormalisation scale introduced 
in eq.~(\ref{chapter_qft:dimensionless_g}).

If we approach the problem from the mathematical side we may view the situation as a complex vector bundle
over a base space, which is parametrised by the kinematic variables $x$.
As we are mainly interested in the local properties and not the global properties,
we may view the base space as an open subset $U$ of ${\mathbb C} {\mathbb P}^{\NB}$.
The open subset $U$ is defined by the requirement
that the master integrals are single-valued and non-degenerate on $U$. 
To get there let's go one step back 
to eq.~(\ref{chapter_basics:def_Feynman_integral_v2})
and start from $\NB+1$ dimensionfull variables $X_1, \dots, X_{\NB+1}$ of the form
\bq
 -p_i \cdot p_j
 & \mbox{or} &
 m_i^2
\eq
of mass dimension $2$ and an arbitrary scale $\mu^2$, again of mass dimension $2$.
Assume that one particular variable $X_j$ is not equal to zero.
Let us now make the choice $\mu^2=X_j$.
Again, no information is lost: $\mu^2$ enters only as a trivial prefactor 
in eq.~(\ref{chapter_basics:def_Feynman_integral_v2}),
and once we know the integral for one particular choice of $\mu^2$, we recover the integral for any other choice
of $\mu^2$ from a scaling relation derived from eq.~(\ref{chapter_basics:def_Feynman_integral_v2}).
Once we made the choice $\mu^2=X_j$, our integral depends on $D$ (or $\eps$) and $X_1, \dots, X_{\NB+1}$.
The dependence on $D$ (or $\eps$) will play no further role in the discussion below
and we focus on the dependence on the variables $X_1, \dots, X_{\NB+1}$.
If we scale all variables $X_1, \dots, X_{\NB+1}$ by a factor $\lambda$ we now have
\bq
\label{chapter_iterated_integrals:projective_scaling_relation}
 I^{\mathrm{chart} \; j}_{\nu_1 \dots \nu_{\ninternal}}\left( \lambda X_1, \dots, \lambda X_{\NB+1}\right)
 & = &
 I^{\mathrm{chart} \; j}_{\nu_1 \dots \nu_{\ninternal}}\left( X_1, \dots, X_{\NB+1}\right).
\eq
The superscript ${\mathrm{chart} \; j}$ indicates that we assumed $X_j \neq 0$ and made the choice $\mu^2=X_j$.
Thus the Feynman integral $I^{\mathrm{chart} \; j}_{\nu_1 \dots \nu_{\ninternal}}$ is invariant
under a simultaneous scaling of all variables $X_1, \dots, X_{\NB+1}$ and defines therefore
a (in general multi-valued) function on the chart $X_j \neq 0$ of ${\mathbb C} {\mathbb P}^{\NB}$.
We have
\bq
\label{chapter_iterated_integrals:projective_function}
 I^{\mathrm{chart} \; j}_{\nu_1 \dots \nu_{\ninternal}}\left( X_1, \dots, X_{j-1}, X_j, X_{j+1}, \dots, X_{\NB+1}\right)
 & = &
 I^{\mathrm{chart} \; j}_{\nu_1 \dots \nu_{\ninternal}}\left( x_1, \dots, x_{j-1}, 1, x_{j+1}, \dots, x_{\NB+1}\right)
 \nonumber \\
\eq
with $x_i=X_i/X_j$.
Since we made the choice $\mu^2=X_j$ we have $x_i=X_i/\mu^2$.
This shows the equivalence of this approach with the previous one
and with the implicit assumption that in the previous approach we always set $x_{\NB+1}$ to one we have
\bq
 I_{\nu_1 \dots \nu_{\ninternal}}\left( x_1, \dots, x_{\NB}\right)
 & = & 
 I^{\mathrm{chart} \; \NB+1}_{\nu_1 \dots \nu_{\ninternal}}\left( x_1, \dots, x_{\NB}, 1 \right).
\eq
We may take
\bq
 \left[ X_1 : X_2 : \dots : X_{\NB+1} \right]
\eq
as homogeneous coordinates on ${\mathbb C} {\mathbb P}^{\NB}$.
Feynman integrals are in general multivalued function of the kinematic variables.
A simple toy example follows directly from eq.~(\ref{chapter_iterated_integrals:one_loop_two_point_unambiguous})
\bq
 I^{\mathrm{chart} \; 2}\left( X_1, X_2 \right)
 \; = \;
 \left( \frac{X_1}{X_2} \right)^{-\eps},
 & &
 I\left(x_1\right)
 \; = \;
 I^{\mathrm{chart} \; 2}\left( x_1, 1 \right)
 \; = \;
 x_1^{-\eps},
\eq
where we ignored an $x$-independent prefactor not relevant to the discussion here.
$I^{\mathrm{chart} \; 2}(X_1, X_2)$ clearly satisfies eq.~(\ref{chapter_iterated_integrals:projective_scaling_relation})
and defines a multi-valued function on ${\mathbb C} {\mathbb P}^{1}$.
The multi-valuedness arises as follows: Consider the chart $X_2=1$ of ${\mathbb C} {\mathbb P}^{1}$
with coordinate $x_1$. If we go anti-clockwise around the origin $x_1=0$, $I(x_1)$ changes by a multiplicative
prefactor
\bq
 e^{-2\pi i \eps},
\eq
which follows from
\bq
 x_1^{-\eps}
 & = & 
 e^{-\eps \ln x_1}
\eq
and the multi-valuedness of the logarithm.
In order to get a single-valued function we restrict to an open subset, 
where we choose a branch of the logarithm which allows us to view
$I(x_1)$ as a single-valued function.
In this example we may choose the open subset as
\bq
 U_2 & = & \left\{ \; x_1 \in {\mathbb C} \; | \; x_1 \notin {\mathbb R}_{\le 0} \; \right\}.
\eq
Let us now return to the general case. Let $U$ be an open subset of ${\mathbb C} {\mathbb P}^{\NB}$, 
where the Feynman integrals are single-valued and where the master integrals are non-degenerate.
We set $U_j$ to be the intersection of $U$ with the set of points of ${\mathbb C} {\mathbb P}^{\NB}$ for which $X_j \neq 0$:
\bq
 U_j & = &
 U \cap
 \left\{ \; \left[ X_1 : \dots : X_{\NB+1} \right] \in {\mathbb C} {\mathbb P}^{\NB} \; | \; X_j \neq 0 \; \right\}.
\eq
Eq.~(\ref{chapter_iterated_integrals:projective_function}) defines then
$I^{\mathrm{chart} \; j}_{\nu_1 \dots \nu_{\ninternal}}$ as a single-valued function on $U_j$.
An analogous definition applies to $I^{\mathrm{chart} \; i}_{\nu_1 \dots \nu_{\ninternal}}$ for $U_i$.
Let us now assume $X_i \neq 0$, $X_j \neq 0$ and $U_i \cap U_j \neq 0$.
We have
\bq
 I^{\mathrm{chart} \; i}_{\nu_1 \dots \nu_{\ninternal}}\left( X_1, \dots, X_i, \dots, X_j, \dots, X_{\NB+1}\right)
 & = &
 \left( \frac{X_i}{X_j} \right)^{\nu-\frac{\loopnumber D}{2}}
 I^{\mathrm{chart} \; j}_{\nu_1 \dots \nu_{\ninternal}}\left( X_1, \dots, X_i, \dots, X_j, \dots, X_{\NB+1}\right).
 \nonumber \\
\eq
For the transition function we have
\bq
 t_{ij}
 & = &
 \left( \frac{X_i}{X_j} \right)^{\nu-\frac{\loopnumber D}{2}}.
\eq
Thus an individual Feynman integral $I_{\nu_1 \dots \nu_{\ninternal}}$ 
defines a complex line bundle over $U$.
A set of $\Nmaster$ master integrals defines a complex vector bundle of rank $\Nmaster$ over $U$.

On $U$ each master integral 
$I_{{\bm{\nu}}_1}, \dots, I_{{\bm{\nu}}_{\Nmaster}}$ defines a local section.
It may happen that for specific values of $x$ some master integrals degenerate and become
linearly dependent.

An example is given by the one-loop two-point function with two unequal internal masses.
For $m_1^2 \neq m_2^2$ we have three master integrals, which may be taken as $I_{11}$, $I_{10}$ and $I_{01}$.
For $m_1^2 = m_2^2$ the two master integrals $I_{10}$ and $I_{01}$ degenerate and are equal to each other
(see the discussion in section~\ref{chapter_iterated_integrals:integration_by_parts}).
For $m_1^2 = m_2^2$ we only have two master integrals, which may be taken as $I_{11}$ and $I_{10}$.

We say that the set of master integrals are {\bf ramified} at a point $x$
if the set of master integrals is linearly dependent for the value $x$. 
We excluded those points from the definition of $U$.
Thus, the set of all master integrals defines a frame on $U$.
This says that on $U$ the $\Nmaster$ master integrals are linearly independent.
The vector bundle is equipped with a connection.
The local connection one-form is given by the matrix-valued one-form $A$ appearing in the differential equation
eq.~(\ref{chapter_iterated_integrals:differential_equation_master_integrals}).

Usually our primary interest is not the global structure, but only a single chart, 
which we may take as $U_{\NB+1}$.
The reason is the following: The chart $U_{\NB+1}$ covers all cases except the ones with $X_{\NB+1} = 0$.
The case $X_{\NB+1} = 0$ has one kinematic variable less and can be considered as simpler.
With these remarks we take in the sequel the base space $M$ to be $U_{\NB+1}$.
$U_{\NB+1}$ is homeomorphic to an open subset of ${\mathbb C}^{\NB}$.

In a mathematical language we are considering a local system on $U_{\NB+1}$ with the corresponding 
Gau{\ss}-Manin connection given by $\nabla = d + A$.
\begin{digression} {\bf Local systems and the Gau{\ss}-Manin connection}
\\
\\
Let $M$ be a topological space.
By a 
\index{local system}
{\bf local system} one understands either:
\begin{enumerate}
\item a vector bundle $\pi : E \rightarrow M$ with parallel transport, i.e. for each homotopy class of paths in $M$ there is
a vector space isomorphism between the fibres. The vector space isomorphisms are compatible with the composition of paths.
\item if $M$ is a differentiable manifold, a vector bundle $\pi : E \rightarrow M$ with a flat connection $\nabla = d + A$.
The connection is flat if $dA + A \wedge A = 0$.
$\nabla$ is called the 
\index{Gau{\ss}-Manin connection}
{\bf Gau{\ss}-Manin connection}.
\item a locally constant sheaf of vector spaces on $M$. (Appendix~\ref{appendix_algebraic_geometry} gives a definition of sheaves.)
\end{enumerate}
The three definitions are equivalent (of course, for definition $2$ we have to assume that $M$ is differentiable).
\end{digression}

\subsection{Fibre transformations}
\label{chapter_iterated_integrals:fibre_transformation}

We have learned that in Feynman integral computations we are considering a vector bundle over a base space
$M$, which is parametrised by the kinematic variables $x_1, \dots, x_{\NB}$.
The master integrals $I_{{\bm{\nu}}_1}(x), \dots, I_{{\bm{\nu}}_{\Nmaster}}(x)$ can be viewed
as local sections, and for each $x$ they define a basis of the vector space in the fibre.
The master integrals 
$\vec{I}=(I_{{\bm{\nu}}_1}, I_{{\bm{\nu}}_2}, \dots, I_{{\bm{\nu}}_{\Nmaster}})^T$
satisfy the differential equation
\bq
 \left(d+A\right) \vec{I} & = & 0.
\eq
The matrix-valued one-form $A$ gives the local connection one-form.
The integrability condition in eq.~(\ref{chapter_iterated_integrals:integrability_condition})
says that the connection is flat:
\bq
 dA + A \wedge A & = & 0.
\eq
Let us now turn back to more practical questions:
We would like to transform a given differential equation (not necessarily in $\eps$-form)
to the $\eps$-form.
In order to achieve this goal, we first have to understand what transformations can be done.
The first transformation, which we discuss in this section, is a change of the master integrals.
This amounts to changing the sections and the frame of the vector bundle.
We consider the transformation
\bq
 \vec{I}'\left(\eps,x\right)
 & = &
 U\left(\eps,x\right) \vec{I}\left(\eps,x\right),
\eq
where $U(\eps,x)$ is an invertible $(\Nmaster \times \Nmaster)$-matrix, which may depend on $\eps$ and $x$.
Let's work out the differential equation satisfies by the new master integrals.
We have 
\bq
 \vec{I} & = & U^{-1} \vec{I}'
\eq
and
\bq
 d \vec{I} 
 \; = \;
 d \left( U^{-1} \vec{I}' \right)
 \; = \;
 U^{-1} d \vec{I}'
 + \left( d U^{-1} \right) \vec{I}'.
\eq
Thus
\bq
 d \vec{I}'
 \; = \;
 U d \vec{I} - \left( U d U^{-1} \right) \vec{I}'
 \; = \;
 - U A \vec{I} - \left( U d U^{-1} \right) \vec{I}'
 \; = \;
 - \left( U A U^{-1} + U d U^{-1} \right) \vec{I}'.
\eq
In summary we have
\begin{tcolorbox}
{\bf Fibre transformation}:
\\
Let $\vec{I}=(I_{{\bm{\nu}}_1}, \dots, I_{{\bm{\nu}}_{\Nmaster}})^T$
be a set of master integrals satisfying the differential equation
\bq
 \left(d+A\right) \vec{I} & = & 0.
\eq
If we change the master integrals according to
\bq
 \vec{I}'
 & = &
 U \vec{I},
\eq
where $U$ is an invertible $(\Nmaster \times \Nmaster)$-matrix, which may depend on $\eps$ and $x$,
the new master integrals satisfy the differential equation
\bq
 \left(d+A'\right) \vec{I}' & = & 0,
\eq
where $A'$ is related to $A$ by
\bq
 A' & = & U A U^{-1} + U d U^{-1}.
\eq
\end{tcolorbox}
We have already seen in example 2 in section~\ref{chapter_iterated_integrals:deriving_the_dgl}
that a suitable fibre transformation may bring the differential equation into an $\eps$-form.
Eq.~(\ref{chapter_iterated_integrals:base_change_box}) may also be written as
\bq
 \vec{I}' & = & U \vec{I},
\eq
with
\bq
 U\left(\eps,x_1,x_2\right) & = &
 \left( \begin{array}{cccc}
  \eps \left(1-2\eps\right) & 0 & 0 & 0 \\ 
  0 & \eps \left(1-2\eps\right) & 0 & 0 \\ 
  0 & 0 & \eps \left(1-2\eps\right) & 0 \\ 
  0 & 0 & 0 & \frac{1}{2} \eps^2 x_1 x_2 \\
 \end{array} \right).
\eq
Let us also consider example 3 from section~\ref{chapter_iterated_integrals:deriving_the_dgl}.
This is the family of the two-loop double box integral with eight master integrals, given by
eq.~(\ref{chapter_iterated_integrals:precanonical_masters_double_box}).
We consider the transformation
\bq
 \vec{I}' & = & U \vec{I},
\eq
where the $(8 \times 8)$-matrix $U$ is given by
\bq
\label{chapter_iterated_integrals:trafo_double_box}
\lefteqn{
 U\left(\eps,x\right) = } & &
 \\
 & &
 \left( \begin{array}{cccccccc} 
 \frac{g_1\left(\eps\right)}{x} & 0 & 0 & 0 & 0 & 0 & 0 & 0 \\
 0 & g_1\left(\eps\right) & 0 & 0 & 0 & 0 & 0 & 0 \\
 0 & 0 & \eps^2\left(1-2\eps\right)^2 & 0 & 0 & 0 & 0 & 0 \\
 0 & 0 & 0 & g_2\left(\eps\right) & 0 & 0 & 0 & 0 \\
 0 & 0 & 0 & g_2\left(\eps\right) & 6 \eps^3\left(1-2\eps\right) x & 0 & 0 & 0 \\
 0 & 0 & 0 & 0 & 0 & 3\eps^4\left(1+x\right) & 0 & 0 \\
 0 & 0 & 0 & 0 & 0 & 0 & \eps^4 x^2 & 0 \\
 -g_1\left(\eps\right) & -g_1\left(\eps\right) x & 0 & g_2\left(\eps\right) x & 6 \eps^3 \left(1-2\eps\right) x^2 & 6 \eps^4 x \left(1+x\right) & 0 & 2 \eps^4 x^2 \\
 \end{array} \right),
 \nonumber
\eq
with
\bq
\label{chapter_iterated_integrals:def_g1_g2}
 g_1\left(\eps\right) & = & - 3 \eps \left(1-2\eps\right) \left(1-3\eps\right) \left(2-3\eps\right), 
 \nonumber \\
 g_2\left(\eps\right) & = & - 3 \eps^2 \left(1-2\eps\right) \left(1-3\eps\right).
\eq
This transforms the differential equation into $\eps$-form. We have
\bq
\label{chapter_iterated_integrals:example_dgl_eps_form_double_box}
 A' & = & 
 \eps \left( C_0 \frac{dx}{x} + C_{-1} \frac{dx}{x+1} \right),
\eq
with
{\footnotesize
\bq
 C_0
 =
 \left( \begin{array}{rrrrrrrr} 
 2 & 0 & 0 & 0 & 0 & 0 & 0 & 0 \\
 0 & 0 & 0 & 0 & 0 & 0 & 0 & 0 \\
 0 & 0 & 2 & 0 & 0 & 0 & 0 & 0 \\
 0 & 0 & 0 & 2 & 0 & 0 & 0 & 0 \\
 0 & 0 & 0 & 1 & 1 & 0 & 0 & 0 \\
 1 & -1 & 0 & 0 & 0 & 2 & 0 & 0 \\
 0 & 0 & 0 & 0 & 0 & 0 & 0 & -1 \\
 0 & 0 & -4 & 0 & 0 & 0 & 4 & 4 \\
 \end{array} \right),
 & &
 C_{-1}
 =
 \left( \begin{array}{rrrrrrrr} 
 0 & 0 & 0 & 0 & 0 & 0 & 0 & 0 \\
 0 & 0 & 0 & 0 & 0 & 0 & 0 & 0 \\
 0 & 0 & 0 & 0 & 0 & 0 & 0 & 0 \\
 0 & 0 & 0 & 0 & 0 & 0 & 0 & 0 \\
 0 & 2 & 0 & -1 & -1 & 0 & 0 & 0 \\
 0 & 0 & 0 & 0 & 0 & -2 & 0 & 0 \\
 -2 & -2 & 0 & 0 & 2 & 4 & 0 & 1 \\
 2 & -2 & 0 & 2 & 0 & 4 & 0 & -1 \\
 \end{array} \right).
 \nonumber
\eq
}

As our last example let's consider example 1 from section~\ref{chapter_iterated_integrals:deriving_the_dgl}.
This is the family of the one-loop two-point function with equal internal masses.
You might be tempted to consider this example to be the easiest example among the three examples
discussed in section~\ref{chapter_iterated_integrals:deriving_the_dgl}, but we will soon see that this example
requires an additional transformation not discussed so far.
We start from the master integrals $\vec{I}=(I_{10}, I_{11})^T$ and the differential equation (see eq.~(\ref{chapter_iterated_integrals:example_A_oneloop_twopoint}))
\bq
\label{chapter_iterated_integrals:dgl_bubble_pre_canonical}
 \left(d+A\right) \vec{I} \; = \; 0,
 & &
 A \; = \; 
 \left(\begin{array}{cc}
 0 & 0 \\
 \frac{1-\eps}{2x} - \frac{1-\eps}{2\left(x+4\right)} & \frac{1}{2x} - \frac{1-2\eps}{2\left(x+4\right)} \\
 \end{array} \right) dx.
\eq
Let's first see if we can transform the differential equation by a fibre transformation such that $\eps$ only appears as a prefactor.
You are welcomed to play around and to convince yourself that there is no transformation rational in $x$ and $\eps$, which achieves
that.
However, if we allow the transformation to be algebraic, we may achieve this goal.
Let's consider the transformation
\bq
\label{chapter_iterated_integrals:trafo_bubble}
 \vec{I}'
 \; = \; U \vec{I},
 & &
 U \; = \;
 \left(\begin{array}{cc}
 2 \eps \left(1-\eps\right) & 0 \\
 2 \eps \left(1-\eps\right) \sqrt{\frac{x}{4+x}} & 2 \eps \left(1-2\eps\right) \sqrt{\frac{x}{4+x}} \\
 \end{array} \right).
\eq
For the transformed system we find
\bq
\label{chapter_iterated_integrals:example_1_eps_form_square_root_singularity}
 \left(d+A'\right) \vec{I}' \; = \; 0,
 & &
 A' \; = \; 
 \eps
 \left(\begin{array}{cc}
 0 & 0 \\
 - \frac{dx}{\sqrt{x\left(4+x\right)}} & \frac{dx}{4+x} \\
 \end{array} \right).
\eq
We have achieved that $\eps$ only appears as a prefactor,
however the condition that the only singularities are simple poles is not met:
The differential one-form
\bq
\label{chapter_iterated_integrals:square_root_singularity}
 \frac{dx}{\sqrt{x\left(4+x\right)}}
\eq
has square root singularities at $x=0$ and $x=-4$.
In the next section we will learn how to transform these away.
As a side remark let us note that we may force eq.~(\ref{chapter_iterated_integrals:square_root_singularity}) 
into a dlog-form:
\bq
\label{chapter_iterated_integrals:example_dlog_square_root}
 \frac{dx}{\sqrt{x\left(4+x\right)}}
 & = &
 d \ln\left(2+x+\sqrt{x\left(4+x\right)} \right).
\eq
We see that in this case the argument of the logarithm is no longer a polynomial, but an algebraic function of $x$.

\subsection{Base transformations}
\label{chapter_iterated_integrals:base_transformation}

In this section we consider coordinate transformation on the base manifold.
Whereas the fibre transformations discussed in the previous section are like gauge transformation in gauge theories,
the coordinate transformations on the base manifold discussed in this section 
are like coordinate transformation in general relativity.

On the base manifold $M$ we perform a change of coordinates: We go from old coordinates $x_1, \dots, x_{\NB}$ 
to new coordinates $x_1', \dots, x_{\NB}'$.
Let's assume that the new coordinates are given in terms of the old coordinates as
\bq
 x_i' & = & f_i\left(x\right), \;\;\;\;\;\;\;\;\; 1 \le i \le \NB.
\eq
If the matrix-valued differential one-form is written in terms of the old coordinates as
\bq
 A & = & \sum\limits_{i=1}^{\NB} A_i dx_i,
\eq
and in terms of the new coordinates as 
\bq
 A & = & \sum\limits_{i=1}^{\NB} A_i' dx_i',
\eq
then $A_i'$ and $A_j$ are related by
\bq
 A_i' & = & \sum\limits_{j}^{\NB} A_j \; \frac{\partial x_j}{\partial x_i'}.
\eq
\begin{tcolorbox}
{\bf Base transformation}:
\\
Let $\vec{I}=(I_{{\bm{\nu}}_1}, \dots, I_{{\bm{\nu}}_{\Nmaster}})^T$
be a set of master integrals satisfying the differential equation
\bq
 \left(d+A\right) \vec{I} & = & 0,
 \;\;\;\;\;\;
 A \; = \; \sum\limits_{i=1}^{\NB} A_i dx_i,
\eq
If we change the coordinates on the base manifold $M$ according to
\bq
 x_i' & = & f_i\left(x\right), \;\;\;\;\;\;\;\;\; 1 \le i \le \NB,
\eq
and write $A$ in terms of the new coordinates as
\bq
 A & = & \sum\limits_{i=1}^{\NB} A_i' dx_i',
\eq
then $A_i'$ and $A_j$ are related by
\bq
 A_i' & = & \sum\limits_{j}^{\NB} A_j \; \frac{\partial x_j}{\partial x_i'}.
\eq
\end{tcolorbox}
Let see how this works in an example:
We continue with example 1 from section~\ref{chapter_iterated_integrals:deriving_the_dgl}.
In the basis $\vec{I}'$ we had
\bq
 \left(d+A'\right) \vec{I}' \; = \; 0,
 & &
 A' \; = \; 
 \eps
 \left(\begin{array}{cc}
 0 & 0 \\
 - \frac{dx}{\sqrt{x\left(4+x\right)}} & \frac{dx}{4+x} \\
 \end{array} \right).
\eq
Let's define $x'$ by
\bq
\label{chapter_iterated_integrals:example_rationalisation}
 x & = & \frac{\left(1-x'\right)^2}{x'}.
\eq
The inverse relation reads
\bq
 x' & = & \frac{1}{2} \left( 2 + x - \sqrt{x\left(4+x\right)} \right),
\eq
where we made a choice for the sign of the square root.
We have
\bq
 \frac{\partial x}{\partial x'} 
 & = & 
 - \frac{\left(1-x'{}^2\right)}{x'^2}
\eq
and
\bq
 \frac{dx}{\sqrt{x\left(4+x\right)}}
 \; = \; - \frac{dx'}{x'},
 & &
 \frac{dx}{4+x}
 \; = \;
 \frac{2 dx'}{x'+1} - \frac{dx'}{x'}.
\eq
Thus in term of the new variable $x'$ we have
\bq
 A' & = &  
 \eps
 \left(\begin{array}{cc}
 0 & 0 \\
 \frac{dx'}{x'} & \frac{2 dx'}{x'+1} - \frac{dx'}{x'} \\
 \end{array} \right).
\eq
The differential equation is now in $\eps$-form: The dimensional regularisation parameter occurs only as a prefactor and
the only singularities of $A'$ are simple poles.
For the case at hand, $A'$ has simple poles at $x'=0$ and $x'=-1$.
 
\section{Cuts of Feynman integrals}
\label{chapter_iterated_integrals:cuts}

Up to now we focused entirely on Feynman integrals, given in the momentum representation by
\bq
 I_{\nu_1 \dots \nu_{\ninternal}}\left(D,x_1,\dots,x_{\NB}\right)
 & = &
 e^{\loopnumber \eps \Eulerconstant} \left(\mu^2\right)^{\nu-\frac{\loopnumber D}{2}}
 \int \prod\limits_{r=1}^{\loopnumber} \frac{d^Dk_r}{i \pi^{\frac{D}{2}}} 
 \prod\limits_{j=1}^{\ninternal} \frac{1}{\left(-q_j^2+m_j^2\right)^{\nu_j}}.
\eq
The integration contour is along the real axes with deformations into the complex domain dictated by
Feynman's $i\delta$-prescription.
The Baikov representation of the Feynman integral reads
\bq
\label{chapter_iterated_integrals:baikov_representation_no_cut}
 I_{\nu_1 \dots \nu_{n}}\left(D,x_1,\dots,x_{\NB}\right) & = &
 C_{\mathrm{pre}}
 \int\limits_{\mathcal C} d^{\NV}z \;
 \left[{\mathcal B}\left(z\right)\right]^{\frac{D-\loopnumber-{\nexternalindependent}-1}{2}}
 \prod\limits_{s=1}^{\NV} z_s^{-\nu_s},
\eq
where the prefactor $C_{\mathrm{pre}}$ is given by
\bq
 C_{\mathrm{pre}} & = &
 \frac{e^{\loopnumber \eps \Eulerconstant} \left(\mu^2\right)^{\nu-\frac{\loopnumber D}{2}} \left[ \det G\left(p_1,...,p_{\nexternalindependent}\right)\right]^{\frac{-D+{\nexternalindependent}+1}{2}}}{\pi^{\frac{1}{2}\left(\NV-\loopnumber\right)} \left( \det C \right) \prod\limits_{j=1}^{\loopnumber} \Gamma\left(\frac{D-\nexternalindependent+1-j}{2}\right)}.
\eq
${\mathcal B}(z)$ denotes the Baikov polynomial and
the Baikov variables are given by $z_j=-q_j^2+m_j^2$.
The domain of integration ${\mathcal C}$ is defined by 
eq.~(\ref{chapter_basics:Baikov_domain_1}) and eq.~(\ref{chapter_basics:Baikov_domain_2}).

In this section we enlarge the set of integrals we are interested in 
and include integrals, which have the same integrands as in eq.~(\ref{chapter_iterated_integrals:baikov_representation_no_cut}),
but are integrated over a different domain.
The new integration domains are not completely arbitrary, but should satisfy the following requirements:
\begin{enumerate}
\item Integration-by-parts identities still hold.
\item The variation of the integral with respect to the kinematic variables comes entirely from the integrand.
\item The symmetries among the integrals are respected.
\end{enumerate}
Condition 1 says that in the language of differential forms
\bq 
 \int\limits_{{\mathcal C}} d\xi \; = \; 
 \int\limits_{\partial {\mathcal C}} \xi
 \; = \; 0,
\eq
either because ${\mathcal C}$ is a cycle and therefore $\partial {\mathcal C}=0$
or because $\xi$ vanishes on $\partial {\mathcal C}$.
Condition 2 is best explained with an example.
Consider 
\bq
 \omega & = & \frac{x}{z-x} dz
\eq
and ${\mathcal C}$ a circle around $z=x$ of radius $\delta$ oriented anti-clockwise.
Then
\bq
 I \; = \; \int\limits_{{\mathcal C}} \omega
 \; = \; 
 2 \pi i x,
 & \mbox{and} &
 \frac{d}{dx} I \; = \; 2 \pi i.
\eq
On the other hand, we have
\bq
 \frac{d}{dx} \omega
 & = &
 \frac{dz}{z-x}
 - x \left( \frac{d}{dz} \frac{1}{\left(z-x\right)} \right) dz,
\eq
and
\bq
 \int\limits_{{\mathcal C}} \frac{dz}{z-x} \; = \; 2 \pi i,
 & &
 - x \int\limits_{{\mathcal C}} \left( \frac{d}{dz} \frac{1}{\left(z-x\right)} \right) dz \; = \; 0,
\eq
in agreement with our previous result.
Although the integration contour moves with $x$ (it is a circle of radius $\delta$ around $z=x$), 
in a small neighbourhood of $x=x_0$ (say of radius $\delta' \ll \delta$) we may deform the integration contour
to a constant integration contour (for example for $x \in [x_0-\delta',x_0+\delta']$ we may deform 
the integration contour
to a circle around $z=x_0$ of radius $\delta$, independently of $x$).
Hence, the derivative of the integral is calculated from the derivative of the integrand.

Also condition 3 is best explained by an example: For the one-loop two-point function with equal internal masses
we have the symmetry $I_{\nu_1 \nu_2}=I_{\nu_2 \nu_1}$.
Changing the integration contour may break this symmetry.
This can be restored by considering a suitable symmetrised contour.
We will require condition 3 in this section, 
but we will dispense ourselves from condition 3 in the next section.

Let us now consider a set of Feynman master integrals $\vec{I}$, satisfying the differential equation
\bq
 \left(d + A \right) \vec{I} & = & 0.
\eq
We now consider a set of new integrals, where the original integration contour ${\mathcal C}$
is replaced by a new integration contour ${\mathcal C}'$, which satisfies conditions 1-3.
Let us denote the new integrals by $\vec{I}'$.
Conditions 1-3 are sufficient to show that the new integrals $\vec{I}'$ satisfy the same differential equation
as the old integrals $\vec{I}$ \cite{Primo:2016ebd}:
\bq
 \left(d + A \right) \vec{I}' & = & 0.
\eq
The proof is rather simple: The differential equations rely only on the forms of the integrands (which are the
same for $\vec{I}$ and $\vec{I}'$) and the properties given by conditions 1-3.

Let us now come to the topic of this section: 
\index{cut}
{\bf Feynman integrals with cuts}.
We define a cut Feynman integral as a special case of the general situation discussed above.
Let us consider the case, where we cut the internal edge $e_j$.
We define the Feynman integral 
with the internal edge $e_j$ 
cut to be given by the Baikov 
representation, where the domain of integration ${\mathcal C}$ is replaced by a modified domain of integration
${\mathcal C}'$ \cite{Frellesvig:2017aai,Bosma:2017ens,Harley:2017qut}.
The modified domain ${\mathcal C}'$ consists of a small anti-clockwise circle around $z_j=0$ in the complex $z_j$-plane.
In all other variables the domain of integration is given by equations similar to 
eq.~(\ref{chapter_basics:Baikov_domain_1}) and eq.~(\ref{chapter_basics:Baikov_domain_2}),
with ${\mathcal B}$ replaced by
\bq
 {\mathcal B}_j\left(z_1,\dots,z_{j-1},z_{j+1},\dots,z_{\NV}\right)
 & = &
 {\mathcal B}\left(z_1,\dots,z_{j-1},0,z_{j+1},\dots,z_{\NV}\right),
\eq
or shortly
\bq
 {\mathcal B}_j
 & = &
 \left. {\mathcal B} \right|_{z_j=0}.
\eq
This is just the intersection of the original integration domain ${\mathcal C}$ with the hyperplane $z_j=0$.

Thus we see that the cut integral is given by an integrand, which is $(2\pi i)$ times the residue at $z_j=0$
of the original integrand, integrated over the remaining variables.
\begin{figure}
\begin{center}
\includegraphics[scale=1.0]{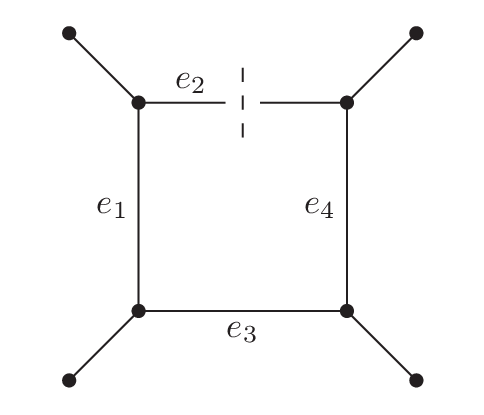}
\end{center}
\caption{
A Feynman graph with a cut.
The edge $e_2$ is cut.
}
\label{chapter_iterated_integrals:cut_graph}
\end{figure}
We draw a cut graph as shown in fig.~\ref{chapter_iterated_integrals:cut_graph}.

The name ``cut Feynman integral'' stems from the fact that
if a propagator occurs only to power one, cutting this propagators corresponds 
in the momentum representation to the replacement
\bq
\label{chapter_iterated_integrals:cut_replacement}
 \frac{1}{-q_j^2 + m_j^2}
 & \rightarrow &
 2 \pi i \; \delta\left(-q_j^2 + m_j^2\right).
\eq
The $\delta$-distribution forces the propagator on-shell.
It is easily seen that for the case where a propagator occurs only to power one
eq.~(\ref{chapter_iterated_integrals:cut_replacement}) is equivalent to taking the residue.
In the Baikov variables eq.~(\ref{chapter_iterated_integrals:cut_replacement}) reads
\bq
 \frac{1}{z_j}
 & \rightarrow &
 2 \pi i \; \delta\left(z_j\right).
\eq
We have for a small anti-clockwise circle $\gamma_j$ around $z_j=0$ and $f(z_j)$ regular at $z_j=0$:
\bq
 2 \pi i \int\limits dz_j f\left(z_j\right) \delta\left(z_j\right)
 \; = \; 
 2 \pi i f\left(0\right)
 \; = \; 
 2 \pi i \; \mathrm{res}\left( \frac{f\left(z_j\right)}{z_j}, z_j=0\right)
 \; = \; 
 \oint\limits_{\gamma_j} dz_j \frac{f\left(z_j\right)}{z_j}.
\eq
Of course we may iterate the procedure and take multiple cuts.
Of particular importance is the 
\index{maximal cut}
{\bf maximal cut}, 
where 
we take for a Feynman integral $I_{\nu_1 \dots \nu_{\ninternal}}$ the cut for all edges
$e_j$ for which $\nu_j>0$.

Let's look at an example:
We consider the one-loop two-point function with equal internal masses 
discussed as example 1 in section~\ref{chapter_iterated_integrals:deriving_the_dgl}.
With $x=-p^2/m^2$ and $\mu^2=m^2=1$ the Baikov polynomial
is given by
\bq
 {\mathcal B}\left(z_1,z_2\right)
 & = &
 - \frac{1}{4}
 \left[ \left(z_1-z_2\right)^2 - 2 x \left(z_1+z_2\right) + x \left(4+x\right)
 \right],
\eq
and the Baikov representation of $I_{11}$ is given by
\bq
 I_{11}
 & = &
 \frac{e^{\eps \Eulerconstant} x^{-\frac{D-2}{2}}}{2 \sqrt{\pi} \Gamma\left(\frac{D-1}{2}\right)}
 \int\limits_{\mathcal C} d^2z \;
 \left[ {\mathcal B}\left(z_1,z_2\right) \right]^{\frac{D-3}{2}} \frac{1}{z_1 z_2}.
\eq
The cut of edge $e_1$ is given by
\bq
 \mathrm{Cut}_{e_1}
 I_{11}
 & = &
 \left( 2 \pi i \right)
 \frac{e^{\eps \Eulerconstant} x^{-\frac{D-2}{2}}}{2 \sqrt{\pi} \Gamma\left(\frac{D-1}{2}\right)}
 \int\limits_{\mathcal C'} dz_2 \;
 \left[ {\mathcal B}\left(0,z_2\right) \right]^{\frac{D-3}{2}} \frac{1}{z_2}.
\eq
We have
\bq
 {\mathcal B}\left(0,z_2\right)
 & = &
 - \frac{1}{4} \left[ z_2 - \left(x - 2 \sqrt{-x}\right) \right] \left[ z_2 - \left(x + 2 \sqrt{-x}\right) \right].
\eq
Let assume for the moment that $x<-4$. Then the integration domain ${\mathcal C'}$ is from $(x - 2 \sqrt{-x})$ to $(x + 2 \sqrt{-x})$
and $0 \notin [x - 2 \sqrt{-x},x + 2 \sqrt{-x}]$.
If $x \not < -4$
we may first pretend that $x < -4$, perform the integration and then continue
analytically to the desired value of $x$.

In a similar way, the cut of the edge $e_2$ is given by
\bq
 \mathrm{Cut}_{e_2}
 I_{11}
 & = &
 \left( 2 \pi i \right)
 \frac{e^{\eps \Eulerconstant} x^{-\frac{D-2}{2}}}{2 \sqrt{\pi} \Gamma\left(\frac{D-1}{2}\right)}
 \int\limits_{\mathcal C''} dz_1 \;
 \left[ {\mathcal B}\left(z_1,0\right) \right]^{\frac{D-3}{2}} \frac{1}{z_1}.
\eq
Cutting the edges $e_1$ and $e_2$ gives the maximal cut.
We have
\bq
 \mathrm{MaxCut} \; I_{11}
 \; = \; 
 \mathrm{Cut}_{e_1,e_2}
 I_{11}
 & = &
 \left( 2 \pi i \right)^2
 \frac{e^{\eps \Eulerconstant} x^{-\frac{D-2}{2}}}{2 \sqrt{\pi} \Gamma\left(\frac{D-1}{2}\right)}
 \left[ {\mathcal B}\left(0,0\right) \right]^{\frac{D-3}{2}}
 \nonumber \\
 & = &
 \left( 2 \pi i \right)^2
 \frac{e^{\eps \Eulerconstant} x^{-\frac{D-2}{2}}}{2 \sqrt{\pi} \Gamma\left(\frac{D-1}{2}\right)}
 \left( - \frac{1}{4} x \left(4+x\right) \right)^{\frac{D-3}{2}}.
\eq
In $D=2-2\eps$ dimensions we have to leading order in the $\eps$-expansion
\bq
 \mathrm{MaxCut} \; I_{11}\left(2-2\eps\right)
 & = &
 - \frac{4 \pi}{\sqrt{-x\left(4+x\right)}}
 + {\mathcal O}\left(\eps\right).
\eq
In eq.~(\ref{chapter_iterated_integrals:trafo_bubble}) we found a fibre transformation, which 
puts the differential equation into a form, where the dimensional regularisation parameter $\eps$ appears only
as a prefactor.
The new master integral in the top sector was
\bq
\label{chapter_iterated_integrals:example_bubble_uniform_weight_4D}
 I_2'
 & = & 
 2 \eps \sqrt{\frac{x}{4+x}} 
 \left[ \left(1-\eps\right) I_{1 0}\left(4-2\eps\right)
      + \left(1-2\eps\right) I_{1 1}\left(4-2\eps\right) 
 \right].
\eq
With the help of the dimensional shift relations (see also exercise~\ref{chapter_iterated_integrals:exercise_dimensional_shift}), 
we may rewrite this expression as
\bq
\label{chapter_iterated_integrals:example_bubble_uniform_weight_2D}
 I_2'
 & = & 
 - \eps \sqrt{x\left(4+x\right)} I_{1 1}\left(2-2\eps\right).
\eq
We see that up to a constant prefactor $I_2'$ is $I_{1 1}\left(2-2\eps\right)$ divided by the leading
term in the $\eps$-expansion of its maximal cut:
\bq
 I_2'
 & = &
 \frac{4 \pi \eps}{\sqrt{-1}} \; \frac{I_{11}\left(2-2\eps\right)}{\mathrm{MaxCut} \; I_{11}\left(2\right)}.
\eq
The cut of the edge $e_j$ of a Feynman integral where the corresponding propagator occurs to a higher power 
(i.e. $\nu_j > 1$)
is obtained by first 
expanding the integrand of the Baikov representation as a Laurent series in the corresponding Baikov variable
and by determining the residue as the coefficient of the $1/z_j$-term.
For example
\bq
 \mathrm{Cut}_{e_1}
 I_{21}
 & = &
 \left(2 \pi i\right)
 \frac{e^{\eps \Eulerconstant} x^{-\frac{D-2}{2}}}{2 \sqrt{\pi} \Gamma\left(\frac{D-3}{2}\right)}
 \int\limits_{\mathcal C'} dz_2 \;
 \left[ {\mathcal B}\left(0,z_2\right) \right]^{\frac{D-5}{2}} \frac{{\mathcal B}'\left(0,z_2\right)}{z_2},
\eq
with 
\bq
 {\mathcal B}'\left(0,z_2\right)
 & = & 
 \left. \frac{\partial}{\partial z_1} {\mathcal B}\left(z_1,z_2\right) \right|_{z_1=0}
 \; = \; 
 \frac{1}{2} \left( z_2+x \right).
\eq
The cut of the edge $e_j$ of a Feynman integral where the corresponding propagator occurs to power $\nu_j \le 0$ is zero,
as there is no residue in this variable.
For example
\bq
 \mathrm{Cut}_{e_1} I_{01}
 \; = \;
 \mathrm{Cut}_{e_1} I_{\left(-1\right)1}
 \; = \;
 0.
\eq
This reveals the power of considering cuts: 
A Feynman integral with $\nu_j \le 0$ corresponds to a
sub-topology, where the edge $e_j$ is pinched.
The corresponding cut integral is zero.
Let us now consider the differential equation for a set of master integrals $\vec{I}$.
Replacing the original integration contour by a contour with a cut in the edge $e_j$ has the effect of 
setting all sub-topologies with edge $e_j$ pinched to zero.
The resulting differential equation is simpler.
In particular, the maximal cut will set all sub-topologies to zero.
\\
\\
\bs
{\it \refstepcounter{exercise}
{\bf Exercise \theexercise}: 
Work out the maximal cut of the double box integral $I_{111111100}$ shown 
\begin{figure}
\begin{center}
\includegraphics[scale=1.0]{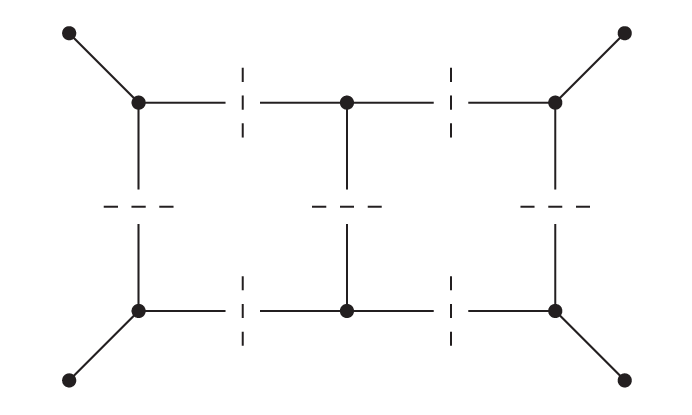}
\end{center}
\caption{
The maximal cut of the double box graph.
}
\label{chapter_iterated_integrals:maximal_cut_graph}
\end{figure}
in fig.~\ref{chapter_iterated_integrals:maximal_cut_graph}.
Use the notation as in example 2 in section~\ref{chapter_iterated_integrals:deriving_the_dgl}.
To work out the maximal cut it is simpler to use the loop-by-loop approach as discussed in 
section~\ref{chapter_basics:sect_Baikov_representation}.
}
\es

\section{Singularities of Feynman integrals}
\label{chapter_iterated_integrals:singularities}

In this section we discuss singularities of Feynman integrals.
There are two aspects to it. 
First there may be singularities, which occur for any values of the kinematic variables.
These are the 
\index{ultraviolet divergence}
{\bf ultraviolet} or 
\index{infrared divergence}
{\bf infrared divergences} of Feynman integrals.
Secondly, there may be singularities, which only occur for specific values of the kinematic variables.
These are called 
\index{Landau singularities}
{\bf Landau singularities}.
A classic textbook on this subject is \cite{Eden}.

The singularities of Feynman integrals are most easily discussed within the Feynman parameter representation.
We denote the kinematic variables by $(x_1,\dots,x_{\NB})$.
We let $U$ be an open subset of ${\mathbb C} {\mathbb P}^{\NB}$ such that
\bq
 (x_1,\dots,x_{\NB}) & \in & U \; \subset \; {\mathbb C} {\mathbb P}^{\NB}.
\eq
The Feynman parameters $[a_1:\dots:a_{\ninternal}]$ denote a point in ${\mathbb C} {\mathbb P}^{\ninternal-1}$
\bq
 [a_1:\dots:a_{\ninternal}] & \in & {\mathbb C} {\mathbb P}^{\ninternal-1}.
\eq
Let us first discuss ultraviolet and infrared singularities.
These manifest themselves as poles in the dimensional regularisation parameter $\eps$ after integration.
Let us assume that $U$ is a subset of the Euclidean region, hence $x_j \ge 0$.
This assumption will simplify the discussion below.

From the Feynman parameter integral in eq.~(\ref{chapter_basics:Feynman_parameter_representation}) we see that
there are three possibilities how poles in $\eps$ can arise:
First of all the gamma function $\Gamma(\nu-\loopnumber D/2)$ of the prefactor can give rise to 
a (single) pole if the argument of this function is close to zero or to a negative integer
value. This divergence is called the 
\index{overall ultraviolet divergence}
{\bf overall ultraviolet divergence}.

Secondly, we consider the polynomial ${\mathcal U}$. 
Depending on the exponent $\nu-(\loopnumber+1)D/2$ of ${\mathcal U}$ 
the vanishing of the polynomial ${\mathcal U}$ 
in some part of the integration region can lead to poles in $\eps$ after integration.
As mentioned in section~\ref{chapter_basics:subsection:Schwinger_parameter}, 
each term of the expanded form of the polynomial ${\mathcal U}$ 
has coefficient $+1$, therefore ${\mathcal U}$ can only vanish if some of the Feynman
parameters are equal to zero.
In other words, ${\mathcal U}$ is non-zero (and positive) inside the integration region, but
may vanish on the boundary of the integration region.
Poles in $\eps$ resulting from the vanishing of ${\mathcal U}$ are related to
\index{ultraviolet sub-divergence}
{\bf ultraviolet sub-divergences}.

Thirdly, we consider the polynomial ${\mathcal F}$. 
In the Euclidean region the polynomial ${\mathcal F}$ 
is also non-zero (and positive) inside the integration region.
Therefore if all kinematic variables are within the Euclidean region
the polynomial ${\mathcal F}$ can only vanish on the boundary of the integration region, 
similar to what has been observed for the the polynomial ${\mathcal U}$.
Depending on the exponent $\nu-\loopnumber D/2$ of ${\mathcal F}$
the vanishing of the polynomial ${\mathcal F}$ on the boundary of
the integration region may lead to poles in $\eps$ after integration.
These poles are related to 
\index{infrared divergence}
{\bf infrared divergences}.

Let us now turn to Landau singularities.
We consider a Feynman integral as a function of the kinematic variables.
We no longer impose any restrictions (like for example within the Euclidean region)
on the kinematic variables.
In particular, the kinematic variables are
now allowed to lie in the physical region. 
Landau's equations give a necessary condition for a singularity to occur in the Feynman integral as
we vary the kinematic variables.

The Feynman integral $I$ as given in eq.~(\ref{chapter_basics:Feynman_parameter_representation})
depends through the polynomial ${\mathcal F}$ on the kinematic variables $x_j$.
As we no longer restrict the kinematic variables $x_j$ to the Euclidean region,
the region where the polynomial ${\mathcal F}$ vanishes is no longer restricted
to the boundary of the Feynman parameter integration region
and we may encounter zeros of the polynomial ${\mathcal F}$ inside the integration region for specific values of the kinematic variables.
The vanishing of ${\mathcal F}$ may in turn result in singularities after integration for specific values of the kinematic variables.
These singularities are called Landau singularities.
Necessary conditions for the occurrence of a Landau singularity are given as follows:
A Landau singularity may occur if there exists a subset $S$ of $\{1,\dots,\ninternal\}$
such that
\bq
\label{chapter_iterated_integrals:Landau_condition}
 a_i & = & 0
 \;\;\; \mbox{for} \; i \in S
 \nonumber \\
 \mbox{and} \;\;\;\;
 \frac{\partial}{\partial a_j} {\mathcal F} & = & 0
 \;\;\; \mbox{for} \; j \in \{1,...,\ninternal\}\backslash S.
\eq
The equations~\ref{chapter_iterated_integrals:Landau_condition} are called 
{\bf Landau equations}.
\\
\\
\bs
{\it \refstepcounter{exercise}
{\bf Exercise \theexercise}: 
Show that the Landau equations imply ${\mathcal F}=0$.
}
\es
\\
\\
The case corresponding to $S = \emptyset$ is called the 
\index{leading Landau singularity}
{\bf leading Landau singularity}, and cases
corresponding to $S \neq \emptyset$ are called 
\index{non-leading Landau singularity}
{\bf non-leading Landau singularities}.
It is sufficient to focus on the leading Landau singularity, since a non-leading singularity
is the leading Landau singularity of a sub-graph of $G$ obtained by contracting the propagators
corresponding to the Feynman parameters $a_i$ with $i \in S$.

Let us now consider the leading Landau singularity of a graph $G$ with $\nexternal$ external lines.
We set
\bq
 A & = & {\mathbb C} {\mathbb P}^{\ninternal-1} \backslash V\left( a_1 \cdot \ldots \cdot a_{\ninternal} {\mathcal U} \right),
\eq
where
\bq
 V\left(f\right)
 & = &
 \left\{
  \; a \in {\mathbb C} {\mathbb P}^{\ninternal-1} \; | \; f\left(a\right) = 0 \;
 \right\}.
\eq
This takes out the regions in Feynman parameter space where we may have sub-leading Landau singularities or
ultraviolet sub-divergences.
Note that if we restrict the Feynman parameters to ${\mathbb R} {\mathbb P}^{\ninternal-1}_{\ge 0}$
we already know that the first graph polynomial can only vanish on the boundary, hence
\bq 
 V\left({\mathcal U}\right) \cap {\mathbb R} {\mathbb P}^{\ninternal-1}_{\ge 0}
 & \subset & 
 V\left( a_1 \cdot \ldots \cdot a_{\ninternal} \right) \cap {\mathbb R} {\mathbb P}^{\ninternal-1}_{\ge 0}.
\eq
Let us consider
\bq
 Y & = &
 \left\{ \, \left(a,x\right) \in A \times U \; | \; \frac{\partial}{\partial a_j} {\mathcal F}=0, \, j=1,\dots,\ninternal \, \right\}
\eq
together with the projection
\bq
 \pi & : & Y \rightarrow U,
 \nonumber \\
 & & \left(a,x\right) \rightarrow x.
\eq
The 
\index{Landau discriminant}
{\bf Landau discriminant} $D_{\mathrm{Landau}}$ 
is defined as the Zariski closure
\bq
 D_{\mathrm{Landau}} & = &
 \overline{\pi\left(Y\right)} \; \subset \; U.
\eq
Essentially, this corresponds to all points $x \in U$, where the equations $\partial {\mathcal F}/\partial a_j=0$ have a solution
in the space $A$.
A computer program to compute the Landau discriminant is described in \cite{Mizera:2021icv}.

Let us look at a simple example.
We consider the one-loop two-point Feynman integral with equal internal masses in $D=4-2\eps$
space-time dimensions (example 1 in section~\ref{chapter_iterated_integrals:deriving_the_dgl}).
This integral is given by
\bq
 I_{11}\left(D,x\right) & = &
 e^{\eps \Eulerconstant}
 \left( m^2 \right)^{\eps}
 \int \frac{d^Dk}{i\pi^{\frac{D}{2}}}\;
 \frac{1}{(-k^2+m^2)(-(k-p)^2+m^2)},
\eq
where we set $\mu^2=m^2$ and
$x=-p^2/m^2$. The second graph polynomial is given by
\bq 
 {\mathcal F}
 & = &
 a_1 a_2 x + \left(a_1+a_2\right)^2.
\eq
The Landau equations for the leading Landau singularities are
\bq
 a_2 x + 2 \left(a_1+a_2\right) \; = \; 0,
 & & 
 a_1 x + 2 \left(a_1+a_2\right) \; = \; 0.
\eq
These equations have a solution with $a_1 \neq 0$ and $a_2 \neq 0$ for
$x \in \left\{ -4, 0 \right\}$,
hence
\bq
 D_{\mathrm{Landau}} & = & \left\{ -4, 0 \right\}.
\eq
The Feynman parameter representation for this integral reads
\bq
 I_{11}\left(4-2\eps,x\right) & = &
 e^{\eps \Eulerconstant}
 \Gamma(\eps) \int\limits_0^1 da_1 \; 
  \left[ 1 + a_1 \left(1-a_1\right) x \right]^{-\eps}.
\eq
Working out this integral we find
\bq 
 I_{11}\left(4-2\eps,x\right)
 & = & 
 \frac{1}{\eps} + 2 
 + \sqrt{\frac{4+x}{x}} \ln \frac{\sqrt{4+x}-\sqrt{x}}{\sqrt{4+x}+\sqrt{x}}
 + {\mathcal O}(\eps).
\eq
The $1/\eps$-term corresponds to an ultraviolet divergence.
As a function of $x$ the Feynman integral has a Landau singularity at $x=-4$ (corresponding to $p^2=4m^2$).
Note that the Feynman integral is finite at $x=0$ and we see that Landau's equations give only a necessary condition
for a Landau singularity, but not a sufficient condition.
The Landau singularity at $x=-4$ is called a 
\index{threshold singularity}
normal 
{\bf threshold singularity}.
The normal threshold manifests itself as a branch point in the complex $x$-plane.
\\
\\
\bs
{\it \refstepcounter{exercise}
{\bf Exercise \theexercise}: 
Work out the Landau discriminant for the double box graph
discussed in exercise~\ref{chapter_iterated_integrals:exercise_doublebox_ibp}.
}
\es

\section{Twisted cohomology}
\label{chapter_iterated_integrals:twisted_cohomology}

We have seen that we can express any Feynman integral
$I_{\nu_1 \dots \nu_{\ninternal}}$ from a family of Feynman integrals 
as a linear combination of master integrals
$I_{{\bm{\nu}}_1}, \dots, I_{{\bm{\nu}}_{\Nmaster}}$.
The coefficients can be obtained by solving a linear system of equations.
The equations themselves are symmetry relations and integration-by-parts identities.
The system of linear equations can systematically be solved with the help of the Laporta algorithm.
There is no principal problem in obtaining the coefficients, after all this is just linear algebra.
However, there is a practical problem: Feynman integrals for cutting-edge precision calculations 
often lead to linear systems which barely can be treated with current computing resources.

We are therefore interested in alternative and more efficient methods to compute the coefficients.
We have seen that the master integrals span a vector space 
and any Feynman integral from the family of Feynman integrals which we are investigating corresponds 
to a specific vector in this vector space.
Writing this Feynman integral as a linear combination of master integrals is nothing else 
than expressing an arbitrary vector 
as a linear combination of basis vectors.
Finding the coefficients is particular easy if the vector space is equipped with an inner product.

This leads directly to the question: Is there an inner product on the vector space of Feynman integrals?
We will see in this section that the answer is almost yes, however we will not work with Feynman integrals, 
but with the integrands of Feynman integrals.
As discussed in section~\ref{chapter_iterated_integrals:integration_by_parts}, the difference is that for Feynman
integrals we take integration-by-parts identities and symmetry relations into account, while for 
the integrands of Feynman integrals we only take integration-by-parts identities into account.
We denote the number of master integrals 
obtained by taking
integration-by-parts identities and symmetries into account by
$\Nmaster$.
If we are only interested in the number of unreduced integrals 
obtained from
integration-by-parts identities alone, we denote this number by 
$\Ncohom$.
The simplest example is the one-loop two-point function with equal internal masses, where we have the symmetry
$I_{\nu 0} = I_{0 \nu}$.
In this case $\Nmaster=2$, but $\Ncohom=3$. The integrands of the two tadpole integrals 
$I_{1 0}$ and $I_{0 1}$ differ, but yield the same result after integration.

In this section we focus on the integrands of Feynman integrals, therefore 
$\Ncohom$ is the relevant quantity.
The mathematical setting is called twisted de Rham cohomology, which we now introduce.

Textbooks on twisted de Rham cohomology are the books by Yoshida \cite{Yoshida:book}
and Aomoto and Kita \cite{Aomoto:book}.
The application towards Feynman integrals started with \cite{Mastrolia:2018uzb,Frellesvig:2019uqt}.
Many examples are provided in \cite{Frellesvig:2019uqt,Frellesvig:2020qot}.
Review articles on this subject are \cite{Mizera:2019ose,Cacciatori:2021nli}.

\subsection{Twisted cocycles}

We start from the $n$-dimensional complex space $\mathbb{C}^n$ and a 
\index{divisor}
{\bf divisor} $D$, on which we will allow singularities.
Instead of $\mathbb{C}^n$ we will later also consider other spaces like $\mathbb{CP}^n$.
For our purpose it is sufficient to think of the divisor $D$ as a union of hypersurfaces, where each hypersurface
is defined by a polynomial equation.
In detail, consider $m$ polynomial equations 
\bq
 p_i(z_1,\dots,z_n) & = & 0,
 \;\;\;\;\;\;
 1 \le i \le m,
\eq
with $p_i \in \mathbb{F}\left[z_1,\dots,z_n\right]$,
where $\mathbb{F}$ is a field, typically $\mathbb{Q}$ or $\mathbb{Q}(x_1,\dots,x_m)$ (the field of rational
functions in $x_1, \dots, x_m$ with rational coefficients).
Each polynomial equation defines a hypersurface
\bq
 D_i \; = \; \{ \left(z_1,\dots,z_n\right) \in \mathbb{C}^n | p_i\left(z_1,\dots,z_n\right) = 0 \},
\eq
and $D$ is the union of the $m$ hypersurfaces:
\bq
 D \; = \; 
 \bigcup\limits_{i=1}^m D_i.
\eq
We consider rational differential $n$-forms $\varphi$ in the variables $z=(z_1,\dots,z_n)$, 
which are holomorphic on ${\mathbb C}^n - D$.
The rational $n$-forms $\varphi$ are of the form
\bq
\label{chapter_iterated_integrals:representative_left}
 \varphi
 & = &
 \frac{q}{p_1^{n_1} \dots p_m^{n_m}} \; dz_n \wedge \dots \wedge dz_1,
 \;\;\;\;\;\;\;\;\;
 q \in \mathbb{F}\left[z_1,\dots,z_n\right],
 \;\;\;
 n_i \in {\mathbb N}_0.
\eq
Using the reversed wedge product $dz_n \wedge \dots \wedge dz_1$ instead of the standard order $dz_1 \wedge \dots \wedge dz_n$
is at this stage just a convention.
$q(z_1,\dots,z_n)$ is a polynomial and the only singularities of $\varphi$ in ${\mathbb C}^n$ are on $D$.

In cohomology theory we call the differential $n$-form $\varphi$ a cocycle.
It is closed on ${\mathbb C}^n - D$, since it is a holomorphic $n$-form:
Obviously we have for $1 \le j \le n$
\bq
 \left( \frac{\partial}{\partial \bar{z}_j} \frac{q}{p_1^{n_1} \dots p_m^{n_m}} \right) \; d\bar{z}_j \wedge dz_n \wedge \dots \wedge dz_1 & = & 0,
\eq
since the derivative in the bracket vanishes,
but also
\bq
 \left( \frac{\partial}{\partial z_j} \frac{q}{p_1^{n_1} \dots p_m^{n_m}} \right) \; dz_j \wedge dz_n \wedge \dots \wedge dz_1 & = & 0,
\eq
since the wedge product contains $dz_j \wedge dz_j$.

Let ${\mathcal C}$ be a $n$-dimensional integration cycle (i.e. an integration domain with no boundary $\partial {\mathcal C}=0$).
We may now consider the integral
\bq
 \left\langle \varphi | {\mathcal C} \right\rangle
 & = &
 \int\limits_{\mathcal C} \varphi.
\eq
This is a pairing between a cycle and a cocycle.
The quantity $\langle \varphi | {\mathcal C} \rangle$ will not change if we add to $\varphi$ the exterior derivative of
a $(n-1)$-form $\xi$:
\bq
\label{chapter_iterated_integrals:invariance_d}
 \varphi & \rightarrow & \varphi + d\xi.
\eq
Due to $\partial {\mathcal C}=0$ and Stokes' theorem
\bq
 \int\limits_{\mathcal C} d\xi
 & = &
 \int\limits_{\partial \mathcal C} \xi
 \; = \; 0,
\eq
we have
\bq
 \left\langle \varphi + d\xi | {\mathcal C} \right\rangle
 & = &
 \left\langle \varphi | {\mathcal C} \right\rangle.
\eq
We call two $n$-forms $\varphi$ and $\varphi'$ equivalent, if they differ 
by the exterior derivative of a $(n-1)$-form $\xi$ as in eq.~(\ref{chapter_iterated_integrals:invariance_d}):
\bq
 \varphi'
 \sim
 \varphi
 & \;\; \Leftrightarrow \;\; &
 \varphi'
 \; = \;
  \varphi
  + d \xi.
\eq
The set of equivalence classes defines the (untwisted) de Rham cohomology group $H^n$.

Let us now introduce the 
\index{twist}
{\bf twist}:
For $m$ complex numbers $\gamma=(\gamma_1,\dots,\gamma_m)$ 
we set
\bq
\label{chapter_iterated_integrals:def_u}
 u
 & = &
 \prod\limits_{i=1}^m p_i^{\gamma_i}.
\eq
Since the exponents $\gamma_i$ of the polynomials $p_i$ are allowed to be complex numbers,
$u$ is in general a multi-valued function on ${\mathbb C}^n - D$.
It will be convenient to define
\bq
\label{chapter_iterated_integrals:def_omega}
 \omega
 & = &
 d \ln u
 \; = \;
 \sum\limits_{i=1}^m \gamma_i d\ln p_i
 \; = \;
 \sum\limits_{j=1}^n  \omega_j dz_j.
\eq
Let us fix a branch of $u$.
We then consider the integral
\bq
\label{chapter_iterated_integrals:def_integral}
 \left\langle \varphi | {\mathcal C} \right\rangle_{\omega}
 & = &
 \int\limits_{\mathcal C} u \; \varphi.
\eq
${\mathcal C}$ is again an integration cycle. 
We may allow ${\mathcal C}$ to have a boundary contained in the divisor $D$: $\partial {\mathcal C} \subset D$.
The integral remains well defined, if we assume that $\mathrm{Re}(\gamma_i)$ is sufficiently large, such that $u \varphi$ vanishes on $D$.
It is not too difficult to see that now the integral remains invariant under
\bq
\label{chapter_iterated_integrals:invariance_nabla_omega}
 \varphi & \rightarrow & \varphi + \nabla_\omega \xi,
\eq
where we introduced the covariant derivative $\nabla_\omega = d + \omega$.
In fact we have
\bq
 \int\limits_{\mathcal C} u \nabla_\omega \xi
 & = &
 \int\limits_{\mathcal C} \left[ u d \xi + u \left( d \ln u \right) \xi \right]
 \; = \;
 \int\limits_{\mathcal C} d \left( u \xi \right)
 \; = \;
 \int\limits_{\partial \mathcal C} u \xi
 \; = \; 0.
\eq
Introducing the twist amounts to going from the normal derivative $d$ in eq.~(\ref{chapter_iterated_integrals:invariance_d}) 
to the covariant derivative $\nabla_\omega = d + \omega$ in eq.~(\ref{chapter_iterated_integrals:invariance_nabla_omega}).
The invariance under eq.~(\ref{chapter_iterated_integrals:invariance_nabla_omega})
motivates the definition of equivalence classes of $n$-forms $\varphi$: 
Two $n$-forms $\varphi'$ and $\varphi$ are called equivalent, if they differ by a 
\index{covariant derivative}
covariant derivative
\bq
 \varphi'
 \sim
 \varphi
 & \;\; \Leftrightarrow \;\; &
 \varphi'
 \; = \;
  \varphi
  + \nabla_\omega \xi
\eq
for some $(n-1)$-form $\xi$.
We denote the equivalence classes by $\langle \varphi |$.
Being $n$-forms, each $\varphi$ is closed with respect to $\nabla_\omega$ and the equivalence classes
define the 
\index{twisted cohomology group}
{\bf twisted cohomology group} $H^n_\omega$:
\bq
 \left\langle \varphi \right|
 & \in &
 H^n_\omega.
\eq
\bs
{\it \refstepcounter{exercise}
{\bf Exercise \theexercise}: 
Show that for $\varphi$ as in eq.~(\ref{chapter_iterated_integrals:representative_left}) 
and $\omega$ as in eq.~(\ref{chapter_iterated_integrals:def_omega})
the differential $n$-form $\varphi$ is closed with respect to $\nabla_\omega$:
\bq
 \nabla_\omega \varphi & = & 0.
\eq
}
\es
\\
\\
The dual twisted cohomology group is given by
\bq
 \left( H^n_\omega \right)^\ast 
 & = &
 H^n_{-\omega}.
\eq
Elements of $( H^n_\omega )^\ast$ are denoted by $| \varphi \rangle$.
We have
\bq
 \left| \varphi' \right\rangle
 =
 \left| \varphi \right\rangle
 & \;\; \Leftrightarrow \;\; &
 \varphi'
 \; = \;
  \varphi
  + \nabla_{-\omega} \xi
\eq
for some $(n-1)$-form $\xi$.
A representative of a dual cohomology class is of the form
\bq
\label{chapter_iterated_integrals:representative_right}
 \varphi
 & = &
 \frac{q}{p_1^{n_1} \dots p_m^{n_m}} \; dz_1 \wedge \dots \wedge dz_n,
 \;\;\;\;\;\;\;\;\;
 q \in {\mathbb F}\left[z_1,\dots,z_n\right],
 \;\;\;
 n_i \in {\mathbb N}_0.
\eq
It will be convenient to use here the order $dz_1 \wedge \dots \wedge dz_n$ in the wedge product.

\begin{digression} {\bf Divisors}
\\
\index{divisor}
In the one-dimensional case the concept of a divisor originates 
from describing the set of zeros and the set of poles of a rational function.
The concept of a divisor can be generalised to higher-dimensional algebraic varieties.
Two different generalisations are in common use: Weil divisors and Cartier divisors.
They agree on non-singular varieties.

Let us start from the one-dimensional case: We consider divisors in the complex plane.
Codimension one sub-varieties are zero-dimensional (i.e. points).

Let $U$ be a connected open sub-set of ${\mathbb C}$.
A divisor is a function
\bq
 D & : & U \rightarrow {\mathbb Z},
\eq
which takes non-zero values $D(z) \neq 0$ at most on a discrete set $\Sigma \subset U$.
We write a divisor as a formal linear combination
\bq
\label{chapter_iterated_integrals:def_divisor}
 D & = & \sum\limits_{z \in U} D(z) \cdot z
\eq
Note that the symbol $\cdot$ in eq.~(\ref{chapter_iterated_integrals:def_divisor}) is a convention for writing a divisor
and has nothing to do with ordinary multiplication.
The 
\index{degree of a divisor}
{\bf degree of a divisor} is the sum of its coefficients:
\bq
 \mathrm{deg}\left(D\right) & = & \sum\limits_{z \in U} D(z).
\eq
A divisor is called 
\index{effective divisor}
{\bf effective}, if all coefficients are non-negative:
\bq
 D\left(z\right) & \ge & 0,
 \;\;\;\;\;\; \forall z \in U.
\eq
In this case we also write
\bq
 D & \ge & 0.
\eq
The divisors form an Abelian group, denoted by $\mathrm{Div}(U)$. 
The zero divisor is given by $D(z)=0$ for all $z$, the addition is defined by
\bq
 D_1 + D_2
 & = &
 \sum\limits_{z \in U} \left( D_1(z) + D_2(z) \right) \cdot z.
\eq
Let us now consider the field of meromorphic functions $K(U)$ on $U$.
We denote by $K^\ast(U)$ the set of meromorphic functions on $U$ without the function $f(z)=0$.
Every $f \in K^\ast(U)$ has a Laurent series
\bq
 f(z) & = & \sum\limits_{j=j_0}^\infty a_j \cdot \left(z-z_0\right)^j
\eq
around $z_0$, starting with $j_0$.
If $f(z_0)$ is finite and non-zero, we have $j_0=0$.
If $f(z_0)=0$, then $j_0$ gives the order of the zero.
If $f(z_0)$ has a pole at $z=z_0$, then $(-j_0)$ denotes the order of the pole.
A divisor $D_f$ is defined by
\bq
 D_f\left(z_0\right) & = & j_0.
\eq
A divisor $D$ is called 
\index{principal divisor}
{\bf principal divisor}, if there is $f\in K^\ast(U)$ such that $D=D_f$.
We have
\bq
 D_{f g} \;\; = \;\; D_f + D_g,
 & &
 D_{\frac{1}{f}} \;\; = \;\; - D_f,
\eq
and therefore the map
\bq
 K^\ast\left(U\right) & \rightarrow & \mathrm{Div}\left(U\right),
 \nonumber \\
 f & \rightarrow & D_f,
\eq
is a group homomorphism.
On ${\mathbb C}$ we have the following statements:
\begin{enumerate}
\item If $D \in \mathrm{Div}(U)$, then there exists a $f \in K^\ast(U)$ such that $D=D_f$.
\item If $f,g \in K^\ast(U)$ with $D_f=D_g$, then $h=f/g$ is a holomorphic function 
with $h(z) \neq 0$ for all $z$.
\end{enumerate}
The first statement says, that on ${\mathbb C}$ every divisor is a principal divisor.
This is true for ${\mathbb C}$, but not for compact Riemann surfaces.
The generalisation is given by the Riemann-Roch theorem. 
\\
\\
\index{Weil divisor}
{\bf Weil divisor}:
We have defined a divisor in the complex plane as a linear combination of points 
(i.e. codimension one sub-varieties) with integer coefficients.
A Weil divisor is the generalisation of this idea of codimension one sub-varieties in higher dimensions

We follow the book of Griffiths and Harris \cite{Griffiths:book}.
Let $X$ be a complex manifold (or algebraic variety) of dimension $n$, not necessarily compact.

A Weil divisor is a locally finite linear combination with integral coefficients of irreducible sub-varieties
of codimension one.
We write
\bq
 D & = & \sum n_i \cdot V_i,
\eq
with $n_i \in {\mathbb Z}$ and $V_i$ irreducible sub-varieties of dimension $(n-1)$.
Locally finite means that for any $p \in M$ there exists a neighbourhood of $p$ meeting only a finite number of the
$V_i$'s appearing in $D$.
\\
\\
\index{Cartier divisor}
{\bf Cartier divisor}:
We have seen that in the complex plane every divisor is a principal divisor and that the associated
function $f \in K^\ast(U)$ is determined up to a multiple of a holomorphic function, everywhere non-zero.
A Cartier divisor is a generalisation of this idea.
For readers familiar with sheaves we give the definition of a Cartier divisor below.
A definition of sheaves is given in appendix~\ref{appendix_algebraic_geometry}.

Let $X$ be a complex manifold (or algebraic variety) of dimension $n$, not necessarily compact.
Further, denote by ${\mathcal O}$ the sheaf of holomorphic functions on $X$ and
by ${\mathcal M}$ the sheaf of meromorphic functions on $X$.
Let ${\mathcal O}^\ast$ denote the sheaf of holomorphic functions which are nowhere zero on $X$
and let ${\mathcal M}^\ast$ denote the sheaf of meromorphic functions on $X$ without the zero function.
The quotient sheaf ${\mathcal D} = {\mathcal M}^\ast / {\mathcal O}^\ast$ is called the sheaf of divisors
and a section of ${\mathcal D}$ is called Cartier divisor.
The set of all sections $\Gamma\left(X, {\mathcal D}\right)$ forms an Abelian group.

\end{digression}

\subsection{Intersection numbers}

There is a non-degenerate bilinear pairing between a cohomology class $\langle \varphi_L|$ 
and a dual cohomology class $| \varphi_R \rangle$, given by the intersection number
\bq
 \left\langle \varphi_L \right. \left| \varphi_R \right\rangle_\omega.
\eq
$\varphi_L$ and $\varphi_R$ are representatives of the cohomology classes $\langle \varphi_L|$
and $| \varphi_R \rangle$, respectively. 
$\varphi_L$ and $\varphi_R$ are differential $n$-forms as 
in eq.~(\ref{chapter_iterated_integrals:representative_left}) and eq.~(\ref{chapter_iterated_integrals:representative_right}), respectively.
It will be convenient to define $\hat{\varphi}_L$ and $\hat{\varphi}_R$ to be the functions obtained by
stripping $dz_n \wedge \dots \wedge dz_1$ and $dz_1 \wedge \dots \wedge dz_n$ off, respectively:
\bq
\label{chapter_iterated_integrals:def_hat_varphi}
 \varphi_L
 \; = \; \hat{\varphi}_L \; dz_n \wedge \dots \wedge dz_1,
 & &
 \varphi_R
 \; = \; \hat{\varphi}_R \; dz_1 \wedge \dots \wedge dz_n.
\eq
The intersection number is defined by \cite{cho1995,Aomoto:book}
\bq
\label{chapter_iterated_integrals:def_intersection_number}
 \left\langle \varphi_L \right. \left| \varphi_R \right\rangle_\omega
 & = &
 \frac{1}{\left(2\pi i\right)^n}
 \int \iota_\omega\left(\varphi_L\right) \wedge \varphi_R
 \; = \;
 \frac{1}{\left(2\pi i\right)^n}
 \int \varphi_L \wedge \iota_{-\omega}\left(\varphi_R\right),
\eq
where $\iota_\omega$ maps $\varphi_L$ to a differential form 
in the same cohomology class as $\varphi_L$ but with compact support.
Similarly, $\iota_{-\omega}$ maps $\varphi_R$ to a differential form 
in the same cohomology class as $\varphi_R$ but with compact support.
That is to say, that $\iota_\omega(\varphi_L)$ and $\iota_{-\omega}(\varphi_R)$
vanish in a tubular neighbourhood of $D$ (and at infinity).
Although we started from differential forms $\varphi_L$ and $\varphi_R$ which are
holomorphic on ${\mathbb C}^n - D$, the compactly supported versions
$\iota_\omega(\varphi_L)$ and $\iota_{-\omega}(\varphi_R)$ are no longer holomorphic on
${\mathbb C}^n - D$.

Please note that the pairing $\langle \varphi | {\mathcal C} \rangle_{\omega}$ 
between an integrand and an integration contour 
denotes the integral defined in eq.~(\ref{chapter_iterated_integrals:def_integral}),
while the pairing $\langle \varphi_L | \varphi_R \rangle_\omega$ between an integrand and a dual integrand 
denotes the intersection number defined in eq.~(\ref{chapter_iterated_integrals:def_intersection_number}).

In order to see how $\iota_\omega(\varphi_L)$ (or $\iota_{-\omega}(\varphi_R)$)
is constructed, let's consider $\mathbb{CP}^1 - D$ \cite{Mizera:2017rqa}.
The divisor is a set of points $D=\{z_1,\dots,z_m\}$. 
Let's assume that none of these points is at infinity.
Around a point $z_j$ we consider two small discs $V_j$ and $U_j$, both centred at $z_j$
and such that the radius of $V_j$ is smaller than the radius of $U_j$.
\begin{figure}
\begin{center}
\includegraphics[scale=1.0]{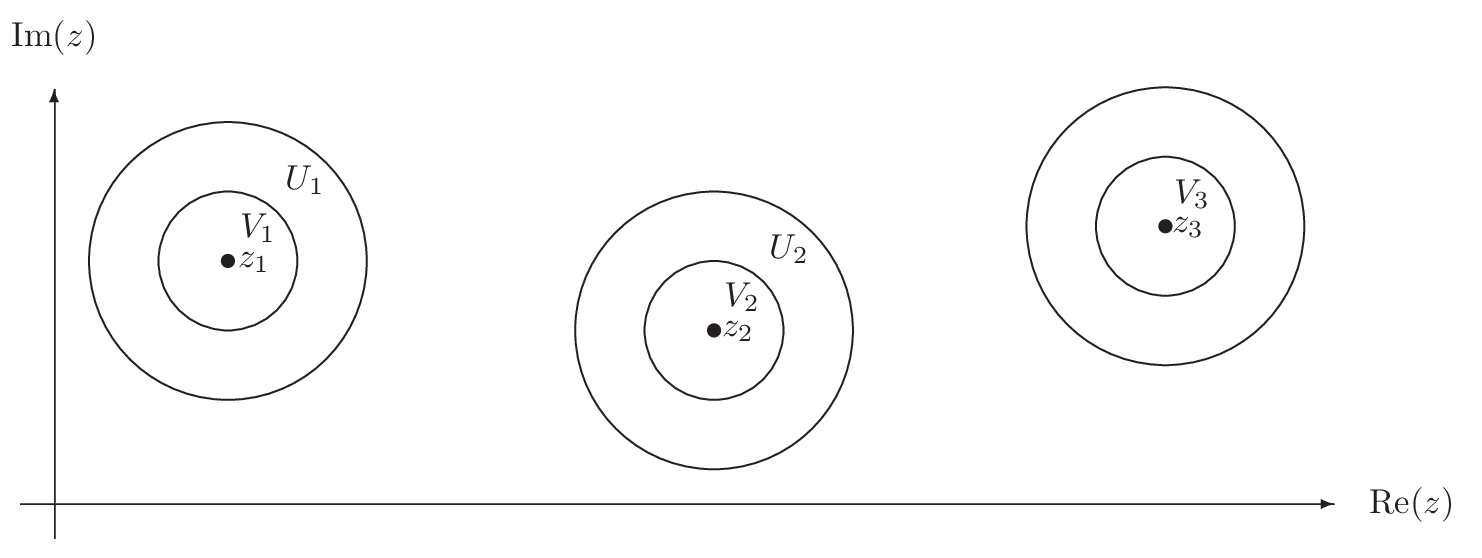}
\end{center}
\caption{
The construction of the differential form with compact support in the one-dimensional case.
The divisor is given by the union of three points: $D=\{z_1\} \cup \{z_2\} \cup \{z_3\}$.
$V_j$ and $U_j$ are small discs around $z_j$ with $V_j \subset U_j$.
}
\label{chapter_iterated_integrals:fig_compact_support}
\end{figure}
This is shown in fig.~\ref{chapter_iterated_integrals:fig_compact_support}.
We assume that the $U_j$'s do not overlap.
We introduce non-holomorphic functions $h_j(z,\bar{z})$ equal to $1$ on $V_j$, equal to $0$
outside $U_j$ and interpolating smoothly in the region $U_j - V_j$.
As $\varphi_L$ and $\iota_\omega(\varphi_L)$ are in the same cohomology class, they differ
by a covariant derivative:
\bq 
 \varphi_L - \iota_\omega\left(\varphi_L\right)
 & = &
 \nabla_\omega \xi.
\eq
Let $\psi_{L,j}$ be a solution of 
\bq
 \nabla_\omega \psi_{L,j} & = & \varphi_L
\eq
on $U_j \backslash \{z_j\}$. We set
\bq
 \xi & = &
 \sum\limits_{j=1}^m h_j \psi_{L,j}.
\eq
By construction, $\iota_\omega(\varphi_L)$ is in the same cohomology class as $\varphi_L$.
Let's verify that $\iota_\omega(\varphi_L)$ has compact support.
We show that $\iota_\omega(\varphi_L)$ vanishes on $V_i$.
We have
\bq
 \iota_\omega\left(\varphi_L\right)
 & = &
 \varphi_L - \nabla_\omega \xi
 \; = \;
 \varphi_L - \sum\limits_{j=1}^m \nabla_\omega \left(h_j \psi_{L,j} \right)
 \nonumber \\
 & = &
 \varphi_L 
 - \sum\limits_{j=1}^m h_j \nabla_\omega \psi_{L,j}
 - \sum\limits_{j=1}^m \left(d h_j \right) \psi_{L,j}
 \nonumber \\
 & = &
 \varphi_L 
 - \sum\limits_{j=1}^m h_j \varphi_L
 - \sum\limits_{j=1}^m \left(d h_j \right) \psi_{L,j}.
\eq
On $V_i$ we have $h_i=1$ and $h_j=0$ for $j \neq i$.
Furthermore we have on $V_i$ that the derivative of all functions $h_j$ vanishes:
$dh_j=0$ for all $j$.
Thus we find on $V_i$
\bq
 \iota_\omega\left(\varphi_L\right)
 & = &
 \varphi_L 
 - h_i \varphi_L
 \; = \;
 \varphi_L 
 - \varphi_L
 \; = \; 
 0.
\eq
Let us now turn to the intersection number for the case $\mathbb{CP}^1 - D$.
We have
\bq
 \left\langle \varphi_L \right. \left| \varphi_R \right\rangle_\omega
 & = &
 \frac{1}{2\pi i}
 \int \iota_\omega\left(\varphi_L\right) \wedge \varphi_R
 \; = \;
 \frac{1}{2\pi i}
 \int \left[ \varphi_L  - \sum\limits_{j=1}^m h_j \varphi_L
 - \sum\limits_{j=1}^m \left(d h_j \right) \psi_{L,j} \right] \wedge \varphi_R
 \nonumber \\
 & = &
 - \frac{1}{2\pi i}
 \sum\limits_{j=1}^m 
 \int \left(d h_j \right) \psi_{L,j} \wedge \varphi_R.
\eq
The first two terms yield $dz \wedge dz$, only the last term yields a non-vanishing wedge product
$dz \wedge d\bar{z}$.
As $d h_j$ is non-zero only on $U_j - V_j$ we obtain
\bq
 \left\langle \varphi_L \right. \left| \varphi_R \right\rangle_\omega
 & = &
 - \frac{1}{2\pi i}
 \sum\limits_{j=1}^m 
 \int\limits_{U_j-V_j} \left(d h_j \right) \psi_{L,j} \wedge \varphi_R
 \nonumber \\
 & = &
 - \frac{1}{2\pi i}
 \sum\limits_{j=1}^m 
 \int\limits_{U_j-V_j} 
  \left[ d \left( h_j \psi_{L,j} \varphi_R \right) 
  - h_j \left(d \psi_{L,j} \right) \wedge \varphi_R
  - h_j \psi_{L,j} d \varphi_R \right]
 \nonumber \\
 & = &
 - \frac{1}{2\pi i}
 \sum\limits_{j=1}^m 
 \int\limits_{U_j-V_j} d \left( h_j \psi_{L,j} \varphi_R \right).
\eq
In the second line the last two terms yield a vanishing contribution, again due to $dz \wedge dz=0$.
We may now use Stokes' theorem and obtain
\bq
\label{chapter_iterated_integrals:intersection_number_psi_phi}
 \left\langle \varphi_L \right. \left| \varphi_R \right\rangle_\omega
 & = &
 \frac{1}{2\pi i}
 \sum\limits_{j=1}^m 
 \int\limits_{\partial V_j} \psi_{L,j} \varphi_R
 \; = \;
 \sum\limits_{j=1}^m 
 \mathrm{Res}_{D_j}\left(\psi_{L,j} \varphi_R\right),
\eq
where we used that $h_j=0$ on $\partial U_j$ and $h_j=1$ on $\partial V_j$.

Alternatively, we could have used $\iota_{-\omega}(\varphi_R)$.
Let $\psi_{R,j}$ be a solution of 
\bq
 \nabla_{-\omega} \psi_{R,j} & = & \varphi_R
\eq
on $U_j \backslash \{z_j\}$. 
Then
\bq
 \left\langle \varphi_L \right. \left| \varphi_R \right\rangle_\omega
 & = &
 \frac{1}{2\pi i}
 \int \varphi_L \wedge \iota_{-\omega}\left(\varphi_R\right)
 \nonumber \\
 & = &
 - 
 \frac{1}{2\pi i}
 \sum\limits_{j=1}^m 
 \int\limits_{\partial V_j} \varphi_L \psi_{R,j} 
 \; = \;
 -
 \sum\limits_{j=1}^m 
 \mathrm{Res}_{D_j}\left(\varphi_L \psi_{R,j}\right).
\eq
Computing the intersection number through the definition in eq.~(\ref{chapter_iterated_integrals:def_intersection_number})
is not the most practical way.
In section~\ref{chapter_iterated_integrals:computation_intersection_numbers} we will learn more efficient methods.

Let's consider an example:
We take $n=1$ and we consider $p_1(z)=z$, $p_2(z)=1-z$.
Thus we have $D_1=\{0\}$ and $D_2=\{1\}$.
We consider
\bq
 u\left(z\right)
 & = &
 z^{\gamma_1} \left(1-z\right)^{\gamma_2}.
\eq
The one-form $\omega$ is then given by
\bq
 \omega
 & = &
 \gamma_1 \frac{dz}{z} - \gamma_2 \frac{dz}{1-z}.
\eq
Let us further consider
\bq
 \varphi_L \; = \; \frac{dz}{z^{1+n_1} \left(1-z\right)^{1+n_2}},
 & &
 \varphi_R \; = \; \frac{dz}{z^{1+n_3} \left(1-z\right)^{1+n_4}},
 \;\;\;\;\;\;
 n_1, n_2, n_3, n_4 \; \in \; {\mathbb Z}.
\eq
Around $z=0$ let $\psi_{L,1}$ be a solution of
\bq
\label{chapter_iterated_integrals:psi_L_1}
 \nabla_\omega \psi_{L,1} & = & \varphi_L.
\eq
For $\psi_{L,1}$ we make the ansatz
\bq
 \psi_{L,1}
 & = &
 \sum\limits_{j=-n_1}^\infty a_j z^j.
\eq
Eq.~(\ref{chapter_iterated_integrals:psi_L_1}) becomes
\bq
 \left( \frac{d}{dz} + \frac{\gamma_1}{z} - \frac{\gamma_2}{1-z} \right)
 \sum\limits_{j=-n_1}^\infty a_j z^j
 & = &
 \frac{1}{z^{1+n_1} \left(1-z\right)^{1+n_2}}.
\eq
This equation can be solved recursively for the unknown coefficients $a_j$, starting with
\bq
 a_{-n_1} & = & 
 \frac{1}{\gamma_1-n_1}.
\eq
In a similar way, we let $\psi_{L,2}$ be a solution of
\bq
\label{chapter_iterated_integrals:psi_L_2}
 \nabla_\omega \psi_{L,2} & = & \varphi_L.
\eq
around $z=1$. We make the ansatz
\bq
 \psi_{L,2}
 & = &
 \sum\limits_{j=-n_2}^\infty b_j \left(1-z\right)^j
\eq
and determine the coefficients $b_j$ recursively.
The intersection number $\langle \varphi_L | \varphi_R \rangle_\omega$
is then obtained with the help of 
eq.~(\ref{chapter_iterated_integrals:intersection_number_psi_phi}) 
as
\bq
 \left\langle \varphi_L \right. \left| \varphi_R \right\rangle_\omega
 & = &
 \mathrm{Res}_{D_1}\left(\psi_{L,1} \varphi_R\right)
 +
 \mathrm{Res}_{D_2}\left(\psi_{L,2} \varphi_R\right).
\eq
We obtain
\bq
\label{chapter_iterated_integrals:intersection_beta_fct}
 \left\langle \varphi_L \right. \left| \varphi_R \right\rangle_\omega
 & = &
 \frac{\left(\gamma_1+\gamma_2\right)}{\gamma_1 \gamma_2}
 \frac{\Gamma\left(1-\gamma_1\right)\Gamma\left(1-\gamma_2\right)}{\Gamma\left(1-\gamma_1-\gamma_2\right)}
 \frac{\Gamma\left(1+\gamma_1\right)\Gamma\left(1+\gamma_2\right)}{\Gamma\left(1+\gamma_1+\gamma_2\right)}
 \nonumber \\
 & &
 \frac{\Gamma\left(1+n_1+n_2-\gamma_1-\gamma_2\right)}{\Gamma\left(1+n_1-\gamma_1\right)\Gamma\left(1+n_2-\gamma_2\right)}
 \frac{\Gamma\left(1+n_3+n_4+\gamma_1+\gamma_2\right)}{\Gamma\left(1+n_3+\gamma_1\right)\Gamma\left(1+n_4+\gamma_2\right)}.
\eq
\bs
{\it \refstepcounter{exercise}
{\bf Exercise \theexercise}: 
Proof eq.~(\ref{chapter_iterated_integrals:intersection_beta_fct}) for the special case $n_1=n_2=n_3=n_4=0$.
}
\es
\\
\\
Under certain assumptions it can be shown \cite{aomoto1975,kita1982,cho:1997,Aomoto:book} that 
the twisted cohomology groups $H^k_\omega$ vanish for $k \neq n$, thus $H^n_\omega$ is the only interesting
twisted cohomology group.
We denote the dimensions of the twisted cohomology groups by
\bq 
 \nu & = & \dim H^n_\omega \; = \; \dim \left( H^n_\omega \right)^\ast.
\eq
Let 
$\langle e_j |$ with $1\le j \le \nu$
be a basis of $H^n_\omega$ and let
$| h_j \rangle$ with $ 1 \le j \le \nu$
be a basis of $( H^n_\omega )^\ast$.
We denote the $(\nu \times \nu)$-dimensional intersection matrix by $C$.
The entries are given by
\bq
 C_{j k}
 & = &
 \left\langle e_{j} \right| \left. h_{k} \right\rangle.
\eq
The matrix $C$ is invertible.
Given a basis $\langle e_j |$ of $H^n_\omega$
we say that a basis $| d_j \rangle$ of $( H^n_\omega )^\ast$ is
the 
\index{dual basis}
{\bf dual basis}
with respect to $\langle e_j |$
if
\bq
\label{chapter_iterated_integrals:def_dual_basis}
 \left\langle e_{j} \right| \left. d_{k} \right\rangle
 & = &
 \delta_{j k}.
\eq
Starting from an arbitrary basis 
$| h_j \rangle$ of $( H^n_\omega )^\ast$
we may always construct the dual basis $| d_j \rangle$ of $( H^n_\omega )^\ast$ with respect to $\langle e_j |$.
The dual basis is given by
\bq
\label{chapter_iterated_integrals:construction_dual_basis}
 \left| d_j \right\rangle
 & = &
 \left| h_k \right\rangle
 \left( C^{-1} \right)_{kj}.
\eq

The dimension of the twisted cohomology groups is related to the number of critical points of 
$f=\ln(u)$.
We have
\bq
 f & = & \ln\left(u\right) \; = \; \sum\limits\limits_{i=1}^m \gamma_i \ln\left(p_i\right).
\eq
A point $z$ is called a 
\index{critical point}
{\bf critical point} if 
\bq
 \left. df \right|_z & = & 0.
\eq
A critical point $z$ is called a 
\index{non-degenerate critical point}
{\bf non-degenerate critical point} if the Hessian matrix
is invertible, i.e.
\bq
 \left. \det \left( \frac{\partial^2 f}{\partial z_i \partial z_j} \right) \right|_{z} & \neq & 0.
\eq
A critical point $z$ is called a 
\index{proper critical point}
{\bf proper critical point} if
\bq
 z & \notin & D.
\eq
By the definition of $\omega$ in eq.~(\ref{chapter_iterated_integrals:def_omega}) we have $df=\omega$. 
Assuming that all critical points are 
proper and non-degenerate we have \cite{Lee:2013hzt,Frellesvig:2019uqt}
\bq
 \dim H^n_\omega & = &
 \left( \# \; \mbox{solutions of } \omega = 0 \;\; \mbox{on} \;\; {\mathbb C}^n - D \right).
\eq
Usually it is not an issue to find a basis.
For completeness, we give here a systematic algorithm to construct a basis for $H^n_\omega$ and $(H^n_\omega)^\ast$
for the case where all critical points are proper and non-degenerate.
We write
\bq
 \omega
 & = &
 \sum\limits_{j=1}^n \omega_j dz_j,
 \;\;\;\;\;\;
 \omega_j \; = \; \frac{P_j}{Q_j},
 \;\;\;\;\;\;
 P_j, Q_j \; \in \; {\mathbb F}\left[z_1,\dots,z_n\right],
 \;\;\;\;\;\;
 \gcd\left(P_j,Q_j\right) \; = \; 1.
 \;\;\;
\eq
We consider the ideal
\bq
 I_n \; = \; \left\lideal P_1, \dots, P_n \right\rideal
 \; \subset \; 
 {\mathbb F}\left[z_1,\dots,z_n\right].
\eq
In the case where all critical points are proper and non-degenerate we have
\bq
 \dim H^n_\omega 
 & = &
 \dim\left( {\mathbb F}\left[z_1,\dots,z_n\right] / I_n \right).
\eq
Let $G_1, \dots, G_r$ be a Gr\"obner basis of $I_n$ with respect to some term order $<$:
\bq
 I_n & = &
 \left\lideal G_1, \dots,G_r \right\rideal.
\eq
For a basis $\langle e_j |$ of $H^n_\omega$ we write as in eq.~(\ref{chapter_iterated_integrals:def_hat_varphi})
\bq
 e_j & = & \hat{e}_j  \; dz_n \wedge \dots \wedge dz_1.
\eq
Similarly, we write for a basis $| h_j \rangle$ of $( H^n_\omega )^\ast$
\bq
 h_j & = & \hat{h}_j \; dz_1 \wedge \dots \wedge dz_n.
\eq
Then $\hat{e}_j$ and $\hat{h}_j$ are given by all monomials
\bq
 \prod\limits_{k=1}^n z_k^{\nu_k},
 \;\;\;\;\;\;\;\;\;
 \nu_k \; \in \; {\mathbb N}_0
\eq
which are not divisible by any leading term of the Gr\"obner basis:
\bq
 \mathrm{lt}\left(G_j\right)
 & \not | &
 \prod\limits_{k=1}^n z_k^{\nu_k}
 \;\;\;\;\;\;\;\;\; \forall \; 0 \; \le \, j \; \le \; r.
\eq
Here, $\mathrm{lt}$ denotes the leading term of a polynomial with respect to the chosen term order.
In general, $\langle e_j |$ and $| h_j \rangle$ defined in this way will not be dual to each other.

Let's look at an example.
We consider a case with two variables $z_1$ and $z_2$ (i.e. $n=2$) and 
three polynomials (i.e. $m=3$)
\bq
 p_1 \; = \; z_1,
 \;\;\;
 p_2 \; = \; z_2,
 \;\;\;
 p_3 \; = \; z_2^2 - 4 z_1^3 + 11 z_1 - 7.
\eq
We set
\bq
 u
 & = &
 \left( p_1 p_2 p_3 \right)^\gamma.
\eq
The differential one-form $\omega$ reads
\bq
 \omega
 & = &
 \gamma \frac{z_2^2 - 16 z_1^3 + 22 z_1 - 7}{p_1 p_3} dz_1
 +
 \gamma \frac{3 z_2^2 - 4 z_1^3 + 11 z_1 - 7}{p_2 p_3} dz_2.
\eq
We therefore have to consider the ideal
\bq
 I_2
 & = &
 \left\lideal
   z_2^2 - 16 z_1^3 + 22 z_1 - 7, 3 z_2^2 - 4 z_1^3 + 11 z_1 - 7
 \right\rideal.
\eq
A Gr\"obner basis with respect to the graded reverse lexicographic order
is given by
\bq
 I_2
 & = &
 \left\lideal
   11 z_2^2 + 22 z_1 -21, 44 z_1^3 - 55 z_1 + 14
 \right\rideal.
\eq
The leading terms of the elements of the Gr\"obner basis are $11 z_2^2$ and $44 z_1^3$.
A basis $\langle e_j |$ of $H^2_\omega$ with $e_j=\hat{e}_j dz_2 \wedge dz_1$ is therefore given by
\bq
 \hat{e}_j & \in & \left\{ 1, z_1, z_2, z_1 z_2 , z_1^2, z_1^2 z_2 \right\}.
\eq
Similarly, a basis $| h_j \rangle$ of $( H^n_\omega )^\ast$ with $h_j=\hat{h}_j dz_1 \wedge dz_2$
is given by
\bq
 \hat{h}_j & \in & \left\{ 1, z_1, z_2, z_1 z_2 , z_1^2, z_1^2 z_2 \right\}.
\eq

\begin{digression} {\bf Gr\"obner bases}
\\
Gr\"obner bases are useful in many situation.
The most prominent application of Gr\"obner bases is probably the simplification of a polynomial
with respect to polynomial siderelations.
A good introduction to Gr\"obner bases is the book by Adams and Loustaunau \cite{Adams_Loustaunau}.

Assume that we have a (possibly rather long) 
expression $f$, which is a polynomial in
several variables $x_1$, \dots, $x_k$.
In addition we have several siderelations of the form
\bq
s_j(x_1,\dots,x_k) & = & 0, \;\;\;1 \le j \le r,
\eq
which are also polynomials in $x_1$, \dots, $x_k$.
A standard task is now to simplify $f$ with respect to the siderelations $s_j$,
e.g. to rewrite $f$ in the form
\bq
 f & = & a_1 s_1 + \dots + a_r s_r + g,
\eq
where $g$ is ``simpler'' than $f$
The precise meaning of ``simpler'' requires the introduction of an order
relation on the multivariate polynomials.
As an example let us consider the expressions
\bq
f_1 = x + 2 y^3, & & 
f_2 = x^2,
\eq
which we would like to simplify with respect to the siderelations
\bq
s_1 & = & x^2 + 2 x y, \nonumber \\
s_2 & = & xy + 2 y^3 -1.
\eq
As an order relation we may 
choose lexicographic ordering, e.g. $x$ is ``more
complicated'' as $y$, and $x^2$ is ``more complicated'' than $x$.
This definition will be made more precise below.
A naive approach would now take each siderelation, determine its ``most
complicated'' element, and replace each occurrence of this element in the 
expression $f$ by the more simpler terms of the siderelation.
As an example let us consider for this approach the simplification
of $f_2$ with respect to the siderelations $s_1$ and $s_2$:
\bq
f_2 = x^2 = s_1 - 2 x y = s_1 - 2 y s_2 + 4 y^4 - 2y,
\eq
and $f_2$ would simplify to $4 y^4 - 2y$. 
In addition, since $f_1$ does not contain $x^2$ nor $x y$, the naive approach would not
simplify $f_1$ at all.
However, this is not the complete
story, since if $s_1$ and $s_2$ are siderelations, any linear
combination of those is again a valid siderelation.
In particular,
\bq
s_3 & = & y s_1 - x s_2 = x
\eq
is a siderelation which can be deduced from $s_1$ and $s_2$.
This implies that $f_2$ simplifies to $0$ with respect to the siderelations
$s_1$ and $s_2$.
Clearly, some systematic approach is needed.
The appropriate tools are ideals in rings, and Gr\"obner bases for these
ideals.
\\
We consider multivariate polynomials in the ring $R[x_1,\dots,x_k]$.
Each element can be written as a sum of monomials of the form
\bq
c x_1^{m_1} \dots x_k^{m_k}.
\eq
We define the 
\index{lexicographic order}
{\bf lexicographic order} of these terms by
\bq
c x_1^{m_1} \dots x_k^{m_k} >
c' x_1^{m_1'} \dots x_k^{m_k'},
\eq
if the leftmost non-zero entry in $(m_1-m_1', \dots, m_k-m_k')$ is positive.
With this ordering we can write any element $f \in R[x_1,\dots,x_k]$ 
as
\bq
f & = & \sum_{i=0}^n h_i
\eq
where the $h_i$ are monomials and 
$h_{i+1} > h_i$ with respect to the lexicographic order.
The term $h_n$ is called the 
\index{leading term}
{\bf leading term} 
and denoted $\mathrm{lt}(f) = h_n$.

Let $ B=\{b_1,\dots,b_r\} \subset R[x_1,\dots,x_k]$ 
be a (finite) set of polynomials.
The set
\bq
\lideal B \rideal = \lideal b_1, \dots, b_r \rideal & = & 
\left\{ \left. 
       \sum\limits_{i=1}^r a_i b_i \right| a_i \in R[x_1,\dots,x_k] \right\}
\eq
is called the 
\index{ideal}
{\bf ideal} 
generated by the set $B$. The set $B$ is also called
a 
\index{basis of an ideal}
{\bf basis} for this ideal.
(In general, given a ring $R$ and a subset $I \subset R$, $I$ 
is called an ideal
if $a+b \in I$ for all $a,b, \in I$ and $r a \in I$ for all $a \in I$ and
$r \in R$. Note the condition for the multiplication:
The multiplication has to be closed with respect
to elements from $R$ and not just $I$.)
\\
Suppose that we have an ideal $I$ and a finite subset $H \subset I$.
We denote by $\mathrm{lt}(H)$ the set of leading terms of $H$ and, 
correspondingly by $\mathrm{lt}(I)$ the set of leading terms of $I$.
Now suppose that the ideal generated by $\mathrm{lt}(H)$ is identical
with the one generated by $\mathrm{lt}(I)$, e.g. $\mathrm{lt}(H)$ is a basis
for $\lideal \mathrm{lt}(I) \rideal$.
Then a mathematical theorem guarantees that $H$ is also a basis for $I$, e.g.
\bq
\lideal \mathrm{lt}(H) \rideal = \lideal \mathrm{lt}(I) \rideal
 & \Rightarrow & 
 \lideal H \rideal = I
\eq
However, the converse is in general not true, e.g. if $H$ is a basis
for $I$ this does not imply that $\mathrm{lt}(H)$ is a basis for
$\lideal \mathrm{lt}(I) \rideal$.
A further theorem (due to Hilbert) states however that there exists
a subset $G \subset I$ such that 
\bq
\lideal G \rideal = I \;\;\;\mbox{and}\;\;\; \lideal \mathrm{lt}(G) \rideal = \lideal \mathrm{lt}(I) \rideal,
\eq
e.g. $G$ is a basis for $I$ and $\mathrm{lt}(G)$ is a basis for
$\lideal \mathrm{lt}(I) \rideal$.
Such a set $G$ is called a 
\index{Gr\"obner basis}
{\bf Gr\"obner basis} 
for $I$.
Buchberger \cite{Buchberger} gave an algorithm to compute $G$, which nowadays
is implemented in many computer algebra systems.

The importance of Gr\"obner bases for simplifications
stems from the following theorem:
Let $G$ be a Gr\"obner basis for an ideal $I \subset R[x_1,\dots,x_k]$ and $f \in R[x_1,\dots,x_k]$.
Then there is a unique polynomial $g \in R[x_1,\dots,x_k]$ with
\bq
f - g \in I
\eq
and no term of $g$ is divisible by any monomial in $\mathrm{lt}(G)$.
\\
In plain text: $f$ is an expression which we would like to simplify
according to the siderelations defined by $I$.
This ideal is originally given 
by a set of polynomials 
$\{s_1,\dots,s_r\}$ and the siderelations are supposed to be of the
form $s_i=0$.
From this set of siderelations a Gr\"obner basis $\{b_1,\dots,b_{r'}\}$
for this ideal is calculated.
This is the natural basis for simplifying the expression $f$.
The result is the expression $g$, from which the  
``most complicated'' terms of $G$ have been eliminated, e.g. the terms $\mathrm{lt}(G)$.
The precise meaning of ``most complicated'' terms depends on the
definition of the order relation.
\\
In our example, $\{s_1,s_2\}$ is not a Gr\"obner basis for
$\lideal s_1, s_2 \rideal$, since $\mathrm{lt}(s_1)=x^2$ and $\mathrm{lt}(s_2)=xy$
and
\bq
\mathrm{lt}\left( y s_1 -x s_2 \right) = x \;\; \notin \;\;
 \lideal \mathrm{lt}(s_1), \mathrm{lt}(s_2) \rideal.
\eq
A Gr\"obner basis for $\lideal s_1, s_2 \rideal$ is given by
\bq
\left\{ x, 2y^3 -1 \right\}.
\eq
With $b_1=x$ and $b_2=2 y^3 -1$ as a Gr\"obner basis, $f_1$ and $f_2$ can 
be simplified as follows:
\bq
f_1 & = & b_1 + b_2 + 1, \nonumber \\
f_2 & = & x b_1 + 0,
\eq
e.g. $f_1$ simplifies to $1$ and $f_2$ simplifies to $0$.

We are not forced to use the lexicographic order introduced above.
We may use any 
\index{term order}
{\bf term order}.
A term order is a total order (this means that between two quantities exactly one of the relations $<$, $=$ or $>$ must be true)
on the monomials $c x_1^{m_1} \dots x_k^{m_k}$ which satisfies
\begin{enumerate}
\item $x_1^{m_1} \dots x_k^{m_k} > 1$ for $(m_1,\dots,m_k) \neq (0,\dots,0)$.
\item $x_1^{m_1} \dots x_k^{m_k} > x_1^{m_1'} \dots x_k^{m_k'}$ implies
\bq
 \left( x_1^{m_1} \dots x_k^{m_k} \right) \left( x_1^{m_1''} \dots x_k^{m_k''} \right) 
 & > & 
 \left( x_1^{m_1'} \dots x_k^{m_k'} \right) \left( x_1^{m_1''} \dots x_k^{m_k''} \right)
\eq
for any monomial $x_1^{m_1''} \dots x_k^{m_k''}$.
\end{enumerate}
Apart from the lexicographic order introduced above other popular choices for a term order are
the 
degree lexicographic order 
and the 
degree reverse lexicographic order.
The 
\index{degree lexicographic order}
{\bf degree lexicographic order} (or graded lexicographic order)
is defined as follows:
We have
\bq
 \left( x_1^{m_1} \dots x_k^{m_k} \right) & > & \left( x_1^{m_1'} \dots x_k^{m_k'} \right) 
\eq
if either
\bq
   \sum\limits_{j=1}^k m_j & > & \sum\limits_{j=1}^k m_j',
 \nonumber
\eq
or in the case that the total degrees are equal
\bq
 \sum\limits_{j=1}^k m_j = \sum\limits_{j=1}^k m_j' 
 \;
 \mbox{and the leftmost non-zero entry in $(m_1-m_1', \dots, m_k-m_k')$ is positive},
 \nonumber
\eq
The
\index{degree reverse lexicographic order}
{\bf degree reverse lexicographic order} (or graded reverse lexicographic order)
is defined as follows:
We have
\bq
 \left( x_1^{m_1} \dots x_k^{m_k} \right) & > & \left( x_1^{m_1'} \dots x_k^{m_k'} \right) 
\eq
if either
\bq
   \sum\limits_{j=1}^k m_j & > & \sum\limits_{j=1}^k m_j',
 \nonumber
\eq
or in the case that the total degrees are equal
\bq
 \sum\limits_{j=1}^k m_j = \sum\limits_{j=1}^k m_j' 
 \;
 \mbox{and the rightmost non-zero entry in $(m_1-m_1', \dots, m_k-m_k')$ is negative}.
 \nonumber
\eq
\end{digression}
\noindent
\bs
{\it \refstepcounter{exercise}
{\bf Exercise \theexercise}: 
Consider the monomials
\bq
 p_1 \; = \; x_1^2 x_2 x_3,
 & &
 p_2 \; = \; x_1 x_2^3.
\eq
Order the two monomials with respect to the degree lexicographic order
and the degree reverse lexicographic order (assuming $x_1>x_2>x_3$).
}
\es
\\
\\
The critical points of 
\bq
 f & = & \ln\left(u\right) \; = \; \sum\limits\limits_{i=1}^m \gamma_i \ln\left(p_i\right)
\eq
allow us to construct a basis $| {\mathcal C}_j \rangle$ of the twisted homology groups $H_n^\omega$ as well:
We first fix a branch of $u$.
As before we assume that all critical points are proper and non-degenerate.
For simplicity let us further assume that $\gamma_i \in {\mathbb R}_{<0}$ and $\gamma_i \notin {\mathbb Z}$.
We split $f$ into the real part and the imaginary part:
\bq
 f\left(z\right)
 & = &
 \sum\limits\limits_{i=1}^m \gamma_i \ln\left|p_i\left(z\right)\right|
 + i
 \sum\limits\limits_{i=1}^m \gamma_i \; \mathrm{arg}\left(p_i\left(z\right)\right).
\eq
We denote the real part by $h(z)$ and the imaginary part by $\phi(z)$.
Thus
\bq
 h\left(z\right)
 \; = \; 
 \sum\limits\limits_{i=1}^m \gamma_i \ln\left|p_i\left(z\right)\right|,
 & &
 \phi\left(z\right)
 \; = \;
 \sum\limits\limits_{i=1}^m \gamma_i \; \mathrm{arg}\left(p_i\left(z\right)\right).
\eq
The value of $\phi$ at a 
critical points $z_{\mathrm{crit}}^{(j)}$ is called a critical phase and denoted by
\bq
 \phi^{(j)} & = & \phi\left(z_{\mathrm{crit}}^{(j)}\right).
\eq
Since we assumed that all critical points are non-degenerate it follows that $h$ is a Morse function.
We will now consider the gradient flow equations for $h$.
If we temporarily introduce $(2n)$ real coordinates $x_1,\dots,x_n,y_1,\dots,y_n$ such that
$z_j=x_j+iy_j$, the gradient flow equations read
\bq
 \frac{dx_j}{d\lambda} \; = \; - \frac{\partial h}{\partial x_j},
 & &
 \frac{dy_j}{d\lambda} \; = \; - \frac{\partial h}{\partial y_j}.
\eq
Changing back to complex coordinates, we may write the gradient flow equations as
\bq
 \frac{d z_j}{d\lambda} & = & - 2 \frac{\partial h}{\partial \bar{z}_j}.
\eq
The gradient flow equations define curves in ${\mathbb C}^n - D$.
We denote by ${\mathcal C}_j$ the union of curves with 
\bq
\label{chapter_iterated_integrals:Lefschetz_thimble_C}
 \lim\limits_{\lambda \rightarrow -\infty} z\left(\lambda\right)
 & = & z_{\mathrm{crit}}^{(j)},
\eq
and by ${\mathcal D}_j$ the union of curves with
\bq
\label{chapter_iterated_integrals:Lefschetz_thimble_D}
 \lim\limits_{\lambda \rightarrow \infty} z\left(\lambda\right)
 & = & z_{\mathrm{crit}}^{(j)}.
\eq
${\mathcal C}_j$ and ${\mathcal D}_j$ are called
\index{Lefschetz thimble} 
{\bf Lefschetz thimbles}.
The curves which make up ${\mathcal D}_j$ start at a point $z \in D$, where $h(z)$ is plus infinity and
approach the critical point $z_{\mathrm{crit}}^{(j)}$ for $\lambda \rightarrow \infty$.
The curves which make up ${\mathcal C}_j$ end at points where $|z|=\infty$ and where $h(z)$ is minus infinity. Tracing back
these curves one reaches the critical point $z_{\mathrm{crit}}^{(j)}$ for $\lambda \rightarrow -\infty$.
Let us reformulate this slightly:
The critical point $z_{\mathrm{crit}}^{(j)}$ is a saddle point of $h(z)$.
The Lefschetz thimble ${\mathcal D}_j$ is the union of all trajectories, which descend by the steepest descent towards $z_{\mathrm{crit}}^{(j)}$.
The Lefschetz thimble ${\mathcal C}_j$ is the union of all trajectories, which descend from $z_{\mathrm{crit}}^{(j)}$ by the steepest descent to minus infinity.
Of interest to us are the Lefschetz thimbles ${\mathcal C}_j$.
Due to the Cauchy-Riemann equations the phase $\phi(z)$ is constant on a
Lefschetz thimbles.
If all critical phases $\phi^{(j)}$ are pairwise distinct, it follows that for $i \neq j$ the
Lefschetz thimbles ${\mathcal C}_i$ and ${\mathcal C}_j$ do not intersect.

The real dimension of ${\mathcal C}_j$ equals the Morse index of the critical point $z_{\mathrm{crit}}^{(j)}$.
For the application towards Feynman integrals we may assume that all critical points have Morse
index $n$. Then ${\mathcal C}_j$ has real dimension $n$ 
and defines a representative for $|{\mathcal C}_j\rangle$.

In summary, each critical point $z_{\mathrm{crit}}^{(j)}$ defines a Lefschetz thimble
${\mathcal C}_j$. The set of all Lefschetz thimbles satisfying eq.~(\ref{chapter_iterated_integrals:Lefschetz_thimble_C}) defines a basis
$|{\mathcal C}_j\rangle$ of $H_n^\omega$.

\begin{figure}
\begin{center}
\includegraphics[scale=1.0]{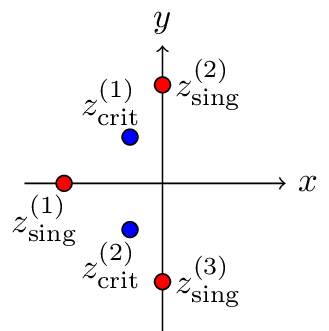}
 \hspace*{10mm}
\includegraphics[scale=1.0]{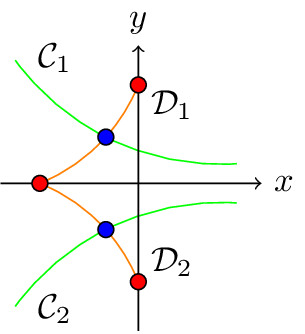}
\end{center}
\caption{\label{chapter_iterated_integrals:fig_singular_and_critical_points}
The left picture shows the location of the singular points 
$z_{\mathrm{sing}}^{(1)}$, $z_{\mathrm{sing}}^{(2)}$ and $z_{\mathrm{sing}}^{(3)}$ (red points)
and the location of the critical points $z_{\mathrm{crit}}^{(1)}$ and $z_{\mathrm{crit}}^{(2)}$ (blue points).
The right picture shows a sketch of the Lefschetz thimbles ${\mathcal C}_1$ and ${\mathcal C}_2$ (green)
as well as the Lefschetz thimbles ${\mathcal D}_1$ and ${\mathcal D}_2$ (orange). 
}
\end{figure}

Let us look at an example. We take $n=1$ (one complex variable $z$) and $m=1$ (one polynomial $p(z)$).
We discuss
\bq
 p\left(z\right) & = & z^3 + z^2 + z + 1.
\eq
We take $\gamma=-\frac{1}{10}$.
Therefore
\bq
 u\left(z\right)
 \; = \; 
 \left[ p\left(z\right) \right]^{-\frac{1}{10}},
 & &
 f\left(z\right)
 \; = \; 
 \ln\left(u\left(z\right)\right)
 \; = \; 
 -\frac{1}{10} \ln\left( p\left(z\right) \right).
\eq
The divisor $D$ is given by the roots of the polynomial $p(z)$:
\bq
 D & = &
 \left\{
  -1, i, -i
 \right\}.
\eq
We call the three roots ``singular points'' and denote them by 
$z_{\mathrm{sing}}^{(1)}=-1$, $z_{\mathrm{sing}}^{(2)}=i$, and $z_{\mathrm{sing}}^{(3)}=-i$. 
We have 
\bq
 p'\left(z\right)
 & = & 3 z^2 + 2 z + 1.
\eq
The critical points are given by the roots of $p'(z)$, hence
we have two critical points $z_{\mathrm{crit}}^{(1)}$ and $z_{\mathrm{crit}}^{(2)}$.
Therefore
\bq
 \mbox{critical points}
 & = &
 \left\{
  z_{\mathrm{crit}}^{(1)}, z_{\mathrm{crit}}^{(2)}
 \right\}
 \; = \; 
 \left\{
  - \frac{1}{3} + \frac{i}{3} \sqrt{2}, 
  - \frac{1}{3} - \frac{i}{3} \sqrt{2}, 
 \right\}.
\eq
The Morse function is given by
\bq
 h\left(x,y\right)
 & = &
 - \gamma \ln\left|p\left(x+iy\right)\right|
 \nonumber \\
 & = &
 - \frac{1}{10} \ln\left| x^6+3x^4y^2+3x^2y^4+y^6+2x^5+4x^3y^2+2xy^4+3x^4+2x^2y^2-y^4
 \right. \nonumber \\ & & \left.
                          +4x^3-4xy^2+3x^2-y^2+2x+1 \right|.
\eq
The left picture of fig.~\ref{chapter_iterated_integrals:fig_singular_and_critical_points} shows the location of the singular points and the location of the critical points.
\begin{figure}
\begin{center}
\includegraphics[scale=0.6]{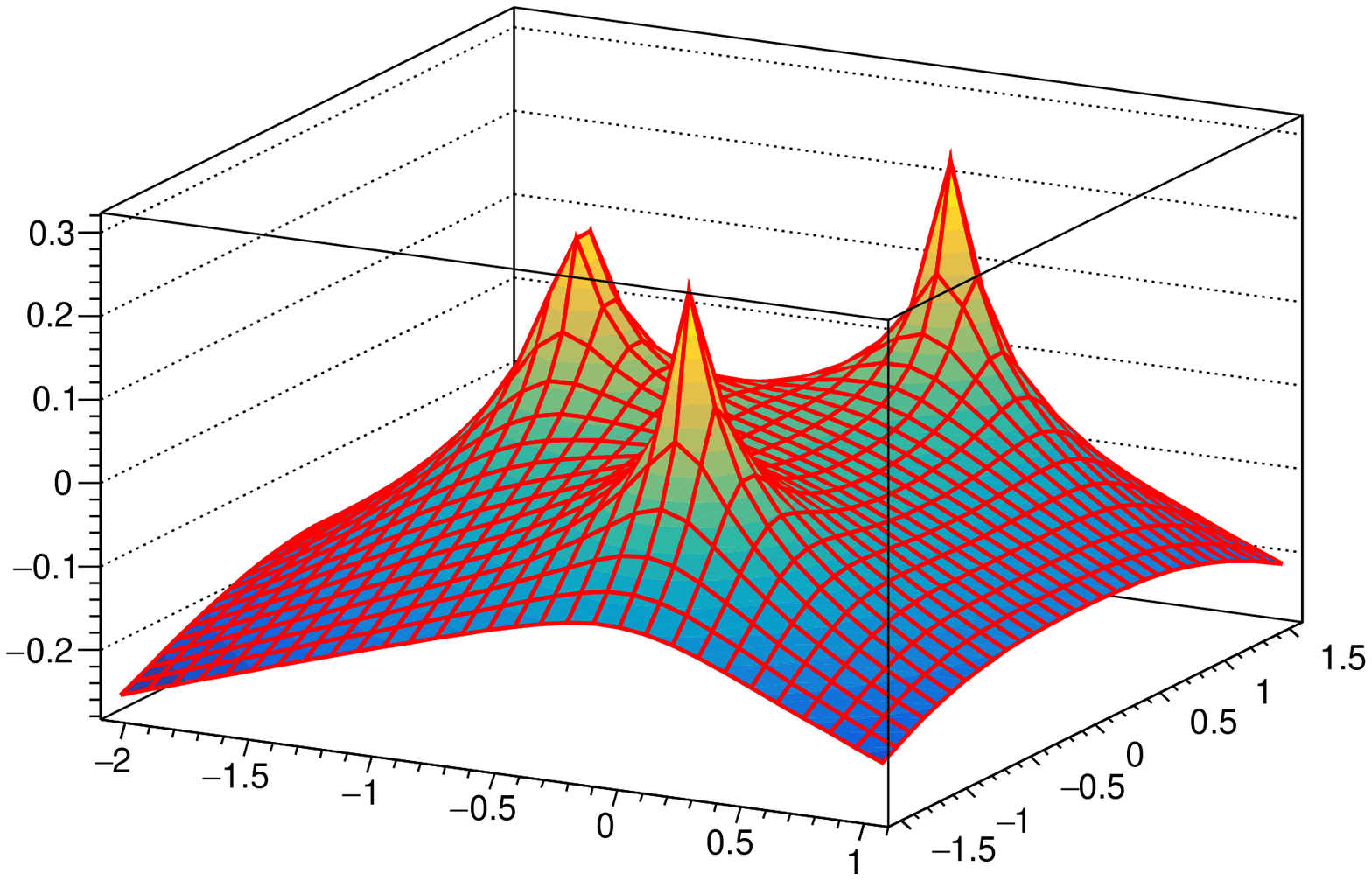}
\end{center}
\caption{\label{chapter_iterated_integrals:fig_morse_root}
The Morse function $h(x,y)$ plotted in the range $x\in [-2,1]$ and $y\in [-1.5,1.5]$.
The Morse function $h$ tends to $+\infty$ at the three singular points 
$z_{\mathrm{sing}}^{(1)}=-1$, $z_{\mathrm{sing}}^{(2)}=i$ and $z_{\mathrm{sing}}^{(3)}=-i$.
}
\end{figure}
Fig.~\ref{chapter_iterated_integrals:fig_morse_root} shows a plot of the Morse function $h$.
The right picture of fig.~\ref{chapter_iterated_integrals:fig_singular_and_critical_points} shows a sketch of the Lefschetz thimbles ${\mathcal C}_1$,
${\mathcal C}_2$, ${\mathcal D}_1$ and ${\mathcal D}_2$.

\begin{digression} {\bf Morse theory}
\\
Morse theory studies the critical points of a real function \cite{Milnor_Morse_Theory}.
There is also a complex analogue, called Picard–Lefschetz theory.

Let $M$ be a compact manifold and $f : M \rightarrow {\mathbb R}$ a smooth real valued function on $M$.
Coordinates on $M$ are denoted by $x_1,\dots,x_n$.

A critical point of $f$ is a point $x \in M$, where all partial derivatives vanish
\bq
 \frac{\partial f}{\partial x_1}
 \; = \;
 \dots
 \; = \; 
 \frac{\partial f}{\partial x_n}
 & = & 
 0,
\eq
or $df=0$ in short.
A critical point $x$ is called non-degenerate, if
\bq
 \det\left( \frac{\partial^2 f}{\partial x_i \partial x_j} \right) & \neq & 0.
\eq
Let $x^{(0)}$ be a non-degenerate critical point. 
In a neighbourhood of $x^{(0)}$ we may choose a coordinate system $(x_1',\dots,x_n')$ such that
\bq
 f
 & = & 
 f\left(x^{(0)}\right)
 - x_{1}'^2
 - \dots
 - x_{\lambda}'^2
 + x_{\lambda+1}'^2
 + \dots
 + x_n'^2
 + {\mathcal O}\left(x'^3\right),
\eq
i.e. there are $\lambda$ downward directions and $(n-\lambda)$ upward directions.
As $x^{(0)}$ is assumed to be a non-degenerate critical point there are no flat directions.
This implies that a non-degenerate critical point is isolated.
The number $\lambda$ of downward directions is called the 
\index{Morse index}
{\bf Morse index} 
of the critical point.
We denote the number of critical points with Morse index $\lambda$ by $C_\lambda$.
The function $f$ is called a 
\index{Morse function}
{\bf Morse function} if all critical points of $f$ are non-degenerate.
Morse theory relates the number of critical points to the topology of $M$.
The relation is provided by the Morse inequalities.
The Morse inequalities for a Morse function $f$ read 
\bq
\label{chapter_iterated_integrals:Morse_inequalities}
 \sum\limits_{k=0}^\lambda \left(-1\right)^{\lambda-k} \dim H_k\left(M\right)
 & \le &
 \sum\limits_{k=0}^\lambda \left(-1\right)^{\lambda-k} C_k.
\eq
The Euler characteristic of $M$ is defined by
\bq
 \chi\left(M\right)
 & = & 
 \sum\limits_{k=0}^n \left(-1\right)^k \dim H_k\left(M\right).
\eq
The Morse inequalities imply
\bq
\label{chapter_iterated_integrals:relation_Euler_characteristic_Morse}
 \chi\left(M\right)
 & = & 
 \sum\limits_{k=0}^n \left(-1\right)^k C_k.
\eq
\bs
{\it \refstepcounter{exercise}
{\bf Exercise \theexercise}: 
Assume $C_{\lambda+1}=C_{\lambda-1}=0$. Show that this implies
\bq
 \dim H_{\lambda+1}\left(M\right) \; = \; 0,
 \;\;\;\;\;\;
 \dim H_{\lambda}\left(M\right) \; = \; C_\lambda,
 \;\;\;\;\;\;
 \dim H_{\lambda-1}\left(M\right) \; = \; 0.
\eq
}
\es

\noindent
\bs
{\it \refstepcounter{exercise}
{\bf Exercise \theexercise}: 
Derive eq.~(\ref{chapter_iterated_integrals:relation_Euler_characteristic_Morse}) from eq.~(\ref{chapter_iterated_integrals:Morse_inequalities}).
}
\es
\end{digression}

\subsection{Computation of intersection numbers}
\label{chapter_iterated_integrals:computation_intersection_numbers}

As the integral appearing in eq.~(\ref{chapter_iterated_integrals:def_intersection_number}) 
in the definition of the intersection number of twisted cocycles is not the most practical way to compute
intersection numbers, let us now turn to a more efficient method to compute intersection numbers.
With a few technical assumptions, outlined in \cite{Mizera:2019gea,Frellesvig:2019uqt,Weinzierl:2020xyy}
we may compute multivariate intersection numbers in $n$ variables $z_1, \dots z_n$ recursively by splitting
the problem into the computation of an intersection number in $(n-1)$ variables $z_1, \dots, z_{n-1}$
and the computation of a (generalised) intersection number in the variable $z_n$.
By recursion, we therefore have to compute only (generalised) intersection numbers in a single variable $z_i$.
This reduces the multivariate problem to an univariate problem.

Let us comment on the word ``generalised'' intersection number:
We only need to discuss the univariate case.
Consider two cohomology classes $\langle \varphi_L |$ and $| \varphi_R \rangle$.
Representatives $\varphi_L$ and $\varphi_R$ for the two cohomology classes
$\langle \varphi_L |$ and $| \varphi_R \rangle$ are in the univariate case differential one-forms and of the form as
in eq.~(\ref{chapter_iterated_integrals:representative_left}) or eq.~(\ref{chapter_iterated_integrals:representative_right}).
We may view the representatives $\varphi_L$ and $\varphi_R$,
the cohomology classes $\langle \varphi_L |$ and $| \varphi_R \rangle$, and the twist $\omega$ 
as scalar quantities.

Consider now a vector of $\nu$ differential one-forms $\varphi_{L,j}$ in the variable $z$, 
where $j$ runs from $1$ to $\nu$.
Similar, consider for the dual space a $\nu$-dimensional vector $\varphi_{R,j}$
and generalise $\omega$ to a $(\nu\times \nu)$-dimensional matrix $\Omega$.
The equivalence classes $\langle \varphi_{L,j} |$ and $| \varphi_{R,j} \rangle$ are now defined
by
\bq
\label{chapter_iterated_integrals:invariance_Omega}
 \varphi_{L,j}' \; = \; \varphi_{L,j} + \partial_z \xi_j + \xi_i \Omega_{i j}
 & \;\; \mbox{and} \;\; &
 \varphi_{R,j}' \; = \; \varphi_{R,j} + \partial_z \xi_j - \Omega_{j i} \xi_i,
\eq
for some zero-forms $\xi_j$ (i.e. functions).
Readers familiar with gauge theories will certainly recognise that the generalisation is exactly the same
step as going from an Abelian gauge theory (like QED) to a non-Abelian gauge theory (like QCD).

Let us now set up the notation for the recursive structure.
We fix an ordered sequence $(z_{\sigma_1},\dots,z_{\sigma_n})$, indicating that we first integrate out $z_{\sigma_1}$,
then $z_{\sigma_2}$, etc..
Without loss of generality we will always consider the order $(z_1,\dots,z_n)$, unless indicated otherwise.

For $i=1,\dots,n$ we consider a fibration $E_i : {\mathbb C}^n \rightarrow B_i$
with total space ${\mathbb C}^n$,
fibre $V_i = {\mathbb C}^i$ parametrised by the coordinates $(z_1,\dots,z_i)$
and base $B_i = {\mathbb C}^{n-i}$ parametrised by the coordinates $(z_{i+1},\dots,z_n)$.
The covariant derivative splits as
\bq
 \nabla_\omega 
 & = &
 \nabla_\omega^{({\bf i}),F} + \nabla_\omega^{({\bf i}),B},
\eq
with
\bq
 \nabla_\omega^{({\bf i}),F}
 \; = \;
 \sum\limits_{j=1}^i dz_j \left( \frac{\partial}{\partial z_j} + \omega_j \right),
 & &
 \nabla_\omega^{({\bf i}),B}
 \; = \;
 \sum\limits_{j=i+1}^n dz_j \left( \frac{\partial}{\partial z_j} + \omega_j \right).
\eq
One sets
\bq
 \omega^{({\bf i})}
 & = &
 \sum\limits_{j=1}^i \omega_j dz_j.
\eq
Clearly, for $i=n$ we have
\bq
 \omega^{({\bf n})}
 \; = \; 
 \omega, 
 & &
 \nabla_\omega^{({\bf n}),F}
 \; = \;
 \nabla_\omega.
\eq
We may now study for each $i$ the twisted cohomology group in the fibre,
defined by replacing $\omega$ with $\omega^{({\bf i})}$.
The additional variables $(z_{i+1},\dots,z_n)$ are treated as parameters, that is to say
we consider all polynomials as polynomials with coefficients in $\tilde{\mathbb F}={\mathbb F}(z_{i+1},\dots,z_n)$.
For each $i$ only the $i$-th cohomology group is of interest and 
for simplicity we write
\bq
 H^{({\bf i})}_\omega
 \; = \;
 H^{i}_{\omega^{({\bf i})}},
 & &
 \left( H^{({\bf i})}_\omega \right)^\ast
 \; = \;
 \left( H^{i}_{\omega^{({\bf i})}} \right)^\ast.
\eq
We denote the dimensions of the twisted cohomology groups by
\bq 
 \nu_{\bf i} & = & \dim H^{({\bf i})}_\omega \; = \; \dim \left( H^{({\bf i})}_\omega \right)^\ast.
\eq
Let 
$\langle e^{({\bf i})}_j |$ with $1\le j \le \nu_{\bf i}$
be a basis of $H^{({\bf i})}_\omega$ and let
$| h^{({\bf i})}_j \rangle$ with $ 1 \le j \le \nu_{\bf i}$
be a basis of $( H^{({\bf i})}_\omega )^\ast$.
We denote the $(\nu_{\bf i} \times \nu_{\bf i})$-dimensional intersection matrix by $C_{\bf i}$.
The entries are given by
\bq
 \left( C_{\bf i} \right)_{j k}
 & = &
 \left\langle e^{({\bf i})}_{j} \right| \left. h^{({\bf i})}_{k} \right\rangle.
\eq
The matrix $C_{\bf i}$ is invertible.
We denote by $| d^{({\bf i})}_j \rangle$ with $ 1 \le j \le \nu_{\bf i}$
the dual basis with respect to $\langle e^{({\bf i})}_j |$.
From eq.~(\ref{chapter_iterated_integrals:construction_dual_basis}) the dual basis is given by
\bq
 \left| d^{({\bf i})}_j \right\rangle
 & = &
 \left| h^{({\bf i})}_k \right\rangle
 \left( C_{\bf i}^{-1} \right)_{kj},
\eq
and satisfies
\bq
 \left\langle e^{({\bf i})}_{j} \right| \left. d^{({\bf i})}_{k} \right\rangle
 & = &
 \delta_{j k}.
\eq
The essential step in the recursive approach is to expand 
the twisted cohomology class $\langle \varphi^{({\bf n})}_L | \in H^{({\bf n})}_\omega$ 
in the basis of $H^{({\bf n-1})}_\omega$
\bq
\label{chapter_iterated_integrals:expansion_varphi_L}
 \left\langle \varphi^{({\bf n})}_L \right|
 & = &
 \sum\limits_{j=1}^{\nu_{\bf n-1}}
 \left\langle \varphi^{({\bf n})}_{L,j} \right| \wedge \left\langle e^{({\bf n-1})}_j \right|,
\eq
and to expand
$| \varphi^{({\bf n})}_R \rangle \in ( H^{({\bf n})}_\omega )^\ast$ in the dual basis of $( H^{({\bf n-1})}_\omega )^\ast$:
\bq
\label{chapter_iterated_integrals:expansion_varphi_R}
 \left| \varphi^{({\bf n})}_R \right\rangle
 & = &
 \sum\limits_{j=1}^{\nu_{\bf n-1}}
 \left| d^{({\bf n-1})}_j \right\rangle
 \wedge
 \left| \varphi^{({\bf n})}_{R,j} \right\rangle.
\eq
Classes in $H^{({\bf n-1})}_\omega$ are represented by rational functions in $z_1, \dots, z_n$ times 
$dz_{n-1} \wedge \dots \wedge dz_1$.
The coefficients $\langle \varphi^{({\bf n})}_{L,j} |$ and $| \varphi^{({\bf n})}_{R,j} \rangle$ are one-forms proportional to $dz_n$
and independent of $z_1,\dots,z_{n-1}$.
They are given by
\bq
 \left\langle \varphi^{({\bf n})}_{L,j} \right|
 \; = \;
 \left\langle \varphi^{({\bf n})}_L \left| d^{({\bf n-1})}_{j} \right. \right\rangle,
 & &
 \left| \varphi^{({\bf n})}_{R,j} \right\rangle
 \; = \;
 \left\langle \left. e^{({\bf n-1})}_{j} \right| \varphi^{({\bf n})}_R \right\rangle.
\eq
The coefficients $\langle \varphi^{({\bf n})}_{L,j}|$ and $| \varphi^{({\bf n})}_{R,j} \rangle$ 
are obtained by computing only
intersection numbers in $(n-1)$ variables.
This is compatible with the recursive approach.
It also shows that the coefficients do not depend on the variables $(z_1,\dots,z_{n-1})$, as these
variables are integrated out.
Let us define a $(\nu_{\bf n-1} \times \nu_{\bf n-1})$-matrix $\Omega^{({\bf n})}$ by
\bq
\label{chapter_iterated_integrals:def_Omega}
 \Omega^{({\bf n})}_{ij}
 & = & 
 \left\langle \left(\partial_{z_n}+\omega_{n}\right) e^{({\bf n-1})}_i \right| \left. d^{({\bf n-1})}_{j} \right\rangle
 \; = \;
 -
 \left\langle e^{({\bf n-1})}_{i} \right| \left. \left(\partial_{z_n}-\omega_{n}\right) d^{({\bf n-1})}_j  \right\rangle.
\eq
The invariance of the original class $\langle \varphi_L |$ under a transformation 
as in eq.~(\ref{chapter_iterated_integrals:invariance_nabla_omega})
translates into the invariance of the vector of coefficients $\langle \varphi_{L,j}|$
as in eq.~(\ref{chapter_iterated_integrals:invariance_Omega}).

The algorithm for computing a multivariate intersection number
consists of three steps:
\begin{enumerate}
\item Recursive approach: The algorithm integrates out one variable at a time.
This part has been outlined above.
It has the advantage to reduce a multivariate problem to a univariate problem.
\item Reduction to simple poles: In general we deal in cohomology with equivalence classes.
We may replace a representative
of an equivalence class with higher poles with an equivalent representative
with only simple poles.
This is similar to integration-by-part reduction.
However, let us stress that the involved systems of linear
equations are usually significantly smaller compared to standard integration-by-part reduction.
\item Evaluation of the intersection number as a global residue. 
Having reduced our objects to simple poles we may use a mathematical theorem 
which states that in this case the intersection number equals a global residue.
The theorem does not hold for higher poles, therefore step 2 is required.
The global residue is easily computed
and does not involve algebraic extensions like square roots.
\end{enumerate}
Let us now fill in the technical details:
We would like to compute the intersection number
\bq
 \left\langle \varphi^{({\bf n})}_L \right| \left. \varphi^{({\bf n})}_R \right\rangle_\omega,
\eq
where $\varphi^{({\bf n})}_L$ and $\varphi^{({\bf n})}_R$ are differential $n$-forms in the variables
$z_1,\dots,z_n$.
Expanding $\langle \varphi^{({\bf n})}_L |$ as in eq.~(\ref{chapter_iterated_integrals:expansion_varphi_L}),
$| \varphi^{({\bf n})}_R \rangle$ as in eq.~(\ref{chapter_iterated_integrals:expansion_varphi_L})
and using the fact that $\langle e^{({\bf n-1})}_j |$ and $| d^{({\bf n-1})}_j \rangle$
are dual bases of $H^{({\bf n-1})}_\omega$ and $( H^{({\bf n-1})}_\omega )^\ast$, respectively, reduces the problem to
\bq
\label{chapter_iterated_integrals:intersection_coeffs}
 \left\langle \varphi^{({\bf n})}_L \right| \left. \varphi^{({\bf n})}_R \right\rangle_\omega
 & = &
 \sum\limits_{j=1}^{\nu_{\bf n-1}}
 \left\langle \varphi^{({\bf n})}_{L,j} \left| \varphi^{({\bf n})}_{R,j} \right. \right\rangle_\Omega.
\eq
The right-hand side is an univariate generalised intersection number in the variable $z_n$.

In the next step we reduce the vector of coefficients
$\langle \varphi^{({\bf n})}_{L,j}|$ and $| \varphi^{({\bf n})}_{R,j}\rangle$
to a form where only simple poles in the variable $z_n$ occur.
A rational function in the variable $z_n$
\bq
 r\left(z_n\right)
 & = &
 \frac{P\left(z_n\right)}{Q\left(z_n\right)},
 \;\;\;\;\;\;\;\;\;
 P, Q \; \in \; {\mathbb F}\left[z_n\right]
 \;\;\;\;\;\;\;\;\;
 \gcd\left(P,Q\right) \; = \; 1,
\eq
has only simple poles if $\deg P < \deg Q$ and if in the partial fraction decomposition each
irreducible polynomial in the denominator occurs only to power $1$.
The condition $\deg P < \deg Q$ ensures that there are no higher poles at infinity.

It is sufficient to discuss the reduction to simple poles for a $\nu$-dimensional vector $\hat{\varphi}_j$
($1 \le j \le \nu$)
which transforms as
\bq
\label{chapter_iterated_integrals:generic_invariance}
 \hat{\varphi}_{j}
 & \rightarrow &
 \hat{\varphi}_{j}
 +
 \left( \delta_{j k} \partial_{z_n} + \Omega_{jk} \right) \xi_k.
\eq
The reduction of $\langle \varphi^{({\bf n})}_{L,j}|$ is then achieved by setting $\Omega = (\Omega^{({\bf n})})^T$,
the reduction of $| \varphi^{({\bf n})}_{R,j}\rangle$ is achieved by setting $\Omega = -\Omega^{({\bf n})}$.
In both case we have $\nu=\nu_{\bf n-1}$.

We first treat poles at infinity:
Assume that $\Omega$ has only simple poles and that
the vector $\hat{\varphi}_{j}$ 
has a pole of order $o > 1$ at infinity.
A transformation as in eq.~(\ref{chapter_iterated_integrals:generic_invariance}) with the seed
\bq
 \xi_j\left(z_n\right) & = & c_j z_n^{o-1},
 \;\;\;\;\;\;\;\;\;
 c_j \; \in \; {\mathbb F}
\eq
reduces the order of the pole at infinity, provided the linear system obtained from 
the condition that the $\nu$ equations
\bq
\label{eq_seed_inf}
 \hat{\varphi}_{j}
 +
 \left( \delta_{j k} \partial_{z_n} + \Omega_{jk} \right) \xi_k
\eq
have only poles of order $(o-1)$ at infinity 
yield a solution for the $\nu$ coefficients $c_j$.
Furthermore, this gauge transformation does not introduce higher poles elsewhere.

The procedure is only slightly more complicated for higher poles at finite points.
Assume that $\Omega$ has only simple poles.
Let $q \in {\mathbb F}[z_n]$ be an irreducible polynomial appearing in the denominator of the partial fraction
decomposition of the $\hat{\varphi}_{j}$'s at worst to the power $o$.
A transformation as in eq.~(\ref{chapter_iterated_integrals:generic_invariance}) with the seed
\bq
 \xi_j\left(z_n\right) & = & \frac{1}{q^{o-1}} \sum\limits_{k=0}^{\mathrm{deg}(q)-1} c_{j,k} \; z_n^k,
 \;\;\;\;\;\;\;\;\;
 c_{j,k} \; \in \; {\mathbb F}.
\eq
reduces the order, provided the linear system obtained from 
the condition that in the partial fraction decomposition of 
\bq
\label{chapter_iterated_integrals:eq_seed_finite_point}
 \hat{\varphi}_{j}
 +
 \left( \delta_{j k} \partial_{z_n} + \Omega_{jk} \right) \xi_k
\eq
terms of the form $z_n^k/q^o$ are absent (with $0 \le k \le \mathrm{deg}(q)-1$)
yield a solution for the $(\nu \cdot \mathrm{deg}(q))$ coefficients $c_{j,k}$.
Furthermore, this gauge transformation does not introduce higher poles elsewhere.

In the third step we relate the intersection number to a global residue.
Let's assume that $\Omega^{({\bf n})}$, $\langle \varphi^{({\bf n})}_{L,j} |$ and $| \varphi^{({\bf n})}_{R,j} \rangle$
have at most only simple poles in $z_n$.
Define two polynomials $P$ and $Q$ by
\bq
\label{chapter_iterated_integrals:def_P_Q}
 \det\left(\Omega^{({\bf n})}\right) & = & \frac{P}{Q},
 \;\;\;\;\;\;\;\;\;
 P, Q \; \in \; {\mathbb F}\left[z_n\right],
 \;\;\;\;\;\;\;\;\;
 \gcd\left(P,Q\right) \; = \; 1,
\eq
and denote by $\mathrm{adj} \; \Omega^{({\bf n})}$ the adjoint matrix of $\Omega^{({\bf n})}$.
This matrix satisfies
\bq
 \Omega^{({\bf n})} \cdot \left( \mathrm{adj} \; \Omega^{({\bf n})} \right)
 \; = \;
 \left( \mathrm{adj} \; \Omega^{({\bf n})} \right) \cdot \Omega^{({\bf n})}
 \; = \;
 \det\left(\Omega^{({\bf n})}\right) \cdot {\bf 1}.
\eq
Let further
\bq
 Y & = & \left\{ \; z_n \in {\mathbb C} \; | \; P\left(z_n\right) = 0 \; \right\}.
\eq
Then
\bq
\label{chapter_iterated_integrals:global_residue_vector_case}
 \left\langle \varphi_L \right| \left. \varphi_R \right\rangle_\omega
 & = &
 - \mathrm{res}_{Y}\left( Q \; \hat{\varphi}_{L,i} \left( \mathrm{adj} \; \Omega^{({\bf n})} \right)_{ i j} \hat{\varphi}_{R,j} \right).
\eq

\begin{digression} {\bf Computation of a global residue}
\\
Consider $n$ meromorphic functions $f_1, f_2, \dots, f_n$ of $n$ variables $x_1,\dots,x_n$
and assume that the system of equations
\bq
 f_1\left(x\right) 
 \; = \; 
 f_2\left(x\right) 
 \; = \; 
 \dots
 f_n\left(x\right) 
 \; = \; 
 0
\eq
has as solutions a finite number of isolated points $x^{(j)}=(x_1^{(j)}, ..., x_n^{(j)})$,
where $j$ labels the individual solutions.
Denote by 
\bq
 Y_j & = & \left\{ x\in \hat{\mathbb C}^n | f_j\left(x\right) = 0 \right\},
 \;\;\;\;\;\;
 1 \; \le \; j \; \le \; n,
\eq
with $\hat{\mathbb C} = {\mathbb C} \cup \{\infty\}$.
In eq.~(\ref{chapter_iterated_integrals:global_residue_2}) we defined the global residue
\bq
 \mathrm{res}_{Y_1,...,Y_n}\left( g \right)
\eq
of a meromorphic function $g$, regular at the solutions $x^{(j)}$, as a sum over the local residues
at the solutions $x^{(j)}$.
Let us now assume that $g$ is a rational function.
The local residues may involve algebraic extensions (i.e. roots), however the global residue does not.
We may compute the global residue without the need to introduce algebraic extensions as follows:

Consider the ring $\mathrm{R}={\mathbb C}[x_1,...,x_n]$
and the ideal $I=\lideal f_1, \dots, f_n \rideal$.
The zero locus of $f_1=\dots=f_n=0$ is a zero-dimensional variety.
It follows that the quotient ring $\mathrm{R}/I$ is a finite-dimensional ${\mathbb C}$-vector space.
Let $\{e_i\}$ be a basis of this vector space and let $P_1, P_2 \in \mathrm{R}/I$
be two polynomials (i.e. vectors) in this vector space.
A theorem of algebraic geometry states that the global residue defines
a {\bf symmetric non-degenerate inner product} \cite{Griffiths:book}:
\bq
 \left\langle P_1, P_2 \right\rangle
 & = &
 \mathrm{res}_{Y_1,...,Y_n}\left( \; P_1 \cdot P_2 \right).
\eq
Since the inner product is non-degenerate there exists a {\bf dual basis} $\{d_i\}$ 
with the property
\bq
 \left\langle e_i, d_j \right\rangle & = & \delta_{ij}.
\eq
To compute the global residue of a polynomial $P(z)$ we therefore obtain the following method:
We express $P$ in the basis $\{e_i\}$ and $1$ in the dual basis $\{d_i\}$:
\bq
 P \; = \; \sum\limits_i \alpha_i e_i, 
 \;\;\;
 1 \; = \; \sum\limits_i \beta_i d_i, 
 \;\;\;
 \;\;\;
 \alpha_i, \beta_i \in {\mathbb C}.
\eq
We then have
\bq
\label{chapter_iterated_integrals:global_residue_polynomial}
 \mathrm{res}_{Y_1,...,Y_n}\left( \; P \; \right)
 \;\; = \;\;
 \mathrm{res}_{Y_1,...,Y_n}\left( \; P \cdot 1 \; \right)
 \;\; = \;\;
 \sum\limits_i \sum\limits_j \alpha_i \beta_j \left\langle e_i, d_j \right\rangle
 \;\; = \;\;
 \sum\limits_i \alpha_i \beta_i.
\eq
Given a basis $\{e_i\}$ and the associated dual basis $\{d_i\}$, 
eq.~(\ref{chapter_iterated_integrals:global_residue_polynomial}) allows us to compute the global residue 
of a polynomial $P$ without knowing the solutions $x^{(j)}$.
Eq.~(\ref{chapter_iterated_integrals:global_residue_polynomial}) simplifies, if the dual basis contains
a constant polynomial $d_{i_0}=c$.
We then have
\bq
 \mathrm{res}_{Y_1,...,Y_n}\left( \; P \; \right)
 & = &
 \frac{\alpha_{i_0}}{c}.
\eq
We would like to compute the global residue of the rational function $g$.
Eq.~(\ref{chapter_iterated_integrals:global_residue_polynomial}) is not yet directly applicable to our problem, 
since $g(x)$ is a rational function, not a polynomial.
We write $g(x)=P(x)/Q(x)$. 
We may assume that $\{f_1,...,f_n,Q\}$ have no common zeros,
since we assumed that $g$ is regular on the solutions $x^{(j)}$.
Hilbert's Nullstellensatz guarantees then 
that there exist polynomials $p_1, ..., p_n, \tilde{Q} \in \mathrm{R}$, such that
\bq
 p_1 f_1 + \dots +p_n f_n + \tilde{Q} Q & = & 1.
\eq
We call $\tilde{Q}$ the polynomial inverse of $Q$ with 
respect to $\lideal f_1, \dots, f_n \rideal$.
For the global residue we have
\bq
 \mathrm{res}_{Y_1,...,Y_n}\left( \; g \; \right)
 \;\; = \;\;
 \mathrm{res}_{Y_1,...,Y_n}\left( \; \frac{P}{Q} \; \right)
 \;\; = \;\;
 \mathrm{res}_{Y_1,...,Y_n}\left( \; P \tilde{Q} \; \right).
\eq
Note $P \tilde{Q}$ is a polynomial. We have therefore reduced the case of a rational function
$g(x)$ to the polynomial case $P(x) \tilde{Q}(x)$.

The above calculations can be carried out with the help of a 
Gr\"obner basis for the ideal $I$ \cite{Cattani:2005,Sogaard:2015dba}.
\end{digression}

Let us apply these ideas to the computation of the global residue
in eq.~(\ref{chapter_iterated_integrals:global_residue_vector_case}).
As we are in the univariate case, the calculation simplifies significantly.
We have to compute the global residue of a rational function in $z_n$.
Let us write
\bq
 \frac{P_g}{Q_g}
 & = &
 Q \; \hat{\varphi}_{L,i} \left( \mathrm{adj} \; \Omega^{({\bf n})} \right)_{ i j} \hat{\varphi}_{R,j},
 \;\;\;\;\;\;
 P_g,Q_g \; \in \; {\mathbb F}[z_n].
\eq
We may assume $\gcd(P, Q_g) \; = \; \gcd(P_g, Q_g) \; = \; 1$.
Let $\nu=\deg P$ and let
$\tilde{Q}_g$ be the polynomial inverse of $Q_g$ with respect to the ideal $\lideal P \rideal$.
Then
\bq
 \mathrm{res}_{Y}\left( \frac{P_g}{Q_g} \right)
 & = &
 \frac{a_\nu}{c_\nu},
\eq
where $a_\nu$ is the coefficient of $z^{\nu-1}$ in the reduction of $P_g \tilde{Q}_g$ modulus $P$
and $c_\nu$ is the coefficient of $z^{\nu}$ of $P$.

\subsection{Inner product for Feynman integrals}
\label{chapter_iterated_integrals:inner_product}

Let us now make contact with Feynman integrals \cite{Mastrolia:2018uzb,Frellesvig:2019kgj,Frellesvig:2019uqt,Mizera:2019vvs,Mizera:2019ose,Frellesvig:2020qot,Chen:2020uyk,Caron-Huot:2021xqj,Caron-Huot:2021iev}.
From section~\ref{chapter_basics:sect_Baikov_representation} we recall the Baikov representation:
\bq
\label{chapter_iterated_integrals:baikov_representation}
 I_{\nu_1 \dots \nu_{n}} & = &
 C
 \int\limits_{\mathcal C} d^{\NV}z \;
 \left[{\mathcal B}\left(z\right)\right]^{\frac{D-\loopnumber-{\nexternalindependent}-1}{2}}
 \prod\limits_{s=1}^{\NV} z_s^{-\nu_s}.
\eq
$C$ is a prefactor (given in eq.~(\ref{chapter_basics:Baikov_representation})) and not relevant for the further discussion.
${\mathcal B}(z)$ denotes the Baikov polynomial. It is obtained from a Gram determinant.
The domain of integration is such that the Baikov polynomial vanishes on the boundary of the integration region.
We note that the indices $\nu_s$ enter only the last factor.
Eq.~(\ref{chapter_iterated_integrals:baikov_representation})
is an integral of the form as in eq.~(\ref{chapter_iterated_integrals:def_integral}) with
\bq
 \varphi & = & \left( \prod\limits_{s=1}^{\NV} z_s^{-\nu_s}\right) d^{\NV}z 
\eq
and
\bq
 u & = & \left[{\mathcal B}\left(z\right)\right]^{\frac{D-\loopnumber-{\nexternalindependent}-1}{2}},
 \;\;\;\;\;\;
 \omega \; = \; d \ln u.
\eq
As the Baikov polynomial vanishes on the boundary of the integration region, the Feynman integral is invariant
under
\bq
 \varphi & \rightarrow & \varphi + \nabla_\omega \xi
\eq
for any $(\NV-1)$-form $\xi$
and we may group the integrands of the Feynman integrals $I_{\nu_1 \dots \nu_n}$ corresponding to different sets of indices $\nu_1, \dots, \nu_n$
into cohomology classes.
The number of independent cohomology classes in $H_\omega^{\NV}$ is finite,
and we may express any $\varphi$ as a linear combination
of a basis of $H_\omega^{\NV}$.
Let $\langle e_j |$ be a basis of $H_\omega^{\NV}$ and $| d_j \rangle$ a basis of the dual cohomology group
$( H^{\NV}_\omega )^\ast$, chosen such that
\bq
 \left\langle e_i | d_j \right\rangle_\omega & = & \delta_{ij}.
\eq
We then have
\bq
 \left\langle \varphi \right|
 & = &
 \sum\limits_j c_j \left\langle e_j \right|,
\eq
where the coefficients are given by the intersection numbers
\bq
 c_j & = & 
 \left\langle \varphi | d_j \right\rangle_\omega.
\eq
This provides an alternative to integration-by-parts reduction.

Note that the dimension of $H_\omega^{\NV}$ can be larger than the number of master integrals, as the latter takes symmetries of integrals
into account, while the former operates on integrands.
This is most easily explained by the simplest example, the one-loop two-point function with two equal internal masses.
This system has two master integrals. A standard choice is $I_{11}$ and $I_{10}$.
By symmetry, the integral $I_{01}$ is identical to $I_{10}$.
At the level of the integrands we have $\dim H_\omega^2=3$.
A basis for $H_\omega^2$ is given by $\langle \varphi_{11} |$, $\langle \varphi_{10} |$, $\langle \varphi_{01} |$
with
\bq
 \varphi_{\nu_1 \nu_2} & = & \frac{dz_2 \wedge dz_1}{z_1^{\nu_1} z_2^{\nu_2}}.
\eq 
The $2$-forms $\varphi_{10}$ and $\varphi_{01}$ 
\bq
 \varphi_{10} \; = \; - \frac{dz_1}{z_1} \wedge dz_2,
 & &
 \varphi_{01} \; = \; - dz_1 \wedge \frac{dz_2}{z_2}.
\eq
are not identical 
(but of course one is obtained from the other up to a sign through the substitution $z_1 \leftrightarrow z_2$),
only the integrals as in eq.~(\ref{chapter_iterated_integrals:def_integral}) give identical results.

Let us now illustrate the technique of intersection numbers by an example.
We consider the double-box integral discussed as example 3 in section~\ref{chapter_iterated_integrals:deriving_the_dgl}.
As a basis of master integrals we take the eight master integrals 
given in eq.~(\ref{chapter_iterated_integrals:precanonical_masters_double_box}).
Suppose we would like to express the integral $I_{1111111(-2)0}$ as a linear combination of the master integrals:
\bq
\label{chapter_iterated_integrals:decomposition_double_box}
 I_{1111111(-2)0}
 & = &
 \sum\limits_{j=1}^8
 c_j I_{{\bm{\nu}}_j},
\eq
with $I_{{\bm{\nu}}_j}$ denoting the master integrals in the order as they appear in
eq.~(\ref{chapter_iterated_integrals:precanonical_masters_double_box}),
e.g. $I_{{\bm{\nu}}_7} = I_{111111100}$ and $I_{{\bm{\nu}}_8} = I_{1111111(-1)0}$.
Let's compute the coefficients $c_7$ and $c_8$ from intersection numbers.
This can be done on the maximal cut, as
\bq
 \mathrm{MaxCut} \; I_{1111111(-2)0}
 & = &
 c_7 \mathrm{MaxCut} \; I_{{\bm{\nu}}_7}
 +
 c_8 \mathrm{MaxCut} \; I_{{\bm{\nu}}_8},
\eq
with the same coefficients $c_7$ and $c_8$ as in eq.~(\ref{chapter_iterated_integrals:decomposition_double_box}).
For simplicity we set $\mu^2=1$.
A Baikov representation of $\mathrm{MaxCut} \; I_{1111111\nu 0}$ is
\bq
\label{chapter_iterated_integrals:maxcut_double_box}
 \mathrm{MaxCut} \; I_{1111111\nu 0}
 & = &
 C
 \int\limits_{\mathcal C} dz_8 \;
 z_8^{-1-2\eps} 
 \left(t-z_8\right)^{-1-\eps}
 \left(s+t-z_8\right)^{\eps}
 z_8^{-\nu}.
\eq
This form of the Baikov representation is obtained within the loop-by-loop approach by first
considering the loop with loop momentum $k_2$ and then the remaining loop with loop momentum $k_1$.
The prefactor $C$ and the integration contour ${\mathcal C}$ are not particularly relevant.
The multi-valued function $u$ is given by
\bq
 u 
 & = & 
 C
 \;
 z_8^{-1-2\eps} 
 \left(t-z_8\right)^{-1-\eps}
 \left(s+t-z_8\right)^{\eps},
\eq
the rational one-form $\varphi_\nu$ (recall that the index $\nu$ in $I_{1111111\nu 0}$ is an integer) by
\bq
 \varphi_\nu
 & = &
 z_8^{-\nu} dz_8.
\eq
The one-form $\omega$ is therefore given by
\bq
 \omega
 & = & d \ln u
 \; = \;
 \left[ - \frac{1+2\eps}{z_8} - \frac{1+\eps}{z_8-t} + \frac{\eps}{z_8-s-t} \right] dz_8.
\eq
The equation $\omega=0$ leads to a quadratic equation for $z_8$, which has two solutions.
We therefore have
\bq
 \dim H^1_\omega & = & 2.
\eq
This is consistent with the fact that there are two master integrals 
($I_{{\bm{\nu}}_7}$ and $I_{{\bm{\nu}}_8}$) in this sector.
The $\varphi$'s corresponding to $I_{{\bm{\nu}}_7}$ and $I_{{\bm{\nu}}_8}$
give us immediately a basis of $H^1_\omega$:
\bq
 e_1 \; = \;  1 \cdot dz_8,
 & &
 e_2 \; = \;  z_8 \cdot dz_8.
\eq
We then compute the dual basis.
We start from an arbitrary basis of $(H^1_\omega)^\ast$, which we take to be
\bq
 h_1 \; = \;  1 \cdot dz_8,
 & &
 h_2 \; = \;  z_8 \cdot dz_8,
\eq
compute the intersection matrix between $(\langle e_1|, \langle e_2|)$ and $(|h_1\rangle, |h_2\rangle)$ 
and obtain the dual basis according to eq.~(\ref{chapter_iterated_integrals:construction_dual_basis}).
We find
\bq
 d_1 
 & = &
 \left[
  \frac{\left(1+\eps\right)\left(2+\eps\right) \left(2z_8-s-2t\right)}{\left(1+2\eps\right)s\left(s+t\right)^2} 
  - \frac{3\left(2z_8-s-2t\right)}{4\left(1+2\eps\right)st^2}
  - \frac{4\left(s+t\right)z_8}{\left(1+\eps\right)s^2t^2}
  + \frac{2\left(2+\eps\right)z_8}{st\left(s+t\right)}
 \right. \nonumber \\
 & & \left.
  + \frac{4\left(t-z_1\right)}{s^2t}
 + \frac{9\left(3+2\eps\right)z_8}{2st^2}
 - \frac{27\left(1+2\eps\right)}{4t^2}
 - \frac{\left(11+30\eps\right)}{2st}
 \right] dz_8,  
 \nonumber \\
 d_2 
 & = &
 \left[
  \frac{3\left(2z_8-s-2t\right)}{2\left(1+2\eps\right)t^2\left(s+t\right)^2}
  - \frac{\left(5+2\eps\right)\left(2z_8-t\right)}{2st\left(s+t\right)^2}
 + \frac{2\left(2+\eps\right)}{st\left(s+t\right)}
 - \frac{6 \eps z_8}{st^2\left(s+t\right)}
 - \frac{\left(3+7\eps\right)z_8}{\left(1+\eps\right)s^2t^2}
 \right. \nonumber \\
 & & \left.
 + \frac{11 z_8}{s^2 t\left(s+t\right)}
 + \frac{9\eps}{st^2}
 - \frac{4}{s^2t}
 + \frac{3}{2st^2}
 \right] dz_8.
\eq
We would like to find the coefficients $c_7$ and $c_8$ in the reduction of $I_{1111111(-2)0}$ to master integrals.
The integral $\mathrm{MaxCut} \; I_{1111111(-2)0}$ corresponds to
\bq
 \varphi_{(-2)}
 & = & 
 z_8^2 \cdot dz_8.
\eq
$c_7$ and $c_8$ are then given by the intersection numbers
\bq
 c_7 
 & = &
 \left\langle \varphi_{(-2)} | d_1 \right\rangle_\omega
 \; = \;
 \frac{2\eps t \left(s+t\right)}{1-2\eps},
 \nonumber \\
 c_8
 & = &
 \left\langle \varphi_{(-2)} | d_2 \right\rangle_\omega
 \; = \;
 \frac{\left(1-4\eps\right)t -3\eps s}{1-2\eps}.
\eq
This agrees with the result obtained from integration-by-parts reduction (compare with exercise~\ref{chapter_iterated_integrals:exercise_doublebox_ibp}).

%% file: transformations.tex
\newpage
\chapter{Transformations of differential equations}
\label{chapter_transformations}

In chapter~\ref{chapter_iterated_integrals} we learned
that the computation of Feynman integrals can be reduced 
to finding appropriate fibre transformations (see section~\ref{chapter_iterated_integrals:fibre_transformation})
and base transformations (see section~\ref{chapter_iterated_integrals:base_transformation}).
However, up to now we didn't discuss methods how to find these fibre and base transformations.

Currently, there is no known method how to do this in full generality.
In this chapter we introduce methods, which allow us to construct the required fibre or base transformation in
special cases.
We focus in this chapter mainly on rational and algebraic transformations.
This covers many Feynman integrals, which evaluate to multiple polylogarithms.
But there are also Feynman integrals, where the required transformations involve transcendental functions.
We discuss an example for this case in chapter~\ref{chapter_elliptics}.

\section{Fibre transformations}
\label{chapter_transformations:fibre_transformation}

We denote by $\vec{I}=(I_{{\bm{\nu}}_1}, \dots, I_{{\bm{\nu}}_{\Nmaster}})^T$
a set of master integrals satisfying the differential equation
\bq
 \left(d+A\right) \vec{I} & = & 0.
\eq
A fibre transformations, given by an
$(\Nmaster \times \Nmaster)$-matrix $U(\eps,x)$, redefines the set of master integrals
as
\bq
 \vec{I}'
 & = &
 U \vec{I},
\eq
and transforms the differential equation to
\bq
 \left(d+A'\right) \vec{I}' & = & 0,
\eq
where $A'$ is related to $A$ by
\bq
 A' & = & U A U^{-1} + U d U^{-1}.
\eq
The goal is to find a transformation $U(\eps,x)$, such that the dependence on $\eps$ of $A'$ is only through
an explicit prefactor $\eps$ as in eq.~(\ref{chapter_iterated_integrals:eps_form}).

In the following subsections we discuss a variety of methods:
We start with block decomposition in subsection~\ref{chapter_transformations:fibre_transformation:block_decomposition}.
This allows us to reduce the original problem involving a $(\Nmaster \times \Nmaster)$-matrix $A$ to matrices of smaller size.
In addition we derive a differential equation for the transformation we are seeking.
In subsection~\ref{chapter_transformations:fibre_transformation:reduction_to_univariate} we reduce a multivariate problem depending on
$\NB$ kinematic variables $x_1,\dots,x_{\NB}$ to an univariate problem depending only on a single kinematic variable $x$.
Obviously, some information is lost in this reduction, but the solution of the simpler univariate problem can be useful
to find a solution for the multivariate problem.
The next three subsections deal with univariate problems:
In subsection~\ref{chapter_transformations:fibre_transformation:picard_fuchs} we convert a system of $\Nmaster$ first-order differential
equations to a higher-order differential equation for one selected master integral. The order of this differential equation
is at most $\Nmaster$. The differential operator of this differential equation is called the Picard-Fuchs operator.
We then study the factorisation properties of the Picard-Fuchs operator when the parameter $D$ denoting the number of space-time dimensions
is an (even) integer.
In subsection~\ref{chapter_transformations:fibre_transformation:magnus} we study the Magnus expansion.
This is particularly useful if the matrix $A$ is linear in $\eps$, i.e. $A=A^{(0)}+\eps A^{(1)}$ with $A^{(0)}$ and $A^{(1)}$ being independent
of $\eps$. 
In subsection~\ref{chapter_transformations:fibre_transformation:moser} we discuss Moser's algorithm.
This algorithm is at the core of several computer programs for finding an appropriate fibre transformation.
In subsection~\ref{chapter_transformations:fibre_transformation:leinartas} we return from the univariate case to the (general)
multivariate case. We discuss the Leinartas decomposition, which can be thought of as a generalisation of partial fraction decomposition
from the univariate case to the multivariate case.
Finally, subsection~\ref{chapter_transformations:maximal_cuts_and_constant_leading_singularities} 
is devoted to maximal cuts and constant leading singularities.
This method allows us often to make an educated guess for a suitable fibre transformation.

\subsection{Block decomposition}
\label{chapter_transformations:fibre_transformation:block_decomposition}

We start with an elementary method \cite{Gehrmann:2014bfa}:
We may order the set of master integrals 
\bq
 \vec{I} & = & (I_{{\bm{\nu}}_1}, \dots, I_{{\bm{\nu}}_{\Nmaster}})^T
\eq
such that $I_{{\bm{\nu}}_1}$ is the simplest integral and $I_{{\bm{\nu}}_{\Nmaster}}$ the most complicated integral.
We may do this with an order criteria as in eq.~(\ref{chapter_iterated_integrals:isp_basis}) or eq.~(\ref{chapter_iterated_integrals:dot_basis}).
Doing so, the matrix $A$ has a lower block-triangular structure.
To give an example, consider the situation with three sectors.
Suppose that the simplest sector has one master integral,
the next sector two master integrals and the most complicated sector one master integral.
The matrix $A$ has then the structure
\bq
\label{chapter_transformations:example_block_matrix}
 A & = &
 \left( \begin{array}{cccc}
 \cellcolor{yellow} A_1                    & 0 & 0                                                              & 0 \\
 \cellcolor{magenta}                       & \multicolumn{2}{>{\columncolor{orange}}c}{}                        & 0 \\
 \cellcolor{magenta} \multirow{-2}{*}{$A_3$} & \multicolumn{2}{>{\columncolor{orange}}c}{ \multirow{-2}{*}{$A_2$} } & 0 \\
 \cellcolor{cyan} A_6                      & \multicolumn{2}{>{\cellcolor{green}}c}{A_5}                        & \cellcolor{red} A_4 \\
 \end{array} \right),
\eq
where only the coloured entries are non-zero.
The blocks on the diagonal, $A_1$, $A_2$ and $A_4$ have the size
$(1 \times 1)$,$(2 \times 2)$ and $(1 \times 1)$, respectively.

In order to find the fibre transformation which transforms the system to an $\eps$-form 
we may split the problem into smaller tasks and first find transformations, which transform a specific block
into an $\eps$-form.
For the example in eq.~(\ref{chapter_transformations:example_block_matrix}) we may do this in the order
$A_1$, $A_2$, $A_3$, $A_4$, $A_5$ and $A_6$.
This has the advantage that for most blocks we may work with matrices of smaller size.
The size of the matrices required for the individual blocks are
\bq
\begin{array}{ll}
 A_1: & 1 \times 1 \\
 A_2: & 2 \times 2 \\
 A_3: & 3 \times 3 \\
 A_4: & 1 \times 1 \\
 A_5: & 3 \times 3 \\
 A_6: & 4 \times 4 \\
\end{array}
\eq
For example, if $\vec{I}=(I_{{\bm{\nu}}_1},I_{{\bm{\nu}}_2},I_{{\bm{\nu}}_3},I_{{\bm{\nu}}_4},)$ the
$(3\times 3)$-system for the block $A_5$ is obtained by using $I_{{\bm{\nu}}_2},I_{{\bm{\nu}}_3},I_{{\bm{\nu}}_4}$
and setting $I_{{\bm{\nu}}_1}$ to zero.
Only for the last block ($A_6$) we need a $(4\times 4)$-system.
We do this bottom-up: We first put the block $A_1$ into an $\eps$-form, then block $A_2$, etc..
In this approach we may assume as we try to find 
a transformation for block $A_i$ that all blocks $A_j$ with $j<i$ have already been put into
an $\eps$-form.

There are two types of blocks: Blocks on the diagonal ($A_1$, $A_2$ and $A_4$ in the example above)
and off-diagonal blocks ($A_3$, $A_5$ and $A_6$ in the example above).
Let's see how a fibre transformation acts on these blocks.
It is sufficient to discuss the case, where $A$ is of the form
\bq
 A
 & = &
 \left( \begin{array}{ccc}
 A_1 & 0 & 0 \\
 A_3 & A_2 & 0 \\
 A_6 & A_5 & A_4 \\
 \end{array} \right)
\eq
and to consider the transformation of the blocks $A_2$ (a diagonal block)
and $A_3$ (an off-diagonal block).
We assume that block $A_1$ is already in $\eps$-form, and the sought-after transformation
should preserve block $A_1$.
The blocks $A_4$, $A_5$ and $A_6$ will be dealt with in a later step.
The transformation for the blocks $A_2$ and $A_3$ is allowed to modify these blocks.

Let's start with block $A_2$.
We consider a transformation of the form
\bq
\label{chapter_transformations:trafo_diagonal_block}
 U
 \; = \;
 \left( \begin{array}{ccc}
 1 & 0 & 0 \\
 0 & U_2 & 0 \\
 0 & 0 & 1 \\
 \end{array} \right),
 & &
 U^{-1}
 \; = \;
 \left( \begin{array}{ccc}
 1 & 0 & 0 \\
 0 & U_2^{-1} & 0 \\
 0 & 0 & 1 \\
 \end{array} \right).
\eq
The transformed $A'$ is given by
\bq
 A'
 & = &
 \left( \begin{array}{ccc}
 A_1 & 0 & 0 \\
 U_2 A_3 & U_2 A_2 U_2^{-1} + U_2 d U_2^{-1} & 0 \\
 A_6 & A_5 U_2^{-1} & A_4 \\
 \end{array} \right).
\eq
Suppose the block $A_2$ contains an unwanted term $F$ and a remainder $R$:
\bq
 A_2 & = & F + R.
\eq
The term $F$ can be removed by a fibre transformation
of the form as in eq.~(\ref{chapter_transformations:trafo_diagonal_block})
with $U_2$ given as a solution of the differential equation
\bq
 d U_2^{-1} & = & - F U_2^{-1}.
\eq
We consider a simple example: Assume that we have only one kinematic variable $x_1=x$ (e.g. $\NB=1$)
and that $A_2$ is of size $(1 \times 1)$ and given by
\bq
 A_2 & = & 
 \left( \frac{1}{x-1} + \frac{2\eps}{x-1} \right) dx.
\eq
We would like to remove the first term $F=dx/(x-1)$ by a fibre transformation.
We have to solve the differential equation
\bq
 \frac{d}{dx} U_2^{-1} + \frac{1}{x-1} U_2^{-1} & = & 0.
\eq
A solution is easily found and given by
\bq
 U_2^{-1} \; = \; \frac{C}{x-1},
 & &
 U_2 \; = \; C^{-1} \left(x-1\right).
\eq
The integration constant $C$ is of no particular relevance, as it corresponds to multiplying a
master integral with a constant prefactor.
We may set $C=1$ and $U_2=x-1$ is the sought-after transformation.

Let us stay with one kinematic variable and $A_2$ of size $(1 \times 1)$.
Let us now consider $F=f(x)dx$ and assume that $f(x)$ is a rational function in $x$.
We have to solve the differential equation
\bq
\label{chapter_transformations:example_dgl_fuchsian}
 \left[ \frac{d}{dx} + f(x) \right] U_2^{-1} & = & 0.
\eq
A solution is easily found and given by
\bq
 U_2^{-1} \; = \; \exp\left[ - \int\limits_{x_0}^x dx' f\left(x'\right) \right],
 & &
 U_2 \; = \; \exp\left[ \int\limits_{x_0}^x dx' f\left(x'\right) \right].
\eq
By using partial fraction decomposition we may write $f(x)$ as a sum of a polynomial, terms with simple poles
and terms with higher poles.
We have with a suitable choice for the integration constant
\begin{align}
\label{chapter_transformations:example_diagonal_block_one_by_one}
 f\left(x\right) 
 & =  
 c_n x^n,
 & 
 U_2\left(x\right)
 & =  
 \exp\left( \frac{c_n x^{n+1}}{n+1} \right),
 \nonumber \\
 f\left(x\right) 
 & = 
 \frac{r_0}{x-z},
 & 
 U_2\left(x\right)
 & = 
 \left(x-z\right)^{r_0},
 \nonumber \\
 f\left(x\right) 
 & = 
 \frac{d_n}{\left(x-z\right)^n},
 & 
 U_2\left(x\right)
 & = 
 \exp\left( - \frac{d_n}{\left(n-1\right) \left(x-z\right)^{n-1}} \right).
\end{align}
A first-order differential equation as in eq.~(\ref{chapter_transformations:example_dgl_fuchsian})
is said to be in 
\index{Fuchsian form}
{\bf Fuchsian form}, 
if $f(x)$ is a rational function in $x$ and if in the partial
fraction decomposition of $f$ polynomial terms and higher poles are absent (i.e. $f$ has only simple poles).
From eq.~(\ref{chapter_transformations:example_diagonal_block_one_by_one}) 
we see that simple poles with integer residues can be removed from a $(1 \times 1)$-block on the diagonal
by a rational fibre transformation.
If the residue is not an integer the fibre transformation is algebraic.

Let us now consider block $A_3$.
At this stage we would like to preserve the blocks $A_1$ and $A_2$.
We consider a transformation of the form
\bq
\label{chapter_transformations:trafo_offdiagonal_block}
 U
 \; = \;
 \left( \begin{array}{ccc}
 1 & 0 & 0 \\
 U_3 & 1 & 0 \\
 0 & 0 & 1 \\
 \end{array} \right),
 & &
 U^{-1}
 \; = \;
 \left( \begin{array}{ccc}
 1 & 0 & 0 \\
 -U_3 & 1 & 0 \\
 0 & 0 & 1 \\
 \end{array} \right).
\eq
The transformed $A'$ is given by
\bq
 A'
 & = &
 \left( \begin{array}{ccc}
 A_1 & 0 & 0 \\
 A_3 - A_2 U_3 + U_3 A_1 - d U_3 & A_2 & 0 \\
 A_6 - A_5 U_3 & A_5 & A_4 \\
 \end{array} \right).
\eq
Suppose the block $A_3$ contains an unwanted term $F$ and a remainder $R$:
\bq
 A_3 & = & F + R.
\eq
The term $F$ can be removed by a fibre transformation
of the form as in eq.~(\ref{chapter_transformations:trafo_offdiagonal_block})
with $U_3$ given as a solution of the differential equation
\bq
\label{chapter_transformations:trafo_offdiagonal_block_diff_eq}
 d U_3 + A_2 U_3 - U_3 A_1 & = & F.
\eq
Let us also consider an example here. We again consider the case of one kinematic variable $x$ (e.g. $\NB=1$).
We further assume that $A_1$ and $A_2$ are both blocks of size $(1 \times 1)$.
Then $A_3$ is also a block of size $(1 \times 1)$.
Assume that $A_1$ and $A_2$ are already in $\eps$-form and given by
\bq
 A_1 \; = \; \frac{\eps dx}{x-1},
 & &
 A_2 \; = \; \frac{2\eps dx}{x-1}.
\eq
Assume further that $F$ is given by
\bq
 F & = & \frac{dx}{\left(x-1\right)^2}.
\eq
We have to solve the differential equation
\bq
 \left[ \frac{d}{dx} + \frac{\eps}{x-1} \right] U_3 & = & \frac{1}{\left(x-1\right)^2}.
\eq
A solution is given by
\bq
 U_3 
 & = &
 \frac{1}{\left(1-\eps\right)\left(1-x\right)}.
\eq

\subsection{Reduction to an univariate problem}
\label{chapter_transformations:fibre_transformation:reduction_to_univariate}

In general our Feynman integrals depend on $\NB$ kinematic variables $x_1, \dots, x_{\NB}$.
If $\NB>1$ we are considering a multivariate problem, if $\NB=1$ we are considering an univariate problem.
Clearly, an univariate problem is simpler than a multivariate problem.

To any multivariate problem we may associate an univariate problem as follows \cite{Adams:2017tga}:
Let $\alpha=[\alpha_1:...:\alpha_{\NB}] \in {\mathbb C} {\mathbb P}^{\NB-1}$ be a point in projective space.
Without loss of generality we work in the chart $\alpha_{\NB}=1$.
We consider a path $\gamma_\alpha : [0,1] \rightarrow {\mathbb C}^n$, indexed by $\alpha$ and parametrised by 
a variable $\lambda$.
Explicitly, we have
\bq
\label{chapter_transformations:def_path}
 x_j\left(\lambda\right) & = & \alpha_j \lambda,
 \;\;\;\;\;\; 1 \le j \le \NB.
\eq
We then view the master integrals as functions of $\lambda$.
In other words, we look at the variation of the master integrals in the direction specified by $\alpha$.
Consider now a set of master integrals $\vec{I}$ with differential equation
\bq
\label{chapter_transformations:reduction_original_system}
 \left( d + A \right) \vec{I} \; = \; 0,
 & &
 A 
 \; = \;
 \sum\limits_{j=1}^{\NB}
 A_{x_j}
 dx_j.
\eq
For the derivative with respect to $\lambda$ we have
\bq
\label{chapter_transformations:diff_eq_lambda}
 \left( \frac{d}{d\lambda} + B_\lambda \right) \vec{I}
 \; = \; 
 0,
 & &
 B_\lambda \; = \;
 \sum\limits_{j=1}^{\NB} \alpha_j A_{x_j}.
\eq
$B$ is a $(\Nmaster \times \Nmaster)$-matrix, whose entries are functions.
Eq.~(\ref{chapter_transformations:diff_eq_lambda}) is now an univariate problem in the variable $\lambda$.
This problem depends on the additional parameters $\alpha=[\alpha_1:\dots:\alpha_{\NB-1}:1]$ specifying the direction
of the path in eq.~(\ref{chapter_transformations:def_path}).
We may now try to find with univariate methods a fibre transformation, which transforms
eq.~(\ref{chapter_transformations:diff_eq_lambda}) into an $\eps$-form.
Let's denote this transformation by $V$:
\bq
 \vec{I}' & = & V \vec{I}.
\eq
$V$ is a function of $\lambda$ and the parameters $\alpha$:
\bq
 V & = & V\left(\alpha_1,...,\alpha_{\NB-1},\lambda\right).
\eq
We recall that we work in the chart $\alpha_{\NB}=1$. 
We may now try to lift the transformation $V$ to the original kinematic space with
coordinates $x_1, \dots, x_{\NB}$.
Let us set
\bq
 U & = & V\left(\frac{x_1}{x_{\NB}},...,\frac{x_{\NB-1}}{x_{\NB}},x_{\NB}\right).
\eq
$U$ defines a transformation in terms of the original variables $x_1$, ..., $x_{\NB}$.

It is important to note that there is no guarantee that the transformation $U$ puts the original system
in eq.~(\ref{chapter_transformations:reduction_original_system}) into an $\eps$-form,
even if the transformation $V$ puts the system in eq.~(\ref{chapter_transformations:diff_eq_lambda}) 
into an $\eps$-form.
The reason is that going from the original system in eq.~(\ref{chapter_transformations:reduction_original_system})
to the simpler univariate system in eq.~(\ref{chapter_transformations:diff_eq_lambda}) we threw away information, 
which we cannot recover by lifting the solution of the univariate system to the multivariate system.
The information we threw away is not so easy to spot, after all we kept the dependence on all directions
in the kinematic space by introducing the parameters $\alpha$.
The information we threw away comes from the specific paths we consider 
in eq.~(\ref{chapter_transformations:def_path}):
We only consider lines through the origin.
Therefore, there might be terms in the original $A$, which map to zero in $B_\lambda$ 
for the class of paths considered
in eq.~(\ref{chapter_transformations:def_path}).
These terms are derivatives of functions being constant on lines through the origin.
An example is given by
\bq 
\label{chapter_transformations:terms_map_to_zero}
 d \ln Z\left(x_1,...,x_{\NB}\right),
\eq
where $Z(x_1,...,x_{\NB})$ is a rational function in $(x_1,...,x_{\NB})$ 
and homogeneous of degree zero in $(x_1,...,x_{\NB})$.

Nevertheless, it can be a promising strategy to first solve the simpler univariate problem and to remove then
any offending terms of the form as in eq.~(\ref{chapter_transformations:terms_map_to_zero})
by a subsequent transformation, which usually is rather easy to find.
\\
\\
\bs
{\it \refstepcounter{exercise}
{\bf Exercise \theexercise}: 
Let $\Nmaster=1$, $\NB=2$ and
\bq
 A & = & 
 d \ln\left(\frac{x_1}{x_1+x_2}\right).
\eq
Show that $B_\lambda$, defined as in eq.~(\ref{chapter_transformations:diff_eq_lambda}), equals zero.
}
\es

\subsection{Picard-Fuchs operators}
\label{chapter_transformations:fibre_transformation:picard_fuchs}

In this section and the two following sections
we investigate differential equations for Feynman integrals, 
which only depend on one kinematic variable $x$.
Thus we consider the case $\NB=1$.
In the previous section we have seen that we may reduce a multivariate problem to an univariate
problem, solve the latter first and finally lift the result to the multivariate case.
Let us therefore consider a set of master integrals $\vec{I}=( I_{{\bm{\nu}}_1}, \dots, I_{{\bm{\nu}}_{\Nmaster}} )^T$ satisfying the differential equation
\bq
\label{chapter_transformations:system_first_order}
 \left( \frac{d}{dx} + A_x \right) \vec{I}
 & = & 0.
\eq
This is a system of $\Nmaster$ coupled first-order differential equations.

In this section we derive a single differential equation, usually of higher order, for one Feynman integral $I$ \cite{MullerStach:2012mp,Adams:2017tga}.
Let $I$ be one of the master integrals $\{I_{{\bm{\nu}}_1}, \dots, I_{{\bm{\nu}}_{\Nmaster}}\}$.
Eq.~(\ref{chapter_transformations:system_first_order}) allows us to express the $k$-th derivative 
of $I$ with respect to $x$ as a linear combination of the original master integrals.
We now determine the largest number $r$, such that the matrix which expresses 
\bq
 I,
 \;\;\;
 \frac{d}{dx} I,
 \;\;\;
 \dots,
 \;\;\;
 \left( \frac{d}{dx} \right)^{r-1} I
\eq
in terms of the original set $\{I_{{\bm{\nu}}_1}, \dots, I_{{\bm{\nu}}_{\Nmaster}}\}$ has full rank.
Obviously, we have $r \le \Nmaster$.
In the case $r < \Nmaster$ we complement the set $I, (d/dx)I, \dots, (d/dx)^{r-1}I$ by $(\Nmaster-r)$ elements
$I_{{\bm{\nu}}_{\sigma_{r+1}}}, \dots, I_{{\bm{\nu}}_{\sigma_{\Nmaster}}} \in \{I_{{\bm{\nu}}_1}, \dots, I_{{\bm{\nu}}_{\Nmaster}}\}$ 
such that the transformation matrix has rank $\Nmaster$.
The elements $I_{{\bm{\nu}}_{\sigma_{r+1}}}, \dots, I_{{\bm{\nu}}_{\sigma_{\Nmaster}}}$ must exist, 
since we assumed that the set $\{I_{{\bm{\nu}}_1}, \dots, I_{{\bm{\nu}}_{\Nmaster}}\}$
forms a basis of master integrals.
The basis 
\bq
\label{chapter_transformations:basis_with_derivatives}
 I,
 \;\;\;
 \frac{d}{dx} I,
 \;\;\;
 \dots,
 \;\;\;
 \left( \frac{d}{dx} \right)^{r-1} I,
 \;\;\;
 I_{{\bm{\nu}}_{\sigma_{r+1}}}, 
 \;\;\;
 \dots, 
 \;\;\;
 I_{{\bm{\nu}}_{\sigma_{\Nmaster}}}
\eq
decouples the system 
into a block of size $r$, which is closed under differentiation 
and a remaining sector of size $(\Nmaster-r)$.

We recall that $r$ is the largest number such that $I, (d/dx)I, ..., (d/dx)^{r-1}I$
are independent.
It follows that $(d/dx)^rI$ can be written as a linear combination of $I, (d/dx)I, ..., (d/dx)^{r-1}I$.
This defines the 
\index{Picard-Fuchs operator}
{\bf Picard-Fuchs operator} $L_r$ for the master integral $I$:
\bq
\label{chapter_transformations:def_picard_fuchs}
 L_{r} I & = & 0,
 \;\;\;\;\;\;
 L_r \; = \; \sum\limits_{k=0}^r R_k \frac{d^k}{dx^k},
\eq
where the coefficients $R_k$ are rational functions in $x$ and we use the normalisation $R_r=1$, hence
\bq
 L_r & = & \frac{d^r}{dx^r} + \sum\limits_{k=0}^{r-1} R_k \frac{d^k}{dx^k}.
\eq
The Picard-Fuchs operator $L_r=L_r(D,x,d/dx)$ depends on $D$, $x$ and $d/dx$.
The Picard-Fuchs operator $L_r$ 
is called 
\index{Fuchsian form}
{\bf Fuchsian}, if $R_k$ has maximally poles of order $r-k$ (including possible poles at infinity).
The Picard-Fuchs operator is a differential operator, which annihilates the Feynman integral $I$.
The Picard-Fuchs operator is easily obtained by a transformation to the basis 
given in eq.~(\ref{chapter_transformations:basis_with_derivatives}).
In this basis the upper-left $r \times r$-block of the transformed matrix $A_x'$ has the form
\bq
 \left( \begin{array}{ccccc}
 0 & -1 & ... & 0 & 0 \\
 && ... && \\
 0 & 0 & ... & 0 & -1 \\
 R_0 & R_1 & ... & R_{r-2} & R_{r-1} \\
 \end{array} \right),
\eq
and the coefficients $R_k$ of the Picard-Fuchs operator can easily be read off.

Let us look at a few examples.
We start with example 1 from section~\ref{chapter_iterated_integrals:deriving_the_dgl}
and work out the Picard-Fuchs operator for the one-loop two-point integral $I_{11}$.
The transformation matrix from the basis $\vec{I}=(I_{10},I_{11})^T$ to the basis $\vec{I}'=(I_{11}, (d/dx) I_{11})^T$ is
\bq
\label{chapter_transformations:basis_change_oneloop_twopoint}
 \left(\begin{array}{c}
 I_{11} \\
 \frac{d}{dx} I_{11}
 \end{array} \right)
 & = &
 \left(\begin{array}{cc}
  0 & 1 \\
  - \frac{D-2}{x\left(4+x\right)} & - \frac{4+\left(4-D\right)x}{2 x\left(4+x\right)} \\
 \end{array} \right)
 \left(\begin{array}{c}
 I_{10} \\
 I_{11} \\
 \end{array} \right).
\eq
Note that the second line of eq.~(\ref{chapter_transformations:basis_change_oneloop_twopoint}) 
is given as the negative of the second line of eq.~(\ref{chapter_iterated_integrals:example_A_oneloop_twopoint}).
In the basis $\vec{I}'$ the differential equation reads
\bq 
 \left( d + A' \right) \vec{I}' & = & 0,
\eq
with
\bq
\label{chapter_transformations:picard_fuchs_oneloop_twopoint}
 A' & = &
 \left( \begin{array}{cc}
 0 & -1 \\
 - \frac{D-4}{2 x \left(4+x\right)} & \frac{12 + \left(8-D\right)x}{2 x \left(4+x\right)}
 \end{array} \right) dx.
\eq
From eq.~(\ref{chapter_transformations:picard_fuchs_oneloop_twopoint}) we may now read off the Picard-Fuchs operator for the integral
$I_{11}$:
\bq
 \left[ 
  \frac{d^2}{dx^2} + \frac{12 + \left(8-D\right)x}{2 x \left(4+x\right)} \frac{d}{dx} - \frac{D-4}{2 x \left(4+x\right)}
 \right] I_{11} & = & 0.
\eq
In this case the Picard-Fuchs operator is a second-order differential operator.
In this example it is a particular simple differential operator, as it factorises
for any $D$ into two first-order differential operators:
\bq
  \frac{d^2}{dx^2} + \frac{12 + \left(8-D\right)x}{2 x \left(4+x\right)} \frac{d}{dx} - \frac{D-4}{2 x \left(4+x\right)}
 & = &
 \left( \frac{d}{dx} + \frac{1}{x} + \frac{1}{4+x} \right)
 \left( \frac{d}{dx} + \frac{1}{2x} - \frac{D-3}{2\left(4+x\right)} \right).
 \nonumber \\
\eq
Let's now look at a second example: We consider the double box integral discussed as 
example 3 in section~\ref{chapter_iterated_integrals:deriving_the_dgl}.
Proceeding along the same lines, we work out the Picard-Fuchs operator for the integral
$I_{111111100}$. This is now a differential operator of order eight, which factorises for
any $D$ as
\bq
 L_{1,1} \; L_{2,1}^3 \; L_{3,1} \; L_{4,1} \; L_{5,2} \; I_{111111100} & = & 0,
\eq
with
\bq
\label{chapter_transformations:picard_fuchs_doublebox}
 L_{1,1}
 & = & 
 \frac{d}{dx} + \frac{7}{x} + \frac{3}{x+1},
 \nonumber \\
 L_{2,1} 
 & = &
 \frac{d}{dx} - \frac{D-10}{x} + \frac{3}{x+1},
 \nonumber \\
 L_{3,1}
 & = &
 \frac{d}{dx} - \frac{D-8}{x} + \frac{D-2}{x+1},
 \nonumber \\
 L_{4,1}
 & = &
 \frac{d}{dx} - \frac{D-10}{2x} + \frac{D-2}{2\left(x+1\right)},
 \nonumber \\
 L_{5,2}
 & = &
  \frac{d^2}{dx^2} 
  + \left(- \frac{2D-13}{x} + \frac{D-2}{2\left(x+1\right)} \right) \frac{d}{dx} 
  + \frac{\left(D-6\right)\left(D-6-3x\right)}{x^2 \left(x+1\right)}.
\eq
For systems with a larger number of master integrals the calculation of the full 
Picard-Fuchs operators becomes soon impractical.
However, the essential information can already be extracted from the Picard-Fuchs
operator for the maximal cut.
Let us therefore consider the maximal cut for the double box integral.
We set
\bq
\label{chapter_transformations:maxcut_double_box_eq_1}
 \vec{J}
 & = &
 \left( \begin{array}{c}
 J_1 \\
 J_2 \\
 \end{array} \right)
 \; = \;
 \left( \begin{array}{l}
 \mathrm{MaxCut} \; I_{111111100} \\
 \mathrm{MaxCut} \; I_{1111111\left(-1\right)0} \\
 \end{array} \right)
\eq
The differential equation for $\vec{J}$ reads
\bq
\label{chapter_transformations:maxcut_double_box_eq_2}
 \left( d + A \right) \vec{J} & = & 0,
\eq
with $A$ given by the lower-right $(2 \times 2)$-block of eq.~(\ref{chapter_iterated_integrals:A_x_precanonical_double_box}):
\bq
\label{chapter_transformations:maxcut_double_box_eq_3}
 A & = &
 \left( \begin{array}{cc}
 \frac{2}{x} & \frac{D-4}{x \left(1+x\right)} \\
  -\frac{D-4}{x} & -\frac{\left(3D-16\right)x+4\left(D-5\right)}{2 x \left(1+x\right)} \\
 \end{array} \right) dx.
\eq
One obtains for the Picard-Fuchs operator for $J_1 = \mathrm{MaxCut} \; I_{111111100}$
the second-order differential operator $L_{5,2}$ defined in eq.~(\ref{chapter_transformations:picard_fuchs_doublebox}):
\bq
 L_{5,2} \; J_1 & = & 0.
\eq
For generic $D$, the second-order differential operator $L_{5,2}$
does not factor into two first-order differential operators.
However, if $D$ equals an even integer, the second-order differential operator $L_{5,2}(D,x,d/dx)$
factorises, for example \cite{MullerStach:2012mp,Tancredi:2015pta,Adams:2017tga}
\bq
 L_{5,2}\left(4,x,\frac{d}{dx}\right)
 & = &
 \left( \frac{d}{dx} + \frac{3}{x} + \frac{1}{x+1} \right)
 \left( \frac{d}{dx} + \frac{2}{x} \right),
 \nonumber \\
 L_{5,2}\left(6,x,\frac{d}{dx}\right)
 & = &
 \left( \frac{d}{dx} + \frac{1}{x} + \frac{2}{x+1} \right)
 \frac{d}{dx}.
\eq
As our last example we look at the two-loop sunrise integral with equal internal masses,
discussed as example 4 in section~\ref{chapter_iterated_integrals:deriving_the_dgl}.
We consider the Picard-Fuchs operator for $\mathrm{MaxCut} \; I_{111}$.
This is a second-order differential operator
\bq
  L_2 & = &
  \frac{d^2}{dx^2} 
  + \left(\frac{D}{2x}  - \frac{D-3}{x+1} - \frac{D-3}{x+9} \right) \frac{d}{dx} 
  + \left(D-3\right) \left( - \frac{D+4}{18x} + \frac{D}{8\left(x+1\right)} - \frac{5D-16}{72\left(x+9\right)} \right)
 \nonumber \\
\eq
which annihilates the integral $\mathrm{MaxCut} \; I_{111}$:
\bq
 L_2 \; \mathrm{MaxCut} \; I_{111} & = & 0.
\eq
This differential operator does not factorise for generic $D$, and remains irreducible in even integer dimensions.
In two space-time dimensions $L_2$ reads
\bq
\label{chapter_transformations:L2_sunrise}
 L_2\left(2,x,\frac{d}{dx}\right)
 & = &
 \frac{d^2}{dx^2} 
  + \left(\frac{1}{x} + \frac{1}{x+1} + \frac{1}{x+9} \right) \frac{d}{dx} 
  + \frac{1}{3x} - \frac{1}{4\left(x+1\right)} - \frac{1}{12\left(x+9\right)}.
\eq
As already mentioned, $L_2(2,x,d/dx)$ does not factorise as a differential operator,
i.e. $L_2$ is an
\index{irreducible differential operator}
{\bf irreducible differential operator}.
In chapter~\ref{chapter_elliptics} we will see that $L_2(2,x,d/dx)$ is also the Picard-Fuchs operator of a family
of elliptic curves, parametrised by the variable $x$.
We will also see in chapter~\ref{chapter_elliptics} that $I_{111}$ cannot be expressed in terms of multiple polylogarithms.
The appearance of an irreducible differential operator of order greater than one in the factorisation
of the Picard-Fuchs operator in even integer space-time dimensions
is an indication that the Feynman integral cannot be expressed 
in terms of multiple polylogarithms.

\begin{digression} {\bf The Frobenius method}
\\
Let us consider the differential equation
\bq
\label{chapter_transformations:def_picard_fuchs_polynomial}
 L_r I & = & 0,
 \;\;\;\;\;\;
 L_r \; = \; \sum\limits_{k=0}^r P_k\left(x\right) \frac{d^k}{dx^k},
\eq
where the coefficients $P_k(x)$ are polynomials in $x$.
This form is equivalent to eq.~(\ref{chapter_transformations:def_picard_fuchs}).
We obtain eq.~(\ref{chapter_transformations:def_picard_fuchs_polynomial})
by multiplying eq.~(\ref{chapter_transformations:def_picard_fuchs}) with the least common multiple of all denominators.
We assume that eq.~(\ref{chapter_transformations:def_picard_fuchs_polynomial}) is a Fuchsian differential equation.
The zeros of $P_r(x)$ (and possibly the point $x=\infty$) are the
\index{singular point} 
{\bf singular points of the differential equation}.

The Frobenius method allows us to construct $r$ independent solutions around a point $x_0 \in {\mathbb C}$ in the form of power series.
The power series converge up to the next nearest singularity.
Without loss of generality we assume $x_0=0$. 
A variable transformation $x'=x-x_0$ (or $x'=1/x$ for the point $x_0=\infty$) will transform to the case $x_0=0$.

Let us introduce the 
\index{Euler operator}
{\bf Euler operator} $\theta$ defined by
\bq
 \theta & = & x \frac{d}{d x}.
\eq
We may rewrite the differential operator $L_r$ in terms of the Euler operator:
\bq
 L_r 
 & = & 
 \sum\limits_{k=0}^r Q_k\left(x\right) \theta^k.
\eq
The conversion between the two forms can be done with the help of
\bq
\label{chapter_transformations:conversion_Euler_operator}
 \frac{d^k}{dx^k} \; = \; x^{-k} \prod\limits_{j=0}^{k-1} \left(\theta - j\right),
 & &
 \theta^k \; = \; \sum\limits_{j=1}^{k} S\left(k,j\right) x^j \frac{d^j}{dx^j},
\eq
where $S(n,k)$ denotes the Stirling numbers of the second kind:
\bq
 S\left(n,k\right)
 & = &
 \frac{1}{k!} \sum\limits_{j=0}^{k} \left(-1\right)^j \left(\begin{array}{c} k \\ j \\ \end{array} \right) \left(k-j\right)^n.
\eq
\bs
{\it \refstepcounter{exercise}
{\bf Exercise \theexercise}: 
Prove the two relations in eq.~(\ref{chapter_transformations:conversion_Euler_operator}).
}
\es
\\
\\
After possibly multiplying by an appropriate power of $x$ we arrive
at
\bq
 \tilde{L}_r 
 & = & 
 \sum\limits_{k=0}^r \tilde{Q}_k\left(x\right) \theta^k,
\eq
where the coefficients $\tilde{Q}_k(x)$ are again polynomials in $x$
and $\tilde{L}_r$ annihilates the integral $I$ as well:
\bq
 \tilde{L}_r I & = & 0.
\eq 
\bs
{\it \refstepcounter{exercise}
{\bf Exercise \theexercise}: 
Rewrite
\bq
 L_2 & = &
 x \left(x+1\right)\left(x+9\right) \frac{d^2}{dx^2}
 + \left(3x^2+20x+9\right) \frac{d}{dx}
 + x + 3
\eq
in Euler operators.
(This is the differential operator of eq.~(\ref{chapter_transformations:L2_sunrise}) multiplied with $x(x+1)(x+9)$).
}
\es
\\
\\
Let us now discuss how the solutions are constructed \cite{Ince:book}.
We consider the 
\index{indicial equation}
{\bf indicial equation}
\bq
 \sum\limits_{k=0}^r \tilde{Q}_k\left(0\right) \alpha^k
 & = & 0.
\eq
This is a polynomial equation of degree $r$ in the variable $\alpha$.
The $r$ solutions for $\alpha$ are called the 
\index{indicial}
{\bf indicials} or
\index{local exponent}
{\bf local exponents} at $x_0=0$.
We denote them by $\alpha_1, \dots, \alpha_r$.

Let us first assume that $\alpha_i-\alpha_j \notin {\mathbb Z}$ for $i \neq j$.
Then the $r$ independent solutions are given by the power series
\bq
 x^{\alpha_i} \sum\limits_{j=0}^\infty c_{i,j} x^j,
 & & 
 c_{i,0} \; = \; 1,
 \;\;\;\;\;\;
 1 \; \le \; i \; \le \; r.
\eq
The coefficients $c_{i,j}$ for $j>0$ can be computed recursively by plugging the ansatz into the differential
equation.
Note that the fact that $\alpha_i$ is a root of the indicial equation ensures that
\bq
 \tilde{L}_r x^{\alpha_i} & = & 0.
\eq
The condition $\alpha_i-\alpha_j \notin {\mathbb Z}$ ensures that
\bq
 \tilde{L}_r x^{\alpha_i+j} & \neq & 0,
 \;\;\;\;\;\;
 j \; \in \; {\mathbb N}.
\eq
Let us now relax the condition $\alpha_i-\alpha_j \notin {\mathbb Z}$.
We now allow that a root $\alpha_i$ occurs with multiplicity $\lambda_i$, but we maintain to condition
$\alpha_i-\alpha_j \notin {\mathbb Z}$ for $i \neq j$. Let $t$ denote the number of distinct roots.
We have
\bq
 \lambda_1 + \dots + \lambda_t & = & r.
\eq
The solutions are now spanned by series of the form
\bq
\label{chapter_transformations:general_Frobenius}
 x^{\alpha_i} \sum\limits_{k=0}^{b} \frac{1}{\left(b-k\right)!} \ln^{\left(b-k\right)}\left(x\right) \sum\limits_{j=0}^\infty c_{i,j,k} x^j,
\eq
where $b \in \{0,1,\dots,\lambda_i-1\}$ and $c_{i,0,k}=\delta_{k 0}$.
\\
\\
\bs
{\it \refstepcounter{exercise}
{\bf Exercise \theexercise}: 
Consider
\bq
 \tilde{L} & = & \left(\theta-\alpha\right)^\lambda.
\eq
Show that the solution space is spanned by
\bq
 x^\alpha, \; x^\alpha \ln\left(x\right), \; \dots, \; 
 \frac{x^\alpha \ln^{\lambda-1}\left(x\right)}{\left(\lambda-1\right)!}.
\eq
}
\es
In the general case we allow $\alpha_i-\alpha_j \in {\mathbb Z}$.
Suppose that the indicials $\alpha_i$ and $\alpha_j$ have multiplicities 
$\lambda_i$ and $\lambda_j$, respectively.
Assume further that $\alpha_i-\alpha_j \in {\mathbb Z}$ and $\mathrm{Re}(\alpha_i) > \mathrm{Re}(\alpha_j)$.
We start with $\lambda_i$ solutions of the form as in eq.~(\ref{chapter_transformations:general_Frobenius}).
These are supplemented by $\lambda_j$ solutions starting with the power $x^{\alpha_j}$.
Up to the power $x^{\alpha_i-1}$ these solutions 
follow the pattern of eq.~(\ref{chapter_transformations:general_Frobenius}).
Starting from the power $x^{\alpha_i}$ we have to allow
logarithms up to the power $(\lambda_i+\lambda_j-1)$.
The following exercise illustrates this:
\\
\\
\bs
{\it \refstepcounter{exercise}
{\bf Exercise \theexercise}: 
Consider the differential operators
\bq
 \tilde{L}_a & = & \left(\theta-1\right) \left(\theta-x\right),
 \nonumber \\
 \tilde{L}_b & = & \left(\theta-x\right) \left(\theta-1\right).
\eq
Construct for both operators two independent solutions around $x_0=0$.
}
\es
\\
\\
Let $x_1,\dots,x_s$ be the set of the singular points of the $r$-th order differential equation, including possibly
the point at infinity. We denote the indicials at the $j$-th singular point
by $\alpha_1^{(j)}, \dots, \alpha_s^{(j)}$.
The 
\index{Riemann $P$-symbol}
{\bf Riemann $P$-symbol} can be viewed as a $(r\times s)$-matrix, which collects the information
on the indicials at all singular points:
\bq
 P\left( \begin{array}{cccc}
  \alpha_1^{(1)} & \alpha_1^{(2)} & \dots & \alpha_1^{(s)} \\
  \vdots & \vdots & & \vdots \\ 
  \alpha_r^{(1)} & \alpha_r^{(2)} & \dots & \alpha_r^{(s)} \\ 
 \end{array} \right).
\eq
The Fuchsian relation states that the sum of all indicials equals
\bq
 \sum\limits_{i=1}^r \sum\limits_{j=1}^s \alpha_i^{(j)}
 & = &
 \frac{1}{2} \left(s-2\right) \left(r-1\right) r.
\eq
A singular point $x_j$, where the indicial equation takes the form
\bq
 \left( \alpha - \alpha_0 \right)^r & = & 0,
\eq
e.g. where there is only one indicial $\alpha_0$ with multiplicity $r$ is called a 
\index{point of maximal unipotent monodromy}
{\bf point of maximal unipotent monodromy}.
A basis of solutions around this point is called a 
\index{Frobenius basis}
{\bf Frobenius basis}.

\end{digression}

\subsection{Magnus expansion}
\label{chapter_transformations:fibre_transformation:magnus}

Let us again consider the univariate problem
\bq
\label{chapter_transformations:magnus_diffenrential_equation}
 \left( \frac{d}{dx} + A_x \right) \vec{I}
 & = & 0.
\eq
A solution to this differential equation with boundary value $\vec{I}_0$ at $x=0$
is given by the infinite series
\bq
\label{chapter_transformations:solution_infinte_series}
 \vec{I}\left(x\right)
 & = &
 \left[
 1
 - \int\limits_0^x dx_1 A_x\left(x_1\right)
 + \int\limits_0^x dx_1 A_x\left(x_1\right)
   \int\limits_0^{x_1} dx_2 A_x\left(x_2\right)
 \right. \nonumber \\
 & & \left.
 - \int\limits_0^x dx_1 A_x\left(x_1\right)
   \int\limits_0^{x_1} dx_2 A_x\left(x_2\right)
   \int\limits_0^{x_2} dx_3 A_x\left(x_3\right)
 + \dots
 \right]
 \vec{I}_0.
\eq
The individual terms of this infinite series are iterated integrals.
If we introduce the 
\index{path ordering operator}
{\bf path ordering operator} 
${\mathcal P}$ by
\bq
 {\mathcal P}\left( A\left(x_1\right) A\left(x_2\right) \dots A\left(x_n\right) \right)
 & = &
 A\left(x_{\sigma_1}\right) A\left(x_{\sigma_2}\right) \dots A\left(x_{\sigma_n}\right),
\eq
where $\sigma$ is the permutation of $\{1,\dots,n\}$ such that
\bq
 x_{\sigma_1} \; > \; 
 x_{\sigma_2} \; > \; 
 \dots \; > \; 
 x_{\sigma_n},
\eq
we may write eq.~(\ref{chapter_transformations:solution_infinte_series}) as
\bq
\label{chapter_transformations:solution_path_ordered_exponential}
 \vec{I}\left(x\right)
 & = &
 {\mathcal P}
 \exp\left( - \int\limits_0^x dx_1 A_x\left(x_1\right) \right)
 \vec{I}_0.
\eq
\bs
{\it \refstepcounter{exercise}
{\bf Exercise \theexercise}: 
Show the equivalence of eq.~(\ref{chapter_transformations:solution_path_ordered_exponential})
with eq.~(\ref{chapter_transformations:solution_infinte_series}).
}
\es
\\
\\
The infinite series in eq.~(\ref{chapter_transformations:solution_infinte_series})
is in general not yet particular useful, as there is no truncation criteria.
If the differential equation is in $\eps$-form, then there is a clear truncation criteria:
We truncate to the desired order in $\eps$ and the solution coincides with the solution discussed in
section~\ref{chapter_iterated_integrals:section:solution_eps_form}.

But let's go on with our formal investigations:
If there is only one master integral ($\Nmaster=1$), $A_x(x_1)$ is a $(1 \times 1)$-matrix
and the path ordering can be ignored, as $(1 \times 1)$-matrices always commute.
In this case the solution is simply
\bq
\label{chapter_transformations:solution_fibre_1dim}
 I\left(x\right)
 & = &
 \exp\left( - \int\limits_0^x dx_1 A_x\left(x_1\right) \right)
 I_0.
\eq
Let us return to the general case, where $A_x$ is a $(\Nmaster \times \Nmaster)$-matrix.
Let us now insist that we write the solution for eq.~(\ref{chapter_transformations:magnus_diffenrential_equation})
in the form of an exponential as in eq.~(\ref{chapter_transformations:solution_fibre_1dim})
as opposed to the form of a path-ordered exponential (as in eq.~(\ref{chapter_transformations:solution_path_ordered_exponential})).
Thus we write
\bq
\label{chapter_transformations:solution_magnus_exponential}
 \vec{I}\left(x\right)
 & = &
 \exp\left( \Omega\left(x\right) \right)
 \vec{I}_0.
\eq
$\Omega(x)$ is in general again given by an infinite series, called
the 
\index{Magnus series}
{\bf Magnus series} \cite{Magnus:1954zz,Argeri:2014qva}.
That's the price we have to pay in order to write the solution in terms of an ordinary exponential.
We write
\bq
\label{chapter_transformations:magnus_expansion}
 \Omega(x) & = & \sum\limits_{n=1}^\infty \Omega_n\left(x\right).
\eq
The first few terms are 
\bq
\label{chapter_transformations:Omega_1_2_3}
 \Omega_1\left(x\right)
 & = &
 - \int\limits_0^x dx_1 A_x\left(x_1\right),
 \\
 \Omega_2\left(x\right)
 & = &
 \frac{1}{2}
 \int\limits_0^x dx_1 \int\limits_0^{x_1} dx_2 \left[ A_x\left(x_1\right), A_x\left(x_2\right)\right],
 \nonumber \\
 \Omega_3\left(x\right)
 & = &
 - \frac{1}{6}
 \int\limits_0^x dx_1 \int\limits_0^{x_1} dx_2 \int\limits_0^{x_2} dx_3 
  \left( 
         \left[ A_x\left(x_1\right), \left[A_x\left(x_2\right),A_x\left(x_3\right)\right]\right] 
         +
         \left[ \left[A_x\left(x_1\right),A_x\left(x_2\right)\right], A_x\left(x_3\right) \right] 
  \right).
 \nonumber
\eq
The higher terms $\Omega_n(x)$ (with $n \ge 2$) correct for the fact that in general $A_x(x_1)$ and $A_x(x_2)$
don't commute.
In general, $\Omega_1(x)$ is given by eq.~(\ref{chapter_transformations:Omega_1_2_3}) and $\Omega_n(x)$ is given for 
$n \ge 2$ by
\bq
 \Omega_n\left(x\right)
 & = &
 \sum\limits_{j=1}^{n-1} \frac{B_j}{j!} \int\limits_0^x dx_1 S_{n}^{(j)}\left(x_1\right),
\eq
where $B_j$ denote the $j$-th 
\index{Bernoulli number}
{\bf Bernoulli number}, 
defined by
\bq
\label{chapter_transformations:def_Bernoulli_number}
 \frac{x}{e^x-1}
 & = &
 \sum\limits_{j=0}^\infty \frac{B_j}{j!} x^j,
\eq
and the $S_{n}^{(j)}$ are defined recursively by
\bq
 S_{n}^{(1)}\left(x\right)
 & = &
 \left[ A_x\left(x\right), \Omega_{n-1}\left(x\right) \right],
 \nonumber \\
 S_{n}^{(j)}\left(x\right)
 & = &
 \sum\limits_{k=1}^{n-j} \left[ \Omega_{k}\left(x\right), S_{n-k}^{(j-1)}\left(x\right) \right],
 \;\;\;\;\;\; 2 \; \le \; j \; \le \; n-1.
\eq
The Magnus expansion is useful when the Magnus series in eq.~(\ref{chapter_transformations:magnus_expansion}) terminates.
This is for example the case, whenever $A_x(x)$ is a diagonal matrix (in which case $\Omega_n=0$ for $n\ge2$)
or a nilpotent matrix.
Let us consider the case where $A_x$ can be decomposed into
a diagonal matrix $D_x$ and a nilpotent matrix $N_x$:
\bq
 A_x & = & D_x + N_x.
\eq
We write
\bq
 \Omega\left[A_x\right]\left(x\right)
\eq
for the Magnus series of $A_x(x)$.
For a diagonal matrix $D_x$ we have
\bq
 \Omega\left[D_x\right]\left(x\right)
 & = & 
 - \int\limits_{0}^x dx_1 D_x\left(x_1\right).
\eq
We set
\bq
 N_x'
 & = &
 e^{-\Omega\left[D_x\right]} N_x e^{\Omega\left[D_x\right]}.
\eq
Since we assumed that $N_x$ is nilpotent, the matrix $N_x'$ is nilpotent as well.
Therefore the Magnus series $\Omega[N_x']$ terminates.
A solution of the differential equation
\bq
 \left( \frac{d}{dx} + D_x + N_x \right) \vec{I} & = & 0
\eq
with boundary value $\vec{I}_0$ at $x=0$ is given by
\bq
\label{chapter_transformations:Magnus_diagonal_nilpotent}
 \vec{I}\left(x\right)
 & = &
 e^{\Omega\left[D_x\right]\left(x\right)} \;
 e^{\Omega\left[N_x'\right]\left(x\right)} \;
 \vec{I}_0.
\eq
\bs
{\it \refstepcounter{exercise}
{\bf Exercise \theexercise}: 
Prove eq.~(\ref{chapter_transformations:Magnus_diagonal_nilpotent}).
}
\es
\\
\\
As an application consider the case where $A_x$ is linear in the dimensional regularisation parameter
$\eps$. We write
\bq
 A_x & = & A_x^{(0)} + \eps A_x^{(1)},
\eq
where $A_x^{(0)}$ and $A_x^{(1)}$ are independent of $\eps$.
Set
\bq
\label{chapter_transformations:Magnus_trafo_A0}
 \vec{I}' & = & U \vec{I},
 \;\;\;\;\;\;\;\;\;
 U \; = \; e^{-\Omega[A_x^{(0)}]\left(x\right)}.
\eq
The matrix $U$ is independent of $\eps$ as well.
The differential equation
\bq
\label{chapter_transformations:Magnus_linear_dgl}
 \left( \frac{d}{dx} + A_x^{(0)} + \eps A_x^{(1)} \right) \vec{I} & = & 0
\eq
transforms under eq.~(\ref{chapter_transformations:Magnus_trafo_A0}) into
\bq
\label{chapter_transformations:Magnus_trafo_A0_eps_form}
 \left( \frac{d}{dx} + \eps U A_x^{(1)} U^{-1} \right) \vec{I}' & = & 0
\eq
The only dependence on $\eps$ is now given by the explicit prefactor in eq.~(\ref{chapter_transformations:Magnus_trafo_A0_eps_form}).
\\
\\
\bs
{\it \refstepcounter{exercise}
{\bf Exercise \theexercise}: 
Show that the transformation in eq.~(\ref{chapter_transformations:Magnus_trafo_A0}) 
transforms the differential equation eq.~(\ref{chapter_transformations:Magnus_linear_dgl}) into
eq.~(\ref{chapter_transformations:Magnus_trafo_A0_eps_form}).
}
\es
\\
\\
As an example let us consider the one-loop two-point function with equal internal masses discussed
in example 1 in section~\ref{chapter_iterated_integrals:deriving_the_dgl}.
The differential equation in $D=4-2\eps$ space-time dimensions is linear in $\eps$ and reads
\bq
 \left( \frac{d}{dx} + A_x^{(0)} + \eps A_x^{(1)} \right) 
 \left( \begin{array}{c} I_{10} \\ I_{11} \\ \end{array} \right) & = & 0
\eq
with
\bq
 A_x^{(0)}
 \; = \;
 \left( \begin{array}{cc} 
  0 & 0 \\
  \frac{2}{x\left(4+x\right)} & \frac{2}{x\left(4+x\right)}  \\
 \end{array} \right),
 & &
 A_x^{(1)}
 \; = \;
 \left( \begin{array}{cc} 
  0 & 0 \\
  -\frac{2}{x\left(4+x\right)} & \frac{1}{4+x}  \\
 \end{array} \right).
\eq
We may write $A_x^{(0)}$ as a sum of a diagonal matrix and a nilpotent matrix:
\bq
 A_x^{(0)}
 & = &
 D_x^{(0)} + N_x^{(0)}
 \; = \;
 \left( \begin{array}{cc} 
  0 & 0 \\
  0 & \frac{2}{x\left(4+x\right)} \\
 \end{array} \right)
 +
  \left( \begin{array}{cc} 
  0 & 0 \\
  \frac{2}{x\left(4+x\right)} & 0 \\
 \end{array} \right).
\eq
The matrix $A_x^{(0)}$ has a singular point at $x=0$. 
For the iterated integrals we take as lower boundary $x_0$.
In the end we take the limit $x_0\rightarrow 0$ and discard any logarithmic divergent terms $\ln^k(x_0)$.
With this prescription we obtain
\bq
 e^{-\Omega[D_x^{(0)}]\left(x\right)}
 & = &
  \left( \begin{array}{cc} 
  1 & 0 \\
  0 & \sqrt{\frac{x}{4+x}} \\
 \end{array} \right),
\eq
and for $N_x^{(0)}{}' = e^{-\Omega[D_x^{(0)}]\left(x\right)} N_x^{(0)} e^{\Omega[D_x^{(0)}]\left(x\right)}$
\bq
 e^{-\Omega[N_x^{(0)}{}']\left(x\right)}
 & = &
  \left( \begin{array}{cc} 
  1 & 0 \\
  \sqrt{\frac{x}{4+x}}  & 1 \\
 \end{array} \right).
\eq
Thus
\bq
 U
 & = &
 e^{-\Omega[A_x^{(0)}]\left(x\right)}
 \; = \;
 e^{-\Omega[N_x^{(0)}{}']\left(x\right)}
 e^{-\Omega[D_x^{(0)}]\left(x\right)}
 \; = \;
  \left( \begin{array}{cc} 
  1 & 0 \\
  \sqrt{\frac{x}{4+x}}  & \sqrt{\frac{x}{4+x}} \\
 \end{array} \right)
\eq
equals up to a prefactor $2\eps(1-\eps)$ the transformation matrix given in eq.~(\ref{chapter_iterated_integrals:trafo_bubble}).
(The prefactor $2\eps(1-\eps)$ comes from the requirement that the non-zero boundary constants have uniform weight zero.)

\subsection{Moser's algorithm}
\label{chapter_transformations:fibre_transformation:moser}

We stay with the univariate problem
\bq
\label{chapter_transformations:moser_diffenrential_equation}
 \left( \frac{d}{dx} + A_x \right) \vec{I}
 & = & 0.
\eq
Suppose that there exists a rational fibre transformation, which brings the differential equation
into $\eps$-form.
Moser's algorithm \cite{Moser:1959,Lee:2014ioa,Lee:2017oca}
allows us to systematically construct such a transformation.
This algorithm has been implemented 
in several computer programs \cite{Prausa:2017ltv,Gituliar:2017vzm,Lee:2020zfb}.

In this section we will always take the complex numbers as ground field.
This is an algebraically closed field.
In particular, any polynomial $p(x) \in {\mathbb C}[x]$ factorises into linear factors.

The entries of $A_x$ are rational functions of the kinematic variable $x$ (and the dimensional regularisation parameter
$\eps$).
Let us denote by $S=\{x_1,x_2,\dots\}$ the set of points, where $A_x$ is singular, including possibly the point
at infinity
and by $S'$ the set of singular points excluding the point at infinity (i.e. the set of finite singular points).

Using partial fraction decomposition in $x$ we may write $A_x$ as
\bq
 A_x
 & = &
 \sum\limits_{j=0}^{o_{\infty}-2}
 M_{\infty,j+2}\left(\eps\right) x^j
 +
 \sum\limits_{x_i \in S'} \sum\limits_{j=1}^{o_{x_i}}
 M_{x_i,j}\left(\eps\right)
 \frac{1}{\left(x-x_i\right)^j}.
\eq
The entries of the $(\Nmaster \times \Nmaster)$-matrices $M_{\infty,j}(\eps)$
and $M_{x_i,j}(\eps)$ are rational functions in $\eps$.
$o_{x_i}$ denotes the order of the pole at $x_i$.

We say that the differential equation in eq.~(\ref{chapter_transformations:moser_diffenrential_equation})
is in 
\index{Fuchsian form}
{\bf Fuchsian form}, if $A_x$ has only simple poles.
In this case, $A_x$ can be written as
\bq
\label{chapter_transformations:moser_A_x_Fuchsian_form}
 A_x
 & = &
 \sum\limits_{x_i \in S'} 
 M_{x_i,1}\left(\eps\right)
 \frac{1}{\left(x-x_i\right)}.
\eq
We call $M_{x_i,1}(\eps)$ the matrix residue at $x=x_i$.
\\
\\
\bs
{\it \refstepcounter{exercise}
\label{chapter_transformations:exercise_moser_Fuchsian_form}
{\bf Exercise \theexercise}: 
Assume that $A_x$ is in Fuchsian form (i.e. of the form as in eq.~(\ref{chapter_transformations:moser_A_x_Fuchsian_form})).
Show that the matrix residue at $x=\infty$ is given by
\bq
 M_{\infty,1}\left(\eps\right)
 & = &
 -
 \sum\limits_{x_i \in S'} 
 M_{x_i,1}\left(\eps\right).
\eq 
}
\es
\\
\\
Moser's algorithm proceeds in three steps: 
In the first step one reduces $A_x$ to a Fuchsian form.
In the second step we treat $\eps$ as an infinitesimal quantity and transform the eigenvalues of the matrices $M_{x_i,1}(\eps)$
into the interval $[-\frac{1}{2},\frac{1}{2}[$.
If all eigenvalues are proportional to $\eps$, the algorithm succeeds and one may factor
out in a third step $\eps$ as a prefactor.

\begin{digression} {\bf Jordan normal form and generalised eigenvectors}
\\
We review a view basic facts from linear algebra.
A quadratic matrix $A \in \mathrm{M}(n \times n,{\mathbb C})$ may or may not be diagonalisable.
However, the matrix can always be put into the Jordan normal form, e.g. there exists 
an invertible matrix $Q \in \mathrm{GL}(n,{\mathbb C})$ 
such that
\bq
 A & = & Q J Q^{-1}
\eq
and $J$ is in the Jordan normal form.
The Jordan normal form consists of Jordan block matrices on the diagonal
\bq
\label{chapter_transformations:Jordan_normal_form}
 J & = &
 \left( \begin{array}{ccccc}
 J_1 & & & & \\
 & & \ddots & & \\
 & & & & J_r \\
 \end{array} \right),
\eq
and the Jordan block matrices $J_i$'s are of the form
\bq
\label{chapter_transformations:Jordan_block}
 J_i & = &
 \left( \begin{array}{cccc}
 \lambda_i & 1 & & \\
 & \lambda_i & \ddots & \\
 & & \ddots & 1 \\
 & & & \lambda_i \\
 \end{array} \right).
\eq
Note that the same eigenvalue $\lambda_i$ may occur in different Jordan blocks.
The Jordan normal form is unique up to permutations of the Jordan blocks.
The number of times the eigenvalue $\lambda_i$ appears on the diagonal in eq.~(\ref{chapter_transformations:Jordan_normal_form}) 
is called the 
\index{algebraic multiplicity of an eigenvalue}
{\bf algebraic multiplicity of the eigenvalue} 
$\lambda_i$.
The number of Jordan blocks corresponding to the eigenvalue $\lambda_i$
is called the 
\index{geometric multiplicity of an eigenvalue}
{\bf geometric multiplicity of the eigenvalue} 
$\lambda_i$.
A right eigenvector $\vec{v}_R$ to the matrix $A$ for the eigenvalue $\lambda_i$ satisfies
\bq
 \left( A - \lambda_i {\bf 1} \right) \cdot \vec{v}_R & = & 0.
\eq
A right 
\index{generalised eigenvector}
{\bf generalised eigenvector} $\vec{v}_R$ to the matrix $A$ for the eigenvalue $\lambda_i$ satisfies
\bq
 \left( A - \lambda_i {\bf 1} \right)^r \cdot \vec{v}_R & = & 0.
\eq
To each Jordan block of size $(r \times r)$ there corresponds a set of $r$ (right) generalised eigenvectors.
A left eigenvector $\vec{v}_L^T$ to the matrix $A$ for the eigenvalue $\lambda_i$ satisfies
\bq
 \vec{v}_L^T \cdot \left( A - \lambda_i {\bf 1} \right) \; = \; 0,
 & \mbox{or} &
 \left( A^T - \lambda_i {\bf 1} \right) \cdot \vec{v}_L \; = \; 0.
\eq
Similar, a left 
generalised eigenvector $\vec{v}_L^T$ to the matrix $A$ for the eigenvalue $\lambda_i$ satisfies
\bq
 \vec{v}_L^T \cdot \left( A - \lambda_i {\bf 1} \right)^r \; = \; 0,
 & \mbox{or} &
 \left( A^T - \lambda_i {\bf 1} \right)^r \cdot \vec{v}_L \; = \; 0.
\eq
For non-diagonalisable matrices we have to distinguish between left and right eigenvectors as the following example shows:
Consider
\bq
 A & = &
 \left( \begin{array}{ccc}
 \lambda & 1 & 0 \\
 0 & \lambda & 1 \\
 0 & 0 & \lambda \\
 \end{array} \right).
\eq
The right eigenvectors are spanned by
\bq
 \vec{v}_R & = & 
 \left( \begin{array}{c} 1 \\ 0 \\ 0 \\ \end{array} \right),
\eq
while the left eigenvectors are spanned by
\bq
 \vec{v}_L^T & = & \left( \begin{array}{ccc} 0 & 0 & 1 \\ \end{array} \right).
\eq
\end{digression}
Let us now look at the technical details of Moser's algorithm.
In the first step we reduce $A_x$ to a Fuchsian form.
The strategy is to remove successively for each singular point $x_i \in S$ the highest pole until only simple poles are left.
A necessary condition for the existence of a fibre transformation which removes the highest pole at $x_i \in S$ is 
that the matrix $M_{x_i,o_{x_i}}(\eps)$ is nilpotent \cite{Moser:1959}.
In the applications towards Feynman integrals this is usually the case and no counter-examples are known.
Very often higher poles in Feynman integral calculations can be removed by a suitable ansatz.
This is usually the most efficient way.
There is also a systematic algorithm based on balance transformations, with projectors constructed from 
generalised eigenvectors \cite{Lee:2014ioa,Barkatou:2007aaa,BARKATOU20091017}.
The balance transformations are discussed below.
This algorithm removes all higher poles and introduces at worst a spurious singularity with a simple pole
at a regular point. The spurious singularity is then removed in the second step.

Let us now assume that $A_x$ is in Fuchsian form:
\bq
\label{chapter_transformations:moser_A_x_Fuchsian_form2}
 A_x
 & = &
 \sum\limits_{x_i \in S'} 
 M_{x_i,1}\left(\eps\right)
 \frac{1}{\left(x-x_i\right)}.
\eq
We treat the dimensional regularisation parameter $\eps$ as an infinitesimal quantity.
We now look at the eigenvalues and the (left and right) eigenvectors of the matrices $M_{x_1,1}$ and $M_{x_2,1}$, where $x_1, x_2 \in S$.
Let $x_1 \in S$ be a singular point such that $M_{x_1,1}$ has an eigenvalue $\lambda_1 \ge \frac{1}{2}$.
Let $\vec{v}_{R,x_1}$ be a right eigenvector of $M_{x_1,1}$ to the eigenvalue $\lambda_1$.
Similar,
let $x_2 \in S$ (with $x_2 \neq x_1$)
be a singular point such that $M_{x_2,1}$ has an eigenvalue $\lambda_2 < -\frac{1}{2}$.
Let $\vec{v}_{L,x_2}^T$ be a left eigenvector of $M_{x_2,1}$ to the eigenvalue $\lambda_2$.
Assume further that
\bq
 \vec{v}_{L,x_2}^T \cdot \vec{v}_{R,x_1}
 & \neq & 0.
\eq
We define a $(\Nmaster \times \Nmaster)$-matrix $P$ by
\bq
\label{chapter_transformations:moser_def_projector}
 P & = &
 \frac{\vec{v}_{R,x_1} \; \vec{v}_{L,x_2}^T}{\left( \vec{v}_{L,x_2}^T \cdot \vec{v}_{R,x_1} \right)}.
\eq
We denote the $(\Nmaster \times \Nmaster)$-unit matrix by ${\bf 1}$.
\\
\\
\bs
{\it \refstepcounter{exercise}
{\bf Exercise \theexercise}: 
Show that $P$ and ${\bf 1}-P$ are projectors, i.e.
\bq
 P^2 \; = \; P,
 & &
 \left({\bf 1}-P\right)^2 \; = \; {\bf 1}-P.
\eq
Show further
\bq
 \left[ \left({\bf 1}-P\right) + \frac{x-x_2}{x-x_1} P \right]
 \left[ \left({\bf 1}-P\right) + \frac{x-x_1}{x-x_2} P \right]
 & = &
 {\bf 1}.
\eq
}
\es
\\
\\
We define a fibre transformation, called a balance transformation, by
\bq
 U^{\mathrm{balance}}\left(x_1,x_2,P\right)
 & = &
 \left\{
  \begin{array}{ll}
    \left({\bf 1}-P\right) + \frac{x-x_1}{x-x_2} P, & x_1,x_2 \neq \infty, \\
    \left({\bf 1}-P\right) - \frac{1}{x-x_2} P, & x_1 = \infty, \\
    \left({\bf 1}-P\right) - \left(x-x_1\right) P, & x_2 = \infty. \\
  \end{array}
 \right.
\eq
The inverse transformation is then given by
\bq
 \left[ U^{\mathrm{balance}}\left(x_1,x_2,P\right) \right]^{-1}
 & = &
 \left\{
  \begin{array}{ll}
    \left({\bf 1}-P\right) + \frac{x-x_2}{x-x_1} P, & x_1,x_2 \neq \infty, \\
    \left({\bf 1}-P\right) - \left(x-x_2\right) P, & x_1 = \infty, \\
    \left({\bf 1}-P\right) - \frac{1}{x-x_1} P, & x_2 = \infty. \\
  \end{array}
 \right.
\eq
The balance transformation lowers the eigenvalue at $x_1$ by one unit and raises the eigenvalue at $x_2$ by one unit.
By a sequence of balance transformations we may try to make all eigenvalues proportional to $\eps$ or reduce them to zero.
This may fail for several reasons.
One reason can be that some eigenvalue is of the form ``half-integer plus $\eps$''.
As we only shift the eigenvalues by units of one, we can never make this eigenvalue proportional to $\eps$.
Another reason can be that there are unbalanced eigenvalues, but the scalar product between the corresponding eigenvectors is zero:
\bq
 \vec{v}_{L,x_2}^T \cdot \vec{v}_{R,x_1}
 & = & 0.
\eq
However, if we succeed we may construct a rational fibre transformation which puts the differential equation into $\eps$-form.

In the third step we factor out $\eps$.
At this stage we may assume that $A_x$ is in Fuchsian form as in eq.~(\ref{chapter_transformations:moser_A_x_Fuchsian_form2})
and that all eigenvalues of the matrices $M_{x_i,1}(\eps)$
are proportional to $\eps$ or are zero.
We seek an $x$-independent fibre transformation $U$, which transforms the differential equation into the $\eps$-form.
Let $V(\eps)$ be such a transformation. This transformation must fulfil
\bq
\label{chapter_transformations:moser_factoring_eps_I}
 V\left(\eps\right) \frac{M_{x_i,1}\left(\eps\right)}{\eps} V\left(\eps\right)^{-1}
 & = &
 N_{x_i},
 \;\;\;\;\;\;
 \forall \; x_i \; \in \; S',
\eq
where the $N_{x_i}$'s are $x$- and $\eps$-independent $(\Nmaster \times \Nmaster)$-matrices.
We don't know the matrix $V(\eps)$ nor do we know matrices $N_{x_i}$.
However, the right-hand side of eq.~(\ref{chapter_transformations:moser_factoring_eps_I}) is independent of $\eps$. 
This implies that
\bq
\label{chapter_transformations:moser_factoring_eps_II}
 V\left(\eps\right) \frac{M_{x_i,1}\left(\eps\right)}{\eps} V\left(\eps\right)^{-1}
 & = &
 V\left(\eps'\right) \frac{M_{x_i,1}\left(\eps'\right)}{\eps'} V\left(\eps'\right)^{-1}
\eq
for any $\eps'$ and all $x_i \in S'$.
Let us now set
\bq
 U\left(\eps,\eps'\right)
 & = &
 V\left(\eps'\right)^{-1} V\left(\eps\right).
\eq
We may re-write eq.~(\ref{chapter_transformations:moser_factoring_eps_II}) as
\bq
\label{chapter_transformations:moser_factoring_eps_III}
 U\left(\eps,\eps'\right) \frac{M_{x_i,1}\left(\eps\right)}{\eps} & = & \frac{M_{x_i,1}\left(\eps'\right)}{\eps'} U\left(\eps,\eps'\right),
 \;\;\;\;\;\;
 \forall \; x_i \; \in \; S', 
\eq
Eq.~(\ref{chapter_transformations:moser_factoring_eps_III}) yield $\Nmaster^2 \cdot |S'|$ linear equations for $\Nmaster^2$ unknowns (i.e. the
entries of $U(\eps,\eps')$).
This system can be solved with standard tools from linear algebra.
This yields the sought-after transformation $U(\eps,\eps')$, which factors out (for any choice of $\eps'$) the dimensional regularisation
parameter $\eps$.
We may then set $\eps'$ to any suitable value.

Let us look at an example. We consider again the maximal cut of the double box integral
(see eq.~(\ref{chapter_transformations:maxcut_double_box_eq_1})-eq.~(\ref{chapter_transformations:maxcut_double_box_eq_3})).
With
\bq
 \vec{J}
 & = &
 \left( \begin{array}{c}
 J_1 \\
 J_2 \\
 \end{array} \right)
 \; = \;
 \left( \begin{array}{l}
 \mathrm{MaxCut} \; I_{111111100} \\
 \mathrm{MaxCut} \; I_{1111111\left(-1\right)0} \\
 \end{array} \right)
\eq
the matrix $A$ of the differential equation $(d+A)\vec{J}=0$ is given by
\bq
\label{chapter_transformations:moser_example_doublebox_A}
 A & = &
 \left( \begin{array}{cc}
 2 & - 2 \eps \\
 2 \eps & 2 + 4 \eps \\
 \end{array} \right)
 \frac{dx}{x}
 +
 \left( \begin{array}{cc}
 0 & 2 \eps \\
 0 & - \eps \\
 \end{array} \right)
 \frac{dx}{x+1}.
\eq
The singular points are $S=\{0,-1,\infty\}$.
The residues are
\bq
 M_{0,1}
 \; = \;
 \left( \begin{array}{cc}
 2 & - 2 \eps \\
 2 \eps & 2 + 4 \eps \\
 \end{array} \right),
 \;\;\;\;
 M_{-1,1}
 \; = \;
 \left( \begin{array}{cc}
 0 & 2 \eps \\
 0 & - \eps \\
 \end{array} \right),
 \;\;\;\;
 M_{\infty,1}
 \; = \;
 \left( \begin{array}{cc}
 -2 & 0 \\
 -2 \eps & -2 - 3 \eps \\
 \end{array} \right),
 \;\;\;\;
\eq
We may read off $M_{0,1}$ and $M_{-1,1}$ directly from eq.~(\ref{chapter_transformations:moser_example_doublebox_A}).
The residue at infinity is given by $M_{\infty,1}=-M_{0,1}-M_{-1,1}$
(see exercise~\ref{chapter_transformations:exercise_moser_Fuchsian_form}).
The matrix $M_{0,1}$ has only the eigenvalue $2+2\eps$ with multiplicity $2$. 
The corresponding right eigenspace is one-dimensional and spanned by $(1,-1)^T$.
The matrix  $M_{-1,1}$ has the eigenvalues $0$ and $-\eps$. These are already as they should be.
The matrix $M_{\infty,1}$ has the eigenvalues $-2$ and $-2-3\eps$.
The left eigenspace for the eigenvalue $-2$ is spanned by $(1,0)$,
the left eigenspace for the eigenvalue $-2-3\eps$ is spanned by $(2,3)$.
We may balance the eigenvalue $2+2\eps$ at $x=0$ against one of the eigenvalues at $x=\infty$. Let us pick the eigenvalue 
$-2$ at $x=\infty$.
Thus we choose 
\bq
 \vec{v}_{R,0} \; = \; 
 \left(\begin{array}{r} 
   1 \\ -1
 \end{array} \right),
 & &
 \vec{v}_{L,\infty} \; = \; 
 \left(\begin{array}{r} 
   1 \\ 0
 \end{array} \right).
\eq
The projector $P$ reads then
\bq
 P
 & = &
 \left(\begin{array}{cc} 
   1 & 0 \\
  -1 & 0 \\
 \end{array} \right)
\eq
and the balance transformation reads
\bq
 U_1
 & = & 
 U^{\mathrm{balance}}\left(0,\infty,P\right)
 \; = \;
 \left(\begin{array}{cc} 
   -x & 0 \\
  x+1 & 1 \\
 \end{array} \right).
\eq
This gives
\bq
 \vec{J}' \; = \; U_1 \vec{J}, 
 \;\;\;\;\;\;
 \left( d + A' \right) \vec{J}'\; = \; 0,
 \;\;\;\;\;\;
 A' \; = \; M_{0,1}' \frac{dx}{x} + M_{-1,1}' \frac{dx}{x+1}
\eq
with
\bq
 M_{0,1}'
 \; = \;
 \left( \begin{array}{cc}
 1+2\eps & 0 \\
 1+ \eps & 2 + 2 \eps \\
 \end{array} \right),
 \;\;\;\;
 M_{-1,1}'
 \; = \;
 \left( \begin{array}{cc}
 0 & 2 \eps \\
 0 & - \eps \\
 \end{array} \right),
 \;\;\;\;
 M_{\infty,1}'
 \; = \;
 \left( \begin{array}{cc}
 -1-2\eps & -2\eps \\
 -1-\eps & -2 - \eps \\
 \end{array} \right).
 \nonumber
\eq
The situation has improved:
$M_{0,1}'$ has now the eigenvalues $1+2\eps$ and $2+2\eps$,
the matrix $M_{\infty,1}'$ has now the eigenvalues $-1$ and $-2-3\eps$.
We may iterated this procedure.
With 
\bq
 U \; = \; U_4 U_3 U_2 U_1,
 \;\;\;\;\;\;
 \vec{J}'' \; = \; U \vec{J}, 
 \;\;\;\;\;\;
 \left( d + A'' \right) \vec{J}''\; = \; 0,
\eq
and
\bq
 U_2
 & = & 
 \left(\begin{array}{cc} 
   -\frac{x-2\eps}{1+2\eps} & \frac{2\eps\left(1+x\right)}{\left(1+\eps\right)\left(1+2\eps\right)} \\
   \frac{\left(1+\eps\right)\left(1+x\right)}{1+2\eps}  & \frac{1-2x\eps}{1+2\eps} \\
 \end{array} \right),
 \nonumber \\
 U_3
 & = & 
 \left(\begin{array}{cc} 
  \frac{1-2\eps x}{1+2\eps} & - \frac{2\eps\left(1+x\right)}{\left(1+\eps\right)\left(1+2\eps\right)} \\
   -\frac{\left(1+\eps\right)\left(1+x\right)}{1+2\eps}  & - \frac{x-2\eps}{1+2\eps} \\
 \end{array} \right),
 \nonumber \\
 U_4
 & = & 
 \left(\begin{array}{cc} 
 1 & 0 \\
 -\left(1+x\right) & -x \\
 \end{array} \right)
\eq
we arrive at
\bq
 A''
 & = &
 \eps
 \left(\begin{array}{rr} 
 0 & -2 \\
 2 & 4 \\
 \end{array} \right)
 \frac{dx}{x} 
 + 
 \eps
 \left(\begin{array}{rr} 
 0 & 2 \\
 0 & -1 \\
 \end{array} \right)
 \frac{dx}{x+1}.
\eq
$A''$ is now already in $\eps$-form and there is nothing to be done in the third step.

Let us now look at the limitations of Moser's algorithm:
We first consider the one-loop two-point function with equal internal masses 
(example 1 in section~\ref{chapter_iterated_integrals:deriving_the_dgl}).
We already know from section~\ref{chapter_iterated_integrals:fibre_transformation}
that in the transformation to the $\eps$-form square roots appear (see eq.~(\ref{chapter_iterated_integrals:trafo_bubble})).
We expect that there is no rational fibre transformation, which brings the differential equation into the $\eps$-form.
Let's see where Moser's algorithm fails: The differential equation in eq.~(\ref{chapter_iterated_integrals:dgl_bubble_pre_canonical})
is already in Fuchsian form, so we may start directly with step $2$ of Moser's algorithm.
We write the quantity $A$ appearing in eq.~(\ref{chapter_iterated_integrals:dgl_bubble_pre_canonical})
as
\bq
 A 
 & = &
 M_{0,1}
 \frac{dx}{x}
 +
 M_{-4,1}
 \frac{dx}{x+4}
\eq
with
\bq
 M_{0,1}
 \; = \;
 \left( \begin{array}{cc}
 0 & 0 \\
 \frac{1-\eps}{2} & \frac{1}{2} \\
 \end{array} \right),
 \;\;\;\;
 M_{-4,1}
 \; = \;
 \left( \begin{array}{cc}
 0 & 0 \\
 -\frac{1-\eps}{2} & -\frac{1}{2} + \eps \\
 \end{array} \right),
 \;\;\;\;
 M_{\infty,1}
 \; = \;
 \left( \begin{array}{cc}
 0 & 0 \\
 0 & -\eps \\
 \end{array} \right).
 \;\;\;\;
\eq
The matrix $M_{0,1}$ has the eigenvalues $0$ and $\frac{1}{2}$, the matrix $M_{-4,1}$ has the eigenvalues $0$ and $-\frac{1}{2}+\eps$.
As we can only shift eigenvalues by units of $1$ with Moser's algorithm, 
there is no way to balance them such that they become zero or proportional to $\eps$.

A a second counter example we look at the two-loop sunrise integral with equal internal masses
(example 4 in section~\ref{chapter_iterated_integrals:deriving_the_dgl}).
We look at the maximal cut in $D=2-2\eps$ dimensions.
With
\bq
 \vec{J}
 \; = \;
 \left( \begin{array}{c}
 I_{111}\left(2-2\eps\right) \\ I_{211}\left(2-2\eps\right) \\
 \end{array} \right),
 \;\;\;\;\;\;
 \left( d + A \right) \vec{J} \; = \; 0,
 \;\;\;\;\;\;
 M_{0,1}
 \frac{dx}{x}
 +
 M_{-1,1}
 \frac{dx}{x+1}
 +
 M_{-9,1}
 \frac{dx}{x+9}
\eq
we have
\begin{align}
 M_{0,1}
 & = 
 \left( \begin{array}{cc}
 1 + 2 \eps & -3 \\
 \frac{1}{3} \left(1+2\eps\right) \left(1+3\eps\right) & -1-3\eps \\
 \end{array} \right),
 &
 M_{-1,1}
 & = 
 \left( \begin{array}{cc}
 0 & 0 \\
 - \frac{1}{4} \left(1+2\eps\right) \left(1+3\eps\right) & 1+2\eps \\
 \end{array} \right),
 \nonumber \\
 M_{-9,1}
 & = 
 \left( \begin{array}{cc}
 0 & 0 \\
 - \frac{1}{12} \left(1+2\eps\right) \left(1+3\eps\right) & 1+2\eps \\
 \end{array} \right),
 & 
 M_{\infty,1}
 & =
 \left( \begin{array}{cc}
 -1-2\eps & 3 \\
 0 & -1-\eps \\
 \end{array} \right).
\end{align}
We do not expect that this differential equation can be put into an $\eps$-form with a rational transformation.
We will discuss this example in more detail in the context of elliptic curves in chapter~\ref{chapter_elliptics}.
Let's see where Moser's algorithm fails:
We find for the eigenvalues
\begin{alignat}{2}
 & \mathrm{Eigenvalues}\left(M_{0,1}\right) & \; = \; & \left\{ 0, -\eps \right\}, 
 \nonumber \\
 & \mathrm{Eigenvalues}\left(M_{-1,1}\right) & \; = \; & \left\{ 0, 1+2\eps \right\}, 
 \nonumber \\
 & \mathrm{Eigenvalues}\left(M_{-9,1}\right) & \; = \; & \left\{ 0, 1+2\eps \right\}, 
 \nonumber \\
 & \mathrm{Eigenvalues}\left(M_{\infty,1}\right) & \; = \; & \left\{ -1-\eps, -1-2\eps \right\}.
\end{alignat}
For example, we may balance the eigenvalue $1+2\eps$ of $M_{-1,1}$ against one of the eigenvalues of $M_{\infty,1}$.
In the next step we would like to balance the eigenvalue $1+2\eps$ of $M_{-9,1}$ against the other eigenvalue of $M_{\infty,1}$.
However, we discover that in this case the corresponding eigenspaces are orthogonal, i.e.
\bq
 \vec{v}_{L,\infty}^T \cdot \vec{v}_{R,-9}
 & = & 0,
\eq
which prohibits the definition of the projector in eq.~(\ref{chapter_transformations:moser_def_projector}).
This observation is independent of the choices we made for the first balance transformation.

\subsection{Leinartas decomposition}
\label{chapter_transformations:fibre_transformation:leinartas}

Up to now we showed in section~\ref{chapter_transformations:fibre_transformation:reduction_to_univariate}
how to reduce a multivariate problem to a univariate problem.
In sections.~\ref{chapter_transformations:fibre_transformation:picard_fuchs}-\ref{chapter_transformations:fibre_transformation:moser}
we treated the univariate case.
Of course, once the univariate case is solved, we have to lift the result to the multivariate case.

In this section we start to treat the multivariate case directly.
The first challenge we have to face is how to represent a rational function.
In the univariate case we may use partial fractioning:
Let $p(x),q(x) \in {\mathbb C}[x]$ and assume that $q(x)$ factorises as
\bq
 q\left(x\right)
 & = & 
 c \prod\limits_{j=1}^r \left(x-x_j\right)^{o_j}.
\eq
We set further $o_\infty=2 +\deg p - \deg q$. 
The quantity $o_\infty$ denotes the order of the pole at infinity.
Using partial fraction decomposition we may write the rational function $p(x)/q(x)$ as
\bq
 \frac{p\left(x\right)}{q\left(x\right)}
 & = &
 \sum\limits_{j=0}^{o_{\infty}-2}
 a_j x^j 
 +
 \sum\limits_{j=1}^r \sum\limits_{k=1}^{o_j}
 \frac{b_{j,k}}{\left(x-x_j\right)^k}.
\eq
If $\deg p < \deg q$ the polynomial part is absent.

Extending partial fractioning iteratively to the multivariate case may introduce spurious poles and may lead to
infinite loops.
Consider as an example the rational function
\bq
 f\left(x_1,x_2\right) & = & \frac{1}{\left(x_1+x_2\right)\left(x_1-x_2\right)}.
\eq
Partial fractioning with respect to $x_1$ leads to
\bq
 f\left(x_1,x_2\right) & = &
 \frac{1}{2 x_2 \left(x_1-x_2\right)} - \frac{1}{2 x_2 \left(x_1+x_2\right)}.
\eq
This introduces the spurious singularity $x_2=0$.
A subsequent partial fraction decomposition with respect to $x_2$ leads to
\bq
 f\left(x_1,x_2\right) & = &
 \frac{1}{2 x_1 \left(x_1+x_2\right)} + \frac{1}{2 x_1 \left(x_1-x_2\right)}.
\eq
This step introduces the spurious singularity $x_1=0$ and spoils the partial fractioning 
with respect to $x_1$.

In the multivariate case we may use the Leinartas decomposition \cite{Leinartas:1978}.
Let ${\mathbb F}$ be a field and $\overline{\mathbb F}$ the 
algebraic closure.
We start with polynomials $p, q \in {\mathbb F}[x_1,\dots,x_n]$ in $n$ variables
$x=(x_1,\dots,x_n)$.
A rational function is a quotient of two polynomials:
\bq
 f\left(x\right)
 & = &
 \frac{p\left(x\right)}{q\left(x\right)}.
\eq
Let's assume that we know the factorisation of the denominator polynomial into
irreducible polynomials:
\bq
 q\left(x\right)
 & = &
 \prod\limits_{j=1}^r \left( q_j\left(x\right) \right)^{o_j}.
\eq
For each irreducible polynomial $q_j(x)$ we denote the corresponding algebraic variety by
\bq
 V_j & = & \left\{ \; x \in \overline{\mathbb F}^n \; | \; q_j\left(x\right) \; = \; 0 \right\}.
\eq
Let $S$ be a subset of $\{1,\dots,r\}$.
We say that the polynomials $q_j(x)$, $j \in S$ have no common zero, if
\bq
 \bigcap\limits_{j \in S} V_j & = & \emptyset.
\eq
In this case 
\index{Hilbert's Nullstellensatz}
{\bf Hilbert's Nullstellensatz} 
guarantees that there are polynomials
$h_j(x) \in {\mathbb F}[x]$, $j \in S$ such that
\bq
\label{chapter_transformations:nullstellensatz_certificate}
 \sum\limits_{j \in S} h_j\left(x\right) q_j\left(x\right) 
 & = & 1.
\eq
Eq.~(\ref{chapter_transformations:nullstellensatz_certificate}) 
is called a 
\index{Nullstellensatz certificate}
{\bf Nullstellensatz certificate}.
An algorithm to compute the polynomials $h_j(x)$ is reviewed 
in appendix~\ref{appendix_algorithms:Nullstellensatz_certificate}.
If the denominator of our rational function contains a set of irreducible polynomials, which
do not share a common zero, we may insert the left-hand side of eq.~(\ref{chapter_transformations:nullstellensatz_certificate}) in the numerator and cancel in each term the common factor $q_j(x)$ 
in the numerator and the denominator.

To give an example consider
\bq
 f_1\left(x\right)
 & = &
 \frac{1}{x_1 x_2 \left(x_1+x_2-1\right)}.
\eq
The polynomials $q_1=x_1$, $q_2=x_2$ and $q_3=x_1+x_2-1$ do not share a common zero
and we have
\bq
 q_1 + q_2 - q_3 & = & 1.
\eq
Thus
\bq
\label{chapter_transformations:example_nullstellensatz_decomposition}
 f_1\left(x\right)
 & = &
 \frac{1}{x_2 \left(x_1+x_2-1\right)}
 +
 \frac{1}{x_1 \left(x_1+x_2-1\right)}
 -
 \frac{1}{x_1 x_2}.
\eq
The polynomials in the denominators of the individual terms in eq.~(\ref{chapter_transformations:example_nullstellensatz_decomposition}) have common zeros, hence a further reduction
with the help of Hilbert's Nullstellensatz is not possible.

We say that a set of polynomials $q_1(x), \dots, q_r(x)$
is 
\index{algebraically independent polynomials}
{\bf algebraically independent} 
if there exists no non-zero polynomial $a$ in $r$ variables
with coefficients in ${\mathbb F}$ such that
\bq
 a\left( q_1, \dots, q_r \right) & = & 0
\eq
in ${\mathbb F}[x]$, otherwise the set of polynomials $q_1(x), \dots, q_r(x)$
is called {\bf algebraically dependent} and the polynomial $a$ is called an
\index{annihilator of a set of polynomials}
{\bf annihilator}.
If $q_1, \dots, q_r$ are algebraically dependent, then also
$q_1^{b_1}, \dots, q_r^{b_r}$ with $b_j \in {\mathbb N}$ are algebraically dependent.
A set of $r$ polynomials $q_1, \dots, q_r$ is always algebraically dependent if $r>n$,
where $n$ denotes the number of variables $x_1, \dots, x_n$.

In appendix~\ref{appendix_algorithms:annihilator} we review an algorithm which allows us to decide
if a given set of polynomials is algebraically dependent, and in the case it is, computes an annihilating polynomial.

We may then reduce the denominators further, until the polynomials in the denominator are algebraically independent.
This is done as follows:
Let us introduce a multi-index notation:
\bq
 x^b \; = \;  \prod\limits_{j=1}^n x_j^{b_j},
 & & 
 q^{\nu} \; = \;  \prod\limits_{j=1}^r q_j^{\nu_j}.
\eq
Let's assume that $q_1, \dots, q_r$ are algebraically dependent
and that the denominator of the rational function is given by
\bq
 q & = & \prod\limits_{j=1}^r q_j^{o_j}.
\eq
We set $Q_j=q_j^{o_j}$. Then also $Q_1, \dots, Q_r$ are algebraically dependent and with
the notation above we write the annihilating polynomial as
\bq
 a\left(Q_1,\dots,Q_r\right)
 & = &
 \sum\limits_{\nu \in I} c_{\nu} Q^{\nu}.
\eq
Let $\nu^{(0)}$ be an $r$-tuple $(\nu^{(0)}_1,\dots,\nu^{(0)}_r)$
with the smallest degree
\bq
 \deg \nu^{(0)} & = & 
 \sum\limits_{j=1}^r \nu^{(0)}_j.
\eq
Then
\bq
 1 & = & 
 \sum\limits_{\nu \in I \backslash\{\nu^{(0)}\} } \frac{c_{\nu}}{c_{\nu^{(0)}}}  Q^{\nu-\nu^{(0)}}.
\eq
As $\nu^{(0)}$ is an $r$-tuple with smallest norm, in each term at least one $Q_j$ in $Q^{\nu-\nu^{(0)}}$
has a positive exponent and removes the corresponding $Q_j$ from the denominator.
As a result, each term will have fewer polynomials in the denominator and repeating this procedure we arrive
at denominators, which are algebraically independent.

Let look at an example:
\bq
 f_1\left(x\right)
 & = &
 \frac{1}{x_1 x_2 \left(x_1+x_2\right)}.
\eq
The polynomials $q_1=x_1$, $q_2=x_2$ and $q_3=x_1+x_2$ share a common zero ($x_1=x_2=0$),
so a decomposition with Hilbert's Nullstellensatz is not possible.
However, they are algebraically dependent. 
An annihilating polynomial is given by
\bq
 a\left(q_1,q_2,q_3\right)
 & = &
 q_1 + q_2 - q_3.
\eq
The three terms ($q_1$, $q_2$ and $(-q_3)$) all are of degree one. Let's pick the last one.
We have
\bq
 1 & = & \frac{q_1}{q_3} + \frac{q_2}{q_3}
\eq
and
\bq
 f_1\left(x\right)
 & = &
 \frac{1}{x_2 \left(x_1+x_2\right)^2}
 +
 \frac{1}{x_1 \left(x_1+x_2\right)^2}.
\eq
Note that the decomposition is not unique, we may picked $q_1$ or $q_2$ as the term $c_{\nu^{(0)}} q^{\nu^{(0)}}$.
Note also that the decomposition may increase the power of the remaining polynomials in the denominator.

Putting Hilbert's Nullstellensatz decomposition and the decomposition based on algebraic dependence together,
we arrive at the Leinartas decomposition:
\begin{tcolorbox}
{\bf Leinartas decomposition}:
A rational function 
\bq
 f
 \; = \; 
 \frac{p}{q},
 & &
 q
 \; = \;
 \prod\limits_{j=1}^r q_j^{o_j},
 \;\;\;\;\;\;\;\;\;
 p, q \; \in \; {\mathbb F}\left[x_1,\dots,x_n\right]
\eq
may be written as 
\bq
 f
 & = &
 \sum\limits_S \frac{p_s}{\prod\limits_{j \in S} q_j^{b_j}},
\eq
where the sum is over subsets $S$ of $\{1,\dots,r\}$ such that 
the polynomials $q_{j}$, $j \in S$ have a common zero and are algebraically independent.
\end{tcolorbox}
Applications towards Feynman integrals have been considered in \cite{Meyer:2016slj,Meyer:2017joq,Heller:2021qkz}.

\subsection{Maximal cuts and constant leading singularities}
\label{chapter_transformations:maximal_cuts_and_constant_leading_singularities}

The study of the maximal cuts is one of the most efficient ways of finding an appropriate fibre transformation,
in particular if the Feynman integrals evaluate to multiple polylogarithms.
Suppose somebody gives us a transformation matrix $U$
\bq
 \vec{I}' & = & U \vec{I}.
\eq
Then it is easy to check if this fibre transformation transforms the differential equation to an $\eps$-form.
We simply calculate
\bq
 A' & = & U A U^{-1} + U d U^{-1}
\eq
and check if $A'$ is in $\eps$-form.
The problem is only to come up initially with the concrete form of the transformation matrix $U$.
This is a situation where a heuristic method may work well: Guessing a suitable $U$ may outperform
any systematic algorithm to construct the matrix $U$.

For the technique discussed below we will focus on the diagonal blocks 
(e.g. the blocks $A_1$, $A_2$ and $A_4$ in eq.~(\ref{chapter_transformations:example_block_matrix})).
The study of the maximal cut allows us to obtain the transformation matrix for this diagonal block up
to an $\eps$-dependent prefactor (i.e. the unknown prefactor may depend on $\eps$, but not on the kinematic variables $x$).

A diagonal block corresponds to the maximal cut of a particular sector \cite{Primo:2016ebd}.
Let us denote the number of master integrals for this sector by $N_{\mathrm{sector}}$
and the integrands of the master integrals by $\varphi_1, \dots, \varphi_{N_\mathrm{sector}}$.
The number $N_{\mathrm{sector}}$ equals the dimension of the diagonal block.
As before, we denote by $N_{\mathrm{prop}}$ the number of propagators having positive indices.
Let's consider a Baikov representation for these integrals.
The number of Baikov variables is denoted by $\NV$. 
For the maximal cut we take a $N_{\mathrm{prop}}$-fold residue.
This leaves us with 
\bq
 \NV - N_{\mathrm{prop}}
\eq
integrations for the maximal cut integrals.
We now choose $N_{\mathrm{sector}}$ independent integration domains
for the remaining integrations.
We denote these integration domains each combined with the $N_{\mathrm{prop}}$-fold residue integration domain
by ${\mathcal C}_1, \dots, {\mathcal C}_{N_\mathrm{sector}}$.
Thus, ${\mathcal C}_j$ defines an $\NV$-dimensional integration domain.
The integration domains are independent, if the $N_{\mathrm{sector}} \times N_{\mathrm{sector}}$-matrix
with entries
\bq
 \left\langle \varphi_i | {\mathcal C}_j \right\rangle
 & = &
 \int\limits_{{\mathcal C}_j} \varphi_i
\eq
has full rank.
We are interested in choosing the integration domains ${\mathcal C}_j$ as simple as possible.
Particular simple integration domains are products of circles around singular points.
These correspond to additional residue calculations.

Having fixed $N_{\mathrm{sector}}$ independent integration domains, we then look for 
$N_{\mathrm{sector}}$ integrands $\varphi_1', \dots, \varphi_{N_\mathrm{sector}}'$
such that the first term in the Laurent expansion in the dimensional regularisation parameter $\eps$ of
\bq
 \left\langle \varphi_i' | {\mathcal C}_j \right\rangle
 & = &
 \int\limits_{{\mathcal C}_j} \varphi_i'
\eq
is a constant (i.e. independent of the kinematic variables $x$) of weight zero for all $j$.
More precisely, let $j_{\min}$ be defined by
\bq
\label{chapter_transformations:def_j_min}
 j_{\min} 
 & = &
 \min\limits_j\left( \mathrm{ldegree}\left(\left\langle \varphi_i' | {\mathcal C}_j \right\rangle,\eps\right)\right),
\eq
where $\mathrm{ldegree}$ denotes the low degree of a Laurent series.
Note that $j_{\min}=j_{\min}(i)$ depends on $i$, two integrands $\varphi_{i_1}'$ and $\varphi_{i_2}'$
may have $j_{\min}(i_1) \neq j_{\min}(i_2)$.
We require that for all $j$ the term of order $\eps^{j_{\min}}$ is a constant of weight zero:
\bq 
\label{chapter_transformations:constant_leading_singularities}
 \mathrm{coeff}\left( \left\langle \varphi_i' | {\mathcal C}_j \right\rangle, \eps^{j_{\min}} \right) \cdot \eps^{j_{\min}}  
 & = &
 \mbox{constant of weight zero},
\eq
where $\mathrm{coeff}(f,\eps^j)$ denotes the coefficient of $\eps^j$ in the Laurent expansion of $f$
around $\eps=0$.
The weight counting is as follows: We define the weight of rational numbers to be zero.
The transcendental constant $\pi$ has weight one, the dimensional regularisation parameter $\eps$ has weight $(-1)$.
The weight of a product is the sum of the weights of its factors.

Let us denote by ${\mathcal C}_{\mathrm{MaxCut}}$ the integration domain for the original maximal cut.
If $\varphi'$ satisfies eq.~(\ref{chapter_transformations:constant_leading_singularities}), we say that
\bq
 \mathrm{MaxCut} \; I
 & = &
 \int\limits_{{\mathcal C}_{\mathrm{MaxCut}}} \varphi'
\eq
has 
\index{constant leading singularities}
{\bf constant leading singularities}.

There is no principal obstruction for restricting us to a diagonal block.
In theory at least we could consider the full system of $\Nmaster$ master integrals.
Let ${\mathcal C}$ denote the original integration domain for the $\Nmaster$ master integrals
and let 
${\mathcal C}_1, \dots, {\mathcal C}_{\Nmaster}$ denote a set of $\Nmaster$ independent integration domains.
If $\varphi'$ satisfies the condition of eq.~(\ref{chapter_transformations:constant_leading_singularities})
for ${\mathcal C}_1, \dots, {\mathcal C}_{\Nmaster}$, we say that
\bq
 I
 & = &
 \int\limits_{{\mathcal C}} \varphi'
\eq
has constant leading singularities \cite{Cachazo:2008vp,Arkani-Hamed:2010pyv}.

Integrals with constant leading singularities are a guess for a basis of master integrals, which puts the differential
equation into an $\eps$-form.
In practice we will be using the requirement of eq.~(\ref{chapter_transformations:constant_leading_singularities}).
We mention that the requirement of eq.~(\ref{chapter_transformations:constant_leading_singularities})
is not a necessary requirement for transforming the differential equation into an $\eps$-form.
We will see in chapter~\ref{chapter_elliptics}
an explicit example, which does not satisfy eq.~(\ref{chapter_transformations:constant_leading_singularities})
but nevertheless puts the differential equation into an $\eps$-form.

Let us now look at an example.
We consider the two-loop double box integral (example 3 in section~\ref{chapter_iterated_integrals:deriving_the_dgl}).
This is a system with eight master integrals.
Suppose we already found suitable master integrals, which puts the sub-system of the first six master integrals into an
$\eps$-form.
A possible choice can be read off from eq.~(\ref{chapter_iterated_integrals:trafo_double_box}):
\bq
 I_{{\bm{\nu}}_1}'
 & = &
 \frac{g_1\left(\eps\right)}{x} I_{001110000},
 \nonumber \\
 I_{{\bm{\nu}}_2}'
 & = &
 g_1\left(\eps\right) I_{100100100},
 \nonumber \\
 I_{{\bm{\nu}}_3}'
 & = &
 \eps^2\left(1-2\eps\right)^2 I_{011011000},
 \nonumber \\
 I_{{\bm{\nu}}_4}'
 & = &
 g_2\left(\eps\right) I_{100111000}
 \nonumber \\
 I_{{\bm{\nu}}_5}'
 & = &
 6 \eps^3\left(1-2\eps\right) I_{111100100} + g_2\left(\eps\right) I_{100111000},
 \nonumber \\
 I_{{\bm{\nu}}_6}'
 & = &
 3\eps^4\left(1+x\right) I_{101110100},
\eq
where $g_1(\eps)$ and $g_2(\eps)$ have been defined in eq.~(\ref{chapter_iterated_integrals:def_g1_g2}).
Thus we are left with finding a fibre transformation, which transforms the last sector, 
consisting of the two master integrals $I_{111111100}$ and $I_{1111111\left(-1\right)0}$
into an $\eps$-form.
We consider the maximal cut of this sector for the integrals $I_{1111111\nu 0}$ (see eq.~(\ref{chapter_iterated_integrals:maxcut_double_box})).
With $\mu^2=t$ we have
\bq
\lefteqn{
 \mathrm{MaxCut} \; I_{1111111\nu 0}
 = } & &
 \nonumber \\
 & &
 \left(2\pi i\right)^7
 \frac{2^{4\eps} \left(s+t\right)^{\eps} t^{3+\nu+3\eps}}{4 \pi^3 \left(\Gamma\left(\frac{1}{2}-\eps\right)\right)^2 s^{2+2\eps}}
 \int\limits_{{\mathcal C}_{\mathrm{MaxCut}}} dz_8 \;
 z_8^{-1-2\eps} 
 \left(t-z_8\right)^{-1-\eps}
 \left(s+t-z_8\right)^{\eps}
 z_8^{-\nu}.
\eq
We now choose two independent integration domains:
\bq
 {\mathcal C}_1 & : & \mbox{small circle around $z_8=0$ for the $z_8$-integration},
 \nonumber \\
 {\mathcal C}_2 & : & \mbox{small circle around $z_8=t$ for the $z_8$-integration}.
\eq
We set
\bq
 \varphi_\nu
 & = &
 \frac{2^{4\eps} \left(s+t\right)^{\eps} t^{3+\nu+3\eps}}{4 \pi^3 \left(\Gamma\left(\frac{1}{2}-\eps\right)\right)^2 s^{2+2\eps}}
 \;
 z_8^{-1-2\eps} 
 \left(t-z_8\right)^{-1-\eps}
 \left(s+t-z_8\right)^{\eps}
 z_8^{-\nu} d^8z.
\eq
With $x=s/t$ we have
\bq
 \left\langle \varphi_0 | {\mathcal C}_1 \right\rangle
 \; = \;
 \frac{64 \pi^4}{x^2}
 + {\mathcal O}\left(\eps\right),
 & &
 \left\langle \varphi_0 | {\mathcal C}_2 \right\rangle
 \; = \;
 - \frac{64 \pi^4}{x^2}
 + {\mathcal O}\left(\eps\right).
\eq
The integral 
\bq
 \mathrm{MaxCut} \; I_{111111100}
 & = &
 \left\langle \varphi_0 | {\mathcal C}_{\mathrm{MaxCut}} \right\rangle
\eq
does not have constant leading singularities, but it is easy to fix this issue:
We multiply the integrand by $x^2$.
If in addition we multiply by $\eps^4$, the leading singularities are constants of weight zero.
Strictly speaking we can only infer from the first term of the $\eps$-expansion of
$\langle \varphi_0 | {\mathcal C}_j \rangle$
that we should multiply by an $\eps$-dependent prefactor, whose $\eps$-expansion starts at $\eps^4$.
In this example we can verify a posteriori that $\eps^4$ is the correct 
$\eps$-dependent prefactor.
We now set
\bq
 \varphi_0' & = & \eps^4 x^2 \varphi_0.
\eq
Then
\bq
 \left\langle \varphi_0' | {\mathcal C}_1 \right\rangle
 \; = \;
 64 \pi^4 \eps^4
 + {\mathcal O}\left(\eps^5\right),
 & &
 \left\langle \varphi_0' | {\mathcal C}_2 \right\rangle
 \; = \;
 - 64 \pi^4 \eps^4
 + {\mathcal O}\left(\eps^5\right).
\eq
Thus
\bq
 \mathrm{MaxCut}\left( \eps^4 x^2 I_{111111100} \right)
 & = &
 \left\langle \varphi_0' | {\mathcal C}_{\mathrm{MaxCut}} \right\rangle
\eq
has constant leading singularities.

As this sector has two master integrals, we need a second master integral.
We consider $\varphi_{-1}$ and compute the leading singularities.
We obtain
\bq
 \left\langle \varphi_{-1} | {\mathcal C}_1 \right\rangle
 \; = \;
 0
 + {\mathcal O}\left(\eps\right),
 & &
 \left\langle \varphi_{-1} | {\mathcal C}_2 \right\rangle
 \; = \;
 - \frac{64 \pi^4}{x^2}
 + {\mathcal O}\left(\eps\right).
\eq
It follows that
\bq
 \mathrm{MaxCut}\left( 2 \eps^4 x^2 I_{1111111\left(-1\right)0} \right)
 & = &
 \left\langle 2 \eps^4 x^2 \varphi_{-1} | {\mathcal C}_{\mathrm{MaxCut}} \right\rangle
\eq
has constant leading singularities.
Including a prefactor of $2$ or not is irrelevant at this stage. We included it to be consistent with eq.~(\ref{chapter_iterated_integrals:trafo_double_box}).

It is easily verified, that the two master integrals
\bq
 \eps^4 x^2 I_{111111100}
 & \mbox{and} &
 2 \eps^4 x^2 I_{1111111\left(-1\right)0} 
\eq
put the $2 \times 2$-diagonal block for this sector into an $\eps$-form.
It remains to treat the off-diagonal block with entries $A_{i,j}$, $i \in \{7,8\}$, $j\in \{1,2,3,4,5,6\}$.
This is most easily done with the methods of section~\ref{chapter_transformations:fibre_transformation:block_decomposition} (see eqs.~(\ref{chapter_transformations:trafo_offdiagonal_block})-(\ref{chapter_transformations:trafo_offdiagonal_block_diff_eq})).
One finds
\bq
 I_{{\bm{\nu}}_7}'
 & = &
 \eps^4 x^2 I_{111111100},
 \nonumber \\
 I_{{\bm{\nu}}_8}'
 & = &
 2 \eps^4 x^2 I_{1111111\left(-1\right)0} 
 + x \left[ 2 I_{{\bm{\nu}}_6}' + I_{{\bm{\nu}}_5}' + I_{{\bm{\nu}}_4}' - I_{{\bm{\nu}}_2}' - I_{{\bm{\nu}}_1}' \right].
\eq

\section{Base transformations}
\label{chapter_transformations:base_transformation}

Let us now discuss base transformations.
We assume that through an appropriate fibre transformation we transformed the differential equation 
\bq
 \left( d + A \right) \vec{I} & = & 0
\eq
into the form
\bq
 A
 & = &
 \eps \sum\limits_{j=1}^{\NL} \; C_j \; \omega_j,
\eq
where the $C_j$'s are $(\Nmaster \times \Nmaster)$-matrices, whose entries are algebraic numbers
and the $\omega_j$'s are dlog-forms 
\bq
 \omega_j & = & d \ln f_j,
\eq
with the $f_j$'s being {\bf algebraic} functions of the kinematic variables $x$.
Eq.~(\ref{chapter_iterated_integrals:example_1_eps_form_square_root_singularity}) is an example.
In this example
the differential one-form
\bq
\label{chapter_transformations:example_algebraic_log}
  d \ln\left(2+x+\sqrt{x\left(4+x\right)} \right)
\eq
appears.
We would like to find a base transformation, which transforms all $f_j$'s to {\bf rational} functions of the new kinematic variables $x'$.
If we achieve this, we may express the Feynman integrals $\vec{I}$ in terms of multiple polylogarithms.

For the example in eq.~(\ref{chapter_transformations:example_algebraic_log})
we have already seen in section~\ref{chapter_iterated_integrals:base_transformation} that the substitution
\bq
 x & = & \frac{\left(1-x'\right)^2}{x'}
\eq
rationalises the argument of the logarithm
\bq
 \ln\left(2+x+\sqrt{x\left(4+x\right)} \right)
 & = &
 \ln\left( \frac{2}{x'} \right).
\eq
We look for a systematic way to find such a transformation.

\subsection{Mathematical set up}

Assume that we have $n$ kinematic variables $x_1, \dots, x_n$.
(In this section we write for simplicity $n=\NB$.)
Consider a polynomial
\bq
 f\left(x_1,\dots,x_n\right) 
 & \in &
 {\mathbb C}\left[x_1,\dots,x_n\right].
\eq
We are interested in
\bq
\label{chapter_transformations:rationalisation_problem}
 \sqrt{f\left(x_1,\dots,x_n\right)}
\eq
and we seek a change of variables from $x_1,\dots,x_n$ to $x_1',\dots,x_n'$ such that
eq.~(\ref{chapter_transformations:rationalisation_problem}) becomes a rational function in the new variables $x_1',\dots,x_n'$.

We first introduce a few concept from algebraic geometry.
An 
\index{affine hypersurface}
{\bf affine hypersurface} 
$V$ is the zero set $V(f)$ of a polynomial $f\in \mathbb{C}[x_1,\dots,x_n]$ in $n$ variables, embedded in $\mathbb{C}^n$: 
\bq
    V(f) & \subset & \mathbb{C}^n.
\eq
The 
\index{degree of a hypersurface}
{\bf degree $d$ of the hypersurface} is the degree of the defining polynomial $f$.

Besides affine hypersurfaces we will also deal with projective hypersurfaces.
These are defined by homogeneous polynomials.
A polynomial $F \in \mathbb{C}[x_0,\dots,x_n]$ in $(n+1)$ variables $x_0,\dots,x_n$
is called {\bf homogeneous of degree $d$} 
if all its terms have the same degree $d$.
In particular, a degree-$d$ homogeneous polynomial satisfies
\bq
 F\left(\lambda x_0,\dots,\lambda x_n\right)
 & = &
 \lambda^d F\left(x_0,\ldots,x_n\right),
 \;\;\;\;\;\;
 \lambda \in \mathbb{C}.
\eq
Note that if a point $(x_0,\dots,x_n) \in \mathbb{C}^{n+1}$ is a zero of a homogeneous polynomial $F$, 
then every point $(\lambda x_0,\dots,\lambda x_n)$ is a zero of $F$.
Thus, the zero set of $F$ is a union of complex lines through the origin in $\mathbb{C}^{n+1}$. 
A 
\index{projective hypersurface}
{\bf projective hypersurface} is the set of zeros
of a homogeneous polynomial $F \in \mathbb{C}[x_0,\dots,x_n]$, embedded in ${\mathbb C} {\mathbb P}^n$:  
\bq
 V(F) & \subset & {\mathbb C} {\mathbb P}^n.
\eq
The 
\index{projective closure}
{\bf projective closure} of an affine hypersurface $V(f)\subset \mathbb{C}^n$ 
is the projective hypersurface $\overline{V} = V(F)\subset {\mathbb C} \mathbb{P}^n$, 
where $F$ is the 
\index{homogenisation}
{\bf homogenisation} of $f$.
We can homogenise a degree-$d$ polynomial $f$ in $n$ variables $x_1,\dots,x_n$ 
to turn it into a degree-$d$ homogeneous polynomial $F$ in $n+1$ variables $x_0,x_1,\dots,x_n$
in the following way:
decompose $f$ into the sum of its homogeneous components of various degrees, $f=g_0+\dots+g_d$, where $g_i$ has degree $i$.
Note that some $g_j$'s may be zero, but $g_d \neq 0$.
We have $g_j \in \mathbb{C}[x_1,\dots,x_n]$.
The homogenisation $F$ of $f$ is defined by
\bq
\label{chapter_transformations:homogenisation}
 F & = & x_0^d g_{0} + x_0^{d-1} g_1 + \dots +x_0 g_{d-1} + g_d
 \; \in \; \mathbb{C}[x_0,x_1,\dots,x_n].
\eq
We call $x_0$ the homogenising variable. 
Note that the restriction of $F$ to the plane $x_0=1$ gives the original polynomial $f$.

As an example, consider the affine parabola $V(y-x^2) \subset \mathbb{C}^2$.
The homogenisation of $f = y-x^2 \in {\mathbb C}[x,y]$ is
\bq
 F & = & z y - x^2 \; \in \; {\mathbb C}\left[x,y,z\right].
\eq
The projective closure $\overline{V} = V(F) \subset {\mathbb C} \mathbb{P}^2$
consists of all points in ${\mathbb C} \mathbb{P}^2$,
which correspond to lines in $\mathbb{C}^3$ that connect the points on the original parabola in the plane $z=1$ 
with the origin plus 
the line $x=z=0$, i.e., the $y$-axis.
The latter line corresponds to the 
``infinitely distant point'' $[0:1:0] \in \mathbb{C} \mathbb{P}^2$.

If $V(f)$ is a hypersurface, affine or projective,
a point $p \in V$ is said to be of 
\index{multiplicity of a point}
{\bf multiplicity} $o \in \mathbb{N}$ if 
all partial derivatives of order $<o$ vanish at $p$
\bq
 \frac{\partial^{i_1+\ldots+i_n}f}{\partial x_1^{i_1}\cdots\partial x_n^{i_n}}(p)
 \; = \; 0
 & &
 \mbox{with} \; i_1+\dots+i_n<o
\eq
and if there exists at least one non-vanishing $o$-th partial derivative
\bq
 \frac{\partial^{i_1+\dots+i_n}f}{\partial x_1^{i_1}\cdots\partial x_n^{i_n}}(p)
 \; \neq \; 0
 & &
 \mbox{with} \; i_1+\dots+i_n=o.
\eq    
We write $\mathrm{mult}_p(V)=o$.
Points of multiplicity $1$ are called 
\index{regular point}
{\bf regular points},
points of multiplicity $o>1$ are called 
\index{singular point}
{\bf singular points} of $V$.

As an example of an (affine) variety with a singular point consider 
the nodal cubic $V(f) \in {\mathbb C}^2$ defined by $f=y^2-x^3-x^2$.
\begin{figure}
\begin{center}
\includegraphics[scale=1.0]{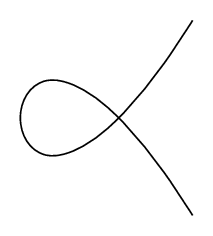}
\end{center}
\caption{
The nodal cubic defined by $y^2-x^3-x^2=0$.
}
\label{chapter_transformations:nodal_cubic}
\end{figure}
The curve is shown in fig.~\ref{chapter_transformations:nodal_cubic}.
The point $p=(0,0)$ is a singular point of multiplicity $o=2$:
One easily verifies 
\bq
 f(p) 
 \; = \; 
 \frac{\partial f}{\partial x}(p)
 \; = \;
 \frac{\partial f}{\partial y}(p)
 \; = \; 0,
 & &
 \frac{\partial^2 f}{\partial x^2}(p) \; \neq \; 0.
\eq
If $d$ denotes the degree of a given hypersurface, we will be particularly interested in the points of multiplicity $d-1$.

To a square root we associate a hypersurface as follows:
Consider a square root $\sqrt{p/q}$ of a rational function, where $p,q \in {\mathbb C}[x_1,\dots,x_n]$ are polynomials.
We introduce a new variable $r$ and set $r = \sqrt{p/q}$. 
After squaring and clearing the denominator we obtain $q r^2 = p$.
Thus we define
\bq
 f & = & q \cdot r^2 - p \; \in \; {\mathbb C}\left[r,x_1,\dots,x_n\right]
\eq
and call $V(f)$ the {\bf associated hypersurface}.
Note that we can also associate a hypersurface to more general algebraic functions such as roots of degree greater than $2$ or nested roots.
For example, $V(r^3-x^3-x^2)$ is associated to $\sqrt[\leftroot{0}\uproot{0}3]{x^3+x^2}$ and
\bq
 V((r^2-x^2)^2-x^4-y^3)
 & \mbox{is associated to} &
 \sqrt{x^2+\sqrt{x^4+y^3}}.
\eq

\subsection{Rationalisation algorithms}

Let us start with an example:
We consider the square root $\sqrt{1-x^2}$ 
and we look for an appropriate transformation $\varphi_x: x' \mapsto \varphi_x(x')$ that turns
\bq
 \sqrt{1-\left(\varphi_x(x')\right)^2} 
\eq
into a rational function of $x'$. One easily checks that the parametrisation
\bq
 \varphi_x(x')
 & = &
 \frac{1-x'{}^2}{1+x'{}^2}
\eq
solves the problem, leading to 
\bq
 \sqrt{1-\left(\varphi_x(x')\right)^2}
 & = &
 \frac{2x'}{1+x'{}^2}.
\eq
There is a systematic way to construct $\varphi_x$.
We start with the associated hypersurface: We introduce a new variable $y$, set $y=\sqrt{1-x^2}$ and arrive after squaring
at the defining equation for the associated hypersurface
\bq
 x^2+y^2-1 & = & 0.
\eq
This equation describes the unit circle. 
We see that asking for a rational change of variables $\varphi_x(x')$ which rationalises the square root $y=\sqrt{1-x^2}$ 
is the same as asking for rational functions $(\varphi_x(x'),\varphi_y(x'))$ which parametrise the unit circle. 
If one can find such rational functions, one would call the circle a {\bf rational algebraic hypersurface}.
For the square root $\sqrt{1-x^2}$, the solution to the rationalisation problem is known since 1500 BC \cite{Ifrah:book}: 
\begin{figure}
\begin{center}
\includegraphics[scale=1.0]{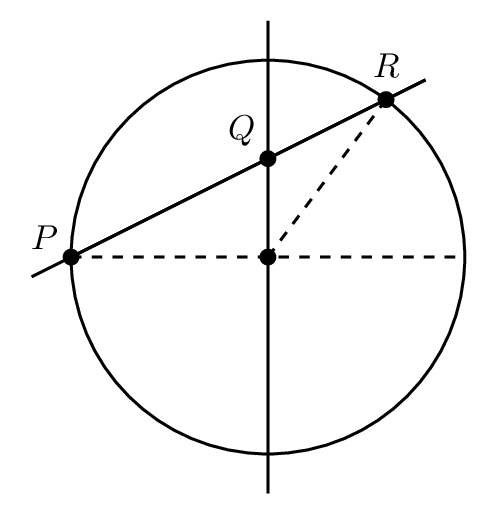}
\end{center}
\caption{
Parametrising the circle by a 1-parameter family of lines.
}
\label{chapter_transformations:parametrisation_circle}
\end{figure}
Consider a fixed point $P$ on the circle and a variable point $Q$ 
moving on a line not passing through $P$ (see fig.~\ref{chapter_transformations:parametrisation_circle}). 
Then look at the second point of intersection $R$ of the line $PQ$ with the circle. 
We observe that, if $Q$ traces its line, then $R$ traces the circle. 
If we take the point $P$ to be $(-1,0)$ and assume $Q$ to move along the $y$-axis, i.e., $Q=(0,x')$, 
then the defining equation of the line $PQ$ is given by $y=x'(1+x)$ from which we find the parametrisation
\bq
 R(x') 
 & = &
 (\varphi_x(x'),\varphi_y(x'))
 \; = \;
 \left(\frac{1-x'{}^2}{1+x'{}^2},\frac{2x'}{1+x'{}^2}\right)
\eq
of the unit circle by a short calculation:
simply determine the intersection points of the line $PQ:y=x'(1+x)$ and the circle $x^2+y^2=1$. 
The first point of intersection is $P$, the second one yields $R(x')$.
Note that, to calculate the expression for $R(x')$, one solely needs rational operations (addition, subtraction, multiplication, division) 
on polynomial expressions with coefficients in $\mathbb{Q}$. 
This is precisely the reason why the above method returns a rational function of $x'$.

We ensure rational coefficients by choosing $P$ to be a point with all coordinate entries lying in $\mathbb{Q}$. 
In principle, nothing prevents us from taking $P \notin \mathbb{Q}^2$,
e.g., choosing $P=\left(-\frac{1}{\sqrt{2}},-\frac{1}{\sqrt{2}}\right)$ as the starting point of our construction. 
Still, the method would return a rational function. 
However, the coefficients of this rational function would no longer be rational, but rather contain factors of $\sqrt{2}$.

This construction generalises to hypersurfaces of degree $d$, whenever the hypersurface possesses a point of multiplicity $(d-1)$:
A generic line through this point will intersect the hypersurface at one other point and
provide a rational parametrisation of the hypersurface.
If we consider affine hypersurfaces, it is not necessary that the affine hypersurface possesses a point of multiplicity
$(d-1)$, it suffices if the projective closure possesses a point of multiplicity $(d-1)$.
In other words, the point of multiplicity $(d-1)$ may be a point at infinity.
This leads to the following algorithm:
\begin{tcolorbox}[breakable]
\begin{myalgorithm}
\label{chapter_transformations:algo_rationalisation_multiplicity}
Rationalisation of an irreducible degree-$d$ hypersurface $V$ defined by $f$ 
whose projective closure $\overline{V}$ has at least one point of multiplicity $d-1$.
\begin{enumerate}
\item Choose a point $p_0$ with $\mathrm{mult}_{p_0}(\overline{V})=d-1$.
\item If $p_0$ is not at infinity, go on with step 3. and finish with step 4.
If on the other hand $p_0$ is at infinity, consider another affine chart $V^\prime$ of the projective closure $\overline{V}$
in which $p_0$ is not at infinity, continue with steps 3., 4., and finish with step 5.
\item With $p_0=(a_0,\dots,a_n)$, compute
\bq
 g(r,x_1,\ldots,x_n) & = & f(r+a_0,x_1+a_1,\ldots,x_n+a_n)
\eq
and write
\bq
 g(r,x_1,\ldots,x_n)
 & = & g_{d}(r,x_1,\ldots,x_n)+g_{d-1}(r,x_1,\ldots,x_n),
\eq
where $g_{d}$ and $g_{d-1}$ are homogeneous components of degree $d$ and $d-1$.
\item Return
\begin{align}
\begin{split}
 \varphi_r(x_0',\dots,x_n')    & = -x_0' \frac{g_{d-1}(x_0',x_1',\dots,x_n')}{g_d(x_0',x_1',\dots,x_n')}+a_0,\\
 \varphi_{x_1}(x_0',\dots,x_n') & = -x_1' \frac{g_{d-1}(x_0',x_1',\dots,x_n')}{g_d(x_0',x_1',\dots,x_n')}+a_1,\\
 & \vdots \\
 \varphi_{x_n}(x_0',\dots,x_n') & = -x_n' \frac{g_{d-1}(x_0',x_1',\dots,x_n')}{g_d(x_0',x_1',\dots,x_n')}+a_n,
\end{split}
\end{align}
where one sets for a single $i\in \{0,\dots,n\}$ the corresponding variable $x_i'=1$.
\item Change coordinates to switch from $V^\prime$ to the original affine chart $V$.         
\end{enumerate}
\end{myalgorithm}
\end{tcolorbox}
As an example let us consider our original problem $\sqrt{x(4+x)}$.
The associated hypersurface $V(f)$ is defined by
\bq
 f\left(r,x\right) & = & r^2 - x\left(4+x\right).
\eq
$f$ is of degree $2$, thus we a need a point $p_0$ of multiplicity $1$, i.e. a regular point.
It is easily checked that $p_0=(r,x)=(0,0)$ is a regular point, as
\bq
 \frac{\partial f}{\partial x}(p) & \neq & 0.
\eq
We then have
\bq
 g_2\left(r,x\right) \; = \; r^2 - x^2,
 & &
 g_1\left(r,x\right) \; = \; - 4 x.
\eq
The rationalisation algorithm with $x_0'=1$ gives
\bq
 \varphi_r \; = \; \frac{4 x_1'}{\left(1-x_1'{}^2\right)},
 & &
 \varphi_x \; = \; \frac{4 x_1'{}^2}{\left(1-x_1'{}^2\right)}.
\eq
This is already a valid rationalisation:
\bq
 x \; = \; \frac{4 x_1'{}^2}{\left(1-x_1'{}^2\right)},
 & &
 \sqrt{x\left(4+x\right)} \; = \; \frac{4 x_1'}{\left(1-x_1'{}^2\right)}.
\eq
We recover the rationalisation of eq.~(\ref{chapter_iterated_integrals:example_rationalisation}) through the substitution
\bq
\label{chapter_transformations:substitution_x1p_xp}
 x_1' \; = \; \frac{1-x'}{1+x'},
 & & 
 x' \; = \; \frac{1-x_1'}{1+x_1'}.
\eq
A slightly more general example is the square root of a quadratic polynomial in one variable
\bq
\label{chapter_transformations:square_root_quadratic_polynomial}
 \sqrt{ \left(x-a\right)\left(x-b\right)},
 \;\;\;\;\;\; \mbox{with} \; a \; \neq \; b
\eq
and where we treat $a$ and $b$ as constants.
The associated hypersurface is now defined by
\bq
 f & = & r^2 - \left(x-a\right) \left(x-b\right).
\eq
This is again a hypersurface of degree two and a point of multiplicity $1$ is given by $p_0=(r,x)=(0,a)$.
We find
\bq
 g_2\left(r,x\right) \; = \; r^2 - x^2,
 & &
 g_1\left(r,x\right) \; = \; \left(b-a\right) x.
\eq
Thus
\bq
 x & = & -\left(b-a\right) \frac{x_1'{}^2}{\left(1-x_1'{}^2\right)} + a
 \; = \;
 \frac{a-b x_1'{}^2}{1-x_1'{}^2}
\eq
is a rationalisation of the square root in eq.~(\ref{chapter_transformations:square_root_quadratic_polynomial}).
Alternatively, we may additionally perform the substitution of eq.~(\ref{chapter_transformations:substitution_x1p_xp}).
This gives another rationalisation
\bq
\label{chapter_transformations:square_root_quadratic_polynomial_rationalisation_II}
 x & = & \frac{\left(a-b\right)}{4 x'} \left( 1 + 2 \frac{\left(a+b\right)}{\left(a-b\right)} x' + x'{}^2 \right).
\eq
\bs
{\it \refstepcounter{exercise}
{\bf Exercise \theexercise}: 
Consider the square roots
\bq
 r_1 \; = \; \sqrt{x\left(4+x\right)}
 & \mbox{and} &
 r_2 \; = \; \sqrt{x\left(36+x\right)}.
\eq
Find a transformation, which simultaneously rationalises $r_1$ and $r_2$.
}
\es
\\
\\
The above algorithm relies on the existence of a point of multiplicity $(d-1)$.
The following theorem allows us under certain conditions to obtain the rationalisation of a degree $d$ hypersurface
from the rationalisation of a hypersurface of lower degree.
To state the theorem, we first have to introduce the concept of
{\bf k-homogenisation}.
This is a generalisation of the homogenisation introduced in eq.~(\ref{chapter_transformations:homogenisation}).
Let $f\in \mathbb{C}[x_1,\dots,x_n]$ be a polynomial of degree $d$
and write $f=g_0+\dots+g_d$, where $g_i$ is homogeneous of degree $i$.
Let $k$ be a positive integer with $k \ge d$.
The $k$-homogenisation of $f$ is the degree-$k$ homogeneous polynomial
\bq
 F & = & x_0^k g_{0} + x_0^{k-1} g_1 + \dots +x_0^{k-(d-1)} g_{d-1} + x_0^{k-d} g_d
 \; \in \; \mathbb{C}[x_0,x_1,\dots,x_n].
\eq
In other words, the $k$-homogenisation of $f$ is $x_0^{k-d}$ times the usual homogenisation of $f$.
To give an example, the $4$-homogenisation of the polynomial $f(x_1,x_2)=x_1x_2$ is given by $F(x_0,x_1,x_2)=x_0^2x_1x_2$.
\begin{theorem}
\label{chapter_transformations:theorem_2}
($F$-decomposition theorem):
Let $V=V(r^2-f_{\frac{d}{2}}^2+4f_{\frac{d}{2}+1}f_{\frac{d}{2}-1})$ be the hypersurface associated to
\bq
 \sqrt{f_{\frac{d}{2}}^2-4f_{\frac{d}{2}+1}f_{\frac{d}{2}-1}},
\eq
where each $f_k\in \mathbb{C}[x_1,\dots,x_n]$ is a polynomial of degree $\deg(f_k) \leq k$.
Then $V$ has a rational parametrisation if $W=V(F_{\frac{d}{2}+1}+F_{\frac{d}{2}}+F_{\frac{d}{2}-1})$ 
has a rational parametrisation with $F_k$ being the $k$-homogenization of $f_k$ using the same homogenising variable for each of the three homogenisations. 
The rational parametrisation of $V$ is obtained from the rational parametrisation 
$(\varphi_{x_0}^W,\varphi_{x_1}^W,\ldots,\varphi_{x_n}^W)$ of $W$ by
\begin{align}
\begin{split}
 \varphi_r^V& = 2\cdot\varphi_{x_0}^W\cdot f_{\frac{d}{2}+1}\left(\varphi_{x_1}^W/\varphi_{x_0}^W,\ldots,\varphi_{x_n}^W/\varphi_{x_0}^W\right)+f_{\frac{d}{2}}\left(\varphi_{x_1}^W/\varphi_{x_0}^W,\ldots,\varphi_{x_n}^W/\varphi_{x_0}^W\right),\\
    \varphi_{x_1}^V& = \frac{\varphi_{x_1}^W}{\varphi_{x_0}^W},\\
    &\vdots\\
    \varphi_{x_n}^V& = \frac{\varphi_{x_n}^W}{\varphi_{x_0}^W}.
\end{split}
\end{align}
\end{theorem}
The proof of this theorem can be found in \cite{Besier:2018jen}.
The following example illustrates many of the facets discussed so far.
Consider the square root
\bq
 \sqrt{x^4+y^3}.
\eq
The associated affine hypersurface $V(f)$ is defined by $f=r^2-x^4-y^3$.
Because $V$ has degree $4$, we need to find a point $p$ with $\mathrm{mult}_p(V)=3$ to apply 
the rationalisation algorithm.
Computing the partial derivatives of the homogenisation $F$ of $f$, 
however, we see that $V$ does not have a point of multiplicity $3$ --- not even at infinity.

We, therefore, use the $F$-decomposition: 
as a first step, we rewrite the square root as  
\bq
 \sqrt{x^4+y^3} & = & \sqrt{f_2^2-4f_3f_1}
\eq
with 
\bq
 f_1(x,y) \; = \; -\frac{1}{4},
 \;\;\;
 f_2(x,y) \; = \; x^2,
 \;\;\;
 f_3(x,y) \; = \; y^3,
\eq
and $k$-homogenisations
\bq
 F_1(x,y,z) \; = \; -\frac{1}{4}z,
 \;\;\;
 F_2(x,y,z) \; = \; x^2,
 \;\;\;
 F_3(x,y,z) \; = \;y^3.
\eq
According to the theorem, $V$ has a rational parametrisation if the hypersurface 
\bq
  W & = & V\left(F_1+F_2+F_3\right)
 \; = \; V\left(-\frac{z}{4}+x^2+y^3\right)
\eq
has a rational parametrisation.
$W$ is an affine hypersurface of degree $3$.
We apply algorithm~\ref{chapter_transformations:algo_rationalisation_multiplicity} 
to $W$.
Because $\mathrm{deg}(W)=3$, we need to find a point of multiplicity $2$.
Looking at the partial derivatives of $F_1+F_2+F_3$, we see that $W$ does not have such a point.
There is, however, a point of multiplicity $2$ at infinity.
We see this by considering the projective closure
\bq
 \overline{W} & = & V\left(v^2F_1+vF_2+F_3\right).
\eq
This projective hypersurface has a single point of multiplicity $2$, namely 
\bq
 p_0 \; = \; [x_0:y_0:z_0:v_0] \; = \; [0:0:1:0].
\eq
Viewed from the affine chart $W$, $p_0$ is at infinity, because $v_0$ is zero.
Therefore, we have to consider a different affine chart $W^\prime$ of $\overline{W}$ in which $p_0$ is not at infinity.
In this particular example, we only have one choice, namely to consider the chart where $z=1$.
Switching from $\overline{W}$ to $W^\prime$ corresponds to the map
\bq
  [x:y:z:v] & \mapsto & \left(x^\prime,y^\prime,v^\prime\right) = \left(x/z,y/z,v/z\right).
\eq
Under this mapping, $p_0\in \overline{W}$ is send to $p_0^\prime = (0,0,0)\in W^\prime$.
The affine hypersurface $W^\prime$ is given by 
\bq
 W^\prime & = & V\left(-\frac{1}{4} \left(v^\prime\right)^2+v^\prime\left(x^\prime\right)^2+\left(y^\prime\right)^3\right).
\eq
Set
\bq
 g(x^\prime,y^\prime,v^\prime)
 & = & -\frac{1}{4}\left(v^\prime+0\right)^2+\left(v^\prime+0\right)\left(x^\prime+0\right)^2+\left(y^\prime+0\right)^3
 \nonumber \\
 & = & g_3(x^\prime,y^\prime,v^\prime)+g_2(x^\prime,y^\prime,v^\prime),
\eq
where 
\bq
 g_3(x^\prime,y^\prime,v^\prime) \; = \; v^\prime\left(x^\prime\right)^2+\left(y^\prime\right)^3
 & \mbox{and} & 
 g_2(x^\prime,y^\prime,v^\prime) \; = \; -\frac{1}{4}\left(v^\prime\right)^2.
\eq
A rational parametrisation of $W'$ is then given by
\bq
 \phi_{x^\prime}(t_1,t_2)
 & = & -\frac{g_{2}(1,t_1,t_2)}{g_3(1,t_1,t_2)}
 \; = \; 
 \frac{t_2^2}{4(t_1^3+t_2)},
 \nonumber \\
 \phi_{y^\prime}(t_1,t_2)
 & = & 
 -t_1\frac{g_{2}(1,t_1,t_2)}{g_3(1,t_1,t_2)}
 \; = \; 
 \frac{t_1t_2^2}{4(t_1^3+t_2)},
 \nonumber \\
 \phi_{v^\prime}(t_1,t_2) 
 & = &
 -t_2\frac{g_{3}(1,t_1,t_2)}{g_4(1,t_1,t_2)}
 \; = \;
 \frac{t_2^3}{4(t_1^3+t_2)}.
\eq
We then translate the rational parametrisation for $W^\prime$ to a rational parametrisation for $W$.
To do this, we solve 
\bq
 \phi_{x^\prime} \; = \; \frac{\phi_{x}}{\phi_{z}},
 \;\;\;
 \phi_{y^\prime} \; = \; \frac{\phi_{y}}{\phi_{z}},
 \;\;\;
 \text{and}
 \;\;\; 
 \phi_{v^\prime} \; = \; \frac{\phi_{v}}{\phi_{z}}
\eq
for $\phi_{x}$, $\phi_{y}$, and $\phi_{z}$ while putting $\phi_{v}=1$.
In this way, we obtain a rational parametrisation of $W$ as
\bq
 \phi_x^W(t_1,t_2)
 \; = \;
 \frac{1}{t_2},
 \;\;\;
 \phi_y^W(t_1,t_2)
 \; = \;
 \frac{t_1}{t_2},
 \;\;\;
 \phi_z^W(t_1,t_2)
 \; = \; 
 \frac{4(t_1^3+t_2)}{t_2^3}.
\eq
In the last step we use the $F$-decomposition theorem to obtain the change of variables that rationalises $\sqrt{x^4+y^3}$:
\bq
\label{chapter_transformations:rationalisation_long_example}
 \phi_x^V(t_1,t_2) \; = \; \frac{\phi_x^W(t_1,t_2)}{\phi_z^W(t_1,t_2)} \; = \; \frac{t_2^2}{4(t_1^3+t_2)},
 & &
 \phi_y^V(t_1,t_2) \; = \; \frac{\phi_y^W(t_1,t_2)}{\phi_z^W(t_1,t_2)} \; = \; \frac{t_1t_2^2}{4(t_1^3+t_2)}.
 \;\;\;
\eq
We may verify that eq.~(\ref{chapter_transformations:rationalisation_long_example})
rationalises the original square root:
\bq
 \sqrt{\left(\phi_x^V(t_1,t_2)\right)^4+\left(\phi_y^V(t_1,t_2)\right)^3}
 & = &
 \frac{t_2^3(2t_1^3+t_2)}{16(t_1^3+t_2)^2}.
\eq
An implementation of these algorithms has been given in \cite{Besier:2019kco}.

\subsection{Theorems on rationalisations}

It is useful to know theorems, which allow us to decide if a given hypersurface
has a rational parametrisation or not.
Proofs of the theorems stated below can be found in \cite{Besier:2020hjf}.
We start with a simple theorem:
\begin{theorem}
Let $r=\sqrt{p/q}$ be the square root of a ratio of two polynomials $p,q \in {\mathbb C}[x_1,\dots,x_n]$ and $q$ non-zero.
Write $p \cdot q = f h^2$, where $f$ is square free.
Then $r$ is rationalisable if and only if $\sqrt{f}$ is.
\end{theorem}
First of all, this theorem reduces the rationalisation of a square root of a rational function to the problem
of the rationalisation of a square root of a polynomial.
Secondly, it states that for the rationalisation of a square root of a polynomial only the square free part is relevant.
Square factors are not relevant.

\begin{theorem}
Let $f \in {\mathbb C}[x_1,\dots,x_n]$ be a non-constant square free polynomial of degree $d$ and denote
by $F \in {\mathbb C}[x_0,x_1,\dots,x_n]$ the homogenisation of $f$.
We have:
\begin{enumerate}
\item If $d$ is even, $\sqrt{f}$ is rationalisable if and only if $\sqrt{F}$ is.
\item If $d$ is odd, $\sqrt{f}$ is rationalisable if and only if $\sqrt{x_0 F}$ is.
\end{enumerate}
\end{theorem}
We may unify the two cases of even degree and odd degree as follows:
Define $h = \lceil d/2 \rceil$ by the ceiling function of $d/2$.
For example, for $d=4$ we have $h=2$ and for $d=3$ we have $h=2$.
Denote by $\tilde{F}$ the $(2h)$-homogenisation of $f$. 
If $d$ is even, this is the usual $d$-homogenisation of $f$, 
if $d$ is odd it is the $(d+1)$-homogenisation of $f$.
The theorem above states that $\sqrt{f}$ is rationalisable if and only if $\sqrt{\tilde{F}}$ is.

Let us now look at square roots in one variable.
We have:
\begin{theorem}
Let $f \in {\mathbb C}[x]$ be a square free polynomial of degree $d$ in one variable.
Then $\sqrt{f}$ is rationalisable if and only if $d \le 2$.
\end{theorem}
For $f \in {\mathbb C}[x]$ as above let $V$ be the affine curve in ${\mathbb C}^2$ defined by $y^2-f(x)=0$ and
$\overline{V}$ its projective closure in ${\mathbb C} {\mathbb P}^2$.
The above theorem is equivalent to the statement that $\sqrt{f}$ is rationalisable if and only if $\overline{V}$ has geometric genus zero.

Similar theorems for cases with more variables are more difficult to state and to obtain.
Already for the case of square roots in two variables we first need to introduce a few technicalities:
Let ${\mathbb F}$ be a field.
In chapter~\ref{chapter_basics} we introduced the projective space 
$\mathrm{P}^n\left({\mathbb F}\right)$ as the set of points in ${\mathbb F}^{n+1} \backslash \{0\}$ modulo
the equivalence relation
\bq
 \left( x_0, x_1, ..., x_n \right) \sim \left( y_0, y_1, ..., y_n \right)
 & \Leftrightarrow &
 \exists \; \lambda \neq 0 : 
 \left( x_0, x_1, ..., x_n \right) = \left( \lambda y_0, \lambda y_1, ..., \lambda y_n \right).
\hspace*{8mm}
\eq
The 
\index{weighted projective space}
{\bf weighted projective space} $\mathrm{P}^n_{w_0,\dots,w_n}\left({\mathbb F}\right)$
with weights $(w_0,w_1,\dots,w_n)$ 
is the set of points in ${\mathbb F}^{n+1} \backslash \{0\}$ modulo
the equivalence relation
\bq
 \left( x_0, x_1, ..., x_n \right) \sim \left( y_0, y_1, ..., y_n \right)
 \; \Leftrightarrow \;
 \exists \; \lambda \neq 0 : 
 \left( x_0, x_1, ..., x_n \right) = \left( \lambda^{w_0} y_0, \lambda^{w_1} y_1, ..., \lambda^{w_n} y_n \right).
\hspace*{8mm}
\eq
We are mainly concerned with the case ${\mathbb F}={\mathbb C}$.
We write
\bq
 {\mathbb C} {\mathbb P}_{w_0,\dots,w_n}^n \; = \; \mathrm{P}_{w_0,\dots,w_n}^n\left({\mathbb C}\right).
\eq
Let $f \in {\mathbb C}[x_1,\dots,x_n]$ be a non-constant square free polynomial of degree $d$
and set $h = \lceil d/2 \rceil$ as above.
We consider the weighted projective space ${\mathbb C} {\mathbb P}_{1,1,\dots,1,h}^{n+1}$ with homogeneous coordinates 
$(x_0,x_1,\dots,x_n,r)$.
The coordinates $x_0, x_1, \dots, x_n$ have weight one, while the coordinate $r$ has weight $h$. 
$x_0$ is the homogenising coordinate, $r$ names the square root.
Denote by 
\bq
 F(x_0,x_1,\dots,x_n) \in {\mathbb C}[x_0,x_1,\dots,x_n]
\eq 
the $(2h)$-homogenisation of $f$.
We associate to $\sqrt{f}$ a hypersurface $\overline{W}$ in the weighted projective space ${\mathbb C} {\mathbb P}_{1,1,\dots,1,h}^{n+1}$.
The hypersurface $\overline{W}$ is defined by
\bq
 r^2 - F\left(x_0,x_1,\dots,x_n\right)
 & = & 0.
\eq
(It is worth thinking about the differences in the definition of $\overline{V}$ and $\overline{W}$:
In defining $\overline{V}$ we start from $\sqrt{f}$ and first consider the affine hypersurface $r^2-f(x_1,\dots,x_n) \in {\mathbb C}^{n+1}$.
We then take the $d$-homogenisation. This gives a projective hypersurface in ${\mathbb C} {\mathbb P}^{n+1}$.
In defining $\overline{W}$ we again start from $\sqrt{f}$ but first consider the $(2h)$-homogenisation 
$F(x_0,x_1,\dots,x_n) \in {\mathbb C}[x_0,x_1,\dots,x_n]$.
In the second step we add the variable $r$ naming the root and consider the hypersurface defined by $r^2-F(x_0,\dots,x_n)$.
This is a hypersurface in the weighted projective space ${\mathbb C} {\mathbb P}_{1,1,\dots,1,h}^{n+1}$.)
One can show that the hypersurface $\overline{W}$ is birationally equivalent to the hypersurface $\overline{V}$ defined
previously.
Some theorems can be formulated more elegantly by referring to $\overline{W}$ instead of $\overline{V}$.

We also need the concept of simple singularities.
We denote by ${\mathbb C}[[x_1,\dots,x_n]]$ the 
\index{ring of formal power series}
{\bf ring of formal power series} in $x_1,\dots,x_n$.
Let $f_1, f_2 \in {\mathbb C}[x_1,\dots,x_n]$ be two polynomials 
and assume that both $V(f_1)$ and $V(f_2)$ have a singular point at the origin.
We say that these two singularities are 
\index{same type singularities}
{\bf of the same type} 
if the two quotient rings
${\mathbb C}[[x_1,\dots,x_n]]/\langle f_1 \rangle$
and
${\mathbb C}[[x_1,\dots,x_n]]/\langle f_2 \rangle$
are isomorphic.
We call ${\mathbb C}[[x_1,\dots,x_n]]/\langle f_i \rangle$ the 
\index{associated quotient ring of a hypersurface}
{\bf associated quotient ring} of $V(f_i)$.

Let $f \in {\mathbb C}[x_1,\dots,x_n]$ be a polynomial and assume that affine hypersurface $V(f)$ has a singular point at the origin
of ${\mathbb C}^n$.
We say that the origin is a 
\index{simple singularity}
{\bf simple singularity} 
(or 
\index{ADE singularity}
{\bf ADE singularity} 
or 
\index{Du Val singularity}
{\bf Du Val singularity}), 
if the associated quotient ring is isomorphic to a quotient ring ${\mathbb C}[[x_1,\dots,x_n]]/\langle g \rangle$,
where $g$ is a polynomial from the following list:
\begin{alignat}{2}
 A_k: \;\;\; & \; x_1^2 + x_2^{k+1} + X, & \;\;\;\;\;\; & k \ge 1, \nonumber \\
 D_k: \;\;\; & \; \left(x_1^2 + x_2^{k-2}\right) x_2 + X, & \;\;\;\;\;\; & k \ge 4, \nonumber \\
 E_6: \;\;\; & \; x_1^3 + x_2^4 + X, & & \nonumber \\
 E_7: \;\;\; & \; x_1\left(x_1^2 + x_2^3\right) + X, & & \nonumber \\
 E_8: \;\;\; & \; x_1^3 + x_2^5 + X, & & 
\end{alignat}
with
\bq
 X & = & x_3^2 + \dots + x_n^2.
\eq
The type of a singularity is invariant under linear coordinate transformations, hence we may always translate any singular point
to the origin.

With these preparations we may now state the theorem on the rationalisation of square roots in two variables:
\begin{theorem}
Let $f \in {\mathbb C}[x_1,x_2]$ be a square free polynomial of degree $d$ in two variables
and assume that the hypersurface $\overline{W} \in {\mathbb C} {\mathbb P}_{1,1,\dots,1,h}^{n+1}$ has at most ADE-singularities.
Then $\sqrt{f}$ is rationalisable if and only if $d \le 4$.
\end{theorem}
We close this section with a theorem on multiple square roots:
\begin{theorem}
Let $f_1, \dots, f_r \in {\mathbb C}[x_1,\dots,x_n]$.
If the set of roots $\{\sqrt{f_1},\dots,\sqrt{f_r}\}$ is simultaneously rationalisable, then for every non-empty
subset $J \subseteq \{1,\dots,r\}$ the square root
\bq
\label{chapter_transformations:product_roots}
 \sqrt{\prod\limits_{j\in J} f_j}
\eq
is rationalisable.
\end{theorem}
The main application of this theorem is to prove that a certain set of square root cannot be rationalised simultaneously
by showing that a specific product as in eq.~(\ref{chapter_transformations:product_roots}) is not rationalisable.

%% file: multiple_polylogs.tex
\newpage
\chapter{Multiple polylogarithms}
\label{chapter_multiple_polylogarithms}

In chapter~\ref{chapter_iterated_integrals} we already encountered multiple polylogarithms.
Multiple polylogarithms are an important class of functions in the context of Feynman integrals.
In this chapter we study them in more detail.

There are two frequently used notations for multiple polylogarithms, either
$G(z_1,\dots,z_r;y)$ 
(which we already introduced in section~\ref{chapter_iterated_integrals:section:solution_eps_form})
or
$\mathrm{Li}_{m_1 \dots m_k}(x_1,\dots,x_k)$.
The former is directly related to the iterated integral representation, while the latter is related
to the nested sum representation.
This reveals already the fact, that multiple polylogarithms may either be defined in terms of
iterated integrals or nested sums.
We will study both cases.

Each of the two representations gives rise to a product: A shuffle product in the case of the iterated integral
representation and a quasi-shuffle product in the case of the nested sum representation.
This turns the vector space spanned by the multiple polylogarithms into an algebra.
This is actually a Hopf algebra. We will discuss the coalgebra properties in more detail in chapter~\ref{chapter_hopf}.

Multiple polylogarithms are generalisations of the logarithms and it comes to no surprise that
these functions have branch cuts. We will study the monodromy around a branch point.

In the last two sections of this chapter we study fibration bases and linearly reducible Feynman integrals.
The linearly reducible Feynman integrals have the property that they evaluate to multiple polylogarithms.
They can be computed efficiently from the Feynman parameter representation.

Multiple polylogarithms surfaced in the work of Kummer \cite{Kummer:1840a,Kummer:1840b,Kummer:1840c}, Poincar\'e \cite{Poincare:1884}
and Lappo-Danilevsky \cite{Lappo-Danilevsky:book}.
Modern references are Goncharov \cite{Goncharov_no_note,Goncharov:2001} and Borwein et al. \cite{Borwein}.
An introductory survey article can be found in \cite{waldschmidt:hal-00416166}.

\section{The integral representation}

In section~\ref{chapter_iterated_integrals:section:solution_eps_form} we introduced multiple polylogarithms as a special case
of iterated integrals.
Let us recall the definition:
If all $z$'s are equal to zero, we define $G(z_1,\dots,z_r;y)$ by
\bq
 G(\underbrace{0,\dots,0}_{r-\mathrm{times}};y)
 & = & 
 \frac{1}{r!} \ln^r\left(y\right).
\eq
This definition includes as a trivial case
\bq
 G(;y)
 & = & 
 1.
\eq
If at least one variable $z$ is not equal to zero we define recursively
\bq
 G\left(z_1,z_2\dots,z_r;y\right)
 & = &
 \int\limits_0^y
 \frac{dy_1}{y_1-z_1}
 G\left(z_2\dots,z_r;y_1\right).
\eq
We have for example
\bq
 G\left(0;y\right) \; = \; \ln\left(y\right),
 & &
 G\left(z;y\right) \; = \; \ln\left(\frac{z-y}{z}\right).
\eq
If one further defines $g(z;y) = 1/(y-z)$, then one has
\bq
\frac{d}{dy} G(z_1,\dots,z_r;y) & = & g(z_1;y) G(z_2,\dots,z_r;y)
\eq
and if at least one variable $z$ is not equal to zero
\bq
\label{chapter_multiple_polylogarithms:Grecursive}
G(z_1,z_2,\dots,z_r;y) & = & \int\limits_0^y dt \; g(z_1;t) G(z_2,\dots,z_r;t).
\eq
The function $G(z_1,\dots,z_r;y)$ is said to have a 
\index{trailing zero}
{\bf trailing zero}, 
if $z_r=0$.
We will soon see that with the help of the shuffle product we can always remove trailing zeros.
Let us therefore focus on multiple polylogarithms $G(z_1,\dots,z_r;y)$ without trailing zeros 
(i.e. $z_r \neq 0$).
For $z_r \neq 0$ the recursive definition translates to 
\bq
\label{chapter_multiple_polylogarithms:Gfuncdef}
 G\left(z_1,\dots,z_r;y\right) 
 & = &
 \int\limits_0^y \frac{dt_1}{t_1-z_1}
 \int\limits_0^{t_1} \frac{dt_2}{t_2-z_2} \ldots
 \int\limits_0^{t_{r-1}} \frac{dt_r}{t_r-z_r}.
\eq
The number $r$ is referred to as the 
\index{depth}
{\bf depth} of the integral representation.
In the case of multiple polylogarithms the number $r$ is also referred to as the
\index{weight}
{\bf weight} of the multiple polylogarithm.
The differential of $G(z_1,\dots,z_r;y)$ is
\bq
\label{chapter_multiple_polylogarithms:differential_Glog}
 d G(z_1,\dots,z_r;y)
 & = &
 \sum\limits_{j=1}^r 
 G(z_1,\dots,\hat{z}_j,\dots,z_r;y) \left[ d\ln\left(z_{j-1}-z_j\right)-d\ln\left(z_{j+1}-z_j\right) \right],
\eq
where we set $z_0=y$ and $z_{r+1}=0$.
A hat indicates that the corresponding variable is omitted.
In addition one uses the convention that for $z_{j+1}=z_j$ the one-form
$d\ln(z_{j+1}-z_j)$ equals zero.
The proof of eq.~(\ref{chapter_multiple_polylogarithms:differential_Glog}) is based on the identity
\bq
 \frac{\partial}{\partial z} \frac{1}{t-z} & = & - \frac{\partial}{\partial t} \frac{1}{t-z}
\eq
and partial integration.

For $z_r \neq 0$ we also have the scaling relation
\bq
\label{chapter_multiple_polylogarithms:G_scaling_relation}
 G\left(z_1,\dots,z_r;y\right) 
 & = & 
 G\left(x z_1, \dots, x z_r; x y\right),
 \;\;\;\;\;\;\;\;\;
 z_r \, \in \, {\mathbb C}\backslash\{0\},
 \;\;\;\;\;
 x \, \in \, {\mathbb C}\backslash\{0\}.
\eq
This allows us to scale the variable $y$ to one:
\bq
 G\left(z_1,\dots,z_r;y\right) 
 & = & 
 G\left(\frac{z_1}{y}, \dots, \frac{z_r}{y}; 1\right),
 \;\;\;\;\;\;\;\;\;
 z_r \, \in \, {\mathbb C}\backslash\{0\},
 \;\;\;\;\;
 y \, \in \, {\mathbb C}\backslash\{0\}.
\eq
Note that the scaling relation does not hold for multiple polylogarithms with trailing zeros.
We have for example
\bq
 G\left(0;xy\right)
 \, = \; \ln\left(x y\right)
 \; = \; \ln\left(x\right) + \ln\left(y\right)
 & \neq &
 \ln\left(y\right)
 \; = \;
 G\left(0;y\right).
\eq
In order to relate the integral representation of the multiple polylogarithms
to the sum representation of the multiple polylogarithms it is convenient
to introduce
the following short-hand notation:
\bq
\label{chapter_multiple_polylogarithms:Gshorthand}
G_{m_1 \dots m_k}(z_1,\dots,z_k;y)
 & = &
 G(\underbrace{0,\dots,0}_{m_1-1},z_1,\dots,z_{k-1},\underbrace{0,\dots,0}_{m_k-1},z_k;y)
\eq
Here, all $z_j$ for $j=1,\dots,k$ are assumed to be non-zero.
For example,
\bq
G_{1 2}(z_1,z_2;y)
 & = &
G(z_1,0,z_2;y).
\eq
The multiply polylogarithm $G_{m_1 \dots m_k}(z_1,\dots,z_k;y)$ has weight $m_1+\dots+m_k$.

\section{The sum representation}

The multiple polylogarithms have also a sum representation.
The standard notation for the sum representation is $\mathrm{Li}_{m_1 \dots m_k}(x_1,\dots,x_k)$.
The sum representation is defined by
\bq 
\label{chapter_multiple_polylogarithms:def_multiple_polylogs_sum}
 \mathrm{Li}_{m_1 \dots m_k}(x_1,\dots,x_k)
  & = & \sum\limits_{n_1>n_2>\ldots>n_k>0}^\infty
     \frac{x_1^{n_1}}{{n_1}^{m_1}}\ldots \frac{x_k^{n_k}}{{n_k}^{m_k}}.
\eq
The sum converges for 
\bq
\label{chapter_multiple_polylogarithms:condconvLi}
\left| x_1 x_2 \dots x_j \right| \le 1 & & \mbox{for all} \; j \in \{1,\dots,k\} \;\; \mbox{and} \;\; (m_1,x_1) \neq (1,1).
\eq
In the following we will always assume that the arguments $x_j$ are such that eq.~(\ref{chapter_multiple_polylogarithms:condconvLi}) is satisfied.
The number $k$ in the definition of the sum representation is referred to as the 
\index{depth}
{\bf depth} of the sum representation of the multiple polylogarithm.
The number $m_1+\dots+m_k$ is referred to the 
\index{weight}
{\bf weight} of the multiple polylogarithm.
Note that for the sum representation of multiple polylogarithms 
the weight and the depth of the sum representation are in general not equal.
Eq.~(\ref{chapter_multiple_polylogarithms:def_multiple_polylogs_sum}) is a nested sum, which we may also write as
\bq
 \mathrm{Li}_{m_1 \dots m_k}(x_1,\dots,x_k)
 & = &
 \sum\limits_{n_1=1}^{\infty} 
 \frac{x_1^{n_1}}{n_1^{m_1}}
 \;
 \sum\limits_{n_2=1}^{n_1-1} 
 \frac{x_2^{n_2}}{n_2^{m_2}}
 \;
 \dots
 \;
 \sum\limits_{n_{k-1}=1}^{n_{k-2}-1}
 \frac{x_{k-1}^{n_{k-1}}}{n_{k-1}^{m_{k-1}}}
 \;
 \sum\limits_{n_k=1}^{n_{k-1}-1}
 \frac{x_k^{n_k}}{n_k^{m_k}},
\eq
with the convention that
\bq
 \sum\limits_{n=a}^b f(n) \; = \; 0,
 & &
 \mbox{for} \;\; b<a.
\eq
The relation between the sum representation in eq.~(\ref{chapter_multiple_polylogarithms:def_multiple_polylogs_sum}) 
and the integral representation in eq.~(\ref{chapter_multiple_polylogarithms:Gshorthand}) is given by
\bq
\label{chapter_multiple_polylogarithms:conversion_Li_to_G}
 \mathrm{Li}_{m_1 \dots m_k}(x_1,\dots,x_k)
 & = & (-1)^k 
 G_{m_1 \dots m_k}\left( \frac{1}{x_1}, \frac{1}{x_1 x_2}, \dots, \frac{1}{x_1 \dots x_k};1 \right),
\eq
and
\bq
\label{chapter_multiple_polylogarithms:conversion_G_to_Li}
 G_{m_1 \dots m_k}(z_1 \dots, z_k;y) 
 & = & 
 (-1)^k \; \mathrm{Li}_{m_1 \dots m_k}\left(\frac{y}{z_1}, \frac{z_1}{z_2}, \dots, \frac{z_{k-1}}{z_k}\right).
\eq
\bs
{\it \refstepcounter{exercise}
{\bf Exercise \theexercise}: 
Prove eq.~(\ref{chapter_multiple_polylogarithms:conversion_Li_to_G}).
}
\es
\\
\\
The multiple polylogarithms include several special cases.
The 
\index{classical polylogarithms}
{\bf classical polylogarithms} are defined by
\bq
 \mathrm{Li}_m(x) & = & \sum\limits_{n=1}^\infty \frac{x^n}{n^m}
\eq
and are the special case of depth one.
The most prominent examples are
\bq
 \mathrm{Li}_1(x) = \sum\limits_{i_1=1}^\infty \frac{x^{i_1}}{i_1} = -\ln(1-x),
 & &
 \mathrm{Li}_2(x) = \sum\limits_{i_1=1}^\infty \frac{x^{i_1}}{i_1^2}.
\eq 
\index{Nielsen polylogarithms}
{\bf Nielsen's generalised polylogarithms} $S_{n,p}(x)$ are defined by \cite{Nielsen} 
\bq
 S_{n,p}(x) 
 & = & \mathrm{Li}_{(n+1) 1 \dots 1}(x,\underbrace{1,\dots,1}_{p-1}),
\eq
Multiple polylogarithms with $x_2=x_3=\ldots=x_k=1$ are a subset of the
\index{harmonic polylogarithms}
{\bf harmonic polylogarithms} $H_{m_1,\dots,m_k}(x)$ \cite{Remiddi:1999ew,Gehrmann:2000zt}
\bq
\label{chapter_iterated_integrals:harmpolylog}
 H_{m_1 \dots m_k}(x) & = & \mathrm{Li}_{m_1 \dots m_k}(x,\underbrace{1,\dots,1}_{k-1}).
\eq
If one restricts in the integral representation $G(z_1,\dots,z_r;y)$
the possible values of $z_j$'s to zero and the $n$-th roots of unity, 
one arrives at the 
\index{cyclotomic harmonic polylogarithms}
$n$-th {\bf cyclotomic harmonic polylogarithms} \cite{Ablinger:2011te}.
The harmonic polylogarithms in eq.~(\ref{chapter_iterated_integrals:harmpolylog})
are just the first cyclotomic harmonic polylogarithms, corresponding to $z_j \in \{0,1\}$.
The word ``harmonic polylogarithms'' is used as a synonym for the second cyclotomic harmonic polylogarithms, 
i.e. multiple polylogarithms with $z_j \in \{-1,0,1\}$.

The values of the multiple polylogarithms at $x_1=\ldots=x_k=1$ are known as 
\index{multiple zeta values}
{\bf multiple $\zeta$-values}:
\bq
\zeta_{m_1 m_2 \dots m_k} \; = \; \mathrm{Li}_{m_1 m_2 \dots m_k}(1,1,\dots,1) 
 \; = \; 
 \sum\limits_{n_1 > n_2 > \dots > n_k > 0}^\infty 
 \;\;\;
 \frac{1}{n_1^{m_1}} \frac{1}{n_2^{m_2}} \dots \frac{1}{n_k^{m_k}},
 \;\;\;\;\;\;\;\;\;
 m_1 \; \neq \; 1.
 \;\;\;\;\;\;
\eq
\begin{digression} {\bf The Clausen and Glaisher functions}
\\
As an excursion let us turn to the Clausen and Glaisher functions. 
These are related to linear combinations of classical polylogarithms.

The 
\index{Clausen function}
{\bf Clausen function} is defined by
\bq
\mathrm{Cl}_n(\theta) & = & 
 \left\{ \begin{array}{cc}
   \mathrm{Im} \; \mathrm{Li}_n\left( e^{i \theta} \right)
   = \frac{1}{2i} \left[ \mathrm{Li}_n\left( e^{i \theta} \right) 
                      -\mathrm{Li}_n\left( e^{-i \theta} \right)
                \right], 
   & n \; \mbox{even}, \\
   & \\
   \mathrm{Re} \; \mathrm{Li}_n\left( e^{i \theta} \right)
   = \frac{1}{2} \left[ \mathrm{Li}_n\left( e^{i \theta} \right) 
                      +\mathrm{Li}_n\left( e^{-i \theta} \right)
                \right],
   & n \; \mbox{odd}, \\
 \end{array} \right.
\eq
the 
\index{Glaisher function}
{\bf Glaisher function} is defined by
\bq
\mathrm{Gl}_n(\theta) & = & 
 \left\{ \begin{array}{cc}
   \mathrm{Re} \; \mathrm{Li}_n\left( e^{i \theta} \right)
   = \frac{1}{2} \left[ \mathrm{Li}_n\left( e^{i \theta} \right) 
                      +\mathrm{Li}_n\left( e^{-i \theta} \right)
                \right], 
   & n \; \mbox{even}, \\
   & \\
   \mathrm{Im} \; \mathrm{Li}_n\left( e^{i \theta} \right)
   = \frac{1}{2i} \left[ \mathrm{Li}_n\left( e^{i \theta} \right) 
                      -\mathrm{Li}_n\left( e^{-i \theta} \right)
                \right],
   & n \; \mbox{odd}. \\
 \end{array} \right.
\eq
From the definition it is clear that
these functions are periodic with period $2\pi$:
\bq
 \mathrm{Cl}_n(\theta+2\pi) \; = \; \mathrm{Cl}_n(\theta),
 & &
 \mathrm{Gl}_n(\theta+2\pi) \; = \; \mathrm{Gl}_n(\theta).
\eq
It is therefore sufficient to restrict the argument to
\bq
 0 \; \le \; \mathrm{Re}\left(\theta\right) \; < \; 2\pi.
\eq
Note that $\mathrm{Cl}_n(\theta)$ and $\mathrm{Gl}_n(\theta)$ are not necessarily continuous as $\theta \rightarrow 2\pi$,
as $\mathrm{Li}_n(x)$ has a branch cut along the positive real axis, starting at $x=1$.

For $l \in {\mathbb N}_0$ we have
\bq
 \mathrm{Li}_{2l}\left( e^{i \theta} \right)
 \; = \; 
 \mathrm{Gl}_{2l}(\theta) + i \mathrm{Cl}_{2l}(\theta),
 & &
 \mathrm{Li}_{2l+1}\left( e^{i \theta} \right)
 \; = \; 
 \mathrm{Cl}_{2l+1}(\theta) + i \mathrm{Gl}_{2l+1}(\theta).
\eq
It is worth knowing the special value
\bq
 \mathrm{Cl}_2\left(\frac{\pi}{2}\right) & = & G,
\eq
where 
\index{Catalan's constant}
{\bf Catalan's constant} $G$ is given by
\bq
G & = & \sum\limits_{n=0}^\infty \frac{(-1)^n}{\left(2n+1\right)^2}.
\eq
It is also worth noting that the Glaisher functions $\mathrm{Gl}_n(\theta)$ are (for $0 \le \mathrm{Re}\left(\theta\right) < 2\pi$) polynomials in $\theta$ of degree $n$.
In detail we have for $0 \le \mathrm{Re}\left(\theta\right) < 2\pi$
\bq
\label{chapter_iterated_integrals:Glaisher_to_Bernoulli}
 \mathrm{Gl}_n(\theta)
 & = &
 \left\{ \begin{array}{cc}
  - \frac{1}{2} \frac{\left(2\pi i\right)^n}{n!} \; B_n\left(\frac{\theta}{2\pi}\right),
   & n \; \mbox{even}, \\
  - \frac{1}{2i} \frac{\left(2\pi i\right)^n}{n!} \; B_n\left(\frac{\theta}{2\pi}\right),
   & n \; \mbox{odd}, \\
 \end{array} \right.
\eq
where $B_n(x)$ is the $n$'th 
\index{Bernoulli polynomial}
{\bf Bernoulli polynomial}
defined by
\bq
 \frac{t e^{x t}}{e^t-1}
 & = &
 \sum\limits_{n=0}^\infty \frac{B_n\left(x\right)}{n!} t^n.
\eq
\end{digression}

We return to the multiple polylogarithms.
The differential of $\mathrm{Li}_{m_1 \dots m_k}(x_1,\dots,x_k)$ with respect to the variables $x_1,\dots,x_k$ is
\bq
\label{chapter_multiple_polylogarithms:dLi}
 d \mathrm{Li}_{m_1 \dots m_k}(x_1,\dots,x_k)
 & = &
 \sum\limits_{j=1}^k \mathrm{Li}_{m_1 \dots m_{j-1} (m_j-1) m_{j+1} \dots m_k}(x_1,\dots,x_k) \cdot d \ln(x_j).
\eq
This follows easily from
\bq
 d \left( \frac{x_j^{n_j}}{n_j^{m_j}} \right)
 & = &
 n_j \frac{x_j^{n_j-1}}{n_j^{m_j}} dx_j
 \; = \;
 \frac{x_j^{n_j}}{n_j^{m_j-1}} \frac{dx_j}{x_j}
 \; = \;
 \frac{x_j^{n_j}}{n_j^{m_j-1}} d\ln(x_j).
\eq
If an index $m_j$ equals one, we obtain in the differential an index with the value zero.
These multiple polylogarithms can be reduced. We have
\bq
 \mathrm{Li}_{0}(x_1)
 & = &
 \frac{x_1}{1-x_1}
\eq
and
\bq
\label{chapter_multiple_polylogarithms:Li_zero_index}
\lefteqn{
 \mathrm{Li}_{m_1 \dots m_{i-1} 0 m_{i+1} \dots m_k}(x_1,\dots,x_{i-1},x_i,x_{i+1},\dots,x_k)
 = } & & \nonumber \\
 & &
 \mathrm{Li}_{0}(x_i)
 \mathrm{Li}_{m_1 \dots m_{i-1} m_{i+1} \dots m_k}(x_1,\dots,x_{i-1},x_{i+1},\dots,x_k)
 \nonumber \\
 & &
 - \sum\limits_{j=i+1}^k
 \mathrm{Li}_{m_1 \dots m_{i-1} m_{i+1} \dots m_j 0 m_{j+1} \dots m_k}(x_1,\dots,x_{i-1},x_{i+1},\dots,x_j,x_i,x_{j+1},\dots,x_k)
 \nonumber \\
 & &
 - \sum\limits_{j=i+1}^k
 \mathrm{Li}_{m_1 \dots m_{i-1} m_{i+1} \dots m_j \dots m_k}(x_1,\dots,x_{i-1},x_{i+1},\dots,x_i \cdot x_j,\dots,x_k)
 \nonumber \\
 & &
 - 
 \left[ 1 + \mathrm{Li}_{0}(x_i) \right]
 \mathrm{Li}_{m_1 \dots m_{i-1} m_{i+1} \dots m_k}(x_1,\dots,x_{i-1}\cdot x_i,x_{i+1},\dots,x_k).
\eq
If $i=1$ the last term is absent.
Eq.~(\ref{chapter_multiple_polylogarithms:Li_zero_index})
allows us to shift recursively the zero index to the last position.
If the zero index is in the last position, the sums from $(i+1)$ to $k$ are empty and the recursion terminates.
We will prove eq.~(\ref{chapter_multiple_polylogarithms:Li_zero_index}) in exercise~\ref{chapter_nested_sums:proof_Li_zero_index}
in chapter~\ref{chapter_nested_sums}, once we learned about the quasi-shuffle product and $Z$-sums.

\section{The shuffle product}
\label{chapter_multiple_polylogarithms:section:shuffle_product}

In this section we introduce the shuffle product for multiple polylogarithms.
The shuffle product is associated with the iterated integral representation.
It is not specific to multiple polylogarithms, but holds for any iterated integral.

We start with the definition of a shuffle algebra.
Consider a finite set of objects, which we will call 
\index{letter}
{\bf letters}.
We denote the letters by $l_1$, $l_2$, $\dots$, 
and the set of all letters the 
\index{alphabet}
{\bf alphabet} $A=\{l_1,l_2,\dots\}$.
A 
\index{word}
{\bf word} is an ordered sequence of letters:
\bq
 w & = & l_1 l_2 ... l_k.
\eq
The word of length zero is denoted by $e$.
Let ${\mathbb F}$ be a field and consider the vector space of words over ${\mathbb F}$.
We may turn this vector space into an algebra by supplying a product for words.
We say that a permutation $\sigma$ is a shuffle of $(1,2,...,k)$ and of $(k+1,...,r)$,
if in 
\bq
 \left( \sigma(1), \sigma(2), \dots, \sigma(r) \right)
\eq
the relative order of $1,2,...,k$ and of $k+1,...,r$ is preserved.
Thus $(1,3,2)$ is a shuffle of $(1,2)$ and $(3)$, while $(2,1,3)$ is not.
The {\bf shuffle product} of two words is defined by
\bq
 l_1 l_2 \dots l_k \shuffle l_{k+1} \dots l_r 
 & = &
 \sum\limits_{\mathrm{shuffles} \; \sigma} l_{\sigma(1)} l_{\sigma(2)} ... l_{\sigma(r)},
\eq
where the sum runs over all permutations $\sigma$ which are shuffles of $(1,\dots,k)$ and $(k+1,\dots,r)$, 
i.e. which preserve the relative order
of $1,2,...,k$ and of $k+1,...,r$.
This product turns the vector space of words into a shuffle algebra ${\mathcal A}$.

The name ``shuffle algebra'' is related to the analogy of shuffling cards: If a deck of cards
is split into two parts and then shuffled, the relative order within the two individual parts
is conserved.
A shuffle algebra is also known under the name ``mould symmetral'' \cite{Ecalle}.

The empty word $e$ is the unit in this algebra:
\bq
 e \shuffle w = w \shuffle e = w.
\eq
A recursive definition of the shuffle product is given by
\bq
\label{chapter_multiple_polylogarithms:def_recursive_shuffle}
 l_1 l_2 ... l_k \shuffle l_{k+1} ... l_r 
 & = &
 l_1 \left( l_2 ... l_k \shuffle l_{k+1} ... l_r \right)
+
 l_{k+1} \left( l_1 l_2 ... l_k \shuffle l_{k+2} ... l_r \right),
\eq
where concatenation of letters is extended on the vector space of words by linearity:
\bq
 l \left( c_1 w_1 + c_2 w_2 \right)
 & = &
 c_1 l w_1 + c_2 l w_2,
 \;\;\;\;\;\;\;\;\;
 c_1, c_2 \in {\mathbb F},
 \;\;\;
 l \in A,
 \;\;\;
 w_1, w_2 \in {\mathcal A}.
\eq
Of course, concatenation of words
would also define a product on the vector space of words, but this is not the product we 
are interested in.
The shuffle product is commutative
\bq
 w_1 \shuffle w_2 & = & w_2 \shuffle w_1,
\eq
while the concatenation product is non-commutative.
A few examples are
\bq
 l_1 l_2 \shuffle l_3
 & = &
 l_1 l_2 l_3
 +
 l_1 l_3 l_2
 +
 l_3 l_1 l_2,
 \nonumber \\
 l_1 l_2 \shuffle l_2
 & = &
 2 l_1 l_2 l_2
 +
 l_2 l_1 l_2,
 \nonumber \\
 l_1 l_2 \shuffle l_3 l_4
 & = & 
 l_1 l_2 l_3 l_4
 +
 l_1 l_3 l_2 l_4
 +
 l_3 l_1 l_2 l_4
 +
 l_1 l_3 l_4 l_2
 +
 l_3 l_1 l_4 l_2
 +
 l_3 l_4 l_1 l_2.
\eq
The shuffle algebra (with the shuffle product as product) is generated by the Lyndon words \cite{Reutenauer}.
If one introduces a lexicographic ordering on the letters of the alphabet
$A$, a 
\index{Lyndon word}
{\bf Lyndon word} 
is defined by the property $w < v$
for any sub-words $u$ and $v$ such that $w= u v$.
To give an example, consider the alphabet $A=\{l_1,l_2\}$ with $l_1<l_2$.
The words
\bq
 w_1 \; = \; l_1 l_1 l_2, 
 & & 
 w_2 \; = \; l_1 l_1 l_2 l_1 l_2 l_2
\eq
are Lyndon words, while $w_3=l_1 l_2 l_1$ is not.
The word $w_3$ may be written as $w_3=u v$ with $u=l_1 l_2$, $v=l_1$ and $v<w_3$.
\\
\\
\bs
{\it \refstepcounter{exercise}
{\bf Exercise \theexercise}: 
Consider the alphabet $A=\{l_1,l_2\}$ with $l_1<l_2$. Write down all Lyndon words of depth $\le 3$.
}
\es

\begin{figure}
\begin{center}
\includegraphics[scale=1.2]{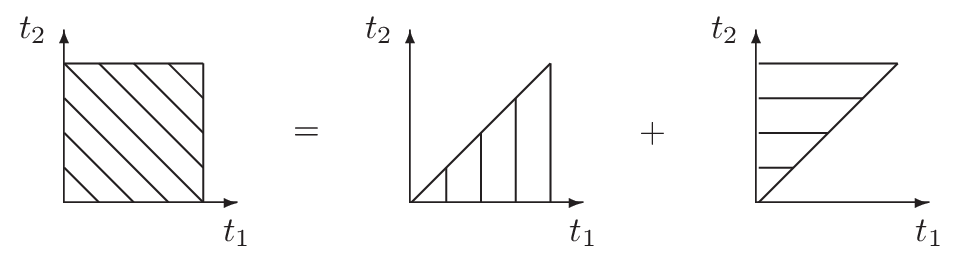}
\caption{\label{chapter_multiple_polylogarithms:fig1} Shuffle algebra from the integral representation: The shuffle product follows from replacing the integral over the square by an integral over the lower triangle and an integral over the upper triangle.}
\end{center}
\end{figure}

Let us now make the connection to multiple polylogarithms $G(z_1,\dots,z_r;y)$.
We take the $z_j$'s as letters.
The alphabet $A$ is given by the distinct $z_j$'s.
A multiple polylogarithm $G(z_1,\dots,z_r;y)$
is therefore specified by a word $w=z_1 z_2\dots z_r$ (i.e. an ordered sequence)
and a value $y$.
The non-trivial statement is the shuffle product for multiple polylogarithms:
\bq
\label{chapter_multiple_polylogarithms:G_shuffle_product}
 G(z_1,z_2,...,z_k;y) \cdot G(z_{k+1},...,z_r; y) = 
 \sum\limits_{\mathrm{shuffles} \; \sigma} G(z_{\sigma(1)},z_{\sigma(2)},...,z_{\sigma(r)};y),
\eq
where the sum runs over all permutations $\sigma$ which are shuffles of $(1,\dots,k)$ and $(k+1,\dots,r)$, 
i.e. which preserve the relative order
of $1,2,...,k$ and of $k+1,...,r$.
An simple example for the shuffle product of two multiple polylogarithms is given by
\bq
\label{chapter_multiple_polylogarithms:example_G_product}
G(z_1;y) \cdot G(z_2;y) 
 & = & 
 G(z_1,z_2;y) + G(z_2,z_1;y).
\eq
The proof that the integral representation of the multiple polylogarithms fulfils the shuffle product formula 
in eq.~(\ref{chapter_multiple_polylogarithms:G_shuffle_product})
is sketched for the example in eq.~(\ref{chapter_multiple_polylogarithms:example_G_product}) in fig.~(\ref{chapter_multiple_polylogarithms:fig1})
and can easily be extended to multiple polylogarithms of higher depth by recursively replacing the 
two outermost integrations by integrations over the upper and lower triangle.

It is clear that the proof does not depend on the specific form of the integrands of the iterated integral, only
the iterated structure is relevant.
This implies that the shuffle product is not specific to the iterated integral representation of multiple polylogarithms,
but holds for any iterated integral of the form as in eq.~(\ref{chapter_iterated_integrals:def_iterated_integral}).

A non-trivial example for the shuffle product of two multiple polylogarithms is given by
\bq
 G(z_1,z_2;y) \cdot G(z_3;y) 
 & = & 
 G(z_1,z_2,z_3;y) + G(z_1,z_3,z_2;y) + G(z_3,z_1,z_2;y).
\eq
For fixed $y$ we may view the multiple polylogarithm $G(z_1,\dots,z_r;y)$ as a function
\bq
 G & : & {\mathbb C}^r \; \rightarrow {\mathbb C},
 \nonumber \\
 & & \left(z_1,\dots,z_r\right) \; \rightarrow \; G(z_1,\dots,z_r;y).
\eq
By linearity this extends to a map from the vector space of words ${\mathcal A}$ to the complex numbers ${\mathbb C}$
\bq
G & : & {\mathcal A} \; \rightarrow \; {\mathbb C},
\eq
e.g. for $w_1=z_1 \dots z_k$, $w_2=z_{k+1} \dots z_r$ and $c_1, c_2 \in {\mathbb C}$
the map is given by
\bq
 G\left( c_1 w_1 + c_2 w_2 \right)
 & = &
 c_1 G\left(w_1\right) + c_2 G\left(w_2\right)
 \nonumber \\
 & = &
 c_1 G(z_1,\dots,z_k;y) + c_2 G(z_{k+1},\dots,z_r;y).
\eq
Eq.~(\ref{chapter_multiple_polylogarithms:G_shuffle_product}) says that this map is an algebra homomorphism, i.e.
\bq
\label{chapter_multiple_polylogarithms:G_algebra_homomorphism}
 G\left( w_1 \shuffle w_2 \right)
 & = &
 G\left(w_1\right) \cdot G\left(w_2\right).
\eq
We may use the shuffle product to remove trailing zeros:
We say that a multiple polylogarithm of the form
\bq
 G(z_1,\dots,z_j,\underbrace{0,\dots,0}_{r-j};y)
\eq
with $z_j \neq 0$
has $(r-j)$ trailing zeroes. 
Multiple polylogarithms with trailing zeroes do not have a Taylor expansion in $y$ around $y=0$, 
but logarithmic singularities at $y=0$.
In removing the trailing zeroes, one explicitly separates these logarithmic terms, such that the rest has a regular expansion
around $y=0$.
The starting point is the shuffle relation
\bq
\lefteqn{
 G(0;y) G(z_1,\dots,z_j,\underbrace{0,\dots,0}_{r-j-1};y) = 
} & & \\
& &
  (r-j) G(z_1,\dots,z_j,\underbrace{0,\dots,0}_{r-j};y)
+ \sum\limits_{(s_1 \dots s_{j}) = (z_1 \dots z_{j-1}) \shuffle (0)} G(s_1,\dots,s_j,z_{j},\underbrace{0,\dots,0}_{r-j-1};y).
\nonumber
\eq
Solving this equation for $G(z_1,\dots,z_j,0,\dots,0;y)$ yields
\bq
\lefteqn{
G(z_1,\dots,z_j,\underbrace{0,\dots,0}_{r-j};y)
 = } & & \\
 & &
 \frac{1}{r-j} \left[
G(0;y) G(z_1,\dots,z_j,\underbrace{0,\dots,0}_{r-j-1};y) 
 -
 \sum\limits_{(s_1 \dots s_{j}) = (z_1 \dots z_{j-1}) \shuffle (0)} G(s_1,\dots,s_j,z_{j},\underbrace{0,\dots,0}_{r-j-1};y)
 \right]. \nonumber 
\eq
In the first term, one logarithm has been explicitly factored out:
\bq
 G(0;y) & = & \ln y.
\eq
All remaining terms have at most $(r-j-1)$ trailing zeroes. 
Using recursion, we may therefore eliminate all trailing zeroes. 
Let's consider an example: Let's assume $z_1 \neq 0$. We have
\bq
 G\left(z_1,0;y\right)
 & = &
 G\left(0;y\right) G\left(z_1;y\right)
 -
 G\left(0,z_1;y\right)
 \nonumber \\
 & = &
 \ln\left(y\right) G\left(z_1;y\right)
 -
 G\left(0,z_1;y\right).
\eq
Both $G(z_1;y)$ and $G(0,z_1;y)$ are free of trailing zeros.
\\
\\
\bs
{\it \refstepcounter{exercise}
{\bf Exercise \theexercise}: 
Express the product
\bq
 G_2\left(z;y\right) \cdot G_3\left(z;y\right)
\eq
as a linear combination of multiple polylogarithms.
}
\es

\section{The quasi-shuffle product}
\label{chapter_multiple_polylogarithms:section:quasi_shuffle_product}

In the previous section we have seen that the iterated integral representation induces
the shuffle product for multiple polylogarithms.
In this section we work out the analogy based on the nested sum representation for multiple polylogarithms.
We will see that the nested sum representation induces a quasi-shuffle product.
Again, it is not specific to multiple polylogarithms, but holds for any nested sum.

We start by considering a generalisation of shuffle algebras. 
Assume that on the alphabet $A$ of letters we have an additional
operation
\bq
\label{chapter_multiple_polylogarithms:additional_operation}
 \circ & : & A \times A \rightarrow A,
 \nonumber \\
       & &  (l_1, l_2) \rightarrow l_1 \circ l_2,
\eq
which is commutative and associative.
Then we can define a new product $\shuffle_q$ of words recursively through
\bq
\label{chapter_multiple_polylogarithms:def_recursive_quasi_shuffle}
 l_1 l_2 ... l_k \; \shuffle_q \; l_{k+1} ... l_r 
 & = &
 l_1 \left( l_2 ... l_k \; \shuffle_q \; l_{k+1} ... l_r \right)
+
 l_{k+1} \left( l_1 l_2 ... l_k \; \shuffle_q \; l_{k+2} ... l_r \right)
 \nonumber \\
 & &
+
(l_1 \circ l_{k+1}) \left( l_2 ... l_k \; \shuffle_q \; l_{k+2} ... l_r \right)
\eq
together with
\bq
 e \; \shuffle_q \; w = w \; \shuffle_q \; e = w.
\eq
This product is a generalisation of the shuffle product and differs from the recursive
definition of the shuffle product in eq.~(\ref{chapter_multiple_polylogarithms:def_recursive_shuffle}) through the extra term in the last line.
This modified product is known under the names quasi-shuffle product \cite{Hoffman},
mixable shuffle product \cite{Guo},
stuffle product \cite{Borwein} or
mould symmetrel \cite{Ecalle}.
This product turns the vector space of words into a quasi-shuffle algebra ${\mathcal A}_q$.

We have for example
\bq
 l_1 l_2 \; \shuffle_q \; l_3
 & = &
 l_1 l_2 l_3
 +
 l_1 l_3 l_2
 +
 l_3 l_1 l_2
 +
 l_1 l_{23}
 +
 l_{13} l_2,
\eq
with $l_{13} = l_1 \circ l_3$ and $l_{23} = l_2 \circ l_3$.

The quasi-shuffle algebra (with the quasi-shuffle product as product) is generated as an algebra by the Lyndon words \cite{Reutenauer}.
This is not too surprising: We already know that the shuffle algebra is generated by the Lyndon words.
Furthermore, the quasi-shuffle product differs from the shuffle product only by terms of lower depth.

Let us now make the connection to multiple polylogarithms $\mathrm{Li}_{m_1 \dots m_k}(x_1,\dots,x_k)$.
As letters we now take pairs $l_j=(m_j,x_j)$.
A multiple polylogarithms $\mathrm{Li}_{m_1 \dots m_k}(x_1,\dots,x_k)$
is uniquely specified by a word $w=l_1 l_2\dots l_k$ in these letters.
We define 
the additional operation in eq.~(\ref{chapter_multiple_polylogarithms:additional_operation}) by
\bq
\label{chapter_multiple_polylogarithms:def_additional_operation}
 \left(m_1,x_1\right)
 \circ
 \left(m_2,x_2\right)
 & = &
 \left( m_1+m_2; x_1 x_2 \right),
\eq
i.e. the first entries are added, while the second entries are multiplied.
We may view the multiple polylogarithm $\mathrm{Li}_{m_1 \dots m_k}(x_1,\dots,x_k)$ as a function
\bq
 \mathrm{Li} & : & {\mathbb N}^k \times {\mathbb C}^k \; \rightarrow {\mathbb C},
 \nonumber \\
 & & \left(m_1,\dots,m_k,x_1,\dots,x_k\right) \; \rightarrow \; \mathrm{Li}_{m_1 \dots m_k}(x_1,\dots,x_k).
\eq
Again, we may extend this by linearity 
to a map from the vector space of words ${\mathcal A}_q$ to the complex numbers ${\mathbb C}$
\bq
 \mathrm{Li} & : & {\mathcal A}_q \; \rightarrow \; {\mathbb C},
\eq
e.g. for $w_1=l_1 \dots l_k$, $w_2=l_{k+1} \dots l_r$, $l_j=(m_j,z_j)$ and $c_1, c_2 \in {\mathbb C}$
we have
\bq
 \mathrm{Li}\left( c_1 w_1 + c_2 w_2 \right)
 & = &
 c_1 \mathrm{Li}\left(w_1\right) + c_2 \mathrm{Li}\left(w_2\right)
 \nonumber \\
 & = &
 c_1 \mathrm{Li}_{m_1 \dots m_k}(x_1,\dots,x_k) + c_2 \mathrm{Li}_{m_{k+1} \dots m_r}(x_{k+1},\dots,x_r).
\eq
This map is again 
an algebra homomorphism, i.e.
\bq
\label{chapter_multiple_polylogarithms:Li_quasishuffle_product}
 \mathrm{Li}\left( w_1 \shuffle_q w_2 \right)
 & = &
 \mathrm{Li}\left(w_1\right) \cdot \mathrm{Li}\left(w_2\right).
\eq
A simple example for the quasi-shuffle product is given by
\bq
\label{chapter_multiple_polylogarithms:example_Li_product}
 \mathrm{Li}_{m_1}(x_1) \mathrm{Li}_{m_2}(x_2) 
& = & 
 \mathrm{Li}_{m_1,m_2}(x_1,x_2) + \mathrm{Li}_{m_2,m_1}(x_2,x_1)
                                 + \mathrm{Li}_{m_1+m_2}(x_1x_2).
\eq
The proof that the sum representation of the multiple polylogarithms 
\begin{figure}
\begin{center}
\includegraphics[scale=1.1]{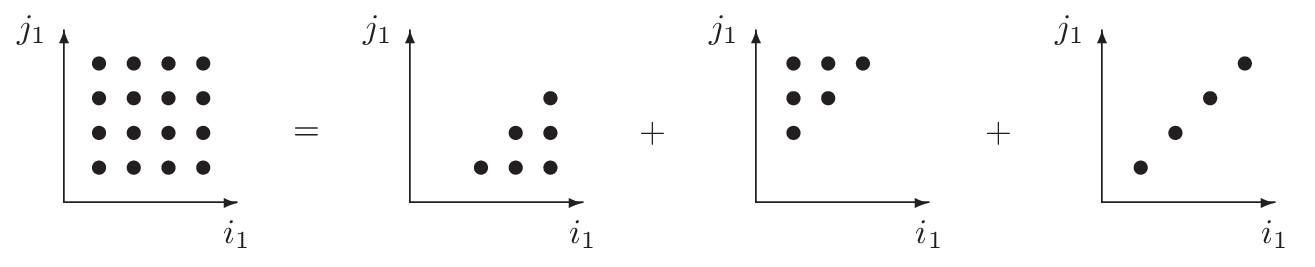}
\caption{\label{chapter_multiple_polylogarithms:fig2} Quasi-shuffle algebra from the sum representation: 
The quasi-shuffle product follows from replacing the sum over the square 
by a sum over the lower triangle, a sum over the upper triangle, and a sum over the diagonal.}
\end{center}
\end{figure}
fulfils the quasi-shuffle product formula in eq.~(\ref{chapter_multiple_polylogarithms:Li_quasishuffle_product})
is sketched for the example in eq.~(\ref{chapter_multiple_polylogarithms:example_Li_product}) 
in fig.~(\ref{chapter_multiple_polylogarithms:fig2})
and can easily be extended to multiple polylogarithms of higher depth by recursively replacing the 
two outermost summations by summations over the upper triangle, the lower triangle, and the diagonal.

Let us provide one further example for the quasi-shuffle product.
Working out the recursive definition of the quasi-shuffle product we obtain
\bq
\lefteqn{
 \mathrm{Li}_{m_1 m_2}(x_1,x_2) \cdot \mathrm{Li}_{m_3}(x_3) =
 } \nonumber \\
 & = &  
 \mathrm{Li}_{m_1 m_2 m_3}(x_1,x_2,x_3) 
 + \mathrm{Li}_{m_1 m_3 m_2}(x_1,x_3,x_2) 
 + \mathrm{Li}_{m_3 m_1 m_2}(x_3,x_1,x_2) 
\nonumber \\ & & 
+ \mathrm{Li}_{m_1 \left(m_2+m_3\right)}(x_1,x_2x_3) 
+ \mathrm{Li}_{\left(m_1+m_3\right) m_2}(x_1 x_3,x_2) 
\eq
\begin{figure}
\begin{center}
\includegraphics[scale=1.2]{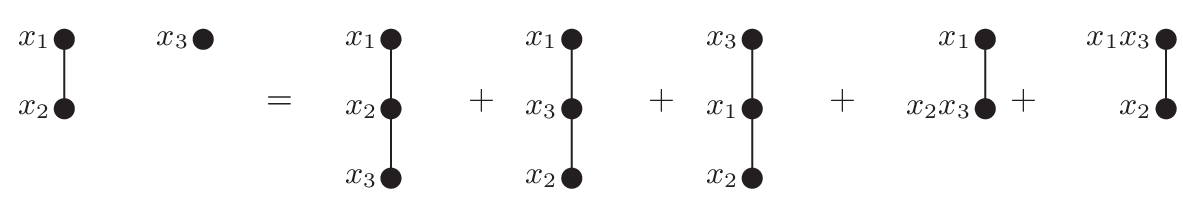}
\caption{\label{chapter_multiple_polylogarithms:fig3} Pictorial representation of the quasi-shuffle multiplication law. The first three terms
on the right-hand side correspond to the ordinary shuffle product, whereas the two last terms are the additional ``stuffle''-terms.}
\end{center}
\end{figure}
This is shown pictorially in fig.~(\ref{chapter_multiple_polylogarithms:fig3}).
The first three terms correspond to the ordinary shuffle product, whereas the two last terms are the additional ``stuffle''-terms.
In fig.~(\ref{chapter_multiple_polylogarithms:fig3}) we show only the $x$-variables, which are multiplied in the stuffle-terms.
Not shown in fig.~(\ref{chapter_multiple_polylogarithms:fig3}) are the indices $m_j$, which are added in the stuffle-terms.
\\
\\
\bs
{\it \refstepcounter{exercise}
{\bf Exercise \theexercise}: 
Work out the quasi-shuffle product
\bq
 \mathrm{Li}_{m_1 m_2}(x_1,x_2) \cdot \mathrm{Li}_{m_3 m_4}(x_3,x_4).
\eq
}
\es

\section{Double-shuffle relations}

We recall that we may denote multiple polylogarithms either as
$G(z_1,\dots,z_r;y)$ (the notation for the integral representation)
or as $\mathrm{Li}_{m_1 \dots m_k}(x_1,\dots,x_k)$ (the notation for the sum representation).
The conversion between these two notations is given 
by eq.~(\ref{chapter_multiple_polylogarithms:conversion_Li_to_G}) 
and eq.~(\ref{chapter_multiple_polylogarithms:conversion_G_to_Li}).

We have seen that shuffle product associated with the integral representation gives
relations among the multiple polylogarithms.
In the notation of section~\ref{chapter_multiple_polylogarithms:section:shuffle_product}
we have for fixed $y$ and 
$w_1=z_1 \dots z_k$, $w_2=z_{k+1} \dots z_r$
\bq
\label{chapter_multiple_polylogarithms:shuffle_relations}
 G\left( w_1 \shuffle w_2 \right)
 & = &
 G\left(w_1\right) \cdot G\left(w_2\right).
\eq
At the same time the quasi-shuffle product associated with the sum representation provides another set
of relations among the multiple polylogarithms.
In the notation of section~\ref{chapter_multiple_polylogarithms:section:quasi_shuffle_product}
we have for $l_j=(m_j,x_j)$ and
$w_1=l_1 \dots l_k$, $w_2=l_{k+1} \dots l_r$,
\bq
\label{chapter_multiple_polylogarithms:quasi_shuffle_relations}
 \mathrm{Li}\left( w_1 \shuffle_q w_2 \right)
 & = &
 \mathrm{Li}\left(w_1\right) \cdot \mathrm{Li}\left(w_2\right).
\eq
The union of the relations given by eq.~(\ref{chapter_multiple_polylogarithms:shuffle_relations}) and eq.~(\ref{chapter_multiple_polylogarithms:quasi_shuffle_relations})
are called the 
\index{double-shuffle relations}
{\bf double-shuffle relations}. 

Multiple zeta values are special values of multiple polylogarithms and as such also have
a sum representation and an integral representation:
\bq
 \zeta_{m_1 \dots m_k}
 & = &
 \mathrm{Li}_{m_1 \dots m_k}\left(1,\dots,1\right),
 \;\;\;\;\;\;\;\;\;\;\;\;\;\;\;\;\;\;\;\;\;\;\;\; m_1 \; \neq \; 1.
 \nonumber \\
 & = &
 \left(-1\right)^k
 G_{m_1 \dots m_k}\left(1,\dots,1;1\right).
\eq
Hence we have double-shuffle relations for multiple zeta values.
Using these, we may for example derive
\bq
\label{chapter_multiple_polylogarithms:example_zeta_31}
 \zeta_{3 1} & = & \frac{1}{4} \zeta_4.
\eq
This follows easily from the shuffle relation
\bq
 \zeta_2^2 & = & \left[ - G\left(0,1;1\right) \right]^2
 \; = \; 2 G\left(0,1,0,1;1\right) + 4 G\left(0,0,1,1;1\right)
 \; = \; 2 \zeta_{2 2} + 4 \zeta_{3 1}
\eq
and the quasi-shuffle relation
\bq
\label{chapter_multiple_polylogarithms:example_zeta_31_quasi_shuffle}
 \zeta_2^2 & = & \left[ \mathrm{Li}_2\left(1\right) \right]^2
 \; = \; 2 \mathrm{Li}_{2 2}\left(1,1\right) + \mathrm{Li}_4\left(1\right)
 \; = \; 2 \zeta_{2 2} + \zeta_4.
\eq
As a second example consider the relation
\bq
 \zeta_{2 1} & = & \zeta_3.
\eq
This relation is due to Euler.
We may derive this relation from the double-shuffle relations in a way similar to what we did above, 
but we have to be careful
since the Riemann zeta function $\zeta(s)$ diverges at $s=1$ (i.e. $\zeta_1$ does not exist).
To do it properly, we are going to use regularised (quasi-) shuffle relations.
Let 
\bq
\label{chapter_multiple_polylogarithms:regularised_zeta_1}
 L & = & - \ln \lambda \; = \; \mathrm{Li}_1\left(1-\lambda\right) \; = \; - G\left(1;1-\lambda\right).
\eq
$L$ is well-defined for $\lambda>0$, but diverges logarithmically for $\lambda \rightarrow 0$.
From the quasi-shuffle product we have
\bq
\label{chapter_multiple_polylogarithms:example_L_zeta_2_v1}
 L \cdot \zeta_2
 & = &
 \mathrm{Li}_1\left(1-\lambda\right) \cdot \mathrm{Li}_{2}\left(1\right)
 \; = \; 
 \mathrm{Li}_{1 2}\left(1-\lambda,1\right)
 +
 \mathrm{Li}_{2 1}\left(1,1-\lambda\right)
 +
 \mathrm{Li}_{3}\left(1-\lambda\right).
\eq
For the shuffle product we consider
\bq
\label{chapter_multiple_polylogarithms:example_L_zeta_2_v2}
 - L \cdot G\left(0,1;1-\lambda\right)
 & = &
 G\left(1;1-\lambda\right) \cdot G\left(0,1;1-\lambda\right)
 \nonumber \\
 & = &
 G\left(1,0,1;1-\lambda\right)
 +
 2 G\left(0,1,1;1-\lambda\right).
\eq
Expressed in the $\mathrm{Li}$-notation eq.~(\ref{chapter_multiple_polylogarithms:example_L_zeta_2_v2}) reads
\bq
\label{chapter_multiple_polylogarithms:example_L_zeta_2_v3}
 L \cdot \mathrm{Li}_{2}\left(1-\lambda\right)
 & = &
 \mathrm{Li}_{1 2}\left(1-\lambda,1\right)
 +
 2 \mathrm{Li}_{2 1}\left(1-\lambda,1\right).
\eq
We now subtract eq.~(\ref{chapter_multiple_polylogarithms:example_L_zeta_2_v1}) from eq.(\ref{chapter_multiple_polylogarithms:example_L_zeta_2_v3}). This yields
\bq
 L \cdot \left[ \mathrm{Li}_{2}\left(1-\lambda\right) - \zeta_2 \right]
 & = &
 2 \mathrm{Li}_{2 1}\left(1-\lambda,1\right)
 -
 \mathrm{Li}_{2 1}\left(1,1-\lambda\right)
 -
 \mathrm{Li}_{3}\left(1-\lambda\right).
\eq
It is easy to see that 
$\mathrm{Li}_{2}\left(1-\lambda\right) - \zeta_2 = {\mathcal O}\left(\lambda\right)$ and therefore
\bq
 \lim\limits_{\lambda \rightarrow 0} \left\{ - \ln \lambda \cdot \left[ \mathrm{Li}_{2}\left(1-\lambda\right) - \zeta_2 \right] \right\}
 & = & 0.
\eq
On the right-hand side all terms are finite and we have
\bq
 \lim\limits_{\lambda \rightarrow 0}
 \left[
 2 \mathrm{Li}_{2 1}\left(1-\lambda,1\right)
 -
 \mathrm{Li}_{2 1}\left(1,1-\lambda\right)
 -
 \mathrm{Li}_{3}\left(1-\lambda\right)
 \right]
 & = &
 \zeta_{2 1} - \zeta_3,
\eq
yielding $\zeta_{2 1} = \zeta_3$.
The relations in eq.~(\ref{chapter_multiple_polylogarithms:example_L_zeta_2_v1}) or eq.~(\ref{chapter_multiple_polylogarithms:example_L_zeta_2_v2}) are examples of
\index{regularised shuffle relation}
regularised (quasi-) shuffle relations.
\\
\\
\bs
{\it \refstepcounter{exercise}
\label{chapter_multiple_polylogarithms:exercise_zeta_4}
{\bf Exercise \theexercise}: 
Use the (regularised) double-shuffle relations to show
\bq
 \zeta_2^2 & = & \frac{5}{2} \zeta_4.
\eq
}
\es
\\
\\
From eq.~(\ref{chapter_one_loop_Li2_numerical_values}) we know that
\bq
 \zeta_2 & = & \mathrm{Li}_2\left(1\right) \; = \; \frac{\pi^2}{6}.
\eq
The above exercise shows that
\bq
 \zeta_4 & = & \frac{\pi^4}{90}.
\eq
In general, the 
\index{zeta value}
even zeta values are powers of $\pi$:
\bq
\label{chapter_multiple_polylogarithms:even_zeta_values}
 \zeta_{n} & = &
 - \frac{B_n}{2 n!} \left(2 \pi i\right)^n,
 \;\;\;\;\;\;\;\;\; 
 n \; = \; 2,4,6,8,\dots,
\eq
where $B_n$ are the Bernoulli numbers defined in eq.~(\ref{chapter_transformations:def_Bernoulli_number}).

A database of relations among multiple zeta values can be found at \cite{Minh:2000aaa,Blumlein:2009cf}.

\section{Monodromy}

In order to motivate the study of monodromies, let us first consider the logarithm $\ln(x)$
for a complex variable $x$.
The logarithm is singular for $x=0$, therefore we consider the logarithm 
on the punctured complex plane ${\mathbb C}\backslash\{0\}$.
It is well-known that the logarithm is a multi-valued function on ${\mathbb C}\backslash\{0\}$.
This is easily seen as follows: By the definition of the logarithm, $y=\ln(x)$ is a number, which fulfils 
\bq
\label{chapter_multiple_polylogarithms:monodromy_logarithm}
 e^y & = & x.
\eq
Now let $y$ be a number, which fulfils eq.~(\ref{chapter_multiple_polylogarithms:monodromy_logarithm}).
Then 
\bq
 y + 2 \pi i n,
 & & 
 n \in {\mathbb Z}
\eq
fulfils eq.~(\ref{chapter_multiple_polylogarithms:monodromy_logarithm}) as well.
We may turn the logarithm into a single-valued function by viewing the logarithm as a function
on a covering space of ${\mathbb C}\backslash\{0\}$,
or by restricting the logarithm to an open subset of ${\mathbb C}\backslash\{0\}$, 
for example ${\mathbb C}\backslash {\mathbb R}_{\le 0}$.
The restriction of the logarithm to an open subset $U$ of ${\mathbb C}\backslash\{0\}$ such that $\ln(x)$ 
is single-valued on $U$ is called a {\bf branch of the logarithm}.

Let us now patch together branches of the logarithm, such that we may analytically continue 
$\ln(x)$ counter clockwise around $x_0=0$.
After analytically continuing $\ln(x)$ around a small loop counter clockwise around $x_0=0$
we do not recover $\ln(x)$ but obtain $\ln(x)+2\pi i$.
This is called the {\bf monodromy}.

In order to prepare for the discussion of the monodromy of the multiple polylogarithms, let us be more explicit.
We consider the analytic continuation of $f(x)=\ln(x)$ around 
a small counter clockwise loop around $x_0=0$.
We parametrise the loop by 
\bq
 x_\eps\left( t \right) & = & x_0 + \eps e^{2 \pi i t},
\eq
with $t \in [0,1]$.
We have
\bq
 \ln\left(x_\eps\left(1\right)\right) - \ln\left(x_\eps\left(0\right)\right)
 & = & 
 2 \pi i.
\eq
We denote by ${\mathcal M}_{x_0} f(x)$ the analytic continuation of $f(x)$ around $x_0$.
Thus
\bq
\label{chapter_multiple_polylogarithms:result_monodromy_logarithm}
 {\mathcal M}_0 \ln\left(x\right)
 & = &
 \ln\left(x\right) + 2 \pi i.
\eq
Let us now turn to the classical polylogarithms.
We first consider
\bq
\mathrm{Li}_1(x)
 & = &
 - \ln\left(1-x\right) 
 \;\; = \;\;
 \int\limits_0^x \frac{dt}{1-t}.
\eq
$\mathrm{Li}_1(x)$ has a branch cut along the positive real axis starting at $x=1$.
Here we find
\bq
 {\mathcal M}_1 \mathrm{Li}_1\left(x\right)
 & = &
 \mathrm{Li}_1\left(x\right) - 2 \pi i.
\eq
The classical polylogarithms are given by
\bq
 \mathrm{Li}_n\left(x\right)
 & = &
 \sum\limits_{j=1}^\infty \frac{x^j}{j^n}
 \;\; = \;\;
 \int\limits_0^x \frac{dt}{t} \mathrm{Li}_{n-1}\left(t\right).
\eq
$\mathrm{Li}_n\left(x\right)$ is analytic at $x_0=0$, therefore
\bq
 {\mathcal M}_0 \mathrm{Li}_n\left(x\right)
 & = &
 \mathrm{Li}_n\left(x\right).
\eq
For the monodromy around $x_1=1$ one finds
\bq
\label{chapter_multiple_polylogarithms:monodromy_Lin}
 {\mathcal M}_1 \mathrm{Li}_n\left(x\right)
 & = &
 \mathrm{Li}_n\left(x\right)
 - 2 \pi i \frac{\ln^{n-1}\left(x\right)}{\left(n-1\right)!}.
\eq
Eq.~(\ref{chapter_multiple_polylogarithms:monodromy_Lin}) is proven by induction \cite{Hain}:
We may write
\bq
 \mathrm{Li}_n\left(x\right)
 & = &
 \int\limits_0^{1-\eps} \frac{dt}{t} \mathrm{Li}_{n-1}\left(t\right)
 +
 \int\limits_{1-\eps}^x \frac{dt}{t} \mathrm{Li}_{n-1}\left(t\right),
\eq
by splitting the integration path into a piece from $0$ to $1-\eps$, followed by second piece from $1-\eps$ to $x$.
The path does not encircle the point $x=1$.
Let's work out ${\mathcal M}_1 \mathrm{Li}_n\left(x\right)$. 
As $\mathrm{Li}_n\left(x\right)$ is given by an integral
from $0$ to $x$, we obtain ${\mathcal M}_1 \mathrm{Li}_n\left(x\right)$ by using an integration path which encircles
the point $x=1$.
We may deform this path into a path, which we split into three pieces:
A first piece from $0$ to $1-\eps$ as above, followed by second piece given by a small circle counter clockwise
around $x=1$ and finally a third piece from $1-\eps$ to $x$.
For the integrand of the third piece we have to use the formula after analytically continuing around a small loop
around $x=1$. This formula is given by the induction hypothesis.
We obtain
\bq
\label{chapter_multiple_polylogarithms:monodromy_Li_n_explicit}
 {\mathcal M}_1 \mathrm{Li}_n\left(x\right)
 -
 \mathrm{Li}_n\left(x\right)
 & = &
 \lim\limits_{\eps \rightarrow 0}
 \left[
 \oint \frac{dx'}{x'} \mathrm{Li}_{n-1}\left(x'\right)
 - \frac{2 \pi i}{\left(n-2\right)!} 
 \int\limits_{1-\eps}^x \frac{dt}{t} \ln^{n-2}\left(t\right)
 \right].
\eq
The first integral is around
\bq
 x'(t) & = & 1 - \eps e^{2\pi i t},
 \;\;\;\;\;\;
 t \in [0,1],
\eq
and corresponds to the second piece (a small circle counter clockwise around $x=1$) mentioned above.
This integral vanishes for $\eps \rightarrow 0$.
For $n>2$ this follows from the fact that $\mathrm{Li}_{n-1}(x)$ is bounded in a neighbourhood of $x=1$.
For $n=2$ we have to consider
\bq
 \lim\limits_{\eps \rightarrow 0}
 \oint \frac{dx'}{x'} \mathrm{Li}_{1}\left(x'\right)
 & = &
 \lim\limits_{\eps \rightarrow 0}
 \left\{
 2 \pi i \eps 
 \int\limits_0^1 dt \frac{e^{2\pi i t}}{1- \eps e^{2\pi i t}}
 \left[ \ln\left( \eps \right) + 2  \pi i t \right]
 \right\}
 =
 0.
\eq
The second integral in eq.~(\ref{chapter_multiple_polylogarithms:monodromy_Li_n_explicit}) 
is along the path $1-\eps$ to $x$ and 
corresponds to the difference of the integrands
\bq
 {\mathcal M}_1 \mathrm{Li}_{n-1}\left(x\right) - \mathrm{Li}_{n-1}\left(x\right).
\eq
We may use the induction hypothesis for this difference.
With
\bq
 \int\limits_{1-\eps}^x \frac{dt}{t} \ln^{n-2}\left(t\right)
 & = &
 \left. \frac{ \ln^{n-1}(t)}{n-1} \right|_{1-\eps}^x
\eq
the claim follows:
\bq
 {\mathcal M}_1 \mathrm{Li}_n\left(x\right)
 & = &
 \mathrm{Li}_n\left(x\right)
 - 2 \pi i \frac{\ln^{n-1}\left(x\right)}{\left(n-1\right)!}.
\eq
\bs
{\it \refstepcounter{exercise}
\label{chapter_multiple_polylogarithms:exercise_monodromy_log}
{\bf Exercise \theexercise}: 
Let
\bq
 f_0\left(x\right) \; = \; \frac{1}{r!} \ln^r\left(x\right),
 & &
 f_1\left(x\right) \; = \; \frac{\left(-1\right)^r}{r!} \ln^r\left(1-x\right).
\eq
Determine
\bq
 {\mathcal M}_0 f_0\left(x\right),
 \;\;\;\;\;\;
 {\mathcal M}_0 f_1\left(x\right),
 \;\;\;\;\;\;
 {\mathcal M}_1 f_0\left(x\right),
 \;\;\;\;\;\;
 {\mathcal M}_1 f_1\left(x\right).
\eq
}
\es
The monodromy of the multiple polylogarithms can be worked out along the same lines.
This is most conveniently done by using the integral representation
$G(z_1,\dots,z_r;y)$.
We consider the case where we analytically continue $y$ around a point $z$.
We assume $(z_1,\dots,z_r) \neq (0,\dots,0)$, as the case $(z_1,\dots,z_r) = (0,\dots,0)$
follows from exercise~\ref{chapter_multiple_polylogarithms:exercise_monodromy_log}.
We then have
\bq
\label{chapter_multiple_polylogarithms:monodromy_Glog}
\lefteqn{
 {\mathcal M}_z G\left(z_1,\dots,z_r;y\right)
 - G\left(z_1,\dots,z_r;y\right)
 = } & &
 \\
 & &
 \lim\limits_{\eps \rightarrow 0}
 \left\{
 \oint \frac{dy'}{y'-z_1} G\left(z_2,\dots,z_r;y'\right)
 + 
 \int\limits_{z+\eps}^y \frac{dy'}{y'-z_1} 
  \left[ {\mathcal M}_z G\left(z_2,\dots,z_r;y'\right) - G\left(z_2,\dots,z_r;y'\right) \right]
 \right\}.
 \nonumber
\eq
The first integral is around
\bq
 y'(t) & = & z + \eps e^{2\pi i t},
 \;\;\;\;\;\;
 t \in [0,1].
\eq
For $z \neq z_1$ the first integral does not contribute, for the same reasons as above: 
For $z \neq z_1$ we obtain from the Jacobian an explicit prefactor $\eps$, which in the limit $\eps \rightarrow 0$
kills any logarithmic singularity which might arise from $G(z_2,\dots,z_r;y)$.

For $z = z_1$ the first integral equals
\bq
\label{chapter_multiple_polylogarithms:monodromy_Glog_z_equals_z1}
 \oint \frac{dy'}{y'-z_1} G\left(z_2,\dots,z_r;y'\right)
 & = &
 2 \pi i \int\limits_0^1 dt G\left(z_2,\dots,z_r;z_1+\eps^{2\pi i t}\right).
\eq
Eq.~(\ref{chapter_multiple_polylogarithms:monodromy_Glog}) allows us to compute the monodromy of the multiple polylogarithms.
\\
\\
\bs
{\it \refstepcounter{exercise}
\label{chapter_multiple_polylogarithms:exercise_monodromy_G11}
{\bf Exercise \theexercise}: 
Compute the monodromy of $G(1,1;y)$ around $y=1$.
}
\es

\section{The Drinfeld associator}

In this section we study the Knizhnik-Zamolodchikov equation \cite{Knizhnik:1984nr}
and the Drinfeld associator \cite{Drinfeld:1990}.
Both topics are related to the first cyclotomic harmonic polylogarithms (i.e. multiple polylogarithms, where
all $z_j$'s are from the set $z_j \in \{0,1\}$).
\index{harmonic polylogarithms}
It is common practice to use the notation
\bq
\label{chapter_multiple_polylogarithms:notation_harmonic_polylogs}
 H\left(z_1,\dots,z_r;x\right)
 & = &
 \left(-1\right)^{n_1} 
 G\left(z_1,\dots,z_r;x\right),
 \;\;\;\;\;\;
 z_j \; \in \; \{0,1\},
\eq
and $n_1$ is the number of times the value $1$ occurs in the sequence $z_1,\dots,z_r$.
We call the functions $H(z_1,\dots,z_r;x)$ harmonic polylogarithms for short.
We also use for harmonic polylogarithms without trailing zeros the notation
\bq
 H_{m_1 \dots m_k}\left(x\right)
 & = &
 \left(-1\right)^{k} 
 G_{m_1 \dots m_k}\left(1,\dots,1;x\right).
\eq
If we write out $H_{m_1 \dots m_k}\left(x\right)$ in the long form $H\left(z_1,\dots,z_r;x\right)$,
the sequence of $z_j$'s so obtained contains exactly $k$ times the letter $1$.
From eq.~(\ref{chapter_iterated_integrals:harmpolylog}) we have
\bq
 H_{m_1 \dots m_k}(x) & = & \mathrm{Li}_{m_1 \dots m_k}(x,\underbrace{1,...,1}_{k-1}).
\eq
We denote by $A=\{0,1\}$ the alphabet corresponding to eq.~(\ref{chapter_multiple_polylogarithms:notation_harmonic_polylogs}).
As in section~\ref{chapter_multiple_polylogarithms:section:shuffle_product}
and section~\ref{chapter_multiple_polylogarithms:section:quasi_shuffle_product}
it is also convenient to denote alternatively harmonic polylogarithms with words.
A word $w=l_1 l_2 ... l_r$ with letters from the alphabet $A=\{0,1\}$
defines a harmonic polylogarithm as follows:
\bq
 H\left(w;x\right)
 & = &
 H\left(l_1,\dots,l_r;x\right).
\eq
In the discussion above we introduced the harmonic polylogarithms as a special cases of the multiple polylogarithms.
We may also define them from scratch.
We introduce two differential one-forms
\bq
 \omega_0\left(x\right) \;\; = \;\; \frac{dx}{x},
 & &
 \omega_1\left(x\right) \;\; = \;\; \frac{dx}{1-x}.
\eq
A word $w=l_1 l_2 \dots l_r$ defines a harmonic polylogarithm as follows: For the empty word $e$ we set
\bq
 H\left( e; x \right) \;\; = \;\; 1.
\eq
For a word consisting only of zeros ($l_1=l_2=\dots=l_r=0$) we set
\bq
\label{chapter_multiple_polylogarithms:H_trailing_zeros}
 H\left( 0^r; x \right) & = & \frac{1}{r!} \ln^r\left(x\right).
\eq
For all other words we define
\bq
 H\left( l w; x\right) & = & \int\limits_0^x \omega_l\left(t\right) H\left( w; t \right).
\eq
Note that we have for all words $w$ not of the form $0^r$
\bq
\label{chapter_multiple_polylogarithms:harmonic_polylog_limit_x_to_zero}
 \lim\limits_{x \rightarrow 0} H\left( w; x \right) & = & 0.
\eq
\begin{digression} {\bf Harmonic polylogarithms up to weight $3$}
\\
It is convenient to know explicit expressions of harmonic polylogarithms of low weight. 
We list here the explicit expressions for the harmonic polylogarithms up to weight
$3$. The tables can be found in \cite{Moch:1999eb}.
At weight $1$ we have
\bq
 H\left( 0; x \right) 
 & = & 
 \ln\left( x \right),
 \nonumber \\
 H\left( 1; x \right) 
 & = & 
 - \ln\left( 1-x \right).
\eq
At weight $2$ we have
\bq
 H\left( 0,0; x \right) 
 & = & 
 \frac{1}{2} \ln^2\left( x \right),
 \nonumber \\
 H\left( 0,1; x \right) 
 & = & 
 \mathrm{Li}_2\left( x \right),
 \nonumber \\
 H\left( 1,0; x \right) 
 & = & 
 - \ln\left( x \right) \ln\left( 1-x \right)
 - \mathrm{Li}_2\left( x \right),
 \nonumber \\
 H\left( 1,1; x \right) 
 & = & 
 \frac{1}{2} \ln^2\left( 1-x \right).
\eq
At weight $3$ we have
\bq
 H\left( 0,0,0; x \right) 
 & = & 
 \frac{1}{6} \ln^3\left( x \right),
 \nonumber \\
 H\left( 0,0,1; x \right) 
 & = & 
 \mathrm{Li}_3\left( x \right),
 \nonumber \\
 H\left( 0,1,0; x \right) 
 & = & 
 \ln\left( x \right) \mathrm{Li}_2\left( x \right)
 - 2 \mathrm{Li}_3\left( x \right),
 \nonumber \\
 H\left( 0,1,1; x \right) 
 & = & 
         \zeta_3
       + \ln\left( 1-x \right) \zeta_2
       - \ln\left( 1-x \right) \mathrm{Li}_2\left( x \right)
       - \frac{1}{2} \ln\left( x \right) \ln^2\left( 1-x \right)
       - \mathrm{Li}_3\left( 1-x \right),
 \nonumber \\
 H\left( 1,0,0; x \right) 
 & = & 
       - \frac{1}{2} \ln^2\left( x \right) \ln\left( 1-x \right)
       - \ln\left( x \right) \mathrm{Li}_2\left( x \right)
       + \mathrm{Li}_3\left( x \right),
 \nonumber \\
 H\left( 1,0,1; x \right) 
 & = & 
       - 2 \zeta_3
       - 2 \ln\left( 1-x \right) \zeta_2
       + \ln\left( 1-x \right) \mathrm{Li}_2\left( x \right)
       + \ln\left( x \right) \ln^2\left( 1-x \right)
       + 2 \mathrm{Li}_3\left( 1-x \right),
 \nonumber \\
 H\left( 1,1,0; x \right) 
 & = & 
         \zeta_3
       + \ln\left( 1-x \right) \zeta_2
       - \mathrm{Li}_3\left( 1-x \right),
 \nonumber \\
 H\left( 1,1,1; x \right) 
 & = & 
       - \frac{1}{6} \ln^3\left( 1-x \right).
\eq
\end{digression}

Let $e_0$ and $e_1$ be two non-commutative variables.
(We may think of $e_0$ and $e_1$ as two generators of a Lie algebra or as two $(N \times N)$-matrices.)
Strings of $e_0$ and $e_1$ are denoted by $e_w$, for example
\bq
 e_{0101} & = & e_0 e_1 e_0 e_1.
\eq
With these preparations, we may now state 
the 
\index{Knizhnik-Zamolodchikov equation}
Knizhnik-Zamolodchikov equation.
\begin{tcolorbox}
{\bf The Knizhnik-Zamolodchikov equation}:
\\
\bq
\label{chapter_multiple_polylogarithms:Knizhnik_Zamolodchikov_equation}
 \frac{d}{dx} L\left(x\right)
 & = & 
 \left(  \frac{e_0}{x} + \frac{e_1}{1-x} \right) L\left(x\right).
\eq
This equation is solved by
\bq
 L\left(x\right)
 & = &
 \sum\limits_w 
 H\left( w; x \right) e_w,
\eq
where the sum runs over all words, which can be formed from the alphabet $A=\{0,1\}$.
The sum includes the empty word.
\end{tcolorbox}
Proof: We have
\bq
 L\left(x\right)
 & = &
 1
 +
 \sum\limits_w H\left( 0, w; x \right) e_{0 \, w}
 +
 \sum\limits_w H\left( 1, w; x \right) e_{1 \, w}
\eq
and therefore
\bq
 \frac{d}{dx}
 L\left(x\right)
 & = &
 \sum\limits_w \frac{H\left( w; x \right)}{x} e_{0 \, w}
 +
 \sum\limits_w \frac{H\left( w; x \right)}{1-x} e_{1 \, w}
 \;\; = \;\;
  \left(  \frac{e_0}{x} + \frac{e_1}{1-x} \right)
 \sum\limits_w 
 H\left( w; x \right) e_w.
\eq
\begin{digression} {\bf Relation to physics}:
\\
The Knizhnik-Zamolodchikov equation is not too far away from physics.
To see this, consider again the two-loop double box Feynman integral discussed as example 3 
in section~\ref{chapter_iterated_integrals:deriving_the_dgl}.
In section~\ref{chapter_iterated_integrals:fibre_transformation} we have shown that a fibre transformation
puts the differential equation into $\eps$-form.
Setting $x'=-x$ we obtain from eq.~(\ref{chapter_iterated_integrals:example_dgl_eps_form_double_box}):
\bq
 \frac{d}{dx'} \vec{I}'\left(x\right) & = & 
 \left( - \frac{\eps C_0}{x'} + \frac{\eps C_{-1}}{1-x'} \right) \vec{I}'\left(x\right),
\eq
where the $(8 \times 8)$-matrices $C_0$ and $C_{-1}$ have been given immediately after 
eq.~(\ref{chapter_iterated_integrals:example_dgl_eps_form_double_box}).
Setting $e_0=-\eps C_0$ and $e_1=\eps C_{-1}$ gives the relation to eq.~(\ref{chapter_multiple_polylogarithms:Knizhnik_Zamolodchikov_equation}).
Note that with the choice $e_0=-\eps C_0$ and $e_1=\eps C_{-1}$ 
the solution $L(x)$ of eq.~(\ref{chapter_multiple_polylogarithms:Knizhnik_Zamolodchikov_equation})
is a $(8 \times 8)$-matrix, while $\vec{I}'$ is a vector of dimension $8$.
In order to obtain a vector-valued solution from the Knizhnik-Zamolodchikov equation, we may always multiply
eq.~(\ref{chapter_multiple_polylogarithms:Knizhnik_Zamolodchikov_equation}) 
by a constant vector $\vec{I}_0'$ from the right.
\end{digression}
Let us return to the general case.
$L(x)$ has for $x \rightarrow 0^+$ the asymptotic value
\bq
 L\left(x\right)
 & = &
 e^{e_0 \ln x} + {\mathcal O}\left(\sqrt{x}\right).
\eq
The exponential term $e^{e_0 \ln x}$ is related to the words consisting of zeros only:
\bq
 e^{e_0 \ln x}
 & = &
 \sum\limits_{n=0}^\infty \frac{\ln^n\left(x\right)}{n!} e_0^n
 \;\; = \;\;
 \sum\limits_{n=0}^\infty H\left( 0^n; x \right) e_{0^n}.
\eq
We further have
\bq
 H\left( 1^n; x \right) 
 & = &
 \left(-1\right)^n \frac{\ln^n\left(1-x\right)}{n!}
\eq
and therefore
\bq
 \sum\limits_{n=0}^\infty H\left( 1^n; x \right) e_{1^n}
 & = &
 e^{-e_1 \ln\left(1-x\right)}.
\eq
For the first terms of $L(x)$ we have
\bq
 L\left(x\right) 
 & = &
 1 
 + H\left(0;x\right) e_0
 + H\left(1;x\right) e_1
 \nonumber \\
 & &
 + H\left(0,0;x\right) e_0^2
 + H\left(0,1;x\right) e_0 e_1
 + H\left(1,0;x\right) e_1 e_0
 + H\left(1,1;x\right) e_1^2
 + ...
\eq
We define the regularised boundary values by
\bq
 C_0 \;\; = \;\; \lim\limits_{x \rightarrow 0} e^{- e_0 \ln x} L\left(x\right),
 & &
 C_1 \;\; = \;\; \lim\limits_{x \rightarrow 1} e^{e_1 \ln\left(1- x\right)} L\left(x\right).
\eq
The first few terms of $C_0$ and $C_1$ are
\bq
 C_0
 & = &
 \lim\limits_{x \rightarrow 0} \left\{
 1 
 + H\left(1;x\right) e_1
 + H\left(1,0;x\right) \left[e_1,e_0\right]
 + H\left(1,1;x\right) e_1^2
 \right. \nonumber \\
 & & \left.
 + H\left(1,0,0;x\right) \left( e_0 \left[e_0,e_1\right] + \left[e_1,e_0\right] e_0 \right)
 + H\left(1,0,1;x\right) \left[e_1,e_0\right] e_1
 + H\left(1,1,0;x\right) \left[e_1^2,e_0\right]
 \right. \nonumber \\
 & & \left.
 + H\left(1,1,1;x\right) e_1^3
 + ...
 \right\}
 \nonumber \\
 & = & 
 1,
 \nonumber \\
 C_1 & = &
 \lim\limits_{x \rightarrow 1} \left\{
 1
 + H\left(0;x\right) e_0
 + H\left(0,0;x\right) e_0^2
 + H\left(0,1;x\right) \left[e_0,e_1\right]
 \right. \nonumber \\
 & & \left. 
 + H\left(0,0,0;x\right) e_0^3
 + H\left(0,0,1;x\right) \left[ e_0^2, e_1 \right]
 + H\left(0,1,0;x\right) \left[ e_0,e_1 \right] e_0
 \right. \nonumber \\
 & & \left. 
 + H\left(0,1,1;x\right) \left( \left[e_0,e_1\right]e_1 + e_1\left[e_1,e_0\right]\right)
 + ...
 \right\}
 \nonumber \\
 & = &
 1
 + \zeta_2 \left[e_0,e_1\right]
 + \zeta_3 \left[ e_0-e_1, \left[ e_0, e_1 \right] \right]
 + ...
\eq
$C_0=1$ follows from eq.~(\ref{chapter_multiple_polylogarithms:harmonic_polylog_limit_x_to_zero}).
$C_1$ and $C_0$ are related by the 
\index{Drinfeld associator}
{\bf Drinfeld associator}
\bq
 C_1 & = & \Phi\left(e_0,e_1\right) C_0.
\eq
The associator is given by
\bq
 \Phi\left(e_0,e_1\right)
 & = &
 \sum\limits_w \zeta\left(w\right) e_{w},
\eq
where $\zeta(w)$ denotes a multiple zeta value in the expanded notation.
\index{multiple zeta values}
We set
\bq
 \zeta\left(e\right) = 1,
 \;\;\;
 \zeta\left(0\right) = 0,
 \;\;\;
 \zeta\left(1\right) = 0.
\eq
The relation with the standard notation $\zeta_{n_1 ... n_k}$ is given for $n_1\ge 2$ by
\bq
 \zeta\left( 0^{n_1-1}, 1, ..., 0^{n_k-1}, 1 \right) & = & \zeta_{n_1 ... n_k}.
\eq
Furthermore we have the shuffle relation
\bq
 \zeta\left(w_1\right) \zeta\left(w_2\right)
 & = &
 \zeta\left(w_1 \shuffle w_2\right).
\eq
Up to weight $3$ we therefore have
\bq
 \Phi\left(e_0,e_1\right)
 & = &
 1
 + \zeta\left(0,0\right) e_0^2
 + \zeta\left(0,1\right) e_0 e_1
 + \zeta\left(1,0\right) e_1 e_0
 + \zeta\left(1,1\right) e_1 e_1
 \nonumber \\
 & &
 + \zeta\left(0,0,0\right) e_0^3
 + \zeta\left(0,0,1\right) e_0^2 e_1
 + \zeta\left(0,1,0\right) e_0 e_1 e_0
 + \zeta\left(0,1,1\right) e_0 e_1^2
 \nonumber \\
 & &
 + \zeta\left(1,0,0\right) e_1 e_0^2
 + \zeta\left(1,0,1\right) e_1 e_0 e_1
 + \zeta\left(1,1,0\right) e_1^2 e_0
 + \zeta\left(1,1,1\right) e_1^3
 + ...
 \nonumber \\
 & = &
 1 + \zeta_2 \left[ e_0, e_1 \right] 
 + \zeta_3 \left[ e_0-e_1, \left[ e_0, e_1 \right] \right]
 + ...
\eq
Here we used 
\bq
 \zeta(0,0) & = & \frac{1}{2} \zeta(0)^2 \;\; = \;\; 0,
 \nonumber \\
 \zeta(1,1) & = & \frac{1}{2} \zeta(1)^2 \;\; = \;\; 0,
 \nonumber \\
 \zeta(1,0) & = & \zeta(0) \zeta(1) - \zeta(0,1) \;\; = \;\; - \zeta(0,1),
\eq
and similar relations at weight $3$.

With the help of the Drinfeld associator we may give the monodromy of $L(x)$ 
(and hence the monodromy of the harmonic polylogarithms) in a compact form as
\bq
 {\mathcal M}_0 L\left(x\right)
 & = &
 L\left(x\right) e^{2\pi i e_0},
 \nonumber \\
 {\mathcal M}_1 L\left(x\right)
 & = &
 L\left(x\right) \Phi^{-1} e^{-2\pi i e_1} \Phi.
\eq
With
\bq
 \Phi^{-1}
 & = &
 1 - \zeta_2 \left[ e_0, e_1 \right] 
 - \zeta_3 \left[ e_0-e_1, \left[ e_0, e_1 \right] \right]
 + ...
\eq
we find
\bq
 {\mathcal M}_1 L\left(x\right) 
 & = &
 1 
 + H\left(0;x\right) e_0
 + \left[ H\left(1;x\right) - 2 \pi i \right] e_1
 + H\left(0,0;x\right) e_0^2
 + \left[ H\left(0,1;x\right) - 2 \pi i H\left(0;x\right) \right] e_0 e_1
 \nonumber \\
 & &
 + H\left(1,0;x\right) e_1 e_0
 + \left[ H\left(1,1;x\right) - 2 \pi i H\left(1;x\right) + \frac{1}{2} \left( 2 \pi i \right)^2 \right] e_1^2
 + ...
\eq
Taking the coefficient of a particular word in $e_0$ and $e_1$ 
on the left-hand side and on the right-hand side, we obtain
the monodromy of the harmonic polylogarithms.
For example, taking the coefficient of $e_1^2$ we find
\bq
 {\mathcal M}_1 H\left(1,1;x\right)
 & = &
 H\left(1,1;x\right) - 2 \pi i H\left(1;x\right) + \frac{1}{2} \left( 2 \pi i \right)^2,
\eq
in agreement with exercise~\ref{chapter_multiple_polylogarithms:exercise_monodromy_G11}.

\section{Fibration bases}

Let's look at
\bq
 f_1\left(x\right)
 & = &
 G\left(1,0;x\right),
 \nonumber \\
 f_2\left(x\right)
 & = &
 -G\left(0,1;x\right) + \ln\left(x\right) \ln\left(1-x\right)
\eq
and
\bq 
 g_1\left(x,y\right)
 & = &
 G\left(0,0;1-x\right)
 - G\left(0;1-x\right) G\left(0;1-y\right)
 + G\left(0,0;1-y\right)
 - G\left(0,1;x\right)
 - G\left(0,1;y\right),
 \nonumber \\
 g_2\left(x,y\right)
 & = &
 G\left(0,1-x;xy-x\right)
 + G\left(0,1-y;xy-y\right)
 - G\left(0,1;xy\right).
\eq
A priori it is not obvious that $f_1(x)=f_2(x)$
and $g_1(x,y)=g_2(x,y)$.
An attentive reader might notice, that a proof of $f_1(x)=f_2(x)$ can be reduced to a shuffle relation, while
a proof of $g_1(x,y)=g_2(x,y)$ can be reduced to the five-term relation for the dilogarithm of eq.~(\ref{chapter_one_loop:dilog_five_term_relation}). Here we are interested in the more general situation: Given two expressions in multiple
polylogarithms, can be prove or disprove that they are equal?
 
In order to prove $f_1(x)=f_2(x)$ we may use the fact that two functions of a variable $x$ are equal,
if their derivatives with respect to $x$ are equal and the two functions agree at one point.
Thus instead of showing $f_1(x)=f_2(x)$ we may show
\bq
 f_1'(x) \; = \; f_2'(x)
 & \mbox{and} &
 f_1(0) \; = \; f_2(0).
\eq
This is simpler, as the derivatives are of lower weight
\bq
 f_1'\left(x\right) \; = \;\frac{\ln\left(x\right)}{x-1},
 & &
 f_2'\left(x\right) \; = \; - \frac{\ln\left(1-x\right)}{x} + \frac{\ln\left(1-x\right)}{x} + \frac{\ln\left(x\right)}{x-1}
 \; = \; \frac{\ln\left(x\right)}{x-1}
\eq
and the equation $f_1(0)=f_2(0)$ has one variable less.
For the case at hand we have
\bq
 f_1\left(0\right) \; = \; f_2\left(0\right) \; = \; 0,
\eq
where we used 
\bq
 \lim\limits_{x\rightarrow 0} \left( x \cdot \ln\left(x\right) \right) & = & 0.
\eq
If $f_1(x)$ and $f_2(x)$ are of higher weight we may iterate this process.

Once we have the derivatives, we may integrate back:
\bq
 f_j\left(x\right) & = & f_j\left(0\right) + \int\limits_0^x d\tilde{x} f_j'\left(\tilde{x}\right),
 \;\;\;\;\;\;
 j \; \in \; \{1,2\}.
\eq
This puts $f_j(x)$ into a standardised form:
\bq
 f_1\left(x\right) \; = \; G\left(1,0;x\right),
 & &
 f_2\left(x\right) \; = \; G\left(1,0;x\right).
\eq
A comparison of $f_1(x)$ and $f_2(x)$ is now straightforward.

We may tackle the proof of $g_1(x,y)=g_2(x,y)$ in a similar way:
We show for example
\bq
 \frac{\partial}{\partial y} g_1\left(x,y\right)
 & = &
 \frac{\partial}{\partial y} g_2\left(x,y\right)
\eq
and
\bq
 g_1\left(x,0\right) & = & g_2\left(x,0\right).
\eq
Note that we have to show $g_1(x,0)=g_2(x,0)$ for all points on the line $y=0$.
For the case at hand we find
\bq
 g_1\left(x,y\right) \; = \; g_2\left(x,y\right)
 & = &
 \frac{1}{y-1} \left[ G\left(1;y\right) - G\left(1;x\right) \right]
 - \frac{1}{y} G\left(1;y\right)
\eq
and
\bq
 g_1\left(x,0\right) \; = \; g_2\left(x,0\right)
 & = &
 - G\left(0,1;x\right)
 + G\left(1,1;x\right).
\eq
Integrating back gives
\bq
 g_j\left(x,y\right)
 & = &
 g_j\left(x,0\right)
 +
 \int\limits_0^y d\tilde{y} \left( \frac{\partial}{\partial \tilde{y}} g_j\left(x,\tilde{y}\right) \right),
 \;\;\;\;\;\;
 j \; \in \; \{1,2\}.
\eq
Thus
\bq
 g_j\left(x,y\right)
 & = &
 - G\left(0,1;x\right)
 + G\left(1,1;x\right)
 - G\left(1;x\right) G\left(1;y\right)
 + G\left(1,1;y\right)
 - G\left(0,1;y\right)
\eq
and a comparison of $g_1(x,y)$ and $g_2(x,y)$ is now straightforward.

Let us now formulate this in generality \cite{Brown:2011b}:
\begin{tcolorbox}
{\bf Fibration basis}:
\\
We consider $n$ variables $x_1,\dots,x_n$ and let $A = \{l_1,l_2,\dots\}$ be a set of rational functions
in the variables $x_1,\dots,x_n$.
We call $A$ an alphabet and denote words by $w=l_1 l_2 \dots l_r$.
We further denote a multiple polylogarithm by
$G(w;z)=G(l_1,\dots,l_r;z)$.
Consider now $G(w;1)$. We may write $G(w;1)$ as
\bq
\label{chapter_multiple_polylogarithms:def_fibration_basis}
 G\left(w;1\right)
 & = &
 \sum\limits_j
 c_j
 \;
 G\left(w_{1,j};x_1\right)
 \dots
 G\left(w_{n-1,j};x_{n-1}\right)
 G\left(w_{n,j};x_n\right),
\eq
where $w_{i,j}$ is a word from an alphabet $A_i$. 
The letters in the alphabet $A_i$ are algebraic functions in the variables $x_1,\dots,x_{i-1}$.
The important point here is that the letters in $A_i$ no longer depend on the variables $x_i,\dots,x_n$.
The $c_j$ are constants with respect to $x_1,\dots,x_n$.

A multiple polylogarithm written as in the right-hand side of eq.~(\ref{chapter_multiple_polylogarithms:def_fibration_basis}) is said to be expressed in the fibration basis with respect to the order $[x_1,\dots,x_n]$.
The expression on the right-hand side of eq.~(\ref{chapter_multiple_polylogarithms:def_fibration_basis}) depends on the order
of the variables $x_1,\dots,x_n$.
\end{tcolorbox}
Please note that although we start with the alphabet $A$ with letters which are rational functions of the
variables $x_1,\dots,x_n$, the letters of the alphabets $A_i$ are in general algebraic functions of the variables
$x_1,\dots,x_{i-1}$.
This can be seen from the following simple example, where the algebraic function $\sqrt{x_1}$ appears.
We assume $x_1,x_2 > 0$ and $x_2^2>x_1$.
\bq
 G\left(\frac{x_2^2}{x_1};1\right)
 & = &
 \ln\left(\frac{x_2^2-x_1}{x_2^2}\right)
 \nonumber \\
 & = &
 G\left(\sqrt{x_1};x_2\right)
 + G\left(-\sqrt{x_1};x_2\right)
 - 2 G\left(0;x_2\right)
 + G\left(0;x_1\right)
 - i \pi.
\eq
This example also shows a second important point: For $x_1,x_2 > 0$ and $x_2^2>x_1$
the function $G(x_2^2/x_1;1)$ gives a real number and there are no singularities on the integration path from $0$ to $1$.
On the other hand we have $\sqrt{x_1}<x_2$ and there is for the function $G(\sqrt{x_1};x_2)$ a singularity 
on the integration path, resulting in a branch cut on the real $x_2$-axis starting at $x_2=\sqrt{x_1}$.
We have to specify how this singularity is avoided.
A standard mathematical convention is that a function with a branch cut starting at a finite point 
and extending to infinity is taken to be continuous as the cut is approached coming
around the finite endpoint of the cut in a counter clockwise direction \cite{C99standard}.
For the case at hand this amounts to deforming the integration contour into the lower complex plane.
Therefore, the function $G(\sqrt{x_1};x_2)$ has an imaginary part for $x_2>\sqrt{x_1}$.
This imaginary part is compensated by the term $(-i\pi)$ and the full result is real.

\section{Linearly reducible Feynman integrals}
\label{chapter_multiple_polylogarithms:linear_reducibility}

In this section we study an algorithm, which allows us for a special class of Feynman integrals to perform
all integrations in the Feynman parameter representation \cite{Brown:2008,Panzer:2014caa}.
The class of Feynman integrals we would like to consider must satisfy two conditions:
\begin{enumerate}
\item The integrand of the Feynman parameter representation is integrable in an integer dimension $\Dint$.
\item The integrand is linearly reducible for at least one ordering $\sigma$ of the Feynman parameters
$a_{\sigma_1}, \dots, a_{\sigma_{\ninternal}}$.
\end{enumerate}
The first condition allows us to expand the integrand in the dimensional regularisation parameter $\eps$:
\bq
 \left[ {\mathcal U}\left(a\right) \right]^{\nu-\frac{\left(\loopnumber+1\right) D}{2}}
 & = &
 \left[ {\mathcal U}\left(a\right) \right]^{\nu-\frac{\left(\loopnumber+1\right) \Dint}{2}}
 \sum\limits_{j=0}^\infty \frac{\left(l+1\right)^j \eps^j}{j!} \left[ \ln\left({\mathcal U}\right)\right]^j,
 \nonumber \\
 \left[ {\mathcal F}\left(a\right) \right]^{\frac{\loopnumber D}{2}-\nu}
 & = &
 \left[ {\mathcal F}\left(a\right) \right]^{\frac{\loopnumber \Dint}{2}-\nu}
 \sum\limits_{j=0}^\infty \frac{\left(-l\right)^j \eps^j}{j!} \left[ \ln\left({\mathcal F}\right)\right]^j.
\eq
The second condition will be explained below.

Let's consider the order $a_1,\dots,a_{\ninternal}$, 
corresponding to the case where we first integrate over $a_{\ninternal}$, then $a_{(\ninternal-1)}$ until $a_1$.
It is convenient to use the Cheng-Wu theorem with the delta distribution $\delta(1-a_{\ninternal})$.
The Feynman integral we are interested in is
\bq
 I
 & = &
 \frac{e^{\loopnumber \eps \Eulerconstant}\Gamma\left(\nu-\frac{\loopnumber D}{2}\right)}{\prod\limits_{j=1}^{\ninternal}\Gamma(\nu_j)}
 \int\limits_0^\infty da_1 
 \dots
 \int\limits_0^\infty da_{\ninternal}
 \; \delta\left(1-a_{\ninternal} \right) \; 
 \cdot R \cdot G.
\eq
where
\bq
 R & = &
 \left( \prod\limits_{j=1}^{\ninternal} a_j^{\nu_j-1} \right)
 \frac{\left[ {\mathcal U}\left(a\right) \right]^{\nu-\frac{\left(\loopnumber+1\right) \Dint}{2}}}{\left[ {\mathcal F}\left(a\right) \right]^{\nu-\frac{\loopnumber \Dint}{2}}},
 \nonumber \\
 G & = &
 \left[ \sum\limits_{j_1=0}^\infty \frac{\left(l+1\right)^{j_1} \eps^{j_1}}{{j_1}!} \left[ \ln\left({\mathcal U}\right)\right]^{j_1} \right]
 \left[ \sum\limits_{j_2=0}^\infty \frac{\left(-l\right)^{j_2} \eps^{j_2}}{{j_2}!} \left[ \ln\left({\mathcal F}\right)\right]^{j_2} \right].
\eq
$R$ is a rational function in the Feynman parameters, while the function $G$ contains logarithms.
The function $G$ has an $\eps$-expansion and we may consider each term in the $\eps$-expansion separately.

Let's consider the integration order $a_1, \dots, a_{\ninternal}$, i.e.
we integrate over $a_{\ninternal}$ first and $a_1$ last.
The integrand for the integration over $a_j$ is then a function of $a_1,\dots,a_j$. 
At this stage the variables $a_{j+1},\dots,a_{\ninternal}$ have already been integrated out.
We say that the integrand is {\bf linearly reducible at stage $j$}, if the integrand at stage $j$
can be written as a sum of terms, where each term
is a product of a rational function and a multiple polylogarithm subject to the conditions that
the denominator of the rational function factorises into linear factors with respect to $a_j$
and the multiple polylogarithm can be cast into a form, where $a_j$ appears only as upper integration limit
and nowhere else.
We say that the integrand is {\bf linearly reducible for the order $1,\dots,\ninternal$}, if it is linearly reducible
at all stages $j$.

Note that also the condition on the multiple polylogarithm is non-trivial: If the letters of the multiple polylogarithm
depend algebraically on $a_j$, but not rationally, eq.~(\ref{chapter_multiple_polylogarithms:def_fibration_basis}) does not apply. 

The two conditions are tailored such that all integrations may be performed within the class of multiple polylogarithms.
The essential integration is
\bq
 \int\limits_0^\infty \frac{da_j}{a_j-l_1} G\left(l_2,\dots,l_r;a_j\right)
 & = &
 \left. G\left(l_1,l_2,\dots,l_r;a_j\right)\right|_0^\infty,
\eq
where $l_1,l_2,\dots,l_r$ may depend on $a_1,\dots,a_{j-1}$, but not on $a_j$.
The condition on the rational function ensures that we may use partial fractioning:
\bq
 R 
 & = &
 P +
 \sum\limits_i
 \sum\limits_n
 \frac{c_{i,n}}{\left(a_j-l_{i}\right)^n},
\eq
where $P$ is a polynomial in $a_j$.
Any terms, which are not simple poles can be reduced by partial integration:
\bq
 \int\limits_0^\Lambda da_j \; a_j^n G\left(w;a_j\right)
 & = &
 \frac{1}{n+1}
 \left[
 \left. a_j^{n+1} G\left(w;a_j\right) \right|_0^\Lambda
 - 
 \int\limits_0^\Lambda da_j \; a_j^{n+1} \frac{\partial}{\partial a_j} G\left(w;a_j\right)
 \right],
 \nonumber \\
 \int\limits_0^\Lambda da_j \; \frac{G\left(w;a_j\right)}{\left(a_j-l\right)^{n+1}}
 & = &
 - \frac{1}{n}
 \left[
 \left. \frac{G\left(w;a_j\right)}{\left(a_j-l\right)^{n}} \right|_0^\Lambda
 - 
 \int\limits_0^\Lambda \frac{da_j}{\left(a_j-l\right)^{n}} \frac{\partial}{\partial a_j} G\left(w;a_j\right)
 \right].
\eq
Let's look at an example: We consider the graph shown in fig.~\ref{chapter_graph_polynomials:fig3} with vanishing
internal masses.
We are interested in $I_{11111}$ in $D=4$ space-time dimensions. This integral is finite
and we may calculate it without regularisation.
We choose the integration order $a_3,a_4,a_5,a_1,a_2$.
We have
\bq
 I_{11111}\left(4,\frac{-p^2}{\mu^2}\right)
 & = &
 \int\limits_{a_j \ge 0} d^5a 
 \frac{\delta\left(1-a_2\right)}{{\mathcal U}{\mathcal F}}
\eq
with
\bq
 {\mathcal U} & = & (a_1+a_4)(a_3+a_5) + (a_1+a_3+a_4+a_5)a_2,
 \nonumber \\
 {\mathcal F} & = & 
      \left[ (a_1+a_5)(a_3+a_4)a_2
      +a_1a_4(a_3+a_5)
      +a_3a_5(a_1+a_4) \right]
      \left( \frac{-p^2}{\mu^2} \right).
\eq
We set $\mu^2=-p^2$.
Thus we have
\bq
 I_{11111}\left(4,1\right)
 & = &
 \int\limits_0^\infty da_3
 \int\limits_0^\infty da_4
 \int\limits_0^\infty da_5
 \int\limits_0^\infty da_1
 \frac{1}{{\mathcal U}_{1} {\mathcal F}_{1}}
 \nonumber \\
 {\mathcal U}_{1}
 & = &
 (a_1+a_4)(a_3+a_5) + a_1+a_3+a_4+a_5,
 \nonumber \\
 {\mathcal F}_{1}
 & = &
 (a_1+a_5)(a_3+a_4) +a_1a_4(a_3+a_5) +a_3a_5(a_1+a_4).
\eq
Partial fractioning with respect to $a_1$ and integration in $a_1$ yields
\bq
\lefteqn{
 \int\limits_0^\infty da_1
 \frac{1}{{\mathcal U}_{1} {\mathcal F}_{1}}
 = } & &
 \nonumber \\
 & &
 \frac{1}{\left[a_3+a_4+\left(a_3+a_5\right)a_4\right]^2}
 \left[ \ln\left(a_3+a_4+a_5+a_3a_4+a_4a_5\right) + \ln\left(a_3+a_4+a_3a_4+a_3a_5+a_4a_5\right)
 \right. \nonumber \\
 & & \left.
 - \ln\left(1+a_3+a_5\right) - \ln\left(a_3+a_4+a_3a_4\right) - \ln\left(a_5\right) \right].
\eq
We then continue with the integration in $a_5$, followed by the integration in $a_4$ and the final integration
in $a_3$.
The final result is
\bq
 I_{11111}\left(4,1\right)
 & = &
 6 \zeta_3.
\eq
It might seem that the integration algorithm for linearly reducible Feynman integrals applies only
to a very narrow set of Feynman integrals (the ones satisfying the two conditions mentioned at the beginning of
this section).
However, this set is larger than one might naively expect.
Let's focus on the first non-trivial integration (in the example above this corresponds to the integration in $a_1$).
Consider for $j \neq 2$ the partial fraction decomposition in $a_j$
of $1/({\mathcal U}(G) {\mathcal F}_0(G))$
(the fact that we set one Feynman parameter to one does not affect the argument).
For ${\mathcal U}(G)$ and ${\mathcal F}_0(G)$ we may use the recursion formulae from eq.~(\ref{chapter_graph_polynomials:recursion_U_F0_summary}).
We thus have
\bq
\lefteqn{
 \frac{1}{\left({\mathcal U}\left(G/e_j\right) + {\mathcal U}\left(G-e_j\right) a_j \right)\left({\mathcal F}_0\left(G/e_j\right) + {\mathcal F}_0\left(G-e_j\right) a_j\right)}
 = } & &
 \nonumber \\
 & &
 \frac{1}{{\mathcal U}\left(G-e_j\right) {\mathcal F}_0\left(G/e_j\right) - {\mathcal U}\left(G/e_j\right) {\mathcal F}_0\left(G-e_j\right)} 
 \left[ \frac{{\mathcal U}\left(G-e_j\right)}{{\mathcal U}\left(G/e_j\right) + {\mathcal U}\left(G-e_j\right) a_j} 
 \right. \nonumber \\
 & & \left.
 - \frac{{\mathcal F}_0\left(G-e_j\right)}{{\mathcal F}_0\left(G/e_j\right) + {\mathcal F}_0\left(G-e_j\right) a_j} \right].
\eq
${\mathcal U}\left(G/e_j\right)$, ${\mathcal U}\left(G-e_j\right)$, ${\mathcal F}_0\left(G/e_j\right)$ and ${\mathcal F}_0\left(G-e_j\right)$ are linear in the remaining Feynman parameters.
We then expect that 
\bq
 {\mathcal U}\left(G-e_j\right) {\mathcal F}_0\left(G/e_j\right) - {\mathcal U}\left(G/e_j\right) {\mathcal F}_0\left(G-e_j\right)
\eq
is quadratic in the remaining Feynman parameters.
However, in the example above we saw that this combination factorises as
\bq
 \left[a_3+a_4+\left(a_3+a_5\right)a_4\right]^2
\eq
and each factor
is again linear in each of the remaining integration variables.
This is no accident: 
Consider the graph $\tilde{G}$ obtained from $G$ by closing the two external edges.
\begin{figure}
\begin{center}
\includegraphics[scale=1.0]{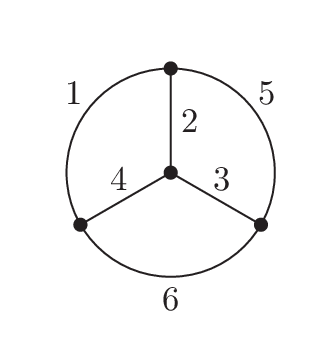}
\end{center}
\caption{\label{chapter_multiple_polylogarithms:fig4} 
The graph $\tilde{G}$ obtained from the two-loop two-point function by closing the two external edges. 
}
\end{figure}    
This gives a three-loop vacuum graph with six edges, as shown in fig.~\ref{chapter_multiple_polylogarithms:fig4}.
It is not too difficult to show that for $\mu^2=-p^2$ and vanishing internal masses
\bq
 {\mathcal U}\left(\tilde{G}\right)
 & = &
 {\mathcal F}_0\left(G\right) + a_6 {\mathcal U}\left(G\right).
\eq
Hence 
\bq
 {\mathcal U}\left(G\right) \; = \; {\mathcal U}\left(\tilde{G}-e_6\right),
 & &
 {\mathcal F}_0\left(G\right) \; = \; {\mathcal U}\left(\tilde{G}/e_6\right)
\eq
and
\bq
\lefteqn{
 {\mathcal U}\left(G-e_j\right) {\mathcal F}_0\left(G/e_j\right) - {\mathcal U}\left(G/e_j\right) {\mathcal F}_0\left(G-e_j\right)
 = } & & 
 \nonumber \\
 & &
 {\mathcal U}\left(\tilde{G}-e_j-e_6\right) {\mathcal U}\left(\tilde{G}/e_j/e_6\right) 
 - {\mathcal U}\left(\tilde{G}/e_j-e_6\right) {\mathcal U}\left(\tilde{G}/e_6-e_j\right).
\eq
For $j \neq 2$ the edges $e_j$ and $e_6$ share one common vertex and the 
factorisation of the expression follows then from Dodgson's identity eq.~(\ref{chapter_graph_polynomials:factorisation_U_U}):
\bq
 {\mathcal U}\left(G-e_j\right) {\mathcal F}_0\left(G/e_j\right) - {\mathcal U}\left(G/e_j\right) {\mathcal F}_0\left(G-e_j\right)
 & = &
 \left( \frac{\Delta_1}{a_j a_6} \right)^2,
\eq
where $\Delta_1$ is defined in eq.~(\ref{chapter_graph_polynomials:def_Delta_1}).

The computer program {\tt HyperInt} implements the algorithm for linearly reducible Feynman integrals \cite{Panzer:2014caa}.
Extensions of the algorithm are discussed in \cite{Hidding:2017jkk,Bourjaily:2018aeq,Bourjaily:2021lnz}.

%% file: nested_sums.tex
\newpage
\chapter{Nested sums}
\label{chapter_nested_sums}

In chapters~\ref{chapter_iterated_integrals} and \ref{chapter_transformations} we developed 
the method of differential equations for the computation of Feynman integrals.
An essential part of this method was the reduction to master integrals.
Only the master integrals need to be calculated, all other Feynman integrals may expressed
as a linear combination of the master integrals.

However, the reduction to master integrals through integration-by-parts identities
is very often the most CPU-time consuming part of an calculation and we are interested in alternatives.
Turning the argument around that any Feynman integral from a family of Feynman integrals can be written
as a linear combination of master integrals, shows that all Feynman integrals in this family
will involve the same final functions (namely the union of all functions appearing in the master integrals).
We now look for algorithms for the computation of Feynman integrals, which can be applied
to all members of a family of Feynman integrals and give the result directly, bypassing the need
for a reduction to master integrals.

Furthermore, there are Feynman integrals which do not depend on any kinematic variable (e.g. $\NB=0$).
A non-trivial example is given by the massless two-loop two-point function discussed as an example 
in section~\ref{chapter_multiple_polylogarithms:linear_reducibility}.
These Feynman integrals cannot be treated directly with the method of differential equations.

Often it is possible to express a Feynman integral with arbitrary values of the space-time dimension $D$ and
the powers of the propagators $\nu_1,\dots,\nu_{\ninternal}$ in terms of generalisations of hypergeometric functions.
In order to arrive at such a representation one may use the Mellin-Barnes representation, close the integration contour 
and sum up the residues as discussed in section~\ref{chapter_basics:Mellin_Barnes}.
In this way one obtains the sum representation of a transcendental function.

For a particular member of the family of Feynman integrals we then specialise 
the indices $\nu_1,\dots,\nu_{\ninternal}$ to the desired integers and $D$ to $\Dint-2\eps$.
We then have to compute the Laurent expansion in $\eps$.
We will discuss algorithms for this task in section~\ref{chapter_nested_sums:expansion_transcendental_functions}.

There are some well-known generalisations of hypergeometric functions:
Appell functions, Lauricella functions, Horn functions and ${\mathcal A}$-hypergeometric functions
(also known as GKZ hypergeometric functions), etc..
We collect useful information on the first three classes of functions in appendix~\ref{appendix_trancendental}.
In section~\ref{chapter_nested_sums:GKZ_hypergeometric_functions}
we discuss ${\mathcal A}$-hypergeometric functions. 
${\mathcal A}$-hypergeometric functions are defined as solutions of a system of partial differential
equations, known as a GKZ system.
We may view any Feynman integral as a special case of an ${\mathcal A}$-hypergeometric function.
At the same time, a GKZ system is holonomic, i.e. 
the solution space is finite dimensional.

\section{Expansion of special transcendental functions}
\label{chapter_nested_sums:expansion_transcendental_functions}

Let us start from two examples to motivate the content of this section.
The first example is the one-loop triangle graph shown in fig.~\ref{chapter_basics:fig_onelooptriangle}.
We are now interested in the case where all internal masses are zero and for the kinematic configuration
$p_2^2=0$, but $p_1^2 \neq 0$ and $p_3^2 \neq 0$.
The integral depends then on one kinematic variable, which we take as $x=p_1^2/p_3^2$.
From the Feynman parameter representation we obtain with $\mu^2=-p_3^2$,
$\nu_{ij}=\nu_i+\nu_j$ and $\nu_{ijk}=\nu_i+\nu_j+\nu_k$
\bq
\label{chapter_nested_sums:example_1_transcendental_function}
I_{\nu_1\nu_2\nu_3}
 & = &
 e^{\eps\Eulerconstant}
 \int \frac{d^{D}k}{i \pi^{\frac{D}{2}}}
 \frac{1}{(-q_1^2)^{\nu_1}}
 \frac{1}{(-q_2^2)^{\nu_2}}
 \frac{1}{(-q_3^2)^{\nu_3}}
 \nonumber \\
 & = & 
 e^{\eps\Eulerconstant}
 \frac{\Gamma(\nu_{123}-\frac{D}{2})}{\Gamma(\nu_1)\Gamma(\nu_2)\Gamma(\nu_3)}
 \int\limits_0^1 da \; a^{\nu_2-1} (1-a)^{\nu_3-1}
 \nonumber \\
 & & 
 \times
 \int\limits_0^1 db \; b^{\frac{D}{2}-\nu_{23}-1} (1-b)^{\frac{D}{2}-\nu_1-1}
 \left[ 1- a \left(1-x\right) \right]^{\frac{D}{2}-\nu_{123}}
 \nonumber \\
 & = &
 e^{\eps\Eulerconstant}
 \frac{\Gamma(\frac{D}{2}-\nu_1)\Gamma(\frac{D}{2}-\nu_{23})}{\Gamma(\nu_1)\Gamma(\nu_2)\Gamma(D-\nu_{123})}
 \sum\limits_{n=0}^\infty
 \frac{\Gamma(n+\nu_2)\Gamma(n-\frac{D}{2}+\nu_{123})}
      {\Gamma(n+1)\Gamma(n+\nu_{23})}
 \left(1-x\right)^n.
\eq
As a second example we consider the two-loop two-point graph already discussed 
in section~\ref{chapter_multiple_polylogarithms:linear_reducibility} and shown in fig.~\ref{chapter_graph_polynomials:fig3}
with vanishing internal masses and $\mu^2=-p^2$.
Using the Mellin-Barnes representation one arrives at the following representation:
To present the result after all residues have been taken in a compact form, we introduce two functions $F_\pm$ with
ten arguments each:
\bq
\lefteqn{
 F_\pm(a_1,a_2,a_3,a_4;b_1,b_2,b_3;c_1,c_2,c_3) 
 =
 \sum\limits_{n=0}^\infty 
 \sum\limits_{j=0}^\infty
  \frac{(-1)^{n+j}}{n! j!}
 }
 \\
 & & 
  \frac{\Gamma(\mp n - j-a_1) \Gamma( \pm n + j+a_2) \Gamma( \pm n + j+a_3)}{\Gamma( \pm n + j+a_4)}
  \frac{\Gamma(\mp n \mp b_1) \Gamma(n+b_2)}{\Gamma(\mp n \mp b_3)}
  \frac{\Gamma(-j-c_1) \Gamma(j+c_2)}{\Gamma(-j-c_3)},
 \nonumber
\eq
together with two operators ${\mathcal L}_{d}$ and ${\mathcal R}_{d}$ acting on the arguments as follows:
\bq
\lefteqn{
{\mathcal L}_{d} F_\pm(a_1,a_2,a_3,a_4;b_1,b_2,b_3;c_1,c_2,c_3)
 = }
 \nonumber \\
 & &
 F_\pm(a_1+d,a_2+d,a_3+d,a_4+d;b_1+2 d,b_2+d,b_3+d;c_1,c_2,c_3),
 \nonumber \\
\lefteqn{
{\mathcal R}_{d} F_\pm(a_1,a_2,a_3,a_4;b_1,b_2,b_3;c_1,c_2,c_3)
 = }
 \nonumber \\
 & &
 F_\pm(a_1+d,a_2+d,a_3+d,a_4+d;b_1,b_2,b_3;c_1+2 d,c_2+d ,c_3+d).
\eq
Then \cite{Bierenbaum:2003ud}
\bq
\lefteqn{
 I_{\nu_1\nu_2\nu_3\nu_4\nu_5}
 = c
 \left( {\bf 1} + {\mathcal L}_{\frac{D}{2}-\nu_{23}} + {\mathcal R}_{\frac{D}{2}-\nu_{25}} 
                + {\mathcal L}_{\frac{D}{2}-\nu_{23}} {\mathcal R}_{\frac{D}{2}-\nu_{25}} 
 \right)
}
 \\
 & &
 F_+\left(\frac{D}{2}-\nu_{14}, \nu_{235}-\frac{D}{2}, \nu_2, D-\nu_{14};
          \nu_{23}-\frac{D}{2}, \frac{D}{2}-\nu_4, -\nu_4;
          \nu_{25}-\frac{D}{2}, \frac{D}{2}-\nu_1, -\nu_1
    \right)
 \nonumber \\
 & &
 + c
 \left( {\bf 1} + {\mathcal R}_{\frac{D}{2}-\nu_{25}} 
 \right)
 \nonumber \\
 & &
 F_-\left( -\nu_1, \frac{D}{2}-\nu_{14}, D-\nu_{1234}, \frac{D}{2}-\nu_1;
          \nu_{12345}-D, \nu_{124}-\frac{D}{2}, \frac{D}{2};
          \nu_{25}-\frac{D}{2}, \frac{D}{2}-\nu_1, -\nu_1
    \right).
 \nonumber 
\eq
Here, ${\bf 1}$ denotes the identity operator with a trivial action on the arguments of the functions $F_\pm$
and the prefactor $c$ is given by
\bq
c & = &  
 \frac{e^{2\eps\Eulerconstant}}{\Gamma(\nu_2)\Gamma(\nu_3)\Gamma(\nu_5)\Gamma(D-\nu_{235})}.
\eq
In both example we obtained for arbitrary indices $\nu_1, \nu_2, \dots$ (multiple) sums.
The task is then to expand all terms in the dimensional regularisation parameter $\eps$ and to re-express
the resulting multiple sums in terms of known functions.
If this can be done we bypass integration-by-parts reduction.
It depends on the form of the multiple sums if this can be done systematically.
The following types of multiple sums occur often and can be evaluated systematically if
all $a_n$, $a_n'$, $b_n$, $b_n'$, $c_n$  and $c_n'$ are 
of the form 
\bq
\label{chapter_nested_sums:condition_arg_Gamma}
 p+q\eps & & \mbox{with} \; p\in{\mathbb Z} \; \mbox{and} \; q\in{\mathbb C}
\eq 
(the typical case is $q\in{\mathbb Z}$ as well) \cite{Moch:2001zr}:
\\
\\
{Type A:}
\bq
\label{chapter_nested_sums:type_A}
     \sum\limits_{i=0}^\infty 
       \frac{\Gamma(i+a_1)}{\Gamma(i+a_1')} ...
       \frac{\Gamma(i+a_k)}{\Gamma(i+a_k')}
       \; x^i
\eq
Up to prefactors the hypergeometric functions $\hypergeometric{J+1}{J}$ fall into this class.
\\
\\
{Type B:}
\bq
\label{chapter_nested_sums:type_B}
     \sum\limits_{i=0}^\infty 
     \sum\limits_{j=0}^\infty 
       \frac{\Gamma(i+a_1)}{\Gamma(i+a_1')} ...
       \frac{\Gamma(i+a_k)}{\Gamma(i+a_k')}
       \frac{\Gamma(j+b_1)}{\Gamma(j+b_1')} ...
       \frac{\Gamma(j+b_l)}{\Gamma(j+b_l')}
       \frac{\Gamma(i+j+c_1)}{\Gamma(i+j+c_1')} ...
       \frac{\Gamma(i+j+c_m)}{\Gamma(i+j+c_m')}
       \; x^i y^j
\eq
An example for a function of this type is given by the first Appell function $F_1$.
\\
\\
{Type C:}
\bq
\label{chapter_nested_sums:type_C}
     \sum\limits_{i=0}^\infty 
     \sum\limits_{j=0}^\infty 
       \left( \begin{array}{c} i+j \\ j \\ \end{array} \right)
       \frac{\Gamma(i+a_1)}{\Gamma(i+a_1')} ...
       \frac{\Gamma(i+a_k)}{\Gamma(i+a_k')}
       \frac{\Gamma(i+j+c_1)}{\Gamma(i+j+c_1')} ...
       \frac{\Gamma(i+j+c_m)}{\Gamma(i+j+c_m')}
       \; x^i y^j
\eq
Here, an example is given by the Kamp\'e de F\'eriet function $S_1$.
\\
\\
{Type D:}
\bq
\label{chapter_nested_sums:type_D}
     \sum\limits_{i=0}^\infty 
     \sum\limits_{j=0}^\infty 
       \left( \begin{array}{c} i+j \\ j \\ \end{array} \right)
       \frac{\Gamma(i+a_1)}{\Gamma(i+a_1')} ...
       \frac{\Gamma(i+a_k)}{\Gamma(i+a_k')}
       \frac{\Gamma(j+b_1)}{\Gamma(j+b_1')} ...
       \frac{\Gamma(j+b_l)}{\Gamma(j+b_l')}
       \frac{\Gamma(i+j+c_1)}{\Gamma(i+j+c_1')} ...
       \frac{\Gamma(i+j+c_m)}{\Gamma(i+j+c_m')}
       \; x^i y^j
\nonumber \\
\eq
An example for a function of this type is the second Appell function $F_2$.

Note that in these examples there are always as many gamma functions in the numerator
as in the denominator.
The task is now to expand these functions systematically into a Laurent series
in $\eps$.
We start with the formula for the expansion of the gamma function $\Gamma(n+\eps)$ with $n\in{\mathbb N}$:
\bq
\label{chapter_nested_sums:expansiongamma}
\lefteqn{
\Gamma(n+\eps)  = 
} & & \\
 & & \Gamma(1+\eps) \Gamma(n)
 \left[
        1 + \eps Z_1(n-1) + \eps^2 Z_{11}(n-1)
          + \eps^3 Z_{111}(n-1) + \dots + \eps^{n-1} Z_{11 \dots 1}(n-1)
 \right],
 \nonumber
\eq
where $Z_{m_1 \dots m_k}(n)$ denotes a
\index{Euler-Zagier sum}
{\bf Euler-Zagier sum}
defined by
\bq
 Z_{m_1 \dots m_k}(n) & = &
  \sum\limits_{n \ge i_1>i_2>\ldots>i_k>0}
     \frac{1}{{i_1}^{m_1}} \ldots \frac{1}{{i_k}^{m_k}}.
\eq
This motivates the following definition of a special form of nested sums, called 
\index{$Z$-sum}
{\bf $\bm{Z}$-sums}:
\bq 
\label{chapter_nested_sums:definition_Zsum}
 Z_{m_1 \dots m_k}(x_1,\dots,x_k;n) 
 & = & 
 \sum\limits_{n\ge i_1>i_2>\ldots>i_k>0}
 \frac{x_1^{i_1}}{{i_1}^{m_1}} \ldots \frac{x_k^{i_k}}{{i_k}^{m_k}}.
\eq
$k$ is called the 
\index{depth}
{\bf depth} 
of the $Z$-sum and $w=m_1+\dots+m_k$ is called the 
\index{weight}
{\bf weight}.
If the sums go to infinity ($n=\infty$) the $Z$-sums are multiple polylogarithms:
\bq
\label{chapter_nested_sums:multipolylog}
 Z_{m_1 \dots m_k}(x_1,\dots,x_k;\infty) 
 & = & 
 \mathrm{Li}_{m_1 \dots m_k}(x_1,\dots,x_k).
\eq
For $x_1=\dots=x_k=1$ the definition reduces to the Euler-Zagier sums:
\bq
 Z_{m_1 \dots m_k}(1,\dots,1;n) 
 & = &
 Z_{m_1 \dots m_k}(n).
\eq
For $n=\infty$ and $x_1=\dots=x_k=1$ the sum is a multiple $\zeta$-value:
\bq
 Z_{m_1 \dots m_k}(1,\dots,1;\infty) 
 & = &
 \zeta_{m_1 \dots m_k}.
\eq
The usefulness of the $Z$-sums lies in the fact, that they interpolate between
multiple polylogarithms and Euler-Zagier sums.
For fixed $n$ the $Z$-sums form a quasi-shuffle algebra in the same way as multiple polylogarithms do.
The letters are pairs $l_j=(m_j,x_j)$. On the alphabet of letter we have an additional operation
``$\circ$''
defined by
\bq
 \left(m_1,x_1\right)
 \circ
 \left(m_2,x_2\right)
 & = &
 \left( m_1+m_2; x_1 x_2 \right).
\eq
This is exactly the same operation as in eq.~(\ref{chapter_multiple_polylogarithms:def_additional_operation}).
On the vector space of words ${\mathcal A}_q$ we have a map
\bq
 Z_n & : & {\mathcal A}_q \; \rightarrow \; {\mathbb C},
\eq
which sends the word $w=l_1 \dots l_r$ with $l_j=(m_j,x_j)$ to
$Z_{m_1 \dots m_r}(x_1,\dots,x_r,n)$.
This map is
an algebra homomorphism, i.e.
\bq
\label{chapter_nested_sums:quasi_shuffle_product_Zsums}
 Z_n\left( w_1 \shuffle_q w_2 \right)
 & = &
 Z_n\left(w_1\right) \cdot Z_n\left(w_2\right),
\eq
where $\shuffle_q$ denotes the quasi-shuffle product introduced in section~\ref{chapter_multiple_polylogarithms:section:quasi_shuffle_product}.
Thus we have for example
\bq
 Z_{m_1}(x_1;n) Z_{m_2}(x_2;n)
 & = & 
 Z_{m_1 m_2}(x_1,x_2;n) + Z_{m_2 m_1}(x_2,x_1;n) + Z_{m_1+m_2}(x_1 x_2;n).
\eq
\bs
{\it \refstepcounter{exercise}
\label{chapter_nested_sums:proof_Li_zero_index}
{\bf Exercise \theexercise}: 
Prove eq.~(\ref{chapter_multiple_polylogarithms:Li_zero_index})
from chapter~\ref{chapter_multiple_polylogarithms}.
}
\es
\\
\\
In addition to $Z$-sums, it is sometimes useful to introduce as well $S$-sums.
A
\index{$S$-sum}
{\bf $\bm{S}$-sum} is defined by
\bq
 S_{m_1 \dots m_k}(x_1,\dots,x_k;n) 
 & = & 
 \sum\limits_{n\ge i_1 \ge i_2\ge \ldots\ge i_k \ge 1}
 \frac{x_1^{i_1}}{{i_1}^{m_1}}\ldots \frac{x_k^{i_k}}{{i_k}^{m_k}}.
\eq
The $S$-sums reduce for $x_1=\dots=x_k=1$ to 
\index{harmonic sum}
{\bf harmonic sums} \cite{Vermaseren:1998uu}:
\bq
 S_{m_1 \dots m_k}(1,\dots,1;n) 
 & = & S_{m_1 \dots m_k}(n).
\eq
The $S$-sums are closely related to the $Z$-sums, the difference being the upper summation boundary
for the nested sums: $(i-1)$ for $Z$-sums, $i$ for $S$-sums.
The introduction of $S$-sums is redundant, since $S$-sums can be expressed in terms
of $Z$-sums and vice versa.
It is however convenient to introduce both $Z$-sums and $S$-sums, since some 
properties are more naturally expressed in terms of
$Z$-sums while others are more naturally expressed in terms of $S$-sums.
An algorithm for the conversion from $Z$-sums to $S$-sums and vice versa can
be found in \cite{Moch:2001zr}.

The quasi-shuffle product of eq.~(\ref{chapter_nested_sums:quasi_shuffle_product_Zsums})
is the essential ingredient to expand functions of type A as in eq.~(\ref{chapter_nested_sums:type_A}) 
in a small parameter.
As a simple example let us consider the function
\bq
\label{chapter_nested_sums:examplehypergeom}
 \sum\limits_{i=0}^\infty
 \frac{\Gamma(i+a_1+t_1\eps)\Gamma(i+a_2+t_2\eps)}{\Gamma(i+1)\Gamma(i+a_3+t_3\eps)}
 x^i.
\eq
Here $a_1$, $a_2$ and $a_3$ are assumed to be integers.
Up to prefactors the expression in eq.~(\ref{chapter_nested_sums:examplehypergeom})
is a hypergeometric function $\hypergeometric{2}{1}$.
We are interested in the Laurent expansion of the function above in the small parameter
$\eps$.

Using $\Gamma(x+1) = x \Gamma(x)$, partial fractioning and an adjustment of the
summation index one can transform eq.~(\ref{chapter_nested_sums:examplehypergeom}) into terms of the form
\bq
\sum\limits_{i=1}^\infty
 \frac{\Gamma(i+t_1\eps)\Gamma(i+t_2\eps)}{\Gamma(i)\Gamma(i+t_3\eps)}
 \frac{x^i}{i^m},
\eq
where $m$ is an integer.
Now using eq.~(\ref{chapter_nested_sums:expansiongamma})
one obtains
\bq
\Gamma(1+\eps) 
\sum\limits_{i=1}^\infty
 \frac{\left(1+\eps t_1 Z_1(i-1)+\dots\right) \left(1+\eps t_2 Z_1(i-1)+\dots\right)}
      {\left(1+\eps t_3 Z_1(i-1)+\dots\right)}
 \frac{x^i}{i^m}.
\eq
Inverting the power series in the denominator and truncating in $\eps$ one obtains
in each order in $\eps$ terms of the form
\bq
\label{chapter_nested_sums:exZ1}
\sum\limits_{i=1}^\infty
 \frac{x^i}{i^{m_0}}
 \; Z_{m_1 \dots m_k}(i-1) \; Z_{m_1' \dots m_l'}(i-1) \; Z_{m_1'' \dots m_n''}(i-1).
\eq
Using the quasi-shuffle product for $Z$-sums the three Euler-Zagier sums
can be reduced to single Euler-Zagier sums and one finally arrives at terms of the form
\bq
\label{chapter_nested_sums:exZ2}
\sum\limits_{i=1}^\infty
 \frac{x^i}{i^{m_0}}
 Z_{m_1 \dots m_k}(i-1),
\eq
which are special cases of multiple polylogarithms, called harmonic polylogarithms $H_{m_0 m_1 \dots m_k}(x)$.
This completes the algorithm for the expansion in $\eps$ for sums of the form as in eq.~(\ref{chapter_nested_sums:examplehypergeom}), and more generally any sum of type A as in eq.~(\ref{chapter_nested_sums:type_A}).

Let us now consider expressions of the form
\bq
\label{chapter_nested_sums:augmented}
\frac{x_0^n}{n^{m_0}} Z_{m_1 \dots m_k}(x_1,\dots,x_k;n),
\eq
e.g. $Z$-sums multiplied by a letter.
Then the following convolution product
\bq
\label{chapter_nested_sums:convolution}
 \sum\limits_{i=1}^{n-1} \; \frac{x^i}{i^{m_0}} Z_{m_1 \dots}(x_1,\dots;i-1)
                         \; \frac{y^{n-i}}{(n-i)^{m_0'}} Z_{m_1' \dots}(x_1',\dots;n-i-1)
\eq
can again be expressed by partial fractioning and relabellings of summation indices
in terms of expressions of the form (\ref{chapter_nested_sums:augmented}).
An example is
\bq
\lefteqn{
 \sum\limits_{i=1}^{n-1} \; \frac{x^i}{i} Z_1(i-1)
                         \; \frac{y^{n-i}}{(n-i)} Z_1(n-i-1)
 = }
 \\
 & &
 \frac{x^n}{n} \left[ 
    Z_{1 1 1}\left(\frac{y}{x},\frac{x}{y},\frac{y}{x};n-1\right)
   +Z_{1 1 1}\left(\frac{y}{x},1,\frac{x}{y};n-1\right)
   +Z_{1 1 1}\left(1,\frac{y}{x},1;n-1\right)
 \right]
 + \left( x \leftrightarrow y \right).
 \nonumber
\eq
Combing this algorithm with the previous algorithms allows to expand any sum of type B as 
in eq.~(\ref{chapter_nested_sums:type_B}).

In addition there is for terms of the form as in eq.~(\ref{chapter_nested_sums:augmented}) 
a conjugation, e.g. sums of the form 
\bq
\label{chapter_nested_sums:conjugation}
 - \sum\limits_{i=1}^n 
       \left( \begin{array}{c} n \\ i \\ \end{array} \right)
       \left( -1 \right)^i
       \; \frac{x^i}{i^{m_0}} S_{m_1 \dots}(x_1,\dots;i)
\eq
can also be reduced to terms of the form (\ref{chapter_nested_sums:augmented}).
Although one can easily convert between the notations for $S$-sums and
$Z$-sums, expressions involving a conjugation tend to be shorter when
expressed in terms of $S$-sums.
The name conjugation stems from the following fact:
To any function $f(n)$ of an integer variable $n$ one can define
a conjugated function $C \ast f(n)$ as the following sum
\bq
C \ast f(n) & = & \sum\limits_{i=1}^n 
       \left( \begin{array}{c} n \\ i \\ \end{array} \right)
       (-1)^i f(i).
\eq
Then conjugation satisfies the following two properties:
\bq
C \ast 1 & = & 1,
 \nonumber \\
C \ast C \ast f(n) & = & f(n).
\eq
An example for a sum involving a conjugation is
\bq
 - \sum\limits_{i=1}^n 
       \left( \begin{array}{c} n \\ i \\ \end{array} \right)
       \left( -1 \right)^i
       \; \frac{x^i}{i} S_1(i)
 & = &
 S_{1 1}\left(1-x, \frac{1}{1-x};n\right)
 -S_{1 1}\left(1-x, 1;n\right).
\eq
Conjugation in combination with the previous algorithms allow us to expand all functions of type C as
in eq.~(\ref{chapter_nested_sums:type_C}).

Finally there is the combination of conjugation and convolution,
e.g. sums of the form 
\bq
\label{chapter_nested_sums:conjugationconvolution}
 - \sum\limits_{i=1}^{n-1} 
       \left( \begin{array}{c} n \\ i \\ \end{array} \right)
       \left( -1 \right)^i
       \; \frac{x^i}{i^{m_0}} S_{m_1 \dots}(x_1,\dots;i)
       \; \frac{y^{n-i}}{(n-i)^{m_0'}} S_{m_1' \dots}(x_1',\dots;n-i)
\eq
can also be reduced to terms of the form (\ref{chapter_nested_sums:augmented}).
An example is given by
\bq
\lefteqn{
 - \sum\limits_{i=1}^{n-1} 
       \left( \begin{array}{c} n \\ i \\ \end{array} \right)
       \left( -1 \right)^i
       \; S_1(x;i)
       \; S_1(y;n-i) =
} \nonumber \\
 & &
 \frac{1}{n} 
 \left\{
 S_1(y;n)
 + (1-x)^n 
   \left[ S_1\left(\frac{x}{x-1};n\right)
        - S_1\left(\frac{x-y}{x-1};n\right)
   \right]
 \right\}
 \nonumber \\
 & &
 + \frac{(-1)^n}{n}
 \left\{
 S_1(x;n)
 + (1-y)^n 
   \left[ S_1\left(\frac{y}{y-1};n\right)
        - S_1\left(\frac{y-x}{y-1};n\right)
   \right]
 \right\}.
 \nonumber \\
\eq
This allows us to expand functions of type D as
in eq.~(\ref{chapter_nested_sums:type_D}).
There are computer packages implementing the algorithms for the expansion of sums 
of type A-D \cite{Weinzierl:2002hv,Moch:2005uc,Maitre:2005uu,Huber:2005yg}.
\\
\\
\bs
{\it \refstepcounter{exercise}
{\bf Exercise \theexercise}: 
Consider $I_{111}$ from eq.~(\ref{chapter_nested_sums:example_1_transcendental_function})
with $\mu^2=-p_3^2$ and $x=p_1^2/p_3^2$ in $D=4-2\eps$ space-time dimensions:
\bq
I_{111} & = &
 e^{\eps\Eulerconstant}
 \frac{\Gamma(-\eps)\Gamma(1-\eps)}{\Gamma(1-2\eps)}
 \sum\limits_{n=0}^\infty
 \frac{\Gamma(n+1+\eps)}
      {\Gamma(n+2)}
 \left(1-x\right)^{n}.
\eq
Expand the sum in $\eps$ and give the first two terms of the $\eps$-expansion for the full expression.
}
\es
\\
\\
Up to now we assumed through eq.~(\ref{chapter_nested_sums:condition_arg_Gamma}) 
that the gamma functions are expanded around an integer value.
The extension to rational numbers is straightforward for sums of type A and B 
if the gamma functions always occur in ratios of the form
\bq
\label{chapter_nested_sums:rational_balanced}
 \frac{\Gamma(n+a-\frac{p}{q} +b \eps)}
      {\Gamma(n+c-\frac{p}{q} +d \eps)},
\eq
where the same rational number $p/q \in {\mathbb Q}$ 
occurs in the numerator and in the denominator \cite{Weinzierl:2004bn}.
The generalisation of eq.~(\ref{chapter_nested_sums:expansiongamma}) reads
\bq
\Gamma\left( n+1-\frac{p}{q}+\eps \right)
 & = &
\frac{\Gamma\left( 1-\frac{p}{q}+\eps\right)\Gamma\left( n+1-\frac{p}{q}\right)}{\Gamma\left( 1-\frac{p}{q} \right)}
 \\
 & & \times
\exp \left( - \frac{1}{q} \sum\limits_{l=0}^{q-1}
            \left( r_q^l \right)^p
            \sum\limits_{k=1}^\infty
             \eps^k \frac{(-q)^k}{k}
             Z_{k}(r_q^l; q \cdot n)
      \right)
 \nonumber
\eq
and introduces the $q$-th roots of unity
\bq
r_q^p & = & \exp \left( \frac{2 \pi i p}{q} \right).
\eq
With the help of the $q$-th roots of unity we may express any $Z$-sum $Z_{m_1 \dots}(x_1,\dots;n)$ as a combination of 
$Z$-sums $Z_{m_1' \dots}(x_1',\dots;q \cdot n)$, where the summation goes now up to $q \cdot n$.

\section{GKZ hypergeometric functions}
\label{chapter_nested_sums:GKZ_hypergeometric_functions}

In this section we show that a Feynman integral can be viewed as a special case of 
a Gelfand-Kapranov-Zelevinsky (GKZ) hypergeometric function.
GKZ hypergeometric functions are solutions to a system of partial differential equations,
called a GKZ system.
In order to present GKZ hypergeometric functions we may either use a geometric language
or an algebraic language.
In both cases we need some preparation.
In preparation for the geometric setting we introduce polytopes in section~\ref{chapter_nested_sums:polytopes}
and in preparation for the algebraic setting we introduce $D$-modules in section~\ref{chapter_nested_sums:D_modules}.
GKZ systems and GKZ hypergeometric functions are then introduced in section~\ref{chapter_nested_sums:GKZ_systems}.
The application towards Feynman integrals is discussed in section~\ref{chapter_nested_sums:Feynman_integrals_GKZ_hypergeometric}.
More information on polytopes can be found in the books by Zieger \cite{Ziegler:book} and by 
De Loera, Rambau and Santos \cite{DeLoera:book}.
Further information on $D$-modules can be found in the books by Bj\"ork \cite{Bjoerk:book},
by Coutinho \cite{Coutinho:book} and
by Saito, Sturmfels and Takayama \cite{Saito:book}.
An  introduction into the topic of GKZ hypergeometric structures can be found in the lecture notes by Stienstra
\cite{Stienstra:2005nr} and by Cattani \cite{Cattani:2006lectures}.
Additional background on GKZ hypergeometric functions can be 
found in the book by Gelfand, Kapranov and Zelevinsky \cite{Gelfand:book}.

In this section it is convenient to use 
\index{multi-index notation}
{\bf multi-index notation}:
For $x=(x_1,\dots,x_n)$ and $\alpha=(\alpha_1,\dots,\alpha_n)$
we set 
\bq
 x^\alpha & = & x_1^{\alpha_1} \dots x_n^{\alpha_n}.
\eq
We denote by $({\mathbb R}^n)^\ast$ the dual space of ${\mathbb R}^n$, 
i.e. the space of linear functionals $\varphi : {\mathbb R}^n \rightarrow {\mathbb R}$.
We may view $a \in {\mathbb R}^n$ as a column vector (or as a ket vector) and $b \in ({\mathbb R}^n)^\ast$ as a row vector (or as a bra vector). 
The row vector $b$ defines a linear functional by
\bq
 \varphi_b & : & a \rightarrow b \cdot a.
\eq

\subsection{Polytopes}
\label{chapter_nested_sums:polytopes}

Let $A=\{a_1,\dots,a_n\} \subset {\mathbb R}^d$ be a non-empty finite set of points in ${\mathbb R}^d$.
The convex hull of these points defines a 
\index{polytope}
{\bf polytope}, denoted by $\mathrm{conv}(A)$.
In a formula
\bq
\label{chapter_nested_sums:def_convex_hull}
 \mathrm{conv}(A)
 & = &
 \left\{
   \alpha_1 a_1 + \dots + \alpha_n a_n | \alpha_j \ge 0, \sum\limits_{j=1}^n \alpha_j = 1
 \right\}.
\eq
A 
\index{cone}
{\bf cone} is defined by
\bq
 \mathrm{cone}(A)
 & = &
 \left\{
   \alpha_1 a_1 + \dots + \alpha_n a_n | \alpha_j \ge 0
 \right\}.
\eq
We define the cone of an empty set to be the set containing the origin, i.e. $\mathrm{cone}(\{\})=\{0\}$.
Every cone contains the origin.
Given two sets $P,Q \subseteq {\mathbb R}^d$ their 
\index{Minkowski sum}
{\bf Minkowski sum}
is defined to be
\bq
 P + Q & = &
 \left\{ a + b | a \in P, b \in Q \right\}.
\eq
\begin{figure}
\begin{center}
\includegraphics[scale=1.0]{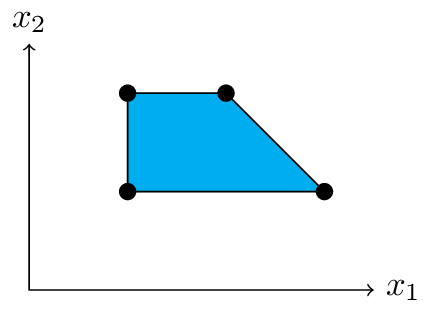}
\hspace*{10mm}
\includegraphics[scale=1.0]{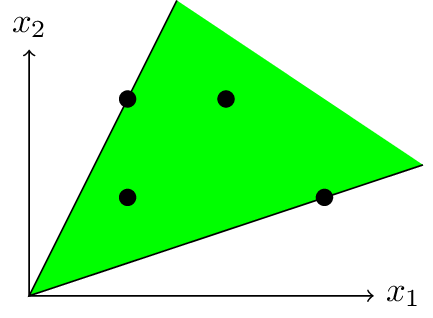}
\hspace*{10mm}
\includegraphics[scale=1.0]{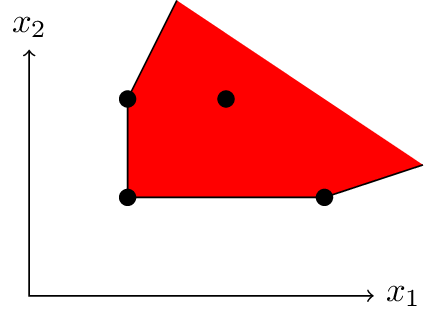}
\end{center}
\caption{
The left picture shows the polytope in ${\mathbb R}^2$ defined by the points
$a_1 = (1,1)^T$, 
$a_2 = (1,2)^T$, 
$a_3 = (2,2)^T$, 
$a_4 = (3,1)^T$.
The middle picture shows the cone defined by these points.
The right picture shows the Minkowski sum of the polytope and the cone.
}
\label{chapter_nested_sums:fig_example_polytope_cone}
\end{figure}
Fig.~\ref{chapter_nested_sums:fig_example_polytope_cone}
shows an example for a polytope and a cone. The figure shows also the Minkowski sum of the polytope and the cone.

A 
\index{polyhedron}
{\bf polyhedron} 
$P$ is the Minkowski sum of a polytope $\mathrm{conv}(A)$ and a cone $\mathrm{cone}(B)$:
\bq
 P & = &
 \mathrm{conv}(A) + \mathrm{cone}(B).
\eq
A hyperplane in ${\mathbb R}^d$ divides the space into two halfspaces.
Let $a \in {\mathbb R}^d$, $b \in ({\mathbb R}^d)^\ast$ and $c \in {\mathbb R}$.
A hyperplane is given by
\bq
 \left\{ \; x \in {\mathbb R}^d \; | \; b \cdot a = c \; \right\}.
\eq
The two closed halfspace are
\bq
 \left\{ \; x \in {\mathbb R}^d \; | \; b \cdot a \ge c \; \right\}
 & \mbox{and} &
 \left\{ \; x \in {\mathbb R}^d \; | \; b \cdot a \le c \; \right\}.
\eq
This can be used to give alternative definitions:
We may define a polyhedron as an intersection of finitely many closed halfspaces
and a polytope as a bounded polyhedron.
A cone is the intersection of finitely many closed halfspaces, where all hyperplanes defining the halfspaces
contain the origin.
The 
\index{lineality space of a cone}
{\bf lineality space of a cone}
is the largest linear subspace contained in the cone.
A cone is called
\index{pointed cone} 
{\bf pointed}, 
if its lineality space is $\{0\}$.
Let us illustrate the last two definitions by an example:
The cone in ${\mathbb R}^2$ generated by
\bq
 \left(\begin{array}{c} 1 \\ 0 \\ \end{array} \right),
 \left(\begin{array}{c} 0 \\ 1 \\ \end{array} \right)
\eq
is pointed. It corresponds to the first quadrant and $\{0\}$ is the largest linear subspace.
On the other hand, the cone in ${\mathbb R}^2$ generated by
\bq
 \left(\begin{array}{r} 1 \\ 0 \\ \end{array} \right),
 \left(\begin{array}{r} 0 \\ 1 \\ \end{array} \right),
 \left(\begin{array}{r} -1 \\ -1 \\ \end{array} \right)
\eq
is not pointed. Any point of ${\mathbb R}^2$ belongs to the cone and hence the largest linear subspace is ${\mathbb R}^2$.

Consider now a hyperplane such that all points of a polytope / cone / polyhedron lie in one closed halfspace
(this includes the points on the hypersurface).
A 
\index{face}
{\bf face} 
is the intersection of a polytope / cone / polyhedron with such a hyperplane.
Faces of dimension zero are called 
\index{vertex}
{\bf vertices}
and faces of dimension $\dim(P)-1$ are called 
\index{facet}
{\bf facets}.
A face $F$ with $\dim(F) < \dim(P)$ is called a proper face.

The dimension of a polytope $P=\mathrm{conv}(A)$ is the dimension of its convex hull.
Let $P$ be a $k$-dimensional polytope. We define the normalised volume $\mathrm{vol}_0(P)$ of $P$
as 
\bq
 \mathrm{vol}_0\left(P\right)
 & = &
 k! \mathrm{vol}\left(P\right),
\eq
where $\mathrm{vol}(P)$ denotes the standard (Euclidean) volume.
\\
\\
\bs
{\it \refstepcounter{exercise}
{\bf Exercise \theexercise}: 
Consider the $k$-dimensional standard simplex in ${\mathbb R}^{k+1}$. This is the polytope with vertices
given by the $(k+1)$ standard unit vectors $e_j \in {\mathbb R}^{k+1}$.
Show that the standard simplex has Euclidean volume $1/k!$ and therefore the normalised volume $1$.
}
\es
\\
\\
Let $\sigma$ be a $k$-dimensional simplex in ${\mathbb R}^{k+1}$, defined by $(k+1)$ points $A=\{a_1,\dots,a_{k+1}\} \in {\mathbb R}^{k+1}$.
By abuse of notation, we also denote by $A$ the $(k+1) \times (k+1)$-matrix, whose columns are given by $a_1, \dots, a_{k+1}$.
Then
\bq
 \mathrm{vol}_0\left(\sigma\right)
 & = &
 \left| \det A \right|.
\eq
A 
\index{triangulation}
{\bf triangulation} 
of a polytope $P=\mathrm{conv}(A) \subset {\mathbb R}^d$
is a set of simplices $\{\sigma_1,\dots,\sigma_r\}$ with vertices from the set $A=\{a_1, \dots, a_{n}\}$
such that all faces of a simplex are contained in this set,
the union of all simplices is the full polytope 
and the intersection of two distinct simplices is a proper face of the two, possibly empty.

A triangulation $\{\sigma_1,\dots,\sigma_r\}$ of a polytope $P=\mathrm{conv}(A) = \mathrm{conv}(a_1,\dots,a_n) \subset {\mathbb R}^d$
is called 
\index{regular triangulation}
{\bf regular},
if there exists a height vector $h \in {\mathbb R}^n$, such that for every simplex $\sigma_i$
of this triangulation there exists a vector $r_i \in {\mathbb R}^d$ satisfying
\bq
 r_i \cdot a_j & = & h_j 
 \;\;\;\;\;\;\;\;\; 
 a_j \; \in \; \sigma_i,
 \nonumber \\
 r_i \cdot a_j & < & h_j 
 \;\;\;\;\;\;\;\;\; 
 a_j \; \notin \; \sigma_i.
\eq
These equations say that if we consider points $\tilde{a}_1,\dots,\tilde{a}_n \in {\mathbb R}^{d+1}$
obtained from $a_1, \dots, a_n \in {\mathbb R}^d$ by adjoining the height as last coordinate
\bq
 \tilde{a}_j & = &
 \left( \begin{array}{c}
  a_{1 j} \\
  \vdots \\
  a_{d j} \\
  h_j \\
 \end{array} \right),
\eq 
the resulting geometrical object obtained from lifting the triangulation to ${\mathbb R}^{d+1}$ is convex.
As an example consider the six points in ${\mathbb R}^1$:
\bq
 a_1 \; = \; \left( 1 \right),
 \;\;\;
 a_2 \; = \; \left( 2 \right),
 \;\;\;
 a_3 \; = \; \left( 3 \right),
 \;\;\;
 a_4 \; = \; \left( 4 \right),
 \;\;\;
 a_5 \; = \; \left( 5 \right),
 \;\;\;
 a_6 \; = \; \left( 6 \right),
\eq
and the triangulation
\bq 
\label{chapter_nested_sums:example_triangulation}
 \left\{ 
  \sigma_{12}, \sigma_{23}, \sigma_{34}, \sigma_{45}, \sigma_{56}, 
  \sigma_1, \sigma_2, \sigma_3, \sigma_4, \sigma_5, \sigma_6
 \right\},
\eq
where $\sigma_{ij}$ denotes the $1$-dimensional simplex defined by $a_i$ and $a_j$, and
$\sigma_j$ denotes the $0$-dimensional simplex defined by $a_j$.
The triangulation in eq.~(\ref{chapter_nested_sums:example_triangulation}) is regular.
A height vector is given by
\bq
 h & = & \left(3,1,0,0,1,3\right)^T,
\eq
\begin{figure}
\begin{center}
\includegraphics[scale=1.0]{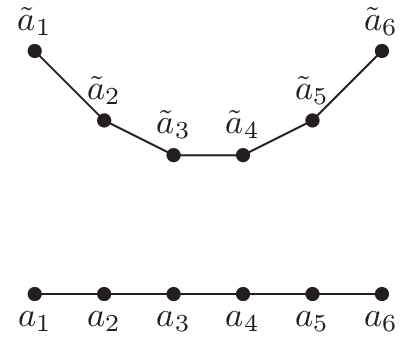}
\end{center}
\caption{
The triangulation of $(a_1,a_2,a_3,a_4,a_5,a_6)$ in ${\mathbb R}^1$ 
and the lift $(\tilde{a}_1,\tilde{a}_2,\tilde{a}_3,\tilde{a}_4,\tilde{a}_5,\tilde{a}_6)$ to ${\mathbb R}^2$.
The lift is convex.
}
\label{chapter_nested_sums:fig_example_triangulation_height_vector}
\end{figure}
see fig.~\ref{chapter_nested_sums:fig_example_triangulation_height_vector}.

As an example for a non-regular triangulations consider the six points
\bq
 A & = & 
 \left\{
  \left(\begin{array}{c} 4 \\ 0 \\ 0 \\ \end{array} \right), 
  \left(\begin{array}{c} 0 \\ 4 \\ 0 \\ \end{array} \right), 
  \left(\begin{array}{c} 0 \\ 0 \\ 4 \\ \end{array} \right), 
  \left(\begin{array}{c} 2 \\ 1 \\ 1 \\ \end{array} \right), 
  \left(\begin{array}{c} 1 \\ 2 \\ 1 \\ \end{array} \right), 
  \left(\begin{array}{c} 1 \\ 1 \\ 2 \\ \end{array} \right)
 \right\}
\eq
in ${\mathbb R}^3$. These points lie in a plane with normal vector $n=(1,1,1)$.
The polytope $P=\mathrm{conv}(A)$ is two-dimensional.
\begin{figure}
\begin{center}
\includegraphics[scale=1.0]{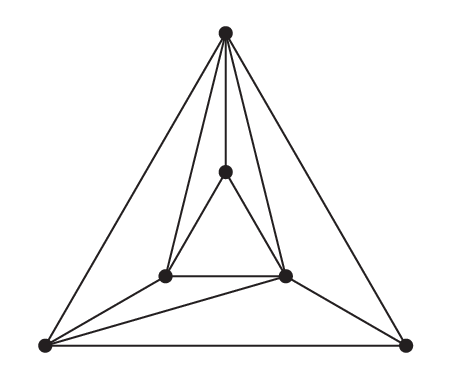}
\hspace*{10mm}
\includegraphics[scale=1.0]{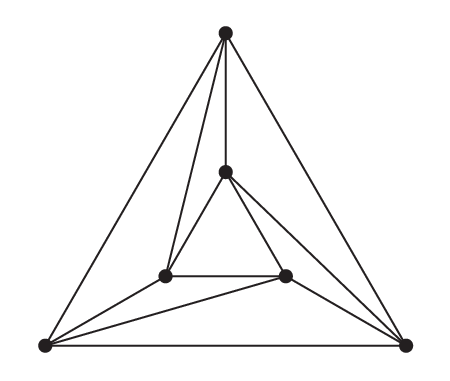}
\end{center}
\caption{
The left picture shows a regular triangulation of the polytope $P$,
the right picture shows a non-regular triangulation of the polytope $P$.
}
\label{chapter_nested_sums:fig_example_regular_triangulation}
\end{figure}
Fig.~\ref{chapter_nested_sums:fig_example_regular_triangulation} shows a regular triangulation of $P$ (left picture)
and a non-regular triangulation of $P$ (right picture).
The right picture of fig.~\ref{chapter_nested_sums:fig_example_regular_triangulation}
is the standard example for a non-regular triangulation.
\\
\\
\bs
{\it \refstepcounter{exercise}
{\bf Exercise \theexercise}: 
Show that the left picture of fig.~\ref{chapter_nested_sums:fig_example_regular_triangulation}
defines a regular triangulation.
}
\es
\\
\\
\bs
{\it \refstepcounter{exercise}
{\bf Exercise \theexercise}: 
Show that the right picture of fig.~\ref{chapter_nested_sums:fig_example_regular_triangulation}
defines a non-regular triangulation.
}
\es
\\
\\
A regular triangulation $\{\sigma_1,\dots,\sigma_r\}$ where all simplices have normalised volume one 
($\mathrm{vol}_0(\sigma_j)=1$) is called a
\index{unimodular triangulation}
{\bf unimodular triangulation}.

A 
\index{fan}
{\bf fan} 
in ${\mathbb R}^d$ is a finite set of cones
\bq
 {\mathcal F} & = & \left\{ C_1, C_2, \dots, C_r \right\}
\eq
such that
\begin{enumerate}
\item Every face of a cone in ${\mathcal F}$ is also a cone in ${\mathcal F}$.
\item The intersection of any two cones in ${\mathcal F}$ is a face of both
\end{enumerate}
The fan is called
\index{complete fan} 
{\bf complete}, if the union of all cones in ${\mathcal F}$ equals ${\mathbb R}^d$:
\bq
 C_1 \cup C_2 \cup \dots \cup C_r
 & = & {\mathbb R}^d.
\eq
A fan is called 
\index{rational fan}
{\bf rational}, if all the cones in the fan are given by inequalities with rational coefficients.

Consider now a polytope $P \in {\mathbb R}^d$ and let $F$ be a face of $P$.
The 
\index{normal cone}
{\bf normal cone} 
$N_F(P)$ of a face $F$ is a cone in the dual space $({\mathbb R}^d)^\ast$ defined by
\bq
 N_F(P)
 & = &
 \left\{
  \; b \in ({\mathbb R}^d)^\ast \; | \; F \subseteq \left\{ \; x \in P \; | \; b \cdot x = \max\limits_{y \in P}\left(b \cdot y\right) \right\}
 \right\}.
\eq
The collection of the normal cones $N_F(P)$ for the faces $F$ of a polytope $P$ forms a complete fan in $({\mathbb R}^d)^\ast$.
This fan is called the
\index{normal fan}
{\bf normal fan} of $P$ and denoted by $N(P)$.
\begin{figure}
\begin{center}
\includegraphics[scale=1.0]{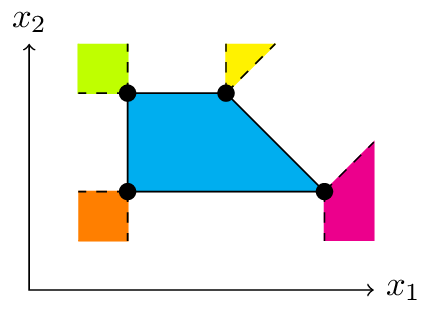}
\hspace*{20mm}
\includegraphics[scale=1.0]{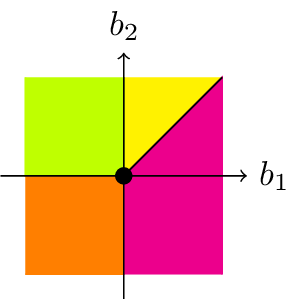}
\end{center}
\caption{
The left picture shows the polytope in ${\mathbb R}^2$ defined by the points
$a_1 = (1,1)^T$, 
$a_2 = (1,2)^T$, 
$a_3 = (2,2)^T$, 
$a_4 = (3,1)^T$.
The right picture shows the normal fan of the polytope.
}
\label{chapter_nested_sums:fig_example_normal_fan}
\end{figure}
An example is shown in fig.~\ref{chapter_nested_sums:fig_example_normal_fan}.

As before consider a non-empty set of $n$ points
$A=\{a_1,\dots,a_n\} \subset {\mathbb R}^d$.
The convex hull of these points defines a polytope $P=\mathrm{conv}(A)$, see eq.~(\ref{chapter_nested_sums:def_convex_hull}).
The polytope $P$ is also called the
\index{primary polytope}
{\bf primary polytope} of $A$.
For any regular triangulation $T=\{\sigma_1,\sigma_2,\dots\}$ of $P$ we define a point $q_T \in {\mathbb R}^n$ with coordinates $q_{1 T},\dots,q_{n T}$ by
\bq
 q_{j T} & = &
 \sum\limits_{\sigma \in T, a_j \in \mathrm{vertices}(\sigma)} \mathrm{vol}_0\left(\sigma\right).
\eq
Let $S$ be the set $\{q_{T_1}, q_{T_2}, \dots\} \subset {\mathbb R}^n$ obtained from all regular triangulations $T_i$ of $P$.
The 
\index{secondary polytope}
{\bf secondary polytope} 
$\Sigma(A)$ of $A$ is defined to be
\bq
 \Sigma(A)
 & = &
 \mathrm{conv}\left(S\right)
 \; \subset \; {\mathbb R}^n.
\eq
The normal fan of the secondary polytope $\Sigma(A)$ 
is called the 
\index{secondary fan}
{\bf secondary fan} 
$N(\Sigma(A))$.

Given a multivariate polynomial $f$ in $d$ variables $x_1, \dots, x_d$, written in multi-index notation 
(e.g. $x^{a_j} = x_1^{a_{j 1}} \dots x_d^{a_{j d}}$)
as
\bq
 f\left(x\right)
 & = & 
 \sum\limits_{j=1}^n c_{a_j} x^{a_j},
 \;\;\;\;\;\;
 c_{a_j} \; \neq \; 0,
\eq
the 
\index{Newton polytope}
{\bf Newton polytope} 
$\Delta_f$ of $f$ is defined by
\bq
 \Delta_f & = & 
 \mathrm{conv}\left(\left\{a_1,\dots,a_n\right\}\right).
\eq
We denote by ${\mathcal V}_f$ the hypersurface defined by $f$:
\bq
 {\mathcal V}_f
 & = &
 \left\{ \; x \in {\mathbb C}^d \; | \; f(x) = 0 \; \right\}.
\eq
The 
\index{amoeba}
{\bf amoeba} 
${\mathcal A}_f$ of ${\mathcal V}_f$ is defined as
\bq
 {\mathcal A}_f
 & = &
 \left\{ \; \left(\ln|x_1|,\dots,\ln|x_d|\right) \in {\mathbb R}^d \; | \; x \in  {\mathcal V}_f \; \right\},
\eq
the 
\index{coamoeba}
{\bf coamoeba} 
${\mathcal A}_f'$ of ${\mathcal V}_f$ is defined as
\bq
 {\mathcal A}_f'
 & = &
 \left\{ \; \left(\arg(x_1),\dots,\arg(x_d)\right) \in {\mathbb R}^d \; | \; x \in  {\mathcal V}_f \; \right\}.
\eq

\subsection{$D$-modules}
\label{chapter_nested_sums:D_modules}

Let ${\mathbb F}$ be a field of characteristic zero and $n\in{\mathbb N}$.
The ring of differential operators $\partial_1, \dots, \partial_n$ (where $\partial_j=\partial/\partial x_j$)
with coefficients in the polynomial ring ${\mathbb F}[x_1,\dots,x_n]$ 
is called the 
\index{Weyl algebra}
{\bf Weyl algebra} $A_n$ in $n$ variables.
The Weyl algebra is generated by
\bq
 x_1, \dots, x_n, \partial_1, \dots, \partial_n
\eq
subject to the commutation relation
\bq
\label{chapter_nested_sums:commutation_relation_Weyl_algebra}
 \left[ \partial_i, x_i \right] & = & 1,
 \;\;\;\;\;\;
 1 \le i \le n.
\eq
All other commutation relations are trivial.
With the multi-index notation
\bq
 x^\alpha \; = \; x_1^{\alpha_1} \dots x_n^{\alpha_n},
 & &
 \partial^{\alpha} \; = \; \partial_1^{\alpha_1} \dots \partial_n^{\alpha_n}
\eq
any $P \in A_n$ can be written uniquely as
\bq
\label{chapter_nested_sums:normal_ordered_differnetial_operator}
 P & = &
 \sum\limits_{\alpha,\beta} c_{\alpha \beta} \; x^\alpha \; \partial^\beta,
 \;\;\;\;\;\;
 c_{\alpha \beta} \; \in \; {\mathbb F}.
\eq
We call $P$ written in the form as in 
eq.~(\ref{chapter_nested_sums:normal_ordered_differnetial_operator}) a {\bf normal ordered expression}.
Let $\xi_1,\dots,\xi_n$ be commutative variables.
It is often useful to replace $\partial_j \rightarrow \xi_j$.
From eq.~(\ref{chapter_nested_sums:normal_ordered_differnetial_operator})
we see that if we view $A_n$ and ${\mathcal F}[x_1,\dots,x_n,\xi_1,\dots,\xi_n]$ 
as vector spaces over ${\mathbb F}$ there is a vector space isomorphism between
\bq
 P(x,\partial) =
 \sum\limits_{\alpha,\beta} c_{\alpha \beta} \; x^\alpha \; \partial^\beta
 & \mbox{and} &
 P(x,\xi) =
 \sum\limits_{\alpha,\beta} c_{\alpha \beta} \; x^\alpha \; \xi^\beta.
\eq
Note that this is not an algebra homomorphism.

Let $D$ be a ring of differential operators.
A 
\index{$D$-module}
{\bf ${\bm D}$-module} 
is a left module over the ring $D$.
If we take $D=A_n$, examples for $D$-modules are the polynomial ring ${\mathbb F}[x_1,\dots,x_n]$ or the field
of rational functions ${\mathbb F}(x_1,\dots,x_n)$.

Let us now consider a system of linear partial differential equations, given by 
differential operators $P_1, \dots, P_r \in A_n$:
\bq
 P_j f\left(x_1,\dots,x_n\right) & = & 0.
\eq
These define an ideal $I=\lideal P_1,\dots,P_r \rideal$ in the Weyl algebra $A_n$, as
\bq
 Q P_j f\left(x_1,\dots,x_n\right) & = & 0,
 \;\;\;\;\;\;
 \mbox{for any} \; Q \; \in \; A_n. 
\eq
A vector $(v,w)=(v_1,\dots,v_n,w_1,\dots,w_n) \in {\mathbb Z}^{2n}$ is called a 
\index{weight vector for the Weyl algebra}
{\bf weight vector} for the Weyl algebra $A_n$ if
\bq
 v_j + w_j & \ge & 0,
 \;\;\;\;\;\;
 1 \; \le \; j \; \le \; n.
\eq
The order of a monomial $x^\alpha \partial^\beta$ with respect to the weight vector $(v,w)$
is
\bq
 \mathrm{ord}_{(v,w)}\left(x^\alpha \partial^\beta\right)
 & = &
 v_1 \alpha_1 + \dots + v_n \alpha_n + w_1 \beta_1 + \dots + w_n \beta_n
\eq
and we define the order of a differential operator $P \in A_n$ as the maximum of the orders
of the individual monomials when $P$ is written in normal form.
A weight vector $(v,w)$ induces a filtration 
\bq
 F_k^{(v,w)}\left(A_n\right)
 & = &
 \left\{
  \; P \; \in \; A_n \; | \; \mathrm{ord}_{(v,w)}\left(P\right) \; \le \; k
 \right\}.
\eq
The 
{\bf associated graded ring} 
is 
\bq
 \mathrm{gr}^{(v,w)}\left(A_n\right)
 & = &
 \bigoplus\limits_{k\in {\mathbb Z}} \mathrm{gr}_k^{(v,w)}\left(A_n\right),
 \;\;\;\;\;\;
 \mathrm{gr}_k^{(v,w)}\left(A_n\right)
 \; = \;
 F_k^{(v,w)}\left(A_n\right) / F_{k-1}^{(v,w)}\left(A_n\right).
\eq
We may think of $\mathrm{gr}^{(v,w)}(A_n)$ as the algebra generated by $x_1, \dots, x_n$ and
for $1\le j \le n$ either $\partial_j$ or $\xi_j$ according to
\bq
 v_j+w_j \; = \; 0 & : &
 \partial_j,
 \nonumber \\
 v_j+w_j \; > \; 0 & : &
 \xi_j.
\eq
The non-trivial commutation relations are the ones given in eq.~(\ref{chapter_nested_sums:commutation_relation_Weyl_algebra}).
If $v_j+w_j>0$ the commutator $[\partial_j,x_j]$ is of lower weight and we may replace $\partial_j$ 
with the commutative variable $\xi$. Only for $v_j+w_j=0$ we have to keep track of order.

Let $P \in A_n$ and assume that $P$ is written in normal form as
in eq.~(\ref{chapter_nested_sums:normal_ordered_differnetial_operator}).
We set $o = \mathrm{ord}_{(v,w)}(P)$.
The 
\index{initial form}
{\bf initial form} 
$\mathrm{in}_{(v,w)}(P)$ of $P$ with respect to the weight vector $(v,w)$ is defined to be
\bq
 \mathrm{in}_{(v,w)}\left(P\right)
 & = &
 \sum\limits_{\substack{\alpha,\beta \\ \alpha v + \beta w = o}} c_{\alpha \beta} \; x^\alpha \; 
 \prod\limits_{v_j+w_j>0} \xi_j^{\beta_j}
 \prod\limits_{v_j+w_j=0} \partial_j^{\beta_j}.
\eq
In other words: We take only the terms of order $o$ and make the replacement $\partial_j \rightarrow \xi_j$, whenever $v_j+w_j>0$.

Consider an ideal $I = \lideal P_1, \dots, P_r \rideal \subset A_n$.
Then
\bq
 \lideal \; \mathrm{in}_{(v,w)}\left(P\right) \; | \; P \in I \; \rideal
\eq
is an ideal in the associated graded ring $\mathrm{gr}^{(v,w)}(A_n)$, 
called the 
\index{initial ideal}
{\bf initial ideal} of $I$ with respect to the weight vector $(v,w)$.

Let us now consider the weight vector
\bq
 \left( {\bm 0}, {\bm 1} \right)
 & = &
 ( \underbrace{0, \dots, 0}_n, \underbrace{1, \dots, 1}_n ).
\eq
For this weight vector the associated graded ring $\mathrm{gr}^{({\bm 0},{\bm 1})}(A_n)$
of the Weyl algebra $A_n$
is the commutative ring ${\mathbb F}[x_1,\dots,x_n,\xi_1,\dots,\xi_n]$.
Let $I$ be an ideal in the Weyl algebra $A_n$.
We define the 
\index{characteristic ideal}
{\bf characteristic ideal} of $I$ to be the initial ideal
\bq
 \mathrm{in}_{({\bm 0},{\bm 1})}\left(I\right)
\eq
with respect to the weight vector $({\bm 0},{\bm 1})$.
The 
\index{characteristic variety}
{\bf characteristic variety} $\mathrm{ch}(I)$ is the zero set of 
$\mathrm{in}_{({\bm 0},{\bm 1})}\left(I\right)$ in the affine space of dimension $2n$ with coordinates
$x_1,\dots,x_n,\xi_1,\dots,\xi_n$.


We say that the ideal $I$ is 
\index{holonomic}
{\bf holonomic}
if the characteristic ideal $\mathrm{in}_{({\bm 0},{\bm 1})}(I)$ of $I$ 
has dimension $n$.
Let us denote by ${\mathbb F}(x)={\mathbb F}(x_1,\dots,x_n)$ the field of rational
functions in $x=(x_1,\dots,x_n)$.
The 
\index{holonomic rank}
{\bf holonomic rank} of $I$ is the dimension of the vector space
${\mathbb F}(x)[\xi]/({\mathbb F}(x)[\xi] \cdot \mathrm{in}_{({\bm 0},{\bm 1})}(I))$:
\bq
 \mathrm{rank}\left(I\right)
 & = &
 \dim\left( \; {\mathbb F}(x)[\xi] \; / \; \left({\mathbb F}(x)[\xi] \cdot \mathrm{in}_{({\bm 0},{\bm 1})}(I)\right) \; \right).
\eq
An important theorem states that if $I$ is holonomic, than $\mathrm{rank}(I)$ is finite.

Let us now consider the weight vector $(-w,w)$ with $w \in {\mathbb Z}^n$.
Consider an ideal $I = \lideal P_1, \dots, P_r \rideal \subset A_n$.
The ideal
\bq
 \mathrm{in}_{(-w,w)}\left(I\right)
 & = &
 \lideal \; \mathrm{in}_{(-w,w)}\left(P\right) \; | \; P \in I \; \rideal
\eq
is called a 
\index{Gr\"obner deformation}
{\bf Gr\"obner deformation} of $I$.
We define the 
\index{Euler operator}
{\bf Euler operators} $\theta_j$ by
\bq
 \theta_j & = & x_j \frac{\partial}{\partial x_j}.
\eq
Let us denote by $\hat{A}_n = {\mathbb F}(x) \cdot A_n$, i.e. the ring of differential operators $\partial_1, \dots, \partial_n$ with coefficients being rational functions of $x_1,\dots,x_n$.
To spell out the difference between $A_n$ and $\hat{A}_n$: In the Weyl algebra the coefficients are
polynomials in $x_1,\dots,x_n$, in $\hat{A}_n$ the coefficients are allowed to be rational functions in $x_1,\dots,x_n$.
Given an ideal $I \subset A_n$ we denote in a similar way by $\hat{I}$  the ideal ${\mathbb F}(x) \cdot I$.
We further denote by ${\mathbb F}[\theta]={\mathbb F}[\theta_1,\dots,\theta_n]$ the (commutative) ring of the Euler
operators.
We may now define the 
\index{indicial ideal}
{\bf indicial ideal} $\mathrm{ind}_w(I)$ of $I \subset A_n$ relative to the weight $w\in{\mathbb Z}^n$:
It is given by
\bq
 \mathrm{ind}_w\left(I\right)
 & = & 
 \widehat{\mathrm{in}}_{(-w,w)}\left(I\right) \; \cap \; {\mathbb F}\left[\theta\right].
\eq

\subsection{GKZ systems}
\label{chapter_nested_sums:GKZ_systems}

A 
\index{GKZ hypergeometric system}
{\bf GKZ hypergeometric system} \cite{Gelfand:1989,Gelfand:1989err}
or 
\index{${\mathcal A}$-hypergeometric system}
{\bf $\bm{\mathcal A}$-hypergeometric system}
is defined by a vector
$c \in {\mathbb C}^{k+1}$ and an $n$-element subset
\bq
\label{chapter_nested_sums:def_GKZ_A}
 {\mathcal A} \; = \; \left\{ a_1, \dots, a_n \right\}
 & \subset &
 {\mathbb Z}^{k+1},
\eq
which satisfies the two conditions
\bq
\label{chapter_nested_sums:conditions_A}
1. & & \hspace*{-5mm} \mbox{${\mathcal A}$ generates ${\mathbb Z}^{k+1}$ as an Abelian group.}
 \\
2. & & \hspace*{-5mm} \mbox{There exists a group homomorphism $h : {\mathbb Z}^{k+1} \rightarrow {\mathbb Z}$ such that
$h(a)=1$ for all $a \in {\mathcal A}$.}
 \nonumber
\eq
We must have $n>k$, otherwise the elements of ${\mathcal A}$ cannot generate ${\mathbb Z}^{k+1}$.
The second condition is equivalent to the statement that all elements $a \in {\mathcal A}$ lie in a hyperplane.
We may think of ${\mathcal A}$ as a $(k+1) \times n$-matrix with integer entries and where the columns are given
by the $a_j$'s. 

We denote by ${\mathbb L} \subset {\mathbb Z}^n$ the lattice of relations in ${\mathcal A}$:
\bq
 {\mathbb L}
 & = &
 \left\{ \; \left(l_1,\dots,l_n\right) \in {\mathbb Z}^n \; | \; l_1 a_1 + \dots + l_n a_n = 0 \; \right\}.
\eq
\begin{tcolorbox}
{\bf GKZ hypergeometric system}:
\\
The GKZ hypergeometric system associated with ${\mathcal A}$ and $c$ is the following system of differential
equations for a function $\phi$ of $n$ variables $x_1,\dots,x_n$:
For every $(l_1,\dots,l_n) \in {\mathbb L}$ one differential equation
\bq
\label{chapter_nested_sums:GKZ_diff_eq_set_1}
 \left[ 
  \prod\limits_{l_j>0} \left( \frac{\partial}{\partial x_j} \right)^{l_j} 
  - 
  \prod\limits_{l_j<0} \left( \frac{\partial}{\partial x_j} \right)^{-l_j} 
 \right] \phi
 & = & 0,
\eq
and $(k+1)$ differential equations
\bq
\label{chapter_nested_sums:GKZ_diff_eq_set_2}
 \left[ \sum\limits_{j=1}^n a_j x_j \frac{\partial}{\partial x_j} - c \right] \phi & = & 0.
\eq
\end{tcolorbox}
The differential equations in eq.~(\ref{chapter_nested_sums:GKZ_diff_eq_set_1})
are called the toric differential equations,
the ones in eq.~(\ref{chapter_nested_sums:GKZ_diff_eq_set_2}) are called the
homogeneity equations.
We may write the system of differential equations in a slightly modified way: Let $l \in {\mathbb L}$ and write
\bq
 l \; = \; u - v, 
 & \mbox{with} &
 u,v \; \in \; {\mathbb N}_0^n.
\eq
Thus $u$ contains all positive entries of $l$, while $v$ is the negative of the negative entries of $l$.
The condition $l_1 a_1 + \dots + l_n a_n = 0$ is equivalent to 
\bq
 {\mathcal A} u & = & {\mathcal A} v.
\eq
Then eq.~(\ref{chapter_nested_sums:GKZ_diff_eq_set_1}) is equivalent to
\bq
\label{chapter_nested_sums:GKZ_diff_eq_set_1_v2}
 \left( \partial^u - \partial^v \right) \phi & = & 0
\eq
for any pair $u,v \in {\mathbb N}_0^n$ with ${\mathcal A} u = {\mathcal A} v$.

Let us denote the vector of Euler operators by
\bq
 {\bm \theta}
 & = &
 \left( \theta_1, \dots, \theta_n \right)^T
 \; = \;
 \left( x_1 \frac{\partial}{\partial x_1}, \dots, x_n \frac{\partial}{\partial x_n} \right)^T.
\eq
Then eq.~(\ref{chapter_nested_sums:GKZ_diff_eq_set_2}) is equivalent to
\bq
\label{chapter_nested_sums:GKZ_diff_eq_set_2_v2}
 \left( {\mathcal A} {\bm \theta} - c \right) \phi & = & 0.
\eq
Let $\gamma \in {\mathbb C}^n$. We define
the 
\index{$\Gamma$-series}
{\bf ${\bm \Gamma}$-series}
associated with ${\mathbb L}$ and $\gamma$ by
\bq
\label{chapter_nested_sums:def_Gamma_series}
 \phi_{{\mathbb L},\gamma}\left(x_1,\dots,x_n\right)
 & = &
 \sum\limits_{(l_1,\dots,l_n) \in {\mathbb L}}
 \prod\limits_{j=1}^n
 \frac{x_j^{l_j+\gamma_j}}{\Gamma\left(l_j+\gamma_j+1\right)}.
\eq
The $\Gamma$-series is in its domain of convergence a solution 
of the GKZ system associated to ${\mathcal A}$ and
\bq
\label{chapter_nested_sums:c_gamma_relation_v1}
 c & = & \sum\limits_{j=1}^n a_j \gamma_j.
\eq
Recall that $c \in {\mathbb C}^{k+1}$, $\gamma \in {\mathbb C}^n$ and ${\mathcal A} \in M(k+1,n,{\mathbb Z})$.
Alternatively we may write eq.~(\ref{chapter_nested_sums:c_gamma_relation_v1}) as
\bq
\label{chapter_nested_sums:c_gamma_relation_v2}
 c & = & {\mathcal A} \gamma.
\eq
Let's look at a simple example: We take $k=2$ and $n=3$.
For the set ${\mathcal A}$ we choose
\bq
 {\mathcal A}
 & = &
 \left\{ 
  \left( \begin{array}{r} 1 \\ 0 \\ \end{array} \right),
  \left( \begin{array}{r} 0 \\ 1 \\ \end{array} \right),
  \left( \begin{array}{r} -1 \\ 2 \\ \end{array} \right)
 \right\}.
\eq
\begin{figure}
\begin{center}
\includegraphics[scale=1.0]{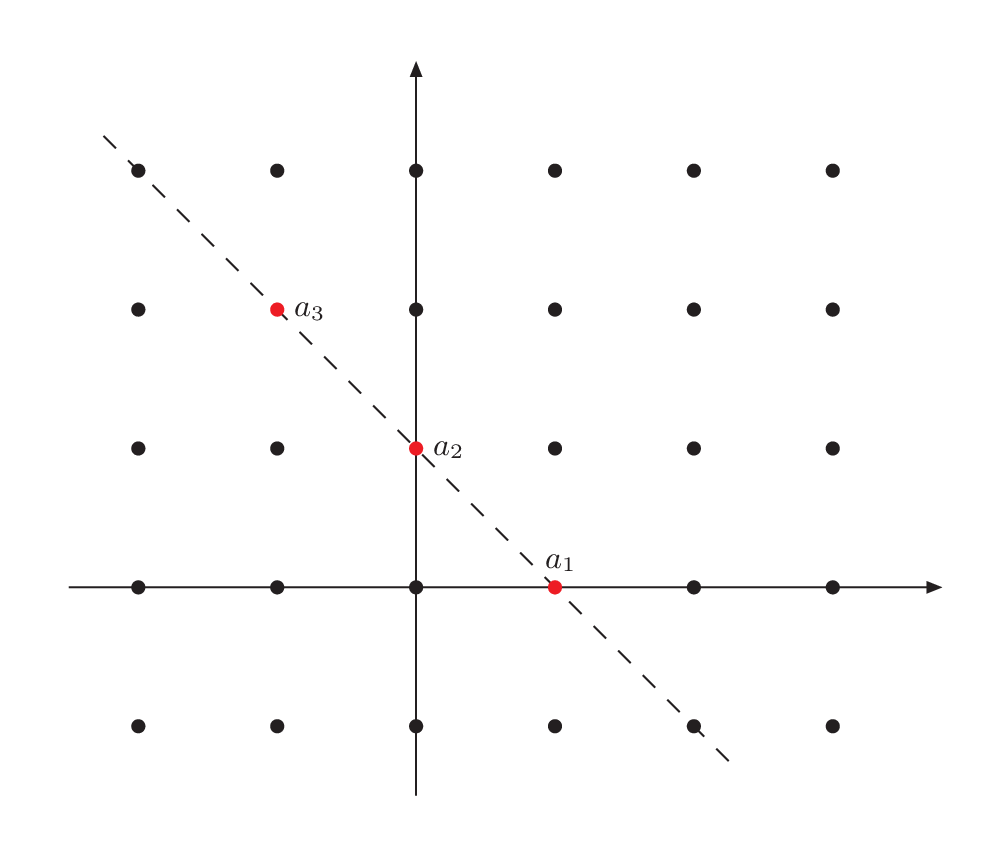}
\end{center}
\caption{
The set ${\mathcal A} = \{a_1,a_2,a_3\}\in {\mathbb Z}^2$.
The vectors $a_1$ and $a_2$ generate ${\mathbb Z}^2$.
Viewed as points, $a_1$, $a_2$ and $a_3$ lie in the hyperplane indicated by the dashed line.
}
\label{chapter_nested_sums:fig_example_GKZ}
\end{figure}
This set spans ${\mathbb Z}^2$ (the first two elements already span ${\mathbb Z}^2$)
and the three points lie in a hyperplane, as shown in fig.~\ref{chapter_nested_sums:fig_example_GKZ}.
The lattice of relations is generated by $(1,-2,1)$, e.g.
\bq
 {\mathbb L}
 & = &
 \left\{ \left(l,-2l,l\right) \in {\mathbb Z}^3 | l \in {\mathbb Z} \right\}.
\eq
The $\Gamma$-series associated to ${\mathcal A}$ and $\gamma$ is
\bq
 \phi_{{\mathbb L},\gamma}\left(x_1,x_2,x_3\right)
 & = &
 \sum\limits_{l \in {\mathbb Z}}
 \frac{x_1^{l+\gamma_1}}{\Gamma\left(l+\gamma_1+1\right)}
 \frac{x_2^{-2l+\gamma_2}}{\Gamma\left(-2l+\gamma_2+1\right)}
 \frac{x_3^{l+\gamma_3}}{\Gamma\left(l+\gamma_3+1\right)}.
\eq
In general, we are interested in the space of local solutions of a GKZ hypergeometric system.
Let us denote by $\Delta_{\mathcal A}$ the polytope in ${\mathbb R}^{k+1}$ defined by 
${\mathcal A} = (a_1,\dots,a_n)$.
If $c$ is generic and $\Delta_{\mathcal A}$ admits a unimodular triangulation, than the dimension of the solution space is
is given by
\bq
 \mathrm{vol}_0\left( \Delta_{\mathcal A} \right).
\eq
We may describe the solution space either in a geometrical language, using polytopes, or in an algebraic language,
using $D$-modules.
Let's start with the geometrical picture:

The key player in the geometrical picture is the secondary fan $N(\Sigma({\mathcal A}))$ of ${\mathcal A}$.
Let $C$ be a maximal cone of the secondary fan $N(\Sigma({\mathcal A}))$.
Such a maximal cone corresponds to a regular triangulation of the primary polytope $\Delta_{\mathcal A}$.
Let us denote by $T_C$ the list of subsets of $\{1,\dots,n\}$ such that each subset denotes the indices of the vertices 
of the maximal simplices in this triangulation.
For example, if $\{a_1,a_4,a_5,a_7\}$ are the vertices of a maximal simplex in the regular triangulation, we would include
in $T_C$ the subset $\{1,4,5,7\}$.
Let us now consider vectors $\gamma \in {\mathbb C}^n$ such that
\bq
\label{chapter_nested_sums:equations_gamma}
 & &
 c \; = \; {\mathcal A} \gamma,
 \nonumber \\
 & &
 \exists \; J \; \in \; T_C 
 \;\;\; 
 \mbox{such that}
 \;\;\;
 \gamma_j \; \in \; {\mathbb Z}_{\le 0}
 \;\;\;
 \mbox{for}
 \;\;\;
 j \; \notin \; J.
\eq
The first condition is the same as in eq.~(\ref{chapter_nested_sums:c_gamma_relation_v2}).
We call $\gamma$ and $\gamma'$ equivalent, if they only differ by $l \in {\mathbb L}$:
\bq
 \gamma \sim \gamma'
 & \Leftrightarrow &
 \gamma - \gamma' \; \in \; {\mathbb L}.
\eq
Given a maximal cone $C$ of the secondary fan and a vector $c \in {\mathbb C}^{k+1}$ we 
say that the vector $c$ is 
{\bf $C$-resonant} 
if the number of equivalence classes of solutions of eq.~(\ref{chapter_nested_sums:equations_gamma})
is less than $\mathrm{vol}_0(\Delta_{\mathcal A})$.
If $c$ is not $C$-resonant, we say that $c$ is generic.

We may now describe the solution space:
We start in a geometric language.
Let's assume that the primary polytope $\Delta_{\mathcal A}$ admits a unimodular triangulation. 
Let $C$ be a maximal cone of the secondary fan $N(\Sigma({\mathcal A}))$ and assume that $c$ is generic.
Then there are $\mathrm{vol}_0(\Delta_{\mathcal A})$ inequivalent solutions of eq.~(\ref{chapter_nested_sums:equations_gamma})
and the corresponding $\Gamma$-series are linearly independent and span the solution space of the GKZ hypergeometric system.

Alternatively, we may give a description in algebraic terms:
We first note that the differential equations of eq.~(\ref{chapter_nested_sums:GKZ_diff_eq_set_1}) and eq.~(\ref{chapter_nested_sums:GKZ_diff_eq_set_1}) generate an ideal $I$ in the Weyl algebra $A_n$.
For the solution of the GKZ system one constructs a basis of logarithmic series. 
A logarithmic series is a series of the form (in multi-index notation)
\bq
 \sum\limits_{\alpha \in A} \sum\limits_{\beta \in B} c_{\alpha \beta} x^\alpha \left( \ln\left(x\right) \right)^\beta,
 \;\;\;\;\;\;
 \left( \ln\left(x\right) \right)^\beta
 \; = \; 
 \left( \ln\left(x_1\right) \right)^{\beta_1} \dots \left( \ln\left(x_n\right) \right)^{\beta_n},
\eq
where $A \subset {\mathbb C}^n$ is a discrete set and $B \subset \{0,\dots,b_{\mathrm{max}}\}^n$ for some $b_{\mathrm{max}} \in {\mathbb N}_0$.
Given a weight vector $w \in {\mathbb Z}^n$ we may define a partial order on the terms of a logarithmic series by
\bq
 x^\alpha \left( \ln\left(x\right) \right)^\beta \; > \; x^{\alpha'} \left( \ln\left(x\right) \right)^{\beta'}
 & \Leftrightarrow &
 \mathrm{Re}\left( w \cdot \alpha \right) \; > \; \mathrm{Re}\left( w \cdot \alpha' \right).
\eq
This partial order can be refined to a total order by using the lexicographic order to break ties.
This total order is denoted by $>_w$. 
The initial term of a logarithmic series is the minimal term with respect to the total order $>_w$.
For a generic weight vector $w$
a basis of solutions of the GKZ system consists of logarithmic series, such that for each logarithmic series
the exponent $\alpha$ of the initial term
\bq
\label{chapter_nested_sums:initial_term_logarithmic_series}
 x^\alpha \left( \ln\left(x\right) \right)^\beta
\eq
is a root of the indicial ideal $\mathrm{ind}_w(I)$, i.e. $\alpha \in V(\mathrm{ind}_w(I))$.
Moreover, for a given root $\alpha$ the number of logarithmic series in the basis with initial term
as in eq.~(\ref{chapter_nested_sums:initial_term_logarithmic_series}) equals the multiplicity of the root $\alpha$ in $V(\mathrm{ind}_w(I))$.
The initial terms of these logarithmic series differ in $\beta$.
Starting from the initial term, it is possible to construct the full logarithmic series.
For an algorithm we refer to the book by
Saito, Sturmfels and Takayama \cite{Saito:book}.

\subsection{Euler-Mellin integrals}
\label{chapter_nested_sums:Euler-Mellin_integrals}

Euler-Mellin integrals are examples of ${\mathcal A}$-hypergeometric functions \cite{Gelfand:1990}.
In order to define Euler-Mellin integrals we start with a function $p(z,x)=p(z_1,\dots,z_k,x_1,\dots,x_n)$ of the form
\bq
\label{chapter_nested_sums:def_Laurent_polynomial}
 p\left(z_1,\dots,z_k,x_1,\dots,x_n\right)
 & = &
 \sum\limits_{j=1}^n x_j \; z_1^{a_{1 j}} \dots z_k^{a_{k j}},
 \;\;\;\;\;\;
 a_{i j} \; \in \; {\mathbb Z}.
\eq
We call a function of this type 
a 
\index{Laurent polynomial}
{\bf Laurent polynomial}.
Note that the exponents $a_{ij}$ are allowed to be negative integers. If for all exponents $a_{i j}$ we have
$a_{ij} \in {\mathbb N}_0$, then the function $p(z,x)$ is a polynomial in $z$ and $x$.
Note further that $p(z,x)$ is linear in each $x_j$ and that there are as many $x_j$'s as there are monomials in the sum
in eq.~(\ref{chapter_nested_sums:def_Laurent_polynomial}).

Let us now consider a GKZ hypergeometric system with
\bq
\label{chapter_nested_sums:def_GKZ_hypergeometric_system_Euler_Mellin_1}
 {\mathcal A} 
 & = &
 \left( \begin{array}{cccc}
 1 & 1 & \dots & 1 \\
 a_{11} & a_{12} & \dots & a_{1n} \\
 \dots & & & \dots \\
 a_{k1} & a_{k2} & \dots & a_{kn} \\
 \end{array} \right),
 \;\;\;\;\;\;
 c
 \; = \;
 \left( \begin{array}{c}
 - \nu_0 \\
 - \nu_1 \\
 \dots \\
 - \nu_k \\
 \end{array} \right).
\eq
${\mathcal A}$ is a $(k+1)\times n$-matrix, $c$ is a vector in ${\mathbb C}^{k+1}$.
${\mathcal A}$ defines a Laurent polynomial $p(x,z)$ through eq.~(\ref{chapter_nested_sums:def_Laurent_polynomial}).
We then consider integrals of the form
\bq
\label{chapter_nested_sums:def_Euler_Mellin_integral}
 \int\limits_{\mathcal C} 
  d^kz \;
 \left( \prod\limits_{j=1}^{k} z_j^{\nu_j-1} \right)
 \left[p\left(z,x\right)\right]^{-\nu_0},
\eq
where ${\mathcal C}$ is a cycle.
An integral of the form as in eq.~(\ref{chapter_nested_sums:def_Euler_Mellin_integral}) is called an 
{\bf Euler-Mellin integral}.

We may extent the definition of Euler-Mellin integrals to integrals involving $m$ Laurent polynomials $p_1(z,x), \dots, p_m(z,x)$
as follows:
We start from a GKZ hypergeometric system of the form
\bq
\label{chapter_nested_sums:def_GKZ_hypergeometric_system_Euler_Mellin_2}
 {\mathcal A} 
 & = &
 \left( \begin{array}{ccccccccccc}
 1 & \dots & 1 & 0 & \dots & 0 & \dots & 0 & \dots & 0 \\
 0 & \dots & 0 & 1 & \dots & 1 & & 0 & \dots & 0 \\
 \dots & & \dots & \dots & & \dots & & \dots & & \dots \\
 0 & \dots & 0 & 0 & \dots & 0 & & 1 & \dots & 1 \\
 a_{111} & \dots & a_{11n_1} & a_{211} & \dots & a_{21n_2} & \dots & a_{m11} & \dots & a_{m1n_m} \\
 \dots & & \dots & \dots & & \dots & & \dots & & \dots \\
 a_{1k1} & \dots & a_{1kn_1} & a_{2k1} & \dots & a_{2kn_2} & \dots & a_{mk1} & \dots & a_{mkn_m} \\
 \end{array} \right),
 \;\;\;\;\;\;
 c
 \; = \;
 \left( \begin{array}{c}
 - \mu_1 \\
 - \mu_2 \\
 \dots \\
 - \mu_m \\
 - \nu_1 \\
 \dots \\
 - \nu_k \\
 \end{array} \right).
 \nonumber \\
\eq
We set $n=n_1+\dots+n_m$.
${\mathcal A}$ is a $(k+m)\times n$-matrix, $c$ is a vector in ${\mathcal C}^{k+m}$.
${\mathcal A}$ defines $m$ Laurent polynomials $p_1(z,x), \dots, p_m(z,x)$ as follows: We set
\bq
 p_i\left(z,x\right)
 & = &
 \sum\limits_{j=1}^{n_i} x_{n_1+\dots+n_{i-1}+j} \; z_1^{a_{i 1 j}} \dots z_k^{a_{i k j}}.
\eq
$p_1(z,x)$ depends on the first $n_1$ variables $x_1, \dots, x_{n_1}$, $p_2(z,x)$ on the next $n_2$ variables
$x_{n_1+1},\dots,x_{n_1+n_2}$ etc..
The associated Euler-Mellin integral is
\bq
\label{chapter_nested_sums:def_general_Euler_Mellin_integral}
 \int\limits_{\mathcal C} 
  d^kz \;
 \left( \prod\limits_{j=1}^{k} z_j^{\nu_j-1} \right)
 \left( \prod\limits_{i=1}^m \left[p_i\left(z,x\right)\right]^{-\mu_i} \right).
\eq

\subsection{Feynman integrals as GKZ hypergeometric functions}
\label{chapter_nested_sums:Feynman_integrals_GKZ_hypergeometric}

Let us now consider a Feynman integral $I_{\nu_1 \dots \nu_{\ninternal}}(D,x_1,\dots,x_{\NB})$.
The Lee-Pomeransky representation reads
\bq
 I_{\nu_1 \dots \nu_{\ninternal}}\left(D,x_1,\dots,x_{\NB}\right)
 & = &
 C
 \int\limits_{z_j \ge 0}  d^{\ninternal}z \;
 \left( \prod\limits_{j=1}^{\ninternal} z_j^{\nu_j-1} \right)
 \left[{\mathcal G}\left(z,x\right)\right]^{-\frac{D}{2}},
\eq
with the prefactor (irrelevant for the discussion here)
\bq
 C & = &
 \frac{e^{\loopnumber \eps \Eulerconstant}\Gamma\left(\frac{D}{2}\right)}{\Gamma\left(\frac{\left(\loopnumber+1\right))D}{2}-\nu\right)\prod\limits_{j=1}^{\ninternal}\Gamma(\nu_j)}
\eq
and the Lee-Pomeransky polynomial 
\bq
 {\mathcal G}\left(z,x\right) = {\mathcal U}\left(z\right) + {\mathcal F}\left(z,x\right).
\eq
The Lee-Pomeransky representation of the Feynman integral is close to an Euler-Mellin integral as in eq.~(\ref{chapter_nested_sums:def_Euler_Mellin_integral}),
but not quite:
First of all, the number of monomials in ${\mathcal G}(z,x)$ will in general not match the number of kinematic
variables $x_1,\dots,x_{\NB}$.
This issue is easily fixed: We consider a generalised Lee-Pomeransky polynomial $G(z,x')=G(z_1,\dots,z_{\ninternal},x_1',\dots,x_n')$ with as many variables $x_j'$ as there are monomials in the original Lee-Pomeransky polynomial ${\mathcal G}(z,x)$.
The original Lee-Pomeransky polynomial ${\mathcal G}(z,x)$ is then recovered as the special case, where the additional
variables take special values.
As an example consider the one-loop two-point function with equal internal masses.
With $x=-p^2/m^2$ the original Lee-Pomeransky polynomial reads
\bq
 {\mathcal G}\left(z_1,z_2,x\right)
 & = &
 z_1 + z_2 + \left(2+x\right) z_1 z_2 + z_1^2 + z_2^2.
\eq
The generalised Lee-Pomeransky polynomial reads
\bq
\label{chapter_nested_sums:example_generalised_Lee_Pomeransky_polynomial}
 G\left(z_1,z_2,x_1',x_2',x_3',x_4',x_5'\right)
 & = &
 x_1' z_1 + x_2' z_2 + x_3' z_1 z_2 + x_4' z_1^2 + x_5' z_2^2.
\eq
We recover the original Lee-Pomeransky polynomial as
\bq
 {\mathcal G}\left(z_1,z_2,x\right)
 & = &
 G\left(z_1,z_2,1,1,2+x,1,1\right).
\eq
The exponents of the monomials in $G(z,x')$ define a $(\ninternal+1)\times n$-matrix ${\mathcal A}$
through eq.~(\ref{chapter_nested_sums:def_GKZ_hypergeometric_system_Euler_Mellin_2}).
For the example from eq.~(\ref{chapter_nested_sums:example_generalised_Lee_Pomeransky_polynomial}) we obtain
\bq
 {\mathcal A}
 & = &
 \left( \begin{array}{ccccc} 
 1 & 1 & 1 & 1 & 1 \\
 1 & 0 & 1 & 2 & 0 \\
 0 & 1 & 1 & 0 & 2 \\
 \end{array} \right).
\eq
Secondly, the matrix ${\mathcal A}$ should satisfy the two conditions of eq.~(\ref{chapter_nested_sums:conditions_A}).
The second condition does not pose any problem: As the first row of ${\mathcal A}$ contains only $1$'s, all points lie in a
hyperplane (with first coordinate equal to one) and the $(\ninternal+1)$-dimensional row vector
\bq
 ( 1, \underbrace{0,\dots,0}_{\ninternal} )
\eq
defines the group homomorphism 
$h : {\mathbb Z}^{k+1} \rightarrow {\mathbb Z}$.
It may happen that ${\mathcal A}$ does not satisfy the first condition (that the columns of ${\mathcal A}$ generate ${\mathbb Z}^{k+1}$
as an Abelian group. In this case we add additional monomials to the generalised Lee-Pomeransky polynomial $G(z,x')$ until
${\mathcal A}$ does satisfy the first condition. These additional monomials come with new variables $x_j'$ and we take the limit $x_j' \rightarrow 0$ in the end.

\begin{theorem}[Feynman integrals and ${\mathcal A}$-hypergeometric functions]
\label{chapter_nested_sums:theorem_A_hypergeometric}
Any Feynman integral 
\bq
 I_{\nu_1 \dots \nu_{\ninternal}}(D,x_1,\dots,x_{\NB})
\eq
depending on the kinematic variables $x_1, \dots, x_{\NB}$
is a special case of a ${\mathcal A}$-hypergeometric function in more variables $x_1', \dots, x_n'$, 
where the variables $x_j'$ take special values.
\end{theorem}
The relation between Feynman integrals and ${\mathcal A}$-hypergeometric functions
has been considered in \cite{Nasrollahpoursamami:2016,Vanhove:2018mto,delaCruz:2019skx,Klausen:2019hrg}.
The above theorem is due to de la Cruz \cite{delaCruz:2019skx}.
\\
\\
\bs
{\it \refstepcounter{exercise}
{\bf Exercise \theexercise}: 
In this exercise we are going to prove theorem~\ref{chapter_nested_sums:theorem_A_hypergeometric}.
Let $G(z,x')=G(z_1,\dots,z_{\ninternal},x_1',\dots,x_n')$ be a generalised Lee-Pomeransky polynomial 
such that the associated $(\ninternal+1)\times n$-matrix ${\mathcal A}$ 
satisfies eq.~(\ref{chapter_nested_sums:conditions_A}).
Consider the integral
\bq
 I
 & = &
 C
 \int\limits_{z_j \ge 0}  d^{\ninternal}z \;
 \left( \prod\limits_{j=1}^{\ninternal} z_j^{\nu_j-1} \right)
 \left[G\left(z,x'\right)\right]^{-\frac{D}{2}}.
\eq
Show that $I$ satisfies the differential equations in eq.~(\ref{chapter_nested_sums:GKZ_diff_eq_set_1}) and eq.~(\ref{chapter_nested_sums:GKZ_diff_eq_set_2}) with
$c=(-D/2,-\nu_1,\dots,-\nu_{\ninternal})^T$.
}
\es
\\
\\
The number of variables $x_1', \dots, x_n'$ for the ${\mathcal A}$-hypergeometric function
can be quite large, as the following example illustrates:
Consider the two-loop double box integral where all internal masses vanish and the external momenta
are light-like: $p_1^2=p_2^2=p_3^2=p_4^2$.
This integral depends only on one kinematic variable, which can be taken as $x=s/t$.
The original Lee-Pomeransky polynomial reads 
\bq
\lefteqn{
 {\mathcal G}\left(z_1,\dots,z_7,x\right)
 = } & &
 \nonumber \\
 & &
 z_1 z_5
 + z_1 z_6
 + z_1 z_7
 + z_2 z_5
 + z_2 z_6
 + z_2 z_7
 + z_3 z_5
 + z_3 z_6
 + z_3 z_7
 + z_1 z_4
 + z_2 z_4
 + z_3 z_4
 + z_4 z_5
 + z_4 z_6
 \nonumber \\
 & &
 + z_4 z_7
 + x z_2 z_3 z_4
 + x z_2 z_3 z_5
 + x z_2 z_3 z_6
 + x z_2 z_3 z_7
 + x z_5 z_6 z_1
 + x z_5 z_6 z_2
 + x z_5 z_6 z_3
 + x z_5 z_6 z_4
 \nonumber \\
 & &
 + x z_2 z_4 z_6
 + x z_3 z_4 z_5
 + z_1 z_4 z_7.
\eq
The original Lee-Pomeransky polynomial is a sum of $26$ monomials.
For the generalised Lee-Pomeransky polynomial $G$ we therefore introduce $26$ variables $x_1', \dots, x_{26}'$.
The generalised Lee-Pomeransky polynomial reads
\bq
\lefteqn{
 G\left(z_1,\dots,z_7,x_{1}',\dots,x_{26}'\right)
 = } & &
 \nonumber \\
 & &
 x_{1}' z_1 z_5
 + x_{2}' z_1 z_6
 + x_{3}' z_1 z_7
 + x_{4}' z_2 z_5
 + x_{5}' z_2 z_6
 + x_{6}' z_2 z_7
 + x_{7}' z_3 z_5
 + x_{8}' z_3 z_6
 + x_{9}' z_3 z_7
 + x_{10}' z_1 z_4
 \nonumber \\
 & &
 + x_{11}' z_2 z_4
 + x_{12}' z_3 z_4
 + x_{13}' z_4 z_5
 + x_{14}' z_4 z_6
 + x_{15}' z_4 z_7
 + x_{16}' z_2 z_3 z_4
 + x_{17}' z_2 z_3 z_5
 + x_{18}' z_2 z_3 z_6
 \nonumber \\
 & &
 + x_{19}' z_2 z_3 z_7
 + x_{20}' z_5 z_6 z_1
 + x_{21}' z_5 z_6 z_2
 + x_{22}' z_5 z_6 z_3
 + x_{23}' z_5 z_6 z_4
 + x_{24}' z_2 z_4 z_6
 + x_{25}' z_3 z_4 z_5
 \nonumber \\
 & &
 + x_{26}' z_1 z_4 z_7.
\eq
Thus we go from the case of one kinematic variable $x$ to a ${\mathcal A}$ hypergeometric function in $26$ 
variables.
At the end we are interested in the limit, where the variables $x_j'$ go either to $1$ or $x$.

\section{The Bernstein-Sato polynomial}
\label{chapter_nested_sums:Bernstein_polynomial}

Given a polynomial $V(x)$ in $n$ variables $x_1, \dots, x_n$, the Bernstein-Sato theorem \cite{Bernstein:1971,Sato:1972}
states that there is an
identity of the form
\bq
 P\left(x,\partial\right) \left[ V\left(x\right) \right]^{\nu+1} & = & B \left[ V\left(x\right) \right]^{\nu},
\eq
where $P(x,\partial)$ is a polynomial of $x$ and $\partial=(\partial_1,\dots,\partial_n)$ with $\partial_j=\partial/\partial_j$.
$B$ and all coefficients of $P$ are polynomials of $\nu$ and of the coefficients of $V(x)$.
$B$ is called the 
\index{Bernstein-Sato polynomial}
{\bf Bernstein-Sato polynomial}.
This can be generalised to several polynomials $V_1(x), \dots, V_r(x)$ in $n$ variables $x_1, \dots, x_n$ \cite{Tkachov:1996wh}:
\bq
 P\left(x,\partial\right) \prod\limits_{j=1}^r \left[ V_j\left(x\right) \right]^{\nu_j+1} 
 & = & 
 B \prod\limits_{j=1}^r \left[ V_j\left(x\right) \right]^{\nu_j},
\eq
where as above 
$P(x,\partial)$ is a polynomial of $x$ and $\partial=(\partial_1,\dots,\partial_n)$.
$B$ and all coefficients of $P$ are polynomials of the $\nu_j$'s and of the coefficients of the $V_j(x)$'s.

As an example consider
\bq
 V\left(x\right)
 & = &
 x^T A x + 2 w^T X + c,
\eq
where $A$ is an invertible $(n\times n)$-matrix and $w$ a $n$-vector.
Then
\bq
 P\left(x,\partial\right) \left[ V\left(x\right) \right]^{\nu+1} & = & B \left[ V\left(x\right) \right]^{\nu},
\eq
with
\bq
 P\left(x,\partial\right) 
 \; = \;
 1- \frac{1}{2\left(\nu+1\right)} \left(x + w^T A^{-1} \right) \partial,
 & &
 B \; = \; c - w^T A^{-1} w.
\eq
The Bernstein-Sato polynomial has been applied in \cite{Tkachov:1996wh,Passarino:2001wv}
to one-loop integrals in the Feynman parameter representation:
For one-loop integrals we have
\bq
 I
 & = &
 \frac{e^{\eps \Eulerconstant}\Gamma\left(\nu-\frac{D}{2}\right)}{\prod\limits_{j=1}^{\ninternal}\Gamma(\nu_j)}
 \int\limits_{a_j \ge 0} d^{\ninternal}a \; \delta\left(1-\sum\limits_{j=1}^{\ninternal} a_j \right) \; 
 \left( \prod\limits_{j=1}^{\ninternal} a_j^{\nu_j-1} \right)
 \left[ {\mathcal F}\left(a\right) \right]^{\frac{D}{2}-\nu},
\eq
and the second graph polynomial ${\mathcal F}(a)$ is quadratic in the Feynman parameters.
Using
\bq
 \left[ {\mathcal F}\left(a\right) \right]^{\frac{D}{2}-\nu}
 & = &
 \frac{1}{B} P\left(x,\partial\right) \left[ {\mathcal F}\left(a\right) \right]^{1+\frac{D}{2}-\nu}
\eq
will raise the exponent of the second graph polynomial.
Using partial integration for the differential operator $P(x,\partial)$ will either produce simpler integrals
on the boundary of the Feynman parameter integration domain or integrals with modified exponents $\nu_j$.
This can be repeated, until the integer part of the exponent of the second graph polynomial is non-negative.
At this point, all poles in the dimensional regularisation parameter $\eps$ originate from the various $1/B$-prefactors.
The remaining integrals are finite and can be performed numerically.

Methods to compute $P(x,\partial)$ and $B$ for a general single polynomial $V(x)$ can be found in the book
by Saito, Sturmfels and Takayama \cite{Saito:book}.

%% file: sector_decomposition.tex
\newpage
\chapter{Sector decomposition}
\label{chapter_sector_decomposition}

Let us consider the Laurent expansion in the dimensional regularisation parameter $\eps$ of a Feynman integral:
\bq
\label{chapter_sector_decomposition:Laurent_expansion}
 I & = &
 \sum\limits_{j=j_{\min}}^\infty
 \eps^j \; I^{(j)}.
\eq
For precision calculations we are interested in the first few terms of this Laurent expansion.
The coefficients $I^{(j)}$ are independent of $\eps$ and we may ask if there is a way to compute
them numerically.
In this section we will discuss the method of sector decomposition, which allows us to compute
the coefficients $I^{(j)}$ by Monte Carlo integration.
The essential step will be to manipulate the original integrand in such a way, that all
$I^{(j)}$'s are given by integrable integrands.
We will start from the Feynman parameter representation of eq.~(\ref{chapter_basics:Feynman_parameter_representation})
\bq
\label{chapter_sector_decomposition:Feynman_parameter_representation}
 I
 & = &
 \frac{e^{\loopnumber \eps \Eulerconstant}\Gamma\left(\nu-\frac{\loopnumber D}{2}\right)}{\prod\limits_{j=1}^{\ninternal}\Gamma(\nu_j)}
 \int\limits_{a_j \ge 0} d^{\ninternal}a \; \delta\left(1-\sum\limits_{j=1}^{\ninternal} a_j \right) \; 
 \left( \prod\limits_{j=1}^{\ninternal} a_j^{\nu_j-1} \right)
 \frac{\left[ {\mathcal U}\left(a\right) \right]^{\nu-\frac{\left(\loopnumber+1\right) D}{2}}}{\left[ {\mathcal F}\left(a\right) \right]^{\nu-\frac{\loopnumber D}{2}}}.
\eq
Depending on the exponents, potential singularities of the integrand come from the regions $a_j=0$,
${\mathcal U}=0$ or ${\mathcal F}=0$.
We already know that ${\mathcal U} \neq 0$ inside the integration, 
but there is the possibility that ${\mathcal U}$ vanishes on the boundary of the integration region.
In order to keep the discussion simple we will assume in this chapter that all kinematic variables
are in the Euclidean region.
Then it follows that ${\mathcal F} \neq 0$ inside the integration,
but ${\mathcal F}$ may vanish on the boundary of the integration region.
It is possible to relax the condition that all kinematic variables should be in the Euclidean region
by an appropriate deformation of the integration contour of the final integration.

In mathematical terms, sector decomposition corresponds to a resolution of singularities by a sequence of blow-ups.
We will discuss the relation between these topics.

Apart from the practical application of being able to compute numerically the coefficients of the Laurent
expansion in $\eps$, we may show that in the case where all kinematic variables are algebraic numbers
in the Euclidean region the coefficients $I^{(j)}$ are numerical periods.

\section{The algorithm of sector decomposition}
\label{chapter_sector_decomposition:algo_sector_decomp}

In this section we discuss the algorithm for iterated
sector decomposition \cite{Binoth:2000ps}.
The starting point is an integral of the form
\bq
\label{chapter_sector_decomposition:basic_integral}
 \int\limits_{x_j \ge 0} d^nx \;\delta(1-\sum_{i=1}^n x_i)
 \left( \prod\limits_{i=1}^n x_i^{\mu_i} \right)
 \prod\limits_{j=1}^r \left[ P_j(x) \right]^{\lambda_j},
\eq
where $\mu_i=a_i+\eps b_i$ and $\lambda_j=c_j+\eps d_j$.
The integration is over the standard simplex.
The $a$'s, $b$'s, $c$'s and $d$'s are integers.
The $P$'s are polynomials in the variables $x_1$, ..., $x_n$.
The polynomials are required to be non-zero
inside the integration region, but
may vanish on the boundaries of the integration region.

The Feynman parameter integral in eq.~(\ref{chapter_sector_decomposition:Feynman_parameter_representation})
is -- apart from a trivial prefactor -- of the form as in eq.~(\ref{chapter_sector_decomposition:basic_integral}).
Eq.~(\ref{chapter_sector_decomposition:basic_integral}) is slightly more general:
We allow more than two polynomials, the polynomials are not required to be homogeneous and the exponents $\mu_i$
are not required to be integers, but are allowed to be of the form $\mu_i=a_i+\eps b_i$, with $a_i, b_i \in {\mathbb Z}$.

The algorithm of sector decomposition consists of the following six steps:
\\
\\
{\bf Step 1}: Convert all polynomials to homogeneous polynomials.
This is easily done as follows:
Due to the presence of the 
Dirac delta distribution in eq.~(\ref{chapter_sector_decomposition:basic_integral})
we have
\bq
 1 & = & x_1 + x_2 + \dots + x_n.
\eq
For each polynomial $P_j$ we determine the highest degree $h_j$ and multiply all terms of lower
degree by an appropriate power of $x_1 + x_2 + \dots + x_n$.
As an example consider $n=2$ and $P=x_1+x_1x_2^2$. The homogenisation of $P$ is
\bq
 x_1 \left(x_1+x_2\right)^2 + x_1 x_2^2.
\eq
After step 1 all polynomials $P_j$ are homogeneous polynomials of degree $h_j$.
\\
\\
{\bf Step 2}: Decompose the integral into $n$ primary sectors.
This is done as follows: We write
\bq
 \int\limits_{x_j \ge 0} d^nx & = &
 \sum\limits_{l=1}^n \int\limits_{x_j \ge 0} d^nx 
     \prod\limits_{i=1, i\neq l}^n \theta(x_l \ge x_i).
\eq
The sum over $l$ corresponds to the sum over the $n$ primary sectors.
In the $l$-th primary sector we make the substitution
\bq
 x_j & = & x_l x_j' \;\;\;\mbox{for} \; j \neq l.
\eq
As after step 1 each polynomial $P_j$ is homogeneous of degree $h_j$ we arrive at
\bq
\lefteqn{
 \int\limits_{x_j \ge 0} d^nx \;\delta(1-\sum_{i=1}^n x_i)
 \left( \prod\limits_{i=1, i\neq l}^n \theta(x_l \ge x_i) \right)
 \left( \prod\limits_{i=1}^n x_i^{a_i+\eps b_i} \right)
 \prod\limits_{j=1}^r \left[ P_j(x) \right]^{c_j+\eps d_j}
= } & & \\
 & & 
 \int\limits_0^1 
 \left( \prod\limits_{i=1, i\neq l}^n dx_i \; x_i^{a_i+\eps b_i} \right)
 \left( 1 + \sum\limits_{j=1, j\neq l}^n x_j \right)^c
 \prod\limits_{j=1}^r \left[ P_j(x_1,...,x_{j-1},1,x_{j+1},...,x_n) \right]^{c_j+\eps d_j},
 \nonumber 
\eq
where
\bq
 c & = & -n - \sum\limits_{i=1}^n \left( a_i+\eps b_i \right) - \sum\limits_{j=1}^r h_j \left( c_j+\eps d_j\right).
\eq
Each primary sector is now a $(n-1)$-dimensional integral over the unit hyper-cube.
Note that in the general case this decomposition introduces an additional polynomial factor
\bq
 \left( 1 + \sum\limits_{j=1, j\neq l}^n x_j \right)^c.
\eq
\bs
{\it \refstepcounter{exercise}
{\bf Exercise \theexercise}: 
Show that for a Feynman integral as 
in eq.~(\ref{chapter_sector_decomposition:Feynman_parameter_representation}) 
we have in any primary sector $c=0$ and therefore the additional factor is absent.
}
\es
\\
\\
The underlying reason for the statement that for any Feynman integral we always have $c=0$ 
is the fact
that the Feynman integral is a projective integral.
In eq.~(\ref{chapter_sector_decomposition:basic_integral}) we consider more general integrals.
In particular we do not require that the integral 
in eq.~(\ref{chapter_sector_decomposition:basic_integral})
descends to an integral on projective space.
Therefore we have in general $c \neq 0$.
In any case, this factor is just an additional polynomial.
After an adjustment $(n-1) \rightarrow n$ and possibly $(r+1) \rightarrow r$
we therefore deal with integrals of the form
\bq
\label{chapter_sector_decomposition:primary_integral}
 \int\limits_{0}^{1} d^{n}x
 \;
 \prod\limits_{i=1}^{n}x_{i}^{a_{i}+\epsilon b_{i}}
 \prod\limits_{j=1}^{r} \left[P_{j}(x)\right]^{c_{j}+\epsilon d_{j}}.
\eq
This is now an integral over the unit hypercube.
The polynomials $P_j$ do not vanish inside the integration region. They may vanish on intersections
of the boundary of the integration region with coordinate subspaces.
\\
\\
{\bf Step 3}: Decompose the sectors iteratively into sub-sectors until each of the polynomials 
is of the form
\bq
\label{chapter_sector_decomposition:monomialised}
 P & = & x_1^{m_1} ... x_n^{m_n} \left( c + P'(x) \right),
\eq
where $c\neq 0$ and $P'(x)$ is a polynomial in the variables $x_j$ without a constant term.
In this case the monomial prefactor $x_1^{m_1} ... x_n^{m_n}$ can be factored out
and the remainder contains a non-zero constant term.
To convert $P$ into the form~(\ref{chapter_sector_decomposition:monomialised}) one chooses a subset
$S=\left\{ \alpha_{1},\,...,\, \alpha_{k}\right\} \subseteq \left\{ 1, \,...\, n \right\}$
according to a strategy discussed in the next section.
One decomposes the $k$-dimensional hypercube into $k$ sub-sectors according to
\bq
\label{chapter_sector_decomposition:decomposition}
 \int\limits_{0}^{1} d^{n}x & = & 
 \sum\limits_{l=1}^{k} 
 \int\limits_{0}^{1} d^{n}x
   \prod\limits_{i=1, i\neq l}^{k}
   \theta\left(x_{\alpha_{l}}\geq x_{\alpha_{i}}\right).
\eq
In the $l$-th sub-sector one makes for each element of $S$ the
substitution
\begin{alignat}{3}
\label{chapter_sector_decomposition:substitution}
 x_{\alpha_{i}} & = x_{\alpha_{l}}' x_{\alpha_{i}}' & \hspace*{3mm} & \mbox{for} & & \;\; i \neq l,
 \nonumber \\
 x_{\alpha_{i}} & = x_{\alpha_{i}}'              & \hspace*{3mm} & \mbox{for} & & \;\; i = l.
\end{alignat}
This procedure is iterated, until all polynomials are of the form~(\ref{chapter_sector_decomposition:monomialised}).
At the end all polynomials contain a constant term.
Each sub-sector integral is of the form as 
in eq.~(\ref{chapter_sector_decomposition:primary_integral}), where every $P_{j}$ is now different from zero in the whole integration domain.
Hence the singular behaviour of the integral depends on the $a_{i}$ and $b_{i}$, the $a_{i}$ being integers.

\begin{figure}
\begin{center}
\includegraphics[scale=0.9]{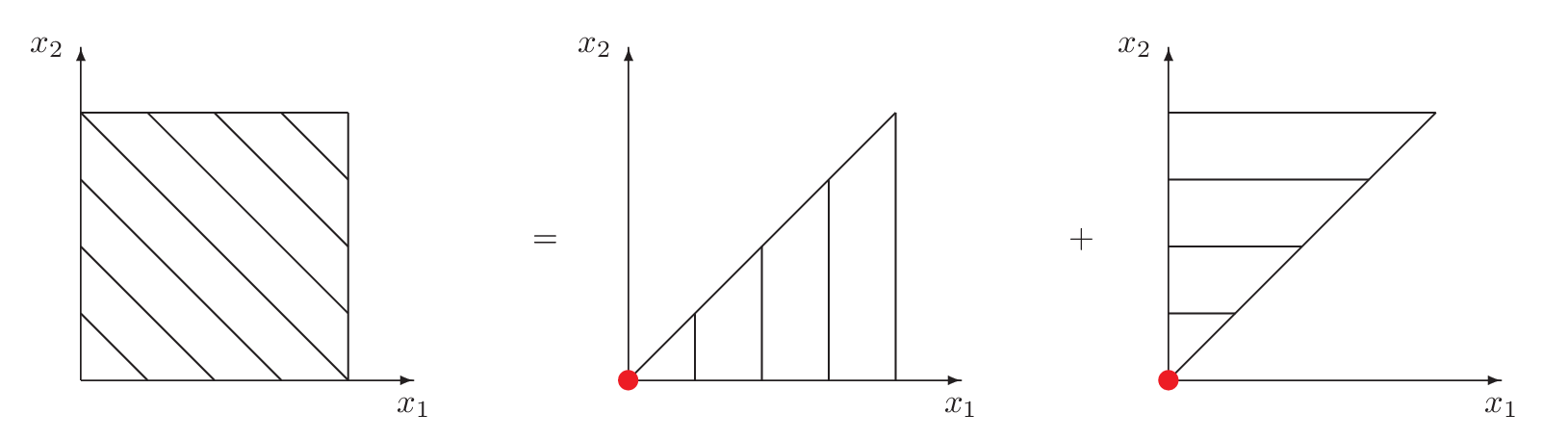}
\includegraphics[scale=0.9]{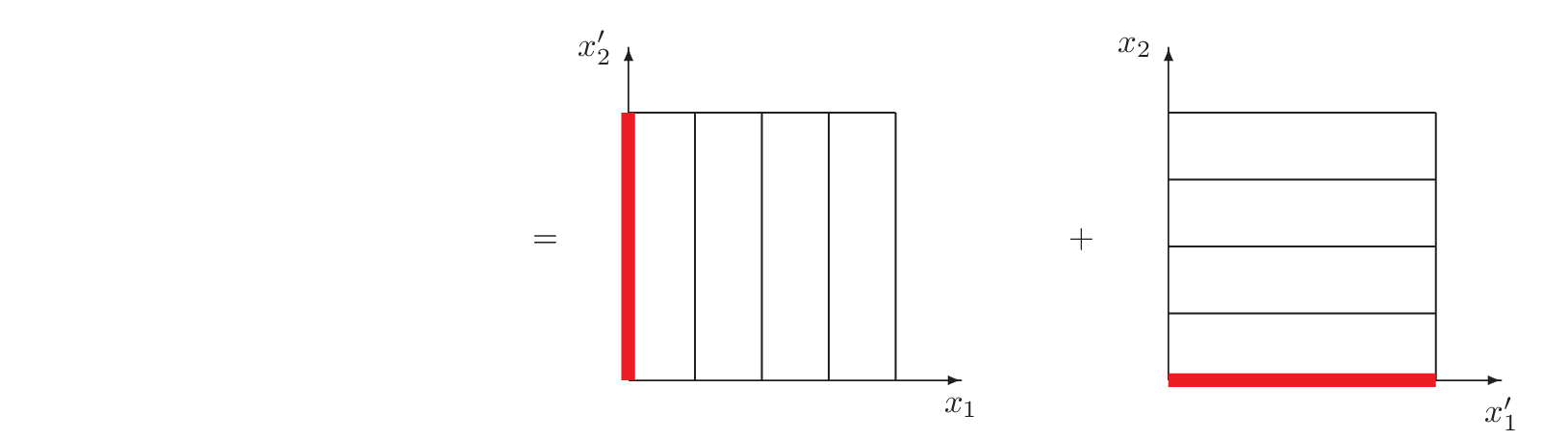}
\end{center}
\caption{
Illustration of sector decomposition and blow-up for a simple example.
}
\label{chapter_sector_decomposition:fig_blow_up}
\end{figure}

Fig.~\ref{chapter_sector_decomposition:fig_blow_up} illustrates this for the simple example $S=\{1,2\}$. 
Eq.~(\ref{chapter_sector_decomposition:decomposition}) gives the decomposition into the two sectors
$x_1>x_2$ and $x_2>x_1$.
Eq.~(\ref{chapter_sector_decomposition:substitution}) transforms the triangles into squares.
This transformation is one-to-one for all points except the origin.
The origin is replaced by the line $x_1=0$ in the first sector and by the line $x_2=0$ in the second sector.
In mathematics this is known as a blow-up.
\\
\\
{\bf Step 4}: The singular behaviour of the integral depends now only on the factor
\bq
  \prod\limits_{i=1}^{n}x_{i}^{a_{i}+\epsilon b_{i}}.
\eq
For every $x_{j}$ with $a_{j}<0$ we perform a Taylor expansion around $x_{j}=0$ 
in order to extract the possible $\epsilon$-poles.
In the variable $x_j$ we write
\bq
 \int\limits_{0}^{1} dx_{j} \; x_{j}^{a_{j}+b_{j}\eps} \mathcal{I}(x_{j})
 = 
 \int\limits_{0}^{1} dx_{j} \; x_{j}^{a_{j}+b_{j}\eps}
   \left(\sum\limits_{p=0}^{\left|a_{j}\right|-1} \frac{x_{j}^{p}}{p!} \mathcal{I}^{(p)} 
         + \mathcal{I}^{(R)}(x_j)
   \right)
\eq
where we defined 
$\mathcal{I}^{(p)} = \left. \left(\frac{\partial}{\partial x_{j}}\right)^{p} \mathcal{I}(x_{j})\right|_{x_{j}=0}$.
The remainder term
\bq
 \mathcal{I}^{(R)}(x_j)
 & = & 
 \mathcal{I}(x_{j}) - \sum\limits_{p=0}^{\left|a_{j}\right|-1} \frac{x_{j}^{p}}{p!} \mathcal{I}^{(p)}
\eq
does not lead to $\eps$-poles in the $x_{j}$-integration.
The integration in the pole part can be carried out analytically:
\bq
 \int\limits_{0}^{1} dx_{j} \; x_{j}^{a_{j}+b_{j}\eps}
 \;
 \frac{x_{j}^{p}}{p!} \mathcal{I}^{(p)} 
  & = &
   \frac{1}{a_{j}+b_{j}\eps+p+1} \frac{\mathcal{I}^{(p)}}{p!}.
\eq
This procedure is repeated for all variables $x_j$ for which $a_j<0$.
\\
\\
{\bf Step 5}: All remaining integrals are now by construction finite.
We can now expand all expressions in a Laurent series in $\eps$
\bq
 \sum\limits_{i=A}^{B}C_{i}\epsilon^{i}+O\left(\epsilon^{B}\right)
\eq
and truncate to the desired order.
\\
\\
{\bf Step 6}: It remains to compute the coefficients of the Laurent series.
These coefficients contain finite integrals, which can be evaluated numerically
by Monte Carlo integration.
This completes the algorithm of sector decomposition.

There are public codes implementing this algorithm \cite{Bogner:2007cr,Smirnov:2008py,Smirnov:2009pb,Carter:2010hi,Borowka:2012yc,Borowka:2015mxa,Borowka:2017idc}.

\begin{digression} {\bf Blow-ups}
\\
Let's define a blow-up in mathematical terms:
We start with the blow-up of a coordinate subspace.
By choosing appropriate coordinates we may always arrange to be in this situation.
Consider ${\mathbb C}^n$ and the submanifold $Z$ defined by
\bq
 x_1 = x_2 = ... = x_k = 0.
\eq
We have $\dim_{\mathbb C} Z = n-k$ and
\bq
 Z & = & 
 \left\{ \left( 0, ..., 0, x_{k+1}, ..., x_n \right) \; | \; x_j \in {\mathbb C}, \; k+1 \le j \le n \right\}.
\eq
Let $P$ be the subset of 
\bq
 {\mathbb C}^n \times {\mathbb C}{\mathbb P}^{k-1},
\eq
which satisfies the equations
\bq
 x_i y_j - x_j y_i & = & 0,
 \;\;\;\;\;\;
 i,j \in \{1,...,k\},
\eq
where $[y_1:y_2:...:y_k]$ are homogeneous coordinates on ${\mathbb C}{\mathbb P}^{k-1}$.
In other words,
\bq
 P & = & 
 \left\{ (x,y) \in {\mathbb C}^n \times {\mathbb C}{\mathbb P}^{k-1}
       | x_i y_j - x_j y_i = 0, \;\; 1 \le i,j \le k \right\}.
\eq
The blow-up map
\bq
 \pi & : & P \rightarrow {\mathbb C}^n
 \nonumber \\
 & &
           \left(x,y\right) \rightarrow x,
\eq
is an isomorphism away from $Z$:
To see this, let $x\in {\mathbb C}^n\backslash Z$. Then there is at least one $x_j$ with 
$x_j\neq 0$ for $1\le j\le k$.
We then have
\bq
 y_i & = & \frac{x_i}{x_j} y_j,
 \;\;\;\;\;\;
 i \in \{1,...,k\}.
\eq
Since $y \in {\mathbb C}{\mathbb P}^{k-1}$ it follows that $y_j\neq 0$
and we find
\bq
 y & = & \left[ x_1 : x_2 : ... : x_k \right].
\eq
On the other hand we have 
for $x \in Z$ that $x_j=0$ for all $1\le j \le k$ and the equations
$x_i y_j - x_j y_i = 0$ are trivially satisfied for all $1\le i \le k$.
Therefore for $x\in Z$ any point $y \in {\mathbb C}{\mathbb P}^{k-1}$ is allowed.
We define
\bq
 E & = & Z \times {\mathbb C}{\mathbb P}^{k-1}
\eq
to be the 
\index{exceptional divisor}
{\bf exceptional divisor}.
The restriction of $\pi$ to $E$ 
\bq
 \left.\pi \right|_E & : & E \rightarrow Z
\eq
gives a fibration with fibre ${\mathbb C}{\mathbb P}^{k-1}$.
\end{digression}

\section{Hironaka's polyhedra game}
\label{chapter_sector_decomposition:hironaka}

In step 2 of the algorithm we choose at each iteration a subset
$S=\left\{ \alpha_{1},\,...,\, \alpha_{k}\right\} \subseteq \left\{ 1, \,...\, n \right\}$,
until we achieve the form of eq.~(\ref{chapter_sector_decomposition:monomialised}) for all polynomials $P_j$.
Up to now we didn't specify how the subset $S$ is chosen.
We will now fill in the details.

Choosing the set $S$ is a non-trivial issue. We have to ensure that we reach
the form of eq.~(\ref{chapter_sector_decomposition:monomialised}) in a finite number of iterations.
In particular, the iteration should not lead to an infinite loop.

To illustrate the problem, let us start with a strategy which does not work:
Suppose we choose $S$ as 
a minimal subset $S=\left\{ \alpha_{1}, \dots, \alpha_{k}\right\} $
such that at least one polynomial $P_j$ vanishes for $x_{\alpha_{1}} = \dots = x_{\alpha_{k}} = 0$.
By a minimal set we mean a set which does not contain a proper subset having this property.
If $S$ contains only one element, $S=\left\{ \alpha\right\}$, then the corresponding Feynman parameter
factorises from $P_j$.
A relative simple example shows, that this procedure may lead to an infinite loop:
If one considers the polynomial
\bq
P\left(x_{1}, x_{2}, x_{3}\right)=x_{1}x_{3}^{2}+x_{2}^{2}+x_{2}x_{3},
\eq
then the subset $S=\left\{ 1, 2 \right\}$ is an allowed choice,
as $P\left(0, 0, x_3\right)=0$ and $S$ is minimal. 
In the first sector the substitution (\ref{chapter_sector_decomposition:substitution}) reads 
$x_{1}=x_{1}^{\prime},\, x_{2}=x_{1}^{\prime}x_{2}^{\prime},\, x_{3}=x_{3}^{\prime}$.
It yields 
\bq
P\left(x_{1}, x_{2}, x_{3}\right)
 =
 x_{1}^{\prime}x_{3}^{\prime2}+x_{1}^{\prime2}x_{2}^{\prime2}+x_{1}^{\prime}x_{2}^{\prime}x_{3}^{\prime}
 =
 x_{1}^{\prime}\left(x_{3}^{\prime2}+x_{1}^{\prime}x_{2}^{\prime2}+x_{2}^{\prime}x_{3}^{\prime}\right)
 =
 x_{1}^{\prime}P\left(x_{1}^{\prime}, x_{3}^{\prime}, x_{2}^{\prime}\right).
\eq
The original polynomial has been reproduced, which leads to an infinite recursion.

To avoid this situation we need a strategy for choosing $S$, for which we can show that the recursion always terminates.
This is a highly non-trivial problem. 
It is closely related to the resolution of singularities 
of an algebraic variety over a field of characteristic zero \cite{Hironaka:1964}.
The polyhedra game was introduced by Hironaka to illustrate the problem of resolution of singularities.
The polyhedra game can be stated with little mathematics.
Any solution to the polyhedra game will correspond to a strategy for choosing the subsets $S$
within sector decomposition, which guarantee termination.

Hironaka's polyhedra game is played by two players, A and B. 
They are given a finite set $M$ of points
$m=\left(m_{1},\,...,\,m_{n}\right) \in \mathbb{N}_{0}^{n}$.
We denote by $\Delta \subset\mathbb{R}_{\ge 0}^{n}$ the positive convex hull of the set $M$.
It is given by the convex hull of the set
\bq
\bigcup\limits_{m\in M}\left(m+\mathbb{R}_{\ge 0}^{n}\right).
\eq
The two players compete in the following game:
\begin{enumerate}
\item Player A chooses a non-empty subset $S\subseteq\left\{ 1,\,...,\, n\right\}$.
\item Player B chooses one element $i$ out of this subset $S$. 
\end{enumerate}
Then, according to the choices of the players,
the components of all $\left(m_{1},\,...,\,m_{n}\right) \in M$
are replaced by new points $\left(m_{1}^{\prime},\,...,\,m_{n}^{\prime}\right)$,
given by:
\bq
\label{chapter_sector_decomposition:update_polyhedron}
m_{j}^{\prime} & = & m_{j}, \;\;\; \textrm{if }j\neq i, \nonumber \\
m_{i}^{\prime} & = & \sum_{j\in S} m_{j} -c,
\eq
where for the moment we set $c=1$. 
This defines the set $M^\prime$.
One then sets $M=M^\prime$ and goes back to step 1.
Player A wins the game if, after a finite number of moves, 
the polyhedron $\Delta$ is of the form
\bq
\label{chapter_sector_decomposition:termination}
 \Delta & = & m+\mathbb{R}_{\ge 0}^{n},
\eq
i.e. generated by one point.
If this never occurs, player $B$ has won.
The challenge of the polyhedra game is to show that player $A$ always has
a winning strategy, no matter how player $B$ chooses his moves.
\begin{figure}
\begin{center}
\includegraphics[scale=0.75]{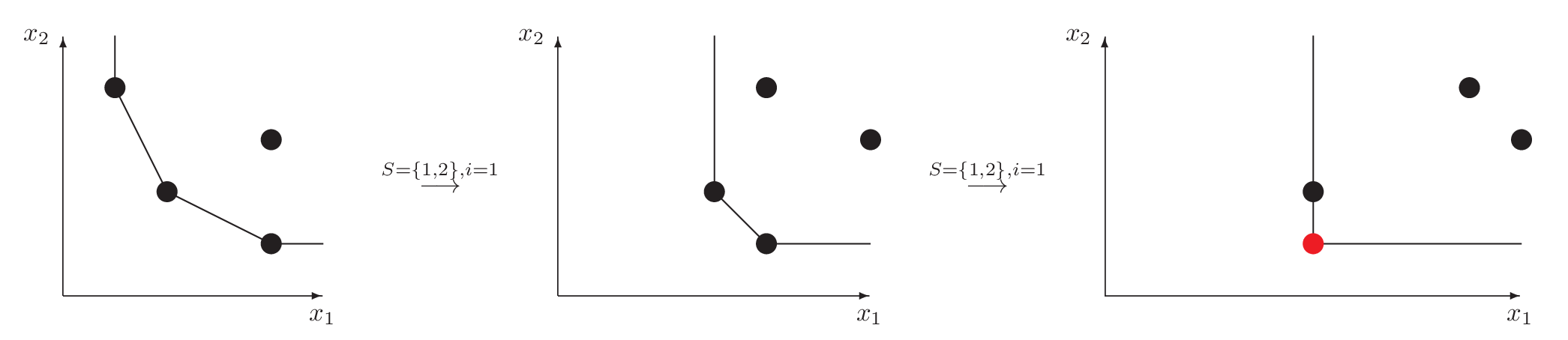}
\end{center}
\caption{
Illustration of Hironaka's polyhedra game.
}
\label{chapter_sector_decomposition:fig_Hironaka_game}
\end{figure}
A simple illustration of Hironaka's polyhedra game in two dimensions is given in
fig.~\ref{chapter_sector_decomposition:fig_Hironaka_game}. Player A always chooses $S=\{1,2\}$.

A winning strategy for Hironaka's polyhedra game
translates directly into a strategy for choosing the sub-sectors within sector decomposition which
guarantees termination.
Without loss of generality we can assume that we have just one polynomial $P$
in eq.~(\ref{chapter_sector_decomposition:basic_integral}).
(If there are several polynomials, we obtain a single polynomial by multiplying them together.
As only the zero-sets of the polynomials are relevant, the exponents $\lambda_j$ can be neglected.)
The polynomial $P$ has the form
\bq
 P & = & 
  \sum\limits_{i=1}^p c_i x_1^{m_1^{(i)}} x_2^{m_2^{(i)}} ... x_n^{m_n^{(i)}}.
\eq
The $n$-tuple $m^{(i)}=\left(m^{(i)}_{1},\,...,\,m^{(i)}_{n}\right)$
defines a point in $\mathbb{N}_{0}^{n}$ and 
$M=\left\{m^{(1)},\dots,m^{(p)} \right\}$ is the set of all such points.
Substituting the parameters $x_{j}$ according to equation~(\ref{chapter_sector_decomposition:substitution}) 
and factoring out a term $x_{i}^c$ yields the same polynomial as replacing the powers
$m_{j}$ according to equation~(\ref{chapter_sector_decomposition:update_polyhedron}). 
In this sense, one iteration of the sector decomposition corresponds to one move in Hironaka's game. 
Reducing $P$ to the form~(\ref{chapter_sector_decomposition:monomialised})
is equivalent to achieving~(\ref{chapter_sector_decomposition:termination}) in the polyhedra game.
Finding a strategy which guarantees termination of the iterated sector decomposition 
corresponds to a winning strategy for player $A$ in the polyhedra game.
Note that we really need a strategy that guarantees player A's victory for every
choice player B can take, because the sector decomposition has to be carried out
in every appearing sector. In other words, we sample over all possible decisions of B.

There are winning strategies for 
Hironaka's polyhedra game \cite{Hironaka:1964,Spivakovsky:1983,Encinas:2002,Hauser:2003,Zeillinger:2006,Smirnov:2008aw,Kaneko:2009qx}.
Common to all strategies is a sequence of positive numbers associated to the polynomials.
All strategies enforce this sequence to decrease with each step in the iteration with respect
to lexicographical ordering.
As the sequence cannot decrease forever, the algorithm is guaranteed to terminate.
The actual construction of this sequence will differ for different strategies.

An an example we discuss here
Spivakovsky's strategy \cite{Spivakovsky:1983}, which was the first solution to Hironaka's polyhedra game.
To state the strategy, we need a few auxiliary definitions:
We define $\omega\left(\Delta\right)\in\mathbb{R}_{\ge 0}^{n}$ as 
the vector given by the minima of the individual coordinates of elements in $\Delta$: 
\bq
\omega_{i} & = & \textrm{min}\left\{ \nu_{i}\mid\nu\in\Delta\right\},
 \;\;\;
 i=1,\dots,n.
\eq
Furthermore we write 
$\tilde{\Delta}=\Delta-\omega\left(\Delta\right)$
and 
$\tilde{\nu}_{i}=\nu_{i}-\omega_{i}$.
For a subset $\Gamma\subseteq\left\{ 1,\,...,\, n\right\} $ we define
\bq
d_{\Gamma}\left(\Delta\right) & = &
 \textrm{min}\left\{ \sum_{j\in\Gamma}\nu_{j}\mid\nu\in\Delta\right\} 
 \;\;\;\textrm{and}\;\;\; 
 d\left(\Delta\right)=d_{\left\{ 1,\,...,\, n\right\} }\left(\Delta\right).
\eq
We then define a sequence of sets
\bq
\label{chapter_sector_decomposition:spivakovsky_set_sequence}
 \left( I_0, \Delta_0, I_1, \Delta_1, ..., I_r, \Delta_r \right)
\eq
starting from
\bq
 I_0 = \left\{ 1,\,...,\, n\right\},
 & &
 \Delta_0 = \Delta.
\eq
For each $\Delta_k$ we define a set 
$H_k$ by
\bq
H_k & = & 
 \left\{ j\in I_k 
         \mid \; \exists \; \nu\in\Delta_k 
              \textrm{ such that }\sum_{i \in I_k} \nu_{i}=d\left(\Delta_k\right)
              \textrm{ and }\tilde{\nu}_{j}\neq0\right\}.
\eq
$I_{k+1}$ is given by
\bq
 I_{k+1} & = & I_{k}\backslash H_k.
\eq
In order to define $\Delta_{k+1}$ we first define 
for the two complementary subsets $H_k$ and $I_{k+1}$ of $I_k$ the set
\bq
M_{H_k} & = &
 \left\{ \nu\in\mathbb{R}_{\ge 0}^{I_k} 
         \mid
         \sum_{j\in H_k}\nu_{j}<1\right\}
\eq
and the projection
\bq
 P_{H_k} & : & M_{H_k} \longrightarrow \mathbb{R}_{\ge 0}^{I_{k+1}},
 \nonumber \\
 & &
 P_{H_k}\left(\alpha,\,\beta\right)=\frac{\alpha}{1-\left|\beta\right|},
 \;\;\;
 \alpha \in \mathbb{R}_{\ge 0}^{I_{k+1}},
 \;\;\;
 \beta \in \mathbb{R}_{\ge 0}^{H_{k}},
 \;\;\;
 \left| \beta \right| = \sum\limits_{j\in H_k} \beta_j.
\eq
Then $\Delta_{k+1}$ is given by
\bq
\label{chapter_sector_decomposition:def_projected_set}
\Delta_{k+1} & = & 
 P_{H_k}
 \left(M_{H_k}
 \cap
  \left(\frac{\tilde{\Delta}_{k}}{d\left(\tilde{\Delta}_{k}\right)}\cup\Delta_{k}\right)
 \right),
\eq
where 
$\tilde{\Delta}_{k} = \Delta_k - \omega\left(\Delta_k\right)$.
The sequence in eq.~(\ref{chapter_sector_decomposition:spivakovsky_set_sequence}) stops if either 
$d\left(\tilde{\Delta}_{r}\right)=0$
or $\Delta_{r}=\emptyset$.
Based on the sequence in eq.~(\ref{chapter_sector_decomposition:spivakovsky_set_sequence}) player A chooses now the
set $S$ as follows:
\begin{enumerate}

\item If $\Delta_{r}=\emptyset$, player A chooses $S=\left\{ 1,\,...,\, n\right\} \backslash I_{r}$.

\item If $\Delta_{r}\neq\emptyset$, player A first chooses a minimal subset $\Gamma_{r}\subseteq I_{r}$,
such that $\sum_{j\in \Gamma_{r}}\nu_{j}\geq1$ for all $\nu\in\Delta_{r}$ and sets
$S=\left(\left\{ 1,\,...,\, n\right\} \backslash I_{r}\right)\cup \Gamma_{r}$.
\end{enumerate}
To each stage of the game (i.e. to each $\Delta$), we can associate
a sequence of $2r+2$ numbers
\bq
\delta\left(\Delta\right) & = &
 \left( d\left(\tilde{\Delta}\right),\,\# I_{1},\, d\left(\tilde{\Delta}_{1}\right),
        \dots,
        \# I_{r},\, d\left(\tilde{\Delta}_{r}\right),\, d\left(\Delta_r\right)\right),
\eq
adopting the conventions $\tilde{\emptyset}=\emptyset$ and $d(\emptyset)=\infty$.
The above strategy forces $\delta(\Delta)$
to decrease with each move with respect to lexicographical ordering.
Further, it can be shown that $\delta(\Delta)$
cannot decrease forever.
Hence player A is guaranteed to win. 
The proof is given in \cite{Spivakovsky:1983}.

\section{Numerical periods}
\label{chapter_sector_decomposition:periods}

As a spin-off of the algorithm of sector decomposition we may prove a theorem
related to the coefficients $I^{(j)}$ of the Laurent expansion in the dimensional regularisation parameter $\eps$
of a Feynman integral $I$.

In order to prepare the ground, 
we start with some sets of numbers:
The natural numbers $\mathbb{N}$,
the integer numbers $\mathbb{Z}$,
the rational numbers $\mathbb{Q}$,
the real numbers $\mathbb{R}$ and 
the complex numbers $\mathbb{C}$
are all well-known. More refined is already the set of algebraic numbers, denoted by $\overline{\mathbb{Q}}$.
An algebraic number is a solution of a polynomial equation with rational
coefficients:
\bq
 x^n + a_{n-1} x^{n-1} + \cdots + a_0 = 0,
 \;\;\; a_j \in \mathbb{Q}.
\eq
As all such solutions lie in $\mathbb{C}$, the set of algebraic numbers $\overline{\mathbb{Q}}$ 
is a sub-set of the complex numbers $\mathbb{C}$.
Numbers which are not algebraic are called transcendental.
The sets $\mathbb{N}$, $\mathbb{Z}$, $\mathbb{Q}$ and $\overline{\mathbb{Q}}$ are countable, whereas
the sets $\mathbb{R}$, $\mathbb{C}$ and the set of transcendental numbers are uncountable.

We now introduce the set of numerical periods ${\mathbb P}$.
The motivation originates in the theory of singly and doubly periodic functions $f(z)$ of a complex variable $z$:
We know that the exponential function is a periodic function with period $(2\pi i)$:
\bq
 \exp\left(z +2 \pi i\right) & = & \exp\left(z\right)
 \;\;\;\;\;\;
 \forall \; z \; \in \; {\mathbb C}.
\eq
In chapter~\ref{chapter_elliptics} we will discuss doubly periodic functions.
A standard example for a doubly periodic function is Weierstrass's $\wp$-function.
Let us denote the two periods of Weierstrass's $\wp$-function by $\psi_1$ and $\psi_2$:
\bq
 \wp\left(z+\psi_1\right)
 \; = \;
 \wp\left(z+\psi_2\right)
 \; = \;
 \wp\left(z\right).
\eq
It is an observation that 
the period of the singly periodic function $\exp(z)$ 
and the periods $\psi_1, \psi_2$
of the doubly periodic function $\wp(z)$ can be expressed as integrals involving only algebraic functions:
For the period of the exponential function we have
\bq
 2 \pi i
 & = &
 2 i \int\limits_{-1}^1 \frac{dt}{\sqrt{1-t^2}}.
\eq
Weierstrass's $\wp$-function is associated to the elliptic curve $y^2=4x^3-g_2x-g_3$.
Assume that the two constants $g_2$ and $g_3$ are algebraic numbers.
The periods of Weierstrass's $\wp$-function can be written as
\bq
 \psi_1 = 2 \int\limits_{t_1}^{t_2} \frac{dt}{\sqrt{4t^3-g_2t-g_3}},
 & &
 \psi_2 = 2 \int\limits_{t_3}^{t_2} \frac{dt}{\sqrt{4t^3-g_2t-g_3}},
\eq
where $t_1$, $t_2$ and $t_3$ are the roots of the cubic equation $4t^3-g_2t-g_3=0$.

This observation motivated Kontsevich and Zagier \cite{Kontsevich:2001}
to define the set of numerical periods ${\mathbb P}$:
\begin{tcolorbox}
\index{numerical period}
{\bf Numerical period}:
\\
\\
A numerical period is
a complex number whose real and imaginary parts are values
of absolutely convergent integrals of rational functions with rational coefficients,
over domains in $\mathbb{R}^n$ given by polynomial inequalities with rational coefficients.
\end{tcolorbox}

We denote the set of numerical periods by $\mathbb{P}$. 
We may replace in the definition above any occurrence of ``rational'' with algebraic, this will not alter the set of numbers.
The algebraic numbers are contained in the set of numerical periods: $\overline{\mathbb{Q}} \in \mathbb{P}$.
In addition, $\mathbb{P}$ contains transcendental numbers, for example the transcendental number $\pi$
\bq
 \pi & = & \iint\limits_{x^2+y^2\le1} dx \; dy,
\eq
or the transcendental number $\ln(2)$
\bq
 \ln\left(2\right)
 & = & 
 \int\limits_1^2 \frac{dx}{x}.
\eq
On the other hand, it is conjectured that the basis of the natural logarithm $e$
and Euler's constant $\Eulerconstant$ are not periods.
The number $(2 \pi i)$ clearly is a period, 
but currently it is not known if the inverse $(2\pi i)^{-1}$ belongs to ${\mathbb P}$
or not.
Although there are uncountably many numbers, which are not periods, only very recently an example
for a number which is not a period has been found \cite{Yoshinaga:2008}.

Periods are a countable set of numbers, lying between $\overline{\mathbb{Q}}$ and $\mathbb{C}$.
The set of periods $\mathbb{P}$ is a $\overline{\mathbb{Q}}$-algebra.
In particular the sum and the product of two periods are again periods.

Let us now turn to Feynman integrals:
\begin{theorem}
\label{chapter_sector_decomposition:Feynman_period}
Consider a Feynman integrals as in eq.~(\ref{chapter_sector_decomposition:Feynman_parameter_representation})
with Laurent expansion as in eq.~(\ref{chapter_sector_decomposition:Laurent_expansion}).
Assume that all kinematic invariants $x_1, \dots, x_{\NB}$ are in the Euclidean region and algebraic:
\bq
 x_j \; \ge \; 0,
 & &
 x_j \; \in \; \overline{\mathbb{Q}},
 \;\;\;\;\;\;
 1 \; \le \; j \; \le \; \NB.
\eq
Then the coefficients $I^{(j)}$ of the Laurent expansion in $\eps$ are numerical periods:
\bq
 I^{(j)} & \in & {\mathbb P}.
\eq
\end{theorem}
The proof of this theorem follows from the algorithm of sector decomposition \cite{Belkale:2003,Bogner:2007mn}:  
If the kinematic variables $x_j$ are in the Euclidean region and algebraic, the integral of each sector
in the sector decomposition is a numerical period.
As the sum of numerical periods is again a numerical period, the theorem follows.

\section{Effective periods and abstract periods}
\label{chapter_sector_decomposition:effective_periods}

There is a more formal definition of periods as follows \cite{Kontsevich:2001}:
Let $X$ be a smooth algebraic variety of dimension $n$ defined over $\mathbb{Q}$ 
and $D \subset X$ a divisor with normal crossings.
(A normal crossing divisor is a subvariety of dimension $n-1$, which looks locally like a union of coordinate hyperplanes.)
Further let $\omega$ be an algebraic differential form on $X$ of degree $n$
and $\Delta$ a singular $n$-chain on the complex manifold $X(\mathbb{C})$ with boundary on the divisor $D(\mathbb{C})$.
We thus have a quadruple $(X,D,\omega,\Delta)$. 
To each quadruple we can associate a complex number $\mathrm{period}(X,D,\omega,\Delta)$ 
called the period of the quadruple and given by the integral
\bq
 \mathrm{period}(X,D,\omega,\Delta)
 & = & 
 \int\limits_\Delta \omega.
\eq
It is clear that the period of the quadruple is an element of $\mathbb{P}$,
and that to any element $p \in \mathbb{P}$ one can find a quadruple,
such that $\mathrm{period}(X,D,\omega,\Delta)=p$.
The period map is therefore surjective. 
The interesting question is whether the period map is also injective.
As it stands above, the period map is certainly not injective for trivial reasons.
For example, a simple change of variables can lead to a different quadruple, but does not change the period.
One therefore considers equivalence classes of quadruples modulo relations induced by 
linearity in $\omega$ and $\Delta$,
changes of variables and Stokes' formula.
The vector space over $\mathbb{Q}$ of the equivalence classes of quadruples $(X,D,\omega,\Delta)$
is called the space of 
\index{effective period}
{\bf effective periods} 
and denoted by ${\mathcal P}$.
${\mathcal P}$ is an algebra.
It is conjectured that the period map from ${\mathcal P}$ to $\mathbb{P}$ is injective and therefore an isomorphism \cite{Grothendieck:1966,Kontsevich:2001,Ayoub:2011}.
This would imply that all relations between numerical periods are due to linearity, change of variables and Stokes' formula. 
Let us summarise:
\begin{tcolorbox}
\index{numerical period}
{\bf Effective period}:
\\
\\
We consider quadruples $(X,D,\omega,\Delta)$,
where $X$ is a smooth algebraic variety of dimension $n$ defined over $\mathbb{Q}$,
$D \subset X$ a divisor with normal crossings,
$\omega$ an algebraic differential form on $X$ of degree $n$
and $\Delta$ a singular $n$-chain on the complex manifold $X(\mathbb{C})$ with boundary on the divisor $D(\mathbb{C})$.

Two quadruples $(X,D,\omega,\Delta)$ and $(X',D',\omega',\Delta')$ are called equivalent, if
\bq
 \int\limits_\Delta \omega
 & = &
 \int\limits_{\Delta'} \omega'
\eq
and this can be derived 
from 
linearity in $\omega$ and $\Delta$,
a change of variables and Stokes' formula.

The equivalence classes are called {\bf effective periods} and the algebra of effective periods is denoted by ${\mathcal P}$.
The period map
\bq 
 \mathrm{period} & : & {\mathcal P} \rightarrow {\mathbb P},
 \nonumber \\
 & & (X,D,\omega,\Delta) \rightarrow \int\limits_\Delta \omega
\eq
maps every effective period to a numerical period.
\end{tcolorbox}

In order to make the definition more concrete, we consider as an example the quadruple given by
$X({\mathbb C}) = {\mathbb C} \backslash \{0\}$, $D=\emptyset$, $\omega = dz/z$ and $\Delta$ the path along the unit circle in the counter-clockwise
direction.
We have
\bq
 \mathrm{period}\left(X,D,\omega,\Delta\right) & = & 2 \pi i.
\eq
As a second example let us consider the quadruple
$X({\mathbb C}) = {\mathbb C}$, $D=\{1,2\}$, $\omega = dz/z$ and $\Delta$ the path from $1$ to $2$ along the real line.
We have
\bq
 \mathrm{period}\left(X,D,\omega,\Delta\right) & = & \ln\left(2\right).
\eq

As in the case of numerical periods it is not known whether there is a quadruple in ${\mathcal P}$, 
whose period is $(2 \pi i)^{-1}$.
One therefore adjoins to ${\mathcal P}$ formally the inverse of the element whose period is $(2\pi i)$ and writes
${\mathcal P}[\frac{1}{2 \pi i}]$ for the so obtained algebra.
Elements of ${\mathcal P}[\frac{1}{2 \pi i}]$ are called 
\index{abstract period}
abstract periods.

\begin{tcolorbox}[breakable]
\index{numerical period}
{\bf Abstract period}:
\\
\\
Adjoin formally to the algebra of effective periods an element $I$ with
\bq
\mathrm{period}\left(I\right)
 & = &
 \frac{1}{2\pi i}.
\eq
The enlarged algebra is called the algebra of abstract periods and denoted by
\bq
 {\mathcal P}\left[\frac{1}{2 \pi i}\right].
\eq
\end{tcolorbox}
\begin{digression} {\bf Algebraic varieties over arbitrary fields}
\\
Let ${\mathbb F}$ be a field and $\overline{\mathbb F}$ its algebraic closure.
Let ${\mathbb F}[t_1,...,t_n]$ be the ring of polynomials
over the field ${\mathbb F}$ in $n$ variables $t_1, \dots, t_n$.
An element $f \in {\mathbb F}[t_1,...,t_n]$ is a polynomial in $t_1,\dots,t_n$ with coefficients
from ${\mathbb F}$.
Let $A \subset {\mathbb F}[t_1,...,t_n]$ be a set of such polynomials.
The corresponding affine algebraic set is given by
\bq
 V\left(A\right) & = &
 \{ \; x \in \overline{\mathbb F}^n \; | \; f(x) = 0 \;\; \forall f \in A \; \}.
\eq
Note that one takes the algebraic closure $\overline{\mathbb F}$ of ${\mathbb F}$.

Let's now specialise to the case where ${\mathbb F}$ is a subfield of ${\mathbb C}$.
The most important example is given by ${\mathbb F}={\mathbb Q}$ (the rational numbers).
The algebraic closure is $\overline{\mathbb F}=\overline{\mathbb Q}$ (the algebraic numbers).
Let $A \subset {\mathbb Q}[t_1,...,t_n]$ and denote
\bq
 X & = &
 \{ \; x \in \overline{\mathbb Q}^n \; | \; f(x) = 0 \;\; \forall f \in A \; \}.
\eq
By $X({\mathbb C})$ we understand the affine algebraic set
\bq
 X\left({\mathbb C}\right) & = &
 \{ \; x \in {\mathbb C}^n \; | \; f(x) = 0 \;\; \forall f \in A \; \}.
\eq
To see the difference between $X$ and $X({\mathbb C})$ consider $A=\{t_1^2+t_2^2-25\}$.
The point
\bq
 \left(x_1,x_2\right) & = & \left(\pi,\sqrt{25-\pi^2}\right)
\eq
is a point of $X({\mathbb C})$, but not of $X$ (the coordinates of any point of $X$ are algebraic numbers).
We may equip $X({\mathbb C})$ with the topology induced from the standard topology on ${\mathbb C}^n$.
(This means in particular that we are not using the Zariski topology, 
which is otherwise widely used in algebraic geometry.)
The set $X({\mathbb C})$ together with this topology becomes then a topological space, 
usually denoted by $X^{\mathrm{an}}$ and called the
\index{analytification}
{\bf analytification} of $X$.
\end{digression}

%% file: hopf.tex
\newpage
\chapter{Hopf algebras, coactions and symbols}
\label{chapter_hopf}

In this chapter we investigate more formal aspects of Feynman integrals.
We first introduce Hopf algebras and discuss where they appear in particle physics.
We then focus on multiple polylogarithms.
There are three Hopf algebras (with different coproducts) associated with multiple polylogarithms.
The first two are combinatorial in nature 
and stem from the shuffle algebra and the quasi-shuffle algebra, respectively.
The third one is motivic. 
We introduce motivic multiple polylogarithms (sounds complicated, but in the end it boils down to the fact that
we introduce objects, which have all the known relations of multiple polylogarithms and no other relations).
We also introduce the de Rham multiple polylogarithms.
The de Rham multiple polylogarithm form a Hopf algebra.
This Hopf algebra coacts on the motivic multiple polylogarithms, turning the motivic multiple polylogarithms
into a comodule.
We then study the symbol and 
the iterated coaction of multiple polylogarithms.
There is a practical application:
The symbol (or the iterated coaction) can be used to simplify long expressions of multiple polylogarithms.

Multiple polylogarithms are in general multi-valued functions. 
In the last section of this chapter we will study a systematic method to associate a single-valued function 
to a multiple polylogarithm.

Textbooks and lecture notes on Hopf algebras can be found in \cite{Sweedler,Kassel,Majid:1990vz,Manchon:2001bf,Frabetti:2008}.

\section{Hopf algebras}
\label{chapter_hopf:hopf_algebras}

Let us start with a brief history of Hopf algebras:
Hopf algebras were introduced in mathematics in 1941 to describe  
similar aspects of groups and algebras in a unified manner \cite{Hopf}.
An article by Woronowicz in 1987 \cite{Woronowicz}, 
which provided explicit examples of non-trivial (non-cocommutative) Hopf algebras, 
triggered the interest of the physics community.
This led to applications of Hopf algebras in the field of integrable systems and quantum groups.
In physics, Hopf algebras received a further boost in 1998, when
Kreimer and Connes re-examined the renormalisation of quantum field theories and showed
that the combinatorial aspects of renormalisation can be described by a Hopf algebra structure \cite{Kreimer:1998dp,Connes:1998qv}.
Since then, Hopf algebras have appeared in several facets of physics.

Let us now consider the definition of a Hopf algebra.
Let $R$ be a commutative ring with unit $1$.
An unitial associative algebra over the ring $R$ 
is an $R$-module together with an associative multiplication $\cdot$ and a unit $e$.
In this chapter we will always assume that the algebra has a unit and that the multiplication is associative.
In this chapter we simply write ``algebra'' whenever we mean a unitial associative algebra.
In physics, the ring $R$ will almost always be a field ${\mathbb F}$
(examples are the rational numbers ${\mathbb Q}$, the real numbers ${\mathbb R}$, or the complex number ${\mathbb C}$). 
In this case the $R$-module will actually be a ${\mathbb F}$-vector space.
Note that the unit $e$ can be viewed as a map from $R$ to $A$ and that the multiplication $\cdot$
can be viewed as a map from the tensor product $A \otimes A$ to $A$ 
(e.g., one takes two elements from $A$, multiplies them, and obtains one element as the outcome):
\begin{align}
\label{chapter_hopf:unit_multiplication}
 & \mbox{Multiplication:} &  \cdot \; : & \;\; A \otimes A \rightarrow A,
 \nonumber \\
 & \mbox{Unit:} & e \; : & \;\; R \rightarrow A.
\end{align}
Instead of multiplication and a unit, a coalgebra has the dual structures,
obtained by reversing the arrows in eq.~(\ref{chapter_hopf:unit_multiplication}):
a comultiplication $\Delta$ and a counit $\bar{e}$.
The 
\index{counit}
{\bf counit} 
$\bar{e}$ is a map from $A$ to $R$, whereas the 
\index{comultiplication}
{\bf comultiplication} 
$\Delta$ is a map from $A$ to
$A \otimes A$:
\begin{align}
 & \mbox{Comultiplication:} &  \Delta \; : & \;\; A \rightarrow A \otimes A, 
 \nonumber \\
 & \mbox{Counit:} & \bar{e} \; : & \;\; A \rightarrow R.
\end{align}
We will always assume that the comultiplication $\Delta$ is 
\index{coassociative}
{\bf coassociative}.
But what does coassociativity mean? We can easily derive it from associativity as follows:
For $a,b,c \in A$ associativity requires
\bq
\label{chapter_hopf:condition_associativity}
 \left( a \cdot b \right) \cdot c
 & = & 
 a \cdot \left( b \cdot c \right).
\eq
We can re-write condition~(\ref{chapter_hopf:condition_associativity}) in the form of a commutative diagram:
\bq
\begin{CD}
A \otimes A \otimes A @>{\mathrm{id} \otimes \cdot}>> A \otimes A \\
@VV{\cdot \otimes \mathrm{id}}V @VV{\cdot}V \\
A \otimes A @>{\cdot}>> A \\
\end{CD}
\eq
We obtain the condition for coassociativity by reversing all arrows and by exchanging multiplication with
comultiplication. We thus obtain the following commutative diagram:
\bq
\begin{CD}
A @>{\Delta}>> A \otimes A \\
@VV{\Delta}V @VV{\Delta \otimes \mathrm{id}}V \\
A \otimes A @>{\mathrm{id} \otimes \Delta}>> A \otimes A \otimes A  \\
\end{CD}
\eq
The general form of the coproduct is
\bq
\Delta(a) & = & \sum\limits_i a_i^{(1)} \otimes a_i^{(2)},
\eq
where $a_i^{(1)}$ denotes an element of $A$ appearing 
in the first slot of $A \otimes A$ and $a_i^{(2)}$ correspondingly denotes an element of $A$ 
appearing in the second slot.
\index{Sweedler's notation}
{\bf Sweedler's notation} \cite{Sweedler} consists of omitting the dummy index $i$ and the summation symbol:
\bq
\Delta(a) & = & 
a^{(1)} \otimes a^{(2)}
\eq 
The sum is implicitly understood. 
This is similar to Einstein's summation convention, except that the dummy summation index $i$ is also dropped. 
The superscripts ${}^{(1)}$ and ${}^{(2)}$ indicate that a sum is involved.
Using Sweedler's notation, coassociativity is equivalent to
\bq
 a^{(1) (1)} \otimes a^{(1) (2)} \otimes a^{(2)}
 & = &
 a^{(1)} \otimes a^{(2) (1)} \otimes a^{(2) (2)}.
\eq
As it is irrelevant whether we apply the second coproduct to the first or the second factor in the tensor product of $\Delta(a)$ ,
we can simply write
\bq
 \Delta^2\left(a\right)
 & = &
 a^{(1)} \otimes a^{(2)} \otimes a^{(3)}.
\eq
If the coproduct of an element $a \in A$ is of the form
\bq
 \Delta\left(a\right) 
 & = & 
 a \otimes a,
\eq
then $a$ is referred to as a 
\index{group-like element in a coalgebra}
{\bf group-like element}.
If the coproduct of $a$ is of the form
\bq
 \Delta\left(a\right)
 & = & 
 a \otimes e + e \otimes a,
\eq
then $a$ is referred to as a 
\index{primitive element in a coalgebra}
{\bf primitive element}.

In an algebra we have for the unit $1$ of the underlying ring $R$ and the unit $e$ of the algebra
the relation
\bq
 a 
 \;\; = \;\;
 1 \cdot a
 \;\; = \;\;
 e \cdot a
 \;\; = \;\;
 a
\eq
for any element $a \in A$ (together with the analogue relation $a = a \cdot 1 = a \cdot e = a$).
In terms of commutative diagrams this is expressed as
\bq
\label{chapter_hopf:axiom_unit}
\begin{CD}
A \otimes A @= A \otimes A \\
 @A{e \otimes \mathrm{id}}AA @VV{\cdot}V \\
R \otimes A @=^{\!\!\!\!\!\! \!\!\!\!\!\! \!\!\!\!\!\! \!\!\!\!\!\! \!\!\!\!\!\! \!\!\!\!\!\! \!\!\! \cong \,\,\,\,\,\, \,\,\,\,\,\, \,\,\,\,\,\, \,\,\,\,\,\, \,\,\,\,\,\,} A \\
\end{CD}
 & &
 \hspace*{15mm}
\begin{CD}
A \otimes A @= A \otimes A \\
 @A{\mathrm{id} \otimes e}AA @VV{\cdot}V \\
A \otimes R @=^{\!\!\!\!\!\! \!\!\!\!\!\! \!\!\!\!\!\! \!\!\!\!\!\! \!\!\!\!\!\! \!\!\!\!\!\! \!\!\! \cong \,\,\,\,\,\, \,\,\,\,\,\, \,\,\,\,\,\, \,\,\,\,\,\, \,\,\,\,\,\,} A \\
\end{CD}
\eq
In a coalgebra we have the dual relations obtained from eq.~(\ref{chapter_hopf:axiom_unit}) by reversing all arrows
and by exchanging multiplication with comultiplication as well as by exchanging the unit $e$ with the counit $\bar{e}$:
\bq
\label{chapter_hopf:axiom_counit}
\begin{CD}
A \otimes A @= A \otimes A \\
 @V{\bar{e} \otimes \mathrm{id}}VV @AA{\Delta}A \\
R \otimes A @=^{\!\!\!\!\!\! \!\!\!\!\!\! \!\!\!\!\!\! \!\!\!\!\!\! \!\!\!\!\!\! \!\!\!\!\!\! \!\!\! \cong \,\,\,\,\,\, \,\,\,\,\,\, \,\,\,\,\,\, \,\,\,\,\,\, \,\,\,\,\,\,} A \\
\end{CD}
 & &
 \hspace*{15mm}
\begin{CD}
A \otimes A @= A \otimes A \\
 @V{\mathrm{id} \otimes \bar{e}}VV @AA{\Delta}A \\
A \otimes R @=^{\!\!\!\!\!\! \!\!\!\!\!\! \!\!\!\!\!\! \!\!\!\!\!\! \!\!\!\!\!\! \!\!\!\!\!\! \!\!\! \cong \,\,\,\,\,\, \,\,\,\,\,\, \,\,\,\,\,\, \,\,\,\,\,\, \,\,\,\,\,\,} A \\
\end{CD}
\eq

A 
\index{bi-algebra}
{\bf bi-algebra} 
is an algebra and a coalgebra at the same time,
such that the two structures are compatible with each other.
In terms of commutative diagrams, the compatibility condition between the product and the
coproduct is expressed as
\bq
\begin{CD}
A \otimes A @>{\cdot}>> A @>{\Delta}>> A \otimes A \\
@VV{\Delta \otimes \Delta}V & & @AA{\cdot \otimes \cdot}A \\
A \otimes A \otimes A \otimes A & @>{\mathrm{id} \otimes \tau \otimes \mathrm{id}}>> & A \otimes A \otimes A \otimes A\\
\end{CD}
\eq
where $\tau : A \otimes A \rightarrow A \otimes A$ is the map, which exchanges the entries in the two slots:
$\tau(a \otimes b) = b \otimes a$.
Using Sweedler's notation, the compatibility between the multiplication and comultiplication is expressed as
\bq
\label{chapter_hopf:bialg}
 \Delta\left( a \cdot b \right)
 & = &
\left( a^{(1)} \cdot b^{(1)} \right)
 \otimes \left( a^{(2)} \cdot b^{(2)} \right).
\eq
It is common practice to write the right-hand side of eq.~(\ref{chapter_hopf:bialg}) as
\bq
\left( a^{(1)} \cdot b^{(1)} \right)
 \otimes \left( a^{(2)} \cdot b^{(2)} \right)
 & = & 
 \Delta\left(a\right) \Delta\left(b\right).
\eq
In addition, there is a compatibility condition between the unit and the coproduct
\bq
\label{chapter_hopf:compatibility_unit_coproduct}
\begin{CD}
 R \otimes R \cong R @>{e}>> A \\
 @V{e \otimes e}VV @VV{\Delta}V \\
 A \otimes A @= A \otimes A \\
\end{CD}
\eq
as well as a compatibility condition between the counit and the product, which is dual to eq.~(\ref{chapter_hopf:compatibility_unit_coproduct}):
\bq
\label{chapter_hopf:compatibility_counit_product}
\begin{CD}
 A @>{\bar{e}}>> R \cong R \otimes R \\
 @A{\cdot}AA @AA{\bar{e} \otimes \bar{e}}A \\
 A \otimes A @= A \otimes A \\
\end{CD}
\eq
The commutative diagrams in eq.~(\ref{chapter_hopf:compatibility_unit_coproduct}) and eq.~(\ref{chapter_hopf:compatibility_counit_product}) are equivalent to
\bq
 \Delta e = e \otimes e,
 \;\;\; \mbox{and} \;\;\; 
 \bar{e}\left(a \cdot b \right) = \bar{e}\left(a\right) \bar{e}\left(b\right),
 \;\;\; \mbox{respectively.}
\eq
An algebra $A$ is {\bf commutative} if for all $a,b \in A$ one has
\bq
\label{chapter_hopf:def_commutative}
 a \cdot b & = & b \cdot a.
\eq
A coalgebra $A$ is 
\index{cocommutative}
{\bf cocommutative}
if for all $a \in A$ one has
\bq
\label{chapter_hopf:def_cocommutative}
 a^{(1)} \otimes a^{(2)} 
 & = &
 a^{(2)} \otimes a^{(1)}.
\eq
With the help of the swap map $\tau$ we may express commutativity and cocommutativity equivalently as
\bq
 \cdot \tau = \cdot,
 \;\;\;\;\;\; \mbox{and} \;\;\;\;\;\;
 \tau \Delta = \Delta,
 \;\;\;\;\;\; \mbox{respectively.}
\eq 
A 
\index{Hopf algebra}
{\bf Hopf algebra}
is a bi-algebra with an additional map from $A$ to $A$, known as the 
\index{antipode}
{\bf antipode} 
$S$, which fulfils
\bq
\label{chapter_hopf:antipode_def1}
\begin{CD}
A @>{\bar{e}}>> R @>{e}>> A \\
@VV{\Delta}V & & @AA{\cdot}A \\
A \otimes A & @>{\mathrm{id} \otimes S}>{S \otimes \mathrm{id}}> & A \otimes A\\
\end{CD}
\eq
An equivalent formulation is
\bq
\label{chapter_hopf:antipode_def2}
 a^{(1)} \cdot S\left( a^{(2)} \right)
 \;\; = \;\;
 S\left(a^{(1)}\right) \cdot a^{(2)} 
 \;\; = \;\;
 e \cdot \bar{e}(a).
\eq
If a bi-algebra has an antipode (satisfying the commutative diagram~(\ref{chapter_hopf:antipode_def1}) 
or eq.~(\ref{chapter_hopf:antipode_def2})), then the antipode is unique.

If a Hopf algebra $A$ is either commutative or cocommutative, then
\bq
 S^2 & = & \mathrm{id}.
\eq
A bi-algebra $A$ is 
\index{graded bi-algebra}
{\bf graded}, if it has a decomposition
\bq
 A & = &
 \bigoplus\limits_{n \ge 0} A_n,
\eq
with
\bq
 A_n \cdot A_m \subseteq A_{n+m},
 & &
 \;\;\;
 \Delta\left(A_n\right) \subseteq \bigoplus\limits_{k+l=n} A_k \otimes A_l.
\eq
Elements in $A_n$ are said to have degree $n$.
The bi-algebra is 
\index{graded connected bi-algebra}
{\bf graded connected}, if in addition one has
\bq
 A_0 & = & R \cdot e.
\eq
It is useful to know that a graded connected bi-algebra is automatically a Hopf algebra \cite{Ehrenborg}.
In a graded Hopf algebra we denote by $\Delta_{i_1,\dots,i_k}(a)$ the projection of $\Delta^n(a)$ 
(where $n=i_1+\dots+i_k$)
onto
\bq
 A_{i_1} \otimes \dots \otimes A_{i_k},
\eq
i.e. we only keep those terms which have degree $i_j$ in the the $j$-th slot.

Let $A$ be a graded connected Hopf algebra.
We have
\bq
\label{chapter_hopf:kernel_counit}
 \mathrm{Ker}\left(\bar{e}\right)
 & = &
 \bigoplus\limits_{n \ge 1} A_n,
\eq
e.g $\bar{e}(a)=0$ if $a\in A_n$ with $n\ge 1$.
For the coproduct one often writes for $a\in A_n$ with $n\ge 1$
\bq
 \Delta\left(a\right) 
 & = &
 a \otimes e
 +
 e \otimes a
 + \tilde{\Delta}\left(a\right).
\eq
$\tilde{\Delta}$ is called the 
\index{reduced coproduct}
{\bf reduced coproduct}.
We have
\bq
 \tilde{\Delta}\left(a\right)
 & \in &
 \bigoplus\limits_{\substack{k+l=n \\ k,l > 0}} A_k \otimes A_l
\eq
For a graded connected Hopf algebra $A$ the antipode is recursively determined by $S(e)=e$ and
\bq
\label{chapter_hopf:antipode_recursive}
 S\left(a\right)
 & = & 
 - a \; - \; \cdot\left(S\otimes \mathrm{id} \right) \tilde{\Delta}\left(a\right)
\eq
for $a\in A_n$ with $n\ge 1$.
In the second term on the right-hand side of eq.~(\ref{chapter_hopf:antipode_recursive})
we first apply the reduced coproduct and apply the antipode to the first tensor slot, while the second is
left as it is. 
By the definition of the reduced coproduct, the weight of the entry in the first tensor slot (as well as the weight
in the second tensor slot) is less than the original weight of $a$, therefore the recursion terminates.
At the end we multiply the two entries in the two tensor slots together, indicated by the little
multiplication sign ``$\cdot$'' just in front of $(S\otimes \mathrm{id})$.

Let us elaborate on the antipode. Let $C$ be a coalgebra over the ring $R$ and $A$ an algebra over the ring $R$.
Both are $R$-modules. Let us denote by $\mathrm{Hom}(C,A)$ the set of linear maps from $C$ to $A$, i.e. for $\varphi \in \mathrm{Hom}(C,A)$
we have
\bq
\label{chapter_hopf:linear_map}
 \varphi\left( \lambda_1 a_1 + \lambda_2 a_2 \right)
 & = &
 \lambda_1 \varphi\left(a_1\right) + \lambda_2 \varphi\left(a_2\right),
 \;\;\;\;\;\;
 \lambda_1, \lambda_2 \; \in \; R,
 \;\;\; 
 a_1, a_2 \; \in \; C.
\eq
As $C$ is a coalgebra and $A$ an algebra we may define a product in $\mathrm{Hom}(C,A)$ as follows:
For $\varphi_1,\varphi_2 \in \mathrm{Hom}(C,A)$
we set
\bq
 \varphi_1 \ast \varphi_2
 & = &
 \cdot \; \left( \varphi_1 \otimes \varphi_2 \right) \Delta.
\eq
This product is called the 
\index{convolution product}
{\bf convolution product}.
$\Delta$ denotes the coproduct in $C$, the multiplication in $A$ is denoted by ``$\cdot$''.
The convolution product is associative.
\\
\\
\bs
{\it \refstepcounter{exercise}
\label{chapter_hopf:convolution_associativity}
{\bf Exercise \theexercise}: 
Show that the convolution product is associative:
\bq
 \left( \varphi_1 \ast \varphi_2 \right) \ast \varphi_3
 & = &
 \varphi_1 \ast \left( \varphi_2 \ast \varphi_3 \right).
\eq
}
\es

\noindent
The convolution product has a neutral element, given by $e \bar{e}$, where $e$ denotes the unit in $A$ and $\bar{e}$ the counit in $C$.
\\
\\
\bs
{\it \refstepcounter{exercise}
\label{chapter_hopf:convolution_neutral}
{\bf Exercise \theexercise}: 
Show that $1_{\mathrm{Hom}} = e \bar{e} \in \mathrm{Hom}(C,A)$ is a neutral element for the convolution product, i.e.
\bq
 \varphi \ast 1_{\mathrm{Hom}} 
 \; = \;
 1_{\mathrm{Hom}} \ast \varphi
 & = & 
 \varphi.
\eq
}
\es

\noindent
Let $H$ be a Hopf algebra and let us specialise to the case $C=H$. 
In particular we now have a product in $H$. 
We therefore restrict our attention to algebra homomorphisms $\mathrm{AlgHom}(H,A)$ from $H$ to $A$ which preserve the unit,
e.g. in additon to eq.~(\ref{chapter_hopf:linear_map}) we require
\bq
 \varphi\left( a_1 \cdot a_2 \right)
 & = &
 \varphi\left(a_1\right) \cdot \varphi\left(a_2\right),
 \;\;\;\;\;\;
 a_1, a_2 \; \in \; H,
 \nonumber \\
 \varphi(e_H) & = & e_A,
\eq
where $e_H$ denotes the unit in $H$ and $e_A$ denotes the unit in $A$.
As before 
$1_{\mathrm{AlgHom}} = e_A \bar{e}_H \in \mathrm{AlgHom}(H,A)$ is a neutral element for the convolution product.
We may now ask, given $\varphi \in \mathrm{AlgHom}(H,A)$ is there an inverse element $\varphi^{-1} \in \mathrm{AlgHom}(H,A)$, such that
\bq
 \varphi \ast \varphi^{-1}
 \; = \;
 \varphi^{-1} \ast \varphi
 & = &
 1_{\mathrm{AlgHom}}.
\eq
There is, and the inverse element is given by
\bq
 \varphi^{-1} & = & \varphi S,
\eq
where $S$ denotes the antipode in $H$.
\\
\\
\bs
{\it \refstepcounter{exercise}
\label{chapter_hopf:convolution_inverse}
{\bf Exercise \theexercise}: 
Show that $\varphi^{-1} = \varphi S$ is an inverse element to $\varphi \in \mathrm{AlgHom}(H,A)$.
}
\es
\\
\\
The exercises~\ref{chapter_hopf:convolution_associativity}, \ref{chapter_hopf:convolution_neutral}
and \ref{chapter_hopf:convolution_inverse} show that the unit-preserving algebra homomorphisms
$\mathrm{AlgHom}(H,A)$ form a group with the convolution product.
If $A$ is in addition commutative, we call
a unit-preserving algebra homomorphism $\varphi \in \mathrm{AlgHom}(H,A)$
a 
\index{character of a Hopf algebra}
{\bf character of the Hopf algebra} $H$.

Now let us specialise even further: We take $A=C=H$ and consider $\mathrm{AlgHom}(H,H)$.
The neutral element with respect to the convolution product is as before 
\bq
 1_{\mathrm{AlgHom}} & = & e \bar{e} \; \in \; \mathrm{AlgHom}(H,H),
\eq
where $e$ denotes the unit in $H$ and $\bar{e}$ denotes the
counit in $H$.
Let us now consider the identity map $\mathrm{id} \in \mathrm{AlgHom}(H,H)$,
\bq
 \mathrm{id}\left(a\right) & = & a,
 \;\;\;\;\;\;
 \forall \; a \; \in \; H.
\eq
Please don't confuse the maps $1_{\mathrm{AlgHom}}$ and $\mathrm{id}$, they are different.
This is most easily seen for a graded connected Hopf algebra. Let $a \in A_n$ and $n \ge 1$.
Then (see eq.~(\ref{chapter_hopf:kernel_counit}))
\bq
 1_{\mathrm{AlgHom}}\left(a\right) & = & e \bar{e}\left(a\right) \; = \; 0,
 \nonumber \\ 
 \mathrm{id}\left(a\right) & = & a.
\eq
The inverse element of the identity map $\mathrm{id}$ with respect to the covolution product is given by
the antipode
\bq
 \mathrm{id}^{-1}
 & = &
 S.
\eq
Let us now consider a few examples of Hopf algebras.

\subsubsection{Example 1: The group algebra}

Let ${\mathbb F}$ be a field and let $G$ be a group.
We denote by ${\mathbb F}[G]$ the vector space with basis $G$ over the field ${\mathbb F}$.
Then ${\mathbb F}[G]$ is an algebra with the multiplication given by the group multiplication. 
The counit, the coproduct, and the antipode are defined for the basis elements $g \in G$ as follows: 
The counit $\bar{e}$ is given by:
\bq
 \bar{e}\left( g\right) 
 & = & 
 1.
\eq
The coproduct $\Delta$ is given by:
\bq
 \Delta\left( g\right) 
 & = & 
 g \otimes g.
\eq
Thus, the basis elements $g \in G$ are goup-like elements in ${\mathbb F}[G]$.
The antipode $S$ is given by:
\bq
 S\left( g \right) 
 & = & 
 g^{-1}.
\eq
Having defined the counit, the coproduct, and the antipode for the basis elements $g \in G$, 
the corresponding definitions for arbitrary vectors in ${\mathbb F}[G]$ 
are obtained by linear extension.
${\mathbb F}[G]$ is a cocommutative Hopf algebra. ${\mathbb F}[G]$ is commutative if $G$ is commutative.

\subsubsection{Example 2: Lie algebras}

A Lie algebra ${\mathfrak g}$ is not necessarily associative nor does it have a unit.
To overcome this obstacle one considers the 
\index{universal enveloping algebra of a Lie algebra}
{\bf universal enveloping algebra} 
$U({\mathfrak g})$,
obtained from the tensor algebra $T({\mathfrak g})$ by factoring out the ideal generated by
\bq
 X \otimes Y - Y \otimes X - \left[ X, Y \right],
\eq
with $X, Y \in {\mathfrak g}$.
The universal enveloping algebra $U({\mathfrak g})$ is a Hopf algebra.
The counit $\bar{e}$ is given by:
\bq
 \bar{e}\left( e\right) = 1,
 & &
 \bar{e}\left( X\right) = 0.
\eq
The coproduct $\Delta$ is given by:
\bq
 \Delta(e) = e \otimes e, 
 & &
 \Delta(X) = X \otimes e + e \otimes X.
\eq
Thus, the elements $X \in {\mathfrak g}$ are primitive elements in $U({\mathfrak g})$.
The antipode $S$ is given by:
\bq
 S(e) = e, 
 & &
 S(X) = -X.
\eq
$U({\mathfrak g})$ is a non-commutative cocommutative Hopf algebra.

\subsubsection{Example 3: Quantum $\mathrm{SU}(2)$}

The Lie algebra $su(2)$ is generated by three generators $H$, $X_\pm$ with
\bq
 \left[ H, X_\pm \right] = \pm 2 X_\pm,
 &  & 
 \left[ X_+, X_- \right] = H. 
\eq
To obtain the deformed universal enveloping algebra $U_q(su(2))$, 
the last relation is replaced with \cite{Majid:1990vz,Schupp:1993hn}
\bq
 \left[ X_+, X_- \right] 
 & = & 
 \frac{q^H - q^{-H}}{q-q^{-1}},
\eq
where $q$ is the deformation parameter.
The undeformed Lie algebra $su(2)$ is recovered in the limit $q \rightarrow 1$. 
The counit $\bar{e}$ is given by:
\bq
 \bar{e}\left( e\right) = 1,
 & &
 \bar{e}\left( H\right) = \bar{e}\left( X_\pm\right) = 0.
\eq
The coproduct $\Delta$ is given by:
\bq
 \Delta(H) 
 & = & 
 H \otimes e + e \otimes H, 
 \nonumber \\
 \Delta(X_\pm) 
 & = & 
 X_\pm \otimes q^{H/2} + q^{-H/2} \otimes X_\pm.
\eq
The antipode $S$ is given by:
\bq
 S(H) = -H, 
 & &
 S(X_\pm) = - q^{\pm 1} X_\pm .
\eq
$U_q(su(2))$ is a non-commutative non-cocommutative Hopf algebra.

\subsubsection{Example 4: Symmetric algebras}

Let $V$ be a finite dimensional vector space with basis $\{v_i\}$.
The symmetric algebra $\mathrm{Sym}(V)$ is the direct sum
\bq
 \mathrm{Sym}(V) 
 & = & 
 \bigoplus\limits_{n=0}^\infty \mathrm{Sym}^n(V),
\eq
where $\mathrm{Sym}^n(V)$ is spanned by elements of the form $v_{i_1} v_{i_2} \dots v_{i_n}$ with
$i_1 \le i_2 \le \dots \le i_n$.
The multiplication is defined by
\bq
 \left( v_{i_1} v_{i_2} \dots v_{i_m} \right) \cdot \left( v_{i_{m+1}} v_{i_{m+2}} \dots v_{i_{m+n}} \right)
 & = &
 v_{i_{\sigma(1)}} v_{i_{\sigma(2)}} \dots v_{i_{\sigma(m+n)}},
\eq
where $\sigma$ is a permutation on $m+n$ elements such that $i_{\sigma(1)} \le i_{\sigma(2)} \le \dots \le i_{\sigma(m+n)}$.
The counit $\bar{e}$ is given by:
\bq
 \bar{e}\left( e\right) = 1, \;\;\;
 & &
 \bar{e}\left( v_1 v_2 \dots v_n\right) = 0.
\eq
The coproduct $\Delta$ is given for the basis elements $v_i$ by:
\bq
 \Delta(v_i) 
 & = & 
 v_i \otimes e + e \otimes v_i.
\eq
Using (\ref{chapter_hopf:bialg})  one obtains for a general element of $\mathrm{Sym}(V)$
\bq
 \Delta\left( v_1 v_2 \dots v_n \right)
 & = &  
  v_1 v_2 \dots v_n \otimes e
 + e \otimes v_1 v_2 \dots v_n
 \nonumber \\
 & &
 + \sum\limits_{j=1}^{n-1} \sum\limits_\sigma
  v_{\sigma(1)} \dots v_{\sigma(j)}
   \otimes 
  v_{\sigma(j+1)} \dots v_{\sigma(n)},
\eq
where $\sigma$ runs over all $(j,n-j)$-shuffles. 
A $(j,n-j)$-shuffle is a permutation $\sigma$ of $(1,\dots,n)$ such that
\bq
 \sigma(1) < \sigma(2) < \dots < \sigma(j)
 & \mbox{and} &
 \sigma(j+1) < \sigma(j+2) < \dots < \sigma(n). \nonumber 
\eq
The antipode $S$ is given by:
\bq
 S( v_{i_1} v_{i_2} \dots v_{i_n}) & = & (-1)^n v_{i_1} v_{i_2} \dots v_{i_n}.
\eq
The symmetric algebra $\mathrm{Sym}(V)$ is a commutative cocommutative Hopf algebra.

\subsubsection{Example 5: Shuffle algebras}

Recall the definition of a shuffle algebra from section~\ref{chapter_multiple_polylogarithms:section:shuffle_product}
(where we denoted the multiplication with the symbol ``$\shuffle$'' instead of ``$\cdot$''):
Consider a set of letters $A$. 
The set $A$ is known as the alphabet.
A word is an ordered sequence of letters:
\bq
 w & = & l_1 l_2 \dots l_k,
\eq
where $l_1, \dots, l_k \in A$.
The word of length zero is denoted by $e$.
The shuffle algebra ${\mathcal A}$ on the vector space spanned by words is defined by
\bq
 \left( l_1 l_2 \dots l_k \right) \shuffle \left( l_{k+1} \dots l_r \right) 
 & = &
 \sum\limits_{\mathrm{shuffles} \; \sigma} l_{\sigma(1)} l_{\sigma(2)} \dots l_{\sigma(r)},
\eq
where the sum runs over all permutations $\sigma$, which preserve the relative order
of $1,2,\dots,k$ and of $k+1,\dots,r$.
The empty word $e$ is the unit in this algebra:
\bq
 e \shuffle w = w \shuffle e = w.
\eq
The shuffle algebra is a (non-cocommutative) Hopf algebra \cite{Reutenauer}.
The counit $\bar{e}$ is given by:
\bq
 \bar{e}\left( e\right) = 1, \;\;\;
 & &
 \bar{e}\left( l_1 l_2 \dots l_n\right) = 0.
\eq
The coproduct $\Delta$ is given by:
\bq
 \Delta\left( l_1 l_2 \dots l_k \right) 
 & = & 
 \sum\limits_{j=0}^k \left( l_{j+1} \dots l_k \right) \otimes \left( l_1 \dots l_j \right).
\eq
This particular coproduct is also known as the 
\index{deconcatenation coproduct}
{\bf deconcatenation coproduct}.
The antipode $S$ is given by:
\bq
\label{chapter_hopf:shuffle_algebra_antipode}
 S\left( l_1 l_2 \dots l_k \right) & = & (-1)^k \; l_k l_{k-1} \dots l_2 l_1.
\eq
The shuffle multiplication is commutative, therefore the antipode satisfies
\bq
 S^2 
 & = & 
 \mathrm{id}.
\eq
From eq.~(\ref{chapter_hopf:shuffle_algebra_antipode}) this is evident.

To summarise,
the shuffle algebra ${\mathcal A}$ is a commutative non-cocommutative Hopf algebra.

\subsubsection{Example 6: Quasi-shuffle algebras}

In section~\ref{chapter_multiple_polylogarithms:section:quasi_shuffle_product} we 
discussed quasi-shuffle algebras ${\mathcal A}_q$ with the quasi-shuffle product $\shuffle_q$.
They are similar to shuffle algebras. 
For a quasi-shuffle algebra we consider as for a shuffle algebra the vector space spanned by words,
but now equipped with the quasi-shuffle product instead of the shuffle product.
The quasi-shuffle product differs from the normal shuffle product only by terms of lower depth.
Quasi-shuffle algebras are Hopf algebras \cite{Hoffman}.

Comultiplication and counit are defined as for the shuffle algebras.
The counit $\bar{e}$ is given by:
\bq
\bar{e}\left( e\right) = 1, \;\;\;
& &
\bar{e}\left( l_1 l_2 \dots l_n\right) = 0.
\eq
The coproduct $\Delta$ is given by:
\bq
\Delta\left( l_1 l_2 \dots l_k \right) 
& = & \sum\limits_{j=0}^k \left( l_{j+1} \dots l_k \right) \otimes \left( l_1 \dots l_j \right).
\eq
The antipode $S$ is recursively defined through
\bq
S\left( l_1 l_2 \dots l_k \right) & = & 
 - l_1 l_2 \dots l_k
 - \sum\limits_{j=1}^{k-1} S\left( l_{j+1} \dots l_k \right) \shuffle_q \left( l_1 \dots l_j \right),
 \;\;\;\;\;\;
 S(e) = e.
\eq
The quasi-shuffle product is commutative, therefore the antipode satisfies
\bq
 S^2 
 & = & 
 \mathrm{id}.
\eq
A quasi-shuffle algebra ${\mathcal A}_q$ is a commutative non-cocommutative Hopf algebra.

\subsubsection{Example 7: Rooted trees}

\begin{figure}
\begin{center}
\includegraphics[scale=0.8]{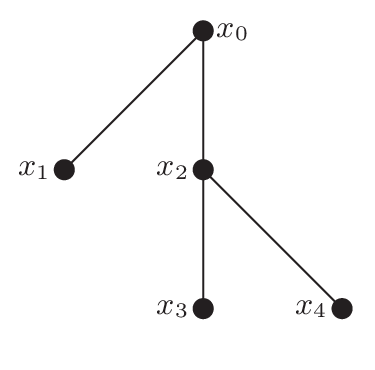}
\caption{\label{chapter_hopf:fig15} Illustration of a rooted tree. The root is drawn at the top and is labeled $x_0$.}
\end{center}
\end{figure}

A rooted tree is a tree where one vertex is marked as the root.
It is common practice in mathematics to draw the root at the top.
(Admittedly, this is a little bit counter-intuitive as a real tree in nature has its root below.) 
An individual rooted tree is shown in figure~(\ref{chapter_hopf:fig15}).
We consider the algebra generated by rooted trees. Elements of this algebra are sets of rooted trees,
conventionally known as forests.
The product of two forests is simply the disjoint union of all trees from the two forests.
The empty forest, consisting of no trees, will be denoted by $e$.
Before we are able to define a coproduct, we first need the definition of an 
\index{admissible cut of a rooted tree}
{\bf admissible cut}.
A single cut is a cut of an edge.
An admissible cut of a rooted tree is any assignment of single cuts such that any path from any vertex of the tree 
to the root has at most one single cut.
An admissible cut $C$ maps a tree $t$ to a monomial in trees $t_1 \cdot \dots \cdot t_{n+1}$. 
Precisely one of these sub-trees $t_j$ will contain the root of $t$. 
We denote this distinguished tree by $R^C(t)$, and the monomial consisting of the $n$ other factors
by $P^C(t)$. 
The counit $\bar{e}$ is given by:
\bq
 \bar{e}(e) = 1, \;\;\;
 & &
 \bar{e}\left(t_1 \cdot \dots \cdot t_k \right) = 0 \;\;\;\mbox{for}\; k \ge 1.
\eq
The coproduct $\Delta$ is given by ($t$ denotes a non-empty tree):
\bq
 \Delta(e) 
 & = & 
 e \otimes e, 
 \nonumber \\
 \Delta(t) 
 & = & 
 t \otimes e + e \otimes t + \sum\limits_{\mathrm{adm. cuts} \; C \mathrm{of} \; t} P^C(t) \otimes R^C(t),
 \nonumber \\
 \Delta\left(t_1 \cdot \dots \cdot t_k\right)
 & = &
 \Delta\left(t_1\right) \; \dots \; \Delta\left(t_k\right).
\eq
The antipode $S$ is given by:
\bq
 S(e) & = & e, 
 \nonumber \\
 S(t) 
 & = & 
 -t - \sum\limits_{\mathrm{adm. cuts} \; C \; \mathrm{of} \; t} S\left( P^C(t) \right) \cdot R^C(t),
 \nonumber \\
 S\left( t_1 \cdot \dots \cdot t_k \right)
 & = &
 S\left(t_1\right) \cdot \dots \cdot S\left(t_k\right).
\eq
The algebra of rooted trees is a commutative non-cocommutative Hopf algebra.
\\
\\
\bs
{\it \refstepcounter{exercise}
{\bf Exercise \theexercise}: 
Which rooted trees are primitive elements in the Hopf algebra of rooted trees?
}
\es
\\
\\
It is possible to classify the examples discussed above into four groups
according to whether they are commutative or cocommutative:
\begin{itemize}
\item Commutative and cocommutative: 
Examples are the group algebra of a commutative group or the symmetric algebras.
\item Non-commutative and cocommutative: 
Examples are the group algebra of a non-com\-mu\-ta\-tive group
or the universal enveloping algebra of a Lie algebra.
\item Commutative and non-cocommutative: 
Examples are the shuffle algebra, the quasi-shuffle algebra or the algebra of rooted trees.
\item Non-commutative and non-cocommutative: 
Examples are given by quantum groups.
\end{itemize}
Let us now turn to a few applications of Hopf algebras in perturbative quantum field theory.

\subsection{Renormalisation}
\label{chapter_hopf:renormalisation}

\begin{figure}
\begin{center}
\includegraphics[scale=0.8]{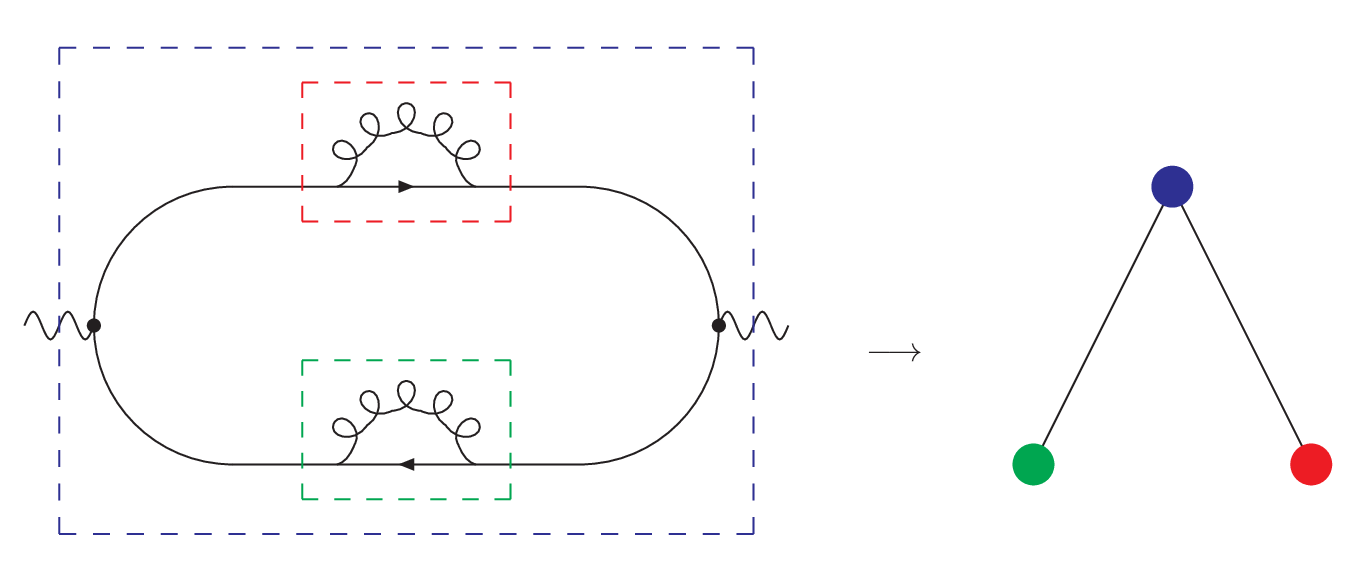}
\caption{\label{chapter_hopf:fig1} A three-loop two-point function with an overall ultraviolet divergence
and two sub-divergences.
We find the corresponding rooted tree by first drawing boxes around all ultraviolet-divergent sub-graphs.
The rooted tree is obtained from the nested structure of these boxes.}
\end{center}
\end{figure}

We start with revisiting the renormalisation of ultraviolet divergences 
in quantum field theory (see section~\ref{chapter_qft:sect:finite}).
A Feynman integral may have ultraviolet (or short-distance) singularities.
These divergences are removed by renormalisation \cite{Zimmermann:1969jj}.
The combinatorics involved in the renormalisation are governed by a Hopf algebra \cite{Kreimer:1998dp,Connes:1998qv}.
The relevant Hopf algebra is the Hopf algebra of decorated rooted trees.
We discussed the Hopf algebra of rooted trees in example 7 above.
The relation between a Feynman integral and a rooted tree is as follows:
A Feynman integral may have nested ultraviolet divergences, i.e. the associated Feynman graph may contain
sub-graphs, which correspond to ultraviolet sub-integrals.
The associated rooted tree of a Feynman graph (or of a Feynman integral)
encodes the nested structure of the sub-divergences.
This is best explained by an example.
Fig.~(\ref{chapter_hopf:fig1}) shows a three-loop two-point function. This Feynman integral has an overall ultraviolet divergence
and two sub-divergences, corresponding to the two fermion self-energy corrections.
We obtain the corresponding rooted tree by drawing boxes around all ultraviolet-divergent sub-graphs.
The rooted tree is obtained from the nested structure of these boxes.
\begin{figure}
\begin{center}
\includegraphics[scale=0.8]{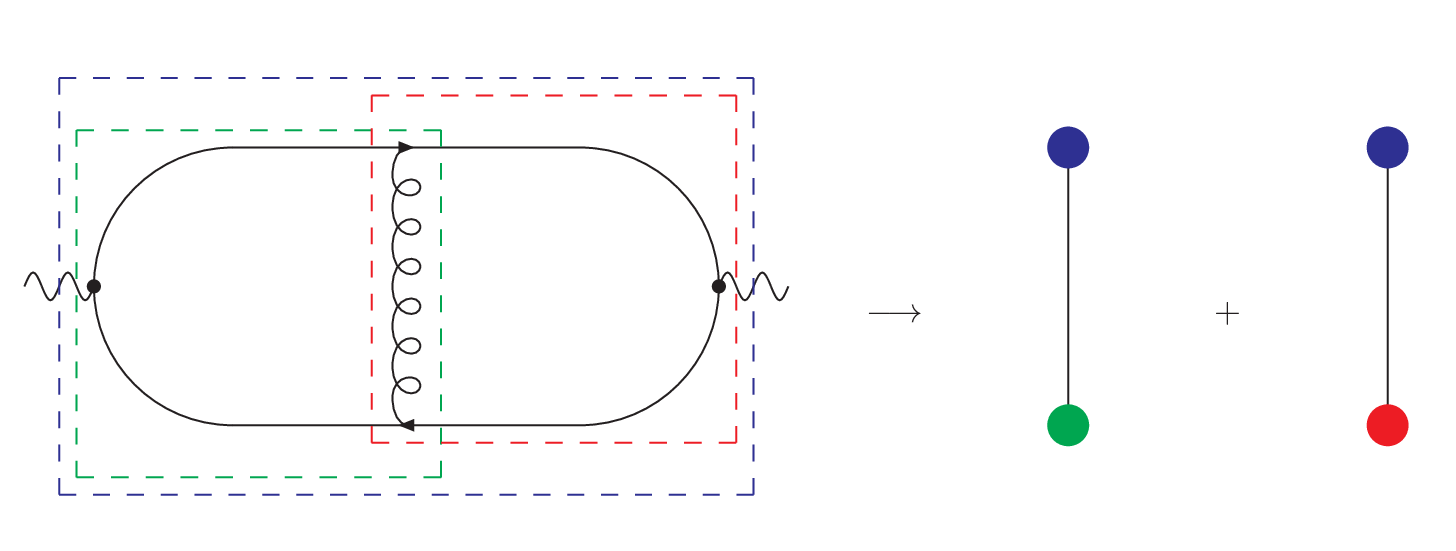}
\caption{\label{chapter_hopf:fig2} Example with overlapping singularities. This graph corresponds to a sum of rooted trees}
\end{center}
\end{figure}
Graphs with overlapping singularities correspond to a sum of rooted trees. This is illustrated for a two-loop example 
with an overlapping singularity in fig.~(\ref{chapter_hopf:fig2}).

Given a Feynman graph $G$ we may associate to $G$ a rooted tree $t$ as above.
In addition, we may associate to each vertex of the rooted tree additional information:
The sub-graph it corresponds to, as well as the momenta $q_j$, masses $m_j$ and powers $\nu_j$ associated with the edges
of the sub-graph.
This ensures that no information is lost in passing from the Feynman graph $G$ to the rooted tree $t$ and we may
recover $G$ from $t$.
A rooted tree with this additional information is called a 
\index{decorated rooted tree}
{\bf decorated rooted tree}.
The decorations do not spoil the Hopf algebra structure.
We denote the Hopf algebra of decorated rooted trees by $H$.
Furthermore we denote by $A$ the (commutative) algebra of Laurent series 
in the dimensional regularisation parameter $\eps$.
The Feynman integral $I(t)$ assigns any decorated rooted tree $t \in H$ a Laurent series.
The map
\bq
 I & : & H \rightarrow A
\eq
is a character of the Hopf algebra $H$ with values in $A$.

We recall that the counit applied to any non-trivial rooted tree $t\neq e$ yields zero:
\bq
 \bar{e}\left(t\right) & = & 0,
 \;\;\;\;\;\; t \neq e.
\eq
If we combine this with the unit $e$ in $A$ we have
\bq
\label{chapter_hopf:renormalisation_unit}
 e \bar{e}\left(t\right) & = & 0,
 \;\;\;\;\;\; t \neq e.
\eq
From the discussion of the convolution product we know that $e \bar{e} \in \mathrm{AlgHom}(H,A)$ is the neutral element
in $\mathrm{AlgHom}(H,A)$ and that the inverse of $I : H \rightarrow A$ is given by
$I^{-1} = I S$.
Thus we may write eq.~(\ref{chapter_hopf:renormalisation_unit}) as
\bq
\label{chapter_hopf:untwisted}
 \left( I^{-1} \ast I \right)\left(t\right) & = & 0,
 \;\;\;\;\;\; t \neq e.
\eq
Eq.~(\ref{chapter_hopf:untwisted}) can also be writtend as
\bq
\label{chapter_hopf:untwisted_v2}
 I^{-1}\left(t^{(1)}\right) \cdot I\left(t^{(2)}\right) & = & 0,
 \;\;\;\;\;\; t \neq e,
\eq
where we used Sweedler's notation.
Eq.~(\ref{chapter_hopf:untwisted}) will be our starting point.
However, rather than obtaining zero on the right-hand side,
we are interested in a finite quantity.
To keep the discussion simple, we only consider Feynman integrals which have ultraviolet but no
infrared divergences. For example, this can be achieved by regulating all infrared divergences with a 
small non-zero mass.

In addition we introduce a map
\bq
\label{chapter_hopf:R_operation}
 R & : & A \rightarrow A
\eq
which does not alter the divergence structure
and which satisfies the Rota-Baxter relation \cite{EbrahimiFard:2006iy}:
\bq
\label{chapter_hopf:rotabaxter}
 R\left( a_1 a_2 \right) + R\left( a_1 \right) R\left( a_2 \right) 
 & = &
 R\left( a_1 R\left( a_2 \right) \right) + R\left( R\left( a_1 \right) a_2 \right).
\eq
The map $R$ defines a 
\index{renormalisation scheme}
{\bf renormalisation scheme}.
An example is given by modified minimal subtraction scheme ($\overline{\mathrm{MS}}$).
With the conventions of this book, the $\overline{\mathrm{MS}}$-scheme is defined by
\bq
\label{chapter_hopf:MSbar}
 R\left( \sum\limits_{k=-L}^\infty c_k \eps^k \right) 
 & = & 
 \sum\limits_{k=-L}^{-1} c_k \eps^k.
\eq
You may wonder, why there are no terms like $\ln(4\pi)-\Eulerconstant$ in eq.~(\ref{chapter_hopf:MSbar}):
We eliminated these terms from the very start by an appropriate choice of the integration measure
in eq.~(\ref{chapter_basics:def_Feynman_integral}).
This motivates a posteriori the choice of the prefactors $e^{\loopnumber \eps \Eulerconstant}$ and $\pi^{-\frac{D}{2}}$
in eq.~(\ref{chapter_basics:def_Feynman_integral}).
\\
\\
\bs
{\it \refstepcounter{exercise}
{\bf Exercise \theexercise}: 
Show that the map $R$ in eq.~(\ref{chapter_hopf:MSbar}) fulfills the Rota-Baxter equation~(\ref{chapter_hopf:rotabaxter}).
}
\es
\\
\\
The notation with the letter $R$ for the map in eq.~(\ref{chapter_hopf:R_operation}) stems from 
\index{Bogoliubov's $R$-operation}
{\bf Bogoliubov's $R$-operation} \cite{Bogoliubov:1957gp}.
One can now twist the map $I^{-1}=I S$ with $R$ and define a new map $I^{-1}_R$ recursively by
\bq
 I_R^{-1}(t) 
 & = & 
 - R \left( I\left(t\right) + \sum\limits_{\mathrm{adm. cuts} \; C \; \mathrm{of} \; t} I_R^{-1}\left( P^C(t) \right) \cdot I\left(R^C(t)\right) \right).
\eq
From the multiplicativity constraint (\ref{chapter_hopf:rotabaxter}) it follows that
\bq
 I_R^{-1}\left(t_1 t_2 \right)
 & = &
 I_R^{-1}\left(t_1 \right)
 I_R^{-1}\left(t_2 \right).
\eq
If we replace $I^{-1}$ by $I_R^{-1}$ in (\ref{chapter_hopf:untwisted_v2})
we no longer obtain zero on the right-hand side, but one may show that
\bq
\label{chapter_hopf:twisted}
 I_R^{-1}\left(t^{(1)}\right) I\left(t^{(2)}\right)
 & = & 
 \mathrm{finite},
 \;\;\;\;\;\; t \neq e.
\eq
This corresponds to the renormalised value of the Feynman integral.
Eq. (\ref{chapter_hopf:twisted}) is equivalent to the forest formula \cite{Zimmermann:1969jj}.
It should be noted that $R$ is not unique and different choices for $R$ correspond
to different renormalisation schemes.
There is certainly more that could be said on the Hopf algebra of renormalisation
and we refer the reader to the original 
literature \cite{Connes:1998qv,Kreimer:1998iv,Krajewski:1998xi,Connes:1999yr,Connes:2000fe,vanSuijlekom:2006fk,Ebrahimi-Fard:2010,Ebrahimi-Fard:2012}.

\subsection{Wick's theorem}
\label{chapter_hopf:wick}

Let us consider bosonic field operators, which we denote by $\phi_i=\phi(x_i)$.
Wick's theorem relates
the time-ordered product of $n$ bosonic field operators to the 
normal product of these operators and contractions.
As an example one has
\bq
\lefteqn{
T\left(\phi_1 \phi_2 \phi_3 \phi_4 \right)
 =  
{:\phi_1 \phi_2 \phi_3 \phi_4:}
+ \left( \phi_1, \phi_2 \right) {:\phi_3 \phi_4:}
} \nonumber \\
& &
+ \left( \phi_1, \phi_3 \right) {:\phi_2 \phi_4:}
+ \left( \phi_1, \phi_4 \right) {:\phi_2 \phi_3:}
+ \left( \phi_2, \phi_3 \right) {:\phi_1 \phi_4:}
 \nonumber \\
 & &
+ \left( \phi_2, \phi_4 \right) {:\phi_1 \phi_3:}
+ \left( \phi_3, \phi_4 \right) {:\phi_1 \phi_2:}
+ \left( \phi_1, \phi_2 \right) \left(\phi_3, \phi_4 \right)
 \nonumber \\
 & &
+ \left( \phi_1, \phi_3 \right) \left(\phi_2, \phi_4 \right)
+ \left( \phi_1, \phi_4 \right) \left(\phi_2, \phi_3 \right),
\eq
where we used the notation
\bq
\label{chapter_hopf:wick1}
\left( \phi_i, \phi_j \right)
 & = & 
 \left\langle 0 \left| T \left( \phi_i \phi_j \right) \right| 0 \right\rangle
\eq
to denote the contraction.
One can use Wick's theorem to define the time-ordered product in terms of the normal product and the contraction.
To establish the connection with Hopf algebras, let $V$ be the vector space with basis $\{\phi_i\}$
and identify the normal product with the symmetric product introduced in example 4 above \cite{Fauser:2000yb,Brouder:2002nu}.
This yields the symmetric algebra $S(V)$.
The contraction defines a bilinear form $V \otimes V \rightarrow {\mathbb C}$.
One extends this pairing to $S(V)$ by
\bq
\label{chapter_hopf:wick2}
\left( {:N_1 N_2:}, M_1 \right)
 & = & \left( N_1, M_1^{(1)} \right) \left( N_2,M_1^{(2)} \right),
 \nonumber \\
\left( N_1, {:M_1 M_2:} \right) 
 & = & \left( N_1^{(1)}, M_1 \right) \left( N_1^{(2)}, M_2 \right).
\eq
Here, $N_1$, $N_2$, $M_1$ and $M_2$ are arbitrary normal products of the $\phi_i$.
With the help of this pairing one defines a new product, called the circle product,
as follows:
\bq
\label{chapter_hopf:wick3}
N \circ M & = &
 \left( N^{(1)}, M^{(1)} \right) \; {:N^{(2)} M^{(2)}:}
\eq
Again, $N$ and $M$ are normal products.
\begin{figure}
\begin{center}
\includegraphics[scale=1.0]{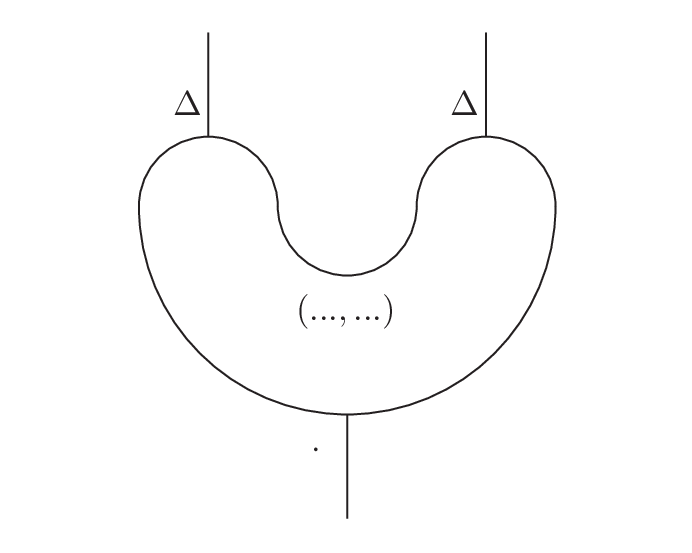}
\caption{\label{chapter_hopf:fig_sausage} The ``sausage tangle'': pictorial representation of the definition of the circle product.}
\end{center}
\end{figure}
Figure~\ref{chapter_hopf:fig_sausage} shows pictorially the definition of the circle product 
involving the coproduct, the pairing $(...,...)$ and the multiplication.
It can be shown that the circle product is associative.
Furthermore, one obtains that
the circle product coincides with the time-ordered product.
For example,
\bq
\label{chapter_hopf:wick4}
\phi_1 \circ \phi_2 \circ \phi_3 \circ \phi_4 
 & = &
T\left(\phi_1 \phi_2 \phi_3 \phi_4 \right).
\eq
The reader is invited to verify the left-hand side of (\ref{chapter_hopf:wick4})
with the help of the definitions (\ref{chapter_hopf:wick1}), (\ref{chapter_hopf:wick2}) and (\ref{chapter_hopf:wick3}).

\subsection{Multiple polylogarithms}
\label{chapter_hopf:polylogs}

Let us now turn to multiple polylogarithms.
In chapter~\ref{chapter_multiple_polylogarithms} we introduced the shuffle algebra and the quasi-shuffle
algebra related to the multiple polylogarithms.
In the examples in section~\ref{chapter_hopf:hopf_algebras} we saw
that a shuffle algebra and a quasi-shuffle algebra are Hopf algebras.
We therefore have two Hopf algebras associated with the multiple polylogarithms,
one associated with the shuffle product and the integral representation, the
other one with the quasi-shuffle product and the sum representation.

Let us start with the shuffle algebra.
Consider an alphabet $A=\{z_1,z_2,\dots\}$ and denote by ${\mathcal A}$ the shuffle algebra of words 
in this alphabet.
From example 5 in section~\ref{chapter_hopf:hopf_algebras} we know that ${\mathcal A}$ is a Hopf algebra.
For fixed $y$ we may view a multiple polylogarithm $G(z_1,\dots,z_r;y)$ as a map
\bq
\label{chapter_hopf:G_character_map}
 G & : & {\mathcal A} \rightarrow {\mathbb C},
 \nonumber \\
 & & w \rightarrow G\left(z_1,\dots,z_r;y\right), 
 \;\;\;\;\;\;
 \mbox{for} \; w = z_1 \dots z_r.
\eq
This map is a character of the Hopf algebra ${\mathcal A}$.
In particular it is an algebra homomorphism (see eq.~(\ref{chapter_multiple_polylogarithms:G_algebra_homomorphism}))
and we have
\bq
 G\left( w_1 \shuffle w_2 \right)
 & = &
 G\left(w_1\right) \cdot G\left(w_2\right).
\eq
The empty word $e$ is the unit in the shuffle algebra ${\mathcal A}$ and mapped to $1$ in ${\mathbb C}$:
\bq
 G\left(e\right) & = & G\left(;y\right) \; = \; 1.
\eq
Let us now see what the antipode gives us:
We start (without any reference to Hopf algebras) with integration-by-parts identities
for the multiple polylogarithms $G(z_1,\dots,z_r;y)$.
The starting point is as follows:
\bq
\lefteqn{
G(z_1,\dots,z_k;y) 
 =
 \int\limits_0^y dt \left( \frac{\partial}{\partial t} G(z_1;t) \right)
   G(z_2,\dots,z_k;t) 
}
 \nonumber \\
 & &
 =
  G(z_1;y) G(z_2,\dots,z_k;y) - \int\limits_0^y dt \; G(z_1;t) 
       g(z_2;t) G(z_3,\dots,z_k;t)
 \nonumber \\ 
 & &
 =
  G(z_1;y) G(z_2,\dots,z_k;y) - \int\limits_0^y dt 
       \left( \frac{\partial}{\partial t} G(z_2,z_1;t) \right) 
       G(z_3,\dots,z_k;t).
\eq
Repeating this procedure one arrives at the following
integration-by-parts identity:
\bq
\label{chapter_hopf:ibp}
 G(z_1,\dots,z_k;y) + (-1)^k G(z_k,\dots,z_1;y)
 & = & 
 G(z_1;y) G(z_2,\dots,z_k;y) - G(z_2,z_1;y) G(z_3,\dots,z_k;y)
\nonumber \\
& &
 + \dots 
 - (-1)^{k-1} G(z_{k-1},\dots,z_1;y) G(z_k;y),
\eq
which relates the combination $G(z_1,\dots,z_k;y) + (-1)^k G(z_k,\dots,z_1;y)$
to $G$-functions of lower depth.
This relation is useful in simplifying expressions.
Eq. (\ref{chapter_hopf:ibp}) can also be derived in a different way.
In the shuffle algebra ${\mathcal A}$ we have for any non-trivial element $w$ 
the following relation involving the antipode:
\bq
\label{chapter_hopf:axiomantipode_v1}
 S\left(w^{(1)}\right) \shuffle w^{(2)} & = & 0.
\eq
Here Sweedler's notation has been used.
Composing eq.~(\ref{chapter_hopf:axiomantipode_v1})
with the map $G : {\mathcal A} \rightarrow {\mathbb C}$ of eq.~(\ref{chapter_hopf:G_character_map})
we obtain
\bq
\label{chapter_hopf:axiomantipode}
 G\left(S\left(w^{(1)}\right) \shuffle w^{(2)}\right) & = & 0.
\eq
Working out the relation (\ref{chapter_hopf:axiomantipode}) for the shuffle 
algebra of the functions
$G(z_1,\dots,$ $z_k;y)$, we recover (\ref{chapter_hopf:ibp}).

Let us now turn to the quasi-shuffle algebra.
We denote letters by $l_j=(m_j,z_j)$ 
and consider an alphabet $A=\{l_1,l_2,\dots\}$. 
We denote by ${\mathcal A}_q$ the quasi-shuffle algebra of words 
in this alphabet as in section~\ref{chapter_multiple_polylogarithms:section:quasi_shuffle_product}.
From example 6 in section~\ref{chapter_hopf:hopf_algebras} we know that ${\mathcal A}_q$ is a Hopf algebra.
We may view a multiple polylogarithm $\mathrm{Li}_{m_1 \dots m_k}(x_1,\dots,x_k)$ as a map
\bq
\label{chapter_hopf:Li_character_map}
 \mathrm{Li} & : & {\mathcal A}_q \; \rightarrow \; {\mathbb C},
 \nonumber \\
 & & w \rightarrow \mathrm{Li}_{m_1 \dots m_k}\left(x_1,\dots,x_k\right),
 \;\;\;\;\;\;
 \mbox{for} \; w = l_1 \dots l_k 
 \;\; \mbox{and} \;\; l_j = (m_j,x_j).
\eq
We may be a little bit more general, fix an integer $n \in {\mathbb N}$
and consider $Z$-sums as in section~\ref{chapter_nested_sums:expansion_transcendental_functions}.
For fixed $n$ we consider the map
\bq
\label{chapter_hopf:Z_sum_character_map}
 \mathrm{Z} & : & {\mathcal A}_q \; \rightarrow \; {\mathbb C},
 \nonumber \\
 & & w \rightarrow Z_{m_1 \dots m_k}\left(x_1,\dots,x_k;n\right),
 \;\;\;\;\;\;
 \mbox{for} \; w = l_1 \dots l_k 
 \;\; \mbox{and} \;\; l_j = (m_j,x_j).
\eq
The maps in eq.~(\ref{chapter_hopf:Li_character_map}) and eq.~(\ref{chapter_hopf:Z_sum_character_map})
are characters of the Hopf algebra ${\mathcal A}_q$.
We may view eq.~(\ref{chapter_hopf:Li_character_map}) as the special case $n=\infty$
of eq.~(\ref{chapter_hopf:Z_sum_character_map}).
Eq.~(\ref{chapter_hopf:Li_character_map}) and eq.~(\ref{chapter_hopf:Z_sum_character_map})
are algebra homomorphisms, therefore
\bq
 \mathrm{Li}\left( w_1 \shuffle_q w_2 \right)
 & = &
 \mathrm{Li}\left( w_1 \right) \cdot \mathrm{Li}\left( w_2 \right),
 \nonumber \\
 Z\left( w_1 \shuffle_q w_2 \right)
 & = &
 Z\left( w_1 \right) \cdot Z\left( w_2 \right).
\eq
The empty word $e$ is the unit in the quasi-shuffle algebra ${\mathcal A}_q$ and mapped to $1$ in ${\mathbb C}$:
\bq
 \mathrm{Li}\left(e\right) & = & \mathrm{Li}\left(\right) \; = \; 1,
 \nonumber \\
 Z\left(e\right) & = & Z\left(;n\right) \; = \; 1.
\eq
We may now proceed and check if the antipode provides
also a non-trivial relation for the quasi-shuffle algebra of $Z$-sums.
This requires first some notation:
A composition of a positive integer $k$ is a sequence $I=(i_1,\dots,i_l)$ of
positive integers such that $i_1+ \dots + i_l = k$.
The set of all composition of $k$ is denoted by ${\mathcal C}(k)$.
Compositions act on words $w=l_1 \dots l_k$ in ${\mathcal A}_q$ as
\bq
\label{chapter_hopf:composition_action}
(i_1,\dots,i_l) \circ \left( l_1 l_2 \dots l_k \right)
 & = & 
 l_1' l_2' \dots l_l',
\eq
with
\bq
\label{chapter_hopf:composition_of_letters}
 l_1' \; = \; l_1 \circ \dots \circ l_{i_1},
 \;\;\;
 l_2' \; = \; l_{i_1+1} \circ \dots \circ l_{i_1+i_2},
 \;\;\;
 \dots
 \;\;\;
 l_l' \; = \; l_{i_1+\dots+i_{l-1}+1} \circ \dots \circ l_{i_1+\dots+i_l},
\eq
where $\circ$ in eq.~(\ref{chapter_hopf:composition_of_letters}) 
denotes the operation defined in eq.~(\ref{chapter_multiple_polylogarithms:def_additional_operation}).
Thus the first $i_1$ letters of the word are combined into one new letter $l_1'$,
the next $i_2$ letters are combined into the second new letter $l_2'$, etc..
To give an example let $l_1=(m_1,x_1), l_2=(m_2,x_2), l_3=(m_3,x_3)$, $w=l_1l_2l_3$  and $I=(2,1)$.
Then
\bq
 I \circ w & = & l_1' l_2',
 \nonumber \\
 & & l_1'=(m_1+m_2,x_1\cdot x_2), 
 \;\;\;
 l_2' = (m_3, x_3).
\eq 
With this notation for compositions one obtains the following closed formula for the
antipode in the quasi-shuffle algebra \cite{Hoffman}:
\bq
 S\left(l_1 l_2 \dots l_k\right)
 & = & (-1)^k \sum\limits_{I \in {\mathcal C}(k)}
 I \circ \left( l_k \dots l_2 l_1 \right).
\eq
The analogue of eq.~(\ref{chapter_hopf:axiomantipode}) reads for $w \neq e$
\bq
 Z\left(S\left(w^{(1)}\right) \shuffle_q w^{(2)}\right) & = & 0.
\eq
Written more explicitly we have
\bq
\lefteqn{
 Z\left(l_1,\dots,l_k\right) + \left(-1\right)^k Z\left(l_k,\dots,l_1\right)
 = } & & 
 \\
 & &
 - \sum\limits_{j=1}^{k-1} Z\left(S\left( l_{j+1} \dots l_k \right)\right) Z\left( l_1 \dots l_j \right)
 - (-1)^k \sum\limits_{I \in {\mathcal C}(k)\backslash (1,1,\dots,1) } Z\left(I \circ \left( l_k \dots l_2 l_1 \right)\right).
 \nonumber
\eq
Again, the combination 
$Z(n;m_1,\dots,m_k;x_1,\dots,x_k) + (-1)^k Z(n;m_k,\dots,m_1;x_k,\dots,x_1)$
reduces to $Z$-sums of lower depth, similar to the integration-by-parts identity in eq.~(\ref{chapter_hopf:ibp}).
We therefore obtained an ``integration-by-parts'' identity for objects, which don't have
an integral representation. 
We first observed, that for the $G$-functions, which have an integral representation,
the integration-by-parts identites are equal to the identities obtained from the antipode.
After this abstraction towards an algebraic formulation, one can translate these relations to cases, which
only have the appropriate algebra structure, but not necessarily a concrete integral representation.
As an example we have
\bq
\lefteqn{
Z(n;m_1,m_2,m_3;x_1,x_2,x_3) - Z(n;m_3,m_2,m_1;x_3,x_2,x_1) 
 = }
 \nonumber \\
 &  & 
Z(n;m_1;x_1) Z(n;m_2,m_3;x_2,x_3)
- Z(n;m_2,m_1;x_2,x_1) Z(n;m_3;x_3)
 \nonumber \\
 & &
- Z(n;m_1+m_2;x_1 x_2) Z(n;m_3;x_3)
+ Z(n;m_2+m_3,m_1;x_2x_3,x_1)
+ Z(n;m_3,m_1+m_2;x_3,x_1x_2)
 \nonumber \\
 & &
+ Z(n;m_1+m_2+m_3;x_1x_2x_3),
\eq
which expresses the combination of the two $Z$-sums of depth $3$ as $Z$-sums
of lower depth.
Taking $n=\infty$ in the equation above we obtain a relation among multiple polylogarithms:
\bq
\lefteqn{
\mathrm{Li}_{m_1 m_2 m_3}(x_1,x_2,x_3) - \mathrm{Li}_{m_3 m_2 m_1}(x_3,x_2,x_1) 
 = }
 \nonumber \\
 &  & 
 \mathrm{Li}_{m_1}(x_1) \mathrm{Li}_{m_2 m_3}(x_2,x_3)
- \mathrm{Li}_{m_2 m_1}(x_2,x_1) \mathrm{Li}_{m_3}(x_3)
 \nonumber \\
 & &
- \mathrm{Li}_{m_1+m_2}(x_1 x_2) \mathrm{Li}_{m_3}(x_3)
+ \mathrm{Li}_{(m_2+m_3) m_1}(x_2x_3,x_1)
+ \mathrm{Li}_{m_3 (m_1+m_2)}(x_3,x_1x_2)
 \nonumber \\
 & &
+ \mathrm{Li}_{m_1+m_2+m_3}(x_1x_2x_3).
\eq
The analog example for the shuffle algebra of the $G$-function reads:
\bq
G(z_1,z_2,z_3;y) - G(z_3,z_2,z_1;y) 
 &= &
G(z_1;y) G(z_2,z_3;y)
- G(z_2,z_1;y) G(z_3;y).
\eq
Multiple polylogarithms obey both the quasi-shuffle algebra and the shuffle algebra.
Therefore we have for multiple polylogarithms two relations, which are in general
independent.

\section{Coactions}
\label{chapter_hopf:coaction}

In the previous section we saw that the shuffle algebra is a Hopf algebra and so is the quasi-shuffle algebra.
When working with multiple polylogarithms we may either use the $G(z_1,\dots,z_r;y)$ notation and work
with shuffle algebra or the $\mathrm{Li}_{m_1 \dots m_k}(x_1,\dots,x_k)$ notation and work with the quasi-shuffle
algebra.
Whatever our choice is, the relations coming from the other algebra are not directly accessible.
We would like to work with a structure, which contains all the relations we know and only those.
As we may view $G$ (for fixed $y$) as a map from the shuffle algebra $A$ to ${\mathbb C}$, and $\mathrm{Li}$
as a map from the quasi-shuffle algebra $A_q$ to ${\mathbb C}$, our first guess might be to look at the
complex numbers ${\mathbb C}$. There we have all the relations we know about.
But it is very hard to prove that there are no additional relations.
(That is to say that one would need to prove that the period map is injective, this is currently a conjecture.)
For this reason we construct a set of objects, called {\bf motivic multiple polylogarithms}, which have exactly
the relations we know about and only those.
We denote the set of motivic multiple polylogarithms by ${\mathcal P}^{\mathfrak m}_{\mathrm{MPL}}$.
The set of motivic multiple polylogarithms is not quite a Hopf algebra, but it is a comodule.
In this section we first introduce coactions and comodules and define then the motivic multiple polylogarithms.
References for this section are \cite{Goncharov:2001,Brown:2011b,Brown:2011ik,Brown:2015bb}.
Applications towards Feynman integrals are considered in \cite{Abreu:2017ptx,Abreu:2017enx,Abreu:2021vhb}.

Let $A$ be a unitial associative algebra over a ring $R$ and $M$ a left $R$-module.
A linear map
\bq
 \cdot & :  & A \otimes M \rightarrow M,
 \nonumber \\
 & & \left(a,v\right) \rightarrow a \cdot v,
\eq
with
\bq
\label{chapter_hopf:condition_action}
 e \cdot v & = & v,
 \nonumber \\
 \left(a_1 \cdot a_2 \right) \cdot v & = & a_1 \cdot \left( a_2 \cdot v \right)
\eq
defines a (left-) 
{\bf action} of $A$ on $M$ (where $e$ denotes the unit in $A$).
This upgrades the left $R$-module $M$ to a left $A$-module.
Please note that we use the multiplication sign ``$\cdot$'' to denote the multiplication in the algebra
(e.g. $a_1\cdot a_2$) as well as for the action of $A$ on $M$ (e.g. $a \cdot v$).

Let $C$ be a coalgebra. We always assume that $C$ is coassociative and that $C$ has a counit $\bar{e}$.
We are now going to define a coaction and a comodule.
This isn't too complicated, we just have to reverse the arrows of all maps.
We start from a linear map
\bq
\label{chapter_hopf:def_coaction}
 \Delta & :  & M \rightarrow C \otimes M ,
 \nonumber \\
 & & v \rightarrow \Delta(v).
\eq
Again, please note that we use the symbol to denote on the one hand the coproduct in $C$ 
(e.g. $\Delta(a)$ for $a\in C$) as well as the new map defined in eq.~(\ref{chapter_hopf:def_coaction})
(e.g. $\Delta(v)$ for $v\in M$).
This is unambiguous, as the argument determines what operation is meant.
(It's like in C++ with operator overloading.)
We will use Sweedler's notation to write
\bq
 \Delta\left(v\right) & = & a^{(1)} \otimes v^{(2)},
 \;\;\;\;\;\;\;\;\;\;\;\;
 a^{(1)} \; \in \; C, 
 \;\;\;\;\;\;
 v, v^{(2)} \; \in \; M.
\eq
Let's now work out the analogue relations of eq.~(\ref{chapter_hopf:condition_action}).
For the map $\Delta$ defined in eq.~(\ref{chapter_hopf:def_coaction})
we require
\bq
\label{chapter_hopf:condition_coaction}
 \cdot \left( \bar{e} \otimes \mathrm{id} \right) \Delta\left(v\right) & = & v,
 \nonumber \\
 \left( \Delta \otimes \mathrm{id} \right) \Delta\left(v\right)
 & = &
 \left( \mathrm{id} \otimes \Delta \right) \Delta\left(v\right).
\eq
\bs
{\it \refstepcounter{exercise}
{\bf Exercise \theexercise}: 
Resolve the operator overloading: In eq.~(\ref{chapter_hopf:condition_coaction})
the symbols ``$\cdot$'', $\bar{e}$ and $\Delta$ appear in various places.
Determine for each occurrence to which operation they correspond.
}
\es
\\
\\
A linear map as in eq.~(\ref{chapter_hopf:def_coaction}) and satisfying eq.~(\ref{chapter_hopf:condition_coaction})
defines a (left-) 
\index{coaction}
{\bf coaction} of $C$ on $M$.
In this case we call $M$ a (left-) 
\index{comodule}
{\bf comodule}.

Let $H$ be a Hopf algebra and $M$ an algebra. We denote the unit in $H$ by $e_H$ and the unit in $M$ by $e_M$.
$M$ is called a 
\index{comodule algebra}
$H$-{\bf comodule algebra} if $M$ is a (left-) H-comodule
and in addition
\bq
 \Delta\left(e_M\right) & = & e_H \otimes e_M,
 \nonumber \\
 \Delta\left(v_1 \cdot v_2\right)
 & = &
 \Delta\left(v_1\right) \Delta\left(v_2\right),
 \;\;\;\;\;\;\;\;\; v_1, v_2 \; \in \; M.
\eq
$M$ is called a 
\index{module algebra}
$H$-{\bf module algebra} if $M$ is a (left-) H-module
and in addition
\bq
  a \cdot e_M & = & \bar{e}\left(a\right) \cdot e_M,
 \nonumber \\
 a \cdot \left(v_1 \cdot v_2 \right)
 & = & 
 \left( a^{(1)} \cdot v_1 \right) \cdot \left( a^{(2)} \cdot v_2 \right),
 \;\;\;\;\;\;\;\;\;
 a \; \in \; H, \;\;\; v_1, v_2 \; \in \; M.
\eq
Note that the definitions of a $H$-comodule algebra and of a $H$-module algebra are not dual to each other,
as $M$ is assumed to be in both cases an algebra.
For $H$ a Hopf algebra and $M$ a coalgebra there are also the notions of
a $H$-module coalgebra and of a $H$-comodule coalgebra.
We will not need them, but their definitions are the duals of the definitions 
of a $H$-comodule algebra and of a $H$-module algebra.

Let us now discuss the application towards multiple polylogarithms.
We first define for $z_1 \neq z_0$ and $z_r \neq z_{r+1}$
\bq
\label{chapter_hopf:def_I_iterated_integral}
 I\left(z_0;z_1,z_2,\dots,z_r;z_{r+1}\right)
 & = &
 \int\limits_{z_0}^{z_{r+1}} \frac{dt_r}{t_r-z_r}
 \int\limits_{z_0}^{t_r} \frac{dt_{r-1}}{t_{r-1}-z_{r-1}}
 \dots
 \int\limits_{z_0}^{t_2} \frac{dt_1}{t_1-z_1},
\eq
together with the convention
\bq
\label{chapter_hopf:def_I_unit}
 I\left(z_0;z_1\right) & = & 1.
\eq
The condition $z_1 \neq z_0$ ensures that there is no divergence at the lower integration boundary,
the condition $z_r \neq z_{r+1}$ ensures that there is no divergence at the upper integration boundary.
We then extend the definition to $z_1 = z_0$ and $z_r = z_{r+1}$ as follows:
For $z_1 = z_0$ we use the shuffle product to isolate all divergences in powers of
$I(z_0;z_0;z_2)$. We then set
\bq
\label{chapter_hopf:shuffle_regularisation_z0}
 I\left(z_0;z_0;z_2\right) & = & \ln\left(z_2-z_0\right).
\eq
In a similar way we handle the case $z_r = z_{r+1}$: We use again the shuffle product and isolate
all divergences in powers of
$I(z_0;z_2;z_2)$. We then set
\bq
\label{chapter_hopf:shuffle_regularisation_z2}
 I\left(z_0;z_2;z_2\right) & = & - \ln\left(z_0-z_2\right).
\eq
The two regularisation prescriptions 
in eq.~(\ref{chapter_hopf:shuffle_regularisation_z0}) and eq.~(\ref{chapter_hopf:shuffle_regularisation_z2})
are compatible with the path decomposition formula:
We have
\bq
 I\left(z_0;z_1;z_2\right)
 & = & 
 I\left(z_0;z_1;z_1\right)
 +
 I\left(z_1;z_1;z_2\right).
\eq
The definition in eq.~(\ref{chapter_hopf:def_I_iterated_integral})
is a slight generalisation of eq.~(\ref{chapter_multiple_polylogarithms:Gfuncdef}), allowing the starting point $z_0$ 
of the integration to be different from zero.
We have
\bq
 G\left(z_1,\dots,z_r;y\right) 
 & = & 
 I\left(0;z_r,\dots,z_1;y\right),
 \;\;\;\;\;\;\;\;\;
 \left( z_1 \neq y, \; z_r \neq 0 \right),
 \nonumber \\
 I\left(z_0;z_1,z_2,\dots,z_r;z_{r+1}\right)
 & = &
 G\left(z_r-z_0,\dots,z_1-z_0;z_{r+1}-z_0\right).
\eq
Let us now consider formal objects
$I^{\mathfrak m}(z_0;z_1,\dots,z_r;z_{r+1})$ (the ${\mathfrak m}$ stands for ``motivic'').
The set of all those objects (modulo an equivalence relation discussed below) will be denoted
by ${\mathcal P}^{\mathfrak m}_{\mathrm{MPL}}$.
We may think of the $I^{\mathfrak m}(z_0;z_1,\dots,z_r;z_{r+1})$'s in the same way we think about words
$w=z_1 \dots z_r \in A$ in the shuffle algebra in the context of multiple polylogarithms.
For fixed $y$ and $z_j \in {\mathbb C}$ we have in the latter case an evaluation map
(see eq.~(\ref{chapter_hopf:G_character_map}))
\bq
 G & : & A \rightarrow {\mathbb C},
 \nonumber \\
 & & w \rightarrow G\left(z_1,\dots,z_r;y\right).
\eq
In the same way we will assume that there is an evaluation map for $I^{\mathfrak m}(z_0;z_1,\dots,z_r;z_{r+1})$:
\bq
\label{chapter_hopf:def_I_period_map}
 \mathrm{period} & : & {\mathcal P}^{\mathfrak m}_{\mathrm{MPL}} \rightarrow {\mathbb C},
 \nonumber \\
 & &
 I^{\mathfrak m}\left(z_0;z_1,\dots,z_r;z_{r+1}\right) \rightarrow I\left(z_0;z_1,\dots,z_r;z_{r+1}\right),
\eq
sending the formal object to the concrete iterated integral of eq.~(\ref{chapter_hopf:def_I_iterated_integral}).
The map in eq.~(\ref{chapter_hopf:def_I_period_map}) will be called the 
\index{period map}
{\bf period map}.

Let's now assume that all $z_j$'s are algebraic: $z_j \in \overline{\mathbb{Q}}$.
In this case the period map takes values in ${\mathbb P}$, the set of numerical periods
(see section~\ref{chapter_sector_decomposition:periods}).
We consider the ${\mathbb Q}$-algebra generated by the $I^{\mathfrak m}(z_0;z_1,\dots,z_r;z_{r+1})$'s
subject to the following relations:
\begin{enumerate}
\item In ${\mathcal P}^{\mathfrak m}_{\mathrm{MPL}}$ we have any relation, which can be derived for
the non-motivic multiple polylogarithms $I(z_0;z_1,\dots,z_r;z_{r+1})$ using linearity, a change of variables
and Stokes' theorem.

\item Shuffle regularisation: An object $I^{\mathfrak m}(z_0;z_1,\dots,z_r; z_{r+1})$ is said to have a trailing zero, if $z_1=z_0$.
It is said to have leading one, if $z_r=z_{r+1}$.
Using the shuffle product, we isolate trailing zeros in powers of $I^{\mathfrak m}(z_0;z_0; z_{r+1})$
and leading ones in powers of $I^{\mathfrak m}(z_0;z_{r+1}; z_{r+1})$.
We then set
\bq
\label{chapter_hopf:I_regularisation}
 I^{\mathfrak m}\left(z_0;z_0; z_{r+1}\right) & = & \ln^{\mathfrak m}\left(z_{r+1}-z_0\right),
 \nonumber \\
 I^{\mathfrak m}\left(z_0;z_{r+1}; z_{r+1}\right) & = & - \ln^{\mathfrak m}\left(z_0-z_{r+1}\right).
\eq
\end{enumerate}
We call the objects $I^{\mathfrak m}(z_0;z_1,\dots,z_r;z_{r+1})$
\index{motivic multiple polylogarithms}
{\bf motivic multiple polylogarithms} and we denote the algebra defined as above 
by ${\mathcal P}^{\mathfrak m}_{\mathrm{MPL}}$.
In point $1$ we impose relations which can be obtained from 
linearity, a change of variables
and Stokes' theorem. This is completely analogue to the definition of effective periods
in section~\ref{chapter_sector_decomposition:effective_periods}.
This includes shuffle relations, i.e. 
for identical start and end points we have
\bq
\label{chapter_hopf:I_shuffle}
 I^{\mathfrak m}(z_0;z_1,\dots,z_k; z_{r+1}) \cdot I^{\mathfrak m}(z_0;z_{k+1},\dots,z_r; z_{r+1}) 
 & = &
 \sum\limits_{\mathrm{shuffles} \; \sigma} I^{\mathfrak m}(z_0; z_{\sigma(1)},\dots,z_{\sigma(r)}; z_{r+1}).
 \nonumber \\
\eq
It also includes path composition: For $y \in \overline{\mathbb{Q}}$ we have
\bq
\label{chapter_hopf:I_path_composition}
 I^{\mathfrak m}(z_0;z_1,\dots,z_r;z_{r+1})
 & = &
 \sum\limits_{k=0}^r
 I^{\mathfrak m}(z_0;z_1,\dots,z_k; y) \cdot I^{\mathfrak m}(y;z_{k+1},\dots,z_r; z_{r+1}),
 \;\;\;\;\;\;
\eq
as well as a relation for a vanishing integration cycle:
For $r \ge 1$ we have
\bq
\label{chapter_hopf:I_zero}
 I^{\mathfrak m}\left(z_0;z_1,\dots,z_r;z_0\right) & = & 0.
\eq
Point $2$ implies that for example for $z \neq 0,1$ we have
\bq
 I^{\mathfrak m}\left(0;0,0,z;1\right)
 & = &
 \frac{1}{2} \left[ I^{\mathfrak m}\left(0;0;1\right) \right]^2 I^{\mathfrak m}\left(0;z;1\right)
 - I^{\mathfrak m}\left(0;0;1\right) I^{\mathfrak m}\left(0;z,0;1\right)
 + I^{\mathfrak m}\left(0;z,0,0;1\right)
 \nonumber \\
 & = &
 \frac{1}{2} \left[ \ln^{\mathfrak m}\left(1\right) \right]^2 I^{\mathfrak m}\left(0;z;1\right)
 - \ln^{\mathfrak m}\left(1\right) I^{\mathfrak m}\left(0;z,0;1\right)
 + I^{\mathfrak m}\left(0;z,0,0;1\right)
 \nonumber \\
 & = &
 I^{\mathfrak m}\left(0;z,0,0;1\right).
\eq
We now define the 
\index{de Rham multiple polylogarithms}
{\bf de Rham multiple polylogarithms}. Roughly speaking, we may think of the de Rham multiple polylogarithms
as the motivic multiple polylogarithms modulo $(2\pi i)$.

Let $r_3=-\frac{1}{2}+\frac{i}{2}\sqrt{3}$ be the third root of unity.
The algebra ${\mathcal P}^{\mathfrak m}_{\mathrm{MPL}}$ contains the element
\bq
 I^{\mathfrak m}\left(1;0;r_3\right) + I^{\mathfrak m}\left(r_3;0;r_3^2\right) + I^{\mathfrak m}\left(r_3^2;0;1\right)
\eq
which we denote by $(2\pi i)^{\mathfrak m}$.
The notation stands for ``the motivic lift of $(2\pi i)$'' (and in particular the super-script ${\mathfrak m}$ stands for ``motivic'', it does not denote an exponent).
We have
\bq
 \mathrm{period}\left( (2\pi i)^{\mathfrak m} \right)
 & = &
 2 \pi i,
\eq
which explains the notation.
We denote by ${\mathcal P}^{\mathfrak{d}\mathfrak{R}}_{\mathrm{MPL}}$ the algebra obtained by factoring out the ideal
$\lideal (2\pi i)^{\mathfrak m} \rideal$:
\bq
 {\mathcal P}^{\mathfrak{d}\mathfrak{R}}_{\mathrm{MPL}}
 & = &
 {\mathcal P}^{\mathfrak m}_{\mathrm{MPL}} / \lideal (2\pi i)^{\mathfrak m} \rideal.
\eq
The super-script $\mathfrak{d}\mathfrak{R}$ stands for ``de Rham''.
Elements in ${\mathcal P}^{\mathfrak{d}\mathfrak{R}}_{\mathrm{MPL}}$ are denoted as
\bq
 I^{\mathfrak{d}\mathfrak{R}}(z_0;z_1,\dots,z_r;z_{r+1})
\eq
and called de Rham multiple polylogarithms.
Note that for de Rham multiple polylogarithms there is no period map, as such a map would be ambiguous by terms
proportional to $(2\pi i)$.
${\mathcal P}^{\mathfrak{d}\mathfrak{R}}_{\mathrm{MPL}}$ is a Hopf algebra \cite{Goncharov:2001,Goncharov:2002b}, the coproduct is given by
\bq
\label{chapter_hopf:I_coproduct}
\lefteqn{
 \Delta
 I^{\mathfrak{d}\mathfrak{R}}\left(z_0;z_1,z_2,\dots,z_r;z_{r+1}\right)
 = 
 \sum\limits_{k=0}^r
 \;\;
 \sum\limits_{0 = i_0 < i_1 < \dots < i_k < i_{k+1} = r+1}
} & & 
 \nonumber \\
 & &
 \prod\limits_{p=0}^k
 I^{\mathfrak{d}\mathfrak{R}}\left(z_{i_p};z_{i_p+1},z_{i_p+2},\dots,z_{i_{p+1}-1};z_{i_{p+1}}\right)
 \otimes
 I^{\mathfrak{d}\mathfrak{R}}\left(z_0;z_{i_1},z_{i_2},\dots,z_{i_k};z_{r+1}\right).
\eq
As ${\mathcal P}^{\mathfrak{d}\mathfrak{R}}_{\mathrm{MPL}}$ is graded connected, the antipode is given 
by (see eq.~(\ref{chapter_hopf:antipode_recursive})):
\bq
\label{chapter_hopf:I_antipode}
 S\left(I^{\mathfrak{d}\mathfrak{R}}\left(z_0;z_1\right)\right)
 & = & 
 I^{\mathfrak{d}\mathfrak{R}}\left(z_0;z_1\right),
 \nonumber \\
 S\left(I^{\mathfrak{d}\mathfrak{R}}\left(z_0;z_1,\dots,z_r;z_{r+1}\right)\right)
 & = & 
 - I^{\mathfrak{d}\mathfrak{R}}\left(z_0;z_1,\dots,z_r;z_{r+1}\right) 
 \nonumber \\
 & &
 - \cdot\left(S\otimes \mathrm{id} \right) \tilde{\Delta}\left(I^{\mathfrak{d}\mathfrak{R}}\left(z_0;z_1,\dots,z_r;z_{r+1}\right)\right).
\eq
On the other hand, ${\mathcal P}^{\mathfrak m}_{\mathrm{MPL}}$ is not a Hopf algebra, it is just a 
${\mathcal P}^{\mathfrak{d}\mathfrak{R}}_{\mathrm{MPL}}$-comodule.
The coaction is given by
\bq
\label{chapter_hopf:I_coaction}
\lefteqn{
 \Delta
 I^{\mathfrak m}\left(z_0;z_1,z_2,\dots,z_r;z_{r+1}\right)
 = 
 \sum\limits_{k=0}^r
 \;\;
 \sum\limits_{0 = i_0 < i_1 < \dots < i_k < i_{k+1} = r+1}
} & & 
 \nonumber \\
 & &
 \prod\limits_{p=0}^k
 I^{\mathfrak{d}\mathfrak{R}}\left(z_{i_p};z_{i_p+1},z_{i_p+2},\dots,z_{i_{p+1}-1};z_{i_{p+1}}\right)
 \otimes
 I^{\mathfrak m}\left(z_0;z_{i_1},z_{i_2},\dots,z_{i_k};z_{r+1}\right).
\eq
This formula is very similar to eq.~(\ref{chapter_hopf:I_coproduct}), 
but note that the entry in the first slot belongs to
${\mathcal P}^{\mathfrak{d}\mathfrak{R}}_{\mathrm{MPL}}$,
while the entry in the second slot belongs to 
${\mathcal P}^{\mathfrak m}_{\mathrm{MPL}}$.
There is a graphical way to represent the formula for the coproduct/coaction.
We may represent $I^{\mathfrak{d}\mathfrak{R}}(z_0; z_1, \dots, z_r; z_{r+1})$ and 
$I^{\mathfrak m}\left(z_0;z_1,\dots,z_r;z_{r+1}\right)$
as polygons drawn on a half-circle 
\begin{figure}
\begin{center}
\includegraphics[scale=0.8]{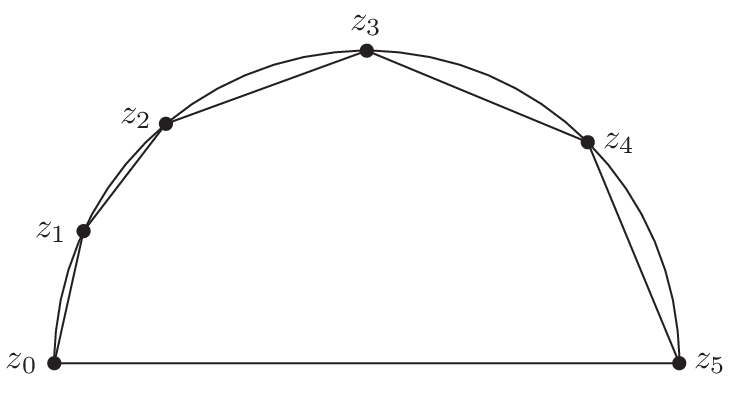}
\hspace*{10mm}
\includegraphics[scale=0.8]{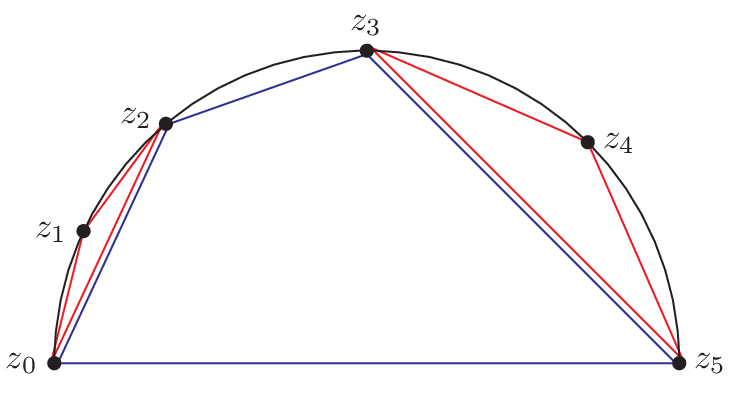}
\caption{\label{chapter_hopf:fig_coproduct_MPL} 
Left figure: We may represent $I^{\mathfrak m}(z_0;z_1,z_2,z_3,z_4;z_5)$ as a polygon on a half-circle.
Right figure: The term $I^{\mathfrak{d}\mathfrak{R}}(z_0;z_1;z_2)\cdot I^{\mathfrak{d}\mathfrak{R}}(z_3;z_4;z_5) \otimes I^{\mathfrak m}(z_0;z_2,z_3;z_5)$
appearing in the coaction.
The polygon corresponding to $I^{\mathfrak m}(z_0;z_2,z_3;z_5)$ is drawn in blue, the polygons
corresponding to $I^{\mathfrak{d}\mathfrak{R}}(z_0;z_1;z_2)$ and $I^{\mathfrak{d}\mathfrak{R}}(z_3;z_4;z_5)$ are drawn in red.
}
\end{center}
\end{figure}
as shown in the left picture in figure~\ref{chapter_hopf:fig_coproduct_MPL}.
The points $z_0$ and $z_{r+1}$ are drawn where the circle segment meets the line, the points $z_1, \dots, z_r$
are drawn in that order on the circle segment, such that $z_1$ is adjacent to $z_0$ (and $z_r$ is adjacent to $z_{r+1}$).
In order to obtain the coproduct or the coaction 
we consider all subsets of $\{i_1,\dots,i_k\} \in \{1,\dots,r\}$ (including the empty set and the full set).
The entry of the second slot is defined by this subset and given by
$I^{\mathfrak{d}\mathfrak{R}/\mathfrak m}(z_0;z_{i_1},z_{i_2},\dots,z_{i_k};z_{r+1})$.
The entry in the first slot is the product of de Rham multiple polylogarithms corresponding to
the smaller polygons, which have been omitted.
This is shown in the right picture of figure~\ref{chapter_hopf:fig_coproduct_MPL} for one specific term obtained
from the coaction on $I^{\mathfrak m}(z_0;z_1,z_2,z_3,z_4;z_5)$.
The full formula reads
\bq
\lefteqn{
 \Delta I^{\mathfrak m}(z_0;z_1,z_2,z_3,z_4;z_5)
 = 
 1 \otimes I^{\mathfrak m}(z_0;z_1,z_2,z_3,z_4;z_5)
 + I^{\mathfrak{d}\mathfrak{R}}(z_0;z_1;z_2) \otimes I^{\mathfrak m}(z_0;z_2,z_3,z_4;z_5)
} \nonumber \\
 & &
 + I^{\mathfrak{d}\mathfrak{R}}(z_1;z_2;z_3) \otimes I^{\mathfrak m}(z_0;z_1,z_3,z_4;z_5)
 + I^{\mathfrak{d}\mathfrak{R}}(z_2;z_3;z_4) \otimes I^{\mathfrak m}(z_0;z_1,z_2,z_4;z_5)
 \nonumber \\
 & &
 + I^{\mathfrak{d}\mathfrak{R}}(z_3;z_4;z_5) \otimes I^{\mathfrak m}(z_0;z_1,z_2,z_3;z_5)
 + I^{\mathfrak{d}\mathfrak{R}}(z_0;z_1,z_2;z_3) \otimes I^{\mathfrak m}(z_0;z_3,z_4;z_5)
 \nonumber \\
 & &
 + I^{\mathfrak{d}\mathfrak{R}}(z_1;z_2,z_3;z_4) \otimes I^{\mathfrak m}(z_0;z_1,z_4;z_5)
 + I^{\mathfrak{d}\mathfrak{R}}(z_2;z_3,z_4;z_5) \otimes I^{\mathfrak m}(z_0;z_1,z_2;z_5)
 \nonumber \\
 & &
 + I^{\mathfrak{d}\mathfrak{R}}(z_0;z_1;z_2) \cdot I^{\mathfrak{d}\mathfrak{R}}(z_2;z_3;z_4) \otimes I^{\mathfrak m}(z_0;z_2,z_4;z_5)
 \nonumber \\
 & &
 + I^{\mathfrak{d}\mathfrak{R}}(z_0;z_1;z_2) \cdot I^{\mathfrak{d}\mathfrak{R}}(z_3;z_4;z_5) \otimes I^{\mathfrak m}(z_0;z_2,z_3;z_5)
 \nonumber \\
 & &
 + I^{\mathfrak{d}\mathfrak{R}}(z_1;z_2;z_3) \cdot I^{\mathfrak{d}\mathfrak{R}}(z_3;z_4;z_5) \otimes I^{\mathfrak m}(z_0;z_1,z_3;z_5)
 + I^{\mathfrak{d}\mathfrak{R}}(z_1;z_2,z_3,z_4;z_5) \otimes I^{\mathfrak m}(z_0;z_1;z_5)
 \nonumber \\
 & &
 + I^{\mathfrak{d}\mathfrak{R}}(z_0;z_1;z_2) \cdot I^{\mathfrak{d}\mathfrak{R}}(z_2;z_3,z_4;z_5) \otimes I^{\mathfrak m}(z_0;z_2;z_5)
 \nonumber \\
 & &
 + I^{\mathfrak{d}\mathfrak{R}}(z_0;z_1,z_2;z_3) \cdot I^{\mathfrak{d}\mathfrak{R}}(z_3;z_4;z_5) \otimes I^{\mathfrak m}(z_0;z_3;z_5)
 \nonumber \\
 & &
 + I^{\mathfrak{d}\mathfrak{R}}(z_0;z_1,z_2,z_3;z_4) \otimes I^{\mathfrak m}(z_0;z_4;z_5)
 + I^{\mathfrak{d}\mathfrak{R}}(z_0;z_1,z_2,z_3,z_4;z_5) \otimes 1.
\eq
The right picture of figure~\ref{chapter_hopf:fig_coproduct_MPL} represents one term in this expression
(the term $I^{\mathfrak{d}\mathfrak{R}}(z_0;z_1;z_2)\cdot I^{\mathfrak{d}\mathfrak{R}}(z_3;z_4;z_5) \otimes I^{\mathfrak m}(z_0;z_2,z_3;z_5)$).

Let us now see the reason why we introduced the motivic multiple polylogarithms and the de Rham multiple polylogarithms.
We start with the coaction on $\mathrm{Li}^{\mathfrak m}_n(x)$:
\bq
 \Delta \mathrm{Li}^{\mathfrak m}_n(x)
 & = &
 - \Delta I^{\mathfrak m}(0;1,\underbrace{0,\dots,0}_{n-1};x)
\eq
Since not all points are distinct, 
many terms in the coaction are zero due to eq.~(\ref{chapter_hopf:I_regularisation}) and eq.~(\ref{chapter_hopf:I_zero}).
We end up with
\bq
\label{chapter_hopf:coaction_Li}
 \Delta \mathrm{Li}^{\mathfrak m}_n(x)
 & = &
 \mathrm{Li}^{\mathfrak{d}\mathfrak{R}}_n(x) \otimes 1
 + \sum\limits_{k=0}^{n-1} \frac{1}{k!} \left[ \ln^{\mathfrak{d}\mathfrak{R}}(x) \right]^k \otimes \mathrm{Li}^{\mathfrak m}_{n-k}(x),
\eq
where
\bq
 \ln^{\mathfrak{d}\mathfrak{R}}(x) & = & I^{\mathfrak{d}\mathfrak{R}}(1;0;x).
\eq
Now let us specialise to $x=1$. Due to eq.~(\ref{chapter_hopf:I_zero}) we have
\bq
 \ln^{\mathfrak m}(1) & = & I^{\mathfrak m}(1;0;1) \; = \; 0,
\eq
and it follows that $\ln^{\mathfrak{d}\mathfrak{R}}(1)=0$ as well.
For $x=1$ eq.~(\ref{chapter_hopf:coaction_Li}) reduces to
\bq
\label{chapter_hopf:coaction_zeta}
 \Delta \zeta^{\mathfrak m}_n
 & = &
 \zeta^{\mathfrak{d}\mathfrak{R}}_n \otimes 1
 + 1 \otimes \zeta^{\mathfrak m}_{n}.
\eq
In exercise~\ref{chapter_multiple_polylogarithms:exercise_zeta_4} you were supposed to show
that
\bq
\label{chapter_hopf:relation_zeta_4}
 \zeta_2^2 & = & \frac{5}{2} \zeta_4.
\eq
This relation prohibits a coproduct for zeta values similar to eq.~(\ref{chapter_hopf:coaction_zeta}).
To see this, assume that $H$ is a Hopf algebra, $\zeta_2, \zeta_4 \in H$ are non-zero elements with coproduct
\bq
 \Delta\left(\zeta_2\right)
 & = & 
 \zeta_2 \otimes 1 + 1 \otimes \zeta_2,
 \nonumber \\
 \Delta\left(\zeta_4\right)
 & = & 
 \zeta_4 \otimes 1 + 1 \otimes \zeta_4,
\eq
and eq.~(\ref{chapter_hopf:relation_zeta_4}) holds in $H$.
We consider $\Delta(\zeta_2^2)$. On the one hand we have
\bq
 \Delta\left(\zeta_2^2\right) 
 & = &
 \frac{5}{2} \Delta\left(\zeta_4\right) 
 \; = \;
 \frac{5}{2} \left[ \zeta_4 \otimes 1 + 1 \otimes \zeta_4 \right]
 \; = \;
 \zeta_2^2 \otimes 1 + 1 \otimes \zeta_2^2,
\eq
on the other hand we obtain using the axiom of compatibility between multiplication and comultiplication in the 
Hopf algebra $H$ 
\bq
 \Delta\left(\zeta_2^2\right) 
 & = &
 \Delta\left( \zeta_2 \cdot \zeta_2 \right)
 \; = \;
 \Delta\left(\zeta_2\right) \cdot \Delta\left(\zeta_2\right)
 \; = \;
 \left[ \zeta_2 \otimes 1 + 1 \otimes \zeta_2 \right] \cdot \left[ \zeta_2 \otimes 1 + 1 \otimes \zeta_2 \right]
 \nonumber \\
 & = &
 \zeta_2^2 \otimes 1 + 2 \zeta_2 \otimes \zeta_2 + 1 \otimes \zeta_2^2.
\eq
We assumed $\zeta_2$ to be a non-zero element of $H$, hence $\zeta_2 \otimes \zeta_2 \neq 0$ and we have a
contradiction.

Now let us return to the motivic zeta values and the de Rham zeta values:
Let $n$ be a positive even integer.
From eq.~(\ref{chapter_multiple_polylogarithms:even_zeta_values})
we know that in this case the zeta value $\zeta_n$ is a rational number times a positive power of $(2\pi i)$.
From the definition of the de Rham multiple polylogarithms ${\mathcal P}^{\mathfrak{d}\mathfrak{R}}_{\mathrm{MPL}}$
it follows that $\zeta_n^{\mathfrak{d}\mathfrak{R}}$ is equivalent to zero in ${\mathcal P}^{\mathfrak{d}\mathfrak{R}}_{\mathrm{MPL}}$.
Thus for positive even integers the coaction on the zeta values reduces to
\bq
 \Delta\left(\zeta^{\mathfrak m}_n\right)
 & = &
 1 \otimes \zeta^{\mathfrak m}_{n},
 \;\;\;\;\;\;\;\;\; 
 n \; = \; 2,4,6,8,\dots.
\eq
We further set
\bq
 \Delta\left( (2\pi i)^{\mathfrak m} \right)
 & = &
 1 \otimes (2\pi i)^{\mathfrak m}.
\eq
\bs
{\it \refstepcounter{exercise}
{\bf Exercise \theexercise}: 
Work out $\Delta(\ln^{\mathfrak m}(x) )$. Note that $\ln^{\mathfrak m}(x) = I^{\mathfrak m}(1;0;x)$.
}
\es
\\
\\
The coaction interacts with derivatives and discontinuities of multiple polylogarithms as follows:
Let $I^{\mathfrak m}_n \in {\mathcal P}^{\mathfrak m}_{\mathrm{MPL}}$ be of weight $n$.
Then
\bq
\label{chapter_hopf:coaction_derivative}
 \Delta\left( \frac{\partial}{\partial z} I^{\mathfrak m}_n \right)
 & = &
 \left( \frac{\partial}{\partial z}  \otimes \mathrm{id} \right)
 \Delta\left( I^{\mathfrak m}_n \right).
\eq
Let's verify this for the example $I^{\mathfrak m}_2 = - I^{\mathfrak m}(0;1,0;x) = \mathrm{Li}^{\mathfrak m}_2(x)$.
On the left-hand side we have
\bq
 \Delta\left( \frac{\partial}{\partial x} \mathrm{Li}^{\mathfrak m}_2(x) \right)
 & = &
 \Delta\left( \frac{1}{x} \mathrm{Li}^{\mathfrak m}_1(x) \right)
 \; = \;
 \frac{1}{x} \left( \mathrm{Li}^{\mathfrak{d}\mathfrak{R}}_1(x) \otimes 1 + 1 \otimes \mathrm{Li}^{\mathfrak m}_1(x) \right).
\eq
On the right-hand side we have
\bq
 \left( \frac{\partial}{\partial x}  \otimes \mathrm{id} \right)
 \Delta\left( \mathrm{Li}^{\mathfrak m}_2(x)\right)
 & = &
 \left( \frac{\partial}{\partial x}  \otimes \mathrm{id} \right)
 \left( \mathrm{Li}^{\mathfrak{d}\mathfrak{R}}_2(x) \otimes 1 + \ln^{\mathfrak{d}\mathfrak{R}}(x) \otimes \mathrm{Li}^{\mathfrak m}_1(x) + 1 \otimes \mathrm{Li}^{\mathfrak m}_2(x) \right)
 \nonumber \\
 & = &
 \left( \frac{\partial}{\partial x} \mathrm{Li}^{\mathfrak{d}\mathfrak{R}}_2(x) \right) \otimes 1 
 + \left( \frac{\partial}{\partial x} \ln^{\mathfrak{d}\mathfrak{R}}(x) \right) \otimes \mathrm{Li}^{\mathfrak m}_1(x)
 \nonumber \\
 & = &
 \frac{1}{x} \left( 
 \mathrm{Li}^{\mathfrak{d}\mathfrak{R}}_1(x) \otimes 1 
 + 1 \otimes \mathrm{Li}^{\mathfrak m}_1(x)
 \right).
\eq
We also have
\bq
\label{chapter_hopf:coaction_derivative_all}
 \frac{\partial}{\partial z} I^{\mathfrak m}_n
 & = &
 \cdot \left( \frac{\partial}{\partial z} \otimes 1 \right) \Delta_{1,n-1}\left(I^{\mathfrak m}_n\right).
\eq
As an alternative to eq.~(\ref{chapter_multiple_polylogarithms:differential_Glog}) 
this formula can be used to calculate the derivative of a multiple polylogarithm.
Suppose we would like to calculate
\bq
 \frac{\partial}{\partial z} G\left(0,z;y\right).
\eq
From eq.~(\ref{chapter_multiple_polylogarithms:differential_Glog})
we have
\bq
 \frac{\partial}{\partial z} G\left(0,z;y\right)
 & = &
 G\left(z;y\right) \frac{\partial}{\partial z} \ln\left(\frac{y}{z}\right)
 +
 G\left(0;y\right) \frac{\partial}{\partial z} \ln\left(\frac{-z}{-z}\right)
 \; = \;
 - \frac{1}{z} G\left(z;y\right).
\eq
We have $G^{\mathfrak m}(0,z;y)= I^{\mathfrak m}(0;z,0;y)$ and from eq.~(\ref{chapter_hopf:coaction_derivative_all}) we obtain
\bq
 \frac{\partial}{\partial z} G^{\mathfrak m}\left(0,z;y\right)
 & = &
 \cdot \left( \frac{\partial}{\partial z} \otimes 1 \right) \Delta_{1,1}\left( G^{\mathfrak m}\left(0,z;y\right) \right)
 \nonumber \\
 & = &
 \cdot \left( \frac{\partial}{\partial z} \otimes 1 \right)
 \left(\ln^{\mathfrak{d}\mathfrak{R}}\left(\frac{y}{z}\right) \otimes G^{\mathfrak m}\left(z;y\right) \right)
 \nonumber \\
 & = &
 \left( \frac{\partial}{\partial z} \ln^{\mathfrak{d}\mathfrak{R}}\left(\frac{y}{z}\right) \right) G^{\mathfrak m}\left(z;y\right)
 \; = \;
 - \frac{1}{z} G^{\mathfrak m}\left(z;y\right).
\eq
\bs
{\it \refstepcounter{exercise}
{\bf Exercise \theexercise}: 
Consider 
\bq
 I\left(0;x,x;y\right)
 & = & 
 G\left(x,x;y\right) 
 \; = \;
 G_{1 1}\left(1,1;\frac{y}{x}\right)
 \; = \;
 \mathrm{Li}_{1 1}\left(\frac{y}{x},1\right)
 \; = \;
 H_{1 1}\left(\frac{y}{x}\right).
\eq
With the techniques of chapter~\ref{chapter_multiple_polylogarithms} it is not too difficult to show that the derivatives
with respect to $x$ and $y$ are
\bq
 \frac{\partial}{\partial x} I\left(0;x,x;y\right)
 & = &
 \frac{y}{x\left(x-y\right)} \ln\left(\frac{x-y}{x}\right),
 \nonumber \\
 \frac{\partial}{\partial y} I\left(0;x,x;y\right)
 & = &
 \frac{1}{y-x} \ln\left(\frac{x-y}{x}\right).
\eq
Re-compute the derivatives using eq.~(\ref{chapter_hopf:coaction_derivative_all}).
}
\es
\\
\\
In section~\ref{chapter_one_loop:sect_unitarity} we defined the discontinuity of a function $f(z)$
across a branch cut as
\bq
 \mathrm{Disc}_z \; f\left(z\right)
 & = &
 f\left(z+i\delta\right)
 -
 f\left(z-i\delta\right),
\eq
where $\delta>0$ is infinitesimal.
Let $I^{\mathfrak m}_n \in {\mathcal P}^{\mathfrak m}_{\mathrm{MPL}}$ be of weight $n$.
Then
\bq
\label{chapter_hopf:coaction_discontinuity}
 \Delta\left( \mathrm{Disc}_z I^{\mathfrak m}_n \right)
 & = &
 \left( \mathrm{id}  \otimes \mathrm{Disc}_z \right)
 \Delta\left( I^{\mathfrak m}_n \right).
\eq
Let us also verify this with an example.
The discontinuity of $\mathrm{Li}_2(z)$ across the branch cut $[1,\infty[$ is
\bq
 \mathrm{Disc}_z \mathrm{Li}_2(z)
 & = & 
 2 \pi i \ln\left(z\right).
\eq
The left-hand side of eq.~(\ref{chapter_hopf:coaction_discontinuity}) gives
\bq
 \Delta\left( \mathrm{Disc}_z \mathrm{Li}^{\mathfrak m}_2(z) \right)
 & = &
 \Delta\left( \left(2 \pi i\right)^{\mathfrak m} \ln^{\mathfrak m}\left(z\right) \right)
 \; = \; 
 \Delta\left( \left(2 \pi i\right)^{\mathfrak m} \right) \cdot \Delta\left( \ln^{\mathfrak m}\left(z\right) \right)
 \nonumber \\
 & = &
 \left( 1 \otimes \left(2 \pi i\right)^{\mathfrak m} \right) \cdot \left( \ln^{\mathfrak{d}\mathfrak{R}}(x) \otimes 1 + 1 \otimes \ln^{\mathfrak m}(x) \right)
 \nonumber \\
 & = &
 \ln^{\mathfrak{d}\mathfrak{R}}(x) \otimes \left(2 \pi i\right)^{\mathfrak m} + 1 \otimes \left(2 \pi i\right)^{\mathfrak m} \cdot \ln^{\mathfrak m}(x)
\eq
The right-hand side of eq.~(\ref{chapter_hopf:coaction_discontinuity}) gives
\bq
 \left( \mathrm{id}  \otimes \mathrm{Disc}_z \right) \Delta\left( \mathrm{Li}^{\mathfrak m}_2(z) \right)
 & = &
 \left( \mathrm{id}  \otimes \mathrm{Disc}_z \right) 
 \left( \mathrm{Li}^{\mathfrak{d}\mathfrak{R}}_2(x) \otimes 1 + \ln^{\mathfrak{d}\mathfrak{R}}(x) \otimes \mathrm{Li}^{\mathfrak m}_1(x) + 1 \otimes \mathrm{Li}^{\mathfrak m}_2(x) \right)
 \nonumber \\
 & = &
 \ln^{\mathfrak{d}\mathfrak{R}}(x) \otimes \mathrm{Disc}_z \mathrm{Li}^{\mathfrak m}_1(x) 
 + 1 \otimes \mathrm{Disc}_z \mathrm{Li}^{\mathfrak m}_2(x) 
 \nonumber \\
 & = &
 \ln^{\mathfrak{d}\mathfrak{R}}(x) \otimes \left(2 \pi i\right)^{\mathfrak m} 
 + 1 \otimes \left(2 \pi i\right)^{\mathfrak m} \cdot \ln^{\mathfrak m}(x),
\eq
where we used $\mathrm{Disc}_z \mathrm{Li}_1(x)=2\pi i$.

Let us summarise:
\begin{tcolorbox}
The {\bf coaction commutes} with the operations of {\bf differentiation} and {\bf taking discontinuities} across branch cuts as follows:
For $I^{\mathfrak m} \in {\mathcal P}^{\mathfrak m}_{\mathrm{MPL}}$ we have
\bq
\label{chapter_hopf:coaction_derivative_and_discontinuity}
 \Delta\left( \frac{\partial}{\partial z} I^{\mathfrak m} \right)
 & = &
 \left( \frac{\partial}{\partial z}  \otimes \mathrm{id} \right)
 \Delta\left( I^{\mathfrak m} \right),
 \nonumber \\
 \Delta\left( \mathrm{Disc}_z I^{\mathfrak m} \right)
 & = &
 \left( \mathrm{id}  \otimes \mathrm{Disc}_z \right)
 \Delta\left( I^{\mathfrak m} \right).
\eq
\end{tcolorbox}
Please note that within the conventions of this book, the first entry of the tensor carries the information
on the derivative, the last entry of the tensor carries the information on the discontinuities.
This is a consequence of the definition of the coaction in eq.~(\ref{chapter_hopf:I_coaction}).
In this book we use the convention that ${\mathcal P}^{\mathfrak m}_{\mathrm{MPL}}$
is a left ${\mathcal P}^{\mathfrak{d}\mathfrak{R}}_{\mathrm{MPL}}$-comodule.

Some authors use a different convention and consider ${\mathcal P}^{\mathfrak m}_{\mathrm{MPL}}$
to be a right ${\mathcal P}^{\mathfrak{d}\mathfrak{R}}_{\mathrm{MPL}}$-comodule,
in which case the roles of the tensor entries are exchanged.

\section{Symbols}
\label{chapter_hopf:symbols}

In this section we introduce symbols and the iterated coaction. Both operations forget information, but they are useful
tools for simplifying expressions.
If two expressions agree, their symbols and their iterated coactions must agree as well.
However, the converse is in general not true.

The symbol is the coarser version, it forgets more. In particular, all constants are mapped to zero.
The symbol is defined for transcendental functions, whose total differential is a linear combination
of dlog-forms times transcendental functions of weight minus one.
The transcendental functions appearing in the total differential are again requested to satisfy the same properties.

If a coaction is available we may use the iterated coaction. This is the finer version.
The iterated coaction keeps the information on transcendental constants, algebraic constants are mapped to zero.

We start with the definition of the symbol for iterated integrals \cite{Goncharov:2010jf,Spradlin:2011wp,Duhr:2011zq,Duhr:2012fh}.
Let $\omega_1, \omega_2, \dots$ be a set of dlog-forms in the kinematic variables $x$.
Thus
\bq
 \omega_j & = & d \ln f_j,
\eq
where $f_j$ is a function of $x$.
In section~\ref{chapter_iterated_integrals:section:solution_eps_form} we associated to any
linear combination of iterated integrals of depth $\le r$
\bq
\label{chapter_hopf:linear_combination_iterated_integrals}
 I
 & = &
 \sum\limits_{j=1}^r \sum\limits_{i_1,\dots,i_j}
 c_{i_1 \dots i_j} I_\gamma\left(\omega_{i_1},\dots,\omega_{i_j};\lambda\right)
\eq
an element in the tensor algebra $T = \bigoplus\limits_{k=0}^\infty (\Omega(X))^{\otimes k}$
\bq
 B
 & = &
 \sum\limits_{j=1}^r \sum\limits_{i_1,\dots,i_j}
 c_{i_1 \dots i_j} \left[\omega_{i_1}|\dots|\omega_{i_j}\right].
\eq
Recall that the bar notation just denotes a tensor product:
\bq
 \left[ \omega_1 | \omega_2 | \dots | \omega_r \right]
 & = &
 \omega_1 \otimes \omega_2 \otimes \dots \otimes \omega_r.
\eq
As we are only considering dlog-forms, we may as well just denote the $f_j$'s instead of the $\omega_j$'s:
\bq
\label{chapter_hopf:def_symbol}
 S
 & = &
 \sum\limits_{j=1}^r \sum\limits_{i_1,\dots,i_j}
 c_{i_1 \dots i_j} \left(f_{i_1} \otimes \dots \otimes f_{i_j}\right).
\eq
$S$ is called the 
\index{symbol}
{\bf symbol} of $I$.
$B$ and $S$ denote the information on the integrand of an iterated integral.
The information on the integration path is not stored in the bar notation nor in the symbol.
We alert the reader that in this section the letter $S$ denotes the symbol, not an antipode.

For the linear combination $I$ of iterated integrals in eq.~(\ref{chapter_hopf:linear_combination_iterated_integrals}) we have the total differential
\bq
\label{chapter_hopf:derivative_iterated_integral}
 d I & = &
 \sum\limits_{j=1}^r \sum\limits_{i_1,\dots,i_j}
 c_{i_1 \dots i_j} \omega_{i_1} I_\gamma\left(\omega_{i_2},\dots,\omega_{i_j};\lambda\right).
\eq
We extend the definition of the symbol from iterated integrals to functions, whose total differential
can be written as
\bq
 d F & = & \sum\limits_i \left(d \ln f_i \right) \; F_i
\eq
with the requirement that the total differential of the function $F_i$ can again be written
in the same fashion.
We then define the symbol recursively through
\bq
\label{chapter_hopf:def_symbol_recursively}
 S\left(F\right) & = & \sum\limits_i f_i \otimes S\left(F_i\right),
 \nonumber \\
 S\left( \ln f \right) & = & f.
\eq
Due to eq.~(\ref{chapter_hopf:derivative_iterated_integral})
the definition in eq.~(\ref{chapter_hopf:def_symbol}) for iterated integrals
agrees with the definition of eq.~(\ref{chapter_hopf:def_symbol_recursively}).

Please note the order in the tensor product: 
In the symbol $(f_{i_1} \otimes \dots \otimes f_{i_r})$ of an iterated integral
$I_\gamma(\omega_{i_1},\dots,\omega_{i_j};\lambda)$
the first entry $f_{i_1}$ corresponds to the outermost integration, while the last entry $f_{i_r}$ corresponds
to the innermost integration.
This notation is consistent with the conventions used in this book: In writing
$G(z_1,\dots,z_r;y)$ or $\mathrm{Li}_{m_1 \dots m_k}(x_1,\dots,x_k)$,
the variables $z_1$ and $x_1$ refer to the outermost integration and the outermost summation, respectively.
The reader should be alerted that most literature on symbols uses the reversed notation.
To make this clear, let's consider the classical polylogarithm
\bq
 \mathrm{Li}_n\left(x\right)
 & = & 
 - G(\underbrace{0,\dots,0}_{n-1},1;x)
 \; = \;
 - I_\gamma(\underbrace{\omega_0,\dots,\omega_0}_{n-1},\omega_1),
\eq
where
$\gamma$ denotes an integration path from zero to $x$
and $\omega_0=d\ln(x)$ and $\omega_1=d\ln(1-x)$. 
Thus $f_0=x$ and $f_1=1-x$.
With the conventions of this book, the symbol of the classical polylogarithm is
\bq
\label{chapter_hopf:symbol_Li_n}
 S\left(\mathrm{Li}_n\left(x\right)\right)
 & = &
 -
 ( \underbrace{x \otimes \dots \otimes x}_{n-1} \otimes \left(1-x\right) ).
\eq
As in the symbol the entries of the individual tensor slots denote arguments of dlog-forms we have the following
rules:
\bq
\label{chapter_hopf:symbol_rules}
 f_1 \otimes \dots \otimes \left( g_a g_b \right) \otimes \dots \otimes f_r
 & = &
 \left( f_1 \otimes \dots \otimes g_a \otimes \dots \otimes f_r \right)
 +
 \left( f_1 \otimes \dots \otimes g_b \otimes \dots \otimes f_r \right),
 \nonumber \\
 f_1 \otimes \dots \otimes \left( c f_j \right) \otimes \dots \otimes f_r
 & = &
 f_1 \otimes \dots \otimes f_j \otimes \dots \otimes f_r,
\eq
where $c$ is a constant (independent of $x$).
Thus we have
\bq
 S\left(\ln\left(2 x\right)\right)
 & = & 
 S\left(\ln\left(x\right)\right)
 \; = \; 
 x.
\eq
Note that the minus sign on the right-hand side of eq.~(\ref{chapter_hopf:symbol_Li_n}) is outside the first tensor slot,
it corresponds to $c_{i_1 \dots i_j}$ in eq.~(\ref{chapter_hopf:def_symbol}).
\\
\\
\bs
{\it \refstepcounter{exercise}
{\bf Exercise \theexercise}: 
Work out the symbols
\bq
 S\left( -\ln\left(x\right) \right) 
 & \mbox{and} &
 S\left( \ln\left(-x\right) \right).
\eq
}
\es
For two iterated integrals 
$f=I_\gamma(\omega_{1},\dots,\omega_{k};\lambda)$ and
$g=I_\gamma(\omega_{k+1},\dots,\omega_{r};\lambda)$ along the same path $\gamma$ we have the shuffle product:
\bq
\label{chapter_hopf:shuffle_product_iterated_integrals}
 I_\gamma\left(\omega_{1},\dots,\omega_{k};\lambda\right) \cdot I_\gamma\left(\omega_{k+1},\dots,\omega_{r};\lambda\right)
 = 
 \sum\limits_{\mathrm{shuffles} \; \sigma} I_\gamma\left(\omega_{\sigma(1)},\omega_{\sigma(1)},\dots,\omega_{\sigma(r)};\lambda\right).
\eq
In eq.~(\ref{chapter_multiple_polylogarithms:G_shuffle_product}) we showed this for the case of multiple polylogarithms.
The proof carries over to iterated integrals.
Taking the symbol on both sides of eq.~(\ref{chapter_hopf:shuffle_product_iterated_integrals}) we find
that 
\bq
 S\left( f \cdot g \right)
 & = & 
 S\left(f\right) \shuffle S\left(g\right),
\eq
where the shuffle product in the tensor algebra $T$ is defined by
\bq
 \left( f_1 \otimes f_2 \otimes \dots \otimes f_k \right) \shuffle \left( f_{k+1} \otimes \dots \otimes f_r \right) 
 & = &
 \sum\limits_{\mathrm{shuffles} \; \sigma} f_{\sigma(1)} \otimes f_{\sigma(2)} \otimes \dots \otimes f_{\sigma(r)}.
 \;\;\;
\eq
Let us now look at a simple example:
We compute
\bq
 S\left(\mathrm{Li}_2\left(x\right)+\mathrm{Li}_2\left(1-x\right)\right)
 & = &
 - \left( x \otimes \left(1-x\right) + \left(1-x\right) \otimes x \right).
\eq
We also have
\bq
 S\left( \ln\left(x\right) \cdot \ln\left(1-x\right) \right)
 & = &
 S\left( \ln\left(x\right) \right) \shuffle S\left( \ln\left(1-x\right) \right)
 \; = \; \left( x \right) \shuffle \left(1-x\right)
 \nonumber \\
 & = &
 x \otimes \left(1-x\right) + \left(1-x\right) \otimes x.
\eq
Thus we obtain
\bq
 S\left(\mathrm{Li}_2\left(x\right)+\mathrm{Li}_2\left(1-x\right)+\ln\left(x\right) \cdot \ln\left(1-x\right) \right)
 & = &
 0.
\eq
This does not imply that
\bq
 \mathrm{Li}_2\left(x\right)+\mathrm{Li}_2\left(1-x\right)+\ln\left(x\right) \cdot \ln\left(1-x\right)
\eq
is zero, but we know that the terms which we are missing are in the kernel of the symbol map.
This could be a weight $2$ constant, or a weight $1$ constant times a logarithm of $x$.
Evaluation the above expression at $x=1$ we find that we should add $(-\zeta_2)$:
Doing so, we already obtain the correct relation
\bq
 \mathrm{Li}_2\left(x\right)+\mathrm{Li}_2\left(1-x\right)+\ln\left(x\right) \cdot \ln\left(1-x\right)
 - \zeta_2 & = & 0.
\eq
We may verify this relation by checking the relation at one point (say $x=1$) and by showing that the 
derivative of the left-hand side
equals zero.
The derivative is of lower weight and repeating this procedure will prove the identity in a finite number of steps.

Transcendental constants like $\pi$ or $\zeta_2$ are in the kernel of the symbol map, and hence not seen
at the level of the symbol.
For multiple polylogarithms we also have a coaction.
In order to get a handle on transcendental constants, we may use a finer variant of the symbol map, the iterated coaction.
We start with $I^{\mathfrak m}_n \in {\mathcal P}^{\mathfrak m}_{\mathrm{MPL}}$ and assume that 
$I^{\mathfrak m}_n$ has homogeneous weight $n$.
We then consider the $(n-1)$-fold iterated coproduct/coaction
\bq
\label{chapter_hopf:iterated_coproduct}
 \Delta^{n-1}\left( I^{\mathfrak m}_n \right)
\eq
Due to coassociativity and eq.~(\ref{chapter_hopf:condition_coaction}) it does not matter
to which tensor slot the second and further coproducts/coactions are applied, the result will be the same.
Eq.~(\ref{chapter_hopf:condition_coaction}) states that
\bq
 \left( \Delta \otimes \mathrm{id} \right) \Delta\left( I^{\mathfrak m}_n \right)
 & = &
 \left( \mathrm{id} \otimes \Delta \right) \Delta\left( I^{\mathfrak m}_n \right),
\eq
and this generalises to higher iterated coproducts/coactions.
We may therefore simply write $\Delta^{n-1}$, as we did in eq.~(\ref{chapter_hopf:iterated_coproduct}).
We then look at $\Delta_{1,\dots,1}(I^{\mathfrak m}_n)$ (with $1+\dots+1=n$, i.e. the number $1$ occurs $n$ times).
We call $\Delta_{1,\dots,1}(I^{\mathfrak m}_n)$ the maximally iterated coaction, any further iteration would produce
tensor slots of weight zero.
In $\Delta_{1,\dots,1}(I^{\mathfrak m}_n)$ the entries of all tensor slots are of weight one.
We have for example
\bq
 \Delta_{1,1}\left( \mathrm{Li}^{\mathfrak m}_2(x) \right)
 & = &
 \ln^{\mathfrak{d}\mathfrak{R}}(x) \otimes \mathrm{Li}^{\mathfrak m}_1(x).
\eq
Up to notation, this is identical to the result from the symbol map.
However, the iterated coaction does not necessarily kill transcendental constants:
\bq
 \Delta_{1,1}\left( \left(2\pi i\right)^{\mathfrak{m}} \cdot \ln^{\mathfrak{m}}(x) \right)
 & = & 
 \ln^{\mathfrak{d}\mathfrak{R}}(x) \otimes \left(2\pi i\right)^{\mathfrak{m}}.
\eq
Here the iterated coaction differs from the symbol map.

If we just look at the maximal iterated coaction we are not sensitive to transcendental constants of weight two
or higher.
For example
\bq
 \Delta_{1,1,1}\left( \zeta_2^{\mathfrak{m}} \cdot \ln^{\mathfrak{m}}(x) \right)
 & = & 0,
\eq
since the coaction does not share out $\zeta_2^{\mathfrak{m}}$ into two weight one pieces.
Hoever, this is easily fixed:
There is actually no need to focus just on the maximally iterated coaction.
Let $i_1+\dots+i_k=n$ with $i_j \in {\mathbb N}$.
We may also look at
\bq
 \Delta_{i_1,\dots,i_k}\left( I^{\mathfrak m}_n \right).
\eq
of the $(k-1)$-fold iterated coproduct/coaction $\Delta^{k-1}(I^{\mathfrak m}_n)$.
We have for example
\bq
 \Delta_{1,2}\left( \zeta_2^{\mathfrak{m}} \cdot \ln^{\mathfrak{m}}(x) \right)
 & = &
 \ln^{\mathfrak{d}\mathfrak{R}}(x) \otimes \zeta_2^{\mathfrak{m}}.
\eq
The general idea is as follows: Suppose we know already relations for weight $<n$ and we would like to establish
a new relation at weight $n$.
Instead of dealing with a single expression of weight $n$, we use the iterated coaction $\Delta_{i_1,\dots,i_k}$.
In each tensor slot the weight is lower than the original weight ($i_j < n$) and we may use in a particular tensor 
slot relations which we already know.
In summary, we may lower the weight of the objects which we would like to manipulate
at the expense of raising the rank of the tensor.

Let's look at an example: We would like to relate the classical polylogarithm $\mathrm{Li}_n(1/x)$ 
to $\mathrm{Li}_n(x)$. 
We may derive the sought-after relation from the integral representation and the substitution $x'=1/x$.
Alternatively, we may derive the relation from the coaction. This derivation nicely illustrates how the coaction
can be applied.
This is an example taken from \cite{Duhr:2014woa,Duhr:2015rjo}.
Let $x \in {\mathbb R}_{>0}$ be a positive real number.
We consider $x-i\delta \in {\mathbb C}$, where $i\delta$ denotes an infinitesimal small imaginary part.
We therefore have
\bq
 \ln\left(-x\right) & = & \ln\left(x\right) + i \pi.
\eq
At weight one we have
\bq
\label{chapter_hopf:example_inversion_Li_1}
 \mathrm{Li}_1\left(\frac{1}{x}\right)
 & = &
 - \ln\left(1-\frac{1}{x}\right)
 \; = \; 
 - \ln\left(1-x\right) + \ln\left(-x\right)
 \nonumber \\
 & = &
 \mathrm{Li}_1\left(x\right) + \ln\left(x\right) + i \pi.
\eq
At weight $2$ we first consider the $\Delta_{1,1}$-part of the coaction
\bq
 \Delta_{1,1}\left(\mathrm{Li}^{\mathfrak m}_2\left(\frac{1}{x}\right)\right)
 & = &
 \ln^{\mathfrak{d}\mathfrak{R}}\left(\frac{1}{x}\right) \otimes \mathrm{Li}^{\mathfrak m}_{1}\left(\frac{1}{x}\right)
 \nonumber \\
 & = &
 - \ln^{\mathfrak{d}\mathfrak{R}}\left(x\right) \otimes \left[ \mathrm{Li}^{\mathfrak m}_{1}\left(x\right) + \ln^{\mathfrak m}\left(x\right) + \left( i \pi \right)^{\mathfrak m} \right]
 \nonumber \\
 & = &
 - \ln^{\mathfrak{d}\mathfrak{R}}\left(x\right) \otimes \mathrm{Li}^{\mathfrak m}_{1}\left(x\right) 
 - \ln^{\mathfrak{d}\mathfrak{R}}\left(x\right) \otimes \ln^{\mathfrak m}\left(x\right) 
 - \ln^{\mathfrak{d}\mathfrak{R}}\left(x\right) \otimes \left( i \pi \right)^{\mathfrak m}
 \nonumber \\
 & = &
 \Delta_{1,1}\left( -\mathrm{Li}^{\mathfrak m}_{2}\left(x\right) 
                   - \frac{1}{2} \left[ \ln^{\mathfrak{m}}\left(x\right) \right]^2
                   - \left( i \pi \right)^{\mathfrak m} \ln^{\mathfrak{m}}\left(x\right)
             \right),
\eq
where we used in the second line the relation eq.~(\ref{chapter_hopf:example_inversion_Li_1}) for $\mathrm{Li}_1(1/x)$.
The $\Delta_{1,1}$-part of the coaction will not detect all terms, in particular we will miss at weight $2$ terms
proportional to $\zeta_2$.
We make the ansatz
\bq
 \mathrm{Li}_2\left(\frac{1}{x}\right)
 & = &
 -\mathrm{Li}_{2}\left(x\right) 
 - \frac{1}{2} \ln^2\left(x\right)
 - i \pi \ln\left(x\right)
 + c \zeta_2,
\eq
with some unknown rational coefficient $c$.
Evaluating the equation at $x=1$ we find $c=2$. It is then easily verified (by taking derivatives and evaluating at special points) that
\bq
\label{chapter_hopf:example_inversion_Li_2}
 \mathrm{Li}_2\left(\frac{1}{x}\right)
 & = &
 -\mathrm{Li}_{2}\left(x\right) 
 - \frac{1}{2} \ln^2\left(x\right)
 - i \pi \ln\left(x\right)
 + 2 \zeta_2
\eq
is the correct relation.

Let us push this example further to weight three: We start with the maximal iterated coaction
\bq
 \Delta_{1,1,1}\left(\mathrm{Li}^{\mathfrak m}_3\left(\frac{1}{x}\right)\right)
 & = &
 \ln^{\mathfrak{d}\mathfrak{R}}\left(\frac{1}{x}\right) \otimes \ln^{\mathfrak{d}\mathfrak{R}}\left(\frac{1}{x}\right) \otimes \mathrm{Li}^{\mathfrak m}_{1}\left(\frac{1}{x}\right)
 \nonumber \\
 & = &
 \ln^{\mathfrak{d}\mathfrak{R}}\left(x\right) \otimes \ln^{\mathfrak{d}\mathfrak{R}}\left(x\right) \otimes \left[ \mathrm{Li}^{\mathfrak m}_{1}\left(x\right) + \ln^{\mathfrak m}\left(x\right) + \left( i \pi \right)^{\mathfrak m} \right]
 \nonumber \\
 & = &
 \Delta_{1,1,1}\left( \mathrm{Li}^{\mathfrak m}_{3}\left(x\right) 
                   + \frac{1}{6} \left[ \ln^{\mathfrak{m}}\left(x\right) \right]^3
                   + \left( i \pi \right)^{\mathfrak m} \frac{1}{2} \left[ \ln^{\mathfrak{m}}\left(x\right) \right]^2
             \right).
\eq
This is not yet the final answer.
In $\Delta_{1,1,1}$ we will not detect terms, which are proportional to $\zeta_2$, $\zeta_3$ or $\pi^3$.
Let's first consider terms proportional to $\zeta_2$. We may detect them in
$\Delta_{1,2}$:
\bq
\lefteqn{
 \Delta_{1,2}\left(\mathrm{Li}^{\mathfrak m}_3\left(\frac{1}{x}\right)
                 - \mathrm{Li}^{\mathfrak m}_{3}\left(x\right) 
                 - \frac{1}{6} \left[ \ln^{\mathfrak{m}}\left(x\right) \right]^3
                 - \left( i \pi \right)^{\mathfrak m} \frac{1}{2} \left[ \ln^{\mathfrak{m}}\left(x\right) \right]^2
 \right)
 = 
 \ln^{\mathfrak{d}\mathfrak{R}}\left(\frac{1}{x}\right) \otimes \mathrm{Li}^{\mathfrak m}_{2}\left(\frac{1}{x}\right)
 } & & 
 \nonumber \\
 & = &
 - \ln^{\mathfrak{d}\mathfrak{R}}\left(x\right) \otimes \mathrm{Li}^{\mathfrak m}_{2}\left(x\right)
 - \ln^{\mathfrak{d}\mathfrak{R}}\left(x\right) \otimes \frac{1}{2} \left[ \ln^{\mathfrak{m}}\left(x\right) \right]^2
 - \ln^{\mathfrak{d}\mathfrak{R}}\left(x\right) \otimes \left( i \pi \right)^{\mathfrak m} \ln^{\mathfrak{m}}\left(x\right)
 \nonumber \\
 & = &
 - \ln^{\mathfrak{d}\mathfrak{R}}\left(x\right) 
   \otimes \left[ \mathrm{Li}^{\mathfrak m}_{2}\left(\frac{1}{x}\right) 
                + \mathrm{Li}^{\mathfrak m}_{2}\left(x\right) 
                + \frac{1}{2} \left[ \ln^{\mathfrak{m}}\left(x\right) \right]^2
                + \left( i \pi \right)^{\mathfrak m} \ln^{\mathfrak{m}}\left(x\right) \right].
\eq
We may now use eq.~(\ref{chapter_hopf:example_inversion_Li_2}) and find
\bq
 \Delta_{1,2}\left(\mathrm{Li}^{\mathfrak m}_3\left(\frac{1}{x}\right)
                 - \mathrm{Li}^{\mathfrak m}_{3}\left(x\right) 
                 - \frac{1}{6} \left[ \ln^{\mathfrak{m}}\left(x\right) \right]^3
                 - \left( i \pi \right)^{\mathfrak m} \frac{1}{2} \left[ \ln^{\mathfrak{m}}\left(x\right) \right]^2
 \right)
 =
 - 2 \Delta_{1,2}\left(\zeta^{\mathfrak{m}}_2 \ln^{\mathfrak{m}}\left(x\right) \right).
 \;\;\;\;\;\;
\eq
Terms proportional to $\zeta_3$ or $\pi^3$ cannot be detected from the coaction, as they do not share out.
We make the ansatz
\bq
 \mathrm{Li}_3\left(\frac{1}{x}\right)
 & = &
 \mathrm{Li}_{3}\left(x\right) 
 + \frac{1}{6} \ln^3\left(x\right)
 + \frac{1}{2} i \pi \ln^2\left(x\right)
 - 2 \zeta_2 \ln\left(x\right)
 + c_1 \zeta_3
 + c_2 i \pi^3.
\eq
Evaluating the expression at $x=1$ yields $c_1=c_2=0$ and we finally obtain
\bq
 \mathrm{Li}_3\left(\frac{1}{x}\right)
 & = &
 \mathrm{Li}_{3}\left(x\right) 
 + \frac{1}{6} \ln^3\left(x\right)
 + \frac{1}{2} i \pi \ln^2\left(x\right)
 - 2 \zeta_2 \ln\left(x\right).
\eq
This relation is then verified by taking derivatives and evaluating at special points.

We may continue in this way and systematically derive inversion relation for $\mathrm{Li}_n(1/x)$.

\section{The single-valued projection}
\label{chapter_hopf:single_valued}

Multiple polylogarithms are in general multi-valued functions.
In relation to Feynman integrals this is what we want: The starting points of branch cuts of multiple polylogarithms 
are related to the thresholds of Feynman integrals.
Nevertheless, we may ask if it is possible to define single-valued multiple polylogarithms.
This is indeed possible and we will define single-valued multiple polylogarithms in this section.
This will also shed some new light on the role of the de Rham multiple polylogarithms.
Up to now we treated them as some formal objects, as we could not associate any numerical value to them.
With the help of the single-valued multiple polylogarithms we may define an evaluation map for 
the de Rham multiple polylogarithms.
References for this section are \cite{Brown:2004,Brown:2013gia,Schnetz:2013hqa,Charlton:2021uhu}.
 
\begin{figure}
\begin{center}
\includegraphics[scale=1.0]{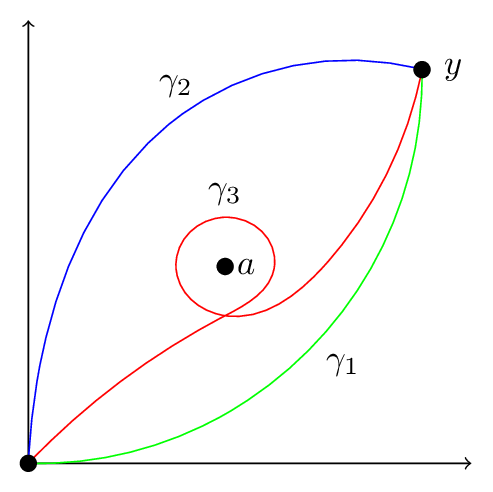}
\caption{\label{chapter_hopf:fig_de_Rham} Consider $I_{\gamma_i}(0;a;y)$. The three integration paths cannot be deformed
continuously into each other without crossing the pole at $a$.
The result for $I_{\gamma_i}(0;a;y)$ will depend on (the homotopy class of) the integration path.
The difference is proportional to $(2\pi i)$, and therefore the de Rham logarithms $I_{\gamma_i}^{\mathfrak{d}\mathfrak{R}}(0;a;y)$ are equivalent.
Phrased differently, the information on the integration path is lost in ${\mathcal P}^{\mathfrak{d}\mathfrak{R}}_{\mathrm{MPL}}$.
}
\end{center}
\end{figure}

In section~\ref{chapter_hopf:coaction} we 
considered the algebras ${\mathcal P}^{\mathfrak m}_{\mathrm{MPL}}$ and
${\mathcal P}^{\mathfrak{d}\mathfrak{R}}_{\mathrm{MPL}}$.
There is a projection
\bq
\label{chapter_hopf:de_Rham_projection}
 \pi^{\mathfrak{d}\mathfrak{R}} & : & {\mathcal P}^{\mathfrak m}_{\mathrm{MPL}} \rightarrow {\mathcal P}^{\mathfrak{d}\mathfrak{R}}_{\mathrm{MPL}}
\eq
whose kernel is the ideal $\lideal (2\pi i)^{\mathfrak m} \rideal$.
We have for example
\bq
 \pi^{\mathfrak{d}\mathfrak{R}}\left( \mathrm{Li}^{\mathfrak m}_{2}\left(x\right) + \left( i \pi \right)^{\mathfrak m} \ln^{\mathfrak{m}}\left(x\right) \right)
 & = &
 \mathrm{Li}^{\mathfrak{d}\mathfrak{R}}_{2}\left(x\right).
\eq
We started from the functions $I(z_0;z_1,\dots,z_r;z_{r+1})$ defined in eq.~(\ref{chapter_hopf:def_I_iterated_integral}).
We implicitly assumed a standard integration path (say a straight line from $z_0$ to $z_{r+1}$, supplemented in the case
of divergent integrals by a tangential base point prescription).
We could have started from an extended definition $I_\gamma(z_0;z_1,\dots,z_r;z_{r+1})$, allowing arbitrary integration paths.
$I_\gamma(z_0;z_1,\dots,z_r;z_{r+1})$ and $I^{\mathfrak m}_\gamma(z_0;z_1,\dots,z_r;z_{r+1})$
would then depend on the integration path.
Let us now restrict our attention to linear combinations, which are homotopy functionals (see figure~\ref{chapter_hopf:fig_de_Rham}).
For those linear combinations 
we may think about the de Rham version as multiple polylogarithms which have lost all information
on the integration path.
If we deform the integration path of a homotopy functional of ordinary multiple polylogarithm across a pole of an integrand, 
we should
compensate by $(2\pi i)$ times the residue at the pole.
However this equals zero in ${\mathcal P}^{\mathfrak{d}\mathfrak{R}}_{\mathrm{MPL}}$ and hence this information is lost
for the de Rham multiple polylogarithms.
For this reason we did not define a period map for de Rham multiple polylogarithms, as a naive attempt would be 
ambiguous by terms of the form $(2\pi i) \times \mbox{functions of weight $(n-1)$}$.

The essential ingredient for the definition of single-valued multiple polylogarithms is the map
\bq
 \mathrm{sv} & : & {\mathcal P}^{\mathfrak{d}\mathfrak{R}}_{\mathrm{MPL}} \rightarrow {\mathcal P}^{\mathfrak m}_{\mathrm{MPL}}
 \nonumber \\
 & & \mathrm{sv}\left( I^{\mathfrak{d}\mathfrak{R}} \right) 
     \; = \; 
     \cdot \left( \mathrm{id} \otimes F_\infty \Sigma \right) \Delta^{\mathfrak m}\left(I^{\mathfrak{d}\mathfrak{R}}\right).
\eq
Let us explain the ingredients. We recall that ${\mathcal P}^{\mathfrak{d}\mathfrak{R}}_{\mathrm{MPL}}$ is a Hopf algebra,
with coproduct $\Delta : {\mathcal P}^{\mathfrak{d}\mathfrak{R}}_{\mathrm{MPL}} \rightarrow {\mathcal P}^{\mathfrak{d}\mathfrak{R}}_{\mathrm{MPL}} \otimes {\mathcal P}^{\mathfrak{d}\mathfrak{R}}_{\mathrm{MPL}}$
and antipode $S : {\mathcal P}^{\mathfrak{d}\mathfrak{R}}_{\mathrm{MPL}} \rightarrow {\mathcal P}^{\mathfrak{d}\mathfrak{R}}_{\mathrm{MPL}}$
defined in eq.~(\ref{chapter_hopf:I_coproduct}) and eq.~(\ref{chapter_hopf:I_antipode}), respectively.

We now define maps
\bq
 \Delta^{\mathfrak m} & : & {\mathcal P}^{\mathfrak{d}\mathfrak{R}}_{\mathrm{MPL}} \rightarrow {\mathcal P}^{\mathfrak m}_{\mathrm{MPL}} \otimes {\mathcal P}^{\mathfrak m}_{\mathrm{MPL}},
 \nonumber \\
 S^{\mathfrak m} & : & {\mathcal P}^{\mathfrak m}_{\mathrm{MPL}} \rightarrow {\mathcal P}^{\mathfrak m}_{\mathrm{MPL}}
\eq
by replacing the superscript $\mathfrak{d}\mathfrak{R}$ with $\mathfrak{m}$ on the right-hand side of eq.~(\ref{chapter_hopf:I_coproduct}) 
and everywhere
in eq.~(\ref{chapter_hopf:I_antipode}).
We then define 
\bq
 \Sigma & : & {\mathcal P}^{\mathfrak m}_{\mathrm{MPL}} \rightarrow {\mathcal P}^{\mathfrak m}_{\mathrm{MPL}}
 \nonumber \\
 & & \Sigma\left(I^{\mathfrak m}\left(z_0;z_1,\dots,z_r;z_{r+1}\right)\right) 
     \; = \;
     \left(-1\right)^r S^{\mathfrak m}\left(I^{\mathfrak m}\left(z_0;z_1,\dots,z_r;z_{r+1}\right)\right).
\eq
The map $F_\infty$ denotes the 
\index{real Frobenius}
{\bf real Frobenius}
\bq
\label{chapter_hopf:real_Frobenius}
 F_\infty & : & {\mathcal P}^{\mathfrak m}_{\mathrm{MPL}} \rightarrow {\mathcal P}^{\mathfrak m}_{\mathrm{MPL}}.
\eq
We may think about $F_\infty$ as complex conjugation. For us the important property will be
\bq
 \mathrm{period}\left(F_\infty\left(I^{\mathfrak m}\right)\right)
 & = & 
 \overline{\mathrm{period}\left(I^{\mathfrak m}\right)},
\eq
where $\overline{z}$ denotes complex conjugation of $z$.
The exact definition of the real Frobenius will be given in chapter~\ref{chapter_motives} in eq.~(\ref{chapter_motives:real_Frobenius_map}) and eq.~(\ref{chapter_motives:induced_real_Frobenius}).

We may now combine the map $\mathrm{sv}$ 
with the period map of eq.~(\ref{chapter_hopf:def_I_period_map}) and obtain a map
\bq
\label{chapter_hopf:single_value_de_Rham}
 \mathrm{sv}^{\mathfrak{d}\mathfrak{R}} & : & {\mathcal P}^{\mathfrak{d}\mathfrak{R}}_{\mathrm{MPL}} \rightarrow {\mathbb C},
 \nonumber \\
 & & \mathrm{sv}^{\mathfrak{d}\mathfrak{R}} \; = \; \mathrm{period} \circ \mathrm{sv}.
\eq
This will assign a complex number to a de Rham multiple polylogarithm.

In a similar way we may combine the map $\pi^{\mathfrak{d}\mathfrak{R}}$ of eq.~(\ref{chapter_hopf:de_Rham_projection})
with the maps $\mathrm{sv}$ and $\mathrm{period}$ and obtain
\bq
\label{chapter_hopf:single_value_projection}
 \mathrm{sv}^{\mathfrak m} & : & {\mathcal P}^{\mathfrak m}_{\mathrm{MPL}} \rightarrow {\mathbb C},
 \nonumber \\
 & & \mathrm{sv}^{\mathfrak m} \; = \; \mathrm{period} \circ \mathrm{sv} \circ \pi^{\mathfrak{d}\mathfrak{R}}.
\eq
Eq.~(\ref{chapter_hopf:single_value_projection}) is called the 
\index{single-value projection}
{\bf single-value projection}.
It can be shown that eq.~(\ref{chapter_hopf:single_value_de_Rham}) 
and eq.~(\ref{chapter_hopf:single_value_projection}) define single-valued functions of $x$.

Let us consider a few examples:
We have
\bq
 \Delta\left( \ln^{\mathfrak{d}\mathfrak{R}}\left(x\right) \right)
 & = &
 1 \otimes \ln^{\mathfrak{d}\mathfrak{R}}\left(x\right) + \ln^{\mathfrak{d}\mathfrak{R}}\left(x\right) \otimes 1,
 \nonumber \\
 S\left( \ln^{\mathfrak{d}\mathfrak{R}}\left(x\right) \right)
 & = &
 -  \ln^{\mathfrak{d}\mathfrak{R}}\left(x\right),
\eq
and therefore
\bq
\label{chapter_hopf:single_value_projection_ln}
 \mathrm{sv}^{\mathfrak m}\left(\ln^{\mathfrak{m}}\left(x\right)\right)
 & = &
 \ln\left(\overline{x}\right) + \ln\left(x\right)
 \; = \; 
 \ln\left( \left|x\right|^2\right).
\eq
For the classical polylogarithms one has
\bq
 \Delta\left(\mathrm{Li}^{\mathfrak{d}\mathfrak{R}}_n(x)\right)
 & = &
 \mathrm{Li}^{\mathfrak{d}\mathfrak{R}}_n(x) \otimes 1
 + \sum\limits_{k=0}^{n-1} \frac{1}{k!} \left[ \ln^{\mathfrak{d}\mathfrak{R}}(x) \right]^k \otimes \mathrm{Li}^{\mathfrak{d}\mathfrak{R}}_{n-k}(x),
 \nonumber \\
 S\left(\mathrm{Li}^{\mathfrak{d}\mathfrak{R}}_n(x)\right)
 & = &
 - \sum\limits_{k=0}^{n-1} \frac{1}{k!} \left[ - \ln^{\mathfrak{d}\mathfrak{R}}(x) \right]^k \mathrm{Li}^{\mathfrak{d}\mathfrak{R}}_{n-k}(x).
\eq
We obtain
\bq
\label{chapter_hopf:single_value_projection_polylogs}
 \mathrm{sv}^{\mathfrak m}\left(\mathrm{Li}^{\mathfrak m}_n(x)\right)
 & = &
 \mathrm{Li}_n\left(x\right)
 - \left(-1\right)^n \sum\limits_{k=0}^{n-1} \frac{1}{k!} \left[ - \ln\left(\left|x\right|^2\right) \right]^k \mathrm{Li}_{n-k}\left(\overline{x}\right),
\eq
e.g.
\bq
 \mathrm{sv}^{\mathfrak m}\left(\mathrm{Li}^{\mathfrak m}_1(x)\right)
 & = &
 \mathrm{Li}_1\left(x\right)
 + \mathrm{Li}_{1}\left(\overline{x}\right),
 \nonumber \\
 \mathrm{sv}^{\mathfrak m}\left(\mathrm{Li}^{\mathfrak m}_2(x)\right)
 & = &
 \mathrm{Li}_2\left(x\right)
 - \mathrm{Li}_{2}\left(\overline{x}\right)
 + \ln\left(\left|x\right|^2\right) \mathrm{Li}_{1}\left(\overline{x}\right),
 \;\;\;\;\;\;\;\;\;\;\;\;
 \mbox{etc..}
\eq
\bs
{\it \refstepcounter{exercise}
{\bf Exercise \theexercise}: 
Fill in the details for the derivation of $\mathrm{sv}^{\mathfrak m}(\mathrm{Li}^{\mathfrak m}_1(x))$
and $\mathrm{sv}^{\mathfrak m}(\mathrm{Li}^{\mathfrak m}_2(x))$.
}
\es
\\
\\
Setting $x=-1$ in eq.~(\ref{chapter_hopf:single_value_projection_ln}) shows that
\bq
 \mathrm{sv}^{\mathfrak m}\left(\left(i \pi\right)^{\mathfrak{m}}\right)
 & = &
 0,
\eq
setting $x=1$ in eq.~(\ref{chapter_hopf:single_value_projection_polylogs}) yields
\bq
\label{chapter_hopf:single_value_zeta_values}
 \mathrm{sv}^{\mathfrak m}\left(\zeta^{\mathfrak m}_n\right)
 & = &
 \mathrm{sv}^{\mathfrak m}\left(\mathrm{Li}^{\mathfrak m}_n\left(1\right)\right)
 \; = \;
 \left\{ \begin{array}{ll}
 2 \zeta_n, & \mbox{$n$ odd}, \\
 0, & \mbox{$n$ even}.
 \end{array}
 \right.
\eq
Eq.~(\ref{chapter_hopf:single_value_zeta_values}) defines the so-called
\index{single-valued zeta values}
{\bf single-valued zeta values}
\bq
 \zeta^{\mathrm{sv}}_n
 & = &
 \mathrm{sv}^{\mathfrak m}\left(\zeta^{\mathfrak m}_n\right).
\eq
Although this may sound like an oxymoron (a zeta value is a number independent of $x$ and therefore certainly
single-valued as a function of $x$), it should be understood as follows:
The ordinary zeta values are the values of the classical polylogarithms at $x=1$.
In exactly the same way we define the single-valued zeta values to be the values of the single-valued classical
polylogarithms at $x=1$.
 
\section{Bootstrap}
\label{chapter_hopf:bootstrap}

Let's look at a practical application of the symbol (or the iterated coaction):
Suppose we expect that a certain Feynman integral can be written as a linear combination of multiple polylogarithms.
Suppose further that we can figure out what the possible arguments of the multiple polylogarithms are.
We can then write down an ansatz for the Feynman integral under consideration as a linear combination of multiple polylogarithms
with the specific arguments and unknown coefficients.
If the Feynman integral is of uniform weight, the unknown coefficients will be algebraic numbers.
In the next step we determine the unknown coefficients. If the symbol of the Feynman integral is known, the symbol of the
ansatz has to match the symbol of the Feynman integral.
This information together boundary data (and possibly other information like information on discontinuities)
can be sufficient to determine all coefficients.
This is the 
\index{bootstrap approach}
{\bf bootstrap approach}.
It is an heuristic approach, as it depends on our original guess of the arguments of the multiple polylogarithms.
However it is quite powerful. In particular it allows in special situations to bypass 
the need to rationalise square roots \cite{Heller:2019gkq,Heller:2021gun}. We will illustrate this with an example below,
taken from \cite{Heller:PhD}.
The bootstrap approach has also been applied to obtain results for scattering amplitudes to an impressive high loop-order,
see for example \cite{Caron-Huot:2016owq,Dixon:2016nkn,Caron-Huot:2019vjl,Caron-Huot:2019bsq}.

Let's now look at an example:
We consider the one-loop two-point function with equal internal masses introduced as example 1 in section~\ref{chapter_iterated_integrals:deriving_the_dgl}.
As master integrals we use (see eq.~(\ref{chapter_iterated_integrals:trafo_bubble}) and eq.~(\ref{chapter_iterated_integrals:example_bubble_uniform_weight_2D}))
\bq
 I_1' & = & -2 \eps I_{10}\left(2-2\eps\right),
 \nonumber \\
 I_2' & = & - \eps \sqrt{x\left(4+x\right)} I_{11}\left(2-2\eps\right).
\eq
The differential equation for $\vec{I}'=(I_1',I_2')^T$ is given by eq.~(\ref{chapter_iterated_integrals:example_1_eps_form_square_root_singularity}). With eq.~(\ref{chapter_iterated_integrals:example_dlog_square_root}) we have
\bq
\label{chapter_hopf:dgl_bubble}
 \left(d+A'\right) \vec{I}' \; = \; 0,
 & &
 A' \; = \; 
 \eps
 \left(\begin{array}{cc}
 0 & 0 \\
 0 & 1 \\
 \end{array} \right) \omega_1
 -
 \eps
 \left(\begin{array}{cc}
 0 & 0 \\
 1 & 0\\
 \end{array} \right) \omega_2,
 \\
 & &
 \omega_1 \; = \;d\ln\left(4+x\right),
 \;\;\;\;\;\;
 \omega_2 \; = \; d \ln\left(2+x+\sqrt{x\left(4+x\right)} \right).
 \nonumber
\eq
The master integral $I_1'$ is a tadpole integral and rather trivial:
\bq
 I_1' & = & -2- \zeta_2 \eps^2 + {\mathcal O}\left(\eps^3\right).
\eq
In section~\ref{chapter_iterated_integrals:base_transformation} we saw that the square root can be rationalised by the transformation
\bq
\label{chapter_hopf:bubble_trafo}
 x \; = \; \frac{\left(1-x'\right)^2}{x'},
 & &
 x' \; = \; \frac{1}{2} \left( 2 + x - \sqrt{x\left(4+x\right)} \right),
\eq
Under this transformation we have
\bq
\label{chapter_hopf:bubble_trafo_dlog}
 \omega_1
 \; = \;
 2 d\ln\left(x'+1\right) - d\ln\left(x'\right),
 \;\;\;\;\;\;
 \omega_2
 \; = \; - d\ln\left(x'\right).
\eq
The master integral $I_2'$ vanishes at $x=0$ (corresponding to $x'=1$).
With this boundary condition we may integrate the differential equation and obtain
\bq
\label{chapter_hopf:result_bubble_v1}
 I_2' & = & 2 \eps G\left(0;x'\right)
 +2 \eps^2 \left[ G\left(0,0;x'\right) - 2 G\left(-1,0;x'\right) - \zeta_2 \right]
 + {\mathcal O}\left(\eps^3\right).
\eq
We have
\bq
 & &
 G\left(0;x'\right) \; = \; \ln\left(x'\right),
 \\
 & &
 G\left(0,0;x'\right) \; = \; \frac{1}{2} \ln^2\left(x'\right),
 \;\;\;\;\;\;
 G\left(-1,0;x'\right) \; = \; \mathrm{Li}_2\left(-x'\right)+\ln\left(x'\right)\ln\left(x'+1\right).
 \nonumber
\eq
Suppose now that we don't know a rationalisation of the square root $\sqrt{x(4+x)}$.
The symbol approach allows us to derive eq.~(\ref{chapter_hopf:result_bubble_v1}) without the need of rationalising the square root
(so we forget eqs.~(\ref{chapter_hopf:bubble_trafo})-(\ref{chapter_hopf:result_bubble_v1}) for the moment).
We will however assume that the result can be expressed in terms of multiple polylogarithms.
We set
\bq
 f_0 \; = \; 2,
 \;\;\;
 f_1 \; = \; x+4,
 \;\;\;
 f_2 \; = \; 2+x+r,
 \;\;\;\;\;\;
 r \; = \; \sqrt{x\left(4+x\right)}.
\eq
The set $\{f_0,f_1,f_2\}$ will be our alphabet. The subset $\{f_0,f_1\}$ is called the 
\index{alphabet, rational part}
{\bf rational part of the alphabet},
the subset $\{f_2\}$ is called the 
\index{alphabet, algebraic part}
{\bf algebraic part}.
For an algebraic letter $f$ we define the 
\index{conjugated letter with respect to a square root}
{\bf conjugated letter} $\bar{f}$ with respect to the root $r$ as the letter obtained
by the substitution $r \rightarrow -r$.
Thus
\bq
 \bar{f}_2 & = & 2+x-r.
\eq
The letters $f_1$and $f_2$ can be directly read off from the differential equation~(\ref{chapter_hopf:dgl_bubble}).
The inclusion of the letter $f_0=2$ seems a little bit artificial, after all $2$ is a constant and we have
\bq
 d \ln 2 & = & 0.
\eq
However, we require that $f_2 \bar{f}_2$ factorises over the rational part of the alphabet.
We have
\bq
 f_2 \bar{f}_2 & = & 4 \; = \; 2^2 \; = \; f_0^2.
\eq
From the differential equation~(\ref{chapter_hopf:dgl_bubble})
we may write down the symbol for $I_2'$:
\bq
\label{chapter_hopf:symbol_bubble}
 S\left(I_2'\right)
 & = & 
 - 2 \eps \left( f_2 \right)
 + 2 \eps^2 \left( f_1 \otimes f_2 \right)
 + {\mathcal O}\left(\eps^3\right).
\eq
We now consider an ansatz in the form of a linear combination of multiple polylogarithms, 
such that the symbol of the ansatz matches the symbol of eq.~(\ref{chapter_hopf:symbol_bubble}).
At order $\eps^1$ this is rather easy:
\bq
 I^{(1)}_{\mathrm{ansatz}} \; = \; - 2 \ln\left(f_2\right)
 & \Rightarrow &
 S\left(I^{(1)}_{\mathrm{ansatz}}\right) \; = \; - 2 \left(f_2\right).
\eq
In the next step we check the total differential
\bq
 d \left( I_2'{}^{(1)} - I^{(1)}_{\mathrm{ansatz}} \right) & = & 0,
\eq
hence $I_2'{}^{(1)}$ and $I^{(1)}_{\mathrm{ansatz}}$ can possibly only differ by a constant.
From the boundary condition we obtain
\bq
 I_2'{}^{(1)}
 & = & 
 - 2 \left[ \ln\left(f_2\right) - \ln\left(f_0\right) \right],
\eq
in agreement with eq.~(\ref{chapter_hopf:result_bubble_v1}). Note that
\bq
 \frac{2}{2+x+r} & = & \frac{2+x-r}{2}.
\eq
The order $\eps^2$ is more interesting: At weight two we expect dilogarithms and products of logarithms.
Let us first consider the possible arguments of the dilogarithm:
We start from a candidate argument of the form as a power product
\bq
\label{chapter_hopf:ansatz_alpha}
 y & = & f_0^{\alpha_0} f_1^{\alpha_1} f_2^{\alpha_2},
 \;\;\;\;\;\;\;\;\;
 \alpha_j \; \in \; {\mathbb Q}.
\eq
Not every combination of $(\alpha_0,\alpha_1,\alpha_2)$ will be an allowed combination.
To find the restrictions, we consider the symbol of the dilogarithm.
We have
\bq
 S\left( \mathrm{Li}_2\left(y\right) \right)
 & = &
 - \left( y \otimes \left(1-y\right) \right).
\eq
In the first tensor slot the power product distributes according to the rules of eq.~(\ref{chapter_hopf:symbol_rules}):
\bq
 \left(y\right)
 & = &
 \alpha_0 \left( f_0 \right) + \alpha_1 \left( f_1 \right) + \alpha_2 \left( f_2 \right).
\eq
The second tensor slot is more problematic: We don't want any additional new dlog-forms. Therefore we require that $(1-y)$
is again a power product:
\bq
\label{chapter_hopf:ansatz_beta}
 1-y & = & f_0^{\beta_0} f_1^{\beta_1} f_2^{\beta_2},
 \;\;\;\;\;\;\;\;\;
 \beta_j \; \in \; {\mathbb Q}.
\eq
Thus the allowed arguments $y$ of $\mathrm{Li}_2$ are such that for a given $y$ in the form 
of eq.~(\ref{chapter_hopf:ansatz_alpha}) there exists
$(\beta_0,\beta_1,\beta_2)$ such that eq.~(\ref{chapter_hopf:ansatz_beta}) holds.
Note that eq.~(\ref{chapter_hopf:ansatz_beta}) is equivalent to
\bq
 \ln\left(1-y\right)
 - \beta_0 \ln\left(f_0\right)
 - \beta_1 \ln\left(f_1\right)
 - \beta_2 \ln\left(f_2\right)
 & = & 0.
\eq
Given $\ln(1-y)$, $\ln(f_0)$, $\ln(f_1)$ and $\ln(f_2)$ we may use the PSLQ algorithm (discussed in chapter~\ref{chapter_numerics})
to check if $(\beta_0,\beta_1,\beta_2)$ exists.

For the case at hand one finds the allowed arguments
\bq
 y & \in &
 \left\{ y_1, y_2 \right\},
 \;\;\;\;\;\;
 y_1 \; = \; f_0^{-\frac{1}{2}} f_1^{-\frac{1}{2}} f_2^{\frac{1}{2}}, 
 \;\;\;\;\;\;
 y_2 \; = \; f_0^{\frac{1}{2}} f_1^{-\frac{1}{2}} f_2^{-\frac{1}{2}}.
\eq
with symbols
\bq
 S\left(\mathrm{Li}_2\left( y_1 \right) \right)
 & = &
 - \left( y_1 \otimes y_2 \right)
 \; = \;
 \frac{1}{4} \left( f_2 \otimes f_2 + f_2 \otimes f_1 - f_1 \otimes f_2 - f_1 \otimes f_1 \right),
 \nonumber \\
 S\left(\mathrm{Li}_2\left( y_2 \right) \right)
 & = &
 - \left( y_2 \otimes y_1 \right)
 \; = \;
 \frac{1}{4} \left( f_2 \otimes f_2 - f_2 \otimes f_1 + f_1 \otimes f_2 - f_1 \otimes f_1 \right).
\eq
Let us now construct an ansatz, which matches the symbol. 
From
\begin{alignat}{6}
 & S\left(-4 \; \mathrm{Li}_2\left( y_1 \right) \right) & = & - & f_2 \otimes f_2 & - & f_2 \otimes f_1 & + & f_1 \otimes f_2 & + & f_1 \otimes f_1, &
 \nonumber \\
 & S\left( \ln\left(f_1\right)\ln\left(f_2\right) \right) & = & & & & f_2 \otimes f_1 & + & f_1 \otimes f_2, & & &
 \nonumber \\
 & S\left( - \frac{1}{2} \ln^2\left(f_1\right) \right) & = & & & & & & & - & f_1 \otimes f_1, &
 \nonumber \\
 & S\left( \frac{1}{2} \ln^2\left(f_2\right) \right) & = & & f_2 \otimes f_2, & & & & & & &
\end{alignat}
it follows that for
\bq
 I^{(2)}_{\mathrm{ansatz}}
 & = &
 - 4 \; \mathrm{Li}_2\left( y_1 \right) +\ln\left(f_1\right)\ln\left(f_2\right) - \frac{1}{2} \ln^2\left(f_1\right) + \frac{1}{2} \ln^2\left(f_2\right)
\eq
we have
\bq
 S\left(I^{(2)}_{\mathrm{ansatz}}\right) & = & 2 f_1 \otimes f_2.
\eq
In the next step we check the derivative: From the differential equation we have
\bq
 d I_2'{}^{(2)}
 & = & - \omega_1 I_2'{}^{(1)}
 \; = \; 
  2 \left[ \ln\left(f_2\right) - \ln\left(f_0\right) \right] d\ln\left(f_1\right).
\eq
However, the derivative of our ansatz is
\bq
 d I^{(2)}_{\mathrm{ansatz}}
 & = & 
 2 \left[ \ln\left(f_2\right) - \frac{1}{2} \ln\left(f_0\right)\right] d\ln\left(f_1\right)
 + \ln\left(f_0\right) d\ln\left(f_2\right).
\eq
This does not match:
\bq
\label{chapter_hopf:diff_derivatives}
 d \left( I_2'{}^{(2)} - I^{(2)}_{\mathrm{ansatz}} \right)
 & = &
  - \ln\left(f_0\right) \left[ d\ln\left(f_1\right) + d\ln\left(f_2\right) \right].
\eq
The difference is proportional to $\ln(f_0)=\ln(2)$. This comes to no surprise: A constant like $\ln(2)$
is in the kernel of the symbol map and terms proportional to $\ln(2)$ are not detected by the symbol.
We can fix this issue as follows:
We add to our ansatz a function, whose derivative is given by the right-hand side of eq.~(\ref{chapter_hopf:diff_derivatives}).
This is a problem of lower weight, as the sought-after function is proportional to the weight one constant $\ln(2)$.
In our case it is rather trivial:
We add
\bq
 - \ln\left(f_0\right) \ln\left(f_1 f_2\right)
\eq
to our ansatz.
Adding this term will not alter the symbol of our ansatz.

Finally, we match the boundary condition at $x=0$ and we arrive at
\bq
\label{chapter_hopf:result_bubble_v2}
 I_2'{}^{(2)}
 & = & 
 - 4 \; \mathrm{Li}_2\left( y_1 \right) + \ln\left(f_1\right)\ln\left(\frac{f_2}{f_0}\right) - \frac{1}{2} \ln^2\left(f_1\right) + \frac{1}{2} \ln^2\left(\frac{f_2}{f_0}\right)
 + 2 \zeta_2
 \nonumber \\
 & = &
 - 4 \; \mathrm{Li}_2\left( y_1 \right) + 2 \ln^2\left(y_2\right) - \ln^2\left(f_1\right) + 2 \zeta_2.
\eq
The result in eq.~(\ref{chapter_hopf:result_bubble_v2}) does not look like our previous result in eq.~(\ref{chapter_hopf:result_bubble_v1}),
but in the next exercise you are supposed to show that these two results are identical:
\\
\\
\bs
{\it \refstepcounter{exercise}
{\bf Exercise \theexercise}: 
Show that eq.~(\ref{chapter_hopf:result_bubble_v1}) and eq.~(\ref{chapter_hopf:result_bubble_v2}) agree in a neighbourhood of $x=0$.
}
\es
\\
\\
In deriving the result of eq.~(\ref{chapter_hopf:result_bubble_v2}) we never had to rationalise the square root
$r = \sqrt{x(4+x)}$.
There are situations, where one can prove that a certain square root is not rationalisable \cite{Besier:2019hqd},
nevertheless the bootstrap approach is able to find a solution in terms of a linear combination of multiple polylogarithms \cite{Heller:2019gkq}.
This shows the power of the bootstrap approach.

Let us make a few more comments:
The representation of the Feynman integral as a linear combination of multiple polylogarithms is not necessarily unique.
This is already obvious from the two representations in eq.~(\ref{chapter_hopf:result_bubble_v1}) and eq.~(\ref{chapter_hopf:result_bubble_v2}).
Within the bootstrap approach we could as well have started with $4 \; \mathrm{Li}_2(y_2)$ instead of $(-4 \; \mathrm{Li}_2(y_1))$
and obtained yet another different representation. Using $y_1+y_2=1$ and the relation~(\ref{chapter_one_loop:dilog_identities})
one may show that the so obtained representation is equivalent to the previous one.

One could argue that the result in eq.~(\ref{chapter_hopf:result_bubble_v1}) is simpler than the result in
eq.~(\ref{chapter_hopf:result_bubble_v2}), as the former contains only a single square root
\bq
 -x' & = & -\frac{1}{2}\left(2+x-\sqrt{x\left(4+x\right)}\right),
\eq
whereas the latter contains a square root of a square root
\bq
 y_1 & = & \sqrt{ \frac{2+x+\sqrt{x\left(4+x\right)}}{2\left(4+x\right)}}.
\eq
This is an artefact of the choice of our alphabet $\{f_0,f_1,f_2\}$.
If we include another constant $(-1)$ in our alphabet, we will find $(-x')$ as an allowed argument of the dilogarithm. 
\\
\\
\bs
{\it \refstepcounter{exercise}
{\bf Exercise \theexercise}: 
Let $f_1, f_2, g_1, g_2$ be algebraic functions of the kinematic variables $x$.
Determine the symbols of
\bq
 \mathrm{Li}_{2 1}\left(f_1, f_2\right)
 & \mbox{and} &
 G_{2 1}\left(g_1,g_2;1\right).
\eq
Assume then $g_1=1/f_1$ and $g_2=1/(f_1 f_2)$. Show that in this case the two symbols agree.

From the two symbols deduce the constraints on the arguments $f_1, f_2$ of $\mathrm{Li}_{2 1}(f_1, f_2)$
and on the arguments $g_1, g_2$ of $G_{2 1}(g_1,g_2;1)$.
}
\es
\\
\\
From the previous exercise and eq.~(\ref{chapter_multiple_polylogarithms:differential_Glog})
we may deduce 
the constraints on the arguments of a multiple polylogarithm in full generality:
\begin{tcolorbox}
{\bf Constraints on the arguments of a multiple polylogarithm}:
\\
Let 
\bq
 A & = & \left\{ f_1, \dots, f_{\NL} \right\}
\eq
be an alphabet. The $f_j$ define dlog-forms $\omega_j=d\ln f_j$.
Consider power products $z_i$ of the form
\bq
\label{chapter_hopf:power_product_z}
 z_i & = & \prod\limits_{j=1}^{\NL} f_j^{\alpha_{i j}}
\eq
and set $Z = \{1,z_1,\dots,z_k\}$.
The symbol of the multiple polylogarithm $G(z_1,\dots,z_k;1)$ can be expressed in the alphabet $A$ if
any difference
\bq
 z_i - z_j,
 \;\;\;\;\;\; z_i, z_j \; \in \; Z
\eq
can again be expressed as a power product in the form of eq.~(\ref{chapter_hopf:power_product_z}).
\end{tcolorbox}
The proof follows directly from eq.~(\ref{chapter_multiple_polylogarithms:differential_Glog}).

In constructing an ansatz it is worth knowing that up to weight four all multiple polylogarithms can be expressed 
in terms of logarithms, $\mathrm{Li}_2(x_1)$, $\mathrm{Li}_3(x_1)$, $\mathrm{Li}_4(x_1)$ and $\mathrm{Li}_{2 2}(x_1,x_2)$ \cite{Frellesvig:2016ske}.
Thus up to weight four it is sufficient to consider only the functions
\bq
 G_{1}\left(z_1;1\right),
 \;\;\;
 G_{2}\left(z_1;1\right),
 \;\;\;
 G_{3}\left(z_1;1\right),
 \;\;\;
 G_{4}\left(z_1;1\right),
 \;\;\;
 G_{2 2}\left(z_1,z_2;1\right).
\eq
At a given weight, products of functions of lower weight may appear.

%% file: cluster.tex
\newpage
\chapter{Cluster algebras}
\label{chapter_cluster}

In order to motivate the content of this chapter let us look again at example 2
of section~\ref{chapter_iterated_integrals:deriving_the_dgl}: The one-loop four-point function
with vanishing internal masses and one non-zero external mass.
In this example we have two kinematic variables $x_1=2 p_1 \cdot p_2/p_4^2$ and $x_2=2 p_2 \cdot p_3 /p_4^2$ 
(see eq.~(\ref{chapter_iterated_integrals:example_box_def_variables})).
By a suitable choice of master integrals we may transform the differential equation into an $\eps$-form
as we did in eq.~(\ref{chapter_iterated_integrals:Aprime_box}).
We obtain an alphabet with five letters. The arguments of the dlog-forms are
\bq
\label{chapter_cluster:example_alphabet}
 x_1,
 \;\;\;
 x_2,
 \;\;\;
 x_1-1,
 \;\;\;
 x_2-1,
  \;\;\;
 x_1+x_2-1.
\eq
We may now ask: Is there a relation between the initial kinematic variables $x_1$, $x_2$, the Feynman graph $G$ and
the alphabet in eq.~(\ref{chapter_cluster:example_alphabet})?

This is where cluster algebras enter the game.
Cluster algebras were introduced in mathematics by Fomin and Zelevinsky in 2001 \cite{Fomin:2001aaa,Fomin:2003aaa}.
A cluster algebra is a commutative ${\mathbb Q}$-algebra generated by the so-called cluster variables.
The cluster variables are grouped into overlapping subsets of fixed cardinality. These subsets are called clusters.
Starting from an initial seed, the clusters are constructed recursively through mutations.

Introductory texts on cluster algebras are \cite{fomin:book_part_I-III,fomin:book_part_IV-V,fomin:book_part_VI,fomin:book_part_VII,Keller:2008}.
The relation between cluster algebras and scattering amplitudes in particle physics appeared for the first time
in \cite{Golden:2013xva} in the context of ${\mathcal N}=4$ supersymmetric Yang-Mills amplitudes.
The relation of Feynman integrals to cluster integrals is an evolving field of research and in this chapter
we merely touch the tip of an iceberg.
A selection of current research literature on this subject is \cite{Drummond:2014ffa,Parker:2015cia,Harrington:2015bdt,Mago:2020kmp,Mago:2020nuv,Mago:2021luw,Chicherin:2020umh,He:2021esx,He:2021eec}.

In section~\ref{chapter_cluster:quivers} we introduce quivers and mutations.
Cluster algebras arising from quivers and without coefficients 
are discussed in section~\ref{chapter_cluster:quiver_cluster_algebras}. These are the simplest ones.
In section~\ref{chapter_cluster:quiver_cluster_algebras_with_coefficients} we introduce coefficients (or frozen vertices).
Cluster algebras may also be defined in terms of matrices.
In section~\ref{chapter_cluster:cluster_algebras} we consider one further generalisation, going
from anti-symmetric matrices to anti-symmetrisable matrices.
In section~\ref{chapter_cluster:Feynman_integrals} we discuss the relation to Feynman integrals.

\section{Quivers and mutations}
\label{chapter_cluster:quivers}

A 
\index{quiver}
{\bf quiver} is an oriented graph.
A quiver therefore consists of a set of vertices $V$, a set of edges $E$ and maps
\bq
 \mathrm{sink} & : & E \rightarrow V,
 \nonumber \\
 \mathrm{source} & : & E \rightarrow V,
\eq
assigning to each edge its sink and source, respectively.
A quiver is called finite if both the sets $V$ and $E$ are finite sets.
We recall that a self-loop is an edge $e$ with
\bq
 \mathrm{sink}\left(e\right) & = & \mathrm{source}\left(e\right).
\eq
A two-cycle is a pair of two distinct edges $e_1$ and $e_2$ with 
\bq
 \mathrm{sink}\left(e_1\right) \; = \;  \mathrm{source}\left(e_2\right)
 & \mbox{and} &
 \mathrm{sink}\left(e_2\right) \; = \;  \mathrm{source}\left(e_1\right).
\eq
We will mainly deal with finite quivers without self-loops and two-cycles.

It is convenient to write $v_i \rightarrow v_j$, if there is an edge $e$ with
\bq
 \mathrm{source}\left(e\right) \; = \; v_i
 & \mbox{and} &
 \mathrm{sink}\left(e\right) \; = \; v_j.
\eq
Let $Q$ be a finite quiver without self-loops and two-cycles.
We denote by $\nverticesquiver=|V|$ the number of vertices of $Q$.
We may associate an anti-symmetric $(\nverticesquiver \times \nverticesquiver)$-matrix $B$ to $Q$ as follows:
The entry $b_{i j}$ is given as the number of edges, which have $v_i$ as source and $v_j$ as sink minus
the number of edges, which have $v_i$ as sink and $v_j$ as source.
Since we exclude two-cycles at least one of these two numbers is zero.
Furthermore, since we exclude self-loops and two-cycles, 
we may reconstruct uniquely the quiver $Q$ from the matrix $B$.
The matrix $B$ is called the 
\index{exchange matrix}
{\bf exchange matrix} of $Q$.

Let $Q$ be a finite quiver without self-loops and two-cycles and $v_k \in V$ a vertex of $Q$.
The 
\index{mutation}
{\bf mutation} of $Q$ at the vertex $v_k$ is a new quiver $Q'$, obtained from $Q$ as follows:
\begin{enumerate}
\item for each path $v_i \rightarrow v_k \rightarrow v_j$ add an edge $v_i \rightarrow v_j$,
\item reverse all arrows on the edges incident with $v_k$,
\item remove any two-cycles that may have formed.
\end{enumerate}
Two quivers $Q$ and $Q'$ are called 
\index{mutation equivalent}
{\bf mutation equivalent}
if $Q'$ can be obtained through a sequence of mutations from $Q$.

Under a mutation at the vertex $v_k$ the matrix $B$ transforms to a matrix $B'$, whose entries are given
by
\bq
\label{chapter_cluster:mutation_B}
 b_{i j}' & = &
 \left\{
  \begin{array}{ll}
    - b_{i j} & \mbox{if $i=k$ or $j=k$}, \\
    b_{i j} + \mathrm{sign}\left(b_{i k}\right) \cdot \max\left(0, b_{i k} b_{k j}\right), & \mbox{otherwise}. \\
  \end{array}
 \right.
\eq
We call two matrices $B$ and $B'$ mutation equivalent, if $B'$ can be obtained through a sequence of mutations from $B$.
\\
\\
\bs
{\it \refstepcounter{exercise}
{\bf Exercise \theexercise}: 
Show that the mutation of the matrix $B$ at a fixed vertex $v_k$ is an involution, i.e. mutating twice at the same vertex
returns the original matrix $B$.
}
\es
\\
\\
Let us now associate a variable $a_j$ to each vertex $v_j$.
We define the mutation of these variables under a mutation at a vertex $v_k$ as
\bq
\label{chapter_cluster:mutation_A}
 a_j' & = &
 \left\{
  \begin{array}{ll}
   a_j & j \neq k, \\
   \frac{1}{a_k} \left( \prod\limits_{i \; | \; b_{i k} > 0} a_i^{b_{i k}} + \prod\limits_{i \; | \; b_{i k} < 0} a_i^{-b_{i k}} \right) & j = k. \\
  \end{array}
 \right.
\eq
We use the standard convention that an empty product equals one.
The variables $a_j$ are called the 
\index{cluster $A$-coordinates}
{\bf cluster $A$-variables} or the
{\bf cluster $A$-coordinates}.

\begin{figure}
\begin{center}
\includegraphics[scale=0.8]{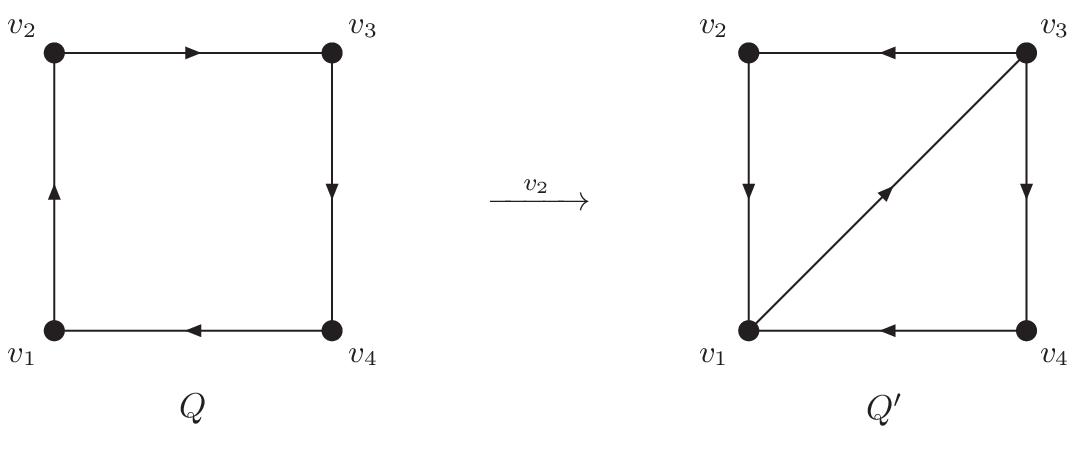}
\caption{\label{chapter_cluster:fig_example_mutation}
The left picture shows the original quiver $Q$, the right picture the quiver $Q'$ obtained through a mutation at the vertex $v_2$. 
}
\end{center}
\end{figure}

Let us set
\bq
\label{chapter_cluster:def_X_coordinates}
 x_j & = & 
 \prod\limits_{i} a_i^{b_{i j}}
\eq
Under a mutation at a vertex $v_k$ the variables $x_j$ transform as
\bq
\label{chapter_cluster:mutation_X}
 x_j' & = &
 \left\{
  \begin{array}{ll}
   x_j \left(1 + x_k^{-\mathrm{sign}(b_{k j})}\right)^{-b_{k j}}  & j \neq k, \\
   \frac{1}{x_k} & j = k. \\
  \end{array}
 \right.
\eq
The variables $x_j$ are called the 
\index{cluster $X$-coordinates}
{\bf cluster $X$-variables} or the
{\bf cluster $X$-coordinates} \cite{Fock2003ClusterEQ}.
\\
\\
\bs
{\it \refstepcounter{exercise}
{\bf Exercise \theexercise}: 
Derive the transformation in eq.~(\ref{chapter_cluster:mutation_X}) 
from eq.~(\ref{chapter_cluster:def_X_coordinates}), eq.~(\ref{chapter_cluster:mutation_A}) and eq.~(\ref{chapter_cluster:mutation_B}).
}
\es
\\
\\
Let's look at an example.
We start with the quiver shown in the left picture of fig.~\ref{chapter_cluster:fig_example_mutation}.
The quiver has four vertices.
The $(4 \times 4)$-matrix $B$ of the quiver $Q$ reads
\bq
 B 
 & = &
 \left(\begin{array}{rrrr}
   0 & 1 & 0 & -1 \\
  -1 & 0 & 1 & 0 \\
   0 & -1 & 0 & 1 \\
   1 & 0 & -1 & 0 \\
 \end{array} \right).
\eq
We consider the mutation at the vertex $v_2$. We obtain the mutated quiver $Q'$ shown in the right picture of 
fig.~\ref{chapter_cluster:fig_example_mutation}.
In $Q'$ the orientation of all edges incident to $v_2$ is reversed. As the original quiver $Q$ contains the edges
$v_1 \rightarrow v_2 \rightarrow v_3$, there is a new edge with the orientation $v_1 \rightarrow v_3$ (according to rule $1$).
The matrix $B'$ associated to $Q'$ reads
\bq
 B' 
 & = &
 \left(\begin{array}{rrrr}
   0 & -1 & 1 & -1 \\
   1 & 0 & -1 & 0 \\
   -1 & 1 & 0 & 1 \\
   1 & 0 & -1 & 0 \\
 \end{array} \right).
\eq
If we assign the variables $(a_1,a_2,a_3,a_4)$ to the vertices $(v_1,v_2,v_3,v_4)$ of the original quiver $Q$, the mutated
variables $(a_1',a_2',a_3',a_4')$ are
\bq
 a_1' \; = \; a_1,
 \;\;\;
 a_2' \; = \; \frac{a_1+a_3}{a_2},
 \;\;\;
 a_3' \; = \; a_3,
 \;\;\;
 a_4' \; = \; a_4.
\eq
The $X$-variables of the original quiver $Q$ are given by
\bq
 x_1 \; = \; \frac{a_4}{a_2},
 \;\;\;
 x_2 \; = \; \frac{a_1}{a_3},
 \;\;\;
 x_3 \; = \; \frac{a_2}{a_4},
 \;\;\;
 x_4 \; = \; \frac{a_3}{a_1},
\eq
the $X$-variables of the mutated quiver $Q'$ are given by
\begin{alignat}{5}
 & x_1' & \;\; = \;\; & x_1\left(1+x_2\right) & \;\; = \;\; & \frac{a_2' a_4'}{a_3'} & \;\; = \;\; & \frac{\left(a_1+a_3\right) a_4}{a_2 a_3},
 \nonumber \\
 & x_2' & \;\; = \;\; & \frac{1}{x_2} & \;\; = \;\; & \frac{a_3'}{a_1'} & \;\; = \;\; & \frac{a_3}{a_1},
 \nonumber \\
 & x_3' & \;\; = \;\; & \frac{x_2 x_3}{1+x_2} & \;\; = \;\; & \frac{a_1'}{a_2' a_4'} & \;\; = \;\; & \frac{a_1 a_2}{\left(a_1+a_3\right) a_4},
 \nonumber \\
 & x_4' & \;\; = \;\; & x_4 & \;\; = \;\; & \frac{a_3'}{a_1'} & \;\; = \;\; & \frac{a_3}{a_1}.
\end{alignat}

\section{Cluster algebras without coefficients}
\label{chapter_cluster:quiver_cluster_algebras}

We start with the simplest cluster algebras: Cluster algebras obtained from a quiver and without coefficients.

A 
\index{seed}
{\bf seed} 
is pair $(Q,a)$, where $Q$ is a finite quiver without self-loops and two-cycles and $a=\{a_1,\dots,a_r\}$ a set
of variables.
The number $r$ equals the number of vertices of the quiver: $r=|V|$.
The set $a=\{a_1,\dots,a_r\}$ is a set of cluster $A$-variables.

We assume the variables $a_1,\dots,a_r$ to be independent, hence they 
generate the field of rational functions ${\mathbb Q}(a_1,\dots,a_r)$.

Let $(Q',a')$ be a pair obtained through a sequence of mutations from the original seed $(Q,a)$ 
(where a mutation transforms the $a_j$'s according to eq.~(\ref{chapter_cluster:mutation_A})).

We call any $a'$ so obtained (including the original $a$) the 
\index{cluster}
{\bf clusters} with respect to $Q$.
We call the union of all variables $a_j'$ of all clusters the 
\index{cluster variables}
{\bf cluster variables}.
Finally, we define the 
\index{cluster algebra}
{\bf cluster algebra} $A_Q$ to be the ${\mathbb Q}$-subalgebra of the field ${\mathbb Q}(a_1,\dots,a_r)$
generated by all the cluster variables.
The cluster algebra is said to be of 
{\bf finite type}, if the number of cluster variables is finite.
Fomin and Zelevinsky \cite{Fomin:2001aaa,Fomin:2003aaa} have classified all cluster algebras of finite type.
A cluster algebra $A_Q$ generated by the seed $(Q,a)$ is of finite type if $Q$ is mutation equivalent 
to an orientation of a simply laced Dynkin diagram (i.e. a Dynkin diagram of type $ADE$).
Dynkin diagrams are reviewed in appendix~\ref{appendix_lie_algebra}.

Let us look at an example. 
\begin{figure}
\begin{center}
\includegraphics[scale=1.0]{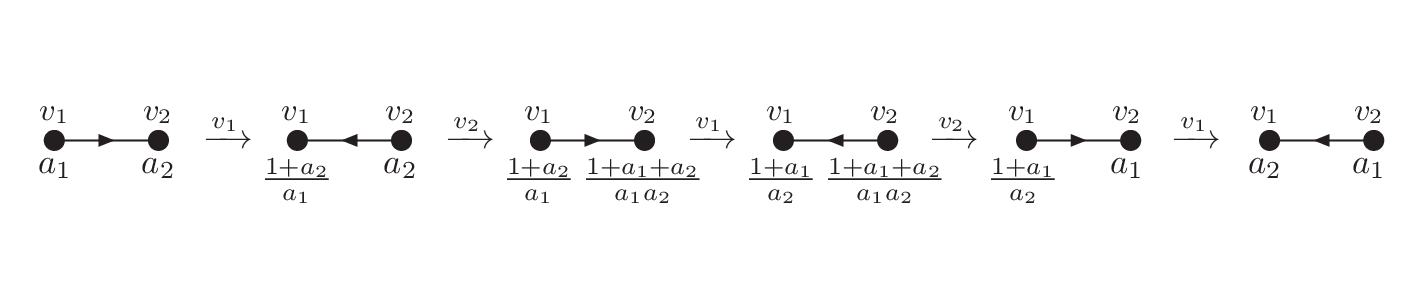}
\caption{\label{chapter_cluster:fig_example_A2_cluster_algebra}
The $A_2$-cluster algebra.
}
\end{center}
\end{figure}
Fig.~\ref{chapter_cluster:fig_example_A2_cluster_algebra} shows the $A_2$-cluster algebra.
We start with the seed shown on the left.
The quiver has two vertices, and we may either mutate at the vertex $v_1$ or the vertex $v_2$.
The mutations of the quiver are not too interesting: Independently of which vertex we choose, the mutation of the quiver
has just the arrow reversed.
The mutation of the cluster variables is more interesting: Performing mutations alternating at the vertices $v_1$ and $v_2$, we obtain
the sequence shown in fig.~\ref{chapter_cluster:fig_example_A2_cluster_algebra}.
We may easily verify that any other mutation reproduces a seed already shown in fig.~\ref{chapter_cluster:fig_example_A2_cluster_algebra}.
There are only a finite number of possibilities to verify.
The cluster variables are therefore
\bq
 a_1, a_2, \frac{1+a_1}{a_2}, \frac{1+a_2}{a_1}, \frac{1+a_1+a_2}{a_1 a_2}.
\eq
This is a finite set and 
the $A_2$-cluster algebra is therefore of finite type.
The $A_2$-cluster algebra is the algebra
\bq
 {\mathbb Q}\left[a_1, a_2, \frac{1+a_1}{a_2}, \frac{1+a_2}{a_1}, \frac{1+a_1+a_2}{a_1 a_2}\right].
\eq

\section{Cluster algebras with coefficients}
\label{chapter_cluster:quiver_cluster_algebras_with_coefficients}

In this section we introduce cluster algebras with coefficients.

Let $1 \le r \le s$ be integers.
An 
\index{ice quiver}
{\bf ice quiver} 
is a quiver with vertex set
\bq
 V & = & \left\{ v_1, \dots, v_r, v_{r+1}, \dots, v_s \right\}
\eq
such that there are no edges between vertices $v_i$ and $v_j$ if $i>r$ and $j>r$.
The vertices $v_{r+1}, \dots, v_s$ are called 
\index{frozen vertex}
{\bf frozen vertices}.
The 
\index{principal part of an ice quiver}
{\bf principal part} of $Q$ is the subquiver consisting of the vertices $v_1,\dots,v_r$ and all the oriented edges 
between them.
For an ice quiver we only allow mutations at vertices $v_k$ with $k \le r$. In addition, in a mutation 
no arrows are drawn
between vertices $v_i$ and $v_j$ if $i>r$ and $j>r$.

To an ice quiver we associate a $(s\times r)$-matrix $\tilde{B}$ with entries $b_{ij}$.
The entry $b_{ij}$ (with $1 \le i \le s$ and $1 \le j \le r$) is given as before:
The number of edges, which have $v_i$ as source and $v_j$ as sink minus
the number of edges, which have $v_i$ as sink and $v_j$ as source.
The matrix $\tilde{B}$ is called the 
\index{extended exchange matrix}
{\bf extended exchange matrix}.
Under a mutation the entries $b_{ij}$ transform as in eq.~(\ref{chapter_cluster:mutation_B}).

The $(r \times r)$-submatrix $B$ with entries $b_{i j}$ with $1  \le i,j \le r$ is called
-- as before -- the exchange matrix.
$B$ is also the exchange matrix of the principal part of $Q$.

In the initial seed we associate the variables
\bq
 \left\{ a_1, \dots, a_r, a_{r+1}, \dots, a_s \right\}
\eq
with the vertices $\{ v_1, \dots, v_r, v_{r+1}, \dots, v_s \}$.
We call $a_1, \dots, a_r$ the cluster variables,
$a_1, \dots, a_r$, $a_{r+1}, \dots, a_s$ the 
{\bf extended cluster variables} 
and $a_{r+1}, \dots, a_s$ the 
{\bf coefficients}.
Under a mutation the extended cluster variables transform as in eq.~(\ref{chapter_cluster:mutation_B}).
Note that the coefficients $a_{r+1}, \dots, a_s$ do not change under mutations.

For $1 \le j \le r$ we define the cluster $X$-variables as in eq.~(\ref{chapter_cluster:def_X_coordinates}).
They transform under mutations as in eq.~(\ref{chapter_cluster:mutation_X}).
Note that there are no cluster $X$-variables with indices $r+1,\dots,s$.

The cluster algebra of an ice quiver is the subalgebra of ${\mathbb Q}(a_1,\dots,a_r,a_{r+1},\dots,a_s)$
generated by the extended cluster variables.
The type of a cluster algebra with coefficients is the type of the cluster algebra generated by the principal part
of the seed.
Thus a cluster algebra with coefficients is of finite type, if its principal part is of finite type.

\begin{figure}
\begin{center}
\includegraphics[scale=0.8]{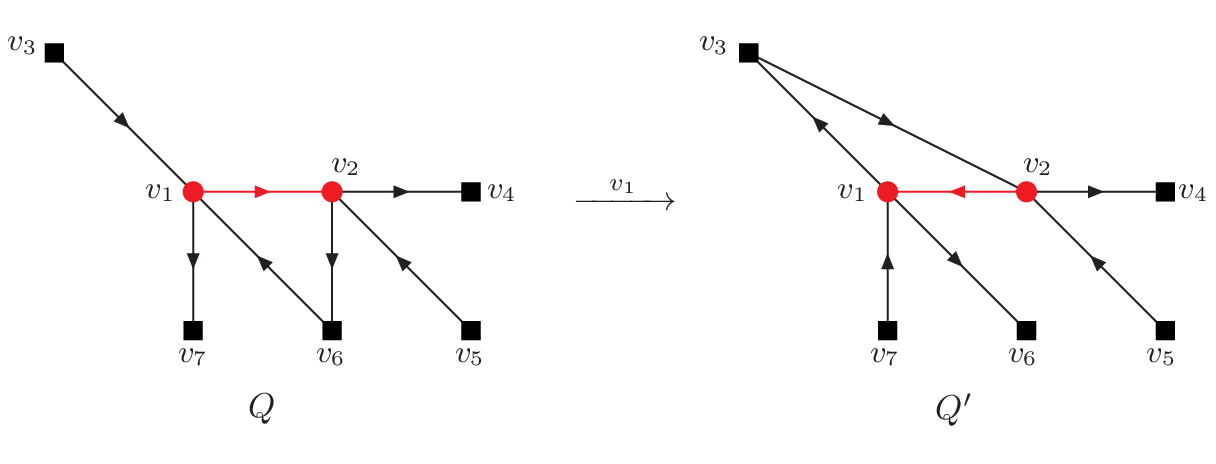}
\caption{\label{chapter_cluster:fig_example_ice_quiver}
The left picture shows an ice quiver $Q$ 
with two unfrozen vertices $(v_1,v_2)$ and four frozen vertices $(v_3,v_4,v_5,v_6)$.
The mutation at vertex $v_1$ gives the ice quiver $Q'$, shown in the right picture.
The principal parts of $Q$ and $Q'$ are drawn in red.
}
\end{center}
\end{figure}

Fig.~\ref{chapter_cluster:fig_example_ice_quiver} shows an example.
It is standard practice to draw frozen vertices as boxes.
The left picture of fig.~\ref{chapter_cluster:fig_example_ice_quiver} shows an ice quiver
with two unfrozen vertices and four frozen vertices.
A mutation at the vertex $v_1$ will produce the quiver $Q'$ shown in the right picture of
fig.~\ref{chapter_cluster:fig_example_ice_quiver}.
The principal part of $Q$ consists of two vertices and one edge shown in red in the left picture of
fig.~\ref{chapter_cluster:fig_example_ice_quiver}.
The cluster algebra generated by $Q$ corresponds to a $A_2$-cluster algebra with coefficients.
\\
\\
\bs
{\it \refstepcounter{exercise}
{\bf Exercise \theexercise}: 
Determine the cluster $A$-variables for the ice quiver $Q'$ of fig.~\ref{chapter_cluster:fig_example_ice_quiver}
in terms of the cluster variables of the ice quiver $Q$.
}
\es
\\
\\
\bs
{\it \refstepcounter{exercise}
{\bf Exercise \theexercise}: 
Mutate the ice quiver $Q'$ of fig.~\ref{chapter_cluster:fig_example_ice_quiver} at the vertex $v_2$ to obtain
an ice quiver $Q''$.
Determine the cluster $A$-variables for the ice quiver $Q''$ 
in terms of the cluster variables of the ice quiver $Q$.
}
\es

\section{Cluster algebras from anti-symmetrisable matrices}
\label{chapter_cluster:cluster_algebras}

In order to arrive at a classification of all cluster algebras of finite type we need one more generalisation:
Up to now we discussed cluster algebras in terms of quivers:
Quivers without frozen vertices for cluster algebras without coefficients and ice quivers for 
cluster algebras with coefficients.
Instead of quivers we could have used the exchange matrix $B$ (for cluster algebras without coefficients)
or the extended exchange matrix $\tilde{B}$ (for cluster algebras with coefficients).
Up to now the exchange matrix $B$ was always an anti-symmetric $(r \times r)$-matrix
\bq
 B & = & - B^T.
\eq
We now relax this condition. We no longer require that $B$ is anti-symmetric,
but only that $B$ is anti-symmetrisable.

A $(r \times r)$-matrix $B$ is called 
{\bf anti-symmetrisable} if there is a diagonal $(r \times r)$-matrix $D$ with positive integers on the diagonal, such
that $D \cdot B$ is anti-symmetric: $D \cdot B = - B^T D$.
Let $D=\mathrm{diag}(d_1,\dots,d_r)$. An anti-symmetrisable matrix satisfies
\bq
 d_i b_{i j} & = & - d_j b_{j i}.
\eq
Example: The matrix
\bq
 B & = &
 \left( \begin{array}{rr}
   0 & -1 \\
   2 & 0 \\
 \end{array} \right)
\eq
is not anti-symmetric. However, it is anti-symmetrisable:
\bq
 \left( \begin{array}{rr}
   2 & 0 \\
   0 & 1 \\
 \end{array} \right)
 \left( \begin{array}{rr}
   0 & -1 \\
   2 & 0 \\
 \end{array} \right)
 & = &
 \left( \begin{array}{rr}
   0 & -2 \\
   2 & 0 \\
 \end{array} \right).
\eq
This generalisation allows us to include cluster algebras, which correspond to non-simply laced Dynkin diagrams
(i.e. Dynkin diagrams of type $B$, $C$, $F$ or $G$).

We now start from an extended exchange matrix $\tilde{B}$ of dimension $(s \times r)$, such that the 
associated exchange matrix $B$
(i.e. the $(r \times r)$-submatrix $B$) is anti-symmetrisable and a set $\{a_1,\dots,a_s\}$ of extended
cluster $A$-variables.
The first $r$ variables $\{a_1,\dots,a_r\}$ are the cluster $A$-variables, the remaining $(s-r)$ variables
$a_{r+1},\dots,a_s$ are the coefficients.
Under a mutation the extended exchange matrix transforms as in eq.~(\ref{chapter_cluster:mutation_B}).
The cluster $A$-variables transform as in eq.~(\ref{chapter_cluster:mutation_A}), the coefficients do not change.
For $1 \le j \le r$ we may define cluster $X$-variables as in eq.~(\ref{chapter_cluster:def_X_coordinates}).
They transform as in eq.~(\ref{chapter_cluster:mutation_X}).
The cluster algebra is the subalgebra of ${\mathbb Q}(a_1,\dots,a_r,a_{r+1},\dots,a_s)$
generated by the extended cluster variables.

In order to state the theorem on the classification of cluster algebras of finite type, we need a little bit more terminology
to establish the relation with Dynkin diagrams.
A $(r \times r)$-matrix $A$ with integer entries is called a 
\index{symmetrisable generalised Cartan matrix}
{\bf symmetrisable generalised Cartan matrix} if
\begin{enumerate}
\item all diagonal entries of $A$ are equal to $2$,
\item all off-diagonal entries of $A$ are non-positive,
\item there exists a diagonal matrix $D$ with positive diagonal entries such that the matrix $D \cdot A$ is symmetric.
\end{enumerate}
A symmetrisable generalised Cartan matrix is called 
{\bf positive}, 
if $D \cdot A$ is positive definite.
This is equivalent to the positivity of all principal minors $|A[I]|$, where $I$ denotes the rows and columns to be deleted.
Now let us choose $I$ as the subset of $(r-2)$ elements, which leaves the $i$-th and $j$-th row and column undeleted.
For a positive symmetrisable generalised Cartan matrix we then have
\bq
 \det\left(\begin{array}{cc}
  2 & a_{ij} \\
  a_{ji} & 2 \\
 \end{array} \right) & > & 0,
\eq
or equivalently
\bq
 a_{i j} a_{j i} & \le & 3.
\eq
A positive symmetrisable generalised Cartan matrix is called a 
\index{Cartan matrix of finite type}
{\bf Cartan matrix of finite type}.

A symmetrisable generalised Cartan matrix is called 
\index{decomposable Cartan matrix}
{\bf decomposable}, if by a simultaneous permutation of rows and columns it can be transformed to a block-diagonal matrix with at least two blocks.
Otherwise it is called 
\index{indecomposable Cartan matrix}
{\bf indecomposable}.
The indecomposable Cartan matrices of finite type
can be classified into four families ($A_n$, $B_n$, $C_n$ and $D_n$) and a finite number of exceptional cases ($E_6$, $E_7$, $E_8$, $F_4$ and $G_2$).
We review this classification in appendix~\ref{appendix_lie_algebra}.
The indecomposable Cartan matrices of finite type
are in one-to-one correspondence with the respective Dynkin diagrams.
Given a Dynkin diagram with $r$ vertices, we obtain the corresponding $r \times r$-Cartan matrix as follows: 
The entries of the Cartan matrix on the diagonal are $a_{ii}=2$ (this is fixed by the definition of a symmetrisable generalised Cartan matrix).
The off-diagonal entries $a_{ij}$ are zero, unless the vertices $i$ and $j$ are connected in the Dynkin diagram.
If they are connected in the Dynkin diagram, they may be connected by one line, two lines or three lines.
In the last two cases there will be in addition an arrow in the Dynkin diagram.
We have:
\begin{center}
\begin{tabular}{lllll}
one line 
& &
\begin{picture}(30,10)(0,0)
\CArc(5,5)(3,0,360)
\CArc(25,5)(3,0,360)
\Line(8,5)(22,5)
\Text(0,5)[r]{$i$}
\Text(30,5)[l]{$j$}
\end{picture}
& &
$a_{ij} \; = \; -1$,
\;\;\;
$a_{ji} \; = \; -1$,
\\
& & & & \\
two lines 
& &
\begin{picture}(30,10)(0,0)
\CArc(5,5)(3,0,360)
\CArc(25,5)(3,0,360)
\Line(8,6)(22,6)
\Line(8,4)(22,4)
\Line(17,5)(12,10)
\Line(17,5)(12,0)
\Text(0,5)[r]{$i$}
\Text(30,5)[l]{$j$}
\end{picture}
& &
$a_{ij} \; = \; -1$,
\;\;\;
$a_{ji} \; = \; -2$,
\\
& & & & \\
three lines 
& &
\begin{picture}(30,10)(0,0)
\CArc(5,5)(3,0,360)
\CArc(25,5)(3,0,360)
\Line(8,7)(22,7)
\Line(8,3)(22,3)
\Line(8,5)(22,5)
\Line(17,5)(12,10)
\Line(17,5)(12,0)
\Text(0,5)[r]{$i$}
\Text(30,5)[l]{$j$}
\end{picture}
& &
$a_{ij} \; = \; -1$,
\;\;\;
$a_{ji} \; = \; -3$,
\\
\end{tabular}
\end{center}
Let $B$ be an anti-symmetrisable
$(r \times r)$-matrix with integer entries.
We associate to $B$ a $(r \times r)$-symmetrisable generalised Cartan matrix $A(B)$ with entries $a_{ij}$ as follows:
\bq
 a_{ij} & = &
 \left\{ \begin{array}{rl}
   2, & \mbox{if} \;\;\; i = j, \\
 -\left|b_{ij}\right|, & \mbox{if} \;\;\; i \neq j. \\
 \end{array}
 \right.
\eq
We may now state the classification theorem for cluster algebras of finite type \cite{Fomin:2003aaa,fomin:book_part_IV-V}:
\begin{tcolorbox}
{\bf Classification of cluster algebras of finite type}:
\begin{theorem}
A cluster algebra is of finite type, if and only if it contains an exchange matrix $B$
such that $A(B)$ is a Cartan matrix of finite type.
\end{theorem}
\end{tcolorbox}
Please note that Dynkin diagrams and quivers are different objects.
This is best seen by an example:
The Dynkin diagram of type $B_2$ corresponds to the exchange matrices $B$
\bq
\begin{picture}(40,10)(0,0)
\CArc(5,5)(3,0,360)
\CArc(25,5)(3,0,360)
\Line(8,6)(22,6)
\Line(8,4)(22,4)
\Line(17,5)(12,10)
\Line(17,5)(12,0)
\Text(0,5)[r]{$1$}
\Text(30,5)[l]{$2$}
\end{picture}
 & \Rightarrow &
 B \; = \; 
 \pm
 \left(\begin{array}{rr}
  0 & -1 \\ 
  2 & 0 \\
 \end{array} \right),
\eq
whereas the quiver $Q$, shown below, corresponds to the exchange matrix $B'$
\bq
\begin{picture}(40,10)(0,0)
\Vertex(5,5){3}
\Vertex(25,5){3}
\ArrowArcn(15,-12.32)(20,120,60)
\ArrowArc(15,22.32)(20,240,300)
\Text(0,5)[r]{$1$}
\Text(30,5)[l]{$2$}
\end{picture}
 & \Rightarrow &
 B' \; = \; 
 \left(\begin{array}{rr}
  0 & 2 \\ 
  -2 & 0 \\
 \end{array} \right).
\eq
The former generates a cluster algebra of finite type, the latter does not.
\\
\\
\bs
{\it \refstepcounter{exercise}
\label{chapter_cluster:example_B_2}
{\bf Exercise \theexercise}: 
The $B_2$-cluster algebra:
Determine the cluster variables from the initial seed
\bq
 B \; = \; 
 \left(\begin{array}{rr}
  0 & -1 \\ 
  2 & 0 \\
 \end{array} \right),
 & &
 a \; = \; \left(a_1,a_2\right).
\eq
}
\es

\section{The relation of cluster algebras to Feynman integrals}
\label{chapter_cluster:Feynman_integrals}

Let us now explore the relation of cluster algebras to Feynman integrals.
In this section we denote the kinematic variables of a Feynman integrals by $x$, the cluster $A$-variables by $a$.
In order to avoid a conflict of notation we will denote in this section cluster $X$-coordinates by $\tilde{x}$.

We define polylogarithmic cluster functions of weight $w$ as follows \cite{Parker:2015cia}: Let us assume that the cluster $A$-variables $a$
(or the cluster $X$-variables $\tilde{x}$) are functions of the kinematic variables $x$.
Polylogarithmic cluster functions of weight $0$ are constants, a polylogarithmic cluster functions $f^{(w)}$ of weight $w$ has
a differential of the form
\bq
 d f^{(w)}
 & = &
 \sum\limits_{j} f^{(w-1)}_j d \ln a_j,
\eq
where the $a_j$'s are cluster $A$-variables and the $f^{(w-1)}_j$'s are polylogarithmic cluster functions of weight $(w-1)$.
A similar definition applies by substituting the cluster $A$-variables with cluster $X$-variables.

Let us now return to the one-loop box integral with one external mass and vanishing internal masses from the introductory remarks 
of this chapter.
The four master integrals $I_1'$, $I_2'$, $I_3'$ and $I_4'$ (defined in eq.~(\ref{chapter_iterated_integrals:base_change_box}))
can be expressed in terms of multiple polylogarithms with the five-letter alphabet
\bq
\label{chapter_cluster:example_alphabet_II}
 x_1,
 \;\;\;
 x_2,
 \;\;\;
 x_1-1,
 \;\;\;
 x_2-1,
  \;\;\;
 x_1+x_2-1.
\eq
Furthermore, the term $I_i^{(j)}{}'$ appearing in the $\eps$-expansion at order $j$
\bq
 I_i' & = & \sum\limits_{j=0}^\infty I_i^{(j)}{}' \cdot \eps^j
\eq
is of uniform weight $j$.
Setting
\bq
 a_1 \; = \; - x_1,
 & &
 a_2 \; = \; - x_2
\eq
shows that $I_i^{(j)}{}'$ is a polylogarithmic cluster functions of weight $j$ for the $A_2$-cluster algebra (compare with fig.~\ref{chapter_cluster:fig_example_A2_cluster_algebra}).
We set
\bq
 a_1 \; = \; a_1,
 \;\;\;
 a_2 \; = \; a_2,
 \;\;\;
 a_3 \; = \; \frac{1+a_2}{a_1},
 \;\;\;
 a_4 \; = \; \frac{1+a_1+a_2}{a_1a_2},
 \;\;\;
 a_5 \; = \; \frac{1+a_1}{a_2}.
\eq
Let us take the definition of $\omega_1$-$\omega_5$ from eq.~(\ref{chapter_iterated_integrals:dlog_form}).
We have
\begin{alignat}{4}
 & \omega_1 & \; = \; & d\ln\left(x_1\right) & & & \; = \; & d\ln\left(a_1\right),
 \\
 & \omega_2 & \; = \; & d\ln\left(x_1-1\right) & \; = \; & d\ln\left(1+a_1\right) & \; = \; & d\ln\left(a_2\right) + d\ln\left(a_5\right),
 \nonumber \\
 & \omega_3 & \; = \; & d\ln\left(x_2\right) & & & \; = \; & d\ln\left(a_2\right),
 \nonumber \\
 & \omega_4 & \; = \; & d\ln\left(x_2-1\right) & \; = \; & d\ln\left(1+a_2\right) & \; = \; & d\ln\left(a_1\right) + d\ln\left(a_3\right),
 \nonumber \\
 & \omega_5 & \; = \; & d\ln\left(x_1+x_2-1\right) & \; = \; & d\ln\left(1+a_1+a_2\right) & \; = \; & d\ln\left(a_1\right) + d\ln\left(a_2\right) + d\ln\left(a_4\right).
 \nonumber 
\end{alignat}
Let us now go to higher loops, keeping the same external kinematic. The planar and non-planar double
box integrals are known \cite{Gehrmann:2000zt,Gehrmann:2001ck}, as well as the planar triple box \cite{DiVita:2014pza}.
These can be expressed as multiple polylogarithms with a six-letter alphabet
\bq
\label{chapter_cluster:example_alphabet_III}
 x_1,
 \;\;\;
 x_2,
 \;\;\;
 x_1-1,
 \;\;\;
 x_2-1,
  \;\;\;
 x_1+x_2-1,
  \;\;\;
 x_1+x_2.
\eq
The sixth letter $(x_1+x_2)$ is not present in the one-loop case.
We may again relate this alphabet to a cluster algebra. 
We need a cluster algebra with six cluster variables. 
The $B_2$-cluster algebra discussed in exercise~\ref{chapter_cluster:example_B_2}
has six cluster variables:
\bq
 a_1,
 \;\;\;
 a_2,
 \;\;\;
 a_3 \; = \; \frac{1+a_2^2}{a_1},
 \;\;\;
 a_4 \; = \; \frac{1+a_1+a_2^2}{a_1a_2},
 \;\;\;
 a_5 \; = \; \frac{1+2a_1+a_1^2+a_2^2}{a_1a_2^2},
 \;\;\;
 a_6 \; = \; \frac{1+a_1}{a_2}.
 \nonumber \\
\eq
Setting \cite{Chicherin:2020umh}
\bq
 x_1 \; = \; - \frac{a_2^2}{1+a_1},
 & & 
 x_2 \; = \; - \frac{1+a_1+a_2^2}{a_1\left(1+a_1\right)}
\eq
allows us to express these Feynman integrals to all order in $\eps$ as polylogarithmic cluster functions.
In detail we have
\begin{alignat}{3}
 & \omega_1 & \; = \; & d\ln\left(x_1\right) & \; = \; & d\ln\left(a_2\right) - d\ln\left(a_6\right),
 \nonumber \\
 & \omega_2 & \; = \; & d\ln\left(x_1-1\right) & \; = \; & d\ln\left(a_1\right) + d\ln\left(a_4\right) - d\ln\left(a_6\right),
 \nonumber \\
 & \omega_3 & \; = \; & d\ln\left(x_2\right) & \; = \; & d\ln\left(a_4\right) - d\ln\left(a_6\right),
 \nonumber \\
 & \omega_4 & \; = \; & d\ln\left(x_2-1\right) & \; = \; & d\ln\left(a_2\right) + d\ln\left(a_5\right) - d\ln\left(a_6\right),
 \nonumber \\
 & \omega_5 & \; = \; & d\ln\left(x_1+x_2-1\right) & \; = \; & d\ln\left(a_2\right) + d\ln\left(a_4\right),
 \nonumber \\
 & \omega_6 & \; = \; & d\ln\left(x_1+x_2\right) & \; = \; & d\ln\left(a_3\right).
\end{alignat}

%% file: elliptics.tex
\newpage
\chapter{Elliptic curves}
\label{chapter_elliptics}

Up to now we discussed mainly Feynman integrals, which can expressed in terms of multiple polylogarithms.
Multiple polylogarithms are an important class of functions for Feynman integrals, 
but not every Feynman integral can be expressed in terms of multiple polylogarithms.
We have already encountered one example:
The two-loop sunrise integral with equal internal masses, discussed as example 4 in section~\ref{chapter_iterated_integrals:deriving_the_dgl}, 
is not expressible in terms of multiple polylogarithms.

This Feynman integral is related to an elliptic curve.
We will see that by a suitable fibre transformation and by a suitable base transformation we may nevertheless transform the 
differential equation for this Feynman integral 
into the $\eps$-form of eq.~(\ref{chapter_iterated_integrals:eps_form}).
The solution is then again given as iterated integrals, in this specific case as iterated integrals of modular forms.

We call the Feynman integrals treated in this chapter ``elliptic Feynman integrals''.
As a rough guide, elliptic Feynman integrals are the next-to-easiest Feynman integrals, with Feynman integrals evaluating
to multiple polylogarithms being the easiest Feynman integrals.
Of course, there are also more complicated Feynman integrals 
beyond these two categories \cite{Brown:2010a,Bourjaily:2018yfy,Bourjaily:2019hmc,Klemm:2019dbm,Bonisch:2020qmm}.

In this chapter we will study elliptic functions, elliptic curves, modular transformations and the moduli space of a genus one curve with $n$ marked points.

Textbooks on elliptic curves are Du Val \cite{Du_Val} and Silverman \cite{Silverman},
textbooks on modular forms are Stein \cite{Stein}, Miyake \cite{Miyake}, Diamond and Shurman \cite{Diamond}
and Cohen and Str\"omberg \cite{Cohen}.

\section{Algebraic curves}
\label{chapter_elliptics:section_algebraic_curves}

We start with the definition of an 
\index{algebraic curve}
{\bf algebraic curve}. 
As ground field we take the complex numbers $\mathbb{C}$.
An algebraic curve in $\mathbb{C}^2$ is defined by the zero set of a polynomial $P(x,y)$ in two variables $x$ and $y$:
\bq
 P\left(x,y\right) & = & 0
\eq
It is more common to consider algebraic curves not in the affine space $\mathbb{C}^2$, but in the projective space
$\mathbb{CP}^2$.
Let $[x:y:z]$ be homogeneous coordinates of $\mathbb{CP}^2$.
An algebraic curve in $\mathbb{CP}^2$ is defined by the zero set of a homogeneous polynomial $P(x,y,z)$ in the three variables $x$, $y$ and $z$:
\bq
\label{chapter_elliptics:def_algebraic_curve}
 P\left(x,y,z\right) & = & 0
\eq
The requirement that $P(x,y,z)$ is a homogeneous polynomial is necessary to have a well-defined zero set on $\mathbb{CP}^2$.

We usually work in the chart $z=1$.
In this chart eq.~(\ref{chapter_elliptics:def_algebraic_curve}) reduces to
\bq
 P\left(x,y,1\right) & = & 0.
\eq
If $d$ is the degree of the polynomial $P(x,y,z)$, the 
\index{arithmetic genus}
{\bf arithmetic genus} 
of the algebraic curve is given by
\bq
\label{chapter_elliptics:def_genus}
 g & = & \frac{1}{2} \left(d-1\right) \left(d-2\right).
\eq
For a smooth curve the arithmetic genus equals the 
\index{geometric genus}
{\bf geometric genus}, 
therefore just using ``genus'' is unambiguous in the smooth case.
Let's look at an example: The equation
\bq
\label{chapter_elliptics:example_elliptic_curve}
 y^2 z - x^3 - x z^2 & = & 0
\eq
defines a smooth algebraic curve of genus $1$.

Let us now turn to elliptic curves:
An 
\index{elliptic curve}
{\bf elliptic curve} 
over $\mathbb{C}$ is a smooth algebraic curve in $\mathbb{CP}^2$ of genus one with one marked point.
It is common practice to work in the chart $z=1$ and to take as the marked point the ``point at infinity''.
Eq.~(\ref{chapter_elliptics:example_elliptic_curve}) reads in the chart $z=1$
\bq
 y^2 - x^3 - x & = & 0,
\eq
The point at infinity, which is not contained in this chart, is given by $[x:y:z]=[0:1:0]$.

Over the complex numbers ${\mathbb C}$ any elliptic curve can be cast into the
\index{Weierstrass normal form}
{\bf Weierstrass normal form}.
In the chart $z=1$ the Weierstrass normal form reads
\bq
 y^2 & = & 4 x^3 - g_2 x - g_3.
\eq
A second important example is to define an elliptic curve by 
a quartic polynomial in the chart $z=1$:
\bq
\label{chapter_elliptics:example_quartic}
 y^2 & = & \left(x-x_1\right) \left(x-x_2\right) \left(x-x_3\right) \left(x-x_4\right).
\eq
If all roots of the quartic polynomial on the right-hand side are distinct, this defines a smooth elliptic curve.
(The attentive reader may ask, how this squares with the genus formula above.
The answer is that the elliptic curve in $\mathbb{CP}^2$ is not given by the homogenisation
$y^2 z^2 = (x-x_1z)(x-x_2z)(x-x_3z)(x-x_4z)$. The latter curve is singular at infinity.
However, there is a smooth elliptic curve, which in the chart $z=1$ is isomorphic to the affine curve defined by
eq.~(\ref{chapter_elliptics:example_quartic}).
)

As one complex dimension corresponds to two real dimensions, we may consider a smooth algebraic curve 
(i.e. an object of complex dimension one) also as a real surface (i.e. an object of real dimension two).
The latter objects are called 
\index{Riemann surfaces}
{\bf Riemann surfaces},
as the real surface inherits the structure of a complex manifold.
We may therefore view an elliptic curve either as a complex one-dimensional smooth algebraic curve in $\mathbb{CP}^2$
with one marked point or as a real Riemann surface of genus one with one marked point.
\begin{figure}
\begin{center}
\includegraphics[align=c,scale=1.0]{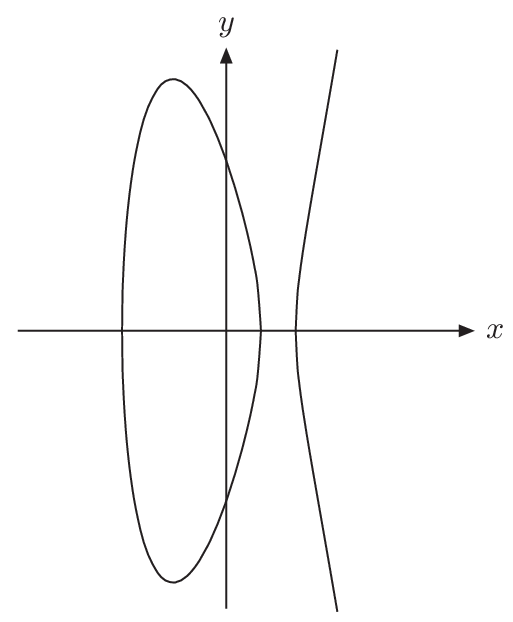}
\hspace*{10mm}
\includegraphics[align=c,scale=1.0]{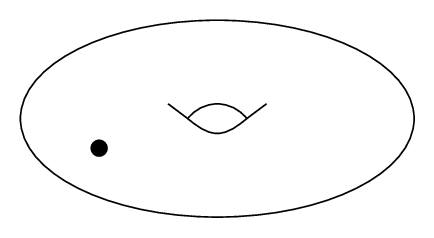}
\end{center}
\caption{
The left picture shows the real part of an elliptic curve in 
the Weierstrass normal form $y^2 = 4 x^3 - g_2 x - g_3$.
The marked point is at infinity.
The right part shows a 
real Riemann surface of genus one with one marked point.
}
\label{chapter_elliptics:fig1}
\end{figure}
This is shown in fig.~\ref{chapter_elliptics:fig1}.
We get from the complex algebraic curve to the real Riemann surface as follows:
Let's consider the curve defined by eq.~(\ref{chapter_elliptics:example_quartic}).
We first note that for $x \notin \{x_1,x_2,x_3,x_4\}$ we have two possible values of $y$:
\bq
 y & = & \pm \sqrt{\left(x-x_1\right) \left(x-x_2\right) \left(x-x_3\right) \left(x-x_4\right)}.
\eq
We denote by $[x_i,x_j]$ the line segment from $x_i$ to $x_j$ in the complex plane.
We may define a single-valued square root for $x \in \mathbb{CP}^1\backslash \mathrm{Cuts}$, where
the cuts remove points between, say, $x_1$ and $x_2$ as well as between $x_3$ and $x_4$:
\bq
 \mathrm{Cuts} & = & [x_1,x_2] \cup [x_3,x_4]
\eq
For the other possible value of $y$ we take a second copy of $\mathbb{CP}^1\backslash \mathrm{Cuts}$.
The two cuts on each copy can be deformed into circles.
We then glue the two copies together along the circles originating from the cuts.
This gives the torus shown in fig.~\ref{chapter_elliptics:fig1}.

\section{Elliptic functions and elliptic curves}
\label{chapter_elliptics:section_elliptic_functions}

Let us now turn to periodic functions and periods.
In chapter~\ref{chapter_sector_decomposition} we already introduced the advanced concepts of numerical periods, effective periods and abstract periods.
We didn't really discuss where the topic of periods originated from. 
We will now close this gap and review periodic functions of a single complex variable $z$.

We consider a non-constant meromorphic function $f$ of a complex variable $z$.
A 
\index{period}
{\bf period}
$\psi$ of the function $f$ is a constant such that
for all $z$:
\bq
 f\left(z+\psi\right) & = & f\left(z\right)
\eq
The set of all periods of $f$ forms a lattice, which is either 
\begin{itemize}
\item trivial (i.e. the lattice consists of $\psi=0$ only),
\item a 
\index{simple lattice}
{\bf simple lattice},
generated by one period $\psi$ : $\Lambda = \left\{ n \psi \; | \; n \in {\mathbb Z} \right\}$,
\item a 
\index{double lattice}
{\bf double lattice},
generated by two periods $\psi_1, \psi_2$ with $\mathrm{Im}(\psi_2/\psi_1) \neq 0$ : 
\bq
\Lambda & = & \left\{ n_1 \psi_1 + n_2 \psi_2 \; | \; n_1, n_2 \in {\mathbb Z} \right\}.
\eq
It is common practice to order these two periods such that $\mathrm{Im}(\psi_2/\psi_1) > 0$.
\end{itemize}
There cannot be more possibilities: Assume that there is a third period $\psi_3$, which is not an element of the lattice
$\Lambda$ spanned by $\psi_1$ and $\psi_2$.
In this case we may construct arbitrary small periods as linear combinations of $\psi_1$, $\psi_2$ and $\psi_3$
with integer coefficients.
In the next step one shows that this implies that the derivative of $f(z)$ vanishes at any point $z$, hence $f(z)$ is a constant.
This contradicts our assumption that $f$ is a non-constant function.

An example for a singly periodic function is given by
\bq
 \exp\left(z\right).
\eq
In this case the simple lattice is generated by $\psi = 2 \pi i$.

Double periodic functions are called 
\index{elliptic functions}
{\bf elliptic functions}.
An example for a doubly periodic function is given by 
\index{Weierstrass's $\wp$-function}
{\bf Weierstrass's $\wp$-function}.
Let $\Lambda$ be the lattice generated by $\psi_1$ and $\psi_2$.
Then
\bq
 \wp\left(z\right)
 & = & 
 \frac{1}{z^2} + \sum\limits_{\psi \in \Lambda \backslash \{0\}} \left( \frac{1}{\left(z+\psi\right)^2} - \frac{1}{\psi^2} \right).
\eq
$\wp(z)$ is periodic with periods $\psi_1$ and $\psi_2$.
Weierstrass's $\wp$-function is an even function, i.e. $\wp(-z)=\wp(z)$.

Of particular interest are also the corresponding inverse functions. These are in general multivalued functions.
In the case of the exponential function $x=\exp(z)$,
the inverse function is given by
\bq
 z 
 & = & 
 \ln\left(x\right).
\eq
The inverse function to Weierstrass's elliptic function $x=\wp(z)$ is an elliptic integral given by
\bq
\label{chapter_elliptics:relation_x_to_z}
 z 
 & = &
 \int\limits_\infty^x \frac{dt}{\sqrt{4t^3-g_2t-g_3}}
\eq
with
\bq
 g_2 = 60 \sum\limits_{\psi \in \Lambda \backslash \{0\}} \frac{1}{\psi^4},
 & \;\;\;\;\;\; &
 g_3 = 140 \sum\limits_{\psi \in \Lambda \backslash \{0\}} \frac{1}{\psi^6}.
\eq
Note that as $\wp(-z)=\wp(z)$, 
\bq
 z 
 & = &
 - \int\limits_\infty^x \frac{dt}{\sqrt{4t^3-g_2t-g_3}}
\eq
is also an inverse function to $x=\wp(z)$.
We may therefore choose any sign of the square root.
In this book we use the convention as in eq.~(\ref{chapter_elliptics:relation_x_to_z})
together with a branch cut of the square root along the negative real axis.

\begin{tcolorbox}
{\bf Elliptic integrals}:
\\
\\
The standard
\index{elliptic integrals}
elliptic integrals 
are classified as
complete or incomplete elliptic integrals and as integrals
of the first, second or third kind.
The complete elliptic integrals
are
\begin{alignat}{2}
 & \mbox{first kind:} 
 & \hspace*{2mm} &
 K\left(x\right) \; = \; \int\limits_0^1 \frac{dt}{\sqrt{\left(1-t^2\right)\left(1-x^2 t^2 \right)}},
 \nonumber \\
 & \mbox{second kind:}
 & &
 E\left(x\right) \; = \; \int\limits_0^1 dt \frac{\sqrt{1-x^2 t^2}}{\sqrt{1-t^2}},
 \nonumber \\
 & \mbox{third kind:}
 & &
 \Pi\left(v,x\right) \; = \; \int\limits_0^1 \frac{dt}{\left(1-vt^2\right)\sqrt{\left(1-t^2\right)\left(1-x^2 t^2 \right)}}.
\end{alignat}
The incomplete elliptic integrals
are
\begin{alignat}{2}
 & \mbox{first kind:} 
 & \hspace*{2mm} &
 F\left(z,x\right) \; = \; \int\limits_0^z \frac{dt}{\sqrt{\left(1-t^2\right)\left(1-x^2 t^2 \right)}},
 \nonumber \\
 & \mbox{second kind:}
 & &
 E\left(z,x\right) \; = \; \int\limits_0^z dt \frac{\sqrt{1-x^2 t^2}}{\sqrt{1-t^2}},
 \nonumber \\
 & \mbox{third kind:}
 & &
 \Pi\left(v,z,x\right) \; = \; \int\limits_0^z \frac{dt}{\left(1-v t^2\right)\sqrt{\left(1-t^2\right)\left(1-x^2 t^2 \right)}}.
\end{alignat}
The complete elliptic integrals are a special case of the incomplete elliptic integrals and obtained
from the incomplete elliptic integrals by setting the variable $z$ to one.
\end{tcolorbox}

The classification of elliptic integrals 
as integrals of the first, second or third kind follows the classification of 
\index{Abelian differential}
{\bf Abelian differentials}:
An Abelian differential $f(z)dz$ is called Abelian differential of the first kind, if $f(z)$ is holomorphic.
It is called an Abelian differential of the second kind, if $f(z)$ is meromorphic, but with all residues vanishing.
It is called an Abelian differential of the third kind, if $f(z)$ is meromorphic with non-zero residues.

In $\mathbb{C}/\Lambda$ with coordinate $z$ the differential $dz$ is clearly 
an Abelian differential of the first kind.
\\
\\
\bs
{\it \refstepcounter{exercise}
{\bf Exercise \theexercise}: 
Consider the elliptic curve $y^2=4x^3-g_2x-g_3$ . Show that
\bq
 dz & = & \frac{dx}{y},
\eq
where $y=\sqrt{4x^3-g_2x-g_3}$. This shows that $dx/y$ is a holomorphic differential. 
}
\es
\\
\\
So far we introduced elliptic curves and elliptic integrals. The link between the two is provided
by the 
\index{period of an elliptic curve}
{\bf periods of an elliptic curve}.
An elliptic curve has one holomorphic differential (i.e. one Abelian differential of the first kind).
If we view the elliptic curve as a genus one Riemann surface (i.e. a torus), we see that there are two independent
\begin{figure}
\begin{center}
\includegraphics[scale=1.0]{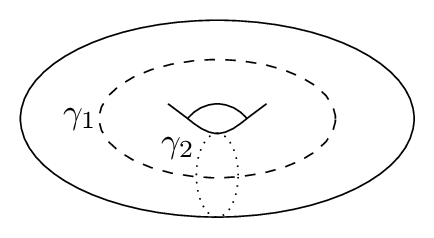}
\end{center}
\caption{
A genus one Riemann surface, where the two independent cycles $\gamma_1$ and $\gamma_2$ are indicated.
}
\label{chapter_elliptics:fig2}
\end{figure}
cycles $\gamma_1$ and $\gamma_2$, as shown in fig.~\ref{chapter_elliptics:fig2}.
A period of an elliptic curve is the integral of the holomorphic differential along a cycle.
As there are two independent cycles, there are two independent periods.
Let's study this for an elliptic curve in the Legendre form
\bq
\label{chapter_elliptics:elliptic_curve_Legendre_form}
 y^2 & = & x \left(x-1\right) \left(x-\lambda\right),
\eq
where $\lambda$ is a parameter not equal to $0$, $1$ or infinity.
We may think of the elliptic curve as two copies of $\mathbb{CP}^1\backslash \mathrm{Cuts}$, where 
the cuts are between $0$ and $\lambda$ as well as between $1$ and $\infty$.
$dx/y$ is the holomorphic differential.
Integrating between $x=0$ and $x=\lambda$ will give a half-period, integrating
from $x=\lambda$ to $x=0$ on the other side of the cut gives another half-period.
In order to obtain two periods, we choose two independent integration paths.
(The integration between $x=0$ and $x=\lambda$ will give a result proportional to the integration between
$x=1$ and $x=\infty$, therefore to obtain two independent periods, the two integrations should have one integration
boundary in common and differ in the other integration boundary.)
A possible choice for the two independent periods is
\bq
 \psi_1 = 2 \int\limits_{0}^{\lambda} \frac{dx}{y} = 4 K\left(\sqrt{\lambda}\right),
 & &
 \psi_2 = 2 \int\limits^{\lambda}_{1} \frac{dx}{y} = 4 i K\left(\sqrt{1-\lambda}\right).
\eq
\bs
{\it \refstepcounter{exercise}
\label{chapter_elliptics:exercise_general_quartic}
{\bf Exercise \theexercise}: 
Determine two independent periods for the elliptic curve defined by
a quartic polynomial:
\bq
 y^2 & = & \left(x-x_1\right) \left(x-x_2\right) \left(x-x_3\right) \left(x-x_4\right).
\eq
}
\es
\noindent
From fig.~\ref{chapter_elliptics:fig2} it is evident that the first homology group $H_1(E)$ of an elliptic curve $E$
is isomorphic to ${\mathbb Z} \times {\mathbb Z}$ and generated by $\gamma_1$ and $\gamma_2$.
The two periods are integrals of the holomorphic differential $dx/y$ along $\gamma_1$ and $\gamma_2$, respectively.
We have $\dim H_1(E) = 2$ and it follows that also the first cohomology group of an elliptic curve is two-dimensional. 
We already know that $dx/y$ is an element of the first de Rham cohomology group
\bq
 \frac{dx}{y} & \in & H^1_{\mathrm{dR}}\left(E\right).
\eq
For the elliptic curve of eq.~(\ref{chapter_elliptics:elliptic_curve_Legendre_form}) we may take
as a second generator
\bq
 \frac{x dx}{y} & \in & H^1_{\mathrm{dR}}\left(E\right).
\eq
Integrating $x dx/y$ over $\gamma_1$ and $\gamma_2$ defines the 
\index{quasi-period}
{\bf quasi-periods} $\phi_1$ and $\phi_2$, respectively.
We obtain
\bq
 \phi_1 \; = \;
 2 \int\limits_0^\lambda \frac{x dx}{y} = 4 K\left(\sqrt{\lambda}\right) - 4 E\left(\sqrt{\lambda}\right),
 & &
 \phi_2 \; = \;
 2 \int\limits_1^\lambda \frac{x dx}{y} = 4 i E\left(\sqrt{1-\lambda}\right).
\eq
The 
\index{period matrix}
{\bf period matrix} is defined by
\bq
 P & = &
 \left( \begin{array}{cc}
 \psi_1 & \psi_2 \\
 \phi_1 & \phi_2 \\
 \end{array} \right)
 =
 \left( \begin{array}{cc}
 4 K & 4 i K' \\
 4 K - 4 E & 4 i E' \\
 \end{array} \right),
\eq
where we used the abbreviations
$K=K(\sqrt{\lambda})$, $E=E(\sqrt{\lambda})$, $K'=K(\sqrt{1-\lambda})$ and $E'=E(\sqrt{1-\lambda})$.
The determinant of the period matrix is given by
\bq
 \det P & = & 8 \pi i.
\eq
This follows from the 
\index{Legendre relation}
{\bf Legendre relation}:
\bq
 K E'
 + E K' 
 - K K'
 & = & 
 \frac{\pi}{2}.
\eq

The elliptic curve $y^2 = x (x-1) (x-\lambda)$ depends on a parameter $\lambda$, 
and so do the periods $\psi_1(\lambda)$ and $\psi_2(\lambda)$.
We may now ask:
How do the periods change, if we change $\lambda$?
The variation is governed by a second-order differential equation:
We have
\bq
\label{chapter_elliptics:Picard_fuchs_equation}
 \left[ 4 \lambda \left(1-\lambda\right) \frac{d^2}{d\lambda^2} 
 + 4 \left(1-2\lambda\right) \frac{d}{d\lambda} - 1 \right] \psi_j & = & 0,
 \;\;\;\;\;\; j = 1,2.
\eq
The differential operator
\bq
 4 \lambda \left(1-\lambda\right) \frac{d^2}{d\lambda^2} 
 + 4 \left(1-2\lambda\right) \frac{d}{d\lambda} - 1
\eq
is called the 
\index{Picard-Fuchs operator}
{\bf Picard-Fuchs operator}
of the elliptic curve $y^2 = x (x-1) (x-\lambda)$.

There is a third possibility to represent an elliptic curve:
We may also represent an elliptic curve as $\mathbb{C}/\Lambda$,
where $\Lambda$ is the double lattice generated by $\psi_1$ and $\psi_2$.
\begin{figure}
\begin{center}
\includegraphics[scale=1.0]{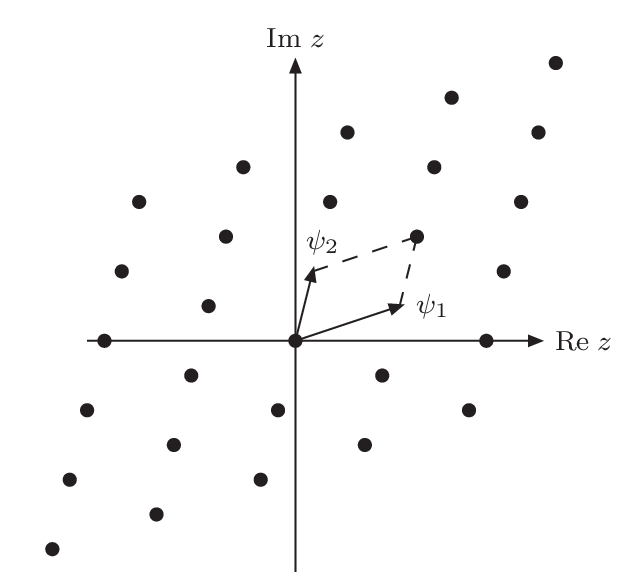}
\end{center}
\caption{
$\mathbb{C}/\Lambda$,
where $\Lambda$ is a double lattice generated by $\psi_1$ and $\psi_2$.
Points inside the fundamental parallelogram correspond to points on the elliptic curve.
A point outside the fundamental parallelogram can always be shifted inside the fundamental parallelogram
through the addition of some lattice vector.
}
\label{chapter_elliptics:fig3}
\end{figure}
This is shown in fig.~\ref{chapter_elliptics:fig3}.
Points, which differ by a lattice vector are considered to be equivalent.
The different equivalence classes are represented by the points inside the fundamental parallelogram,
as shown in fig.~\ref{chapter_elliptics:fig3}.
They correspond to points on the elliptic curve.
Before we go into the details, let us first remark that this is not too surprising:
If we start from the representation of an elliptic curve as a genus one Riemann surface and cut open this surface
along the two cycles $\gamma_1$ and $\gamma_2$ shown in fig.~\ref{chapter_elliptics:fig2}, we obtain
a parallelogram.

Let's now fill in the technical detail: We would like to map a point on an elliptic curve, defined by a
polynomial $P$, to a point in $\mathbb{C}/\Lambda$ and vice versa.
For simplicity we assume that the elliptic curve is given in the Weierstrass normal form
$y^2-4x^3+g_2x+g_3=0$.

We start with the direction from the 
Weierstrass normal form to $\mathbb{C}/\Lambda$:
Given a point $(x,y)$ with $y^2-4x^3+g_2x+g_3=0$ the corresponding point $z \in \mathbb{C}/\Lambda$ is given by the incomplete elliptic integral
\bq
 z & = &
\int\limits_\infty^x \frac{dt}{\sqrt{4t^3-g_2t-g_3}}.
\eq
Let's now consider the reverse direction from $z \in \mathbb{C}/\Lambda$ to
a point on the curve defined by the Weierstrass normal form.
Given a point $z \in \mathbb{C}/\Lambda$ the corresponding point $(x,y)$ on $y^2-4x^3+g_2x+g_3=0$
is given by
\bq
 \left(x,y\right)
 & = &
 \left( \wp\left(z\right), \wp'\left(z\right) \right).
\eq
$\wp(z)$ denotes Weierstrass's $\wp$-function.

Let us now introduce some additional notation and conventions:
It is common practise to normalise one period to one:
$(\psi_2,\psi_1) \rightarrow (\tau, 1)$, where
\bq
\label{chapter_elliptics:def_tau}
 \tau & = & \frac{\psi_2}{\psi_1}.
\eq
In addition one requires $\mathrm{Im}(\tau) > 0$.
This is always possible: If $\mathrm{Im}(\tau) < 0$ simply  exchange $\psi_1$ and $\psi_2$ and proceed as above.
The possible values of $\tau$ lie therefore in the
\index{complex upper half-plane}
{\bf complex upper half-plane},
defined by
\bq
 \mathbb{H}
 & = &
 \left\{ \; \tau \in \mathbb{C} \; | \; \mathrm{Im}(\tau) > 0 \; \right\}.
\eq
Let us now consider an elliptic curve as being given by ${\mathbb C}/\Lambda$, where $\Lambda$ is a lattice.
Two elliptic curves $E = {\mathbb C}/\Lambda$ and $E' = {\mathbb C}/\Lambda'$ are called
\index{isomorphic elliptic curves}
{\bf isomorphic}, if there is a complex number $c$
such that
\bq
 c \Lambda & = & \Lambda'.
\eq
For example, the two elliptic curves defined by the lattices with the periods $(\psi_2,\psi_1)$ and $(\tau,1)=(\psi_2/\psi_1,1)$ are isomorphic. 
Two elliptic curves are 
\index{isogenic elliptic curves}
{\bf isogenic}, if there is a complex number $c$ such that
\bq
\label{def_isogeny}
 c \Lambda & \subset & \Lambda',
\eq
i.e. $c \Lambda$ is a sub-lattice of $\Lambda'$.

\subsection{Calculations with elliptic curves}
\label{chapter_elliptics:section_calculations_elliptic_curves}

Let us now turn to practicalities of doing calculations with elliptic curves.
We consider the generic quartic case
\bq
\label{chapter_elliptics:calculation_quartic_curve}
 E
 & : &
 v^2 - \left(u-u_1\right) \left(u-u_2\right) \left(u-u_3\right) \left(u-u_4\right)
 \; = \; 0.
\eq
where the roots $u_j$ may depend on variables $x=(x_1,\dots,x_{\NB})$:
\bq
 u_j & = & u_j\left(x\right),
 \;\;\;\;\;\;
 j \in \{1,2,3,4\}.
\eq
In practical applications the $x_j$'s will be the kinematic variables the Feynman integral depends on.
In eq.~(\ref{chapter_elliptics:calculation_quartic_curve}) we used the variables $(u,v)$ instead of $(x,y)$ 
to avoid a clash of notation with the kinematic variables.
In this section we consider an elliptic curve together with a fixed choice of two independent periods $\psi_1, \psi_2$.
In mathematical terms we are considering a 
\index{framed elliptic curve}
{\bf framed elliptic curve}.
In section~\ref{chapter_elliptics:section_modular_transformations} we will remove the framing and discuss arbitrary choices
for the two periods.

But let us now proceed and define a standard choice for the two periods:
We set
\bq
 U_1 \; = \; \left(u_3-u_2\right)\left(u_4-u_1\right),
 \;\;\;\;
 U_2 \; = \; \left(u_2-u_1\right)\left(u_4-u_3\right),
 \;\;\;\;
 U_3 \; = \; \left(u_3-u_1\right)\left(u_4-u_2\right).
 \;
\eq 
Note that we have
\bq
 U_1 + U_2 & = & U_3.
\eq
We define the 
\index{modulus of an elliptic curve}
{\bf modulus} $k$ and the 
\index{complementary modulus of an elliptic curve}
{\bf complementary modulus} $\bar{k}$ of the elliptic curve $E$ by
\bq
\label{chapter_elliptics:def_modulus}
 k^2 
 \; = \; 
 \frac{U_1}{U_3},
 & &
 \bar{k}^2 
 \; = \;
 1 - k^2 
 \; = \;
 \frac{U_2}{U_3}.
\eq
Note that there are six possibilities of defining $k^2$ (compare with exercise~\ref{chapter_elliptics:exercise_general_quartic}).
Our standard choice for the periods and quasi-periods is
\bq
\label{chapter_elliptics:def_generic_periods}
 \psi_1 
 \; = \; 
 \frac{4 K\left(k\right)}{U_3^{\frac{1}{2}}},
 & &
 \psi_2
 \; = \; 
 \frac{4 i K\left(\bar{k}\right)}{U_3^{\frac{1}{2}}},
 \nonumber \\
 \phi_1 
 \; = \; 
 \frac{4 \left[ K\left(k\right) - E\left(k\right) \right]}{U_3^{\frac{1}{2}}},
 & &
 \phi_2
 \; = \; 
 \frac{4 i E\left(\bar{k}\right)}{U_3^{\frac{1}{2}}}.
\eq
This defines the framing of the elliptic curve.
The Legendre relation for the periods and the quasi-periods reads
\bq
 \psi_1 \phi_2 - \psi_2 \phi_1
 & = &
 \frac{8 \pi i}{U_3}.
\eq
As in eq.~(\ref{chapter_elliptics:def_tau}) we define the modular parameter $\tau$ by
\bq
\label{chapter_elliptics:def_tau_quartic_case}
 \tau & = & \frac{\psi_2}{\psi_1}.
\eq
In addition we define the 
\index{nome}
{\bf nome} $q$ and the 
{\bf nome squared} $\bar{q}$ by
\bq
\label{chapter_elliptics:def_nome}
 q \; = \; \exp\left(i \pi \tau\right),
 & &
 \bar{q} \; = \; \exp\left(2 i \pi \tau\right).
\eq
(In the literature the letter $q$ is either used for the nome or the nome squared.
In this book we denote the nome by $q$ and the nome squared by $\bar{q}$. 
Obviously we have $\bar{q}=q^2$. We will mainly use the nome squared $\bar{q}$.)

We assumed that the roots $u_1, u_2, u_3, u_4$ depend on the variables $x=(x_1,\dots,x_{\NB})$,
hence also the periods and quasi-periods will depend on $x$.
We would like to know how the periods and quasi-periods vary with $x$.
The answer is provided by the following system of first-order differential equations
\bq
\label{chapter_elliptics:dgl_periods}
 d 
 \left( \begin{array}{c} \psi_i \\ \phi_i \end{array} \right)
 & = &
 \left( \begin{array}{cc}
 - \frac{1}{2} d \ln U_2 & \frac{1}{2} d \ln \frac{U_2}{U_1} \\
 - \frac{1}{2} d \ln \frac{U_2}{U_3} & \frac{1}{2} d \ln \frac{U_2}{U_3^2} \\
 \end{array}
 \right)
 \left( \begin{array}{c} \psi_i \\ \phi_i \end{array} \right),
 \;\;\;\;\;\; i \in \{1,2\},
\eq
where $d$ denotes the differential with respect to the variables $x_1, \dots, x_{\NB}$, e.g.
\bq
 d f\left(x\right)
 & = &
 \sum\limits_{j=1}^{\NB} \left( \frac{\partial f}{\partial x_j} \right) dx_j.
\eq
We further have
\bq
 2 \pi i \; d\tau
 \; = \; 
 d\ln \bar{q}
 & = &
 \frac{2 \pi i}{\psi_1^2} \; \frac{4 \pi i}{U_3} d\ln\frac{U_2}{U_1}.
\eq
In general, our base space is $\NB$-dimensional (with coordinates $x_1,\dots,x_{\NB}$).
Sometimes we want to restrict to a one-dimensional subspace.
To this aim consider a path
$\gamma : [0,1] \rightarrow {\mathbb C}^{\NB}$ such that $x_i=x_i(\lambda)$, where the variable $\lambda$ parametrises the path.
For a path $\gamma$ we may view the periods $\psi_1$ and $\psi_2$ as functions of the path variable $\lambda$.
We may then write down a second-order Picard-Fuchs equation for the variation of the periods along the path $\gamma$
(as we did in eq.~(\ref{chapter_elliptics:Picard_fuchs_equation})):
\bq
\label{chapter_elliptics:Picard_Fuchs_along_gamma}
 \left[ \frac{d^2}{d\lambda^2} + p_{1,\gamma} \frac{d}{d\lambda} + p_{0,\gamma} \right] \psi_i & = & 0,
 \;\;\;\;\;\;\;\;\;\;\;\;\;\;\;
 i \in \{1,2\}.
\eq
The coefficients $p_{1,\gamma}$ and $p_{0,\gamma}$ are given by
\bq
 p_{1,\gamma} & = & 
 \frac{d}{d\lambda} \ln U_3 
 - \frac{d}{d\lambda} \ln\left( \frac{d}{d\lambda} \ln \frac{U_2}{U_1} \right),
 \\
 p_{0,\gamma} & = &
 \frac{1}{2}
 \left( \frac{d}{d\lambda} \ln U_1 \right)
 \left( \frac{d}{d\lambda} \ln U_2 \right)
 - \frac{1}{2} \frac{ \left(\frac{d}{d\lambda} U_1\right)\left(\frac{d^2}{d\lambda^2} U_2\right)
                     -\left(\frac{d^2}{d\lambda^2} U_1\right)\left(\frac{d}{d\lambda} U_2\right) }
                    { U_1\left(\frac{d}{d\lambda} U_2\right) - U_2\left(\frac{d}{d\lambda} U_1\right)}
 \nonumber \\
 & &
 +
 \frac{1}{4U_3} 
 \left[ 
   \frac{1}{U_1} \left( \frac{d}{d\lambda} U_1 \right)^2
   +
   \frac{1}{U_2} \left( \frac{d}{d\lambda} U_2 \right)^2
 \right].
 \nonumber
\eq
This defines the Picard-Fuchs operator along the path $\gamma$:
\bq
\label{chapter_elliptics:Picard_Fuchs_operator_along_gamma}
 L_\gamma & = & \frac{d^2}{d\lambda^2} + p_{1,\gamma} \frac{d}{d\lambda} + p_{0,\gamma}.
\eq
Note that eq.~(\ref{chapter_elliptics:Picard_Fuchs_along_gamma}) and eq.~(\ref{chapter_elliptics:Picard_Fuchs_operator_along_gamma})
follow from eq.~(\ref{chapter_elliptics:dgl_periods}) by restricting to $\gamma$ and by eliminating the quasi-periods $\phi_i$.

The 
\index{Wronskian}
{\bf Wronskian} of eq.~(\ref{chapter_elliptics:Picard_Fuchs_along_gamma}) is defined by
\bq
\label{chapter_elliptics:def_Wronskian}
 W_\gamma & = & 
 \psi_1 \frac{d}{d\lambda} \psi_2 - \psi_2 \frac{d}{d\lambda} \psi_1,
\eq
and given by
\bq
 W_\gamma & = & 
 \frac{4 \pi i}{U_3} 
   \frac{d}{d\lambda} \ln \frac{U_2}{U_1}.
\eq
We further have
\bq
\label{chapter_elliptics:derivative_Wronskian}
 \frac{d}{d\lambda} W_\gamma & = & - p_{1,\gamma} W_\gamma,
 \nonumber \\
 2\pi i d\tau & = & \frac{2\pi i \; W_\gamma}{\psi_1^2} d\lambda.
\eq
Let us now consider a one-parameter family of elliptic curves, which we parametrise by a variable $x$.
This occurs, if either we just have one kinematic variable $x$ (i.e. $\NB=1$) or if we restrict to a one-dimensional
subspace as above.
The discussion from above (eqs.~(\ref{chapter_elliptics:Picard_Fuchs_along_gamma})-(\ref{chapter_elliptics:derivative_Wronskian})) carries over, with $\lambda$ replaced by $x$.

In this situation we may want 
to perform a base transformation as in section~\ref{chapter_transformations:base_transformation} and change variables from $x$ to $\tau$ (or from $x$ to $\bar{q}$, the change from $\tau$ to $\bar{q}$ is rather trivial).
Eqs.~(\ref{chapter_elliptics:def_tau_quartic_case}) and (\ref{chapter_elliptics:def_nome})
gives us $\tau$ and $\bar{q}$ as a function of $x$.
From eq.~(\ref{chapter_elliptics:derivative_Wronskian}) we have for the Jacobian of the transformation
\bq
 \frac{d\tau}{dx} & = & \frac{W}{\psi_1^2},
\eq
where $W$ denotes the Wronskian, defined as in eq.~(\ref{chapter_elliptics:def_Wronskian}) and with $\lambda$ replaced by $x$.
However, what we really need is not $\tau$ or $\bar{q}$ as a function of $x$,
but $x$ as a function of $\tau$ or $\bar{q}$.
In this context it is useful to know about the Jacobi theta functions and the Dedekind eta function.
\begin{digression} 
\index{Jacobi theta functions}
\index{Dedekind eta function}
{\bf Jacobi theta functions and Dedekind eta function}
\\
Dedekind's eta function $\eta(\tau)$ is defined for $\tau \in {\mathbb H}$ by
\bq
 \eta\left(\tau\right) 
 & = & 
 e^{\frac{i \pi \tau}{12}} \prod\limits_{n=1}^{\infty} \left(1-e^{2\pi i n \tau}\right).
\eq
With $\bar{q}=\exp(2 \pi i \tau)$ this becomes
\bq
 \eta\left(\tau\right)
 & = &
 \bar{q}^{\frac{1}{24}} \prod\limits_{n=1}^\infty \left( 1 - \bar{q}^{n} \right).
\eq
We have
\bq
 \eta\left(\tau\right)
 & = & 
 \sum\limits_{n=-\infty}^\infty \left(-1\right)^n \bar{q}^{\frac{\left(6n-1\right)^2}{24}}
 \; = \;
 \bar{q}^{\frac{1}{24}} \left\{ 1 + \sum\limits_{n=1}^\infty \left(-1\right)^n \left[ \bar{q}^{\frac{1}{2}\left(3n-1\right)n} + \bar{q}^{\frac{1}{2}\left(3n+1\right)n} \right] \right\}. \;
\eq
Dedekind's eta function is related to the Jacobi theta function $\theta_2$ (defined below):
\bq
 \eta\left(\tau\right)
 & = & 
 \frac{1}{\sqrt{3}} \; \theta_2\left( \frac{\pi}{6}, \bar{q}^{\frac{1}{6}} \right).
\eq
Under modular transformations we have
\bq
 \eta\left(\tau + 1\right) \; = \; e^{\frac{2\pi i}{24}} \eta\left(\tau\right),
 & &
 \eta\left(-\frac{1}{\tau}\right) \; = \; \left(-i\tau\right)^{\frac{1}{2}} \eta\left(\tau\right).
\eq
Dedekind's eta function is related to the 
\index{modular discriminant}
{\bf modular discriminant} $\Delta(\tau)$ through
\bq
 \Delta\left(\tau\right) 
 & = &
 \left(2\pi i\right)^{12} \eta\left(\tau\right)^{24}.
\eq
Let us now turn to the Jacobi theta functions.
For historical reasons, they are defined through the nome $q=\exp(i\pi \tau)$ and 
a variable $z$, which within the conventions of this book we later always will rescale as
$z \rightarrow \pi z$.
But for the moment, we follow the standard (historical) notation.
The general theta function is defined for $a,b \in {\mathbb R}$ by
\bq
 \theta\left[a,b\right]\left(z,q\right)
 & = & 
 \theta\left[a,b\right]\left(z|\tau\right)
 =
 \sum\limits_{n=-\infty}^\infty
 q^{\left(n+\frac{1}{2}a\right)^2} e^{2i\left(n+\frac{1}{2}a\right)\left(z-\frac{1}{2} \pi b\right)}.
\eq
The theta function satisfies
\bq
 4 i \frac{\partial}{\partial \tau} \theta\left[a,b\right]\left(z|\tau\right)
 & = &
 \pi \frac{\partial^2}{\partial z^2} \theta\left[a,b\right]\left(z|\tau\right).
\eq
For $n\in {\mathbb Z}$ we have
\bq
 \theta\left[a+2n,b\right]\left(z|\tau\right) 
 =
 \theta\left[a,b\right]\left(z|\tau\right),
 & &
 \theta\left[a,b+2n\right]\left(z|\tau\right) 
 =
 e^{- n \pi i a}
 \theta\left[a,b\right]\left(z|\tau\right).
\eq
If $a,b \in {\mathbb Z}$ it is therefore sufficient to consider the four cases
\bq
 \theta_1\left(z,q\right) = \theta\left[1,1\right]\left(z,q\right),
 & &
 \theta_2\left(z,q\right) = \theta\left[1,0\right]\left(z,q\right),
 \nonumber \\
 \theta_3\left(z,q\right) = \theta\left[0,0\right]\left(z,q\right),
 & &
 \theta_4\left(z,q\right) = \theta\left[0,1\right]\left(z,q\right).
\eq
Explicitly the four theta functions are defined by
\begin{alignat}{4}
\label{chapter_elliptics:def_Jacobi_theta_functions}
 &
 \theta_1\left(z,q\right) 
 & \; = \; &
 \theta_1\left(z | \tau \right)
 & \; = \; &
 -i \sum\limits_{n=-\infty}^\infty \left(-1\right)^n q^{\left(n+\frac{1}{2}\right)^2} e^{i\left(2n+1\right)z}
 & \; = \; &
 2 \sum\limits_{n=0}^\infty \left(-1\right)^n q^{\left(n+\frac{1}{2}\right)^2} \sin\left(\left(2n+1\right)z\right),
 \nonumber \\
 &
 \theta_2\left(z,q\right) 
 & \; = \; & 
 \theta_2\left(z | \tau \right)
 & \; = \; &
 \sum\limits_{n=-\infty}^\infty q^{\left(n+\frac{1}{2}\right)^2} e^{i\left(2n+1\right)z}
 & \; = \; &
 2 \sum\limits_{n=0}^\infty q^{\left(n+\frac{1}{2}\right)^2} \cos\left(\left(2n+1\right)z\right),
 \nonumber \\
 & 
 \theta_3\left(z,q\right) 
 & \; = \; &
 \theta_3\left(z | \tau \right)
 & \; = \; &
 \sum\limits_{n=-\infty}^\infty q^{n^2} e^{2 i n z}
 & \; = \; &
 1 + 2 \sum\limits_{n=1}^\infty q^{n^2} \cos\left(2 n z\right),
 \nonumber \\
 &
 \theta_4\left(z,q\right) 
 & \; = \; & 
 \theta_4\left(z | \tau \right)
 & \; = \; &
 \sum\limits_{n=-\infty}^\infty \left(-1\right)^n q^{n^2} e^{2 i n z}
 & \; = \; &
 1 + 2 \sum\limits_{n=1}^\infty \left(-1\right)^n q^{n^2} \cos\left(2 n z\right).
\end{alignat}
The theta functions have a representation as infinite products:
\bq
\theta_1\left(z,q\right) & = & 
 2 q^{\frac{1}{4}} \sin z \prod\limits_{n=1}^\infty \left( 1 - q^{2n} \right) \left( 1 - 2 q^{2n} \cos\left(2z\right) + q^{4n} \right),
 \nonumber \\
\theta_2\left(z,q\right) & = & 
 2 q^{\frac{1}{4}} \cos z \prod\limits_{n=1}^\infty \left( 1 - q^{2n} \right) \left( 1 + 2 q^{2n} \cos\left(2z\right) + q^{4n} \right),
 \nonumber \\
\theta_3\left(z,q\right) & = & 
 \prod\limits_{n=1}^\infty \left( 1 - q^{2n} \right) \left( 1 + 2 q^{2n-1} \cos\left(2z\right) + q^{4n-2} \right),
 \nonumber \\
\theta_4\left(z,q\right) & = & 
 \prod\limits_{n=1}^\infty \left( 1 - q^{2n} \right) \left( 1 - 2 q^{2n-1} \cos\left(2z\right) + q^{4n-2} \right).
\eq
The functions $\theta_1$ and $\theta_2$ are periodic in $z$ with period $2\pi$, 
the functions $\theta_3$ and $\theta_4$ are periodic in $z$ with period $\pi$.
\bq
 \theta_1\left(z+2\pi,q\right) = \theta_1\left(z,q\right), & & \theta_3\left(z+\pi,q\right) = \theta_3\left(z,q\right),
 \nonumber \\
 \theta_2\left(z+2\pi,q\right) = \theta_2\left(z,q\right), & & \theta_4\left(z+\pi,q\right) = \theta_4\left(z,q\right).
\eq
$\pi \tau$ is a quasi-period of the theta functions with periodicity factor $\pm \left( q e^{2 i z} \right)^{-1}$:
\bq
 \theta_1\left(z+\pi \tau,q\right) = - \left( q e^{2 i z} \right)^{-1} \theta_1\left(z,q\right), 
 & &
 \theta_3\left(z+\pi \tau,q\right) = \left( q e^{2 i z} \right)^{-1} \theta_3\left(z,q\right), 
 \nonumber \\
 \theta_2\left(z+\pi \tau,q\right) = \left( q e^{2 i z} \right)^{-1} \theta_2\left(z,q\right), 
 & &
 \theta_4\left(z+\pi \tau,q\right) = - \left( q e^{2 i z} \right)^{-1} \theta_4\left(z,q\right).
\eq
A prime denotes the derivative with respect to the first variable $z$
\bq 
 \theta_i'\left(z,q\right)
 & = & 
 \frac{\partial}{\partial z} \theta_i\left(z,q\right),
 \;\;\;\;\;\;\;\;\;
 i \in \{1,2,3,4\}.
\eq
Useful relations are
\bq
 \theta_1'\left(0,q\right)
 & = &
 \theta_2\left(0,q\right)
 \theta_3\left(0,q\right)
 \theta_4\left(0,q\right),
 \nonumber \\
 \theta_3^4\left(0,q\right)
 & = & 
 \theta_2^4\left(0,q\right) + \theta_4^4\left(0,q\right).
\eq
The theta functions can be used to express the modulus $k$, 
the complementary modulus $k'$, 
the complete elliptic integral of the first kind $K$
and 
the complete elliptic integral of the second kind $E$
as functions of the nome $q$:
\begin{alignat}{3}
\label{chapter_elliptics:Jacobi_theta_relation_1}
 k & = \frac{\theta_2^2\left(0,q\right)}{\theta_3^2\left(0,q\right)},
 & \;\;\;\;\;\;\;\;\; & &
 k' & = \frac{\theta_4^2\left(0,q\right)}{\theta_3^2\left(0,q\right)},
 \nonumber \\
 K & = \frac{\pi }{2} \theta_3^2\left(0,q\right),
 & \;\;\;\;\;\;\;\;\; & &
 E & = \frac{\pi }{2} \left( 1 - \frac{\theta_4''\left(0,q\right)}{\theta_3^4\left(0,q\right) \theta_4\left(0,q\right)} \right) \theta_3^2\left(0,q\right).
\end{alignat}
We have the following relations with Dedekind's eta function:
\bq
\label{chapter_elliptics:Jacobi_theta_relation_2}
 \theta_2\left(0,q\right) 
 =
 2 \frac{\eta\left(2\tau\right)^2}{\eta\left(\tau\right)},
 \;\;\;\;\;\;
 \theta_3\left(0,q\right) 
 =
 \frac{\eta\left(\tau\right)^5}{\eta\left(\frac{\tau}{2}\right)^2 \eta\left(2\tau\right)^2},
 \;\;\;\;\;\;
 \theta_4\left(0,q\right) 
 =
 \frac{\eta\left(\frac{\tau}{2}\right)^2}{\eta\left(\tau\right)}.
\eq
\end{digression}

\noindent
\bs
{\it \refstepcounter{exercise}
\label{chapter_elliptics:exercise_modulus_squared}
{\bf Exercise \theexercise}: 
Express the modulus squared $k^2$ and the complementary modulus squared $k'{}^2$ as a quotient of eta
functions.
}
\es
\\
\\
Equipped with the Jacobi theta functions and the Dedekind eta function we now return
to our original problem, expressing $x$ as a function of $\tau$ or $\bar{q}$.
Let's look at the modulus squared $k^2$, defined in eq.~(\ref{chapter_elliptics:def_modulus}) 
for the elliptic curve $E$ of eq.~(\ref{chapter_elliptics:calculation_quartic_curve}).
On the one hand we have from eq.~(\ref{chapter_elliptics:def_modulus})
\bq
 k^2 & = & \frac{U_1}{U_3} \; = \; \frac{\left(u_3-u_2\right)\left(u_4-u_1\right)}{\left(u_3-u_1\right)\left(u_4-u_2\right)}.
\eq
The right-hand side is a (known) function of $x$.
On the other hand, we learned in exercise~\ref{chapter_elliptics:exercise_modulus_squared}
that
\bq
\label{chapter_elliptics:def_modular_lambda}
 k^2
 & = &
 16 
 \frac{\eta\left(\frac{\tau}{2}\right)^8 \eta\left(2\tau\right)^{16}}{\eta\left(\tau\right)^{24}}.
\eq
Here, the right-hand side has an expansion in $\bar{q}^{\frac{1}{2}}$ (the square root originates from the argument $\tau/2$):
\bq
 16 
 \frac{\eta\left(\frac{\tau}{2}\right)^8 \eta\left(2\tau\right)^{16}}{\eta\left(\tau\right)^{24}}
 & = & 
 16 \bar{q}^{\frac{1}{2}}
 - 128 \bar{q}
 + 704 \bar{q}^{\frac{3}{2}}
 - 3072 \bar{q}^2
 + 11488 \bar{q}^{\frac{5}{2}}
 + {\mathcal O}\left(\bar{q}^3\right).
\eq
Thus
\bq
\label{chapter_elliptics:relation_modulus_squared}
 \frac{U_1}{U_3}
 & = &
 16 
 \frac{\eta\left(\frac{\tau}{2}\right)^8 \eta\left(2\tau\right)^{16}}{\eta\left(\tau\right)^{24}}.
\eq
We may then solve eq.~(\ref{chapter_elliptics:relation_modulus_squared}) for $x$ as a power series in $\bar{q}_N=\bar{q}^{\frac{1}{N}}$
for an appropriate $N$.
It is usually possible with the help of computer algebra programs
to obtain a large number of terms of this power series.
In a second step we try to find a closed form for this power series.
There are several heuristic methods how this can be done:
\begin{enumerate}
\item If we expect that the result should lie within a certain function space, we can start from an ansatz
with unknown coefficients and determine the coefficients by comparing sufficient many terms in the power series.
\item If we expect that the result can be expressed as an eta quotient, we may use dedicated computer programs
to find this eta quotient \cite{Garvan:1998}.
\item We may use the ``On-Line Encyclopedia of Integer Sequences'' \cite{oeis} by typing the first few coefficients of the power
series into the web interface.
\end{enumerate}
Let us remark that the first method can be turned into a strict method by first proving that the result must lie 
within a finite-dimensional function space.
Once this is established, we know that the result can be written as a finite linear combination of certain
basis functions and it suffices to determine the coefficients by comparing sufficient many terms in the power series.

Let us now look at an example: We consider a family of elliptic curves
\bq
\label{chapter_elliptics:def_example_elliptic_curve}
 E
 & : &
 v^2 - u
       \left(u + 4 \right) 
       \left[u^2 + 2 \left(1+x\right) u + \left(1-x\right)^2 \right]
 \; = \; 0,
\eq
depending on a parameter $x$.
We denote the roots of the quartic polynomial in eq.~(\ref{chapter_elliptics:def_example_elliptic_curve}) by
\bq
 u_1 \; = \; -4,
 \;\;\;
 u_2 \; = \; -\left(1+\sqrt{x}\right)^2,
 \;\;\;
 u_3 \; = \; -\left(1-\sqrt{x}\right)^2,
 \;\;\;
 u_4 \; = \; 0.
\eq
We have
\bq
 k^2 
 & = & \frac{U_1}{U_3}
 \; = \;
 \frac{16 \sqrt{x}}{\left(1+\sqrt{x}\right)^3 \left(3-\sqrt{x}\right)},
\eq
and therefore
\bq
 \frac{16 \sqrt{x}}{\left(1+\sqrt{x}\right)^3 \left(3-\sqrt{x}\right)}
 & = &
 16 
 \frac{\eta\left(\frac{\tau}{2}\right)^{8} \eta\left(2\tau\right)^{16}}{\eta\left(\tau\right)^{24}}.
\eq
We first solve this equation for $\sqrt{x}$ as a power series in $\bar{q}_2=\bar{q}^{\frac{1}{2}}$:
\bq
 \sqrt{x}
 & = &
 3 \bar{q}_2
 - 6 \bar{q}_2^3
 + 9 \bar{q}_2^5
 - 12 \bar{q}_2^7
 + 21 \bar{q}_2^9
 - 36 \bar{q}_2^{11}
 + 51 \bar{q}_2^{13}
 + {\mathcal O}\left(\bar{q}_2^{15}\right).
\eq
Squaring the left-hand side and the right-hand side we obtain
\bq
 x & = &
 9 \bar{q}
 - 36 \bar{q}^2
 + 90 \bar{q}^3
 - 180 \bar{q}^4
 + 351 \bar{q}^5
 - 684 \bar{q}^6
 + 1260 \bar{q}^7
 + {\mathcal O}\left(\bar{q}^8\right).
\eq
In the last step we convert the power series to a closed form:
\bq
\label{chapter_elliptics:hauptmodul_0}
 x
 & = &
 9
 \frac{\eta\left(\tau\right)^4 \eta\left(6\tau\right)^8}
      {\eta\left(3\tau\right)^4 \eta\left(2\tau\right)^8}.
\eq
We therefore obtained an expression for $x$ as a function of $\tau$ (or $\bar{q}$).

\section{Modular transformations and modular forms}
\label{chapter_elliptics:section_modular_transformations}

In the previous section we considered a framed elliptic curve, i.e. an elliptic curve together with 
a fixed choice of periods $\psi_1$ and $\psi_2$, which generate the lattice $\Lambda$.
In this section we investigate the implications of our freedom of choice for the periods $\psi_1$ and $\psi_2$.
This will lead us to modular transformations.

We recall that we may represent an elliptic curve as 
$\mathbb{C}/\Lambda$,
where $\Lambda$ is a double lattice generated by $\psi_1$ and $\psi_2$.
\begin{figure}
\begin{center}
\includegraphics[scale=1.0]{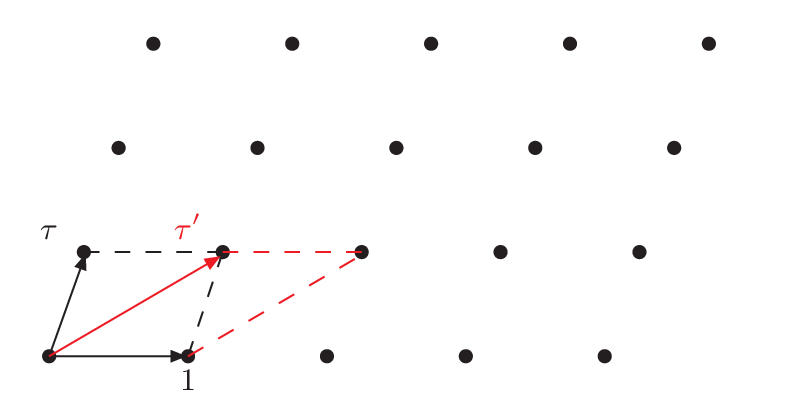}
\end{center}
\caption{
The generators $\tau$ and $1$ generate the same lattice as the generators $\tau'$and $1$.
}
\label{chapter_elliptics:fig4}
\end{figure}
As only the lattice $\Lambda$ matters, but not the specific generators, we may consider a different pair
of periods $(\psi_2',\psi_1')$,
which generate the same lattice $\Lambda$. An example is shown in fig.~\ref{chapter_elliptics:fig4}:
The generators $\tau$ and $1$ generate the same lattice as the generators $\tau'$and $1$.

Let's return to the general case
and consider a change of basis from the pair of periods $(\psi_2,\psi_1)$
to the pair of periods $(\psi_2',\psi_1')$.
The new pair of periods $(\psi_2',\psi_1')$ is again a pair of lattice vectors,
so it can be written as
\bq
\label{chapter_elliptics:transformation_periods}
 \left( \begin{array}{c}
 \psi_2' \\
 \psi_1' \\
 \end{array} \right)
 & = &
 \left( \begin{array}{cc}
 a & b \\
 c & d \\
 \end{array} \right)
 \left( \begin{array}{c}
 \psi_2 \\
 \psi_1 \\
 \end{array} \right),
\eq
with $a, b, c, d \in {\mathbb Z}$.
The transformation should be invertible and $(\psi_2,\psi_1)$
and $(\psi_2',\psi_1')$ should generate the same lattice $\Lambda$.
This implies
\bq
 \left( \begin{array}{cc}
 a & b \\
 c & d \\
 \end{array} \right)
 & \in & \mathrm{SL}_2\left({\mathbb Z}\right).
\eq
The group $\mathrm{SL}_2({\mathbb Z})$ is called the 
\index{modular group}
{\bf modular group}. 
It is generated by the two matrices
\bq
 T
 \; = \;
 \left( \begin{array}{rr}
         1 & 1 \\ 
         0 & 1
        \end{array}  \right)
 & \mbox{and} &
 S
 \; = \;
 \left( \begin{array}{rr}
         0 & -1 \\ 
         1 & 0
        \end{array}  \right).
\eq
In terms of $\tau$ and $\tau'$ we have
\bq
\label{chapter_elliptics:trafo_modular}
 \tau' & = & \frac{a \tau +b}{c \tau +d}.
\eq
A transformation of the form
as in eq.~(\ref{chapter_elliptics:trafo_modular}) is called a 
\index{modular transformation}
{\bf modular transformation}.

We may then look at functions $f(\tau)$, which transform under modular transformations in a particular way.
This will lead us to modular forms.
A meromorphic function $f: \mathbb{H} \rightarrow \mathbb{C}$ 
is a 
\index{modular form}
{\bf modular form} 
of modular weight $k$ for $\mathrm{SL}_2(\mathbb{Z})$ if
\begin{enumerate}
\item $f$ transforms under modular transformations as
\bq
\label{chapter_elliptics:trafo_modular_form}
 f\left( \dfrac{a\tau+b}{c\tau+d} \right)
 =
 (c\tau+d)^k \cdot f(\tau) 
 \qquad \text{for} \;\;
 \gamma = \left( \begin{array}{cc}
 a & b \\ 
 c & d
 \end{array} \right)
 \in \mathrm{SL}_2(\mathbb{Z}),
\eq
\item $f$ is holomorphic on $\mathbb{H}$,
\item $f$ is holomorphic at $i \infty$.
\end{enumerate}
The prefactor $(c\tau+d)^k$ in eq.~(\ref{chapter_elliptics:trafo_modular_form}) is called
\index{automorphic factor}
{\bf automorphic factor}
and equals
\bq
 (c\tau+d)^k
 & = &
 \left( \frac{\psi_1'}{\psi_1} \right)^k.
\eq
It is convenient to introduce the
\index{$\slashoperator{\gamma}{k}$ operator}
{\bf $\slashoperator{\gamma}{k}$ operator},
defined by
\bq
(f \slashoperator{\gamma}{k})(\tau) & = & (c\tau+d)^{-k} \cdot f(\gamma(\tau)).
\eq
\bs
{\it \refstepcounter{exercise}
{\bf Exercise \theexercise}: 
Show that
\bq
 \left(f \slashoperator{\gamma_1}{k}\right) \slashoperator{\gamma_2}{k}
 & = &
 f \slashoperator{\left(\gamma_1 \gamma_2\right)}{k}.
\eq
}
\es
\\
\\
With the help of the $\slashoperator{\gamma}{k}$ operator
we may rewrite eq.~(\ref{chapter_elliptics:trafo_modular_form}) as
\bq
\label{chapter_elliptics:trafo_modular_form_v2}
 (f \slashoperator{\gamma}{k})
 & = &
 f
 \qquad \text{for} \;\; 
 \gamma  \in \mathrm{SL}_2(\mathbb{Z}).
\eq
A meromorphic function $f : {\mathbb H} \rightarrow {\mathbb C}$,
which only satisfies eq.~(\ref{chapter_elliptics:trafo_modular_form})
(or equivalently only eq.~(\ref{chapter_elliptics:trafo_modular_form_v2})) is called
\index{weakly modular}
{\bf weakly modular}
of weight $k$ for $\mathrm{SL}_2(\mathbb{Z})$.
It is clear that $f$ is weakly modular of weight $k$ for $\mathrm{SL}_2(\mathbb{Z})$ 
if eq.~(\ref{chapter_elliptics:trafo_modular_form_v2}) holds for the two generators of $\mathrm{SL}_2(\mathbb{Z})$: 
\bq
 f\left(T\left(\tau\right)\right) 
 \; = \;
 f\left(\tau+1\right) 
 \; = \; 
 f\left(\tau\right) 
 & \mbox{and} & 
 f\left(S\left(\tau\right)\right) 
 \; = \; 
 f\left(\frac{-1}{\tau}\right) 
 \; = \; \tau^k f\left(\tau\right). 
\eq
From the periodicity $f(\tau+1)=f(\tau)$ and the holomorphicity at the cusp $\tau=i\infty$ it follows that
a modular form $f(\tau)$ of $\mathrm{SL}_2(\mathbb{Z})$ has a $\bar{q}$-expansion
\bq
 f\left(\tau\right)
 & = &
 \sum\limits_{n=0}^{\infty} a_n \bar{q}^n \qquad \text{with} \qquad \bar{q}=e^{2\pi i \tau}.
\eq
A modular form for $\mathrm{SL}_2(\mathbb{Z})$ is called a 
\index{cusp form}
{\bf cusp form} 
of $\mathrm{SL}_2(\mathbb{Z})$, if it vanishes at the cusp $\tau=i \infty$.
This is the case if $a_0=0$ in the $\bar{q}$-expansion of $f(\tau)$.
The set of modular forms of weight $k$ for $\mathrm{SL}_2(\mathbb{Z})$ 
is denoted by $\mathcal{M}_k(\mathrm{SL}_2(\mathbb{Z}))$, 
the set of cusp forms of weight $k$ for $\mathrm{SL}_2(\mathbb{Z})$ 
is denoted by $\mathcal{S}_k(\mathrm{SL}_2(\mathbb{Z}))$.

Apart from $\mathrm{SL}_2({\mathbb Z})$ we may also look at congruence subgroups.
The 
{\bf standard congruence subgroups}
are defined by
\begin{align}
 \Gamma_0(N) & =
 \left\{ \left( \begin{array}{cc}
                a & b \\ 
                c & d \\
                \end{array}  \right)
 \in \mathrm{SL}_2(\mathbb{Z}): c \equiv 0\ \text{mod}\ N \right\},
 \nonumber \\
\Gamma_1(N) & =
 \left\{ \left( \begin{array}{cc}
                a & b \\ 
                c & d \\
                \end{array}  \right)
 \in \mathrm{SL}_2(\mathbb{Z}): a,d \equiv 1\ \text{mod}\ N, \; c \equiv 0\ \text{mod}\ N  \right\},
 \nonumber \\
 \Gamma(N) & =
 \left\{ \left( \begin{array}{cc}
                a & b \\ 
                c & d \\
                \end{array}  \right)
 \in \mathrm{SL}_2(\mathbb{Z}): a,d \equiv 1\ \text{mod}\ N, \; b,c \equiv 0\ \text{mod}\ N \right\}.
\nonumber
\end{align}
$\Gamma(N)$ is called the 
\index{principle congruence subgroup}
{\bf principle congruence subgroup} 
of level $N$.
The principle congruence subgroup $\Gamma(N)$ is a normal subgroup of $\mathrm{SL}_2({\mathbb Z})$.
In general, a subgroup $\Gamma$ of $\mathrm{SL}_2({\mathbb Z})$ is called a 
\index{congruence subgroup}
{\bf congruence subgroup},
if there exists an $N$ such that
\bq
 \Gamma\left(N\right) & \subseteq & \Gamma.
\eq
The smallest such $N$ is called the
\index{level of a congruence subgroup}
{\bf level of the congruence subgroup}.

We may now define modular forms for a congruence subgroup $\Gamma$, by relaxing the transformation law
in eq.~(\ref{chapter_elliptics:trafo_modular_form}) to hold only for
modular transformations from the subgroup $\Gamma$, plus holomorphicity on ${\mathbb H}$ and at the cusps.
In detail:
A meromorphic function $f: \mathbb{H} \rightarrow \mathbb{C}$ 
is a modular form of modular weight $k$ for the congruence subgroup $\Gamma$ if
\begin{enumerate}
\item $f$ transforms as
\bq
 (f \slashoperator{\gamma}{k}) & = & f
 \qquad \text{for} \;\; 
 \gamma 
 \in \Gamma,
\eq
\item $f$ is holomorphic on $\mathbb{H}$,
\item $f \slashoperator{\gamma}{k}$ is holomorphic at $i \infty$ for all $\gamma \in \mathrm{SL}_2({\mathbb Z})$.
\end{enumerate}
Let $\Gamma$ be a congruence subgroup of level $N$. 
Modular forms for $\Gamma$ are invariant under $\tau'=\tau+N$, since
\bq
 T_N & = & 
\left( \begin{array}{cc}
1 & N \\ 
0 & 1
\end{array} \right) \; \in \; \Gamma.
\eq
In other words, they are periodic with period $N$: $f(\tau+N)=f(\tau)$.
Depending on $\Gamma$, there might even be a smaller $N'$ with $N' | N$ such that $T_{N'} \in \Gamma$.
For example for $\Gamma_0(N)$ and $\Gamma_1(N)$ we have
\bq
\label{chapter_elliptics:translation_Gamma_0_Gamma_1}
\left( \begin{array}{cc}
1 & 1 \\ 
0 & 1
\end{array} \right) \; \in \; \Gamma_0\left(N\right)
 & \mbox{and} &
\left( \begin{array}{cc}
1 & 1 \\ 
0 & 1
\end{array} \right) \; \in \; \Gamma_1\left(N\right).
\eq
Let now $N'$ be the smallest positive integer such that $T_{N'} \in \Gamma$.
It follows that modular forms for $\Gamma$ have a Fourier expansion in $\bar{q}_{N'}=\bar{q}^{\frac{1}{N'}}$:
\bq
 f\left(\tau\right)
 & = &
 \sum\limits_{n=0}^{\infty} a_n \bar{q}_{N'}^n.
\eq
We remark that eq.~(\ref{chapter_elliptics:translation_Gamma_0_Gamma_1}) 
implies that modular forms for $\Gamma_0(N)$ and $\Gamma_1(N)$ have a 
Fourier expansion in $\bar{q}$:
\bq
 f\left(\tau\right)
 & = &
 \sum\limits_{n=0}^{\infty} a_n \bar{q}^n.
\eq
In the following we will use frequently the notation
\bq
 \tau_N \; = \; \frac{\tau}{N},
 \;\;\;\;\;\;
 \bar{q}_N \; = \; e^{\frac{2\pi i \tau}{N}}
 & \mbox{and therefore} &
 \bar{q}_N \; = \; \exp\left(2\pi i \tau_N\right)
 \; = \; \bar{q}^{\frac{1}{N}}.
\eq
A modular form $f(\tau)$ for $\Gamma$ is called a 
\index{cusp form}
{\bf cusp form},
if $a_0=0$ in the Fourier expansion of $f \slashoperator{\gamma}{k}$ for all $\gamma \in \mathrm{SL}_2({\mathbb Z})$.

For a congruence subgroup $\Gamma$ of $\mathrm{SL}_2({\mathbb Z})$ we denote by
${\mathcal M}_k(\Gamma)$ the space of modular forms of weight $k$,
and by $ \mathcal{S}_k(\Gamma)$ the space of cusp forms of weight $k$.
The space $\mathcal{M}_k(\Gamma)$ is a finite dimensional $\mathbb{C}$-vector space.
Furthermore, $\mathcal{M}_k(\Gamma)$ is the direct sum of two finite dimensional $\mathbb{C}$-vector spaces:
the space of cusp forms $ \mathcal{S}_k(\Gamma)$ and the Eisenstein subspace $\mathcal{E}_k(\Gamma)$.

From the inclusions
\begin{align}
 \Gamma(N)
 \subseteq \Gamma_1(N)
 \subseteq \Gamma_0(N)
 \subseteq \text{SL}_2(\mathbb{Z})
\end{align}
follow the inclusions
\bq
\label{chapter_elliptics:inclusion}
 \mathcal{M}_k(\mathrm{SL}_2(\mathbb{Z}))
 \subseteq \mathcal{M}_k(\Gamma_0(N))
 \subseteq \mathcal{M}_k(\Gamma_1(N))
 \subseteq \mathcal{M}_k(\Gamma(N)).
\eq
For a given $N$, the space $\mathcal{M}_k(\Gamma(N))$ of modular forms of weight $k$ for the principal congruence
subgroup $\Gamma(N)$ is the largest one among the spaces listed in eq.~(\ref{chapter_elliptics:inclusion}).
By definition we have for $f \in \mathcal{M}_k(\Gamma(N))$ and $\gamma \in \Gamma(N)$
\begin{align}
 f \slashoperator{\gamma}{k} & = f,
 & 
 \gamma & \in \Gamma(N).
\end{align}
We may ask what happens if we transform by a $\gamma \in \mathrm{SL}_2(\mathbb{Z})$, which does not belong
to the congruence subgroup $\Gamma(N)$.
One may show that in this case we have
\begin{align}
\label{chapter_elliptics:modular_form_covariance_Gamma_N}
 f \slashoperator{\gamma}{k} & \in \mathcal{M}_k(\Gamma(N)),
 & 
 \gamma & \in \mathrm{SL}_2(\mathbb{Z}) \backslash \Gamma(N),
\end{align}
i.e. $f \slashoperator{\gamma}{k}$ is again a modular form
of weight $k$ for $\Gamma(N)$, although not necessarily identical to $f$.
The proof relies on the fact that $\Gamma(N)$ is a normal subgroup of $\mathrm{SL}_2(\mathbb{Z})$.
This is essential: If $\Gamma$ is a non-normal congruence subgroup of $\mathrm{SL}_2(\mathbb{Z})$ one has in general
$f \slashoperator{\gamma}{k} \notin \mathcal{M}_k(\Gamma)$.
\\
\\
\bs
{\it \refstepcounter{exercise}
{\bf Exercise \theexercise}: 
Let $f \in \mathcal{M}_k(\Gamma(N))$ and $\gamma \in \mathrm{SL}_2(\mathbb{Z}) \backslash \Gamma(N)$.
Show that $f\slashoperator{\gamma}{k} \in \mathcal{M}_k(\Gamma(N))$.
}
\es
\\
\\
Let $N'$ be a divisor of $N$. We have
\bq
 \Gamma_0\left(N\right) \subseteq \Gamma_0\left(N'\right),
 \;\;\;\;\;\;
 \Gamma_1\left(N\right) \subseteq \Gamma_1\left(N'\right),
 \;\;\;\;\;\;
 \Gamma\left(N\right) \subseteq \Gamma\left(N'\right),
\eq
and therefore
\bq
 \mathcal{M}_k\left(\Gamma_0\left(N'\right)\right) \subseteq \mathcal{M}_k\left(\Gamma_0\left(N\right)\right),
 \;\;\;
 \mathcal{M}_k\left(\Gamma_1\left(N'\right)\right) \subseteq \mathcal{M}_k\left(\Gamma_1\left(N\right)\right),
 \;\;\;
 \mathcal{M}_k\left(\Gamma\left(N'\right)\right) \subseteq \mathcal{M}_k\left(\Gamma\left(N\right)\right).
 \nonumber \\
\eq
In other words, a modular form $f \in \mathcal{M}_k(\Gamma_0(N))$ is also a modular form for $\Gamma_0(K \cdot N)$
where $K \in {\mathbb N}$ (and similar for $\Gamma_1(N)$ and $\Gamma(N)$).

There is one more generalisation which we can do:
We may consider for $\Gamma_0(N)$ modular forms with a character $\chi$.
In appendix~\ref{appendix_dirichlet} we review in detail Dirichlet characters $\chi(n)$.
In essence, a Dirichlet character of modulus $N$ is a function
$\chi: \mathbb{Z} \rightarrow \mathbb{C}$ satisfying
\begin{alignat}{3}
 & (i) & \hspace*{3mm} & \chi(n) = \chi(n+N) & \hspace*{5mm} & \forall\ n \in \mathbb{Z}, \nonumber \\
 & (ii) & & \chi(n) = 0 & & \mbox{if} \;\; \gcd(n,N) > 1, \nonumber \\
 &      & & \chi(n) \neq 0 & & \mbox{if} \;\; \gcd(n,N) = 1, \nonumber \\ 
 & (iii) & & \chi(nm) = \chi(n) \chi(m) & & \forall\ n,m \in \mathbb{Z}.
\end{alignat}
Let $N$ be a positive integer and let $\chi$ be a Dirichlet character modulo $N$.
A meromorphic function $f: \mathbb{H} \rightarrow \mathbb{C}$ is a modular form of weight $k$ for $\Gamma_0(N)$ 
with character $\chi$ if
\begin{enumerate}
\item $f$ transforms as
\bq
 f \left( \dfrac{a\tau+b}{c\tau+d}\right) 
 & = & \chi(d) (c\tau+d)^k f(\tau)
 \qquad \text{for} \;\; 
 \left( \begin{array}{cc}
         a & b \\ 
         c & d
        \end{array}  \right)
 \in \Gamma_0(N),
\eq
\item $f$ is holomorphic on $\mathbb{H}$,
\item $f \slashoperator{\gamma}{k}$ is holomorphic at $i \infty$ for all $\gamma \in \mathrm{SL}_2({\mathbb Z})$.
\end{enumerate}
The space of modular forms of weight $k$ and character $\chi$ for the congruence subgroup $\Gamma_0(N)$
is denoted by $\mathcal{M}_k(N,\chi)$,
the associated space of cusp forms by $\mathcal{S}_k(N,\chi)$ and the Eisenstein subspace
by $\mathcal{E}_k(N,\chi)$.

Introducing modular forms with characters is useful due to the following theorem:
\begin{theorem}
\label{chapter_elliptics:theorem_Eisenstein}
The space $\mathcal{M}_k(\Gamma_1(N))$ is a direct sum of spaces of modular forms with characters:
\bq
\label{chapter_elliptics:mod_form_space_decomp}
 \mathcal{M}_{k}(\Gamma_1(N)) 
 & = &
 \bigoplus\limits_{\chi}\ \mathcal{M}_k(N,\chi),
\eq
where the sum runs over all Dirichlet characters modulo $N$.
Similar decompositions hold for the space of cusp forms and the Eisenstein subspaces:
\bq
\label{chapter_elliptics:mod_form_space_decomp_cusp_and_Eisenstein}
 \mathcal{S}_{k}(\Gamma_1(N))
 \; = \;
 \bigoplus\limits_{\chi}\ \mathcal{S}_k(N,\chi),
 & &
 \mathcal{E}_{k}(\Gamma_1(N)) 
 \; = \; 
 \bigoplus\limits_{\chi}\ \mathcal{E}_k(N,\chi).
\eq
\end{theorem}
The following exercise shows, why we only consider modular forms with characters for the congruence
subgroup $\Gamma_0(N)$ (and subgroups thereof):
\\
\\
\bs
{\it \refstepcounter{exercise}
{\bf Exercise \theexercise}: 
Let $\chi$ be a Dirichlet character with modulus $N$ and
$f \in \mathcal{M}_k(N,\chi)$.
Let further $\gamma_1, \gamma_2 \in \Gamma_0(N)$ and set $\gamma_{12} = \gamma_1 \gamma_2$.
Show that
\bq
 f\left(\gamma_1\left(\gamma_2\left(\tau\right)\right)\right)
 & = &
 f\left(\gamma_{12}\left(\tau\right)\right).
\eq
}
\es
\\
\\
We now introduce 
\index{iterated integrals of modular forms}
{\bf iterated integrals of modular forms}.
Let $f_1, \dots, f_n$ be modular forms 
(for $\mathrm{SL}_2(\mathbb{Z})$ or some congruence subgroup $\Gamma$ of level $N$).
For a modular form $f \in \mathcal{M}_k(\Gamma)$ of level $N$ let $N'$ be the smallest positive integer such
that 
\bq
\left( \begin{array}{cc}
1 & N' \\ 
0 & 1
\end{array} \right) \in \Gamma.
\eq
We set
\bq
\label{chapter_elliptics:def_omega_modular}
 \omega^{\mathrm{modular}}\left(f\right)
 & = &
 2 \pi i \; f\left(\tau\right) \frac{d\tau}{N'}
 \;\; = \;\;
 2 \pi i \; f\left(\tau\right) d\tau_{N'}
 \;\; = \;\;
 f\left(\tau\right) \frac{d\bar{q}_{N'}}{\bar{q}_{N'}}.
\eq
If the modular form $f(\tau)$ has the $\bar{q}_{N'}$-expansion
\bq
 f\left(\tau\right)
 & = &
 \sum\limits_{n=0}^\infty a_n \bar{q}_{N'}^n,
\eq
we have
\bq
\label{chapter_elliptics:differential_form_modular_form}
 \omega^{\mathrm{modular}}\left(f\right)
 & = &
 \sum\limits_{n=0}^\infty a_n \bar{q}_{N'}^{n-1} d\bar{q}_{N'}.
\eq
Let $\gamma : [a,b] \rightarrow {\mathbb H}$ be a path with $\gamma(a)=\tau_0$ and $\gamma(b)=\tau$.
We set
\bq
 I\left(f_1,\dots,f_n; \tau \right) 
 & = &
 I_\gamma\left(\omega^{\mathrm{modular}}\left(f_1\right),\dots,\omega^{\mathrm{modular}}\left(f_n\right);b\right),
\eq
where the right-hand side refers to the general definition 
of an iterated integral given in eq.~(\ref{chapter_iterated_integrals:def_iterated_integral}).
Explicitly
\bq
 I\left(f_1,f_2,\dots,f_n; \tau \right) 
 & = &
 \left( \frac{2 \pi i}{N'} \right)^n
 \int\limits_{\tau_0}^{\tau} d\tau_1
 f_1\left(\tau_1\right)
 \int\limits_{\tau_0}^{\tau_1} d\tau_2
 f_2\left(\tau_2\right)
 \dots
 \int\limits_{\tau_0}^{\tau_{n-1}} d\tau_n
 f_n\left(\tau_n\right).
\eq
As base point we usually take $\tau_0=i\infty$.
Please note that an integral over a modular form is in general not a modular form. 
This is not surprising if we consider the following analogy:
An integral over a rational function is in general not a rational function.

We usually like iterated integrals appearing in solutions of Feynman integrals to have at worst simple poles.
Let's study iterated integrals of modular forms. As modular forms are holomorphic in the complex upper half-plane,
there are no poles there. So the only interesting points are the cusps.
Let's consider as an example modular forms $f \in \mathcal{M}_k(\mathrm{SL}_2(\mathbb{Z}))$, so the only cusp is at $\tau=i\infty$.
By definition a modular form $f(\tau)$ is holomorphic at the cusp and has a $\bar{q}$-expansion
\bq
 f(\tau)
 & = &
 a_0 + a_1 \bar{q} + a_2 \bar{q}^2 + \dots,
 \;\;\;\;\;\;\;\;\;\;\;\;
 \bar{q}=\exp(2\pi i \tau).
\eq
The transformation $\bar{q}=\exp(2\pi i \tau)$ transforms the point $\tau=i\infty$ to $\bar{q}=0$ and we have
\bq
 2 \pi i \; f(\tau) d\tau
 & = &
 \frac{d\bar{q}}{\bar{q}} \left( a_0 + a_1 \bar{q} + a_2 \bar{q}^2 + \dots \right).
\eq
Thus a modular form non-vanishing at the cusp $\tau=i\infty$ has a simple pole at $\bar{q}=0$.

\subsection{Eisenstein series}
\label{chapter_elliptics:section_Eisenstein}

In applications towards Feynman integrals we will need the $\bar{q}_{N'}$-expansions of modular
forms.
A complete treatment is beyond the scope of this book and we refer the reader 
to textbooks on modular forms \cite{Stein,Miyake,Diamond,Cohen}.

We limit ourselves to the aspects most relevant to Feynman integrals.
In this section we will look at the $\bar{q}_{N'}$-expansions of modular forms spanning the Eisenstein
subspace.
We will study Eisenstein series for $\mathrm{SL}_2({\mathbb Z})$, $\Gamma_1(N)$ and $\Gamma(N)$.

The case of the full modular group $\mathrm{SL}_2({\mathbb Z})$ is rather simple and the main result
is that any modular form for $\mathrm{SL}_2({\mathbb Z})$ can be written as a polynomial in two Eisenstein series
$e_4(\tau)$ and $e_6(\tau)$ of modular weight $4$ and $6$, respectively.

Eisenstein series for $\Gamma_1(N)$ appear in the simplest elliptic Feynman integrals.
A basis for $\mathcal{E}_k(\Gamma_1(N))$ can be given explicitly.

For a modular form $f$ of a congruence subgroup $\Gamma$ of level $N$
we also would like to know the transformation behaviour under
$\gamma \in \mathrm{SL}_2({\mathbb Z}) \backslash \Gamma$.
We first note that by the definition of a congruence subgroup we have $f \in \mathcal{M}_k(\Gamma(N))$.
If in addition $f \in \mathcal{E}_k(\Gamma(N))$ we may answer this question explicitly.

\subsubsection{Eisenstein series for $\mathrm{SL}_2({\mathbb Z})$}

The $z$-dependent Eisenstein series $E_k(z,\tau)$ are defined by
\bq
\label{chapter_elliptics:z_dependent_Eisenstein_series}
 E_k\left(z,\tau\right)
 & = &
 \sideset{}{_e}\sum\limits_{(n_1,n_2) \in {\mathbb Z}^2} \frac{1}{\left(z+n_1 + n_2\tau \right)^k}.
\eq
The subscript $e$ at the summation sign denotes the 
Eisenstein summation prescription defined by
\bq
 \sideset{}{_e}\sum\limits_{(n_1,n_2) \in {\mathbb Z}^2} f\left(z+n_1 + n_2\tau \right)
 & = &
 \lim\limits_{N_2\rightarrow \infty} \sum\limits_{n_2=-N_2}^{N_2}
 \left(
 \lim\limits_{N_1\rightarrow \infty} \sum\limits_{n_1=-N_1}^{N_1}
 f\left(z + n_1 + n_2 \tau \right)
 \right).
\eq
The series in eq.~(\ref{chapter_elliptics:z_dependent_Eisenstein_series}) is absolutely convergent for $k \ge 3$.
For $k=1$ and $k=2$ the Eisenstein summation depends on the choice of generators. 
One further sets
\bq
\label{chapter_elliptics:def_e_k}
 e_k\left(\tau\right)
 & = &
 \sideset{}{_e}\sum\limits_{(n_1,n_2) \in {\mathbb Z}^2\backslash (0,0)} \frac{1}{\left(n_1 + n_2\tau \right)^k}.
\eq
We have $e_k(\tau)=0$ whenever $k$ is odd.
The $\bar{q}$-expansions of the first few Eisenstein series are
\bq
 e_{2}\left(\tau\right)
 & = &
 2 \left(2 \pi i\right)^2 \left[
 - \frac{1}{24} + \bar{q} + 3 \bar{q}^2 + 4 \bar{q}^3 + 7 \bar{q}^4 + 6 \bar{q}^5 + 12 \bar{q}^6
 \right]
 + {\mathcal O}\left(\bar{q}^7\right),
 \\
 e_{4}\left(\tau\right)
 & = &
 \frac{\left(2 \pi i\right)^4}{3} \left[
 \frac{1}{240} + \bar{q} + 9 \bar{q}^2 + 28 \bar{q}^3 + 73 \bar{q}^4 + 126 \bar{q}^6
 \right]
 + {\mathcal O}\left(\bar{q}^7\right),
 \nonumber \\
 e_{6}\left(\tau\right)
 & = &
 \frac{\left(2 \pi i\right)^6}{60} \left[
 - \frac{1}{504} + \bar{q} + 33 \bar{q}^2 + 244 \bar{q}^3 + 1057 \bar{q}^4 + 3126 \bar{q}^5 + 8052 \bar{q}^6
 \right]
 + {\mathcal O}\left(\bar{q}^7\right).
 \nonumber 
\eq
For $k \ge 4$ the Eisenstein series $e_k(\tau)$ are modular forms of ${\mathcal M}_k(\mathrm{SL}_2({\mathbb Z}))$.
The space ${\mathcal M}_k(\mathrm{SL}_2({\mathbb Z}))$ has a basis of the form
\bq
\label{chapter_elliptics:basis_SL2Z}
 \left( e_4\left(\tau\right) \right)^{\nu_4} \left(e_6\left(\tau\right)\right)^{\nu_6},
\eq
where $\nu_4$ and $\nu_6$ run over all non-negative integers with $4 \nu_4 + 6 \nu_6 = k$.

As an example, let us give the cusp form of modular weight $12$ for $\mathrm{SL}_2({\mathbb Z})$:
\bq
\label{chapter_elliptics:def_Delta}
 \Delta\left(\tau\right)
 & = &
 \left(2\pi i\right)^{12} \eta\left(\tau\right)^{24}
 \;\; = \;\;
 10800 \left( 20 \left(e_4\left(\tau\right)\right)^3 - 49 \left(e_6\left(\tau\right)\right)^2 \right).
\eq
Note that $e_2(\tau)$ is not a modular form. Under modular transformations $e_2(\tau)$ transforms as
\bq
 e_2\left(\frac{a\tau+b}{c\tau+d}\right)
 & = &
 \left( c\tau+d \right)^2 e_2\left(\tau\right)
 - 2 \pi i c \left( c\tau+d \right).
\eq
Modularity is spoiled by the second term on the right-hand side.
\\
\\
\bs
{\it \refstepcounter{exercise}
{\bf Exercise \theexercise}: 
Consider
\bq
 f\left(\tau\right) & = & 
 e_2\left(\tau\right) - 2 e_2\left(2\tau\right)
\eq
and work out the transformation properties under $\gamma \in \Gamma_0(2)$.
}
\es
 
\subsubsection{Eisenstein series for $\Gamma_1(N)$}

Let $\Gamma$ be a congruence subgroup of $\mathrm{SL}_2({\mathbb Z})$.
By definition there exists an $N$, such that
\bq
 \Gamma\left(N\right) & \subseteq & \Gamma.
\eq
This implies
\bq
 {\mathcal M}_k\left(\Gamma\right) & \subseteq & {\mathcal M}_k\left(\Gamma\left(N\right)\right) 
\eq
and this reduces in a first step the study of modular forms for an arbitrary congruence subgroup $\Gamma$
to the study of modular forms of the principal congruence subgroup $\Gamma\left(N\right)$.
Now let $\eta(\tau) \in {\mathcal M}_k(\Gamma(N))$.
Then \cite{Miyake}
\bq
 \eta\left(N\tau\right)
 & \in &
 {\mathcal M}_k\left(\Gamma_1\left(N^2\right)\right),
\eq
which reduces in a second step the study of modular forms for an arbitrary congruence subgroup $\Gamma$
to the study of modular forms of the congruence subgroup $\Gamma_1\left(N\right)$.

Let us therefore consider modular forms for the congruence subgroups $\Gamma_1(N)$, and here
in particular the Eisenstein subspace $\mathcal{E}_k(\Gamma_1(N))$.
Let us first note that
\bq
 T_1 \; = \;
 \left( \begin{array}{cc} 1 & 1 \\ 0 & 1 \\ \end{array} \right)
 & \in &
 \Gamma_1\left(N\right),
\eq
and therefore the modular forms $f \in {\mathcal M}_k(\Gamma_1(N))$ have an expansion in $\bar{q}$.
From eq.~(\ref{chapter_elliptics:mod_form_space_decomp_cusp_and_Eisenstein}) 
we know that $\mathcal{E}_k(\Gamma_1(N))$ decomposes into Eisenstein spaces $\mathcal{E}_k(N,\chi)$.
\bq
 \mathcal{E}_{k}(\Gamma_1(N)) 
 & = &  
 \bigoplus\limits_{\chi}\ \mathcal{E}_k(N,\chi).
\eq
A basis for the Eisenstein subspace $\mathcal{E}_k(N,\chi)$ can be given explicitly.
To this aim we first define 
\index{generalised Eisenstein series}
{\bf generalised Eisenstein series}.
Let $\chi_a$ and $\chi_b$ be primitive Dirichlet characters with conductors $d_a$ and $d_b$, respectively.
We set
\bq
\label{chapter_elliptics:def_Eisenstein_E}
 E_{k}\left(\tau,\chi_a,\chi_b\right)
 & = &
 a_0 + \sum\limits_{n=1}^{\infty} \left( \sum\limits_{d|n} \chi_a(n/d) \cdot \chi_b(d) \cdot d^{k-1} \right) \bar{q}^{n},
\eq
The normalisation is such that the coefficient of $\bar{q}$ is one.
The constant term $a_0$ is given by
\begin{align}
 a_0 &
 = 
 \begin{cases}
 -\frac{B_{k,\chi_b}}{2k}, \qquad &\text{if}\ d_a=1, \\
0, \qquad &\text{if}\ d_a>1.
\end{cases}
\end{align}
Note that the constant term $a_0$ depends on $\chi_a$ and $\chi_b$.
The 
\index{generalised Bernoulli numbers}
{\bf generalised Bernoulli numbers}
$B_{k,\chi_b}$ are defined by
\bq
\label{chapter_elliptics:def_generalised_Bernoulli}
 \sum\limits_{n=1}^{d_b} \chi_b(n) \dfrac{xe^{nx}}{e^{d_b x}-1}
 & = &
 \sum\limits_{k=0}^{\infty} B_{k,\chi_b} \dfrac{x^k}{k!}. 
\eq
Note that in the case of the trivial character $\chi_b=1$,
eq.~(\ref{chapter_elliptics:def_generalised_Bernoulli}) reduces to
\bq
 \dfrac{xe^{x}}{e^{x}-1}
 & = &
 \sum\limits_{k=0}^{\infty} B_{k,1} \dfrac{x^k}{k!},
\eq
yielding $B_{1,1}=1/2$.
The ordinary Bernoulli numbers $B_k$ are generated by $x/(e^x-1)$ 
(i.e. without an extra factor $e^x$ in the numerator)
and yield $B_1=-1/2$.

Let now $\chi_a$, $\chi_b$ and $k$ be such that 
\bq
\label{chapter_elliptics:condition_k}
 \chi_a\left(-1\right) \chi_b\left(-1\right) & = & \left(-1\right)^k
\eq
and if $k=1$ one requires in addition
\bq
\label{chapter_elliptics:condition_k_eq_1}
 \chi_a\left(-1\right) \; = \; 1,
 & &
 \chi_b\left(-1\right) \; = \; -1.
\eq
Let $K$ be a positive integer.
We then set
\bq
\label{chapter_elliptics:def_Eisenstein_E_B}
 E_{k,K}\left(\tau,\chi_a,\chi_b\right) 
 & = & 
 \left\{
  \begin{array}{lll}
    E_{k}\left(K\tau,\chi_a,\chi_b\right), & (k,\chi_a,\chi_b) \neq (2,1,1), & K \ge 1, \\ 
    E_{2}\left(\tau,1,1\right) - K E_{2}\left(K\tau,1,1\right), & (k,\chi_a,\chi_b) = (2,1,1), & K > 1. \\
  \end{array}
 \right.
 \nonumber \\
\eq
Let $N$ be an integer multiple of $(K \cdot d_a \cdot d_b)$.
Then $E_{k,K}(\tau,\chi_a,\chi_b)$ is a modular form for $\Gamma_1(N)$ of modular weight $k$ and level $N$:
\bq
 E_{k,K}\left(\tau,\chi_a,\chi_b\right)
 & \in & 
 \mathcal{E}_k(\Gamma_1(N)),
 \;\;\;\;\;\;\;\;\;\;\;\; 
 \left(K \cdot d_a \cdot d_b\right) \; | \; N.
\eq
In more detail we have that 
for $(k,\chi_a,\chi_b) \neq (2,1,1)$
\bq
 E_{k,K}\left(\tau,\chi_a,\chi_b\right)
 & \in & 
 \mathcal{E}_k(N,\tilde{\chi}),
\eq
where
$\tilde{\chi}$ is the Dirichlet character with modulus $N$ induced by $\chi_a \chi_b$.
Furthermore, for $(k,\chi_a,\chi_b) = (2,1,1)$ and $K > 1$ we have
\bq
 E_{k,K}\left(\tau,\chi_a,\chi_b\right)
 & \in & 
 \mathcal{E}_k(N,\tilde{1})
 \; = \; 
 \mathcal{E}_k(\Gamma_0(N)).
\eq
\begin{theorem}
Let $(K \cdot d_a \cdot d_b)$ be a divisor of $N$.
The $E_{k,K}(\tau,\chi_a,\chi_b)$ subject to the conditions outlined above 
(eqs.~(\ref{chapter_elliptics:condition_k}) - (\ref{chapter_elliptics:condition_k_eq_1}))
form a basis
of $\mathcal{E}_k(N,\tilde{\chi})$,
where $\tilde{\chi}$ is the Dirichlet character with modulus $N$ induced by $\chi_a \chi_b$.
\end{theorem}
Of particular interest are characters which are obtained from the Kronecker symbol.
These characters take the values $\{-1,0,1\}$.
In general, the value of a Dirichlet character is a root of unity or zero.
The restriction to Dirichlet characters obtained from the Kronecker symbol
has the advantage that the $\bar{q}$-expansion of the Eisenstein series can be computed 
within the rational numbers.

Let $a$ be an integer, which is either one or the discriminant of a quadratic field.
In appendix~\ref{appendix_dirichlet:kronecker_symbol} we give a criteria for $a$ being the
discriminant of a quadratic field.
The Kronecker symbol, also defined in appendix~\ref{appendix_dirichlet:kronecker_symbol}, 
then defines a primitive Dirichlet character
\bq
\label{chapter_elliptics:primitive_Dirichlet_character}
 \chi_a\left(n\right) & = & 
 \left( \frac{a}{n} \right)
\eq
of conductor $|a|$.
On the right-hand side of eq.~(\ref{chapter_elliptics:primitive_Dirichlet_character})
we have the Kronecker symbol. 
Be alert that the right-hand side of eq.~(\ref{chapter_elliptics:primitive_Dirichlet_character})
does not denote a fraction.
Let $a$ and $b$ be integers, which are either one or the discriminant of a quadratic field.
For 
\bq
 \chi_a\left(n\right) \; = \; 
 \left( \frac{a}{n} \right),
 & &
 \chi_b\left(n\right) \; = \; 
 \left( \frac{b}{n} \right)
\eq
we introduce the short-hand notations
\bq
\label{chapter_elliptics:Eisenstein_short_hand_notation}
 E_{k,a,b}\left(\tau\right)
 & = &
 E_{k}\left(\tau,\chi_a,\chi_b\right)
 \nonumber \\
 E_{k,a,b,K}\left(\tau\right)
 & = & 
 E_{k,K}\left(\tau,\chi_a,\chi_b\right).
\eq
These Eisenstein series have a $\bar{q}$-expansion with rational coefficients.

Remark: For $k$ even and $k \ge 4$ the relation between the Eisenstein series $e_k(\tau)$ defined in eq.~(\ref{chapter_elliptics:def_e_k}) and the Eisenstein series with a trivial character is
\bq
\label{chapter_elliptics:relation_e_k_to_E_k_1_1}
 e_k\left(\tau\right)
 & = &
 2 \frac{\left(2 \pi i \right)^k}{\left(k-1\right)!}
 E_{k,1,1}\left(\tau\right).
\eq

\subsubsection{Eisenstein series for $\Gamma(N)$}

Let us now turn to Eisenstein series for $\Gamma(N)$. 
For these Eisenstein series we may give the transformation law for any $\gamma \in \mathrm{SL}_2({\mathbb Z})$ and not just $\gamma \in \Gamma(N)$.

Let $r, s$ be integers with $0 \le r,s <N$.
Following \cite{Broedel:2018iwv,Duhr:2019rrs} we set
\bq
\label{chapter_elliptics:def_Eisenstein_h}
 h_{k,N,r,s}\left(\tau\right)
 & = & \sum\limits_{n=0}^\infty a_n \bar{q}^n_N.
\eq
For $n \ge 1$ the coefficients are given by
\bq
\label{coeff_h_a_n}
 a_n
 & = &
 \frac{1}{2 N^k}
 \sum\limits_{d|n}
 \sum\limits_{c_1=0}^{N-1}
 d^{k-1}
 \left[
   e^{\frac{2\pi i}{N} \left(r \frac{n}{d} - \left(s-d\right) c_1 \right)}
   + \left(-1\right)^k
   e^{-\frac{2\pi i}{N} \left(r \frac{n}{d} - \left(s+d\right) c_1 \right)}
 \right].
\eq
The constant term is given for $k \ge 2$ by
\bq
 a_0
 & = & 
 - \frac{1}{2k} B_k\left(\frac{s}{N}\right),
\eq
where $B_k(x)$ is the $k$'th 
\index{Bernoulli polynomial}
{\bf Bernoulli polynomial}
defined by
\bq
 \frac{t e^{x t}}{e^t-1}
 & = &
 \sum\limits_{k=0}^\infty \frac{B_k\left(x\right)}{k!} t^k.
\eq
For $k=1$ the constant term is given by
\bq
 a_0
 & = &
 \left\{
 \begin{array}{ll}
  \frac{1}{4} - \frac{s}{2N}, & s \neq 0, \\
 0, & (r,s) = (0,0), \\
 \frac{i}{4} \cot\left(\frac{r}{N} \pi \right),
 & \mbox{otherwise}.
 \end{array}
 \right.
\eq
We have 
\bq
\label{chapter_elliptics:def_Eisenstein_h_k_N_r_s}
 h_{k,N,r,s}\left(\tau\right)
 & = &
 \frac{1}{2} \frac{\left(k-1\right)!}{\left(2\pi i\right)^k}
 \sideset{}{_e}\sum\limits_{(n_1,n_2) \in {\mathbb Z}^2\backslash (0,0)} 
 \frac{e^{\frac{2\pi i}{N} \left(n_1s-n_2r\right)}}{\left(n_1+n_2\tau\right)^k}.
\eq
\bs
{\it \refstepcounter{exercise}
{\bf Exercise \theexercise}: 
Show eq.~(\ref{chapter_elliptics:def_Eisenstein_h_k_N_r_s}).
}
\es
\\
\\
The normalisation is compatible with the normalisation of the previous subsection.
For example we have
\bq
 h_{k,1,0,0}\left(\tau\right)
 & = &
 E_{k,1,1}\left(\tau\right).
\eq
With the exception of $(k,r,s) \neq (2,0,0)$ the $h_{k,N,r,s}(\tau)$ are Eisenstein series for $\Gamma(N)$:
\bq
 h_{k,N,r,s}\left(\tau\right) 
 & \in & 
 \mathcal{E}_k\left(\Gamma\left(N\right)\right).
\eq
For $(k,N,r,s)=(2,1,0,0)$ we have
\bq
 h_{2,1,0,0}\left(\tau\right)
 & = &
 E_{2,1,1}\left(\tau\right)
 \; = \;
 \frac{1}{2\left(2\pi i\right)^2} e_2\left(\tau\right),
\eq
which is not a modular form.

The most important property of the Eisenstein series $h_{k,N,r,s}(\tau)$ is their transformation behaviour
under the full modular group:
The Eisenstein series $h_{k,N,r,s}(\tau)$
transform under modular transformations 
\bq
 \gamma \; = \;
 \left(\begin{array}{cc} a & b \\ c & d \\ \end{array} \right)
 & \in &
 \mathrm{SL}_2\left({\mathbb Z}\right)
\eq
of the full modular group $\mathrm{SL}_2({\mathbb Z})$ as
\bq
\label{chapter_elliptics:trafo_Eisenstein_h_1}
 h_{k,N,r,s}\left(\frac{a\tau+b}{c\tau+d}\right)
 & = &
 \left(c\tau +d\right)^k
 h_{k,N,\left(rd+sb\right) \bmod N,\left(rc+sa\right) \bmod N}\left(\tau\right),
\eq
or equivalently with the help of the $\slashoperator{\gamma}{k}$ operator
\bq
\label{chapter_elliptics:trafo_Eisenstein_h_2}
 \left(h_{k,N,r,s}\slashoperator{\gamma}{k}\right)\left(\tau\right)
 & = &
 h_{k,N,\left(rd+sb\right) \bmod N,\left(rc+sa\right) \bmod N}\left(\tau\right).
\eq
\bs
{\it \refstepcounter{exercise}
{\bf Exercise \theexercise}: 
Prove eq.~(\ref{chapter_elliptics:trafo_Eisenstein_h_2}) for the case $k \ge 3$.
}
\es

\subsection{The modular lambda function and Klein's $j$-invariant}
\label{chapter_elliptics:section_modular_lambda}

In this paragraph we discuss the modular lambda function and Klein's $j$-invariant.
The latter allows us to decide, if two elliptic curves are isomorphic.

In eq.~(\ref{chapter_elliptics:def_modular_lambda}) we have seen that
\bq
 k^2
 & = &
 16 
 \frac{\eta\left(\frac{\tau}{2}\right)^8 \eta\left(2\tau\right)^{16}}{\eta\left(\tau\right)^{24}}.
\eq
The right-hand side defines a function of $\tau$, called the 
\index{modular lambda function}
{\bf modular lambda function}:
\bq
 \lambda\left(\tau\right)
 & = &
 16 
 \frac{\eta\left(\frac{\tau}{2}\right)^8 \eta\left(2\tau\right)^{16}}{\eta\left(\tau\right)^{24}}.
\eq
The function $\lambda(\tau)$ is invariant 
under $\Gamma(2)$.
The congruence subgroup $\Gamma(2)$ is generated by
\bq
 \left( \begin{array}{rr}
  1 & 2 \\
  0 & 1 \\
 \end{array} \right),
 \;\;\;\;\;\;
 \left( \begin{array}{rr}
  1 & 0 \\
  2 & 1 \\
 \end{array} \right),
 \;\;\;\;\;\;
 \left( \begin{array}{rr}
  -1 & 0 \\
  0 & -1 \\
 \end{array} \right).
\eq
Thus $\lambda(\tau)$ is invariant under
\bq
 \tau' \; = \; \tau + 2,
 & \mbox{and} &
 \tau' \; = \;
 \frac{\tau}{1+2\tau}.
\eq
Under the generators of $\mathrm{SL}_2({\mathbb Z})$ the modular lambda function transforms as
\bq
 \tau' \; = \; \tau +1:
 & &
 \lambda\left(\tau'\right) 
 \; = \; 
 - \frac{\lambda\left(\tau\right)}{1-\lambda\left(\tau\right)},
 \nonumber \\
 \tau' \; = \; \frac{-1}{\tau}:
 & &
 \lambda\left(\tau'\right) 
 \; = \; 
 1-\lambda\left(\tau\right).
\eq
\index{Klein's $j$-invariant}
{\bf Klein's $j$-invariant} is defined by
\bq
\label{chapter_elliptics:def_j_invariant}
 j\left(\tau\right)
 & = &
 \frac{1728 g_2\left(\tau\right)^3}{g_2\left(\tau\right)^3-27g_3\left(\tau\right)^2},
\eq
where
\bq
 g_2\left(\tau\right) \; = \; 60 e_4\left(\tau\right),
 & &
 g_3\left(\tau\right) \; = \; 140 e_6\left(\tau\right).
\eq
Note that the denominator in eq.~(\ref{chapter_elliptics:def_j_invariant}) is the modular discriminant
\bq
 \Delta\left(\tau\right) 
 & = & 
 g_2\left(\tau\right)^3-27g_3\left(\tau\right)^2
 \; = \;
 \left(2\pi i\right)^{12} \eta\left(\tau\right)^{24}.
\eq
Two elliptic curves are isomorphic if their $j$-invariants agree.

The relation of the modular lambda function to Klein's $j$-invariant is
\bq
 j\left(\tau\right)
 & = &
 256 \frac{\left(1-\lambda+\lambda^2\right)^3}{\lambda^2\left(1-\lambda\right)^2}.
\eq

\section{Moduli spaces}
\label{chapter_elliptics:section_moduli_spaces}

For elliptic Feynman integrals, which only depend on one kinematic variable (e.g. $\NB=1$) we now have all
necessary tools: In essence, we perform in addition to a fibre transformation also
a base transformation and change from 
the original kinematic variable $x$ to the modular parameter $\tau$. We then express the Feynman integrals
as iterated integrals of modular forms.
We will see in an example later on, how this is done in practice.

However, there are elliptic Feynman integrals, which depend on more than one kinematic variable
(e.g. $\NB>1$)
and for those we need one more ingredient:
We have to introduce the moduli space ${\mathcal M}_{1,n}$ of a smooth genus one curve with $n$ marked points
and we will see that an elliptic Feynman integral can be expressed as a linear combination of iterated integrals 
on a covering space of the moduli space ${\mathcal M}_{1,n}$ with integrands having only simple poles.
This comes to no surprise, as a ``simple'' Feynman integral, which evaluates to multiple polylogarithms, can
be expressed as a linear combination of iterated integrals 
on a covering space of the moduli space ${\mathcal M}_{0,n}$ of
a genus zero curve with $n$ marked points, again with integrands having only simple poles.

In appendix~\ref{appendix_moduli_space} we give a detailed introduction into the moduli space ${\mathcal M}_{g,n}$
of a smooth algebraic curve of genus $g$ with $n$ marked points.
Here in this section we briefly summarise the main points and proceed then to the relevant aspects for computing
elliptic Feynman integrals.

Let's start with the short summary:
Let $X$ be a topological space. The configuration space of $n$ ordered points in $X$ is
\bq
 \mathrm{Conf}_n\left(X\right)
 & = &
 \left\{ \left. \left(x_1,\dots,x_n\right) \in X^n \right| x_i \neq x_j \; \mbox{for} \; i \neq j \right\}.
\eq
The non-trivial ingredient is the requirement that the points are distinct: $x_i \neq x_j$.
Without this requirement we would simply look at $X^n$.

An example is the configuration space of $n$ points on the Riemann sphere ${\mathbb C}{\mathbb P}^1$:
\bq
 \mathrm{Conf}_n\left({\mathbb C}{\mathbb P}^1\right)
 & = &
 \left\{ \left. \left(z_1,\dots,z_n\right) \in \left({\mathbb C}{\mathbb P}^1\right)^n \right| z_i \neq z_j \; \mbox{for} \; i \neq j \right\}.
\eq
A M\"obius transformation
\bq
 z'
 & = &
 \frac{az+b}{cz+d}
\eq
transforms the Riemann sphere into itself.
These transformations form a group $\mathrm{PSL}\left(2,{\mathbb C}\right)$.
Usually we are not interested in configurations 
\bq
 (z_1,\dots,z_n) \in \mathrm{Conf}_n\left({\mathbb C}{\mathbb P}^1\right)
 & \mbox{and} &
 (z_1',\dots,z_n') \in \mathrm{Conf}_n\left({\mathbb C}{\mathbb P}^1\right),
\eq
which differ only by a M\"obius transformation:
\bq
 z_j'
 & = &
 \frac{az_j+b}{cz_j+d},
 \;\;\;\;\;\;
 j \; \in \; \{1,\dots,n\}.
\eq
This brings us to the definition of the moduli space of the Riemann sphere with $n$ marked points:
\bq 
 {\mathcal M}_{0,n}
 & = &
 \mathrm{Conf}_n\left({\mathbb C}{\mathbb P}^1\right) / \mathrm{PSL}\left(2,{\mathbb C}\right).
\eq
We may use the freedom of M\"obius transformations to fix three points (usually $0$, $1$ and $\infty$).
Therefore
\bq
\label{chapter_elliptics:dim_M_0_n}
 \dim\left( \mathrm{Conf}_n\left({\mathbb C}{\mathbb P}^1\right) \right) & = & n,
 \nonumber \\
 \dim\left({\mathcal M}_{0,n}\right) & = & n-3.
\eq
Let's generalise this:
We are interested in the situation, where the topological space $X$ 
is a smooth algebraic curve $C$ in ${\mathbb C}{\mathbb P}^2$.
This implies that there exists a homogeneous polynomial $P(z_1,z_2,z_3)$ such that
\bq
 C & = & \left\{ \left. \left[z_1:z_2:z_3\right] \in {\mathbb C}{\mathbb P}^2 \right| P\left(z_1,z_2,z_3\right) = 0 \right\}. 
\eq
If $d$ is the degree of the polynomial $P(z_1,z_2,z_3)$, the genus $g$ of $C$ is given by
eq.~(\ref{chapter_elliptics:def_genus}).
\begin{figure}
\begin{center}
\includegraphics[scale=0.85]{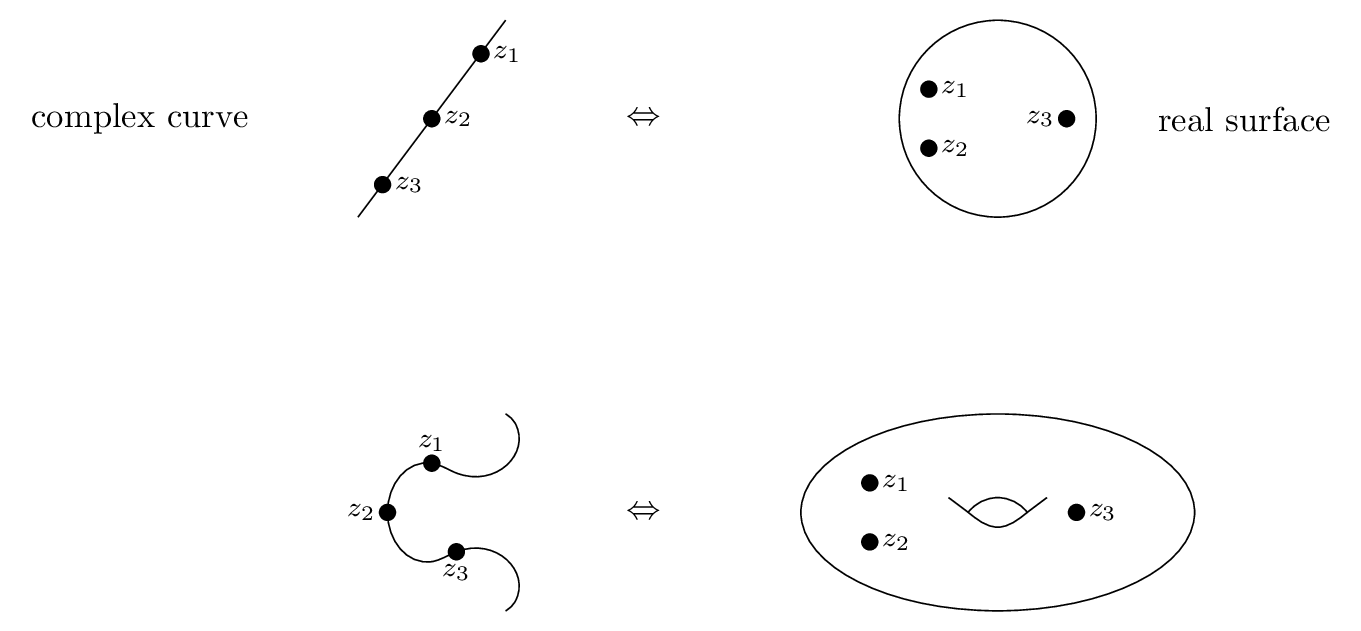}
\end{center}
\caption{
The upper left figure shows a configuration of three marked points on a complex curve of genus zero,
the upper right figure shows the corresponding configuration when the complex curve is viewed as a real
Riemann surface.
The lower figures show the analogous situation for a complex curve of genus one.
}
\label{chapter_elliptics:fig5}
\end{figure}
Note that we may view $C$ either as a complex curve (of complex dimension one) or as a real surface (of real dimension two).
This is illustrated in fig.~\ref{chapter_elliptics:fig5}.

Let's pause for a second and let us convince ourselves that this set-up is a generalisation of the previous case: 
The special case $C={\mathbb C}{\mathbb P}^1$ is obtained for example by the choice
$P(z_1,z_2,z_3)=z_3$. The genus formula gives us that ${\mathbb C}{\mathbb P}^1$ has genus zero.

Let us now consider a smooth curve $C$ of genus $g$ with $n$ marked points.
Two such curves $(C;z_1,\dots,z_n)$ and $(C';z_1',\dots,z_n')$ are isomorphic if there is an isomorphism
\bq
 \phi & : & C \rightarrow C'
 \;\;\;\;\;\;
 \mbox{such that} \;\;
 \phi\left(z_i\right) = z_i'.
\eq
The moduli space
\bq
 {\mathcal M}_{g,n}
\eq
is the space of isomorphism classes of smooth curves of genus $g$ with $n$ marked points.
For $g \ge 1$ the isomorphism classes do not only depend on the positions of the marked points,
but also on the ``shape'' of the curve.
For $g=0$ there is only one ``shape'', the Riemann sphere.
For $g=1$ the shape of the torus is described by the modular parameter $\tau$.

The dimension of ${\mathcal M}_{g,n}$ is
\bq
\label{chapter_elliptics:dim_M_g_n}
 \dim\left({\mathcal M}_{g,n}\right) & = & 3g + n -3,
\eq
for $g=0$ this formula agrees with the previous result in eq.~(\ref{chapter_elliptics:dim_M_0_n}).

Let us now focus on the moduli spaces ${\mathcal M}_{0,n}$ and ${\mathcal M}_{1,n}$ and
work out natural choices for coordinates on ${\mathcal M}_{0,n}$ and ${\mathcal M}_{1,n}$.
\begin{itemize}
\item
We start with genus $0$. We have $\dim {\mathcal M}_{0,n} = n - 3$.
As mentioned above, the sphere has a unique shape. 
We may use M\"obius transformations to fix three points,
say $z_{n-2}=1$, $z_{n-1}=\infty$, $z_n=0$. This leaves 
\bq
 (z_1,\dots,z_{n-3})
\eq
as coordinates on ${\mathcal M}_{0,n}$.
\item
We now turn to genus $1$.
From eq.~(\ref{chapter_elliptics:dim_M_g_n}) we have $\dim {\mathcal M}_{1,n} = n$.
We need one coordinate to describe the shape of the elliptic curve (or the shape of the torus or the shape
of the parallelogram). We may take $\tau$ as defined in eq.~(\ref{chapter_elliptics:def_tau}) for this.
We may use translation transformations to fix one marked point, say $z_n=0$.
This gives
\bq
 (\tau,z_1,\dots,z_{n-1})
\eq
as coordinates on ${\mathcal M}_{1,n}$.
\end{itemize}
We then consider iterated integrals on ${\mathcal M}_{0,n}$ and ${\mathcal M}_{1,n}$.
We recall from section~\ref{chapter_iterated_integrals:section:solution_eps_form}
that iterated integrals on a manifold $M$ are defined by a set $\omega_1$, \dots, $\omega_k$ of differential 1-forms on $M$
and a path $\gamma : [0,1] \rightarrow M$ as
\bq
 I_{\gamma}\left(\omega_1,\dots,\omega_k;\lambda\right)
 & = &
 \int\limits_0^{\lambda} d\lambda_1 f_1\left(\lambda_1\right)
 \int\limits_0^{\lambda_1} d\lambda_2 f_2\left(\lambda_2\right)
 \dots
 \int\limits_0^{\lambda_{k-1}} d\lambda_k f_k\left(\lambda_k\right),
\eq
where the pull-back of $\omega_j$ to the interval $[0,1]$ is denoted by
\bq
 f_j\left(\lambda\right) d\lambda & = & \gamma^\ast \omega_j.
\eq
Let us briefly discuss iterated integrals on ${\mathcal M}_{0,n}$.
We are interested in differential one-forms, which have only simple poles.
Thus we consider
\bq
 \omega & = &
 d\ln\left(z_i-z_j\right)
 \; = \; 
 \frac{dz_i-dz_j}{z_i-z_j}.
\eq
Keeping $z_1, \dots, z_{n-4}$ fixed and integrating along $y=z_{n-3}$ leads to
\bq
 \omega^{\mathrm{mpl}} & = & \frac{dy}{y-z_j}.
\eq
The iterated integrals constructed from these differential one-forms are the multiple polylogarithms:
\bq
G(z_1,\dots,z_k;y) & = & \int\limits_0^y \frac{dy_1}{y_1-z_1}
 \int\limits_0^{y_1} \frac{dy_2}{y_2-z_2} \dots
 \int\limits_0^{y_{k-1}} \frac{dy_k}{y_k-z_k},
 \;\;\;\;\;\;
 z_k \neq 0,
\eq
discussed in detail in section~\ref{chapter_multiple_polylogarithms}.
As discussed in section~\ref{chapter_multiple_polylogarithms} we may relax the condition $z_k\neq 0$
and allow trailing zeros.

Let's now consider iterated integrals on ${\mathcal M}_{1,n}$.
We recall that we may take $(\tau,z_1,\dots,z_{n-1})$ as coordinates on ${\mathcal M}_{1,n}$.
We may decompose an arbitrary integration path into pieces along $d\tau$ (with $z_1=\dots =z_{n-1}=\mathrm{const}$) 
and pieces along the $dz_j$'s (with $\tau=\mathrm{const}$).
Thus we obtain two classes of standardised iterated integrals:
Iterated integrals on ${\mathcal M}_{1,n}$ with integration along $d\tau$ and
iterated integrals on ${\mathcal M}_{1,n}$ with integration along the $dz_j$'s.

In addition we have to specify the differential one-forms we want to integrate.
The differential one-forms which we want to consider in the case of ${\mathcal M}_{1,n}$
are derived from the Kronecker function.
The Kronecker function $F(x,y,\tau)$ is defined in terms of the first Jacobi theta function by
\bq
 F\left(x,y,\tau\right)
 & = &
 \pi
 \theta_1'\left(0,q\right) \frac{\theta_1\left( \pi\left(x+y\right), q \right)}{\theta_1\left( \pi x, q \right)\theta_1\left( \pi y, q \right)},
\eq
where $q=\exp(\pi i \tau)$
and the first Jacobi theta function $\theta_1(z,q)$ is defined in eq.~(\ref{chapter_elliptics:def_Jacobi_theta_functions}).
$\theta_1'$ denotes the derivative with respect to the first argument.
Please note that in order to make contact with the standard notation for the Jacobi
theta functions we used here the nome $q=\exp(\pi i \tau)$ and not the nome squared
$\bar{q}=q^2=\exp(2 \pi i \tau)$. The definition of the Kronecker function is cleaned up
if we define
\bq
 \bar{\theta}_1\left(z,\bar{q}\right) 
 & = &
 \theta_1\left(\pi z,\bar{q}^{\frac{1}{2}} \right).
\eq
Then
\bq
 F\left(x,y,\tau\right)
 & = &
 \bar{\theta}_1'\left(0,\bar{q}\right) \frac{\bar{\theta}_1\left(x+y, \bar{q}\right)}{\bar{\theta}_1\left(x, \bar{q}\right)\bar{\theta}_1\left(y, \bar{q}\right)}.
\eq
\begin{digression}
{\bf Properties of the Kronecker function}:
\\
\\
From the definition it is obvious that
the Kronecker function is symmetric in the variables $(x,y)$:
\bq
 F\left(y,x,\tau\right)
 & = &
 F\left(x,y,\tau\right).
\eq
The (quasi-) periodicity properties are
\bq
 F\left(x+1,y,\tau\right)
 \; = \;
 F\left(x,y,\tau\right),
 & &
 F\left(x+\tau,y,\tau\right)
 \; = \;
 e^{-2\pi i y}
 F\left(x,y,\tau\right).
\eq
The function
\bq
 \Omega\left(x,y,\tau\right) 
 & = &
 \exp\left(2\pi i \frac{\left(y\mathrm{Im}\left(x\right)+x\mathrm{Im}\left(y\right)\right)}{\mathrm{Im}\left(\tau\right)} \right)
 F\left(x,y,\tau\right)
\eq
is symmetric in $x$ and $y$ and doubly periodic
\bq
 \Omega\left(x+1,y,\tau\right) 
 \; = \;
 \Omega\left(x+\tau,y,\tau\right) 
 \; = \;
 \Omega\left(x,y,\tau\right),
\eq
but no longer meromorphic (that is to say that $\Omega$ also depends on $\bar{x}$, $\bar{y}$ and $\bar{\tau}$,
the dependency on the anti-holomorphic variables enters through $2 i \, \mathrm{Im}(x)= x -\bar{x}$).
\\
\\
The Fay identity reads
\bq
\lefteqn{
 F\left(x_1,y_1,\tau\right) F\left(x_2,y_2,\tau\right)
 = } & &
 \\
 & &
 F\left(x_1,y_1+y_2,\tau\right) F\left(x_2-x_1,y_2,\tau\right)
 +
 F\left(x_2,y_1+y_2,\tau\right) F\left(x_1-x_2,y_1,\tau\right).
 \nonumber
\eq
\end{digression}
We recall that the Kronecker function is symmetric in $x$ and $y$.
We are interested in the Laurent expansion in one of these variables. We define functions
$g^{(k)}(z,\tau)$ through
\bq
\label{chapter_elliptics:def_g_n}
 F\left(z,\alpha,\tau\right)
 & = &
 \sum\limits_{k=0}^\infty g^{(k)}\left(z,\tau\right) \alpha^{k-1}.
\eq
We are primarily interested in the coefficients $g^{(k)}(z,\tau)$ of the Kronecker function.
Let us recall some of their properties \cite{Zagier:1991,Brown:2011,Broedel:2018qkq}.
\begin{tcolorbox}[breakable]
{\bf Properties of the coefficients $g^{(k)}(z,\tau)$ of the Kronecker function}:
\begin{enumerate}

\item The functions $g^{(k)}(z,\tau)$ have the symmetry
\bq
 g^{(k)}(-z,\tau)
 & = &
 \left(-1\right)^k g^{(k)}(z,\tau).
\eq

\item When viewed as a function of $z$, the function $g^{(k)}(z,\tau)$ has only simple poles.
More concretely, the function $g^{(1)}(z,\tau)$ has a simple pole with unit residue at every point of the lattice.
For $k>1$ the function $g^{(k)}(z,\tau)$ has a simple pole only at those lattice points 
that do not lie on the real axis.

\item The (quasi-) periodicity properties are
\bq
\label{chapter_elliptics:quasi_periodicity_g_k}
 g^{(k)}\left(z+1,\tau\right) & = &  g^{(k)}\left(z,\tau\right),
 \nonumber \\
 g^{(k)}\left(z+\tau,\tau\right) & = &  
 \sum\limits_{j=0}^k \frac{\left(-2\pi i\right)^j}{j!} g^{(k-j)}\left(z,\tau\right).
\eq
We see that $g^{(k)}(z,\tau)$ is invariant under translations by $1$, but not by $\tau$.
The translation invariance by $\tau$ is only spoiled by the terms with $j \ge 1$ in eq.~(\ref{chapter_elliptics:quasi_periodicity_g_k}).

\item Under modular transformations the functions $g^{(k)}(z,\tau)$ transform as
\bq
\label{chapter_elliptics:modularity_g_k}
 g^{(k)}\left(\frac{z}{c\tau+d},\frac{a\tau+b}{c\tau+d}\right)
 \; = \;
 \left(c\tau +d \right)^k
 \sum\limits_{j=0}^k
 \frac{\left(2\pi i\right)^j}{j!}
 \left( \frac{c z}{c\tau+d} \right)^j
 g^{(k-j)}\left(z,\tau\right).
 \;\;\;\;\;\;\;\;\;\;\;\;
\eq
Modular invariance is only spoiled by the terms with $j \ge 1$ in eq.~(\ref{chapter_elliptics:modularity_g_k}).

\item The $\bar{q}$-expansion of the $g^{(k)}(z,\tau)$ functions is given by 
(with $\bar{q}=\exp(2\pi i\tau)$ and $\bar{w}=\exp(2\pi i z)$)
\bq
\label{chapter_elliptics:qbar_expansion_Kronecker_coefficients}
 g^{(0)}\left(z,\tau\right)
 & = & 1,
 \nonumber \\
 g^{(1)}\left(z,\tau\right)
 & = &
 - 2 \pi i \left[
                  \frac{1+\bar{w}}{2 \left(1-\bar{w}\right)}
                  + \overline{\mathrm{E}}_{0,0}\left(\bar{w};1;\bar{q}\right)
 \right],
 \nonumber \\
 g^{(k)}\left(z,\tau\right)
 & = &
 - \frac{\left(2\pi i\right)^k}{\left(k-1\right)!} 
 \left[
 - \frac{B_k}{k}
       + \overline{\mathrm{E}}_{0,1-k}\left(\bar{w};1;\bar{q}\right)
 \right],
 \;\;\;\;\;\;\;\;\;
 k > 1,
\eq
where $B_k$ denotes the 
\index{Bernoulli number}
$k$-th Bernoulli number, defined in eq.~(\ref{chapter_transformations:def_Bernoulli_number})
and
\bq
 \overline{\mathrm{E}}_{n;m}\left(\bar{u};\bar{v};\bar{q}\right) 
 & = &
  \mathrm{ELi}_{n;m}\left(\bar{u};\bar{v};\bar{q}\right)
  - \left(-1\right)^{n+m} \mathrm{ELi}_{n;m}\left(\bar{u}^{-1};\bar{v}^{-1};\bar{q}\right),
 \nonumber \\
 \mathrm{ELi}_{n;m}\left(\bar{u};\bar{v};\bar{q}\right) & = & 
 \sum\limits_{j=1}^\infty \sum\limits_{k=1}^\infty \; \frac{\bar{u}^j}{j^n} \frac{\bar{v}^k}{k^m} \bar{q}^{j k}.
\eq

\item We may relate $g^{(k)}(z, K \tau)$ (with $K \in {\mathbb N}$)
to functions with argument $\tau$ according to
\bq
\label{chapter_elliptics:Kronecker_g_K_multiple}
 g^{(k)}\left(z, K \tau\right)
 & = &
 \frac{1}{K} \sum\limits_{l=0}^{K-1} 
  g^{(k)}\left(\frac{z+l}{K},\tau\right).
\eq

\end{enumerate}
\end{tcolorbox}
Having defined the functions $g^{(k)}(z,\tau)$, we may now state the differential one-forms
which we would like to integrate on ${\mathcal M}_{1,n}$.
To keep the discussion simple, we focus on ${\mathcal M}_{1,2}$ with coordinates $(\tau,z)$.
(The general case ${\mathcal M}_{1,n}$ is only from a notational perspective more cumbersome.)
We consider
\bq
\label{chapter_elliptics:omega_Kronecker}
 \omega^{\mathrm{Kronecker}}_{k}
 & = &
 \left(2\pi i\right)^{2-k}
 \left[
  g^{(k-1)}\left( z-c_j, \tau\right) d z + \left(k-1\right) g^{(k)}\left( z-c_j, \tau\right) \frac{d\tau}{2\pi i}
 \right],
 \;\;\;
\eq
with $c_j$ being a constant.
The differential one-form $\omega^{\mathrm{Kronecker}}_{k}$ is closed
\bq
 d \omega^{\mathrm{Kronecker}}_{k} & = & 0.
\eq

Let us first consider the integration along $d\tau$ (i.e. $z=\mathrm{const}$).
Here, the part
\bq
\label{chapter_elliptics:omega_Kronecker_dtau}
 \omega^{\mathrm{Kronecker},\tau}_{k}
 & = &
 \left(2\pi i\right)^{2-k}
 \left(k-1\right) g^{(k)}\left( z-c_j, \tau\right) \frac{d\tau}{2\pi i}
 \nonumber \\
 & = &
 \frac{\left(k-1\right)}{\left(2\pi i\right)^{k}} g^{(k)}\left(z-c_j, \tau\right) \frac{d\bar{q}}{\bar{q}}
\eq
is relevant.
This is supplemented by $z$-independent differential one-forms constructed from modular forms:
Let $f_k(\tau) \in \mathcal{M}_k(\Gamma)$ be a modular form of weight $k$ and level $N$
for the congruence subgroup $\Gamma$. Let $N'$ be the smallest positive integer such that $T_{N'} \in \Gamma$.
We set as in eq.~(\ref{chapter_elliptics:def_omega_modular})
\bq
\label{chapter_elliptics:omega_modular}
 \omega^{\mathrm{modular}}_{k}
 & = &
 \left( 2 \pi i \right) f_k\left(\tau\right) \frac{d\tau}{N'}.
\eq
Let us now assume for simplicity $N'=1$ 
(otherwise we should use the variable $\bar{q}_{N'}$ instead of the variable $\bar{q}$
in the formulae below).
Let $\omega_j$ with weight $k_j$ be as 
in eq.~(\ref{chapter_elliptics:omega_Kronecker_dtau}) or as in eq.~(\ref{chapter_elliptics:omega_modular})
and $\gamma$ the path from $\tau=i\infty$ to $\tau$, corresponding in $\bar{q}$-space to a path from $\bar{q}=0$ to $\bar{q}$.
We then consider in $\bar{q}$-space the iterated integrals 
\bq
 I_\gamma\left( \omega_1, \dots, \omega_r; \bar{q} \right).
\eq
The integrands have no poles in $0 < | \bar{q} |< 1$.
A simple pole at $\bar{q}=0$ is possible and allowed.
If $\omega_r$ has a simple pole at $\bar{q}=0$ we say that the iterated integral has a trailing zero.
We may split $\omega_r$ into a part proportional to $d\bar{q}/\bar{q}$ and a regular remainder.
The singular part of a trailing zero can be treated in exactly the same way as we did in the case of multiple polylogarithms.

To summarise:
\begin{tcolorbox}
{\bf Iterated integrals along $d\tau$}:
\\
\\
For the integration along $d\tau$ we consider in $\bar{q}$-space the iterated integrals 
\bq
\label{chapter_elliptics:def_iterated_integral_dtau}
 I_\gamma\left( \omega_1, \dots, \omega_r; \bar{q} \right),
\eq
where $\omega_j$ is of the form
\bq
\label{chapter_elliptics:def_iterated_integral_omega_Kronecker_and_omega_modular}
 \omega^{\mathrm{Kronecker},\tau}_{k_j}
 \; = \;
 \frac{\left(k_j-1\right)}{\left(2\pi i\right)^{k_j}} g^{(k_j)}\left(z-c_j, \tau\right) \frac{d\bar{q}}{\bar{q}}
 & \;\; \mbox{or} \;\; &
 \omega^{\mathrm{modular}}_{k_j}
 \; = \;
 f_{k_j}\left(\tau\right) \frac{d\bar{q}}{\bar{q}},
 \;\;\;\;
\eq
with $f_{k_j}(\tau)$ being a modular form of weight $k_j$.
\end{tcolorbox}

Let us now consider the integration along $dz$ (i.e. $\tau=\mathrm{const}$).
For the integration along $dz$ the part
\bq
\label{chapter_elliptics:omega_Kronecker_dz}
 \omega^{\mathrm{Kronecker},z}_{k}
 & = &
 \left(2\pi i\right)^{2-k}
 g^{(k-1)}\left( z-c_j, \tau\right) d z
\eq
is relevant.
The iterated integrals of the differential one-forms in eq.~(\ref{chapter_elliptics:omega_Kronecker_dz})
along a path $\gamma$ from $z=0$ to $z$ are the elliptic multiple polylogarithms $\widetilde{\Gamma}$,
as defined in ref.~\cite{Broedel:2017kkb}:
\begin{tcolorbox}
\index{elliptic multiple polylogarithms}
{\bf The elliptic multiple polylogarithms $\widetilde{\Gamma}$ (i.e. iterated integrals along $dz$)}:
\bq
\label{chapter_elliptics:Gammatilde}
\lefteqn{
 \widetilde{\Gamma}\!\left({\begin{smallmatrix} n_1 & \dots & n_r \\ c_1 & \dots & c_r \\ \end{smallmatrix}}; z; \tau \right)
 = } & &
 \nonumber \\
 & &
 \left( 2 \pi i\right)^{n_1+\dots+n_r-r}
 I_\gamma\left( \omega^{\mathrm{Kronecker},z}_{n_1+1}\left(c_1,\tau\right), \dots, \omega^{\mathrm{Kronecker},z}_{n_r+1}\left(c_r,\tau\right); z \right).
\eq
\end{tcolorbox}
Let us stress that this is one possibility to define elliptic multiple polylogarithms.
In the literature there exist various definitions of elliptic multiple polylogarithms due to the following
problem:
It is not possible that the differential one-forms $\omega$ entering 
the definition of elliptic multiple polylogarithms
have at the same time the following three properties:
\begin{description}
\item{(i)} $\omega$ is double-periodic,
\item{(ii)} $\omega$ is meromorphic,
\item{(iii)} $\omega$ has only simple poles.
\end{description}
We can only require two of these three properties.
The definition of the $\widetilde{\Gamma}$-functions
selects meromorphicity and simple poles.
Meromorphicity means that $\omega$ does not depend on the anti-holomorphic variables.
The differential one-forms are not double-periodic. 
(This is spoiled by the quasi-periodicity of $g^{(k)}( z, \tau)$ with respect to $z \rightarrow z+\tau$.)
However, this is what physics (i.e. the evaluation of Feynman integrals)
dictates us to choose.
The integrands are then either multi-valued functions on ${\mathcal M}_{1,n}$
or single-valued functions
on a covering space,
in the same way as $\ln(z)$ is a multi-valued function on ${\mathbb C}^\times$ or a single-valued function
on a covering space of ${\mathbb C}^\times$.
Of course, in mathematics one might also consider alternative definitions, which prioritise other properties.
A definition of elliptic multiple polylogarithms, which implements properties (i) and (ii), but gives up property (iii)
can be found in \cite{Levin:2007},
a definition, which implements properties (i) and (iii), but gives up (ii) can be found in \cite{Brown:2011}.
It is a little bit unfortunate that these different function are all named elliptic multiple polylogarithms.
The reader is advised to carefully check what is meant by the name ``elliptic multiple polylogarithm'', this also concerns
the definitions in \cite{Passarino:2017EPJC,Remiddi:2017har}.

It is not unusual in Feynman integral calculations 
that we end up with an integration involving a square root of a quartic polynomial in combination with
integrands known from multiple polylogarithms.
There is a systematic way to convert these integrals to the elliptic multiple polylogarithms
$\widetilde{\Gamma}$ \cite{Broedel:2017kkb,Broedel:2018qkq}.
The square root defines an elliptic curve
\bq
 E
 & : &
 v^2 - \left(u-u_1\right) \left(u-u_2\right) \left(u-u_3\right) \left(u-u_4\right)
 \; = \; 0,
\eq
and we consider as integrands the field of rational functions of the elliptic curve $E$,
i.e. rational functions in $u$ and $v$ subject to the relation $v^2=(u-u_1)(u-u_2)(u-u_3)(u-u_4)$.
Primitives, which are not in this function field originate from the integrands
\bq
 \frac{du}{v},
 \;\;\;
 \frac{u \, du}{v},
 \;\;\;
 \frac{u^2 \, du}{v},
 \;\;\;
 \frac{du}{u-c},
 \;\;\;
 \frac{du}{\left(u-c\right)v},
\eq
where $c$ is a constant. This list includes the integrands $du/(u-c)$ for the multiple polylogarithms.
Note that in the quartic case $u \, du/v$ is a differential of the third kind, a differential
of the second kind can be constructed from a linear combination of $u^2 \, du/v$ and $u \, du/v$:
\bq
\label{chapter_elliptics:differential_second_kind_quartic_case}
 \left( u^2 - \frac{1}{2} s_1 u \right) \frac{du}{v},
 & &
 s_1 = u_1+u_2+u_3+u_4.
\eq
In section~\ref{chapter_elliptics:section_calculations_elliptic_curves} we discussed the general quartic case
and defined a pair of periods (see eq.~(\ref{chapter_elliptics:def_generic_periods}), we use the notation as in section~\ref{chapter_elliptics:section_calculations_elliptic_curves})
\bq
 \psi_1 
 \; = \; 
 2 \int\limits_{u_2}^{u_3} \frac{du}{v}
 \; = \;
 \frac{4 K\left(k\right)}{U_3^{\frac{1}{2}}},
 & &
 \psi_2
 \; = \; 
 2 \int\limits_{u_4}^{u_3} \frac{du}{v}
 \; = \; 
 \frac{4 i K\left(\bar{k}\right)}{U_3^{\frac{1}{2}}}.
\eq
It is convenient to normalise one period to one, hence we define $\tau=\psi_2/\psi_1$ and we consider
the lattice $\Lambda$ generated by $(1,\tau)$.
Abel's map
\bq
 z & = & \frac{1}{\psi_1} \int\limits_{u_1}^u \frac{du}{v}
\eq
relates a point $(u,v)$ on the elliptic curve to a point $z \in {\mathbb C}/\Lambda$.
We have
\bq
\label{chapter_elliptics:quartic_case_weight_1}
 \frac{2\pi i}{\psi_1} \frac{du}{v} 
 & = & 2 \pi i \; dz 
 \; = \;\omega^{\mathrm{Kronecker},z}_{1}.
\eq
Let us also discuss the case of modular weight $k=2$.
This includes the integrands for the multiple polylogarithms $du/(u-c)$.
These integrands have a pole at $u=c$ and a pole at $u=\infty$.
We define the images of $u=c$ and $u=\infty$ under Abel's map by
\bq
 z_c \; = \; \frac{1}{\psi_1} \int\limits_{u_1}^c \frac{du}{v},
 & &
 z_\infty \; = \; \frac{1}{\psi_1} \int\limits_{u_1}^\infty \frac{du}{v}.
\eq
We then have
\bq
\label{chapter_elliptics:quartic_case_weight_2_case_1}
 \frac{du}{u-c}
 & = &
 \left[ 
        g^{(1)}\left(z-z_c,\tau\right)
      + g^{(1)}\left(z+z_c,\tau\right)
      - g^{(1)}\left(z-z_\infty,\tau\right)
      - g^{(1)}\left(z+z_\infty,\tau\right) \right] dz.
 \;\;\;\;\;\;
\eq
The integrand $du/((u-c)v)$ translates to
\bq
\label{chapter_elliptics:quartic_case_weight_2_case_2}
 \frac{v_c du}{\left(u-c\right)v}
 & = &
 \left[ 
        g^{(1)}\left(z-z_c,\tau\right)
      - g^{(1)}\left(z+z_c,\tau\right)
      + g^{(1)}\left(z_c-z_\infty,\tau\right)
      + g^{(1)}\left(z_c+z_\infty,\tau\right) \right] dz.
 \;\;\;\;\;\;\;\;\;\;\;\;\;\;\;
\eq
where $v_c^2=(c-u_1)(c-u_2)(c-u_3)(c-u_4)$.
For the integrand $u\,du/v$ one finds
\bq
\label{chapter_elliptics:quartic_case_weight_2_case_3}
 \frac{\left(u-u_1\right) \, du}{v}
 & = &
 \left[ 
      g^{(1)}\left(z+z_\infty,\tau\right)
      - g^{(1)}\left(z-z_\infty,\tau\right)
      - 2g^{(1)}\left(z_\infty,\tau\right)
 \right] dz.
\eq
The term $u_1 du/v$ can be translated with formula~(\ref{chapter_elliptics:quartic_case_weight_1}).
It remains to treat the integrand $u^2\,du/v$ or equivalently the linear combination appearing
in eq.~(\ref{chapter_elliptics:differential_second_kind_quartic_case}).
This integrand has a double pole at $u=\infty$.
We would like to have integrands with only simple poles.
This can be enforced by introducing a primitive.
To illustrate this with a simple example consider
\bq
 \frac{dx}{x^2} & = & - d\left(\frac{1}{x}\right),
\eq
where $1/x$ has only a simple pole.
In order to enforce simple poles, we therefore introduce 
\bq
 Z_4
 & = &
 - 
 \int\limits_{u_1}^u \frac{du}{v} \left[ u^2 - \frac{1}{2} s_1 u + \frac{1}{2}\left(u_1u_2+u_3u_4\right) - \frac{1}{2} U_3 \frac{\phi_1}{\psi_1} \right],
\eq
where $\phi_1$ is the quasi-period defined in eq.~(\ref{chapter_elliptics:def_generic_periods}).
The first two terms of the integrand are proportional to
eq.~(\ref{chapter_elliptics:differential_second_kind_quartic_case}), 
the remainder of the integrand is proportional to the holomorphic one-form $du/v$.
It can be shown that $Z_4$ has as a function of $u$ only simple poles.
One then finds
\bq
\label{chapter_elliptics:quartic_case_weight_2_case_4}
 \frac{Z_4 \, du}{v}
 & = &
 \left[ 
        g^{(1)}\left(z-z_\infty,\tau\right)
      + g^{(1)}\left(z+z_\infty,\tau\right)
 \right] dz.
\eq
However, enforcing simple poles with the introduction of $Z_4$ has a price: We now have to consider
higher powers of $Z_4$ as well. This will give rise to an infinite tower of integrands, corresponding
to modular weight $k>2$.
For the details we refer to \cite{Broedel:2017kkb,Broedel:2018qkq}.
Eqs.~(\ref{chapter_elliptics:quartic_case_weight_2_case_1}),
(\ref{chapter_elliptics:quartic_case_weight_2_case_2}), 
(\ref{chapter_elliptics:quartic_case_weight_2_case_3}) and 
(\ref{chapter_elliptics:quartic_case_weight_2_case_4})
are the complete set of formulae at modular weight $2$.

\section{Elliptic Feynman integrals}

With the background in mathematics on elliptic curves, modular forms and moduli spaces we are now
in a position to tackle the first Feynman integrals, which cannot be expressed in terms of multiple polylogarithms.

We do this in two steps: We start with elliptic Feynman integrals, which depend only on one kinematic variable $x$.
The essential trick is to change from the variable $x$ to a new variable, the modular parameter $\tau$.
We will find that the Feynman integrals can be expressed as iterated integrals on ${\mathcal M}_{1,1}$
(i.e. iterated integrals of modular forms).

In a second step we generalise to elliptic Feynman integrals which depend on 
several kinematic variables $x_1, x_2, \dots$.
This will lead us to iterated integrals on ${\mathcal M}_{1,n}$.

\subsection{Feynman integrals depending on one kinematic variable}
\label{chapter_elliptics:feynman_integrals_one_kinematic_variable}

%
%
In section~\ref{chapter_iterated_integrals:deriving_the_dgl}
we introduced already the two-loop sunrise integral with equal internal masses:
\bq
\label{chapter_elliptics:def_equal_mass_sunrise}
 I_{\nu_1 \nu_2 \nu_3}\left(D,x\right)
 & = &
 e^{2 \eps \Eulerconstant} \left(m^2\right)^{\nu_{123}-D}
 \int \frac{d^Dk_1}{i \pi^{\frac{D}{2}}} \frac{d^Dk_2}{i \pi^{\frac{D}{2}}} 
 \frac{1}{\left(-q_1^2+m^2\right)^{\nu_1} \left(-q_2^2+m^2\right)^{\nu_2} \left(-q_3^2+m^2\right)^{\nu_3}},
 \nonumber \\
\eq
with $x=-p^2/m^2$ and $q_1=k_1$, $q_2=k_2-k_1$, $q_3=-k_2-p$.
We have set $\mu^2=m^2$.
This is the simplest example of an elliptic Feynman integral.
It has been studied intensively in the literature \cite{Sabry:1962,Broadhurst:1993mw,Laporta:2004rb,Bloch:2013tra,Adams:2015ydq,Adams:2017ejb,Broedel:2017siw,Adams:2018yfj,Honemann:2018mrb}.

\begin{figure}
\begin{center}
\includegraphics[align=c,scale=1.0]{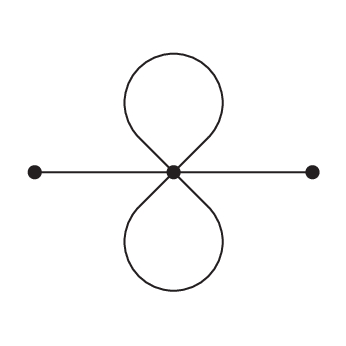}
\hspace*{10mm}
\includegraphics[align=c,scale=1.0]{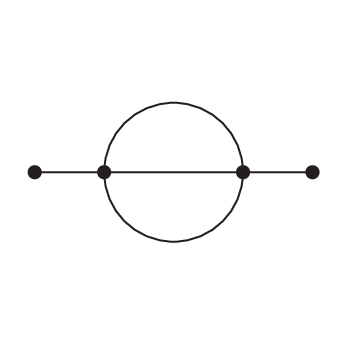}
\hspace*{10mm}
\includegraphics[align=c,scale=1.0]{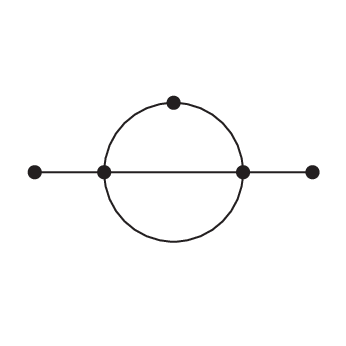}
\end{center}
\caption{
The three master integrals for the family of the equal mass sunrise integral.
}
\label{chapter_elliptics:fig_masters_equal_sunrise}
\end{figure}

There are three master integrals
and we start from the basis 
\bq
 \vec{I} & = & \left( I_{110}, I_{111}, I_{211} \right)^T.
\eq
The three master integrals are shown in fig.~\ref{chapter_elliptics:fig_masters_equal_sunrise}.
The differential equation for this basis has been given in eq.~(\ref{chapter_iterated_integrals:dgl_equal_mass_sunrise}).

It is simpler to analyse this system in $D=2-2\eps$ dimensions. 
This is no restriction: With the help of the dimensional shift relations we may always relate integrals
in $(2-2\eps)$ dimensions to integrals in $(4-2\eps)$ dimensions.
We have for example
\bq
\lefteqn{
 I_{110}\left(2-2\eps,x\right)
 = 
 I_{220}\left(4-2\eps,x\right),
 } & & 
\nonumber \\
\lefteqn{
 I_{111}\left(2-2\eps,x\right)
 = 
 \frac{3}{\left(x+1\right)\left(x+9\right)}
 \left[
  \left(3+x\right) I_{220}\left(4-2\eps,x\right)
 \right.
} & & \nonumber \\
 & &
 \left.
  +
  2 \left(1-2\eps\right) \left(2-3\eps\right) I_{111}\left(4-2\eps,x\right)
  +
 2 \left(1-2\eps\right) \left(3-x\right) I_{211}\left(4-2\eps,x\right)
 \right],
 \nonumber \\
\lefteqn{
 I_{211}\left(2-2\eps,x\right)
 = 
} & & \nonumber \\
 & & 
 \frac{1}{\left(x+1\right)^2\left(x+9\right)^2}
 \left\{
  \left[ 3 \left(x+1\right) \left(x+9\right) + \eps \left(2x^3+34x^2+54x+54\right) \right] I_{220}\left(4-2\eps,x\right)
 \right. \nonumber \\
 & & \left.
  + 2 \left(1-2\eps\right) \left(2-3\eps\right) \left[ \left(x+1\right) \left(x+9\right) - 2 \eps \left(x-3\right)\left(x+3\right) \right] I_{111}\left(4-2\eps,x\right)
 \right. \nonumber \\
 & & \left.
 + 2 \left(1-2\eps\right) \left[ 3 \left(x+1\right)\left(x+9\right) - \eps \left(x^3+36x^2+45x-54\right) \right]  I_{211}\left(4-2\eps,x\right)
 \right\}.
\eq
The equal mass sunrise integral is the simplest Feynman integral related to an elliptic curve.
The first question which we should address is how to obtain the elliptic curve associated to this integral.
For the sunrise integral there are two possibilities, 
we may either obtain an elliptic curve from the Feynman graph polynomial or from the maximal cut.
The sunrise integral has three propagators, hence we need three Feynman parameters, which we denote by
$a_1,a_2,a_3$.
The Feynman parameter representation for $I_{111}$ reads
\bq
 I_{111}\left(2-2\eps,x\right)
 & = &
 e^{2 \eps \Eulerconstant}\Gamma\left(1+2\eps\right)
 \int\limits_{a_j \ge 0} d^{3}a \; \delta\left(1-\sum\limits_{j=1}^{3} a_j \right) \; 
 \frac{\left[{\mathcal U}\left(a\right)\right]^{3\eps}}{\left[ {\mathcal F}\left(a\right) \right]^{1+2\eps}},
 \nonumber \\
 {\mathcal U}\left(a\right)
 & = & 
 a_1 a_2 + a_2 a_3 + a_3 a_1,
 \nonumber \\
 {\mathcal F}\left(a\right)
 & = &
 a_1 a_2 a_3 x + \left( a_1 + a_2 + a_3 \right) \left( a_1 a_2 + a_2 a_3 + a_3 a_1 \right).
\eq
The second graph polynomial defines an elliptic curve
\bq
 E^{\mathrm{Feynman}}
 & : &
 a_1 a_2 a_3 x + \left( a_1 + a_2 + a_3 \right) \left( a_1 a_2 + a_2 a_3 + a_3 a_1 \right) 
 \; = \; 0,
\eq
in $\mathbb{CP}^2$, with $[a_1:a_2:a_3]$ being the homogeneous coordinates of $\mathbb{CP}^2$.
The elliptic curve varies with the kinematic variable $x$.
In general, the Feynman parameter space can be viewed as $\mathbb{CP}^{n-1}$, with $n$ being the number of propagators
of the Feynman integral.
It is clear that this approach does not generalise in a straightforward way to other elliptic Feynman integrals
with more than three propagators. (For an elliptic curve we want the zero set of a single polynomial in $\mathbb{CP}^2$).

We therefore turn to the second method of obtaining the elliptic curve, which generalises easily:
We study the maximal cut of the sunrise integral.
Within the loop-by-loop approach
\bq
\label{chapter_elliptics:maxcut}
 \mathrm{MaxCut} \; I_{111}\left(2-2\eps,x\right)
 & = &
 \frac{\left(2\pi i\right)^3}{\pi^2}
 \int\limits_{{\mathcal C}_{\mathrm{MaxCut}}} 
 \frac{dz}{\sqrt{z \left(z + 4 \right) \left[z^2 + 2 \left(1-x\right) z + \left(1+x\right)^2 \right]}}
 + {\mathcal O}\left(\eps\right),
 \nonumber \\
\eq
Thus we obtain an elliptic curve
as a quartic polynomial $P(u,v)=0$:
\bq
\label{chapter_elliptics:def_elliptic_curve_maximal_cut}
 E^{\mathrm{cut}}
 & : &
 v^2 - u
       \left(u + 4 \right) 
       \left[u^2 + 2 \left(1-x\right) u + \left(1+x\right)^2 \right]
 \; = \; 0.
\eq
Also this elliptic curve varies with the kinematic variable $x$.
Please note that these two elliptic curves $E^{\mathrm{Feynman}}$ and $E^{\mathrm{cut}}$ are not isomorphic,
but only isogenic.
For the sunrise integral we may work with either one of the two. In the following we will use $E^{\mathrm{cut}}$.

Let us therefore consider $E^{\mathrm{cut}}$, defined by the quartic polynomial in eq.~(\ref{chapter_elliptics:def_elliptic_curve_maximal_cut}).
We denote the roots by
\bq
 u_1 \; = \; -4,
 \;\;\;
 u_2 \; = \; -\left(1+\sqrt{-x}\right)^2,
 \;\;\;
 u_3 \; = \; -\left(1-\sqrt{-x}\right)^2,
 \;\;\;
 u_4 \; = \; 0.
\eq
We may then determine two independent periods $\psi_1$ and $\psi_2$ as in section~\ref{chapter_elliptics:section_calculations_elliptic_curves}.
Therefore, let $\psi_1$ and $\psi_2$ be defined by eq.~(\ref{chapter_elliptics:def_generic_periods}).
We set $\tau = \psi_2/\psi_1$
and we denote the Wronskian by
\bq
\label{chapter_elliptics:def_Wronskian_example}
 W & = & \psi_{1} \frac{d}{dx} \psi_{2} - \psi_{2} \frac{d}{dx} \psi_{1}
 \; = \;
 - \frac{6 \pi i}{x\left(x+1\right)\left(x+9\right)}.
\eq
We then perform a change of the basis of the master integrals from the pre-canonical basis $(I_{110},I_{111},I_{211})$ to
\bq
\label{chapter_elliptics:def_basis}
 J_1
 & = &
 4 \eps^2 \; I_{110}\left(2-2\eps,x\right),
 \nonumber \\
 J_2
 & = &
 \eps^2 \frac{\pi}{\psi_1} \; I_{111}\left(2-2\eps,x\right),
 \nonumber \\
 J_3
 & = &
 \frac{1}{\eps} \frac{\psi_1^2}{2 \pi i W} \frac{d}{dx} J_2 
 + \frac{\psi_1^2}{2 \pi i W} \frac{\left(3x^2+10x-9\right)}{2x\left(x+1\right)\left(x+9\right)} J_2.
\eq
This transformation is not rational or algebraic in $x$, as can be seen from the prefactor $1/\psi_1$ in the definition of $J_2$.
The period $\psi_1$ is a transcendental function of $x$.

The fibre transformation in eq.~(\ref{chapter_elliptics:def_basis}) can be understood and motivated as follows:
The definition of $J_1$ is straightforward and follows from eq.~(\ref{chapter_basics:result_tadpole}).
As far as $J_2$ is concerned, we first note that $\psi_1$ and $\psi_2$ are obtained by integrating the holomorphic
one-form $du/v$ of the elliptic curve along two cycles $\gamma_1$ and $\gamma_2$, respectively (see also fig.~\ref{chapter_elliptics:fig2}).
If we replace in eq.~(\ref{chapter_elliptics:maxcut})
${\mathcal C}_{\mathrm{MaxCut}}$ by $\gamma_1$ we obtain
\bq
\label{chapter_elliptics:def_P11}
 \frac{\left(2\pi i\right)^3}{\pi^2}
 \int\limits_{\gamma_1} 
 \frac{dz}{\sqrt{z \left(z + 4 \right) \left[z^2 + 2 \left(1-x\right) z + \left(1+x\right)^2 \right]}}
 & = &
 - 8 i \pi^2
 \frac{\psi_1}{\pi}.
\eq
$J_2$ is obtained by dividing $I_{111}$ by $\psi_1$ and by adjusting powers of $\pi$ and $\eps$.
Prefactors consisting of algebraic numbers (for example $(-8i)$) are not relevant.

Let us turn to $J_3$: It is well-known in mathematics, that the first cohomology group for
a family of elliptic curves $E_x$, parametrised by $x$, is generated by the holomorphic one-form $du/v$ and its $x$-derivative.
This motivates an ansatz, consisting of $J_2$ and its $\tau$-derivative:
\bq
 J_3 & = & c_2\left(x\right) \frac{1}{2\pi i} \frac{d}{d\tau} J_2 + c_3\left(x\right) J_2,
\eq
with unknown functions $c_2(x)$ and $c_3(x)$.
We determine $c_2(x)$ and $c_3(x)$ such that this ansatz transforms the differential equation into an $\eps$-form.
One finds
\bq
 c_2\left(x\right) \; = \; \frac{1}{\eps}
 & \mbox{and} &
 c_3\left(x\right)
 \; = \;
 \frac{1}{24} \left(3x^2+10x-9\right) \frac{\psi_1^2}{\pi^2},
\eq
and therefore
\bq
 J_3
 & = &
 \frac{1}{\eps} \frac{1}{2\pi i} \frac{d}{d\tau} J_2
 +
 \frac{1}{24} \left(3x^2+10x-9\right) \frac{\psi_1^2}{\pi^2} J_2.
\eq
This agrees with eq.~(\ref{chapter_elliptics:def_basis}):
From eq.~(\ref{chapter_elliptics:derivative_Wronskian}) it follows that
\bq
\label{chapter_elliptics:sunrise_jacobian}
 \frac{1}{2\pi i} \frac{d}{d\tau}
 & = &
 \frac{\psi_1^2}{2 \pi i W} \frac{d}{dx}.
\eq
In the basis $J=(J_1,J_2,J_3)^T$ we have now
\bq
 \left( \frac{d}{dx} + A_x \right) \vec{J} & = & 0
\eq
with
\bq
\label{chapter_elliptics:sunrise_A_x}
 A_x
 & = &
 \eps 
 \left( \begin{array}{rrr}
 0 & 0 & 0 \\
 0 & \frac{\left(3x^2+10 x - 9 \right)}{2 x \left(x+1\right) \left(x+9\right)} & -\frac{2 \pi i W}{\psi_1^2} \\
 \frac{3i}{4} \frac{\psi_{1}}{W} \frac{1}{x \left(x+1\right)\left(x+9\right)} 
 & -\frac{i}{288} \frac{W \psi_{1}^2}{\pi^3} \; \left(3-x\right)^4  
 & \frac{\left(3x^2+10 x - 9 \right)}{2 x \left(x+1\right) \left(x+9\right)} \\
 \end{array} \right).
\eq
Through the fibre transformation we have managed that the dimensional regularisation parameter $\eps$ 
is factored out, however the matrix $A_x$ is not yet in a particular nice and suitable form.
The situation is similar to eq.~(\ref{chapter_iterated_integrals:example_1_eps_form_square_root_singularity}), where
a fibre transformation allowed us to factor out $\eps$, but left us with a square root singularity.
In eq.~(\ref{chapter_elliptics:sunrise_A_x}) we have a transcendental function (i.e. $\psi_1$) appearing in the matrix $A_x$.

In order to put the matrix $A$ into a nice form, we perform a base transformation and change the variable from $x$ to $\tau$.
With the help of eq.~(\ref{chapter_elliptics:sunrise_jacobian}) we obtain
\bq
 \left( \frac{1}{2\pi i} \frac{d}{d\tau} + A_{\tau} \right) \vec{J} & = & 0
\eq
with
\bq
\label{chapter_elliptics:sunrise_A_tau}
 A_{\tau}
 & = &
 \eps
 \left( \begin{array}{rrr}
 0 & 0 & 0 \\
 0 & \eta_2 & \eta_0 \\
 \eta_3 & \eta_4 & \eta_2 \\
 \end{array} \right)
\eq
and
\bq
\label{chapter_elliptics:integration_kernels}
 \eta_{0}
 & = & 
 -1,
 \nonumber \\
 \eta_{2} 
 & = &
 \frac{1}{24} \ \frac{\psi_1^2}{\pi^2} \ \left(3x^2+10 x - 9 \right),
 \nonumber \\
 \eta_{3} 
 & = &
 - \frac{1}{96} \ \frac{\psi_1^3}{\pi^3} \ x \left(x+1\right) \left(x+9\right),
 \nonumber \\
 \eta_{4}
 & = &
 - \frac{1}{576} \ \frac{\psi_{1}^4}{\pi^4} \ \left(3-x\right)^4.
\eq
It remains to express $\eta_2$, $\eta_3$ and $\eta_4$ as a function of $\tau$.
To this aim we first express $x$ as a function of $\tau$.
Noting that upon the substitution $x \rightarrow (-x)$
the elliptic curve in eq.~(\ref{chapter_elliptics:def_elliptic_curve_maximal_cut})
is identical to the one discussed in eq.~(\ref{chapter_elliptics:def_example_elliptic_curve}) 
it follows that
\bq
 x
 & = &
 - 9
 \frac{\eta\left(\tau\right)^4 \eta\left(6\tau\right)^8}
      {\eta\left(3\tau\right)^4 \eta\left(2\tau\right)^8}.
\eq
We may then obtain the $\bar{q}$-expansions of $\eta_2$, $\eta_3$ and $\eta_4$.
(For the complete elliptic integral $K$ appearing in $\psi_1$ we use eq.~(\ref{chapter_elliptics:Jacobi_theta_relation_1}).)
For example, for $\eta_2$ we obtain
\bq
 \eta_2 & = &
- \frac{1}{2} - 8 \bar{q} - 4 \bar{q}^2 - 44 \bar{q}^3 + 4 \bar{q}^4 - 48 \bar{q}^5 - 40 \bar{q}^6
 +{\mathcal O}\left(\bar{q}^7\right).
\eq
By comparing the $\bar{q}$-expansions 
one checks that $\eta_0$, $\eta_2$, $\eta_3$ and $\eta_4$ are modular forms of $\Gamma_1(6)$ of modular weight $0$, $2$, $3$ and $4$, respectively.
In order to get an explicit expression we introduce a basis $\{b_1,b_2\}$ for the modular forms of modular weight $1$ 
for the Eisenstein subspace ${\mathcal E}_1(\Gamma_1(6))$ as follows:
We define two primitive Dirichlet characters $\chi_1$ and $\chi_{(-3)}$ with conductors $1$ and $3$, respectively,
through the Kronecker symbol 
\bq
 \chi_1
 \; = \; 
 \left( \frac{1}{n} \right), 
 & &
 \chi_{(-3)} 
 \; = \; 
 \left( \frac{-3}{n} \right).
\eq
Explicitly we have
\bq
 \chi_1\left(n\right) 
 & = & 
 1, 
 \;\;\;\;\;\; 
 \forall n \in {\mathbb Z},
 \nonumber \\
 \chi_{-3}\left(n\right) 
 & = & 
 \left\{ \begin{array}{rl}
  0, & n = 0 \mod 3, \\
  1, & n = 1 \mod 3, \\
 -1, & n = 2 \mod 3, \\
 \end{array} \right.
\eq
We then set
\bq
\label{chapter_elliptics:def_e1_e2}
 b_1 & = & E_1\left(\tau;\chi_1,\chi_{(-3)}\right)
 \; = \; 
 E_{1,1,-3,1}\left(\tau\right),
 \nonumber \\
 b_2 & = & E_1\left(2\tau;\chi_1,\chi_{(-3)}\right)
 \; = \; E_{1,1,-3,2}\left(\tau\right).
\eq
The generalised Eisenstein series $E_k(\tau,\chi_a,\chi_b)$ have been defined in eq.~(\ref{chapter_elliptics:def_Eisenstein_E}), the generalised Eisenstein series $E_{k,a,b,K}(\tau)$ have been defined in eq.~(\ref{chapter_elliptics:Eisenstein_short_hand_notation}).
The integration kernels may be expressed as polynomials in $b_1$ and $b_2$:
\bq
 \eta_2
 & = &
 -6 \left( b_1^2 + 6 b_1 b_2 - 4 b_2^2 \right),
 \nonumber \\
 \eta_3
 & = &
 - 9 \sqrt{3} \left( b_1^3 - b_1^2 b_2 - 4 b_1 b_2^2 + 4b_2^3 \right),
 \nonumber \\
 \eta_4 
 & = &
 - 324 b_1^4.
\eq
Let us summarise: Through a fibre transformation and a base transformation we obtained the differential equation
\bq
\label{chapter_elliptics:sunrise_final_v0}
\left( d + A \right) \vec{J} & = & 0
\eq
with
\bq
\label{chapter_elliptics:sunrise_final}
 A & = &
 2 \pi i \; \eps
 \left( \begin{array}{ccc}
 0 & 0 & 0 \\
 0 & \eta_2\left(\tau\right) & \eta_0\left(\tau\right) \\
 \eta_3\left(\tau\right) & \eta_4\left(\tau\right) & \eta_2\left(\tau\right) \\
 \end{array} \right)
 d\tau,
\eq
where $\eta_k(\tau)$ denotes a modular form of modular weight $k$ for $\Gamma_1(6)$.
The differential equation for the equal mass sunrise system is now in $\eps$-form and the kinematic variable
matches the standard coordinate on ${\mathcal M}_{1,1}$.
With the additional information of a boundary value, the differential equation is now easily solved order by order
in $\eps$ in terms of iterated integrals of modular forms.
One finds for example
\bq
\label{chapter_elliptics:sunrise_final_qbar_expansion}
 J_2 & = &
 \left[ 3 \, \mathrm{Cl}_2\left(\frac{2\pi}{3}\right) 
 + 
 4 I\left(\eta_0,\eta_3;\tau\right)
 \right] \eps^2
 + {\mathcal O}\left(\eps^3\right).
\eq
The Clausen value $\mathrm{Cl}_2(2\pi/3)$ comes from the boundary value.
The first few terms of the $\bar{q}$-expansion read
\bq
 J_2^{(2)}
 & = &
 3 \, \mathrm{Cl}_2\left(\frac{2\pi}{3}\right)
 - 3 \sqrt{3}
   \left[
         \bar{q}-\frac{5}{4}\,\bar{q}^{2}+\bar{q}^{3}-{\frac {11}{16}}\,\bar{q}^{4}+{\frac {24}{25}}\,\bar{q}^{5}
         -\frac{5}{4}\,\bar{q}^{6}+{\frac {50}{49}}\,\bar{q}^{7}-{\frac {53}{64}}\,\bar{q}^{8}+\bar{q}^{9}
   \right]
 \nonumber \\
 & &
 + {\mathcal O}\left(\bar{q}^{10}\right).
\eq

\subsubsection{Analytic continuation}

In the calculation above we solved the differential equation in a neighbourhood of $x=0$ (corresponding to $\bar{q}=0$)
and the result is valid in a neighbourhood of $x=0$.
We are interested in extending this result to all $x \in {\mathbb R}-i\delta$, where the infinitesimal imaginary part originates
from Feynman's $i\delta$-prescription.
We have to ensure that the periods $\psi_1$ and $\psi_2$ vary smoothly, as $x$ varies in ${\mathbb R}-i\delta$ \cite{Bogner:2017vim}.

Let's look at the details:
The expressions for the the modulus $k$ and the complementary modulus $k'$ of the elliptic curve $E^{\mathrm{cut}}$ are
\bq
 k^2 
 \; = \;
 \frac{16 \sqrt{-x}}{\left(1+\sqrt{-x}\right)^3 \left(3-\sqrt{-x}\right)},
 & &
 k'{}^2
 \; = \;
 \frac{\left(1-\sqrt{-x}\right)^3 \left(3+\sqrt{-x}\right)}{\left(1+\sqrt{-x}\right)^3 \left(3-\sqrt{-x}\right)},
\eq
where Feynman's $i\delta$-prescription ($x\rightarrow x-i\delta$) is understood.
We define the periods $\psi_1$ and $\psi_2$ for all $x \in {\mathbb R}-i\delta$ by
\bq
\label{chapter_elliptics:def_periods_choice_0}
 \left( \begin{array}{c}
  \psi_2 \\
  \psi_1 \\
 \end{array} \right)
 & = &
 \frac{4}{\left(1+\sqrt{-x}\right)^{\frac{3}{2}} \left(3-\sqrt{-x}\right)^{\frac{1}{2}}}
 \; \gamma \;
 \left( \begin{array}{c}
  i K\left( k' \right) \\
  K\left( k \right) \\
 \end{array} \right),
\eq
where $K(x)$ denotes the complete elliptic integral of the first kind.
The essential new ingredient is the 
$2 \times 2$-matrix $\gamma$ given by
\bq
\label{chapter_elliptics:def_periods_choice_0_gamma}
 \gamma
 & = &
 \left\{ \begin{array}{ccrcccl}
  \left( \begin{array}{rr}
   1 & 0 \\
   2 & 1 \\
  \end{array} \right),
  & & -\infty & < & x & < & -1.
  \\
  \left( \begin{array}{rr}
   1 & 0 \\
   0 & 1 \\
  \end{array} \right),
  & & -1 & < & x & < & -3 + 2 \sqrt{3}, 
  \\
  \left( \begin{array}{rr}
   1 & 0 \\
   2 & 1 \\
  \end{array} \right),
  & & -3 + 2 \sqrt{3} & < & x & < & \infty, 
  \\
 \end{array} \right.
\eq
The matrix $\gamma$ ensures that the periods $\psi_1$ and $\psi_2$ vary smoothly as the kinematic variable $x$ varies smoothly
in $x \in {\mathbb R} - i \delta$ \cite{Bogner:2017vim}.
The complete elliptic integral $K(k)$ can be viewed as a function of $k^2$: 
We set $\tilde{K}(k^2)=K(k)$.
The function $\tilde{K}(k^2)$ has a branch cut at $[1,\infty[$ in the complex $k^2$-plane.
The matrix $\gamma$ compensates for the discontinuity when we cross this branch cut.
It is relatively easy to see that $k^2$ as a function of $x$ crosses this branch cut at the point
$x=-3+2\sqrt{3} \approx 0.46$, the corresponding value in the $k^2$-plane is 
$k^2=2$.
The point $x=-1$ is a little bit more subtle. Let us parametrise a small path around $x=-1$ by
\bq
 x\left(\phi\right) & = & -1 - \delta e^{i \phi},
 \;\;\;\;\;\;
 \phi \in [0,\pi],
\eq
then
\bq
 k^2
 & = &
 1 + \frac{1}{32} \delta^3 e^{3 i \phi}
 + {\mathcal O}\left(\delta^4\right),
\eq
and the path in $k^2$-space winds around the point $k^2=-1$ by an angle $3\pi$ as 
the path in $x$-space winds around the point $x=1$ by the angle $\pi$.

Eq.~(\ref{chapter_elliptics:def_periods_choice_0}) defines the periods $\psi_1$ and $\psi_2$
for all values $x \in {\mathbb R} - i \delta$.
The periods take values in ${\mathbb C} \cup \left\{ \infty \right\}$.

The original differential equation (eq.~(\ref{chapter_iterated_integrals:dgl_equal_mass_sunrise}))
has singular points at
\bq
 x & \in & \left\{ -9, -1, 0, \infty \right\}.
\eq
The point $x=-9$ is called the threshold, the point $x=-1$ is called the pseudo-threshold.
The singular points of the differential equation are mapped in $\tau$-space to
\bq
 \tau\left(x=-9\right) \; = \; \frac{1}{3},
 \;\;\;
 \tau\left(x=-1\right) \; = \; 0,
 \;\;\;
 \tau\left(x=0\right) \; = \; i \infty,
 \;\;\;
 \tau\left(x=\infty\right) \; = \; \frac{1}{2}.
\eq
The points 
\bq
 \left\{ 0, \frac{1}{3}, \frac{1}{2}, i \infty \right\}
\eq
in $\tau$-space are the cusps of $\Gamma_1(6)$.

\begin{figure}
\begin{center}
\includegraphics[scale=1.0]{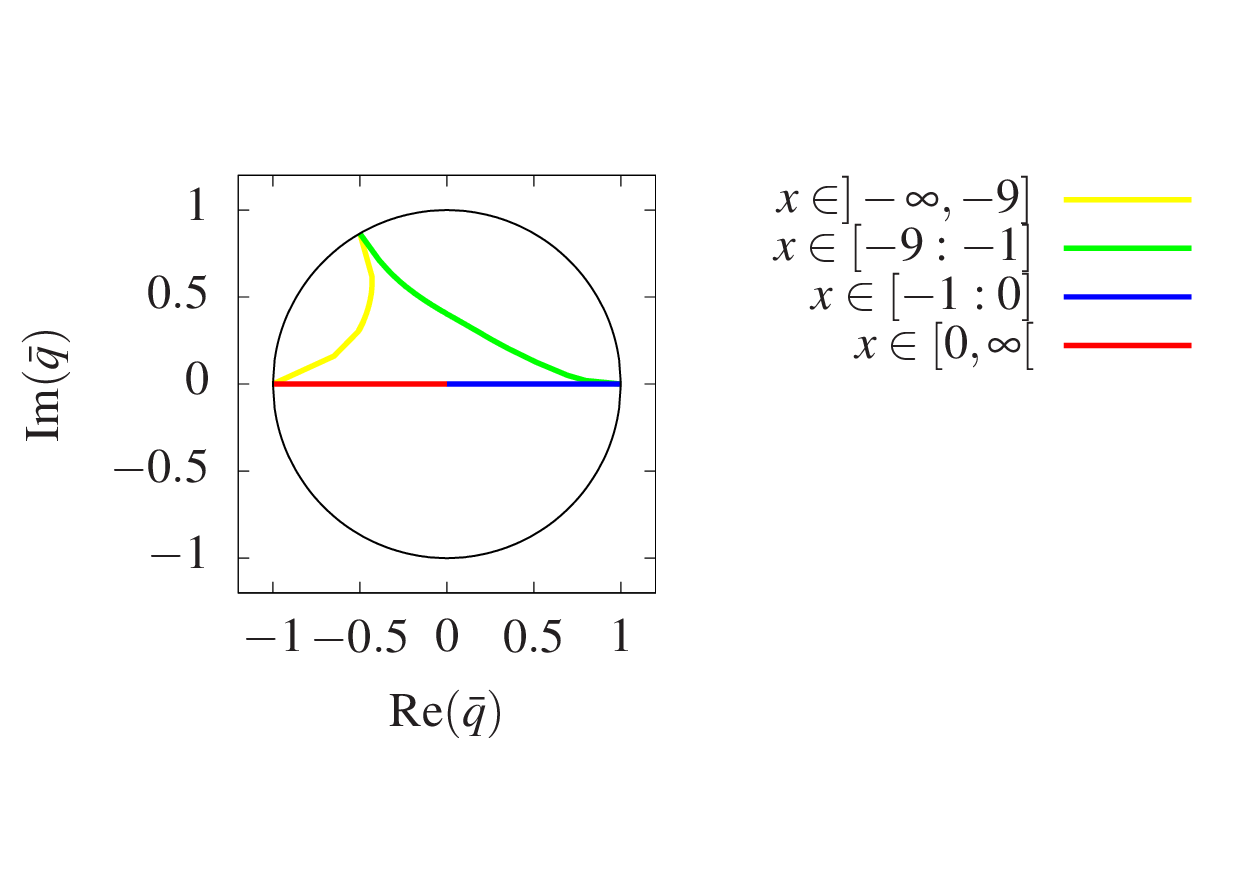}
\end{center}
\caption{\label{chapter_elliptics:fig_qbar_path}
The path in $\bar{q}$-space as $x$ ranges over ${\mathbb R}$.
We always have $|\bar{q}| \le 1$ and $|\bar{q}| = 1$ only at $x \in \{-9,-1,\infty\}$. 
}
\end{figure}

In fig.~\ref{chapter_elliptics:fig_qbar_path} we plot the values of the variable $\bar{q}$ as $x$ ranges over
${\mathbb R}$.
We see that all values of $\bar{q}$ are inside the unit disc 
with the exception of the three points $x \in \{-9,-1,\infty\}$, where the corresponding $\bar{q}$-values
are on the boundary of the unit disc.
Once the periods $\psi_1$ and $\psi_2$ are defined
as in eq.~(\ref{chapter_elliptics:def_periods_choice_0}), 
the $\bar{q}$-expansion in eq.~(\ref{chapter_elliptics:sunrise_final_qbar_expansion}) 
gives the correct result for $x \in {\mathbb R} \backslash \{-9,-1,\infty\}$.
Note that although the $\bar{q}$-expansion corresponds to an expansion around $x=0$, it gives also the correct result
for 
\bq
 x \; \in \; ]-\infty,-9[ & \mbox{and} & x \; \in \; ]-9,-1[.
\eq
These intervals correspond to the yellow and green segments in fig.~\ref{chapter_elliptics:fig_qbar_path}.
The $\bar{q}$-expansion in eq.~(\ref{chapter_elliptics:sunrise_final_qbar_expansion}) does not converge for
$x \in \{-9,-1,\infty\}$ and in a neighbourhood of these points the convergence is slow.
In the next subsection we will see how to improve the convergence in a neighbourhood of these points.

\subsubsection{Modular transformations}

In this paragraph we address two questions:
In the calculation of the equal mass two-loop sunrise integral we started with a specific choice of periods $\psi_1$ and $\psi_2$.
Our first question is: What happens if we make a different choice $\psi_1'$ and $\psi_2'$?

We have seen that we may express the equal mass two-loop sunrise integral as iterated integrals of modular forms.
We may evaluate these integrals through their $\bar{q}$-expansion. The convergence of $\bar{q}$-series is fast,
if $|\bar{q}| \ll 1$, however it is rather slow if $|\bar{q}| \lesssim 1$.
The sunrise integral depends only on one kinematic variable, which we take as the modular parameter $\tau$.
By a modular transformation $\tau' = \gamma(\tau)$ with $\gamma \in \mathrm{SL}_2({\mathbb Z})$ we may transform $\tau'$
into the fundamental domain ${\mathcal F}$ 
\begin{figure}
\begin{center}
\includegraphics[align=c,scale=1.4]{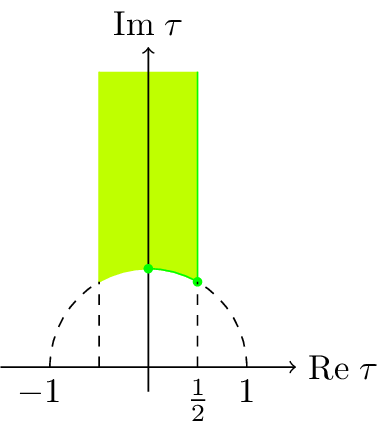}
\end{center}
\caption{
The fundamental domain ${\mathcal F}$ for $\tau$.
}
\label{chapter_elliptics:fig_fundamental_domain}
\end{figure}
shown in fig.~\ref{chapter_elliptics:fig_fundamental_domain} and defined by
\bq
\lefteqn{
 {\mathcal F} 
 = } & & \\
 & &
 \left\{ \tau' \in {\mathbb H} \left| \left| \tau' \right| > 1 \;\mbox{and}\;  -\frac{1}{2} < \mathrm{Re}\left(\tau'\right) \le \frac{1}{2} \right. \right\}
 \cup
 \left\{ \tau' \in {\mathbb H} \left| \left| \tau' \right| = 1 \;\mbox{and}\; 0 \le \mathrm{Re}\left(\tau'\right) \le \frac{1}{2} \right. \right\}.
 \nonumber
\eq
Such a transformation
achieves that
\bq
 \left| \bar{q}' \right|
 & \le &
 e^{- \pi \sqrt{3}}
 \; \approx \; 0.0043.
\eq
This is a small expansion parameter.
Our second question is: How do elliptic Feynman integrals transform under modular transformations?

The two questions are of course related: A modular transformation corresponds exactly to the transformation
(see eq.~(\ref{chapter_elliptics:transformation_periods}))
\bq
 \left( \begin{array}{c}
 \psi_2' \\
 \psi_1' \\
 \end{array} \right)
 & = &
 \gamma
 \left( \begin{array}{c}
 \psi_2 \\
 \psi_1 \\
 \end{array} \right).
\eq
There are two complications we have to take into account:
First of all, in order to transform $\tau$ into the fundamental domain ${\mathcal F}$ we need a
$\gamma \in \mathrm{SL}_2({\mathbb Z})$. Already in the simplest example of an elliptic Feynman integral we have seen
that the modular forms appearing in the solution belong to a congruence subgroup $\Gamma$ (e.g. $\Gamma_1(6)$ in the case of the
sunrise integral).
For $f \in {\mathcal M}_k(\Gamma)$ and $\gamma \in \mathrm{SL}_2({\mathbb Z})$ we have in general
\bq
 f \slashoperator{\gamma}{k} & \notin & {\mathcal M}_k(\Gamma).
\eq
The second complication is as follows: Suppose our Feynman integral is expressed as iterated integrals of modular forms
for a congruence subgroup $\Gamma$ and -- in order to avoid the first complication --
suppose that $\gamma \in \Gamma$.
By changing variables from $\tau$ to $\tau'=\gamma(\tau)$ 
we leave the class of functions we started with and generate new integrands
with additional appearances of $\ln(\bar{q})$.
To see this in more detail, let's look at an example:
Let $f$ be a modular form of weight $k$ for $\Gamma$.
For simplicity we assume that $f$ vanishes at the cusp $\tau=i\infty$.
Let $N'$ be the smallest positive integer such that $T_{N'} \in \Gamma$.
The modular form has then the Fourier expansion
\bq
 f
 & = &
 \sum\limits_{n=1}^\infty a_n \tilde{q}_{N'}^n.
\eq
Let $\gamma \in \Gamma$.
We then have
\bq
 I\left(f;\tau\right)
 & = &
 \frac{2\pi i}{N'}
 \int\limits_{i \infty}^{\tau} f\left(\tilde{\tau}\right) d\tilde{\tau}
 \; = \;
 \sum\limits_{n=1}^\infty 
 \int\limits_{0}^{\bar{q}} a_n \tilde{q}_{N'}^n \frac{d\tilde{q}_{N'}}{\tilde{q}_{N'}}
 \; = \;
 \sum\limits_{n=1}^\infty 
  \frac{a_{n}}{n}
  \bar{q}_{N'}^{n}.
\eq
Let us now consider a coordinate transformation
\bq
 \tau & = & \gamma^{-1}\left(\tau'\right) \; = \;
 \frac{a\tau'+b}{c\tau'+d},
 \;\;\;\;\;\;\;\;\;\;\;\;
 \gamma^{-1} \; \in \; \Gamma.
\eq
It is simpler to consider the inverse transformation here.
We have
\bq
 I\left(f;\tau\right)
 & = &
 \frac{2\pi i}{N'}
 \int\limits_{i \infty}^{\tau} f\left(\tilde{\tau}\right) d\tilde{\tau}
 \; = \;
 \frac{2\pi i}{N'}
 \int\limits_{\gamma\left(i \infty\right)}^{\gamma\left(\tau\right)} 
  f\left(\gamma^{-1}\left(\tilde{\tau}'\right)\right) \frac{d\tilde{\tau}'}{\left(c\tilde{\tau}'+d\right)^2}
 \nonumber \\
 & = &
 \frac{2\pi i}{N'}
 \int\limits_{\gamma\left(i \infty\right)}^{\gamma\left(\tau\right)} 
 \left(c\tilde{\tau}'+d\right)^{k-2} 
 (f \slashoperator{\gamma^{-1}}{k})(\tilde{\tau}')
 \; d\tilde{\tau}'.
\eq
As we only consider $\gamma \in \Gamma$,
the expression $(f \slashoperator{\gamma^{-1}}{k})(\tilde{\tau}')$ is again a modular form for $\Gamma$, this is fine.
However, we picked up a factor $(c\tilde{\tau}'+d)^{k-2}$.
Only for the modular weight $k=2$ this factor is absent.
In general we leave the class of integrands constructed purely from modular forms
and end up with integrands which contain in addition powers of the automorphic factor $(c\tau'+d)$.

The two complications are solved as follows \cite{Weinzierl:2020fyx}: 
For the first complication we note that if $f \in {\mathcal M}_k(\Gamma)$ is of level $N$ we have
$f \in {\mathcal M}_k(\Gamma(N))$. For any $\gamma \in \mathrm{SL}_2({\mathbb Z})$ we then have
(see eq.~(\ref{chapter_elliptics:modular_form_covariance_Gamma_N}))
\bq
 f \slashoperator{\gamma}{k}
 & \in & 
 {\mathcal M}_k\left(\Gamma\left(N\right)\right).
\eq
For the second complication we note
that we may view the transformation $\tau'=\gamma(\tau)$ as a base transformation.
In order to stay within the class of iterated integrals of modular forms we should accompany
the base transformation by a fibre transformation.

Let us illustrate the two aspects with an example.
We consider again the system of the equal mass sunrise integral with the master integrals
$\vec{J}=(J_1,J_2,J_3)^T$ defined in eq.~(\ref{chapter_elliptics:def_basis})
and the differential equation (see eqs.~(\ref{chapter_elliptics:sunrise_final_v0})-(\ref{chapter_elliptics:sunrise_final}))
\bq
\left( d + A \right) J \; = \; 0,
 & &
 A \; = \;
 2 \pi i \; \eps
 \left( \begin{array}{ccc}
 0 & 0 & 0 \\
 0 & \eta_2\left(\tau\right) & \eta_0\left(\tau\right) \\
 \eta_3\left(\tau\right) & \eta_4\left(\tau\right) & \eta_2\left(\tau\right) \\
 \end{array} \right)
 d\tau.
\eq
For this particular example, the $\eta_k(\tau)$'s are modular forms of $\Gamma_1(6)$ 
and therefore also modular forms of $\Gamma(6)$.
Let us now consider for
\bq
 \gamma\left(\tau\right)
 & = & 
 \frac{a\tau+b}{c\tau+d},
 \;\;\;\;\;\;\;\;\;\;\;\;
 \gamma \; \in \; \mathrm{SL}_2({\mathbb Z})
\eq
the combined transformation
\bq
\label{chapter_elliptics:combined_trafo_sunrise}
 J' 
 & = &
 \left( \begin{array}{ccc}
 1 & 0 & 0 \\
 0 & \left(c\tau+d\right)^{-1} & 0 \\
 0 & \frac{c}{2\pi i \eps \eta_0} & \left(c\tau+d\right) \\
 \end{array} \right) J,
 \nonumber \\
 \tau'
 & = &
 \frac{a\tau+b}{c\tau+d}.
\eq
Working out the transformed differential equation we obtain
\bq
\left( d + A' \right) J' & = & 0
\eq
with
\bq
 A' & = &
 2 \pi i \; \eps
 \left( \begin{array}{ccc}
 0 & 0 & 0 \\
 0 & (\eta_2 \slashoperator{\gamma^{-1}}{2})(\tau') & (\eta_0 \slashoperator{\gamma^{-1}}{0})(\tau') \\
 (\eta_3 \slashoperator{\gamma^{-1}}{3})(\tau') & (\eta_4 \slashoperator{\gamma^{-1}}{4})(\tau') & (\eta_2 \slashoperator{\gamma^{-1}}{2})(\tau') \\
 \end{array} \right)
 d\tau'.
\eq
We have
\bq
 \eta_k \slashoperator{\gamma^{-1}}{k}
 & \in & 
 \mathcal{M}_k(\Gamma(6))
\eq
and therefore we don't leave the space of modular forms with the combined transformation of eq.~(\ref{chapter_elliptics:combined_trafo_sunrise}).
The transformed system may therefore again be solved 
for any $\gamma \in \mathrm{SL}_2({\mathbb Z})$
in terms of iterated integrals of modular forms.
In particular, we achieved that terms with additional automorphic factors do not occur.

The fact that we need to redefine the master integrals is not too surprising. Let's look at $J_2$.
We originally defined $J_2$ by
\bq
 J_2
 & = &
 \eps^2 \frac{\pi}{\psi_1} \; I_{111}\left(2-2\eps,x\right),
\eq
i.e. we rescaled $I_{111}$ (up to a constant) by $1/\psi_1$.
This definition is tied to our initial choice of periods.
Noting that the automorphic factor $(c\tau+d)$ is nothing than the ratio of two periods
\bq
 c \tau + d & = & 
 \frac{\psi_1'}{\psi_1},
\eq
we find that $J_2'$ is given by
\bq
 J_2'
 & = &
 \eps^2 \frac{\pi}{\psi_1'} \; I_{111}\left(2-2\eps,x\right).
\eq

\subsubsection{The maximal cut and the period matrix}

We have seen that the basis $J_1$, $J_2$ and $J_3$ defined in eq.~(\ref{chapter_elliptics:def_basis})
puts the differential equation for the system into an $\eps$-form.
In section~\ref{chapter_transformations:maximal_cuts_and_constant_leading_singularities} we introduced
a technique, which guesses a basis of master integrals from a set of master integrands $\varphi_i$ and
a set of master contours ${\mathcal C}_j$.
The various master integrands integrated against the various master contours define a period matrix $P$ with
entries
\bq
 P_{ij} & = & \left\langle \varphi_i | {\mathcal C}_j \right\rangle.
\eq
In the $i$-th row of this matrix we then look at the term of order $j_{\min}$ in the $\eps$-expansion
(with $j_{\min}$ defined as in eq.~(\ref{chapter_transformations:def_j_min})).
This defines a matrix $P^{\mathrm{leading}}$ with entries
\bq
\label{chapter_elliptics:def_P_leading}
 P^{\mathrm{leading}}_{ij}
 & = & 
 \mathrm{coeff}\left( \left\langle \varphi_i | {\mathcal C}_j \right\rangle, \eps^{j_{\min}} \right) \cdot \eps^{j_{\min}}  
\eq
Our strategy in section~\ref{chapter_transformations:maximal_cuts_and_constant_leading_singularities}
was to look for integrands $\varphi_i$ such that all entries of $P^{\mathrm{leading}}$ are constants of weight zero.
We already know that this is not a sufficient condition, as we may always multiply $\varphi_i$ by an
$\eps$-dependent prefactor with leading term $1$.
We now show that it is neither a necessary condition.
(But the condition is helpful in practice.)

We will focus on the maximal cut, hence only the integrals $J_2$ and $J_3$ are relevant.
We work with the loop-by-loop Baikov representations, where we have four integration variables $z_1-z_4$.
Let ${\mathcal C}'$ be the integration domain selecting the maximal cut, 
i.e. a small counter-clockwise circle around $z_1=0$, 
a small counter-clockwise circle around $z_2=0$ and
a small counter-clockwise circle around $z_3=0$.
We set $z_4=z$ in accordance with the notation used in eq.~(\ref{chapter_elliptics:def_P11}).
We denote by $\gamma_1$ and $\gamma_2$ the two cycles of the elliptic curve. They define the integration domain
in the variable $z$.
We define
\bq
 {\mathcal C}_2 \; = \; {\mathcal C}' \cup \gamma_1,
 & &
 {\mathcal C}_3 \; = \; {\mathcal C}' \cup \gamma_2.
\eq
We denote the integrands of $J_2$ and $J_3$ by $\varphi_2$ and $\varphi_3$, respectively.
Let us look at the period matrix
\bq
 P
 & = &
 \left( \begin{array}{cc}
  \left\langle \varphi_2 | {\mathcal C}_2 \right\rangle & \left\langle \varphi_2 | {\mathcal C}_3 \right\rangle \\
  \left\langle \varphi_3 | {\mathcal C}_2 \right\rangle & \left\langle \varphi_3 | {\mathcal C}_3 \right\rangle \\
 \end{array} \right)
\eq
and the matrix $P^{\mathrm{leading}}$ defined as in eq.~(\ref{chapter_elliptics:def_P_leading}).
We compute the entries of $P^{\mathrm{leading}}$. To this aim we first express $J_2$ and $J_3$ 
as a linear combination of $I_{111}$ and $I_{211}$.
One finds
\bq
 J_2
 & = &
 \eps^2\frac{\pi}{\psi_1} I_{111},
 \nonumber \\
 J_3 
 & = &
 \frac{\eps^2}{24} \left(7x^2+50x+27\right) \frac{\psi_1}{\pi} I_{111}
 + \frac{\eps}{12} \left(x+1\right)\left(x+9\right) \left( \frac{\psi_1}{\pi} + x \frac{d}{dx} \frac{\psi_1}{\pi} \right) I_{111}
 \nonumber \\
 & &
 - \frac{\eps}{4} \left(x+1\right)\left(x+9\right) \frac{\psi_1}{\pi} I_{211}.
\eq
The derivative $d\psi_1/dx$ may be expressed with the help of eq.~(\ref{chapter_elliptics:dgl_periods}) as a linear combination
of $\psi_1$ and $\phi_1$:
\bq
 \frac{d}{dx} \psi_1
 & = &
 - \frac{1}{2} \psi_1 \frac{d}{dx} \ln U_2 + \frac{1}{2} \phi_1 \frac{d}{dx} \ln \frac{U_2}{U_1},
 \\
 & &
 U_1 \; = \; 16 \sqrt{-x},
 \;\;\;
 U_2 \; = \; \left(1-\sqrt{-x}\right)^3 \left(3+\sqrt{-x}\right).
 \nonumber 
\eq
Let $\varphi_{111}$ and $\varphi_{211}$ denote the integrands of $I_{111}$ and $I_{211}$, respectively.
Replacing ${\mathcal C}_{\mathrm{MaxCut}}$ by $\gamma_1$ or $\gamma_2$ in eq.~(\ref{chapter_elliptics:maxcut})
we immediately have
\bq
 \left\langle \varphi_{111} | {\mathcal C}_2 \right\rangle
 \; = \; - 8 i \pi \psi_1 + {\mathcal O}\left(\eps\right),
 & &
 \left\langle \varphi_{111} | {\mathcal C}_3 \right\rangle
 \; = \; - 8 i \pi \psi_2 + {\mathcal O}\left(\eps\right),
\eq
For $\varphi_{211}$ the following two integrals are helpful
($v=\sqrt{(u-u_1)(u-u_2)(u-u_3)(u-u_4)}$):
\bq
 2 \int\limits_{u_2}^{u_3} \frac{du}{\left(u-u_1\right) v}
 & = &
 \frac{1}{u_2-u_1} \psi_1 - \frac{u_4-u_2}{\left(u_2-u_1\right)\left(u_4-u_1\right)} \phi_1,
 \nonumber \\
 2 \int\limits_{u_4}^{u_3} \frac{du}{\left(u-u_1\right) v}
 & = &
 \frac{1}{u_2-u_1} \psi_2 - \frac{u_4-u_2}{\left(u_2-u_1\right)\left(u_4-u_1\right)} \phi_2.
\eq
One finds
\bq
 P^{\mathrm{leading}}
 & = &
 2 i
 \left( \begin{array}{cc}
 \left(2\pi i \eps\right)^2 & \left(2\pi i \eps\right)^2 \tau \\
 0 & - \left(2\pi i \eps\right) \\ 
 \end{array} \right).
\eq
We see that the second entry in the first row $P^{\mathrm{leading}}_{1 2}$ is not constant, this entry depends on the kinematic variable $\tau$.

\subsubsection{Further examples}

There are more Feynman integrals depending on one kinematic variable $x=-p^2/m^2$ and evaluating 
to iterated integrals of modular forms.
\begin{figure}
\begin{center}
\includegraphics[align=c,scale=1.0]{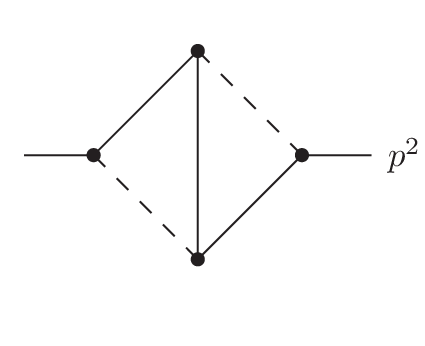}
\includegraphics[align=c,scale=1.0]{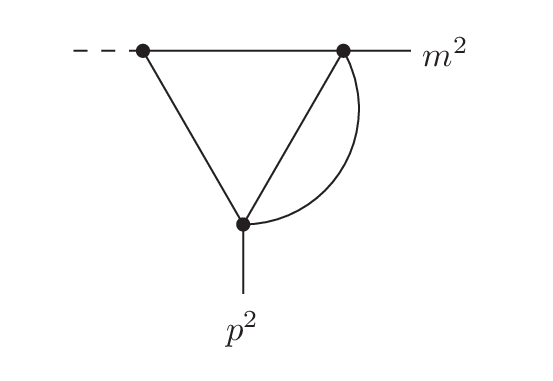}
\includegraphics[align=c,scale=1.0]{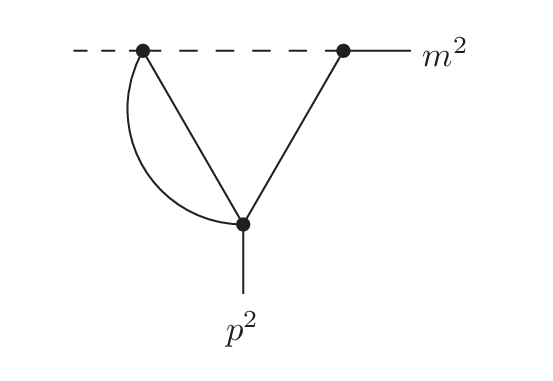}
\end{center}
\caption{
Further examples of Feynman integrals evaluating to iterated integrals of modular forms.
Internal solid lines correspond to a propagator with mass $m^2$, internal dashed lines to a massless 
propagator. External dashed lines indicate a light-like external momentum.
}
\label{chapter_elliptics:fig_further_examples_M_1_1}
\end{figure}
Fig.~\ref{chapter_elliptics:fig_further_examples_M_1_1} shows some additional examples \cite{Sabry:1962,Remiddi:2016gno,Adams:2016xah,Adams:2018bsn,Adams:2018kez,Bezuglov:2020ywm}.
In computing these integrals, we encounter two additional integration kernels
\bq
\label{chapter_elliptics:polylog_kernels}
 \omega_0 \; = \; \frac{dx}{x},
 \;\;\;\;\;\;
 \omega_1 \; = \; \frac{dx}{x+1}.
\eq
If we only would have those integration kernels in the differential equation, we would be able
to express the result in terms of multiple polylogarithms.
However, we have the ones appearing in eq.~(\ref{chapter_elliptics:sunrise_final})
combined with the ones from eq.~(\ref{chapter_elliptics:polylog_kernels}).
We therefore have to rewrite all integration kernels in terms of the variable $\tau$.
We may express $\omega_0$ and $\omega_1$ in terms of modular forms:
\bq
 \omega_0 \; = \; 2 \pi i \; \eta_{2,0} \; d\tau,
 \;\;\;\;\;\;
 \omega_1 \; = \; 2 \pi i \; \eta_{2,1} \; d\tau.
\eq
The modular forms $\eta_{2,0}$ and $\eta_{2,1}$, both of modular weight $2$, are given by
\bq
 \eta_{2,0} 
 & = &
 \frac{1}{2 i \pi} \frac{\psi_{1}^2}{W} \frac{1}{x}
 \; =  \;
 - 12 \left( b_1^2 - 4 b_2^2 \right),
 \nonumber \\
 \eta_{2,1} 
 & = &
 \frac{1}{2 i \pi} \frac{\psi_{1}^2}{W} \frac{1}{x+1}
 \; = \;
 - 18 \left( b_1^2 + b_1 b_2 - 2 b_2^2 \right),
\eq
where $b_1$ and $b_2$ are the two Eisenstein series defined in eq.~(\ref{chapter_elliptics:def_e1_e2}).

\subsection{Feynman integrals depending on several kinematic variables}
\label{chapter_elliptics:feynman_integrals_several_kinematic_variables}

Let us now turn our attention to elliptic Feynman integrals, which depend on more than one kinematic variable
(e.g. $\NB>1$).
In the case of just one kinematic variable ($\NB=1$) we may think of the base space as 
a covering space of ${\mathcal M}_{1,1}$
with coordinate $\tau$.
In the case of more kinematic variables this generalises to a covering space of ${\mathcal M}_{1,n}$
with coordinates $(\tau,z_1,\dots,z_{n-1})$.

In section~\ref{chapter_elliptics:section_moduli_spaces} we introduced in eq.~(\ref{chapter_elliptics:omega_Kronecker})
the differential one-form $\omega^{\mathrm{Kronecker}}_{k}$.
We can be a little bit more general than eq.~(\ref{chapter_elliptics:omega_Kronecker}):
Let $K \in {\mathbb N}$ and $L(z)$ a linear function of $z_1, \dots, z_{n-1}$:
\bq
\label{chapter_elliptics:linear_function}
 L\left(z\right)
 & = &
 \sum\limits_{j=1}^{n-1} \alpha_j z_j + \beta.
\eq
The generalisation of eq.~(\ref{chapter_elliptics:omega_Kronecker}) which we would like to consider is 
\bq
\label{chapter_elliptics:omega_kronecker_general}
 \omega_k\left( L\left(z\right), K \tau\right)
 = 
 \left(2\pi i\right)^{2-k}
 \left[
  g^{(k-1)}\left( L\left(z\right), K \tau\right) d L\left(z\right)
  + K \left(k-1\right) g^{(k)}\left( L\left(z\right), K \tau\right) \frac{d\tau}{2\pi i}
 \right].
 \;\;\;
\eq
The differential one-form $\omega_k(L(z), K \tau)$ is closed
\bq
 d \omega_k\left(L\left(z\right), K \tau\right) & = & 0.
\eq
We may always reduce the case $K>1$ to the case $K=1$ with help of (compare with eq.~(\ref{chapter_elliptics:Kronecker_g_K_multiple}))
\bq
\label{chapter_elliptics:omega_reduce_K}
 \omega_k\left(L\left(z\right),K \tau\right)
 & = &
 \sum\limits_{l=0}^{K-1} \omega_k\left(\frac{L\left(z\right)+l}{K},\tau\right).
\eq
It is therefore sufficient to focus on the case $K=1$.

Let us study the case of elliptic Feynman integrals depending on several kinematic variables 
with a concrete example.
We don't have to go very far, we may generalise the equal mass sunrise integral
to the unequal mass sunrise integral.
We now take the three masses squared $m_1^2$, $m_2^2$ and $m_3^2$ in the propagators
to be pairwise distinct.
We consider
\bq
\label{chapter_elliptics:def_unequal_mass_sunrise}
\lefteqn{
 I_{\nu_1 \nu_2 \nu_3}\left(D,x,y_1,y_2\right)
 = } & &
 \nonumber \\
 & &
 e^{2 \eps \Eulerconstant} \left(m_3^2\right)^{\nu_{123}-D}
 \int \frac{d^Dk_1}{i \pi^{\frac{D}{2}}} \frac{d^Dk_2}{i \pi^{\frac{D}{2}}} 
 \frac{1}{\left(-q_1^2+m_1^2\right)^{\nu_1} \left(-q_2^2+m_2^2\right)^{\nu_2} \left(-q_3^2+m_3^2\right)^{\nu_3}},
 \nonumber \\
\eq
with 
$x = -p^2/m_3^2$,
$y_1 = m_1^2/m_3^2$,
$y_2 = m_2^2/m_3^2$.
and as before $q_1=k_1$, $q_2=k_2-k_1$, $q_3=-k_2-p$.
We have set $\mu^2=m_3^2$.
Also this integral has been studied intensively in the literature \cite{Berends:1993ee,Caffo:1998du,MullerStach:2011ru,Adams:2013nia,Remiddi:2013joa,Adams:2014vja,Adams:2015gva,Bloch:2016izu,Broedel:2017siw,Bogner:2019lfa}.

There are now seven master integrals
and we may start from the basis 
\bq
 \vec{I} & = & \left( I_{110}, I_{101}, I_{011}, I_{111}, I_{211}, I_{121}, I_{112} \right)^T.
\eq
In mathematical terms we are looking at a rank $7$ vector bundle over ${\mathcal M}_{1,3}$.

Finding the elliptic curve proceeds exactly in the same way as discussed in the equal mass case.
The second graph polynomial defines an elliptic curve
\bq
 E^{\mathrm{Feynman}}
 & : &
 a_1 a_2 a_3 x + \left( a_1 y_1 + a_2 y_2 + a_3 \right) \left( a_1 a_2 + a_2 a_3 + a_3 a_1 \right) 
 \; = \; 0,
\eq
in $\mathbb{CP}^2$, with $[a_1:a_2:a_3]$ being the homogeneous coordinates of $\mathbb{CP}^2$.

Alternatively, the loop-by-loop approach for the maximal cut gives
\bq
\label{chapter_elliptics:def_elliptic_curve_maximal_cut_unequal}
 E^{\mathrm{cut}}
 & : &
 v^2 - \left[u^2 + 2 \left(y_1+y_2\right) u  + \left(y_1-y_2\right)^2 \right] 
       \left[u^2 + 2 \left(1-x\right) u + \left(1+x\right)^2 \right]
 \; = \; 0.
 \;\;\;\;\;\;
\eq
As in the equal mass case the two elliptic curves $E^{\mathrm{Feynman}}$ and $E^{\mathrm{cut}}$ are not isomorphic,
but only isogenic.
We may work with either of the two curves.
In the following we will use $E^{\mathrm{cut}}$.

In the next step we would like to change the kinematic variables from $(x,y_1,y_2)$ to the standard
coordinates $(\tau,z_1,z_2)$ on ${\mathcal M}_{1,3}$.
This raises the question: How to express the new coordinates in terms of the old coordinates and vice versa?
For $\tau$ the answer is straightforward: $\tau$ is again the ratio of the two periods 
\bq
\label{chapter_elliptics:def_tau_unequal_sunrise}
 \tau & = & \frac{\psi_2}{\psi_1},
\eq
and $\psi_1$ and $\psi_2$ are functions of $x$, $y_1$ and $y_2$, given by eq.~(\ref{chapter_elliptics:def_generic_periods}).

Also for $z_1$ and $z_2$ there is a simple geometric interpretation:
In the Feynman parameter representation there are two geometric objects of interest:
the domain of integration $\Delta$ (the simplex $a_1, a_2, a_3 \ge 0$, $a_1+a_2+a_3\le 1$)
and the elliptic curve $E^{\mathrm{Feynman}}$ (the zero set of the second graph polynomial).
\begin{figure}
\begin{center}
\includegraphics[scale=0.9]{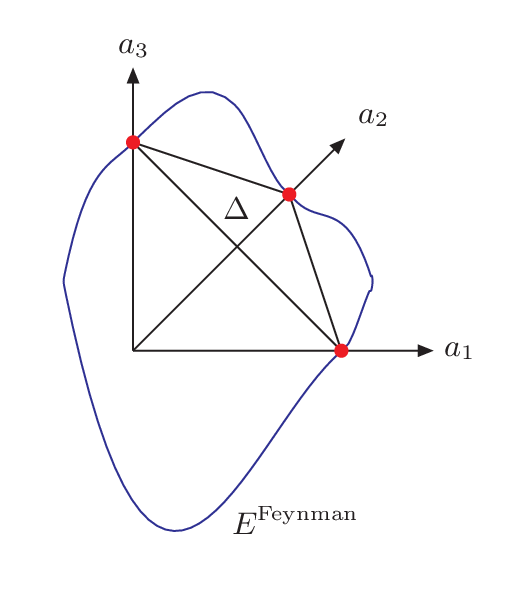}
\includegraphics[scale=0.9]{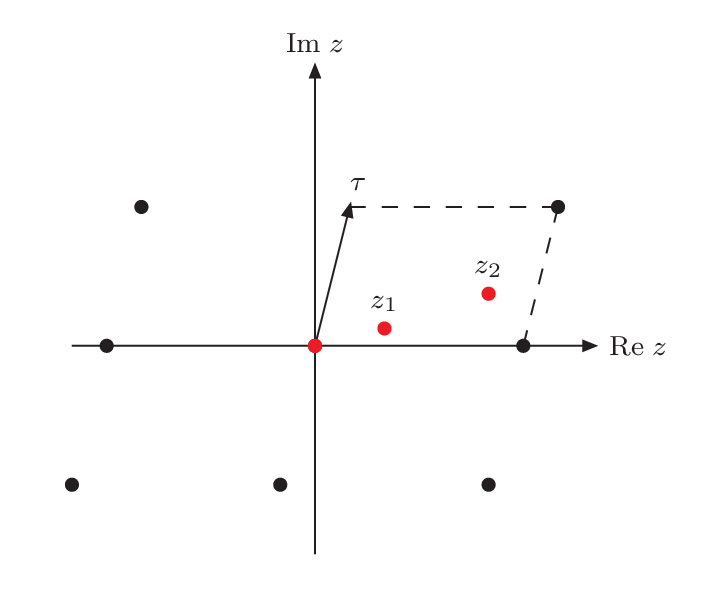}
\end{center}
\caption{
$E^{\mathrm{Feynman}}$ and $\Delta$ intersect at three points, the images of these three points in ${\mathbb C}/\Lambda$ are $0, z_1, z_2$.
}
\label{chapter_elliptics:fig6}
\end{figure}
The two objects
$E^{\mathrm{Feynman}}$ and $\Delta$ intersect at three points, as shown in fig.~\ref{chapter_elliptics:fig6}.
The images of these three points in ${\mathbb C}/\Lambda^{\mathrm{Feynman}}$ are 
\bq
 0, \;\; z_1^{\mathrm{Feynman}}, \;\; z_2^{\mathrm{Feynman}}, 
\eq
where we used a translation transformation to fix one point at $0$.
The elliptic curves $E^{\mathrm{Feynman}}$ and $E^{\mathrm{cut}}$ are related by $\tau=\tau^{\mathrm{cut}}=\frac{1}{2}\tau^{\mathrm{Feynman}}$
and assuming that the points $z_1^{\mathrm{Feynman}}$ and $z_2^{\mathrm{Feynman}}$ are inside the fundamental parallelogram of $E^{\mathrm{cut}}$ we have
\bq
 z_i \; = \; z_i^{\mathrm{cut}} \; = \; z_i^{\mathrm{Feynman}},
 \;\;\;\;\;\; 
 i \; \in \; \{1,2\}.
\eq
Working out the details we find
\bq
\label{chapter_elliptics:def_z_i_unequal_sunrise}
 z_{i} & = &
 \frac{F\left(u_{i},k\right)}{2 K\left(k\right)},
 \;\;\;\;\;\; 
 i \; \in \; \{1,2\},
\eq
where $K(x)$ denotes the complete elliptic integral of the first kind, 
$F(z,x)$ denotes the incomplete elliptic integral of the first kind and $k$ denotes 
the modulus of the elliptic curve $E^{\mathrm{cut}}$ as defined by eq.~(\ref{chapter_elliptics:def_modulus}).
The variables $u_1$ and $u_2$ are given by
\bq
 u_{1}
 & = &
 \frac{\sqrt{
4 y_1 y_2 - 4x\left(y_1+y_2\right) - \left(1+x-y_1-y_2\right)^2 + 8 \sqrt{-xy_1y_2}
}}{
2
\left( \sqrt{-x y_1} + \sqrt{y_2} \right)
},
 \nonumber \\
 u_{2}
 & = &
 \frac{\sqrt{
4 y_1 y_2 - 4x\left(y_1+y_2\right) - \left(1+x-y_1-y_2\right)^2 + 8 \sqrt{-xy_1y_2}
}}{
2
\left( \sqrt{-x y_2} + \sqrt{y_1} \right)
}.
\eq
Eq.~(\ref{chapter_elliptics:def_tau_unequal_sunrise}) and eq.~(\ref{chapter_elliptics:def_z_i_unequal_sunrise})
define the new coordinates $(\tau,z_1,z_2)$ as functions of the old coordinates $(x,y_1,y_2)$.
We also need the inverse relation, which gives us $(x,y_1,y_2)$ as functions of $(\tau,z_1,z_2)$.
One finds
\bq
\label{chapter_elliptics:lambda_to_x}
 x
 & = &
 \frac{\left(1-\kappa_1\right)\left(1-\kappa_2\right) \kappa_1 \kappa_2 \lambda^2}{\left(1-\lambda\kappa_1\right) \left(1-\lambda\kappa_2\right)},
 \nonumber \\
 y_1
 & = &
 \frac{\kappa_1 \left(1-\kappa_1\right)}{\left(1-\lambda\kappa_1\right) \left(\kappa_1-\kappa_2\right)^2\left(1-\kappa_1-\kappa_2+\lambda\kappa_1\kappa_2\right)^2} R,
 \nonumber \\
 y_2
 & = &
 \frac{\kappa_2 \left(1-\kappa_2\right)}{\left(1-\lambda\kappa_2\right) \left(\kappa_1-\kappa_2\right)^2\left(1-\kappa_1-\kappa_2+\lambda\kappa_1\kappa_2\right)^2} R,
\eq
with
\bq
 R
 & = &
 \left( 1+\kappa_1^{3}\kappa_2^{3}\lambda^{3} \right)  \left( \kappa_1+\kappa_2 \right) 
 - \left( 1+\kappa_1^{2}\kappa_2^{2}\lambda^{2} \right)  \left( \kappa_1^{2}+\kappa_2^{2} \right)  \left( 1+\lambda \right) 
- 8 \left( 1+\kappa_1^{2}\kappa_2^{2}\lambda^{2} \right) \kappa_1\,\kappa_2
 \nonumber \\
 & &
+ \left( 1+\lambda\,\kappa_1\,\kappa_2 \right)  \left( \kappa_1+\kappa_2 \right) ^{3}\lambda
+ 3 \left( 1+\lambda\,\kappa_1\,\kappa_2 \right) \kappa_1\,\kappa_2\, \left( \kappa_1+\kappa_2 \right) \lambda
 \nonumber \\
 & &
+ 8 \left( 1+\lambda\,\kappa_1\,\kappa_2 \right) \kappa_1\,\kappa_2\, \left( \kappa_1+\kappa_2 \right) 
+4\,\kappa_1^{2}\kappa_2^{2} \lambda \left( 1-\lambda \right) 
-8\,\kappa_1\,\kappa_2\, \left( \kappa_1+\kappa_2 \right) ^{2}\lambda-8\,\kappa_1^{2}\kappa_2^{2}
 \nonumber \\
 & &
 - 2 \left(1-2\kappa_1+\lambda\kappa_1^2\right) \left(1-2\kappa_2+\lambda\kappa_2^2\right)
  \sqrt{ \kappa_1 \kappa_2 \left(1-\kappa_1\right)\left(1-\kappa_2\right) \left(1-\lambda\kappa_1\right) \left(1-\lambda\kappa_2\right)}.
 \nonumber \\
\eq
and
\bq
\label{chapter_elliptics:tau_to_lambda}
 \lambda
 & = &
 \frac{\theta_2^4\left(0,q\right)}{\theta_3^4\left(0,q\right)},
 \nonumber \\
 \kappa_1
 & = &
 \frac{\theta_3^2\left(0,q\right)}{\theta_2^2\left(0,q\right)} 
 \frac{\theta_1^2\left(\frac{\pi z_{1}}{2}, q\right)}{\theta_4^2\left(\frac{\pi z_{1}}{2}, q\right)},
 \nonumber \\
 \kappa_2
 & = &
 \frac{\theta_3^2\left(0,q\right)}{\theta_2^2\left(0,q\right)} 
 \frac{\theta_1^2\left(\frac{\pi z_{2}}{2}, q\right)}{\theta_4^2\left(\frac{\pi z_{2}}{2}, q\right)}.
\eq
Eq.~(\ref{chapter_elliptics:lambda_to_x}) and eq.~(\ref{chapter_elliptics:tau_to_lambda}) 
together with $q=\exp(i\pi \tau)$ 
allow us to express $(x,y_1,y_2)$ in terms of $(\tau,z_1,z_2)$.

The system of differential equations can again be transformed into an $\eps$-form
by a redefinition of the master integrals 
and a change of coordinates from $(x,y_1,y_2)$ to $(\tau,z_1,z_2)$ \cite{Bogner:2019lfa,Weinzierl:2020fyx}.
Let's look at the fibre transformation: We seek a matrix $U$ relating the new master integrals $\vec{J}$
to the old master integrals $\vec{I}$
\bq
 \vec{J} & = & U \vec{I},
\eq
such that in the transformed differential equation the dimensional regularisation parameter $\eps$ appears only
as a prefactor.
We construct $U$ in two steps:
\bq
 U & = & U_2 U_1,
\eq
Let us set
\bq
 \vec{J}_1 \; = \; U_1 \vec{I},
 & &
 d \vec{J}_1 \; = \; \hat{A}_1 \vec{J}_1.
\eq
The entries of $U_1$ are constructed such that $\hat{A}_1$ is linear in $\eps$ and the $\eps^0$-part is strictly lower triangular, i.e.
\bq
 \hat{A}_1 & = & \hat{A}^{(0)}_1 + \eps \hat{A}^{(1)}_1,
\eq
where $\hat{A}^{(0)}_1$ and $\hat{A}^{(1)}_1$ are independent of $\eps$ and
$\hat{A}^{(0)}_1$ is strictly lower triangular.
This can be done with a transformation,
where the entries of $U_1$ are rational functions of $\eps$, $x$, $y_1$, $y_2$, $\psi_1$ and $\partial_x \psi_1$
(i.e. compared to the full transformation $U$ the entries of $U_1$ do not involve incomplete elliptic integrals.
They only involve complete elliptic integrals related to $\psi_1$ and $\partial_x \psi_1$).
The entries of $U_1$ are determined from an ansatz and 
with the help of the methods of section~\ref{chapter_transformations:fibre_transformation}.
Explicitly, $U_1$ is given by setting $F_{54}=F_{64}=F_{74}=0$ in the formula~(\ref{chapter_elliptics:def_J}) below.

In a second step $U_2$ is constructed. $U_2$ eliminates the non-zero entries of $\hat{A}^{(0)}_1$.
As $\hat{A}^{(0)}_1$ is strictly lower triangular,
this can be done systematically by integration and will lead to incomplete elliptic integrals.
In a final clean-up (and after the change of coordinates on the base manifold) we trade the incomplete elliptic integrals
for $d\ln(y_1)/d\tau$ and $d\ln(y_2)/d\tau$.

In order to present the explicit formulae, we first introduce a few abbreviations:
We introduce the monomial symmetric polynomials $M_{\lambda_1 \lambda_2 \lambda_3}(a_1,a_2,a_3)$ in the variables $a_1$, $a_2$ and $a_3$.
These are defined by
\bq
\label{chapter_elliptics:def_notation_M_ijk}
 M_{\lambda_1 \lambda_2 \lambda_3}\left(a_1,a_2,a_3\right) & = &
 \sum\limits_{\sigma} \left( a_1 \right)^{\sigma\left(\lambda_1\right)} \left( a_2 \right)^{\sigma\left(\lambda_2\right)} \left( a_3 \right)^{\sigma\left(\lambda_3\right)},
\eq
where the sum is over all distinct permutations $\sigma$ of $\left(\lambda_1,\lambda_2,\lambda_3\right)$.
A few examples are
\bq
 M_{100}\left(a_1,a_2,a_3\right)
 & = &
 a_1 + a_2 + a_3,
 \nonumber \\
 M_{111}\left(a_1,a_2,a_3\right)
 & = &
 a_1 a_2 a_3,
 \nonumber \\
 M_{210}\left(a_1,a_2,a_3\right)
 & = &
 a_1^2 a_2 + a_2^2 a_3 + a_3^2 a_1 + a_2^2 a_1 + a_3^2 a_2 + a_1^2 a_3.
\eq
As an abbreviation we then set
\bq
 M_{\lambda_1 \lambda_2 \lambda_3}
 & = &
 M_{\lambda_1 \lambda_2 \lambda_3}\left(y_1,y_2,1\right).
\eq
As an example we have
\bq
 M_{110} & = & M_{110}\left(y_1,y_2,1\right) 
 \; = \; 
 y_1 y_2 + y_1 + y_2.
\eq
In addition, we introduce the abbreviation
\bq
\label{def_delta}
\Delta & = & 
 2 M_{110} - M_{200}
 \; = \;
 2 y_1 y_2 + 2 y_1 + 2 y_2 - y_1^2 - y_2^2 - 1.
\eq
We further set
\bq
 W_x & = & 
 \psi_1 \frac{d}{dx} \psi_2 - \psi_2 \frac{d}{dx} \psi_1.
\eq
Let us now present the basis $\vec{J}$:
\bq
\label{chapter_elliptics:def_J}
 J_1
 & = &
 \eps^2 I_{101},
 \\
 J_2
 & = &
 \eps^2 I_{011},
 \nonumber \\
 J_3
 & = &
 \eps^2 I_{110},
 \nonumber \\
 J_4 
 & = &
 \eps^2 \frac{\pi}{\psi_1} I_{111},
 \nonumber \\
 J_5 
 & = &
 \eps \left[ 
             \left(y_1+y_2-2\right) I_{111}
             + \left(3-x-y_1-3y_2\right) y_1 I_{211}
             + \left(3-x-3y_1-y_2\right) y_2 I_{121}
             + 2 \left(1+x\right) I_{112}
      \right]
 \nonumber \\
 & &
 + \frac{2 \eps^2}{\left( 3 x^2+2 M_{100} x + \Delta \right)}
   \left[
          7 \left(y_1+y_2-2\right) x^2 
          + 2 \left(3y_1^2+3y_2^2 -6 + y_1 + y_2 - 2 y_1 y_2 \right) x
 \right. \nonumber \\
 & & \left.
          + \left(y_1+y_2 - 2 \right) \Delta
   \right]
   I_{111}
 + F_{54} J_4,
 \nonumber \\
 J_6 
 & = &
 \eps \left[ 
             \left(y_1-y_2\right) I_{111}
             - \left(1+x+y_1-y_2\right) y_1 I_{211}
             + \left(1+x-y_1+y_2\right) y_2 I_{121}
             - 2 \left(y_1-y_2\right) I_{112}
      \right]
 \nonumber \\
 & &
 + \frac{2 \eps^2 \left(y_1-y_2\right) }{\left( 3 x^2+2 M_{100} x + \Delta \right)}
   \left[
          7 x^2 
          + 2 \left(3y_1+3y_2 -1\right) x 
          + \Delta
   \right]
   I_{111}
 + F_{64} J_4,
 \nonumber \\
 J_7 
 & = &
 \frac{1}{\eps} \frac{\psi_1^2}{2\pi i W_x} \frac{d}{dx} J_4
 + \frac{\eps^2}{8}  \frac{1}{\left(3x^2+2M_{100}x+\Delta \right)^2} 
 \left[ 
        9 x^6
        + 22 M_{100} x^5
        + \left( 50 M_{110} - M_{200} \right) x^4
 \right. \nonumber \\
 & & \left.
        - \left( 44 M_{300} - 76 M_{210} + 216 M_{111} \right) x^3
        - \left( 41 M_{400} - 84 M_{310} + 86 M_{220} + 52 M_{211} \right) x^2
 \right. \nonumber \\
 & & \left.
        + 2 \Delta \left( 5 M_{300} - 5 M_{210} + 2 M_{111} \right) x
        - \Delta^3
 \right] \frac{\psi_1}{\pi} I_{111}
 - \frac{1}{8} F_{64} J_6
 - \frac{1}{24} F_{54} J_5
 + F_{74} J_4.
 \nonumber 
\eq
The three functions $F_{54}$, $F_{64}$, $F_{74}$, appearing in the definition of $J_5$, $J_6$ and $J_7$
are given by
\bq
\lefteqn{
 F_{54}
 = } & &
 \nonumber \\
 & &
 \frac{6 i}{\left(3x^2+2M_{100}x+\Delta\right) \psi_1}
 \left[
       \left(1+x+y_1-y_2\right) \frac{1}{y_1} \frac{dy_1}{d\tau}
       +
       \left(1+x-y_1+y_2\right) \frac{1}{y_2} \frac{dy_2}{d\tau}
 \right],
 \nonumber \\
 & & \nonumber \\
\lefteqn{
 F_{64}
 = } & &
 \nonumber \\
 & &
 \frac{2 i}{\left(3x^2+2M_{100}x+\Delta\right) \psi_1}
 \left[
       \left(3y_1+y_2-1 + 3x\right) \frac{1}{y_1} \frac{dy_1}{d\tau}
       -
       \left(y_1+3y_2-1 + 3x\right) \frac{1}{y_2} \frac{dy_2}{d\tau}
 \right],
 \nonumber \\
 & & \nonumber \\
\lefteqn{
 F_{74}
 = } & &
 \nonumber \\
 & &
 - \frac{1}{\left(3x^2+2M_{100}x+\Delta\right)^2 \psi_1^2}
 \left[
  \left( 3 y_1^2 + y_2^2 + 1 - 2 y_2 + 6 y_1 x + 3 x^2 \right) \left(\frac{1}{y_1} \frac{dy_1}{d\tau}\right)^2
 \right. \nonumber \\
 & & \left.
  - \left( 3 y_1^2 + 3 y_2^2 - 1 + 2 y_1 y_2 - 2 y_1 - 2 y_2 + 6 \left( y_1 +y_2 - 1 \right) x + 3 x^2 \right) 
    \left( \frac{1}{y_1} \frac{dy_1}{d\tau} \right) 
    \left( \frac{1}{y_2} \frac{dy_2}{d\tau} \right)
 \right. \nonumber \\
 & & \left.
  + \left( y_1^2 + 3 y_2^2 + 1 - 2 y_1 + 6 y_2 x + 3 x^2 \right) \left( \frac{1}{y_2} \frac{dy_2}{d\tau} \right)^2
 \right].
\eq
In this basis the differential equation reads
\bq
\label{chapter_elliptics:diff_eq_sunrise_unequal}
 \left( d + A \right) J & = & 0,
\eq
with
\bq
 A & = &
 \eps
 \left( \begin{array}{ccccccc}
 a_{11} & 0 & 0 & 0 & 0 & 0 & 0 \\
 0 & a_{22} & 0 & 0 & 0 & 0 & 0 \\
 0 & 0 & a_{33} & 0 & 0 & 0 & 0 \\
 0 & 0 & 0 & a_{44} & a_{45} & a_{46} & a_{47} \\
 a_{51} & a_{52} & a_{53} & a_{54} & a_{55} & a_{56} & a_{57} \\
 a_{61} & a_{62} & a_{63} & a_{64} & a_{65} & a_{66} & a_{67} \\
 a_{71} & a_{72} & a_{73} & a_{74} & a_{75} & a_{76} & a_{77} \\
 \end{array} \right).
\eq
In order to present the entries of $A$ in a compact form we 
introduce $z_3=-z_1-z_2$ and a constant $\beta=1$.
We define (for arbitrary $\beta$)
\bq
 \Omega_k\left(z,\beta,\tau\right)
 & = &
 \frac{1}{2} \omega_k\left(z,\tau\right)
 + \frac{1}{4} \omega_k\left(z-\beta,\tau\right)
 + \frac{1}{4} \omega_k\left(z+\beta,\tau\right)
 \\
 & &
 - 2 \left(k-1\right) \left[
 \omega_k\left(\frac{z}{2},\tau\right)
 + \frac{1}{2} \omega_k\left(\frac{z-\beta}{2},\tau\right)
 + \frac{1}{2} \omega_k\left(\frac{z+\beta}{2},\tau\right)
 \right],
 \nonumber
\eq
where $\omega_k(z,\tau)$ denotes the one-form defined in eq.~(\ref{chapter_elliptics:omega_kronecker_general}).
For $\beta=1$ we have
\bq
 \Omega_k\left(z,1,\tau\right)
 & = &
 \omega_k\left(z,\tau\right)
 - 2 \left(k-1\right) \omega_k\left(z,2\tau\right).
\eq
We will encounter $\Omega_2(z,\beta,\tau)$ and $\Omega_3(z,\beta,\tau)$.
For the entries $a_{ij}$ we also need two differential forms $\eta_2(\tau)$ and $\eta_4(\tau)$, which depend on $\tau$, but not on the $z_i$'s.
These are defined by
\bq
 \eta_2\left(\tau\right)
 \; = \;
 \left[ e_2\left(\tau\right) - 2  e_2\left(2\tau\right) \right] \frac{d\tau}{2\pi i},
 & &
 \eta_4\left(\tau\right)
 \; = \;
 \frac{1}{\left(2\pi i\right)^2} e_4\left(\tau\right) \frac{d\tau}{2\pi i},
\eq
where $e_k(\tau)$ denotes the standard Eisenstein series, defined in eq.~(\ref{chapter_elliptics:def_e_k}).
We have
\bq
 e_2\left(\tau\right) - 2  e_2\left(2\tau\right) \; \in \; {\mathcal M}_2(\Gamma_0(2)),
 & &
 e_4\left(\tau\right) \; \in \; {\mathcal M}_4(\mathrm{SL}_2({\mathbb Z})).
\eq
For the entries of $A$ we have the following relations
\begin{align}
 a_{45} & = \frac{1}{24} a_{57},
 &
 a_{46} & = \frac{1}{8} a_{67},
 & 
 a_{33} & = a_{11} + a_{22},
 \nonumber \\
 a_{53} & = a_{11} + a_{22} - a_{51} - a_{52},
 &
 a_{56} & = 3 a_{65},
 &
 a_{77} & = a_{44},
 \nonumber \\
 a_{61} & = 2 a_{11} - a_{51},
 &
 a_{62} & = -2 a_{11} + a_{51},
 &
 a_{63} & = a_{11}-a_{22},
 \nonumber \\
 a_{75} & = \frac{1}{24} a_{54},
 & 
 a_{76} & = \frac{1}{8} a_{64},
 & &
\end{align}
and the following symmetries
\begin{align}
 a_{22}\left(z_1,z_2,z_3\right) & = a_{11}\left(z_2,z_1,z_3\right),
 &
 a_{52}\left(z_1,z_2,z_3\right) & = a_{51}\left(z_2,z_1,z_3\right),
 \nonumber \\
 a_{72}\left(z_1,z_2,z_3\right) & = a_{71}\left(z_2,z_1,z_3\right),
 &
 a_{73}\left(z_1,z_2,z_3\right) & = a_{71}\left(z_1,z_3,z_2\right).
\end{align}
Thus we need to specify only a few entries. 
We group them by modular weight.
\\
\\
Modular weight $0$:
\bq
 a_{4,7}
 & = &
 \omega_0\left(z,\tau\right) \; = \; -2\pi i d\tau.
\eq
Modular weight $1$:
\bq
 a_{5,7}
 & = &
 6 i \left[ \omega_1\left(z_1,\tau\right) + \omega_1\left(z_2,\tau\right) \right], 
 \nonumber \\
 a_{6,7}
 & = &
 2 i \left[ \omega_1\left(z_1,\tau\right) - \omega_1\left(z_2,\tau\right) \right].
\eq
Note that $\omega_1(z,\tau)=2\pi i dz$ is independent of $\tau$.
\\
\\
Modular weight $2$:
\bq
 a_{1,1}
 & = &
 -2 \left[ \Omega_2\left(z_1,\beta,\tau\right) - \Omega_2\left(z_3,\beta,\tau\right) \right],
 \nonumber \\
 a_{4,4}
 & = &
 \omega_2\left(z_1,\tau\right) + \omega_2\left(z_2,\tau\right) + \omega_2\left(z_3,\tau\right) 
 - \Omega_2\left(z_1,\beta,\tau\right) - \Omega_2\left(z_2,\beta,\tau\right) + 3 \Omega_2\left(z_3,\beta,\tau\right) 
 \nonumber \\
 & &
 - 6 \eta_2\left(\tau\right),
 \nonumber \\
 a_{5,1}
 & = &
 - 2 \left[ \Omega_2\left(z_1,\beta,\tau\right) - \Omega_2\left(z_2,\beta,\tau\right) - 2 \Omega_2\left(z_3,\beta,\tau\right) \right],
 \nonumber \\
 a_{5,5}
 & = &
 - 3 \omega_2\left(z_3,\tau\right) 
 - \Omega_2\left(z_1,\beta,\tau\right) - \Omega_2\left(z_2,\beta,\tau\right) + 3 \Omega_2\left(z_3,\beta,\tau\right) 
 - 6 \eta_2\left(\tau\right),
 \nonumber \\
 a_{6,5}
 & = &
 - \omega_2\left(z_1,\tau\right) + \omega_2\left(z_2,\tau\right),
 \nonumber \\
 a_{6,6}
 & = &
 - 2 \omega_2\left(z_1,\tau\right) - 2 \omega_2\left(z_2,\tau\right) + \omega_2\left(z_3,\tau\right) 
 - \Omega_2\left(z_1,\beta,\tau\right) - \Omega_2\left(z_2,\beta,\tau\right) + 3 \Omega_2\left(z_3,\beta,\tau\right) 
 \nonumber \\
 & &
 - 6 \eta_2\left(\tau\right),
\eq
Modular weight $3$:
\bq
 a_{5,4}
 & = &
 12 i \left[ \omega_3\left(z_1,\tau\right) + \omega_3\left(z_2,\tau\right) - 2 \omega_3\left(z_3,\tau\right) \right],
 \nonumber \\
 a_{6,4}
 & = &
 12 i \left[ \omega_3\left(z_1,\tau\right) - \omega_3\left(z_2,\tau\right) \right],
 \nonumber \\
 a_{7,1}
 & = &
 i \left[ 
  \Omega_3\left(z_1,\beta,\tau\right) - \Omega_3\left(z_2,\beta,\tau\right) + \Omega_3\left(z_3,\beta,\tau\right) 
 \right].
\eq
Modular weight $4$:
\bq
 a_{7,4}
 & = &
 12 \left[ \omega_4\left(z_1,\tau\right) + \omega_4\left(z_2,\tau\right) + \omega_4\left(z_3,\tau\right) - 6 \eta_4\left(\tau\right) \right].
\eq
We have managed to transform the differential equation for the family of the unequal mass sunrise integral 
(e.g. an elliptic Feynman integral depending on three kinematic variables)
into an $\eps$-form.
This differential equation can be solved in terms of the iterated integrals 
discussed in section~\ref{chapter_elliptics:section_moduli_spaces}.

Let us close this section with a discussion of
the behaviour of the system under a modular transformation
\bq
 \gamma
 \; = \;
 \left(\begin{array}{cc}
   a & b \\
   c & d \\
 \end{array} \right)
 & \in &
 \mathrm{SL}_2({\mathbb Z}).
\eq
The coordinate transform as
\bq
\label{chapter_elliptics:trafo_base_unequal_sunrise}
 z_1' \; = \; \frac{z_1}{c\tau+d},
 \;\;\;\;\;\;
 z_2' \; = \; \frac{z_2}{c\tau+d},
 \;\;\;\;\;\;
 \tau' \; = \; \frac{a\tau+b}{c\tau+d}.
\eq
The constant $\beta=1$ transforms as
\bq
\label{chapter_elliptics:trafo_constants_unequal_sunrise}
 \beta' \; = \; \frac{\beta}{c\tau+d},
\eq
so in general we will have $\beta' \neq 1$.
We may view $\beta$ as being a further marked point in a higher dimensional space
$\mathcal{M}_{1,n'}$ with $n'>n$.
We set again $z_3'=-z_1'-z_2'$.
We also need to redefine the master integrals.
We set
\bq
\label{chapter_elliptics:trafo_fibre_unequal_sunrise}
 J' & = U_\gamma \; J,
\eq
where $U_\gamma$ is given by
\bq
\label{chapter_elliptics:def_U_trafo_fibre_unequal_sunrise}
 U_\gamma & = &
 \left( \begin{array}{ccccccc}
 1 & 0 & 0 & 0 & 0 & 0 & 0 \\
 0 & 1 & 0 & 0 & 0 & 0 & 0 \\
 0 & 0 & 1 & 0 & 0 & 0 & 0 \\
 0 & 0 & 0 & \frac{1}{c\tau+d} & 0 & 0 & 0 \\
 0 & 0 & 0 & \frac{6 i c \left(z_1+z_2\right)}{c\tau+d} & 1 & 0 & 0 \\
 0 & 0 & 0 & \frac{2 i c \left(z_1-z_2\right)}{c\tau+d} & 0 & 1 & 0 \\
 0 & 0 & 0 & -\frac{c}{2\pi i \eps} + \frac{c^2\left(z_1^2+z_1z_2+z_2^2\right)}{c\tau+d} & -\frac{i c \left(z_1+z_2\right)}{4} & -\frac{i c \left(z_1-z_2\right)}{4} & c\tau+d \\
 \end{array} \right).
 \;\;\;\;\;\;
\eq
The transformation matrix $U$ is not too difficult to construct,
if one starts from the assumption that the first elliptic master integral (i.e. $J_4$) should be rescaled as
\bq
 J_4' & = & 
 \frac{\omega_1}{\omega_1'} J_4
 \; = \;
 \frac{1}{c\tau+d} J_4.
\eq
Under this combined transformation 
the differential equation for the transformed system reads then
\bq
 \left( d + A' \right) J' & = & 0,
\eq
where $A'$ is obtained (with one exception) from $A$ by replacing all unprimed variables with primed variables.
For example $a_{7,1}'$ is given by
\bq
 a_{7,1}'
 & = &
 i \left[ 
  \Omega_3\left(z_1',\beta',\tau'\right) - \Omega_3\left(z_2',\beta',\tau'\right) + \Omega_3\left(z_3',\beta',\tau'\right) 
 \right].
\eq
The only exception is $\eta_2(\tau)$.
For $\gamma \in \Gamma_0(2)$ the differential one-form $\eta_2(\tau)$ transforms into
$\eta_2(\tau')$.
For a general $\gamma \in \mathrm{SL}_2({\mathbb Z})$ let us set $b_2(\tau)=e_2(\tau) - 2  e_2(2\tau)$.
Then $\eta_2(\tau)$ is replaced by
\bq
 (b_2 \slashoperator{\gamma^{-1}}{2})(\tau') \; \frac{d\tau'}{2\pi i}.
\eq
$(b_2 \slashoperator{\gamma^{-1}}{2})(\tau')$ is again a modular form for $\Gamma(2)$,
but not necessarily identical to $b_2(\tau')$.

It remains to work out $(b_2 \slashoperator{\gamma^{-1}}{2})(\tau')$.
To this aim we first express $b_2(\tau)$ in terms of Eisenstein series for $\Gamma(2)$.
We find
\bq
 b_2\left(\tau\right)
 & = &
 4 \left(2\pi i\right)^2 h_{2,2,0,1}\left(\tau\right),
\eq
where the Eisenstein series $h_{k,N,r,s}(\tau)$ have been defined in eq.~(\ref{chapter_elliptics:def_Eisenstein_h}).
The transformation law for $b_2(\tau)$ follows then from the transformation law for $h_{k,N,r,s}(\tau)$
given in eq.~(\ref{chapter_elliptics:trafo_Eisenstein_h_1}).
We obtain
\bq
 (b_2 \slashoperator{\gamma^{-1}}{2})(\tau')
 & = &
 4 \left(2\pi i\right)^2 
 h_{2,2,b \bmod 2,d \bmod 2}\left(\tau'\right),
 \;\;\;\;\;\;\;\;\;
 \gamma^{-1} \; = \;
 \left( \begin{array}{rr}
  d & -b \\
 -c & a \\
 \end{array}
 \right).
 \;\;\;
\eq
Further examples of elliptic Feynman integrals can be found in \cite{Sogaard:2014jla,Bonciani:2016qxi,vonManteuffel:2017hms,Ablinger:2017bjx,Bourjaily:2017bsb,Broedel:2019hyg,Abreu:2019fgk,Kniehl:2019vwr,Campert:2020yur,Kristensson:2021ani}.

%% file: motives.tex
\newpage
\chapter{Motives and mixed Hodge structures}
\label{chapter_motives}

In most chapters of this book we followed the pattern to show how known facts in mathematics
may help a physicist (to compute Feynman integrals).
In this chapter we reverse the pattern.
We would like to illustrate how known facts in physics (Feynman integrals we know how to calculate)
may help a mathematician.
This concerns the theory of motives.
In part, such a theory is conjectural (hence the interest of mathematicians).
Feynman integrals provide non-trivial examples such a theory should contain.
In this chapter we introduce the mathematical language and the main ideas behind motives.

Motives were introduced by Alexander Grothendieck to unify cohomology theories.
In order to see at least two different cohomology theories we discuss 
in section~\ref{chapter_motives:cohomology} Betti cohomology and de Rham cohomology.
Unification will require a more abstract language,
and in section~\ref{chapter_motives:categories} we introduce categories as the appropriate language.
In section~\ref{chapter_motives:motives} we try to give a glimpse on what a theory of motives should be about.
As remarked earlier, this is still partly a conjectural theory
and sections~\ref{chapter_motives:categories} and \ref{chapter_motives:motives} mainly serve 
to acquaint the reader with the main ideas and the language of this field.

We can be more concrete if we look at the Hodge realisation of a motive (also called an H-motive).
Here we focus on the interplay between Betti cohomology and de Rham cohomology.
The theory is based on Hodge structures, which we discuss in section~\ref{chapter_motives:hodge_structures}.
Finally, in section~\ref{chapter_motives:examples_Feynman_integrals} we discuss examples from Feynman integrals.
We will also see how this formalism offers us an alternative way to compute the differential equation for a Feynman
integral.

Readers interested in the topics of this chapter would almost certainly consult additional text books.
Introductory texts for categories are the review article by Deligne and Milne \cite{Deligne1982}
and the books by
Mac Lane \cite{maclane:71}, Leinster \cite{leinster_2014} and Etingof, Gelaki, Nikshych and Ostrik \cite{Etingof_book}.
More information on motives can be found in the book by Andr\'e \cite{Andre_book}.
Hodge structures are treated in the books by Voisin \cite{Voisin_book} and Peters and Steenbrink \cite{Peters_Steenbrink_book}.
The proceedings of a summer school on Hodge structures \cite{Cattani_book} give also very useful information.
Review articles related to the content of this chapter are \cite{Brown:2015aa,Brown:2015bb,Rella:2020ivo}.

\section{Cohomology}
\label{chapter_motives:cohomology}

In this section we review Betti cohomology and de Rham cohomology.
We also discuss relative Betti cohomology and relative de Rham cohomology,
as well as algebraic de Rham cohomology.

\subsection{Betti cohomology}

\index{Betti cohomology}
{\bf Betti cohomology} 
(also known as 
\index{singular cohomology}
{\bf singular cohomology}) is based on a triangulation of a topological space by simplices.
We already encountered simplices in section~\ref{chapter_nested_sums:polytopes}.
Let $a_0, a_1, \dots, a_n$ be $(n+1)$ linear independent vectors in ${\mathbb R}^d$ (with $d \ge n+1$).
The $n$-simplex $\sigma_n=\langle a_0,a_1, \dots, a_n \rangle$ is 
the polytope
\bq
 \sigma_n \; = \; 
 \langle a_0,a_1, \dots, a_n \rangle
 & = &
 \left\{
   \alpha_0 a_0 + \alpha_1 a_1 + \dots + \alpha_n a_n | \alpha_j \ge 0, \sum\limits_{j=0}^n \alpha_j = 1
 \right\}.
\eq
For example a 0-simplex is a point, a 1-simplex is a line interval, 
a 2-simplex is a triangle and a 3-simplex is a tetrahedron.
The $a_i$ are called the vertices of the simplex $\sigma_n$.
Faces and facets (we recall that a facet is a face of codimension one) of simplices are defined as for polytopes (see section~\ref{chapter_nested_sums:polytopes}).
They correspond to $k$-dimensional simplices defined by a subset of $(k+1)$ vertices of $\sigma_n$. 
We denote the $i$-th facet of the simplex $\sigma_n= \langle a_0,a_1, \dots, a_n \rangle$ by
\bq
 \sigma_{n-1}^i
 & = &
 \langle a_0, \dots, \hat{a}_i, \dots, a_n \rangle,
\eq
where the hat denotes that the corresponding vertex is omitted.
The 
\index{incident number of simplices}
{\bf incident number} $[\sigma_n : \sigma_{n-1}^i ]$ is defined as
\bq
 [\sigma_n : \sigma_{n-1}^i ]
 & = &
 \left\{
 \begin{array}{ll}
 +1 & \mbox{if $(i,0,\dots,i-1,i+1,\dots,n)$ is an even permutation of $(0,1,\dots,n)$},\\
 -1 & \mbox{if $(i,0,\dots,i-1,i+1,\dots,n)$ is an odd permutation of $(0,1,\dots,n)$}.\\
 \end{array}
 \right.
 \nonumber \\
\eq
A 
\index{simplicial complex}
{\bf simplicial complex} $B$ is a set of a finite number of simplexes in ${\mathbb R}^d$ satisfying the following two properties:
\begin{enumerate}
\item an arbitrary face of a simplex $\sigma$ of $B$ belongs to $B$, 
\item if $\sigma$ and $\sigma'$ are two simplexes of $B$, the intersection $\sigma \cap \sigma'$
is either empty or a face of $\sigma$ and $\sigma'$.
\end{enumerate}
The union of all simplices of $B$ defines a subset of ${\mathbb R}^d$, which we denote by $|B|$.
Let $X$ be a topological space. If there exists a simplicial complex $B$ and a 
homeomorphism $t : |B| \rightarrow X$ then $X$ is said to be triangulable and the pair $(B,t)$ is called a 
\index{triangulation of a topological space}
{\bf triangulation} of $X$.

For a simplex $\sigma_n$ we define the boundary of the simplex by
\bq
 \partial \sigma_n & = & \sum\limits_{i=0}^n [\sigma_n : \sigma_{n-1}^i ] \sigma_{n-1}^i.
\eq
In other words, the boundary of a simplex is a linear combination of its facets with the coefficients being given by
the incident numbers.

Let $B$ be a simplicial complex. An 
\index{integral $k$-chain}
{\bf integral $k$-chain} of $B$ is a function mapping the set
of $k$-simplexes $\{\sigma_{k,1},\sigma_{k,2},\dots\}$ into the integers ${\mathbb Z}$, such that $\sigma_{k,j} \rightarrow n_j$.
An integral $k$-chain is denoted as
\bq
 c_k
 & = & 
 \sum\limits_j n_j \; \sigma_{k,j}.
\eq
The integral $k$-chains form a group with respect to addition. We denote this group by $C_k(B)$.
The 
\index{boundary of a $k$-chain}
{\bf boundary of a $k$-chain} is given by
\bq
 \partial c_k
 & = & 
 \sum\limits_j \sum\limits_{i=0}^k n_j \; [\sigma_{k,j} : \sigma_{k-1,j}^i ] \sigma_{k-1,j}^i,
\eq
where $\sigma_{k-1,j}^i$ denotes the $i$-th facet of the $k$-dimensional simplex $\sigma_{k,j}$.

A 
\index{cycle in singular homology}
{\bf $k$-cycle} $z_k$ is defined by
\bq
 \partial z_k 
 & = &
 0.
\eq
A $k$-chain $b_k$ is called a 
{\bf boundary} if there exists a $(k+1)$-chain $c_{k+1}$ such that
\bq
 \partial c_{k+1} & = & b_{k}.
\eq
Due to the alternating signs originating from the incidence numbers 
we have $\partial \partial c_k = 0$ for any chain, so each boundary $b_k=\partial c_{k+1}$ is also a cycle 
(since $\partial b_k=\partial \partial c_{k+1}=0$).
The sets of all $k$-chains, $k$-cycles and $k$-boundaries form Abelian groups, denoted by
$C_k(B)$, $Z_k(B)$ and $B_k(B)$, respectively. 
The boundary operator $\partial : C_k(B) \rightarrow C_{k-1}(B)$, which maps $k$-chains into $(k-1)$-chains, is
also a group homomorphism. The group of boundaries $B_k(B)$ forms a subgroup of the group of
cycles $Z_k(B)$. Since all groups are Abelian, the factor group
\bq
\label{chapter_motives:def_Betti_homology}
 H_k(B) & = & \frac{Z_k(B)}{B_k(B)}
\eq
is well defined and is called the 
{\bf $k$-th homology group} $H_k(B)$ of the complex $B$.

A sequence of Abelian groups $\dots, C_0, C_1, C_2, \dots$ together with homomorphisms $\partial_k : C_k \rightarrow C_{k-1}$ such that
$\partial_{k-1} \circ \partial_k = 0$ in an example of a 
\index{chain complex}
{\bf chain complex}.
A chain complex is denoted as
\bq
 \cdots 
 \stackrel{\partial_3}{\longrightarrow}
 C_2 
 \stackrel{\partial_2}{\longrightarrow}
 C_1
 \stackrel{\partial_1}{\longrightarrow}
 C_0 
 \stackrel{\partial_0}{\longrightarrow}
 \cdots.
\eq
The chain groups $C_k(B)$ form a chain complex.

The cohomology groups are defined as follows:
Given an Abelian group $K$ and the $k$-chain groups $C_k(B)$ of a simplicial complex $B$
we consider the 
\index{cochain}
{\bf $k$-cochain group} 
\bq
 C^k(B,K) & = & \mathrm{Hom}( C_k(B), K ).
\eq 
The group $K$ is called the 
\index{coefficient group of cochains}
{\bf coefficient group}.
A typical choice would be $K={\mathbb Z}$.
We define the 
\index{coboundary operator}
{\bf coboundary operator} $d : C^k(B, K) \rightarrow C^{k+1}(B, K)$ by
\bq
 \left( d f^k \right) \left( c_{k+1} \right) 
 & = & 
 f^k\left( \partial c_{k+1} \right)
\eq
for all $f^k \in C^k(B,K)$ and $c_{k+1} \in C_{k+1}(B)$.
\index{cocycle}
{\bf Cocycles} 
and 
\index{coboundary}
{\bf coboundaries} are then defined in the usual way : 
A $k$-cochain $f^k$ is called a $k$-cocylce if $d f^k = 0 $, 
it is called a $k$-coboundary if there is a $(k-1)$-cochain $g^{k-1}$ such that $f^k = d g^{k-1}$.
The $k$-cohomology group $H^k(B,K)$ is defined as
\bq
\label{chapter_motives:def_Betti_cohomology}
 H^k(B,K)
 & = &
 \frac{Z^k(B,K)}{B^k(B,K)},
\eq
where $Z^k(B,K)$ is the group of $k$-cocycles and $B^k(B,K)$ is the group of $k$-coboundaries.

Let us now define Betti homology/cohomology for a triangulable topological space.
For a triangulable topological space $X$ one first chooses a triangulation $(B,t)$ 
and defines the Betti homology/cohomology (or singular homology/cohomology) by 
eq.~(\ref{chapter_motives:def_Betti_homology}) and eq.~(\ref{chapter_motives:def_Betti_cohomology}), respectively.
One can show that this is independent of the chosen triangulation.
We denote the $k$-th cohomology group by
\bq
 H^k_B(X)
\eq
The subscript $B$ stands for ``Betti'', not the simplicial complex $B$. As we mentioned above, $H^k_B(X)$ is independent of the chosen
triangulation $(B,t)$. 
If we want to emphasise the coefficient group $K$, we write $H^k_B(X,K)$.

We may also define 
\index{relative homology groups}
{\bf relative homology groups} and 
\index{relative cohomology groups}
{\bf relative cohomology groups}:
Whereas a $k$-cycle of a complex $B$ is a chain with no boundary at all, we can relax this condition
by requiring that the boundary lies only within some specified subcomplex $A$.
If $A$ is a subcomplex of $B$, we define the relative chain group as
\bq
 C_k(B,A)
 & = &
 \frac{C_k(B)}{C_k(A)}.
\eq
The relative cycle group $Z_k(B,A)$ and the relative boundary group $B_k(B,A)$ are defined in a similar way:
A chain $z_k \in C_k(B)$ defines a cycle $[z_k] \in C_k(B,A)$ if 
\bq
 \partial z_k & \in & A.
\eq
A chain $b_k \in C_k(B)$ defines a boundary $[b_k] \in C_k(B,A)$ if there is a chain $c_{k+1} \in C_{k+1}(B)$ with
\bq
 b_k - \partial c_{k+1} & \in & A.
\eq
It can be shown that $B_k(B,A)$ is a subgroup of $Z_k(B,A)$ and thus the
relative homology group
\bq
 H_k(B,A)
 & = &
 \frac{Z_k(B,A)}{B_k(B,A)}
\eq
is well defined.
\\
\\
\bs
{\it \refstepcounter{exercise}
{\bf Exercise \theexercise}: 
Show that a relative boundary is a relative cycle.
}
\es
\\
\\
The relative cohomology groups are obtained from the 
relative homology groups
in complete analogy to the way the 
cohomology groups are obtained from the 
homology groups, starting with the relative $k$-cochain group
\bq
 C^k(B,A,K) & = & \mathrm{Hom}( C_k(B,A), K ).
\eq 

\subsection{De Rham cohomology}
\label{chapter_motives:de_Rham_cohomology}

If $X$ is a differentiable manifold, 
we may consider de Rham cohomology (as we did in section~\ref{chapter_iterated_integrals:section:eps_form}):
The 
\index{de Rham cohomology}
$k$-th {\bf de Rham cohomology group} 
$H^k_{\mathrm{dR}}(X)$ is the set of equivalence classes of closed $k$-forms modulo exact $k$-forms.
The group law is the addition of $k$-forms.

Let us now look at the relation between de Rham cohomology and Betti cohomology:
Let $X$ be a differentiable manifold.
To this manifold we can on the one hand associate the de Rham cohomology $H_{\mathrm{dR}}^k(X)$, as well as
the Betti cohomology $H_{\mathrm{B}}^k(X)$.
There is an isomorphism between the de Rham and Betti cohomology:
\bq
 \mathrm{comparison}
 & : & 
 H_{\mathrm{dR}}^k(X) \otimes \mathbb{C}
 \rightarrow
 H_{\mathrm{B}}^k(X) \otimes \mathbb{C},
 \nonumber \\
 & & \omega \rightarrow \left( \gamma \rightarrow \int\limits_\gamma \omega \right).
\eq
If we fix a basis for $H_{\mathrm{dR}}^k(X)$ and $H_{\mathrm{B}}^k(X)$
the isomorphism is given explicitly as follows:
Let $n = \dim H_{\mathrm{dR}}^k(X) = \dim H_{\mathrm{B}}^k(X)$.
Let us denote by $\omega_1, \dots, \omega_n$ a basis for $H_{\mathrm{dR}}^k(X)$ 
and by $\gamma_1, \dots, \gamma_n$ a basis for the Betti homology $H^{\mathrm{B}}_k(X)$.
A basis for the Betti cohomology $H_{\mathrm{B}}^k(X)$ is then given by the duals $\gamma_1^\ast, \dots, \gamma_n^\ast$ 
and satisfies $\gamma_i^\ast(\gamma_j) = \delta_{ij}$. 
We then have
\bq
\label{chapter_motives:def_comparison}
 \omega_i = \sum\limits_j p_{ij} \gamma_j^\ast,
 & &
 p_{ij} = \int\limits_{\gamma_j} \omega_i.
\eq
The coefficients $p_{ij}$ are called 
{\bf periods} and the $n \times n$-matrix $P$ with entries $p_{ij}$ is called the 
{\bf period matrix}.

If $Y$ is a closed submanifold of $X$, we may consider 
{\bf relative de Rham cohomology}.
We first define $\Omega^k(X,Y)$ to be the space of differential $k$-forms on $X$, whose restriction to $Y$
is zero.
The relative de Rham cohomology group $H_{\mathrm{dR}}^k(X,Y)$ is then given by the equivalence classes
of the closed forms in $\Omega^k(X,Y)$ modulo the exact ones.

Let's look at an example: We take $X={\mathbb C}^\ast = {\mathbb C}\backslash\{0\}$ and
$Y=\{1,2\}$.
For the non-relative cohomology we have
\bq
 \dim H_{\mathrm{B}}^1(X) \; = \; 1,
 & &
 \dim H_{\mathrm{dR}}^1(X) \; = \; 1.
\eq
A basis for $H^{\mathrm{B}}_1(X)$ is given by an anti-clockwise circle $\gamma_1$ around $z=0$,
a basis for $H_{\mathrm{dR}}^1(X)$ is given by $\omega_1=dz/z$.
For the relative cohomology we have
\bq
 \dim H_{\mathrm{B}}^1(X,Y) \; = \; 2,
 & &
 \dim H_{\mathrm{dR}}^1(X,Y) \; = \; 2.
\eq
A basis for $H^{\mathrm{B}}_1(X,Y)$ is now given by $\gamma_1$ as above and the line segment $\gamma_2$
from $z=1$ to $z=2$. The boundary of this line segments are the points $z=1$ and $z=2$, which are in $Y$.
A basis for the relative de Rham cohomology $H_{\mathrm{dR}}^1(X,Y)$ is given by $\omega_1=dz/z$
and $\omega_2=dz$.
Note that in the relative case $dz$ is no longer an exact form, as the zero form (i.e. the function)
$f(z) = z$ does not belong to $\Omega^0(X,Y)$. 
All functions from $\Omega^0(X,Y)$ are required to vanish on $Y$ and $f(z)$ does not
(for example $f(1)=1\neq 0$).
As a side-remark, let us note that the restriction of a $k$-form to a $(k-1)$-dimensional submanifold
always vanishes, there are simply not enough linear independent tangential vectors to contract into the
$k$-form. Thus $\omega_1,\omega_2 \in \Omega^1(X,Y)$.
There is again an isomorphism between the de Rham and Betti cohomology:
\bq
 H_{\mathrm{dR}}^k(X,Y) \otimes \mathbb{C}
 \rightarrow
 H_{\mathrm{B}}^k(X,Y) \otimes \mathbb{C}.
\eq
The period matrix is given by
\bq
 P 
 & = &
 \left(\begin{array}{cc}
 2 \pi i & \ln 2 \\
 0 & 1 \\
 \end{array} \right).
\eq

\subsection{Algebraic de Rham cohomology}

On a differential manifold $M$ we have the complex of differential forms $\Omega^\bullet(M)$.
Let $U \subset M$ be an open subset. On $U$ we may write any differential $k$-form as
\bq
 \omega & = &
 \sum\limits_{i_1 < \dots < i_k} f_{i_1 \dots i_k}\left(x\right) dx^{i_1} \wedge \dots \wedge dx^{i_k},
 \;\;\;\;\;\;
 f_{i_1 \dots i_k} \; \in \; {\mathcal C}^\infty\left(U,{\mathbb C}\right).
\eq
Let us stress that there is no point in singling out polynomial or rational functions: A polynomial or rational function
in one coordinate system will in general not be a polynomial or rational function in another coordinate system.

However, the situation changes if we consider algebraic varieties. 
Algebraic varieties are defined by polynomial equations and we may single out the variables defining the variety.
For an algebraic variety $X$ we may consider regular functions on $X$, these are functions which
we may write locally on an open set $U$ as a rational function such that the denominator polynomial 
is nowhere vanishing on $U$.
One denotes the regular functions on $X$ by ${\mathcal O}(X)$.
An 
\index{algebraic form}
{\bf algebraic form} on $X$ is given by
\bq
\label{chapter_motives:algebraic_form}
 \omega & = &
 \sum\limits_{i_1 < \dots < i_k} f_{i_1 \dots i_k}\left(x\right) dx^{i_1} \wedge \dots \wedge dx^{i_k},
 \;\;\;\;\;\;
 f_{i_1 \dots i_k} \; \in \; {\mathcal O}\left(X\right).
\eq
We denote the complex of algebraic forms on $X$ by $\Omega_{\mathrm{alg}}^\bullet(X)$.
Differentiation is defined in the usual way.

If ${\mathbb F}$ is a subfield of ${\mathbb C}$ and $X$ is defined over ${\mathbb F}$,
we denote by $X^{\mathrm{an}}$ the analytification of $X$ (see section~\ref{chapter_sector_decomposition:effective_periods}).
This means that the coordinates of the points in $X^{\mathrm{an}}$ may be complex (and not just in $\overline{\mathbb F}$)
and $X^{\mathrm{an}}$ is equipped with the standard topology (the one which is Hausdorff).

Algebraic de Rham cohomology is the cohomology obtained by restricting ourselves to the algebraic forms of eq.~(\ref{chapter_motives:algebraic_form}).
We denote algebraic de Rham cohomology groups by
\bq
 H_{\mathrm{alg \; dR}}^k\left(X\right).
\eq
Any cohomology class in $H_{\mathrm{alg \; dR}}^k(X)$ defines a cohomology class in $H_{{\mathrm dR}}^k(X^{\mathrm{an}})$
and we trivially have
\bq
 H_{\mathrm{alg \; dR}}^k\left(X\right) & \subseteq & H_{\mathrm{dR}}^k\left(X^{\mathrm{an}}\right).
\eq
Non-trivial is the following theorem \cite{Grothendieck:1966}:
\begin{theorem}
\label{chapter_motives:theorem_Grothendieck}
\bq
 H_{\mathrm{alg \; dR}}^k\left(X\right) \otimes {\mathbb C}
 & = & H_{\mathrm{dR}}^k\left(X^{\mathrm{an}}\right).
\eq
\end{theorem}
This theorem states that every cohomology class of $H_{{\mathrm dR}}^k(X^{\mathrm{an}})$ has a representative as an algebraic
cohomology class.
It is therefore sufficient to consider just $H_{\mathrm{alg \; dR}}^k(X)$.
This allows us to work just with rational functions.

Let's look at an example. We consider
\bq
 x y - 1 & \in & {\mathbb Q}\left[x,y\right].
\eq
This defines a variety over ${\mathbb Q}$:
\bq
 X & = & \left\{ \, \left(x,y\right) \in \overline{\mathbb Q}^2 \; | \; x y - 1 = 0 \, \right\}.
\eq
For $x$ we may choose any algebraic number not equal to zero, $y$ is then given by $y=1/x$.
Hence $X$ is isomorphic to
\bq
 X & \cong & \overline{\mathbb Q} \backslash \{0\}.
\eq
The analytification of $X$ is given
by
\bq
 X^{\mathrm{an}} & \cong & {\mathbb C} \backslash \{0\} \; = \; {\mathbb C}^\ast. 
\eq
We have
\bq
 \Omega_{\mathrm{alg}}^0\left(X\right)
 & = & 
 {\mathbb Q}\left[x,y\right]/\left\langle xy-1 \right\rangle 
 \; = \; 
 {\mathbb Q}\left[x,\frac{1}{x}\right],
 \nonumber \\
 \Omega_{\mathrm{alg}}^1\left(X\right)
 & = & 
 {\mathbb Q}\left[x,\frac{1}{x}\right] \cdot dx,
 \nonumber \\
 \Omega_{\mathrm{alg}}^2\left(X\right)
 & = & 
 0.
\eq
An algebraic one-form is a sum of terms $x^n \cdot dx$ with $n \in {\mathbb Z}$. The only term which is not exact is the one with $n=-1$.
Hence, $H_{\mathrm{alg \; dR}}^1(X)$ is generated by $dx/x$.

There is a generalisation of theorem~\ref{chapter_motives:theorem_Grothendieck} to the relative case:
\begin{theorem}
\label{chapter_motives:theorem_Grothendieck_relative}
\bq
 H_{\mathrm{alg \; dR}}^k\left(X,Y\right) \otimes {\mathbb C}
 & = & H_{\mathrm{dR}}^k\left(X^{\mathrm{an}},Y^{\mathrm{an}}\right).
\eq
\end{theorem}
As $Y$ is allowed to intersect itself, the definition of $H_{\mathrm{alg \; dR}}^k(X,Y)$ is more involved.
We illustrate here the construction of the relative algebraic de Rham cohomology groups 
for the case where $X$ is a smooth affine algebraic
variety and $Y$ a simple normal crossing divisor.
For the general case, where $X$ is a smooth variety and $Y$ a closed subvariety we refer to the book by
Huber and M\"uller-Stach \cite{Huber_book}.

A codimension $1$ closed subvariety $Y \subset X$ is called a 
\index{normal crossing divisor}
{\bf normal crossing divisor}, if for every
point $x \in Y$ there is an open neighbourhood $U \subseteq X$ of $x$ with local coordinates
$x_1, \dots, x_n$ such that $Y$ is locally given by
\bq
\label{chapter_motives:coordinate_system_normal_crossing}
 x_1 \cdot x_2 \cdot \ldots \cdot x_k & = & 0,
 \;\;\;\;\;\;
 \mbox{for some} \; 1 \le k \le n.
\eq
$Y$ is called a 
\index{simple normal crossing divisor}
{\bf simple normal crossing divisor} if in addition the irreducible components of $Y$ are smooth.
In other words, $Y$ looks locally like a union of coordinate hyperplanes.

Let us assume that $Y$ is a union of $r$ irreducible components $Y_i$:
\bq
 Y & = & Y_1 \cup Y_2 \cup \dots \cup Y_r.
\eq
Let $I \subseteq \{1,\dots,r\}$. We set
\bq
 Y_I \; = \; \bigcap\limits_{i \in I} Y_i,
 & &
 Y^p \; = \;
 \left\{
  \begin{array}{ll}
    X, & p \; = \; 0, \\
    \bigsqcup\limits_{\left|I\right|=p} Y_I, & p \ge 1. \\
  \end{array}
 \right.
\eq
Given a subset $I=\{i_0,\dots,i_p\} \subseteq \{1,\dots,r\}$ with $(p+1)$ elements, we define $I_l$ to be the subset with
$p$ elements, obtained from $I$ by removing the $l$-th element:
\bq
 I_l & = & \left\{ i_0, \dots, \hat{i}_l, \dots, i_p \right\}.
\eq 
As $Y_{I_l}$ is the intersection of $p$ irreducible components and $Y_I$ is the intersection of $Y_{I_l}$ with $Y_{i_l}$
we have
\bq
 Y_I &\subset & Y_{I_l}.
\eq
With these definitions we may now look at the double complex $K^{p,q} = \Omega_{\mathrm{alg}}^q(Y^p)$.
In the double complex $K^{\bullet,\bullet}$ we have two differentials: 
The first one is defined by
\bq
 d_{\mathrm{vertical}} & : & K^{p,q} \rightarrow K^{p,q+1}, 
 \nonumber \\
 & & 
 d_{\mathrm{vertical}} \; = \; \left(-1\right)^p d,
\eq
where $d$ is the ordinary exterior derivative. The alternating sign is required to turn $K^{p,q}$ into a complex.
The second one is given by
\bq
 d_{\mathrm{horizontal}} & : & K^{p,q} \rightarrow K^{p+1,q}, 
 \nonumber \\
 & & 
 d_{\mathrm{horizontal}} \; = \; \bigoplus\limits_{\left|I\right|=p+1} \bigoplus\limits_{l=0}^p \left(-1\right)^l r_{I_l I},
\eq
and
\bq
 r_{I_l I} & : & \Omega_{\mathrm{alg}}^q\left(Y_{I_l}\right) \rightarrow \Omega_{\mathrm{alg}}^q\left(Y_I\right)
\eq
is the restriction map.
In the next step one considers the total complex
\bq
 \Omega_{\mathrm{alg}}^n\left(X,Y\right) & = & 
 \bigoplus\limits_{p+q=n} K^{p,q}
\eq
together with the differential
\bq
 d_{\mathrm{total}} & = & d_{\mathrm{vertical}} + d_{\mathrm{horizontal}}.
\eq
The relative algebraic de Rham cohomology group $H^k_{\mathrm{alg \; dR}}(X,Y)$ is then the $k$-th cohomology group of 
$\Omega_{\mathrm{alg}}^k(X,Y)$ with respect to $d_{\mathrm{total}}$.

We may visualise the double complex $K^{p,q}$ as follows:
\bq
\begin{CD}
 \dots && \dots && \dots \\
 @A{d_{\mathrm{vertical}}}AA @A{d_{\mathrm{vertical}}}AA @A{d_{\mathrm{vertical}}}AA \\
\Omega_{\mathrm{alg}}^2\left(X\right) @>{d_{\mathrm{horizontal}}}>> \bigoplus\limits_{i} \Omega_{\mathrm{alg}}^2\left(Y_i\right) @>{d_{\mathrm{horizontal}}}>> \bigoplus\limits_{i < j}\Omega_{\mathrm{alg}}^2\left(Y_i \cap Y_j \right) @>{d_{\mathrm{horizontal}}}>> \dots \\
 @A{d_{\mathrm{vertical}}}AA @A{d_{\mathrm{vertical}}}AA @A{d_{\mathrm{vertical}}}AA \\
\Omega_{\mathrm{alg}}^1\left(X\right) @>{d_{\mathrm{horizontal}}}>> \bigoplus\limits_{i} \Omega_{\mathrm{alg}}^1\left(Y_i\right) @>{d_{\mathrm{horizontal}}}>> \bigoplus\limits_{i < j}\Omega_{\mathrm{alg}}^1\left(Y_i \cap Y_j \right) @>{d_{\mathrm{horizontal}}}>> \dots \\
 @A{d_{\mathrm{vertical}}}AA @A{d_{\mathrm{vertical}}}AA @A{d_{\mathrm{vertical}}}AA \\
\Omega_{\mathrm{alg}}^0\left(X\right) @>{d_{\mathrm{horizontal}}}>> \bigoplus\limits_{i} \Omega_{\mathrm{alg}}^0\left(Y_i\right) @>{d_{\mathrm{horizontal}}}>> \bigoplus\limits_{i < j}\Omega_{\mathrm{alg}}^0\left(Y_i \cap Y_j \right) @>{d_{\mathrm{horizontal}}}>> \dots \\
\end{CD}
\eq
The total complex $\Omega_{\mathrm{alg}}^{\bullet}(X,Y)$ is obtained by summing up the diagonals.
For example
\bq
 \Omega_{\mathrm{alg}}^2\left(X,Y\right)
 & = &
 \Omega_{\mathrm{alg}}^2\left(X\right)
 \oplus
 \left( \bigoplus\limits_{i} \Omega_{\mathrm{alg}}^1\left(Y_i\right) \right)
 \oplus
 \left( \bigoplus\limits_{i < j}\Omega_{\mathrm{alg}}^0\left(Y_i \cap Y_j \right) \right).
\eq
The construction above works the same way if we drop the restriction to algebraic forms.
It is instructive to see how this definition reduces to the one we had previously in section~\ref{chapter_motives:de_Rham_cohomology}.
To this aim we consider the case where $Y$ has only one irreducible component without self-intersections.
In this case
\bq
 \Omega^k\left(X,Y\right)
 & = &
 \Omega^k\left(X\right)
 \oplus
 \Omega^{k-1}\left(Y\right).
\eq
We may therefore write an element in 
$\Omega^k(X,Y)$ as a pair $(\varphi,\xi)$ with 
$\varphi \in \Omega^k(X)$ and $\xi \in \Omega^{k-1}(Y)$.
The differential $d_{\mathrm{total}}$ works out as
\bq
 d_{\mathrm{total}} \left(\varphi,\xi\right)
 & = &
 \left( d\varphi, \iota^\ast \varphi - d \xi \right),
\eq
where $\iota : Y \rightarrow X$ denotes the inclusion.
In~\ref{chapter_motives:de_Rham_cohomology} we defined the relative de Rham cohomology as equivalence
classes of closed differential forms, which vanish on $Y$.
Let us denote a representative of such a class by $\omega$. We would like to show that the two definitions are equivalent:
Any $\omega \in H^k_{\mathrm{dR}}(X,Y)$ (according to the definition in section~\ref{chapter_motives:de_Rham_cohomology}) defines
\bq
 \left(\varphi, \xi \right) & = & \left(\omega, 0 \right).
\eq
$\omega$ is closed and vanishes on $Y$, hence $\iota^\ast \omega = 0$.
We therefore have
\bq
 d_{\mathrm{total}} \left(\omega,0\right)
 & = &
 \left( d\omega, \iota^\ast \omega \right) \; = \; \left(0,0\right).
\eq
On the other hand, any $(\varphi,\xi) \in H^k_{\mathrm{dR}}(X,Y)$ (according to the definition of this sub-section)
defines
\bq
 \omega & = & \varphi - d \left( h \pi^\ast \xi \right).
\eq
This requires some explanation. We start from a tubular neighbourhood $T$ of $Y$ in $X$ and denote
by $\pi : T \rightarrow Y$ the projection.
Furthermore $h$ is a bump function. Let $T_1$ and $T_2$ be two further tubular neighbourhoods of $Y$ in $X$ with
$T_1 \subset T_2 \subset T$.
The bump function $h$ is equal to $1$ on $T_1$ and equal to zero on $X\backslash T_2$.
One may show that the cohomology class defined by $\omega$ is independent of the choices made for $T$ and $h$.
If $(\varphi,\xi)$ is closed, we have $d\varphi=0$ and $\iota^\ast \varphi = d \xi$.
It follows that $\omega$ is closed ($d\omega=d\varphi=0$) and vanishes when restricted to $Y$:
\bq
 \iota^\ast \omega
 & = &
 \iota^\ast \varphi - \iota^\ast d \left( h \pi^\ast \xi \right)
 \; = \; 
 d \xi - d\xi \; = \; 0.
\eq
Therefore $\omega$ defines a cohomology class according to the definition of section~\ref{chapter_motives:de_Rham_cohomology}.

\section{Categories}
\label{chapter_motives:categories}

In mathematics it is quite common to consider for example vector spaces together with linear maps 
between vector spaces
(i.e. vector space homomorphisms)
or groups together with group homomorphisms.
There are proofs which work the same way for vector spaces as they do for groups.
It is therefore useful to introduce another layer of abstraction, where the common properties are treated
in a unified way. This brings us to category theory.

Let us expand on the examples of vector spaces and groups:
We may view the vector spaces (or groups) as objects 
and we denote morphisms between two objects $X$ and $Y$ (with $X$ as source and $Y$ as target)
by an arrow $X \stackrel{\alpha}{\rightarrow} Y$.
\begin{tcolorbox}[breakable]
A 
\index{category}
{\bf category} 
consists of
\begin{itemize}

\item a class of {\bf objects} denoted by 
$\gls{Objectscategory}({\mathcal C})$,

\item for every pair $X,Y \in \Obj({\mathcal C})$ 
a class of {\bf morphisms} $X \stackrel{\alpha}{\rightarrow} Y$
from $X$ to $Y$ denoted by 
$\gls{Morphismscategory}_{{\mathcal C}}(X,Y)$
or simply $\Hom(X,Y)$ if no confusion arises,

\item for every ordered triple of objects $X,Y,Z$ a map from
$\Hom(X,Y) \times \Hom(Y,Z)$ to $\Hom(X,Z)$ called {\bf composition}, 
the composition of $\alpha \in \Hom(X,Y)$ with $\beta \in \Hom(Y,Z)$ is denoted by
$\beta \circ \alpha$,

\item for $\alpha \in \Hom(W,X)$, $\beta \in \Hom(X,Y)$  and $\gamma \in \Hom(Y,Z)$ 
we have the {\bf associativity} law 
$\gamma \circ \left( \beta \circ \alpha \right) = \left( \gamma \circ \beta \right) \circ \alpha$,
\item an {\bf identity morphism} $\mathrm{id} : X \rightarrow X$
for every object $X$, such that
$\alpha \circ \mathrm{id}_X = \alpha$ and $\mathrm{id}_Y \circ \alpha = \alpha$ for any $\alpha \in \Hom(X,Y)$.
\end{itemize}
\end{tcolorbox}
A morphism $\alpha \in \Hom(X,Y)$ is called an 
\index{isomorphism}
{\bf isomorphism} if there exists a morphism $\alpha^{-1} \in \Hom(Y,X)$ such that
\bq
 \alpha^{-1} \circ \alpha \; = \; \mathrm{id}_X
 & \mbox{and} &
 \alpha \circ \alpha^{-1} \; = \; \mathrm{id}_Y.
\eq
Examples of categories are 
the category of sets, denoted by 
$\gls{CategorySet}$,
where the objects are sets and the morphisms are maps between sets,
the category of groups, denoted by 
$\gls{CategoryGrp}$,
where the objects are groups and the morphisms are group homomorphisms
and 
the category of finite-dimensional ${\mathbb F}$-vector spaces, denoted 
$\gls{CategoryVectF}$,
where the objects
are finite-dimensional vector space over the field ${\mathbb F}$ and the morphisms are linear maps between vector spaces.
A less standard example of a category is a quiver, where every vertex has a self-loop attached to it and for any two oriented edges
$v_i \rightarrow v_j$ and $v_j \rightarrow v_k$ there is an oriented edge $v_i \rightarrow v_k$, subject to the associativity law.
The objects are the vertices and the morphisms are the oriented edges.

In the definition of a category we wrote ``class of objects'' and ``class of morphisms''. 
This wording avoids Russell's ``set of all sets''-contradiction.
A category ${\mathcal C}$ is called a 
\index{small category}
{\bf small category} if the class of all morphisms of ${\mathcal C}$ is a set.
(This implies automatically that $\Obj({\mathcal C})$ is a set too, as objects are in one-to-one
correspondence with the identity maps.)
A category ${\mathcal C}$ is called a 
\index{locally small category}
{\bf locally small category} if for each $X,Y \in \Obj({\mathcal C})$
the class $\Hom(X,Y)$ is a set.

Given a category ${\mathcal C}$ the 
\index{dual category}
{\bf dual category} 
${\mathcal C}^\ast$ is given by
\bq
 \Obj\left({\mathcal C}^\ast\right)
 & = &
 \Obj\left({\mathcal C}\right),
 \nonumber \\
 \Hom_{{\mathcal C}^\ast}(X,Y)
 & = & 
 \Hom_{{\mathcal C}}(Y,X).
\eq
In other words, the dual category ${\mathcal C}^{\ast}$ is
obtained from the category ${\mathcal C}$ by reversing all arrows.
 
Maps between categories which preserve composition and identities are called functors.
In detail:
\begin{tcolorbox}
A 
\index{covariant functor}
{\bf covariant functor} $T : \mathcal{C} \rightarrow \mathcal{D}$ from a category $\mathcal{C}$ to a category $\mathcal{D}$
consists of
\begin{itemize}
\item a map $T : \Obj({\mathcal C}) \rightarrow \Obj({\mathcal D})$, and
\item maps $T=T_{XY} : \Hom(X,Y) \rightarrow \Hom(TX,TY)$, which preserve
composition and identities, i.e. 
\begin{align}
\label{chapter_motives:def_covariant_functor}
 & T(\beta \circ \alpha) = (T\beta) \circ (T \alpha)
 & &
 \mbox{for all morphisms $X \stackrel{\alpha}{\rightarrow} Y \stackrel{\beta}{\rightarrow} Z$ in ${\mathcal C}$},
 \nonumber \\
 & T(\mathrm{id}_X) = \mathrm{id}_{TX}
 & &
 \mbox{for all $X \in \Obj({\mathcal C})$}.
\end{align}
\end{itemize}
\end{tcolorbox}
If no confusion arises we will simply write $T$ for $T_{XY}$ (as we did already above).

A functor is called 
\index{faithful functor}
{\bf faithful}, if for any $X,Y \in \Obj({\mathcal C})$ the map
\bq
 T_{XY} & : & \Hom_{{\mathcal C}}(X,Y) \rightarrow \Hom_{{\mathcal D}}(TX,TY)
\eq 
is injective, the functor is called
\index{full functor}
{\bf full}, if the map is surjective
and the functor is called
\index{fully faithful functor}
{\bf fully faithful}, if the map is bijective.

There is also the notion of a contravariant functor: A
contravariant functor $T : {\mathcal C} \rightarrow {\mathcal D}$ is a covariant functor 
${\mathcal C} \rightarrow {\mathcal D}^\ast$ (or equivalently 
a covariant functor 
${\mathcal C}^\ast \rightarrow {\mathcal D}$).
Spelled out as in eq.~(\ref{chapter_motives:def_covariant_functor}) we have
\begin{tcolorbox}
A 
\index{contravariant functor}
{\bf contravariant functor} $T : \mathcal{C} \rightarrow \mathcal{D}$ from a category $\mathcal{C}$ to a category $\mathcal{D}$
consists of
\begin{itemize}
\item a map $T : \Obj({\mathcal C}) \rightarrow \Obj({\mathcal D})$, and
\item maps $T=T_{XY} : \Hom(X,Y) \rightarrow \Hom(TY,TX)$, which preserve
composition and identities, i.e. 
\begin{align}
\label{chapter_motives:def_contravariant_functor}
 & T(\beta \circ \alpha) = (T\alpha) \circ (T \beta)
 & &
 \mbox{for all morphisms $X \stackrel{\alpha}{\rightarrow} Y \stackrel{\beta}{\rightarrow} Z$ in ${\mathcal C}$},
 \nonumber \\
 & T(\mathrm{id}_X) = \mathrm{id}_{TX}
 & &
 \mbox{for all $X \in \Obj({\mathcal C})$}.
\end{align}
\end{itemize}
A contravariant functor reverses all arrows.
\end{tcolorbox}
A 
\index{natural transformation of functors}
{\bf natural transformation} 
$\Phi : T_1 \rightarrow T_2$ between two functors 
$T_1 : {\mathcal C}_1 \rightarrow {\mathcal C}_2$ 
and 
$T_2 : {\mathcal C}_1 \rightarrow {\mathcal C}_2$
(between the same categories)
is a map that assigns to each object $X \in \Obj({\mathcal C_1})$
a morphism $\Phi_X \in \Hom_{{\mathcal C}_2}(T_1(X),T_2(X))$ such that for
any morphism $\alpha \in \Hom_{{\mathcal C}_1}(X,Y)$
\bq
 \Phi_Y \circ T_1(\alpha) & = & T_2(\alpha) \circ \Phi_X.
\eq
In terms of a commutative diagram:
\bq
 \begin{CD}
   T_1\left(X\right) @>{\Phi_X}>> T_2\left(X\right) \\
   @V{T_1\left(\alpha\right)}VV  @VV{T_2\left(\alpha\right)}V \\
   T_1\left(Y\right) @>{\Phi_Y}>> T_2\left(Y\right) \\
 \end{CD}
\eq
A natural transformation $\Phi$ is called a 
\index{natural equivalence of functors}
{\bf natural equivalence of functors}
if each map $\Phi_X$ is an isomorphism.
$\Phi_X$ is called a 
\index{functorial isomorphism}
{\bf functorial isomorphism}.

For a category ${\mathcal C}$ we denote by $I_{\mathcal C} : {\mathcal C} \rightarrow {\mathcal C}$ the identity functor, which assigns 
each object and morphism to itself. 
Two categories ${\mathcal C}_1$ and ${\mathcal C}_2$ are called 
\index{equivalent categories}
{\bf equivalent categories}, 
if there exists two functors 
$T_{21} : {\mathcal C}_1 \rightarrow {\mathcal C}_2$
and
$T_{12} : {\mathcal C}_2 \rightarrow {\mathcal C}_1$
such that $T_{12} \circ T_{21} : {\mathcal C}_1 \rightarrow {\mathcal C}_1$ and $I_{{\mathcal C}_1} : {\mathcal C}_1 \rightarrow {\mathcal C}_1$ are natural equivalent
as well as $T_{21} \circ T_{12} : {\mathcal C}_2 \rightarrow {\mathcal C}_2$ and $I_{{\mathcal C}_2} : {\mathcal C}_2 \rightarrow {\mathcal C}_2$.

We are interested in categories, which have additional structures.
We therefore introduce various specialisations: Monoidal categories,
tensor categories, Abelian categories and finally Tannakian categories.
It is the last one (Tannakian categories) which is most relevant to us.

\subsection{Monoidal categories}

A monoidal category ${\mathcal C}$ is a category
equipped with a functor from ${\mathcal C} \times {\mathcal C}$ into ${\mathcal C}$,
denoted by $\otimes$, a unit object ${\bf 1}$ and an isomorphism $\iota : {\bf 1} \otimes {\bf 1} \rightarrow {\bf 1}$, 
subject to the following constraints:

For any three objects $X,Y,Z \in \Obj({\mathcal C})$
there is a functorial isomorphism 
\bq
 \Phi_{X,Y,Z} & : & X \otimes ( Y \otimes Z) \rightarrow ( X \otimes Y ) \otimes Z
\eq
satisfying the pentagon identity:
\bq
 (\Phi_{X,Y,Z} \otimes \mathrm{id}) \circ \Phi_{X,Y \otimes Z, W} \circ (\mathrm{id} \otimes \Phi_{Y,Z,W} )
 & = &
 \Phi_{X \otimes Y, Z,W} \circ \Phi_{X, Y, Z \otimes W}.
\eq
$\Phi_{X,Y,Z}$ is also called an 
\index{associativity constraint in a monoidal category}
{\bf associativity constraint}.
The pentagon identity reads in terms of a commutative diagram
\bq
 \begin{CD}
  X \otimes \left( Y \otimes \left( Z \otimes W \right) \right) 
   @>{\Phi_{X, Y, Z \otimes W}}>> 
 \left( X \otimes Y \right) \otimes \left( Z \otimes W \right)
   @>{\Phi_{X \otimes Y, Z,W}}>>
 \left( \left( X \otimes Y \right) \otimes Z \right) \otimes W 
 \\
 @VV{\mathrm{id}_X \otimes \Phi_{Y,Z,W}}V & & @AA{\Phi_{X,Y,Z} \otimes \mathrm{id}_W}A
 \\
 X \otimes \left( \left( Y \otimes Z \right) \otimes W \right)
 &
  @>{\Phi_{X,Y \otimes Z, W}}>>
 &
 \left( X \otimes \left( Y \otimes Z \right) \right)\otimes W
 \\
 \end{CD}
\eq
For the unit object ${\bf 1}$ one requires that there are functorial isomorphisms
\bq
 l_X \; : \; {\bf 1} \otimes X \; \rightarrow \; X
 & \mbox{and} &
 r_X \; : \; X \otimes {\bf 1} \; \rightarrow \; X,
\eq
satisfying the triangle diagram
\bq
\label{chapter_motives:monoidal_category_triangle_diagram}
\begin{tikzcd}
 X \otimes \left({\bf 1} \otimes Y \right) \arrow[rr, "\Phi_{X,{\bf 1},Y}"] \arrow[rd, "\mathrm{id}_X \otimes l_Y"] & & \left(X \otimes {\bf 1} \right) \otimes Y \arrow[ld, "r_X \otimes \mathrm{id}_Y"] \\
 & X \otimes Y & \\
\end{tikzcd}
\eq
Instead of eq.~(\ref{chapter_motives:monoidal_category_triangle_diagram}) we may require that the functors
$L_{\bf 1} : {\mathcal C} \rightarrow {\mathcal C}$ and 
$R_{\bf 1} : {\mathcal C} \rightarrow {\mathcal C}$ acting on the objects of ${\mathcal C}$ as
\bq
 L_{\bf 1} & : & X \; \rightarrow {\bf 1} \otimes X,
 \nonumber \\
 R_{\bf 1} & : & X \; \rightarrow X \otimes {\bf 1}
\eq
are autoequivalences of ${\mathcal C}$ \cite{Etingof_book}.
We denote a monoidal category by $({\mathcal C},\otimes,\Phi,{\bf 1},\iota)$.
The functorial isomorphisms $l_X$ and $r_X$ are related to this data as
\bq
 L_{\bf 1}\left(l_X\right) & = & \left( \iota \otimes {\mathrm id}_X \right) \circ \phi_{{\bf 1}, {\bf 1}, X},
 \nonumber \\
 R_{\bf 1}\left(r_X\right) & = & \left( {\mathrm id}_X \otimes \iota \right) \circ \phi^{-1}_{X, {\bf 1}, {\bf 1}}.
\eq
One of the simplest examples of a monoidal category is $\mathrm{\bf Vect}_{{\mathbb F}}$,
the category of finite-dimensional vector spaces over the field ${\mathbb F}$.
The unit object ${\bf 1}$ in this category is the one-dimensional vector space isomorphic to ${\mathbb F}$.

A monoidal category ${\mathcal C}$ is called
\index{rigid monoidal category}
(left) {\bf rigid} (or has left duals), if for each
object $X$ there is an object $X^{\ast}$ and morphisms
$\mathrm{ev}_X : X^{\ast} \otimes X \rightarrow {\bf 1}$,
$\mathrm{coev}_X : {\bf 1} \rightarrow X \otimes X^{\ast}$ such that
\bq
 \begin{CD}
 X
 @>{\mathrm{coev}_X}>>
 (X \otimes X^{\ast}) \otimes X
 @>{\Phi^{-1}_{X,X^\ast,X}}>>
 X \otimes ( X^{\ast} \otimes X )
 @>{\mathrm{ev}_X}>>
 X,
 \\
 X^{\ast}
 @>{\mathrm{coev}_X}>>
 X^{\ast} \otimes (X \otimes X^{\ast})
 @>{\Phi_{X^\ast,X,X^\ast}}>>
 (X^{\ast} \otimes X) \otimes X
 @>{\mathrm{ev}_X}>>
 X^{\ast}
 \end{CD}
\eq
compose to $\mathrm{id}_X$ and $\mathrm{id}_{X^{\ast}}$, respectively.

\subsection{Tensor categories}

A tensor category is a monoidal category such that the two functors
${\mathcal C} \times {\mathcal C} \rightarrow {\mathcal C}$ given by
$(X, Y) \rightarrow X \otimes Y$ and $(X,Y) \rightarrow Y \otimes X$
are naturally equivalent, i.e. there exists a functorial isomorphism
\bq
 \Psi_{X,Y} & : & X \otimes Y \rightarrow Y \otimes X.
\eq 
The functional isomorphism $\Psi_{X,Y}$ is required to satisfy
\bq
 \Psi_{Y,X} \circ \Psi_{X,Y} & = & \mathrm{id}_{X \otimes Y}
\eq
and the hexagon identity
\bq
 (\Psi_{X,Z} \otimes \mathrm{id}) \circ \Phi_{X,Z,Y} \circ (\mathrm{id} \otimes \Psi_{Y,Z})
 & = &
 \Phi_{Z,X,Y} \circ \Psi_{X \otimes Y, Z} \circ \Phi_{X,Y,Z}.
\eq
$\Psi_{X,Y}$ is also called a 
\index{commutativity constraint in a tensor category}
{\bf commutativity constraint}.
The hexagon identity ensures that the associativity constraint and the commutativity constraint
are compatible.
In terms of a commutative diagram we have
\bq
 \begin{CD}
  X \otimes \left( Y \otimes Z \right)
   @>{\Phi_{X,Y,Z}}>>
  \left( X \otimes Y \right) \otimes Z
   @>{\Psi_{X \otimes Y, Z}}>>
  Z \otimes \left( X \otimes Y \right)
  \\
  @VV{\mathrm{id}_X \otimes \Psi_{Y,Z}}V & & @VV{\Phi_{Z,X,Y}}V
  \\
  X \otimes \left( Z \otimes Y \right)
   @>{\Phi_{X,Z,Y}}>>
  \left( X \otimes Z \right) \otimes Y
  @>{\Psi_{X,Z} \otimes \mathrm{id}_Y}>>
  \left( Z \otimes X \right) \otimes Y
  \\
 \end{CD}
\eq
Let ${\mathcal C}$ and ${\mathcal C}'$ be two tensor categories. 
A 
\index{tensor functor}
{\bf tensor functor}
is a pair $(T,c)$ consisting of a functor $T : {\mathcal C} \rightarrow {\mathcal C}'$ and 
functorial isomorphisms $c_{X,Y} : T(X) \otimes T(Y) \rightarrow T(X \otimes Y)$ satisfying
\begin{enumerate}
\item for all $X, Y, Z \in \Obj({\mathcal C})$ the following diagram commutes:
\bq
 \begin{CD}
  T\left(X\right) \otimes \left( T\left(Y\right) \otimes T\left(Z\right) \right)
   @>{\mathrm{id}_{T(X)} \otimes c_{Y,Z}}>>
  T\left(X\right) \otimes T\left( Y \otimes Z \right)
   @>{c_{X,Y \otimes Z}}>>
  T\left(X \otimes \left( Y \otimes Z \right) \right)
  \\
  @VV{\Phi'_{T(X),T(Y),T(Z)}}V & & @VV{T(\Phi_{X,Y,Z})}V
  \\
  \left( T\left(X\right) \otimes T\left(Y\right) \right) \otimes T\left(Z\right)
   @>{c_{X,Y} \otimes\mathrm{id}_{T(Z)}}>>
  T\left(X \otimes Y\right) \otimes T\left(Z\right)
  @>{c_{X \otimes Y, Z}}>>
  T\left( \left(X \otimes Y \right) \otimes Z \right)
  \\
 \end{CD}
 \nonumber \\
\eq

\item for all $X, Y \in \Obj({\mathcal C})$ the following diagram commutes:
\bq
 \begin{CD}
  T\left(X\right) \otimes T\left(Y\right)
   @>{c_{X,Y}}>>
  T\left( X \otimes Y \right)
  \\
  @VV{\Psi'_{T(X),T(Y)}}V @VV{T(\Psi_{X,Y})}V
  \\
  T\left(Y\right) \otimes T\left(X\right)
   @>{c_{Y,X}}>>
  T\left(Y \otimes X\right)
  \\
 \end{CD}
\eq

\item we have ${\bf 1}' = T({\bf 1})$ and $\iota' = T(\iota)$ up to a unique isomorphism.

\end{enumerate}

\subsection{Abelian categories}

An 
\index{additive category}
{\bf additive category} ${\mathcal C}$ is a locally small category where
\begin{enumerate}
\item every set $\Hom_{\mathcal C}(X,Y)$ is an Abelian group (written additively) such that for
$\alpha_1, \alpha_2 \in \Hom(X,Y)$ and $\beta_1, \beta_2 \in \Hom(Y,Z)$ we have
\bq
 \left( \beta_1 + \beta_2 \right) \circ \left( \alpha_1 + \alpha_2 \right)
 & = & 
 \beta_1 \circ \alpha_1
 + 
 \beta_1 \circ \alpha_2
 + 
 \beta_2 \circ \alpha_1
 + 
 \beta_2 \circ \alpha_2.
\eq
\item There exists a zero object ${\bf 0} \in \Obj({\mathcal C})$ such that $\Hom_{\mathcal C}({\bf 0}, {\bf 0}) = 0$.
\item For all objects $X_1, X_2 \in \Obj({\mathcal C})$ there exists an object $Y \in \Obj({\mathcal C})$
and morphisms $i_1 \in \Hom(X_1,Y)$, $i_2 \in \Hom(X_2,Y)$, $p_1 \in \Hom(Y,X_1)$ and $p_2 \in \Hom(Y,X_2)$
such that
\bq
 p_1 \circ i_1 \; = \; \mathrm{id}_{X_1},
 \;\;\;\;\;\;
 p_2 \circ i_2 \; = \; \mathrm{id}_{X_2},
 \;\;\;\;\;\;
 i_1 \circ p_1 + i_2 \circ p_2 \; = \; \mathrm{id}_{Y}.
\eq
\end{enumerate}
It can be shown that the object $Y$ is unique up to a unique isomorphism.
One denotes this element as $Y = X_1 \oplus X_2$ and calls it the 
\index{direct sum}
{\bf direct sum} 
of $X_1$ and $X_2$.
This defines a bifunctor $\oplus : {\mathcal C} \times {\mathcal C} \rightarrow {\mathcal C}$.

One of the simplest examples of an additive category is again $\mathrm{\bf Vect}_{{\mathbb F}}$,
the category of finite-dimensional vector spaces over the field ${\mathbb F}$.
The zero object ${\bf 0}$ in this category is the zero-dimensional vector space consisting only of the zero vector.

Let ${\mathbb F}$ be a field.
An additive category ${\mathcal C}$ is said to be
\index{${\mathbb F}$-linear category}
{\bf ${\mathbb F}$-linear} if 
every set $\Hom_{\mathcal C}(X,Y)$ is a ${\mathbb F}$-vector space, such that the composition of morphisms is ${\mathbb F}$-linear.

Let us now turn to the definition of an Abelian category: 
An Abelian category is an additive category, where kernels and cokernels exist.
In order to define kernels and cokernels in an additive category, we have to do some gymnastics:

Let ${\mathcal C}$ be an additive category and $\alpha \in \Hom_{\mathcal C}(X,Y)$ a morphism.
Suppose there exists an object $K \in \Obj({\mathcal C})$ and a morphism $k \in \Hom_{\mathcal C}(K,X)$ such that
$ \alpha \circ k = 0$ and if $k' \in \Hom_{\mathcal C}(K',X)$ is such that $\alpha \circ k' = 0$ then there exists a unique $l \in \Hom_{\mathcal C}(K',K)$ such that
$ k l = k'$.
The pair $(K,k)$ is called the 
\index{kernel of a morphism}
{\bf kernel} of $\alpha$ and denoted $\mathrm{Ker}(\alpha)$.

The cokernel is defined in a similar way:
Suppose there exists an object $C \in \Obj({\mathcal C})$ and a morphism $c \in \Hom_{\mathcal C}(Y,C)$ such that
$ c \circ \alpha = 0$ and if $c' \in \Hom_{\mathcal C}(Y,C')$ is such that $c' \circ \alpha = 0$ then there exists a unique $l \in \Hom_{\mathcal C}(C, C')$ such that
$ l c = c'$.
The pair $(C,c)$ is called the
\index{cokernel of a morphism} 
{\bf cokernel} of $\alpha$ and denoted $\mathrm{Coker}(\alpha)$.

An 
\index{Abelian category}
{\bf Abelian category} ${\mathcal C}$ is an additive category, where 
for every $\alpha \in \Hom_{\mathcal C}(X,Y)$ there exists a sequence
\bq
 K \stackrel{k}{\rightarrow} X \stackrel{i}{\rightarrow} I \stackrel{j}{\rightarrow} Y \stackrel{c}{\rightarrow} C
\eq
with $\alpha = j \circ i$ and 
\begin{alignat}{3}
 \mathrm{Ker}\left(\alpha\right) & = \left(K,k\right), & \hspace*{5mm} & & \mathrm{Coker}\left(\alpha\right) & = \left(C,c\right),
 \nonumber \\
 \mathrm{Ker}\left(c\right) & = \left(I,j\right), & & & \mathrm{Coker}\left(k\right) & = \left(I,i\right).
\end{alignat}
A functor is called 
\index{exact functor}
{\bf exact}, if it preserves short exact sequences, e.g.
\bq
 {\bf 0} \longrightarrow X \stackrel{\alpha}{\longrightarrow} Y \stackrel{\beta}{\longrightarrow} Z \longrightarrow {\bf 0}
\eq
implies
\bq
 {\bf 0} \longrightarrow T(X) \stackrel{T(\alpha)}{\longrightarrow} T(Y) \stackrel{T(\beta)}{\longrightarrow} T(Z) \longrightarrow {\bf 0}.
\eq

\subsection{Tannakian categories}

We now have all ingredients to define Tannakian categories.
As before, we denote by ${\mathbb F}$ a field 
and by ${\bf Vect}_{\mathbb F}$ the category of finite-dimensional 
${\mathbb F}$-vector spaces.

Let $R$ be a ring.
We denote by 
$\gls{CategoryModR}$
the category of finitely generated $R$-modules and by 
$\gls{CategoryProjR}$
the category of finitely generated projective $R$-modules.

Let ${\mathcal C}$ be a category. For $X \in \Obj({\mathcal C})$ we denote by 
$\mathrm{End}(X)=\Hom(X,X)$ the endomorphisms of $X$, and in particular $\mathrm{End}({\bf 1})=\Hom({\bf 1},{\bf 1})$.

A 
\index{neutral Tannakian category}
{\bf neutral Tannakian category} ${\mathcal C}$ over ${\mathbb F}$ is a rigid Abelian tensor category 
with $\mathrm{End}({\bf 1}) = {\mathbb F}$
and a ${\mathbb F}$-linear exact faithful tensor functor 
\bq
 \omega & : & {\mathcal C} \; \rightarrow \; {\bf Vect}_{\mathbb F},
\eq
called the 
\index{fibre functor}
{\bf fibre functor}.
We say that the fibre functor $\omega$ takes values in ${\mathbb F}$.

We may generalise the definition of a neutral Tannakian category towards a Tannakian category as follows:
Let $R$ be a non-zero ${\mathbb F}$-algebra.
We define a fibre functor on ${\mathcal C}$ which takes values in $R$ as a 
${\mathbb F}$-linear exact faithful tensor functor $\eta : {\mathcal C} \rightarrow \mathrm{\bf Mod}_R$ that takes
values in the subcategory $\mathrm{\bf Proj}_R$ of $\mathrm{\bf Mod}_R$.

With these preparations we finally arrive at the definition of a Tannakian category:
\begin{tcolorbox}
A 
\index{Tannakian category}
{\bf Tannakian category} ${\mathcal C}$ over ${\mathbb F}$ is a rigid Abelian tensor category 
with $\mathrm{End}({\bf 1}) = {\mathbb F}$
and a fibre functor with values in the ${\mathbb F}$-algebra $R$.
\end{tcolorbox}
Let's look at an example: 
Let $V$ be a finite-dimensional vector space and $\mathrm{GL}(V)$ the group of automorphisms of $V$.
For example $V={\mathbb C}^n$ and $\mathrm{GL}(V) = \mathrm{GL}(n,{\mathbb C})$.
Let $G$ be a group.
A representation of $G$ is a homomorphism $\rho : G \rightarrow \mathrm{GL}(V)$.
Since $\rho$ is a homomorphism we have
\bq
 \rho\left( g_1 g_2 \right) & = & \rho\left(g_1\right) \rho\left(g_2\right).
\eq
By abuse of notation we denote by $\rho$ also the map $\rho : G \times V \rightarrow V$,
$v \rightarrow \rho(g) v$.
Let's now fix the ground field to be ${\mathbb C}$.
The finite-dimensional representations of $G$ form a category, which we denote by
$\mathrm{\bf Rep}_{\mathbb C}(G)$.
The objects in this category are pairs $(\rho,V) \in \Obj(\mathrm{\bf Rep}_{\mathbb C}(G))$.
The morhpisms in this category are as follows:
Let $(\rho_1,V_1)$ and $(\rho_2,V_2)$ be two objects in $\mathrm{\bf Rep}_{\mathbb C}(G)$.
The class of morphisms $\Hom_{\mathrm{\bf Rep}_{\mathbb C}(G)}((\rho_1,V_1),(\rho_2,V_2))$ 
consists of maps $\alpha : V_1 \rightarrow V_2$
such that for all $g \in G$ the following diagram commutes:
\bq
 \begin{CD}
  G \times V_1
   @>{\rho_1}>>
  V_1
  \\
  @VV{\mathrm{id}_G \times \alpha}V @VV{\alpha}V
  \\
  G \times V_2
   @>{\rho_2}>>
  V_2
  \\
 \end{CD}
\eq
$\mathrm{\bf Rep}_{\mathbb C}(G)$ is a (neutral) Tannakian category.
The fibre functor is given by the forgetful functor, which associates 
to any object $(\rho,V) \in \Obj(\mathrm{\bf Rep}_{\mathbb C}(G))$ 
the object $V \in \Obj({\bf Vect}_{\mathbb C})$.

A second example is given by the category of mixed Hodge structures
$\gls{CategoryMHS}$,
discussed in section~\ref{chapter_motives:hodge_structures}.

\section{Motives}
\label{chapter_motives:motives}

Motives are a conjectured framework to unify different cohomology theories.
For an algebraic variety $X$ there is more than one cohomology theory.
Most relevant to us are de Rham cohomology or Betti cohomology.
Other cohomology theories are for example 
$l$-adic cohomology or crystalline cohomology.
Before outlining the main ideas behind motives, we have to introduce correspondences, which will play the role
of morphisms in the category of motives.

\begin{digression} {\bf Correspondences and adequate equivalence relations}
\\
Let $X$ be an algebraic variety. 
The group of cycles $Z(X)$ is the free Abelian group generated by the set of subvarieties of $X$.
We write a cycle as a formal linear combination
\bq
 Z & = & \sum\limits_j n_j Y_j,
\eq
where $n_j \in {\mathbb Z}$ and $Y_j$ a subvariety of $X$.

Let $X$ and $Y$ be two algebraic varieties.
A 
\index{correspondence}
{\bf correspondence} between $X$ and $Y$ is a cycle of $Z(X \times Y)$.

Two cycles $Z_1$ and $Z_2$ are called 
{\bf rational equivalent}, if there is a cycle $W$ on ${\mathbb P}^1 \times X$ (i.e. a correspondence between ${\mathbb P}^1$ and $X$) 
and $t_1, t_2 \in {\mathbb P}^1$
such that
\bq
 Z_1 - Z_2 & = & W \cap \left(\{t_1\} \times X\right) - W \cap \left(\{t_2\} \times X\right).
\eq
We write
\bq
 Z_1 & \sim_{\mathrm{rat}} & Z_2.
\eq
The group
\bq
 Z(X) / \sim_{\mathrm{rat}}
\eq
is called the 
\index{Chow group}
{\bf Chow group} of $X$.

Rational equivalence is the statement that we may interpolate between the cycles $Z_1$ and $Z_2$ with a parameter $t$, being the coordinate of a curve
of genus zero (i.e. ${\mathbb P}^1$). This can be generalised to curves of higher genus:
Two cycles $Z_1$ and $Z_2$ are called 
{\bf algebraic equivalent}, if there is an irreducible curve ${\mathcal C}$ and a cycle $W$ on ${\mathcal C} \times X$ and $t_1, t_2 \in {\mathcal C}$
such that
\bq
 Z_1 - Z_2 & = & W \cap \left(\{t_1\} \times X\right) - W \cap \left(\{t_2\} \times X\right).
\eq
We write
\bq
 Z_1 & \sim_{\mathrm{alg}} & Z_2.
\eq
Clearly, rational equivalence implies algebraic equivalence.

Two cycles $Z_1$ and $Z_2$ are called 
{\bf numerical equivalent}, if 
$\deg(Z_1 \cap W) = \deg(Z_2 \cap W)$ for any cycle with $\dim W = \mathrm{codim} \, Z$.
We write
\bq
 Z_1 & \sim_{\mathrm{num}} & Z_2.
\eq
$\deg(Z_1 \cap Z_2)$ is the intersection number of $Z_1$ and $Z_2$. If $\dim Z_1 = \mathrm{codim}\, Z_2$ and
if the intersection is a set of points (counted with multiplicities)
\bq
 Z_1 \cap Z_2 & = & \sum\limits_j n_j P_j,
 \;\;\;\;\;\; P_j \; \in \; X
\eq
one has
\bq
 \deg(Z_1 \cap Z_2) & = & \sum\limits_j n_j.
\eq
We have the implications
\bq
 \sim_{\mathrm{rat}}
 \;\;\; \Rightarrow \;\;\;
 \sim_{\mathrm{alg}}
 \;\;\; \Rightarrow \;\;\;
 \sim_{\mathrm{num}},
\eq
hence rational equivalence is the finest equivalence relation and numerical equivalence is the coarsest equivalence relation.
\end{digression}

Let us now give a short summary on the conjectured theory of motives:
Let us denote by 
$\gls{CategoryVarQ}$,
the category of algebraic varieties defined over ${\mathbb Q}$ and by 
$\gls{CategorySmProjQ}$
the subcategory of smooth projective varieties over ${\mathbb Q}$.
It is conjectured that there exists a Tannakian category of mixed motives 
$\gls{CategoryMixMot}$
and a functor 
\bq
 h & : & \mathrm{\bf Var}_{\mathbb Q} \rightarrow \mathrm{\bf MixMot},
\eq
such that the cohomologies $H_{\mathrm{dR}}$ and $H_B$ factor through $h$, e.g. there exist commutative diagrams
\bq
\begin{tikzcd}
 \mathrm{\bf Var}_{\mathbb Q} \arrow[r, "h"] \arrow[rd, "H_{\mathrm{dR}}"] & \mathrm{\bf MixMot} \arrow[d, "\eta_{\mathrm{dR}}"] \\
 & \mathrm{\bf Vect}_{\mathbb Q} \\
\end{tikzcd},
 & &
\begin{tikzcd}
 \mathrm{\bf Var}_{\mathbb Q} \arrow[r, "h"] \arrow[rd, "H_{B}"] & \mathrm{\bf MixMot} \arrow[d, "\eta_{B}"] \\
 & \mathrm{\bf Vect}_{\mathbb Q} \\
\end{tikzcd}.
\eq
$\eta_{\mathrm{dR}}$ and $\eta_{B}$ are fibre functors in $\mathrm{\bf MixMot}$.
If we just consider the subcategory of smooth projective varieties $\mathrm{\bf SmProj}_{\mathbb Q}$ one expects 
\bq
 h & : & \mathrm{\bf SmProj}_{\mathbb Q} \rightarrow \mathrm{\bf PureMot},
\eq
where 
$\gls{CategoryPureMot}$
denotes the subcategory of pure motives.

Morphisms in $\mathrm{\bf MixMot}$ are given by correspondences.

One further expects that motives extend to the relative setting, i.e. to any pair $(X,Y)$ with $X$ a smooth algebraic variety
and $Y$ a closed subvariety there is a motive $h(X,Y)$.

Motives are often studied through their Hodge realisation. The Hodge realisation is a functor
\bq
 \mathrm{\bf MixMot} & \rightarrow & \mathrm{\bf MHS},
\eq
where $\mathrm{\bf MHS}$ denotes the category of mixed Hodge structures (again a Tannakian category).
Restricted to the category of pure motives we have
\bq
 \mathrm{\bf PureMot} & \rightarrow & \mathrm{\bf HS},
\eq
where 
$\gls{CategoryHS}$
denotes the category of (pure) Hodge structures.

With this short interlude on motives we now leave the field of conjectural mathematics and return to solid grounds.
In the next section we introduce Hodge structures, followed by examples from Feynman integrals.

\section{Hodge structures}
\label{chapter_motives:hodge_structures}

Hodge structures have their origin in the study of compact K\"ahler manifolds \cite{Ballmann}.
Let $M$ be a complex manifold with complex structure $J$.
A Riemannian metric $g$ on $M$ is called Hermitian,
if it is compatible with the complex structure $J$, in other words
for vector fields $X, Y$ on $M$ we have
\bq
 g\left( J X, J Y \right)
 & = &
 g\left( X, Y \right).
\eq
For a Hermitian manifold one defines an associated differential two-form by
\bq
 K\left(X,Y\right) & = & g\left( J X, Y \right).
\eq
$K$ is called the K\"ahler form.
A Hermitian manifold is called a
\index{K\"ahler manifold} 
{\bf K\"ahler manifold}, if the two-form $K$ is closed:
\bq
 d K & = & 0.
\eq
Examples of compact K\"ahler manifolds are provided by compact Riemann surfaces.
Riemann surfaces are complex manifolds of complex dimension one 
and the K\"ahler form of any Hermitian metric is necessarily closed.
A second example is given by the complex projective space $\mathbb{P}^n(\mathbb{C})$.
As a third example we mention complex submanifolds of K\"ahler manifolds. These submanifolds are again K\"ahler.

On a compact K\"ahler manifold we have the following decomposition of the cohomology groups
\bq
 H^k\left(X\right) \otimes {\mathbb C}& = & \bigoplus\limits_{p+q=k} H^{p,q}(X),
 \;\;\;\;\;\;
 \overline{H^{p,q}(X)} = H^{q,p}(X).
\eq
For a fixed $k$ this provides an example of a pure Hodge structure of weight $k$.

\subsection{Pure Hodge structures}

Let us now define pure Hodge structures.
Let $V_{\mathbb Z}$ be a $\mathbb Z$-module of finite rank and
$V_{\mathbb C} = V_{\mathbb Z} \otimes_{\mathbb Z} {\mathbb C}$ its complexification.
\begin{tcolorbox}
A 
\index{pure Hodge structure}
{\bf pure Hodge structure} of weight $k$ on the $\mathbb Z$-module $V_{\mathbb Z}$
is a direct sum decomposition
\bq
 V_{\mathbb C} & = & 
 \bigoplus\limits_{p+q=k} V^{p,q}
 \;\;\;
 \mbox{with}
 \;\;\;
 \overline{V^{p,q}} = V^{q,p}.
\eq
\end{tcolorbox}
If one replaces $\mathbb Z$ by $\mathbb Q$ or $\mathbb R$, one speaks about a rational or real Hodge structure, respectively.
The bar in $\overline{V^{q,p}}$ denotes complex conjugation 
with respect to the real structure $V_{\mathbb C} = V_{\mathbb R} \otimes_{\mathbb R} {\mathbb C}$
(if we start from $V_{\mathbb Z}$ we have $V_{\mathbb R} = V_{\mathbb Z} \otimes_{\mathbb Z} {\mathbb R}$).

The numbers
\bq
 h^{p,q}(V) & = & \mbox{dim}\;V^{p,q}
\eq
are called the 
\index{Hodge numbers}
{\bf Hodge numbers}.

There is a second definition of a pure Hodge structure, which is more adapted for generalisations.
The second definition is based on a 
\index{Hodge filtration}
{\bf Hodge filtration}: Let $F^\bullet V_{\mathbb C}$ be 
a finite decreasing filtration:
\bq
 V_{\mathbb C} \supseteq ... \supseteq F^{p-1} V_{\mathbb C} \supseteq F^{p} V_{\mathbb C} \supseteq F^{p+1} V_{\mathbb C} \supseteq ... \supseteq (0)
\eq
such that
\bq
 V_{\mathbb C} & = &
 F^p V_{\mathbb C} \oplus \overline{F^{k-p+1} V_{\mathbb C}}.
\eq
Then $V$ carries a pure Hodge structure of weight $k$.
These two definitions are equivalent: Given the Hodge decomposition, we can define the
corresponding Hodge filtration by
\bq
 F^{p} V_{\mathbb C} & = & \bigoplus\limits_{j \ge p} V^{j,k-j}.
\eq
Conversely, given a Hodge filtration we obtain the Hodge decomposition by
\bq
 V^{p,q} & = & F^p V_{\mathbb C} \cap \overline{F^q V_{\mathbb C}}.
\eq
Hodge structures behave under the operations of direct sums, tensor products and
duality as follows:
\begin{itemize}

\item Direct sum:
If $V$ and $W$ are Hodge structures of weight $k$, then also $V \oplus W$ is a Hodge structure of weight $k$.

\item Tensor product: If $V$ is a Hodge structure of weight $k$, and $W$ is a Hodge structure of weight $l$, 
then the tensor product $V \otimes W$ is a Hodge structure of weight $(k \cdot l)$.

\item Duality: If $V$ is a Hodge structure of weight $k$, then
$\Hom\left(V, {\mathbb Z}\right)$ 
is a Hodge structure of weight $(-k)$.
\end{itemize}
Let look at a few examples:

Example 1: Let $(e_1,e_2)$ be a basis of ${\mathbb R}^2$ and consider
$V_{\mathbb Z} = {\mathbb Z} e_1 \oplus {\mathbb Z} e_2$. 
We can define a Hodge structure of weight $1$ with the decomposition
\bq
 V_{\mathbb C} & = & V^{1,0} \oplus V^{0,1},
\eq
by setting
\bq
 V^{1,0} = {\mathbb C} \left( e_1 - i e_2 \right),
 & &
 V^{0,1} = {\mathbb C} \left( e_1 + i e_2 \right).
\eq
Note that the definition $V^{1,0} = {\mathbb C} e_1$, $V^{0,1} = {\mathbb C} e_2$ would not work:
Since $\overline{e_1} = e_1$ and $\overline{e_2} = e_2$, we have
$\overline{V^{1,0}} = {\mathbb C} e_1 \neq V^{0,1} = {\mathbb C} e_2$.

Example 2: 
The 
\index{Tate Hodge structure}
{\bf Tate Hodge structure} 
${\mathbb Z}(1)$ is the Hodge structure with underlying
${\mathbb Z}$-module given by $V_{\mathbb Z} = {\mathbb Z}(1) = 2\pi i \; {\mathbb Z}$.
One sets
\bq
 V_{\mathbb C} & = & {\mathbb Z}(1) \otimes_{\mathbb Z} {\mathbb C} = {\mathbb C}.
\eq
For the Hodge decomposition one sets
\bq
 V_{\mathbb C} & = & V^{-1,-1},
\eq
hence $V^{p,q}=0$ for $(p,q)\neq(-1,-1)$.
The Tate Hodge structure ${\mathbb Z}(1)$ is a pure Hodge structure of weight $-2$.

One further defines 
\bq
 {\mathbb Z}(m) & = & {\mathbb Z}^{\otimes m}.
\eq
We therefore have
\bq
 {\mathbb Z}(m) & = & \left(2 \pi i \right)^m {\mathbb Z}.
\eq
${\mathbb Z}(m)$ is a pure Hodge structure of weight $(-2m)$ with the decomposition
\bq
 {\mathbb Z}(m) \otimes {\mathbb C} = V^{-m,-m}.
\eq
Given a Hodge structure on $V_{\mathbb Z}$ of weight $k$, one defines the 
\index{Tate twist}
{\bf Tate twist} 
$V(m)$ as the Hodge structure
of weight $k-2m$ with underlying ${\mathbb Z}$-module
\bq
 V(m)_{\mathbb Z} & = & \left(2 \pi i \right)^m \otimes V_{\mathbb Z}
\eq
and Hodge decomposition
\bq
 V(m)_{\mathbb C} & = & \bigoplus\limits_{p+q=k-2m} V(m)^{p,q}
\;\;\;\;\;\;
 \mbox{with}\;\;\;
 V(m)^{p,q} = V^{p+m,q+m}.
\eq
A 
\index{polarisation of a pure Hodge structure}
{\bf polarisation} of a pure Hodge structure $V$ of weight $k$ is a non-degenerate bilinear form
\bq
 Q & : & V_{\mathbb Z} \otimes V_{\mathbb Z} \rightarrow {\mathbb Z},
\eq
with $Q(v,w) = (-1)^k Q(w,v)$.
For the complex extension $Q : V_{\mathbb C} \otimes V_{\mathbb C} \rightarrow {\mathbb C}$ 
one requires for $v \in V^{p,q}$ and $w \in V^{p',q'}$
\bq
 Q(v,w) = 0 & & \mbox{for} \;\;\; (p',q') \; \neq \; (k-p,k-q)
\eq
and
\bq
 i^{p-q} Q(v, \bar{v} ) &  > & 0
\eq
for $v \neq 0$.

As an example let us consider polarised Hodge structures of dimension $2$ and weight $1$:
Let $V_{\mathbb Z}$ be generated by $e_1, e_2$.
For $v, w \in V_{\mathbb Z}$ we write $v=v_1 e_1 + v_2 e_2$, $w=w_1 e_1 + w_2 e_2$.
Let us assume that the bilinear form defining the polarisation $Q : V_{\mathbb Z} \otimes V_{\mathbb Z} \rightarrow {\mathbb Z}$
is given by
\bq
\label{chapter_motives:polarisation_dim_2_weight_1}
 Q\left(v,w\right)
 & = &
 \left(w_2, w_1 \right)
 \left(\begin{array}{rr}
 0 & 1 \\
 -1 & 0 \\
 \end{array} \right)
 \left(\begin{array}{c}
 v_2 \\
 v_1 \\
 \end{array} \right)
\eq
We assume that $V_{\mathbb Z}$ is a Hodge structure of weight $1$, hence
\bq
 V_{\mathbb C}
 & = &
 V^{1,0} \otimes V^{0,1}
\eq
and $h^{1,0}=h^{0,1}=1$. Let $\psi=\psi_1 e_1 + \psi_2 e_2$ with $\psi_1,\psi_2 \in {\mathbb C}$
be a basis of $V^{1,0}$.
From 
\bq
\label{chapter_motives:positive_polarisation}
 i Q\left(\psi,\bar{\psi}\right) & > & 0
\eq
it follows that
\bq
 i \left( \bar{\psi}_2 \psi_1 - \bar{\psi}_1 \psi_2 \right)
 & > & 0.
\eq
This implies in particular $\psi_1 \neq 0$.
We may therefore rescale the generator of $V^{1,0}$ by $1/\psi_1$.
We then have
\bq
 V^{1,0} & = & \left\langle e_1 + \tau e_2 \right\rangle,
 \;\;\;\;\;\;
 \tau \; = \; \frac{\psi_2}{\psi_1}
\eq
and
\bq
 i Q\left(e_1 + \tau e_2,e_1 + \bar{\tau} e_2\right)
 & = & 
 - i \left( \tau - \bar{\tau} \right)
 \; = \; 
 2 \mathrm{Im} \tau
 \; > \; 0.
\eq
This shows that all Hodge structures of dimension $2$ and weight $1$ with the polarisation form as in eq.~(\ref{chapter_motives:polarisation_dim_2_weight_1})
are parametrised by $\tau \in {\mathbb H}$.
$V^{0,1}$ is generated by
\bq
 V^{0,1} & = & \overline{V^{1,0}} \; = \; \left\langle e_1 + \bar{\tau} e_2 \right\rangle.
\eq
It is instructive to discuss this concretely for elliptic curves:
Let 
\bq
 E & : & y^2 \; = \; 4 x \left(x-1\right) \left(x-\lambda\right)
\eq
be an elliptic curve.
We denote by $\gamma_1$ and $\gamma_2$ two independent cycles. 
As they are independent, they form a basis of the first Betti homology group $H_1^{\mathrm{B}}(E)$.
We then consider Betti cohomology.
Let $\gamma_1^\ast, \gamma_2^\ast \in H^1_{\mathrm{B}}(E)$ be the dual basis, i.e. the basis which satisfies
\bq
\label{chapter_motives:def_duality_Betti}
 \left\langle \gamma_i^\ast, \gamma_i \right\rangle & = & \delta_{i j}.
\eq
$\gamma_1^\ast$ and $\gamma_2^\ast$ correspond to $e_1$ and $e_2$ in the discussion above.
From section~\ref{chapter_elliptics:section_elliptic_functions} we know that a basis of $H^1_{\mathrm{dR}}(E)$
is given by
\bq
 \omega_1 \; = \; \frac{dx}{y},
 & & 
 \omega_2 \; = \; \frac{x \, dx}{y}.
\eq
We denote the period matrix by
\bq
 P & = &
 \left( \begin{array}{cc}
  \left\langle \omega_1, \gamma_1 \right\rangle & \left\langle \omega_1, \gamma_2 \right\rangle \\
  \left\langle \omega_2, \gamma_1 \right\rangle & \left\langle \omega_2, \gamma_2 \right\rangle \\
 \end{array} \right)
 \; = \; 
 \left( \begin{array}{cc}
 \psi_1 & \psi_2 \\
 \phi_1 & \phi_2 \\
 \end{array} \right).
\eq
From the Legendre relation we have
\bq
 \det P & = & 2 \pi i.
\eq
We recall that instead of $(x,y)$ we may use a complex coordinate $z$ through
\bq
 z \; = \; \int\limits_\infty^x \frac{dt}{\sqrt{4 t\left(t-1\right)\left(t-\lambda\right)}},
 & &
 \left(x,y\right) \; = \; \left(\wp\left(z\right), \wp'\left(z\right) \right).
\eq
In terms of the complex coordinate $z$, the one-form $\omega_1$ is given by $\omega_1=dz$.
On the other hand, we may express $\omega_1$ as a linear combination of $\gamma_1^\ast$ and $\gamma_2^\ast$:
We make the ansatz $\omega_1 = c_1 \gamma_1^\ast + c_2 \gamma_2^\ast$, contract with $\gamma_1$ and $\gamma_2$
and find
\bq
 \omega_1 & = & \psi_1 \gamma_1^\ast + \psi_2 \gamma_2^\ast.
\eq
This corresponds to $\psi=\psi_1 e_1 + \psi_2 e_2$ just before 
eq.~(\ref{chapter_motives:positive_polarisation}).
Thus we have
\bq
 H^1_{\mathrm{B}}(E)_{\mathbb C} & = &
 H^{1,0} \otimes H^{0,1},
\eq
with $H^{1,0}$ being generated by $\omega_1=dz$ and $H^{0,1}$ being generated by $\overline{\omega}_1=d\bar{z}$.
\\
\\
\bs
{\it \refstepcounter{exercise}
{\bf Exercise \theexercise}: 
We now have two bases of $H^1_{\mathrm{dR}}(E)$: on the one hand $(\omega_1,\omega_2)$, on the other hand $(dz,d\bar{z})$. We already know $\omega_1=dz$. Work out the full relation between the two bases.
}
\es

\subsection{Mixed Hodge structures}
\label{chapter_motives:section_mixed_hodge_structures}

Pure Hodge structures are relevant for smooth projective algebraic varieties, these are necessarily compact.
If one gives up the requirement of smoothness or compactness one is lead to a 
generalisation called mixed Hodge structure \cite{Deligne:1970,Deligne:1971,Deligne:1974}.
\begin{tcolorbox}[breakable]
A 
\index{mixed Hodge structure}
{\bf mixed Hodge structure} is given by
a $\mathbb Z$-module $V_{\mathbb Z}$ of finite rank,
a finite increasing filtration on 
$V_{\mathbb Q} = V_{\mathbb Z} \otimes_{\mathbb Z} {\mathbb Q}$, called the 
\index{weight filtration}
{\bf weight filtration}:
\bq
 (0) \subseteq ... \subseteq W_{k-1} V_{\mathbb Q} \subseteq W_{k} V_{\mathbb Q} \subseteq W_{k+1} V_{\mathbb Q} \subseteq ... \subseteq V_{\mathbb Q},
\eq
and a finite decreasing filtration on
$V_{\mathbb C} = V_{\mathbb Z} \otimes_{\mathbb Z} {\mathbb C}$, called the 
\index{Hodge filtration}
{\bf Hodge filtration}:
\bq
 V_{\mathbb C} \supseteq ... \supseteq F^{p-1} V_{\mathbb C} \supseteq F^{p} V_{\mathbb C} \supseteq F^{p+1} V_{\mathbb C} \supseteq ... \supseteq (0),
\eq
such that $F^\bullet$ induces a pure Hodge structure of weight $k$ on
\bq
 \mbox{Gr}_k^W V_{\mathbb Q} & = & {W_k V_{\mathbb Q}} / {W_{k-1} V_{\mathbb Q}}.
\eq
\end{tcolorbox}
A mixed Hodge structure is called a 
\index{mixed Tate Hodge structure}
{\bf mixed Tate Hodge structure} if
\bq
 h^{p,q} & = & 0
 \;\;\;\;\;\;
 \mbox{for $p \neq q$}.
\eq
Example 1:
Let us fix two independent vectors $e_0$ and $e_{-1}$ and a complex number $x$.
We set
\bq
 V_{\mathbb Q} & = & \left\langle e_0 + \ln x \cdot e_{-1}, 2 \pi i e_{-1} \right\rangle.
\eq
On $V_{\mathbb Q}$ we define the weight filtration by
\bq
\label{chapter_motives:example_1_weight_filtration}
 W_0 V_{\mathbb Q} & = & V_{\mathbb Q} = \left\langle e_0 + \ln x \cdot e_{-1}, 2 \pi i e_{-1} \right\rangle,
 \nonumber \\
 W_{-1} V_{\mathbb Q} & = & W_{-2} V_{\mathbb Q} = \left\langle 2 \pi i e_{-1} \right\rangle,
 \nonumber \\
 W_{-3} V_{\mathbb Q} & = & 0.
\eq
On $V_{\mathbb C}$ we define the Hodge filtration by
\bq
 F^1 V_{\mathbb C} = 0,
 \;\;\;
 F^0 V_{\mathbb C} = \left\langle e_0 \right\rangle,
 \;\;\;
 F^{-1} V_{\mathbb C} = \left\langle e_0, e_{-1} \right\rangle.
\eq
We have
\bq
 \mathrm{Gr}_{0}^W V_{\mathbb C}
 & = & 
 \left\langle e_0 + \ln x \cdot e_{-1}, 2 \pi i e_{-1} \right\rangle / \left\langle 2 \pi i e_{-1} \right\rangle
 =
 \left\langle e_0, e_{-1} \right\rangle / \left\langle e_{-1} \right\rangle
 \cong 
 \left\langle e_0 \right\rangle,
 \nonumber \\
 \mathrm{Gr}_{-1}^W V_{\mathbb C} & = & 0,
 \nonumber \\
 \mathrm{Gr}_{-2}^W V_{\mathbb C} & = & \left\langle 2 \pi i e_{-1} \right\rangle.
\eq
In the decomposition
\bq
 \mathrm{Gr}_{0}^W V_{\mathbb C} & = & \bigoplus\limits_{p} V^{p,-p}
\eq
one easily finds that $V^{p,-p}=0$ for $p \ge 1$, since $F^1 V_{\mathbb C} = 0$.
From $V^{-p,p}=\overline{V^{p,-p}}$ it follows then that also $V^{-p,p}=0$ for $p\ge 1$.
Therefore 
\bq
 \mathrm{Gr}_{0}^W V_{\mathbb C} & = & V^{0,0} \cong \left\langle e_0 \right\rangle.
\eq
In a similar way one finds
\bq
 \mathrm{Gr}_{-2}^W V_{\mathbb C} & = & V^{-1,-1} = \left\langle 2 \pi i e_{-1}  \right\rangle.
\eq
Therefore the Hodge structure of $\mathrm{Gr}_{0}^W V_{\mathbb Q}$ is isomorph to ${\mathbb Q}(0)$,
and the Hodge structure of $\mathrm{Gr}_{-2}^W V_{\mathbb Q}$ is isomorph to ${\mathbb Q}(1)$.
Thus, $V_{\mathbb Q}$ defines a mixed Tate Hodge structure with $V^{0,0}$ and $V^{-1,-1}$ non-zero.

Note that $V_{\mathbb Q}$ is spanned by the columns of 
\bq
 P & = &
 \left( \begin{array}{cc}
 1 & 0 \\
 \ln x & 2 \pi i \\
 \end{array} \right).
\eq
With 
\bq
 C_0 & = &
 \left( \begin{array}{cc}
 0 & 0 \\
 1 & 0 \\
 \end{array} \right)
\eq
we have
\bq
 \frac{d}{dx} P \; = \; \frac{C_0}{x} P,
 & &
 {\mathcal M}_0 P \; = \; P \exp\left(C_0\right),
\eq
where ${\mathcal M}_0$
denotes the monodromy operator around $x=0$ (see eq.~(\ref{chapter_multiple_polylogarithms:result_monodromy_logarithm})).
\\
\\
Example 2:
In a similar spirit let us consider the $(n+1)\times(n+1)$-matrix
\bq
\label{chapter_motives:example_polylog_def_P}
 P & = &
 \left( \begin{array}{ccccc}
 1 & 0 & 0 & \cdots & 0 \\
 -\mathrm{Li}_1(x) & 2 \pi i & 0 & \cdots & 0 \\
 -\mathrm{Li}_2(x) & 2 \pi i \ln(x) & (2 \pi i)^2 & \cdots & 0 \\
 \cdots & \cdots & \cdots & \cdots & \cdots \\
 -\mathrm{Li}_n(x) & 2 \pi i \frac{\ln^{n-1}(x)}{(n-1)!} & (2 \pi i)^2 \frac{\ln^{n-2}(x)}{(n-2)!} & \cdots & (2 \pi i)^n \\
 \end{array} \right)
\eq
For later use we note that with
\bq
\label{chapter_motives:example_polylog_def_C0_C1}
 C_0
 \; = \;
 \left( \begin{array}{ccccc}
 0 & 0 & \cdots & \cdots & 0 \\
 0 & 0 & \cdots & \cdots & 0 \\
 0 & 1 & \ddots & & 0 \\
 \vdots & \vdots & \ddots & \ddots & \vdots \\
 0 & 0 & \cdots & 1 & 0 \\
 \end{array} \right),
 & &
 C_1
 \; = \;
 \left( \begin{array}{ccccc}
 0 & 0 & 0 & \cdots & 0 \\
 1 & 0 & 0 & \cdots & 0 \\
 0 & 0 & 0 & \cdots & 0 \\
 \vdots & \vdots & \vdots &  & \vdots \\
 0 & 0 & 0 & \cdots & 0 \\
 \end{array} \right)
\eq
we have
\bq
\label{chapter_motives:example_polylog_def_dgl}
 \frac{d}{dx} P \; = \; \left( \frac{C_0}{x} + \frac{C_1}{x-1} \right) P,
 & &
 {\mathcal M}_0 P \; = \; P \exp\left(C_0\right),
 \;\;\;
 {\mathcal M}_1 P \; = \; P \exp\left(C_1\right).
\eq 
${\mathcal M}_0$ and ${\mathcal M}_1$ denote the monodromy operators around $x=0$ and $x=1$, respectively.

We introduce independent vectors $e_0, e_{-1}, \dots, e_{-n}$ and set
\begin{alignat}{6}
\label{chapter_motives:example_polylog_v_e}
 v_0 & = &  e_0 & & - \mathrm{Li}_1(x) e_{-1} & & - \mathrm{Li}_2(x) e_{-2} & & - \dots & & - \mathrm{Li}_n(x) e_{-n},
 \nonumber \\
 v_{1} & = & & & 2 \pi i e_{-1} & & + 2 \pi i \ln(x) e_{-2} & & + \cdots & & + 2 \pi i \frac{\ln^{n-1}(x)}{(n-1)!} e_{-n},
 \nonumber \\
 v_{2} & = & & & & & \left(2 \pi i\right)^2 e_{-2} & & + \cdots & & + \left(2 \pi i\right)^2 \frac{\ln^{n-2}(x)}{(n-2)!} e_{-n},
 \nonumber \\
 & & \cdots & & & & & & & &
 \nonumber \\
 v_{n} & = & & & & & & & & & (2 \pi i)^n e_{-n}.
\end{alignat}
We then consider
\bq
 V_{\mathbb Q} & = &
 \left\langle v_0, v_1, \dots, v_n \right\rangle.
\eq
$V_{\mathbb Q}$ is a mixed Hodge structure with the weight filtration
\begin{alignat}{3}
 & & W_0 V_{\mathbb Q}  & \; = \; & \left\langle v_0, v_1, v_2, \dots, v_n  \right\rangle, &
\label{chapter_motives:example_2_weight_filtration}
 \nonumber \\
 W_{-1} V_{\mathbb Q} & \; = \; & W_{-2} V_{\mathbb Q} & \; = \; & \left\langle v_1, v_2, \dots, v_n  \right\rangle, &
 \nonumber \\
 W_{-3} V_{\mathbb Q} & \; = \; & W_{-4} V_{\mathbb Q} & \; = \; & \left\langle v_2, \dots, v_n  \right\rangle, &
 \nonumber \\
 \cdots & & \cdots && \cdots &
 \nonumber \\
 W_{-2n+1} V_{\mathbb Q} & \; = \; & W_{-2n} V_{\mathbb Q} & \; = \; & \left\langle v_n  \right\rangle, &
 \nonumber \\
 W_{-2n-1} V_{\mathbb Q} & \; = \; & && 0, &
\end{alignat}
and the Hodge filtration
\bq
 F^1 V_{\mathbb C} & = & 0,
 \nonumber \\
 F^0 V_{\mathbb C} & = & \left\langle e_0 \right\rangle,
 \nonumber \\
 F^{-1} V_{\mathbb C} & = &\left\langle e_0, e_{-1} \right\rangle,
 \nonumber \\
 & & \cdots
 \nonumber \\
 F^{-n} V_{\mathbb C} & = &\left\langle e_0, e_{-1}, \dots, e_{-n} \right\rangle \; = \; V_{\mathbb C}.
\eq
$V_{\mathbb Q}$ is a mixed Tate Hodge structure.
\\
\\
\bs
{\it \refstepcounter{exercise}
{\bf Exercise \theexercise}: 
Work out all $V^{p,q}$ and show that $V_{\mathbb Q}$ is mixed Tate.
}
\es

\subsection{Variations of Hodge structures}

We are in particular interested in families of (mixed) Hodge structures, parametrised by a manifold $B$ (the $B$ stands for ``base'').
We assume that for every point $x \in B$ we have a mixed Hodge structure.
This will lead us to a variation of mixed Hodge structures \cite{Griffiths:1968i,Griffiths:1968ii}.
In detail:
\begin{tcolorbox}
A 
{\bf variation of mixed Hodge structure} 
on the manifold $B$ consists of
\begin{itemize}
\item a local system ${\mathcal L}_{\mathbb Z}$ of ${\mathbb Z}$-modules of finite rank,
\item a finite increasing filtration ${\mathcal W}$ of ${\mathcal L}_{\mathbb Q} = {\mathcal L}_{\mathbb Z} \otimes {\mathbb Q}$ by 
sublocal systems of rational vector spaces,
\item a finite decreasing filtration ${\mathcal F}$ of ${\mathcal L}_{{\mathcal O}_B} = {\mathcal L}_{\mathbb Z} \otimes {\mathcal O}_B$, satisfying
Griffiths' transversality condition:
\bq
 \nabla\left({\mathcal F}^p\right) & \subset & {\mathcal F}^{p-1} \otimes \Omega^1_B.
\eq
\item the filtrations ${\mathcal W}$ and ${\mathcal F}$ define a mixed Hodge structure on each fibre 
$({\mathcal L}_{{\mathcal O}_B}(x), {\mathcal W}(x), {\mathcal F}(x))$ of the bundle ${\mathcal L}_{{\mathcal O}_B}(x)$ at point $x$.
\end{itemize}
\end{tcolorbox}
Here, ${\mathcal O}_B$ denotes the sheaf of holomorphic functions on $B$, and $\Omega^1_B$ denotes
the sheaf of differential one-forms on $B$.

Let's continue with example 2 from the previous section:
$V_{\mathbb Q}$ is generated by  $\langle v_0, v_1, \dots, v_n \rangle$, where the vectors $v_j$ are given by
the columns of the matrix $P$ in eq.~(\ref{chapter_motives:example_polylog_def_P}).
From eq.~(\ref{chapter_motives:example_polylog_def_dgl}) we deduce
\bq
 \left[ d - C_0 \; d\ln\left(x\right) - C_1 \; d\ln\left(x-1\right) \right] v_j & = & 0,
\eq
with the $(n+1)\times(n+1)$-matrices $C_0$ and $C_1$ defined in eq.~(\ref{chapter_motives:example_polylog_def_C0_C1}).
The connection $\nabla$ is therefore given by
\bq
 \nabla & = & d - C_0 \; d\ln\left(x\right) - C_1 \; d\ln\left(x-1\right)
\eq
and we write
\bq
 \nabla v_j & = & 0.
\eq
In physics jargon we call $\nabla$ a covariant derivative and say that the $v_j$'s are covariantly constant
(or parallel transported with respect to $\nabla$).
In the mathematical language one says that the $v_j$'s define a locally constant sheaf.

Let's consider $U = {\mathbb C}\backslash (]-\infty,0] \cup [1,\infty[)$. 
In this region $P$ is single-valued and from section~\ref{chapter_motives:section_mixed_hodge_structures}
we know already that for any $x \in U$ the columns of $P$ define a mixed Tate Hodge structure.
It remains to verify Griffiths' transversality condition.
We need to determine $\nabla e_{-j}$ for $j \in \{0,\dots,n\}$.
We find 
\bq
\label{chapter_motives:example_Griffiths}
 \nabla e_0 & = & - \frac{dx}{x-1} e_{-1},
 \nonumber \\
 \nabla e_{-j} & = & - \frac{dx}{x} e_{-j-1},
 \;\;\;\;\;\;
 1 \; \le \; j \; < \; n,
 \nonumber \\
 \nabla e_{-n} & = & 0.
\eq
With eq.~(\ref{chapter_motives:example_Griffiths}) it follows that
\bq
 \nabla\left({\mathcal F}^p\right) & \subset & {\mathcal F}^{p-1} \otimes \Omega^1_U.
\eq
\bs
{\it \refstepcounter{exercise}
{\bf Exercise \theexercise}: 
Derive eq.~(\ref{chapter_motives:example_Griffiths}).
}
\es

\subsection{Mixed Hodge structures on cohomology groups}
\label{chapter_motives:mixed_Hodge_structure_on_cohomology_groups}

Let $X$ be a complex algebraic variety and $Y$ a (possibly empty) closed subvariety.
It has been shown by Deligne  \cite{Deligne:1971,Deligne:1974}
that the relative cohomology group $H^k(X,Y)$ carries a mixed Hodge structure.
If $Y$ is empty, this reduces to the (non-relative) cohomology $H^k(X)$.

Let us look at the details: 
To simplify life we assume that $X$ is smooth and $Y$ a simple normal crossing divisor.
If $Y$ is not a simple normal crossing divisor, one uses first the resolution of singularities (see chapter~\ref{chapter_sector_decomposition}) to achieve this condition.

For the mixed Hodge structure we have to define the weight filtration and the Hodge filtration.
This is best explained with differential forms and de Rham cohomology.
Due to theorem~\ref{chapter_motives:theorem_Grothendieck_relative} we may replace 
de Rham cohomology with algebraic de Rham cohomology.
But let us stick in this section with de Rham cohomology.
We denote by $\Omega^p_X$ the sheaf of holomorphic differential forms of degree $p$ on $X$
and by $\Omega^p_X(\log Y)$ the sheaf of meromorphic differential forms of degree $p$ on $X$ 
with at most logarithmic poles along $Y$.
A meromorphic differential form $\omega$ has at most logarithmic poles along $Y$ 
if both $\omega$ and $d\omega$ have at most a simple pole along $Y$.

This implies that in the coordinate system of eq.~(\ref{chapter_motives:coordinate_system_normal_crossing})
a differential one-form $\omega \in \Omega^1_X$ can locally be written
as
\bq
 \omega & = & 
 \sum\limits_{j=1}^n f_j\left(x\right) dx_j,
\eq
with $f_j$ holomorphic, while a differential one-form $\omega \in \Omega^1_X(\log Y)$ can locally be written
as
\bq
 \omega & = & 
 \sum\limits_{j=1}^k f_j\left(x\right) \frac{dx_j}{x_j}
 + \sum\limits_{j=k+1}^n f_j\left(x\right) dx_j,
\eq
again with $f_j$ holomorphic.
The differential $p$-forms in $\Omega^p_X$ and $\Omega^p_X(\log Y)$
are then obtained from the $p$-fold wedge product of forms in $\Omega^1_X$ and $\Omega^1_X(\log Y)$,
respectively.

Let's now consider $\Omega^\bullet_X(\log Y)$.
There is a trivial filtration:
\bq
 F^p \Omega^\bullet_X(\log Y)
 & = &
 \bigoplus\limits_{r \ge p} \Omega^p_X(\log Y).
\eq
This filtration induces the Hodge filtration on the cohomology.

For the weight filtration we count the number of dlog-forms.
One sets
\bq
\label{chapter_motives:weight_filtration_cohomology}
 W_m \Omega^p_X(\log Y)
 & = &
 \left\{
  \begin{array}{ll}
   0 & \mbox{for} \; m < 0, \\
   \Omega^{p-m}_X \wedge \Omega^m_X(\log Y) & \mbox{for} \; 0 \le m \le p, \\
   \Omega^p_X(\log Y) & \mbox{for} \; p < m. \\
  \end{array}
 \right.
\eq
This filtration induces the weight filtration on the cohomology.
Without going into the details let us note that in the $k$-th cohomology group differential forms with $m$
dlog-one-forms will contribute at weight $m+k$.

Let us illustrate the weight filtration with an example.
We take $X={\mathbb C}^3$ and
\bq
 Y & = & \left\{x_1=0\right\} \cup \left\{x_2=0\right\} \cup \left\{x_3=1\right\}.
\eq
We consider the three-forms
\bq
 \omega_1 & = & 
 - \frac{dx_1}{x_1} \wedge \frac{dx_2}{x_2} \wedge \frac{dx_3}{1-x_3},
 \nonumber \\
 \omega_2 & = & 
 - \frac{dx_1}{x_1} \wedge dx_2 \wedge \frac{dx_3}{1-x_3}.
\eq
As both are three-forms, we have
\bq 
 \omega_1, \omega_2 & \in & F^3 \Omega^\bullet_X(\log Y).
\eq
In a local coordinate system around $(x_1,x_2,x_3)=(0,0,1)$ the three-form $\omega_1$ has
three dlog forms, while $\omega_2$ has only two.
Hence
\bq
 \omega_1 \; \in \; W_3 \Omega^3_X(\log Y),
 & &
 \omega_2 \; \in \; W_2 \Omega^3_X(\log Y).
\eq
The counting of the dlog forms as weight is illustrated as follows:
Consider the integration domain $\gamma : 0 \le x_3 \le x_2 \le x_1 \le 1$.
We then have
\bq
 \int\limits_\gamma \omega_1 & = & \zeta_3,
 \nonumber \\
 \int\limits_\gamma \omega_2 & = & - \zeta_2 + 2.
\eq
Thus $\langle \omega_1 | \gamma \rangle$ is pure of (polylogarithmic) weight $3$, while
$\langle \omega_2 | \gamma \rangle$ is mixed with highest (polylogarithmic) weight $2$.

The attentive reader might have noticed that we already discussed weight filtrations of mixed Hodge structures
in two examples. From eq.~(\ref{chapter_motives:example_1_weight_filtration}) and eq.~(\ref{chapter_motives:example_2_weight_filtration}) we would expect that the weight of $\zeta_n$ within Hodge theory is $(-2n)$.
This deserves some explanation.
Let's look at $\zeta_3$: We start from the three-form $\omega_1$, hence we are interested in 
the relative cohomology group $H^3(X,Y)$.
The form $\omega_1$ is a wedge product of three dlog-one-forms.
We therefore have $k=3$ and $m=3$. The comment after eq.~(\ref{chapter_motives:weight_filtration_cohomology})
implies that $\omega_1$ contributes at weight $m+k=6$ in cohomology.
Instead of looking at cohomology we may also look at homology, which is just the dual.
Under dualisation the weight of the weight filtration changes sign, in our example $6 \rightarrow -6$,
in agreement with the examples of eq.~(\ref{chapter_motives:example_1_weight_filtration}) and eq.~(\ref{chapter_motives:example_2_weight_filtration}).

\subsection{Motivic periods}

We may now bring the various pieces together and define motivic periods.
We remind the reader that we discussed effective periods in chapter~\ref{chapter_sector_decomposition}
and that we already discussed motivic periods in chapter~\ref{chapter_hopf}.
The definition which we provide now is a slight generalisation of the definition of effective periods discussed
in section~\ref{chapter_sector_decomposition:effective_periods}.
The motivic periods discussed in
chapter~\ref{chapter_hopf} correspond to the subset of motivic mixed Tate periods.

Let $X$ be a smooth variety defined by polynomials with coefficients in ${\mathbb Q}$.
Let $Y$ be a closed subvariety.
We denote by $X^{\mathrm{an}}$ and $Y^{\mathrm{an}}$ the analytifications.
Let $\omega \in H^k_{\mathrm{alg \; dR}}\left(X,Y\right)$ be a class of the $k$-th relative algebraic de Rham cohomology  
and let $\gamma \in H_k^{\mathrm{B}}\left(X^{\mathrm{an}},Y^{\mathrm{an}},{\mathbb Q}\right)$ be a class in the $k$-th relative Betti homology.

From the previous section we know that we have mixed Hodge structure on $H^k(X,Y)$.
That is to say that $H^k_{\mathrm{alg \; dR}}(X,Y)$ is equipped with a Hodge filtration $F^\bullet$ 
and a weight filtration $W_\bullet$, 
that the Betti cohomology $H^k_{\mathrm{B}}(X^{\mathrm{an}},Y^{\mathrm{an}},{\mathbb Q})$ is equipped with a weight filtration 
$W_\bullet$ and there is a comparison isomorphism
\bq
 \mathrm{comparison}
 & : & 
 H^k_{\mathrm{alg \; dR}}\left(X,Y\right) \otimes {\mathbb C}
 \rightarrow
 H^k_{\mathrm{B}}\left(X^{\mathrm{an}},Y^{\mathrm{an}},{\mathbb Q}\right) \otimes {\mathbb C}
\eq
compatible with the weight filtration.

The triple
\bq
 M^k\left(X,Y\right)
 & = &
 \left[
  H^k_{\mathrm{alg \; dR}}\left(X,Y\right), H^k_{\mathrm{B}}\left(X^{\mathrm{an}},Y^{\mathrm{an}},{\mathbb Q}\right),
  \mathrm{comparison}
 \right]^{\mathfrak m}
\eq
is called an 
\index{H-motive}
{\bf H-motive} (or ``motive'' for short, the long official name is ``Hodge realisation of the motive'').

The 
\index{real Frobenius}
{\bf real Frobenius} is a linear involution 
\bq
\label{chapter_motives:real_Frobenius_map}
 F_\infty & : & 
 H^k_{\mathrm{B}}\left(X^{\mathrm{an}},Y^{\mathrm{an}},{\mathbb Q}\right) \rightarrow H^k_{\mathrm{B}}\left(X^{\mathrm{an}},Y^{\mathrm{an}},{\mathbb Q}\right)
\eq
defined as follows: 
We denote by $\mathrm{conj}_{\mathrm{alg \; dR}}$ the ${\mathbb C}$-antilinear involution on
$H^k_{\mathrm{alg \; dR}}\left(X,Y\right) \otimes {\mathbb C}$ given by
\bq
 \mathrm{conj}_{\mathrm{alg \; dR}}\left( \omega \otimes z \right)
 & = &
 \omega \otimes \bar{z}
\eq
and by $\mathrm{conj}_{\mathrm{B}}$ the analogous ${\mathbb C}$-antilinear involution on
$H^k_{\mathrm{B}}\left(X^{\mathrm{an}},Y^{\mathrm{an}}\right) \otimes {\mathbb C}$.
Then $F_\infty$ is defined such that the following diagram commutes:
\bq
\label{chapter_motives:real_Frobenius_commutative_diagram}
\begin{CD}
 H^k_{\mathrm{alg \; dR}}\left(X,Y\right) \otimes {\mathbb C} @>{\mathrm{comparison}}>> H^k_{\mathrm{B}}\left(X^{\mathrm{an}},Y^{\mathrm{an}}\right) \otimes {\mathbb C} \\
 @V{\mathrm{conj}_{\mathrm{alg \; dR}} }VV @VV{\left(F_\infty \otimes \mathrm{id}\right) \; \mathrm{conj}_{\mathrm{B}}}V \\
 H^k_{\mathrm{alg \; dR}}\left(X,Y\right) \otimes {\mathbb C} @>{\mathrm{comparison}}>> H^k_{\mathrm{B}}\left(X^{\mathrm{an}},Y^{\mathrm{an}}\right) \otimes {\mathbb C} \\
\end{CD}
\eq
The following exercise illustrates the action of the real Frobenius:
\\
\\
\bs
{\it \refstepcounter{exercise}
{\bf Exercise \theexercise}: 
Consider $X={\mathbb C} \backslash \{0\}$ and $Y=\emptyset$.
Take $\omega=dx/x$ as a basis of $H^1_{\mathrm{alg \; dR}}(X)$
and let $\gamma$ be a small counter-clockwise circle around $x=0$.
$\gamma$ is a basis of $H_1^{\mathrm{B}}(X)$. Denote by $\gamma^\ast$ the dual basis of $H^1_{\mathrm{B}}(X)$.
Work out
\bq
 F_\infty\left(\gamma^\ast\right).
\eq
}
\es
\\
Let us then consider the ${\mathbb Q}$-vector space of equivalence classes of triples
\bq
 \left[ M^k\left(X,Y\right), \omega, \gamma \right]^{\mathfrak m}
\eq
modulo the relations induced by linearity in $\omega$ and $\gamma$,
changes of variables and Stokes' formula.
Linearity states that for $c_1, c_2 \in {\mathbb Q}$
\bq
\label{chapter_motives:linearity}
 \left[ M^k\left(X,Y\right), c_1 \omega_1 + c_2 \omega_2, \gamma \right]^{\mathfrak m}
 & = &
 c_1 \left[ M^k\left(X,Y\right), \omega_1, \gamma \right]^{\mathfrak m}
 +
 c_2 \left[ M^k\left(X,Y\right), \omega_2, \gamma \right]^{\mathfrak m},
 \nonumber \\
 \left[ M^k\left(X,Y\right), \omega, c_1 \gamma_1 + c_2 \gamma_2 \right]^{\mathfrak m}
 & = &
 c_1 \left[ M^k\left(X,Y\right), \omega, \gamma_1 \right]^{\mathfrak m}
 +
 c_2 \left[ M^k\left(X,Y\right), \omega, \gamma_2 \right]^{\mathfrak m}.
 \;\;\;\;\;\;\;\;\;
\eq
Let $f : X \rightarrow X'$ be a regular map and $Y'=f(Y)$. A change of variables implies
\bq
\label{chapter_motives:change_of_variables}
 \left[ M^k\left(X,Y\right), f^\ast \omega', \gamma \right]^{\mathfrak m}
 & = &
 \left[ M^k\left(X',Y'\right), \omega', f_\ast \gamma \right]^{\mathfrak m}.
\eq
For Stokes' theorem consider $X \supset Y \supset Z$ and denote the 
connecting morphism by $d : H^{k-1}(Y,Z) \rightarrow H^k(X,Y)$.
Then
\bq
\label{chapter_motives:Stokes}
 \left[ M^k\left(X,Y\right), d\omega, \gamma \right]^{\mathfrak m}
 & = &
 \left[ M^{k-1}\left(Y,Z\right), \omega, \partial \gamma \right]^{\mathfrak m}.
\eq
The equivalence classes with respect to these relations are called 
\index{effective motivic periods}
{\bf effective motivic periods} 
and denoted as ${\mathcal P}^{\mathfrak m}$.
To each effective motivic period we can associate a numerical period.
In other words, there is a map
\bq
\label{chapter_motives:effective_motivic_period_map}
 \mathrm{period} & : & {\mathcal P}^{\mathfrak m} \rightarrow {\mathbb P},
 \nonumber \\
 & &
 \left[ M^k\left(X,Y\right), \omega, \gamma \right]^{\mathfrak m}
 \rightarrow \int\limits_\gamma \omega.
\eq
Every motivic period comes with a mixed Hodge structure (the one discussed in section~\ref{chapter_motives:mixed_Hodge_structure_on_cohomology_groups})
and in particular a weight filtration.
If we forget about the extra information on the mixed Hodge structure we have an effective period as introduced
in chapter~\ref{chapter_sector_decomposition}, which can be specified by the quadruple 
$(X, Y, \omega, \gamma)$.

The real Frobenius $F_\infty : H^k_{\mathrm{B}}(X^{\mathrm{an}},Y^{\mathrm{an}},{\mathbb Q}) \rightarrow H^k_{\mathrm{B}}(X^{\mathrm{an}},Y^{\mathrm{an}},{\mathbb Q})$ of eq.~(\ref{chapter_motives:real_Frobenius_map})
induces a map
\bq
\label{chapter_motives:induced_real_Frobenius}
 F_\infty & : & {\mathcal P}^{\mathfrak m} \rightarrow {\mathcal P}^{\mathfrak m}
\eq
through
\bq
 \left[ M^k\left(X,Y\right), \omega, \gamma \right]^{\mathfrak m}
 \rightarrow 
 \left[ M^k\left(X,Y\right), \omega, F_\infty\left(\gamma\right) \right]^{\mathfrak m}.
\eq
By abuse of notation we also denote this map by $F_\infty$.
The latter map we already encountered in eq.~(\ref{chapter_hopf:real_Frobenius}).

\subsection{The motivic Galois group}

In order not to raise false expectations let us state from the beginning that although despite the title of this section
we will be dealing with the motivic Galois group only indirectly through its dual, which is a Hopf algebra.
But that is o.k., as this is the way we would like to apply it anyway.
Of course, it is possible to construct the group from its dual.

We start by introducing algebraic groups.
An 
\index{affine algebraic group}
{\bf affine algebraic group $G$} is a group defined by polynomial equations such that the group law is polynomial.

To give an example, consider $\mathrm{SL}_2(\overline{\mathbb Q})$, the set of $(2 \times 2)$-matrices with entries
from $\overline{\mathbb Q}$ and determinant $1$:
\bq
 g & = &
 \left( \begin{array}{cc}
 z_{11} & z_{12} \\
 z_{21} & z_{22} \\
 \end{array} \right),
 \;\;\;\;\;\;\;\;\;
 z_{11}, z_{12}, z_{21}, z_{22} \; \in \; \overline{\mathbb Q},
 \;\;\;\;\;\;\;\;\;
 \det g \; = \; 1.
\eq
Alternatively, we may view $\mathrm{SL}_2(\overline{\mathbb Q})$ as an affine algebraic variety:
\bq
 \mathrm{SL}_2(\overline{\mathbb Q})
 & = &
 \left\{ \,
  \left(z_{11}, z_{12}, z_{21}, z_{22} \right) \in \overline{\mathbb Q}^4
  \; | \;
  z_{11} z_{22} - z_{12} z_{21} - 1 = 0 
 \, \right\}
\eq
\bs
{\it \refstepcounter{exercise}
{\bf Exercise \theexercise}: 
Let ${\mathbb F}$ be a sub-field of ${\mathbb C}$. Show that $\mathrm{GL}_n({\mathbb F})$
(the group of $(n \times n)$-matrices with entries from ${\mathbb F}$ and non-zero determinant)
can be defined by a polynomial equation.
}
\es
\\
\\
Let us now consider functions on the group, i.e. maps
\bq
 G \rightarrow {\mathbb F},
\eq
where we take ${\mathbb F}$ to be a sub-field of ${\mathbb C}$.
If we specialise to matrix groups, like $\mathrm{GL}_n({\mathbb F})$, the entry of the $i$-th row and $j$-th column
provides a function on the group:
\bq
 a_{ij} & : & \mathrm{GL}_n({\mathbb F}) \rightarrow {\mathbb F},
 \nonumber \\
 & &  g \rightarrow z_{ij}.
\eq
A polynomial function on the group is a function on the group, which is a polynomial in the $a_{ij}$'s.
We denote by $H$ the set of polynomial functions on the group.
$H$ is a Hopf algebra.
The coproduct comes from the multiplication in the group.
Let $g_1, g_2 \in \mathrm{GL}_n({\mathbb F})$ with entries $z_{ij}^{(1)}$ and $z_{ij}^{(2)}$, respectively, and consider
\bq
 a_{ij}\left(g_1 \cdot g_2\right)
 & = &
 \sum\limits_{k=1}^n z_{ik}^{(1)} z_{kj}^{(2)}
 \; = \; 
 \sum\limits_{k=1}^n a_{ik}\left(g_1\right) a_{kj}\left(g_2\right)
 \nonumber \\
 & = &
 \cdot \left( \sum\limits_{k=1}^n a_{ik} \otimes a_{kj} \right) \left(g_1 \otimes g_2\right).
\eq
This gives us the coproduct
\bq
 \Delta & : & H \rightarrow H \otimes H,
 \nonumber \\
 & &
 a_{ij} \rightarrow \sum\limits_{k=1}^n a_{ik} \otimes a_{kj}.
\eq
It can be verified that $H$ is indeed a Hopf algebra.

Let us summarise: Given a group $G=\mathrm{GL}_n({\mathbb F})$ we obtain a Hopf algebra $H$ by considering 
the polynomial functions on $G$.
This also goes in the reverse direction: One may show that one can construct the group $G$ from the Hopf algebra $H$.

A group may act on a vector space.
Also in this case we may consider the dual picture.
For concreteness, we take again $G=\mathrm{GL}_n({\mathbb F})$ and consider $V={\mathbb F}^n$.
Let
\bq
 v & = & \left( \begin{array}{c} x_1 \\ \vdots \\ x_n \end{array} \right)
 \; \in \; V.
\eq
We consider functions on $V$. We start from the coordinate functions
\bq
 b_i & : & {\mathbb F}^n \rightarrow {\mathbb F},
 \nonumber \\
 & & v \rightarrow x_i.
\eq
Let $M$ denote the set of polynomial functions on $V$ (i.e. functions which are given as polynomials in the $b_i$'s).
The action of $G$ on $V$ induces a coaction of $H$ on $M$:
Consider
\bq
 b_i\left( g \cdot v \right)
 & = &
 \sum\limits_{k=1}^n z_{ik} x_k
 \; = \; 
 \sum\limits_{k=1}^n a_{ik}\left(g\right) b_k\left(v\right)
 \nonumber \\
 & = &
 \cdot \left( \sum\limits_{k=1}^n a_{ik} \otimes b_k \right) \left( g \otimes v \right).
\eq
This gives us the coaction (which we also denote by $\Delta$)
\bq
 \Delta & : & M \rightarrow H \otimes M,
 \nonumber \\
 & & 
 b_i \rightarrow \sum\limits_{k=1}^n a_{ik} \otimes b_k.
\eq
Let's now apply these ideas to motivic periods:
We first define motivic de Rham periods. 
These are triples of the form
\bq
 \left[ M^k\left(X,Y\right), \omega, \omega^\ast \right]^{\mathfrak{d}\mathfrak{R}},
\eq
where $\omega \in H^k_{\mathrm{alg \; dR}}\left(X,Y\right)$ and 
$\omega^\ast \in (H^k_{\mathrm{alg \; dR}}\left(X,Y\right))^\ast$
(the dual). 
As usual, we consider these triples modulo the relations 
of linearity eq.~(\ref{chapter_motives:linearity}), change of variables eq.~(\ref{chapter_motives:change_of_variables}) 
and Stokes eq.~(\ref{chapter_motives:Stokes}).
We call the set of equivalence classes 
\index{motivic de Rham periods}
{\bf motivic de Rham periods} 
and denote this set by
${\mathcal P}^{\mathfrak{d}\mathfrak{R}}$.

Let us denote by $\omega_1, \dots, \omega_r \in H^k_{\mathrm{alg \; dR}}(X,Y)$ a basis of $H^k_{\mathrm{alg \; dR}}(X,Y)$
and by
$\omega_1^\ast, \dots, \omega_r^\ast \in (H^k_{\mathrm{alg \; dR}}(X,Y))^\ast$ the dual basis of $(H^k_{\mathrm{alg \; dR}}(X,Y))^\ast$.
On ${\mathcal P}^{\mathfrak{d}\mathfrak{R}}$ we have a coproduct
\bq
 \Delta & : & {\mathcal P}^{\mathfrak{d}\mathfrak{R}} \rightarrow {\mathcal P}^{\mathfrak{d}\mathfrak{R}} \otimes {\mathcal P}^{\mathfrak{d}\mathfrak{R}}
\eq
which is given by
\bq
 \Delta \left[ M^k\left(X,Y\right), \omega_i, \omega_k^\ast \right]^{\mathfrak{d}\mathfrak{R}}
 & = &
 \sum\limits_{l=1}^r
 \left[ M^k\left(X,Y\right), \omega_i, \omega_l^\ast \right]^{\mathfrak{d}\mathfrak{R}}
 \otimes
 \left[ M^k\left(X,Y\right), \omega_l, \omega_k^\ast \right]^{\mathfrak{d}\mathfrak{R}}.
 \;\;\;\;\;\;
\eq
${\mathcal P}^{\mathfrak{d}\mathfrak{R}}$ is a Hopf algebra.
The 
\index{motivic Galois group}
{\bf motivic Galois group} $G^{\mathfrak{d}\mathfrak{R}}$ is the dual of the Hopf algebra ${\mathcal P}^{\mathfrak{d}\mathfrak{R}}$.
(This is the definition of the motivic Galois group. As mentioned in the introduction of this section, it is an indirect definition.)
${\mathcal P}^{\mathfrak{d}\mathfrak{R}}$ coacts on ${\mathcal P}^{\mathfrak m}$, this gives us the motivic coaction:
\begin{tcolorbox}[breakable]
\index{motivic coaction}
{\bf Motivic coaction}:
\\
We have a coaction of the motivic de Rham periods
${\mathcal P}^{\mathfrak{d}\mathfrak{R}}$ on
the effective motivic periods ${\mathcal P}^{\mathfrak m}$
\bq
 \Delta & : & {\mathcal P}^{\mathfrak m} \rightarrow {\mathcal P}^{\mathfrak{d}\mathfrak{R}} \otimes {\mathcal P}^{\mathfrak m}
\eq
given by
\bq
\label{chapter_motives:motivic_coaction}
 \Delta \left[ M^k\left(X,Y\right), \omega, \gamma \right]^{\mathfrak m}
 & = &
 \sum\limits_{l=1}^r
 \left[ M^k\left(X,Y\right), \omega, \omega_l^\ast \right]^{\mathfrak{d}\mathfrak{R}}
 \otimes
 \left[ M^k\left(X,Y\right), \omega_l, \gamma \right]^{\mathfrak m}.
 \;\;\;\;\;\;
\eq
Eq.~(\ref{chapter_motives:motivic_coaction}) is called the 
{\bf motivic coaction}.
\end{tcolorbox}

Let's look at an example. We elaborate on the example from section~\ref{chapter_motives:de_Rham_cohomology}.
We take $X={\mathbb C}^\ast = {\mathbb C}\backslash\{0\}$ and $Y=\{1,x\}$ 
(in section~\ref{chapter_motives:de_Rham_cohomology} we considered the special case $x=2$).
A basis for $H^{\mathrm{B}}_1(X,Y)$ is given by an anti-clockwise circle $\gamma_1$ around $z=0$
and the line segment $\gamma_2$ from $z=1$ to $z=x$.
A basis for $H_{\mathrm{dR}}^1(X,Y)$ is given by $\omega_1=dz/z$
and $\omega_2=dz/(x-1)$.
The period matrix is given by
\bq
 P 
 & = &
 \left(\begin{array}{cc}
 2 \pi i & \ln x \\
 0 & 1 \\
 \end{array} \right).
\eq
We have chosen the normalisation of $\omega_2$ such that 
\bq
\label{chapter_motives:example_motivic_coaction_normalisation}
 \int\limits_{\gamma_2} \omega_2 & = & 1.
\eq
We set
\bq
 \ln^{\mathfrak m}(x)
 & = &
 \left[ M^1\left(X,Y\right), \omega_1, \gamma_2 \right]^{\mathfrak m}.
\eq
The motivic coaction gives us
\bq
 \Delta\left(\ln^{\mathfrak m}(x)\right)
 & = &
 \left[ M^1\left(X,Y\right), \omega_1, \omega_1^\ast \right]^{\mathfrak{d}\mathfrak{R}}
 \otimes
 \left[ M^1\left(X,Y\right), \omega_1, \gamma_2 \right]^{\mathfrak m}
 \nonumber \\
 & &
 +
 \left[ M^1\left(X,Y\right), \omega_1, \omega_2^\ast \right]^{\mathfrak{d}\mathfrak{R}}
 \otimes
 \left[ M^1\left(X,Y\right), \omega_2, \gamma_2 \right]^{\mathfrak m}.
\eq
Let's look at the individual expressions:
The period map of eq.~(\ref{chapter_motives:effective_motivic_period_map})
sends $[ M^1\left(X,Y\right), \omega_2, \gamma_2 ]^{\mathfrak m}$ to $1$.
This can be read off from eq.~(\ref{chapter_motives:example_motivic_coaction_normalisation}).
Assuming that the period map is injective, we set
\bq
 \left[ M^1\left(X,Y\right), \omega_2, \gamma_2 \right]^{\mathfrak m}
 & = & 1^{\mathfrak m}.
\eq
$[ M^1\left(X,Y\right), \omega_1, \omega_1^\ast ]^{\mathfrak{d}\mathfrak{R}}$ 
is an element of ${\mathcal P}^{\mathfrak{d}\mathfrak{R}}$.
The motivic de Rham periods form a Hopf algebra and it can be shown that the counit
maps $[ M^1\left(X,Y\right), \omega_1, \omega_1^\ast ]^{\mathfrak{d}\mathfrak{R}}$ to $1$.
We denote
\bq
 \left[ M^1\left(X,Y\right), \omega_1, \omega_1^\ast \right]^{\mathfrak{d}\mathfrak{R}}
 & = & 1^{\mathfrak{d}\mathfrak{R}}.
\eq
If we finally define
\bq
 \ln^{\mathfrak{d}\mathfrak{R}}(x)
 & = &
 \left[ M^1\left(X,Y\right), \omega_1, \omega_2^\ast \right]^{\mathfrak{d}\mathfrak{R}}
\eq
we arrive at
\bq
 \Delta\left(\ln^{\mathfrak m}(x)\right)
 & = &
 1^{\mathfrak{d}\mathfrak{R}} \otimes \ln^{\mathfrak m}(x)
 +
 \ln^{\mathfrak{d}\mathfrak{R}}(x) \otimes 1^{\mathfrak m},
\eq
in agreement with eq.~(\ref{appendix_solutions:coaction_logarithm}).
This provides the link with the coaction defined for multiple polylogarithms in section~\ref{chapter_hopf:coaction}.

\section{Examples from Feynman integrals}
\label{chapter_motives:examples_Feynman_integrals}

Feynman integrals provide ample examples for motivic periods.
Throughout this section we assume that all kinematic variables satisfy
\bq
\label{chapter_motives:condition_euclidean}
 x_j & \ge & 0,
 \;\;\;\;\;\;
 \left( \mbox{Euclidean region} \right)
\eq
and
\bq
\label{chapter_motives:condition_rational}
 x_j & \in & {\mathbb Q}.
\eq
Condition~(\ref{chapter_motives:condition_euclidean}) 
ensures that in the Feynman parametrisation the singularities of the integrand are 
at worst on the boundary of the integration domain, but not inside.
Condition~(\ref{chapter_motives:condition_rational}) ensures that the variety where the integrand is singular
is defined over ${\mathbb Q}$.

\subsection{Feynman motives depending only on one graph polynomial}
\label{chapter_motives:motive_one_graph_polynomial}

We recall that with these assumptions 
we may use the algorithm of sector decomposition (see chapter~\ref{chapter_sector_decomposition}) and express 
any term $I^{(j)}$ in the $\eps$-expansion of any Feynman integral $I$
\bq
 I & = &
 \sum\limits_{j=j_{\mathrm{min}}}^\infty \eps^j I^{(j)}
\eq
as an absolute convergent integral.
This will also ensure that the singularities have normal crossings.
From theorem~\ref{chapter_sector_decomposition:Feynman_period} we know that each $I^{(j)}$ gives a numerical period.
Denoting by $\eta$ the integrand (a $k$-form), $\gamma$ the integration domain (a relative $k$-cycle with boundary contained in $B$), 
$P$ the space containing $\gamma$ and $Y$
the variety where $\eta$ is singular we obtain a motive
\bq
 M^k\left( P \backslash Y, B \backslash \left(B \cap Y\right) \right)
\eq
and the motivic period
\bq
 \left[ M^k\left( P \backslash Y, B \backslash \left(B \cap Y\right) \right), \eta, \gamma \right]^{\mathfrak m}.
\eq
Let us now specialise to two cases, where $Y$ is determined by a single graph polynomial.
(The case where $Y$ is determined by both graph polynomials is only from a notational perspective more cumbersome.)
We start from the Feynman parameter representation eq.~(\ref{chapter_basics:Feynman_representation_differential_forms})
and set from the beginning $\nu_1=\cdots=\nu_{\ninternal}=1$:
\bq
 I
 & = &
 e^{\loopnumber \eps \Eulerconstant}\Gamma\left(\ninternal-\frac{\loopnumber D}{2}\right)
 \int\limits_{\Delta} 
 \frac{{\mathcal U}^{\ninternal-\frac{\left(\loopnumber+1\right) D}{2}}}{{\mathcal F}^{\ninternal-\frac{\loopnumber D}{2}}} \omega.
\eq
The differential $(\ninternal-1)$-form $\omega$ is defined in eq.~(\ref{chapter_basics:def_omega}).
Let us further assume that the integration over the Feynman parameters is well-defined 
without regularisation.
This implies that the Feynman integral is either finite or has at most an overall ultraviolet
singularity, which manifests itself in the prefactor $\Gamma(\ninternal-\loopnumber D/2)$.
Let us further assume that either the exponent of the ${\mathcal F}$-polynomial is zero
or the exponent of the ${\mathcal U}$-polynomial is zero.
In the former case we end up with (apart from prefactors)
\bq
\label{chapter_motives:example_U}
 \int\limits_{\Delta} 
 \frac{\omega}{{\mathcal U}^{\frac{D}{2}}},
\eq
in the latter case with
\bq
\label{chapter_motives:example_F}
 \int\limits_{\Delta} 
 \frac{\omega}{{\mathcal F}^{\frac{D}{2}}}.
\eq
In both cases the integrand depends only on one graph polynomial. For $D$ an even integer, the integrand
is a rational function.

Let's see if there are interesting examples matching our assumptions:
Setting $D=2$ in eq.~(\ref{chapter_motives:example_U}) yields $\ninternal = \loopnumber$ 
(since the exponent of the ${\mathcal F}$-polynomial has to vanish).
This only allows for a product of tadpole integrals and is not so interesting.
However, $D=4$ leads to $\ninternal = 2 \loopnumber$ and there are interesting graphs.
There is the family of the wheel with $l$ spokes graphs
and the family of zigzag graphs \cite{Bloch:2006,Doryn:2010zz,Brown:2012ia}.
\begin{figure}
\begin{center}
\includegraphics[scale=1.0]{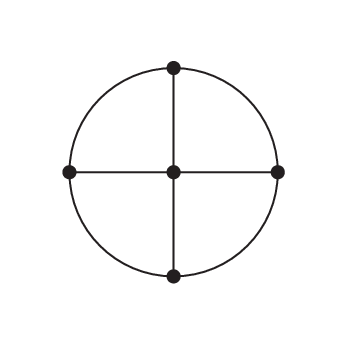}
 \hspace*{25mm}
\includegraphics[scale=1.0]{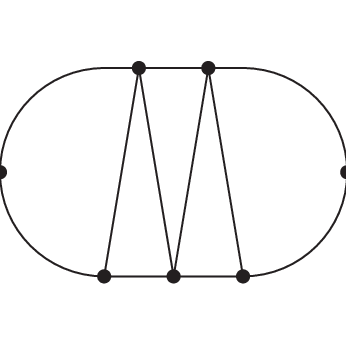}
\end{center}
\caption{
The $4$-loop wheel with four spokes graph (left) and the $6$-loop zigzag graph (right).
}
\label{chapter_motives:fig_wheel_spoke_zigzag}
\end{figure}
Two examples are shown in fig.~\ref{chapter_motives:fig_wheel_spoke_zigzag}.

In the case of eq.~(\ref{chapter_motives:example_F}) we may set $D=2$. We obtain then $\ninternal=(\loopnumber+1)$.
\begin{figure}
\begin{center}
\includegraphics[scale=1.0]{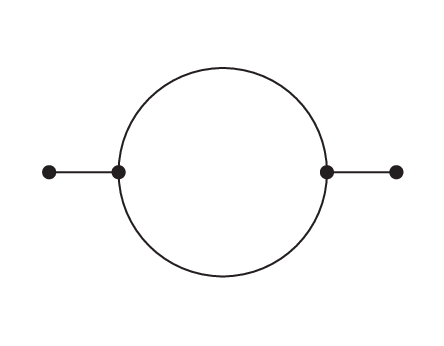}
\includegraphics[scale=1.0]{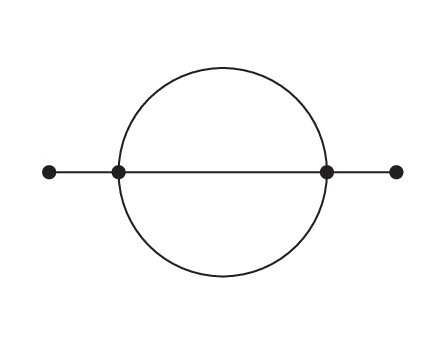}
\includegraphics[scale=1.0]{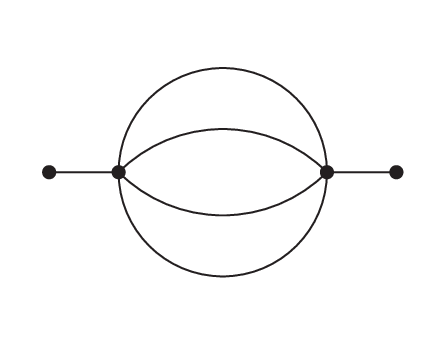}
\end{center}
\caption{
The first three members of the family of banana graphs.
}
\label{chapter_motives:fig_banana}
\end{figure}
This gives the family of banana graphs shown in fig.~\ref{chapter_motives:fig_banana} \cite{Aluffi:2008sy,MullerStach:2011ru}.
For non-zero internal masses these integrals are finite (in $D=2$ space-time dimensions).

Thus the examples which we are going to consider are
\bq
\label{chapter_motives:example_final_U_F}
 \int\limits_{\Delta} 
 \frac{\omega}{{\mathcal U}^2} \;\;\; \mbox{(wheel with $l$ spokes graphs, zigzag graphs)},
 & &
 \int\limits_{\Delta} 
 \frac{\omega}{{\mathcal F}} \;\;\; \mbox{(banana graphs)}.
 \;\;\;\;\;\;
\eq
In both cases the integrand is a rational differential form on ${\mathbb C}{\mathbb P}^{\ninternal-1}$ and $\Delta={\mathbb R} {\mathbb P}^{\ninternal-1}_{\ge 0}$.
The integration region has a boundary, which we denote by $\partial \Delta$.
Despite the fact that we assumed the integrals are finite, we cannot conclude that the integrand has no singularities: 
There might be (and will be) integrable singularities.
This will happen, whenever the relevant graph polynomial vanishes.
Let us set ${\mathcal X}={\mathcal U}$ for the wheel with $l$ spokes graphs and the zigzag graphs
and ${\mathcal X}={\mathcal F}$ for the banana graphs.
We define
\bq
 X & = &
 \left\{ \; a \in {\mathbb C}{\mathbb P}^{\ninternal-1} \; | \;
 {\mathcal X}\left(a\right) \; = \; 0 \;
 \right\}.
\eq
$X$ is an algebraic variety, defined as the zero set of the graph polynomial ${\mathcal X}$.
In the integration, we encounter a singularity wherever $X$ and $\Delta$ intersect.
It can be shown for the first graph polynomial ${\mathcal U}$ that
\bq
\label{chapter_motives:eq_singularities_boundary}
 X \cap \Delta & \subset & \partial \Delta,
\eq
e.g. intersections happens only on the boundary $\partial \Delta$ of the integration region $\Delta$.
This is the case because ${\mathcal U}$ is a sum of monomials with all coefficients equal to $1$.
Eq.~(\ref{chapter_motives:eq_singularities_boundary}) also holds true for the second graph polynomial ${\mathcal F}$, 
if we restrict ourselves for the kinematics to the Euclidean region (e.g. all kinematic variables $x_j \ge 0$).
In this case ${\mathcal F}$ is a sum of monomials with all coefficients positive.
This is the reason why we imposed assumption eq.~(\ref{chapter_motives:condition_euclidean}).

If $X \cap \partial \Delta$ is non-empty, we blow-up ${\mathbb C}{\mathbb P}^{\ninternal-1}$ in this region.
Let us denote the blow-up by $P$.
We further denote the strict transform of $X$ by $Y$ and 
we denote the total transform of the set 
$\{ x_1 \cdot x_2 \cdot \ldots \cdot x_n = 0 \}$ by $B$.

We view the integrals of eq.~(\ref{chapter_motives:example_final_U_F}) as periods of a mixed Hodge structures,
obtained from two geometric objects: The algebraic variety $X$ and the domain of integration $\Delta$.
As the domain of integration $\Delta$ has a boundary we have to consider relative cohomology.
In order to avoid (integrable) singularities of the integrand we consider the blow-up.
This leads us to the mixed Hodge structure given by the relative cohomology group \cite{Bloch:2006,Bloch:2008jk,Connes:2005zp,Aluffi:2009aa,Brown:2009a,Schnetz:2008mp}
\bq
 H^{\ninternal-1}\left(P \backslash Y, B \backslash \left( B \cap Y \right) \right).
\eq
The Feynman integral is then a period of this cohomology class.
The
\index{motive associated to a Feynman integral}
{\bf motive associated to the Feynman integral}
is given by
\bq
\label{chapter_motives:feynman_motive}
\lefteqn{
 M^{\ninternal-1}\left(P \backslash Y, B \backslash \left( B \cap Y \right) \right)
 = } & &
 \\ 
 & &
 \left[
  H^{\ninternal-1}_{\mathrm{alg \; dR}}\left(P \backslash Y, B \backslash \left( B \cap Y \right) \right),
  H^{\ninternal-1}_{\mathrm{B}}\left( \left(P \backslash Y\right)^{\mathrm{an}}, \left(B \backslash \left( B \cap Y \right)\right)^{\mathrm{an}}, {\mathbb Q} \right),
  \mathrm{comparison}
 \right]^{\mathfrak m}.
 \nonumber
\eq

\subsection{The sunrise motive}

Let us now look at a concrete example: The two-loop sunrise integral with unequal masses.
We encountered this Feynman integral already in eq.~(\ref{chapter_elliptics:def_unequal_mass_sunrise}).
With the notation as in eq.~(\ref{chapter_elliptics:def_unequal_mass_sunrise}) we consider
\bq
 I_{111}\left(2,x,y_1,y_2\right)
 & = &
 \int \frac{d^2k_1}{i \pi} \frac{d^2k_2}{i \pi} 
 \frac{m_3^2}{\left(-q_1^2+m_1^2\right) \left(-q_2^2+m_2^2\right) \left(-q_3^2+m_3^2\right)}.
\eq
In the Feynman parameter representation we have
\bq
 I_{111}\left(2,x,y_1,y_2\right)
 & = &
 \int\limits_{\Delta} 
 \frac{\omega}{{\mathcal F}},
\eq
with
\bq
 {\mathcal F} & = &
 a_1 a_2 a_3 x + \left( a_1 y_1 + a_2 y_2 + a_3 \right) \left( a_1 a_2 + a_2 a_3 + a_3 a_1 \right),
 \nonumber \\
 \omega & = & 
 a_1 da_2 \wedge da_3
 - a_2 da_1 \wedge da_3
 + a_3 da_1 \wedge da_2,
 \nonumber \\
 \Delta & = & 
 {\mathbb R} {\mathbb P}^{2}_{\ge 0}.
\eq
We define the variety where ${\mathcal F}$ vanishes:
\bq
 X & = & 
 \left\{ \; \left[a_1:a_2:a_3\right] \, \in \, {\mathbb C} {\mathbb P}^{2} \; | \; {\mathcal F}(a) \, = \, 0 \; \right\}.
\eq
$X$ and $\Delta$ intersect in the three points
\bq
 \left[1:0:0\right],
 \;\;\;
 \left[0:1:0\right],
 \;\;\;
 \left[0:0:1\right].
\eq
Let $P$ be the blow-up of ${\mathbb C} {\mathbb P}^{2}$ in these three points.
The exceptional divisors in these three points are denoted by $E_1$, $E_2$ and $E_3$, respectively.
We denote the strict transform of $X$ by $Y$ and we denote by $B$ the total transform of $a_1 \cdot a_2 \cdot a_3 = 0$. 
The mixed Hodge structure (or motive) associated to the Feynman integral $I_{111}(2,x,y_1,y_2)$ is then
\bq
 H^2\left(P \backslash Y, B \backslash B \cap Y \right).
\eq 
The Feynman integral is a period of this motive.

What can be said about this motive?
One can show that there is a short exact sequence of mixed Hodge structures \cite{MullerStach:2011ru}
\bq
\label{chapter_motives:short_exact_sequence}
\begin{CD}
 0 
 @>>> 
 {\mathbb Z}\left(-1\right) 
 @>>> 
 H^2\left(P \backslash Y, B \backslash B \cap Y \right) 
 @>>> 
 H^2\left(P \backslash Y \right) 
 @>>> 
 0,
\end{CD}
\eq
and for $H^2(P \backslash Y)$ we have the short exact sequence
\bq
\label{chapter_motives:gysin}
\begin{CD}
 0 
 @>>> 
 {\mathbb Z} E_1 \oplus {\mathbb Z} E_2 \oplus {\mathbb Z} E_3
 @>>> 
 H^2\left(P \backslash Y \right) 
 @>{\mathrm{res}}>> 
 H^1\left( Y \right) 
 @>>> 
 0,
\end{CD}
\eq
This sequence is split as a sequence of mixed Hodge structures via
\bq
\begin{CD}
 H^2\left(P \backslash Y \right) @>{\mathrm{res}}>> H^1\left( Y \right) \\
@A{\pi^{*}}AA		@VV{\cong}V \\
H^2( {\mathbb C}{\mathbb P}^2 \backslash X) @>{\mathrm{res}}>{\cong}> H^1\left(X\right).
\end{CD}
\eq
\begin{digression} {\bf Exact sequences}
\\
We start from a category where kernels and cokernels are defined.
Typical examples are
the category of groups,
the category of vector spaces or the category of modules.
We denote the objects by $O_i$ and the morphisms by $f_j$.
Consider a sequence of morphisms
\bq
 \begin{CD}
 O_0 @>{f_1}>> O_1 @>{f_2}>> O_2 @>{f_3}>> \dots @>{f_n}>> O_n.
 \end{CD}
\eq
The sequence is said to be 
\index{exact sequence}
{\bf exact}, if
\bq
 \mathrm{im}\left(f_i\right) & = & \mathrm{ker}\left(f_{i+1}\right).
\eq
An exact sequence of the form
\bq
 \begin{CD}
 0 @>>> O_1 @>{f}>> O_2
 \end{CD}
\eq
states that $f$ is injective, i.e. $f$ is a monomorphism.
The image of $0$ under the first map is $0$, which by the assumption of exactness is the kernel of $f$. 
Hence the kernel of $f$ is trivial and $f$ is injective.

By a similar reasoning, the exact sequence
\bq
 \begin{CD}
 O_1 @>{f}>> O_2 @>>> 0
 \end{CD}
\eq
expresses that $f$ is surjective, i.e. $f$ is an epimorphism: The kernel of the rightmost morphism is $O_2$.
Since the sequence is supposed to be exact, this equals the image of $f$, hence $f$ is surjective.

A 
\index{short exact sequence}
{\bf short exact sequence} is an exact sequence of the form
\bq
 \begin{CD}
 0 @>>> O_1 @>{f}>> O_2 @>{g}>> O_3 @>>> 0.
 \end{CD}
\eq
This implies that $f$ is injective and $g$ is surjective.
A short exact sequence is said to be 
\index{split short exact sequence}
{\bf split}, if there is a morphism $h : O_3 \rightarrow O_2$ such that $g \circ h$ is the identity map on $O_3$.

In order to distinguish a general exact sequence from the special case of a short exact sequence, the term 
\index{long exact sequence}
{\bf long exact sequence} is also used for the former. 
\end{digression}

Now let us discuss how these considerations can be turned into a practical tool.
Suppose we would like to compute the differential equation of $I_{111}(2,x,y_1,y_2)$ with respect to $x$.
Of course, we already know one possibility how to do this: 
We could start with integration-by-parts identities in $D$ dimensions
as in section~\ref{chapter_iterated_integrals:integration_by_parts}, derive a coupled system
of first-order differential equations as in section~\ref{chapter_iterated_integrals:differential_equations}
and convert the coupled system of first-order differential equations to a single higher order differential
equation as in section~\ref{chapter_transformations:fibre_transformation:picard_fuchs}.
Doing so, we would expect an inhomogeneous fourth order differential equation for $I_{111}(D,x,y_1,y_2)$.
(We expect a fourth order differential equation because there are four master integrals in the top sector.
We expect an inhomogeneous differential equation because there are sub-topologies.)
For generic $D$ we would indeed obtain an inhomogeneous fourth order differential equation.
However we are interested in the finite integral $I_{111}(2,x,y_1,y_2)$ in two space-time dimensions.
This raises the question if we can carry out the calculation directly in $D=2$ space-time dimensions
without the need of introducing an additional symbolic variable $D$.
Secondly, we will see shortly that in two space-time dimensions there are only 
two independent master integrals in the top sector
(instead of four master integrals for generic $D$).
Thus we would like to avoid to work in intermediate stages with four master integrals in the top sector, if only
two master integrals are required.

These questions are particularly important for cutting-edge calculations, where additional variables
or additional master integrals can push the required computing resources (memory and/or CPU time) beyond
the available resources.
Therefore we would like to calculate only these parts which are strictly necessary.
Motivic methods allow us to do this \cite{MullerStach:2011ru}.

Let's look at the details. We would like to derive the differential equation
for $I_{111}(2,x,y_1,y_2)$ with respect to the kinematic variable $x$. 
We treat the two additional kinematic variables $y_1$ and $y_2$ as (fixed) parameters.
We seek
\bq
 L \; I_{111}(2,x,y_1,y_2)
 & = &
 Q\left(x\right),
\eq
where $L$ is a differential operator
\bq
 L & = & \sum\limits_{j=0}^r P_j\left(x\right) \frac{d^j}{dx^j}
\eq
and $Q(x)$ the inhomogeneous term.
The order $r$ of the differential operator $L$ is a priori unknown.

Let's focus first on the differential operator $L$.
We start from the algebraic variety $X$ defined by the second Symanzik polynomial ${\mathcal F}$:
\bq
 a_1 a_2 a_3 x + \left( a_1 y_1 + a_2 y_2 + a_3 \right) \left( a_1 a_2 + a_2 a_3 + a_3 a_1 \right) & = & 0.
\eq
This defines for generic values of the parameters $x$, $y_1$ and $y_2$ an elliptic curve.
The elliptic curve varies smoothly with the parameters $x$, $y_1$ and $y_2$.
By a birational change of coordinates this equation can brought into the Weierstrass normal form
(instead of the common $x, y, z$ we use $u, v, w$ as coordinates in ${\mathbb C}{\mathbb P}^2$)
\bq
 v^2 w - u^3 - f_2(x) u w^2 - f_3(x) w^3 & = & 0.
\eq
$f_2$ and $f_3$ are functions of the kinematic variables $x$, $y_1$ and $y_2$.
As we are mainly interested in the dependence on $x$, we suppress in the notation the dependence on $y_1$ and $y_2$.
In the chart $w=1$ the above equation reduces to
\bq
\label{chapter_motives:weierstrass_normal_form}
 v^2 - u^3 - f_2(x) u - f_3(x) & = & 0.
\eq
In these coordinates $H^1(X)$ is generated by
\bq
 \eta = \frac{du}{v}
 & \mbox{and} &
 \eta' = \frac{d}{dx} \eta.
\eq
Since $H^1(X)$ is two-dimensional it follows that $\eta''=\frac{d^2}{dx^2} \eta$ 
must be a linear combination of $\eta$ and $\eta'$.
In other words we must have a relation of the form
\bq
\label{chapter_motives:eq_eta}
 \eta'' +  R_1(x) \eta' + R_0(x) \eta & = & 0.
\eq
It is convenient to bring this equation onto a common denominator. 
Doing so and carrying out the derivatives with respect to $x$ we have
\bq
 \eta & = & \left( u^3 + f_2 u + f_3 \right)^2 \frac{du}{v^5},
 \nonumber \\
 \eta' & = & - \frac{1}{2} \left( f_2' u + f_3' \right) \left( u^3 + f_2 u + f_3 \right) \frac{du}{v^5},
 \nonumber \\
 \eta'' & = & \left[ - \frac{1}{2} \left( f_2'' u + f_3'' \right) \left( u^3 + f_2 u + f_3 \right)
                          + \frac{3}{4} \left( f_2' u + f_3' \right)^2 \right] \frac{du}{v^5}.
\eq
The numerator of eq.~(\ref{chapter_motives:eq_eta}) is then a polynomial of degree $6$ in the single variable $u$.
Since we work in $H^1(X)$, we can simplify the expression by adding an exact form
\bq
 d \left( \frac{u^n}{v^3} \right)
 & = & u^{n-1} \left[ \left(n-\frac{9}{2}\right)u^3 + \left( n-\frac{3}{2} \right) f_2 u + n f_3 \right] \frac{du}{v^5}.
\eq
This allows us to reduce the numerator polynomial from degree six to a linear polynomial.
The two coefficients of this linear polynomial have to vanish, on account of eq.~(\ref{chapter_motives:eq_eta}).
We obtain therefore two equations for the two unknowns $R_1$ and $R_0$. 
Solving for $R_1$ and $R_0$ we find
\bq
 R_1\left(x\right) = \frac{P_1\left(x\right)}{P_2\left(x\right)},
 & &
 R_0\left(x\right) = \frac{P_0\left(x\right)}{P_2\left(x\right)},
\eq
with
\bq
\label{chapter_motives:res_p012}
 P_2\left(x\right) & = &
  -x
 \left[ 3 x^2 + 2 M_{100} x - M_{200} + 2 M_{110} \right]
 \left[ 
  x^4
  + 4 M_{100} x^3
  + 2 \left( 3 M_{200} + 2 M_{110} \right) x^2
 \right. \nonumber \\
 & & \left.
  + 4 \left( M_{300} - M_{210} + 10 M_{111} \right) x
  + \left(M_{200} - 2 M_{110} \right)^2
 \right],
 \nonumber \\
 P_1\left(x\right) & = & 
  - 9 x^6
  - 32 M_{100} x^5
  - \left( 37 M_{200} + 70 M_{110} \right) x^4
  - \left( 8 M_{300} + 56 M_{210} + 144 M_{111} \right) x^3
 \nonumber \\
 & &
  + \left( 13 M_{400} - 36 M_{310} + 46 M_{220} - 124 M_{211} \right) x^2
 \nonumber \\
 & &
  - \left( -8 M_{500} + 24 M_{410} - 16 M_{320} - 96 M_{311} + 144 M_{221} \right) x
 \nonumber \\
 & &
  + \left( M_{600} - 6 M_{510} + 15 M_{420} - 20 M_{330} + 18 M_{411} - 12 M_{321} - 6 M_{222} \right),
 \nonumber \\
 P_0\left(x\right) & = &
  - 3 x^5
  - 7 M_{100} x^4
  - \left( 2 M_{200} + 16 M_{110} \right) x^3
  + \left( 6 M_{300} - 14 M_{210} \right) x^2
 \nonumber \\
 & &
  + \left( 5 M_{400} - 8 M_{310} + 6 M_{220} - 8 M_{211} \right) x
  + \left( M_{500} - 3 M_{410} + 2 M_{320} + 8 M_{311} 
 \right. \nonumber \\
 & & \left.
 - 10 M_{221} \right).
\eq
Here we used the same notation as in section~\ref{chapter_elliptics:feynman_integrals_several_kinematic_variables}: 
\bq
 M_{\lambda_1 \lambda_2 \lambda_3}
 & = &
 M_{\lambda_1 \lambda_2 \lambda_3}\left(y_1,y_2,1\right)
\eq
and $M_{\lambda_1 \lambda_2 \lambda_3}\left(a_1,a_2,a_3\right)$ is defined in eq.~(\ref{chapter_elliptics:def_notation_M_ijk}).

We therefore obtain the Picard-Fuchs operator for $\eta \in H^1(X)$ as
\bq
 L
 & = & 
 P_2\left(x\right) \frac{d^2}{dx^2}
 +
 P_1\left(x\right) \frac{d}{dx}
 +
 P_0\left(x\right),
\eq
with $P_2$, $P_1$ and $P_0$ defined in eq.~(\ref{chapter_motives:res_p012}).
This is also the Picard-Fuchs operator of the Feynman form 
\bq
 \varphi & = & \frac{\omega}{{\mathcal F}}
 \; \in \; H^2\left(P \backslash Y\right)
\eq 
due to the splitting of the sequence in eq.~(\ref{chapter_motives:gysin})
and the flatness of the system ${\mathbb Z} E_1\oplus {\mathbb Z} E_2\oplus {\mathbb Z} E_3$. 
So for any cycle $\gamma \in H_2(P \backslash Y)$ we have
\bq
 L \left( \int\limits_\gamma \varphi \right) & = & 0.
\eq
Let us now turn to the inhomogeneous part $Q(x)$.
The integration domain for the Feynman integral $I_{111}$ is not a cycle $\gamma \in H_2(P \backslash Y)$, but the relative cycle
$\Delta \in H_2(P \backslash Y, B \backslash B \cap Y )$.
From the short exact sequence in eq.~(\ref{chapter_motives:short_exact_sequence}) we deduce that $L \varphi$ is exact, i.e. there is a one-form $\beta$ such that
\bq
\label{chapter_motives:eq_beta}
 L \; \varphi & = & d \beta.
\eq
We make the ansatz \cite{Griffiths:1969}
\bq
 \beta & = & 
 \frac{1}{{\mathcal F}^2} 
 \left[ \left( a_2 q_3 - a_3 q_2 \right) da_1 + \left( a_3 q_1 - a_1 q_3 \right) da_2 + \left( a_1 q_2 - a_2 q_1 \right) da_3 \right],
\eq
where $q_1$, $q_2$ and $q_3$ are polynomials of degree $4$ in the variables $a_1$, $a_2$ and $a_3$.
The most general form is
\bq
\lefteqn{
 q_i = } & & \nonumber \\
 & &
 c^{(i)}_{400} a_1^4 + c^{(i)}_{040} a_2^4 + c^{(i)}_{004} a_3^4
 + 
 c^{(i)}_{310} a_1^3 a_2 + c^{(i)}_{301} a_1^3 a_3 + c^{(i)}_{130} a_1 a_2^3 + c^{(i)}_{103} a_1 a_3^3 + c^{(i)}_{031} a_2^3 a_3 + c^{(i)}_{013} a_2 a_3^3
 \nonumber \\
 & &
 +
 c^{(i)}_{211} a_1^2 a_2 a_3 + c^{(i)}_{121} a_1 a_2^2 a_3 + c^{(i)}_{112} a_1 a_2 a_3^2
 +
 c^{(i)}_{220} a_1^2 a_2^2 + c^{(i)}_{202} a_1^2 a_3^2 + c^{(i)}_{022} a_2^2 a_3^2.
\eq   
We would like $\beta$ to be finite on the boundary $\partial \sigma$. This implies
\bq
 c^{(1)}_{040} = c^{(1)}_{004} = c^{(2)}_{400} = c^{(2)}_{004} = c^{(3)}_{400} = c^{(3)}_{040} = 0.
\eq
The remaining 39 coefficients $c^{(i)}_{jkl}$ are found by solving the linear system of equations obtained from inserting the
ansatz into eq.~(\ref{chapter_motives:eq_beta}).
The solution of this linear system is not unique, corresponding to the fact that $\beta$ can be changed by a closed one-form.
The solutions for the coefficients $c^{(i)}_{jkl}$ are rather lengthy and not listed here.
In the next step we integrate $\beta$ along the boundary $\partial \Delta$ to get $Q(x)$:
\bq
 Q(x) & = & \int\limits_{\partial \Delta} \beta.
\eq
Note that the integration is in the blow-up $P$ of ${\mathbb P}^2$. 
This is most easily done as follows: We start from the integration domain
${\mathbb R} {\mathbb P}^{2}_{\ge 0}$, which we take as the union of three squares
\bq
 \Delta_{12,3} \cup \Delta_{23,1} \cup \Delta_{31,2}
\eq
with
\bq
 \Delta_{ij,k}
 & = & 
 \left\{ \; \left[a_1:a_2:a_3\right] \; | \; 0 \le a_i, a_j \le 1, \; a_k=1 \; \right\}.
\eq
Each square contains one point, which needs to be blown-up. In the square $\Delta_{ij,k}$ this is the point $a_i=a_j=0, a_k=1$.
The blow-up can be done as described in section~\ref{chapter_sector_decomposition:algo_sector_decomp}.
The blow-up of each square can be covered with two charts. Thus we get six charts in total.
\begin{figure}
\begin{center}
\includegraphics[scale=1.2]{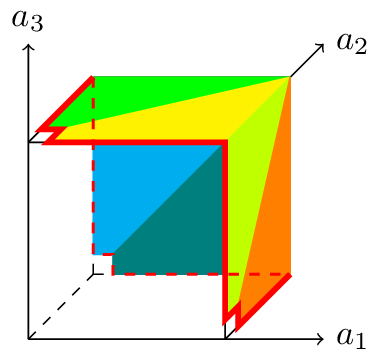}
\end{center}
\caption{
Sketch of the integration domain $\Delta$ covered by six charts. Each chart is drawn in a different colour.
The boundary $\partial \Delta$ is shown in red.
}
\label{chapter_motives:fig_integration_contour_sunrise}
\end{figure}
The integration contour $\partial \Delta$ is sketched in fig.~\ref{chapter_motives:fig_integration_contour_sunrise}.
Performing the integration we obtain
\bq
\label{res_p3}
 Q\left(x\right) & = & 
 -18 x^4
 - 24 M_{100} x^3
 + \left( 4 M_{200} - 40 M_{110} \right) x^2
 - \left( - 8 M_{300} + 8 M_{210} + 48 M_{111} \right) x
 \nonumber \\
 & & 
 + \left( - 2 M_{400} + 8 M_{310} - 12 M_{220} - 8 M_{211} \right)
 \nonumber \\
 & &
 + 2 c\left(x,y_1,y_2,1\right)  \ln y_1
 + 2 c\left(x,y_2,1,y_1\right)  \ln y_2
\eq
and
\bq
\lefteqn{
c\left(x,y_1,y_2,y_3\right) = } & &
 \nonumber \\
 & &
 \left( 2 y_1 - y_2 - y_3 \right) x^3
 + \left( 6 y_1^2 - 3 y_2^2 - 3 y_3^2 - 7 y_1 y_2 - 7 y_1 y_3 + 14 y_2 y_3 \right) x^2
 \nonumber \\
 & &
 - \left( -6 y_1^3 + 3 y_2^3 + 3 y_3^3 + 11 y_1^2 y_2 + 11 y_1^2 y_3 - 8 y_1 y_2^2 - 8 y_1 y_3^2 - 3 y_2^2 y_3 - 3 y_2 y_3^2 \right) x
 \nonumber \\
 & & 
 + \left( 2 y_1^4 - y_2^4 - y_3^4 - 5 y_1^3 y_2 - 5 y_1^3 y_3 + y_1 y_2^3 + y_1 y_3^3 + 4 y_2^3 y_3 + 4 y_2 y_3^3 
 \right. \nonumber \\
 & & \left.
        + 3 y_1^2 y_2^2 + 3 y_1^2 y_3^2 - 6 y_2^2 y_3^2 
        + 2 y_1^2 y_2 y_3 - y_1 y_2^2 y_3 - y_1 y_2 y_3^2 \right).
\eq
The coefficients $c(x,y_i,y_j,y_k)$ of the logarithms of the masses vanish for equal masses.

Putting everything together the sought-after differential equation reads
\bq
 \left[
 P_2\left(x\right) \frac{d^2}{dx^2}
 +
 P_1\left(x\right) \frac{d}{dx}
 +
 P_0\left(x\right)
 \right]
 I_{111}\left(2,x,y_1,y_2\right)
 & = & 
 Q\left(x\right).
\eq
This is an inhomogeneous second-order differential equation for the sunrise integral with unequal masses
in two space-time dimensions.
As it is second-order, we have two master integrals in the top sector in two space-time dimensions.
We recall that for generic $D$ we have four master integrals in the top sector.
The two additional master integrals can be chosen such that they vanish in the limit $D \rightarrow 2$ \cite{Remiddi:2013joa}.
\\
\\
\bs
{\it \refstepcounter{exercise}
\label{chapter_motives:exercise_motivic_bubble}
{\bf Exercise \theexercise}: 
Consider the one-loop two-point function with equal internal masses
\bq
 I_{11}\left(2,x\right)
 & = &
 \int \frac{d^2k}{i \pi}
 \frac{m^2}{\left(-q_1^2+m_1^2\right) \left(-q_2^2+m_2^2\right)},
 \;\;\;\;\;\;
 x \; = \; - \frac{p^2}{m^2}.
\eq
Derive with the methods of this section the differential equation for $I_{11}(2,x)$ with respect to the kinematic variable $x$.
}
\es

\subsection{Banana motives}

In the previous section we considered the one-loop bubble integral (in exercise~\ref{chapter_motives:exercise_motivic_bubble}) and
the two-loop sunrise integral, both in two space-time dimensions.
These are the first two members of the family of banana graphs.
The first three members of this family are shown in fig.~\ref{chapter_motives:fig_banana}.
 
The $l$-loop banana integral has $(l+1)$ propagators and is given in $D$ space-time dimensions by
\bq
 I_{\nu_1 \dots \nu_{l+1}}\left(D\right)
 = 
 e^{\loopnumber \eps \Eulerconstant} \left(\mu^2\right)^{\nu-\frac{\loopnumber D}{2}}
 \int \prod\limits_{a=1}^{\loopnumber} \frac{d^Dk_a}{i \pi^{\frac{D}{2}}} 
 \delta^D\left(p-\sum\limits_{b=1}^{\loopnumber+1} k_b \right)
 \left( \prod\limits_{c=1}^{\loopnumber+1} \frac{1}{\left(-k_c^2+m_c^2\right)^{\nu_j}} \right).
 \;\;\;
\eq
We introduce
\bq
 x \; = \; \frac{-p^2}{\mu^2},
 \;\;\;\;\;\;
 y_1 \; = \; \frac{m_1^2}{\mu^2},
 \;\;\;\;\;\;
 y_2 \; = \; \frac{m_2^2}{\mu^2},
 \;\;\;\;\;\;
 \dots,
 \;\;\;\;\;\;
 y_{\loopnumber+1} \; = \; \frac{m_{\loopnumber+1}^2}{\mu^2}.
\eq
As usual we may set one variable to one. 
The $l$-loop banana integral depends therefore in the generic case 
(i.e. all internal masses pairwise distinct and $(-p^2)$ non-zero and not equal to any internal mass squared)
on $\NB=\loopnumber+1$ kinematic variables.
Of particular interest is also the equal mass case $m_1=\dots=m_{\loopnumber+1}=m$, in which case the
$l$-loop banana integral depends on one kinematic variable, which we may take as $x=-p^2/m^2$.

The Feynman parameter representation of the $l$-loop banana integral is given by
\bq
\label{chapter_motives:Feynman_representation_banana}
 I_{\nu_1 \dots \nu_{l+1}}\left(D\right)
 & = &
 \frac{e^{\loopnumber \eps \Eulerconstant}
 \Gamma\left(\nu-\frac{\loopnumber D}{2}\right)}{\prod\limits_{j=1}^{\loopnumber+1}\Gamma(\nu_j)}
 \int\limits_{\Delta} \omega \;
 \left( \prod\limits_{j=1}^{\loopnumber+1} a_j^{\nu_j-1} \right)
 \frac{{\mathcal U}^{\nu-\frac{\left(\loopnumber+1\right) D}{2}}}{{\mathcal F}^{\nu-\frac{\loopnumber D}{2}}},
\eq
with $\Delta={\mathbb R} {\mathbb P}^{\loopnumber}_{\ge 0}$ and
\bq
 \omega & = & \sum\limits_{j=1}^{l+1} (-1)^{j-1}
  \; a_j \; da_1 \wedge ... \wedge \widehat{da_j} \wedge ... \wedge da_n.
\eq
The hat indicates that the corresponding term is omitted.
The graph polynomials are given by
\bq
 {\mathcal U}
 \; = \;
 \left( \prod\limits_{i=1}^{\loopnumber+1} a_i \right) \cdot \left( \sum\limits_{j=1}^{\loopnumber+1} \frac{1}{a_j} \right),
 & &
 {\mathcal F}
 \; = \;
 x \left( \prod\limits_{i=1}^{\loopnumber+1} a_i \right)
 + \left( \sum\limits_{i=1}^{\loopnumber+1} a_i y_i \right) {\mathcal U}.
\eq
At one, two and three loops we have
\bq
 l=1 & : & 
 {\mathcal F} = a_1 a_2 x + \left( a_1 + a_2 \right) \left( a_1 y_1 + a_2 y_2 \right),
 \\
 l=2 & : & 
 {\mathcal F} = a_1 a_2 a_3 x + \left( a_1 a_2 + a_1 a_3 + a_2 a_3 \right) \left( a_1 y_1 + a_2 y_2 + a_3 y_3 \right),
 \nonumber \\
 l=3 & : & 
 {\mathcal F} = a_1 a_2 a_3 a_4 x + \left( a_1 a_2 a_3 + a_1 a_2 a_4 + a_1 a_3 a_4 + a_2 a_3 a_4 \right) \left( a_1 y_1 + a_2 y_2 + a_3 y_3 + a_4 y_4 \right).
 \nonumber
\eq
Any sub-topology of the $l$-loop banana integral is a product of $l$ one-loop tadpole integrals.
If all masses are distinct and $(-p^2)$ and $D$ generic, we have
for the $l$-loop banana family
\bq
 \Nmaster & = & 2^{\loopnumber+1} - 1
\eq
master integrals.
We have $(\loopnumber+1)$ sub-topologies and there is one master integral for each sub-topology.
These may be taken as
\bq
 I_{011 \dots 11}, \; I_{101 \dots 11}, \; \dots, \;I_{111 \dots 10}.
\eq 
The top sector has
\bq
 2^{\loopnumber+1} - \loopnumber - 2
\eq
master integrals.
A basis for the top sector is given by
\bq
 \nu_j \; \in \; \left\{ 1,2 \right\},
 & &
 \loopnumber+1 \; \le \; \nu \; \le \; 2 \loopnumber,
\eq
where as usual $\nu=\nu_1+\dots+\nu_{\loopnumber+1}$.
At one, two and three loops a basis for the top sector is given by
\bq
 l=1 & : & 
 I_{11},
 \\
 l=2 & : &
 I_{111}, \; I_{211}, \; I_{121}, \; I_{112}, 
 \nonumber \\
 l=3 & : & 
 I_{1111}, \; 
 I_{2111}, \; I_{1211}, \; I_{1121}, \; I_{1112}, \;  
 I_{2211}, \; I_{2121}, \; I_{2112}, \; I_{1221}, \; I_{1212}, \; I_{1122}.
 \nonumber
\eq
In the equal mass case the number of master integrals is
\bq
 \Nmaster & = & \loopnumber + 1.
\eq
There is only one sub-sector with one master integral.
The top sector has $\loopnumber$ master integrals in the equal mass case,
which can be taken as
\bq
 I_{1 1 \dots 1}, \; I_{2 1 \dots 1}, \; I_{3 1 \dots 1}, \; \dots, \; I_{\left(\loopnumber-1\right) 1 \dots 1}, \; I_{\loopnumber 1 \dots 1}.
\eq
Let us now study the $l$-loop banana integral in $D=2$ space-time dimensions.
We go back to the unequal mass case.
As already observed in the two-loop case, the master integrals may be ramified
for special values of $D$. For $D=2$ one has in the unequal mass case in the top sector only
\bq
 2^{\loopnumber+1}
 -
 \left(\begin{array}{c}
  \loopnumber+2 \\
  \lfloor \frac{\loopnumber+2}{2} \rfloor
 \end{array} \right)
\eq
master integrals.
For example, there are $1, 2, 6$ master integrals in the top sector for $1, 2, 3$ loops, respectively.

We are in particular interested in the case
$\nu_1=\dots=\nu_{\loopnumber+1}=1$ and $D=2$.
Eq.~(\ref{chapter_motives:Feynman_representation_banana}) simplifies in this case to
\bq
\label{chapter_motives:Feynman_representation_banana_D_2}
 I_{1 \dots 1}\left(2\right)
 & = &
 \int\limits_{\Delta} \frac{\omega}{{\mathcal F}}.
\eq
As a sideremark we note that there exists 
a one-dimensional integral representation for $I_{1 \dots 1}(2)$
involving Bessel functions \cite{Groote:2005ay,Groote:2012pa}: 
\bq
 I_{1 \dots 1}\left(2\right)
 & = &
 2^\loopnumber \int\limits_0^\infty dt \; t \; J_0\left( t \sqrt{x} \right)
 \prod\limits_{i=1}^{\loopnumber+1} K_0\left( t \sqrt{y_i} \right).
\eq
However, our main interest here is the geometry underlying the $l$-loop banana integral.
The geometry is determined by the variety where ${\mathcal F}$ vanishes:
\bq
 X & = & 
 \left\{ \; \left[a_1:a_2:\dots:a_{\loopnumber+1}\right] \, \in \, {\mathbb C} {\mathbb P}^{\loopnumber} \; | \; {\mathcal F}(a) \, = \, 0 \; \right\}.
\eq
The second graph polynomial is a homogeneous polynomial of degree $(\loopnumber+1)$.
For generic kinematic variables the hypersurface $X \in {\mathbb C} {\mathbb P}^{\loopnumber}$ 
is smooth and defines by theorem~\ref{chapter_motives:theorem_example_Calabi_yau} 
below a Calabi-Yau $(\loopnumber-1)$-fold.
In particular we have at two-loops an elliptic curve and at three-loops a K3 surface.

The motive associated to the Feynman integral $I_{1 \dots 1}(2)$ is obtained along the lines of section~\ref{chapter_motives:motive_one_graph_polynomial}.
We denote the blow-up of ${\mathbb C} {\mathbb P}^{\loopnumber}$ by $P$, the strict transform of $X$ by $Y$ and 
the total transform of the set 
$\{ x_1 \cdot x_2 \cdot \ldots \cdot x_n = 0 \}$ by $B$.
The motive associated to the Feynman integral $I_{1 \dots 1}(2)$ is then
\bq
 M^{\loopnumber}\left(P \backslash Y, B \backslash \left( B \cap Y \right) \right).
\eq
The banana integrals are the simplest example, where higher-dimensional algebraic varieties (i.e. Calabi-Yau $(\loopnumber-1)$-folds) enter the computation of Feynman integrals.
The banana integrals have been studied 
in \cite{Aluffi:2008sy,MullerStach:2011ru,Bloch:2014qca,Vanhove:2014wqa,Primo:2017ipr,Broedel:2019kmn,Klemm:2019dbm,Bonisch:2020qmm,Bonisch:2021yfw,Broedel:2021zij},
other Feynman integrals related to Calabi-Yau manifolds have been studied 
in \cite{Brown:2010a,Bourjaily:2018yfy,Bourjaily:2019hmc}.

\begin{digression} {\bf Calabi-Yau manifolds}
\\
\index{Calabi-Yau manifold}
A Calabi-Yau manifold of complex dimension $n$ (or a {\bf Calabi-Yau $n$-fold} for short)
is a compact K\"ahler manifold $M$ of complex dimension $n$, satisfying one of the following equivalent 
conditions:
\begin{enumerate}
\item \label{chapter_motives:CY_cond_1} The first Chern class of $M$ vanishes (over ${\mathbb R}$).
\item \label{chapter_motives:CY_cond_2} $M$ has a K\"ahler metric with vanishing Ricci curvature.
\item \label{chapter_motives:CY_cond_3} $M$ has a K\"ahler metric with local holonomy $\mathrm{Hol}(p) \subseteq \mathrm{SU}(n)$ (where $p \in M$ denotes a point).
\item \label{chapter_motives:CY_cond_4} A positive power of the canonical bundle of $M$ is trivial.
\item \label{chapter_motives:CY_cond_4b} $M$ has a finite cover that is a product of a torus and a simply connected manifold with trivial canonical bundle.
\end{enumerate}
If $M$ is simply connected, the following conditions are equivalent to the ones above:
\begin{enumerate}
\setcounter{enumi}{5}
\item \label{chapter_motives:CY_cond_5} $M$ has a holomorphic $n$-form that vanishes nowhere.
\item \label{chapter_motives:CY_cond_6} The canonical bundle of $M$ is trivial.
\end{enumerate}
Note that the conditions $(\ref{chapter_motives:CY_cond_1})-(\ref{chapter_motives:CY_cond_4b})$ are in general weaker than the conditions $(\ref{chapter_motives:CY_cond_5})-(\ref{chapter_motives:CY_cond_6})$.
The exact definition of Calabi-Yau manifolds varies slightly in the literature.
Some authors define Calabi-Yau manifolds by conditions $(\ref{chapter_motives:CY_cond_5})-(\ref{chapter_motives:CY_cond_6})$, such a definition excludes for example Enriques surfaces as Calabi-Yau manifolds.
Other authors are even more restrictive and require
\begin{enumerate}
\setcounter{enumi}{7}
\item \label{chapter_motives:CY_cond_7} $M$ has a K\"ahler metric with local holonomy $\mathrm{Hol}(p) = \mathrm{SU}(n)$.
\end{enumerate}
Defining a Calabi-Yau manifold by condition $(\ref{chapter_motives:CY_cond_7})$ excludes for example Abelian surfaces as Calabi-Yau manifolds.
On the other hand, condition $(\ref{chapter_motives:CY_cond_7})$ ensures that the Hodge numbers $h^{j,0}$ vanish for $0 < j < n$.
It can be shown that condition $(\ref{chapter_motives:CY_cond_7})$ implies condition $(\ref{chapter_motives:CY_cond_5})$ \cite{Gross:book}.
Condition $(\ref{chapter_motives:CY_cond_5})$ is equivalent to condition $(\ref{chapter_motives:CY_cond_6})$ and
condition $(\ref{chapter_motives:CY_cond_6})$ clearly implies condition $(\ref{chapter_motives:CY_cond_4})$.
Therefore requiring condition $(\ref{chapter_motives:CY_cond_7})$ defines a Calabi-Yau manifold according to 
$(\ref{chapter_motives:CY_cond_1})$ - $(\ref{chapter_motives:CY_cond_4b})$.

A simply connected Calabi-Yau $2$-fold is called a 
\index{K3 surface}
{\bf K3 surface} (after Kummer, K\"ahler and Kodaira).
The requirement of simple connectedness in the definition of a K3 surface excludes complex tori.

In table~\ref{chapter_motives:Hodge_diamonds}
\begin{table}
\bq
\begin{array}{ccccc}
 \begin{array}{ccccc}
  & & 1 & & \\
  & 0 & & 0 & \\
  1 & & 20 & & 1 \\
  & 0 & & 0 & \\
  & & 1 & & \\
 \end{array}
 & \hspace*{5mm} &
 \begin{array}{ccccc}
  & & 1 & & \\
  & 2 & & 2 & \\
  1 & & 4 & & 1 \\
  & 2 & & 2 & \\
  & & 1 & & \\
 \end{array}
 & \hspace*{5mm} &
 \begin{array}{ccccc}
  & & 1 & & \\
  & 0 & & 0 & \\
  0 & & 10 & & 0 \\
  & 0 & & 0 & \\
  & & 1 & & \\
 \end{array}
 \\
 \mbox{K3 surface} && \mbox{Abelian surface} && \mbox{Enriques surface} \\
\end{array}
 \nonumber 
\eq
\caption{
The Hodge diamonds for a K3 surface, an Abelian surface and an Enriques surface.
}
\label{chapter_motives:Hodge_diamonds}
\end{table}
we show the Hodge diamonds for a K3 surface, an Abelian surface and an Enriques surface.
These Hodge diamonds illustrate that condition $(\ref{chapter_motives:CY_cond_7})$ ensures
\bq
 h^{n,0} \; = \; 1,
 & &
 h^{j,0} \; = \; 0,
 \;\;\;\;\;\;
 0 \; < \; j \; < \; n.
\eq
Examples of Calabi-Yau manifolds can be obtained from the following theorem:
\begin{theorem}
\label{chapter_motives:theorem_example_Calabi_yau}
Let $X$ be the hypersurface defined by a homogeneous polynomial $P$ of degree $d=n+2$ in ${\mathbb C} {\mathbb P}^{n+1}$.
If $X$ is smooth, than $X$ is a Calabi-Yau $n$-fold.
\end{theorem}
\noindent
In the definitions above we used some technical terms, which we now explain:

We start from a K\"ahler manifold $M$. First of all, a K\"ahler manifold is a special case of a Riemannian manifold
of real dimension $(2n)$,
so $M$ comes with a Riemannian metric $g$ and the associated Levi-Civita connection $\nabla$.
$M$ is also a complex manifold, so there is a complex structure $J$.
The manifold $M$ is further a Hermitian manifolds, this means that the complex structure $J$ is compatible with the metric $g$:
\bq
 g\left(J X, J Y\right) & = & g\left(X,Y\right)
\eq
for $X,Y \in T_p M$.
Finally, the fact that $M$ is K\"ahler implies that the complex structure satisfies
\bq
 \nabla J & = & 0.
\eq
From the Levi-Civita connection one defines the Riemann curvature tensor $R$ and the Ricci tensor $\mathrm{Ric}$.
The 
\index{Ricci form}
{\bf Ricci form} is then defined by
\bq
 \rho\left(X,Y\right) & = & \mathrm{Ric}\left(J X, Y\right).
\eq
The Ricci form of a K\"ahler manifold is a real closed $(1,1)$-form and can be written locally as
\bq
 \rho & = & -i \partial \bar{\partial} \ln \det g.
\eq
As $\rho$ is closed, it defines a class $[\rho] \in H^{1,1}(M,{\mathbb R}) \subset H^{2}(M,{\mathbb R})$.
The 
\index{Chern class}
{\bf first Chern class} of $M$ is given by
\bq
 c_1 & = & \frac{1}{2\pi} \left[\rho\right].
\eq
The connection $\nabla$ naturally defines a transformation group at each tangent space $T_p M$ as follows:
Let $p \in M$ be a point and consider the set of closed loops at $p$,
$\{\gamma(t) | 0 \le t \le 1, \gamma(0)=\gamma(1)=p \}$. 
Take a vector $X \in T_p M$ and parallel transport $X$ along the curve $\gamma$. 
After a trip along $\gamma$ we end up with a new vector $X_{\gamma} \in T_p M$. 
Thus the loop $\gamma$ and the connection $\nabla$ induces a linear transformation
\bq
 P_{\gamma} & : & T_p M \rightarrow T_p M.
\eq
The set of these transformations is denoted $\mathrm{Hol}(p)$ and is called the 
\index{holonomy group}
{\bf holonomy group}
at $p$.
The holonomy of a Riemannian manifold of real dimension $(2n)$ is contained in $\mathrm{O}(2n)$. 
If the manifold is orientable this becomes $\mathrm{SO}(2n)$. 
If $M$ is flat, the holonomy group consists only of the identity element.
If $M$ is K\"ahler, the holonomy group is contained in $\mathrm{U}(n)$ and $M$ is Calabi-Yau if the 
the holonomy group is contained in $\mathrm{SU}(n)$.

The 
\index{canonical bundle}
{\bf canonical bundle} of a smooth algebraic variety $X$ of (complex) dimension $n$ is the line bundle of holomorphic $n$-forms
\bq
 K_X & = & \bigwedge\limits^n \Omega_X^1.
\eq
The $k$-th tensor product $K_X^k$ of the canonical bundle is again a line bundle.
The 
\index{plurigenus}
{\bf plurigenus} of $X$ is the dimension of the vector space of global sections of $K_X^k$:
\bq
 P_k & = & \dim H^0\left(X,K_X^k\right).
\eq
The 
\index{Kodaira dimension}
{\bf Kodaira dimension} $\kappa$ of $X$ is defined to be $(-\infty)$, if all plurigenera $P_k$ are zero for $k>0$, otherwise 
it is defined as the minimum such that $P_k/k^\kappa$ is bounded.
In other words, the Kodaira dimension $\kappa$ gives the rate of growth of the plurigenera:
\bq
 P_k & = & {\mathcal O}\left(k^\kappa\right).
\eq
Let's look at an example in (complex) dimension $1$: The smooth algebraic curves are classified by their genus $g$.
One has
\begin{align}
 g & = 0 \; : & \kappa & = - \infty, & P_k & = 0, & k & > 0, \nonumber \\
 g & = 1 \; : & \kappa & = 0,        & P_k & = 1, & k & \ge 0, \nonumber \\
 g & \ge 2 \; : & \kappa & = 1,        & P_k & = \left(2k-1\right)\left(g-1\right), & k & \ge 2.
\end{align}

An Abelian surface is an Abelian variety of (complex) dimension two.
An 
\index{Abelian variety}
{\bf Abelian variety} is a projective algebraic variety that is also an algebraic group.
The group law is necessarily commutative.
It can be shown that the Abelian varieties are exactly those complex tori, which can be embedded into projective space.
To give an example:
An elliptic curve is an Abelian variety of (complex) dimension one.

An 
\index{Enriques surface}
{\bf Enriques surface} 
is a smooth projective minimal algebraic surface of Kodaira dimension $0$ with Betti numbers $b_1=0$ and $b_2=10$ \cite{Dolgachev:2014}.

\end{digression}

%% file: numerics.tex
\newpage
\chapter{Numerics}
\label{chapter_numerics}

At the end of the day of an analytic calculation of Feynman integrals we would like to get a number.
This requires methods for the numerical evaluation of all functions appearing in the final result
for a Feynman integral.
In this chapter we discuss methods how this can be done for multiple polylogarithms and the 
elliptic generalisations discussed in chapter~\ref{chapter_elliptics}.
All these methods can be pushed to obtain numerical results with a precision of hundred or thousand digits.
For phenomenological applications in physics this is certainly overkill, but there is one application, where
high-precision numerics is extremely useful: The PSLQ algorithm allows us to find relations among 
dependent transcendental constants. This algorithm is often used in the context of Feynman integral calculations
to find a simple form for the boundary constants.
We discuss this algorithm in section~\ref{chapter_numerics:pslq}.

\section{The dilogarithm}
\label{chapter_numerics:dilogarithm}

As a warm-up example let us start with the numerical evaluation of the dilogarithm \cite{'tHooft:1979xw}:
The dilogarithm is defined by
\bq
 \mathrm{Li}_{2}(x)
 & = &
 - \int\limits_{0}^{x} \frac{dt}{t} \ln\left(1-t\right),
\eq
and has a branch cut along the positive real axis, starting at the point $x=1$.
For $\left|x\right| \le 1$ one has the convergent power series expansion
\bq
\label{chapter_numerics:Li2power}
 \mathrm{Li}_{2}(x)
 & = &
 \sum\limits_{n=1}^{\infty} \frac{x^{n}}{n^{2}}.
\eq
The first step for a numerical evaluation consists in mapping an arbitrary (complex) argument $x$ into
the region, where the power series in eq.~(\ref{chapter_numerics:Li2power}) converges.
This can be done with the help of the reflection identity (see eq.~(\ref{chapter_one_loop:dilog_identities})
\bq
\label{chapter_numerics:Li2inversion}
 \mathrm{Li}_2(x)
 & = &
 -\mathrm{Li}_2\left(\frac{1}{x}\right) -\frac{\pi^2}{6} -\frac{1}{2} \left( \ln(-x) \right)^2,
\eq
which is used to map the argument $x$, lying outside the unit circle into the unit circle.
The function $\ln(-x)$ appearing on the right-hand side of eq.~(\ref{chapter_numerics:Li2inversion}) is considered 
to be ``simpler'', e.g. it is assumed that a numerical evaluation routine for this function is known.
In addition we can shift the argument into the range
$-1 \leq \mathrm{Re}(x) \leq 1/2$ with the help of
\bq
\label{chapter_numerics:Li2reflection}
 \mathrm{Li}_2(x)
 & = &
 -\mathrm{Li}_2(1-x) + \frac{\pi^2}{6} -\ln(x) \ln(1-x).
\eq
Although one can now attempt a brute force evaluation of the power series in eq.~(\ref{chapter_numerics:Li2power}), it is more
efficient to rewrite the dilogarithm 
as a series involving the Bernoulli numbers $B_j$ (defined in eq.~(\ref{chapter_transformations:def_Bernoulli_number})):
\bq
\label{chapter_numerics:Li2Bernoulli}
 \mathrm{Li}_2(x)
 & = &
 \sum\limits_{j=0}^\infty \frac{B_j}{(j+1)!} z^{j+1},
 \;\;\;\;\;\;\;\;\;
 z = - \ln(1-x).
\eq
Therefore the numerical evaluation of the dilogarithm consists 
in using eqs.~(\ref{chapter_numerics:Li2inversion}) and (\ref{chapter_numerics:Li2reflection})
to map any argument $x$ into the unit circle with the additional condition $\mathrm{Re}(x) \le 1/2$.
One then uses the series expansion in terms of Bernoulli numbers eq.~(\ref{chapter_numerics:Li2Bernoulli}).
\\
\\
\bs
{\it \refstepcounter{exercise}
{\bf Exercise \theexercise}: 
Derive eq.~(\ref{chapter_numerics:Li2Bernoulli}).
}
\es

\section{Multiple polylogarithm}
\label{chapter_numerics:multiple_polylogarithms}

Let us now consider multiple polylogarithms.
We used several notations for them (see chapter~\ref{chapter_multiple_polylogarithms}):
A long notation related to the integral representation
\bq
\label{chapter_numerics:G_long}
 G\left(z_1,z_2\dots,z_r;y\right),
\eq
where the $z_j$'s are allowed to be zero, a short notation related to the integral representation
\bq
\label{chapter_numerics:G_short}
 G_{m_1 \dots m_k}(z_1,\dots,z_k;y),
\eq
where all $z_j$'s are assumed to be non-zero and 
a representation related to the sum representation
\bq
\label{chapter_numerics:Li}
 \mathrm{Li}_{m_1 \dots m_k}(x_1,\dots,x_k).
\eq
These notations are related by eq.~(\ref{chapter_multiple_polylogarithms:Gshorthand}) and
eqs.~(\ref{chapter_multiple_polylogarithms:conversion_Li_to_G})-(\ref{chapter_multiple_polylogarithms:conversion_G_to_Li}).
We call $k$ the depth of the multiple polylogarithm.

We would like to evaluate numerically the multiple polylogarithms in eq.~(\ref{chapter_numerics:G_long})
for arbitrary complex arguments.
Let us first note that with the help of the shuffle product we may always remove trailing zeros
(as discussed in section~\ref{chapter_multiple_polylogarithms:section:shuffle_product}).
If $z_r \neq 0$ we may use the scaling relation eq.~(\ref{chapter_multiple_polylogarithms:G_scaling_relation})
to scale $y$ to $1$ (or a positive real number).
Let us therefore assume without loss of generality that $y$ is a positive real number.
Our integration path is then the line segment from $0$ to $y$ along the positive real axis.
The $z_j'$ are assumed not to lie on this line segment (but they are allowed to be infinitesimal close to this line segment):
\bq
 z_j & \in & {\mathbb C} \backslash \left[0,y\right].
\eq
As we already removed trailing zeros, we may use the notation as in eq.~(\ref{chapter_numerics:G_short}) or eq.~(\ref{chapter_numerics:Li}).
The principal ideas for the algorithm for the numerical evaluation of multiple polylogarithms are very similar
to the example of the dilogarithm discussed above.
We first use the integral representation
to transform all arguments into a region, where the sum representation converges.
Truncating the sum representation to an appropriate order provides a numerical evaluation.
In addition, there are methods which can be used to accelerate the convergence 
for the series representation of the multiple polylogarithms.

Let's look at the details:
In most physical applications, the $z_j$'s appearing in the integral representation will be real numbers.
To distinguish if the integration contour runs above or below a cut, we define the abbreviations
$z_\pm$, meaning that a small positive, respectively negative imaginary part is to be added to the value of the
variable:
\bq
z_+ = z + i \delta, & & z_- = z- i \delta,
 \;\;\;\;\;\;\;\;\; \delta \; > \; 0.
\eq
The sum representation $\mathrm{Li}_{m_1,\dots,m_k}(x_1,\dots,x_k)$ is convergent, if
\bq
\label{chapter_numerics:condconvLi}
 \left| x_1 x_2 \dots x_j \right| \le 1 & & \mbox{for all} \; j \in \{1,\dots,k\} \; \mbox{and} \; (m_1,x_1) \neq(1,1).
\eq
Therefore the function $G_{m_1,\dots,m_k}\left(z_1,\dots,z_k;y\right)$
has a convergent series representation
if
\bq
\label{chapter_numerics:condconv}
\left| y \right| \le \left| z_j \right| \;\;\; \mbox{for all}\;j,
\eq
e.g. no element in the set $\{|z_1|,\dots,|z_k|,|y|\}$ is smaller than $|y|$ and in addition if $m_1=1$ we have $y/z_1 \neq 1$.

\subsection{Transformation into the region where the sum representation converges}

If eq.~(\ref{chapter_numerics:condconv}) is not satisfied, we first transform into the domain, where
the sum representation is convergent.
This transformation is based on the integral representation.
We start from the function
\bq
\label{chapter_numerics:generalcaseordering}
 G_{m_1,\dots,m_k}\left(z_1,\dots,z_{j-1},s,z_{j+1},\dots,z_k;y\right),
\eq
with the assumption that $|s|$ is the smallest element in the set
$\{|z_1|,...,|z_{j-1}|,|s|,|z_{j+1}|,...,|z_k|,|y|\}$.
The algorithm goes by induction and introduces the more general structure
\bq
\label{chapter_numerics:generalstructure}
 \int\limits_0^{y_1} \frac{ds_1}{s_1-b_1}
 \dots
 \int\limits_0^{s_{r-1}}  \frac{ds_r}{s_r-b_r}  G(a_1,\dots,s_r,\dots,a_w;y_2),
\eq
where $|y_1|$ is the smallest element in the set
$\{|y_1|, |b_1'|, \dots, |b_r'|, |a_1'|, \dots, |a_w'|, |y_2| \}$.
The prime indicates that only the non-zero elements of $a_i$ and $b_j$ are considered.
If the integrals over $s_1$ to $s_r$ are absent, we recover the original $G$-function in eq. (\ref{chapter_numerics:generalcaseordering}).
Since we can always remove trailing zeroes with the help of the algorithm in section \ref{chapter_multiple_polylogarithms:section:shuffle_product}, we can assume that
$a_w \neq 0$.
We first consider the case where the $G$-function is of depth one, e.g.
\bq
 \int\limits_0^{y_1} \frac{ds_1}{s_1-b_1}
 \dots
 \int\limits_0^{s_{r-1}}  \frac{ds_r}{s_r-b_r}  G(\underbrace{0,\dots,0}_{m-1},s_r;y_2)
 & = &
 \int\limits_0^{y_1} \frac{ds_1}{s_1-b_1}
 \dots
 \int\limits_0^{s_{r-1}}  \frac{ds_r}{s_r-b_r}  G_m(s_r;y_2),
 \;\;\;\;\;\;
\eq
and show that we can relate the function $G_m(s_r;y_2)$ to $G_m\left(y_2;s_r\right)$,
powers of $\ln(s_r)$ and functions, which do not depend on $s_r$.
For $m=1$ we have
\bq
 G_1\left(\left.s_r\right._{\pm};y_2\right)
 & = & 
 G_1\left(\left.y_2\right._{\mp};s_r\right) - G(0;s_r) + \ln\left(-\left.y_2\right._{\mp}\right).
\eq
For $m \ge 2$ one can use the transformation $1/y$ and one obtains:
\bq
\label{chapter_numerics:oneoverxforclasspolylog}
 G_m\left(\left.s_r\right._{\pm};y_2\right)
 & = &
 -\zeta_m + \int\limits_0^{y_2} \frac{dt}{t} G_{m-1}\left(\left.t\right._{\pm};y_2\right)
          - \int\limits_0^{s_r} \frac{dt}{t} G_{m-1}\left(\left.t\right._{\pm};y_2\right).
\eq
One sees that the first and second term in eq. (\ref{chapter_numerics:oneoverxforclasspolylog}) yield functions
independent of $s_r$.
The third term has a reduced weight and we may therefore use recursion.
This completes the discussion for $G_m\left(s_r;y_2\right)$. 
We now turn to the general case with a $G$-function of depth greater than one in eq. (\ref{chapter_numerics:generalstructure}). 
Here we first consider the sub-case,
that $s_r$ appears in the last place in the parameter list and $(m-1)$ zeroes precede $s_r$, e.g.
\bq
 \int\limits_0^{y_1} \frac{ds_1}{s_1-b_1}
 \dots
 \int\limits_0^{s_{r-1}}  \frac{ds_r}{s_r-b_r}  G(a_1,\dots,a_{k},\underbrace{0,\dots,0}_{m-1},s_r;y_2).
\eq
Since we assumed that the $G$-function has a depth greater than one, we have $a_k \neq 0$.
Here we use the shuffle relation to relate this case to the case where $s_r$ does not appear in the last place:
\bq
\label{chapter_numerics:shuffleG}
\lefteqn{
G(a_1,\dots,a_{k},\underbrace{0,\dots,0}_{m-1},s_r;y_2)
 = } & &  \\
 & &
 G(a_1,\dots,a_{k};y_2) G(\underbrace{0,\dots,0}_{m-1},s_r;y_2)
 - \sum\limits_{\mathrm{shuffles}'} G(\alpha_1,\dots,\alpha_{k+m};y_2),
 \nonumber
\eq
where the sum runs over all shuffles of $(a_1,\dots,a_{k})$ with $(0,\dots,0,s_r)$ and the prime indicates that
$(\alpha_1,\dots,\alpha_{k+m}) = (a_1,\dots,a_{k},0,\dots,0,s_r)$ is to be excluded from this sum.
In the first term on the right-hand side of eq. (\ref{chapter_numerics:shuffleG}) the factor $G(a_1,\dots,a_{k};y_2)$
is independent of $s_r$ , whereas the second factor $G(0,\dots,0,s_r;y_2)$ is of depth one
and can be treated with the methods discussed above.
The terms corresponding to the sum over the shuffles in eq. (\ref{chapter_numerics:shuffleG}) have either $s_r$ not appearing in the last place
in the parameter list or a reduced number of zeroes preceding $s_r$. In the last case we may use recursion to remove $s_r$ from
the last place in the parameter list.
It remains to discuss the case, where the $G$-function has depth greater than one and $s_r$ does not appear in the last
place in the parameter list, e.g.
\bq
 \int\limits_0^{y_1} \frac{ds_1}{s_1-b_1}
 \dots
 \int\limits_0^{s_{r-1}}  \frac{ds_r}{s_r-b_r}  G(a_1,\dots,a_{i-1},s_r,a_{i+1},\dots,a_w;y_2),
\eq
with $a_w \neq 0$.
Obviously, we have
\bq
\lefteqn{
 G\left(a_1,\dots,a_{i-1},s_r,a_{i+1},\dots,a_{w};y_2\right)
 = } & & \\
& &
  G\left(a_1,\dots,a_{i-1},0,a_{i+1},\dots,a_{w};y_2\right)
  + \int\limits_0^{s_r} ds_{r+1} \frac{\partial}{\partial s_{r+1}}  
    G\left(a_1,\dots,a_{i-1},s_{r+1},a_{i+1},\dots,a_{w};y_2\right). \nonumber 
\eq
The first term $G\left(a_1,\dots,a_{i-1},0,a_{i+1},\dots,a_{w};y_2\right)$ does no longer depend on $s_r$ and has a reduced depth.
For the second term we first write out the integral representation of the $G$-function. We then use
\bq
\frac{\partial}{\partial s} \frac{1}{t-s} 
 & = & 
 - \frac{\partial}{\partial t} \frac{1}{t-s},
\eq
followed by partial integration in $t$ and finally partial fraction decomposition
according to
\bq
 \frac{1}{\left(t-\alpha\right)\left(t-s\right)}
 & = & 
 \frac{1}{s-\alpha} \left( \frac{1}{t-s} - \frac{1}{t-\alpha} \right).
\eq
If $s_r$ is not in the first place of the parameter list, we obtain
\bq
\lefteqn{
\int\limits_0^{s_r} ds_{r+1} \frac{\partial}{\partial s_{r+1}}  
    G\left(a_1,\dots,a_{i-1},s_{r+1},a_{i+1},\dots,a_{w};y_2\right)
} \nonumber \\
 & = &
 - \int\limits_0^{s_r} \frac{ds_{r+1}}{s_{r+1}-a_{i-1}}
   G\left(a_1,\dots,a_{i-2},s_{r+1},a_{i+1},\dots,a_w;y_2\right)
 \nonumber \\
 & &
 + \int\limits_0^{s_r} \frac{ds_{r+1}}{s_{r+1}-a_{i-1}}
   G\left(a_1,\dots,a_{i-2},a_{i-1},a_{i+1},\dots,a_w;y_2\right)
 \nonumber \\
 & &
 + \int\limits_0^{s_r} \frac{ds_{r+1}}{s_{r+1}-a_{i+1}}
   G\left(a_1,\dots,a_{i-1},s_{r+1},a_{i+2},\dots,a_w;y_2\right)
 \nonumber \\
 & &
 - \int\limits_0^{s_r} \frac{ds_{r+1}}{s_{r+1}-a_{i+1}}
   G\left(a_1,\dots,a_{i-1},a_{i+1},a_{i+2},\dots,a_w;y_2\right).
\eq
Each $G$-function has a weight reduced by one unit and we may use recursion.
If $s_r$ appears in the first place we have the following special case:
\bq
\lefteqn{
\int\limits_0^{s_r} ds_{r+1} \frac{\partial}{\partial s_{r+1}}  
    G\left(s_{r+1},a_{i+1},\dots,a_{w};y_2\right)
 = 
 \int\limits_0^{s_r} \frac{ds_{r+1}}{s_{r+1}-y_2}
   G\left(a_{i+1},\dots,a_w;y_2\right)
} & & \\
 & &
 + \int\limits_0^{s_r} \frac{ds_{r+1}}{s_{r+1}-a_{i+1}}
   G\left(s_{r+1},a_{i+2},\dots,a_w;y_2\right)
 - \int\limits_0^{s_r} \frac{ds_{r+1}}{s_{r+1}-a_{i+1}}
   G\left(a_{i+1},a_{i+2},\dots,a_w;y_2\right).
 \nonumber 
\eq
There is however a subtlety: If $\alpha_{i-1}$ or $\alpha_{i+1}$ are zero, the algorithm generates
terms of the form
\bq
 \int\limits_0^y \frac{ds}{s} F(s) 
- \int\limits_0^y \frac{ds}{s} F(0).
\eq
Although the sum of these two terms is finite, individual pieces diverge at $s=0$.
We regularise the individual contributions with a lower cut-off $\lambda$:
\bq
 \int\limits_\lambda^y \frac{ds}{s} F(s) 
- \int\limits_\lambda^y \frac{ds}{s} F(0).
\eq
In individual contributions we therefore obtain at the end of the day powers of $\ln\lambda$ from integrals of the form
\bq
\int\limits_\lambda^y \frac{ds_1}{s_1} \int\limits_\lambda^{s_1} \frac{ds_2}{s_2} 
 & = & \frac{1}{2} \ln^2 y - \ln y \ln \lambda + \frac{1}{2} \ln^2 \lambda.
\eq
In the final result, all powers of $\ln \lambda$ cancel, and we are left with $G$-functions with trailing zeros.
These are then converted by standard algorithms to $G$-functions without trailing zeros.
The $G$-functions without trailing zeros can then be evaluated numerically by their power series expansion.

In addition, the algorithms may introduce in intermediate steps $G$-functions with leading ones,
e.g. $G(1,\dots,z_k;1)$.
These functions are divergent, but the divergence can be factorised and expressed in terms
of the basic divergence $G(1;1)$. 
The algorithm is very similar to the one for the extraction of trailing zeroes.
In the end all divergences cancel.

\subsection{Series acceleration}

The $G$-function $G_{m_1,\dots,m_k}(z_1,\dots,z_k;y)$ has a convergent sum representation if the conditions
in eq. (\ref{chapter_numerics:condconv}) are met.
This does not necessarily imply, that the convergence is sufficiently fast, such
that the power series expansion can be used in a straightforward way.
In particular, if $z_1$ is close to $y$ the convergence is rather poor.
In this paragraph we consider methods to improve the convergence.
The main tool will be the H\"older convolution.

The multiple polylogarithms satisfy the 
\index{H\"older convolution}
{\bf H\"older convolution} \cite{Borwein}.
For $z_1 \neq 1$ and $z_r \neq 0$ this identity reads
\bq
\label{chapter_numerics:defhoelder}
G\left(z_1,\dots,z_r; 1 \right) 
 & = & 
 \sum\limits_{j=0}^r \left(-1\right)^j 
  G\left(1-z_j, 1-z_{j-1},\dots,1-z_1; 1 - \frac{1}{p} \right)
  G\left( z_{j+1},\dots, z_r; \frac{1}{p} \right).
 \nonumber \\
\eq
The H\"older convolution can be used to improve the rate of convergence for the series
representation of multiple polylogarithms.

Let us see how this is done: We consider $G_{m_1,\dots,m_k}(z_1,\dots,z_k;y)$ and assume that
the conditions of eq. (\ref{chapter_numerics:condconv}) are met 
(i.e. the multiple polylogarithm has a convergent sum representation).
By assumption we have $z_k \neq 0$, and therefore we can normalise $y$ to one.
We are therefore considering $G_{m_1,\dots,m_k}(z_1,\dots,z_k;1)$.
Convergence implies then, that we have $|z_j| \ge 1$ and $(z_1,m_1) \neq (1,1)$.
If some $z_j$ is close to the unit circle, say,
\bq
\label{chapter_numerics:condhoelder}
1 \le \left| z_j \right| \le 2,
\eq
we use the H\"older convolution eq. (\ref{chapter_numerics:defhoelder}) with $p=2$ to rewrite the $G$-functions as
\bq
\label{chapter_numerics:localhoelder}
\lefteqn{
G\left(z_1,\dots,z_r; 1 \right) 
 =  
 G\left(2 z_1,\dots, 2 z_r; 1 \right)
 + (-1)^r G\left( 2(1-z_r), 2(1-z_{r-1}),\dots, 2(1-z_1); 1 \right)
 } & &
 \nonumber \\
 & & 
 + \sum\limits_{j=1}^{r-1} \left(-1\right)^j 
  G\left( 2(1-z_j), 2(1-z_{j-1}),\dots, 2(1-z_1); 1 \right)
  G\left( 2 z_{j+1},\dots, 2 z_r; 1 \right).
 \;\;\;\;\;\;\;\;\;
\eq
Here, we normalised the right-hand side to one and explicitly wrote the first and last term of the sum.
We observe, that 
the first term
$G\left(2 z_1,\dots, 2 z_r; 1 \right)$ has all arguments outside $\left| 2 z_j \right| \ge 2$.
This term has therefore a better convergence.
Let us now turn to the second term in eq, (\ref{chapter_numerics:localhoelder}).
If some $z_j$ lies within $\left|z_j-1\right| < 1/2$, the H\"older convolution transforms the arguments out of the
region of convergence. In this case, we repeat the steps above, e.g. transformation into the region of convergence,
followed by a H\"older convolution, if necessary.
While this is a rather simple recipe to implement into a computer program, it is rather tricky to proof that this procedure does
not lead to an infinite recursion, and besides that, does indeed lead to an improvement in the convergence.
For the proof we have to understand how the algorithms for the transformation into the region of convergence act
on the arguments of a $G$-function with length $r$.
In particular we have to understand how in the result the $G$-functions of length $r$ are related to the original
$G$-function. Products of $G$-functions of lower length are ``simpler'' and not relevant for the argument here.
We observe, that this algorithm for the $G$-function $G\left(z_1,\dots,z_r;y\right)$ substitutes $y$ by the element
with the smallest non-zero modulus from the set $\{ \left|z_1\right|, \dots, \left|z_r\right|, \left|y\right| \}$,
permutes the remaining elements into an order, which is of no relevance here and possibly substitutes
some non-zero elements by zero.
The essential point is, that it does not introduce any non-trivial new arguments (e.g. new non-zero arguments).
The details can be found in \cite{Vollinga:2004sn}.

\subsection{Series expansion}

With the preparations of the previous paragraphs we may now assume that we have a multiple polylogarithm
$G_{m_1,\dots,m_k}(z_1,\dots,z_k;y)$, which has a sufficient fast converging sum representation.
With the help of eq.~(\ref{chapter_multiple_polylogarithms:conversion_G_to_Li}) we switch to the $\mathrm{Li}$-notation
\bq 
 \mathrm{Li}_{m_1 \dots m_k}(x_1,\dots,x_k)
  & = & \sum\limits_{n_1>n_2>\ldots>n_k>0}^\infty
     \frac{x_1^{n_1}}{{n_1}^{m_1}}\ldots \frac{x_k^{n_k}}{{n_k}^{m_k}}.
\eq
Let us write
\bq
 \mathrm{Li}_{m_1 \dots m_k}(x_1,\dots,x_k)
  & = & 
 \sum\limits_{n_1=1}^\infty d_{n_1},
 \;\;\;\;\;\;\;\;\;
 d_{n_1} \; = \; 
 \frac{x_1^{n_1}}{n_1^{m_1}}
 \;
 \sum\limits_{n_2=1}^{n_1-1} 
 \frac{x_2^{n_2}}{n_2^{m_2}}
 \;
 \dots
 \;
 \sum\limits_{n_k=1}^{n_{k-1}-1}
 \frac{x_k^{n_k}}{n_k^{m_k}}.
\eq
We may approximate $\mathrm{Li}_{m_1 \dots m_k}(x_1,\dots,x_k)$ by
\bq
 I^{\mathrm{approx}}(N) 
 & = &
 \sum\limits_{n_1=1}^N d_{n_1}
\eq
for some $N \in {\mathbb N}$. This is a finite sum and can be evaluated on a computer.
Choosing $N$ large enough, such that the neglected terms contribute below the numerical precision
gives the numerical evaluation of the iterated integral.

In more detail, let us define for two numbers $a$ and $b$ an equivalence relation.
We say $a \sim b$, if they have exactly the same floating-point representation
within a given numerical precision.
A reasonable truncation criteria is as follows: We truncate the iterated integral at $N$ if
\bq
\label{chapter_numerics:truncation_criterion}
 I^{\mathrm{approx}}\left(N\right)
 \; \sim \;
 I^{\mathrm{approx}}\left(N-1\right)
 & \mbox{and} &
 d_N \; \neq \; 0.
\eq
This gives reliable results in most cases.

\subsection{Examples}

The algorithms for the numerical evaluation of multiple polylogarithm are implemented in the 
computer algebra program {\tt GiNaC} \cite{Vollinga:2004sn}.
{\tt GiNaC} is a C++ library and allows symbolic calculations as well as numerical calculations with arbitrary precision 
within C++.
Alternatively, {\tt GiNaC} offers also a small interactive shell called {\tt ginsh}.

Let us consider as a first example
\bq
 \mathrm{Li}_{3 1}\left(\frac{1}{2},\frac{3}{4}\right).
\eq
This multiple polylogarithm is evaluated numerically in 
{\tt ginsh} as follows:
\begin{verbatim}
> Li({3,1},{0.5,0.75});
0.029809219570239646653
\end{verbatim}
We may change the number of digits:
\begin{verbatim}
> Digits=30;
30
> Li({2,2,1},{3.0,2.0,0.2});
0.0298092195702396466595180002639066394709
\end{verbatim}
We may also use the $G(z_1,z_2\dots,z_r;y)$-notation:
We have
\bq
 \mathrm{Li}_{3 1}\left(\frac{1}{2},\frac{3}{4}\right)
 \; = \; G_{3 1}\left(6,8;3\right) 
 \; = \;
 G\left(0,0,6,8;3\right).
\eq
The $G$-function is evaluated as follows:
\begin{verbatim}
> Digits=30;
30
> G({0,0,6.0,8.0},3.0);
0.0298092195702396466595180002639066394709
\end{verbatim}
Let us also consider
\bq
 G\left(\frac{1}{2} \pm i \delta; 1 \right)
 & = & \pm i \pi,
\eq
where $\delta$ is an infinitesimal small positive number.
Here we have to specify a small imaginary part, which indicates, if the pole at $\frac{1}{2}$ lies above or below the integration path.
In {\tt ginsh} we may include the signs of the small imaginary parts of the $z_j$'s as an optional second list:
\begin{verbatim}
> G({0.5},{1},1.0);
3.1415926535897932385*I
> G({0.5},{-1},1.0);
-3.1415926535897932385*I
\end{verbatim}
The default choice is a small positive imaginary part for the $z_j$'s:
\begin{verbatim}
> G({0.5},1.0);
3.1415926535897932385*I
\end{verbatim}
Implementations, which work with floating-point data types are {\tt handyG} \cite{Naterop:2019xaf} and {\tt FastGPL} \cite{Wang:2021imw}.
These programs offer only a fixed precision, but are significantly faster 
and therefore better suited to be used in situations where multiple polylogarithms need to be evaluated several million times
(like in Monte Carlo integrations).

Furthermore, there are dedicated implementations for the subclass of harmonic polylogarithms \cite{Gehrmann:2001pz,Maitre:2005uu,Ablinger:2018sat}.

\section{Iterated integrals in the elliptic case}
\label{chapter_numerics:iterated_integrals}

Let us now turn to the numerical evaluation of iterated integrals
related to elliptic Feynman integrals. 
We introduced these integrals in section~\ref{chapter_elliptics:section_moduli_spaces}.
Let us recall that these are iterated integrals on a covering space of the moduli space ${\mathcal M}_{1,n}$.
Standard coordinates on ${\mathcal M}_{1,n}$ are $(\tau,z_1,\dots,z_{n-1})$ 
(see section~\ref{chapter_elliptics:section_moduli_spaces})
and we may decompose an arbitrary integration path into pieces along $d\tau$ (with $z_1=\dots =z_{n-1}=\mathrm{const}$) 
and pieces along the $dz_j$'s (with $\tau=\mathrm{const}$).
By choosing appropriate boundary values it is sufficient to limit ourselves to 
iterated integrals with integration along $d\tau$.
Thus we consider in this section iterated integrals of the form as in eq.~(\ref{chapter_elliptics:def_iterated_integral_dtau}).

Comparing the numerical evaluation of these iterated integrals
to the numerical evaluation of multiple polylogarithms, there are several similarities, but also 
two fundamental differences.
We recall that the essential steps for the evaluation of multiple polylogarithms were
(i) removal of trailing zeros,
(ii) transformation into a region, where the series representation converges,
(iii) series acceleration and
(iv) evaluation of the truncated series.

The first difference of the numerical evaluation of 
iterated integrals in $d\tau$ with the numerical evaluation of
multiple polylogarithms is actually good news:
There are no poles in $\tau$-space along the integration path. 
There might be poles at the starting point of the integration path (``trailing zeros'')
or at the endpoint, but not in between.
This implies that the iterated integrals in $\tau$-space always have a convergent series representation
except for a few points.
These few points correspond in $\bar{q}$-space to an integration up to $|\bar{q}|=1$.
In physical terms, this corresponds to a threshold or (more generally) to a singularity of the differential equation.
Thus, for iterated integrals in $d\tau$ we do not need to consider point (ii) from the list above
(i.e. no need for a transformation into a region, where the series representation converges).
Let us emphasize, that this is not true for the iterated integrals in $dz$, 
e.g. the elliptic multiple polylogarithms $\widetilde{\Gamma}$ 
defined in eq.~(\ref{chapter_elliptics:Gammatilde}): As in the case of ordinary multiple polylogarithms 
we may integrate in the case of elliptic multiple polylogarithms $\widetilde{\Gamma}$ past poles.
However, as already mentioned above, there is no need for the functions $\widetilde{\Gamma}$.
With appropriate boundary values we may always integrate along $d\tau$.

On the other hand, the second difference is not so pleasant:
For endpoints of the integration path in $\bar{q}$-space close to $|\bar{q}| \lesssim 1$ 
the series expansion of the iterated integral converges rather slowly and 
we would like to apply methods to accelerate the series convergence.
A modular transformation is the natural candidate.
By a modular transformation $\tau' = \gamma(\tau)$ with $\gamma \in \mathrm{SL}_2({\mathbb Z})$ we may transform $\tau'$
into the fundamental domain 
\bq
\lefteqn{
 {\mathcal F} 
 = } & & \\
 & &
 \left\{ \tau' \in {\mathbb H} \left| \left| \tau' \right| > 1 \;\mbox{and}\;  -\frac{1}{2} < \mathrm{Re}\left(\tau'\right) \le \frac{1}{2} \right. \right\}
 \cup
 \left\{ \tau' \in {\mathbb H} \left| \left| \tau' \right| = 1 \;\mbox{and}\; 0 \le \mathrm{Re}\left(\tau'\right) \le \frac{1}{2} \right. \right\}
 \nonumber
\eq
and achieve that
\bq
 \left| \bar{q}' \right|
 & \le &
 e^{- \pi \sqrt{3}}
 \; \approx \; 0.0043.
\eq
This is a small expansion parameter.
So far, so good.
However, as discussed in section~\ref{chapter_elliptics:feynman_integrals_one_kinematic_variable}
and section~\ref{chapter_elliptics:feynman_integrals_several_kinematic_variables}
individual iterated integrals of the form as in eq.~(\ref{chapter_elliptics:def_iterated_integral_dtau})
do in general not transform nicely under modular transformations.
In general, we will leave through a modular transformation the space of iterated integrals 
of the form as in eq.~(\ref{chapter_elliptics:def_iterated_integral_dtau}).
We will stay inside the space of iterated integrals 
of the form as in eq.~(\ref{chapter_elliptics:def_iterated_integral_dtau})
if we perform simultaneously a fibre transformation and change the basis of our master integrals.
(We have seen explicit examples in eq.~(\ref{chapter_elliptics:combined_trafo_sunrise}) and eq.~(\ref{chapter_elliptics:trafo_fibre_unequal_sunrise}).)
Unfortunately, this implies that the acceleration techniques are tied to the specific family of
Feynman integrals under consideration
and that we cannot implement acceleration techniques based on modular transformations
into a black-box algorithm for iterated integrals 
of the form as in eq.~(\ref{chapter_elliptics:def_iterated_integral_dtau}).

What can be done, is the following: Assuming that the iterated integral under consideration
has a sufficiently fast converging series expansion, we may implement points (i) (removal of trailing zeros)
and (iv) (evaluation of the truncated series) into a black-box algorithm.
On the positive side, this can be done with a generality 
which exceeds the specific forms of eq.~(\ref{chapter_elliptics:def_iterated_integral_omega_Kronecker_and_omega_modular}).
We will now discuss this in more detail.

\subsubsection{Setup}

Let $M$ be a one-dimensional complex manifold with coordinate $x$ and
let $\omega_1$, ..., $\omega_r$ be differential $1$-forms on $M$.
Let $\lambda_0 \in {\mathbb R}_{>0}$ and denote by $U$ the domain $U=\{ x \in {\mathbb C} | |x| \le \lambda_0 \}$.
Let us assume that all $\omega_j$ are holomorphic in $U\backslash\{0\}$ and have at most a simple pole
at $x=0$.
In other words
\bq
\label{chapter_numerics:Laurent_expansion_one_forms}
 \omega_j
 & = &
 f_j\left(x\right) dx
 \; = \;
 \sum\limits_{n=0}^\infty c_{j,n} \; x^{n-1} dx,
 \;\;\;\;\;\;\;\;\;
 c_{j,n} \; \in \; {\mathbb C}.
\eq
We say that $\omega_j$ has a 
\index{trailing zero}
{\bf trailing zero}, if $c_{j,0} \neq 0$.
We denote by
\bq
\label{chapter_numerics:def_L0}
 L_0 & = & d\ln(x) \;\; = \;\; \frac{dx}{x}
\eq
the logarithmic form with $c_{0}=1$ and $c_n=0$ for $n>0$.

We set
\bq
\label{chapter_numerics:only_trailing_zeros}
 I(\underbrace{L_0,\dots,L_0}_{r};x_0)
 & = &
 \frac{1}{r!} \ln^r\left(x_0\right)
\eq
and define recursively
\bq
\label{chapter_numerics:recursive_definition}
 I\left(\omega_1,\omega_2,\dots,\omega_r;x_0\right)
 & = &
 \int\limits_0^{x_0} dx_1 f_1\left(x_1\right)
 I\left(\omega_2,\dots,\omega_r;x_1\right).
\eq
We say that the iterated integral $I(\omega_1,\dots,\omega_r;x_0)$ has a 
\index{trailing zero}
{\bf trailing zero},
if $\omega_r$ has a trailing zero.
If $\omega_r$ has a trailing zero, we may always write
\bq
\label{chapter_numerics:def_regularised_form}
 \omega_r
 & = &
 c_{r,0} L_0 + \omega_r^{\mathrm{reg}},
\eq
with
\bq
 \omega_r^{\mathrm{reg}}
 & = &
 \sum\limits_{n=1}^\infty c_{j,n} \; x^{n-1} dx
\eq
having no trailing zero.
In the case where $I(\omega_1,\omega_2,\dots,\omega_r;x_0)$ has no trailing zero, 
the definition in eq.~(\ref{chapter_numerics:recursive_definition}) agrees with the previous
definition of iterated integrals in eq.~(\ref{chapter_iterated_integrals:def_iterated_integral}).
Furthermore, we do not need to specify the path: As all $\omega_j$'s are holomorphic in $U$ and $\dim M = 1$, the iterated
integral is path-independent.

\subsubsection{Shuffle product and trailing zeros}

Iterated integrals always come with a shuffle product (compare with section~\ref{chapter_multiple_polylogarithms:section:shuffle_product}):
\bq
 I(\omega_{1},\dots,\omega_{k};x_0)
 \cdot
 I(\omega_{k+1},\dots,\omega_{r};x_0)
 & = &
 \sum\limits_{\mathrm{shuffles} \; \sigma}
 I(\omega_{\sigma(1)},\dots,\omega_{\sigma(r)};x_0),
\eq
where the sum runs over all shuffles $\sigma$ of $(1,\dots,k)$ with $(k+1,\dots,r)$.
The proof of this formula is identical to the proof of the shuffle product formula for multiple polylogarithms
given in section~\ref{chapter_multiple_polylogarithms:section:shuffle_product}.
We may use the shuffle product and eq.~(\ref{chapter_numerics:def_regularised_form}) to remove trailing zeros,
for example if $c_{1,0}=0$ and $c_{2,0}=1$ we have
\bq
 I(\omega_{1},\omega_{2};x_0)
 & = &
 I(\omega_{1},L_0;x_0)
 +
 I(\omega_{1},\omega_{2}^{\mathrm{reg}};x_0)
 \nonumber \\
 & = &
 I(L_0;x_0) I(\omega_{1};x_0)
 -
 I(L_0,\omega_{1};x_0)
 +
 I(\omega_{1},\omega_{2}^{\mathrm{reg}};x_0).
\eq
This isolates all trailing zeros in integrals of the form~(\ref{chapter_numerics:only_trailing_zeros}),
for which we may use the explicit formula in eq.~(\ref{chapter_numerics:only_trailing_zeros}).
It is therefore sufficient to focus on iterated integrals with no trailing zeros.
For
\bq
 I(\omega_{1},\dots,\omega_{r};x_0)
\eq
this means $c_{r,0}=0$.
Please note that $c_{k,0} \neq 0$ is allowed for $k<r$
and in particular that the form $L_0$ is allowed in positions $k<r$.

For integrals with no trailing zeros we introduce the notation
\bq
 I_{m_1,\dots,m_r}(\omega_{1},\dots,\omega_{r};x_0)
 & = &
 I(\underbrace{L_0,\dots,L_0}_{m_1-1},\omega_{1},\dots,\omega_{r-1},\underbrace{L_0,\dots,L_0}_{m_r-1},\omega_{r};x_0),
\eq
where we assumed that $\omega_k \neq L_0$ and $(m_k-1)$ $L_0$'s precede $\omega_k$.
This notation resembles the notation of multiple polylogarithms.
The motivation for this notation is as follows:
The iterated integrals $I_{m_1,\dots,m_r}(\omega_{1},\dots,\omega_{r};x_0)$ have just a $r$-fold series expansion,
and not a $(m_1+\dots+m_r)$-fold one.

\subsubsection{Series expansion}

With the same assumptions as in the previous subsection
(all $\omega_j$ are holomorphic in $U\backslash\{0\}$ and have at most a simple pole at $x=0$)
an iterated integral with no trailing zero has a convergent series expansion
in $U$:
\bq
\label{chapter_numerics:iter_int_series_expansion}
 I_{m_1,\dots,m_r}(\omega_{1},\dots,\omega_{r};x_0)
 & = &
 \sum\limits_{i_1=1}^\infty \sum\limits_{i_2=1}^{i_1} \dots \sum\limits_{i_r=1}^{i_{r-1}}
  x_0^{i_1}
  \frac{c_{1,i_1-i_2} \dots c_{r-1,i_{r-1}-i_r} c_{r,i_r}}
       {i_1^{m_1} i_2^{m_2} \cdot \dots \cdot i_r^{m_r}},
\eq
where the Laurent expansion around $x=0$ of the differential one-forms $\omega_j$ is given
by eq.~(\ref{chapter_numerics:Laurent_expansion_one_forms}).
Eq.~(\ref{chapter_numerics:iter_int_series_expansion})
can be used for the numerical evaluation of the iterated integral:
We truncate the outer sum over at $i_1=N$. 
Let us write eq.~(\ref{chapter_numerics:iter_int_series_expansion}) as
\bq
 I_{m_1,\dots,m_r}(\omega_{1},\dots,\omega_{r};x_0)
 & = &
 \sum\limits_{i_1=1}^\infty d_{i_1},
 \nonumber \\
 d_{i_1} & = &
  x_0^{i_1}
  \sum\limits_{i_2=1}^{i_1} \dots \sum\limits_{i_r=1}^{i_{r-1}}
  \frac{c_{1,i_1-i_2} \dots c_{r-1,i_{r-1}-i_r} c_{r,i_r}}
       {i_1^{m_1} i_2^{m_2} \cdot \dots \cdot i_r^{m_r}}.
\eq
This gives a numerical approximation $I^{\mathrm{approx}}(N)$ of the iterated integral
\bq
 I^{\mathrm{approx}}(N)
 & = &
 \sum\limits_{i_1=1}^N d_{i_1}.
\eq
As truncation criteria we may again use eq.~(\ref{chapter_numerics:truncation_criterion}).

\subsubsection{Iterated integrals along $d\tau$}

The discussion of the previous paragraphs applies to 
the iterated integrals along $d\tau$, introduced in eq.~(\ref{chapter_elliptics:def_iterated_integral_dtau}).
For the integration along $d\tau$ we consider in $\bar{q}$-space the iterated integrals 
\bq
 I_\gamma\left( \omega_1, \dots, \omega_r; \bar{q} \right),
\eq
where $\omega_j$ is of the form
\bq
 \omega^{\mathrm{Kronecker},\tau}_{k_j}
 \; = \;
 \frac{\left(k_j-1\right)}{\left(2\pi i\right)^{k_j}} g^{(k_j)}\left(z-c_j, \tau\right) \frac{d\bar{q}}{\bar{q}}
 & \;\; \mbox{or} \;\; &
 \omega^{\mathrm{modular}}_{k_j}
 \; = \;
 f_{k_j}\left(\tau\right) \frac{d\bar{q}}{\bar{q}},
 \;\;\;\;
\eq
with $f_{k_j}(\tau)$ being a modular form of weight $k_j$.
The $\bar{q}$-expansion of $\omega^{\mathrm{Kronecker},\tau}_{k_j}$ is given in eq.~(\ref{chapter_elliptics:qbar_expansion_Kronecker_coefficients}),
the $\bar{q}$-expansion of Eisenstein series has been discussed in section~\ref{chapter_elliptics:section_Eisenstein}.
Note that if a modular form is non-vanishing at the cusp $\tau=i\infty$, then it has a simple
pole at $\bar{q}=0$ in $\bar{q}$-space (and no further poles inside the unit disk $|\bar{q}|<1$).
The simple pole at $\bar{q}=0$ comes from the Jacobian of the transformation from $\tau$ to $\bar{q}$:
\bq
 2 \pi i d\tau & = & \frac{d\bar{q}}{\bar{q}}.
\eq

\subsubsection{Example}

We may use {\tt GiNaC} to evaluate numerically iterated integrals 
of the form as in eq.~(\ref{chapter_elliptics:def_iterated_integral_dtau}) \cite{Walden:2020odh}.
Let us see how this works in a full example.
We consider the equal mass sunrise integral in two space-time dimensions:
\bq
 I_{111}\left(2,x\right)
 & = &
 \frac{m^2}{\pi^2}
 \int d^2k_1 \int d^2k_2 \int d^2k_3
 \frac{\delta^2\left(p-k_1-k_2-k_3\right)}{\left(k_1^2-m^2\right)\left(k_1^2-m^2\right)\left(k_1^2-m^2\right)}
\eq
with $x=-p^2/m^2$.
Feynman's $i\delta$-prescription translates into an infinitesimal small negative imaginary part of $x$.
In section~\ref{chapter_elliptics:feynman_integrals_one_kinematic_variable} we worked out this integral
and found
\bq
 I_{111}\left(2,x\right)
 & = &
 \frac{\psi_1}{\pi} 
 \left[ 3 \, \mathrm{Cl}_2\left(\frac{2\pi}{3}\right)  +  4 I\left(\eta_0,\eta_3;\tau\right) \right].
\eq
This involves the iterated integral
\bq
 I\left(\eta_0,\eta_3;\tau\right),
\eq
with
\bq
 \eta_0 & = & -1,
 \nonumber \\
 \eta_3 & = & 
 - 9 \sqrt{3} \left( b_1^3 - b_1^2 b_2 - 4 b_1 b_2^2 + 4b_2^3 \right)
\eq
and 
\bq
 b_1 & = & E_1\left(\tau;\chi_1,\chi_{(-3)}\right)
 \; = \; 
 E_{1,1,-3,1}\left(\tau\right),
 \nonumber \\
 b_2 & = & E_1\left(2\tau;\chi_1,\chi_{(-3)}\right)
 \; = \; E_{1,1,-3,2}\left(\tau\right).
\eq
The following C++ code computes the Feynman integral $I_{111}(2,x)$
for $x \in {\mathbb R}\backslash\{-9,-1,0\}$:
\begin{verbatim}
#include <ginac/ginac.h>

int main()
{
  using namespace std;
  using namespace GiNaC;

  Digits = 30;

  // input x = -p^2/m^2, x real and not equal to {-9,-1,0}
  numeric x = numeric(-1,100);

  numeric sqrt_3  = sqrt(numeric(3));
  numeric sqrt_mx = sqrt(-x);
  numeric k2      = 16*sqrt_mx/pow(1+sqrt_mx,numeric(3))/(3-sqrt_mx);
 
  ex  pre = 4*pow(1+sqrt_mx,numeric(-3,2))*pow(3-sqrt_mx,numeric(-1,2));
  if (x < -9) pre = -pre;
  ex psi1 = pre*EllipticK(sqrt(k2));
  ex psi2 = pre*I*EllipticK(sqrt(1-k2));
  if ((x < -1) || (x > -3+2*sqrt_3)) psi1 += 2*psi2;
  if ((x > -9) && (x < -1)) psi1 += 2*psi2;
  ex tau  = psi2/psi1;
  ex qbar = exp(2*Pi*I*tau);

  ex L0   = basic_log_kernel();
  ex b1   = Eisenstein_kernel(1, 6, 1, -3, 1);
  ex b2   = Eisenstein_kernel(1, 6, 1, -3, 2);
  ex eta3 = modular_form_kernel(3, -9*sqrt_3*(pow(b1,3)-pow(b1,2)*b2
             -4*b1*pow(b2,2)+4*pow(b2,3)));

  ex Cl2  = numeric(1,2)/I*(Li(2,exp(2*Pi*I/3))-Li(2,exp(-2*Pi*I/3)));
  ex I111 = psi1/Pi*(3*Cl2-4*iterated_integral(lst{L0,eta3},qbar));

  cout << "I111 = " << I111.evalf() << endl;

  return 0;
}
\end{verbatim}
Let us explain the code: The input is given in the line
\begin{verbatim}
  numeric x = numeric(-1,100);
\end{verbatim}
One may change this value to any other value except $x \notin \{-9,-1,0\}$.
The program computes then the two periods $\psi_1$ and $\psi_2$, the modular parameter $\tau$
and the variable $\bar{q}$.
The line 
\begin{verbatim}
  if ((x < -1) || (x > -3+2*sqrt_3)) psi1 += 2*psi2;
\end{verbatim}
corresponds to eq.~(\ref{chapter_elliptics:def_periods_choice_0_gamma}).

There is a convention how mathematical software should evaluate a function on a branch cut:
Implementations shall map a cut so the function is continuous as the cut is approached coming
around the finite endpoint of the cut in a counter clockwise direction \cite{C99standard}.  
{\tt GiNaC} follows this convention. 
In physics, Feynman's $i\delta$-prescription dictates how a function should be evaluated on a branch cut.
The lines
\begin{verbatim}
  if (x < -9) pre = -pre;
  if ((x < -9) && (x < -1)) psi1 += 2*psi2;
\end{verbatim}
correct for a mismatch between the standard convention for mathematical software and
Feynman's $i\delta$-prescription.

We then define the modular forms. \verb|basic_log_kernel()| represents
\bq
 2 \pi i \; d\tau & = & \frac{d\bar{q}}{\bar{q}}.
\eq
The Eisenstein series $E_{k,a,b,K}(\tau)$ for $\Gamma_1(N)$ are defined by
\begin{verbatim}
  Eisenstein_kernel(k, N, a, b, K);
\end{verbatim}
Finally,
\begin{verbatim}
  iterated_integral(lst{L0,eta3},qbar);
\end{verbatim}
defines the iterated integral $I(1,\eta_3;\tau)=-I(\eta_0,\eta_3;\tau)$.

Running the code for
\bq 
 x & = & -0.01
\eq
yields
\begin{verbatim}
I111 = 2.34505440991241557114658013997777317976
\end{verbatim}

\section{The PSLQ algorithm}
\label{chapter_numerics:pslq}

The possibility to evaluate numerically transcendental functions to high precision allows us to simplify
boundary constants.
Suppose the boundary value of a Feynman integral at a certain kinematic point is given by a linear combination
of harmonic polylogarithms at $x=1$:
\bq
 G\left(z_1,\dots,z_r;1\right),
 & &
 z_j \in \{-1,0,1\}.
\eq
These are just transcendental numbers. 
Excluding trailing zeros and leading ones we have at weight $r \ge 2$
\bq
 4 \cdot 3^{r-2}
\eq
transcendental numbers.
However, they are not linearly independent.

The PSLQ algorithm \cite{Ferguson:1979,Ferguson:1992,Ferguson:1999,Bailey:1999nv}
may be used to find relations among a set of transcendental numbers.
The input to the PSLQ algorithm are high-precision numerical values for the transcendental numbers.
The PSLQ algorithm then finds integer coefficients, such that the linear combination with these integer coefficients
is close to zero within the numerical precision.
This does not provide a strict mathematical proof that the linear combination is indeed zero.
However, we may increase the numerical precision and if the coefficients stay constant we may be confident that the relation is
correct.

Let us start with an example:
We consider 
\bq
 G\left(-1,0,-1,-1;1\right)
\eq
and we ask if there is a relation between $G(-1,0,-1,-1;1)$ and the simpler constants
\bq
 \mathrm{Li}_4\left(\frac{1}{2}\right),
 \;\;\;
 \zeta_4,
 \;\;\;
 \zeta_3 \ln(2),
 \;\;\;
 \zeta_2 \ln^2(2),
 \;\;\;
 \ln^4(2).
\eq
We evaluate numerically all quantities to $50$ digits:
\bq
 G\left(-1,0,-1,-1;1\right) 
 & \approx & 
 0.0328931951943560413263595656028689325387,
 \nonumber \\
 \mathrm{Li}_4\left(\frac{1}{2}\right)
 & \approx &
 0.517479061673899386330758161898862945618,
 \nonumber \\
 \zeta_4 
 & \approx &
 1.082323233711138191516003696541167902776,
 \nonumber \\
 \zeta_3 \ln(2)
 & \approx &
 0.833202353297691993445762529661560103894,
 \nonumber \\
 \zeta_2 \ln^2(2)
 & \approx &
 0.790313530113954608772917335680644104204,
 \nonumber \\
 \ln^4(2)
 & \approx &
 0.230835098583083451887497717767812771517.
\eq
The PSLQ algorithm gives then
\bq
 8 G\left(-1,0,-1,-1;1\right) 
 - 24 \mathrm{Li}_4\left(\frac{1}{2}\right)
 + 24 \zeta_4 
 - 22 \zeta_3 \ln(2)
 + 6 \zeta_2 \ln^2(2)
 - \ln^4(2)
 & = & 0
 \nonumber
\eq
and therefore
\bq
 G\left(-1,0,-1,-1;1\right) 
 & = &
 3 \mathrm{Li}_4\left(\frac{1}{2}\right)
 -3 \zeta_4 
 + \frac{11}{4} \zeta_3 \ln(2)
 - \frac{3}{4} \zeta_2 \ln^2(2)
 + \frac{1}{8} \ln^4(2).
 \;\;\;
\eq
We may repeat the calculation with a higher number of digits. 
The empirical relation will stay the same.
This gives us confidence that the relation is correct.

On the other hand, if no relation exists, 
the PSLQ algorithm will tell us that no relation with integer coefficients smaller than a certain
bound exists.
The bound depends on the numerical precision.

The input to the PSLQ algorithm is a vector $x=(x_1,\dots,x_n) \in {\mathbb R}^n$ of real numbers, 
given as floating-point numbers to a certain precision.
An integer relation is given by a vector $m=(m_1,\dots,m_n) \in {\mathbb Z}^n$ of integer numbers such that
\bq
 m_1 x_1 + m_2 x_2 + \dots + m_n x_n & = & 0.
\eq
The name of the PSLQ algorithm derives from the {\bf partial sums}
\bq
\label{chapter_numerics:pslq_def_partial_sums}
 s_k & = & \sqrt{\sum\limits_{j=k}^n x_j^2}
\eq
and the
\index{LQ-decomposition of matrices} 
{\bf LQ-decomposition of matrices}:
Any $(n \times m)$-matrix $M$ may be written as 
\bq
 M & = & L \cdot Q,
\eq
where $L$ is a lower trapezoidal $(n \times m)$-matrix
and $Q$ is an orthogonal $(m \times m)$-matrix (i.e. $Q^{-1}=Q^T$).
A $(n\times m)$-matrix $L$ is called a 
\index{lower trapezoidal matrix}
{\bf lower trapezoidal matrix},
if 
\bq
 L_{ij} & = & 0
 \;\;\;\;\;\; \mbox{for} \;\;\; i < j.
\eq
The PSLQ algorithm is not too complicated to state.
For the proof why the algorithm works we refer to the literature \cite{Ferguson:1992,Ferguson:1999}. 
In order to state the algorithm we denote for a real number $x$ by $[x]$ the rounded value to the nearest integer.
The PSQL algorithm depends on two parameters $\gamma$ and $\delta$.
The first parameter $\gamma$ is chosen as
\bq
 \gamma & \ge & \sqrt{\frac{4}{3}}
\eq
and determines the weighting of the diagonal elements of a matrix $H$ in the first iteration step below.
The second parameter $\delta$ defines the detection threshold for an integer relation.
As a rule of thumb, if we expect an integer relation between $n$ input number $x_1, \dots,x_n$
with integer coefficients of maximum size $d$ digits, we should work with a precision of $(n \cdot d)$ digits
(i.e. the input numbers $x_j$ have to be given with this precision, and all internal arithmetic has to be carried
out with this precision).
In order to tolerate numerical rounding errors, we set the detection threshold a few orders of magnitude greater
than $10^{-n d}$, e.g.
\bq
 \delta & = & 10^{-n d + o},
\eq
where $o$ is a small positive integer.
Let us now state the algorithm:
\begin{tcolorbox}[breakable]
\begin{myalgorithm}
\label{chapter_numerics:algo_pslq}
The PSLQ algorithm
\\
\\
The algorithm is divided into an initialisation phase and an iteration phase.
\begin{description}
\item{\bf Initialisation:}
\begin{enumerate}
\item Initialise two integer $(n \times n)$-matrices $A$ and $B$ by
\bq
 A \; = \; {\bf 1}_{n \times n},
 & &
 B \; = \; {\bf 1}_{n \times n}.
\eq
\item Initialise $s_k$ by eq.~(\ref{chapter_numerics:pslq_def_partial_sums}) and set
\bq
 y_k & = & \frac{x_k}{s_1},
 \;\;\;\;\;\; 1 \le k \le n.
\eq
\item Initialise a lower trapezoidal $n \times (n-1)$-matrix $H$ by
\bq
 H_{ij}
 & = & 
 \left\{ \begin{array}{rl}
 - \frac{x_i x_j}{s_j s_{j+1}}, & i > j, \\
 \frac{s_{j+1}}{s_j}, & i = j, \\
 0, & i < j. \\
 \end{array}
 \right.
\eq
\item Reduce $H$:
\begin{algorithmic}[]
 \For{$i=2$ to $n$}
  \For{$j=i-1$ to $1$ step $-1$}
   \State{$t \gets [H_{ij}/H_{jj}]$}
   \State{$y_j \gets y_j + t y_i$}
   \For{$k=1$ to $j$}
    \State{$H_{ik} \gets H_{ik} - t H_{jk}$}
   \EndFor 
   \For{$k=1$ to $n$}
    \State{$A_{ik} \gets A_{ik} - t A_{jk}$}
    \State{$B_{kj} \gets B_{kj} - t B_{ki}$}
   \EndFor 
  \EndFor 
 \EndFor 
\end{algorithmic}
\end{enumerate}

\item{\bf Iteration:}
\begin{enumerate}

\item Select $l$ such that $\gamma^i |H_{ii}|$ is maximal for $i=l$.

\item Exchange the entries of $y$ indexed $l$ and $(l+1)$, the corresponding rows of $A$ and $H$, and the corresponding columns of $B$.

\item Remove corner:
\begin{algorithmic}[]
 \If{$l \le n-2$}
  \State{$t_0 \gets \sqrt{H_{l l}^2+H_{l (l+1)}^2}$}
  \State{$t_1 \gets \frac{H_{l l}}{t_0}$}
  \State{$t_2 \gets \frac{H_{l (l+1)}}{t_0}$}
  \For{$i=l$ to $n$}
   \State{$t_3 \gets H_{i l}$}
   \State{$t_4 \gets H_{i (l+1)}$}
   \State{$H_{i l} \gets t_1 t_3 + t_2 t_4$}
   \State{$H_{i (l+1)} \gets -t_2 t_3 + t_1 t_4$}
  \EndFor 
 \EndIf
\end{algorithmic}

\item Reduce $H$:
\begin{algorithmic}[]
 \For{$i=l+1$ to $n$}
  \For{$j=\min(i-1,l+1)$ to $1$ step $-1$}
   \State{$t \gets [H_{ij}/H_{jj}]$}
   \State{$y_j \gets y_j + t y_i$}
   \For{$k=1$ to $j$}
    \State{$H_{ik} \gets H_{ik} - t H_{jk}$}
   \EndFor 
   \For{$k=1$ to $n$}
    \State{$A_{ik} \gets A_{ik} - t A_{jk}$}
    \State{$B_{kj} \gets B_{kj} - t B_{ki}$}
   \EndFor 
  \EndFor 
 \EndFor 
\end{algorithmic}

\item Termination test:
If the largest entry of $A$ exceeds the numerical precision, 
then no relation exists where the Euclidean norm of the vector $m$ is less than $1/\max_j|H_{jj}|$.
If the smallest entry of the vector $y$ is less than the detection threshold $\delta$, return the corresponding
column of $B$.
Otherwise go back to step 1 of the iteration.

\end{enumerate}
\end{description}
\end{myalgorithm}
\end{tcolorbox}
The PSLQ algorithm is implemented in many commercial computer algebra systems.

%% file: final_project.tex
\newpage
\chapter{Final project}
\label{chapter_final_project}

In the last chapter of this book, let's do a final project.
In this way we are going to review many techniques introduced in the previous chapters.

\section{A two-loop penguin integral}

\begin{figure}
\begin{center}
\includegraphics[scale=1.0]{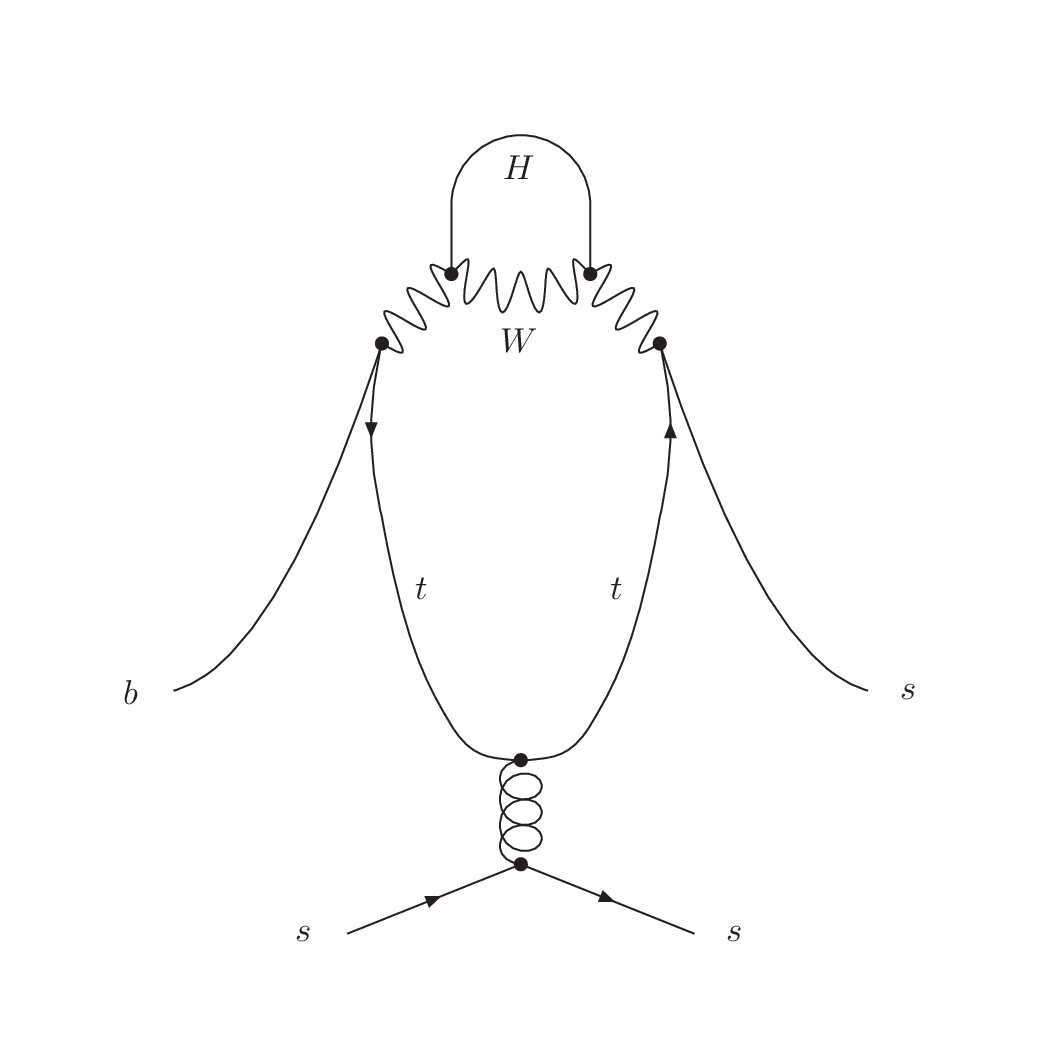}
\end{center}
\caption{
A two-loop penguin diagram.
}
\label{chapter_final_project:fig_penguin_1}
\end{figure}

We are going to compute the Feynman integral 
of the penguin graph shown in
fig.~\ref{chapter_final_project:fig_penguin_1}.
We are going to neglect all light quark masses ($m_s=m_b=0$) and only keep
the heavy masses $m_W, m_H, m_t$ non-zero.
The integral corresponding to the graph shown in 
fig.~\ref{chapter_final_project:fig_penguin_1} has six loop propagators 
(and one tree-like propagator: the gluon propagator drawn by a curly line at the bottom).
We notice that two of the six loop propagators are identical: These are the $W$-boson propagators
making up the left and the right shoulder of the penguin. 
The momenta flowing through these lines is the same, as is the internal mass (i.e. $m_W$).
Thus we actually only need to consider a two-loop integral with five loop propagators, where one propagator
is raised to the power two.
\begin{figure}
\begin{center}
\includegraphics[scale=1.0]{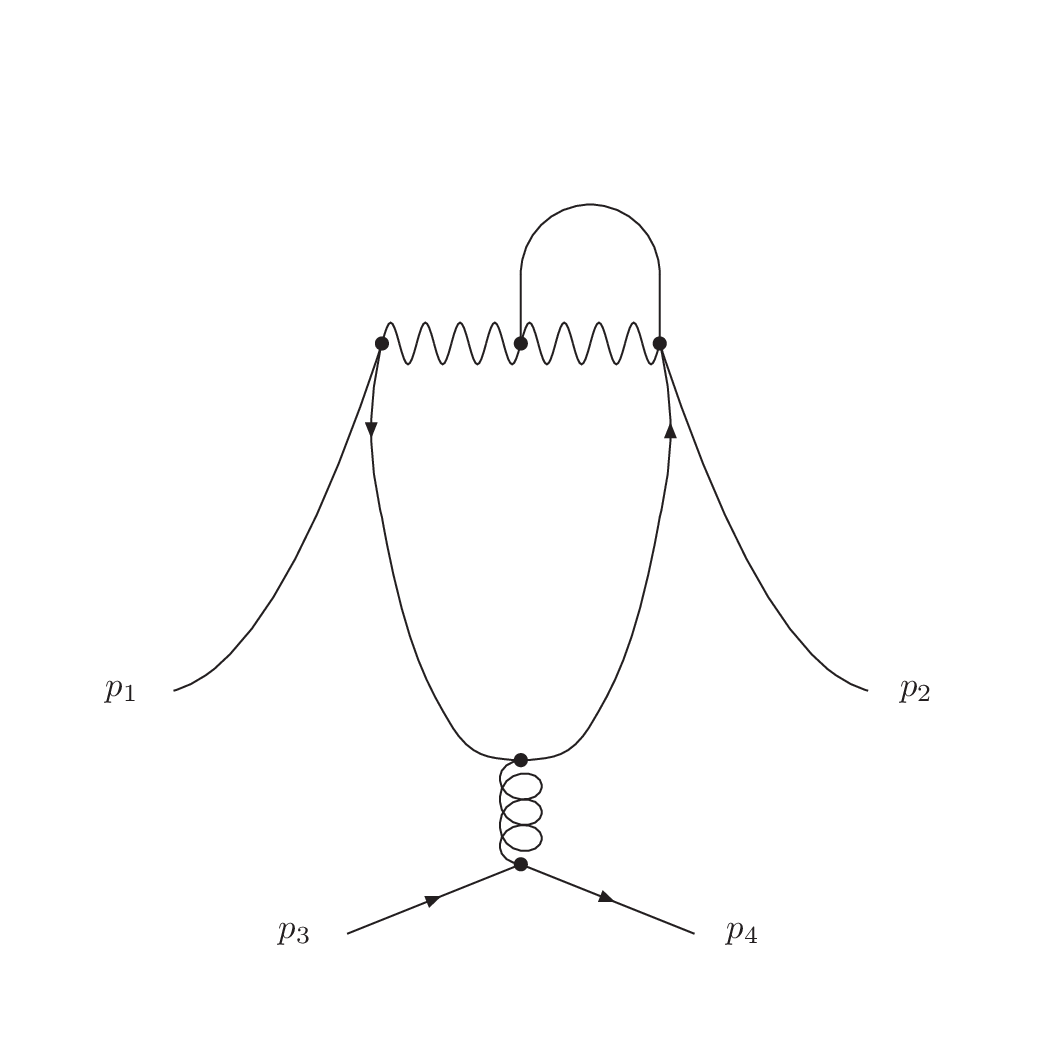}
\end{center}
\caption{
The distorted penguin:
A two-loop diagram with five loop propagators.
}
\label{chapter_final_project:fig_penguin_2}
\end{figure}
The corresponding diagram (a distorted penguin) is shown in fig~\ref{chapter_final_project:fig_penguin_2}.
In fig.~\ref{chapter_final_project:fig_penguin_2} we also indicated the external momenta.
As usual our convention is to take all momenta as outgoing, therefore momentum conservation reads
\bq
 p_1 + p_2 + p_3 + p_4 & = & 0.
\eq
The external particles are assumed to be on the mass-shell, and since we assumed $m_b=m_s=0$ we have
\bq
 p_1^2 \; = \; m_b^2 \; = \; 0,
 & &
 p_2^2 \; = \; p_3^2 \; = \; p_4^2 \; = \; m_s^2 \; = \; 0.
\eq
It is clear from the diagram that the scalar Feynman integrals will only depend on
$p_1, p_2, (p_3+p_4)$, but not on $p_3$ nor $p_4$ individually.
Thus the only non-vanishing Lorentz invariant is
\bq
 s & = & \left(p_1+p_2\right)^2 \; = \; \left(p_3+p_4\right)^2.
\eq
Therefore our kinematic variables are
\bq
 x_1 \; = \; \frac{-s}{\mu^2},
 \;\;\;\;\;\;
 x_2 \; = \; \frac{m_W^2}{\mu^2},
 \;\;\;\;\;\;
 x_3 \; = \; \frac{m_H^2}{\mu^2},
 \;\;\;\;\;\;
 x_4 \; = \; \frac{m_t^2}{\mu^2}.
\eq
In our calculation we may set (temporarily) $\mu$ to any value we want, and in particular
the choice $\mu=m_t$ sets the last variable equal to one: $x_4=1$.
As $\mu$ enters only as a trivial prefactor the definition of Feynman integrals, the $\mu$ dependence
can be restored at the end of the calculation.
We therefore have a problem with three kinematic variables $x_1, x_2, x_3$ and $\NB=3$
(see the discussion in section~\ref{chapter_basics:momentum_representation_of_Feynman_integrals}
and in section~\ref{chapter_iterated_integrals:fibre_bundles_in_physics}).

\begin{figure}
\begin{center}
\includegraphics[scale=1.0]{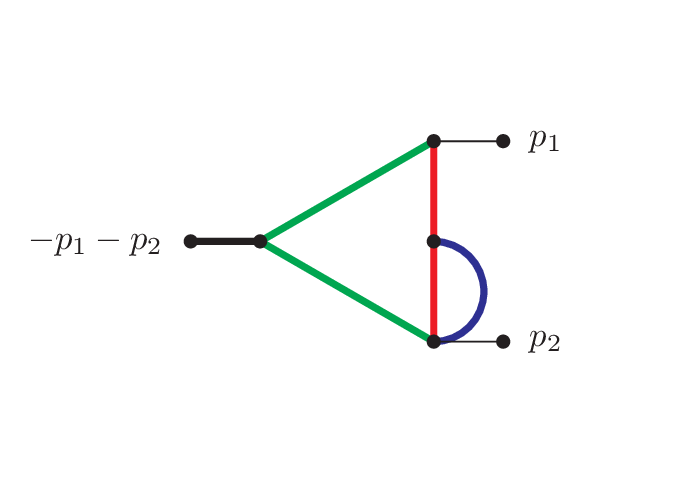}
\end{center}
\caption{
A two-loop three-point function.
A green line denotes a propagator with mass $m_t$, 
a red line denotes a propagator with mass $m_W$,
a blue line denotes a propagator with mass $m_H$.
A thick black external line indicates that $p^2=s$, a thin black external line indicates that $p^2=0$.
}
\label{chapter_final_project:fig_scalar_diagram}
\end{figure}

We recall from section~\ref{chapter_qft:sect:tensor_reduction} that all tensor integrals 
can be reduced to scalar integrals.
Therefore we only need to consider the relevant scalar integrals.
In fig.~\ref{chapter_final_project:fig_scalar_diagram}
we draw the relevant Feynman graph in a more standard way: We are interested in a two-loop three-point
function with five propagators as shown in fig.~\ref{chapter_final_project:fig_scalar_diagram}.

\begin{figure}
\begin{center}
\includegraphics[scale=1.0]{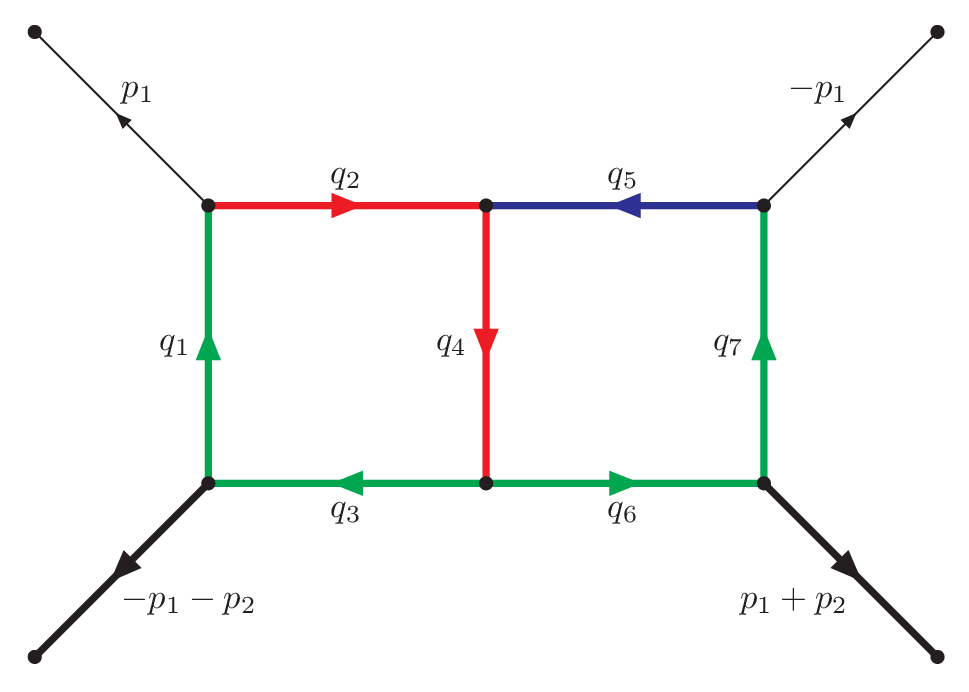}
\end{center}
\caption{
The auxiliary graph $\tilde{G}$ with seven propagators. The arrow indicate the momentum flow.
The colour coding is as in fig.~\ref{chapter_final_project:fig_scalar_diagram}.
}
\label{chapter_final_project:fig_auxiliary_graph}
\end{figure}

We are going to use integration-by-parts identities.
We have two linear independent external momenta ($p_1$ and $p_2$) and two independent loop momenta 
(which we label $k_1$ and $k_2$).
Therefore we have (see section~\ref{chapter_basics:sect_Baikov_representation})
\bq
 \NV & = &
 \frac{1}{2} \loopnumber \left(\loopnumber+1\right) + \nexternalindependent \loopnumber
 \; = \;
 7
\eq
linear independent scalar products involving the loop momenta.
We therefore consider an auxiliary graph with seven loop propagators, such that any
scalar product involving the loop momenta can be expressed as a linear combination of the propagators and a constant
and vice versa.
This is to say that we seek an auxiliary graph such that eq.~(\ref{chapter_basics:baikov_condition2}) holds.
This is not too complicated, the double-box graph $\tilde{G}$ shown in fig.~\ref{chapter_final_project:fig_auxiliary_graph} will do the job.
Note that this double-box graph depends only on two linear independent external momenta.

We set $\mu = m_t$ and we consider the family of Feynman integrals
\bq
 I_{\nu_1 \nu_2 \nu_3 \nu_4 \nu_5 \nu_6 \nu_7}\left(D,x_1,x_2,x_3\right)
 & = &
 e^{2 \eps \Eulerconstant} \left(m_t^2\right)^{\nu-D}
 \int 
 \frac{d^Dk_1}{i \pi^{\frac{D}{2}}} 
 \frac{d^Dk_2}{i \pi^{\frac{D}{2}}} 
 \prod\limits_{j=1}^{7} \frac{1}{\left(-q_j^2+m_j^2\right)^{\nu_j}},
\eq
with
\begin{align}
 q_1 & = k_1 + p_1 + p_2, &   m_1 & = m_t,
 \nonumber \\
 q_2 & = k_1 + p_2,       &   m_2 & = m_W,
 \nonumber \\
 q_3 & = k_1,             &   m_3 & = m_t,
 \nonumber \\
 q_4 & = k_1 + k_2,       &   m_4 & = m_W,
 \nonumber \\
 q_5 & = k_2 - p_2,       &   m_5 & = m_H,
 \nonumber \\
 q_6 & = k_2,             &   m_6 & = m_t,
 \nonumber \\
 q_7 & = k_2 - p_1 - p_2, &   m_7 & = m_t.
\end{align}
We are interested in the integrals with $\nu_6 \le 0, \nu_7 \le 0$, these correspond to the
topology shown in fig.~\ref{chapter_final_project:fig_scalar_diagram}.

\section{Deriving the differential equation}

We first determine the two graph polynomials ${\mathcal U}$ and ${\mathcal F}$
of the graph $\tilde{G}$ with the help
of the methods from chapter~\ref{chapter_graph_polynomials}.
The first graph polynomial can actually be copied directly from eq.~(\ref{chapter_basics:result_U_and_F_double_box}),
for the second graph polynomial we have to take into account that our kinematic configuration is
different.
We obtain
\bq
{\mathcal U} & = & \left( a_1+a_2+a_3 \right) \left( a_5+a_6+a_7 \right) + a_4 \left( a_1+a_2+a_3+a_5+a_6+a_7 \right),
 \nonumber \\
{\mathcal F} & = &
 x_1 \left[ a_1 a_3 \left( a_4+a_5+a_6+a_7 \right)
          + a_6 a_7 \left( a_1+a_2+a_3+a_4 \right)
          + a_1 a_4 a_6 + a_3 a_4 a_7 \right] 
 \nonumber \\
 & &
 + \left[ a_1 + a_3 + a_6 + a_7 + x_2 \left( a_2 + a_4 \right) + x_3 a_5 \right] {\mathcal U}.
\eq
In the next step we generate the integration-by-parts identities.
We do this with the help of a computer program, like
{\tt Fire} \cite{Smirnov:2008iw,Smirnov:2019qkx},
{\tt Reduze} \cite{Studerus:2009ye,vonManteuffel:2012np} or
{\tt Kira} \cite{Maierhoefer:2017hyi,Klappert:2020nbg}.
We may modify the set-up from exercise~\ref{chapter_iterated_integrals:exercise_doublebox_ibp}.
For example, if we are using {\tt Kira} the file
{\tt integralfamilies.yaml} should now read
{\footnotesize
\begin{verbatim}
integralfamilies:
  - name: "doublebox"
    loop_momenta: [k1, k2]
    top_level_sectors: [31]
    propagators:
      - [ "k1+p1+p2", "mt2" ]
      - [ "k1+p2", "mW2" ]
      - [ "k1", "mt2" ]
      - [ "k1+k2", "mW2" ]
      - [ "k2-p2", "mH2" ]
      - [ "k2", "mt2" ]
      - [ "k2-p1-p2", "mt2" ]
\end{verbatim}
}
\noindent
and the file {\tt kinematics.yaml} should read
{\footnotesize
\begin{verbatim}
kinematics : 
  incoming_momenta: []
  outgoing_momenta: [p1, p2, p3]
  momentum_conservation: [p3,-p1-p2]
  kinematic_invariants:
    - [s,  2]
    - [mW2,2]
    - [mH2,2]
    - [mt2,2]
  scalarproduct_rules:
    - [[p1,p1],  0]
    - [[p2,p2],  0]
    - [[p1+p2,p1+p2],  s]
  symbol_to_replace_by_one: mt2
\end{verbatim}
}
\noindent
Running an integration-by-parts reduction program we also obtain a list of master integrals.
This list is not unique and will depend on the chosen ordering criteria for the Laporta algorithm
(see section~\ref{chapter_iterated_integrals:integration_by_parts}).
A possible basis of master integrals is tabulated in table~\ref{chapter_final_project:table_master_integrals}.
\begin{table}
\begin{center}
\begin{tabular}{|c|r|r|l|l|c|}
\hline
 number of   & block & sector & master integrals & master integrals & kinematic \\
 propagators &       &        & basis $\vec{I}$  & basis $\vec{J}$  & dependence \\
\hline
\hline
 $2$ & $1$ &  $9$ & $I_{1001000}$ & $J_{1}$ & $x_2$ \\
     & $2$ & $10$ & $I_{0101000}$ & $J_{2}$ & $x_2$ \\
     & $3$ & $17$ & $I_{1000100}$ & $J_{3}$ & $x_3$ \\
     & $4$ & $18$ & $I_{0100100}$ & $J_{4}$ & $x_2,x_3$ \\
\hline
\hline
 $3$ & $5$ & $13$ & $I_{1011000}$ & $J_{5}$ & $x_1,x_2$ \\
     & $6$ & $21$ & $I_{1010100}$ & $J_{6}$ & $x_1,x_3$ \\
     & $7$ & $25$ & $I_{1001100}$ & $J_{7}$ & $x_2,x_3$ \\
     & $8$ & $26$ & $I_{0101100}$ & $J_{8}$ & $x_2,x_3$ \\
\hline
\hline
 $4$ &  $9$ & $15$ & $I_{1111000}$ & $J_{9}$ & $x_1,x_2$ \\
     & $10$ & $23$ & $I_{1110100}$ & $J_{10}$ & $x_1,x_2,x_3$ \\
     & $11$ & $29$ & $I_{1011100}, I_{1(-1)11100},$ & $J_{11},J_{12},J_{13},J_{14}$ & $x_1,x_2,x_3$ \\
     &      &      & $I_{10111(-1)0}, I_{1(-2)11100}$ & & \\
\hline
\hline
 $5$ & $12$ & $31$ & $I_{1111100}$ & $J_{15}$ & $x_1,x_2,x_3$ \\
\hline
\end{tabular}
\end{center}
\caption{Overview of the set of master integrals.
The first column denotes the number of propagators, the second column labels consecutively the sectors or topologies,
the third column gives the sector id $N_{\mathrm{id}}$ (defined in eq.~(\ref{chapter_iterated_integrals:def_notation_sectors})),
the fourth column lists the master integrals in the basis $\vec{I}$,
the fifth column the corresponding ones in the basis $\vec{J}$.
The last column denotes the kinematic dependence. 
}
\label{chapter_final_project:table_master_integrals}
\end{table}
We have $15$ master integrals, hence $\Nmaster=15$.
There are $12$ sectors. 
\begin{figure}
\begin{center}
\includegraphics[scale=0.75]{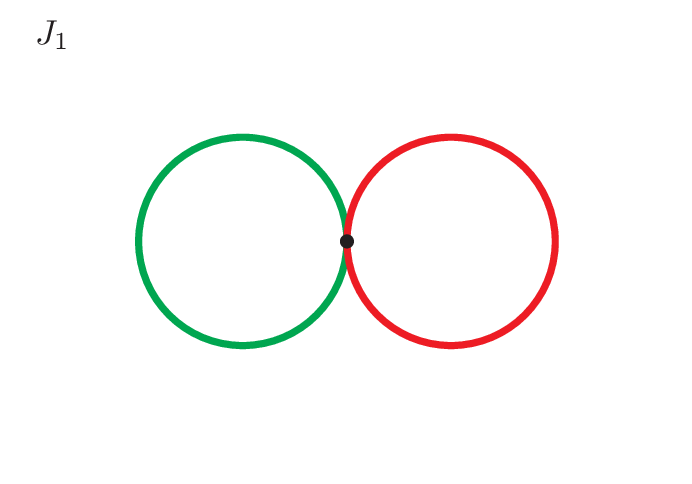}
\includegraphics[scale=0.75]{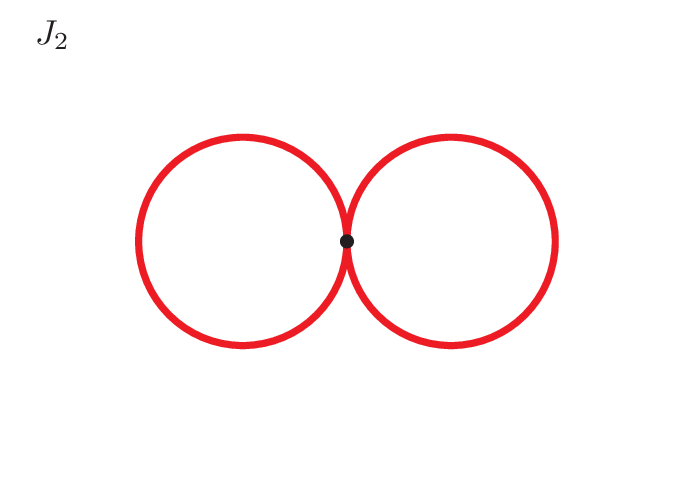}
\includegraphics[scale=0.75]{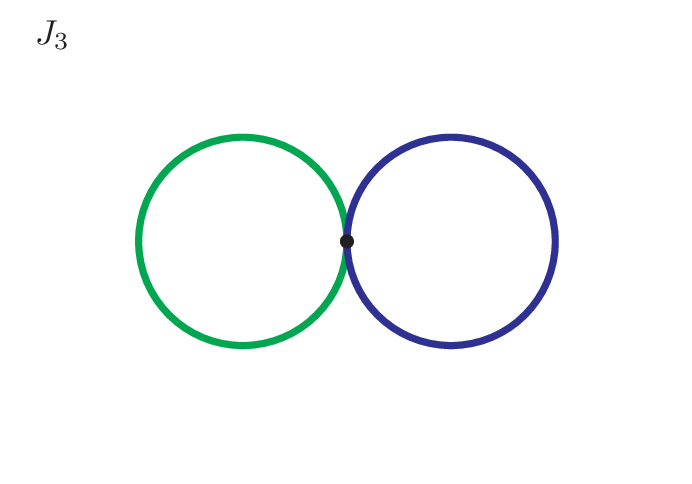}
\includegraphics[scale=0.75]{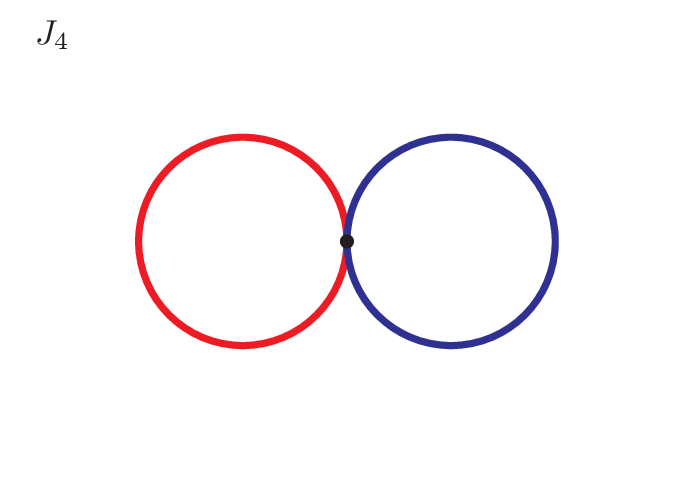}
\includegraphics[scale=0.75]{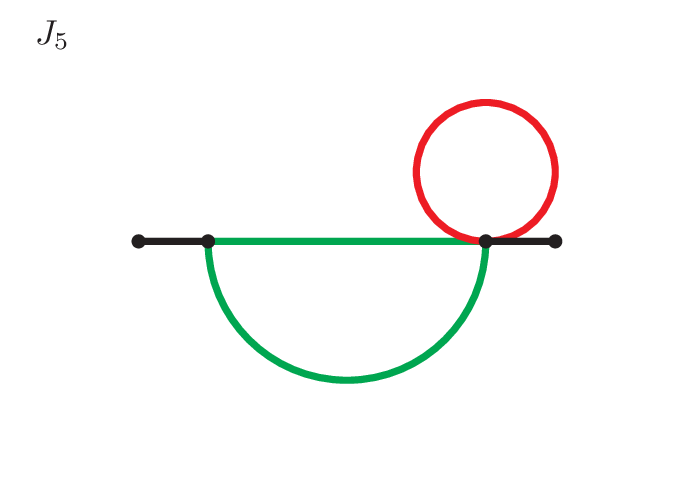}
\includegraphics[scale=0.75]{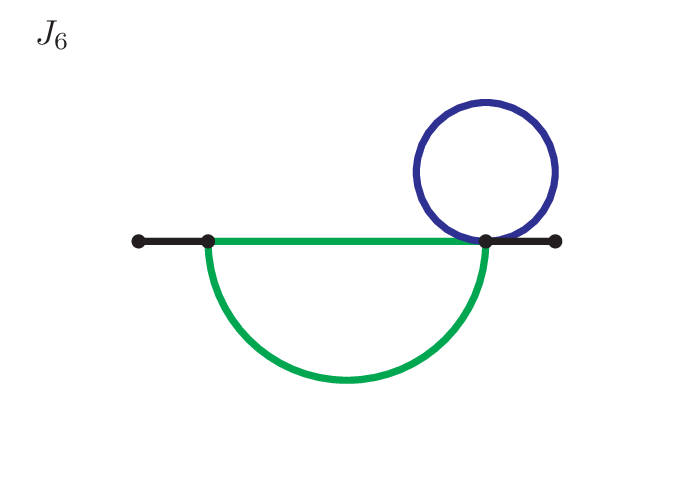}
\includegraphics[scale=0.75]{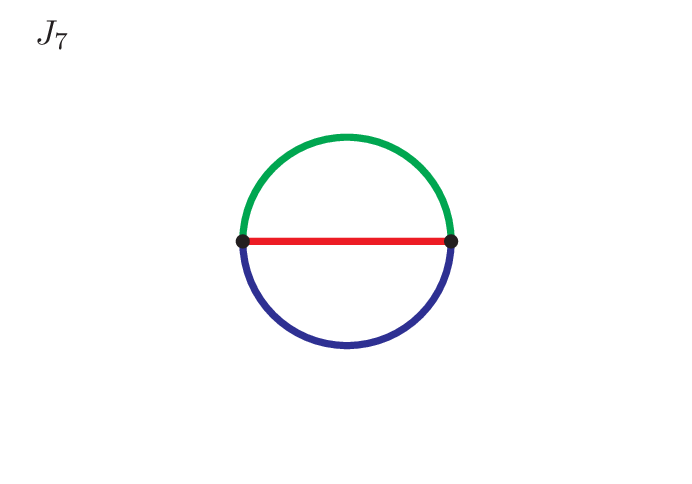}
\includegraphics[scale=0.75]{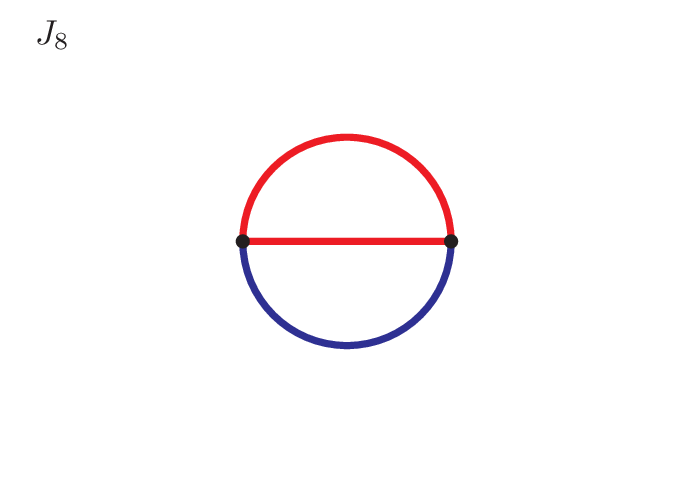}
\includegraphics[scale=0.75]{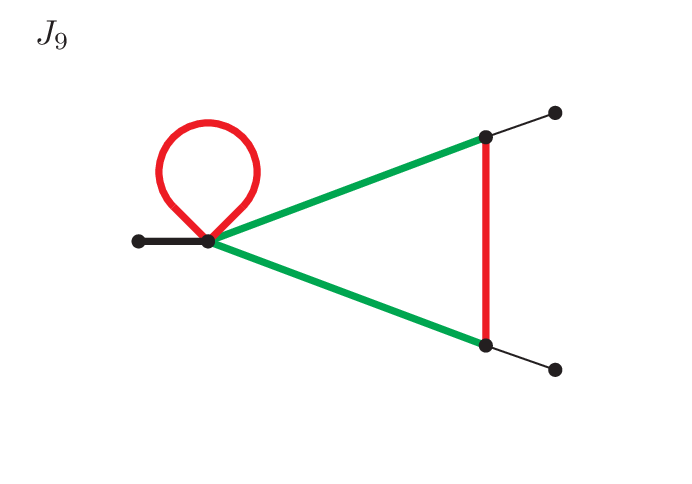}
\includegraphics[scale=0.75]{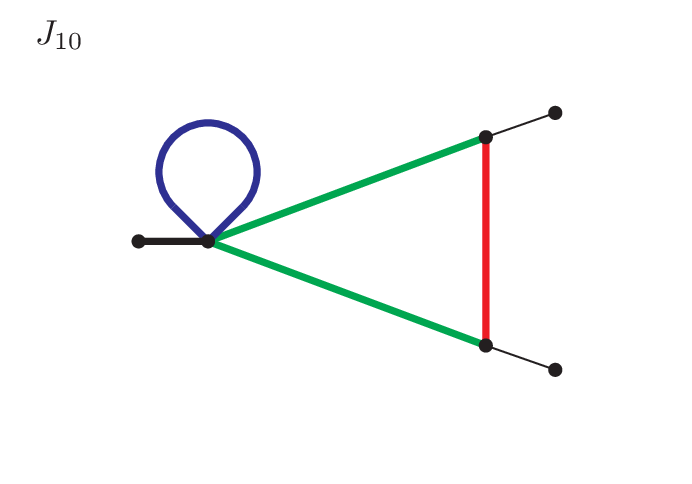}
\includegraphics[scale=0.75]{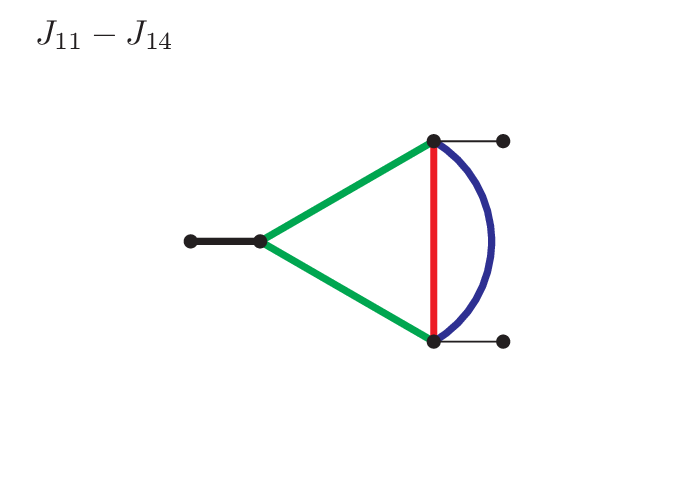}
\includegraphics[scale=0.75]{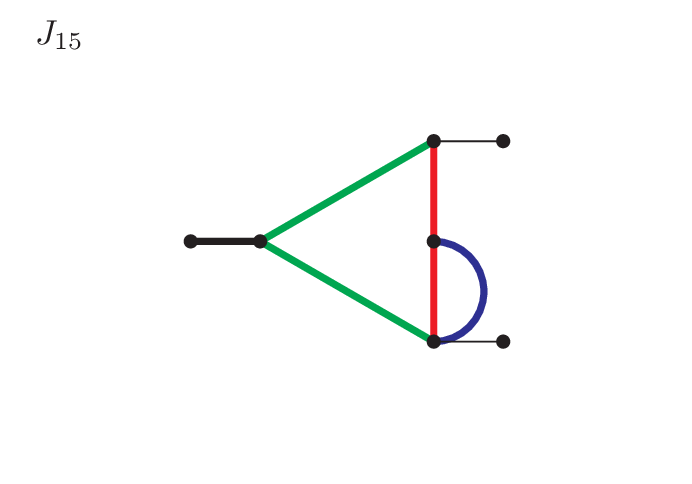}
\end{center}
\caption{
The master topologies.
}
\label{chapter_final_project:fig_master_topologies}
\end{figure}
The sectors (or master topologies) are shown in fig.~\ref{chapter_final_project:fig_master_topologies}.
Eleven sectors have only one master integral per sector, 
while one sector (with sector id $N_{\mathrm{id}}=29$) has four master integrals.

We set
\bq
 \vec{I}
 & = &
 \left(
 I_{1001000},
 I_{0101000},
 I_{1000100},
 I_{0100100},
 I_{1011000},
 I_{1010100},
 I_{1001100},
 I_{0101100},
 \right. \nonumber \\
 & & \left.
 I_{1111000},
 I_{1110100},
 I_{1011100}, I_{1(-1)11100},
 I_{10111(-1)0}, I_{1(-2)11100},
 I_{1111100}
 \right)^T.
\eq
In the next step one derives the differential equation for $\vec{I}$, as explained in section~\ref{chapter_iterated_integrals:differential_equations}.
For the derivatives with respect to $x_1, x_2, x_3$ we may use eq.~(\ref{chapter_iterated_integrals:partial_derivative}).
As we are considering a two-loop integral we get from ${\mathcal F}_{x_j}'({\bf 1}^+,\dots,{\bf \ninternal}^+)$ three raising operators 
and we therefore need integration-by-parts reduction identities for up to three dots.
We also need the dimensional shift relations, 
which involves ${\mathcal U}({\bf 1}^+,\dots,{\bf \ninternal}^+)$. 
As for a two-loop graph ${\mathcal U}$ is homogeneous of degree $2$, this leads to two raising operators
and requires integration-by-parts reduction identities for up to two dots. These are a subset of the ones required for up to three dots.

For this particular example the full differential equation can actually be derived with reduction identities for only one dot.
This saves quite some computer memory and CPU time. 
The trick is as follows: We do not set $\mu=m_t$ in the beginning and first compute the derivatives with respect to
$x_2, x_3, x_4$. These are the derivatives with respect to the internal masses and we may use
eq.~(\ref{chapter_iterated_integrals:partial_derivative_internal_mass_generalised}).
This involves only one raising operator (and no dimensional shift), therefore reduction identities for one dot are sufficient.
The derivative with respect to $x_1$ is then obtained from the scaling relation eq.~(\ref{chapter_iterated_integrals:differential_scaling_relation}).
Having obtained the derivatives with respect to $x_1, x_2, x_3$ we may set $\mu=m_t$.

In this way we obtain the differential equation
\bq
 \left( d + A \right) \vec{I} & = & 0,
\eq
where 
\bq
 d \; = \;
 \sum\limits_{j=1}^3
 dx_j \frac{\partial}{\partial x_j},
 & &
 A \; = \;
 \sum\limits_{j=1}^3
 dx_j \; A_{x_j},
\eq
and the $A_{x_j}$ are $(15 \times 15)$-matrices, whose entries are rational functions of $x_1,x_2,x_3$ and $\eps$.
These matrices are not in a form to be printed here, 
but they may be computed in a straightforward way and stored on a computer.
The matrices $A_{x_1}, A_{x_2}, A_{x_3}$ have to satisfy the integrability condition of eq.~(\ref{chapter_iterated_integrals:integrability_condition}). Spelled out in components we must have
\bq
 \partial_{x_1} A_{x_2} - \partial_{x_2} A_{x_1} + \left[ A_{x_1}, A_{x_2} \right] & = & 0,
 \nonumber \\
 \partial_{x_1} A_{x_3} - \partial_{x_3} A_{x_1} + \left[ A_{x_1}, A_{x_3} \right] & = & 0,
 \nonumber \\
 \partial_{x_2} A_{x_3} - \partial_{x_3} A_{x_2} + \left[ A_{x_2}, A_{x_3} \right] & = & 0,
\eq
where $[A,B]=A \cdot B - B \cdot A$ denotes the commutator of the two matrices $A$ and $B$. 
It is highly recommended to check these relations at this stage.

\section{Fibre transformation}

Let us set $D=4-2\eps$.
In the next step we perform a fibre transformation, e.g. we redefine the master integrals
\bq
 \vec{J} & = & U\left(\eps,x\right) \vec{I}.
\eq
We seek a transformation $U$ such that in the transformed differential equation
\bq
 \left( d + A' \right) \vec{J} & = & 0,
 \;\;\;\;\;\;
 A' \; = \; U A U^{-1} + U d U^{-1}
\eq
the dimensional regularisation parameter $\eps$ appears only as a prefactor of $A'$.
To this aim we define a new basis of master integrals
\bq
 \vec{J} & = & \left( J_1, J_2, \dots, J_{15} \right)^T
\eq
such that the $J_i$'s are of uniform weight. Expressing the new $J_i$'s as a linear combination of the old basis
$\vec{I}=(I_1,\dots,I_{15})^T$ defines the matrix $U$:
\bq
 J_i & = & \sum\limits_{j=1}^{15} U_{i j} I_j.
\eq
We may use the methods of chapter~\ref{chapter_transformations} to construct the $J_i$'s.
However, it is usually the case that the first few $J_i$'s may already be obtained from known examples.
This is also the case here.
The first four master integrals are each products of two one-loop tadpole integrals.
The tadpole integral was the first Feynman integral we calculated 
and from eq.~(\ref{chapter_basics:result_tadpole_T1_2D_uniform_weight}) we know that
$\eps \; T_1(2-2\eps)$ is of uniform weight.
We have set $D=4-2\eps$, hence we may write
\bq
 \eps \; T_1\left(2-2\eps\right)
 & = &
 \eps \; {\bf D}^- T_1\left(4-2\eps\right).
\eq
From the dimensional shift relation eq.~(\ref{chapter_basics:tadpole_dimensional_shift}) 
we have $T_1(2-2\eps) = T_2(4-2\eps)$ and therefore
\bq
 \eps \; T_1\left(2-2\eps\right)
 & = &
 \eps \; {\bf D}^- T_1\left(4-2\eps\right)
 \; = \;
 \eps \; T_2\left(4-2\eps\right).
\eq
Thus the first four master integrals of uniform weight are
\bq
 J_{1}
 & = &
 \eps^2 \; {\bf D}^- I_{1001000}
 \; = \;
 \eps^2 \; I_{2002000},
 \nonumber \\
 J_{2}
 & = &
 \eps^2 \; {\bf D}^- I_{0101000}
 \; = \;
 \eps^2 \; I_{0202000},
 \nonumber \\
 J_{3}
 & = &
 \eps^2 \; {\bf D}^- I_{1000100}
 \; = \;
 \eps^2 \; I_{2000200},
 \nonumber \\
 J_{4}
 & = &
 \eps^2 \; {\bf D}^- I_{0100100}
 \; = \;
 \eps^2 \; I_{0200200}.
\eq
The next two sectors (sectors $13$ and $21$) are again products of one-loop integrals, in this case the
product of a tadpole integral and a bubble integral.
The bubble integral is the one discussed as example 1 in section~\ref{chapter_iterated_integrals:deriving_the_dgl}.
In eq.~(\ref{chapter_iterated_integrals:example_bubble_uniform_weight_2D}) we have given the corresponding master
integral of uniform weight.
This involves the root
\bq
 r_1 & = & \sqrt{x_1\left(4+x_1\right)}
 \; = \; \frac{1}{m_t^2} \sqrt{-s \left(4m_t^2-s\right)}.
\eq
Thus
\bq
 J_{5}
 & = &
 - \eps^2 r_1 \; {\bf D}^- I_{1011000}
 \; = \;
 2 \eps^2 \frac{r_1}{4+x_1} 
 \left[ \left(1-\eps\right) I_{1002000}
      + \left(1-2\eps\right) I_{1012000}
 \right],
 \nonumber \\
 J_{6}
 & = &
 - \eps^2 r_1 \; {\bf D}^- I_{1010100}
 \; = \;
 2 \eps^2 \frac{r_1}{4+x_1} 
 \left[ \left(1-\eps\right) I_{1000200}
      + \left(1-2\eps\right) I_{1010200}
 \right].
\eq
The sectors $25$ and $26$ are genuine two-loop topologies.
In order to find master integrals of uniform weight we look at maximal cuts and constant
leading singularities (see section~\ref{chapter_transformations:maximal_cuts_and_constant_leading_singularities}).
For the sector $25$ the maximal cut in $D=2$ space-time dimensions is given by
\bq
 \mathrm{MaxCut} \; I_{1001100}(2)
 & = &
 \frac{\left(2 \pi i\right)^3 m_t^2}{\pi^2}
 \int\limits_{{\mathcal C}_{\mathrm{MaxCut}}} dz_2 \;
 \frac{1}{\left(z_2+m_t^2-m_W^2\right)\sqrt{4m_H^2m_W^2-\left(z_2+m_H^2\right)^2}}
 \nonumber \\
 & = & 
 \int\limits_{{\mathcal C}_{\mathrm{MaxCut}}} \varphi.
\eq
The last equation defines the integrand $\varphi$.
We now replace the integration domain ${\mathcal C}_{\mathrm{MaxCut}}$ by a simpler
integration domain ${\mathcal C}$, given by a small anti-clockwise circle around $z_2=m_W^2-m_t^2$.
This gives 
\bq
 \left\langle \varphi | {\mathcal C} \right\rangle
 & = &
 \frac{16 \pi^2}{\sqrt{-\lambda\left(x_2,x_3,1\right)}},
\eq
where $\lambda(x,y,z)$ denotes the K\"all\'en function
\bq
 \lambda\left(x,y,z\right)
 & = &
 x^2 + y^2 + z^2 - 2 x y - 2 y z - 2 z x.
\eq
The analysis for sector $26$ is similar and gives instead of $\lambda(x_2,x_3,1)$ the expression $\lambda(x_2,x_3,x_2)$.
We introduce two new square roots
\bq
 r_2 & = & 
 \sqrt{-\lambda\left(x_2,x_3,1\right)}
 \; = \; 
 \sqrt{2x_2x_3 + 2x_2 + 2x_3 - x_2^2 - x_3^2 - 1}
 \; = \; 
 \frac{1}{m_t^2} \sqrt{-\lambda\left(m_W^2,m_H^2,m_t^2\right)},
 \nonumber \\
 r_3 & = & 
 \sqrt{-\lambda\left(x_2,x_3,x_2\right)}
 \; = \; 
 \sqrt{4 x_2 x_3 - x_3^2}
 \; = \;
 \frac{1}{m_t^2} \sqrt{m_H^2 \left(4m_W^2-m_H^2\right)}.
\eq
Our tentative guess for $J_7$ and $J_8$ is
\bq
 J_{7}
 & = &
 \frac{1}{2} \eps^2 r_2 \; {\bf D}^- I_{1001100},
 \nonumber \\
 J_{8}
 & = &
 \frac{1}{2} \eps^2 r_3 \; {\bf D}^- I_{0101100}.
\eq
This guess is obtained (apart from irrelevant rational prefactors)
by dividing $I_7$ and $I_8$ by the appropriate period $\langle \varphi | {\mathcal C} \rangle$ and by replacing $\pi$ by $\eps^{-1}$.
There may be additional terms proportional to sub-topologies. In these two examples we verify a posteriori that there are no additional
terms and $J_7$ and $J_8$ define master integrals of uniform weight.

The sectors $15$ and $23$ are again products of one-loop integrals.
We need the one-loop triangle of uniform weight, which again can be obtained from the maximal cut.
Proceeding as above we obtain
\bq
 J_{9}
 & = &
 \eps^3 x_1\; I_{1112000},
 \nonumber \\
 J_{10}
 & = &
 \eps^3 x_1\; I_{1110200}.
\eq
We now come to sector $29$. This is the most challenging sector, as it has four master integrals.
We started from an ISP-basis, where we chose
\bq
 I_{1011100}, I_{1(-1)11100},
 I_{10111(-1)0}, I_{1(-2)11100}
\eq
as a basis for this sector.
We could have chosen a dot-basis, in which case
\bq
 I_{1011100}, I_{1011200},
 I_{1012100}, I_{1021100}
\eq
would be an appropriate basis.
Our strategy is to put first the $(4 \times 4)$-diagonal block into an $\eps$-form and to treat the
off-diagonal blocks (corresponding to sub-topologies) in a second stage.
For the diagonal block we may work on the maximal cut.
For the maximal cut we use the loop-by-loop approach (see section~\ref{chapter_basics:sect_Baikov_representation}),
this yields a one-fold integral representation in the Baikov variable $z_2$ for the maximal cut.
We look at the maximal cuts for various sets of indices $\nu_j$ and various values of the space-time dimension.
In these integral representations we recognise in the denominators a few recurring expressions:
\bq
\label{chapter_final_project:example_polynomials_punctures}
 P_1 & = & z_2 - m_W^2,
 \nonumber \\
 P_2 & = & \left(z_2+m_t^2-m_W^2\right)^2 - s \left( z_2 - m_W^2 \right),
 \nonumber \\
 R & = & \sqrt{4m_W^2m_H^2-\left(z_2+m_H^2\right)^2}.
\eq
$P_1$ is a linear polynomial in $z_2$, $P_2$ is a quadratic polynomial in $z_2$ and $R$ is the square
root of a quadratic polynomial in $z_2$.
In terms of maximal cuts we have for example
\bq
\label{chapter_final_project:example_maxcuts}
 \mathrm{MaxCut} \; I_{1012100}\left(4\right)
 & = &
 4 \pi^2 i \int\limits_{{\mathcal C}_{\mathrm{MaxCut}}} \frac{m_t^2 \left(m_H^2-z_2\right)dz_2}{s P_1 R},
 \nonumber \\
 \mathrm{MaxCut} \; I_{1011200}\left(4\right)
 & = &
 4 \pi^2 i \int\limits_{{\mathcal C}_{\mathrm{MaxCut}}} \frac{m_t^2 \left(2m_W^2-m_H^2-z_2\right)dz_2}{s P_1 R},
 \nonumber \\
 \frac{1}{\eps} \mathrm{MaxCut} \; I_{1011100}\left(2\right)
 & = &
 - 8 \pi^2 i \int\limits_{{\mathcal C}_{\mathrm{MaxCut}}} \frac{m_t^4  dz_2}{P_2 R},
\eq
where the $z_2$-dependent expressions in the denominator are $P_1$, $P_2$ and $R$.
We have four master integrals for this sector, hence we look for four master contours ${\mathcal C}_1,\dots, {\mathcal C}_4$.
The four master contours have to be independent.
We recall that on a Riemann sphere with $5$ punctures we may define four independent contours as small circles
around four of the five punctures. The small circle around the fifth puncture is equivalent to minus the sum of the first four
contours.

\begin{figure}
\begin{center}
\includegraphics[scale=1.0]{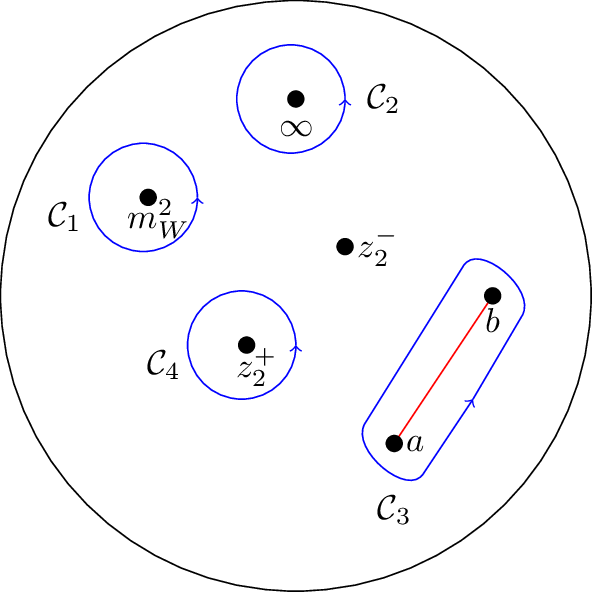}
\end{center}
\caption{
The integration contours ${\mathcal C}_1, \dots, {\mathcal C}_4$ on the Riemann sphere. 
Note that a small counter clockwise circle
around $z_2^-$ is equivalent to $(-{\mathcal C}_1 - \dots - {\mathcal C}_4)$.
}
\label{chapter_final_project:fig_contours}
\end{figure}

We may also count a square root $\sqrt{(z-a)(z-b)}$ as a deformed puncture:
Consider for simplicity the square root $f=\sqrt{x}\sqrt{4+x}$ with the usual branch cut of the square root along the negative
real axis. The function $f$ is single valued for $x \in {\mathbb C} \backslash [-4,0]$.
The essential point is that $f$ is single valued in a neighbourhood of the negative real axis for $x<-4$, the two sign ambiguities cancel
each other. Thus we may define $\sqrt{(z-a)(z-b)}$ as a single-valued function on ${\mathbb C}\backslash [a,b]$, where
$[a,b]$ denotes a slit between $a$ and $b$.
The situation is shown in fig.~\ref{chapter_final_project:fig_contours}.
Let us now investigate which integrands give constant residues. We start with the square root $\sqrt{(z-a)(z-b)}$.
Let ${\mathcal C}$ be an anticlockwise closed contour around the slit $[a,b]$. Then
\bq
\label{chapter_final_project:square_root_contour}
 \int\limits_{{\mathcal C}} \frac{dz}{\sqrt{(z-a)(z-b)}}
 & = &
 2 \pi i.
\eq
For the combination of the square root $\sqrt{(z-a)(z-b)}$ with a simple pole $(z-c)$ in the denominator
one finds that the integrand
\bq
\label{chapter_final_project:square_root_simple_pole_contour}
 \frac{1}{2\pi i} 
 \frac{\sqrt{\left(c-a\right)\left(c-b\right)}}{\left(z-c\right)\sqrt{\left(z-a\right)\left(z-b\right)}}
\eq
gives $\pm 1$, when integrated around a small circle at $z=c$ 
or along a cycle around the slit $[a,b]$.

Let us now turn to quadratic polynomials in the denominator.
We consider the quadratic polynomial
\bq
 z^2 - 2 c z + c^2 - r^2 & = & \left(z-c-r\right)\left(z-c+r\right),
\eq
which we may factorise at the expense of introducing the $z$-independent square root $r$.
We consider integrals of the form
\bq
 \int\limits_{{\mathcal C}} \frac{\left(a z+b\right) dz}{\left(z-c-r\right)\left(z-c+r\right)}
\eq
where ${\mathcal C}=n_+{\mathcal C}_++n_-{\mathcal C}_-$ is a ${\mathbb Z}$-linear combination 
of small anticlockwise circles around the two poles
$z=c+r$ (contour ${\mathcal C}_+$)
and $z=c-r$ (contour ${\mathcal C}_-$).
For ${\mathcal C}={\mathcal C}_++{\mathcal C}_-$ the square root $r$ will not show up in the final result:
\bq
 \int\limits_{{\mathcal C}_++{\mathcal C}_-} \frac{\left(a z+b\right) dz}{\left(z-c-r\right)\left(z-c+r\right)}
 & = &
 2 \pi i \; a. 
\eq
However, on a punctured Riemann sphere this contour is equivalent to minus the sum of all other contours and hence not independent
of those.
We need the integral around one pole (or the difference between them).
We have
\bq
 \int\limits_{{\mathcal C}_+-{\mathcal C}_-} \frac{\left(a z+b\right) dz}{\left(z-c-r\right)\left(z-c+r\right)}
 & = &
 2 \pi i \; \frac{ac+b}{r}.
\eq
Dividing the integrand on the left-hand side by the right-hand side we obtain an integrand with unit residue.

Let us now return to sector $29$ of our example.
We denote the two roots of $P_2$ by
\bq
 z_2^\pm & = & \frac{1}{2}\left[ 2m_W^2 - 2m_t^2 + s \pm \sqrt{-s\left(4m_t^2-s\right)}\right].
\eq
We will need the value of $R$ at $z_2=z_2^\pm$.
We introduce
\bq
 r_4 & = & 
 \sqrt{-\lambda\left(x_2,x_3,1\right) - \frac{x_1^2}{2} - x_1 \left(2-x_2-x_3\right) + r_1 \left(1 + \frac{x_1}{2} - x_2 -x_3\right)},
 \nonumber \\
 r_5 & = & 
 \sqrt{-\lambda\left(x_2,x_3,1\right) - \frac{x_1^2}{2} - x_1 \left(2-x_2-x_3\right) - r_1 \left(1 + \frac{x_1}{2} - x_2 -x_3\right)}.
\eq
Then $R(z_2^+)= m_t^2 r_4$ and $R(z_2^-)= m_t^2 r_5$.
With these preparations it is now clear from eq.~(\ref{chapter_final_project:example_polynomials_punctures})
and eq.~(\ref{chapter_final_project:example_maxcuts}) that a suitable set of master contours is
\bq
 {\mathcal C}_1 & : & \mbox{small anticlockwise circle around $z_2=m_W^2$}.
 \\
 {\mathcal C}_2 & : & \mbox{small anticlockwise circle around $z_2=\infty$}.
 \nonumber \\
 {\mathcal C}_3 & : & \mbox{anticlockwise cycle around the slit $[-m_H^2-2m_Hm_W,-m_H^2+2m_Hm_W]$}.
 \nonumber \\
 {\mathcal C}_4 & : & \mbox{small anticlockwise circle around $z_2=z_2^+$}.
 \nonumber 
\eq
The integration contours are sketched in fig.~\ref{chapter_final_project:fig_contours}.
Inspecting eq.~(\ref{chapter_final_project:example_maxcuts}) we define 
(note that kinematic independent algebraic prefactors don't matter here)
\bq
\label{chapter_final_project:differential_forms_sector_29}
 \varphi_1
 & = &
 i \pi^2 \eps^3 \frac{\left(m_H^2-z_2\right)}{P_1 R} dz_2,
 \nonumber \\
 \varphi_2
 & = &
 i \pi^2 \eps^3 \frac{\left(2m_W^2-m_H^2-z_2\right)}{P_1 R} dz_2,
 \nonumber \\
 \varphi_3
 & = &
 \pi^2  \eps^3 \frac{m_t^2 r_4 \left(z_2-z_2^-\right)}{P_2 R} dz_2,
 \nonumber \\
 \varphi_4
 & = &
 \pi^2  \eps^3 \frac{m_t^2 r_5 \left(z_2-z_2^+\right)}{P_2 R} dz_2.
\eq
Note that the singularities of $\varphi_i$ are all as in eq.~(\ref{chapter_final_project:square_root_contour}) and eq.~(\ref{chapter_final_project:square_root_simple_pole_contour}).
In particular we have $P_2=(z_2-z_2^+)(z_2-z_2^-)$.
Hence the factor $(z_2-z_2^-)$ in the numerator in the definition
of $\varphi_3$ cancels the same factor of $P_2$ in the denominator, leaving just a single factor $(z_2-z_2^+)$ in the denominator.

We verify that $\langle \varphi_i | {\mathcal C}_j \rangle$ is a constant of weight zero
and that the so defined $(4 \times 4)$-matrix is invertible.
Indeed, we have
\bq
 \left\langle \varphi_i | {\mathcal C}_j \right\rangle
 & = &
 2 i \pi^3 \eps^3 
 \left( \begin{array}{rrrr}
 1  &  1 & -2 & 0 \\
 -1 &  1 &  0 & 0 \\
 0  &  0 & -1 & 1 \\
 0  &  0 & -1 & 0 \\
 \end{array} \right).
\eq
As the matrix is invertible, $\varphi_1, \dots, \varphi_4$ form a basis.
We now look for Feynman integrals, whose maximal cut on the sector $29$ 
gives the differential forms of eq.~(\ref{chapter_final_project:differential_forms_sector_29})
up to irrelevant kinematic independent algebraic prefactors (that is to say we do not care about prefactors
like $2$, $i$ or $\sqrt{3}$). We do however care about kinematic dependent prefactors (like $x_1$).
Comparing eq.~(\ref{chapter_final_project:differential_forms_sector_29}) with eq.~(\ref{chapter_final_project:example_maxcuts}) we see that $\varphi_1$ matches $\mathrm{MaxCut} \; I_{1012100}(4)$
and $\varphi_2$ matches $\mathrm{MaxCut} \; I_{1011200}(4)$.
The last two cases ($\varphi_3$ and $\varphi_4$) are only slightly more complicated:
We first see that the denominator of $\varphi_3$ and $\varphi_4$ matches the one of 
$\mathrm{MaxCut} \; I_{1011100}(2)$.
A factor $z_2$ in the numerator of $\varphi_3$ or $\varphi_4$
corresponds to an irreducible scalar product in the Feynman integral and translates to $\nu_2=-1$.
We therefore deduce
\bq
 J_{11} & = & \eps^3 x_1 I_{1012100},
 \nonumber \\
 J_{12} & = & \eps^3 x_1 I_{1011200},
 \nonumber \\
 J_{13} & = & \eps^2 r_4 \left[ {\bf D}^- I_{1(-1)11100} + \left(1 + \frac{1}{2} x_1 - x_2 + \frac{1}{2} r_1\right) {\bf D}^- I_{1011100} \right], 
 \nonumber \\
 J_{14} & = & \eps^2 r_5 \left[ {\bf D}^- I_{1(-1)11100} + \left(1 + \frac{1}{2} x_1 - x_2 - \frac{1}{2} r_1\right) {\bf D}^- I_{1011100} \right].
\eq
Note that
\bq
 z_2^\pm & = & - m_t^2 \left( 1 + \frac{1}{2} x_1 - x_2 \mp \frac{1}{2} r_1 \right).
\eq
As we work on the maximal cut, there might be corrections corresponding to sub-topologies.
Also in this case we are lucky and we verify (by inspecting $A'$) that no correction terms are required.

It remains to treat the last sector.
The sector $31$ has again only one master integral, and we first consider the maximal cut.
The steps are similar to what we did for $J_7$ and $J_8$.
From the maximal cut we obtain the tentative guess
\bq
\label{chapter_final_project:J_15_tentative}
 J_{15}^{\mathrm{tentative}} & = & \eps^2 r_3 \left[ \left(1-x_2\right)^2 -x_1 x_2 \right] {\bf D}^- I_{1111100}.
\eq
Again, there might be corrections due to sub-topologies (which are not detected at the maximal cut).
In this case there are actually corrections.
We could use the methods of section~\ref{chapter_transformations:fibre_transformation}
to systematically obtain them.
However, with an educated guess we might reach our goal faster:
From the differential equation for $J_{15}^{\mathrm{tentative}}$ we see that we need correction terms
proportional to the master integrals of sector $29$.
In eq.~(\ref{chapter_final_project:J_15_tentative}) we recognise $(1-x_2)^2-x_1x_2$ as the 
constant part of the polynomial $P_2$:
\bq
 \left. P_2 \right|_{z_2=0} & = & m_t^4 \left[ \left(1-x_2\right)^2 -x_1 x_2 \right].
\eq
Terms proportional to $z_2$ cancel the second propagator of $I_{1111100}$ and correspond to Feynman
integrals from sector $29$.
Therefore we take for the educated guess the terms proportional to $z_2$ and $z_2^2$ of $P_2$ into account:
\bq
\label{chapter_final_project:J_15_educated_guess}
 J_{15} & = & \eps^2 r_3 
 \left\{ \left[ \left(1-x_2\right)^2 -x_1 x_2 \right] {\bf D}^- I_{1111100}
  + \left(2+x_1-2x_2\right) {\bf D}^- I_{1011100}
  + {\bf D}^- I_{1(-1)11100}
 \right\}.
 \nonumber \\
\eq
Plugging this ansatz into the differential equation we verify that this guess factorises $\eps$ out.
We obtain
\bq
\label{chapter_final_project:penguin_dgl_eps_form}
 \left( d + A' \right) \vec{J} & = & 0
\eq
with
\bq
 A' & = & \eps \sum\limits_{j=1}^{25} C_j \omega_j,
 \;\;\;\;\;\;
 \omega_j \; = \; d \ln f_j.
\eq
The $f_j$ are algebraic functions of the kinematic variables.
Some of them are polynomials in the kinematic variables.
These define the rational part of the alphabet:
\begin{align}
\label{chapter_final_project:rational_alphabet}
 f_1 & = x_1,
 &
 f_2 & = x_1+4,
 &
 f_3 & = x_2,
 \\
 f_4 & = x_2-1,
 &
 f_5 & = x_3,
 &
 f_6 & = \left(1-x_2\right)^2-x_1x_2,
 \nonumber \\
 f_7 & = 4x_2-x_3,
 &
 f_8 & = -\lambda\left(x_2,x_3,1\right),
 &
 f_9 & = \left[ x_1\left(x_2+x_3\right) - \lambda\left(x_2,x_3,1\right) \right] ^2 - 4 x_1\left(4+x_1\right) x_2 x_3.
 \nonumber 
\end{align}
The algebraic part of the alphabet can be taken as
\begin{align}
\label{chapter_final_project:algebraic_alphabet}
 f_{10} & = 2+x_1-r_1,
 & 
 f_{11} & = 2+x_1-2x_2-r_1,
 \\
 f_{12} & = -\lambda\left(x_2,x_3,1\right) - x_1 -\left(1+\frac{x_1}{2}-x_2-x_3\right)\left(x_1-r_1\right),
 &
 f_{13} & = 1+x_2-x_3-i r_2,
 \nonumber \\
 f_{14} & = 1-x_2+x_3-i r_2,
 &
 f_{15} & = x_3-i r_3,
 \nonumber \\
 f_{16} & = 1-x_2+i r_2+i r_3,
 &
 f_{17} & = 1+\frac{x_1}{2}+x_2-x_3-\frac{r_1}{2}-i r_4,
 \nonumber \\
 f_{18} & = 1+\frac{x_1}{2}-x_2+x_3-\frac{r_1}{2}-i r_4,
 &
 f_{19} & = 1+\frac{x_1}{2}+x_2-x_3+\frac{r_1}{2}-i r_5,
 \nonumber \\
 f_{20} & = 1+\frac{x_1}{2}-x_2+x_3+\frac{r_1}{2}-i r_5,
 &
 f_{21} & = 
 \left(r_1-ir_4-ir_5\right)\left(r_1+ir_4+ir_5\right),
 \nonumber \\
 f_{22} & = x_1-r_1-2 i r_2-2 i r_4,
 &
 f_{23} & = 1+\frac{x_1}{2}-x_2-\frac{r_1}{2}+i r_3-i r_4,
 \nonumber \\
 f_{24} & = x_1+r_1-2 i r_2-2 i r_5,
 &
 f_{25} & = 1+\frac{x_1}{2}-x_2+\frac{r_1}{2}+i r_3-i r_5.
 \nonumber
\end{align}
Note that the choice of the $\omega_j$'s and of the $f_j$'s is not unique. For example, we may always 
transform the $\omega_j$'s by a $\mathrm{GL}_{\NL}(\mathbb{Q})$-transformation.

Note further that all roots are included in our alphabet:
\bq
\label{chapter_final_project:roots_in_alphabet}
 r_1^2 \; = \; f_1 f_2,
 \;\;\;\;\;\;
 r_2^2 \; = \; f_8,
 \;\;\;\;\;\;
 r_3^2 \; = \; f_5 f_7,
 \;\;\;\;\;\;
 r_4^2 \; = \; f_{12},
 \;\;\;\;\;\;
 r_5^2 \; = \; \frac{f_9}{f_{12}}.
\eq
Rewriting the $\omega_j$'s as dlog-forms is not entirely straightforward.
The exercises~\ref{chapter_final_project:exercise_polynomial} (for the polynomial case) and \ref{chapter_final_project:exercise_algebraic} (for the algebraic case) show how to do this.
\\
\\
\bs
{\it \refstepcounter{exercise}
\label{chapter_final_project:exercise_polynomial}
{\bf Exercise \theexercise}: 
Rewrite the differential one-form 
\bq
 \omega & = &
 - \frac{\left(1-x_2\right)^2 dx_1}{x_1 \left[\left(1-x_2\right)^2-x_1x_2\right]}
 - \frac{x_1\left(1+x_2\right)dx_2}{\left(1-x_2\right) \left[\left(1-x_2\right)^2-x_1x_2\right]}
\eq
as a dlog-form.
}
\es
\\
\\
The $C_j$ are $15 \times 15$ matrices, whose entries are of the form ${\mathbb Q} + i {\mathbb Q}$.
They are usually sparse matrices.
As an example we have
\bq
\label{chapter_final_project:C_0}
 C_1 
 & = &
 \left( \begin{array}{rrrrrrrrrrrrrrr}
 0 & 0 & 0 & 0 & 0 & 0 & 0 & 0 & 0 & 0 & 0 & 0 & 0 & 0 & 0 \\
 0 & 0 & 0 & 0 & 0 & 0 & 0 & 0 & 0 & 0 & 0 & 0 & 0 & 0 & 0 \\
 0 & 0 & 0 & 0 & 0 & 0 & 0 & 0 & 0 & 0 & 0 & 0 & 0 & 0 & 0 \\
 0 & 0 & 0 & 0 & 0 & 0 & 0 & 0 & 0 & 0 & 0 & 0 & 0 & 0 & 0 \\
 0 & 0 & 0 & 0 & 0 & 0 & 0 & 0 & 0 & 0 & 0 & 0 & 0 & 0 & 0 \\
 0 & 0 & 0 & 0 & 0 & 0 & 0 & 0 & 0 & 0 & 0 & 0 & 0 & 0 & 0 \\
 0 & 0 & 0 & 0 & 0 & 0 & 0 & 0 & 0 & 0 & 0 & 0 & 0 & 0 & 0 \\
 0 & 0 & 0 & 0 & 0 & 0 & 0 & 0 & 0 & 0 & 0 & 0 & 0 & 0 & 0 \\
 0 & 0 & 0 & 0 & 0 & 0 & 0 & 0 & 1 & 0 & 0 & 0 & 0 & 0 & 0 \\
 0 & 0 & 0 & 0 & 0 & 0 & 0 & 0 & 0 & 1 & 0 & 0 & 0 & 0 & 0 \\
 0 & 0 & 0 & 0 & 0 & 0 & 0 & 0 & 0 & 0 & 1 & 0 & 0 & 0 & 0 \\
 0 & 0 & 0 & 0 & 0 & 0 & 0 & 0 & 0 & 0 & 0 & 1 & 0 & 0 & 0 \\
 0 & 0 & 0 & 0 & 0 & 0 & -2 & 0 & 0 & 0 & 0 & 0 & \frac{1}{2} & \frac{1}{2} & 0 \\
 0 & 0 & 0 & 0 & 0 & 0 & -2 & 0 & 0 & 0 & 0 & 0 & \frac{1}{2} & \frac{1}{2} & 0 \\
 0 & 0 & 0 & 0 & 0 & 0 & 0 & -2 & 0 & 0 & 0 & 0 & 0 & 0 & 1 \\
 \end{array} \right).
\eq
The matrix $C_1$ accompanies the differential one-form $\omega_1=dx_1/x_1$.
From the lists in eq.~(\ref{chapter_final_project:rational_alphabet}) 
and eq.~(\ref{chapter_final_project:algebraic_alphabet}) one may check that
we have chosen the remaining $\omega_j$'s such that they do not have a pole on $x_1=0$. 
For the family of Feynman integrals under consideration we do not expect for the master integrals
$J_1$-$J_{15}$ any logarithmic singularities
in the limit $x_1\rightarrow 0$.
This translates into the requirement that there should be no trailing zeros with respect to the
integration in $x_1$.
From eq.~(\ref{chapter_final_project:C_0}) we see that trailing zeros in the $x_1$-integration are absent
if
\bq
\label{chapter_final_project:limits_from_C_0}
 0 & = & 
 \lim\limits_{x_1\rightarrow 0} J_9
 \; = \;
 \lim\limits_{x_1\rightarrow 0} J_{10}
 \; = \;
 \lim\limits_{x_1\rightarrow 0} J_{11}
 \; = \;
 \lim\limits_{x_1\rightarrow 0} J_{12}
 \nonumber \\
 & = &
 \lim\limits_{x_1\rightarrow 0} \left( J_{13} + J_{14} - 4 J_7 \right)
 \; = \;
 \lim\limits_{x_1\rightarrow 0} \left( J_{15} - 2 J_8 \right).
\eq

With our choice of the $\omega_j$'s as in eq.~(\ref{chapter_final_project:rational_alphabet}) 
and eq.~(\ref{chapter_final_project:algebraic_alphabet})
only $\omega_1$ has a pole along $x_1=0$.
We may always choose the remaining $\omega_j$'s such that they do not have a pole along $x_1=0$.
The following exercise shows that this is not entirely trivial:
\\
\\
\bs
{\it \refstepcounter{exercise}
{\bf Exercise \theexercise}: 
Let
\bq
 \tilde{f}_1 & = & \lambda\left(x_2,x_3,1\right) + 8x_3 - x_1 \left(x_2-x_3\right) - r_4 r_5,
 \nonumber \\
 \tilde{f}_2 & = & \lambda\left(x_2,x_3,1\right) + 8x_2 + x_1 \left(x_2-x_3\right) - r_4 r_5,
 \nonumber \\
 \tilde{f}_3 & = & \lambda\left(x_2,x_3,1\right) - x_1 \left(x_2-x_3\right) + r_4 r_5,
 \nonumber \\
 \tilde{f}_4 & = & \lambda\left(x_2,x_3,1\right) + x_1 \left(x_2-x_3\right) + r_4 r_5,
\eq
Show that
\bq
 \tilde{\omega} & = & d\ln\left(\frac{\tilde{f}_1\tilde{f}_2}{\tilde{f}_3\tilde{f}_4}\right)
\eq
has a pole along $x_1=0$, while
\bq
 \omega & = & d\ln\left(x_1^2 \frac{\tilde{f}_1\tilde{f}_2}{\tilde{f}_3\tilde{f}_4}\right)
\eq
does not.
}
\es

\section{Base transformation}
\label{chapter_final_project:base_transformation}

The differential equation eq.~(\ref{chapter_final_project:penguin_dgl_eps_form}) 
for the Feynman integrals involve dlog-forms with algebraic
arguments (square roots).
A sufficient but not necessary criteria to express the result in terms of multiple polylogarithms
is the possibility to rationalise simultaneously all occurring square roots.
Unfortunately, we are not so lucky here:
With the methods of section~\ref{chapter_transformations:base_transformation}
we may rationalise four out the five square roots $r_1$-$r_5$.
One square root (either $r_4$ or $r_5$) remains unrationalised.
Nevertheless it is instructive to see how four of the square roots are rationalised.

The root $r_1$ occurs frequently in Feynman integrals and we encountered this root already before
in sections.~\ref{chapter_iterated_integrals:fibre_transformation} and \ref{chapter_iterated_integrals:base_transformation}.
The transformation (see eq.~(\ref{chapter_iterated_integrals:example_rationalisation}))
\bq
 x_1 & = & \frac{\left(1-x_1'\right)^2}{x_1'}
\eq
rationalises the root $r_1$:
\bq
 r_1 & = & \frac{1-x_1'^2}{x_1'}.
\eq
We note that in terms of the new variable $x_1'$ the roots $r_4$ and $r_5$ are given by
\bq
\label{chapter_final_project:r4_r5_simplified}
 r_4 \; = \; \sqrt{-\lambda\left(x_2,x_3,x_1'\right)},
 & &
 r_5 \; = \; \sqrt{-\lambda\left(x_2,x_3,\frac{1}{x_1'}\right)}.
\eq
The roots $r_2$ and $r_3$ depend only on $x_2$ and $x_3$.
With the help of the methods from section~\ref{chapter_transformations:base_transformation} one finds that these roots are rationalised 
simultaneously for example by
\bq
\label{chapter_final_project:rationalisation_r2_r3}
 x_2
 \; = \;
 \frac{\left(1+x_2'^2\right)\left(2-x_3'\right)}{x_3'\left(3-2x_3'-x_2'^2\right)},
 & &
 x_3
 \; = \;
 \frac{4 \left(2-x_3'\right)}{x_3'\left(3-2x_3'-x_2'^2\right)}.
\eq
We have
\bq
 r_2 \; = \;
 \frac{2 i \left(3-4x_3'-x_2'^2+x_3'^2\right)}{x_3'\left(3-2x_3'-x_2'^2\right)},
 & &
 r_3 \; = \;
 \frac{4 x_2' \left(2-x_3'\right)}{x_3'\left(3-2x_3'-x_2'^2\right)}.
\eq
This leaves us with the roots $r_4$ and $r_5$. From eq.~(\ref{chapter_final_project:r4_r5_simplified}) it is clear that the argument of the square root 
$r_4$ or $r_5$ is quadratic in $x_1'$.
It is therefore straightforward to rationalise in addition either $r_4$ or $r_5$ (but not both).
For example, $r_4$ is rationalised by
\bq
 x_1'
 \; = \; 
 \frac{\left[x_4'^2-\left(x_2-x_3\right)^2\right]}{2\left(x_4'-x_2-x_3\right)},
 & &
 r_4
 \; = \;
 i
 \frac{\lambda\left(x_2,x_3,x_4'\right)}{2\left(x_4'-x_2-x_3\right)},
\eq
while $r_5$ is rationalised by
\bq
 x_1'
 \; = \; 
 \frac{2\left(x_5'-x_2-x_3\right)}{\left[x_5'^2-\left(x_2-x_3\right)^2\right]},
 & &
 r_5
 \; = \;
 i
 \frac{\lambda\left(x_2,x_3,x_5'\right)}{2\left(x_5'-x_2-x_3\right)}.
\eq
The variables $(x_2',x_3',x_4')$ rationalise simultaneously the roots $\{r_1,r_2,r_3,r_4\}$, the variables
$(x_2',x_3',x_5')$ rationalise simultaneously the roots $\{r_1,r_2,r_3,r_5\}$.

Rationalisations are also helpful to convert the $\omega_j$'s to dlog-forms, as the following exercise shows:
\\
\\
\bs
{\it \refstepcounter{exercise}
\label{chapter_final_project:exercise_algebraic}
{\bf Exercise \theexercise}: 
Rewrite the differential one-form 
\bq
 \omega & = &
 - \frac{\left(2-2x_2-x_1x_2\right)r_1 dx_1}{x_1\left(4+x_1\right) \left[\left(1-x_2\right)^2-x_1x_2\right]}
 - \frac{r_1 dx_2}{\left[\left(1-x_2\right)^2-x_1x_2\right]}
\eq
where $r_1=\sqrt{x_1(4+x_1)}$
as a dlog-form.
}
\es

\section{Boundary values}

In order to solve the differential equation we need boundary values for the master integrals $\vec{J}$.
A convenient boundary point is the point
\bq
 x_1 \; = \; 0,
 \;\;\;\;\;\;
 x_2 \; = \; 1,
 \;\;\;\;\;\;
 x_3 \; = \; 1.
\eq
At this point we have
\bq
 r_1 \; = \; 0,
 & &
 r_2
 \; = \;
 r_3
 \; = \;
 r_4
 \; = \;
 r_5
 \; = \;
 \sqrt{3}.
\eq
At the boundary point, the four tadpole integrals $J_1$-$J_4$ are all equal and given by
\bq
 J_1
 \;\; = \;\; 
 J_2
 \;\; = \;\; 
 J_3
 \;\; = \;\; 
 J_4
 & = & 
  1 + \zeta_2 \eps^2 - \frac{2}{3} \zeta_3 \eps^3 + \frac{7}{4} \zeta_4 \eps^4 
 + {\mathcal O}\left(\eps^5\right).
\eq
The master integrals $J_5$ and $J_6$ vanish at the boundary point (due to the prefactor $r_1$):
\bq
 J_5
 \;\; = \;\; 
 J_6
 & = & 0.
\eq
The master integrals $J_7$ and $J_8$ are equal at the boundary point.
The value is given by the value of the equal mass sunrise integral at zero external momentum squared
(times a trivial prefactor). 
Let us introduce the following linear combination of harmonic polylogarithms
\bq
 \overline{H}_{m_1 \dots m_k}\left(x\right)
 & = &
 H _{m_1 \dots m_k}\left(x\right)
 -
 H _{m_1 \dots m_k}\left(x^{-1}\right).
\eq
Then \cite{Adams:2015ydq}
\bq
\label{chapter_final_project:boundary_J7_J8}
 J_7
 \;\; = \;\; 
 J_8
 & = & 
 \frac{3}{2i}
 \left\{
  \eps^2 \overline{H}_2\left(e^{\frac{2\pi i}{3}}\right)
  + \eps^3 \left[
                 - 2 \overline{H}_{2,1}\left(e^{\frac{2\pi i}{3}}\right) - \overline{H}_3\left(e^{\frac{2\pi i}{3}}\right) 
                 - \ln\left(3\right) \overline{H}_2\left(e^{\frac{2\pi i}{3}}\right) 
          \right]
 \right. \nonumber \\
 & & \left.
 + \eps^4 \left[
                4 \overline{H}_{2,1,1}\left(e^{\frac{2\pi i}{3}}\right) - 2 \overline{H}_{3,1}\left(e^{\frac{2\pi i}{3}}\right) + \overline{H}_4\left(e^{\frac{2\pi i}{3}}\right)
                + \frac{2}{9} \pi^2 \overline{H}_2\left(e^{\frac{2\pi i}{3}}\right)
 \right. \right. \nonumber \\
 & & \left. \left.
                + \ln\left(3\right) \left[ 2 \overline{H}_{2,1}\left(e^{\frac{2\pi i}{3}}\right) + \overline{H}_3\left(e^{\frac{2\pi i}{3}}\right) \right]
                + \frac{1}{2} \ln^2\left(3\right) \overline{H}_2\left(e^{\frac{2\pi i}{3}}\right)
          \right]
 \right\}
 \nonumber \\
 & &
 + {\mathcal O}\left(\eps^5\right).
\eq
The master integrals $J_9$ and $J_{10}$ vanish again at the boundary point and so do the master integrals $J_{11}$ and $J_{12}$:
\bq
 J_9
 \;\; = \;\; 
 J_{10}
 & = & 0,
 \nonumber \\
 J_{11}
 \;\; = \;\; 
 J_{12}
 & = & 0.
\eq
From the definition of the master integrals $J_{13}$, $J_{14}$ and $J_{15}$ it follows that they are equal at the boundary point:
\bq
 J_{13}
 \;\; = \;\; 
 J_{14}
 \;\; = \;\; 
 J_{15} 
 & = &
 \eps^2 \sqrt{3} \; {\bf D}^- I_{1(-1)11100}.
\eq
The integral $I_{1(-1)11100}$ reduces at the boundary point to the equal mass sunrise integral at zero external momentum squared
and we find
\bq
 J_{13}
 \;\; = \;\; 
 J_{14}
 \;\; = \;\; 
 J_{15} 
 & = &
 2 J_7,
\eq
with $J_7$ given by eq.~(\ref{chapter_final_project:boundary_J7_J8}).

Note that boundary information on the master integrals $J_{9}-J_{15}$ can already be extracted from the matrix $C_1$
in eq.~(\ref{chapter_final_project:C_0}).
In fact, this matrix does not only give information on the boundary point $(x_1,x_2,x_3)=(0,1,1)$, 
but on the complete hyperplane $x_1=0$.
Assuming that the master integrals do not have any logarithmic singularities at $x_1=0$ and using the fact
that $J_{13}$ and $J_{14}$ are equal for $x_1=0$ (this follows from the definition of $J_{13}$ and $J_{14}$),
it follows that in the hyperplane $x_1=0$ we have (compare with eq.~(\ref{chapter_final_project:limits_from_C_0}))
\bq
 J_9 \; = \; J_{10} \; = \; J_{11} \; = \; J_{12} \; = \; 0,
 \;\;\;\;\;\;
 J_{13} \; = \; J_{14} \; = \; 2 J_{7},
 \;\;\;\;\;\;
 J_{15} \; = \; 2 J_{8}.
\eq

\section{Integrating the differential equation}

With the differential equation and the boundary values we have everything at hand to solve
for the master integrals $\vec{J}$ in terms of iterated integrals.
Let $\gamma : [0,1] \rightarrow {\mathbb C}^3$ be a path from the boundary point
$\gamma(0)=(0,1,1)$ to the point of interest $\gamma(1)=(x_1,x_2,x_3)$.
We integrate the differential equation order by order in $\eps$ as described in section~\ref{chapter_iterated_integrals:section:solution_eps_form}.
Thus we may express any master integral as a linear combination of iterated integrals along the path $\gamma$.
For example
\bq
 J_1 & = &
 1 - I_\gamma\left(\omega_3\right) \eps
 + \left[ I_\gamma\left(\omega_3,\omega_3\right) + \zeta_2 \right] \eps^2
 + {\mathcal O}\left(\eps^3\right).
\eq
It remains to express these iterated integrals in terms of more commonly used functions.

For the master integrals $J_1$-$J_{10}$ this is straightforward: The roots $r_4$ and $r_5$ do not enter these master integrals.
The master integrals $J_1$-$J_{10}$ depend only on the roots $r_1$, $r_2$ and $r_3$.
These roots can be rationalised simultaneously and we may express the master integrals
$J_1$-$J_{10}$ to any order in $\eps$ in terms of multiple polylogarithms.
A trivial example is
\bq
 J_1 & = &
 1 - \ln\left(x_2\right) \eps
 + \left[ \frac{1}{2} \ln^2\left(x_2\right) + \zeta_2 \right] \eps^2
 + {\mathcal O}\left(\eps^3\right).
\eq
The situation is more complicated for the master integrals $J_{11}$-$J_{15}$.
These involve all five roots
\begin{align}
 r_1 & = \sqrt{x_1\left(4+x_1\right)},
 &
 r_2 & = \sqrt{-\lambda\left(x_2,x_3,1\right)},
 &
 r_3 & = \sqrt{-\lambda\left(x_2,x_3,x_2\right)},
 \nonumber \\
 r_4 & = \sqrt{-\lambda\left(x_2,x_3,x_1'\right)},
 &
 r_5 & = \sqrt{-\lambda\left(x_2,x_3,\frac{1}{x_1'}\right)},
 & 
 x_1' & = \frac{1}{2}\left( 2+x_1-r_1\right).
\end{align}
For $x_1=0$ (corresponding to $x_1'=1$) we have
\bq
 r_1 \; = \; 0,
 & &
 r_4 \; = \; r_5 \; = \; r_2.
\eq
This means that on the hyperplane $x_1=0$ we only deal with two square roots $r_2$ and $r_3$, which can be rationalised
simultaneously.
Thus for the special kinematic configuration $x_1=0$ we may express all master integrals 
in terms of multiple polylogarithms.
For a generic kinematic configuration with $x_1 \neq 0$ we are left with an (iterated) integration along the $x_1$-direction.
From section~\ref{chapter_final_project:base_transformation} we know that we may rationalise four of the five square roots.
We may treat the variable $x_1$ for the variable $x_4'$.
In the variables $(x_2',x_3',x_4')$ only the square root $r_5$ remains unrationalised.
In these variables $r_5$ is the square root of a quartic polynomial in $x_4'$.
Thus we are in the situation that we have an integration in a single variable ($x_4'$) involving 
a single square root of a quartic polynomial.

With the methods of section~\ref{chapter_elliptics:section_moduli_spaces}
we may transform these iterated integrals to elliptic multiple polylogarithms.
This does not exclude the possibility that the (first few terms of the $\eps$-expansion of the) master integrals can be expressed in terms 
of simpler functions (i.e. multiple polylogarithms).

We may search for a representation in terms of multiple polylogarithms with the help of the bootstrap approach
described in section~\ref{chapter_hopf:bootstrap}.
\\
\\
\bs
{\it \refstepcounter{exercise}
{\bf Exercise \theexercise}: 
Let
\bq
 g & = & \frac{2+x_1-r_1}{2+x_1+r_1}.
\eq
Express $g$ and $(1-g)$ as a power product in the letters of the alphabet 
defined by eq.~(\ref{chapter_final_project:rational_alphabet}), eq.~(\ref{chapter_final_project:algebraic_alphabet}) 
and the constant $f_0=2$.
}
\es
\\
\\
\bs
{\it \refstepcounter{exercise}
{\bf Exercise \theexercise}: 
The master integral $J_{15}$ starts at order ${\mathcal O}(\eps^2)$.
The weight two term of $J_{15}$ is given in terms of iterated integrals 
by
\bq
 J_{15}^{(2)}
 & = &
 2 i I_\gamma\left(2\omega_{15}-\omega_{3}-\omega_{5},\omega_{5}-\omega_{3};1\right)
 + 2 J_7^{(2)}(0,1,1),
\eq
where $J_7^{(2)}(0,1,1)$ denotes the boundary value of eq.~(\ref{chapter_final_project:boundary_J7_J8}):
\bq
 J_7^{(2)}(0,1,1) & = & \frac{3}{2i} \overline{H}_2\left(e^{\frac{2\pi i}{3}}\right)
 \; = \; 
 \frac{3}{2i} \left[ \mathrm{Li}_2\left(e^{\frac{2\pi i}{3}}\right) - \mathrm{Li}_2\left(e^{-\frac{2\pi i}{3}}\right) \right].
\eq
Express $J_{15}^{(2)}$ in terms of multiple polylogarithms.
}
\es

\section{Final result}

Let's now return to our original problem:
The scalar Feynman integral corresponding to
the penguin diagram in fig.~\ref{chapter_final_project:fig_penguin_1}
is $I_{1211100}$.
As a pedagogical example we
work out the first non-vanishing term in the $\eps$-expansion of this Feynman integral.
This illustrates the general procedure.
(Of course, if we are only interested in this particular coefficient, there are simpler ways to obtain the result.)

Using integration-by-parts identities we may express the integral $I_{1211100}$
as a linear combination of the master integrals $\vec{I}$.
By using
\bq
 \vec{I} & = & U^{-1} \vec{J}
\eq
we express $I_{1211100}$ as a linear combination of the master integrals $\vec{J}$.
The integral $I_{1211100}(4-2\eps)$ has a Laurent expansion in the dimensional regularisation
parameter starting with $\eps^{-1}$:
\bq
 I_{1211100}\left(4-2\eps,x_1,x_2,x_3\right)
 & = &
 \sum\limits_{j=-1}^\infty
 \eps^j \; I_{1211100}^{(j)}\left(x_1,x_2,x_3\right).
\eq
Let's focus on the first non-vanishing coefficient $I_{1211100}^{(-1)}$.
For the master integrals we have a similar expansion in $\eps$
\bq
 J_i\left(4-2\eps,x_1,x_2,x_3\right) & = &
 \sum\limits_{j=0}^\infty \eps^j \; J_i^{(j)}\left(x_1,x_2,x_3\right).
\eq
In terms of the master integrals $\vec{J}$ we obtain for $I_{1211100}^{(-1)}$
\bq
\label{chapter_final_project:I_1211100_reduction_to_J}
 I_{1211100}^{(-1)}
 & = & 
 \frac{1}{\left[\left(1-x_2\right)^2-x_1x_2\right]}
  \left\{
   \frac{\left(1+x_2\right)\left(2x_2-x_3\right)}{2x_2\left(1-x_2\right)} \left( J_2^{(1)} - J_1^{(1)} \right)
   + \frac{\left(1+x_2\right)x_3}{2x_2\left(1-x_2\right)} \left( J_4^{(1)} - J_3^{(1)} \right)
 \right. \nonumber \\
 & & \left.
   + \frac{\left(2x_2-x_3\right)r_1}{4x_1x_2} J_5^{(1)}
   + \frac{x_3r_1}{4x_1x_2} J_6^{(1)}
  \right\}
   + \frac{\left(x_2-x_3\right)}{2x_1x_2^2} \left( J_9^{(2)} - J_{10}^{(2)} \right)
 \nonumber \\
 & & 
   + \frac{\left(3x_2-x_3\right)r_3}{2x_1x_2^2\left(4x_2-x_3\right)} \left(J_{15}^{(2)}-2J_8^{(2)} \right)
\eq
We then substitute the results for $\vec{J}$ and obtain
\bq
 I_{1211100}^{(-1)}
 & = & 
 \frac{1}{\left[\left(1-x_2\right)^2-x_1x_2\right]}
 \left\{
 \sqrt{\frac{4+x_1}{x_1}} I_\gamma\left(\omega_{10}\right)
 - \frac{1+x_2}{1-x_2} I_\gamma\left(\omega_3\right)
 \right\}.
\eq
With
\bq
 \omega_3 \; = \; d\ln\left(x_2\right),
 & &
 \omega_{10} \; = \; 
 d\ln\left(2+x_1-r_1\right)
 \; = \;
 \frac{1}{2} d\ln\left(\frac{2+x_1-r_1}{2+x_1+r_1}\right)
\eq
we finally obtain
\bq
\label{chapter_final_project:final_result}
\lefteqn{
 I_{1211100}\left(4-2\eps,x_1,x_2,x_3\right)
 = } & & 
 \\
 & &
 \frac{1}{\eps}
 \frac{1}{\left[\left(1-x_2\right)^2-x_1x_2\right]}
 \left\{
 \frac{1}{2} \sqrt{\frac{4+x_1}{x_1}} \ln\left(\frac{2+x_1-r_1}{2+x_1+r_1}\right)
 - \frac{1+x_2}{1-x_2} \ln\left(x_2\right)
 \right\}
 + {\mathcal O}\left(\eps^0\right).
 \nonumber
\eq
Note that $I_{1211100}^{(-1)}$ is finite in the $(x_1 \rightarrow 0)$-limit and in the $(x_2 \rightarrow 1)$-limit.
Note further that eq.~(\ref{chapter_final_project:final_result}) does not contain any weight two terms.
Although we might expect from eq.~(\ref{chapter_final_project:I_1211100_reduction_to_J})
terms of weight two, they cancel out in this order.
Finally note that $I_{1211100}^{(-1)}$ is independent of $x_3$.
The $1/\eps$-term $I_{1211100}^{(-1)}$ originates from the ultraviolet divergence of the sub-graph formed
by propagators $4$ and $5$. The ultraviolet divergence is independent of the masses propagating in the
sub-graph.
As $x_3$ (or $m_H^2$) enters only this sub-graph, the result for $I_{1211100}^{(-1)}$ is independent of 
$x_3$.
Essentially, $I_{1211100}^{(-1)}$ is given by the pole term of the sub-graph times
a one-loop three-point function obtained by contracting the sub-graph to a point.

Let's plug in some numbers: With
\bq
 \left(x_1,x_2,x_3\right)
 & = &
 \left(\frac{5^2}{173^2},\frac{80^2}{173^2},\frac{125^2}{173^2}\right)
\eq
we obtain
\bq
\label{chapter_final_project:numerical_result}
 I_{1211100}^{(-1)}\left(\frac{25}{29929},\frac{6400}{29929},\frac{15625}{29929}\right)
 & \approx &
 0.61751141179432938382213384.
\eq
It is always recommended to perform an independent cross check.
Sector decomposition offers the possibility to check a Feynman integral at a specific kinematic point.
The following C++ code uses the program \verb|sector_decomposition| \cite{Bogner:2007cr}:
{\footnotesize
\begin{verbatim}
#include <iostream>
#include <stdexcept>
#include <vector>
#include <ginac/ginac.h>
#include "sector_decomposition/sector_decomposition.h"

int main()
{
  using namespace sector_decomposition;
  using namespace GiNaC;

  symbol eps("eps");

  int n            =  5;
  int loops        =  2;
  int order        = -1;
  int D_int_over_2 =  2;

  std::vector<ex> nu = {1,2,1,1,1};

  ex x1  = numeric(25,29929);
  ex x2  = numeric(6400,29929);
  ex x3  = numeric(15625,29929);
	
  // --------------------------------------------------------------

  int verbose_level = 0;

  CHOICE_STRATEGY = STRATEGY_C;

  monte_carlo_parameters mc_parameters = monte_carlo_parameters( 5, 15, 100000, 1000000 );

  // --------------------------------------------------------------

  symbol a1("a1"), a2("a2"), a3("a3"), a4("a4"), a5("a5");
  std::vector<ex> parameters = { a1, a2, a3, a4, a5 };

  ex U = a1*a4+a1*a5+a4*a3+a5*a3+a2*a4+a5*a4+a5*a2;
  ex F = a1*a3*(a4+a5)*x1 + U*( (a2+a4)*x2 + a5*x3 + (a1+a3) );

  std::vector<ex> poly_list = {U,F};

  std::vector<exponent> nu_minus_1(n);
  for (int i1=0; i1<n; i1++) nu_minus_1[i1] = exponent(nu[i1]-1,0);

  std::vector<exponent> c(poly_list.size());
  c[0] = exponent( n-(loops+1)*D_int_over_2, loops+1 );
  c[1] = exponent( -n+loops*D_int_over_2, -loops );
  for (int k=0; k<n; k++) 
  {
    c[0].sum_up(nu_minus_1[k]);
    c[1].subtract_off(nu_minus_1[k]);
  }

  integrand my_integrand = integrand(nu_minus_1, poly_list, c);

  // --------------------------------------------------------------

  integration_data global_data(parameters, eps, order);

  monte_carlo_result res = 
	  do_sector_decomposition(global_data, my_integrand, mc_parameters, verbose_level);

  std::cout << "Order " << pow(eps,order) << ": " << res.get_mean() 
	    << " +/- " << res.get_error() << std::endl;
    
  return 0;
}
\end{verbatim}
}
\noindent
Running this program will print out
\begin{verbatim}
Order eps^(-1): 0.617517 +/- 9.57571e-06
\end{verbatim}
in agreement with eq.~(\ref{chapter_final_project:numerical_result}).

%% file: spinor.tex
\newpage
\chapter{Spinors}
\label{appendix_spinors}

In this appendix we summarise properties of spinors in four space-time
dimensions.
Although we use dimensional regularisation throughout this book, four-dimensional
formulae can be used for the external kinematic and within some variants of dimensional
regularisation like the FDH-scheme.
Note that some formulae (like the Schouten identity) are specific to four space-time dimensions.

\section{The Dirac equation}

The Lagrange density for a Dirac field in four space-time dimensions
depends on four-component spinors $\psi_\alpha(x)$ $(\alpha=1,2,3,4)$ and
$\bar{\psi}_\alpha(x) = \left( \psi^\dagger(x) \gamma^0 \right)_\alpha$:
\bq
{\mathcal L}(\psi, \bar{\psi}, \partial_\mu \psi ) & = & i \bar{\psi}(x) \gamma^\mu \partial_\mu \psi(x) - m \bar{\psi}(x) \psi(x)
\eq
Here, the $(4 \times 4)$-Dirac matrices satisfy the anti-commutation rules
\bq
 \{ \gamma^\mu, \gamma^\nu \} = 2 g^{\mu\nu} {\bf 1},
 \;\;\;
 \{ \gamma^\mu, \gamma_5 \} = 0,
 \;\;\;
 \gamma_5 = i \gamma^0 \gamma^1 \gamma^2 \gamma^3
          = \frac{i}{24} \eps_{\mu\nu\rho\sigma} \gamma^\mu \gamma^\nu \gamma^\rho \gamma^\sigma.
\eq
The Dirac equations read
\bq
 \left( i \gamma^\mu \partial_\mu - m \right) \psi(x) \;\; = \;\; 0,
 & &
 \bar{\psi}(x) \left( i \gamma^\mu \stackrel{\leftarrow}{\partial}_\mu + m \right ) \;\; = \;\; 0.
\eq
It is useful to have an explicit representation of the Dirac matrices.
There are several widely used representations. A particular useful one is the {\bf Weyl representation} of the Dirac matrices:
\bq
 \gamma^{\mu} = \left(\begin{array}{cc}
 {\bf 0} & \sigma^{\mu} \\
 \bar{\sigma}^{\mu} & {\bf 0} \\
\end{array} \right),
& &
\gamma_{5} = i \gamma^0 \gamma^1 \gamma^2 \gamma^3 
           = \left(\begin{array}{cc}
 {\bf 1} & {\bf 0} \\
 {\bf 0} & -{\bf 1} \\
\end{array} \right)
\eq
Here, ${\bf 0}$ denotes the $(2\times 2)$-zero matrix and ${\bf 1}$ denotes the $(2\times 2)$-unit matrix.
The 4-dimensional $\sigma^{\mu}$-matrices are defined by
\bq
\sigma_{A \dot{B}}^{\mu} = \left( {\bf 1}, - \vec{\sigma} \right), & &
\bar{\sigma}^{\mu \dot{A} B} = \left( {\bf 1},  \vec{\sigma} \right) .
\eq
and $\vec{\sigma}=(\sigma_x,\sigma_y,\sigma_z)$ are the standard
Pauli matrices:
\bq
\sigma_x = \left(\begin{array}{cc}
 0 & 1\\
 1 & 0 \\
\end{array} \right),
&
\sigma_y = \left(\begin{array}{cc}
 0 & -i\\
 i & 0 \\
\end{array} \right),
&
\sigma_z = \left(\begin{array}{cc}
 1 & 0\\
 0 & -1 \\
\end{array} \right).
\eq
The $\sigma^{\mu}$-matrices satisfy the Fierz identities
\bq
 \sigma^{\mu}_{A\dot{A}} \bar{\sigma}_{\mu}^{\dot{B}B} 
 = 
 2 \delta_{A}^{\; B} \delta_{\dot{A}}^{\; \dot{B}},
 \;\;\;\;\;\;\;\;\;
 \sigma^{\mu}_{A\dot{A}} \sigma_{\mu B \dot{B}} 
 = 
 2 \varepsilon_{AB} \varepsilon_{\dot{A}\dot{B}},
 \;\;\;\;\;\;\;\;\;
 \bar{\sigma}^{\mu\dot{A}A} \bar{\sigma}_{\mu}^{\dot{B}B} 
 = 
 2 \varepsilon^{\dot{A}\dot{B}} \varepsilon^{AB}.
\eq
Let us now look for plane wave solutions of the Dirac equation.
We make the ansatz
\bq
 \psi(x) & = & 
 \left\{ \begin{array}{llll}
  u(p) e^{-i p x}, & p^0>0, & p^2=m^2, & \mbox{incoming fermion}, \\
  v(p) e^{+i p x}, & p^0>0, & p^2=m^2, & \mbox{outgoing anti-fermion}. \\
  \end{array}
 \right.
\eq
$u(p)$ describes incoming particles, $v(p)$ describes outgoing anti-particles.
Similar,
\bq
 \bar{\psi}(x) & = & 
 \left\{ \begin{array}{llll}
  \bar{u}(p) e^{+i p x}, & p^0>0, & p^2=m^2, & \mbox{outgoing fermion}, \\
  \bar{v}(p) e^{-i p x}, & p^0>0, & p^2=m^2, & \mbox{incoming anti-fermion}, \\
  \end{array}
 \right.
\eq
where
\bq
 \bar{u}(p) = u^\dagger(p) \gamma^0,
 & &
 \bar{v}(p) = v^\dagger(p) \gamma^0.
\eq
$\bar{u}(p)$ describes outgoing particles, $\bar{v}(p)$ describes incoming anti-particles.
Then
\bq 
\label{appendix_spinors_Dirac_equations}
\left( p\!\!\!/ - m \right) u(p) = 0, & & \left( p\!\!\!/ + m \right) v(p) = 0, \nonumber \\
\bar{u}(p) \left( p\!\!\!/ - m \right) = 0, & & \bar{v}(p) \left( p\!\!\!/ + m \right) = 0,  
\eq
There are two solutions for $u(p)$ (and the other spinors $\bar{u}(p)$, $v(p)$, $\bar{v}(p)$).
We will label the various solutions with $\lambda \in \{+,-\}$ and we will use the notation $\bar{\lambda}=-\lambda$.
The degeneracy is related to the additional spin degree of freedom.
We require that the two solutions satisfy 
the orthogonality relations
\bq
\label{appendix_spinors_spinor_orthogonality}
\bar{u}(p,\bar{\lambda}) u(p,\lambda) & = & 2 m \delta_{\bar{\lambda}\lambda}, \nonumber \\
\bar{v}(p,\bar{\lambda}) v(p,\lambda) & = & -2 m \delta_{\bar{\lambda}\lambda}, \nonumber \\
\bar{u}(p,\bar{\lambda}) v(p,\lambda) & = & \bar{v}(\bar{\lambda}) u(\lambda) \;\; = \;\; 0,
\eq
and the completeness relations
\bq
\label{appendix_spinors_spinor_completeness_relations}
\sum\limits_{\lambda} u(p,\lambda) \bar{u}(p,\lambda) \;\; = \;\; p\!\!\!/ + m, 
 & &
\sum\limits_{\lambda} v(p,\lambda) \bar{v}(p,\lambda) \;\; = \;\; p\!\!\!/ - m.
\eq

\section{Massless spinors in the Weyl representation}
\label{appendix_spinors_sect_massless_spinors}

Let us now try to find explicit solutions for the spinors $u(p)$, $v(p)$, $\bar{u}(p)$ and $\bar{v}(p)$.
The simplest case is the one of a massless fermion:
\bq
 m & = & 0.
\eq
In this case the Dirac equation for the $u$- and the $v$-spinors are identical and it is sufficient to consider
\bq
 p\!\!\!/ u(p) = 0, & & 
\bar{u}(p) p\!\!\!/ = 0.
\eq
In the Weyl representation $p\!\!\!/$ is given by
\bq 
 p\!\!\!/ & = &
 \left(\begin{array}{cc}
 {\bf 0} & p_\mu \sigma^{\mu} \\
 p_\mu \bar{\sigma}^{\mu} & {\bf 0} \\
\end{array} \right),
\eq
therefore the $4 \times 4$-matrix equation for $u(p)$ (or $\bar{u}(p)$) decouples into two
$2 \times 2$-matrix equations.
We introduce the following notation:
Four-component Dirac spinors are constructed out of two {\bf Weyl spinors} as follows:
\bq
\label{appendix_spinors_Weyl_ket}
u(p) & = & \left(\begin{array}{c} \left| p + \right\rangle \\ \left| p - \right\rangle \\ \end{array} \right) 
       =   \left(\begin{array}{c} \left| p \right\rangle \\ \left| p \right] \\ \end{array} \right)
       =   \left(\begin{array}{c} p_A \\ p^{\dot{B}} \\ \end{array} \right)
       =   \left(\begin{array}{c} u_+(p) \\ u_-(p) \\ \end{array} \right).
\eq
Bra-spinors are given by
\bq
\label{appendix_spinors_Weyl_bra}
\overline{u}(p) & = & \left( \; \left\langle p - \right|, \; \left\langle p + \right| \; \right)
                  =   \left( \; \left\langle p \right|, \; \left[ p \right| \; \right)
                  =   \left( \; p^A, \; p_{\dot{B}} \; \right)
                  =   \left( \; \bar{u}_-(p), \; \bar{u}_+(p) \; \right).
\eq
In the literature there exists various notations for Weyl spinors.
Eq.~(\ref{appendix_spinors_Weyl_ket}) and eq.~(\ref{appendix_spinors_Weyl_bra}) show four of them and the way how to translate from one
notation to another notation.
By a slight abuse of notation we will in the following not distinguish between 
a two-component Weyl spinor and a Dirac spinor, where either the upper two components or the lower two components
are zero.
If we define the chiral projection operators
\bq
P_+ \;\; = \;\; \frac{1}{2} \left( 1 + \gamma_5 \right) 
 = \left( \begin{array}{cc} {\bf 1} & {\bf 0} \\ {\bf 0} & {\bf 0} \\ \end{array} \right),
 & &
P_- \;\; = \;\; \frac{1}{2} \left( 1 - \gamma_5 \right) 
 = \left( \begin{array}{cc} {\bf 0} & {\bf 0} \\ {\bf 0} & {\bf 1} \\ \end{array} \right),
\eq
then (with the slight abuse of notation mentioned above)
\bq
u_\pm(p) = P_\pm u(p),
 & &
\bar{u}_\pm(p) = \bar{u}(p) P_\mp.
\eq
The two solutions of the Dirac equation
\bq
 p\!\!\!/ u(p,\lambda) & = & 0 
\eq
are then
\bq
 u(p,+) = u_+(p),
 & & 
 u(p,-) = u_-(p).
\eq
We now have to solve
\bq
p_\mu \bar{\sigma}^\mu \left| p+ \right\rangle = 0,
 & &
p_\mu \sigma^\mu \left| p- \right\rangle = 0,
 \nonumber \\
\left\langle p+ \right| p_\mu \bar{\sigma}^\mu  = 0,
 & &
\left\langle p- \right| p_\mu \sigma^\mu  = 0. 
\eq
It it convenient to express the four-vector $p^\mu=(p^0,p^1,p^2,p^3)$ in terms of light-cone
coordinates:
\bq
\label{appendix_spinors_def_contravariant_lightcone_coordinates}
 p^+=
 \frac{1}{\sqrt{2}}
 \left(p^0+p^3\right), 
 \;\;\;
 p^-=
 \frac{1}{\sqrt{2}}
 \left(p^0-p^3\right), 
 \;\;\; 
 p^\bot=
 \frac{1}{\sqrt{2}}
 \left(p^1+ip^2\right), 
 \;\;\;
 p^{\bot\ast}=
 \frac{1}{\sqrt{2}}
 \left(p^1-ip^2\right).
 \nonumber
\eq
Note that $p^{\bot\ast}$ does not involve a complex conjugation of $p^1$ or $p^2$.
For null-vectors one has
\bq
p^{\bot\ast} p^{\bot} & = & p^+ p^-.
\eq
Then the equation for the ket-spinors becomes
\bq
\left( \begin{array}{cc}
        p^- & -p^{\bot\ast} \\
        -p^\bot & p^+ \\ 
       \end{array}
\right)
 \left| p+ \right\rangle = 0,
 & &
\left( \begin{array}{cc}
        p^+ & p^{\bot\ast} \\
        p^\bot & p^- 
       \end{array}
\right)
 \left| p- \right\rangle = 0,
\eq
and similar equations can be written down for the bra-spinors.
This is a problem of linear algebra.
Solutions for ket-spinors are
\bq
 \left| p+ \right\rangle = p_A =
  c_1
  \left( \begin{array}{c} p^{\bot\ast} \\ p^- \end{array} \right),
 & &
 \left| p- \right\rangle = p^{\dot{A}} =
 c_2
 \left( \begin{array}{c} p^- \\ -p^\bot \end{array} \right),
\eq
with some yet unspecified multiplicative constants $c_1$ and $c_2$.
Solutions for bra-spinors are 
\bq
 \left\langle p+ \right| = p_{\dot{A}} =
 c_3
 \left( p^\bot, p^- \right),
 & &
 \left\langle p- \right| = p^A =
 c_4
 \left( p^-, -p^{\bot\ast} \right),
\eq
with some further constants $c_3$ and $c_4$.
Let us now introduce the 2-dimensional antisymmetric tensor:
\bq
\varepsilon_{AB} = \left(\begin{array}{cc}
 0 & 1\\
 -1 & 0 \\
\end{array} \right),
& &
\varepsilon_{BA} = - \varepsilon_{AB} 
\eq
Furthermore we set
\bq
\varepsilon^{AB} = \varepsilon^{\dot{A}\dot{B}} = \varepsilon_{AB} = \varepsilon_{\dot{A}\dot{B}}.
\eq
Note that these definitions imply
\bq
 \eps^{A C} \eps_{B C} = \delta^{A}_{\;B},
 & &
 \eps^{\dot{A} \dot{C}} \eps_{\dot{B} \dot{C}} = \delta^{\dot{A}}_{\;\dot{B}}.
\eq
We would like to have the following relations for raising and lowering a spinor index $A$ or $\dot{B}$:
\bq
\label{appendix_spinors_raising_and_lowering_spinor_indices}
 p^A = \varepsilon^{AB} p_B,
  & & 
 p^{\dot{A}} = \varepsilon^{\dot{A}\dot{B}} p_{\dot{B}},
  \nonumber\\
 p_{\dot{B}} = p^{\dot{A}} \varepsilon_{\dot{A}\dot{B}}, 
  & & 
 p_B = p^A \varepsilon_{AB}.
\eq
Note that raising an index is done by left-multiplication, whereas 
lowering is performed by right-multiplication.
Postulating these relations implies
\bq
 c_1 = c_4, & & c_2=c_3.
\eq
In addition we normalise the spinors according to
\bq
\label{appendix_spinors_normalisation_massless_spinors}
\langle p \pm | \gamma^{\mu} | p \pm \rangle = 2 p^{\mu}.
\eq
This implies
\bq
 c_1 c_3 = \frac{\sqrt{2}}{p^-},
 & &
 c_2 c_4 = \frac{\sqrt{2}}{p^-}.
\eq
Eq.~(\ref{appendix_spinors_raising_and_lowering_spinor_indices}) and eq.~(\ref{appendix_spinors_normalisation_massless_spinors})
determine the spinors only up to a scaling
\bq
 p_A \rightarrow \lambda p_A,
 & &
 p_{\dot{A}} \rightarrow \frac{1}{\lambda} p_{\dot{A}}.
\eq
This scaling freedom is referred to as {\bf little group scaling}.
Keeping the scaling freedom, we define the spinors as
\bq
\label{appendix_spinors_def_spinors}
\left| p+ \right\rangle = p_A = 
  \frac{\lambda_p 2^{\frac{1}{4}}}{\sqrt{p^-}} 
  \left( \begin{array}{c} p^{\bot\ast} \\ p^- \end{array} \right),
 & &
\left| p- \right\rangle = p^{\dot{A}} = 
 \frac{2^{\frac{1}{4}}}{\lambda_p\sqrt{p^-}} 
 \left( \begin{array}{c} p^- \\ -p^\bot \end{array} \right),
 \nonumber \\
\left\langle p+ \right| = p_{\dot{A}} = 
 \frac{2^{\frac{1}{4}}}{\lambda_p\sqrt{p^-}} 
 \left( p^\bot, p^- \right),
 & &
\left\langle p- \right| = p^A =
 \frac{\lambda_p 2^{\frac{1}{4}}}{\sqrt{p^-}} 
 \left( p^-, -p^{\bot\ast} \right).
\eq
Popular choices for $\lambda_p$ are
\bq
 \lambda_p = 1 & : & \mbox{symmetric},
 \nonumber \\
 \lambda_p = 2^{\frac{1}{4}} \sqrt{p^-} & : & \mbox{$p_A$ linear in $p^\mu$},
 \nonumber \\
 \lambda_p = \frac{1}{2^{\frac{1}{4}} \sqrt{p^-}} & : & \mbox{$p_{\dot{A}}$ linear in $p^\mu$}.
\eq
Note that all formulae in this sub-section~(\ref{appendix_spinors_sect_massless_spinors}) 
work not only for real momenta $p^\mu$
but also for complex momenta $p^\mu$.
This will be useful later on, where we encounter situations with complex momenta.
However there is one exception:
The relations
$p_A^\dagger = p_{\dot{A}}$ and $p^A{}^\dagger = p^{\dot{A}}$ (or equivalently $\bar{u}(p) = u(p)^\dagger \gamma^0$)
encountered in previous sub-sections are valid only for real momenta $p^\mu=(p^0,p^1,p^2,p^3)$.
If on the other hand the components $(p^0,p^1,p^2,p^3)$ are complex, these relations will in general not hold.
In the latter case $p_A$ and $p_{\dot{A}}$ are considered to be independent quantities.
The reason, why the relations $p_A^\dagger = p_{\dot{A}}$ and $p^A{}^\dagger = p^{\dot{A}}$
do not hold in the complex case lies in the definition of $p^{\bot\ast}$:
We defined $p^{\bot\ast}$ as $p^{\bot\ast}=(p^1-ip^2)/\sqrt{2}$, and not as $(p^1-i(p^2)^\ast)/\sqrt{2}$.
With the former definition $p^{\bot\ast}$ is a holomorphic function of 
$p^1$ and $p^2$.
There are applications where holomorphicity is more important than nice properties under
hermitian conjugation.

\section{Spinor products}

Let us now make the symmetric choice $\lambda_p=1$.
Spinor products are defined by
\bq
\label{appendix_spinors_spinor_products}
 \langle p q \rangle 
 & = & 
 \langle p - | q + \rangle 
 = 
 p^A q_A
 =
 \frac{\sqrt{2}}{\sqrt{p^-} \sqrt{q^-}} \left( p^- q^{\bot\ast} - q^- p^{\bot\ast} \right),
 \nonumber \\
 \left[ q p \right] 
 & = & 
 \langle q + | p - \rangle 
 = 
 q_{\dot{A}} p^{\dot{A}}
 =
 \frac{\sqrt{2}}{\sqrt{p^-} \sqrt{q^-}} \left( p^- q^\bot - q^- p^\bot \right),
\eq 
where the last expression in each line used the choice $\lambda_p=\lambda_q=1$.
We have
\bq
 \langle p q \rangle \left[ q p \right]
 & = & 
 2 p q.
\eq
If $p^\mu$ and $q^\mu$ are real we have 
\bq
\left[ q p \right] 
 & = & \left\langle p q \right\rangle^\ast \; \mathrm{sign}(p^0) \; \mathrm{sign}(q^0).
\eq
The spinor products are anti-symmetric
\bq
 \langle q p \rangle = - \langle p q \rangle,
 & &
 [ p q ] = - [ q p ].
\eq
From the Schouten identity for the 2-dimensional antisymmetric tensor 
\bq
 \eps_{A B} \eps_{C D}
 +
 \eps_{B C} \eps_{A D}
 +
 \eps_{C A} \eps_{B D}
 & = &
 0.
\eq
one derives
\bq
 \langle p_1 p_2 \rangle \langle p_3 p_4 \rangle
 +
 \langle p_2 p_3 \rangle \langle p_1 p_4 \rangle
 +
 \langle p_3 p_1 \rangle \langle p_2 p_4 \rangle
 & = & 0,
 \nonumber \\
 \left[ p_1 p_2 \right] \left[ p_3 p_4 \right]
 +
 \left[ p_2 p_3 \right] \left[ p_1 p_4 \right]
 +
 \left[ p_3 p_1 \right] \left[ p_2 p_4 \right]
 & = & 0.
\eq
The Fierz identity reads
\bq
 \langle p_1+ | \gamma_{\mu} | p_2+ \rangle \langle p_3- | \gamma^{\mu} | p_4- \rangle 
 & = &
 2 [ p_1 p_4 ] \langle p_3 p_2 \rangle.
\eq
Note that with our slight abuse of notation we identify a two-component Weyl spinor
with a Dirac spinor, where the other two components are zero.
Therefore
\bq
 \langle p_1+ | \gamma_{\mu} | p_2+ \rangle
 \;\; = \;\;
 \langle p_1+ | \bar{\sigma}_{\mu} | p_2+ \rangle,
 & &
 \langle p_3- | \gamma^{\mu} | p_4- \rangle
 \;\, = \;\;
 \langle p_3- | \sigma^{\mu} | p_4- \rangle.
\eq
We further have the reflection identities
\bq
 \langle p\pm|\gamma^{\mu_1} ... \gamma^{\mu_{2n+1}} | q\pm \rangle 
 & = &
 \langle q\mp|\gamma^{\mu_{2n+1}} ... \gamma^{\mu_1} | p\mp \rangle,
 \nonumber \\
 \langle p\pm|\gamma^{\mu_1} ... \gamma^{\mu_{2n}} | q\mp \rangle 
 & = &
 - \langle q\pm|\gamma^{\mu_{2n}} ... \gamma^{\mu_1} | p\mp \rangle.
\eq

\section{Massive spinors}
\label{appendix_spinors_sect_massive_spinors}

As in the massless case, a massive spinor satisfying the Dirac equation has
a two-fold degeneracy. We will label the two different eigenvectors by ``+'' and
``-''.
Let $p$ be a massive four-vector with $p^2=m^2$, and let $q$ be an arbitrary
light-like four-vector.
With the help of $q$ we can construct a light-like vector $p^\flat$ associated to $p$:
\bq
 p^\flat & = & p - \frac{p^2}{2 p \cdot q} q.
\eq
We define \cite{Kleiss:1986qc,Schwinn:2005pi,Rodrigo:2005eu}
\bq
\label{appendix_spinors_def_u_spinor}
u(p,+) = \frac{1}{\langle p^\flat + | q - \rangle} \left( p\!\!\!/ + m \right) | q - \rangle,
& & 
v(p,-) = \frac{1}{\langle p^\flat + | q - \rangle} \left( p\!\!\!/ - m \right) | q - \rangle, \nonumber \\
u(p,-) = \frac{1}{\langle p^\flat - | q + \rangle} \left( p\!\!\!/ + m \right) | q + \rangle,
& & 
v(p,+) = \frac{1}{\langle p^\flat - | q + \rangle} \left( p\!\!\!/ - m \right) | q + \rangle.
\eq
For the conjugate spinors we have
\bq
\label{appendix_spinors_def_ubar_spinor}
\bar{u}(p,+) = \frac{1}{\langle q - | p^\flat + \rangle} \langle q - | \left( p\!\!\!/ + m \right), 
& & 
\bar{v}(p,-) = \frac{1}{\langle q - | p^\flat + \rangle}\langle q - | \left( p\!\!\!/ - m \right), \nonumber \\
\bar{u}(p,-) = \frac{1}{\langle q + | p^\flat - \rangle} \langle q + | \left( p\!\!\!/ + m \right),
& & 
\bar{v}(p,+) = \frac{1}{\langle q + | p^\flat - \rangle} \langle q + | \left( p\!\!\!/ - m \right). 
\eq
These spinors satisfy the Dirac equations of eq.~(\ref{appendix_spinors_Dirac_equations}),
the orthogonality relations of eq.~(\ref{appendix_spinors_spinor_orthogonality})
and the completeness relations of eq.~(\ref{appendix_spinors_spinor_completeness_relations}).
We further have
\bq
 \bar{u}(p,\bar{\lambda}) \gamma^\mu u(p,\lambda) = 2 p^\mu \delta_{\bar{\lambda} \lambda},
 & &
 \bar{v}(p,\bar{\lambda}) \gamma^\mu v(p,\lambda) = 2 p^\mu \delta_{\bar{\lambda} \lambda}.
\eq
In the massless limit the definition reduces to
\bq
 u(p,+) = v(p,-) = | p+ \rangle,
 &&
 \bar{u}(p,+) = \bar{v}(p,-) = \langle p+ |, 
 \nonumber \\
 u(p,-) = v(p,+) = | p- \rangle,
 & &
 \bar{u}(p,-) = \bar{v}(p,+) = \langle p- |,
\eq
and the spinors are in the massless limit independent of the reference spinors $|q+\rangle$ and $\langle q+|$.

%% file: oneloopintegrals.tex
\newpage
\chapter{Scalar one-loop integrals}
\label{appendix_one_loop_integrals}

In this appendix we list the basic scalar one-loop integrals for massless theories in 
$D = 4 - 2\eps$ dimensions as an expansion in the dimensional regularisation parameter $\eps$
up to and including the ${\mathcal O}(\eps^0)$ term.
The basic scalar integrals consist of the scalar two-point, the scalar three-point and the
scalar four-point functions.
The scalar one-point function vanishes in dimensional regularisation.
Since we restrict ourselves to massless quantum field theories, 
all internal propagators are massless and we only
have to distinguish the momentum configurations of the external momenta.
We call an external momentum $p$ ``massive'' if $p^2 \neq 0$. 
All scalar integrals have been known for a long time in the literature.
Classical papers on scalar integrals are \cite{'tHooft:1979xw,Denner:1991qq}.
Scalar integrals within dimensional regularization are treated in \cite{Bern:1993em,Bern:1994kr}.
Useful information on the three-mass triangle can be found in \cite{Lu:1992ny,Ussyukina:1993jd,Bern:1997ka}.
The scalar boxes have been recalculated in \cite{Duplancic:2000sk,Duplancic:2002dh}.
The compilation given here is based on \cite{vanHameren:2005ed}.

The basic scalar integrals with internal masses constitute a longer list.
These integrals can be found in the literature \cite{Ellis:2007qk,Denner:2010tr}.

\section{Massless scalar one-loop integrals}

We recall the notation for massless one-loop integrals from eq.~(\ref{chapter_one_loop:basic_one_loop_scalar_int}):
\bq
I_{\nexternal} & = &
 e^{\eps \Eulerconstant} \mu^{2\eps} 
  \int \frac{d^Dk}{i \pi^{\frac{D}{2}}}
  \frac{1}{(-k^2) (-(k-p_1)^2) ... (-(k-p_1-...p_{\nexternal-1})^2)}.
\eq

\subsection*{The two-point function}

The scalar two-point function is given by
\bq
I_2(p_1^2,\mu^2) & = & 
 \frac{1}{\eps} + 2 - \ln\left(\frac{-p_1^2}{\mu^2}\right) 
 + {\cal O}(\eps).
\eq

\subsection*{Three-point functions}

For the three-point functions we have three different cases: One external mass, two external masses
and three external masses.
The one-mass scalar triangle with $p_1^2\neq 0$, $p_2^2=p_3^2=0$ is given by
\bq
I^{1m}_3(p_1^2,\mu^2) & = &
 \frac{1}{\eps^2 p_1^2} 
 - \frac{1}{\eps p_1^2} \ln\left(\frac{-p_1^2}{\mu^2}\right)
 + \frac{1}{2 p_1^2} \ln^2\left(\frac{-p_1^2}{\mu^2}\right)
 - \frac{1}{2 p_1^2} \zeta_2
 + {\cal O}(\eps).
\eq
The two-mass scalar triangle with $p_1^2\neq 0$, $p_2^2\neq 0$ and $p_3^2=0$ is given by
\bq
I^{2m}_3(p_1^2,p_2^2,\mu^2) & = & 
 \frac{1}{\eps} 
 \frac{1}{\left( p_1^2-p_2^2 \right)}
 \left[ - \ln\left(\frac{-p_1^2}{\mu^2}\right) + \ln\left(\frac{-p_2^2}{\mu^2}\right) \right]
\nonumber \\
 & &
 + \frac{1}{2(p_1^2-p_2^2)} 
   \left[ \ln^2\left(\frac{-p_1^2}{\mu^2}\right) - \ln^2\left(\frac{-p_2^2}{\mu^2}\right) \right]
 + {\cal O}(\eps).
\eq
The three-mass scalar triangle with $p_1^2\neq 0$, $p_2^2\neq 0$ and $p_3^2\neq 0$:
This integral is finite and we have
\bq
 I_3^{3m}\left(p_1^2,p_2^2,p_3^2,\mu^2\right) & = &
    - \int\limits_0^1 d^3 \alpha 
      \frac{\delta\left(1-\alpha_1-\alpha_2-\alpha_3\right)}
           {-\alpha_1\alpha_2 p_1^2 - \alpha_2 \alpha_3 p_2^2 - \alpha_3 \alpha_1 p_3^2}
 + {\cal O}(\eps).
\eq
With the notation
\bq
 & &
 \delta_1 = p_1^2 - p_2^2 - p_3^2,
 \;\;\;
 \delta_2 = p_2^2 - p_3^2 - p_1^2,
 \;\;\;
 \delta_3 = p_3^2 - p_1^2 - p_2^2,
 \nonumber \\
 & &
 \Delta_3 = \left(p_1^2\right)^2 + \left(p_2^2\right)^2 + \left(p_3^2\right)^2 
               - 2 p_1^2 p_2^2 - 2 p_2^2 p_3^2 - 2 p_3^2 p_1^2,
\eq
the three-mass triangle $I_3^{3m}$ is expressed in the region
$p_1^2,p_2^2,p_3^2 < 0$ and $\Delta_3 < 0$  by 
\bq
\lefteqn{
 I_3^{3m} = 
   -\frac{2}{\sqrt{-\Delta_3}} 
} & & \nonumber \\
 & & 
   \times \left[
   \mathrm{Cl}_2\left( 2 \arctan \left( \frac{\sqrt{-\Delta_3}}{\delta_1}\right)\right)
   +\mathrm{Cl}_2\left( 2 \arctan \left( \frac{\sqrt{-\Delta_3}}{\delta_2}\right)\right) 
   +\mathrm{Cl}_2\left( 2 \arctan \left( \frac{\sqrt{-\Delta_3}}{\delta_3}\right)\right)
   \right]
 \nonumber \\
 & & + {\cal O}(\eps).
\eq
The Clausen function $\mathrm{Cl}_2(\theta)$ is defined 
by
\bq
\mathrm{Cl}_2(\theta) & = & 
   \frac{1}{2i} \left[ \mathrm{Li}_n\left( e^{i \theta} \right) 
                      -\mathrm{Li}_n\left( e^{-i \theta} \right)
                \right] 
\eq
and discussed in more detail in chapter~\ref{chapter_multiple_polylogarithms}.

In  the region $p_1^2,p_2^2,p_3^2 < 0$ and $\Delta_3 > 0$ as well as in the region
$p_1^2, p_3^2 < 0$, $p_2^2 > 0$ (for which $\Delta_3$ is always positive) the integral
$I_3^{3m}$ is given by 
\bq
 I_3^{3m} & = & \frac{1}{\sqrt{\Delta_3}} \mathrm{Re} \left[
       2 \left( \mathrm{Li}_2(-\rho x)+\mathrm{Li}_2(-\rho y) \right) + \ln(\rho x) \ln(\rho y) 
       + \ln \left( \frac{y}{x} \right) \ln \left( \frac{1+\rho x}{1+\rho y} \right) + \frac{\pi^2}{3} \right] 
 \nonumber \\
 & & 
       + \frac{i \pi \theta(p_2^2)}{\sqrt{\Delta_3}} \ln \left( 
         \frac{\left(\delta_1+\sqrt{\Delta_3}\right) \left(\delta_3+\sqrt{\Delta_3}\right)}
              {\left(\delta_1-\sqrt{\Delta_3}\right) \left(\delta_3-\sqrt{\Delta_3}\right)} \right)
 + {\cal O}(\eps),
\eq
where
\bq
 x = \frac{p_1^2}{p_3^2}, \;\;\;\; y = \frac{p_2^2}{p_3^2}, \;\;\;\;
 \rho = \frac{2 p_3^2}{\delta_3+\sqrt{\Delta_3}}.
\eq
The step function $\theta(x)$ is defined as $\theta(x)=1$ for $x>0$ and $\theta(x)=0$ otherwise.

\subsection*{Four-point functions}

For the four-point function we use the invariants
\bq
s = \left(p_1+p_2\right)^2,
 & &
t = \left(p_2+p_3\right)^2
\eq
together with the external masses $m_i^2=p_i^2$.
\\
The zero-mass box ($m_1^2=m_2^2=m_3^2=m_4^2=0$):
\bq
I^{0m}_4\left(s,t,\mu^2\right)  & = & 
 \frac{4}{\eps^2 s t} - \frac{2}{\eps s t} 
                      \left[ \ln\left(\frac{-s}{\mu^2}\right) + \ln\left(\frac{-t}{\mu^2}\right) \right]
 \nonumber \\
 & &
 + \frac{1}{s t} \left[ \ln^2\left(\frac{-s}{\mu^2}\right) + \ln^2\left(\frac{-t}{\mu^2}\right)
                       - \ln^2\left(\frac{-s}{-t}\right) - 8 \zeta_2
                 \right]
 + {\cal O}(\eps).
\eq
The one-mass box ($m_1^2=m_2^2=m_3^2=0$):
\bq
\label{appendix_one_loop_integrals:one_mass_box}
\lefteqn{
I^{1m}_4\left(s,t,m_4^2,\mu^2\right) = 
 \frac{2}{\eps^2 s t} - \frac{2}{\eps s t} 
                      \left[ \ln\left(\frac{-s}{\mu^2}\right) + \ln\left(\frac{-t}{\mu^2}\right) 
                             - \ln\left(\frac{-m_4^2}{\mu^2}\right) \right]
 + \frac{1}{s t} \left[ \ln^2\left(\frac{-s}{\mu^2}\right) 
 \right.
 } & &
 \nonumber \\
 & & \left.
 + \ln^2\left(\frac{-t}{\mu^2}\right)
                       - \ln^2\left(\frac{-m_4^2}{\mu^2}\right)
                       - \ln^2\left(\frac{-s}{-t}\right) 
                       - 2 \; \mathrm{Li}_2\left(1- \frac{(-m_4^2)}{(-s)}\right)
                       - 2 \; \mathrm{Li}_2\left(1- \frac{(-m_4^2)}{(-t)}\right)
 \right.
 \nonumber \\
 & & \left.
                       - 3 \zeta_2
                 \right]
 + {\cal O}(\eps).
\eq
The easy two-mass box ($m_1^2=m_3^2=0$):
\bq
\lefteqn{
I^{2me}_4\left(s,t,m_2^2,m_4^2,\mu^2\right) = 
 } & &
 \nonumber \\
 & &
 - \frac{2}{\eps \left( s t - m_2^2 m_4^2 \right)} 
                      \left[ \ln\left(\frac{-s}{\mu^2}\right) + \ln\left(\frac{-t}{\mu^2}\right) 
                             - \ln\left(\frac{-m_2^2}{\mu^2}\right) - \ln\left(\frac{-m_4^2}{\mu^2}\right) \right]
 \nonumber \\
 & &
 + \frac{1}{s t - m_2^2 m_4^2} 
     \left[ \ln^2\left( \frac{-s}{\mu^2}\right) + \ln^2\left(\frac{-t}{\mu^2}\right)
                        - \ln^2\left(\frac{-m_2^2}{\mu^2}\right) - \ln^2\left(\frac{-m_4^2}{\mu^2}\right)
                       - \ln^2\left(\frac{-s}{-t}\right) 
 \right. \nonumber \\
 & & \left.
                       - 2 \; \mathrm{Li}_2\left(1- \frac{(-m_2^2)}{(-s)}\right)
                       - 2 \; \mathrm{Li}_2\left(1- \frac{(-m_2^2)}{(-t)}\right)
                       - 2 \; \mathrm{Li}_2\left(1- \frac{(-m_4^2)}{(-s)}\right)
 \right. \nonumber \\
 & & \left.
                       - 2 \; \mathrm{Li}_2\left(1- \frac{(-m_4^2)}{(-t)}\right)
                       + 2 \; \mathrm{Li}_2\left(1- \frac{(-m_2^2)}{(-s)} \frac{(-m_4^2)}{(-t)}\right)
                 \right]
 + {\cal O}(\eps).
\eq
The hard two-mass box ($m_1^2=m_2^2=0$):
\bq
\lefteqn{
I^{2mh}_4\left(s,t,m_3^2,m_4^2,\mu^2\right) = 
 \frac{1}{\eps^2 s t} - \frac{1}{\eps s t} 
                      \left[ \ln\left(\frac{-s}{\mu^2}\right) + 2 \ln\left(\frac{-t}{\mu^2}\right) 
                             - \ln\left(\frac{-m_3^2}{\mu^2}\right) - \ln\left(\frac{-m_4^2}{\mu^2}\right) \right]
 } & &
 \nonumber \\
 & &
 + \frac{1}{s t} 
     \left[ \frac{3}{2} \ln^2\left( \frac{-s}{\mu^2}\right) + \ln^2\left(\frac{-t}{\mu^2}\right)
                        - \frac{1}{2} \ln^2\left(\frac{-m_3^2}{\mu^2}\right) 
                        - \frac{1}{2} \ln^2\left(\frac{-m_4^2}{\mu^2}\right)
                        - \ln^2\left(\frac{-s}{-t}\right) 
 \right. \nonumber \\
 & & \left.
                        - \ln\left(\frac{-s}{\mu^2}\right) \ln\left(\frac{-m_3^2}{\mu^2}\right) 
                        - \ln\left(\frac{-s}{\mu^2}\right) \ln\left(\frac{-m_4^2}{\mu^2}\right) 
                        + \ln\left(\frac{-m_3^2}{\mu^2}\right) \ln\left(\frac{-m_4^2}{\mu^2}\right) 
 \right. \nonumber \\
 & & \left.
                       - 2 \; \mathrm{Li}_2\left(1- \frac{(-m_3^2)}{(-t)}\right)
                       - 2 \; \mathrm{Li}_2\left(1- \frac{(-m_4^2)}{(-t)}\right)
                       - \frac{1}{2} \zeta_2
                 \right]
 + {\cal O}(\eps).
\eq
The three-mass box ($m_1^2=0$):
\bq
\lefteqn{
I^{3m}_4\left(s,t,m_2^2,m_3^2,m_4^2,\mu^2\right) = 
 } & &
 \\
 & &
 - \frac{1}{\eps \left( s t - m_2^2 m_4^2 \right)} 
                      \left[ \ln\left(\frac{-s}{\mu^2}\right) + \ln\left(\frac{-t}{\mu^2}\right) 
                             - \ln\left(\frac{-m_2^2}{\mu^2}\right) - \ln\left(\frac{-m_4^2}{\mu^2}\right) \right]
 \nonumber \\
 & &
 + \frac{1}{s t - m_2^2 m_4^2} 
     \left[   \frac{3}{2} \ln^2\left( \frac{-s}{\mu^2}\right) 
            + \frac{3}{2} \ln^2\left(\frac{-t}{\mu^2}\right)
            - \frac{1}{2} \ln^2\left(\frac{-m_2^2}{\mu^2}\right) 
            - \frac{1}{2} \ln^2\left(\frac{-m_4^2}{\mu^2}\right)
            - \ln^2\left(\frac{-s}{-t}\right) 
 \right. \nonumber \\
 & & \left.
            - \ln\left(\frac{-s}{\mu^2}\right) \ln\left(\frac{-m_3^2}{\mu^2}\right) 
            - \ln\left(\frac{-s}{\mu^2}\right) \ln\left(\frac{-m_4^2}{\mu^2}\right) 
            + \ln\left(\frac{-m_3^2}{\mu^2}\right) \ln\left(\frac{-m_4^2}{\mu^2}\right) 
 \right. \nonumber \\
 & & \left.
            - \ln\left(\frac{-t}{\mu^2}\right) \ln\left(\frac{-m_2^2}{\mu^2}\right) 
            - \ln\left(\frac{-t}{\mu^2}\right) \ln\left(\frac{-m_3^2}{\mu^2}\right) 
            + \ln\left(\frac{-m_2^2}{\mu^2}\right) \ln\left(\frac{-m_3^2}{\mu^2}\right) 
 \right. \nonumber \\
 & & \left.
            - 2 \; \mathrm{Li}_2\left(1- \frac{(-m_2^2)}{(-s)}\right)
            - 2 \; \mathrm{Li}_2\left(1- \frac{(-m_4^2)}{(-t)}\right)
            + 2 \; \mathrm{Li}_2\left(1- \frac{(-m_2^2)}{(-s)} \frac{(-m_4^2)}{(-t)}\right)
                 \right]
 + {\cal O}(\eps).
 \nonumber
\eq
The four-mass box:
\bq
I^{4m}_4\left(s,t,m_1^2,m_2^2,m_3^2,m_4^2,\mu^2\right) & = & I_3^{3m}(s t, m_1^2 m_3^2, m_2^2 m_4^2,\mu^2) + K(s, t, m_1^2, m_3^2, m_2^2, m_4^2),
\eq
where
\bq
\lefteqn{
K(s_1,t_1,s_2,t_2,s_3,t_3) = -\frac{2\pi i}{\lambda}
 \sum\limits_{i=1}^{3} \theta(-s_i) \theta(-t_i) 
} & & \nonumber \\
 & & \times
 \left[  
       \ln\left( \sum\limits_{j \neq i} s_j t_j - \left( s_i t_i - \lambda \right) (1+i \delta) \right)
     - \ln\left( \sum\limits_{j \neq i} s_j t_j - \left( s_i t_i + \lambda \right) (1+i \delta) \right)
 \right],
\eq
and
\bq
 \lambda & = & \sqrt{ \left( s_1 t_1 \right)^2 + \left( s_2 t_2 \right)^2 + \left( s_3 t_3 \right)^2
                      - 2 s_1 t_1 s_2 t_2 - 2 s_2 t_2 s_3 t_3 - 2 s_3 t_3 s_1 t_1 }.
\eq
$\delta > 0$ is an infinitesimal quantity.


\section{Analytic continuation}

In one-loop integrals the functions
\bq
 \ln\left(\frac{-s}{-t}\right), & &
 \mathrm{Li}_2\left(1- \frac{(-s)}{(-t)} \right)
\eq
and generalizations thereof
occur. The analytic continuation is defined by giving all quantities a small
imaginary part, e.g.
\bq 
 s \rightarrow s + i \delta,
\eq
with $\delta > 0$ being an infinitesimal quantity.
Explicitly, the imaginary parts of the logarithm and the dilogarithm are given by
\bq
 \ln\left( \frac{-s}{-t} \right) & = & 
    \ln\left(\left|\frac{s}{t}\right|\right) -i \pi \left[ \theta(s) - \theta(t) \right],
 \nonumber \\
 \mathrm{Li}_2\left(1 - \frac{(-s)}{(-t)} \right) & = & 
    \mathrm{Re} \mathrm{Li}_2\left(1 - \frac{s}{t} \right) 
    - i \theta\left(-\frac{s}{t}\right)
        \ln\left(1-\frac{s}{t}\right) 
        \mathrm{Im} \ln \left(\frac{-s}{-t}\right).
\eq
This generalizes as follows:
\bq
 \ln\left( \frac{(-s_1)}{(-t_1)} \frac{(-s_2)}{(-t_2)} \right) & = & 
    \ln\left(\left|\frac{s_1 s_2}{t_1 t_2}\right|\right) 
    -i \pi \left[ \theta(s_1) + \theta(s_2) - \theta(t_1) - \theta(t_2) \right],
 \nonumber \\
 \mathrm{Li}_2\left(1 - \frac{(-s_1)}{(-t_1)} \frac{(-s_2)}{(-t_2)} \right) & = & 
    \mathrm{Re} \mathrm{Li}_2\left(1 - \frac{s_1 s_2}{t_1 t_2} \right) 
    - i 
        \ln\left(1-\frac{(-s_1)}{(-t_1)} \frac{(-s_2)}{(-t_2)} \right) 
        \mathrm{Im} \ln \left(\frac{(-s_1)}{(-t_1)} \frac{(-s_2)}{(-t_2)} \right), 
 \nonumber 
\eq
where
\bq
 \ln\left(1-\frac{(-s_1)}{(-t_1)} \frac{(-s_2)}{(-t_2)} \right)
 & = &
 \ln\left|1-\frac{s_1 s_2}{t_1 t_2} \right| 
 - \frac{1}{2} i \pi
   \left[ \theta(s_1) + \theta(s_2) - \theta(t_1) - \theta(t_2) \right]
   \theta\left( \frac{s_1 s_2}{t_1 t_2} - 1 \right).
 \nonumber
\eq

%% file: transcendental.tex
\newpage
\chapter{Transcendental functions}
\label{appendix_trancendental}

In this appendix we summarise definitions and properties of a few 
transcendental functions.
We start with hypergeometric functions in one variable in section~\ref{appendix_trancendental:hypergeometric}.
Appell functions are generalisations to two variables and discussed in section~\ref{appendix_trancendental:appell_functions}.
Lauricella functions are particular generalisations to $n$ variables and briefly 
discussed in section~\ref{appendix_trancendental:Lauricella}.
The general case with $n$ variables is known as Horn functions. These are introduced in section~\ref{appendix_trancendental:Horn}.

In defining these functions, the Pochhammer symbol occurs frequently.
The Pochhammer symbol $(a)_n$ is defined by
\bq
(a)_n & = & \frac{\Gamma(a+n)}{\Gamma(a)}.
\eq

\section{Hypergeometric functions}
\label{appendix_trancendental:hypergeometric}

The generalised hypergeometric function ${}_AF_B$ (or hypergeometric function for short)
is defined by
\bq
\label{appendix_trancendental:def_hypergeometric}
 \hypergeometric{A}{B}(a_1,\dots,a_A;b_1,\dots,b_B;x)
 & = &
 \sum\limits_{n=0}^{\infty} 
 \frac{(a_1)_n \dots (a_A)_n}{(b_1)_n \dots (b_B)_n}
 \frac{x^{n}}{n!}.
\eq
We are mainly interested in the case $\hypergeometric{A+1}{A}$.
The case $\hypergeometric{1}{0}$ is trivial
\bq
 \hypergeometric{1}{0}(a;;x) 
 & = &
 \left(1-x\right)^{-a}.
\eq
The first non-trivial case in the family $\hypergeometric{A+1}{A}$ is the function 
\bq
 \hypergeometric{2}{1}(a_1,a_2;b_1;x) 
 & = & 
 \sum\limits_{n=0}^{\infty} \frac{(a_1)_n (a_2)_n}{(b_1)_n} \frac{x^{n}}{n!}.
\eq
This function is sometimes referred to as ``the'' hypergeometric function or the Gau{\ss} hypergeometric function.
In this book we will call any function of the form as in eq.~(\ref{appendix_trancendental:def_hypergeometric})
a hypergeometric function.
From the definition it is easy to verify that
\bq
\lefteqn{
 \hypergeometric{A+1}{B+1}(a_1,\dots,a_A,a_{A+1};b_1,\dots,b_B,b_{B+1};x) 
 \; = \; 
 \frac{\Gamma(b_{B+1})}{\Gamma(a_{A+1}) \Gamma(b_{B+1}-a_{A+1})}
 } & & \nonumber \\
 & &
 \times
 \int\limits_0^1dt \; t^{a_{A+1}-1} (1-t)^{b_{B+1}-a_{A+1}-1} \hypergeometric{A}{B}(a_1,\dots,a_A;b_1,\dots,b_B;tx).
\eq
This allows us to deduce a $A$-fold integral representation for $\hypergeometric{A+1}{A}$.
In particular
\bq
 \hypergeometric{2}{1}(a_1,a_2;b_1;x)
 & = &
 \frac{\Gamma(b_{1})}{\Gamma(a_{2}) \Gamma(b_{1}-a_{2})}
 \int\limits_0^1dt \; t^{a_{2}-1} (1-t)^{b_{1}-a_{2}-1} (1-tx)^{-a_1},
 \\
 \hypergeometric{3}{2}(a,b_1,b_2;c_1,c_2;x)
 & = & 
 \frac{\Gamma(c_1)}{\Gamma(b_1) \Gamma(c_1-b_1)}
 \frac{\Gamma(c_2)}{\Gamma(b_2) \Gamma(c_2-b_2)}
 \nonumber \\
 &&
 \int\limits_0^1 du \int\limits_0^1 dv \; u^{b_1-1} (1-u)^{c_1-b_1-1} v^{b_2-1} (1-v)^{c_2-b_2-1} (1-uvx)^{-a}.
 \nonumber 
\eq

\section{Appell functions}
\label{appendix_trancendental:appell_functions}

In this section we discuss the four Appell functions and the Kamp\'e de F\'eriet function \cite{Appell:1880,Appell}.
They are generalisations of the hypergeometric function from one variable $x$ to two variables $x_1$ and $x_2$.
 
The Appell function of the first kind is defined by
\bq
 F_1(a,b_1,b_2;c;x_1,x_2)  
 & = & 
 \sum\limits_{m_1=0}^\infty \sum\limits_{m_2=0}^\infty 
 \frac{(a)_{m_1+m_2} (b_1)_{m_1} (b_2)_{m_2}}{(c)_{m_1 + m_2}}
 \frac{x_1^{m_1}}{m_1!} \frac{x_2^{m_2}}{m_2!}.
\eq
The Appell function of the second kind is defined by
\bq
 F_2(a,b_1,b_2;c_1,c_2;x_1,x_2)  
 & = &
 \sum\limits_{m_1=0}^\infty \sum\limits_{m_2=0}^\infty 
 \frac{(a)_{m_1+m_2} (b_1)_{m_1} (b_2)_{m_2}}{(c_1)_{m_1} (c_2)_{m_2}}
 \frac{x_1^{m_1}}{m_1!} \frac{x_2^{m_2}}{m_2!}.
\eq
The Appell function of the third kind is defined by
\bq
 F_3(a_1,a_2,b_1,b_2;c;x_1,x_2)  
 & = &
 \sum\limits_{m_1=0}^\infty \sum\limits_{m_2=0}^\infty 
 \frac{(a_1)_{m_1} (a_2)_{m_2} (b_1)_{m_1} (b_2)_{m_2}}{(c)_{m_1 + m_2}}
 \frac{x_1^{m_1}}{m_1!} \frac{x_2^{m_2}}{m_2!}.
\eq
The Appell function of the fourth kind is defined by
\bq
 F_4(a,b;c_1,c_2;x_1,x_2)  
 & = &
 \sum\limits_{m_1=0}^\infty \sum\limits_{m_2=0}^\infty 
 \frac{(a)_{m_1+m_2} (b)_{m_1+m_2}}{(c_1)_{m_1} (c_2)_{m_2}}
 \frac{x_1^{m_1}}{m_1!} \frac{x_2^{m_2}}{m_2!}.
\eq
The generalized Kamp\'e de F\'eriet function $S_1$ is defined by
\bq
 S_1(a_1,a_2,b_1;c,c_1;x_1,x_2)  
 & = &
 \sum\limits_{m_1=0}^\infty \sum\limits_{m_2=0}^\infty 
 \frac{(a_1)_{m_1+m_2} (a_2)_{m_1+m_2} (b_1)_{m_1} }{(c)_{m_1 + m_2} (c_1)_{m_1} }
 \frac{x_1^{m_1}}{m_1!} \frac{x_2^{m_2}}{m_2!}.
\eq
We have the following integral representations:
\bq
\lefteqn{
 F_1(a,b_1,b_2;c;x_1,x_2) =
 \frac{\Gamma(c)}{\Gamma(a) \Gamma(c-a) } 
 \int\limits_0^1 dy \; y^{a-1} (1-y)^{c-a-1} \left(1-x_1 y \right)^{-b_1}\left( 1- x_2 y\right)^{-b_2}
 } & & \nonumber \\  
 & = & 
 \frac{\Gamma(c)}{\Gamma(b_1) \Gamma(b_2) \Gamma(c-b_1-b_2)}
 \int d^3y \;\delta(1-\sum_{j=1}^3 y_j) \; y_1^{b_1-1} y_2^{b_2-1} y_3^{c-b_1-b_2-1} \left(1-x_1 y_1 - x_2 y_2\right)^{-a},
 \nonumber \\
\lefteqn{
 F_2(a,b_1,b_2;c_1,c_2;x_1,x_2)
 =
 \frac{\Gamma(c_1)}{\Gamma(b_1) \Gamma(c_1-b_1)}
 \frac{\Gamma(c_2)}{\Gamma(b_2) \Gamma(c_2-b_2)}
 } & & \nonumber \\
 & & 
 \times
 \int\limits_0^1 du \int\limits_0^1 dv \; 
 u^{b_1-1} \left(1-u\right)^{c_1-b_1-1} v^{b_2-1} \left(1-v\right)^{c_2-b_2-1} \left(1-ux_1-vx_2\right)^{-a},
 \nonumber \\
\lefteqn{
 F_3(a_1,a_2,b_1,b_2;c;x_1,x_2)
 =
 \frac{\Gamma(c)}{\Gamma(b_1) \Gamma(b_2) \Gamma(c-b_1-b_2)}
 } & & \nonumber \\  
 & & 
 \times
 \int d^3y \; \delta(1-\sum_{j=1}^3 y_j) \; y_1^{b_1-1} y_2^{b_2-1} y_3^{c-b_1-b_2-1}
 \left( 1-x_1 y_1\right)^{-a_1} \left(1 - x_2 y_2\right)^{-a_2},
 \nonumber \\
\lefteqn{
 F_4(a,b;c_1,c_2;x_1(1-x_2),x_2(1-x_1)) 
 =
\frac{\Gamma(c_1)}{\Gamma(a) \Gamma(c_1-a)}
\frac{\Gamma(c_2)}{\Gamma(b) \Gamma(c_2-b)}
 } & & \nonumber \\  
 & & 
 \times
 \int\limits_0^1 du \int\limits_0^1 dv \;
 u^{a-1} \left(1-u\right)^{c_1-a-1} v^{b-1} \left(1-v\right)^{c_2-b-1}
 \left(1-ux_1\right)^{a-c_1-c_2+1} \left(1-vx_2\right)^{b-c_1-c_2+1}
 \nonumber \\
 & &
 \cdot
 \left(1-ux_1-vx_2\right)^{c_1+c_2-a-b-1},
 \nonumber \\
\lefteqn{
 S_1(a_1,a_2,b_1;c,c_1;x_1,x_2)
 =
 \frac{\Gamma(c)}{\Gamma(a_1) \Gamma(c-a_1)}
 \frac{\Gamma(c_1)}{\Gamma(b_1) \Gamma(c_1-b_1)}
 } & & \nonumber \\
 & & 
 \int\limits_0^1 du \int\limits_0^1 dv \;
 u^{a_1-1} \left(1-u\right)^{c-a_1-1} v^{b_1-1} \left(1-v\right)^{c_1-b_1-1} \left(1-uvx_1-ux_2\right)^{-a_2}.
\eq
Note that in the equation above the arguments of the fourth Appell function are $x_1(1-x_2)$ and $x_2(1-x_1)$.

\section{Lauricella functions}
\label{appendix_trancendental:Lauricella}

The Lauricella functions are generalisations to $n$ variables $x_1, \dots, x_n$.
The four Lauricella functions are defined by \cite{Lauricella:1893}
\bq
\lefteqn{
 F_A(a,b_1,\dots,b_n;c_1,\dots,c_n;x_1,\dots,x_n) 
 = } & & 
 \nonumber \\
 & &
 \sum\limits_{m_1=0}^\infty \dots \sum\limits_{m_n=0}^\infty 
 \frac{(a)_{m_1+\dots+m_n} (b_1)_{m_1} \dots (b_n)_{m_n}}{(c_1)_{m_1} \dots (c_n)_{m_n}}
 \frac{x_1^{m_1}}{m_1!} \dots \frac{x_n^{m_n}}{m_n!},
 \nonumber \\
\lefteqn{
 F_B(a_1,\dots,a_n,b_1,\dots,b_n;c;x_1,\dots,x_n) 
 = 
 } & & 
 \nonumber \\
 & &
 \sum\limits_{m_1=0}^\infty \dots \sum\limits_{m_n=0}^\infty 
 \frac{(a_1)_{m_1} \dots (a_n)_{m_n} (b_1)_{m_1} \dots (b_n)_{m_n}}{(c)_{m_1+\dots+m_n} }
 \frac{x_1^{m_1}}{m_1!} \dots \frac{x_n^{m_n}}{m_n!},
 \nonumber \\
\lefteqn{
 F_C(a,b;c_1,\dots,c_n;x_1,\dots,x_n) 
 = } & & 
 \nonumber \\
 & &
 \sum\limits_{m_1=0}^\infty \dots \sum\limits_{m_n=0}^\infty 
 \frac{(a)_{m_1+\dots+m_n} (b)_{m_1+\dots+m_n}}{(c_1)_{m_1} \dots (c_n)_{m_n}}
 \frac{x_1^{m_1}}{m_1!} \dots \frac{x_n^{m_n}}{m_n!},
 \nonumber \\
\lefteqn{
 F_D(a,b_1,\dots,b_n;c;x_1,\dots,x_n) 
 = } & & 
 \nonumber \\
 & &
  \sum\limits_{m_1=0}^\infty \dots \sum\limits_{m_n=0}^\infty 
 \frac{(a)_{m_1+\dots+m_n} (b_1)_{m_1} \dots (b_n)_{m_n}}{(c)_{m_1 +\dots+ m_n}}
 \frac{x_1^{m_1}}{m_1!} \dots \frac{x_n^{m_n}}{m_n!}.
\eq

\section{Horn functions}
\label{appendix_trancendental:Horn}

In his original publication, Jakob Horn extended the list of hypergeometric functions in two variables 
(examples are the four Appell functions and the Kamp\'e de F\'eriet function from section~\ref{appendix_trancendental:appell_functions})
to 34 functions \cite{Horn:1931}. 
There is no point in listing them all here.

In the modern literature the name ``Horn-type hypergeometric function'' is used for a function of the following type:
With the multi-index notation
\bq
 {\bf x} \; = \; (x_1,\dots,x_n),
 \;\;\;
 {\bf i} \; = \; (i_1,\dots,i_n),
 \;\;\;
 {\bf x}^{\bf i} \; = \; x_1^{i_1} \cdot \dots \cdot x_n^{i_n}
\eq
a Horn-type hypergeometric function is defined by
\bq
\label{appendix_trancendental:def_horn_function}
 H
 & = &
 \sum\limits_{{\bf i} \in {\mathbb N}_0^n} C_{{\bf i}} {\bf x}^{\bf i},
 \nonumber \\
 C_{{\bf i}}
 & = &
 \frac{\prod\limits_{j=1}^p \Gamma\left(\sum\limits_{k=1}^n A_{jk} i_k + u_j\right)}{\prod\limits_{j=1}^q \Gamma\left(\sum\limits_{k=1}^n B_{jk} i_k + v_j\right)},
\eq
with $A_{jk}, B_{jk} \in {\mathbb Z}$ and $u_j, v_j \in {\mathbb C}$.
Let us denote
\bq
 {\bf i} + {\bf e}_j & = & \left(i_1,\dots,i_{j-1},i_j+1,i_{j+1},\dots,i_n\right).
\eq
It follows from $\Gamma(z+1)=z\Gamma(z)$ that the ratio $C_{{\bf i}+{\bf e}_j}/C_{{\bf i}}$ is a rational function
in $i_1,\dots,i_n$:
\bq
\label{appendix_trancendental:horn_ratio}
 \frac{C_{{\bf i}+{\bf e}_j}}{C_{{\bf i}}}
 & = &
 \frac{P_j\left({\bf i}\right)}{Q_j\left({\bf i}\right)},
\eq
with $P_j({\bf i})$ and $Q_j({\bf i})$ being polynomials in $(i_1,\dots,i_n)$.
Let us denote the Euler operators by
\bq
 {\bm \theta}
 & = &
 \left( \theta_1, \dots, \theta_n \right)^T
 \; = \;
 \left( x_1 \frac{\partial}{\partial x_1}, \dots, x_n \frac{\partial}{\partial x_n} \right)^T.
\eq
The function $H({\bf x})$ satisfies the differential equation
\bq
\label{appendix_trancendental:dgl_horn_function}
 \left(1+\theta_j\right)
 \left[ 
  Q_j\left({\bf \theta}\right) \frac{1}{x_j} 
  - P_j\left({\bf \theta}\right) 
 \right] H\left({\bf x}\right)
 & = & 0,
 \;\;\;\;\;\;
 j \; = \; 1,\dots,n.
\eq
\bs
{\it \refstepcounter{exercise}
{\bf Exercise \theexercise}: 
Prove eq.~(\ref{appendix_trancendental:dgl_horn_function}).
}
\es
\\
\\
As $\sum\limits_{k=1}^n A_{jk} i_k$ is an integer we may use the reflection identity eq.~(\ref{chapter_basics:Gamma_function_reflection_identity}) to write
\bq
 \Gamma\left(\sum\limits_{k=1}^n A_{jk} i_k + u_j\right)
 & = &
 \left(-1\right)^{\sum\limits_{k=1}^n A_{jk} i_k}
 \frac{\Gamma\left(u_j\right)\Gamma\left(1-u_j\right)}{\Gamma\left(-\sum\limits_{k=1}^n A_{jk} i_k + 1 - u_j\right)}.
\eq
This converts the series representation of the Horn functions of eq.~(\ref{appendix_trancendental:def_horn_function})
into a form similar to the $\Gamma$-series defined in eq.~(\ref{chapter_nested_sums:def_Gamma_series}).

Note that the Horn functions define a rather large class of functions: If we consider the system of differential equations they satisfy,
this system may not be holonomic \cite{Dickenstein:2003BivariateHD}.
This implies that the space of local solutions of the system of differential equations may be infinite dimensional.

The relation of Feynman integrals with Horn-type hypergeometric function is discussed 
in \cite{Bytev:2013gva,Bytev:2017jmx,Kalmykov:2020cqz}.

%% file: lie_algebra.tex
\newpage
\chapter{Lie groups and Lie algebras}
\label{appendix_lie_algebra}

In this appendix we give a short introduction to the theory of Lie groups and Lie algebras.
Lie groups figure prominently as symmetry groups in particle physics, 
and we assume that most readers have already come across Lie groups and Lie algebras.
The main point of this appendix is the classification of the simple Lie algebras.

Lie groups and Lie algebras are treated in many textbooks, examples are the books by
Helgason \cite{Helgason}, Fulton and Harris \cite{Fulton}, 
Bourbaki \cite{Bourbaki:Lie} and Weyl \cite{Weyl}.

\section{Definitions}

We start with the definitions:
\begin{tcolorbox}
{\bf Lie group}:
\\
A Lie group $G$ is a group which is also an analytic manifold such that the mapping $(a,b) \rightarrow
ab^{-1}$ of the product manifold $G \times G$ into $G$ is analytic.
\end{tcolorbox}

\begin{tcolorbox}
{\bf Lie algebra}:
\\
A Lie algebra $\mathfrak{g}$ over a field ${\mathbb F}$ is a vector space together with a bilinear mapping
$[\cdot,\cdot] : \mathfrak{g} \times \mathfrak{g} \rightarrow \mathfrak{g}$,
$(X,Y) \rightarrow [X,Y]$  such that for $X,Y,Z \in \mathfrak{g}$:
\bq
\label{appendix_lie_algebra:def_Lie_algebra}
\left[X,X\right] & = & 0,
  \nonumber \\
   \left[ \left[ X, Y \right], Z \right] 
 + \left[ \left[ Y, Z \right], X \right] 
 + \left[ \left[ Z, X \right], Y \right] & = & 0.
\eq
\end{tcolorbox}
\bs
{\it \refstepcounter{exercise}
{\bf Exercise \theexercise}: 
Show that $[X, X]=0$ implies the anti-symmetry of the Lie bracket $[X,Y]=-[Y,X]$.
Show further that also the converse is true, provided $\mathrm{char}\; {\mathbb F} \neq 2$.
Explain, why the argument does not work for $\mathrm{char}\; {\mathbb F} = 2$.
}
\es
\\
\\
We are mainly interested in the case where the ground field 
are the real numbers ${\mathbb R}$ or the complex numbers ${\mathbb C}$.

Let $\mathfrak{g}$ be a Lie algebra and $X^1,...,X^n$ a basis of $\mathfrak{g}$ as a vector space.
$[X^a,X^b]$ is again in $\mathfrak{g}$ and can be expressed as a linear combination of the basis
vectors $X^k$:
\bq
 \left[ X^a, X^b \right] & = & \sum\limits_{c=1}^n c^{abc} X^c.
\eq
The coefficients $c^{abc}$ are called the 
\index{structure constants of a Lie algebra}
{\bf structure constants} of the Lie algebra.
For matrix algebras the $X^a$'s are anti-hermitian matrices.

The notation above is mainly used in the mathematical literature. In physics a slightly different
convention is often used (which corresponds for matrix algebras to having hermitian matrices as a basis): 
Denote by $T^1,...,T^n$ a basis of $\mathfrak{g}$ as a (complex) vector space and write
\bq
 \left[ T^a, T^b \right] & = & i \sum\limits_{c=1}^n f^{abc} T^c.
\eq
We can get from one convention to the other one by letting
\bq
 T^a & = & i X^a.
\eq
In this case we have
\bq
 f^{abc} & = & c^{abc}.
\eq
The standard normalisation of the generators is
\bq
 \mathrm{Tr}\left( T^a T^b \right) & = & \frac{1}{2} \delta^{ab}.
\eq
The relation between Lie groups and Lie algebras is as follows:
Let $G$ be a Lie group. 
It is therefore a manifold and a group.
Let $n$ be the dimension of $G$ as a manifold.
Choose a local coordinate system with coordinates $(\theta_1,\dots,\theta_n)$, 
such that the identity element $e$ is given
by
\bq
 e & = & g(0,\dots,0).
\eq
In a neighbourhood of $e$ we may write
\bq
 g(0,\dots,\theta_a,\dots,0) & = & g(0,\dots,0,\dots,0) + \theta_a X^a + {\mathcal O}(\theta^2)
 \nonumber \\
 & = & g(0,\dots,0,\dots,0) - i \theta_a T^a + {\mathcal O}(\theta^2),
\eq
where
\bq
\label{appendix_lie_algebra:def_generators}
 X^a & = & \lim\limits_{\theta_a\rightarrow 0} \frac{g(0,\dots,\theta_a,\dots,0)-g(0,\dots,0,\dots,0)}{\theta_a},
 \nonumber \\
 T^a & = & i \lim\limits_{\theta_a\rightarrow 0} \frac{g(0,\dots,\theta_a,\dots,0)-g(0,\dots,0,\dots,0)}{\theta_a}.
\eq
The $T^a$'s (and the $X^a$'s) are called the 
\index{generators of a Lie group}
{\bf generators} of the Lie group $G$.
\begin{theorem}
The generators $T^a$ of a Lie group are a basis of a Lie algebra $\mathfrak{g}$, in particular
the commutators of the generators are linear combinations
of the generators:
\bq
\label{appendix_lie_algebra:commutation_relation}
 \left[ T^a, T^b \right] & = & i \sum\limits_{c=1}^n f^{abc} T^c.
\eq
$\mathfrak{g}$ is called the Lie algebra of the Lie group $G$.
\end{theorem}
We will often use Einstein's summation convention and write eq.~(\ref{appendix_lie_algebra:commutation_relation}) as
\bq
 \left[ T^a, T^b \right] & = & i f^{abc} T^c.
\eq
We have seen that given a Lie group $G$ we obtain its Lie algebra from eq.~(\ref{appendix_lie_algebra:def_generators}).
We may now ask if the converse is also possible:
Given a Lie algebra $\mathfrak{g}$, can we reconstruct the Lie group $G$ ?
The answer is that this can almost be done.
Note that a Lie group need not be connected.
Given a Lie algebra we have information about the connected component in which the identity lies.
The exponential map takes us from the Lie algebra into the group, more precisely into the connected component in which the
identity lies.
In a neighbourhood of the identity we have
\bq
 g(\theta_1,\dots,\theta_n) & = & \exp\left( - i \sum\limits_{a=1}^n \theta_a T^a \right).
\eq
A few examples of Lie groups are:
\begin{enumerate}

\item $\mathrm{SU}(n,{\mathbb C})$: The group of special unitary $(n \times n)$-matrices defined through
\bq
 U U^\dagger \; = \; {\bf 1} & \mbox{and} & \det\left(U\right) \; = \; 1.
\eq
The group $\mathrm{SU}(n,{\mathbb C})$ has $n^2-1$ real parameters.

\item $\mathrm{SO}(n,{\mathbb R})$: 
The group of special orthogonal $(n \times n)$-matrices defined through
\bq
 R R^T \; = \; {\bf 1} & \mbox{and} & \det\left(R\right) \; = \; 1.
\eq
The group $\mathrm{SO}(n,{\mathbb R})$ has $n(n-1)/2$ real parameters.

\item $\mathrm{Sp}(n,\mathbb{R})$: The symplectic group is the group of $(2n\times 2n)$-matrices satisfying
\bq
 M^T \left( \begin{array}{rr} {\bf 0} & {\bf 1} \\ - {\bf 1} & {\bf 0} \\ \end{array} \right) M
 & = & 
 \left( \begin{array}{rr} {\bf 0} & {\bf 1} \\ - {\bf 1} & {\bf 0} \\ \end{array} \right),
\eq
where ${\bf 0}$ denotes the $(n \times n)$-zero matrix and ${\bf 1}$ denotes the $(n \times n)$-identity matrix.  
The group $\mathrm{Sp}(n,\mathbb{R})$ has $(2n+1)n$ real parameters.
The group $\mathrm{Sp}(n,\mathbb{R})$ can also be defined as the transformation group 
of a real $(2n)$-dimensional vector space with coordinates $(x_1,\dots,x_n,x_{n+1},\dots,x_{2n})$, 
which preserves the inner product
\bq
 \sum\limits_{j=1}^n \left( x_j y_{j+n} - x_{j+n} y_j \right).
\eq
\end{enumerate}
The corresponding Lie algebras are denotes ${\mathfrak s}{\mathfrak u}(n)$, ${\mathfrak s}{\mathfrak o}(n)$  
and ${\mathfrak s}{\mathfrak p}(n)$, respectively.

Let us now focus on Lie algebras. Let $\mathfrak{g}$ be the Lie algebra of a Lie group $G$.
\begin{definition}
The 
\index{rank of a Lie algebra}
{\bf rank of a Lie algebra} is the number of simultaneously diagonalisable generators.
\end{definition}
\begin{definition}
A 
\index{Casimir operator}
{\bf Casimir operator} is an operator, which commutes with all the generators of the group.
\end{definition}
\begin{theorem}
The number of independent Casimir operators is equal to the rank of the Lie algebra.
\end{theorem}
\begin{definition}
A Lie algebra is called 
\index{simple Lie Algebra}
{\bf simple} if it is non-Abelian and has no non-trivial ideals.
($\mathfrak{g}$ and $\{0\}$ are the two trivial ideals every Lie algebra has.)
\end{definition}
\begin{definition}
A Lie algebra is called 
\index{semi-simple Lie algebra}
{\bf semi-simple} if it has no non-trivial Abelian ideals.
\end{definition}
\begin{theorem}
A Lie algebra $\mathfrak{g}$ is semi-simple if and only if
\bq
 \det\left(g\right) & \neq & 0,
\eq
where
\bq
 g^{ab} & = & f^{acd} f^{bcd}.
\eq
\end{theorem}
\begin{definition}
A Lie algebra is called 
\index{reductive Lie algebra}
{\bf reductive} if it is the sum of a semi-simple and an Abelian Lie algebra.
\end{definition}
A simple Lie algebra is also semi-simple and a semi-simple Lie algebra is also reductive.
Let us look at a few examples:
The Lie algebras 
\bq
 {\mathfrak s}{\mathfrak u}(n), 
 \;\;\;
 {\mathfrak s}{\mathfrak o}(n), 
 \;\;\;
 {\mathfrak s}{\mathfrak p}(n)
\eq
are simple.
Semi-simple Lie algebras are sums of simple Lie algebras, for example
\bq
 {\mathfrak s}{\mathfrak u}(n_1) \oplus {\mathfrak s}{\mathfrak u}(n_2).
\eq
Reductive Lie algebras may have in addition an Abelian part, for example
\bq
 {\mathfrak u}(1) \oplus {\mathfrak s}{\mathfrak u}(2) \oplus {\mathfrak s}{\mathfrak u}(3).
\eq
The Abelian Lie algebras are rather trivial, they only have
one-dimensional irreducible representations.
Therefore the classification of all reductive Lie algebras essentially boils down
to the classification of all simple Lie algebras.

\section{The Cartan basis and root systems}

We now take the complex numbers ${\mathbb C}$ as the ground field.
Consider $A,X \in \mathfrak{g}$ with $X \neq 0$ and assume that
\bq
\label{appendix_lie_algebra:def_root}
 \left[ A, X \right] & = & \rho X,
 \;\;\;\;\;\;
 \rho \; \in \; {\mathbb C}.
\eq
$\rho$ is called a 
\index{root of a Lie algebra}
{\bf root} of the Lie algebra $\mathfrak{g}$.
We write
\bq
 A \; = \; \sum\limits_{a=1}^n c_a T^a,
 & & 
 X \; = \; \sum\limits_{a=1}^n x_a T^a.
\eq
For a non-trivial solution $X \neq 0$ of eq.~(\ref{appendix_lie_algebra:def_root}) we must have
\bq
\label{appendix_lie_algebra:secular_equation}
 \det\left( c_a i f^{abc} - \rho \delta^{bc} \right) & = & 0.
\eq
Eq.~(\ref{appendix_lie_algebra:secular_equation}) is called the
\index{secular equation}
{\bf secular equation}.
In general the secular equation will give a $n$-th order polynomial in $\rho$.
Solving for $\rho$ one obtains $n$ roots.
One root may occur more than once.
The degree of degeneracy is called the 
\index{multiplicity of a root of a Lie algebra}
{\bf multiplicity of the root}.
\\
\\
\bs
{\it \refstepcounter{exercise}
{\bf Exercise \theexercise}: 
Derive eq.~(\ref{appendix_lie_algebra:secular_equation}) from eq.~(\ref{appendix_lie_algebra:def_root}).
}
\es
\begin{theorem}
If $A$ is chosen such that the secular equation has the maximum number of distinct roots,
then only the root $\rho=0$ is degenerate.
Further if $r$ is the multiplicity of that root, there exist $r$ linearly independent
generators $H_i$, which mutually commute
\bq
 \left[ H_i, H_j \right] & = & 0, 
 \;\;\;\;\;\; i,j \in \left\{1,\dots,r\right\}.
\eq
The multiplicity $r$ of the root $\rho=0$ equals the rank of the Lie algebra.
\end{theorem}
The generators $H_1, \dots, H_r$ generate an Abelian sub-algebra of $\mathfrak{g}$.
This sub-algebra is called the 
\index{Cartan sub-algebra}
{\bf Cartan sub-algebra} of $\mathfrak{g}$.

\begin{figure}
\begin{center}
\includegraphics[scale=1.0]{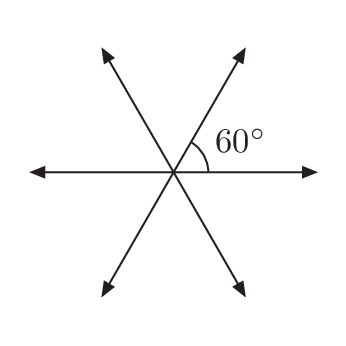}
\caption{\label{appendix_lie_algebra:fig_root_system_su3}
The root system of ${\mathfrak s}{\mathfrak u}(3)$ (or equivalently of $A_2$).
}
\end{center}
\end{figure}

It is a standard convention to use Latin indices $i \in \{1,\dots,r\}$ to denote the $r$ mutually commuting
generators $H_i$ and greek indices and the letter $E$ to denote the remaining $(n-r)$ generators $E_\alpha$.
\begin{theorem}
For any semi-simple Lie algebra, non-zero roots occur in pairs of opposite sign and are denoted
$E_\alpha$ and $E_{-\alpha}$ with $\alpha\in\{1,\dots,\frac{1}{2}(n-r)\}$.
\end{theorem}
We thus have the 
\index{Cartan standard form}
{\bf Cartan standard form} or {\bf Cartan basis}:
\bq
 \left[ H_i, H_j \right] & = & 0,
 \nonumber \\
 \left[ H_i, E_\alpha \right] & = & \rho\left(\alpha,i\right) E_\alpha.
\eq
We write $\alpha_i = \rho(\alpha,i)$. With this notation
the last equation may be written as
\bq
 \left[ H_i, E_\alpha \right] & = & \alpha_i E_\alpha.
\eq
The standard normalisation for the Cartan basis is
\bq
 \sum\limits_{\alpha=1}^{\frac{1}{2}(n-r)} \alpha_i \alpha_j & = & \delta_{ij}.
\eq
For a fixed generator $E_\alpha$ we collect the $r$ numbers $\alpha_1,\dots,\alpha_r$ into one vector
\bq
 \vec{\alpha}
 & = &
 \left( \alpha_1, \dots, \alpha_r \right)^T.
\eq
The vector $\vec{\alpha}$ is called the 
\index{root vector}
{\bf root vector} of $E_\alpha$.
The set of root vectors $\vec{\alpha}$ for all $E_\alpha$ is called the 
\index{root system}
{\bf root system} of $\mathfrak{g}$.
\\
\\
\bs
{\it \refstepcounter{exercise}
{\bf Exercise \theexercise}: 
Consider the Lie algebra ${\mathfrak s}{\mathfrak u}(2)$:
Start from the generators
\bq
 & &
 I^1 =
 \frac{1}{2} 
 \left( \begin{array}{rr}
 0 & 1 \\
 1 & 0 \\
 \end{array} \right),
\;\;\;
 I^2 =
 \frac{1}{2} 
 \left( \begin{array}{rr}
 0 & -i \\
 i & 0 \\
 \end{array} \right),
\;\;\;
 I^3 =
 \frac{1}{2} 
 \left( \begin{array}{rr}
 1 & 0 \\
 0 & -1 \\
 \end{array} \right).
\eq
These generators are proportional to the Pauli matrices and normalised as
\bq
 \mathrm{Tr}\left( I^a I^b \right) & = & \frac{1}{2} \delta^{ab}.
\eq
The commutators are given by 
\bq
 \left[ I^a, I^b \right] & = & i \eps^{abc} I^c,
\eq
where $\eps^{abc}$ denotes the totally antisymmetric tensor.
Start from $A=I^3$. Determine for this choice the roots, the Cartan standard form and the root vectors.
}
\es
\\
\\
Let us look at an example: The Cartan basis for ${\mathfrak s}{\mathfrak u}(3)$ is
\begin{alignat}{6}
 & H_1 
 & = &
 \frac{1}{\sqrt{6}} 
 \left( \begin{array}{ccc}
 1 & 0 & 0 \\
 0 & -1 & 0 \\
 0 & 0 & 0 \\
 \end{array} \right),
 \;\;\;
 & &
 H_2 
 & = &
 \frac{1}{3\sqrt{2}} 
 \left( \begin{array}{ccc}
 1 & 0 & 0 \\
 0 & 1 & 0 \\
 0 & 0 & -2 \\
 \end{array} \right),
 \;\;\;
 & &
 & &
 \\
 & E_{1}
 & = &
 \frac{1}{\sqrt{3}} 
 \left( \begin{array}{ccc}
 0 & 1 & 0 \\
 0 & 0 & 0 \\
 0 & 0 & 0 \\
 \end{array} \right),
 & &
 E_{2}
 & = &
 \frac{1}{\sqrt{3}} 
 \left( \begin{array}{ccc}
 0 & 0 & 1 \\
 0 & 0 & 0 \\
 0 & 0 & 0 \\
 \end{array} \right),
 & &
 E_{3}
 & = &
 \frac{1}{\sqrt{3}} 
 \left( \begin{array}{ccc}
 0 & 0 & 0 \\
 0 & 0 & 1 \\
 0 & 0 & 0 \\
 \end{array} \right),
 \nonumber \\
 & E_{-1}
 & = &
 \frac{1}{\sqrt{3}} 
 \left( \begin{array}{ccc}
 0 & 0 & 0 \\
 1 & 0 & 0 \\
 0 & 0 & 0 \\
 \end{array} \right),
 & &
 E_{-2}
 & = &
 \frac{1}{\sqrt{3}} 
 \left( \begin{array}{ccc}
 0 & 0 & 0 \\
 0 & 0 & 0 \\
 1 & 0 & 0 \\
 \end{array} \right),
 & &
 E_{-3}
 & = &
 \frac{1}{\sqrt{3}} 
 \left( \begin{array}{ccc}
 0 & 0 & 0 \\
 0 & 0 & 0 \\
 0 & 1 & 0 \\
 \end{array} \right).
 \nonumber
\end{alignat}
The roots of $E_1$, $E_2$ and $E_3$ are
\begin{alignat}{4}
 & \left[ H_1, E_1 \right] 
 & \; = \; &
 \frac{1}{3} \sqrt{6} E_1,
 & &
 \left[ H_2, E_1 \right] 
 & \; = \; &
 0,
\nonumber \\
 & \left[ H_1, E_2 \right] 
 & \; = \; &
 \frac{1}{6} \sqrt{6} E_2,
 & &
 \left[ H_2, E_2 \right] 
 & \; = \; &
 \frac{1}{2} \sqrt{2} E_2
\nonumber \\
 & \left[ H_1, E_3 \right] 
 & \; = \; &
 -\frac{1}{6} \sqrt{6} E_3,
 \;\;\;
 & &
 \left[ H_2, E_3 \right] 
 & \; = \; &
 \frac{1}{2} \sqrt{2} E_3,
\end{alignat}
and similar for $E_{-1}$, $E_{-2}$ and $E_{-3}$.
For the root vectors we obtain
\begin{alignat}{6}
 &
 \vec{\alpha}(E_1) 
 & = & \left( \begin{array}{c} \frac{1}{3} \sqrt{6} \\ 0 \\ \end{array} \right),
 & &
 \vec{\alpha}(E_2) 
 & = & \left( \begin{array}{c} \frac{1}{6} \sqrt{6} \\ \frac{1}{2} \sqrt{2} \\ \end{array} \right),
 & &
 \vec{\alpha}(E_3) 
 & = & \left( \begin{array}{c} - \frac{1}{6} \sqrt{6} \\ \frac{1}{2} \sqrt{2} \\ \end{array} \right),
 \nonumber \\
 &
 \vec{\alpha}(E_{-1}) 
 & = & \left( \begin{array}{c} -\frac{1}{3} \sqrt{6} \\ 0 \\ \end{array} \right),
 \;\;\;
 & &
 \vec{\alpha}(E_{-2}) 
 & = & \left( \begin{array}{c} -\frac{1}{6} \sqrt{6} \\ -\frac{1}{2} \sqrt{2} \\ \end{array} \right),
 \;\;\;
 & &
 \vec{\alpha}(E_{-3}) 
 & = & \left( \begin{array}{c} \frac{1}{6} \sqrt{6} \\ -\frac{1}{2} \sqrt{2} \\ \end{array} \right).
\end{alignat}
Figure~\ref{appendix_lie_algebra:fig_root_system_su3}
shows the root system for ${\mathfrak s}{\mathfrak u}(3)$.

There are a few theorems on root vectors:
\begin{theorem}
\label{appendix_lie_algebra:theorem_root_vectors_1}
If $\vec{\alpha}$ is a root vector, so is $-\vec{\alpha}$, 
(since roots always occur in pairs of opposite sign).
\end{theorem}
\begin{theorem}
\label{appendix_lie_algebra:theorem_root_vectors_2}
If $\vec{\alpha}$ and $\vec{\beta}$ are root vectors then
\bq
\frac{2 \vec{\alpha} \cdot \vec{\beta}}{|\vec{\alpha}|^2} & \mbox{and} &  
\frac{2 \vec{\alpha} \cdot \vec{\beta}}{|\vec{\beta}|^2} 
\eq
are integers.
\end{theorem}
\begin{theorem}
\label{appendix_lie_algebra:theorem_root_vectors_3}
If $\vec{\alpha}$ and $\vec{\beta}$ are root vectors so is
\bq
\vec{\gamma} & = & \vec{\beta} - \frac{2 \vec{\alpha} \cdot \vec{\beta}}{|\vec{\alpha}|^2} \vec{\alpha}
\eq
\end{theorem}
Let us now investigate the implications of these theorems.
We start with theorem~\ref{appendix_lie_algebra:theorem_root_vectors_2}.
Denote the two integers by $p$ and $q$, i.e.
\bq
\frac{2 \vec{\alpha} \cdot \vec{\beta}}{|\vec{\alpha}|^2} \; = \; p,
 & &  
\frac{2 \vec{\alpha} \cdot \vec{\beta}}{|\vec{\beta}|^2} \; = \; q.
\eq
Then
\bq
 \frac{\left(\vec{\alpha}\cdot \vec{\beta}\right)^2}{|\vec{\alpha}|^2 |\vec{\beta}|^2}
 & = & 
 \frac{p q }{4} 
 \; = \; \cos^2 \theta 
 \; \le \; 
 1,
\eq
where $\theta$ denotes the angle between the root vectors $\vec{\alpha}$ and $\vec{\beta}$.
Therefore
\bq 
 p q & \le & 4.
\eq
As $p$ and $q$ are integers, this puts strong constraints on the angle between $\vec{\alpha}$ and $\vec{\beta}$
and the ratio of their lengths.
We have
\bq 
 \cos^2 \theta & \in & \left\{ 0, \frac{1}{4}, \frac{1}{2}, \frac{3}{4}, 1 \right\}.
\eq
This restricts the angle between two root vectors to
\bq
 0^\circ, 30^\circ, 45^\circ, 60^\circ, 90^\circ, 120^\circ, 135^\circ, 150^\circ, 180^\circ.
\eq
This is a finite list. Let's go through all possibilities:
\begin{itemize}

\item Case $\cos^2 \theta = 1$: This implies $\cos \theta = \pm 1$ and the angle is either $\theta=0^\circ$ or
$\theta=180^\circ$. We further have $|\vec{\alpha}|=|\vec{\beta}|$ and therefore either
$\vec{\alpha}=\vec{\beta}$ or $\vec{\alpha}=-\vec{\beta}$.

\item Case $\cos^2 \theta = \frac{3}{4}$: This implies $\cos \theta = \pm \frac{1}{2} \sqrt{3}$ and the angle is either $\theta=30^\circ$ or
$\theta=150^\circ$.
We have $p q =3$ and therefore either $p=1, q=3$ or $p=3, q=1$.
Let us first discuss $p=1, q=3$.
This means
\bq
\frac{2 \vec{\alpha} \cdot \vec{\beta}}{|\vec{\alpha}|^2} = 1,
 & & 
\frac{2 \vec{\alpha} \cdot \vec{\beta}}{|\vec{\beta}|^2} = 3.
\eq
Therefore
\bq
 \frac{|\vec{\alpha}|^2}{|\vec{\beta}|^2} & = & 3.
\eq
The case $p=3, q=1$ is similar and in summary we obtain
\bq
 \frac{|\vec{\alpha}|^2}{|\vec{\beta}|^2} & \in & \left\{ \frac{1}{3}, 3 \right\}.
\eq

\item Case $\cos^2 \theta = \frac{1}{2}$: This implies $\cos \theta = \pm \frac{1}{2} \sqrt{2}$ and the angle is either $\theta=45^\circ$ or
$\theta=135^\circ$.
We have $p q =2$ and either $p=1, q=2$ or $p=2, q=1$.
It follows
\bq
 \frac{|\vec{\alpha}|^2}{|\vec{\beta}|^2} & \in & \left\{ \frac{1}{2}, 2 \right\}.
\eq

\item Case $\cos^2 \theta = \frac{1}{4}$: This implies $\cos \theta = \pm \frac{1}{2}$ and the angle is either $\theta=60^\circ$ or
$\theta=120^\circ$.
We have $p q =1$ and hence $p=1, q=1$.
It follows
\bq
 \frac{|\vec{\alpha}|^2}{|\vec{\beta}|^2} & = & 1.
\eq

\item Case $\cos^2 \theta = 0$: This implies $\cos \theta = 0$ and the angle is $\theta=90^\circ$.
In this case we have $p=0$ and $q=0$. This leaves the ratio $|\vec{\alpha}|^2/|\vec{\beta}|^2$ undetermined.

\end{itemize}
Let us now explore theorem~\ref{appendix_lie_algebra:theorem_root_vectors_3}.
We have already seen that if 
$\vec{\alpha}$ and $\vec{\beta}$ are root vectors so is
\bq
\vec{\gamma} & = & \vec{\beta} - \frac{2 \vec{\alpha} \cdot \vec{\beta}}{|\vec{\alpha}|^2} \vec{\alpha}.
\eq
Let us now put this a little bit more formally.
For any root vector $\alpha$ we define a mapping $W_\alpha$ from the set of root vectors
to the set of root vectors by
\bq
 W_\alpha(\beta) 
 & = &
 \vec{\beta} - \frac{2 \vec{\alpha} \cdot \vec{\beta}}{|\vec{\alpha}|^2} \vec{\alpha}.
\eq
$W_\alpha$ can be described as the reflection by the plane $\Omega_\alpha$ perpendicular to $\alpha$.
It is clear that this mapping is an involution: After two reflections one obtains the original
root vector again.
The set of all these mappings $W_\alpha$ generates a group, which is called the 
\index{Weyl group}
{\bf Weyl group}.

Since $W_\alpha$ maps a root vector to another root vector, we have the following corollary:
\begin{corollary}
The set of root vectors is invariant under the Weyl group.
\end{corollary}
For root vectors we define an 
\index{ordering of root vectors}
{\bf ordering} as follows:
$\vec{\alpha}$ is said to be higher than $\vec{\alpha}'$ if the $r^{\mbox{th}}$ component of 
$(\vec{\alpha}-\vec{\alpha}')$ is positive (if zero look at the $(r-1)^{\mbox{th}}$ component etc.).
If $\vec{\alpha}$ is higher than $\vec{\alpha}'$ we write $\vec{\alpha} > \vec{\alpha}'$.
\begin{definition}
A root vector $\vec{\alpha}$ is called 
\index{positive root vector}
{\bf positive}, 
if $\vec{\alpha}>\vec{0}$.
\end{definition}
Therefore the set of non-zero root vectors $R$ decomposes into
\bq
 R & = & R^+ \cup R^-,
\eq
where $R^+$ denotes the positive roots and $R^-$ denotes the negative roots.
\begin{definition}
The 
\index{Weyl chamber}
(closed) {\bf Weyl chamber} relative to a given ordering
is the set of points $\vec{x}$ in the $r$-dimensional space of root vectors,
such that
\bq
 2 \frac{\vec{x} \cdot \vec{\alpha}}{|\vec{\alpha}|^2} \ge 0
 & & \forall \; \vec{\alpha} \in R^+.
\eq
\end{definition}
An example of a root system, the positive roots and the Weyl chamber 
is shown for the Lie algebra ${\mathfrak s}{\mathfrak u}(3)$ in
\begin{figure}
\begin{center}
\includegraphics[scale=1.0]{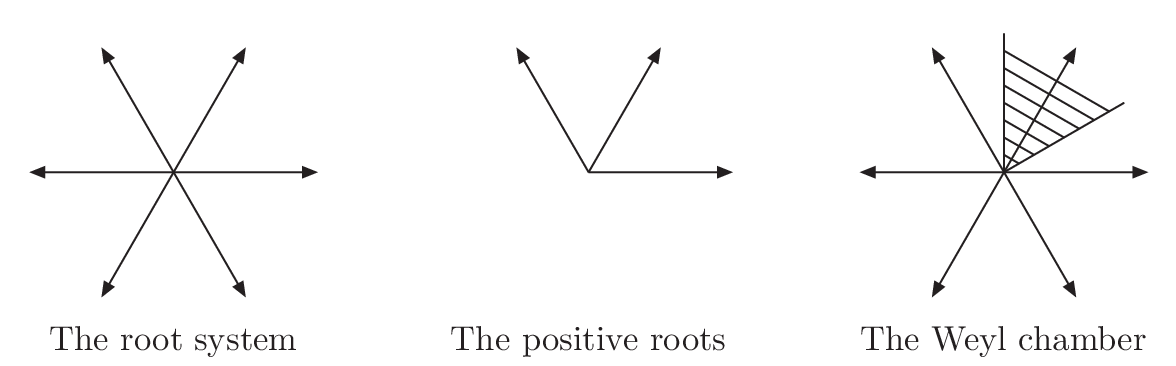}
\caption{\label{appendix_lie_algebra:fig_weyl_chamber_su3}
The root system, the positive roots and the Weyl chamber for ${\mathfrak s}{\mathfrak u}(3)$ (or equivalently $A_2$).
}
\end{center}
\end{figure}
fig.~\ref{appendix_lie_algebra:fig_weyl_chamber_su3}.

Let us summarise: The root system $R$ of a Lie algebra ${\mathfrak g}$ has the following properties:
\begin{enumerate}
\item $R$ is a finite set.
\item If $\vec{\alpha}\in R$, then also $-\vec{\alpha}\in R$.
\item For any $\vec{\alpha}\in R$ the reflection $W_\alpha$ maps $R$ to itself.
\item If $\vec{\alpha}$ and $\vec{\beta}$ are root vectors then
$2 \vec{\alpha} \cdot \vec{\beta}/|\alpha|^2$
is an integer.
\end{enumerate}
This puts strong constraints on the geometry of a root system.
Before we embark on the general classification, let us investigate the possible root systems of 
rank $1$ and $2$.
For rank $1$ the root vectors are one-dimensional
and the only possibility is
\begin{figure}
\begin{center}
\includegraphics[scale=1.0]{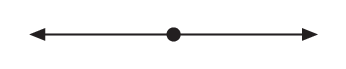}
\caption{\label{appendix_lie_algebra:fig_A1}
The root system $A_1$.
}
\end{center}
\end{figure}
the one shown in fig.~\ref{appendix_lie_algebra:fig_A1}.
This is the root system of ${\mathfrak s}{\mathfrak u}(2)$.

For rank $2$ we first note that due to property (3) the angle between two roots must be the same
for any pair of adjacent roots.
It will turn out that any of the four angles $90^\circ$, $60^\circ$, $45^\circ$ and $30^\circ$ can occur.
Once this angle is specified, the relative lengths of the roots are fixed except for the case of right angles.

Let us start with the case $\theta=90^\circ$. 
Up to rescaling the root system is
\begin{figure}
\begin{center}
\includegraphics[scale=1.0]{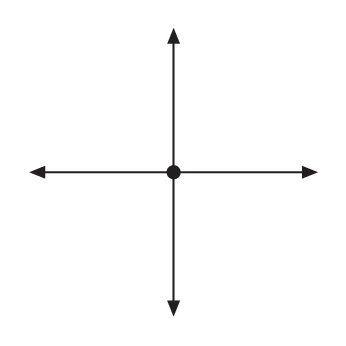}
\caption{\label{appendix_lie_algebra:fig_A1_A1}
The root system $A_1 \times A_1$.
}
\end{center}
\end{figure}
the one shown in fig.~\ref{appendix_lie_algebra:fig_A1_A1}.
This corresponds to ${\mathfrak s}{\mathfrak u}(2) \oplus {\mathfrak s}{\mathfrak u}(2)$. 
This Lie algebra is semi-simple, but not simple.

\begin{figure}
\begin{center}
\includegraphics[scale=1.0]{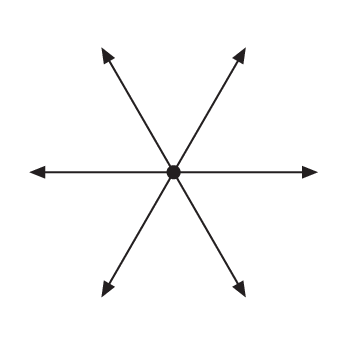}
\caption{\label{appendix_lie_algebra:fig_A2}
The root system $A_2$.
}
\end{center}
\end{figure}

For the angle $\theta=60^\circ$ we have the root system shown in fig.~\ref{appendix_lie_algebra:fig_A2}.
This is the root system of ${\mathfrak s}{\mathfrak u}(3)$.

\begin{figure}
\begin{center}
\includegraphics[scale=1.0]{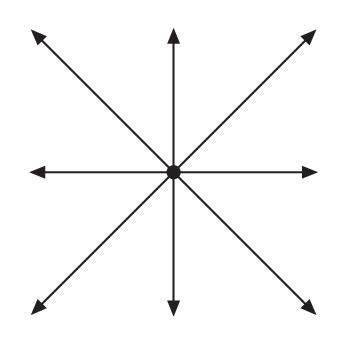}
\caption{\label{appendix_lie_algebra:fig_B2}
The root system $B_2$.
}
\end{center}
\end{figure}

For the angle $\theta=45^\circ$ we have
the root system shown in fig.~\ref{appendix_lie_algebra:fig_B2}.
This is the root system of ${\mathfrak s}{\mathfrak o}(5)$.

\begin{figure}
\begin{center}
\includegraphics[scale=1.0]{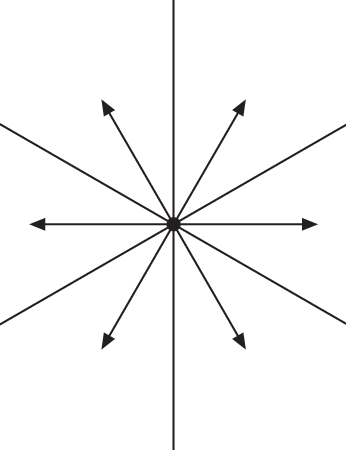}
\caption{\label{appendix_lie_algebra:fig_G2}
The root system $G_2$.
}
\end{center}
\end{figure}

Finally, for $\theta=30^\circ$ we have
the root system shown in fig.~\ref{appendix_lie_algebra:fig_G2}.
This is the root system of the exceptional Lie group $G_2$.

\section{Dynkin diagrams}

Let us try to reduce further the data of a root system. We already learned that with the help of an ordering
we can divide the non-zero root vectors into a disjoint union of positive and negative roots:
\bq
 R & = & R^+ \cup R^-.
\eq
\begin{definition}
A positive root vector is called 
\index{simple root vector}
{\bf simple} if it is not the sum of two other positive roots.
\end{definition}
We illustrate this for the Lie algebra ${\mathfrak s}{\mathfrak u}(3)$
\begin{figure}
\begin{center}
\includegraphics[scale=1.0]{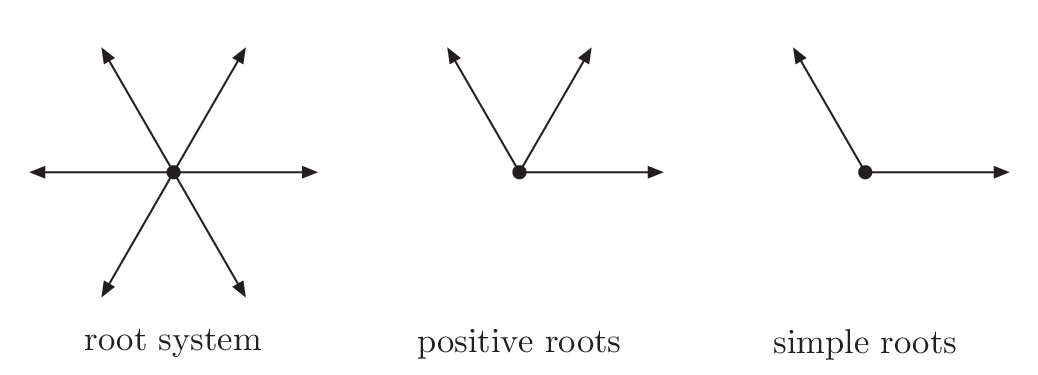}
\caption{\label{appendix_lie_algebra:fig_example_simple_roots}
The root system, the positive roots and the simple roots for ${\mathfrak s}{\mathfrak u}(3)$ (or equivalently $A_2$).
}
\end{center}
\end{figure}
in fig.~\ref{appendix_lie_algebra:fig_example_simple_roots}.
The angle between the two simple roots is $\theta=120^\circ$.

The 
\index{Dynkin diagram}
{\bf Dynkin diagram} 
of the root system is constructed by drawing one vertex $\circ$ for each simple root
and joining two vertices by a number of lines depending on the angle $\theta$ between the two roots:
\begin{center}
\begin{tabular}{lllll}
no lines 
& &
\begin{picture}(30,10)(0,0)
\CArc(5,5)(3,0,360)
\CArc(25,5)(3,0,360)
\end{picture}
& &
if $\theta=90^\circ$ 
\\
& & & & \\
one line 
& &
\begin{picture}(30,10)(0,0)
\CArc(5,5)(3,0,360)
\CArc(25,5)(3,0,360)
\Line(8,5)(22,5)
\end{picture}
& &
if $\theta=120^\circ$ 
\\
& & & & \\
two lines 
& &
\begin{picture}(30,10)(0,0)
\CArc(5,5)(3,0,360)
\CArc(25,5)(3,0,360)
\Line(8,6)(22,6)
\Line(8,4)(22,4)
\Line(17,5)(12,10)
\Line(17,5)(12,0)
\end{picture}
& &
if $\theta=135^\circ$ 
\\
& & & & \\
three lines 
& &
\begin{picture}(30,10)(0,0)
\CArc(5,5)(3,0,360)
\CArc(25,5)(3,0,360)
\Line(8,7)(22,7)
\Line(8,3)(22,3)
\Line(8,5)(22,5)
\Line(17,5)(12,10)
\Line(17,5)(12,0)
\end{picture}
& &
if $\theta=150^\circ$ 
\\
\end{tabular}
\end{center}
When there is one line, the roots have the same length. If two roots are connected by two or three
lines, an arrow is drawn pointing from the longer to the shorter root.
\\
\\
Example: The Dynkin diagram of ${\mathfrak s}{\mathfrak u}(3)$ is 
\begin{center}
\begin{picture}(30,10)(0,0)
\CArc(5,5)(3,0,360)
\CArc(25,5)(3,0,360)
\Line(8,5)(22,5)
\end{picture}
\end{center}
We have the following theorems:
\begin{theorem}
Two complex semi-simple Lie algebras are isomorphic if and only if they have the same Dynkin diagram.
\end{theorem}
\begin{theorem}
A complex semi-simple Lie algebra is simple if and only if its Dynkin diagram is connected.
\end{theorem}
\begin{theorem}
\label{appendix_lie_algebra:Dynkin_classification}
Classification of simple Lie algebras:
The connected Dynkin diagrams can be grouped into four families ($A_n$, $B_n$, $C_n$, $D_n$)
and a set of five exceptional Dynkin diagrams.
The four families are
\begin{itemize}
\item $A_n$ for $n \ge 1$
\begin{center}
\begin{picture}(110,10)(0,0)
\CArc(5,5)(2,0,360)
\CArc(25,5)(2,0,360)
\CArc(45,5)(2,0,360)
\CArc(85,5)(2,0,360)
\CArc(105,5)(2,0,360)
\Line(7,5)(23,5)
\Line(27,5)(43,5)
\Line(47,5)(53,5)
\Line(77,5)(83,5)
\Line(87,5)(103,5)
\Text(65,5)[t]{$\cdots$}
\Text(5,2)[t]{$\alpha_1$}
\Text(25,2)[t]{$\alpha_2$}
\Text(45,2)[t]{$\alpha_3$}
\Text(85,2)[t]{$\alpha_{n-1}$}
\Text(105,2)[t]{$\alpha_n$}
\end{picture}
\end{center}
\item $B_n$ for $n \ge 2$
\begin{center}
\begin{picture}(110,10)(0,0)
\CArc(5,5)(2,0,360)
\CArc(25,5)(2,0,360)
\CArc(45,5)(2,0,360)
\CArc(85,5)(2,0,360)
\CArc(105,5)(2,0,360)
\Line(7,5)(23,5)
\Line(27,5)(43,5)
\Line(47,5)(53,5)
\Line(77,5)(83,5)
\Line(87,6)(103,6)
\Line(87,4)(103,4)
\Line(95,5)(93,3)
\Line(95,5)(93,7)
\Text(65,5)[t]{$\cdots$}
\Text(5,2)[t]{$\alpha_1$}
\Text(25,2)[t]{$\alpha_2$}
\Text(45,2)[t]{$\alpha_3$}
\Text(85,2)[t]{$\alpha_{n-1}$}
\Text(105,2)[t]{$\alpha_n$}
\end{picture}
\end{center}
\item $C_n$ for $n \ge 3$
\begin{center}
\begin{picture}(110,10)(0,0)
\CArc(5,5)(2,0,360)
\CArc(25,5)(2,0,360)
\CArc(45,5)(2,0,360)
\CArc(85,5)(2,0,360)
\CArc(105,5)(2,0,360)
\Line(7,5)(23,5)
\Line(27,5)(43,5)
\Line(47,5)(53,5)
\Line(77,5)(83,5)
\Line(87,6)(103,6)
\Line(87,4)(103,4)
\Line(95,5)(97,3)
\Line(95,5)(97,7)
\Text(65,5)[t]{$\cdots$}
\Text(5,2)[t]{$\alpha_1$}
\Text(25,2)[t]{$\alpha_2$}
\Text(45,2)[t]{$\alpha_3$}
\Text(85,2)[t]{$\alpha_{n-1}$}
\Text(105,2)[t]{$\alpha_n$}
\end{picture}
\end{center}
\item $D_n$ for $n \ge 4$
\begin{center}
\begin{picture}(110,10)(0,0)
\CArc(5,5)(2,0,360)
\CArc(25,5)(2,0,360)
\CArc(45,5)(2,0,360)
\CArc(85,5)(2,0,360)
\CArc(105,0)(2,0,360)
\CArc(105,10)(2,0,360)
\Line(7,5)(23,5)
\Line(27,5)(43,5)
\Line(47,5)(53,5)
\Line(77,5)(83,5)
\Line(87,5)(103,10)
\Line(87,5)(103,0)
\Text(65,5)[t]{$\cdots$}
\Text(5,2)[t]{$\alpha_1$}
\Text(25,2)[t]{$\alpha_2$}
\Text(45,2)[t]{$\alpha_3$}
\Text(85,2)[t]{$\alpha_{n-2}$}
\Text(109,10)[l]{$\alpha_{n-1}$}
\Text(109,0)[l]{$\alpha_{n}$}
\end{picture}
\end{center}
\end{itemize}
The exceptional Dynkin diagrams are
\begin{itemize}
\item $E_6$
\begin{center}
\begin{picture}(110,10)(0,0)
\CArc(5,5)(2,0,360)
\CArc(25,5)(2,0,360)
\CArc(45,5)(2,0,360)
\CArc(65,5)(2,0,360)
\CArc(85,5)(2,0,360)
\CArc(45,-5)(2,0,360)
\Line(7,5)(23,5)
\Line(27,5)(43,5)
\Line(47,5)(63,5)
\Line(67,5)(83,5)
\Line(45,3)(45,-3)
\Text(5,8)[b]{$\alpha_1$}
\Text(25,8)[b]{$\alpha_2$}
\Text(45,8)[b]{$\alpha_3$}
\Text(65,8)[b]{$\alpha_5$}
\Text(85,8)[b]{$\alpha_6$}
\Text(49,-5)[l]{$\alpha_4$}
\end{picture}
\end{center}
\item $E_7$
\begin{center}
\begin{picture}(110,10)(0,0)
\CArc(5,5)(2,0,360)
\CArc(25,5)(2,0,360)
\CArc(45,5)(2,0,360)
\CArc(65,5)(2,0,360)
\CArc(85,5)(2,0,360)
\CArc(105,5)(2,0,360)
\CArc(45,-5)(2,0,360)
\Line(7,5)(23,5)
\Line(27,5)(43,5)
\Line(47,5)(63,5)
\Line(67,5)(83,5)
\Line(87,5)(103,5)
\Line(45,3)(45,-3)
\Text(5,8)[b]{$\alpha_1$}
\Text(25,8)[b]{$\alpha_2$}
\Text(45,8)[b]{$\alpha_3$}
\Text(65,8)[b]{$\alpha_5$}
\Text(85,8)[b]{$\alpha_6$}
\Text(105,8)[b]{$\alpha_7$}
\Text(49,-5)[l]{$\alpha_4$}
\end{picture}
\end{center}
\item $E_8$
\begin{center}
\begin{picture}(110,10)(0,0)
\CArc(5,5)(2,0,360)
\CArc(25,5)(2,0,360)
\CArc(45,5)(2,0,360)
\CArc(65,5)(2,0,360)
\CArc(85,5)(2,0,360)
\CArc(105,5)(2,0,360)
\CArc(125,5)(2,0,360)
\CArc(45,-5)(2,0,360)
\Line(7,5)(23,5)
\Line(27,5)(43,5)
\Line(47,5)(63,5)
\Line(67,5)(83,5)
\Line(87,5)(103,5)
\Line(107,5)(123,5)
\Line(45,3)(45,-3)
\Text(5,8)[b]{$\alpha_1$}
\Text(25,8)[b]{$\alpha_2$}
\Text(45,8)[b]{$\alpha_3$}
\Text(65,8)[b]{$\alpha_5$}
\Text(85,8)[b]{$\alpha_6$}
\Text(105,8)[b]{$\alpha_7$}
\Text(125,8)[b]{$\alpha_8$}
\Text(49,-5)[l]{$\alpha_4$}
\end{picture}
\end{center}
\item $F_4$
\begin{center}
\begin{picture}(110,10)(0,0)
\CArc(5,5)(2,0,360)
\CArc(25,5)(2,0,360)
\CArc(45,5)(2,0,360)
\CArc(65,5)(2,0,360)
\Line(7,5)(23,5)
\Line(27,4)(43,4)
\Line(27,6)(43,6)
\Line(35,5)(37,7)
\Line(35,5)(37,3)
\Line(47,5)(63,5)
\Text(5,8)[b]{$\alpha_1$}
\Text(25,8)[b]{$\alpha_2$}
\Text(45,8)[b]{$\alpha_3$}
\Text(65,8)[b]{$\alpha_4$}
\end{picture}
\end{center}
\item $G_2$
\begin{center}
\begin{picture}(110,10)(0,0)
\CArc(25,5)(2,0,360)
\CArc(45,5)(2,0,360)
\Line(27,3)(43,3)
\Line(27,5)(43,5)
\Line(27,7)(43,7)
\Line(35,5)(39,9)
\Line(35,5)(39,1)
\Text(25,8)[b]{$\alpha_1$}
\Text(45,8)[b]{$\alpha_2$}
\end{picture}
\end{center}
\end{itemize}
\end{theorem}
Up to now we considered Lie algebras over the complex numbers ${\mathbb C}$.
We are also interested in real Lie groups, whose Lie algebra is a real Lie algebra.
Starting from a real simple Lie algebra $\mathfrak{g}$ we consider the complexification $\mathfrak{g}^{\mathbb C}$
of $\mathfrak{g}$. The latter is classified by theorem~\ref{appendix_lie_algebra:Dynkin_classification}.
Thus a classification of all real simple Lie algebras amounts to a classification of all real forms 
of the complex simple Lie algebras $\mathfrak{g}^{\mathbb C}$ from theorem~\ref{appendix_lie_algebra:Dynkin_classification}.

The types of the Lie algebras of the classical real compact simple Lie groups are
\bq
 \mathrm{SU}\left(n+1\right) & : & A_n,
 \nonumber \\
 \mathrm{SO}\left(2n+1\right) & : & B_n,
 \nonumber \\
 \mathrm{Sp}\left(n\right) & : & C_n,
 \nonumber \\
 \mathrm{SO}\left(2n\right) & : & D_n.
\eq
In order to prove theorem~\ref{appendix_lie_algebra:Dynkin_classification} 
we have to show that the only possible connected Dynkin diagrams are the ones mentioned in 
theorem~\ref{appendix_lie_algebra:Dynkin_classification}
and that for every connected Dynkin diagrams from the list of theorem~\ref{appendix_lie_algebra:Dynkin_classification}
there is a simple Lie algebra corresponding to it.

For the first part of the proof it is sufficient to consider only the angles between the simple roots, 
the relative lengths do not enter the proof.
We may therefore drop the arrows in the Dynkin diagrams.
Such diagrams, without the arrows to indicate the relative lengths, are called 
\index{Coxeter diagrams}
{\bf Coxeter diagrams}.
Let us now consider a Coxeter diagram with $n$ vertices. Two vertices are connected by
either $0$, $1$, $2$ or $3$ lines.
We call a Coxeter diagram
\index{admissible Coxeter diagram}
{\bf admissible} if there are $n$ independent unit vectors $\vec{e}_1$, \dots, $\vec{e}_n$
in an Euclidean space with the angle $\theta$ between $\vec{e}_i$ and $\vec{e}_j$ as follows:
\begin{center}
\begin{tabular}{lllll}
no lines 
& &
\begin{picture}(30,10)(0,0)
\CArc(5,5)(3,0,360)
\CArc(25,5)(3,0,360)
\end{picture}
& &
if $\theta=90^\circ$ 
\\
& & & & \\
one line 
& &
\begin{picture}(30,10)(0,0)
\CArc(5,5)(3,0,360)
\CArc(25,5)(3,0,360)
\Line(8,5)(22,5)
\end{picture}
& &
if $\theta=120^\circ$ 
\\
& & & & \\
two lines 
& &
\begin{picture}(30,10)(0,0)
\CArc(5,5)(3,0,360)
\CArc(25,5)(3,0,360)
\Line(8,6)(22,6)
\Line(8,4)(22,4)
\end{picture}
& &
if $\theta=135^\circ$ 
\\
& & & & \\
three lines 
& &
\begin{picture}(30,10)(0,0)
\CArc(5,5)(3,0,360)
\CArc(25,5)(3,0,360)
\Line(8,7)(22,7)
\Line(8,3)(22,3)
\Line(8,5)(22,5)
\end{picture}
& &
if $\theta=150^\circ$ 
\\
\end{tabular}
\end{center}
We prove theorem~\ref{appendix_lie_algebra:Dynkin_classification}
with the help of the following proposition:
\begin{proposition}
\label{appendix_lie_algebra:proposition_admissible_Coxeter_diagrams}
The only connected admissible Coxeter graphs are the ones listed in theorem~\ref{appendix_lie_algebra:Dynkin_classification} (without the arrows).
\end{proposition}
To prove this proposition, we will first prove the following four lemmata:
\begin{lemma}
\label{appendix_lie_algebra:lemma_Coxeter_1}
Any sub-diagram of an admissible diagram, obtained by removing some vertices
and all lines attached to them, will also be admissible.
\end{lemma}
\begin{proof}
Suppose we have an admissible diagram with $n$ vertices. By definition there are $n$ vectors $\vec{e}_j$, such that
the angle between a pair of vectors is in the set
\bq
 \{ 90^\circ, 120^\circ, 135^\circ, 150^\circ \}
\eq
Removing some of the vectors $\vec{e}_j$ does not change the angles between the remaining ones.
Therefore any sub-diagram of an admissible diagram is again admissible.
\end{proof}
\begin{lemma}
\label{appendix_lie_algebra:lemma_Coxeter_2}
There are at most $(n-1)$ pairs of vertices that are connected by lines. The diagram has no loops.
\end{lemma}
\begin{proof}
We have
\bq
 2 \vec{e}_i \cdot \vec{e}_j \in \{ 0, -1, -\sqrt{2}, -\sqrt{3} \}
\eq
Therefore if $\vec{e}_i$ and $\vec{e}_j$ are connected we have $\theta>90^\circ$ and
\bq
  2 \vec{e}_i \cdot \vec{e}_j & \le & -1.
\eq
Now
\bq
 0 & < & \left( \sum\limits_i \vec{e}_i \right) \cdot \left( \sum\limits_j \vec{e}_i \right)
 \;\; = \;\;n + 2 \sum\limits_{i<j} \vec{e}_i \cdot \vec{e}_j
 \;\;< \;\;n - \# \;\mbox{connected pairs}.
\eq
Therefore
\bq
\# \;\mbox{connected pairs} & < & n.
\eq
Connecting $n$ vertices with $(n-1)$ connections (of either $1$, $2$ or $3$ lines)
implies that there are no loops.
\end{proof}
\begin{lemma}
\label{appendix_lie_algebra:lemma_Coxeter_3}
No vertex has more than three lines attached to it.
\end{lemma}
\begin{proof}
We first note that
\bq
 \left( 2 \vec{e}_i \cdot \vec{e}_j \right)^2 & = & \#\;\mbox{number of lines between}\;\vec{e}_i\;\mbox{and}\;\vec{e}_j.
\eq
Consider the vertex $\vec{e}_1$ and let $\vec{e}_2, \dots \vec{e}_j$, be the vertices connected to $\vec{e}_1$.
We want to show
\bq
 \sum\limits_{i=2}^j \left( 2 \vec{e}_1 \cdot \vec{e}_i \right)^2 & < & 4.
\eq
Since there are no loops, no pair of $\vec{e}_2$,\dots,$\vec{e}_j$ is connected.
Therefore $\vec{e}_2$, \dots, $\vec{e}_j$ are perpendicular unit vectors.
Further, by assumption $\vec{e}_1$, $\vec{e}_2$,\dots,$\vec{e}_j$ are linearly independent vectors.
Therefore $\vec{e}_1$ is not in the span of $\vec{e}_2$,\dots,$\vec{e}_j$.
It follows
\bq
 1 & = & \left( \vec{e}_1 \cdot \vec{e}_1 \right)^2
 \;\; > \;\;
 \sum\limits_{i=2}^j \left( \vec{e}_1 \cdot \vec{e}_i \right)^2
\eq
and therefore
\bq
 \sum\limits_{i=2}^j \left( \vec{e}_1 \cdot \vec{e}_i \right)^2 < 1
 & \mbox{and} &
 \sum\limits_{i=2}^j \left( 2 \vec{e}_1 \cdot \vec{e}_i \right)^2 < 4.
\eq
\end{proof}
\begin{lemma}
\label{appendix_lie_algebra:lemma_Coxeter_4}
In an admissible diagram, any chain of vertices connected to each other by one line, with none
but the ends of the chain connected to any other vertices, can be collapsed to one vertex, 
and the resulting diagram remains admissible.
\end{lemma}
\begin{proof}
Let us consider a chain of $r$ vertices:
\bq
 & & \nonumber \\
\begin{picture}(110,10)(0,0)
\CArc(5,5)(2,0,360)
\CArc(25,5)(2,0,360)
\CArc(45,5)(2,0,360)
\CArc(65,5)(2,0,360)
\CArc(85,5)(2,0,360)
\Line(7,5)(23,5)
\Line(27,5)(43,5)
\Line(47,5)(63,5)
\Line(67,5)(83,5)
\DashLine(-15,-15)(3,3){4}
\DashLine(-15,25)(3,7){4}
\DashLine(87,7)(105,25){4}
\DashLine(87,3)(105,-15){4}
\Text(5,9)[b]{$1$}
\Text(25,9)[b]{$2$}
\Text(85,9)[b]{$r$}
\end{picture}
& \rightarrow &
\begin{picture}(110,10)(0,0)
\CArc(45,5)(2,0,360)
\DashLine(25,-15)(43,3){4}
\DashLine(25,25)(43,7){4}
\DashLine(47,7)(65,25){4}
\DashLine(47,3)(65,-15){4}
\end{picture}
\\ \nonumber 
\eq
If $\vec{e}_1$, \dots, $\vec{e}_r$ are the unit vectors corresponding to the chain of vertices as indicated above,
then
\bq
 \vec{e}' & = & \vec{e}_1 + \dots + \vec{e}_r
\eq
is a unit vector since
\bq
 \vec{e}' \cdot \vec{e}'
 & = & 
 \left( \vec{e}_1 + \dots + \vec{e}_r \right)^2
 = r + 2 \vec{e}_1 \cdot \vec{e}_2 + 2 \vec{e}_2\cdot \vec{e}_3 + \dots + 2 \vec{e}_{r-1} \cdot \vec{e}_r
 \nonumber \\
 & = & r - (r-1) = 1.
\eq
Furthermore, $\vec{e}'$ satisfies the same conditions with respect to the other vectors since
$\vec{e}' \cdot \vec{e}_j$ is either
$\vec{e}_1 \cdot \vec{e}_j$ or $\vec{e}_r \cdot \vec{e}_j$.
\end{proof}
With the help of these lemmata we can now prove proposition~\ref{appendix_lie_algebra:proposition_admissible_Coxeter_diagrams}:
\begin{proof}
From lemma~\ref{appendix_lie_algebra:lemma_Coxeter_3} it follows that the only connected diagram with a triple line is $G_2$.
\\
\\
Furthermore we cannot have a diagram with two double lines, otherwise we would have a sub-diagram,
which we could contract as
\bq
\begin{picture}(110,10)(0,0)
\CArc(5,5)(2,0,360)
\CArc(25,5)(2,0,360)
\CArc(45,5)(2,0,360)
\CArc(65,5)(2,0,360)
\CArc(85,5)(2,0,360)
\CArc(105,5)(2,0,360)
\Line(7,6)(23,6)
\Line(7,4)(23,4)
\Line(27,5)(43,5)
\Text(55,5)[c]{$\dots$}
\Line(67,5)(83,5)
\Line(87,6)(103,6)
\Line(87,4)(103,4)
\end{picture}
& \rightarrow &
\begin{picture}(110,10)(0,0)
\CArc(25,5)(2,0,360)
\CArc(45,5)(2,0,360)
\CArc(65,5)(2,0,360)
\Line(27,6)(43,6)
\Line(27,4)(43,4)
\Line(47,6)(63,6)
\Line(47,4)(63,4)
\end{picture}
\eq
contradicting again lemma~\ref{appendix_lie_algebra:lemma_Coxeter_3}.
By the same reasoning we cannot have a diagram with a double line and a vertex with three single lines attached to it:
\bq
\begin{picture}(110,10)(0,0)
\CArc(5,5)(2,0,360)
\CArc(25,5)(2,0,360)
\CArc(45,5)(2,0,360)
\CArc(65,5)(2,0,360)
\CArc(85,5)(2,0,360)
\CArc(105,10)(2,0,360)
\CArc(105,0)(2,0,360)
\Line(7,6)(23,6)
\Line(7,4)(23,4)
\Line(27,5)(43,5)
\Text(55,5)[c]{$\dots$}
\Line(67,5)(83,5)
\Line(87,6)(103,10)
\Line(87,4)(103,0)
\end{picture}
& \rightarrow &
\begin{picture}(110,10)(0,0)
\CArc(25,5)(2,0,360)
\CArc(45,5)(2,0,360)
\CArc(65,10)(2,0,360)
\CArc(65,0)(2,0,360)
\Line(27,6)(43,6)
\Line(27,4)(43,4)
\Line(47,6)(63,10)
\Line(47,4)(63,0)
\end{picture}
\eq
Again this contradicts lemma~\ref{appendix_lie_algebra:lemma_Coxeter_3}.
\\
\\
To finish the case with double lines, we rule out the diagram
\bq
\begin{picture}(110,15)(0,0)
\CArc(5,5)(2,0,360)
\CArc(25,5)(2,0,360)
\CArc(45,5)(2,0,360)
\CArc(65,5)(2,0,360)
\CArc(85,5)(2,0,360)
\Line(7,5)(23,5)
\Line(27,4)(43,4)
\Line(27,6)(43,6)
\Line(47,5)(63,5)
\Line(67,5)(83,5)
\Text(5,9)[b]{$1$}
\Text(25,9)[b]{$2$}
\Text(45,9)[b]{$3$}
\Text(65,9)[b]{$4$}
\Text(85,9)[b]{$5$}
\end{picture}
\eq
Consider the vectors
\bq
 \vec{v} = \vec{e}_1 + 2 \vec{e}_2, 
 & & 
 \vec{w} = 3 \vec{e}_3 + 2 \vec{e}_4 + \vec{e}_5.
\eq
We find
\bq
 \left( \vec{v} \cdot \vec{w} \right)^2 = 18,
\;\;\;
 \left| \vec{v} \right|^2 = 3,
\;\;\;
 \left| \vec{w} \right|^2 = 6.
\eq
This violates the Cauchy-Schwarz inequality
\bq
 \left( \vec{v} \cdot \vec{w} \right)^2 & < & \left| \vec{v} \right|^2 \cdot \left| \vec{w} \right|^2.
\eq
By a similar reasoning one rules out the following (sub-) graphs with single lines:
\bq
\begin{picture}(110,50)(0,0)
\CArc(5,25)(2,0,360)
\CArc(25,25)(2,0,360)
\CArc(45,25)(2,0,360)
\CArc(65,15)(2,0,360)
\CArc(65,35)(2,0,360)
\CArc(85,5)(2,0,360)
\CArc(85,45)(2,0,360)
\Line(7,25)(23,25)
\Line(27,25)(43,25)
\Line(47,26)(63,34)
\Line(67,36)(83,44)
\Line(47,24)(63,16)
\Line(67,14)(83,6)
\end{picture}
\eq
\bq
\begin{picture}(110,10)(0,0)
\CArc(5,5)(2,0,360)
\CArc(25,5)(2,0,360)
\CArc(45,5)(2,0,360)
\CArc(65,5)(2,0,360)
\CArc(85,5)(2,0,360)
\CArc(105,5)(2,0,360)
\CArc(125,5)(2,0,360)
\CArc(65,-5)(2,0,360)
\Line(7,5)(23,5)
\Line(27,5)(43,5)
\Line(47,5)(63,5)
\Line(67,5)(83,5)
\Line(87,5)(103,5)
\Line(107,5)(123,5)
\Line(65,3)(65,-3)
\end{picture}
\eq
\bq
\begin{picture}(110,10)(0,0)
\CArc(5,5)(2,0,360)
\CArc(25,5)(2,0,360)
\CArc(45,5)(2,0,360)
\CArc(65,5)(2,0,360)
\CArc(85,5)(2,0,360)
\CArc(105,5)(2,0,360)
\CArc(125,5)(2,0,360)
\CArc(145,5)(2,0,360)
\CArc(45,-5)(2,0,360)
\Line(7,5)(23,5)
\Line(27,5)(43,5)
\Line(47,5)(63,5)
\Line(67,5)(83,5)
\Line(87,5)(103,5)
\Line(107,5)(123,5)
\Line(127,5)(143,5)
\Line(45,3)(45,-3)
\end{picture}
 \\ \nonumber
\eq
These sub-diagrams rules out all graphs not in the list of theorem~\ref{appendix_lie_algebra:Dynkin_classification}.
To finish the proof of the proposition it remains to show that all graphs in the list are admissible.
This is equivalent to show that for each Dynkin diagram in the list there exists a corresponding
Lie algebra. (The simple root vectors of such a Lie algebra will then have automatically the corresponding
angles of the Coxeter diagram.)
\\
\\
To prove the existence it is sufficient to give for each Dynkin diagram an example
of a Lie algebra corresponding to it.
For the four families $A_n$, $B_n$, $C_n$ and $D_n$  we have already seen
that they correspond to the Lie algebras of ${\mathfrak s}{\mathfrak u}(n+1)$, ${\mathfrak s}{\mathfrak o}(2n+1)$, ${\mathfrak s}{\mathfrak p}(n)$ and ${\mathfrak s}{\mathfrak o}(2n)$
In addition one can write down explicit matrix representations for the Lie algebras corresponding
to the five exceptional groups $E_6$, $E_7$, $E_8$, $F_4$ and $G_2$.
\end{proof}

%% file: dirichlet.tex
\newpage
\chapter{Dirichlet characters}
\label{appendix_dirichlet}

\section{Definition}
\label{appendix_dirichlet:sect_definition}

Let $N$ be a positive integer. 
We denote by ${\mathbb Z}_N^\ast$ the set of invertible elements in ${\mathbb Z}_N$. 
These are all elements $a \in {\mathbb Z}_N$ with $\mathrm{gcd}(a,N)=1$.
The set ${\mathbb Z}_N^\ast$ is an Abelian group with respect to multiplication.
We further denote $\mathbb{C}^\ast = \mathbb{C}\backslash\{0\}$.
A 
\index{Dirichlet character}
{\bf Dirichlet character} modulo $N$
is a function 
\bq
 \chi & : & {\mathbb Z}_N^\ast \rightarrow \mathbb{C}^\ast
\eq
that is a homomorphism of groups, i.e. 
\bq
 \chi(nm) 
 & = & 
 \chi(n) \chi(m) 
 \;\;\;\;\;\; 
 \mbox{for all} \;\;  n,m \in {\mathbb Z}_N^\ast.
\eq
We may extend $\chi$ to a function $\chi: {\mathbb Z}_N \rightarrow \mathbb{C}$ by setting
$\chi(n)=0$ if $\gcd(n,N) > 1$ and then further extend to a function $\chi: \mathbb{Z} \rightarrow \mathbb{C}$
by setting $\chi(n) = \chi(n \mod N)$.
By abuse of notation we denote both extensions again by $\chi$.
This function satisfies
\begin{alignat}{3}
 & (i) & \hspace*{3mm} & \chi(n) = \chi(n+N) & \hspace*{5mm} & \forall\ n \in \mathbb{Z}, \nonumber \\
 & (ii) & & \chi(n) = 0 & & \mbox{if} \;\; \gcd(n,N) > 1, \nonumber \\
 &      & & \chi(n) \neq 0 & & \mbox{if} \;\; \gcd(n,N) = 1, \nonumber \\ 
 & (iii) & & \chi(nm) = \chi(n) \chi(m) & & \forall\ n,m \in \mathbb{Z}.
\end{alignat}
Property $(ii)$ and $(iii)$ imply for any Dirichlet character $\chi$:
\bq
 \chi(1) & = &1.
\eq
Let's look at a few examples of Dirichlet characters with modulus $N \in \{1,2,3,4,5,6\}$.
\begin{description}
\item{Modulus $1$}: For $N=1$ there is only the trivial character
\begin{center}
\begin{tabular}{c|r}
 $n$ & $0$ \\
\hline
 $\chi_{1,1}(n)$ & $1$ \\
\end{tabular}
\end{center}
\item{Modulus $2$:} There is one character modulo $2$
\begin{center}
\begin{tabular}{c|r|r}
 $n$ & $0$ & $1$ \\
\hline
 $\chi_{2,1}(n)$ & $0$ & $1$ \\
\end{tabular}
\end{center}
\item{Modulus $3$:} There are two characters modulo $3$
\begin{center}
\begin{tabular}{c|r|r|r}
 $n$ & $0$ & $1$ & $2$ \\
\hline
 $\chi_{3,1}(n)$ & $0$ & $1$ & $1$ \\
 $\chi_{3,2}(n)$ & $0$ & $1$ & $-1$ \\
\end{tabular}
\end{center}
\item{Modulus $4$:} There are two characters modulo $4$
\begin{center}
\begin{tabular}{c|r|r|r|r}
 $n$ & $0$ & $1$ & $2$ & $3$ \\
\hline
 $\chi_{4,1}(n)$ & $0$ & $1$ & $0$ & $1$ \\
 $\chi_{4,2}(n)$ & $0$ & $1$ & $0$ & $-1$ \\
\end{tabular}
\end{center}
\item{Modulus $5$}: There are four characters modulo $5$
\begin{center}
\begin{tabular}{c|r|r|r|r|r}
 $n$ & $0$ & $1$ & $2$ & $3$ & $4$ \\
\hline
 $\chi_{5,1}(n)$ & $0$ & $1$ & $1$ & $1$ & $1$ \\
 $\chi_{5,2}(n)$ & $0$ & $1$ & $i$ & $-i$ & $-1$ \\
 $\chi_{5,3}(n)$ & $0$ & $1$ & $-1$ & $-1$ & $1$ \\
 $\chi_{5,4}(n)$ & $0$ & $1$ & $-i$ & $i$ & $-1$ \\
\end{tabular}
\end{center}
\item{Modulus $6$}: There are two characters modulo $6$
\begin{center}
\begin{tabular}{c|r|r|r|r|r|r}
 $n$ & $0$ & $1$ & $2$ & $3$ & $4$ & $5$ \\
\hline
 $\chi_{6,1}(n)$ & $0$ & $1$ & $0$ & $0$ & $0$ & $1$ \\
 $\chi_{6,2}(n)$ & $0$ & $1$ & $0$ & $0$ & $0$ & $-1$ \\
\end{tabular}
\end{center}
\end{description}
We denote by $\chi_{N,1}$ the trivial character modulo $N$ 
(with $\chi_{N,1}(n) = 1$ if $\gcd(n,N)=1$ and $\chi_{N,1}(n)=0$ otherwise).
If no confusion with the modulus arises, we simply write $\chi_{1}$ instead of $\chi_{N,1}$.
The trivial character modulo $1$ is denoted by $1=\chi_{1,1}$ 
(we have $\chi_{1,1}(n)=1$ for all $n$, hence the notation).

The 
\index{conductor}
{\bf conductor} of $\chi$ is the smallest positive divisor $d|N$ such that there is a character $\chi'$ modulo $d$
with
\bq
 \chi(n) & = & \chi'(n) \qquad \forall\ n \in \mathbb{Z} \ \mbox{with}\ \gcd(n,N) = 1.
\eq
A Dirichlet character is called 
\index{primitive Dirichlet character}
{\bf primitive},
if its modulus equals its conductor.

To give an example, the Dirichlet character $\chi_{4,1}$ of modulus $4$ from the examples above 
has conductor $1$, since
\bq
 \chi_{4,1}(1) \; = \; \chi_{1,1}(1),
 & &
 \chi_{4,1}(3) \; = \; \chi_{1,1}(3).
\eq
On the other hand, $\chi_{4,2}$ has conductor $4$ and is therefore a primitive Dirichlet character.

If $\chi$ is a Dirichlet character modulo $N$ and $M$ a positive integer, 
$\chi$ induces a Dirichlet character $\tilde{\chi}$ with modulus $(M \cdot N)$ by setting
\bq
 \tilde{\chi}(n)
 & = &
 \left\{\begin{array}{rl}
  \chi(n), & \mbox{if} \; \mbox{gcd}(n,M \cdot N) = 1, \\
  0, & \mbox{if} \; \mbox{gcd}(n,M \cdot N) \neq 1. \\
 \end{array} \right.
\eq
We call $\tilde{\chi}$ the
\index{induced character}
{\bf induced character} of modulus $(M \cdot N)$ induced by $\chi$.
In the examples above $\chi_{N,1}$ (the trivial character modulo $N$) 
is the induced character of modulus $N$
induced by $\chi_{1,1}$ (the trivial character modulo $1$).

In the other direction we may associate to a Dirichlet character $\chi$ 
with modulus $N$ and conductor $d$ a primitive Dirichlet character $\bar{\chi}$ 
with modulus $d$ as follows:
We first note if $\mbox{gcd}(n,d)=1$ there exists an integer $n'$ such that $\mbox{gcd}(n',N)=1$ 
and $n' \equiv n \mod d$.
We set
\bq
 \bar{\chi}(n)
 & = &
 \left\{ \begin{array}{rl}
  \chi(n'), & \mbox{if} \; \mbox{gcd}(n,d) = 1, \\
  0, & \mbox{if} \; \mbox{gcd}(n,d) \neq 1. \\
 \end{array} \right.
\eq
$\bar{\chi}$ is called the 
\index{associated primitive character}
{\bf primitive character associated with $\chi$}.

\section{The Kronecker symbol}
\label{appendix_dirichlet:sect_kronecker_symbol}

In this section we introduce the
Kronecker symbol
\bq
\label{appendix_dirichlet:kronecker_symbol}
 \left( \frac{a}{n} \right).
\eq
(There is an overloading of the name ``Kronecker symbol'': In this section we mean the symbol
as in eq.~(\ref{appendix_dirichlet:kronecker_symbol}), not $\delta_{ij}$. 
The two symbols are not related.)
The Kronecker symbol defines a Dirichlet character, which takes values $\{-1,0,1\}$.
In addition, we may give a criteria under which condition this 
Dirichlet character is primitive.

Let $a$ be an integer and $n$ a non-zero integer with prime factorisation
$n = u p_1^{\alpha_1} p_2^{\alpha_2} ... p_k^{\alpha_k}$,
where $u \in \{1,-1\}$ is a unit and the $p_j$'s are prime numbers.
The 
\index{Kronecker symbol}
{\bf Kronecker symbol} 
is defined by
\bq
 \left( \frac{a}{n} \right)
 & = & 
 \left( \frac{a}{u} \right)
 \left( \frac{a}{p_1} \right)^{\alpha_1}
 \left( \frac{a}{p_2} \right)^{\alpha_2}
 ...
 \left( \frac{a}{p_k} \right)^{\alpha_k}.
\eq
The individual factors are defined as follows:
For a unit $u$ we define
\bq
 \left( \frac{a}{u} \right)
 & = &
 \left\{ \begin{array}{rl}
 1, & u=1, \\
 1, & u=-1, \; a \ge 0, \\
 -1, & u=-1, \; a<0. \\
 \end{array} \right.
\eq
For $p=2$ we define
\bq
 \left( \frac{a}{2} \right)
 & = &
 \left\{ \begin{array}{rl}
 1, & a \equiv \pm 1 \mod 8, \\
 -1, & a \equiv \pm 3 \mod 8, \\
 0, & a \;\; \mbox{even}.  \\
 \end{array} \right.
\eq
For an odd prime $p$ we have
\bq
 \left( \frac{a}{p} \right)
 & = & 
 a^{\frac{p-1}{2}} \mod p
 \;\; = \;\;
 \left\{ \begin{array}{rl}
 1, & a \equiv b^2 \mod p, \\
 -1, & a \not\equiv b^2 \mod p, \\
 0, & a \equiv 0 \mod p. \\
 \end{array} \right.
\eq
We further set
\bq
 \left( \frac{a}{0} \right)
 & = &
 \left\{ \begin{array}{rl}
 1, & a = \pm 1 \\
 0, & \mbox{otherwise}.  \\
 \end{array} \right.
\eq
For any non-zero integer $a$ the mapping 
\bq
 n & \rightarrow & 
 \left( \frac{a}{n} \right)
\eq
is a Dirichlet character, which we denote by $\chi_a$:
\bq
 \chi_a\left(n\right)
 & = &
 \left( \frac{a}{n} \right).
\eq
If $a$ is the discriminant of a quadratic field,
then it is a primitive Dirichlet character with conductor $|a|$.
One may give a condition for $a$ being the discriminant of a quadratic field \cite{Miyake}.
We first set for $p$ being a prime number, $-1$ or $-2$
\bq
 p^\ast
 & = &
 \left\{ \begin{array}{rl}
 p, & \mbox{if} \quad p \equiv 1 \mod 4, \\
 -p, & \mbox{if} \quad p \equiv -1 \mod 4 \quad \mbox{and} \quad p \neq -1, \\
 -4, & \mbox{if} \quad p = -1, \\
 8, & \mbox{if} \quad p = 2, \\
-8, & \mbox{if} \quad p = -2. \\
 \end{array} \right.
\eq
Then an integer $a$ is the discriminant of a quadratic field if and only if
$a$ is a product of distinct $p^\ast$'s.

Including the trivial character (for which $a=1$) the possible values for $a$ with smallest absolute value
are
\bq
 1, -3, -4, 5, -7, 8, -8, -11, 12, \dots
\eq

%% file: moduli_space.tex
\newpage
\chapter{The moduli space ${\mathcal M}_{g,n}$}
\label{appendix_moduli_space}

In this appendix we discuss the moduli space of a smooth algebraic curve of genus $g$ with $n$ marked points.
This moduli space is denoted by  ${\mathcal M}_{g,n}$.

We start in section~\ref{appendix_moduli_space:config_spaces} 
from configuration spaces of $n$ points in a topological space $X$ and mod out configurations 
which are isomorphic. This gives us the moduli space.
We are in particular interested in the case, where the space $X$ is a smooth complex algebraic curve of genus $g$.
These complex curves, and their equivalence to real Riemann surfaces are discussed in section~\ref{appendix_moduli_space:riemann_surfaces}.

With these preparations, we 
specialise in section~\ref{appendix_moduli_space:moduli_space_M_g_n} to the main topic
of this appendix: The moduli space ${\mathcal M}_{g,n}$
of a smooth algebraic curve of genus $g$ with $n$ marked points.
This moduli space is non-compact, and one is interested in its compactification $\overline{\mathcal M}_{g,n}$.
The compactification includes configurations, where points and/or the algebraic curve degenerates.

The cases of genus zero and genus one are the most important ones for applications towards Feynman integrals.
We discuss the moduli space ${\mathcal M}_{0,n}$ in section~\ref{appendix_moduli_space:genus_zero_case}
and the moduli space ${\mathcal M}_{1,n}$ in section~\ref{appendix_moduli_space:genus_one_case}


\section{Configuration spaces}
\label{appendix_moduli_space:config_spaces}

Let $X$ be a topological space. The 
\index{configuration space}
{\bf configuration space} of $n$ ordered points
in $X$ is
\bq
 \mathrm{Conf}_n\left(X\right)
 & = &
 \left\{ \left. \left(x_1,\dots,x_n\right) \in X^n \right| x_i \neq x_j \; \mbox{for} \; i \neq j \right\}.
\eq
Please note that we require that the points are distinct: $x_i \neq x_j$.
As a simple example consider the configuration space of $2$ ordered points in ${\mathbb R}$:
\bq
 \mathrm{Conf}_2\left({\mathbb R}\right)
 & = &
 \left\{ \left. \left(x_1,x_2\right) \in {\mathbb R}^2 \right| x_1 \neq x_2\right\}.
\eq
$\mathrm{Conf}_2({\mathbb R})$ is the plane ${\mathbb R}^2$ with the diagonal $x_1=x_2$ removed.
It is a two-dimensional space.

As a second example consider the configuration space of $2$ ordered points in the complex projective space
${\mathbb C}{\mathbb P}^1$ (i.e. the Riemann sphere):
\bq
 \mathrm{Conf}_2\left({\mathbb C}{\mathbb P}^1\right)
 & = &
 \left\{ \left. \left(z_1,z_2\right) \in \left({\mathbb C}{\mathbb P}^1\right)^2 \right| z_1 \neq z_2\right\}.
\eq
This is again a two-dimensional space.
A M\"obius transformation
\bq
 z' & = & \frac{az+b}{cz+d}
\eq
transforms the Riemann sphere into itself.
These transformations form the group $\mathrm{PSL}\left(2,{\mathbb C}\right)$.
Usually we are not interested in configurations 
\bq
 (z_1,\dots,z_n) \in \mathrm{Conf}_n\left({\mathbb C}{\mathbb P}^1\right)
 & \mbox{and} &
 (z_1',\dots,z_n') \in \mathrm{Conf}_n\left({\mathbb C}{\mathbb P}^1\right),
\eq
which differ only by a M\"obius transformation.
This brings us to the definition of the {\bf moduli space} of the Riemann sphere with $n$ marked points:
\bq 
 {\mathcal M}_{0,n}
 & = &
 \mathrm{Conf}_n\left({\mathbb C}{\mathbb P}^1\right) / \mathrm{PSL}\left(2,{\mathbb C}\right).
\eq
We may use the freedom of M\"obius transformations to fix three points (usually $0$, $1$ and $\infty$).
Therefore
\bq
 \dim\left( \mathrm{Conf}_n\left({\mathbb C}{\mathbb P}^1\right) \right) & = & n,
 \nonumber \\
 \dim\left({\mathcal M}_{0,n}\right) & = & n-3.
\eq


\section{Complex algebraic curves and Riemann surfaces}
\label{appendix_moduli_space:riemann_surfaces}

We are mainly interested in the situation, where the topological space $X$ is a Riemann surface $C$.
Let us start with a compact, connected and smooth Riemann surface $C$.

On the one hand, we may view $C$ as a two-dimensional real surface (hence Riemann surface) with a complex structure.

On the other hand we may view $C$ as an algebraic curve (i.e. of complex dimension one) in ${\mathbb C}{\mathbb P}^2$:
There exists a homogeneous polynomial $P(z_1,z_2,z_3)$ such that
\bq
 C & : & \left\{ \left. \left[z_1:z_2:z_3\right] \in {\mathbb C}{\mathbb P}^2 \right| P\left(z_1,z_2,z_3\right) = 0 \right\} 
\eq
If $d$ is the degree of the polynomial $P(z_1,z_2,z_3)$, the 
\index{arithmetic genus}
{\bf arithmetic genus} of $C$ is given by
\bq
 g & = & \frac{1}{2} \left(d-1\right) \left(d-2\right).
\eq
Example:
\bq
 y^2 z - x^3 - x z^2 & = & 0
\eq
is a smooth curve of genus $1$.
\begin{figure}
\begin{center}
\includegraphics[scale=1.0]{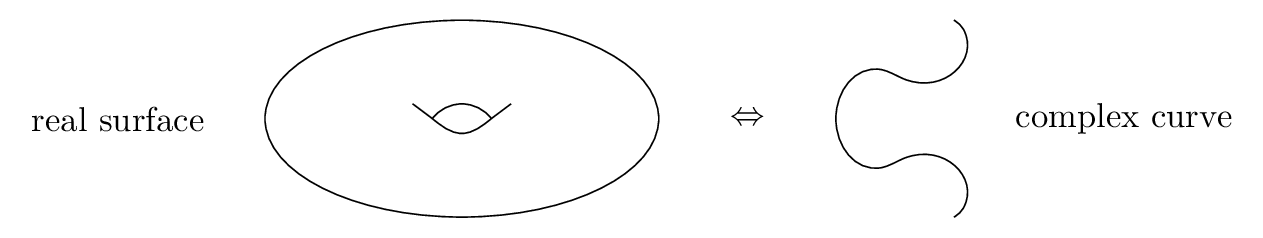}
\end{center}
\caption{
We may view $C$ either as a Riemann surface (a two-dimensional real surface) or as a complex algebraic curve
(of complex dimension one, corresponding to two real dimensions). 
}
\label{appendix_moduli_space:fig_torus}
\end{figure}
The fact that we may view $C$ either as a Riemann surface (of real dimension two) or as a complex algebraic curve
is illustrated in fig.~\ref{appendix_moduli_space:fig_torus}.

The requirement that the curve is smooth is a little bit too restrictive and we consider the generalisation towards nodal curves.
A 
\index{node}
{\bf node} of a curve is a singularity isomorphic to
\bq
 x y & = & 0
 \;\;\; \mbox{in} \;\; {\mathbb C}^2.
\eq
A 
\index{nodal curve}
{\bf nodal curve} is a compact, connected curve which is smooth except for a finite number of points, which are nodes.
\begin{figure}
\begin{center}
\includegraphics[scale=1.0]{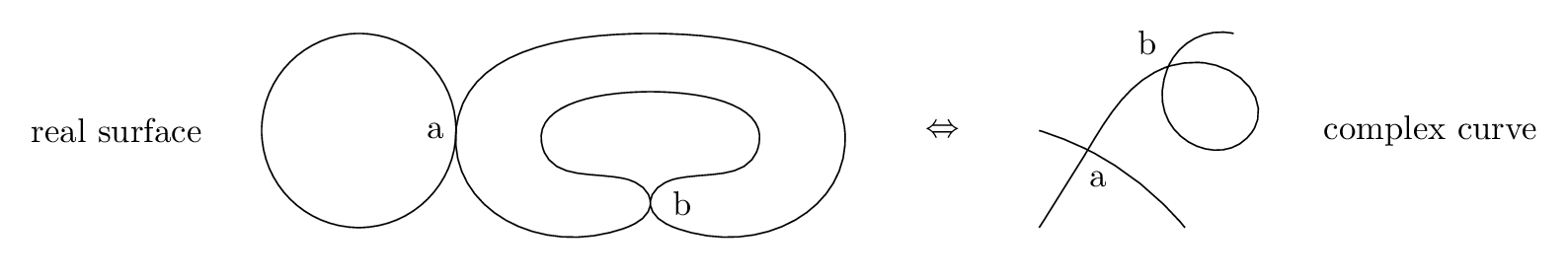}
\end{center}
\caption{
A nodal curve, shown as a real surface (left) or as a complex curve (right).
}
\label{appendix_moduli_space:fig_nodal_curve}
\end{figure}
Fig.~\ref{appendix_moduli_space:fig_nodal_curve} shows an example.

There are two operations, which we may perform on a node:
We may 
\index{ungluing of a node}
{\bf unglue} a node or we may 
\index{smoothing of a node}
{\bf smoothen} a node.
\begin{figure}
\begin{center}
\includegraphics[scale=1.0]{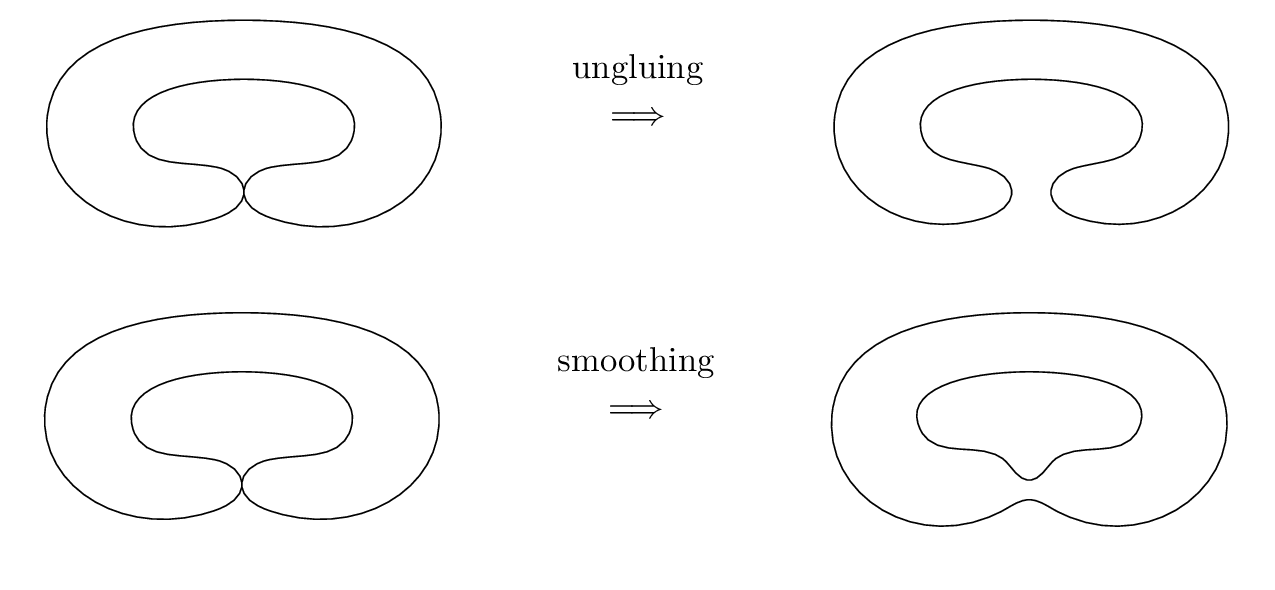}
\end{center}
\caption{
On a node we may perform the operations of 
ungluing (top) and smoothing (bottom).
}
\label{appendix_moduli_space:fig_gluing_smoothing}
\end{figure}
These two operations are illustrated in fig.~\ref{appendix_moduli_space:fig_gluing_smoothing}.

For a nodal curve we have to distinguish the geometric genus and the arithmetic genus.
The 
\index{geometric genus}
{\bf geometric genus} of an irreducible nodal curve is its genus once all of the nodes are unglued, and
the geometric genus of a (reducible) nodal curve is the sum of the geometric genera of the irreducible components.
The 
\index{arithmetic genus}
{\bf arithmetic genus} of a nodal curve is the genus of the curve obtained by smoothing.
\begin{figure}
\begin{center}
\includegraphics[scale=1.0]{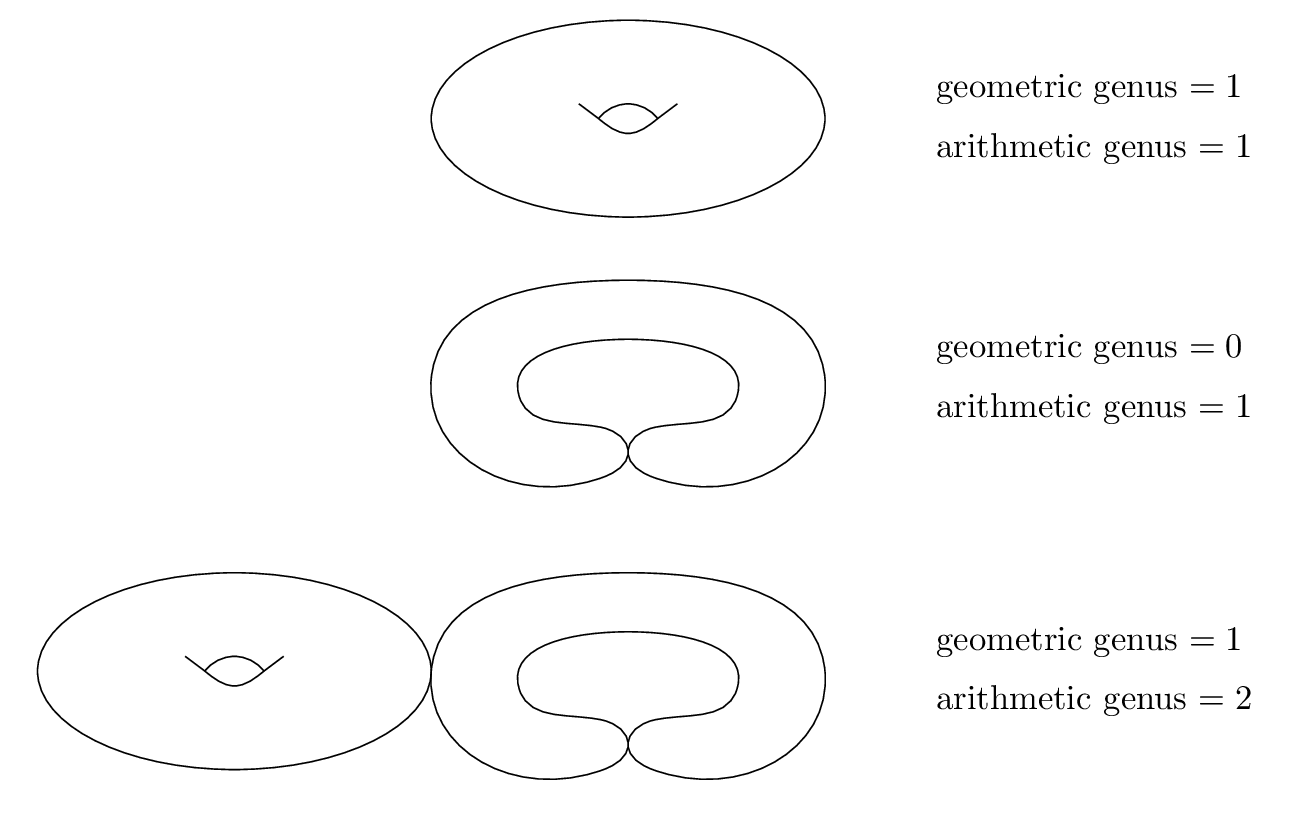}
\end{center}
\caption{
The arithmetic genus and the geometric genus for various examples.
}
\label{appendix_moduli_space:fig_garithmetic_geometric_genus}
\end{figure}
Fig.~\ref{appendix_moduli_space:fig_garithmetic_geometric_genus} shows a few examples.

\begin{figure}
\begin{center}
\includegraphics[scale=1.0]{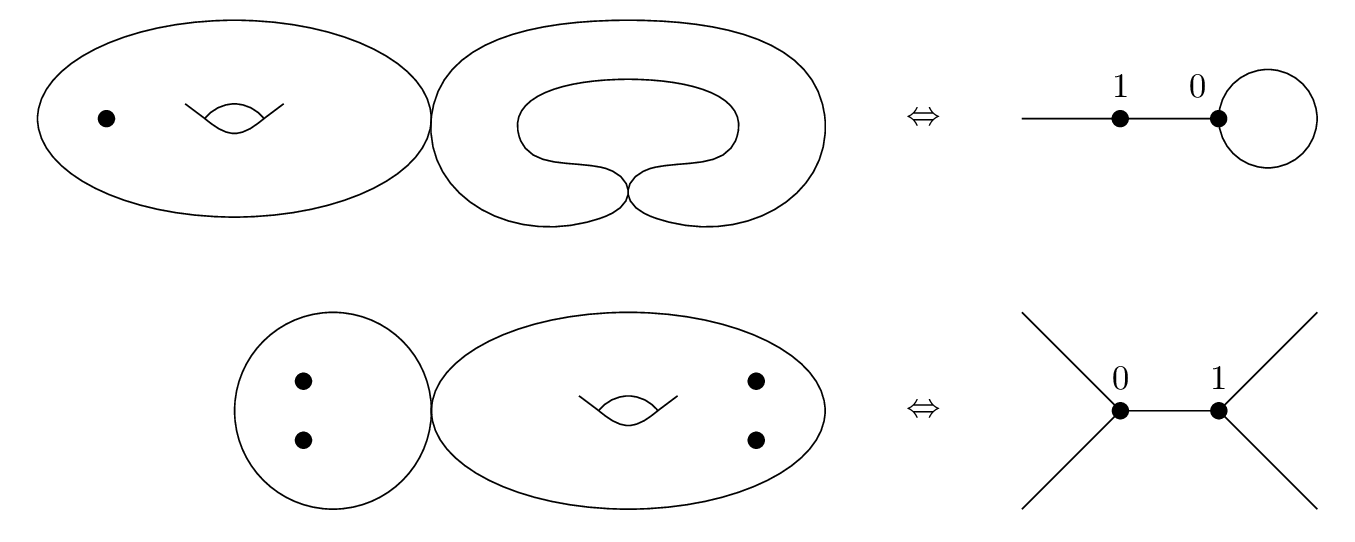}
\end{center}
\caption{
The correspondence between a nodal curve and its dual graph.
}
\label{appendix_moduli_space:fig_nodal_curve_dual_graph}
\end{figure}

If the curve $C$ has $s$ nodes and $k$ irreducible components 
the relation between the arithmetic genus $g_{\mathrm{arithm}}$ and the geometric genus $g_{\mathrm{geom}}$ is
\bq
 g_{\mathrm{arithm}} & = & g_{\mathrm{geom}} + 1 + s - k.
\eq
Let us now consider nodal curves with $n$ marked points.
\begin{figure}
\begin{center}
\includegraphics[scale=1.0]{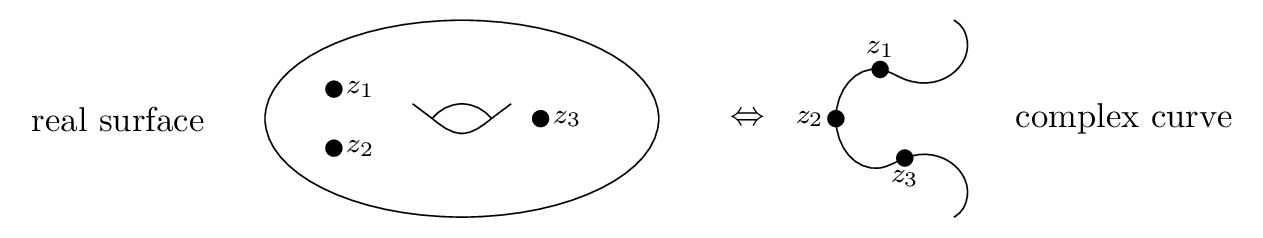}
\end{center}
\caption{
A (smooth) curve with three marked points.
The left picture shows the real surface, the right picture the complex curve. 
}
\label{appendix_moduli_space:fig_curve_with_marked_points}
\end{figure}
An example with three marked points is shown in fig.~\ref{appendix_moduli_space:fig_curve_with_marked_points}.
To a nodal curve we may associate a 
\index{dual graph of a nodal curve}
{\bf dual graph} as follows:
\begin{itemize}
\item The irreducible components of the curve $C$ are drawn as vertices, labelled with their geometric genera.
\item The nodes are drawn as edges.
\item The marked points are drawn as half-edges.
\end{itemize}
Fig.~\ref{appendix_moduli_space:fig_nodal_curve_dual_graph} shows a few examples.
We call a nodal curve with marked points a 
\index{stable curve}
{\bf stable curve}, if in the dual graph
\begin{itemize}
\item each genus $0$ vertex has valence $\ge 3$,
\item each genus $1$ vertex has valence $\ge 1$.
\end{itemize}
This definition implies that 
smooth curves of genus $g$ with $n$ marked points are not stable if
\bq
 \left(g,n\right)
 & \in &
 \left\{
  \left(0,0\right),
  \left(0,1\right),
  \left(0,2\right),
  \left(1,0\right)
 \right\}.
\eq
The smooth stable curves with $n$ marked points are the ones with
\bq 
 \chi 
 \; = \; 2 - 2g - n & < & 0.
\eq
$\chi$ is the 
\index{Euler characteristic of a smooth curve}
{\bf Euler characteristic} of the smooth curve with $n$ marked points.

Let us remark that for a smooth curve the arithmetic genus equals the geometric genus, 
therefore just using ``genus'' is unambiguous in the smooth case.


\section{The moduli space of a curve of genus $g$ with $n$ marked points}
\label{appendix_moduli_space:moduli_space_M_g_n}

Let us now consider a smooth curve $C$ of genus $g$ with $n$ marked points.
Two such curves $(C;z_1,\dots,z_n)$ and $(C';z_1',\dots,z_n')$ are 
\index{isomorphic curves}
{\bf isomorphic} if there is an isomorphism
\bq
 \phi & : & C \rightarrow C'
 \;\;\;\;\;\;
 \mbox{such that} \;\;
 \phi\left(z_i\right) = z_i'.
\eq
The {\bf moduli space} 
\bq
 {\mathcal M}_{g,n}
\eq
is the space of isomorphism classes of smooth curves of genus $g$ with $n$ marked points.

For $g \ge 1$ the isomorphism classes do not only depend on the positions of the marked points,
but also on the ``shape'' of the curve.
For $g=0$ there is only one ``shape'', the Riemann sphere.

The dimension of ${\mathcal M}_{g,n}$ is
\bq
 \dim\left({\mathcal M}_{g,n}\right) & = & 3g + n -3.
\eq
The Euler characteristic of ${\mathcal M}_{g,n}$ is
\bq
 \chi\left({\mathcal M}_{g,n}\right)
 & = &
 \left(-1\right)^n \frac{\left(2g+n-3\right)!}{2g \left(2g-2\right)!} B_{2g},
\eq
where $B_j$ are the Bernoulli numbers.
\\
\\
The moduli space is not compact. We are interested in a specific compactification, the 
\index{Deligne-Mumford-Knudsen compactification}
{\bf Deligne-Mumford-Knudsen compactification} \cite{Deligne:1969,Knudsen:1976,Knudsen:1983,Knudsen:1983a}, 
which we denote by $\overline{\mathcal M}_{g,n}$.
The space $\overline{\mathcal M}_{g,n}$ is the moduli space of stable nodal curves of arithmetic genus $g$ with
$n$ marked points.
${\mathcal M}_{g,n}$ is an open subset in $\overline{\mathcal M}_{g,n}$.
A generic element of $\overline{\mathcal M}_{g,n}$ is drawn in the dual graph picture as
\begin{center}
\begin{tikzpicture}
 \draw (1.0,1.0) -- (0.0,1.0);
 \draw (1.0,1.0) -- (0.5,0.13);
 \draw (1.0,1.0) -- (1.5,0.13);
 \draw (1.0,1.0) -- (2.0,1.0);
 \draw (1.0,1.0) -- (1.5,1.87);
 \draw (1.0,1.0) -- (0.5,1.87);
 \draw[fill=gray] (1.0,1.0) circle (0.2);
 \draw (1.0,1.0) circle (0.2);
 \node [above left] at (0.8,1.0) {$g$};
\end{tikzpicture}
\end{center}
If the curve is smooth, the grey blob corresponds to a single genus $g$ vertex.
If the curve is non-smooth, the grey blob represents the appropriate sub-graph.
\\
\\
\index{stratification}
{\bf Stratification}:
Let $\Gamma$ be the dual graph of a stable curve.
We denote by ${\mathcal M}_\Gamma$ the subset of $\overline{\mathcal M}_{g,n}$ 
of stable curves with dual graph $\Gamma$.
The various ${\mathcal M}_\Gamma$'s give a stratification of $\overline{\mathcal M}_{g,n}$.
The dense open set ${\mathcal M}_{g,n}$ is one stratum,
all other strata are called 
\index{boundary strata}
{\bf boundary strata}.
The closure of the codimension $1$ strata are called 
\index{boundary divisor of a moduli space}
{\bf boundary divisors}.
The divisors meet transversely along smaller strata.
Let us denote the number of internal edges of $\Gamma$ by $n_e$. Then
\bq
 \dim\left({\mathcal M}_\Gamma\right) & = & 3g + n -3 - n_e.
\eq
Let us look at a few examples:
\begin{center}
\begin{tikzpicture}
 \node [left] at (0.0,1.0) {$\Gamma_1 = $};
 \draw (0.5,1.5) -- (1.0,1.0);
 \draw (0.5,0.5) -- (1.0,1.0);
 \draw (1.0,1.0) -- (2.0,1.0);
 \draw (2.0,1.0) -- (2.5,0.5);
 \draw (2.0,1.0) -- (2.5,1.5);
 \draw[fill] (1.0,1.0) circle (0.08);
 \draw[fill] (2.0,1.0) circle (0.08);
 \node [above] at (1.0,1.1) {$0$};
 \node [above] at (2.0,1.1) {$0$};
 \node at (5.0,1.0) {$\Rightarrow$};
 \node [right] at (6.0,1.0) {$\dim\left({\mathcal M}_{\Gamma_1}\right) = 0$};
\end{tikzpicture}
\vspace*{5mm}
\\
\begin{tikzpicture}
 \node [left] at (0.0,1.5) {$\Gamma_2 = $};
 \draw (1.0,1.0) -- (2.0,1.0);
 \draw (1.0,2.0) -- (2.0,2.0);
 \draw (1.0,1.0) -- (1.0,2.0);
 \draw (2.0,1.0) -- (2.0,2.0);
 \draw (0.5,2.5) -- (1.0,2.0);
 \draw (0.5,0.5) -- (1.0,1.0);
 \draw (2.0,1.0) -- (2.5,0.5);
 \draw (2.0,2.0) -- (2.5,2.5);
 \draw[fill] (1.0,1.0) circle (0.08);
 \draw[fill] (2.0,1.0) circle (0.08);
 \draw[fill] (1.0,2.0) circle (0.08);
 \draw[fill] (2.0,2.0) circle (0.08);
 \node [above] at (1.0,2.1) {$0$};
 \node [above] at (2.0,2.1) {$0$};
 \node [below] at (1.0,0.9) {$0$};
 \node [below] at (2.0,0.9) {$0$};
 \node at (5.0,1.5) {$\Rightarrow$};
 \node [right] at (6.0,1.5) {$\dim\left({\mathcal M}_{\Gamma_2}\right) = 0$};
\end{tikzpicture}
\end{center}
\begin{center}
\begin{tikzpicture}
 \node [left] at (0.0,1.5) {$\Gamma_3 = $};
 \draw (1.0,1.5) -- (2.0,1.0);
 \draw (1.0,1.5) -- (2.0,2.0);
 \draw (2.0,1.0) -- (2.0,2.0);
 \draw (0.5,1.75) -- (1.0,1.5);
 \draw (0.5,1.25) -- (1.0,1.5);
 \draw (2.0,1.0) -- (2.5,0.75);
 \draw (2.0,2.0) -- (2.5,2.25);
 \draw[fill] (1.0,1.5) circle (0.08);
 \draw[fill] (2.0,1.0) circle (0.08);
 \draw[fill] (2.0,2.0) circle (0.08);
 \node [above] at (1.0,1.6) {$0$};
 \node [above] at (2.0,2.1) {$0$};
 \node [below] at (2.0,0.9) {$0$};
 \node at (5.0,1.5) {$\Rightarrow$};
 \node [right] at (6.0,1.5) {$\dim\left({\mathcal M}_{\Gamma_3}\right) = 1$};
\end{tikzpicture}
\vspace*{5mm}
\\
\begin{tikzpicture}
 \node [left] at (0.0,1.5) {$\Gamma_4 = $};
 \draw (1.5,1.5) circle (0.5);
 \draw (0.5,1.75) -- (1.0,1.5);
 \draw (0.5,1.25) -- (1.0,1.5);
 \draw (2.0,1.5) -- (2.5,1.75);
 \draw (2.0,1.5) -- (2.5,1.25);
 \draw[fill] (1.0,1.5) circle (0.08);
 \draw[fill] (2.0,1.5) circle (0.08);
 \node [above] at (0.9,1.6) {$0$};
 \node [above] at (2.1,1.6) {$0$};
 \node at (5.0,1.5) {$\Rightarrow$};
 \node [right] at (6.0,1.5) {$\dim\left({\mathcal M}_{\Gamma_4}\right) = 2$};
\end{tikzpicture}
\end{center}
Of course we may also consider dual graphs with vertices of higher genus:
\begin{center}
\begin{tikzpicture}
 \node [left] at (0.0,1.0) {$\Gamma_5 = $};
 \draw (0.5,1.5) -- (1.0,1.0);
 \draw (0.5,0.5) -- (1.0,1.0);
 \draw (1.0,1.0) -- (2.0,1.0);
 \draw (2.0,1.0) -- (2.5,0.5);
 \draw (2.0,1.0) -- (2.5,1.5);
 \draw[fill] (1.0,1.0) circle (0.08);
 \draw[fill] (2.0,1.0) circle (0.08);
 \node [above] at (1.0,1.1) {$0$};
 \node [above] at (2.0,1.1) {$1$};
 \node at (5.0,1.0) {$\Rightarrow$};
 \node [right] at (6.0,1.0) {$\dim\left({\mathcal M}_{\Gamma_5}\right) = 3$};
\end{tikzpicture}
\end{center}
The 
\index{forgetful morphism}
{\bf forgetful morphism}: 
Consider a stable nodal curve with $n$ marked points (and assume $n>0$, $(g,n) \neq (0,3), (1,1)$).
We may forget the $n$-th point. This gives a nodal curve with $(n-1)$ marked points.
This curve may not be stable.
One stabilises the curve by contracting all components to a point, which correspond to genus $0$ vertices
of valency $2$.
This gives the forgetful morphism
\bq
 \overline{\mathcal M}_{g,n} & \rightarrow & \overline{\mathcal M}_{g,n-1}.
\eq
Example:
\begin{center}
\begin{tikzpicture}
 \genuszero{0.0}{8.0}
 \genusone{2.0}{8.0}
 \draw[fill] (0.8,9.3) circle (0.08);
 \draw[fill] (0.8,8.7) circle (0.08);
 \draw[fill] (5.1,9.3) circle (0.08);
 \draw[fill] (5.1,8.7) circle (0.08);
 \node [right] at (5.1,9.3) {$z_1$};
 \node [right] at (5.1,8.7) {$z_2$};
 \node [left] at (0.8,8.7) {$z_3$};
 \node [left] at (0.8,9.3) {$z_4$};
 \node at (7.0,9.0) {$\Leftrightarrow$};
 \draw (8.0,10.0) -- (9.0,9.0);
 \draw (8.0,8.0) -- (9.0,9.0);
 \draw (9.0,9.0) -- (10.0,9.0);
 \draw (10.0,9.0) -- (11.0,8.0);
 \draw (10.0,9.0) -- (11.0,10.0);
 \draw[fill] (9.0,9.0) circle (0.08);
 \draw[fill] (10.0,9.0) circle (0.08);
 \node [above] at (9.0,9.1) {$0$};
 \node [above] at (10.0,9.1) {$1$};
 \node [right] at (11.0,10.0) {$z_1$};
 \node [right] at (11.0,8.0) {$z_2$};
 \node [left] at (8.0,8.0) {$z_3$};
 \node [left] at (8.0,10.0) {$z_4$};
 \node at (4.0,7.0) {$\Downarrow$};
 \node at (7.0,7.0) {forget $z_4$};
 \node at (9.5,7.0) {$\Downarrow$};
 \genuszero{0.0}{4.0}
 \genusone{2.0}{4.0}
 \draw[fill] (0.8,4.7) circle (0.08);
 \draw[fill] (5.1,5.3) circle (0.08);
 \draw[fill] (5.1,4.7) circle (0.08);
 \node [right] at (5.1,5.3) {$z_1$};
 \node [right] at (5.1,4.7) {$z_2$};
 \node [left] at (0.8,4.7) {$z_3$};
 \node at (7.0,5.0) {$\Leftrightarrow$};
 \draw (8.0,5.0) -- (9.0,5.0);
 \draw (9.0,5.0) -- (10.0,5.0);
 \draw (10.0,5.0) -- (11.0,4.0);
 \draw (10.0,5.0) -- (11.0,6.0);
 \draw[fill] (9.0,5.0) circle (0.08);
 \draw[fill] (10.0,5.0) circle (0.08);
 \node [above] at (9.0,5.1) {$0$};
 \node [above] at (10.0,5.1) {$1$};
 \node [right] at (11.0,6.0) {$z_1$};
 \node [right] at (11.0,4.0) {$z_2$};
 \node [left] at (8.0,5.0) {$z_3$};
 \node at (4.0,3.0) {$\Downarrow$};
 \node at (7.0,3.0) {stabilise};
 \node at (9.5,3.0) {$\Downarrow$};
 \genusone{2.0}{0.0}
 \draw[fill] (2.8,0.7) circle (0.08);
 \draw[fill] (5.1,1.3) circle (0.08);
 \draw[fill] (5.1,0.7) circle (0.08);
 \node [right] at (5.1,1.3) {$z_1$};
 \node [right] at (5.1,0.7) {$z_2$};
 \node [left] at (2.8,0.7) {$z_3$};
 \node at (7.0,1.0) {$\Leftrightarrow$};
 \draw (9.0,1.0) -- (10.0,1.0);
 \draw (10.0,1.0) -- (11.0,0.0);
 \draw (10.0,1.0) -- (11.0,2.0);
 \draw[fill] (10.0,1.0) circle (0.08);
 \node [above] at (10.0,1.1) {$1$};
 \node [right] at (11.0,2.0) {$z_1$};
 \node [right] at (11.0,0.0) {$z_2$};
 \node [left] at (9.0,1.0) {$z_3$};
\end{tikzpicture}
\end{center}
\index{gluing morphisms}
{\bf Gluing morphisms}: There are two gluing morphisms
\bq
 \overline{\mathcal M}_{g_1,n_1+1} \times \overline{\mathcal M}_{g_2,n_2+1} & \rightarrow & \overline{\mathcal M}_{g_1+g_2,n_1+n_2},
 \nonumber \\
 \overline{\mathcal M}_{g,n+2} & \rightarrow & \overline{\mathcal M}_{g+1,n}.
\eq
We may represent them graphically as
\begin{center}
\begin{tikzpicture}
 \draw (1.0,1.0) -- (0.0,1.0);
 \draw (1.0,1.0) -- (0.5,0.13);
 \draw (1.0,1.0) -- (1.5,0.13);
 \draw (1.0,1.0) -- (2.0,1.0);
 \draw (1.0,1.0) -- (1.5,1.87);
 \draw (1.0,1.0) -- (0.5,1.87);
 \draw[fill=gray] (1.0,1.0) circle (0.2);
 \draw (1.0,1.0) circle (0.2);
 \node [above left] at (0.8,1.0) {$g_1$};
 \node at (2.5,1.0) {$\times$};
 \draw (4.0,1.0) -- (3.0,1.0);
 \draw (4.0,1.0) -- (3.5,0.13);
 \draw (4.0,1.0) -- (4.5,0.13);
 \draw (4.0,1.0) -- (5.0,1.0);
 \draw (4.0,1.0) -- (4.5,1.87);
 \draw (4.0,1.0) -- (3.5,1.87);
 \draw[fill=gray] (4.0,1.0) circle (0.2);
 \draw (4.0,1.0) circle (0.2);
 \node [above left] at (3.8,1.0) {$g_2$};
 \node at (6.0,1.0) {$\rightarrow$};
 \draw (8.0,1.0) -- (7.0,1.0);
 \draw (8.0,1.0) -- (7.5,0.13);
 \draw (8.0,1.0) -- (8.5,0.13);
 \draw (8.0,1.0) -- (9.0,1.0);
 \draw (8.0,1.0) -- (8.5,1.87);
 \draw (8.0,1.0) -- (7.5,1.87);
 \draw[fill=gray] (8.0,1.0) circle (0.2);
 \draw (8.0,1.0) circle (0.2);
 \node [above left] at (7.8,1.0) {$g_1$};
 \draw (10.0,1.0) -- (9.0,1.0);
 \draw (10.0,1.0) -- (9.5,0.13);
 \draw (10.0,1.0) -- (10.5,0.13);
 \draw (10.0,1.0) -- (11.0,1.0);
 \draw (10.0,1.0) -- (10.5,1.87);
 \draw (10.0,1.0) -- (9.5,1.87);
 \draw[fill=gray] (10.0,1.0) circle (0.2);
 \draw (10.0,1.0) circle (0.2);
 \node [above left] at (9.8,1.0) {$g_2$};
\end{tikzpicture}
\\
\begin{tikzpicture}
 \draw (1.0,1.0) -- (1.0,2.0);
 \draw (1.0,1.0) -- (0.13,1.5);
 \draw (1.0,1.0) -- (0.13,0.5);
 \draw (1.0,1.0) -- (1.0,0.0);
 \draw (1.0,1.0) -- (1.87,0.5);
 \draw (1.0,1.0) -- (1.87,1.5);
 \draw[fill=gray] (1.0,1.0) circle (0.2);
 \draw (1.0,1.0) circle (0.2);
 \node [left] at (0.7,1.0) {$g$};
 \node at (3.0,1.0) {$\rightarrow$};
 \draw (5.0,1.0) -- (5.0,2.0);
 \draw (5.0,1.0) -- (4.13,1.5);
 \draw (5.0,1.0) -- (4.13,0.5);
 \draw (5.0,1.0) -- (5.0,0.0);
 \draw (5.0,1.0) to [curve through={(5.433,1.25)(5.87,1.5)(6.5,1.0)(5.87,0.5)(5.433,0.75)}] (5.0,1.0);
 \draw[fill=gray] (5.0,1.0) circle (0.2);
 \draw (5.0,1.0) circle (0.2);
 \node [left] at (4.7,1.0) {$g$};
\end{tikzpicture}
\end{center}
Note that the images of these morphisms are necessarily non-smooth.
\\
\\
Remark: We are not restricted to forget the last marked point in the forgetful morphism, nor
are we restricted to glue the last two points together in the gluing morphisms.
We therefore have various forgetful morphisms and gluing morphisms, which we may index by the point,
which we forget, or by the two points which are glued together, respectively.


\section{The genus zero case}
\label{appendix_moduli_space:genus_zero_case}

Let us now consider a Riemann surface of genus $0$, i.e. the Riemann sphere.
Such a surface is isomorphic to ${\mathbb C}{\mathbb P}^1$ and the group of M\"obius transformations $\mathrm{PSL}\left(2,{\mathbb C}\right)$
\bq
 z & \rightarrow & \frac{az+b}{cz+d}
\eq
acts as an automorphism.
We mark on the Riemann sphere $n$ points.
The moduli space is denoted
by ${\mathcal M}_{0,n}$.
\bq
 {\mathcal M}_{0,n}
 & = &
 \left\{ \left(z_1,\dots,z_n\right) \; | \; z_i \in {\mathbb C} {\mathbb P}^1, z_i \neq z_j \right\} / \mathrm{PSL}\left(2,{\mathbb C}\right).
\eq
This is an affine variety of dimension
\bq
 \mathrm{dim} \; {\mathcal M}_{0,n}
 & = &
 n-3.
\eq
We may use the freedom of $\mathrm{PSL}(2,{\mathbb C})$-transformations to fix three points.
The standard choice will be $z_1=0$, $z_{n-1}=1$ and $z_n=\infty$.
Thus
\bq
\label{appendix_moduli_space:def_simplicial_coordinates}
 {\mathcal M}_{0,n}
 & = &
 \left\{ (z_2,\dots,z_{n-2}) \in {\mathbb C}^{n-3} \; : \; z_i \neq z_j, \; z_i \neq 0, \; z_i \neq 1 \right\}.
\eq
The variables $z_2,\dots,z_{n-2}$ are called 
\index{simplicial coordinates}
{\bf simplicial coordinates}.
We denote the 
\index{set of real points of ${\mathcal M}_{0,n}$}
{\bf set of real points} by ${\mathcal M}_{0,n}({\mathbb R})$:
\bq
 {\mathcal M}_{0,n}\left({\mathbb R}\right)
 & = &
 \left\{ (z_2,\dots,z_{n-2}) \in {\mathbb R}^{n-3} \; : \; z_i \neq z_j, \; z_i \neq 0, \; z_i \neq 1 \right\}.
\eq
Let us look at a few examples.
\begin{enumerate}
\item For $n=3$ we have $\mathrm{dim} \; {\mathcal M}_{0,3} = 0$ and ${\mathcal M}_{0,3}$ consists of a single point.
We may use $\mathrm{PSL}\left(2,{\mathbb C}\right)$-invariance to take
\bq
 \left(z_1,z_2,z_3\right) 
 & = &
 \left(0,1,\infty\right)
\eq
as a representative of this point.

\item For $n=4$ we have $\mathrm{dim} \; {\mathcal M}_{0,4} = 1$ and elements of ${\mathcal M}_{0,4}$ 
can be represented by
\bq
 \left(z_1,z_2,z_3,z_4\right)
 & = &
 \left(0,z,1,\infty\right)
\eq
with
\bq
 z & \in & {\mathbb C} {\mathbb P}^1 \backslash \left\{ 0,1,\infty \right\}.
\eq
Thus
\bq
 {\mathcal M}_{0,4} 
 \;\; \simeq \;\;
 {\mathbb C} {\mathbb P}^1 \backslash \left\{ 0,1,\infty \right\}
 \;\; \simeq \;\;
 {\mathbb C} \backslash \left\{ 0,1 \right\}.
\eq

\item For $n=5$ we have $\mathrm{dim} \; {\mathcal M}_{0,5} = 2$ and ${\mathcal M}_{0,5}$ 
is isomorphic to
\bq
 {\mathcal M}_{0,5} 
 \;\; \simeq \;\;
 \left\{ \left( z_1, z_2 \right) \; | \; z_i \in \mathbb{C}, \; z_i \neq 0,1,  \;\; z_1 \neq z_2 \right\}.
\eq
Thus ${\mathcal M}_{0,5}$ is isomorphic to the complement of the five lines $z_1=0$, $z_1=1$, $z_2=0$, $z_2=1$ and $z_1=z_2$
in ${\mathbb C}^2$.

In fig.~(\ref{appendix_moduli_space:fig_M_0_5}) we sketch the moduli space ${\mathcal M}_{0,5}({\mathbb R})$.
\begin{figure}
\begin{center}
\includegraphics[scale=0.8]{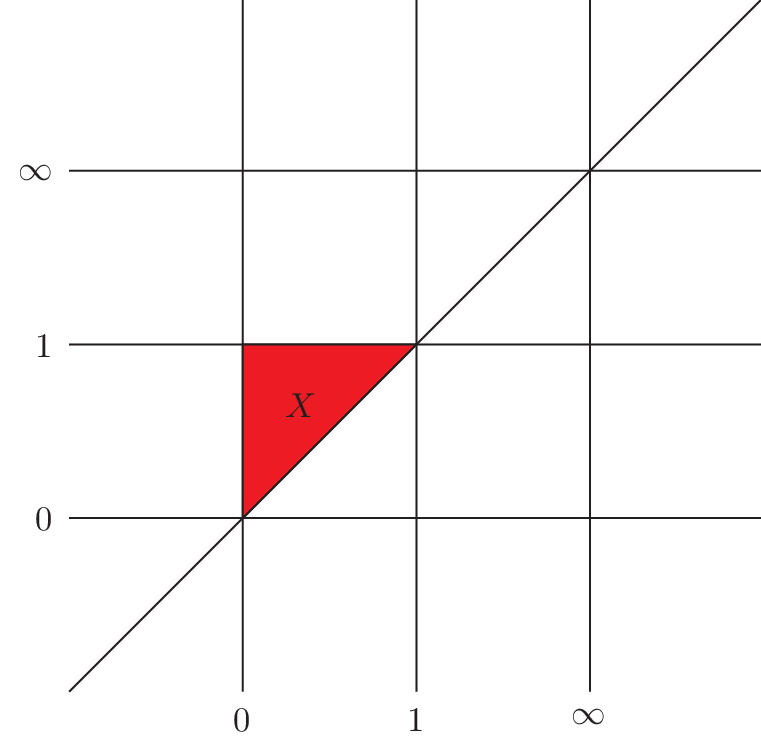}
\hspace*{10mm}
\includegraphics[scale=0.8]{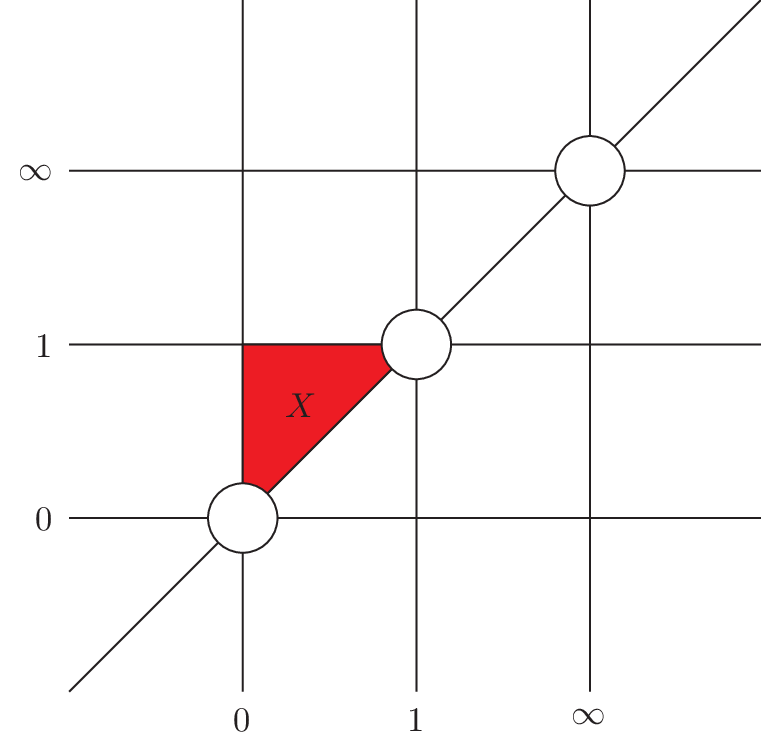}
\end{center}
\caption{
The moduli space ${\mathcal M}_{0,5}({\mathbb R})$ (left).
The region $X$ is bounded by $z_2=0$, $z_3=1$ and $z_2=z_3$.
The right figure shows $\overline{\mathcal M}_{0,5}({\mathbb R})$, obtained from ${\mathcal M}_{0,5}({\mathbb R})$ by blowing up the points
$(z_2,z_3)=(0,0)$, $(z_2,z_3)=(1,1)$ and $(z_2,z_3)=(\infty,\infty)$.
}
\label{appendix_moduli_space:fig_M_0_5}
\end{figure}
In fig.~\ref{appendix_moduli_space:fig_M_0_5} we indicated in red a region $X$ of ${\mathcal M}_{0,5}({\mathbb R})$.
This region is bounded by $z_2=0$, $z_3=1$ and $z_2=z_3$.
In general there will be points, where the boundaries do not cross normally.
For the region $X$ in the example above this occurs for $(z_2,z_3)=(0,0)$ and $(z_2,z_3)=(1,1)$. 
We denote by $\overline{\mathcal M}_{0,n}$ the blow-up of ${\mathcal M}_{0,n}$ in all those points, such that
in $\overline{\mathcal M}_{0,n}$ all boundaries cross normally.
In this way the region $X$ of our example transforms from a triangle in ${\mathcal M}_{0,5}({\mathbb R})$ into
a pentagon in $\overline{\mathcal M}_{0,5}({\mathbb R})$.
\end{enumerate}
For a set of points $\{z_i, z_j, z_k, z_l \} \subseteq \{z_1,z_2,\dots,z_n\}$ we define a
\index{cross-ratio}
{\bf cross-ratio} as follows
\bq
 \left[ i,j | k,l \right]
 & = &
 \frac{\left(z_i-z_k\right)\left(z_j-z_l\right)}{\left(z_i-z_l\right)\left(z_j-z_k\right)}.
\eq
The cross-ratios are invariant under M\"obius transformations $\mathrm{PSL}\left(2,{\mathbb C}\right)$.
We have the relations
\bq
 \left[ i,j | k,l \right]
 & = &
 \left[ i,j | l,k \right]^{-1},
 \nonumber \\
 \left[ i,j | k,l \right]
 & = &
 \left[ j,i | l,k \right]^{-1},
 \nonumber \\
 \left[ i,j | k,l \right]
 & = &
 \left[ k,l | i,j \right],
 \nonumber \\
 \left[ i,j | k,l \right]
 & = &
 1 - \left[ i,k | j,l \right].
\eq

\subsection{The Deligne-Mumford-Knudsen compactification}

Let us now review a systematic way to construct $\overline{\mathcal M}_{0,n}$.
There is a smooth compactification 
\bq
 {\mathcal M}_{0,n} \subset \overline{\mathcal M}_{0,n},
\eq
known as the Deligne-Mumford-Knudsen compactification \cite{Deligne:1969,Knudsen:1976,Knudsen:1983,Knudsen:1983a},
such that $\overline{\mathcal M}_{0,n} \backslash \mathcal M_{0,n}$ is a smooth normal
crossing divisor.
In order to describe $\overline{\mathcal M}_{0,n}$ we follow ref.~\cite{Brown:2006}.

Let $\pi=(\pi_1,\dots,\pi_n)$ be a permutation of $(1,\dots,n)$.
A 
\index{cyclic order}
{\bf cyclic order} is defined as a permutation modulo cyclic permutations $(\pi_1,\pi_2,\dots,\pi_n) \rightarrow (\pi_2,\dots,\pi_n,\pi_1)$.
We may represent a cyclic order by an $n$-gon, where the edges of the $n$-gon are indexed clockwise
by $\pi_1$, $\pi_2$, \dots, $\pi_n$.
A 
\index{dihedral structure}
{\bf dihedral structure} is defined as a permutation modulo cyclic permutations and
reflection $(\pi_1,\pi_2,\dots,\pi_n) \rightarrow (\pi_n,\dots,\pi_2,\pi_1)$.
We may represent a dihedral structure by an $n$-gon, where the edges of the $n$-gon are indexed either clockwise or anti-clockwise
by $\pi_1$, $\pi_2$, \dots, $\pi_n$.

The construction of $\overline{\mathcal M}_{0,n}$ proceeds through intermediate spaces $\mathcal M_{0,n}^\pi$, labelled by a dihedral structure $\pi$,
such that
\bq
 \mathcal M_{0,n} \subset \mathcal M_{0,n}^\pi \subset \overline{\mathcal M}_{0,n}.
\eq
Let $z=(z_1,\dots,z_n)$ denote the (ordered) set of the $n$ marked points on the curve.
In the following we will use the notation 
\bq
 \mathcal M_{0,z}
\eq
for $\mathcal M_{0,n}$.
This notation allows us to distinguish $\mathcal M_{0,z'}$ from $\mathcal M_{0,z''}$ if $z'$ and $z''$ are two non-identical subsets of $z$
with $k$ elements each (i.e. $z' \neq z''$ but $|z'|=|z''|=k$).
Let $\pi$ denote a permutation of $(1,\dots,n)$, which defines a dihedral structure.
We may draw a 
\index{regular $n$-gon}
{\bf regular $n$-gon}, where the edges are labelled by $z_{\pi_1}$, $z_{\pi_2}$, \dots, $z_{\pi_n}$ in this order.
In order to keep the notation simple let us assume that $\pi=(1,2,\dots,n)$. Then the edges are labelled by $z_1$, $z_2$, \dots, $z_n$.

\begin{figure}
\begin{center}
\includegraphics[scale=0.6]{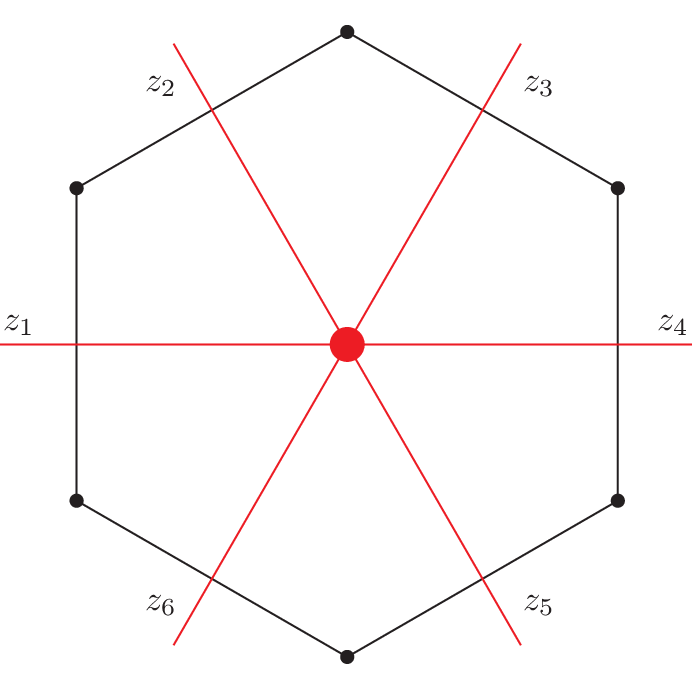}
\end{center}
\caption{
An example of the relation between the dual graph (red) and the $n$-gon (black).
}
\label{appendix_moduli_space:fig_dual_graph_dual_graph}
\end{figure}

We may think of the $n$-gon as the ``dual graph of the dual graph'', as shown in fig.~\ref{appendix_moduli_space:fig_dual_graph_dual_graph}.
The words ``of the dual graph'' refers to the dual graph of a nodal curve introduced in section~\ref{appendix_moduli_space:riemann_surfaces}, the words ``dual graph of'' refer to the construction of a dual graph from
a planar graph as discussed in section~\ref{chapter_graph_polynomials:sect_duality}.
In this example one starts from a Riemann sphere with six marked points $z_1$-$z_6$.
The dual graph is given by a genus-$0$ vertex with six external legs.
Choosing a dihedral structure defines in particular a cyclic order and we draw the dual graph with this cyclic order in the plane.
We then construct the dual graph of the dual graph, this gives the hexagon.
Please note that the term ``dual graph'' is used with two different meanings:
Once we mean the graph dual to a Riemann sphere with marked points, the second meaning refers to the dual of a graph drawn in
a plane.

A 
\index{chord}
{\bf chord} of the polygon connects two non-adjacent vertices and may be specified 
by giving the two edges preceding the two vertices in
the clockwise orientation.
Thus $(i,j)$ denotes the chord from the vertex between edge $z_i$ and $z_{i+1}$ to the vertex between the edge $z_j$ and $z_{j+1}$.
There are
\bq
 \frac{1}{2} n \left( n-3 \right)
\eq
chords for a regular $n$-gon.
We denote by $\chi(z,\pi)$ the set of all chords of the $n$-gon defined by the set $z$ and the dihedral structure $\pi$.
Each chord defines a cross-ratio as follows (for the dihedral structure $\pi=(1,2,\dots,n)$):
\bq
 u_{i,j} 
 & = &
 \left[ i,i+1 | j+1,j \right]
 \;\; = \;\;
 \frac{\left(z_i-z_{j+1}\right)\left(z_{i+1}-z_j\right)}{\left(z_i-z_j\right)\left(z_{i+1}-z_{j+1}\right)}
\eq
For an arbitrary dihedral structure $\pi$ we set
\bq
 u_{i,j}^\pi 
 & = &
 \left[ \pi_i,\pi_{i+1} | \pi_{j+1},\pi_j \right]
 \;\; = \;\;
 \frac{\left(z_{\pi_i}-z_{\pi_{j+1}}\right)\left(z_{\pi_{i+1}}-z_{\pi_j}\right)}{\left(z_{\pi_i}-z_{\pi_j}\right)\left(z_{\pi_{i+1}}-z_{\pi_{j+1}}\right)}.
\eq
As already mentioned, 
the cross-ratio is invariant under 
$\mathrm{PSL}(2,{\mathbb C})$-transformations.
Each cross-ratio defines a function
\bq
 \mathcal M_{0,z}
 & \rightarrow &
 {\mathbb C} {\mathbb P}^1 \backslash \{0,1,\infty\},
\eq
or equivalently
\bq
 \mathcal M_{0,z}
 & \rightarrow &
 {\mathbb C} \backslash \{0,1\}.
\eq
If a cross-ratio takes a value from the set $\{0,1,\infty\}$, one can show that two points of the $z_i$'s coincide
(which contradicts the assumption that all marked points are distinct).
The set of all cross-ratios for a given dihedral structure $\pi$ defines an embedding
\bq
 \mathcal M_{0,z}
 & \rightarrow &
 {\mathbb C}^{n (n-3)/2}.
\eq
We may now consider the Zariski closure of the image of this embedding
and take the Zariski closure as a chart of the dihedral extension $\mathcal M_{0,z}^\pi$.
This defines the dihedral extension $\mathcal M_{0,z}^\pi$.
Since the chart and the dihedral extension $\mathcal M_{0,z}^\pi$ are homeomorphic, one usually does not distinguish between the two.
The Deligne-Mumford-Knudsen compactification is obtained by gluing these charts together:
\bq
 \overline{\mathcal M}_{0,z}
 & = &
 \bigcup\limits_{\pi} \; \mathcal M_{0,z}^\pi,
\eq
where $\pi$ ranges over the $(n-1)!/2$ inequivalent dihedral structures.
\\
\\
Example 1: The simplest case is $n=4$ and
from $(z_1,z_2,z_3,z_4) = (0,z,1,\infty)$ we concluded that
\bq
 {\mathcal M}_{0,4} 
 \;\; \simeq \;\;
 {\mathbb C} {\mathbb P}^1 \backslash \left\{ 0,1,\infty \right\}
 \;\; \simeq \;\;
 {\mathbb C} \backslash \left\{ 0,1 \right\}.
\eq
For $n=4$ there are three inequivalent dihedral structures, which we may take as
\bq
 \pi_1 \; = \; \left(1,2,3,4\right),
 \;\;\;
 \pi_2 \; = \; \left(2,3,1,4\right),
 \;\;\;
 \pi_3 \; = \; \left(3,1,2,4\right).
\eq
For each dihedral structure there are two chords.
Let's start with $\pi_1$.
We have
\bq
 u_{1,3}^{\pi_1}
 \;\; = \;\;
 \left[ 1,2 | 4,3 \right]
 \;\; = \;\;
 1-z,
 & &
 u_{2,4}^{\pi_1}
 \;\; = \;\;
 \left[ 2,3 | 1,4 \right]
 \;\; = \;\;
 z.
\eq
The embedding is given by
\bq
 {\mathcal M}_{0,4}
 \; = \; 
 {\mathbb C} \backslash \left\{ 0,1 \right\}
 & \rightarrow &
 {\mathbb C}^2,
 \nonumber \\
 z & \rightarrow & \left( \begin{array}{c} 1-z \\ z \\ \end{array} \right).
\eq
Taking the closure of the image, we add the points
\bq
 \left( \begin{array}{c} 1 \\ 0 \\ \end{array} \right),
 & &
 \left( \begin{array}{c} 0 \\ 1 \\ \end{array} \right).
\eq
Therefore
\bq
 {\mathcal M}_{0,4}^{\pi_1}
 & = &
 {\mathbb C} 
 \;\; = \;\; 
 {\mathbb C} {\mathbb P}^1 \backslash \left\{ \infty \right\}.
\eq
The point at infinity has not been added.
In order to get the point at infinity we look at the other dihedral extensions.
We have
\bq
 u_{1,3}^{\pi_2}
 \;\; = \;\;
 \left[ 2,3 | 4,1 \right]
 \;\; = \;\;
 \frac{1}{z},
 & &
 u_{2,4}^{\pi_2}
 \;\; = \;\;
 \left[ 3,1 | 2,4 \right]
 \;\; = \;\;
 1 - \frac{1}{z},
\eq
and therefore
\bq
 {\mathcal M}_{0,4}^{\pi_2}
 & = &
 {\mathbb C} {\mathbb P}^1 \backslash \left\{ 0 \right\}.
\eq
Furthermore
\bq
 u_{1,3}^{\pi_3}
 \;\; = \;\;
 \left[ 3,1 | 4,2 \right]
 \;\; = \;\;
 1-\frac{1}{1-z},
 & &
 u_{2,4}^{\pi_3}
 \;\; = \;\;
 \left[ 1,2 | 3,4 \right]
 \;\; = \;\;
 \frac{1}{1-z},
\eq
and hence
\bq
 {\mathcal M}_{0,4}^{\pi_3}
 & = &
 {\mathbb C} {\mathbb P}^1 \backslash \left\{ 1 \right\}.
\eq
In total we find
\bq
 \overline{\mathcal M}_{0,4}
 \;\; = \;\;
 {\mathbb C} {\mathbb P}^1,
 & \;\;\; &
 \overline{\mathcal M}_{0,4} \backslash {\mathcal M}_{0,4}
 \;\; = \;\;
 \left\{ 0, 1, \infty \right\}.
\eq
We have constructed three charts for $\overline{\mathcal M}_{0,4}$, indexed by the dihedral structures
$\pi_1$, $\pi_2$ and $\pi_3$.
A single chart does not cover $\overline{\mathcal M}_{0,4}$, we need at least two of them.
The boundary divisor $\overline{\mathcal M}_{0,4} \backslash {\mathcal M}_{0,4}$ consists of three points.
This is a particular simple example, where blow-ups do not yet enter.
\\
\\
Example 2:
In order to see how blow-ups enter the game, we discuss the next more complicated example given by
$n=5$.
With $(z_1,z_2,z_3,z_4,z_5)=(0,z_2,z_3,1,\infty)$ we have
\bq
 {\mathcal M}_{0,5} 
 \;\; \simeq \;\;
 \left\{ \left( z_2, z_3 \right) \; | \; z_i \in \mathbb{C}, \; z_i \neq 0,1,  \;\; z_2 \neq z_3 \right\}.
\eq
There are now $12$ inequivalent dihedral structures. Let us take $\pi=(1,2,3,4,5)$.
For each dihedral structure we have $5$ chords.
We have
\bq
 u_{1,3} 
 & = & 
 \left[ 1,2 | 4,3 \right]
 \;\; = \;\;
 \frac{z_3-z_2}{\left(1-z_2\right) z_3},
 \nonumber \\
 u_{1,4} 
 & = & 
 \left[ 1,2 | 5,4 \right]
 \;\; = \;\;
 1 -z_2,
 \nonumber \\
 u_{2,4} 
 & = & 
 \left[ 2,3 | 5,4 \right]
 \;\; = \;\;
 \frac{1-z_3}{1-z_2},
 \nonumber \\
 u_{2,5} 
 & = & 
 \left[ 2,3 | 1,5 \right]
 \;\; = \;\;
 \frac{z_2}{z_3},
 \nonumber \\
 u_{3,5} 
 & = & 
 \left[ 3,4 | 1,5 \right]
 \;\; = \;\;
 z_3.
\eq
The five cross-ratios define the embedding
\bq
 {\mathcal M}_{0,5} & \rightarrow & {\mathbb C}^5,
\eq
the image of this embedding is contained in a plane.
The image does not include five lines, given by the intersection of the plane with one of
the hyperplanes defined by
\bq
 u_{1,3} \; = \; 0,
 \;\;\;
 u_{1,4} \; = \; 0,
 \;\;\;
 u_{2,4} \; = \; 0,
 \;\;\;
 u_{2,5} \; = \; 0,
 \;\;\;
 u_{3,5} \; = \; 0.
\eq
One checks, that the intersection of the plane with one of the hyperplanes defined by
\bq
 u_{1,3} \; = \; 1,
 \;\;\;
 u_{1,4} \; = \; 1,
 \;\;\;
 u_{2,4} \; = \; 1,
 \;\;\;
 u_{2,5} \; = \; 1,
 \;\;\;
 u_{3,5} \; = \; 1
\eq
does not give new lines, for example $u_{3,5}=1$ is equivalent to $u_{2,4}=0$.
For the Zariski closure we add these five lines back.
Let us now understand how we get from the red triangle $X$ in the left picture of fig.~\ref{appendix_moduli_space:fig_M_0_5}
to the pentagon in the right picture.
It is clear that we have away from critical values the correspondence
\bq
 u_{1,3} \; = \; 0 & \Rightarrow & z_2 = z_3,
 \nonumber \\
 u_{2,5} \; = \; 0 & \Rightarrow & z_2 = 0,
 \nonumber \\
 u_{2,4} \; = \; 0 & \Rightarrow & z_3 = 1.
\eq
This gives us three edges of the pentagon.
\begin{figure}
\begin{center}
\includegraphics[scale=0.9]{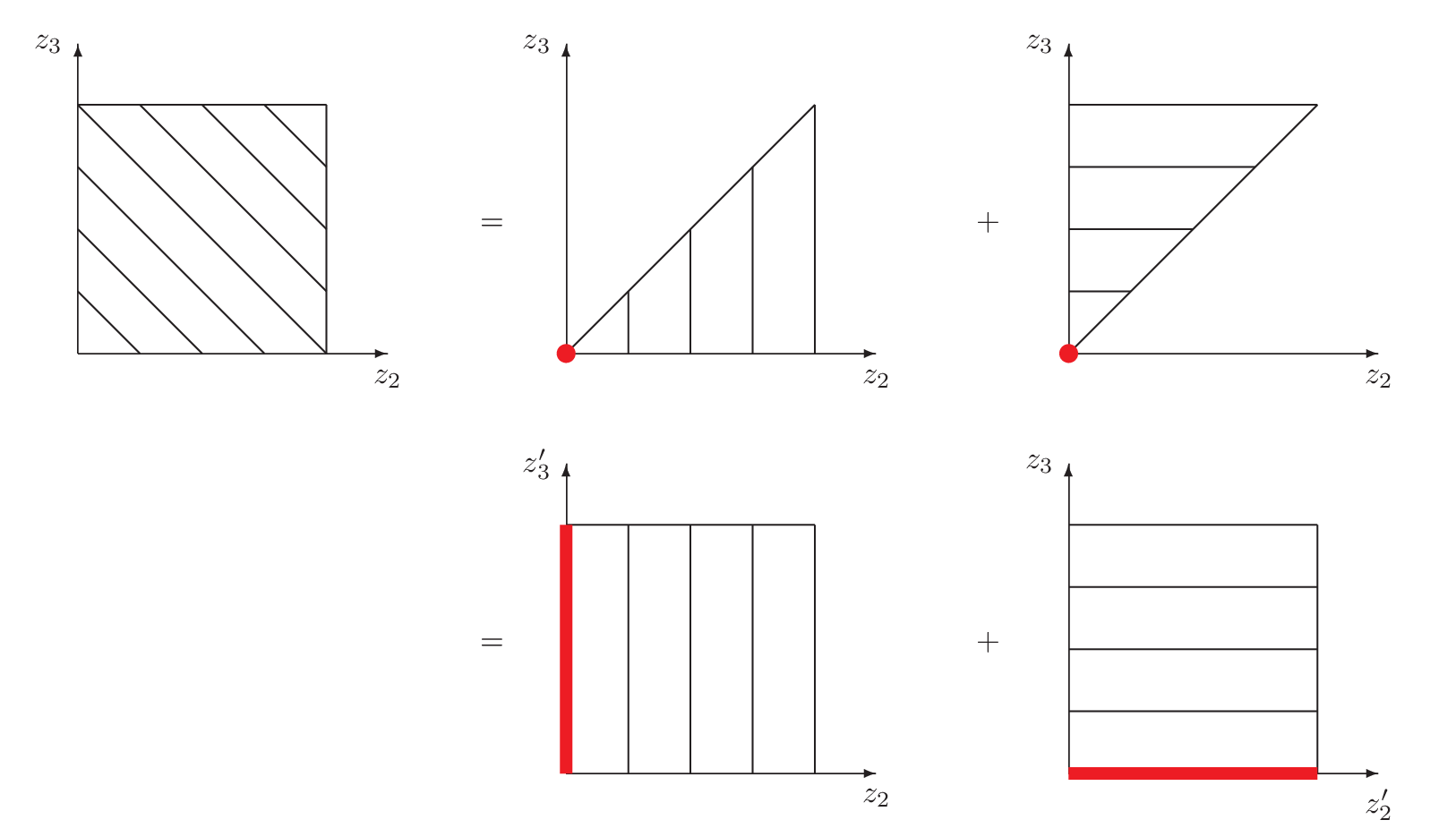}
\end{center}
\caption{
The blow-up of the point $(z_2,z_3)=(0,0)$.
}
\label{appendix_moduli_space:fig_blowup}
\end{figure}
Let us now see what happens, if we blow-up the point $(z_2,z_3)=(0,0)$ by 
${\mathbb C}{\mathbb P}^1$. For ${\mathbb C}{\mathbb P}^1$ we need two charts.
In the first chart ($z_2 \neq 0$) we use coordinates $(z_2,z_3')$, where $z_3'$ is related to the old coordinates
by
\bq
 z_3 \;\; = \;\; z_2 z_3'
 & \Leftrightarrow &
 z_3' \;\; = \;\; \frac{z_3}{z_2}.
\eq
In the second chart ($z_3 \neq 0$) we use coordinates $(z_2',z_3)$, where $z_2'$ is related to the old coordinates
by
\bq
 z_2 \;\; = \;\; z_3 z_2'
 & \Leftrightarrow &
 z_2' \;\; = \;\; \frac{z_2}{z_3}.
\eq
Graphically this is shown in fig.~\ref{appendix_moduli_space:fig_blowup}.
Note that the procedure for the blow-up of this point is completely analogous to the discussion
in chapter~\ref{chapter_sector_decomposition}.
The condition $u_{3,5}=0$ gives in the second chart the line $z_3=0$, i.e. the blow-up of the point $(0,0)$.

A similar argumentation holds for the blow-up of the point $(z_2,z_3)=(1,1)$
and the condition $u_{1,4}=0$.

\subsection{The associahedron}

\begin{figure}
\begin{center}
\includegraphics[scale=0.6]{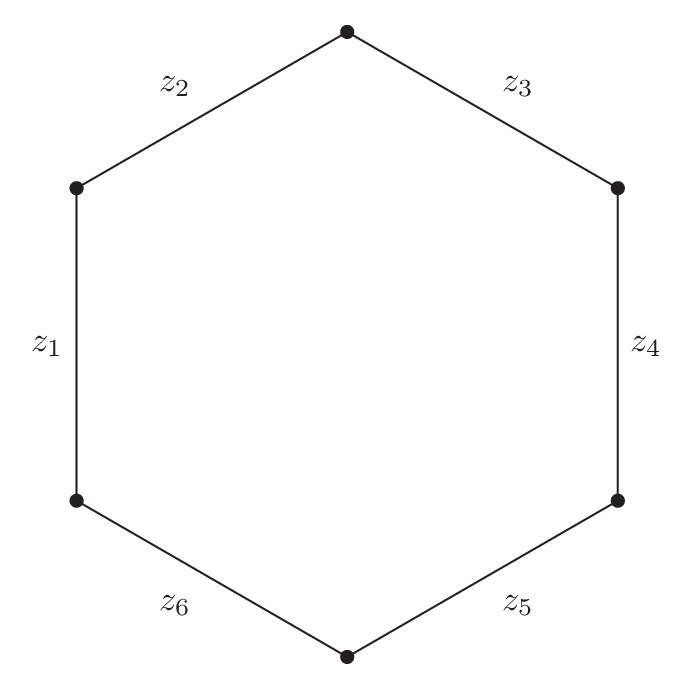}
\hspace*{10mm}
\includegraphics[scale=0.6]{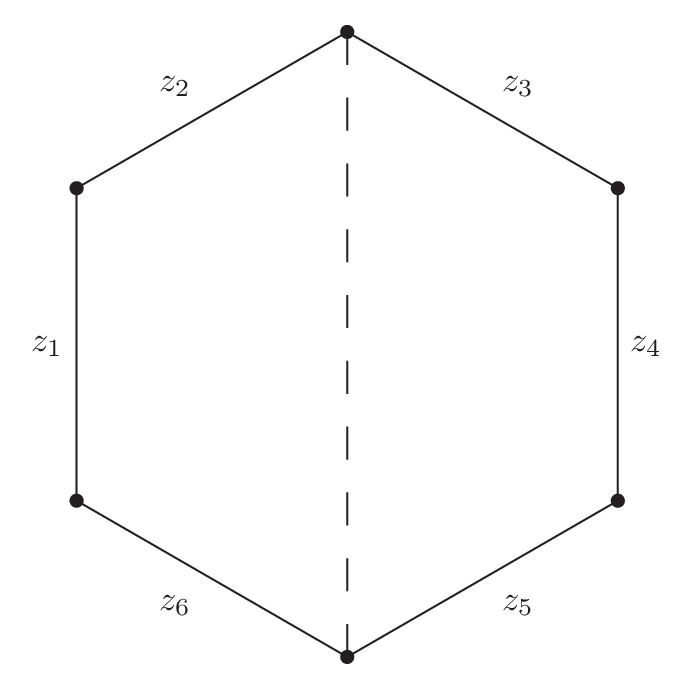}
\hspace*{10mm}
\includegraphics[scale=0.6]{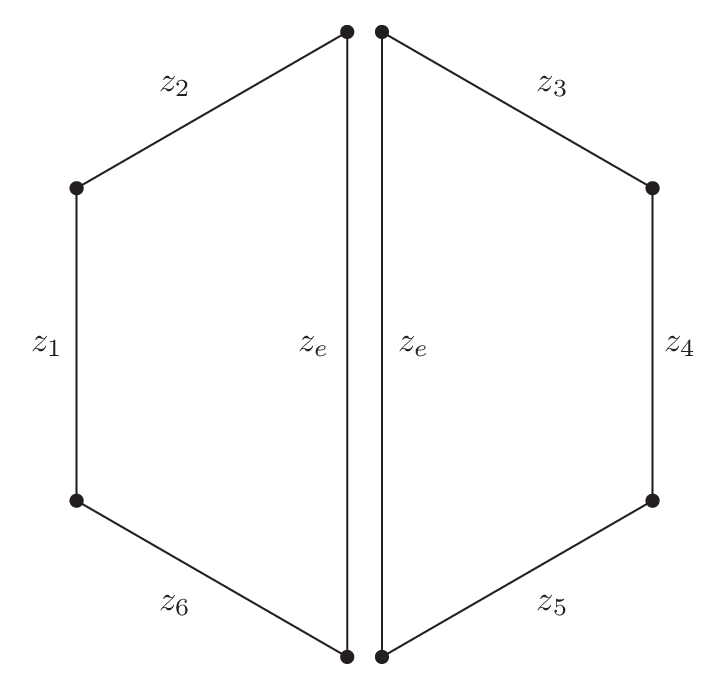}
\end{center}
\caption{
A hexagon, where the edges are labelled by the cyclic ordered variables $(z_1,z_2,\dots,z_6)$ (left picture).
The middle picture shows the chord $(2,5)$.
Right picture: A chord divides the hexagon into two lower $n$-gons, in this case two quadrangles.
}
\label{appendix_moduli_space:fig_polygon}
\end{figure}

\begin{figure}
\begin{center}
\includegraphics[scale=0.6]{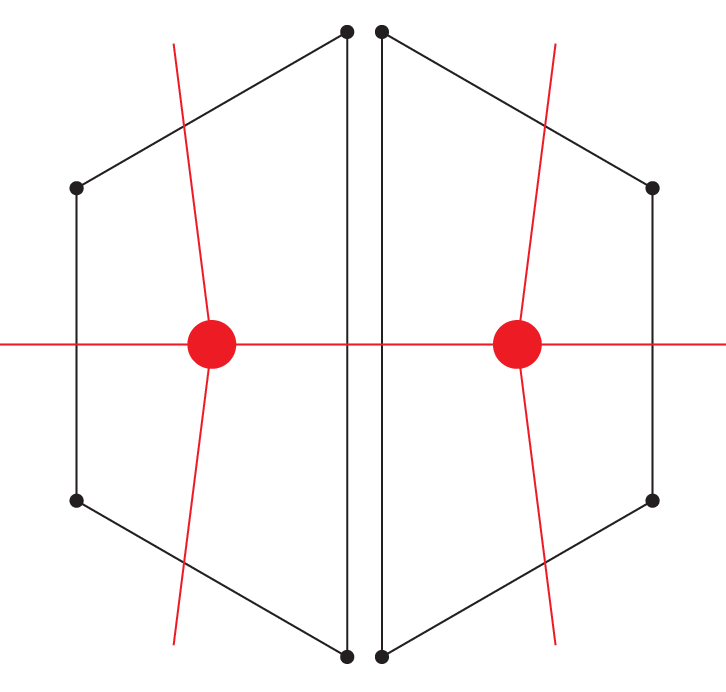}
\hspace*{20mm}
\includegraphics[scale=0.6]{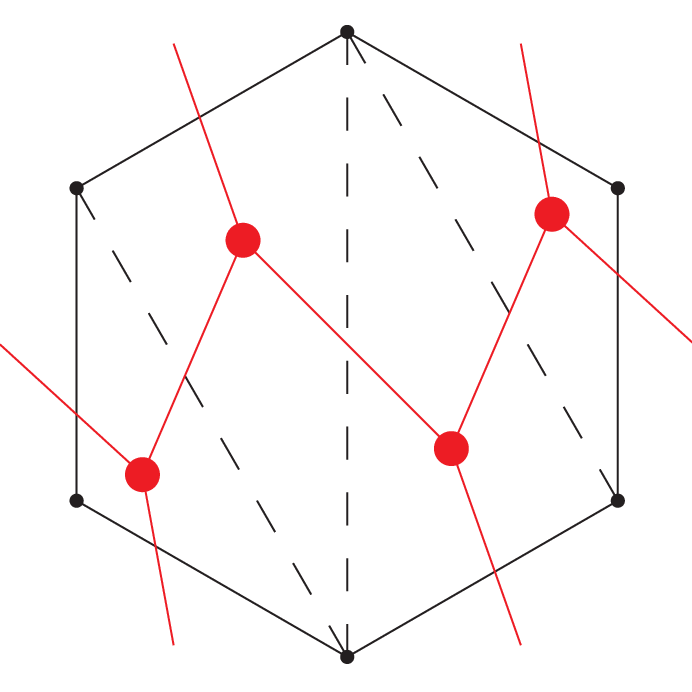}
\end{center}
\caption{
The limit $u_{i,j}\rightarrow 0$ leads to a factorisation of the dual graph (left).
A complete triangulation of the $n$-gon leads to a dual graph with trivalent vertices only (right).
}
\label{appendix_moduli_space:fig_factorisation}
\end{figure}

Let us discuss the dihedral extension $\mathcal M_{0,z}^\pi$ in more detail.
We recall that the construction of $\mathcal M_{0,z}^\pi$ requires the specification
of a dihedral structure $\pi$ (i.e. a permutation up to cyclic permutations and reflection).
We will need a few properties of the dihedral extension $\mathcal M_{0,z}^\pi$ \cite{Brown:2006}:
\begin{enumerate}
\item The complement $\mathcal M_{0,z}^\pi \backslash \mathcal M_{0,z}$ is a normal crossing divisor, whose irreducible components are
\bq
\label{appendix_moduli_space:def_divisors}
 D_{i j} & = & \left\{ \; u_{i,j} \; = \; 0 \; \right\},
\eq
indexed by the chords $(i,j) \in \chi(z,\pi)$. 
\item Each divisor is again a product of spaces of the same type:
Let us consider a chord $(i,j)$. This chord decomposes the original polygon $(z,\pi)$ into two smaller polygons,
as shown in fig.~\ref{appendix_moduli_space:fig_polygon}.
We denote the new edge by $z_e$.
The set of edges for the two smaller polygons are
$z' \cup \{z_e\}$ and $z'' \cup \{z_e\}$, where $z = z' \cup z''$ and $z' \cap z'' = \emptyset$.
The two smaller polygons inherit their dihedral structures $\pi'$ and $\pi''$ from $\pi$ and the chord $(i,j)$.
We have
\bq
\label{appendix_moduli_space:product_structure}
 D_{i j} 
 & \cong &
 \mathcal M_{0,z' \cup \{z_e\}}^{\pi'}
 \times
 \mathcal M_{0,z'' \cup \{z_e\}}^{\pi''}.
\eq
This factorisation translates to the dual graphs, as shown in fig.~\ref{appendix_moduli_space:fig_factorisation}.
Iteration of this procedure corresponds to a triangulation of the $n$-gon or equivalently to a dual tree graph with three-valent vertices only.
\end{enumerate}
Let us now consider the space of real points.
For a given set $z$ and dihedral structure $\pi$ we set
\bq
 X_{0,z}^\pi
 & = &
 \left\{
  u_{i,j} > 0 \; : \; (i,j) \in \chi(z,\pi)
 \right\}
\eq
and
\bq
 \overline{X}_{0,z}^\pi
 & = &
 \left\{
  u_{i,j} \ge 0 \; : \; (i,j) \in \chi(z,\pi)
 \right\}.
\eq
One has 
\bq
 \mathcal M_{0,n}({\mathbb R})
 & = &
 \bigsqcup\limits_{\pi} \; X_{0,z}^\pi,
\eq
where 
$\pi$ ranges again over the $(n-1)!/2$ inequivalent dihedral structures.

For a given set $z$ and dihedral structure $\pi$ the cell $\overline{X}_{0,z}^\pi$ is called 
a Stasheff polytope or associahedron \cite{Stasheff:1963a,Stasheff:1963b,Devadoss:1998,Devadoss:2004}.
The associahedron has the properties
\begin{enumerate}
\item Its facets (i.e. codimension one faces) 
\bq
 F_{i j} & = & \left\{ \; u_{i,j} \; = \; 0 \; \right\},
\eq
are indexed by the chords 
$(i,j) \in \chi(z,\pi)$.
\item From eq.~(\ref{appendix_moduli_space:product_structure}) it follows that each facet is a product
\bq
 F_{ij} & = &
 \overline{X}_{0,z' \cup \{z_e\}}^{\pi'}
 \times
 \overline{X}_{0,z'' \cup \{z_e\}}^{\pi''}.
\eq
\item Two facets $F_{i j}$ and $F_{k l}$ meet if and only if the chords $(i,j)$ and $(k,l)$ do not cross.
\item Faces of codimension $k$ are given by sets of $k$ non-crossing chords.
In particular, the set of vertices of $\overline{X}_{0,z}^\pi$ are in one-to-one
correspondence with the set of triangulations of the $n$-gon defined by the set $z$ and the
dihedral structure $\pi$.
\end{enumerate}
Properties (1) and (2) are the analogues of eq.~(\ref{appendix_moduli_space:def_divisors}) 
and eq.~(\ref{appendix_moduli_space:product_structure}), respectively.
\begin{figure}
\begin{center}
\includegraphics[scale=1.0]{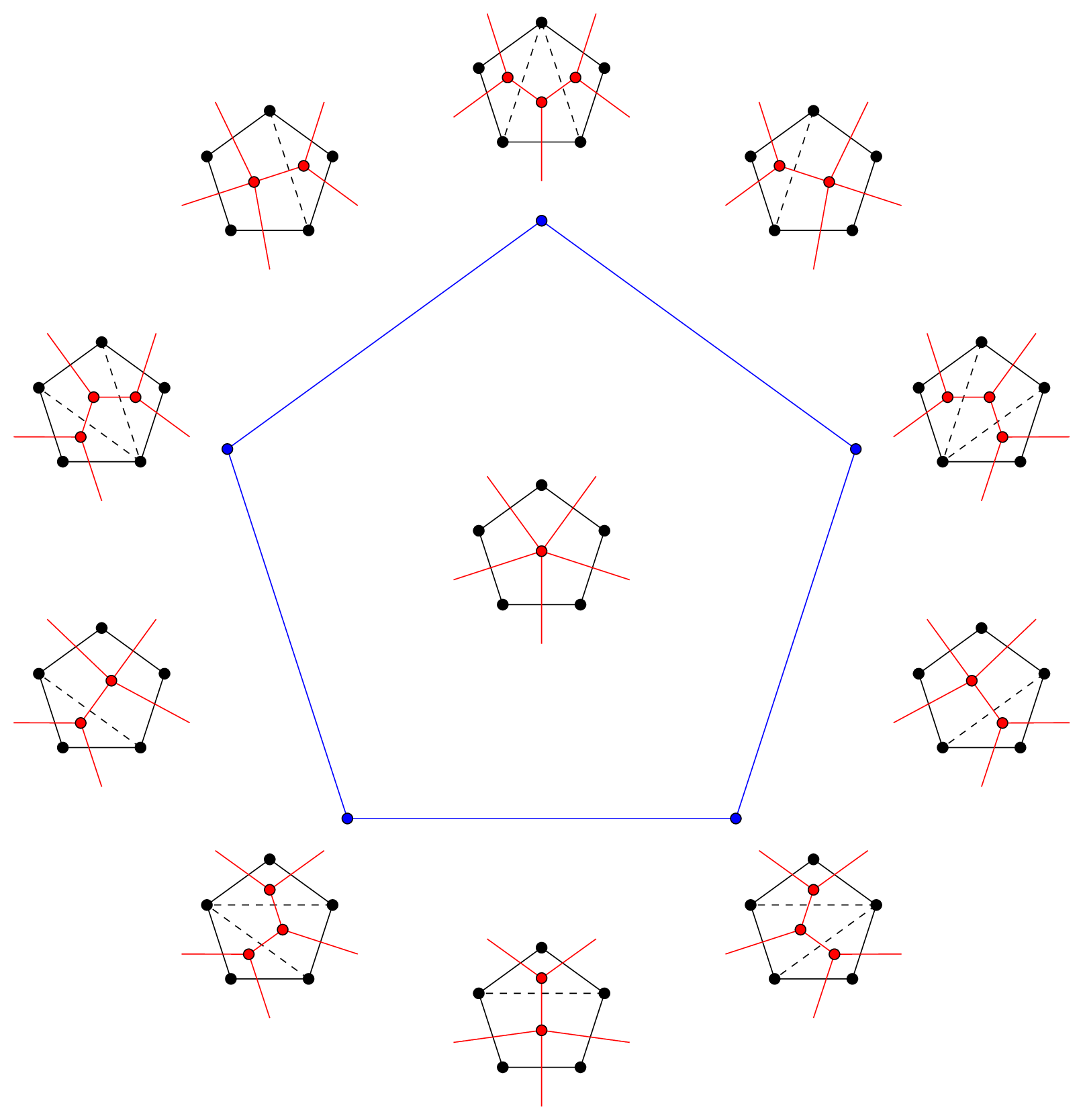}
\end{center}
\caption{
The associahedron for $n=5$ (blue). 
For the Riemann sphere with $n$ marked points the associahedron is a $(n-3)$-dimensional object.
The codimension $k$ faces are either indexed by an $n$-gon with $k$ non-crossing chords (black)
or by dual graphs with $k$ internal edges (red).
}
\label{appendix_moduli_space:fig_associahedron}
\end{figure}
The associahedron for $n=5$ is shown in fig.~\ref{appendix_moduli_space:fig_associahedron}.

\begin{figure}
\begin{center}
\includegraphics[scale=0.6]{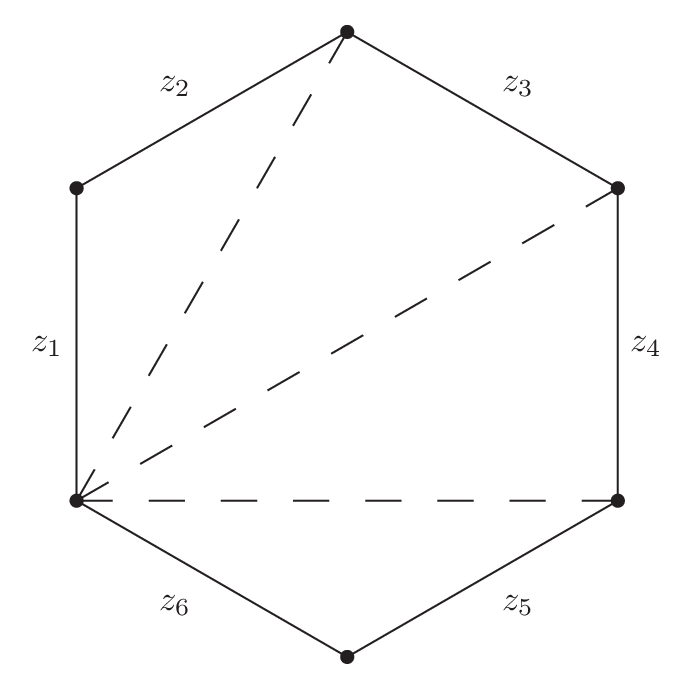}
\end{center}
\caption{
The three chords $(2,6)$, $(3,6)$ and $(4,6)$ define cross-ratios $u_{2,6}$, $u_{3,6}$ and $u_{4,6}$, which may be used as coordinates
on $\mathcal M_{0,6}^\pi$ (for $\pi=(1,2,3,4,5,6)$).
}
\label{appendix_moduli_space:fig_coordinates}
\end{figure}

Let us now have a closer look at coordinates on $\mathcal M_{0,z}^\pi$.
We already introduced the simplicial coordinates $(z_2,\dots,z_{n-2})$
in eq.~(\ref{appendix_moduli_space:def_simplicial_coordinates}).
Let us fix a dihedral structure $\pi$. Without loss of generality we may take the cyclic order to be
$(1,2,\dots,n)$.
Let us consider a chord from $\chi(z,\pi)$. Due to cyclic invariance we may limit ourselves to chords
of the form $(i,n)$.
With the gauge choice $z_1=0$, $z_{n-1}=1$ and $z_n=\infty$ we have
\bq
 u_{2,n} \;\; = \;\; \frac{z_2}{z_3},
 \;\;\;\;\;\;
 \dots
 \;\;\;\;\;\;
 u_{(n-3),n} \;\; = \;\; \frac{z_{n-3}}{z_{n-2}},
 \;\;\;\;\;\;
 u_{(n-2),n} \;\; = \;\; z_{n-2},
\eq
and hence
\bq
\label{appendix_moduli_space:eq_z_coordinates}
 z_i & = &
 \prod\limits_{j=i}^{n-2} u_{j,n},
 \;\;\;\;\;\;\;\;\;\;\;\;
 i \in \{2,\dots,n-2\}.
\eq
Thus we may use as coordinates on $\mathcal M_{0,z}^\pi$ instead of the $(n-3)$ coordinates
$(z_2,\dots,z_{n-2})$ the $(n-3)$ cross-ratios $(u_{2,n},\dots,u_{n-2,n})$.
For $n=6$ this is illustrated in fig.~\ref{appendix_moduli_space:fig_coordinates}.
We have
\bq
 d^{n-3}z & = & \left( \prod\limits_{j=3}^{n-2} u_{j,n}^{j-2} \right) d^{n-3}u.
\eq
Let us further set
\bq
\label{appendix_moduli_space:eq_x_coordinates}
 x_j & = &
 u_{j,n}^{-1}, 
 \;\;\;\;\;\;\;\;\;
 2 \le j \le n-2.
\eq
The $x_j$'s are called 
\index{cubical coordinates}
{\bf cubical coordinates}.
\\
\\
\bs
{\it \refstepcounter{exercise}
{\bf Exercise \theexercise}: 
Let $(z_2,\dots,z_{n-2})$ be simplicial coordinates and $(x_2,\dots,x_{n-2})$ the corresponding cubical coordinates.
Show that
\bq
 \mathrm{Li}_{m_{n-2} \dots m_3 m_2}\left(x_{n-2},\dots,x_3,x_2\right)
 & = &
 \left(-1\right)^{n-3}
 G_{m_{n-2} \dots m_3 m_2}\left(z_{n-2},\dots,z_3,z_2;1\right)
\eq
}
\es
Let us now fix $i_0 \in \{2,\dots,n-2\}$. We will study the limit $u_{i_0,n} \rightarrow 0$.
The chord $(i_0,n)$ splits the polygon into two smaller polygons. 
We set $z'=(z_1,z_2,\dots,z_{i_0})$ and $z''=(z_{i_0+1},\dots,z_n)$.
As before we label the new edge by $z_e$.
One of the two smaller polygons has the edges $z' \cup \{z_e\}$ 
and the dihedral structure $\pi'=(1,2,\dots,i_0,e)$,
the other smaller polygon has the edges $z'' \cup \{z_e\}$ and
the dihedral structure $\pi''=(e,i_0+1,i_0+2,\dots,n)$.
In the limit $u_{i_0,n} \rightarrow 0$ we have
\bq
 \lim\limits_{u_{i_0,n} \rightarrow 0} u_{i,j} & = & 1
\eq
for any chord $(i,j) \in \chi(z,\pi)$ which crosses the chord $(i_0,n) \in \chi(z,\pi)$.


\section{The genus one case}
\label{appendix_moduli_space:genus_one_case}

Let us now turn to the genus one case.
We use the genus one case to introduce fine and coarse moduli spaces.

\subsection{Fine and coarse moduli spaces}

A moduli space is a space (or a scheme or a stack), whose points represent isomorphism classes of algebro-geometric objects.

Let us now introduce the concepts of a fine moduli space and of a coarse moduli space.
For a 
\index{fine moduli space}
{\bf fine moduli space} ${\mathcal M}$ we require that 
\begin{itemize}
\item there is a universal family of objects ${\mathcal C} \rightarrow {\mathcal M}$, 
such that the fibre over $z \in {\mathcal M}$ is the object the point $z$ is parametrising.
\\
Example: If we consider the moduli space ${\mathcal M}_g$ of genus $g$ curves, the fibre over $z$ would be given by the corresponding curve $C$.
\item for any family of objects, parametrised by some base $B$, say ${\mathcal C}_B \rightarrow B$, we require that there is a map
\bq
 f & : & B \rightarrow {\mathcal M},
\eq
and ${\mathcal C}_B$ is isomorphic to $f^\ast {\mathcal C}$.
\bq
\begin{CD}
 {\mathcal C}_B @. {\mathcal C} \\
 @VVV @VVV \\
 B @>{f}>> {\mathcal M} \\
\end{CD}
\eq
\end{itemize}
For a 
\index{coarse moduli space}
{\bf coarse moduli space} $M$ we require that 
\begin{itemize}
\item for any family of objects, parametrised by some base $B$, say ${\mathcal C}_B \rightarrow B$, we require that there is a map
\bq
 f & : & B \;\; \rightarrow \;\; M,
\eq
which sends the fibres of $B$ to their isomorphism classes.
\end{itemize}
A coarse moduli space does not necessarily carry any family of appropriate objects, let alone a universal one.
In other words, a fine moduli space includes both a base space ${\mathcal M}$ and an universal family ${\mathcal C} \rightarrow {\mathcal M}$,
while a coarse moduli space only has the base space $M$.

\subsection{Framed elliptic curves}

We recall that an elliptic curve is a cubic curve in ${\mathbb C}{\mathbb P}^2$ with a marked (rational) point.
Equivalently, we may represent an elliptic curve as the quotient of ${\mathbb C}$ by a lattice $\Lambda$:
\bq
 {\mathbb C} / \Lambda.
\eq
The origin of ${\mathbb C}$ corresponds to the marked point
and we denote the curve together with the marked point by $({\mathbb C} / \Lambda,0)$.

A 
\index{framed elliptic curve}
{\bf framed elliptic curve} is an elliptic curve $E$ together with an ordered basis $\gamma_1$, $\gamma_2$ of $H_1(E,{\mathbb Z})$
such that the intersection number $\gamma_1 \cdot \gamma_2 = 1$.
A framing of a lattice $\Gamma$ in ${\mathbb C}$ is an ordered basis $\psi_1$, $\psi_2$ such that
\bq
\label{appendix_moduli_space:condition_framing}
 \mathrm{Im}\; \left(\frac{\psi_2}{\psi_1}\right) & > & 0.
\eq
We may think of a framed elliptic curve as ${\mathbb C} / \Lambda$ together with the choice of an
ordered basis $\psi_1$, $\psi_2$ satisfying eq.~(\ref{appendix_moduli_space:condition_framing}). 
Two elliptic curves are isomorphic if there is a $c \in {\mathbb C}^\ast$ such that
\bq
 \Lambda' & = & c \Lambda.
\eq
Let $\Lambda$ be generated by $(\psi_1,\psi_2)$.
We may therefore rescale the lattice such that $\psi_1=1$ and $\mathrm{Im}(\psi_2) > 0$.
We label the basis vectors $(1,\tau)$. 
The same lattice is generated if we perform a $\mathrm{SL}\left(2,{\mathbb Z}\right)$-transformation.
Thus we are tempted to consider as a set
\bq
 M_{1,1}
 & \cong &
 {\mathbb H} / \mathrm{SL}\left(2,{\mathbb Z}\right).
\eq
We will later see that this is just a coarse moduli space, not a fine one.
In order to get a fine moduli space we have to consider the 
\index{orbifold}
{\bf orbifold}
\bq
 {\mathcal M}_{1,1}
 & \cong &
 {\mathbb H} \sslash \mathrm{SL}\left(2,{\mathbb Z}\right).
\eq
We will define orbifolds in a second.
But let us first discuss why the set $M_{1,1}$ is just a coarse moduli space.
All points of $M_{1,1}$ have a non-trivial stabiliser group.
The stabiliser group (or isotropy group or little group) of the point $\tau$ is
\bq
 \left\{ \gamma \in \mathrm{SL}\left(2,{\mathbb Z}\right) | \gamma\left(\tau\right) = \tau \right\}.
\eq
There is an isomorphism between the stabilizer group at the point $\tau$ 
and the automorphism group of the corresponding elliptic curve:
\bq
 \mathrm{Aut}\left({\mathbb C}/\Lambda_\tau, 0 \right)
 & \cong &
 \left\{ \gamma \in \mathrm{SL}\left(2,{\mathbb Z}\right) | \gamma\left(\tau\right) = \tau \right\}.
\eq
Each point in ${\mathbb H}$ is invariant under
\bq
 \gamma & = &
 \left( \begin{array}{rr}
  -1 & 0 \\ 
  0 & -1
 \end{array}  \right).
\eq
In addition, 
the point $\tau=i$ is invariant under $S$, while
the point $\tau=r_3=\exp(2\pi i/3)$ is invariant under $U = S T$, where $S$ and $T$ are the usual generators of $\mathrm{SL}\left(2,{\mathbb Z}\right)$:
\bq
 T 
 \;\; = \;\;
 \left( \begin{array}{rr}
  1 & 1 \\ 
  0 & 1
 \end{array}  \right),
 & &
 S
 \;\; = \;\;
 \left( \begin{array}{rr}
  0 & -1 \\ 
  1 & 0
 \end{array}  \right),
\eq
and hence
\bq
 U 
 \;\; = \;\;
 \left( \begin{array}{rr}
  0 & -1 \\ 
  1 & 1
 \end{array}  \right).
\eq
The element $S$ is of order $4$, the element $U$ is of order $6$. Furthermore, $S^2=U^3=-I$.
Thus
\bq
 \mathrm{SL}\left(2,{\mathbb Z}\right)
 & = &
 \left\langle S,T | S^2 = \left(ST\right)^3, S^4=I \right\rangle.
\eq
We therefore have
\bq
 \mathrm{Aut}\left({\mathbb C}/\Lambda_\tau, 0 \right)
 & \cong &
 {\mathbb Z}_2,
 \;\;\;\;\;\;
 \tau \neq i, r_3,
\eq
and
\bq
 \mathrm{Aut}\left({\mathbb C}/\Lambda_i, 0 \right)
 \;\; \cong \;\; 
 {\mathbb Z}_4,
 & &
 \mathrm{Aut}\left({\mathbb C}/\Lambda_{r_3}, 0 \right)
 \;\; \cong \;\; 
 {\mathbb Z}_6.
\eq

\subsection{The universal family of framed elliptic curves}

We first discuss framed elliptic curves, i.e. elliptic curves
together with a fixed choice of an ordered basis $\psi_1, \psi_2$.
We may construct a
\index{universal family of framed elliptic curves}
{\bf universal family of framed elliptic curves}:
Let us consider ${\mathbb C} \times {\mathbb H}$ with an ${\mathbb Z}^2$-action given by
\bq
 \left(n_2,n_1\right) & : &
 \left(z,\tau\right) \;\; \rightarrow \;\; 
 \left( z + n_2 \tau + n_1, \tau \right).
\eq
The ${\mathbb Z}^2$-action corresponds to the translation of $z$ by a lattice vector.
We set 
\bq
 {\mathcal C}_{\mathbb H} & = &
 \left( {\mathbb C} \times {\mathbb H} \right) / {\mathbb Z}^2.
\eq
There is a projection
\bq
 \pi & : & {\mathcal C}_{\mathbb H} \;\; \rightarrow \;\; {\mathbb H},
 \nonumber \\
 & & \left(z,\tau\right) \;\; \rightarrow \;\; \tau 
\eq
such that the fibre over $\tau$ is ${\mathbb C}/\Lambda_\tau$.

A 
\index{framed family of elliptic curves}
family of elliptic curves is {\bf framed}, if it has a locally constant framing.
This means that the cycles $\gamma_1, \gamma_2 \in H_1(E,{\mathbb Z})$ defining the framing vary smoothly.
The family ${\mathcal C}_{\mathbb H} \rightarrow {\mathbb H}$ is framed.
Let ${\mathcal C}_B \rightarrow B$ be a family of framed elliptic curves.
We have a function
\bq
 f & : & B \;\; \rightarrow {\mathbb H},
 \nonumber \\
 & & t \;\; \rightarrow \;\;
 \frac{\int\limits_{\gamma_2(t)} \omega_t}{\int\limits_{\gamma_1(t)} \omega_t},
\eq
where $\omega_t$ is any non-zero holomorphic differential one-form on ${\mathcal C}_t$.
This mapping is called the 
\index{period mapping}
{\bf period mapping}.
The period mapping is holomorphic.
Furthermore, ${\mathcal C}_B$ is isomorphic to $f^\ast {\mathcal C}_{\mathbb H}$.
Thus ${\mathbb H}$ is a fine moduli space for families of framed elliptic curves.
\\
\\
Let us now try to remove the framing.
In other words, we not only allow for $z$ translations by lattice vectors, but also allow 
a change of basis of the lattice vectors by a modular transformation.
We consider again ${\mathbb C} \times {\mathbb H}$, but now with the action of the 
semi-direct product $\mathrm{SL}\left(2,{\mathbb Z}\right) \ltimes {\mathbb Z}^2$.
We denote elements of this group by 
\bq
 \left( \gamma, \vec{n} \right),
 & &
 \gamma \in \mathrm{SL}\left(2,{\mathbb Z}\right),
 \;\;\;\;\;\;
 \vec{n} = (n_2,n_1) \in {\mathbb Z}^2.
\eq
The group composition is given by
\bq
 \left( \gamma_1, \vec{n}_1 \right)
 \left( \gamma_2, \vec{n}_2 \right),
 & = &
 \left( \gamma_1 \gamma_2, \vec{n}_1 \gamma_2 + \vec{n}_2 \right).
\eq
The group $\mathrm{SL}\left(2,{\mathbb Z}\right) \ltimes {\mathbb Z}^2$ acts on 
${\mathbb C} \times {\mathbb H}$ as
\bq
 z' \;\; = \;\; \frac{z+n_2 \tau + n_1}{c \tau + d},
 & & 
 \tau' \;\; = \;\; \frac{a\tau +b}{c\tau +d}.
\eq
In order to grasp the point of what follows, the following simple exercise is helpful:
\\
\\
\bs
{\it \refstepcounter{exercise}
\label{appendix_moduli_space:exercise_action}
{\bf Exercise \theexercise}: 
Let
\bq
 \gamma \; = \; \left(\begin{array}{rr} -1 & 0 \\ 0 & -1 \\ \end{array}\right),
 & &
 \vec{n} \; = \; \left( 0, 0 \right).
\eq
Work out $z'$ and $\tau'$.
}
\es
\\
\\
Let us now consider
\bq
 {\mathcal C}_{M_{1,1}} & = & \left({\mathbb C} \times {\mathbb H}\right) / \left(\mathrm{SL}\left(2,{\mathbb Z}\right) \ltimes {\mathbb Z}^2\right),
 \nonumber \\
 M_{1,1} & = & {\mathbb H} / \mathrm{SL}\left(2,{\mathbb Z}\right).
\eq
There is a projection $\pi : {\mathcal C}_{M_{1,1}} \rightarrow M_{1,1}$ given by $(z,\tau)\rightarrow \tau$, but
${\mathcal C}_{M_{1,1}} \rightarrow M_{1,1}$ is not a universal elliptic curve.
To see this, let us consider the fibre above $\tau \in M_{1,1}$.
We have
\bq
 \pi^{-1}\left(\tau \right)
 & = &
 \left( {\mathbb C}/\Lambda_\tau, 0 \right) / \mathrm{Aut}\left({\mathbb C}/\Lambda_\tau, 0 \right).
\eq
For a generic value of $\tau$ we have $\mathrm{Aut}\left({\mathbb C}/\Lambda_\tau, 0 \right) \cong {\mathbb Z}_2$ and
${\mathbb Z}_2$ acts on $z$ (see exercise~\ref{appendix_moduli_space:exercise_action}) by
\bq
 z' & = & -z.
\eq
This additional symmetry makes $\pi^{-1}(\tau)$ isomorphic to the Riemann sphere ${\mathbb C}{\mathbb P}^1$.
In order to see this, consider first the parallelogram spanned by $1$ and $\tau$.
The additional ${\mathbb Z}_2$-symmetry identifies the points 
\bq
 z
 & \mbox{and} &
 1 + \tau -z.
\eq
Thus, we have to consider only half of the points, i.e. for example just the triangle as shown in the figure below.
\begin{center}
\begin{tikzpicture}
\draw [->] (0.0,0.0) -- (2.0,0.0);
\draw [->] (0.0,0.0) -- (0.5,1.5);
\draw [dashed] (0.5,1.5) -- (2.5,1.5);
\draw [dashed] (2.0,0.0) -- (2.5,1.5);
\draw [->] (4.0,0.0) -- (6.0,0.0);
\draw [->] (4.0,0.0) -- (4.5,1.5);
\draw [dashed] (4.5,1.5) -- (6.0,0.0);
\draw [->] (8.0,0.0) -- (10.0,0.0);
\draw [->] (8.0,0.0) -- (8.5,1.5);
\draw [dashed] (8.5,1.5) -- (10.0,0.0);
\draw[fill=red] (8.33,0.0) circle (0.08);
\draw[fill=red] (9.67,0.0) circle (0.08);
\draw[fill=green] (8.1667,0.5) circle (0.08);
\draw[fill=green] (8.3333,1.0) circle (0.08);
\draw[fill=blue] (9.0,1.0) circle (0.08);
\draw[fill=blue] (9.5,0.5) circle (0.08);
\end{tikzpicture}
\end{center}
Along the edges of the triangle we then identify points which are symmetric around the mid-point, as shown
by the points of the same colour in the figure above.
This gives a sphere.
\\
\\
In particular, no fibre of ${\mathcal C}_{M_{1,1}}$ is an elliptic curve.
Thus, ${\mathcal C}_{M_{1,1}} \rightarrow M_{1,1}$ is not a universal elliptic curve
and $M_{1,1}$ is just a coarse moduli space.

\subsection{Orbifolds}

In order to get a fine moduli space, we have to introduce 
\index{orbifold}
orbifolds.
We start with the definition of an 
\index{orbifold chart}
{\bf orbifold chart} (also called {\bf uniformising system}) \cite{Chen_Ruan:2001aaa,Pronk:2016aaa}.
Let $U_i$ be a non-empty connected topological space.
An orbifold chart of dimension $n$ for $U_i$ is a quadruple $(V_i, \Gamma_i, \rho_i, \phi_i)$, where
\begin{itemize}
\item $V_i$ is a connected and simply connected open subset of ${\mathbb R}^n$,
\item $\Gamma_i$ is a {\bf finite group},
\item $\rho_i : \Gamma_i \rightarrow \mathrm{Aut}\left(V_i\right)$ 
is a ({\bf not necessarily injective}) homomorphism from $\Gamma_i$ to the group of smooth automorphisms of $V_i$.
We set
\bq
 \mathrm{Ker}\left(\Gamma_i\right) \; = \; \mathrm{Ker}\left(\rho_i\right) \; \subseteq \; \Gamma_i,
 & &
 \Gamma_i^{\mathrm{red}} \; = \; \rho_i\left(\Gamma_i\right) \; \subseteq \; \mathrm{Aut}\left(V_i\right).
\eq
\item $\phi_i : V_i \rightarrow U_i$ is a continuous and surjective map from $V_i$ to $U_i$ invariant under $\Gamma_i$, which defines a homeomorphism
\bq
 V_i / \Gamma_i^{\mathrm{red}} & \rightarrow & U_i.
\eq
\end{itemize}
Note that we do not require that $\Gamma_i$ acts effectively on $V_i$. 
\\
The chart $(V_i,\Gamma_i,\rho_i,\phi_i)$ is called {\bf linear}, if the $\Gamma_i$-action on ${\mathbb R}^n$ is linear.
\\
\\
Next we define 
\index{embedding of orbifold charts}
{\bf embeddings} \cite{Henriques:2003aaa}: Let us consider $U_i \subset U_j$, this defines an inclusion $\iota : U_i \rightarrow U_j$.
Let $(V_i, \Gamma_i, \rho_i, \phi_i)$ be a chart for $U_i$ and 
let $(V_j, \Gamma_j, \rho_j, \phi_j)$ be a chart of $U_j$.
An embedding is given by a pair $(\psi_{ji}, \rho_{ji})$, where
\begin{itemize}
\item $\rho_{ji} : \Gamma_i \rightarrow \Gamma_j$ is an injective group homomorphism,
which induces a group isomorphism $\rho_{ji} : \mathrm{Ker}\left(\Gamma_i\right) \rightarrow \mathrm{Ker}\left(\Gamma_j\right)$,
\item $\psi_{ji} : V_i \rightarrow V_j$ is a homeomorphism, called {\bf gluing map}, of $V_i$ onto an open subset of $V_j$,
\item the gluing map is compatible with the chart:
\bq
\begin{CD}
 V_i @>{\psi_{ji}}>> V_j \\
 @V{\phi_i}VV @VV{\phi_j}V \\
 U_i @>{\iota}>> U_j \\
\end{CD}
 & \hspace*{10mm} &
 \phi_j \circ \psi_{ji} \; = \; \iota \circ \phi_i.
\eq
\item the gluing map is equivariant:
\bq
\begin{CD}
 V_i @>{\psi_{ji}}>> V_j \\
 @V{\rho_i(\gamma_i)}VV @VV{\rho_j(\rho_{ji}(\gamma_i))}V \\
 V_i @>{\psi_{ji}}>> V_j \\
\end{CD}
 & \hspace{20mm} &
 \psi_{ji}\left( \rho_i\left(\gamma_i\right) \cdot x \right) 
 \; = \; 
 \rho_j\left(\rho_{ji}\left(\gamma_i\right)\right) \cdot \psi_{ji}\left(x\right)
\eq
In order to simplify the notation we drop the maps $\rho_i$ and $\rho_j$ (which are understood implicitly), thus
\bq
 \psi_{ji}\left( \gamma_i \cdot x \right) 
 & = &
 \rho_{ji}\left(\gamma_i\right) \cdot \psi_{ji}\left(x\right).
\eq
\item the gluing map $\psi_{ji}$ is unique up to a right action of $\Gamma_i$ and a left action of $\Gamma_j$.
\end{itemize}
An 
\index{orbifold atlas}
{\bf orbifold atlas} on $X$ is a family of orbifold charts, which cover $X$ and are compatible in the following sense:
Given a chart $(V_i, \Gamma_i, \rho_i, \phi_i)$ of $U_i$ and a chart 
$(V_j, \Gamma_j, \rho_j, \phi_j)$ of $U_j$, and given any $x \in U_i \cap U_j$, there exists an open neighbourhood 
$U_k \subset U_i \cap U_j$ and a chart $(V_k, \Gamma_k, \rho_k, \phi_k)$ of $U_k$, such there are embeddings
\bq
 \left(V_k, \Gamma_k, \rho_k, \phi_k\right) & \rightarrow & \left(V_i, \Gamma_i, \rho_i, \phi_i\right),
 \nonumber \\
 \left(V_k, \Gamma_k, \rho_k, \phi_k\right) & \rightarrow & \left(V_j, \Gamma_j, \rho_j, \phi_j\right).
\eq
Two such atlases are said to be equivalent, if they have a common refinement.
\\
\\
An 
\index{orbifold}
{\bf orbifold} $O$ is the space $X$ together with an equivalence class of orbifold atlases.
The space $X$ is called the {\bf underlying space}.
An orbifold contains more information than just its underlying space $X$.
\\
\\
Let $M$ be a manifold and $\Gamma$ a finite group acting properly on $M$.
This defines an orbifold, which we denote by $M \sslash \Gamma$ with underlying space $M / \Gamma$.
\\
\\
If $x \in X$ and $v \in \phi^{-1}(x)$ is a point in the inverse image of $x$ in some local chart,
then the stabiliser group at $v$ is independent of the chart.
We call this group the 
\index{local group}
{\bf local group} at $x$ and denote it by $\Gamma_x$.
The 
\index{singular points of an orbifold}
{\bf singular points} of an orbifold are the points $x \in X$ with a non-trivial local group $\Gamma_x$.
\\
\\
Example 1: The circle $S^1$, defined by
\bq
 x^2 + y^2 & = & 1,
\eq
together with the group $\Gamma : y \rightarrow -y$.
Then $S^1 \sslash \Gamma$ is an orbifold.
The singular points are $(-1,0)$ and $(1,0)$.
\\
\\
Example 2: The complex upper-half plane ${\mathbb H}$ with the group $\mathrm{PSL}\left(2,{\mathbb Z}\right)$.
Then
\bq
 {\mathbb H} \sslash \mathrm{PSL}\left(2,{\mathbb Z}\right)
\eq
is an orbifold.
The singular points are $\tau=i$ and $\tau=r_3$.
\\
\\
Example 3: The complex upper-half plane ${\mathbb H}$ with the group $\mathrm{SL}\left(2,{\mathbb Z}\right)$.
Then
\bq
 {\mathbb H} \sslash \mathrm{SL}\left(2,{\mathbb Z}\right)
\eq
is an orbifold.
All points of ${\mathbb H} / \mathrm{SL}\left(2,{\mathbb Z}\right)$ are singular points of the orbifold, since
\bq
 \left(\begin{array}{rr}
 -1 & 0 \\
 0 & -1 \\
 \end{array} \right)
\eq
acts trivially.
\\
\\
Let $O_1$ and $O_2$ be two orbifolds.
A 
\index{smooth map of orbifolds}
{\bf smooth map} between $O_1$ and $O_2$ is defined as follows:
There is a continuous map
\bq
 f & : & X_1 \; \rightarrow \; X_2
\eq
between the underlying spaces 
such that for any point $x_1 \in X_1$ there is a chart $(V_1,\Gamma_1,\rho_1,\phi_1)$ around $x_1$
and a chart $(V_2,\Gamma_2,\rho_2,\phi_2)$ around $f(x_1)$ together with a group homomorphism $\rho : \Gamma_1 \rightarrow \Gamma_2$
with the properties that
$f$ maps $\phi_1(V_1)$ into $\phi_2(V_2)$ and can be lifted to a smooth map
\bq
 \tilde{f} & : & V_1 \; \rightarrow \; V_2
\eq
such that
\bq
 \phi_2 \circ \tilde{f} & = & f \circ \phi_1,
 \nonumber \\
 \tilde{f}\left( \gamma_1 \cdot x \right) & = & \rho\left(\gamma_1\right) \cdot \tilde{f}\left(x\right).
\eq
In terms of commutative diagrams:
\bq
\begin{CD}
 V_1 @>{\tilde{f}}>> V_2 \\
 @V{\phi_1}VV @VV{\phi_2}V \\
 U_1 @>{f}>> U_2 \\
\end{CD}
 & \hspace*{20mm}&
\begin{CD}
 V_1 @>{\tilde{f}}>> V_2 \\
 @V{\gamma_1}VV @VV{\rho(\gamma_1)}V \\
 V_1 @>{\tilde{f}}>> V_2 \\
\end{CD}
\eq
Let us discuss an example:
We consider
$O_1 = \left({\mathbb C} \times {\mathbb C}\right) \sslash \left( C_2 \times C_2 \right)$
and
$O_2 = {\mathbb C} \sslash C_2$,
where $C_2 = \{1,-1\}$ denotes the multiplicative group with two elements,
together with the group actions
\bq
 O_1 & : &
 \left(\lambda_1,\lambda_2\right) \cdot \left(z_1,z_2\right) 
 \; = \;
 \left( \lambda_1 z_1, \lambda_2 z_2 \right),
 \nonumber \\
 O_2 & : &
 \lambda \cdot z \; = \; \lambda z.
\eq
An orbifold map $O_1 \rightarrow O_2$ is defined by
\bq
\label{appendix_moduli_space_def_orbifold_map_f}
 f & : &  \left({\mathbb C} \times {\mathbb C}\right) / \left( C_2 \times C_2 \right)
          \; \rightarrow \;
          {\mathbb C} \sslash C_2,
 \nonumber \\
 & & \left(z_1,z_2\right) \; \rightarrow \; z_1,
\eq
and
\bq
\label{appendix_moduli_space_def_orbifold_map_rho}
 \rho & : & C_2 \times C_2 \; \rightarrow \; C_2,
 \nonumber \\
 & & \left(\lambda_1,\lambda_2\right) \; \rightarrow \; \lambda_1.
\eq
This is a projection and the fibre above the point $[\pm z_1]$ is given by the second factor
$O_3 = {\mathbb C} \sslash C_2$.
\\
\\
Now let us discuss a slight modification of this example:
We consider the case that some groups act trivially:
\bq
 \tilde{O}_1 & : &
 \left(\lambda_1,\lambda_2\right) \cdot \left(z_1,z_2\right) 
 \; = \;
 \left( z_1, \lambda_2 z_2 \right),
 \nonumber \\
 \tilde{O}_2 & : &
 \lambda \cdot z \; = \; z.
\eq
Note that $\tilde{O}_1$ is not identical to $O_1$ and $\tilde{O}_2$ is not identical to $O_2$, as the group actions are different.
We consider an orbifold map $\tilde{O}_1 \rightarrow \tilde{O}_2$
with $f$ and $\rho$ as in eq.~(\ref{appendix_moduli_space_def_orbifold_map_f}) and eq.~(\ref{appendix_moduli_space_def_orbifold_map_rho}), respectively.
This is again a projection, 
where the fibre above the point $[z_1]$ is given by the second factor
$\tilde{O}_3 = {\mathbb C} \sslash C_2$.
Note that $\lambda_1$ acts trivially on $\tilde{O}_1$ and (through $\rho$) trivially on $\tilde{O}_2$, but does not act on $\tilde{O}_3$.

\subsection{The universal family of elliptic curves}

With the preparation on orbifolds we are now in a position to present 
the universal family of elliptic curves (without any framing).
We set
\bq
 {\mathcal C} & = & \left({\mathbb C} \times {\mathbb H}\right) \sslash \left(\mathrm{SL}\left(2,{\mathbb Z}\right) \ltimes {\mathbb Z}^2\right),
 \nonumber \\
 {\mathcal M}_{1,1}
 & = &
 {\mathbb H} \sslash \mathrm{SL}\left(2,{\mathbb Z}\right).
\eq
The projection ${\mathbb C} \times {\mathbb H} \rightarrow {\mathbb H}$,
given by $(z,\tau) \rightarrow \tau$, induces an orbifold morphism
\bq
 {\mathcal C} & \rightarrow & {\mathcal M}_{1,1}.
\eq
In more detail, this orbifold morphism is given by
\bq
 f & : & \left({\mathbb C} \times {\mathbb H}\right) / \left(\mathrm{SL}\left(2,{\mathbb Z}\right) \ltimes {\mathbb Z}^2\right)
 \; \rightarrow 
 {\mathbb H} / \mathrm{SL}\left(2,{\mathbb Z}\right),
 \nonumber \\
 & &\left(z,\tau\right) \; \rightarrow \; \tau,
\eq
and
\bq
 \rho & : & \mathrm{SL}\left(2,{\mathbb Z}\right) \ltimes {\mathbb Z}^2 
 \; \rightarrow \;
 \mathrm{SL}\left(2,{\mathbb Z}\right),
 \nonumber \\
 & &
 \left( \gamma, \vec{n} \right) \; \rightarrow \; \gamma.
\eq
The fibre above $\tau$ is ${\mathbb C} \sslash {\mathbb Z}^2$, i.e. an elliptic curve,
and ${\mathcal C} \rightarrow {\mathcal M}_{1,1}$ is a fine moduli space
of smooth genus one curves with one marked point.

\subsection{Compactification of ${\mathcal M}_{1,1}$}

The moduli space ${\mathcal M}_{1,1}$ parametrises equivalence classes of smooth genus one curves with one marked point,
i.e. smooth elliptic curves.
For the compactification $\overline{\mathcal M}_{1,1}$ we add one point, corresponding to a nodal genus one curve
with one marked point (the marked point does not coincide with the node).
Technically this is done as follows:
We denote by ${\mathbb D}$ the (open) unit disc
\bq
 {\mathbb D}
 & = &
 \left\{
   \; \bar{q} \in {\mathbb C} \; | \; \left|\bar{q}\right| < 1 \;
 \right\}
\eq
and by ${\mathbb D}^\ast$ the punctured unit disc
\bq
 {\mathbb D}^\ast
 & = &
 {\mathbb D} \backslash \left\{ 0 \right\}.
\eq
We construct $\overline{\mathcal M}_{1,1}$ with the help of two charts.
The first chart 
\bq
 {\mathcal M}_{1,1}
 & = &
 {\mathbb H} \sslash \mathrm{SL}\left(2,{\mathbb Z}\right)
\eq
covers ${\mathcal M}_{1,1}$ and has been discussed above.
The second chart is given by
\bq
 {\mathbb D} \sslash C_2,
\eq
where $C_2$ acts trivially on ${\mathbb D}$ and is the relict of the modular transformation
\bq
 \left(\begin{array}{rr}
  -1 & 0 \\
  0 & -1 \\
 \end{array} \right)
\eq
The mapping between the coordinate $\tau$ on ${\mathcal M}_{1,1}$ and $\bar{q}$ on ${\mathbb D} \sslash C_2$
is given by
\bq
 \bar{q} & = & \exp\left(2\pi i \tau\right).
\eq
The two charts overlap on ${\mathbb D}^\ast \sslash C_2$.
\\
\\
We call an action of a group $\Gamma$ on a space $X$ 
\index{virtually free group action}
{\bf virtually free},
if $\Gamma$ has a finite index subgroup $\Gamma'$, that acts freely.
\\
\\
Let $\Gamma$ be a discrete group which acts virtually free and properly discontinuous on $X$.
Let $\Gamma'$ be the finite index normal subgroup, which acts freely.
The 
\index{orbifold Euler characteristic}
{\bf orbifold Euler characteristic} is defined by
\bq
 \chi\left(X \sslash \Gamma \right)
 & = & 
 \frac{1}{\left[ \Gamma : \Gamma' \right]}
 \chi\left( X / \Gamma' \right).
\eq
For the two moduli spaces ${\mathcal M}_{1,1}$ and $\overline{\mathcal M}_{1,1}$ one finds 
the orbifold Euler characteristics 
\bq
 \chi\left({\mathcal M}_{1,1}\right) \; = \; - \frac{1}{12},
 & &
 \chi\left(\overline{\mathcal M}_{1,1}\right) \; = \; \frac{5}{12}.
\eq

%% file: algebraic_geometry.tex
\newpage
\chapter{Algebraic geometry}
\label{appendix_algebraic_geometry}

In the main part of this book we made no reference to sheaves or schemes,
although they are fundamental concepts in algebraic geometry.
As mathematicians always aim to state theorems as general as possible, 
this inevitably involves generalisations and abstraction, leading to sheaves and schemes.
It is therefore no surprise that one will encounter these terms 
in the mathematical literature quite frequently.
While physicists are certainly familiar with concrete examples of sheaves and schemes,
the abstract language makes it sometimes difficult to read mathematical literature.
In this appendix we give a short introduction to sheaves and schemes.
This appendix may serve as a survival kit for reading the 
mathematical literature.

References for sheaves and schemes are the books of 
Hartshorne~\cite{Hartshorne_book},
Eisenbud and Harris~\cite{Eisenbud_schemes:book} and
Holme~\cite{Holme:book}.

\section{Topology}
\label{appendix_algebraic_geometry:sect_topology}

We start with the definition of a topology:
\begin{tcolorbox}
Let $X$ be a set and ${\mathcal T}$ a collection of subsets of $X$.
${\mathcal T}$ is called a 
\index{topology}
{\bf topology} of $X$ if
\begin{enumerate}
\item $\emptyset \in {\mathcal T}$ and $X \in {\mathcal T}$,
\item $U_1, U_2 \in {\mathcal T} \Rightarrow U_1 \cap U_2 \in {\mathcal T}$,
\item $U_\alpha \in {\mathcal T}$ for all $\alpha \in I \Rightarrow \bigcup\limits_{\alpha \in I} U_\alpha \in {\mathcal T}$.
\end{enumerate}
\end{tcolorbox}
The pair $(X,{\mathcal T})$ is called a 
\index{topological space}
{\bf topological space}.
A subset $U \subseteq X$ is called 
\index{open set}
{\bf open} if $U \in {\mathcal T}$. 
A subset $A \subseteq X$ is called 
\index{closed set}
{\bf closed} if the complement $X\backslash A$ is open.
For closed sets we have
\begin{enumerate}
\item $\emptyset$ and $X$ are closed sets,
\item if $A$ and $B$ are closed, then $A \cup B$ is closed,
\item if $A_\alpha$ is closed, where $\alpha \in I$, then $\bigcap\limits_{\alpha \in I} A_\alpha$ is closed.
\end{enumerate}
The closure of an open set $U$ is denoted by $\overline{U}$.
We defined a topological space by its open sets. Alternatively, we may define a topological space by
its closed sets and the requirement that the closed sets satisfy items $1$-$3$ for closed sets.

A map between topological spaces is called 
\index{continuous map}
{\bf continuous} if the pre-image of any open set is again open.
A bijective map which is continuous in both directions is called a 
\index{homeomorphism}
{\bf homeomorphism}.

A topological space is called 
\index{Hausdorff space}
{\bf Hausdorff} if for any two distinct points $x_1, x_2 \in X$ there exist open sets
$U_1, U_2 \in {\mathcal T}$ with
\bq
 p_1 \in U_1, 
 \;\;\;
 p_2 \in U_2, 
 \;\;\;
 U_1 \cap U_2 = \emptyset.
\eq

For a subset $X' \subseteq X$ we define the 
\index{induced topology}
{\bf induced topology} by
\bq
 {\mathcal T}' & = & \left\{ U' | U' = U \cap X', U \in {\mathcal T} \right\}.
\eq
Note that $U'$ is open in $X$ only if $X'$ is open in $X$.

A topological space $X$ is called 
\index{irreducible topological space}
{\bf irreducible}, 
if it cannot be expressed as the union of two proper closed subsets,
otherwise the topological space is called reducible.

Let $Y \subseteq X$. $Y$ is called an 
\index{irreducible component}
{\bf irreducible component} of $X$ 
if $Y$ is irreducible (within the relative topology ${\cal T}'$)
and if $Y$ is maximal (i.e. $Y \subseteq Y'$ and $Y'$ irreducible implies $Y = Y'$).
Note that if $Y$ is an irreducible component of $X$, then $Y = \overline{Y}$.

Let $X$ be a topological space.
The 
\index{dimension of a topological space}
{\bf dimension} of $X$ is defined to be the supremum of all integers $n$ such that there
exists a chain
\bq
 Y_0 \subset Y_1 \subset ... \subset Y_n
\eq
of distinct irreducible subsets of $X$.
\begin{theorem}
A topological space $X$ is the union of irreducible components $Y_\alpha$:
\bq
X & = & \bigcup\limits_\alpha Y_\alpha
\eq
\end{theorem}
A topological space $X$ is called a 
\index{Noetherian space}
{\bf Noetherian space}, 
if each ascending chain of open sets $U_1 \subseteq U_2 \subseteq ...$ becomes stationary,
i.e. there exists a $k \in {\mathbb N}$ with
$U_j = U_k$ for all $j \ge k$.
This is equivalent to the requirement that each descending chain of closed sets $A_1 \supseteq A_2 \supseteq ...$ becomes stationary. 
A third equivalent
formulation is, that each non-empty set of open sets contains a maximal element.
\begin{theorem}
Let $X$ be a Noetherian topological space. 
Then $X$ has only a finite number of irreducible components $Y_1,Y_2,...,Y_r$:
\bq
 X & = & Y_1 \cup Y_2 \cup ... \cup Y_r.
\eq
If we require $Y_i \not\subseteq Y_j$,
this decomposition is unique up to ordering.
\end{theorem}

\section{Rings}

Rings play an essential part in the the theory of schemes.
In this section we recall the most important facts and definitions.
\begin{tcolorbox}[breakable]
A set $(R,+,\cdot)$ is called a 
\index{ring}
{\bf ring} if
\begin{description}
\item{(R1)\; :} \; $(R,+)$ is an {\bf Abelian group},
\item{(R2)\; :} \; the operation $\cdot$ is {\bf associative}: $a \cdot ( b \cdot c ) = ( a \cdot b ) \cdot c$,
\item{(R3)\; :} \; the operation $\cdot$ is {\bf distributive} with respect to the operation $+$:
\bq
 a \cdot ( b+ c ) &=  & (a \cdot b ) + ( a \cdot c ), 
 \nonumber \\
 (a + b ) \cdot c & = & (a \cdot c) + ( a \cdot b).
\eq
\end{description}
\end{tcolorbox}
The 
\index{trivial ring}
{\bf trivial ring} (or zero ring) is the ring consisting of one element $0$
with $0+0=0$ and $0 \cdot 0 = 0$.
The trivial ring is denoted by $\{0\}$.

If there is a neutral element $1$ for the operation $\cdot$, the ring is called a 
\index{ring with $1$}
\index{unital ring}
{\bf ring with $1$}.
Other names for a ring with $1$ are unital ring, ring with unity or ring with identity.

A ring $(R,+,\cdot)$ is called 
\index{commutative ring}
{\bf commutative} if the operation $\cdot$ is commutative :
\bq 
 a \cdot b & = & b \cdot a.
\eq
Let $R$ be a ring with $1$. An element $a \in R$ is called 
\index{invertible element in a ring}
{\bf invertible} or 
\index{unit in a ring}
{\bf unit} in $R$, if
$a$ has a left-inverse and a right-inverse with respect to multiplication.
We denote the set of invertible elements by $R^\ast$. $R^\ast$ is a group with respect to multiplication.

An element $a \in R$, $a \neq 0$ is called 
\index{zero-divisor}
{\bf zero-divisor}, if there is a $b \in R$,
$b \neq 0$ such that $a b = 0$ or $b a =0$.

A non-trivial commutative ring with $1$ and with no zero-divisors is called an 
\index{integral domain}
{\bf integral domain}.

Of particular importance are ideals of a ring:
\begin{tcolorbox}
\index{ideal}
A sub-group $I$ of $(R,+)$ is called an {\bf ideal} (or two-sided ideal), if
\begin{description}
\item{(I1)\; :} \; $r \cdot a \in I$ for all $r \in R$ and $a \in I$ (or short $R I \subseteq I$),
\item{(I2)\; :} \; $a \cdot r \in I$ for all $r \in R$ and $a \in I$ (or short $I R \subseteq I$).
\end{description}
If a sub-group $I$ of $(R,+)$ satisfies only $R I \subseteq I$, we call $I$ a left-ideal.
Similar, if a sub-group $I$ of $(R,+)$ satisfies only $I R \subseteq I$, we call $I$ a right-ideal.
\end{tcolorbox}
Every ring $R$ has the ideals $\{0\}$ and $R$. 
We call a non-trivial ring 
\index{simple ring}
{\bf simple}, if these are the only ideals.
An ideal $I$ is called a 
\index{proper ideal}
{\bf proper ideal}, if $I \neq R$.

An ideal $I$ is called a 
\index{principal ideal}
{\bf principal ideal}, if it is generated by one element:
\bq
 I & = & \left\lideal a \right\rideal.
\eq
$R$ is called a 
\index{principal ideal ring}
{\bf principal ideal ring}
if every ideal $I$ of $R$ is a principal ideal.
If $R$ is an integral domain and every ideal $I$ of $R$ is a principal ideal, then $R$ is called a 
\index{principal ideal domain}
{\bf principal ideal domain}.

Let $R$ be a commutative ring with $1$.
An ideal $P$ is called a 
\index{prime ideal}
{\bf prime ideal}, if $P$ is a proper ideal and
\bq
 a \cdot b \in P & \Rightarrow &
 a \in P \;\; \mbox{or} \;\; b \in P.
\eq
An ideal $P$ in a commutative ring $R$ with $1$ is prime if and only if $R / P$ is an integral domain.

An ideal $M$ is called a 
\index{maximal ideal}
{\bf maximal ideal}, if $M$ is a proper ideal and for any other ideal $I$ with
\bq
 M \; \subseteq \; I \; \subseteq \; R
\eq
it follows that $I = M$ or $I=R$.
Every ring with $1$ contains a maximal ideal.
An ideal $M$ in a commutative ring $R$ with $1$ is maximal if and only if $R / M$ is a field.
Every maximal ideal $M$ in a commutative ring $R$ with $1$ is a prime ideal.

The 
\index{height of a prime ideal}
{\bf height} of a prime ideal $P$ is the supremum of all integers $n$ such that there exists a chain
\bq
 P_0 \subset P_1 \subset ... \subset P_n = P
\eq
of distinct prime ideals.
The dimension or 
\index{Krull dimension}
{\bf Krull dimension} of the ring $R$ is defined to be the supremum of the heights of all prime ideals.

Let $R$ be a commutative ring with $1$.
$R$ is called a 
\index{local ring}
{\bf local ring} if $R$ contains exactly one maximal ideal.
In order to check if a ring is local, the following theorem is useful:
\begin{theorem}
\label{appendix_algebraic_geometry:theorem_local_ring}
Let $R$ be a commutative ring with $1$. $R$ is a local ring if and only if the set of
non-invertible elements $N=R\backslash R^\ast$ is an ideal in $R$.
If $N$ is an ideal in $R$ then $N$ is a maximal ideal and the only maximal ideal in $R$.
\end{theorem}
The concept of 
\index{localisation}
{\bf localisation} is of central importance for the theory of schemes:
Let $R$ be a commutative ring with $1$ and $S$ a subset of $R$ closed under multiplication 
together with the conditions $1 \in S$ and $0 \notin S$.
One defines the quotient ring
\bq
 R_S
 & = &
 \left\{ \; \left[\frac{r}{s}\right] \; | \; r \in R, s \in S \; : \; \frac{r_1}{s_1} \sim \frac{r_2}{s_2} \Leftrightarrow 
 \; \exists \; s \in S \; \mbox{such that} \; s \left(s_2r_1-s_1r_2\right) = 0 \; \right\}
\eq
The ring $R$ is a subring of $R_S$ via the identification 
\bq
 r & = & \left[ \frac{r}{1} \right].
\eq
If $P$ is a prime ideal of $R$, we may consider $S_P = R \backslash P$.
By definition of a prime ideal this set is closed under multiplication.
The quotient ring $R_{S_P}$ is a local ring. We denote the maximal ideal by $P_{S_P}$.
One often simplifies the notation and writes
\bq
 R_P \; = \; R_{S_P},
 & &
 P \; = \; P_P \; = \; P_{S_P}.
\eq
\bs
{\it \refstepcounter{exercise}
{\bf Exercise \theexercise}: 
Let $R$ be a commutative ring with $1$ and $P$ a prime ideal. Set $S_P = R \backslash P$.
Show that $S_P$ is closed under multiplication.
}
\es
\\
\\
\bs
{\it \refstepcounter{exercise}
\label{appendix_algebraic_geometry:exercise_Z_localised_at_5}
{\bf Exercise \theexercise}: 
Consider the commutative ring ${\mathbb Z}$ and the prime ideal $P=\lideal 5 \rideal$.
Define $S_P = {\mathbb Z}\backslash \lideal 5 \rideal$.
Describe the quotient ring $R_{S_P}$ and its maximal ideal $P_{S_P}$.
}
\es
\\
\\
Finally, let us introduce Noetherian rings and Artinian rings:
\begin{tcolorbox}
A ring $R$ is called a 
\index{Noetherian ring}
{\bf Noetherian ring} if one of the following conditions is satisfied:
\begin{enumerate}
\item Every ascending sequence $I_1 \subseteq I_2 \subseteq ...$ of ideals $I_j$ of $R$ becomes stationary, i.e. there exists a $k \in {\mathbb N}$ with
$I_j = I_k$ for all $j \ge k$.
\item Every non-empty set of ideals of $R$, partially ordered by inclusion, has a maximal element with respect to set inclusion.
\item Every ideal $I$ of $R$ is finitely generated. 
\end{enumerate}
\end{tcolorbox}
\begin{tcolorbox}
A ring $R$ is called a 
\index{Artinian ring}
{\bf Artinian ring} if
every descending sequence $I_1 \supseteq I_2 \supseteq ...$ of ideals $I_j$ of $R$ becomes stationary, i.e. there exists a $k \in {\mathbb N}$ with
$I_j = I_k$ for all $j \ge k$.
\end{tcolorbox}

\section{Algebraic varieties}

\subsubsection*{Affine algebraic varieties}

Let ${\mathbb A}$ be an algebraically closed field. The set of all $n$-tuples of elements of ${\mathbb A}$ defines the
affine $n$-space over ${\mathbb A}$, which we denote by ${\mathbb A}^n$.
An element $x \in {\mathbb A}^n$ will be called a point, and if $x=(x_1,...,x_n)$ with $x_i \in {\mathbb A}$, then
the $x_i$ will be called the coordinates of $x$.

Let $T = (t_1,...,t_n)$ be an $n$-tuple of independent variables, ${\mathbb A}[T]= {\mathbb A}[t_1,...,t_n]$ the ring of polynomials
over the field ${\mathbb A}$ and let $A \subset {\mathbb A}[T]$. 
An 
\index{affine algebraic set}
{\bf affine algebraic set} is given by
\bq
 V(A) & = & \{ \; x \in {\mathbb A}^n \; | \; f(x) = 0 \;\; \forall f \in A \; \}.
\eq
Note that $V(A) = V({\mathcal A})$, where ${\mathcal A} = \langle A \rangle$ is the ideal generated by $A$ in ${\mathbb A}[T]$.
Note further that ${\mathbb A}[T]$ is a Noetherian ring, therefore every ideal is generated by a finite set of polynomials.
If ${\mathcal A} = \langle f \rangle$ is a principal ideal, e.g. generated by one element $f$, then $V(f)$ is called a 
\index{hyperplane}
{\bf hyperplane} in ${\mathbb A}^n$.
$V$ is antiton, i.e.
\bq
 {\mathcal A} \subseteq {\mathcal B} 
 & \Rightarrow & 
 V({\mathcal B}) \subseteq V({\mathcal A}).
\eq
The empty set $\emptyset$ and the whole space ${\mathbb A}^n$ are algebraic sets:
\bq
 \emptyset \; = \; V\left(\left\langle 1 \right\rangle \right),
 & &
 {\mathbb A}^n \; = \; V\left( \left\{ 0 \right\} \right).
\eq
We further have
\bq
 & & 
 V({\mathcal A}) \cup V({\mathcal B}) 
 \; = \;
 V({\mathcal A B}) 
 \; = \; V({\mathcal A} \cap {\mathcal B}),
 \nonumber \\
 & & 
 \bigcap\limits_{i \in  I} V({\mathcal A}_i) 
 \; = \; 
 V( \sum\limits_{i \in I} {\mathcal A}_i).
\eq
Thus the algebraic sets can be viewed as closed sets. 
The open sets are then given as the complements of the closed sets.
This defines the 
\index{Zariski topology}
{\bf Zariski topology}.
\begin{tcolorbox}
An 
\index{affine algebraic variety}
{\bf affine algebraic variety} is an irreducible closed subset of ${\mathbb A}^n$ (within the induced topology).
An open subset of an affine algebraic variety is called a 
\index{quasi-affine algebraic variety}
{\bf quasi-affine algebraic variety}.
\end{tcolorbox}
Note that the terminology differs in the literature: Some authors use the term ``affine algebraic variety'' 
for an affine algebraic set and indicate explicitly if they refer to an irreducible set.

Let's look at an example: We take ${\mathbb A}={\mathbb C}$ and consider the affine line ${\mathbb A}^1 = {\mathbb C}^1$.
The proper irreducible closed subsets of ${\mathbb C}^1$ are sets $\{x\}$ consisting of a single point $x$.
The non-trivial open sets are the complements: ${\mathbb C}^1\backslash\{x\}$.
Please note that the affine line ${\mathbb C}^1$ together with the Zariski topology is not a Hausdorff space: It is impossible
to find open sets $U_1$ and $U_2$ with $x_1 \in U_1$, $x_2 \in U_2$ and $U_1 \cap U_2 = \emptyset$.

The {\bf annihilation ideal} of a set $Y \subseteq {\mathbb A}^n$ is defined by
\bq
 I(Y) 
 & = & 
 \{ \; f \in {\mathbb A}[T] \; | \; f(x) = 0 \;\; \forall x \in Y \; \}.
\eq
We have
\bq
 I( \emptyset ) \; = \; {\mathbb A}[T], 
 & &
 I( {\mathbb A}^n ) \; = \; \left\{ 0 \right\},
\eq
and
\bq
 Y_1 \subseteq Y_2 & \Rightarrow & I(Y_2) \subseteq I(Y_1).
\eq
For any two subsets $Y_1, Y_2 \subseteq {\mathbb A}^n$ we have
\bq
 I\left( Y_1 \cup Y_2 \right)
 & = &
 I\left(Y_1\right) \cap I\left(Y_2\right).
\eq
We have for any ideal ${\mathcal A}$
\bq
 I\left(V\left({\mathcal A}\right)\right)
 & = & 
 \mathrm{Rad} {\mathcal A},
\eq
where $\mathrm{Rad} {\mathcal A}$ denotes the 
\index{radical}
{\bf radical} of ${\mathcal A}$:
\bq
 \mathrm{Rad} {\mathcal A}
 & = &
 \left\{ \; f \in {\mathbb A}[T] \; | \; f^r \in {\mathcal A} \; \mbox{for some} \; r \in {\mathbb N} \; \right\}.
\eq
We further have for any set $Y \subseteq {\mathbb A}^n$
\bq
 V\left(I\left(Y\right)\right)
 & = &
 \overline{Y}.
\eq
There is a one-to-one inclusion-reversing correspondence between algebraic sets in ${\mathbb A}^n$
and radical ideals (i.e. ideals which are equal to their own radical) in ${\mathbb A}[T]$, given by
\bq
 Y \rightarrow I(Y), & &
 {\mathcal A} \rightarrow V\left({\mathcal A}\right).
\eq
Furthermore, an algebraic set is irreducible if and only if its ideal is a prime ideal.

Let $Y$ be an affine algebraic set.
The 
\index{affine coordinate ring}
{\bf affine coordinate ring} of $Y$ is defined by
\bq
 {\mathbb A}[T] / I\left(Y\right).
\eq
If $Y$ is an affine algebraic set, then the dimension of $Y$ is equal to the dimension of its affine coordinate ring.

\subsubsection*{Projective algebraic varieties}

Let's now consider the projective case.
We denote by $P^n({\mathbb A})$ (or ${\mathbb P}^n$ for short)
the $n$-dimensional projective space over the field ${\mathbb A}$.
Points $x \in {\mathbb P}^n$ will be denoted by $x=[x_0:x_1:...:x_n]$.
We now consider {\bf homogeneous} polynomials $f \in {\mathbb A}[t_0,...,t_n]$.
We denote by ${\mathbb A}^h[T]$ the set of homogeneous polynomials in ${\mathbb A}[t_0,...,t_n]$.
Let $A \subseteq {\mathbb A}^h[T]$.
A 
\index{projective algebraic set}
{\bf projective algebraic set} is given by
\bq
 V(A) & = & \{ \; x \in {\mathbb P}^n \; | \; f(x) = 0 \;\; \forall f \in A \; \}.
\eq
\begin{tcolorbox}
A {\bf projective algebraic variety} is an irreducible projective algebraic set.
An open subset of a projective algebraic variety is a {\bf quasi-projective variety}.
\end{tcolorbox}
As in the affine case the algebraic sets are the closed sets within the Zariski topology.
The 
\index{annihilation ideal}
{\bf annihilation ideal} of a set $Y \subseteq {\mathbb P}^n$ is defined by
\bq
 I(Y) 
 & = & 
 \{ \; f \in {\mathbb A}^h[T] \; | \; f(x) = 0 \; \forall x \in Y \; \}.
\eq
Note that all $f$'s are homogeneous polynomials.

Let $Y$ be a projective algebraic set.
The 
\index{homogeneous coordinate ring}
{\bf homogeneous coordinate ring} of $Y$ is defined by
\bq
 {\mathbb A}[T] / I\left(Y\right).
\eq
If $Y$ is a projective algebraic set, then the dimension of $Y$ is equal to the dimension of its homogeneous coordinate ring minus one.

Every affine variety is also a quasi-projective variety.
Furthermore the complement of an algebraic set in an affine variety is a quasi-projective variety.
In the following we will take 
\index{variety}
``{\bf variety}'' to mean either an affine, a quasi-affine, a projective or a quasi-projective variety.

\subsubsection*{Regular functions}

Let $X$ be a {\bf quasi-affine} algebraic variety in ${\mathbb A}^n$.
We consider functions
\bq
 f & : & X \rightarrow {\mathbb A}.
\eq
A function $f : X \rightarrow {\mathbb A}$ is 
\index{regular function}
{\bf regular} at a point $x$ if there is an open neighbourhood $U$ with $x \in U \subseteq X$,
and polynomials $p,q \in {\mathbb A}[T]$, such that $q$ is nowhere zero on $U$, and
\bq
 f & = & \frac{p}{q}
\eq
on $U$.
The function $f$ is regular on $X$ if it is regular at every point of $X$.

Let $X$ be now a {\bf quasi-projective} algebraic variety in ${\mathbb P}^n$.
A function $f : X \rightarrow {\mathbb A}$ is {\bf regular} at a point $x$ if there is an open neighbourhood $U$ with $x \in U \subseteq X$,
and homogeneous polynomials $p,q \in {\mathbb A}[t_0,t_1,...,t_n]$ of the same degree, 
such that $q$ is nowhere zero on $U$, and
\bq
 f & = & \frac{p}{q}
\eq
on $U$.
The function $f$ is regular on $X$ if it is regular at every point of $X$.

Let $X$ be a variety. We denote by ${\mathcal O}(X)$ the {\bf ring of all regular functions} on $X$.

Let $x$ be a point of $X$.
We define the {\bf local ring} ${\mathcal O}_x$ of $x$ on $X$ to be the ring of germs of regular functions
on $X$ near $x$.
In other words, an element of ${\mathcal O}_x$ is a pair $(U,f)$, where $U$ is an open subset of $X$ containing $x$, and
$f$ is a regular function on $U$, and we identify two such pairs $(U_1,f_1)$ and $(U_2,f_2)$ if $f_1=f_2$ on $U_1 \cap U_2$.
${\mathcal O}_x$ is a local ring: its maximal ideal $M$ is the set of germs of regular functions which vanish at $x$.
The residue field ${\mathcal O}_x/M$ is isomorphic to ${\mathbb A}$.

The {\bf function field} ${\mathbb A}(X)$ of $X$ is defined as follows:
an element of ${\mathbb A}(X)$ is an equivalence class of pairs $(U,f)$, where $U$ is a non-empty open subset of $X$, 
$f$ is a regular function on $U$, and where we identify two pairs $(U_1,f_1)$ and $(U_2,f_2)$ if $f_1=f_2$ on
$U_1 \cap U_2$.
The elements of ${\mathbb A}(X)$ are called {\bf rational functions} on $X$.

\section{Sheaves}

The law of physics are often given locally, i.e. as differential equations.
Prominent examples are the equations of motion for a particle or -- in the context of this book --
the differential equations for a family of Feynman integrals.
We almost never state it explicitly, but the first step is usually to study these systems
in an open neighbourhood of a point of interest.
Thus we have open sets $U$ 
(with a time coordinate $t$ in the case of equation of motions 
and with kinematic coordinates $x$ in the case of Feynman integrals)
and we are interested in local sections with values in some space
(position space in the case of a particle or a vector space with dimension equal to the number of master integrals
in the case of Feynman integrals), 
giving us the trajectory of a particle as a function of $t$ 
or the values of the Feynman integrals as a function of $x$.
Of course, the result should not change if we restrict to a slightly smaller open set $U' \subset U$.
Furthermore, if we have two overlapping open sets $U_1$ and $U_1$ the results on the intersection
$U_1 \cap U_2$ should be compatible.

Sheaves formalise this concept. We may think of a sheaf as data attached to open sets of a topological space,
compatible with restriction to smaller open set and compatible with intersections of open sets.

Let's now consider the definition:
We start with the definition of a 
\index{presheaf}
{\bf presheaf}.
Let $X$ be a topological space, and let ${\mathcal C}$ be a category.
Usually ${\mathcal C}$ is taken to be the category of sets, the category of groups, the category of Abelian groups or
the category of commutative rings.
A presheaf ${\mathcal F}$ on $X$ assigns to each open set $U \subset X$ an object ${\mathcal F}(U) \in {\mathcal C}$,
called the 
\index{section, sheaf}
{\bf sections} of ${\mathcal F}$ over $U$, 
and to each inclusion of open sets $U \subset V$ 
a morphism $r_{V,U} : {\mathcal F}(V) \rightarrow {\mathcal F}(U)$, 
called the 
\index{restriction map}
{\bf restriction map}, 
satisfying:
\begin{itemize}

\item For every open set $U \in X$, the restriction morphism $r_{U,U} : {\mathcal F}(U) \rightarrow {\mathcal F}(U)$
is the identity.

\item For any triple $U \subset V \subset W$ of open sets,
\bq
r_{W,U} & = & r_{V,U} \circ r_{W,V}.
\eq
By virtue of this relation, we may write $\left. \sigma \right|_U$ for $r_{V,U}(\sigma)$ without loss of information.

\end{itemize}
A {\bf sheaf} is a presheaf, which in addition satisfies the following two conditions:
\begin{itemize}

\item {\bf Locality}: 
Let $\left( U_i \right)$ be an open covering of an open set $U$, 
and $\sigma, \tau \in {\mathcal F}(U)$, such that
$\left. \sigma \right|_{U_i} = \left. \tau \right|_{U_i}$ for each subset $U_i$ of the covering, then
\bq
 \sigma & = & \tau.
\eq
This means that a section over $U$ is determined by all its restrictions to subsets $U_i$ of
$U$.

\item {\bf Gluing}: 
Let $\left( U_i \right)$ be an open covering of an open set $U$.
If for each $i$ a section $\sigma_i \in {\mathcal F}(U_i)$ is given with the property
that for any pair $i,j$ we have
\bq
 \left. \sigma_i \right|_{U_i \cap U_j}
 & = &
 \left. \sigma_j \right|_{U_i \cap U_j},
\eq
then there exists a section $\sigma \in {\mathcal F}(U)$ such that
\bq
 \left. \sigma \right|_{U_i}
 & = &
 \sigma_i
\eq
for all $i$.
This allows the passage from local data to global data: A section $\rho$
on $U$ may be assembled from the local data on the subsets $U_i$ of $U$.

\end{itemize}
Let ${\mathcal F}$ be a sheaf on $X$ and $x \in X$ a point.
We define the
\index{stalk}
{\bf stalk} ${\mathcal F}_x$ at $x$
to be the direct limit of ${\mathcal F}(U)$ for all open sets $U$ containing $x$
\bq
{\mathcal F}_x & = & \varinjlim_{U \ni x} {\mathcal F}(U).
\eq
The direct limit is taken over all open subsets of $X$ containing the point $x$ with the restriction map.
An element $\sigma \in {\mathcal F}_x$ is called a 
\index{germ}
{\bf germ}.

Let's look at a few examples of sheaves which occur frequently.
Example of sheaves, which take values in the category of Abelian groups, are:
\begin{center}
\begin{tabular}{ll}
 ${\mathcal O}(U)$ & holomorphic functions on $U$ with addition, \\
 ${\mathcal O}^\ast(U)$ & holomorphic functions on $U$ which are nowhere zero with multiplication, \\
 ${\mathcal M}(U)$ & meromorphic functions on $U$ with addition, \\
 ${\mathcal M}^\ast(U)$ & meromorphic functions on $U$ without the zero function with multiplication, \\
 $\Lambda^p(U)$ & $p$-forms on $U$ with addition, \\
 $\Omega^p(U)$ & holomorphic $p$-forms on $U$ with addition, \\
 ${\mathbb Z}(U)$ & locally constants ${\mathbb Z}$-valued functions on $U$ with addition. \\
\end{tabular}
\end{center}
Example of sheaves, which take values in the category of commutative rings with $1$, are:
\begin{center}
\begin{tabular}{ll}
 ${\mathcal O}(U)$ & holomorphic functions on $U$ with addition and multiplication, \\
 ${\mathbb Z}(U)$ & locally constants ${\mathbb Z}$-valued functions on $U$ with addition and multiplication. \\
\end{tabular}
\end{center}
Example of sheaves, which take values in the category of rings with $1$, are:
\begin{center}
\begin{tabular}{ll}
 $\Lambda^\bullet(U)$ & differential forms on $U$ with addition and the wedge product, \\
 $\Omega^\bullet(U)$ & holomorphic differential forms on $U$ with addition and the wedge product. \\
\end{tabular}
\end{center}
The notation ${\mathcal F}_U$ instead of ${\mathcal F}(U)$ is also used frequently.

\section{The spectrum of a ring}

Let $R$ be a commutative ring with $1$. 
The spectrum $\mathrm{Spec}(R)$ of $R$ is a pair $(S,{\mathcal O})$, where $S$ is a set with a topology
defined on it (hence a topological space) and ${\mathcal O}$ a sheaf on $S$.
The sheaf ${\mathcal O}$ is called the 
\index{structure sheaf}
{\bf structure sheaf}.
By abuse of notation, the set $S$ is also often denoted as $\mathrm{Spec}(R)$.
It should be clear from the context if the set $S$ or the pair $(S,{\mathcal O})$ is meant.

The set $S$ is given as the set of all prime ideals of $R$:
\bq
 S \; = \; \mathrm{Spec}\left(R\right)
 & = &
 \left\{ \; P \; | \; P \; \mbox{prime ideal of} \; R \; \right\}.
\eq
Example:
\bq
 \mathrm{Spec}\left({\mathbb Z}\right)
 & = &
 \left\{ \left(0\right), \left(2\right), \left(3\right), \left(5\right), \left(7\right), \left(11\right), ... \right\}.
\eq
Let $I$ be an ideal of $R$.
We set
\bq
 V\left(I\right)
 & = &
 \left\{ \, P \, \in \, \mathrm{Spec}\left(R\right) \; | \; I \, \subseteq \, P \, \right\}.
\eq
We have $V(R)=\emptyset$, $V(\{0\})=\mathrm{Spec}\left(R\right)$ and
\bq
 V\left(I_1 \cdot I_2\right) & = & V\left(I_1\right) \cup V\left(I_2\right),
 \nonumber \\
 V\left(\sum\limits_\alpha I_\alpha\right) & = & \bigcap\limits_\alpha V\left(I_\alpha\right).
\eq
This defines a topology on $S$, where the $V(I)$ are the closed sets.

Now let $P$ be a prime ideal of $R$ and denote by $R_P$ the localisation of $R$ at $P$.
We now define the structure sheaf ${\mathcal O}$.
Let $U \subseteq \mathrm{Spec}(R)$ be an open set. The sheaf ${\mathcal O}(U)$ consists of functions
\bq
 \sigma & : & U \rightarrow \bigsqcup\limits_{P \in U} R_P,
\eq
where $\bigsqcup$ denotes the disjoint union, such that
\begin{enumerate}
\item
\bq
 \sigma\left(P\right) \in R_P
\eq
\item $\sigma$ is locally a quotient of elements from $R$: 
This means that for all $P \in U$ there exists a $V$ with $P \in V$ and $V \subseteq U$ as well as $a,b \in R$
such that for all $Q \in V$ we have $b \notin Q$ and
\bq
 \sigma\left(Q\right)
 \; = \; \frac{a}{b} 
\eq 
in $R_Q$.
\end{enumerate}
Let's look at an example: The structure sheaf of $\mathrm{Spec}({\mathbb Z})$ has ${\mathbb Q}$ as stalk
in the point $(0)$. For a prime number $p$ the stalk at $(p)$ is ${\mathbb Z}$ localised at $(p)$
(see exercise~\ref{appendix_algebraic_geometry:exercise_Z_localised_at_5}).

\section{Schemes}

A pair $(X, {\mathcal O}_X)$ of a topological space $X$ and a sheaf of commutative rings with $1$ on $X$ is called a 
\index{ringed space}
{\bf ringed space}.
The sheaf ${\mathcal O}_X$ is called the 
\index{structure sheaf}
{\bf structure sheaf} of the space.
If all the stalks of the structure sheaf are local rings, the pair $(X, {\mathcal O}_X)$
is called a 
\index{locally ringed space}
{\bf locally ringed space}.

Examples:
\begin{enumerate}
\item An arbitrary topological space $X$ can be considered a locally ringed space 
by taking ${\mathcal O}_X$ to be the sheaf of real-valued (or complex-valued) continuous functions 
on open subsets of $X$.
The stalk at a point $x$ can be thought of as the set of all germs of continuous functions at $x$; 
this is a local ring with maximal ideal consisting of those germs whose value at $x$ is $0$.

Remark: There may exist continuous functions over open subsets of $X$ that are not the restriction of any continuous function over $X$.

\item If $X$ is a differentiable manifold, we may take the sheaf of differentiable functions.
If $X$ is a complex manifold, we may take the sheaf of holomorphic functions.
Both of these give rise to locally ringed spaces.

\item If $X$ is an algebraic variety with the Zariski topology, 
we can define a locally ringed space by taking ${\mathcal O}_X(U)$ 
to be the ring of rational mappings defined on the open set $U$ that do not become infinite on $U$.

\end{enumerate}
We now have all ingredients to define a scheme. We do this in two steps: We first define an affine scheme and then a (general) scheme.
You may think about the relation between an affine scheme and a scheme as being similar to the relation between a coordinate patch and a manifold,
the latter being described by a collection of coordinate patches.
\begin{tcolorbox}
An 
\index{affine scheme}
{\bf affine scheme} is a locally ringed space $(X, {\mathcal O}_X)$ which is isomorphic (as a locally ringed space) to the spectrum of some ring $R$.
\end{tcolorbox}
\begin{tcolorbox}
A 
\index{scheme}
{\bf scheme} is a locally ringed space $(X, {\mathcal O}_X)$ such that every point $x \in X$ has an open neighbourhood $U$ such that
$(U, {\mathcal O}_X|_U)$ is an affine scheme.
\end{tcolorbox}

%% file: algorithms.tex
\newpage
\chapter{Algorithms for polynomial rings}
\label{appendix_algorithms}

In this appendix we review a few basic algorithms related to polynomial rings.
We discuss algorithms for computing a Gr\"obner basis, a Nullstellensatz certificate, an annihilator and the syzygy module.

\section{Computing a Gr\"obner basis}
\label{appendix_algorithms:Groebner_basis}

One of the essential tools is the computation of a Gr\"obner basis.
Most computer algebra systems offer implementations to do this.
Here we review the basics. A standard reference is the book by
Adams and Loustaunau \cite{Adams_Loustaunau}.

Let ${\mathbb F}$ be a field 
and ${\mathbb F}[x_1,\dots,x_n]$ the ring of polynomials in $n$ variables
$x_1, \dots, x_n$ with coefficients from the field ${\mathbb F}$.
We fix a term order, which we denote by $<$.

Consider now $f, q_1, \dots, q_r \in {\mathbb F}[x_1,\dots,x_n]$.
Using long division (almost as in primary school) we may write
\bq
\label{appendix_algorithms:polynomial_division}
 f
 & = &
 \sum\limits_{j=1}^r h_j q_j + r,
\eq
with $h_j q_j \le f$, $r \le f$ and no term in $r$ is divisible by any leading term $\mathrm{lt}(q_j)$.
$r$ is called a remainder for $f$ with respect to $\{q_1,\dots,q_r\}$.
The division algorithm proceeds as follows:
We start with one polynomial from the set $q_1, \dots, q_r$, say $q_1$ and check if 
$\mathrm{lt}(q_1)$ divides $\mathrm{lt}(f)$. 
If this is the case, we reduce $f$, if this is not the case, we try the
next polynomial $q_2$. 
If none of the leading terms $\mathrm{lt}(q_j)$ divides $\mathrm{lt}(f)$, we move 
$\mathrm{lt}(f)$ from $f$ to the remainder and continue with the next term of $f$.
Note that the result of the division algorithm is not necessarily unique.
The result may depend on the order in which we try the polynomials $q_1, \dots, q_r$.

For two polynomials $f_i, f_j \in {\mathbb F}[x_1,\dots,x_n]$, the $S$-polynomial is defined by
\bq
 S\left(f_i,f_j\right)
 & = &
 \frac{l_{i j}}{\mathrm{lt}\left(f_i\right)} f_i - \frac{l_{i j}}{\mathrm{lt}\left(f_j\right)} f_j,
 \;\;\;\;\;\;
 l_{i j} \; = \; \mathrm{lcm}\left(\mathrm{lt}\left(f_i\right),\mathrm{lt}\left(f_j\right)\right).
\eq
Consider now the ideal $I$ generated by $q_1, \dots, q_r$:
\bq
 I & = & \left\lideal q_1, \dots, q_r \right\rideal.
\eq
A basic algorithm for the computation of a Gr\"obner basis $G$ of the ideal $I$ is as follows: 
One starts from $G=\{q_1,\dots,q_r\}$
and one computes for a pair $f_i,f_j \in G$ the $S$-polynomial $S(f_i,f_j)$ and reduces 
$S(f_i,f_j)$ with the multivariate division algorithm relative to $G$.
If the remainder is non-zero, it is added to $G$.
This process is iterated until for any pair $f_i,f_j \in G$ the 
$S$-polynomial $S(f_i,f_j)$ reduces to zero relative to $G$.

A Gr\"obner basis $G$ for $I$ is not necessarily unique.
There are two places were choices are made: 
The order of the polynomials in the multivariate division algorithm already discussed above and 
the order in which pairs $f_i, f_j \in G$ are selected.

A Gr\"obner basis $G$ is called a 
\index{reduced Gr\"obner basis}
{\bf reduced Gr\"obner basis}, 
if for any $f_i \in G$ the coefficient of $\mathrm{lt}(f_i)$ is one and no term of $f_i$
is divisible by $\mathrm{lt}(f_j)$ for any $j \neq i$.
For a given term order, the reduced Gr\"obner basis is unique.

\section{Computing a Nullstellensatz certificate}
\label{appendix_algorithms:Nullstellensatz_certificate}

Let ${\mathbb F}$ be a field 
and ${\mathbb F}[x_1,\dots,x_n]$ the ring of polynomials in $n$ variables
$x_1, \dots, x_n$ with coefficients from the field ${\mathbb F}$.
If a set of $r$ polynomials $q_1, \dots, q_r \in {\mathbb F}[x_1,\dots,x_n]$
have a common zero, Hilbert's Nullstellensatz guarantees that there exist
$r$ polynomials $h_1, \dots, h_r \in {\mathbb F}[x_1,\dots,x_n]$ such that
\bq
\label{appendix_algorithms:Hilbert_Nullstellensatz}
 \sum\limits_{j=1}^r h_j q_j & = & 1.
\eq
We may compute the $h_j$'s as follows:
Consider the ideal
\bq
 I
 & = &
 \left\lideal q_1, \dots, q_r \right\rideal.
\eq
Eq.~(\ref{appendix_algorithms:Hilbert_Nullstellensatz}) states that $1 \in I$.
Hence, a Gr\"obner basis $G$ for the ideal $I$ is given by
\bq
 G & = &
 \left\{ 1 \right\}.
\eq
Thus, we may just compute a Gr\"obner basis for the ideal $I$ together
with the transformation matrix, which expresses any element of the Gr\"obner
basis as a linear combination of the input polynomials $q_1,\dots,q_r$.
We already know that the Gr\"obner basis will be $\{1\}$, and we are only interested
in the transformation matrix, which expresses the generator $1$
as a linear combination of the polynomials $q_1,\dots,q_r$.
The coefficients are the sought-after polynomials $h_1,\dots,h_r$.

\section{Computing an annihilator}
\label{appendix_algorithms:annihilator}

Let ${\mathbb F}$ be a field 
and ${\mathbb F}[x_1,\dots,x_n]$ the ring of polynomials in $n$ variables
$x_1, \dots, x_n$ with coefficients from the field ${\mathbb F}$.
A set of $r$ polynomials $q_1, \dots, q_r \in {\mathbb F}[x_1,\dots,x_n]$
is said to be algebraically dependent, if there is a 
non-zero polynomial $a(y_1,\dots,y_r) \in {\mathbb F}[y_1,\dots,y_r]$ in $r$ variables $y_1,\dots,y_r$
with coefficients in ${\mathbb F}$ such that
\bq
\label{appendix_algorithms:annihilating_polynomial}
 a\left( q_1, \dots, q_r \right) & = & 0.
\eq
The polynomial $a(y_1,\dots,y_r)$ is called an 
\index{annihilator of a set of polynomials}
{\bf annihilating polynomial} of $q_1, \dots, q_r$.
In order to compute an annihilating polynomial we start from the ideal
\bq
 I 
 & = &
 \left\lideal y_1-q_1, \dots, y_r-q_r \right\rideal
 \; \in \; 
 {\mathbb F}[x_1,\dots,x_n,y_1,\dots,y_r].
\eq
Let $G$ be a Gr\"obner basis of $I$ with respect to the lexicographic order
\bq
 x_1 
 \; > \; 
 \dots 
 \; > \; 
 x_n
 \; > \; 
 y_1
 \; > \; 
 \dots
 \; > \; 
 y_r.
\eq
Set
\bq
 G_Y & = & G \cap {\mathbb F}[y_1,\dots,y_r].
\eq
$G_Y$ is a Gr\"obner basis for the ideal $I \cap {\mathbb F}[y_1,\dots,y_r]$
and any $a \in G_Y$ is an annihilating polynomial.

\section{Computing the syzygy module}
\label{appendix_algorithms:syzygy}

Let ${\mathbb F}$ be a field 
and ${\mathbb F}[x_1,\dots,x_n]$ the ring of polynomials in $n$ variables
$x_1, \dots, x_n$ with coefficients from the field ${\mathbb F}$.
Consider a set of $r$ polynomials $q_1, \dots, q_r \in {\mathbb F}[x_1,\dots,x_n]$
and the ideal
\bq
 I
 & = &
 \left\lideal q_1, \dots, q_r \right\rideal.
\eq
A syzygy is a relation
\bq
\label{appendix_algorithms:def_syzygy}
 \sum\limits_{j=1}^r h_j q_j & = & 0,
\eq
with $h_1, \dots, h_r \in {\mathbb F}[x_1,\dots,x_n]$.
In comparison with eq.~(\ref{appendix_algorithms:Hilbert_Nullstellensatz})
note that eq.~(\ref{appendix_algorithms:Hilbert_Nullstellensatz}) has a one on the right-hand side,
while eq.~(\ref{appendix_algorithms:def_syzygy}) has a zero on the right-hand side.

The difference with an annihilating polynomial as in eq.~(\ref{appendix_algorithms:annihilating_polynomial})
is as follows: A syzygy is linear in the $q_j$'s, while an annihilating polynomial is allowed to be
polynomial in the $q_j$'s. On the other hand, the coefficients $h_j$ of a syzygy are allowed to be in ${\mathbb F}[x_1,\dots,x_n]$,
while the coefficients of an annihilating polynomial are in ${\mathbb F}$.

The set of syzygies form a module over ${\mathbb F}[x_1,\dots,x_n]$. We may compute a basis of this module as follows:
We first compute a Gr\"oebner basis for the ideal $I$ (as outlined in section~\ref{appendix_algorithms:Groebner_basis}).
Let therefore
\bq
 G & = & \left\{ f_1, f_2, \dots, f_s \right\}
\eq
be a Gr\"obner basis for $I$.
For each pair $(f_i,f_j)$ with $1 \le i,j \le s$ we consider the $S$-polynomial $S(f_i,f_j)$.
As $G$ is a Gr\"obner basis, multivariate division with remainder reduces $S(f_i,f_j)$ to zero relative to $G$.
That is to say, we find by multivariate division polynomials $g_{ijk} \in {\mathbb F}[x_1,\dots,x_n]$
such that
\bq
 S\left(f_i,f_j\right)
 & = &
 \sum\limits_{k=1}^s
 g_{ijk} f_k.
\eq
We set
\bq
 r_{ij}
 & = &
 \frac{l_{i j}}{\mathrm{lt}\left(f_i\right)} y_i - \frac{l_{i j}}{\mathrm{lt}\left(f_j\right)} y_j
 -
 \sum\limits_{k=1}^s
 g_{ijk} y_k,
 \;\;\;\;\;\;
 l_{i j} \; = \; \mathrm{lcm}\left(\mathrm{lt}\left(f_i\right),\mathrm{lt}\left(f_j\right)\right).
\eq
The syzygy module is generated by all relations $r_{ij}$ with $1 \le i < j \le s$.
Substituting $f_k$ for $y_k$ 
gives a syzygy relation of the form as in eq.~(\ref{appendix_algorithms:def_syzygy}).
More refined algorithms to compute the syzygy module are given in \cite{Erocal:2016}.

Syzygies are of interest in extending the unitarity-based methods 
discussed in section~\ref{chapter_one_loop:sect_amplitude_methods} 
from one-loop to higher loops \cite{Gluza:2010ws,Schabinger:2011dz,Ita:2015tya,Larsen:2015ped,Zhang:2016kfo}.

%% file: finite_fields.tex
\newpage
\chapter{Finite fields methods}
\label{appendix_finite_fields}

Integration-by-parts identities and the reduction to master integrals 
are at the core of many Feynman integral computations.
On the positive side we note that this only involves linear algebra and rational functions in the kinematic
variables $x$ and the dimension of space-time $D$.
However, the simplification of the rational functions (i.e. cancelling common factors in the numerator
and in the denominator) is actually the bottle-neck.
Finite-field methods can be used to improve the performance.
In this appendix we first discuss Euclid's algorithm for the greatest common divisor and gradually
turn to finite field methods.

\section{The greatest common divisor and the Euclidean algorithm}

It is often required to simplify rational functions
by cancelling common factors in the numerator and denominator.
As an example let us consider
\bq
 \frac{(x+y)^2 (x-y)^3}{(x+y)(x^2-y^2)} 
 & = &
 (x-y)^2.
\eq
One factor of $(x+y)$ is trivially removed,
the remaining factors are cancelled once we noticed that $(x^2-y^2) = (x+y)(x-y)$.
For the implementation in a computer algebra system this is however not the way to proceed.
The factorization of the numerator and the denominator into irreducible polynomials
is a very expensive calculation and actually not required.
To cancel the common factors in the numerator and in the denominator it is sufficient to 
calculate the 
\index{greatest common divisor}
{\bf greatest common divisor (gcd)} of the two expressions.
The efficient implementation of an algorithm for the calculation of the greatest common divisor is essential
for many other algorithms.
Like in the example above, most gcd calculations are done in polynomial rings.
It is therefore useful to recall first some basic definitions from ring theory:

A
\index{commutative ring} 
{\bf commutative ring} $(R,+,\cdot)$ is a set $R$ with two operations $+$ and $\cdot$, such 
that $(R,+)$ is an Abelian group and $\cdot$ is associative, distributive and commutative.
In addition we always assume that there is a unit element for the multiplication.
An example for a commutative ring would be ${\mathbb Z}_8$,
i.e. the set of integers modulo 8.
In this ring one has for example $3+7 = 2$ and $2 \cdot 4 = 0$.
From the last equation one sees that it is possible to obtain zero
by multiplying two non-zero elements.

An 
\index{integral domain}
{\bf integral domain} $D$ is a commutative ring with the additional requirement
\bq
 a \cdot b = 0 & \Rightarrow & a=0 \;\; \mbox{or} \;\; b=0 \;\;\; \mbox{(no zero divisors)}.
\eq
Sometimes an integral domain $D$ is defined by requiring
\bq
a \cdot b = a \cdot c \;\;\mbox{and}\;\; a \neq 0 
 & \Rightarrow & b=c \;\;\; \mbox{(cancellation law)}.
\eq
It can be shown that these two requirements are equivalent.
An example for an integral domain would be the subset of the complex numbers
defined by
\bq
\label{appendix_finite_fields:exampleintdomain}
S & = & \left\{ \left. a + b i \sqrt{5} \; \right| a,b \in {\mathbb Z} \right\}
\eq
An element $u \in D$ is called 
\index{unit element in an integral domain}
{\bf unit} or {\bf invertible} if $u$ has a
multiplicative inverse in $D$.
The only units in the example eq.~(\ref{appendix_finite_fields:exampleintdomain}) are $1$ and $(-1)$.
We further say that $a$ divides $b$ if there is an element $x \in D$ such that
$b = a x$. In that case one writes $a | b$.
Two elements $a,b \in D$ are called 
\index{associates in an integral domain}
{\bf associates} if $a$ divides $b$ and $b$ divides $a$.
In the integral domain $S$ defined in eq.~(\ref{appendix_finite_fields:exampleintdomain})
the elements $1$ and $(-1)$ are associates.

We can now define the 
\index{greatest common divisor}
{\bf greatest common divisor}: An element $c \in D$ is called
the greatest common divisor of $a$ and $b$ if $c | a$ and $c | b$ and $c$ is a multiple
of every other element which divides both $a$ and $b$.
Closely related to the greatest common divisor is the
\index{least common multiple} 
{\bf least common multiple (lcm)} of two elements $a$ and $b$:
$d$ is called least common multiple of $a$ and $b$ if $a | d$ and $b | d$ and $d$ is a divisor
of every other element which is a multiple of both $a$ and $b$.
Since gcd and lcm are related by
\bq
\mathrm{lcm}(a,b) & = & \frac{a b}{\mathrm{gcd}(a,b)}
\eq
it is sufficient to focus on an algorithm for the calculation of the greatest common divisor. 

An element $p \in D\backslash\{0\}$ is called 
\index{prime elements}
{\bf prime} if
from $p | a b$ it follows that $p | a$ or $p | b$.
An element $p \in D\backslash\{0\}$ is called 
\index{irreducible elements}
{\bf irreducible} if
$p$ is not a unit and whenever
$p=a b$ either $a$ or $b$ is a unit.
In an integral domain, any prime element is automatically also an irreducible
element.
However, the reverse is in general not true.
This requires some additional properties in the ring.

Let us now turn to these additional properties:
An integral domain $D$ is called a 
\index{unique factorization domain}
{\bf unique factorization domain} 
if for all $a\in D\backslash\{0\}$, either $a$ is a unit or else $a$ can be expressed as a finite product
of irreducible elements such that this factorization into irreducible elements is unique 
up to associates and reordering.
It can be shown that in an unique factorization domain the notions of irreducible element and
prime element are equivalent.
In a unique factorization domain the greatest common divisor exists and is unique (up to associates and reordering).
The integral domain $S$ in eq.~(\ref{appendix_finite_fields:exampleintdomain}) is not a unique
factorization domain, since for example
\bq
 21 & = & 3 \cdot 7 = \left( 1 - 2 i \sqrt{5} \right) \left( 1 + 2 i \sqrt{5} \right)
\eq 
are two factorizations into irreducible elements. An example for a unique
factorization domains is the polynomial ring ${\mathbb Z}[x]$ in one variable with integer
coefficients.

An 
\index{Euclidean domain}
{\bf Euclidean domain} is an integral domain $D$ 
with a valuation map $v: D\backslash\{0\} \rightarrow {\mathbb N}_0$ into the non-negative integer numbers, such that
$v(a b) \ge v(a)$ for all $a,b \in D\backslash\{0\}$, and for all
$a,b \in D$ with $b \neq 0$, there exist elements $q,r \in D$ 
such that
\bq
a = bq +r,
\eq
where either $r=0$ or $v(r) < v(b)$.
This means that in an Euclidean domain division with remainder is possible.
An example for an Euclidean domain is given by the integer numbers ${\mathbb Z}$.

Finally, a 
\index{field}
{\bf field} is a commutative ring in which every non-zero element has a multiplicative inverse,
e.g. $R\backslash\{0\}$ is an Abelian group.
Any field is automatically an Euclidean domain.
Examples for fields are given by the rational numbers ${\mathbb Q}$, 
the real numbers ${\mathbb R}$, the complex numbers ${\mathbb C}$
or ${\mathbb Z}_p$, the integers modulo $p$ with $p$ a prime number.
${\mathbb Z}_p$ is a 
\index{finite field}
{\bf finite field}, it has $p$ elements $0,1,\dots,(p-1)$. 

\begin{figure}
\begin{center}
\includegraphics[scale=1.0]{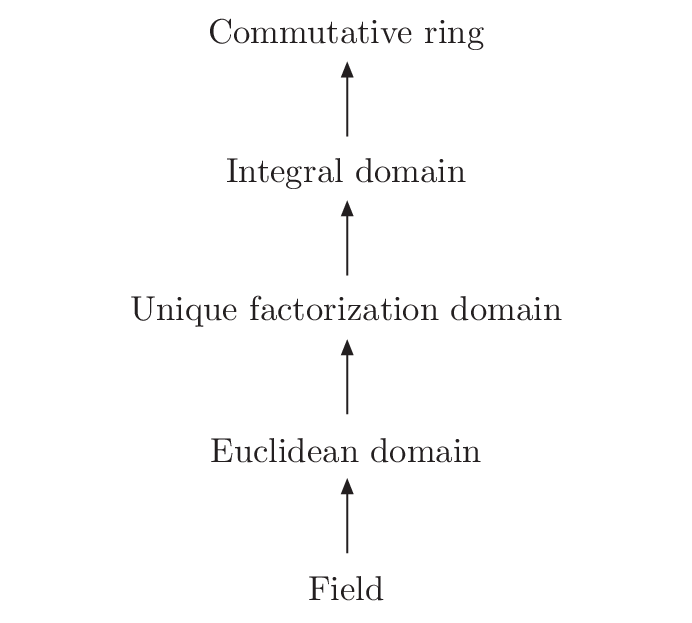}
\caption{\label{appendix_finite_fields:domains} \it Hierarchy of domains. Arrows $A \rightarrow B$
indicate that $A$ is a specialisation of $B$ (the notation is borrowed from derived classes in C++).}
\end{center}
\end{figure}

Fig.~(\ref{appendix_finite_fields:domains}) summarises the relationships between the various domains.
Of particular importance are polynomial rings in one or several variables.
Fig.~(\ref{appendix_finite_fields:unipolring}) summarises the structure of these
domains.
\begin{figure}
\begin{center}
\begin{tabular}{l|l|l}
 $R$ & $R[x]$ & $R[x_1,x_2,\dots,x_n]$ \\
\hline
 & \\
 commutative ring & commutative ring & commutative ring \\
 integral domain  & integral domain  & integral domain  \\
 unique factorization domain & unique factorization domain & unique factorization domain \\
 Euclidean domain & unique factorization domain & unique factorization domain \\
 field            & Euclidean domain & unique factorization domain \\
\end{tabular}
\caption{\label{appendix_finite_fields:unipolring} \it Structure of polynomial rings in one variable and several variables
depending on the
underlying coefficient ring $R$.}
\end{center}
\end{figure}
Note that a multivariate polynomial ring $R[x_1,\dots,x_n]$ can always be viewed
as an univariate polynomial ring in one variable $x_n$ with coefficients in the
ring $R[x_1,\dots,x_{n-1}]$.

The algorithm for the calculation of the gcd in an Euclidean domain dates back
to Euclid \cite{Euclid}. 
It is based on the fact that if $a = b q + r$, then
\bq
\mathrm{gcd}(a,b) & = & \mathrm{gcd}(b,r).
\eq
This is easily seen as follows: Let $c=\mathrm{gcd}(a,b)$ and $d=\mathrm{gcd}(b,r)$.
Since $r=a-b q$ we see that $c$ divides $r$, therefore it also divides $d$.
On the other hand $d$ divides $a=b q + r$ and therefore it also divides $c$.
We now have $c|d$ and $d|c$ and therefore $c$ and $d$ are associates.
\\
It is clear that for $r=0$, e.g. $a = b q$ we have $\mathrm{gcd}(a,b)=b$.
Let us denote the remainder as $r=\mathrm{rem}(a,b)$.
We can now define a sequence
$r_0 = a$, $r_1 = b$ and $r_i=\mathrm{rem}(r_{i-2},r_{i-1})$ for $i \ge 2$.
Then there is a finite index $k$ such that $r_{k+1}=0$ 
(since the valuation map applied to the remainders is a strictly decreasing function).
We have
\bq
\mathrm{gcd}(a,b) & = & \mathrm{gcd}(r_0,r_1) = \mathrm{gcd}(r_1,r_2)
 = \dots = \mathrm{gcd}(r_{k-1},r_k) = r_k.
\eq
This is the Euclidean algorithm.
We briefly mention that as a side product one can find elements $s$, $t$ such that
\bq
s a + t b & = & \mathrm{gcd}(a,b).
\eq
This is called the 
\index{extended Euclidean algorithm}
{\bf extended Euclidean algorithm}.
For the extended Euclidean algorithm one defines three sequences $r_i$, $s_i$ and $t_i$, starting from
\begin{align}
 r_0 & = a, & s_0 & = 1, & t_0 & = 0,
 \nonumber \\
 r_1 & = b, & s_1 & = 0, & t_1 & = 1,
\end{align}
and updates these sequences with $q_i = \lfloor \frac{r_{i-2}}{r_{i-1}} \rfloor$ as
\begin{align}
 r_i & =
 \mathrm{rem}(r_{i-2},r_{i-1}) 
 =
 r_{i-2} - q_i r_{i-1},
 &
 s_i & =
 s_{i-2} - q_i s_{i-1},
 &
 t_i & =
 t_{i-2} - q_i t_{i-1}.
\end{align}
As before, the algorithm stops whenever $r_{k+1}=0$. Then $\mathrm{gcd}(a,b)=r_k$, $s=s_k$ and $t=t_k$.
This allows the solution of the
\index{Diophantine equation} 
{\bf Diophantine equation}
\bq
s a + t b & = & c,
\eq
for $s$ and $t$ whenever $\mathrm{gcd}(a,b)$ divides $c$.

We are primarily interested in gcd computations in polynomial rings.
However, polynomial rings are usually only unique factorization domains,
but not Euclidean domains, e.g. division with remainder is in general not possible.
As an example consider the polynomials $a(x) = x^2+2x+3$ and $b(x)=5x+7$ in ${\mathbb Z}[x]$. 
It is not possible to write $a(x)$ in the form
$a(x) = b(x) q(x) + r(x)$, where the polynomials $q(x)$ and $r(x)$ have
integer coefficients.
However in ${\mathbb Q}[x]$ we have
\bq
x^2+2x+3 & = & \left( 5 x + 7 \right) \left( \frac{1}{5} x + \frac{3}{25} \right)
 + \frac{54}{25}
\eq
and we see that the obstruction arises from the leading coefficient of $b(x)$.
It is therefore appropriate to introduce a
\index{pseudo-division with remainder} 
{\bf pseudo-division with remainder}.
Let $D[x]$ be a polynomial ring over a unique factorization domain $D$.
For $ a(x) = a_n x^n + \dots + a_0$, $b(x) = b_m x^m + \dots + b_0$ with $n \ge m$ 
and $b(x) \neq 0$ there exists $q(x), r(x) \in D[x]$ such that
\bq
b_m^{n-m+1} a(x) & = & b(x) q(x) + r(x)
\eq
with $\mathrm{deg}(r(x)) < \mathrm{deg}(b(x))$.
This pseudo-division property is sufficient to extend the Euclidean algorithm to
polynomial rings over unique factorization domains.

Unfortunately, the Euclidean algorithm as well as the extended algorithm with pseudo-division
have a severe drawback: Intermediate expressions can become quite long.
This can be seen in the following example, where we would like to calculate the 
gcd of the polynomials
\bq
 a(x)
 & = &
 x^8 + x^6 - 3 x^4 - 3 x^3 + 8 x^2 + 2 x -5,
 \nonumber \\
 b(x)
 & = &
 3 x^6 + 5 x^4 - 4 x^2 - 9 x + 21,
\eq
in ${\mathbb Z}[x]$.
Calculating the pseudo-remainder sequence $r_i(x)$ we obtain
\bq
 r_2(x)
 & = &
 -15x^4 + 3x^2 -9,
 \nonumber \\
 r_3(x)
 & = &
 15795 x^2 + 30375 x - 59535,
 \nonumber \\
 r_4(x)
 & = &
 1254542875143750 x - 1654608338437500,
 \nonumber \\
 r_5(x)
 & = &
 12593338795500743100931141992187500.
\eq
This implies that $a(x)$ and $b(x)$ are relatively prime, but the numbers which occur
in the calculation are large.
An analysis of the problem shows, that the large numbers can be avoided if each
polynomial is split into a content part and a primitive part.
The 
\index{content of a polynomial}
{\bf content of a polynomial} 
is the gcd of all it's coefficients.
For example we have
\bq
 15795 x^2 + 30375 x - 59535
 & = &
 1215 \left( 13 x^2 + 25 x + 49 \right)
\eq
and $1215$ is the content and $13 x^2 + 25 x + 49$ the 
\index{primitive part of a polynomial}
{\bf primitive part}.
Taking out the content of a polynomial in each step requires a gcd calculation in the
coefficient domain and avoids large intermediate expressions in the example above.
However the extra cost for the gcd calculation in the coefficient domain is prohibitive
for multivariate polynomials.
The art of gcd calculations consists in finding an algorithm which keeps intermediate
expressions at reasonable size and which at the same time does not involve too much
computational overhead.
An acceptable algorithm is given by the 
\index{subresultant algorithm}
{\bf subresultant algorithm} \cite{Collins:srm,Brown:srm}:
Similar to the methods discussed above, one calculates a polynomial remainder 
sequence $r_0(x), r_1(x), \dots, r_k(x)$.
This sequence is obtained through
$r_0(x) = a(x)$, $r_1(x)=b(x)$ and
\bq
 c_i^{\delta_i+1} r_{i-1}(x)
 & = &
 q_i(x) r_i(x) + d_i r_{i+1}(x),
\eq
where $c_i$ is the leading coefficient of $r_i(x)$, 
$\delta_i = \mathrm{deg}(r_{i-1}(x)) - \mathrm{deg}(r_{i}(x))$
and
$d_1 = (-1)^{\delta_1+1}$,
$d_i = - c_{i-1} \psi_i^{\delta_i}$ for $2 \le i \le k$.
The $\psi_i$ are defined by $\psi_1 = -1$ and
\bq
\psi_i & = & \left( - c_{i-1} \right)^{\delta_{i-1}} \psi_{i-1}^{1-\delta_{i-1}}.
\eq
Then the primitive part of the last non-vanishing remainder equals the primitive
part of the greatest common divisor $\mathrm{gcd}(a(x),b(x))$.

\subsection{Heuristic methods}

In order to further speed up the calculation of polynomial gcds
one may resort to heuristic algorithms \cite{Char}.
In general a heuristic algorithm maps a problem to a simpler problem, solves
the simpler problem and tries to reconstruct the solution of the original problem
from the solution of the simpler problem.
\begin{figure}
\begin{center}
\includegraphics[scale=1.0]{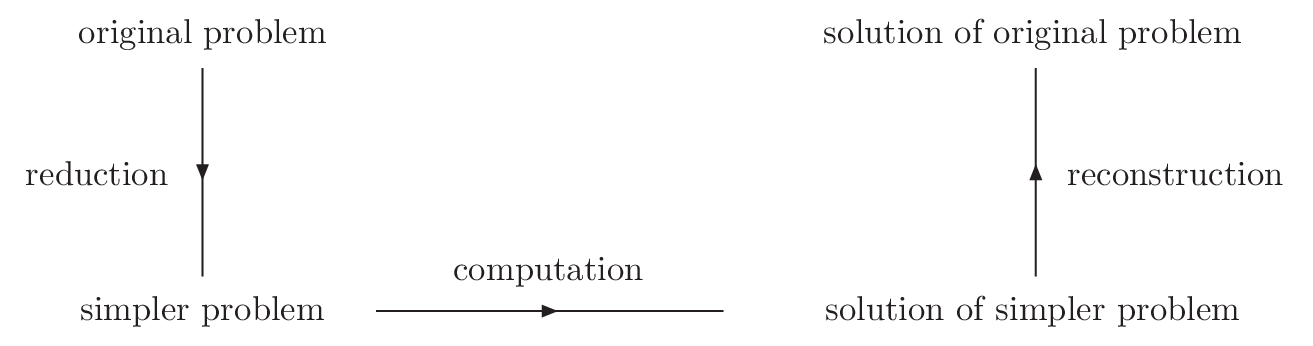}
\caption{\label{appendix_finite_fields:modularappr} 
The modular approach: Starting from the original problem,
one first tries to find a related simpler problem. The solution of the simpler problem
is used to reconstruct a solution of the original problem.}
\end{center}
\end{figure}
This is illustrated in fig.~\ref{appendix_finite_fields:modularappr}.
This approach is at the core of finite field methods.

Let's see how this works with a concrete example:
For the calculation of polynomial gcds one evaluates the polynomials at a specific point
and one considers the gcd of the results in the coefficient domain.
Since gcd calculations in the coefficient domain are cheaper, this can lead to a
sizeable speed-up, if both the evaluation of the polynomial and the reconstruction
of the polynomial gcd can be done at reasonable cost.
Let us consider the polynomials
\bq
 a(x)
 & = &
 6 x^4 + 21 x^3 + 35 x^2 + 27 x + 7,
 \nonumber \\
 b(x)
 & = &
 12 x^4 - 3 x^3 -17 x^2 - 45 x + 21.
\eq
Evaluating these polynomials at the point $\xi = 100$ yields
$a(100) = 621352707$ and $b(100)=1196825521$.
The gcd of theses two numbers is
\bq
c & = & \mathrm{gcd}(621352707,1196825521) = 30607.
\eq
To reconstruct the polynomial gcd one writes $c$ in the
$\xi$-adic representation
\bq
c & = & c_0 + c_1 \xi + \dots + c_n \xi^n, \;\;\; - \frac{\xi}{2} < c_i \le \frac{\xi}{2}.
\eq
Then the candidate for the polynomial gcd is
\bq
g(x) & = & c_0 + c_1 x + \dots + c_n x^n.
\eq
In our example we have
\bq
30607 & = & 7 + 6 \cdot 100 + 3 \cdot 100^2
\eq
and the candidate for the polynomial gcd is $g(x) = 3 x^2 +6 x +7$.
A theorem guarantees now if $\xi$ is chosen such that
\bq
 \xi 
 & > & 
 1 + 2 \; \mathrm{min}\left( ||a(x)||_\infty, ||b(x)||_\infty \right),
\eq
then $g(x)$ is the greatest common divisor of $a(x)$ and $b(x)$ if and only if
$g(x)$ divides $a(x)$ and $b(x)$.
This can easily be checked by a trial division.
In the example above, $g(x) = 3 x^2 +6 x +7$ divides both $a(x)$ and $b(x)$ and is therefore
the gcd of the two polynomials.

Note that there is no guarantee that the heuristic algorithm will succeed in finding
the gcd.
But if it does, this algorithm is usually faster than the subresultant algorithm discussed
previously.
Therefore, a strategy for a computer algebra system could be to
try first a few times the heuristic algorithm with various evaluation points and to fall
back onto the subresultant algorithm, if the greatest common divisor has not been found by the 
heuristic algorithm.

\section{The Chinese remainder theorem}

An important theorem is the Chinese remainder theorem, which allows us to use several (cheap)
calculations in the rings ${\mathbb Z}_{n_1}, {\mathbb Z}_{n_2}, \dots, {\mathbb Z}_{n_k}$ to obtain
the corresponding result in the larger ring ${\mathbb Z}_{n_1 \cdot \ldots \cdot n_k}$, provided no pair $(n_i,n_j)$ 
has a common factor greater than one.
This can be advantageous, as the computational cost for the calculations in 
the rings ${\mathbb Z}_{n_1}, \dots, {\mathbb Z}_{n_k}$ plus the computational cost for the reconstruction can be
significantly lower than the computational cost for a direct calculation in the ring 
${\mathbb Z}_{n_1 \cdot \ldots \cdot n_k}$. 

Let's see how this is done:
Two integers $n_1, n_2 \in {\mathbb Z}$ are called 
\index{coprime}
{\bf coprime}, 
if $\mathrm{gcd}(n_1,n_2)=1$.
Let $n_1,\dots,n_k$ be a set of natural numbers, which are pairwise coprime.
Set
\bq
 n & = & n_1 \cdot n_2 \cdot \ldots \cdot n_k.
\eq
Let $N \in {\mathbb Z}$ be an (unknown) integer. Assume that we know all remainders of $N \bmod n_j$:
\bq
 r_j & = & N \bmod n_j,
 \;\;\;\;\;\;
 1 \; \le \; j \; \le \; k.
\eq
We have $0 \le r_j < n_j$.
Given the remainders $r_1,\dots,r_k$ the Chinese remainder theorem allows 
us to compute the remainder $r = N \bmod n$:
We first set
\bq
 \tilde{n}_j & = & \prod\limits_{\substack{i=1 \\ i \neq j}}^k n_i \; = \; \frac{n}{n_j}.
\eq
The integers $n_j$ and $\tilde{n}_j$ are coprime $\mathrm{gcd}(n_j,\tilde{n}_j)=1$, 
hence there exist integers $s_j$ and $t_j$ such that
\bq
 s_j \tilde{n}_j + t_j n_j & = & 1.
\eq
$s_j$ and $t_j$ can be obtained with the extended Euclidean algorithm.
Then
\bq
\label{appendix_finite_fields:chinese_remainder_theorem_r}
 r & = &
 \sum\limits_{j=1}^k r_j s_j \tilde{n}_j \bmod n.
\eq
In mathematical terms, the Chinese remainder theorem says that there is a ring homomorphism
\bq
 {\mathbb Z}_{n_1} \times {\mathbb Z}_{n_2} \times \dots \times {\mathbb Z}_{n_k}
 & \cong &
 {\mathbb Z}_{n}.
\eq
In one direction, this isomorphism is given by eq.~(\ref{appendix_finite_fields:chinese_remainder_theorem_r}).
In the other direction, the isomorphism is trivially given by
\bq
 r_j & = & r \bmod n_j.
\eq
\bs
{\it \refstepcounter{exercise}
{\bf Exercise \theexercise}: 
Let $r$ be defined by eq.~(\ref{appendix_finite_fields:chinese_remainder_theorem_r}). Determine
\bq
 r \bmod n_i.
\eq
}
\es

\section{Black box reconstruction}

Suppose we are given a routine, which returns the numerical value of a rational function $f$ in several variables
$x=(x_1,\dots,x_n)$
for a choice of the input variables $x$.
The black box reconstruction problem asks, if it is possible to find the analytic form of the rational
function $f$ by evaluating numerically the function $f$ for sufficiently many distinct input values $x$.

A typical application in the context of Feynman integrals would be the following \cite{Peraro:2016wsq,Peraro:2019svx,Klappert:2019emp}:
Integration-by-parts reduction expresses any Feynman integral as a linear combination of master integrals.
The coefficients of this linear combination are rational functions in the number of space-time dimensions $D$
and the kinematic variables $x$.
It is much faster to run the integration-by-parts reduction algorithm for specific values of the kinematic variables $x$
(and $D$). Doing this several times we may reconstruct the coefficients.

The numerical evaluation is usually done with finite field arithmetic. This is exact (i.e. avoids rounding errors)
and limits the size of the numbers at intermediate stages (a finite field contains only finitely many numbers).
On the other side, there is some loss of information, as in a finite field there are relations
(like $1+1=0$ in ${\mathbb Z}_2$), which do not hold in a field of characteristic $0$.

A 
\index{finite field}
{\bf finite field} is a field with finitely many elements.
It can be shown that the number of elements must always be a power of a prime number: A finite field has
$p^k$ elements, where $p$ is a prime number and $k \in {\mathbb N}$.
For us it is sufficient to focus on the finite fields ${\mathbb F}_p={\mathbb Z}_p$, where $p$ is a prime number.
The field ${\mathbb F}_p$ has $p$ elements
\bq
 \left\{ 0,1,\dots,p-1 \right\}.
\eq
Addition, subtraction and multiplication are done modulo $p$.
For example
\bq
 2 \cdot 3 & = & 1 
 \;\;\;\;\;\;
 \mbox{in}
 \;\;\; 
 {\mathbb Z}_5.
\eq
Division is only slightly more complicated: First of all, $a$ divided by $b$ is nothing than the multiplication
of $a$ with the inverse of $b$. As in characteristic zero, we assume that $b \neq 0$.
As $p$ is a prime number and $1 \le b < p$ we have $\mathrm{gcd}(b,p)=1$.
From the extended Euclidean algorithm we find $s$ and $t$ such that
\bq
 s \cdot b + t \cdot p & = & 1.
\eq
$s$ is the inverse of $b$ in ${\mathbb Z}_p$, since
\bq
 s \cdot b & = & 1\bmod p.
\eq
Let us remark that $b^{-1}$ exists in ${\mathbb Z}_n$ (with $n$ not necessarily prime) if 
$b$ and $n$ are coprime.

Let $\frac{p}{q} \in {\mathbb Q}$ be a rational number and assume that we know the image
\bq
 c & = & \frac{p}{q} \bmod n
 \;\;\;\;\;\; \mbox{with} \;\;\;
 \mathrm{gcd}(q,n) \; = \; 1.
\eq
In general, there is no inverse mapping from ${\mathbb Z}_n$ to ${\mathbb Q}$.
(${\mathbb Z}_n$ has finitely many elements, while ${\mathbb Q}$ has countable many elements.)
We may use Wang's algorithm \cite{10.1145/800206.806398} to obtain a guess for $(p,q)$:
We use the extended Euclidean algorithm for $a=c$ and $b=n$ and monitor the sequence $(r_i,s_i)$.
We run the extended Euclidean algorithm until $r_i^2 \le n/2$.
If in addition $s_i^2 \le n/2$ and $\mathrm{gcd}(r_i,s_i)=1$ one returns $(p,q)=(r_i,s_i)$.
If on the other hand $s_i^2 > n/2$ or $\mathrm{gcd}(r_i,s_i)=1$, the reconstruction failed.

Let us look at an example: We consider the image of $\frac{2}{3} \in {\mathbb Q}$ in ${\mathbb Z}_{37}$.
From the extended Euclidean algorithm we obtain
\bq
 25 \cdot 3 - 2 \cdot 37 & = & 1,
\eq
and hence $3^{-1} = 25$ in ${\mathbb Z}_{37}$.
Thus the image of $\frac{2}{3} \in {\mathbb Q}$ in ${\mathbb Z}_{37}$ is
\bq
 \left( 2 \cdot 25 \right) \bmod 37 & = &
 50 \bmod 37 
 \; = \; 
 13.
\eq
Let us now consider the other direction: Assume we know $c=13 \in {\mathbb Z}_{37}$.
Can we find a rational number $\frac{p}{q} \in {\mathbb Q}$ such that the image of $\frac{p}{q}$
in ${\mathbb Z}_{37}$ is $c=13$?
We use Wang's algorithm from above. We first note that
\bq
 \left\lfloor \sqrt{\frac{37}{2}} \right\rfloor
 & = &
 4.
\eq
From the extended Euclidean algorithm with $a=13$ and $b=37$ we obtain
\begin{align}
 r_0 & = 13, & s_0 & = 1, & t_0 & = 0,
 \nonumber \\
 r_1 & = 37, & s_1 & = 0, & t_1 & = 1,
 \nonumber \\
 r_2 & = 13, & s_2 & = 1, & t_2 & = 0, 
 \nonumber \\
 r_3 & = 11, & s_3 & = -2, & t_3 & = 1, 
 \nonumber \\
 r_4 & = 2, & s_4 & = 3, & t_4 & = -1, 
 \nonumber \\
 r_5 & = 1, & s_5 & = -17, & t_5 & = 6, 
 \nonumber \\
 r_6 & = 0, & s_6 & = 37, & t_6 & = -13. 
\end{align}
We stop with $r_4$, as 
\bq
 r_4 & \le & \left\lfloor \sqrt{\frac{37}{2}} \right\rfloor.
\eq
In addition we have $s_4=3 \le \lfloor \sqrt{\frac{37}{2}} \rfloor$ and $\mathrm{gcd}(r_4,s_4)=\mathrm{gcd}(2,3)=1$.
Hence Wang's algorithm returns $(p,q)=(2,3)$, which is the correct result.

\subsection{Univariate polynomials}

Let us start with the reconstruction of a polynomial $f(x)$ in one variable $x$.
A degree $d$ polynomial is uniquely specified by the values $y_i$ at $(d+1)$ points $x_i$:
\bq
 y_i & = & f\left(x_i\right), 
 \;\;\;\;\;\;
 0 \; \le i \; \le \; d.
\eq
There is only a small problem: We do not know the degree of the polynomial a priori.
In this situation, a representation in terms of Newton polynomials is convenient.
We write
\bq
\label{appendix_finite_fields:Newton_representation}
 f\left(x\right)
 & = &
 \sum\limits_{j=0}^d a_j \prod\limits_{i=0}^{j-1} \left(x-x_i\right)
 \nonumber \\
 & = &
 a_0
 + \left(x-x_0\right) 
   \left( a_1 + \left(x-x_1\right) 
                \left( a_2 + \left(x-x_2\right)
                             \left( \dots + \left(x-x_{d-1}\right) a_d \right) 
                \right) 
   \right).
\eq
The coefficients $a_i$ are computed recursively as
\bq
\label{appendix_finite_fields:Newton_coefficients}
 a_0 & = & y_0,
 \nonumber \\
 a_1 & = & \frac{y_1-a_0}{x_1-x_0},
 \nonumber \\
 a_2 & = & \left( \frac{y_2-a_0}{x_2-x_0} - a_1 \right) \frac{1}{x_2-x_1},
 \nonumber \\
 & \dots &
 \nonumber \\
 a_d & = & 
 \left( \left( \frac{y_d-a_0}{x_d-x_0} - a_1 \right) \frac{1}{x_d-x_1} - \dots - a_{d-1} \right) \frac{1}{x_d-x_{d-1}}.
\eq
This representation has the advantage that additional evaluation points will not change the values
of the already computed coefficients $a_j$.
If $f$ is of degree $d$ and if we probe more than $(d+1)$ points, the coefficients $a_j$ with $j>d$ will be zero.
This can be used as a termination criteria: We fix a positive integer $\eta$ and terminate the algorithm
if 
\bq
 a_{d+1} \; = \; a_{d+2} \; = \; \dots \; = \; a_{d+\eta} \; = \; 0.
\eq
This is a heuristic algorithm: It may fail if for example $f$ is of degree $(d+\eta+1)$ and we accidentally choose the points $x_i$ such that
$a_{d+1}=\dots=a_{d+\eta}=0$.
However, it can be shown that the probability that we obtain the correct polynomial in ${\mathbb Z}_p$ is no less than
\cite{DBLP:conf/issac/2001,DBLP:journals/jsc/KaltofenL03}
\bq
 1 - \left(d+1\right) \left( \frac{d}{p} \right)^\eta.
\eq

\subsection{Univariate rational functions}

For the reconstruction of a rational function $f$ in one variable $x$ we may use Thiele's formula \cite{abramowitz+stegun},
which is based on a continued fraction
\bq
 f\left(x\right)
 & = &
 a_0 + \frac{x-x_0}{a_1+\frac{x-x_1}{a_2+\frac{x-x_2}{\dots+\frac{x-x_{d-1}}{a_d}}}}
 \\
 & = &
 a_0
 + \left(x-x_0\right) 
   \left( a_1 + \left(x-x_1\right) 
                \left( a_2 + \left(x-x_2\right)
                             \left( \dots + \frac{\left(x-x_{d-1}\right)}{a_d} \right)^{-1} 
                \right)^{-1} 
   \right)^{-1}.
 \nonumber
\eq
The coefficients $a_i$ are computed recursively as
\bq
 a_0 & = & y_0,
 \nonumber \\
 a_1 & = & \frac{x_1-x_0}{y_1-a_0},
 \nonumber \\
 a_2 & = & \left( \frac{x_2-x_0}{y_2-a_0} - a_1 \right)^{-1} \left(x_2-x_1\right),
 \nonumber \\
 & \dots &
 \nonumber \\
 a_d & = & 
 \left( \left( \frac{x_d-x_0}{y_d-a_0} - a_1 \right)^{-1} \left(x_d-x_1\right) - \dots - a_{d-1} \right)^{-1} \left(x_d-x_{d-1}\right).
\eq
As in the case of Newton interpolation, the coefficient $a_j$ depends only on the evaluations $0, 1, \dots, j$
and will not change if we add additional evaluations.

It should be noted that this algorithm may lead to spurious singularities (for example if $y_1=a_0$) in which case
the algorithm fails.
In this case one may re-try with different values of the $x_j$'s.

\subsection{Multivariate polynomials}

Let us now turn to multivariate polynomials.
Let $f$ be a polynomial in the variables $x=(x_1,\dots,x_n)$.
We denote the values of the $i$-th variable where we probe the function $f$ 
by 
\bq
 x_{i,0}, x_{i,1}, x_{i,2}, \dots.
\eq
We may view a multivariate polynomial $f(x_1,\dots,x_n)$ as a univariate polynomial
in the variable $x_n$ with coefficients in the ring ${\mathbb F}[x_1,\dots,x_{n-1}]$.
Thus we may reconstruct $f$ recursively:
We start with Newton interpolation in the variable $x_n$:
\bq
\label{appendix_finite_fields:multivariate_Newton_representation}
 f\left(x_1,\dots,x_n\right)
 & = &
 \sum\limits_{j=0}^d a_j\left(x_1,\dots,x_{n-1}\right) \prod\limits_{i=0}^{j-1} \left(x_n-x_{n,i}\right)
\eq
The coefficients $a_j(x_1,\dots,x_{n-1})$ are polynomials in $(n-1)$ variables, i.e. one variable less.
Given a numerical black box routine for $f$ and choices $x_{n,0}, x_{n,1}, \dots$ for the last variable
$x_n$, we immediately have a numerical black box routine for $a_0(x_1,\dots,x_{n-1})$ by choosing
$x_n=x_{n,0}$. We therefore first reconstruct $a_0(x_1,\dots,x_{n-1})$.
Once $a_0(x_1,\dots,x_{n-1})$ is known, we obtain a numerical black box routine for $a_1(x_1,\dots,x_{n-1})$
by using
\bq
 a_1\left(x_1,\dots,x_{n-1}\right)
 & = &
 \frac{f\left(x_1,\dots,x_n\right)-a_0\left(x_1,\dots,x_{n-1}\right)}{x_n-x_{n,0}}.
\eq
By using eq.~(\ref{appendix_finite_fields:Newton_coefficients}) we may continue this process to 
reconstruct $a_2(x_1,\dots,x_{n-1})$, $a_3(x_1,\dots,x_{n-1})$, etc..
Each $a_j(x_1,\dots,x_{n-1})$ is a polynomial in $(n-1)$ variables.

In order to reconstruct $a_j(x_1,\dots,x_{n-1})$ we repeat the ansatz and view
$a_j(x_1,\dots,x_{n-1})$ as an univariate polynomial in $x_{n-1}$ with coefficients in ${\mathbb F}[x_1,\dots,x_{n-2}]$.
This recursion terminates with the reconstruction of univariate polynomials in the variable $x_1$ with coefficients
in the field ${\mathbb F}$.

For sparse multivariate polynomials the algorithm above can be improved.
A multivariate polynomial in $n$ variables and total degree $d$ has only a finite number of coefficients.
A (multivariate) polynomial is called 
\index{sparse polynomial}
{\bf sparse}, 
if most of its coefficients are zero.
The contrary of a sparse polynomial is a 
\index{dense polynomial}
{\bf dense polynomial}, where most of its coefficients are non-zero.
We may always convert a polynomial from Newton's representation as in 
eq.~(\ref{appendix_finite_fields:Newton_representation}) and eq.~(\ref{appendix_finite_fields:multivariate_Newton_representation})
to the standard monomial representation
\bq
 f\left(x\right)
 & = &
 \sum\limits_\alpha c_\alpha x^\alpha,
\eq
where we use the multi-index notation $x^\alpha = x_1^{\alpha_1} \cdot \dots \cdot x_n^{\alpha_n}$.
The main idea of Zippel's algorithm \cite{10.1007/3-540-09519-5_73,10.1016/S0747-7171(08)80018-1}
for sparse polynomials is the following:
Suppose we are in the recursive algorithm above
at a stage where we reconstruct a polynomial in the variables $x_1, \dots, x_k$ (with $k < n$).
Assume further that the coefficient $c_{\alpha_1 \dots \alpha_k}(x_{k+1,j_{k+1}},\dots,x_{n,j_n})$ of the monomial
\bq
\label{appendix_finite_fields:Zippel_assumption}
 c_{\alpha_1 \dots \alpha_k} x_1^{\alpha_1} \cdot \dots \cdot x_k^{\alpha_k}
\eq
turns out to be zero
for the chosen numerical values of the remaining variables $x_{k+1}, \dots, x_{n}$.
Zippel's algorithm assumes that this remains true in ${\mathbb F}[x_{k+1}, \dots, x_{n}]$, e.g.
the final polynomial will not contain a term of the form as in eq.~(\ref{appendix_finite_fields:Zippel_assumption})
with a coefficient $c_{\alpha_1 \dots \alpha_k} \in {\mathbb F}[x_{k+1}, \dots, x_{n}]$.
This can lead to a significant speed-up. 
Let us stress that this is a guess (which may or may not be true).
In order to minimise the risk of 
accidental zeros, one chooses the numerical values $x_{k+1,j_{k+1}},\dots,x_{n,j_n}$
with care.
A typical choice is given by random values for the variables
$x_{i,0}$ and the values
\bq
 x_{i,j} & = & \left( x_{i,0} \right)^{j+1},
\eq
for the remaining variables.

\subsection{Multivariate rational functions}

We finally turn to the reconstruction of multivariate rational functions.
We may write a multivariate rational function as
\bq
 f\left(x\right)
 & = &
 \frac{p\left(x\right)}{q\left(x\right)},
\eq
where $p(x)$ and $q(x)$ are multivariate polynomials in $n$ variables $x_1,\dots,x_n$.
In order to make this representation unique, one may require that the coefficient of the smallest monomial of $q(x)$ 
with respect to a chosen term order equals one.
If $q(x)$ contains a constant term, this implies that the constant term equals one.

Let us now discuss the main ideas of the algorithm of Cuyt and Lee \cite{10.1145/1823931.1823938}
for the reconstruction of a multivariate rational function.
We focus on the ideas how this can be done in principle.
For the details how this is implemented efficiently we refer to the 
literature \cite{10.1145/1823931.1823938,Peraro:2016wsq,Klappert:2019emp}.

We introduce an auxiliary variable $t$ 
and we consider the function $g(t,x)$ defined by
\bq
 g\left(t,x\right)
 & = &
 f\left(t x_1,\dots,t x_n\right).
\eq
We view $g$ as a function of $t$, depending in addition on parameters $x_1,\dots,x_n$.
Clearly
\bq
 f\left(x\right) & = & g\left(1,x\right).
\eq
Let us first assume that the denominator polynomial $q(x)$ of the rational function $f(x)$ contains a constant term.
We choose the normalisation where this constant term equals one.
Then $g(t,x)$ can be written as
\bq
 g\left(t,x\right)
 & = &
 \frac{\sum\limits_{r=0}^{d_p} p_r\left(x\right) t^r}{1 + \sum\limits_{r'=1}^{d_q} q_{r'}\left(x\right) t^{r'}},
\eq
where the $p_r$ and $q_r$ are homogeneous polynomials of degree $r$ in the variables $x=(x_1,\dots,x_n)$.

For chosen numerical values $x_{1,j_1}, \dots, x_{n,j_n}$ for the variables $x_1,\dots,x_n$
we may use Thiele's algorithm for the reconstruction of the 
univariate rational function $g$ in $t$.
This reconstruction will provide
\begin{enumerate}
\item a verification (or falsification)
of our assumption that the denominator contains a constant term,
\item the degrees $d_p$ and $d_q$ of the numerator polynomial and of the denominator polynomial, respectively,
\item numerical black box routines for the polynomials $p_r$ and $q_r$ (given as the coefficient of $t^r$ in the
numerator and denominator, respectively).
\end{enumerate}
With the numerical black box routines for the polynomials $p_r$ and $q_r$ at hand, we may use any algorithm
for the reconstruction of multivariate polynomials (for example the algorithm discussed above).
As we know that $p_r$ and $q_r$ are homogeneous of degree $r$, one usually uses a dedicated algorithm
for homogeneous polynomials.

It may happen that our initial assumption that the denominator polynomial $q(x)$ contains a constant term is not
justified. In this case one considers first a modified function
\bq
 \tilde{f}\left(x_1,\dots,x_n\right)
 & = &
 f\left(x_1+s_1,\dots,x_n+s_n\right)
\eq
obtained by a random shift $(s_1,\dots,s_n)$.
One reconstructs $\tilde{f}$ and obtains $f(x_1,\dots,x_n)=\tilde{f}(x_1-s_1,\dots,x_n-s_n)$.
The denominator polynomials $\tilde{q}$ of the rational function $\tilde{f}$ will in general have a constant
term (if accidentally this is not the case, one may try a different shift).
This completes the algorithm for the reconstruction of a multivariate rational function.
However, the last step comes with a caveat: If the numerator polynomial $p$ and the denominator polynomial $q$
of the original rational function $f$ are sparse polynomials, a random shift will in general lead to dense
polynomials $\tilde{p}$ and $\tilde{q}$ of the shifted rational function $\tilde{f}$: 
For example, $x^5$ is a sparse polynomial in one variable $x$, while
\bq
 \left(x+1\right)^5
 & = & 
 x^5 + 5 x^4 + 10 x^3 + 10 x^2 + 5 x +1
\eq
is a dense polynomial.

%% file: solutions.tex
\newpage
\chapter{Solutions to the exercises}
\label{appendix_solutions}

\setcounter{exercise}{0}

%
%
\bs
{\it \refstepcounter{exercise}
{\bf Exercise \theexercise}: 
Consider a connected graph $G$ with the notation as above.
Show that momentum conservation at each vertex of valency $>1$ implies momentum conservation
of the external momenta:
\bq
 \sum\limits_{e_j \in E^{\mathrm{in}}} q_j 
 & = &
 \sum\limits_{e_j \in E^{\mathrm{out}}} q_j.
\eq
If we choose an orientation such that all external edges have a vertex of valency $1$ as sink 
(e.g. $E^{\mathrm{in}} = \emptyset$) this translates to
\bq
 \sum\limits_{j=1}^{\nexternal} p_j & = & 0.
\eq
{\bf Solution}: 
We proof the claim by induction on the number of internal edges $\ninternal$.
For $\ninternal=0$ our graph looks like
\bq
 \begin{picture}(100,100)(0,0)
 \Vertex(50,50){2}
 \Vertex(84.64,70){2}
 \Vertex(84.64,30){2}
 \Vertex(56.95,89.39){2}
 \Vertex(56.95,10.61){2}
 \Vertex(24.29,80.64){2}
 \Vertex(24.29,19.36){2}
 \Vertex(10,50){2}
 \ArrowLine(50,50)(84.64,70)
 \ArrowLine(84.64,30)(50,50)
 \ArrowLine(50,50)(56.95,89.39)
 \ArrowLine(50,50)(56.95,10.61)
 \ArrowLine(24.29,80.64)(50,50)
 \ArrowLine(24.29,19.36)(50,50)
 \ArrowLine(50,50)(10,50)
 \Text(60,50)[l]{$v$}
\end{picture} 
\eq
and has exactly one vertex $v$ of valency $>1$.
Momentum conservation at this vertex reads
\bq
 \sum\limits_{e_j \in E^{\mathrm{in}}} q_j 
 & = &
 \sum\limits_{e_j \in E^{\mathrm{out}}} q_j.
\eq
and corresponds exactly to the claim.

Let us now assume that the claim is correct for $(\ninternal-1)$.
Consider now a graph with $\ninternal$ internal edges. 
Pick one internal edge $e_i$.
Denote by $v_a$ its source and by $v_b$ its sink.
Let us write down momentum conservation at $v_a$ and $v_b$:
\begin{alignat}{3}
 v_a & : & \;\;\; && q_j + \sum\limits_{e_r \in E^{\mathrm{source}}(v_a)\backslash\{e_j\}} q_r & = \sum\limits_{e_r \in E^{\mathrm{sink}}(v_a)} q_r,
 \nonumber \\
 v_b & : & \;\;\; && \sum\limits_{e_r \in E^{\mathrm{source}}(v_b)} q_r & = q_j + \sum\limits_{e_r \in E^{\mathrm{sink}}(v_b)\backslash\{e_j\}} q_r.
\end{alignat}
We may eliminate $q_i$ from these two equations. We obtain a single equation
\begin{alignat}{3}
 v & : & \;\;\; && 
 \sum\limits_{e_r \in E^{\mathrm{source}}(v_a)\backslash\{e_j\}} q_r + \sum\limits_{e_r \in E^{\mathrm{source}}(v_b)} q_r 
 & = 
 \sum\limits_{e_r \in E^{\mathrm{sink}}(v_a)} q_r + \sum\limits_{e_r \in E^{\mathrm{sink}}(v_b)\backslash\{e_j\}} q_r.
\end{alignat}
The momentum $q_j$ appears at no other vertex.
As far as momentum conservation is concerned, we may replace the graph $G$ with a new graph $\tilde{G}$, where
the edge $e_j$ has been contracted (e.g. the edge $e_j$ is removed and the vertices $v_a$ and $v_b$ are merged to a new vertex $v$. Pictorially we have
\bq
 \begin{picture}(150,50)(0,50)
 \Vertex(50,50){2}
 \Vertex(100,50){2}
 \ArrowLine(100,50)(134.64,70)
 \ArrowLine(134.64,30)(100,50)
 \ArrowLine(100,50)(106.95,89.39)
 \ArrowLine(100,50)(106.95,10.61)
 \ArrowLine(24.29,80.64)(50,50)
 \ArrowLine(24.29,19.36)(50,50)
 \ArrowLine(50,50)(10,50)
 \ArrowLine(50,50)(100,50)
 \Text(75,55)[b]{$e_j$}
 \Text(52,45)[tl]{$v_a$}
 \Text(98,45)[tr]{$v_b$}
\end{picture} 
 & \rightarrow &
 \begin{picture}(100,50)(0,50)
 \Vertex(50,50){2}
 \ArrowLine(50,50)(84.64,70)
 \ArrowLine(84.64,30)(50,50)
 \ArrowLine(50,50)(56.95,89.39)
 \ArrowLine(50,50)(56.95,10.61)
 \ArrowLine(24.29,80.64)(50,50)
 \ArrowLine(24.29,19.36)(50,50)
 \ArrowLine(50,50)(10,50)
 \Text(60,50)[l]{$v$}
\end{picture} 
 \\
 \nonumber \\
 \nonumber \\
 \nonumber
\eq
The new graph $\tilde{G}$ has $(\ninternal-1)$ internal edges and we may use the induction hypothesis.
As $G$ and $\tilde{G}$ only differ by the contraction of an internal edge, the claim holds for the graph $G$ as well.
\\
\\
If we choose an orientation such that all external edges have a vertex of valency $1$ as sink 
we have
\bq
 E^{\mathrm{in}} \; = \; \emptyset,
 & &
 E^{\mathrm{out}} \; = \; \left\{ e_{\ninternal+1}, \dots, e_{\ninternal+\nexternal}\right\}
\eq
and
\bq
 0
 \; = \; 
 \sum\limits_{e_j \in E^{\mathrm{in}}} q_j 
 \; = \;
 \sum\limits_{e_j \in E^{\mathrm{out}}} q_j
 \; = \;
 \sum\limits_{j=1}^{\nexternal} p_j.
\eq
}
\es
\\
\\
\bs
{\it \refstepcounter{exercise}
{\bf Exercise \theexercise}: 
Consider the one-loop graph shown in fig.~\ref{chapter_basics:fig_oneloopbox}.
Write down the equations expressing momentum conservation at each vertex of valency $>1$.
Use $p_1,p_2,p_3$ as independent external momenta and $k_1=q_4$ as the independent loop momentum.
Express all other momenta as linear combinations of these.
\\
\\
{\bf Solution}: 
Momentum conservation at the four vertices $v_1,\dots,v_4$ reads
\begin{alignat}{3}
 v_1 & : & \;\;\; && p_1 + q_1 & = q_4,
 \nonumber \\
 v_2 & : & \;\;\; && p_2 + q_2 & = q_1,
 \nonumber \\
 v_3 & : & \;\;\; && p_3 + q_3 & = q_2,
 \nonumber \\
 v_4 & : & \;\;\; && p_4 + q_4 & = q_3.
\end{alignat}
With $k_1=q_4$ we express $q_1$, $q_2$, $q_3$ and $p_4$ in terms 
of $k_1$, $p_1$, $p_2$ and $p_3$ as
\bq
 q_1 & = & k_1 - p_1,
 \nonumber \\
 q_2 & = & k_1 - p_1 - p_2,
 \nonumber \\
 q_3 & = & k_1 - p_1 - p_2 - p_3,
 \nonumber \\
 p_4 & = & - p_1 - p_2 - p_3.
\eq
}
\es
\\
\\
\bs
{\it \refstepcounter{exercise}
{\bf Exercise \theexercise}: 
Prove
\bq
 T_\nu\left(D\right)
 & = &
 \nu T_{\nu+1}\left(D+2\right).
\eq
{\bf Solution}: 
The tadpole integral is given by eq.~(\ref{chapter_basics:result_tadpole}).
The left-hand side of our equation equals
\bq
 T_\nu\left(D\right)
 & = &
 \frac{e^{\eps \Eulerconstant} \Gamma\left(\nu-\frac{D}{2}\right)}{\Gamma\left(\nu\right)}
 \left( \frac{m^2}{\mu^2} \right)^{\frac{D}{2}-\nu}.
\eq
Let's work out the right-hand side:
\bq
 \nu T_{\nu+1}\left(D+2\right)
 & = &
 \nu \frac{e^{\eps \Eulerconstant} \Gamma\left(\left(\nu+1\right)-\frac{D+2}{2}\right)}{\Gamma\left(\nu+1\right)}
 \left( \frac{m^2}{\mu^2} \right)^{\frac{D+2}{2}-\left(\nu+1\right)}
 \nonumber \\
 & = &
 \frac{e^{\eps \Eulerconstant} \Gamma\left(\nu-\frac{D}{2}\right)}{\Gamma\left(\nu\right)}
 \left( \frac{m^2}{\mu^2} \right)^{\frac{D}{2}-\nu},
\eq
where we used $\Gamma(\nu+1) = \nu \Gamma(\nu)$. 
}
\es
\\
\\
\bs
{\it \refstepcounter{exercise}
{\bf Exercise \theexercise}: 
Derive eq.~(\ref{chapter_basics:master_one_loop_v1}).
\\
\\
{\bf Solution}: 
We repeat the steps from the calculation of the tadpole integral.
Starting from
\bq
 \tilde{T}
 & = &
 e^{\eps \Eulerconstant} \left(\mu^2\right)^{\nu-\frac{D}{2}-a}
 \int \frac{d^Dk}{i \pi^{\frac{D}{2}}} 
\frac{\left(-k^2\right)^a}{\left( -U k^2 + F\right)^\nu}
\eq
we perform a Wick rotation and obtain
\bq
 \tilde{T}
 & = &
 e^{\eps \Eulerconstant} \left(\mu^2\right)^{\nu-\frac{D}{2}-a}
 \int \frac{d^DK}{\pi^{\frac{D}{2}}} 
\frac{\left(K^2\right)^a}{\left( U K^2 + F\right)^\nu}.
\eq
We then introduce spherical coordinates and integrate over the angles. This yields
\bq
 \tilde{T}
 & = &
 \frac{e^{\eps \Eulerconstant} \left(\mu^2\right)^{\nu-\frac{D}{2}-a}}{\Gamma\left( \frac{D}{2} \right)}
 \int\limits_0^\infty dK^2 
\frac{\left(K^2\right)^{\frac{D}{2}+a-1}}{\left( U K^2 + F\right)^\nu}.
\eq
We then substitute $t=U K^2/F$. We obtain
\bq
 \tilde{T}
 & = &
 \frac{e^{\eps \Eulerconstant} \left(\mu^2\right)^{\nu-\frac{D}{2}-a}}{\Gamma\left( \frac{D}{2} \right)}
 \frac{U^{-\frac{D}{2}-a}}{F^{\nu-\frac{D}{2}-a}}
 \int\limits_0^\infty dt 
\frac{t^{\frac{D}{2}+a-1}}{\left( t + 1 \right)^\nu}.
\eq
The remaining integral is again Euler's beta integral and we finally obtain
\bq
 \tilde{T}
 & = &
 e^{\eps \Eulerconstant} \left(\mu^2\right)^{\nu-\frac{D}{2}-a}
 \frac{\Gamma\left(\frac{D}{2}+a\right)}{\Gamma\left(\frac{D}{2}\right)}
 \frac{\Gamma\left(\nu-\frac{D}{2}-a\right)}{\Gamma\left(\nu\right)} 
 \frac{U^{-\frac{D}{2}-a}}{F^{\nu-\frac{D}{2}-a}}.
\eq
}
\es
\\
\\
\bs
{\it \refstepcounter{exercise}
{\bf Exercise \theexercise}: 
Derive eq.~(\ref{chapter_basics:k2_eps_numerator}).
\\
\\
Hint: Split the $D$-dimensional integration into a $\Dint$-dimensional part and a
$(-2\eps)$-dimensional part.
Eq.~(\ref{chapter_basics:k2_eps_numerator}) can be derived by just considering the $(-2\eps)$-dimensional part.
\\
\\
{\bf Solution}: 
$f(k_{(\Dint)},k_{(-2\eps)}^2)$ may depend arbitrarily on $k^0$, $k^1$, ..., $k^{\Dint-1}$,
but the dependence on $k^{\Dint}$, $k^{\Dint+1}$, ..., $k^{D-1}$ is only
through $k_{(-2\eps)}^2$.
We split the integration into a $\Dint$-dimensional part and a
$(-2\eps)$-dimensional part. We write
\bq
 \frac{d^Dk}{i \pi^{\frac{D}{2}}}
 & = &
 \frac{d^{\Dint}k}{i \pi^{\frac{\Dint}{2}}} 
 \frac{d^{(-2\eps)}k}{i \pi^{-\eps}}.
\eq
From eq.~(\ref{chapter_basics:def_master_one_loop}) and eq.~(\ref{chapter_basics:master_one_loop_v1}) we have
\bq
 \int \frac{d^{(-2\eps)}k}{i \pi^{-\eps}} 
 \left(-k_{(-2\eps)}^2\right)^r 
 f\left(k_{(\Dint)},k_{(-2\eps)}^2\right) 
 & = &
 \frac{\Gamma(r-\eps)}{\Gamma(-\eps)}
 \int \frac{d^{(-2\eps+2r)}k}{i \pi^{-\eps+r}}
 f\left(k_{(\Dint)},k_{(-2\eps)}^2\right).
\eq
Integrating then also over the $\Dint$-dimensional part gives
\bq
 \int \frac{d^Dk}{i \pi^{\frac{D}{2}}} 
 \left(-k_{(-2\eps)}^2\right)^r 
 f\left(k_{(\Dint)},k_{(-2\eps)}^2\right) 
 & = &
 \frac{\Gamma(r-\eps)}{\Gamma(-\eps)}
 \int \frac{d^{D+2r}k}{i \pi^{\frac{D+2r}{2}}} 
 f\left(k_{(\Dint)},k_{(-2\eps)}^2\right).
\eq
}
\es
\\
\\
\bs
{\it \refstepcounter{exercise}
{\bf Exercise \theexercise}: 
Derive eq.~(\ref{chapter_basics:basic_eq_negative_dimensions}).
\\
\\
Hint: Consider the mass dimension of the integral to prove the statement for $D/2+a\neq0$
and the normalisation of the integral measure in eq.~(\ref{chapter_basics:normalisation_D_int}) to prove the statement for $D/2+a = 0$.
\\
\\
{\bf Solution}: 
We start with the case $D/2+a\neq0$.
The integral
\bq
 \int \frac{d^Dk}{i \pi^{\frac{D}{2}}} 
 \left( - k^2 \right)^a  
\eq
has mass dimension $(D+2a)$, so it must be proportional to some scale $\mu$ raised to the power
of the mass dimension:
\bq
 \left( \mu^2 \right)^{\frac{D}{2}+a}.
\eq
However, the integral is a scaleless integral.
Thus there is no such scale available, hence the integral must be zero.

Let us now turn to the case $D/2+a = 0$.
Let us consider $a\in{\mathbb N}$. The space-time dimension is then necessarily 
even and negative: $D=-2a$.
For this reason, the use of eq.~(\ref{chapter_basics:basic_eq_negative_dimensions}) is sometimes
called the 
\index{negative dimension approach}
{\bf negative dimension approach}.
After Wick rotation we have to show
\bq 
\label{appendix_solutions:negative_dimension_Wick_rotated_integer_value}
 \int \frac{d^DK}{\pi^{\frac{D}{2}}} 
 \left( K^2 \right)^a & = & 
 \left(-1\right)^{\frac{D}{2}} \Gamma\left(1-\frac{D}{2}\right), 
 \;\;\;\;\;\; \mbox{for}
 \;\;\; 
 \frac{D}{2}+a=0
 \;\;\;
 \mbox{and}
 \;\;\; 
 a\in{\mathbb N}.
\eq
The normalisation of the integral measure reads
\bq
 \int \frac{d^DK}{\pi^{\frac{D}{2}}} \exp \left( - K^2 \right) 
 & = &
 1.
\eq
We expand the left-hand side
\bq
 \int \frac{d^DK}{\pi^{\frac{D}{2}}} \exp \left( - K^2 \right) 
 & = &
 \sum\limits_{a=0}^\infty 
 \frac{\left(-1\right)^a}{a!}
 \int \frac{d^DK}{\pi^{\frac{D}{2}}} 
 \left(K^2\right)^a.
\eq
We already know that integrals with $D/2+a \neq 0$ vanish, thus only the term
$a=-D/2$ survives:
\bq
 \int \frac{d^DK}{\pi^{\frac{D}{2}}} \exp \left( - K^2 \right) 
 & = &
 \frac{\left(-1\right)^{-\frac{D}{2}}}{\left(-\frac{D}{2}\right)!}
 \int \frac{d^DK}{\pi^{\frac{D}{2}}} 
 \left(K^2\right)^{-\frac{D}{2}}.
\eq
This should be equal to $1$, hence
\bq
 \int \frac{d^DK}{\pi^{\frac{D}{2}}} 
 \left(K^2\right)^{-\frac{D}{2}}
 & = &
 \left(-1\right)^{\frac{D}{2}}
 \Gamma\left(1-\frac{D}{2}\right).
\eq
This proves eq.~(\ref{appendix_solutions:negative_dimension_Wick_rotated_integer_value}).
Analytic continuation in $D$ (or the parameter $a$) on the variety $D/2+a=0$ completes the proof.
}
\es
\\
\\
\bs
{\it \refstepcounter{exercise}
{\bf Exercise \theexercise}: 
Consider again the one-loop box graph shown in fig.\ref{chapter_basics:fig_oneloopbox}.
Assume first that all internal masses are non-zero and pairwise distinct and that the external momenta
are as generic as possible.
How many kinematic variables are there?
\\
\\
Secondly, assume that all internal masses are zero and that the external momenta satisfy
$p_1^2=p_2^2=p_3^2=p_4^2=0$.
How many kinematic variables are there now?
\\
\\
{\bf Solution}: We start with the case where all internal masses are non-zero and pairwise distinct and the external momenta
are as generic as possible.
The external momenta still satisfy momentum conservation. If we take all momenta outgoing, momentum conservation reads
\bq
 p_1 + p_2 + p_3 + p_4 & = & 0.
\eq
Momentum conservation allows us to eliminate $p_4$.
Thus we have
\bq
 \frac{-p_1^2}{\mu^2}, 
 \;\;\;
 \frac{-p_1 \cdot p_2}{\mu^2},
 \;\;\;
 \frac{-p_1 \cdot p_3}{\mu^2},
 \;\;\;
 \frac{-p_2^2}{\mu^2}, 
 \;\;\;
 \frac{-p_2 \cdot p_3}{\mu^2},
 \;\;\;
 \frac{-p_3^2}{\mu^2},
 \;\;\;
 \frac{m_1^2}{\mu^2},
 \;\;\;
 \frac{m_2^2}{\mu^2},
 \;\;\;
 \frac{m_3^2}{\mu^2},
 \;\;\;
 \frac{m_4^2}{\mu^2}
\eq
as kinematic variables, where $m_1,\dots,m_4$ denote the internal masses.
Due to the scaling relation in eq.~(\ref{chapter_basics:kinematic_variables_scaling_relation}) we may set one
kinematic variable to one (say the last one $m_4^2/\mu^2$). This gives us $\NB=9$ for the most general one-loop box graph.
Quite often one uses instead of the scalar products $p_1 \cdot p_2$, $p_2 \cdot p_3$, $p_1 \cdot p_3$ the
\index{Mandelstam variables}
{\bf Mandelstam variables}
\bq
 s \; = \; \left(p_1+p_2\right)^2,
 \;\;\;\;\;\;
 t \; = \; \left(p_2+p_3\right)^2,
 \;\;\;\;\;\;
 u \; = \; \left(p_1+p_3\right)^2.
\eq
These satisfy the relation (inherited from momentum conservation)
\bq
 s + t + u & = & p_1^2 + p_2^2 + p_3^2 + p_4^2.
\eq
Thus we may either eliminate $p_4^2$ (as we did above) or $u$ (another popular choice).
\\
\\
Let us now discuss the second part, where we assume that all internal masses are zero and that the external momenta satisfy
$p_1^2=p_2^2=p_3^2=p_4^2=0$.
The relations $p_1^2=p_2^2=p_3^2=m_1^2=m_2^2=m_3^2=m_4^2=0$ leave only 
\bq
 \frac{-p_1 \cdot p_2}{\mu^2},
 \;\;\;
 \frac{-p_1 \cdot p_3}{\mu^2},
 \;\;\;
 \frac{-p_2 \cdot p_3}{\mu^2}
\eq
from the list above. However, we haven't used $p_4^2=0$ yet, which allows to eliminate another kinematic variable.
In the massless case we have $s=2p_1\cdot p_2$, $t=2p_2\cdot p_3$ and $u=2p_1\cdot p_3$
and the Mandelstam relation simplifies to
\bq
 s + t + u & = & 0.
\eq
Thus we may trade $p_4^2=0$ to eliminate $u$ (e.g. $p_1\cdot p_3/\mu^2$).
This brings us down to two kinematic variables, which can be taken as
$-2p_1\cdot p_2/\mu^2$ and $-2p_2\cdot p_3/\mu^2$.
As above we may set one kinematic variable to one, giving us $\NB=1$ in the massless case.
It is quite common to use as kinematic variable
\bq
 x & = & \frac{s}{t}
\eq
in this case.
}
\es
\\
\\
\bs
{\it \refstepcounter{exercise}
{\bf Exercise \theexercise}: 
Determine with the method above the graph polynomials ${\mathcal U}$ and ${\mathcal F}$ for
the graph shown in fig.~\ref{chapter_basics:fig_nonplanar_vertex} for the case where
all internal masses are zero.
\\
\\
{\bf Solution}: 
Let us first note that the graph shown in fig.~\ref{chapter_basics:fig_nonplanar_vertex} may equally well be drawn as in fig.~\ref{chapter_solutions:fig_nonplanar_vertex_v2}.
\begin{figure}
\begin{center}
\includegraphics[scale=1.0]{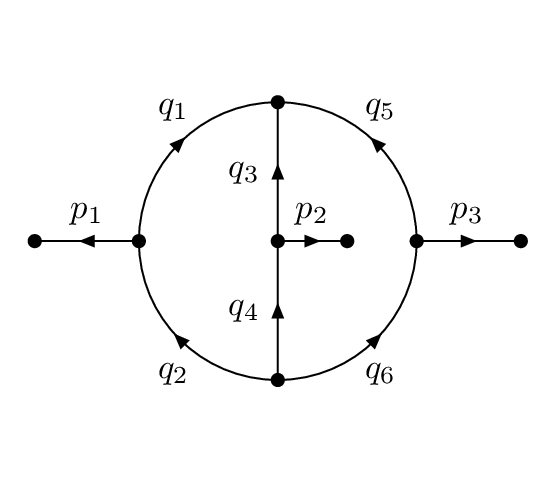}
\end{center}
\caption{
The two-loop non-planar vertex graph of fig.~\ref{chapter_basics:fig_nonplanar_vertex} drawn in an alternative way.
}
\label{chapter_solutions:fig_nonplanar_vertex_v2}
\end{figure}
As independent loop momenta we take
\bq
 k_1 \; = \; q_1, & & k_2 \; = \; q_5.
\eq
Then
\bq
 q_2 \; = \; k_1 + p_1,
 \;\;\;\;\;\;
 q_3 \; = \; - k_1 - k_2,
 \;\;\;\;\;\;
 q_4 \; = \; - k_1 - k_2 + p_2,
 \;\;\;\;\;\;
 q_6 \; = \; k_2 + p_3,
\eq
and $p_3=-p_1-p_2$.
We work out
\bq
 \sum\limits_{j=1}^6 \alpha_j \left(-q_j^2\right) & = &
 - \left(\alpha_1+\alpha_2+\alpha_3+\alpha_4\right) k_1^2 
 - 2 \left( \alpha_3 + \alpha_4 \right) k_1 \cdot k_2 - \left( \alpha_3+\alpha_4+\alpha_5+\alpha_6\right) k_2^2
 \nonumber \\
 & &
 + 2 \left( \alpha_4 p_2 - \alpha_2 p_1 \right) \cdot k_1
 + 2 \left( \alpha_4 p_2 - \alpha_6 p_3 \right) \cdot k_2
 - \alpha_2 p_1^2 - \alpha_4 p_2^2 - \alpha_6 p_3^2.
\eq
In comparing with eq.~(\ref{chapter_basics:eq_poly_calc_1})
we find
\bq
 M & = & \left( \begin{array}{cc}
 \alpha_1+\alpha_2+\alpha_3+\alpha_4 & \alpha_3+\alpha_4 \\
 \alpha_3+\alpha_4 & \alpha_3+\alpha_4+\alpha_5+\alpha_6 \\
 \end{array} \right),
 \nonumber \\
 v & = & \left( \begin{array}{c}
          \alpha_4 p_2 - \alpha_2 p_1 \\
          \alpha_4 p_2 - \alpha_6 p_3 \\
          \end{array} \right),
 \nonumber \\
 J & = & \alpha_2 \left(-p_1\right)^2 + \alpha_4 \left(-p_2\right)^2 + \alpha_6 \left(-p_3\right)^2.
\eq
From momentum conservation we have $(p_1+p_2)^2=p_3^2$ and hence
\bq
 2 p_1 \cdot p_2 & = & p_3^2 - p_1^2 - p_2^2,
 \nonumber \\
 2 p_2 \cdot p_3 & = & p_1^2 - p_2^2 - p_3^2,
 \nonumber \\
 2 p_3 \cdot p_1 & = & p_2^2 - p_3^2 - p_1^2.
\eq
Using this and eq.~(\ref{chapter_basics:eq_poly_calc_2}) we finally obtain
\bq
\label{appendix_solutions:graph_polynomials_crossed_vertex}
 {\mathcal U} & = & 
 \left( \alpha_1+\alpha_2\right) \left( \alpha_3+\alpha_4\right) 
 + \left( \alpha_1+\alpha_2\right) \left( \alpha_5+\alpha_6\right) 
 + \left( \alpha_3+\alpha_4\right) \left( \alpha_5+\alpha_6\right),
 \nonumber \\
{\mathcal F} & = & 
 \left[ \alpha_1 \alpha_4 \alpha_6 + \alpha_2 \alpha_3 \alpha_5 + \alpha_1 \alpha_2 \left( \alpha_3 + \alpha_4 + \alpha_5 \alpha_6 \right) \right] 
 \left( \frac{-p_1^2}{\mu^2} \right)
 \nonumber \\
 & &
 + \left[ \alpha_3 \alpha_2 \alpha_6 + \alpha_4 \alpha_1 \alpha_5 + \alpha_3 \alpha_4 \left( \alpha_1 + \alpha_2 + \alpha_5 \alpha_6 \right) \right] 
 \left( \frac{-p_2^2}{\mu^2} \right)
 \nonumber \\
 & &
 + \left[ \alpha_5 \alpha_2 \alpha_4 + \alpha_6 \alpha_1 \alpha_3 + \alpha_5 \alpha_6 \left( \alpha_1 + \alpha_2 + \alpha_3 \alpha_4 \right) \right] 
 \left( \frac{-p_3^2}{\mu^2} \right).
\eq
}
\es
\\
\\
\bs
{\it \refstepcounter{exercise}
{\bf Exercise \theexercise}: 
Prove eq.~(\ref{chapter_basics:Feynman_trick}).
\\
\\
{\bf Solution}: 
The attentive reader might have noticed that we (implicitly) already gave a proof of eq.~(\ref{chapter_basics:Feynman_trick})
when we we derived the Feynman parameter representation from the Schwinger parameter representation.
For clarity, let's distil the proof here:
For $A_j>0$ and $\mathrm{Re}(\nu_j)>0$ we have
\bq
 \frac{1}{A_j^{\nu_j}} & = &
 \frac{1}{\Gamma\left(\nu_j\right)} \int\limits_0^\infty d\alpha_j \; \alpha_j^{\nu_j-1} \; e^{-\alpha A_j},
\eq
and therefore
\bq
 \prod\limits_{j=1}^{n} \frac{1}{A_{j}^{\nu_{j}}} 
 & = &
 \frac{1}{\prod\limits_{j=1}^{n} \Gamma(\nu_{j})} \;\;
 \int\limits_{\alpha_j \ge 0}  d^n\alpha \;
 \left( \prod\limits_{j=1}^n \alpha_j^{\nu_j-1} \right)
 \exp\left(-\sum\limits_{j=1}^n \alpha_j A_j \right).
\eq
Clearly, $\sum\limits_{j=1}^n \alpha_j \ge 0$. We insert
\bq
 1 & = &
 \int\limits_{0}^\infty dt \; \delta\left(t - \sum\limits_{j=1}^n \alpha_j \right)
\eq
and change variables according to $a_j = \alpha_j/t$.
This gives
\bq
 \prod\limits_{j=1}^{n} \frac{1}{A_{j}^{\nu_{j}}} 
 & = &
 \frac{1}{\prod\limits_{j=1}^{n} \Gamma(\nu_{j})} \;\;
 \int\limits_{a_j \ge 0}  d^na \;
 \delta\left(1 - \sum\limits_{j=1}^n a_j \right)
 \left( \prod\limits_{j=1}^n a_j^{\nu_j-1} \right)
 \int\limits_{0}^\infty dt \; t^{\nu-1}
 e^{-t \sum\limits_{j=1}^n a_j A_j}.
\eq
A further change of variable $t \rightarrow t/(\sum\limits_{j=1}^n a_j A_j)$ yields
\bq
 \prod\limits_{j=1}^{n} \frac{1}{A_{j}^{\nu_{j}}} 
 & = &
 \frac{1}{\prod\limits_{j=1}^{n} \Gamma(\nu_{j})} \;\;
 \int\limits_{a_j \ge 0}  d^na \;
 \delta\left(1 - \sum\limits_{j=1}^n a_j \right)
 \left( \prod\limits_{j=1}^n a_j^{\nu_j-1} \right)
 \frac{1}{\left(\sum\limits_{j=1}^n a_j A_j\right)^{\nu}}
 \int\limits_{0}^\infty dt \; t^{\nu-1}
 e^{-t}.
\eq
With
\bq
 \int\limits_{0}^\infty dt \; t^{\nu-1}
 e^{-t}
 & = &
 \Gamma\left(\nu\right)
\eq
we arrive at
\bq
 \prod\limits_{j=1}^{n} \frac{1}{A_{j}^{\nu_{j}}} 
 & = &
 \frac{\Gamma\left(\nu\right)}{\prod\limits_{j=1}^{n} \Gamma(\nu_{j})} \;\;
 \int\limits_{a_j \ge 0}  d^na \;
 \delta\left(1 - \sum\limits_{j=1}^n a_j \right)
 \left( \prod\limits_{j=1}^n a_j^{\nu_j-1} \right)
 \frac{1}{\left(\sum\limits_{j=1}^n a_j A_j\right)^{\nu}}.
\eq
}
\es
\\
\\
\bs
{\it \refstepcounter{exercise}
{\bf Exercise \theexercise}: 
Calculate with the help of the Feynman parameter representation 
the 
one-loop triangle integral 
\bq
 I_{\nu_1\nu_2\nu_3}
 & = &
 e^{\eps \Eulerconstant} \left(\mu^2\right)^{\nu-\frac{D}{2}}
 \int \frac{d^Dk}{i \pi^{\frac{D}{2}}} 
 \frac{1}{\left(-q_1^2\right)^{\nu_1} \left(-q_2^2\right)^{\nu_2} \left(-q_3^2\right)^{\nu_3}},
\eq
shown in fig.~\ref{chapter_basics:fig_onelooptriangle}
for the case where all internal masses are zero ($m_1=m_2=m_3=0$) and 
for the kinematic configuration $p_1^2=p_2^2=0$, $p_3^2 \neq 0$.
\\
\\
{\bf Solution}: 
The second graph polynomial is given by
\bq 
 {\mathcal F} 
 & = &
 a_1 a_3 \left( \frac{-p_3^2}{\mu^2} \right)
\eq
and the Feynman parameter representation reads
\bq
 I_{\nu_1\nu_2\nu_3}
 & = &
 \frac{e^{\eps \Eulerconstant}\Gamma\left(\nu-\frac{D}{2}\right)}{\Gamma(\nu_1) \Gamma(\nu_2) \Gamma(\nu_3)}
 \int\limits_{a_j \ge 0} d^3a \; \delta\left(1-a_1-a_2-a_3\right) \;  
 \frac{a_1^{\nu_1-1} a_2^{\nu_2-1} a_3^{\nu_3-1}}{\left[ {\mathcal F}\left(a\right) \right]^{\nu-\frac{D}{2}}}
 \\
 & = &
 \frac{e^{\eps \Eulerconstant}\Gamma\left(\nu-\frac{D}{2}\right)}{\Gamma(\nu_1) \Gamma(\nu_2) \Gamma(\nu_3)}
 \left( \frac{-p_3^2}{\mu^2} \right)^{\frac{D}{2}-\nu}
 \int\limits_{a_j \ge 0} d^3a \; \delta\left(1-a_1-a_2-a_3\right) \;  
 a_1^{\frac{D}{2}-\nu_{23}-1} a_2^{\nu_2-1} a_3^{\frac{D}{2}-\nu_{12}-1}.
 \nonumber 
\eq
The Feynman parameter integral is a generalisation of Euler's beta function: For $n \in {\mathbb N}$ we have
\bq
 \int\limits_{a_j \ge 0} d^{n}a \; \delta\left(1-\sum\limits_{j=1}^{n} a_j \right) \; 
 \left( \prod\limits_{j=1}^{n} a_j^{\nu_j-1} \right)
 & = & 
 \frac{\prod\limits_{j=1}^{n}\Gamma(\nu_j)}{\Gamma(\nu_1+...+\nu_{n})}
\eq
and therefore
\bq
 I_{\nu_1\nu_2\nu_3}
 & = &
 \frac{e^{\eps \Eulerconstant}\Gamma\left(\nu-\frac{D}{2}\right)\Gamma\left(\frac{D}{2}-\nu_{12}\right)\Gamma\left(\frac{D}{2}-\nu_{23}\right)}{\Gamma(\nu_1) \Gamma(\nu_3) \Gamma\left(D-\nu\right)}
 \left( \frac{-p_3^2}{\mu^2} \right)^{\frac{D}{2}-\nu}.
\eq
}
\es
\\
\\
\bs
{\it \refstepcounter{exercise}
{\bf Exercise \theexercise}: 
Consider again the one-loop box graph in fig.~\ref{chapter_basics:fig_oneloopbox},
this time for the kinematic configuration
\bq
 p_2^2 \; = \; p_4^2 \; = \; 0,
 & &
 m_1 \; = \; m_2 \; = \; m_3 \; = \; m_4 \; = \; 0.
\eq
Write down the Feynman parameter representation as in eq.~(\ref{chapter_basics:Feynman_parameter_representation}).
Obtain a second integral representation by first combining propagators $1$ and $2$ with a
pair of Feynman parameters, then combining propagators $3$ and $4$ with a second pair of
Feynman parameters and finally the two results with a third pair of Feynman parameters.
\\
\\
{\bf Solution}: 
Let us start with the standard (democratic) Feynman parameter representation.
The graph polynomials are
\bq
 {\mathcal U} & = & a_1 + a_2 + a_3 + a_4,
 \nonumber \\
 {\mathcal F} & = & 
 a_2 a_4 \left( \frac{-s}{\mu^2} \right)
 + a_1 a_3 \left( \frac{-t}{\mu^2} \right)
 + a_1 a_4 \left( \frac{-p_1^2}{\mu^2} \right)
 + a_2 a_3 \left( \frac{-p_3^2}{\mu^2} \right).
\eq
The democratic Feynman parameter representation is then
\bq
 I
 & = &
 \frac{e^{\eps \Eulerconstant}\Gamma\left(\nu-\frac{D}{2}\right)}{\prod\limits_{j=1}^4\Gamma(\nu_j)}
 \int\limits_{a_j \ge 0} d^4a \; \delta\left(1-\sum\limits_{j=1}^4 a_j \right) \; 
 \left( \prod\limits_{j=1}^4 a_j^{\nu_j-1} \right)
 \frac{1}{{\mathcal F}^{\nu-\frac{D}{2}}}.
\eq
Note that we may ignore the ${\mathcal U}$-polynomial due to the delta distribution.

Let us now follow a hierarchical approach: It will be convenient to use the following 
notation: $\bar{a}=1-a$, $\bar{b}=1-b$, $\bar{c}=1-c$ and $\nu_{i_1 \dots i_k}=\nu_{i_1} + \dots + \nu_{i_k}$.
We first combine propagators $1$ and $2$
\bq
 \frac{1}{\left(-q_1^2\right)^{\nu_1} \left(-q_2^2\right)^{\nu_1}}
 & = &
 \frac{\Gamma\left(\nu_{12}\right)}{\Gamma\left(\nu_1\right)\Gamma\left(\nu_2\right)}
 \int\limits_0^1 da \; a^{\nu_1-1} \bar{a}^{\nu_2-1}
 \frac{1}{\left( - a q_1^2 - \bar{a} q_2^2 \right)^{\nu_{12}}},
\eq
then propagators $3$ and $4$
\bq
 \frac{1}{\left(-q_3^2\right)^{\nu_1} \left(-q_4^2\right)^{\nu_1}}
 & = &
 \frac{\Gamma\left(\nu_{34}\right)}{\Gamma\left(\nu_3\right)\Gamma\left(\nu_4\right)}
 \int\limits_0^1 db \; b^{\nu_3-1} \bar{b}^{\nu_4-1}
 \frac{1}{\left( - b q_3^2 - \bar{b} q_4^2 \right)^{\nu_{34}}},
\eq
and finally the two intermediate results:
\bq
 \frac{1}{\left( - a q_1^2 - \bar{a} q_2^2 \right)^{\nu_{12}} \left( - b q_3^2 - \bar{b} q_4^2 \right)^{\nu_{34}}}
 & = &
 \frac{\Gamma\left(\nu_{1234}\right)}{\Gamma\left(\nu_{12}\right)\Gamma\left(\nu_{34}\right)}
 \int\limits_0^1 dc \; 
 \frac{c^{\nu_{12}-1} \bar{c}^{\nu_{34}-1}}{\left( - a c q_1^2 - \bar{a} c q_2^2 - b \bar{c} q_3^2 - \bar{b} \bar{c} q_4^2 \right)^{\nu_{1234}}}.
 \hspace*{12mm}
\eq
Let's work out the denominator:
\bq
\lefteqn{
 - a c q_1^2 - \bar{a} c q_2^2 - b \bar{c} q_3^2 - \bar{b} \bar{c} q_4^2
= } & & 
 \\
 & &
 - \left(k-a c p_1 - \bar{a} c \left(p_1+p_2\right) - b \bar{c} \left(p_1+p_2+p_3\right)\right)^2
 + c \bar{c} \left[ \bar{a} \bar{b} \left(-s\right) + a b \left(-t\right)
                    + a \bar{b} \left(-p_1^2\right) + \bar{a} b \left(-p_3^2\right)
 \right].
 \nonumber
\eq
We see that the $c$-dependence in the second term factors out.
We may now use eq.~(\ref{chapter_basics:master_one_loop}) and obtain
\bq
 I 
 & = &
 \frac{e^{\eps \Eulerconstant}\Gamma\left(\nu-\frac{D}{2}\right)}{\prod\limits_{j=1}^4\Gamma(\nu_j)}
 \int\limits_0^1 da \; a^{\nu_1-1} \bar{a}^{\nu_2-1}
 \int\limits_0^1 db \; b^{\nu_3-1} \bar{b}^{\nu_4-1}
 \frac{1}{\left[\bar{a} \bar{b} \left(-s\right) + a b \left(-t\right)
                    + a \bar{b} \left(-p_1^2\right) + \bar{a} b \left(-p_3^2\right)\right]^{\nu-\frac{D}{2}}}
 \nonumber \\
 & &
 \times
 \int\limits_0^1 dc \; c^{\frac{D}{2}-\nu_{34}-1} \bar{c}^{\frac{D}{2}-\nu_{12}-1}.
\eq
The integral over $c$ is trivial and gives Euler's beta function. We obtain
\bq
\lefteqn{
 I 
 = } & & \\
 & &
 \frac{e^{\eps \Eulerconstant}\Gamma\left(\nu-\frac{D}{2}\right)\Gamma\left(\frac{D}{2}-\nu_{12}\right)\Gamma\left(\frac{D}{2}-\nu_{34}\right)}{\Gamma\left(D-\nu\right)\prod\limits_{j=1}^4\Gamma(\nu_j)}
 \int\limits_0^1 da  
 \int\limits_0^1 db 
 \frac{a^{\nu_1-1} \bar{a}^{\nu_2-1} b^{\nu_3-1} \bar{b}^{\nu_4-1}}{\left[\bar{a} \bar{b} \left(-s\right) + a b \left(-t\right)
                    + a \bar{b} \left(-p_1^2\right) + \bar{a} b \left(-p_3^2\right)\right]^{\nu-\frac{D}{2}}}.
 \nonumber
\eq
This leaves two non-trivial integrations, compared to three non-trivial integrations within the standard Feynman parameter representations.
}
\es
\\
\\
\bs
{\it \refstepcounter{exercise}
{\bf Exercise \theexercise}: 
Prove eq.~(\ref{chapter_basics:dgl_integrand}).
\\
\\
{\bf Solution}: 
We have to prove
\bq
 \sum\limits_{j=1}^{\ninternal} a_j \frac{\partial}{\partial a_j} f\left(a\right) 
 & = & - \ninternal f\left(a\right).
\eq
$f(a)$ is a product of three factors, each factor is a homogeneous polynomial raised
to some power.
The derivatives act by the product rules. It is therefore sufficient to prove
\bq
 \sum\limits_{j=1}^{\ninternal} a_j \frac{\partial}{\partial a_j} \left(p\left(a\right)\right)^\nu 
 & = & \nu h p\left(a\right)
\eq
for a homogeneous polynomial $p(a)$ of degree $h$.
We may simplify the task further by noting that it is sufficient to prove
\bq
 \sum\limits_{j=1}^{\ninternal} a_j \frac{\partial}{\partial a_j} p\left(a\right) 
 & = & h p\left(a\right).
\eq
Let us now write
\bq
 p\left(a\right)
 & = &
 \sum_i c_i \prod\limits_{j=1}^{\ninternal} a_j^{\nu_{ij}}
\eq
We assumed that $p(a)$ is homogeneous of degree $h$, therefore we have for all $i$
\bq
 \sum\limits_{j=1}^{\ninternal} \nu_{ij} & = & h.
\eq
It is easy to show that for each term we have
\bq
 \sum\limits_{j=1}^{\ninternal} a_j \frac{\partial}{\partial a_j} 
 \left(c_i \prod\limits_{j=1}^{\ninternal} a_j^{\nu_{ij}}\right)
 & = &
 \left( \sum\limits_{j=1}^{\ninternal} \nu_{ij} \right)
 \left(c_i \prod\limits_{j=1}^{\ninternal} a_j^{\nu_{ij}}\right)
 \; = \;
 h
 \left(c_i \prod\limits_{j=1}^{\ninternal} a_j^{\nu_{ij}}\right),
\eq
which completes the proof.
}
\es
\\
\\
\bs
{\it \refstepcounter{exercise}
{\bf Exercise \theexercise}: 
Show explicitly that eq.~(\ref{chapter_basics:Feynman_representation_differential_forms})
is equivalent to eq.~(\ref{chapter_basics:Feynman_parameter_representation}).
\\
\\
{\bf Solution}: 
Let us denote by
\bq
 \tilde{\Delta} & = & \left\{ \left(a_1,\dots,a_{{\ninternal}-1}\right) \in \left. {\mathbb R}^{{\ninternal}-1} \right| \sum\limits_{j=1}^{{\ninternal}-1} a_j \le 1, a_j \ge 0 \right\}.
\eq
$\tilde{\Delta}$ is a coordinate patch for the standard simplex.
Let us further agree that in this exercise we always define
\bq
\label{appendix_solutions:def_a_n}
 a_{\ninternal} & = & 1 - \sum\limits_{j=1}^{{\ninternal}-1} a_j.
\eq
The Feynman parameter representation from eq.~(\ref{chapter_basics:Feynman_parameter_representation}) is
then
\bq
 I
 & = &
 \frac{e^{\loopnumber \eps \Eulerconstant}\Gamma\left(\nu-\frac{\loopnumber D}{2}\right)}{\prod\limits_{j=1}^{\ninternal}\Gamma(\nu_j)}
 \int\limits_{\tilde{\Delta}} d^{{\ninternal}-1}a \; 
 \left( \prod\limits_{j=1}^{\ninternal} a_j^{\nu_j-1} \right)
 \frac{\left[ {\mathcal U}\left(a\right) \right]^{\nu-\frac{\left(\loopnumber+1\right) D}{2}}}{\left[ {\mathcal F}\left(a\right) \right]^{\nu-\frac{\loopnumber D}{2}}}.
\eq
In other words, we have used the Dirac delta distribution to integrate out $a_{\ninternal}$.

Let us now consider eq.~(\ref{chapter_basics:Feynman_representation_differential_forms}).
We work out $\omega$ for our coordinate chart:
From eq.~(\ref{appendix_solutions:def_a_n}) we have
\bq
 d a_{\ninternal} & = & - \sum\limits_{j=1}^{{\ninternal}-1} d a_j.
\eq
and hence
\bq
 \omega & = &
 \sum\limits_{j=1}^{\ninternal} (-1)^{{\ninternal}-j}
  \; a_j \; da_1 \wedge ... \wedge \widehat{da_j} \wedge ... \wedge da_{\ninternal}
 \nonumber \\
 & = &
 \left(-1\right)^{\ninternal-1} a_1 da_2 \wedge \dots \wedge da_{\ninternal -1} \wedge \left(-da_1\right)
 \; + \; \left(-1\right)^{\ninternal-2} a_2 da_1 \wedge da_3 \wedge \dots \wedge da_{\ninternal -1} \wedge \left(-da_2\right)
 \; + \; \dots 
 \nonumber \\
 & &
 \; + \; \left(-1\right)^{1} a_{\ninternal-1} da_1 \wedge \dots \wedge da_{\ninternal -2} \wedge \left(-da_{\ninternal-1}\right)
 \; + \; \left(1-\sum\limits_{j=1}^{{\ninternal}-1} a_j\right) da_1 \wedge \dots \wedge da_{\ninternal -2} \wedge da_{\ninternal-1}
 \nonumber \\
 & = &
 da_1 \wedge \dots \wedge da_{\ninternal-1}
 \; = \;
 d^{{\ninternal}-1}a.
\eq
}
\es
\\
\\
\bs
{\it \refstepcounter{exercise}
{\bf Exercise \theexercise}: 
Prove eq.~(\ref{chapter_basics:interior_product}).
\\
\\
{\bf Solution}: 
Let us write with Einstein's summation convention
\bq
 \omega
 \; = \;
 \frac{1}{\left(\ninternal-1\right)!} \omega_{i_2 \dots i_{\ninternal}} da_{i_2} \wedge \dots \wedge da_{i_{\ninternal}},
 & \mbox{with} &
 \omega_{i_2 \dots i_{\ninternal}}
 \; = \; 
 \left(-1\right)^{\ninternal -1} \eps_{i_1 i_2 \dots i_{\ninternal}} a_{i_1},
\eq
where $\eps_{i_1 i_2 \dots i_{\ninternal}}$ denotes the totally antisymmetric tensor with $\eps_{1 2 \dots \ninternal}=1$.
For
\bq
 X & = & \lambda_i e_i
\eq
the interior product is given by
\bq
 \iota_X \omega
 & = &
 \frac{1}{\left(\ninternal-2\right)!}
 \lambda_{i_2}
 \omega_{i_2 i_3 \dots i_{\ninternal}} da_{i_3} \wedge \dots \wedge da_{i_{\ninternal}}.
\eq
Integrating along the radial direction we have $a_i=\lambda_i t$, where $t$ is the curve parameter.
We then have
\bq
 \lambda_{i_2} \omega_{i_2 i_3 \dots i_{\ninternal}}
 & = &
 \left(-1\right)^{\ninternal -1} a_{i_1} \lambda_{i_2} \eps_{i_1 i_2 i_3 \dots i_{\ninternal}} 
 \; = \;
 \left(-1\right)^{\ninternal -1} \lambda_{i_1} \lambda_{i_2} \eps_{i_1 i_2 i_3 \dots i_{\ninternal}} t.
\eq
This vanishes, as $\lambda_{i_1} \lambda_{i_2}$ is symmetric under the exchange of $i_1$ and $i_2$,
while $\eps_{i_1 i_2 i_3 \dots i_{\ninternal}}$ is antisymmetric.
}
\es
\\
\\\bs
{\it \refstepcounter{exercise}
{\bf Exercise \theexercise}: 
Prove eq.~(\ref{chapter_basics:f_omega_closed}).
\\
\\
{\bf Solution}: 
From Leibniz's rule we have
\bq
 d \left( f \omega \right) & = & \left(df\right) \wedge \omega + f \left(d\omega\right).
\eq
We work out $d\omega$ first:
\bq
 d\omega
 & = &
 \sum\limits_{j=1}^{\ninternal} (-1)^{{\ninternal}-j}
  \; da_j \wedge da_1 \wedge ... \wedge \widehat{da_j} \wedge ... \wedge da_{\ninternal}
 \nonumber \\
 & = &
 \ninternal (-1)^{{\ninternal}-1} da_1 \wedge \dots da_{\ninternal}.
\eq
$df$ is given by
\bq
 df & = &
 \sum\limits_{j=1}^{\ninternal}
 \frac{\partial f}{\partial a_j} da_j.
\eq
In the wedge product with $\omega$ only a single sum survives:
\bq
 \left(df\right) \wedge \omega
 & = &
 \left( \sum\limits_{j_1=1}^{\ninternal} \frac{\partial f}{\partial a_{j_1}} da_{j_1} \right)
 \wedge
 \left( \sum\limits_{j_2=1}^{\ninternal} (-1)^{{\ninternal}-j_2} \; a_{j_2} \wedge da_1 \wedge ... \wedge \widehat{da_{j_2}} \wedge ... \wedge da_{\ninternal} \right)
 \nonumber \\
 & = &
 \sum\limits_{j=1}^{\ninternal} (-1)^{{\ninternal}-j} a_{j} \frac{\partial f}{\partial a_{j}} da_{j} 
 \wedge da_1 \wedge ... \wedge \widehat{da_{j}} \wedge ... \wedge da_{\ninternal}
 \nonumber \\
 & = &
 (-1)^{{\ninternal}-1} \left( \sum\limits_{j=1}^{\ninternal} a_{j} \frac{\partial f}{\partial a_{j}} \right)
 da_1 \wedge ... \wedge da_{\ninternal}.
\eq
With eq.~(\ref{chapter_basics:dgl_integrand}) we have
\bq
 \left(df\right) \wedge \omega
 & = &
 - \ninternal (-1)^{{\ninternal}-1} f
 da_1 \wedge ... \wedge da_{\ninternal}.
\eq
}
\es
\\
\\
\bs
{\it \refstepcounter{exercise}
{\bf Exercise \theexercise}: 
An alternative proof of the Cheng-Wu theorem: Prove the Cheng-Wu theorem directly from the
Schwinger parameter representation by inserting
\bq
 1 & = & 
 \int\limits_{-\infty}^\infty dt \; \delta\left(t - \sum\limits_{j \in S} \alpha_j \right)
 \;\; = \;\;
 \int\limits_{0}^\infty dt \; \delta\left(t - \sum\limits_{j \in S} \alpha_j \right),
\eq
where in the last step we used again the fact that the sum of the Schwinger parameters is non-negative.
\\
\\
{\bf Solution}: 
We start from the Schwinger parameter representation
\bq
I  & = &
 \frac{e^{\loopnumber \eps \Eulerconstant}}{\prod\limits_{j=1}^{\ninternal}\Gamma(\nu_j)}
 \;
 \int\limits_{\alpha_j \ge 0}  d^{\ninternal}\alpha \;
 \left( \prod\limits_{j=1}^{\ninternal} \alpha_j^{\nu_j-1} \right)
 \left[ {\mathcal U}\left(\alpha\right) \right]^{-\frac{D}{2}}
 \exp\left( - \frac{{\mathcal F}\left(\alpha\right)}{{\mathcal U}\left(\alpha\right)}\right)
\eq
and insert
\bq
 1 & = & 
 \int\limits_{0}^\infty dt \; \delta\left(t - \sum\limits_{j \in S} \alpha_j \right).
\eq
We then change variables as
$a_j = \alpha_j/t$ (for all $j \in \{1,\dots,\ninternal\}$)
and obtain:
\bq
I  \; = \;
 \frac{e^{\loopnumber \eps \Eulerconstant}}{\prod\limits_{j=1}^{\ninternal}\Gamma(\nu_j)}
 \;
 \int\limits_{a_j \ge 0}  d^{\ninternal}a \;
 \delta\left(1 - \sum\limits_{j \in S} a_j \right)
 \left( \prod\limits_{j=1}^{\ninternal} a_j^{\nu_j-1} \right)
 \left[ {\mathcal U}\left(a\right) \right]^{-\frac{D}{2}}
 \int\limits_{0}^\infty dt \; t^{\nu-\frac{\loopnumber D}{2}-1} \exp\left( - t \frac{{\mathcal F}\left(a\right)}{{\mathcal U}\left(a\right)}\right).
 \hspace*{5mm}
\eq
The remaining steps are as in the derivation of the Feynman parameter representation from the Schwinger
parameter representation and yield the result
\bq
I  & = &
 \frac{e^{\loopnumber \eps \Eulerconstant}\Gamma\left(\nu-\frac{\loopnumber D}{2}\right)}{\prod\limits_{j=1}^{\ninternal}\Gamma(\nu_j)}
 \int\limits_{a_j \ge 0} d^{\ninternal}a \; \delta\left(1-\sum\limits_{j \in S} a_j \right) \; 
 \left( \prod\limits_{j=1}^{\ninternal} a_j^{\nu_j-1} \right)
 \frac{\left[ {\mathcal U}\left(a\right) \right]^{\nu-\frac{\left(\loopnumber+1\right) D}{2}}}{\left[ {\mathcal F}\left(a\right) \right]^{\nu-\frac{\loopnumber D}{2}}}.
\eq
}
\es
\\
\\
\bs
{\it \refstepcounter{exercise}
{\bf Exercise \theexercise}: 
Prove eq.~(\ref{chapter_basics:k_orthogonal_squared}).
\\
\\
{\bf Solution}: 
We have to show
\bq
 \det G^{\mathrm{eucl}}\left(K,P_1,...,P_{\nexternalindependent}\right)
 & = &
 K_\bot^2 \det G^{\mathrm{eucl}}\left(P_1,...,P_{\nexternalindependent}\right).
\eq
We start with the following lemma:
\bq
 \det G^{\mathrm{eucl}}\left(K+P_j,P_1,...,P_{\nexternalindependent}\right)
 & = &
 \det G^{\mathrm{eucl}}\left(K,P_1,...,P_{\nexternalindependent}\right),
 \;\;\;\;\;\;
 1 \; \le \; j \; \le \; \nexternalindependent.
\eq
The $(\nexternalindependent+1)$ vectors $K, P_1,\dots, P_{\nexternalindependent}$ 
span at most a vector space
of dimension $(\nexternalindependent+1)$.
Let $V$ be a vector space of dimension $(\nexternalindependent+1)$
containing these vectors and define (with respect to a basis of $V$)
\bq
 J\left(K,P_1,...,P_{\nexternalindependent}\right)
 & = &
 \left(
 \begin{array}{cccc}
  K^0 & K^1 & \dots & K^{\nexternalindependent} \\
  P_1^0 & P_1^1 & \dots & P_1^{\nexternalindependent} \\
  & & \dots & \nonumber \\
  P_{\nexternalindependent}^0 & P_{\nexternalindependent}^1 & \dots & P_{\nexternalindependent}^{\nexternalindependent} \\
 \end{array}
 \right),
\eq
where we labelled the coordinates with respect to the basis of $V$ from $0$ to $\nexternalindependent$.
It is clear that
\bq
 \det J\left(K+P_j,P_1,...,P_{\nexternalindependent}\right)
 & = &
 \det J\left(K,P_1,...,P_{\nexternalindependent}\right),
\eq
since we may always add a linear dependent row inside a determinant.
We have
\bq
 \det G^{\mathrm{eucl}}\left(K,P_1,...,P_{\nexternalindependent}\right)
 & = &
 \det\left( J\left(K,P_1,...,P_{\nexternalindependent}\right)
 \cdot
 J\left(K,P_1,...,P_{\nexternalindependent}\right)^T \right)
\eq
and the lemma 
$\det G^{\mathrm{eucl}}(K+P_j,P_1,...,P_{\nexternalindependent}) = \det G^{\mathrm{eucl}}(K,P_1,...,P_{\nexternalindependent})$ follows.

Thus we have to show
\bq
 \det G^{\mathrm{eucl}}\left(K_\bot,P_1,...,P_{\nexternalindependent}\right)
 & = &
 K_\bot^2 \det G^{\mathrm{eucl}}\left(P_1,...,P_{\nexternalindependent}\right).
\eq
We are free to choose an appropriate basis of $V$.
We choose a basis $V_0,V_1, \dots V_{\nexternalindependent}$ such that
\bq
 \left\langle K_\bot \right\rangle
 \; = \;
 \left\langle V_0 \right\rangle,
 & &
 \left\langle P_1, \dots P_{\nexternalindependent} \right\rangle
 \; = \;
 \left\langle V_1, \dots, V_{\nexternalindependent} \right\rangle.
\eq
Then
\bq
 J\left(K_\bot,P_1,...,P_{\nexternalindependent}\right)
 & = &
 \left(
 \begin{array}{cccc}
  K_\bot^0 & 0 & \dots & 0 \\
  0 & P_1^1 & \dots & P_1^{\nexternalindependent} \\
  & & \dots & \nonumber \\
  0 & P_{\nexternalindependent}^1 & \dots & P_{\nexternalindependent}^{\nexternalindependent} \\
 \end{array}
 \right)
\eq
and the claim follows.
}
\es
\\
\\
\bs
{\it \refstepcounter{exercise}
{\bf Exercise \theexercise}: 
Perform the integration in eq.~(\ref{chapter_basics:tadpole_Baikov}).
\\
\\
{\bf Solution}: 
We have to compute
\bq
 T_{\nu}\left( D, x \right)
 & = &
 \frac{e^{\Eulerconstant \eps} \left(\mu^2\right)^{\nu-\frac{D}{2}}}{\Gamma\left(\frac{D}{2}\right)}
 \int\limits_{m^2}^\infty dz_1
 \left[ z_1-m^2 \right]^{\frac{D}{2}-1}
 \frac{1}{z_1^\nu}.
\eq
We first set $s=(z_1-m^2)/\mu^2$:
\bq
 T_{\nu}\left( D, x \right)
 & = &
 \frac{e^{\Eulerconstant \eps}}{\Gamma\left(\frac{D}{2}\right)}
 \int\limits_{0}^\infty ds \;
 s^{\frac{D}{2}-1}
 \frac{1}{\left(s + x \right)^\nu}.
\eq
A further substitution $s=t x$ gives
\bq
 T_{\nu}\left( D, x \right)
 & = &
 \frac{e^{\Eulerconstant \eps}}{\Gamma\left(\frac{D}{2}\right)}
 x^{\frac{D}{2}-\nu}
 \int\limits_{0}^\infty dt \;
 t^{\frac{D}{2}-1}
 \frac{1}{\left(t + 1 \right)^\nu}.
\eq
The integral gives Euler's beta function:
\bq
 \int\limits_{0}^\infty dt \;
 t^{\frac{D}{2}-1}
 \frac{1}{\left(t + 1 \right)^\nu}
 & = &
 \frac{\Gamma\left(\frac{D}{2}\right)\Gamma\left(\nu-\frac{D}{2}\right)}{\Gamma\left(\nu\right)}
\eq
and we obtain
\bq
 T_{\nu}\left( D, x \right)
 & = &
 e^{\Eulerconstant \eps}
 \frac{\Gamma\left(\nu-\frac{D}{2}\right)}{\Gamma\left(\nu\right)}
 x^{\frac{D}{2}-\nu}
\eq
in agreement with eq.~(\ref{chapter_basics:result_tadpole}).
}
\es
\\
\\
\bs
{\it \refstepcounter{exercise}
{\bf Exercise \theexercise}: 
Derive the Baikov representation of the graph shown in fig.~\ref{chapter_basics:fig_sunrise}
within the democratic approach and within the loop-by-loop approach. 
Assume that all internal masses are non-zero and equal.
\\
\\
{\bf Solution}: 
We start with the democratic approach.
We have
\bq
 \nexternalindependent & = &
 \dim \left\langle p, -p \right\rangle
 \; = \; 1,
\eq
and thus
\bq
 \NV & = &
 \frac{1}{2} \loopnumber \left(\loopnumber+1\right) + \nexternalindependent \loopnumber
 \; = \;
 5.
\eq
For the democratic approach we need a graph $\tilde{G}$ with $\ninternal=5$ internal propagators, which allows us to express 
any scalar products involving the loop momenta 
\begin{figure}
\begin{center}
\includegraphics[scale=1.0]{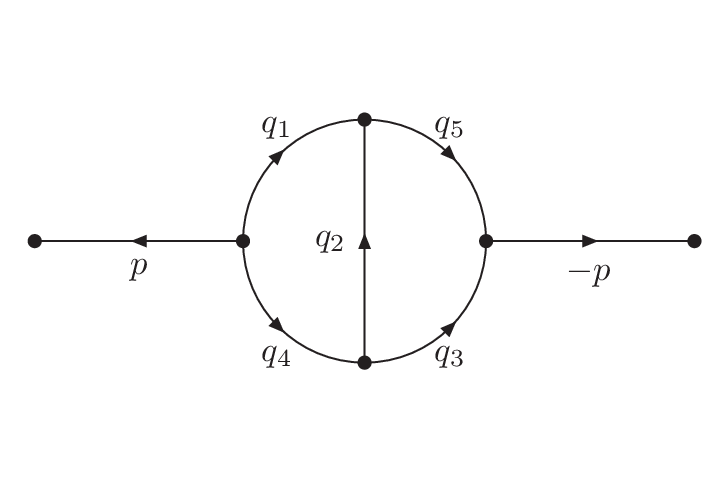}
\end{center}
\caption{
The two-loop kite diagram.
}
\label{chapter_basics:fig_kite}
\end{figure}
in terms of inverse propagators and terms independent of the loop momenta. 
The graph $\tilde{G}$ shown in fig.~\ref{chapter_basics:fig_kite} has this property.
This graph is known as the kite graph.
We denote
\bq
 q_1 \; = \; k_1,
 \;\;\;
 q_2 \; = \; k_2 - k_1,
 \;\;\;
 q_3 \; = \; -k_2 - p,
 \;\;\;
 q_4 \; = \; -k_1 - p,
 \;\;\;
 q_5 \; = \; k_2.
\eq
Propagators four and five are auxiliary propagators. We may associate any mass to them.
The simplest choice is that propagators four and five are massless propagators.
The momentum representation of the Feynman integral for the kite graph is
\bq
 I_{\nu_1\nu_2\nu_3\nu_4\nu_5}
 & = &
 e^{2 \eps \Eulerconstant} \left(\mu^2\right)^{\nu-D}
 \int \frac{d^Dk_1}{i \pi^{\frac{D}{2}}} \frac{d^Dk_2}{i \pi^{\frac{D}{2}}} 
 \frac{1}{\left(-q_1^2+m^2\right)^{\nu_j} \left(-q_2^2+m^2\right)^{\nu_j} \left(-q_3^2+m^2\right)^{\nu_j} \left(-q_4^2\right)^{\nu_j} \left(-q_5^2\right)^{\nu_j}}.
 \nonumber \\
\eq
The original sunrise integral $S_{\nu_1\nu_2\nu_3}$ is simply
\bq
 S_{\nu_1\nu_2\nu_3}
 & = &
 I_{\nu_1\nu_2\nu_3 0 0}.
\eq
The detour through the kite integral is necessary to satisfy the condition that we 
may express any internal inverse propagator as a linear combination
of the linear independent scalar products involving the loop momenta and terms independent of the loop momenta.
The democratic Baikov representation of the kite integral reads
\bq
 I_{\nu_1\nu_2\nu_3\nu_4\nu_5}
 & = &
 \frac{e^{2 \eps \Eulerconstant} \left(\mu^2\right)^{\nu-D} \left(-p^2\right)^{1-\frac{D}{2}}}{8 \pi^{\frac{3}{2}} \Gamma\left(\frac{D-1}{2}\right) \Gamma\left(\frac{D-2}{2}\right)}
 \int\limits_{\mathcal C} d^5z \;
 \left[{\mathcal B}\left(z\right)\right]^{\frac{D}{2}-2}
 \prod\limits_{s=1}^{5} z_s^{-\nu_s},
\eq
Here we used
\bq
 \det C \; = \; 8,
 & &
 \det G\left(p\right) \; = \; - p^2.
\eq
The Baikov polynomial reads
\bq
 {\mathcal B}\left(z\right)
 & = &
 \det G\left(k_1,k_2,p\right) 
 \nonumber \\
 & = &
 \frac{1}{4}
 \left\{
        \left( z_1 z_3 - z_4 z_5 \right) \left( z_4 + z_5 - z_1 - z_3 \right)
        + z_2 \left( z_1 - z_4 \right) \left( z_3 - z_5 \right)
 \right. \nonumber \\
 & & \left.
+ \left[
        z_1 \left( z_1 - z_2 - z_4 \right)
        + z_3 \left( z_3 - z_2 - z_5 \right)
        +3\,z_1\,z_3
        +z_2 \left( z_4+z_5\right)
        -3\,z_5\,z_4
  \right] m^2
 \right. \nonumber \\
 & & \left.
+ \left[
        z_2 \left( z_2 - z_1 - z_3 - z_4 - z_5 \right)
        -\left(z_1-z_5\right) \left( z_3 - z_4 \right)
  \right] p^2
-z_2\,\left(p^2\right)^{2}
+ 2 \left( z_1+z_3 \right) m^2 p^2
 \right. \nonumber \\
 & & \left.
+ \left( z_2-2\,z_1-2\,z_3 \right) \left(m^2\right)^2
+ m^2 \left(m^2-p^2\right)^2
 \right\}.
\eq
Let us now consider the loop-by-loop approach. 
We start with the loop formed by the internal edges $e_1$ and $e_2$ in the sunrise graph.
The external momenta with respect to this loop are $k_2$ and $-k_2$.
Thus we need for the first loop only the auxiliary edge $e_5$, but not $e_4$.
Re-writing the measure gives us
\bq
 \frac{d^Dk_1}{i \pi^{\frac{D}{2}}} 
 & = &
 \frac{1}{2 \sqrt{\pi} \Gamma\left(\frac{D-1}{2}\right)}
 \left[\det G\left(k_2\right) \right]^{1-\frac{D}{2}}
 \left[\det G\left(k_1,k_2\right)\right]^{\frac{D-3}{2}}
 dz_1 dz_2.
\eq
Having done the first loop, we turn to the second loop. We still have the edge $e_3$.
In addition, we introduced the auxiliary edge $e_5$ in the previous step. Thus we deal again
with a one-loop two-point function.
The loop momentum is $k_2$, the external momenta are $p$ and $-p$.
Re-writing the second measure in terms of Baikov variables gives
\bq
 \frac{d^Dk_2}{i \pi^{\frac{D}{2}}} 
 & = &
 \frac{1}{2 \sqrt{\pi} \Gamma\left(\frac{D-1}{2}\right)}
 \left[\det G\left(p\right) \right]^{1-\frac{D}{2}}
 \left[\det G\left(k_2,p\right)\right]^{\frac{D-3}{2}}
 dz_3 dz_5.
\eq
Putting everything together we arrive at the Baikov representation within the loop-by-loop approach:
\bq
\lefteqn{
 S_{\nu_1\nu_2\nu_3}
 = } & & 
 \\
 &&
 \frac{e^{2 \eps \Eulerconstant} \left(\mu^2\right)^{\nu-D} \left(-p^2\right)^{1-\frac{D}{2}}}{4 \pi \left[\Gamma\left(\frac{D-1}{2}\right)\right]^2}
 \int\limits_{\mathcal C} dz_1 dz_2 dz_3 dz_5 \;
 \left[\det G\left(k_2,p\right)\right]^{\frac{D-3}{2}}
 \left[\det G\left(k_1,k_2\right)\right]^{\frac{D-3}{2}}
 z_5^{1-\frac{D}{2}}
 \prod\limits_{s=1}^{3} z_s^{-\nu_s}.
 \nonumber
\eq
Here, we already used
\bq
 \det G\left(p\right) \; = \; -p^2,
 & &
 \det G\left(k_2\right) \; = \; -k_2^2 \; = \; z_5.
\eq
The remaining Gram determinants, expressed in terms of the Baikov variables, read
\bq
 \det G\left(k_2,p\right) 
 & = &
 \frac{1}{4} \left[ - \left(m^2-p^2-z_3\right)^2 + 2 \left( z_3 - p^2 - m^2 \right) z_5 - z_5^2 \right],
 \nonumber \\
 \det G\left(k_1,k_2\right)
 & = &
 \frac{1}{4} \left[ - \left(z_1-z_2\right)^2 + 2 \left(z_1+z_2-2m^2\right) z_5 - z_5^2 \right].
\eq
It is worth noting that within the loop-by-loop approach we only have four integration variables
($z_1$, $z_2$, $z_3$, $z_5$), compared to five integration variables in the democratic approach.
}
\es
\\
\\
%
%
\bs
{\it \refstepcounter{exercise}
{\bf Exercise \theexercise}: 
Re-compute the first graph polynomial ${\mathcal U}$
for the graph shown in fig.~\ref{chapter_basics:fig_nonplanar_vertex} 
from the set of spanning trees.
\\
\\
{\bf Solution}: 
The graph shown in fig.~\ref{chapter_basics:fig_nonplanar_vertex} can alternatively be drawn as
shown in fig.~\ref{chapter_solutions:fig_nonplanar_vertex_v3}.
We have to find all spanning trees for this graph.
The graph has three chains, which we may take as
\bq
 C_1 \; = \; \left\{ e_1, e_2 \right\},
 \;\;\;\;\;\;
 C_2 \; = \; \left\{ e_3, e_4 \right\},
 \;\;\;\;\;\;
 C_3 \; = \; \left\{ e_5, e_6 \right\}.
\eq
In order to obtain a spanning tree, we have to delete from two chains one edge each.
There are three possibilities to pick two chains out of three.
For any choice of these two chains there are four possibilities to delete one edge
from each chain.
Thus there are in total $3 \cdot 4 = 12$ possibilities.
This is the number of spanning trees.
The first graph polynomial ${\mathcal U}$ contains for each spanning tree
a monomial corresponding to the edges, which have been deleted to obtain the spanning
tree.
Thus
\bq
 {\mathcal U} & = & 
 \left( \alpha_1+\alpha_2\right) \left( \alpha_3+\alpha_4\right) 
 + \left( \alpha_1+\alpha_2\right) \left( \alpha_5+\alpha_6\right) 
 + \left( \alpha_3+\alpha_4\right) \left( \alpha_5+\alpha_6\right),
\eq
in agreement with eq.~(\ref{appendix_solutions:graph_polynomials_crossed_vertex}).
}
\es
\\
\\
\bs
{\it \refstepcounter{exercise}
{\bf Exercise \theexercise}: 
Re-compute the first graph polynomial ${\mathcal U}$
for the graph shown in fig.~\ref{chapter_basics:fig_nonplanar_vertex} 
from the Laplacian of the graph.
\\
\\
{\bf Solution}: 
Let us label the internal vertices 
\begin{figure}
\begin{center}
\includegraphics[scale=1.0]{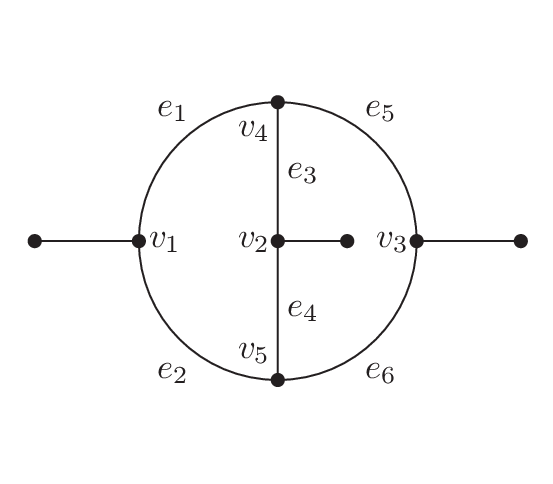}
\end{center}
\caption{
The labelling of the internal vertices for the 
two-loop non-planar vertex graph of fig.~\ref{chapter_basics:fig_nonplanar_vertex}.
}
\label{chapter_solutions:fig_nonplanar_vertex_v3}
\end{figure}
as shown in fig.~\ref{chapter_solutions:fig_nonplanar_vertex_v3}.
The Laplacian for this graph 
is given by
\bq
 L_{\mathrm{int}} 
 & = & \left( \begin{array}{ccccc}
 a_1+a_2 & 0 & 0 & -a_1 & -a_2 \\
 0 & a_3+a_4 & 0 & -a_3 & -a_4 \\
 0 & 0 & a_5+a_6 & -a_5 & -a_6 \\
 -a_1 & -a_3 & -a_5 & a_1+a_3+a_5 & 0 \\
 -a_2 & -a_4 & -a_6 & 0 & a_2+a_4+a_6 \\
                \end{array} \right).
\eq
We obtain the Kirchhoff polynomial by deleting row and column $j$ and taking the determinant afterwards.
Here, $j$ can be any number $j \in \{1,2,3,4,5\}$. Let's take $j=5$:
\bq
 {\mathcal K}_{\mathrm{int}}\left(a_1,a_2,a_3,a_4,a_5\right) & = & \det \; L_{\mathrm{int}}[5]
 \; = \;
         \left| \begin{array}{cccc}
 a_1+a_2 & 0 & 0 & -a_1 \\
 0 & a_3+a_4 & 0 & -a_3 \\
 0 & 0 & a_5+a_6 & -a_5 \\
 -a_1 & -a_3 & -a_5 & a_1+a_3+a_5 \\
                \end{array} \right|.
\eq
We then obtain the first graph polynomial from eq.~(\ref{chapter_graph_polynomials:convert_U_K}):
\bq
 {\mathcal U}(a_1,a_2,a_3,a_4,a_5) 
 & = &
 a_1 a_2 a_3 a_4 a_5 \; {\mathcal K}_{\mathrm{int}}\left(\frac{1}{a_1},\frac{1}{a_2},\frac{1}{a_3},\frac{1}{a_4},\frac{1}{a_5}\right)
 \\
 & = &
 \left( \alpha_1+\alpha_2\right) \left( \alpha_3+\alpha_4\right) 
 + \left( \alpha_1+\alpha_2\right) \left( \alpha_5+\alpha_6\right) 
 + \left( \alpha_3+\alpha_4\right) \left( \alpha_5+\alpha_6\right).
 \nonumber 
\eq
Again, we find agreement with eq.~(\ref{appendix_solutions:graph_polynomials_crossed_vertex}).
}
\es
\\
\\
\bs
{\it \refstepcounter{exercise}
{\bf Exercise \theexercise}: 
Consider a massless theory.
Show that in this case the 
Lee-Pomeransky polynomial ${\mathcal G}$ satisfies for any regular edge $e_k$ the recursion
\bq
 {\mathcal G}(G) & = & {\mathcal G}(G/e_k) + a_k {\mathcal G}(G-e_k).
\eq
\\
\\
{\bf Solution}: 
In a massless theory we have
\bq
 {\mathcal F} & = & {\mathcal F}_0.
\eq
From eq.~(\ref{chapter_graph_polynomials:recursion_U_F0}) we have for any regular edge $e_k$ the recursion
\bq
 {\mathcal U}(G) & = & {\mathcal U}(G/e_k) + a_k {\mathcal U}(G-e_k), 
 \nonumber \\
 {\mathcal F}_0(G) & = & {\mathcal F}_0(G/e_k) + a_k {\mathcal F}_0(G-e_k).
\eq
The Lee-Pomeransky polynomial ${\mathcal G}$ is given by
\bq
 {\mathcal G} & = & 
 {\mathcal U} + {\mathcal F}
 \; = \;
 {\mathcal U} + {\mathcal F}_0,
\eq
and hence the claim follows for any regular edge $e_k$:
\bq
 {\mathcal G}(G) & = & {\mathcal G}(G/e_k) + a_k {\mathcal G}(G-e_k).
\eq
}
\es
\\
\\
\bs
{\it \refstepcounter{exercise}
{\bf Exercise \theexercise}: 
Let $G$ be a graph with $\ninternal$ edges and $\nexternal$ edges and set $\nedges=\ninternal+\nexternal$.
Label the edges as 
\bq
 \mbox{internal edges} & : & \{ e_1, e_2, ..., e_{\ninternal} \},
 \nonumber  \\
 \mbox{external edges} & : & \{ e_{\ninternal+1}, e_{\ninternal+2}, ..., e_{\ninternal+\nexternal} \}.
\eq
Let $G_{\mathrm{int}}$ be the internal graph of $G$.
Define ${\mathcal U}$, ${\mathcal K}$ and ${\mathcal K}_{\mathrm{int}}$ as before.
Define $\tilde{{\mathcal U}}$ by
\bq
 \tilde{{\mathcal U}}\left(a_1,...,a_{\nedges}\right) 
 & = & 
 a_1 ... a_{\nedges} \; {\mathcal K}\left(\frac{1}{a_1},...,\frac{1}{a_{\nedges}}\right).
\eq
Show
\bq
 \tilde{{\mathcal U}}\left(a_1,...,a_{\nedges}\right) 
 & = & 
 {\mathcal U}\left(a_1,...,a_{\ninternal}\right),
 \nonumber \\
 {\mathcal K}\left(a_1,...,a_{\nedges}\right) 
 & = & 
 a_{\ninternal+1} \dots a_{\nedges}
 {\mathcal K}\left(a_1,...,a_{\ninternal}\right).
\eq
{\bf Solution}: 
The key to the solution is to realise that there is a one-to-one correspondence between the spanning trees
of $G$ and $G_{\mathrm{int}}$. The internal graph $G_{\mathrm{int}}$ is obtained from $G$ by deleting all external vertices
and edges.
Now consider a spanning tree of $G$. A spanning tree $T$ of $G$ is obtained from $G$ by deleting $\loopnumber$ edges,
such that $T$ is connected and a tree.
However, we cannot delete an external edge: If we delete an external edge, the resulting graph is disconnected.
Thus, any spanning tree of $G$ can be mapped onto a spanning tree of $G_{\mathrm{int}}$ and vice versa.
From eq.~(\ref{chapter_graph_polynomials:def_Kirchhoff_polynomial})
\bq
 {\mathcal K}\left(a_1,...,a_{\nedges}\right)
 & = & 
 \sum\limits_{T\in {\mathcal T}_1} \;
     \prod\limits_{e_j \in T} a_j
\eq
and ${\mathcal K}_{\mathrm{int}}\left(G\right) = {\mathcal K}\left(G_{\mathrm{int}}\right)$ it follows that
\bq
 {\mathcal K}\left(a_1,...,a_{\nedges}\right) 
 & = & 
 a_{\ninternal+1} \dots a_{\nedges}
 {\mathcal K}\left(a_1,...,a_{\ninternal}\right).
\eq
The second equation
\bq
 \tilde{{\mathcal U}}\left(a_1,...,a_{\nedges}\right) 
 & = & 
 {\mathcal U}\left(a_1,...,a_{\ninternal}\right)
\eq
follows then from eq.~(\ref{chapter_graph_polynomials:convert_U_K}) and the definition of $\tilde{{\mathcal U}}$.
}
\es
\\
\\
\bs
{\it \refstepcounter{exercise}
{\bf Exercise \theexercise}: 
Determine the number of loops for $K_{5}$ and $K_{3,\,3}$.
\\
\\
{\bf Solution}: 
The loop number is given by (see eq.~(\ref{chapter_basics:def_loop_number}))
\bq
 \loopnumber & = & \nedges-\nvertices+k,
\eq
where $\nedges$ denotes the number of edges, 
$\nvertices$ denotes the number of vertices and $k$ denotes the number of connected components.
Both $K_{5}$ and $K_{3,\,3}$ are connected, hence $k=1$ in both cases.
It remains to count for each graph the number of edges and vertices.
The graph $K_{5}$ has $10$ edges and $5$ vertices. We therefore find
\bq
 \loopnumber_{K_{5}} & = & 10 - 5 + 1 \; = \; 6.
\eq
The graph $K_{3,\,3}$ has $9$ edges and $6$ vertices. We therefore find
\bq
 \loopnumber_{K_{3,\,3}} & = & 9 - 6 + 1 \; = \; 4.
\eq
}
\es
\\
\\
\bs
{\it \refstepcounter{exercise}
{\bf Exercise \theexercise}: 
Consider the two graphs $G_1$ and $G_2$ shown in fig.~\ref{chapter_graph_polynomials:fig1735}, which differ by a self-loop.
For each of the two graphs, give the Kirchhoff polynomial $\mathcal{K}$ and the first graph polynomial $\mathcal{U}$.
Show that the cycle matroids are not isomorphic.
\\
\\
{\bf Solution}: We may work out the Kirchhoff polynomial and the first graph polynomial from the spanning trees
(or alternatively from the Laplacian). The result is:
\bq
 \mathcal{K}\left(G_1\right)
 & = & 
 \left(a_1+a_2\right)\left(a_3+a_4\right) + a_3 a_4,
 \nonumber \\
 \mathcal{K}\left(G_2\right)
 & = & 
 \left(a_1+a_2\right)\left(a_3+a_4\right) + a_3 a_4,
 \nonumber \\
 \mathcal{U}\left(G_1\right)
 & = & 
 \left(a_1+a_2\right)\left(a_3+a_4\right) + a_1 a_2,
 \nonumber \\
 \mathcal{U}\left(G_2\right)
 & = & 
 \left[ \left(a_1+a_2\right)\left(a_3+a_4\right) + a_1 a_2 \right] a_5.
\eq
The two graphs $G_1$ and $G_2$ have the same set of spanning trees.
Hence the Kirchhoff polynomials are equal $\mathcal{K}(G_1)=\mathcal{K}(G_2)$. 
On the other hand, in the first graph polynomial $\mathcal{U}$ the deleted edges enter.
In the graph $G_2$ we have to delete $3$ edges to obtain a tree graph, whereas in the graph $G_1$
we only have to deleted two edges.
Hence the first graph polynomials differ:
$\mathcal{U}(G_1) \neq \mathcal{U}(G_2)$.

Now let's look at the cycle matroids: It is clear that we cannot transform $G_2$ by a sequence of vertex identifications,
vertex cleavings and twistings into $G_1$. We may detach the self-loop formed by $e_5$ from the rest of the graph by 
vertex cleaving, but we cannot delete the self-loop.
From Whitney's theorem it follows that the cycle matroids are not isomorphic.

It is instructive to go back to the definition of a matroid:
A matroid is specified by a ground set $E$ and the set of independent sets $\mathcal{I}$.
We get the ground set $E$ and the set of independent sets $\mathcal{I}$ from the incidence matrix:
\bq
 B_{\mathrm{incidence}}\left(G_1\right)
 \; = \;
 \left(\begin{array}{cccc}
 1\, & \,1\, & \,1\, & 0\\
 0 & 0 & 1 & 1\\
 1 & 1 & 0 & 1
 \end{array}\right),
 & &
 B_{\mathrm{incidence}}\left(G_2\right)
 \; = \;
 \left(\begin{array}{ccccc}
 1\, & \,1\, & \,1\, & \,0\, & 0\\
 0 & 0 & 1 & 1 & 0 \\
 1 & 1 & 0 & 1 & 0
 \end{array}\right).
\eq
The column $j$ in the incidence matrix corresponds to the edge $e_j$. 
We have
\bq
 E\left(G_1\right)
 & = &
 \left\{ e_1, e_2, e_3, e_4 \right\},
 \nonumber \\
 \mathcal{I}\left(G_1\right)
 & = & 
 \left\{ \emptyset, \, \left\{ e_{1}\right\}, \, \left\{ e_{2}\right\}, \,
         \left\{ e_{3}\right\}, \, \left\{ e_{4}\right\},\right. \nonumber \\
 &  & \left.\left\{ e_{1}, \, e_{3}\right\}, \, \left\{ e_{1}, \, e_{4}\right\}, \,
      \left\{ e_{2}, \, e_{3}\right\}, \, \left\{ e_{2}, \, e_{4}\right\}, \,
      \left\{ e_{3}, \, e_{4}\right\} \right\},
 \nonumber \\
 E\left(G_2\right)
 & = &
 \left\{ e_1, e_2, e_3, e_4, e_5 \right\},
 \nonumber \\
 \mathcal{I}\left(G_2\right)
 & = & 
 \left\{ \emptyset, \, \left\{ e_{1}\right\}, \, \left\{ e_{2}\right\}, \,
         \left\{ e_{3}\right\}, \, \left\{ e_{4}\right\},\right. \nonumber \\
 &  & \left.\left\{ e_{1}, \, e_{3}\right\}, \, \left\{ e_{1}, \, e_{4}\right\}, \,
      \left\{ e_{2}, \, e_{3}\right\}, \, \left\{ e_{2}, \, e_{4}\right\}, \,
      \left\{ e_{3}, \, e_{4}\right\} \right\}.
\eq
Although
\bq
 \mathcal{I}\left(G_1\right) & = & \mathcal{I}\left(G_2\right),
\eq
the two cycle matroids are not isomorphic, as there is no bijection from $E(G_1)$ to $E(G_2)$.
$E(G_2)$ has one element more than $E(G_1)$.
}
\es
\\
\\
\bs
{\it \refstepcounter{exercise}
{\bf Exercise \theexercise}: 
Consider first the two graphs $G_1$ and $G_2$ shown in fig.~\ref{chapter_graph_polynomials:fig_twisting},
both with three external legs.
Assume that all internal masses vanish.
Show that
\bq
 \mathcal{U}\left(G_1\right) \; = \; \mathcal{U}\left(G_2\right),
 & &
 \mathcal{F}\left(G_1\right) \; = \; \mathcal{F}\left(G_2\right).
\eq
Consider then the graphs $G_3$ and $G_4$ with four external legs.
Show that
\bq
 \mathcal{U}\left(G_3\right) & = & \mathcal{U}\left(G_4\right),
\eq
but
\bq
 \mathcal{F}\left(G_3\right) & \neq & \mathcal{F}\left(G_4\right).
\eq
{\bf Solution}: 
Let us first 
\begin{figure}
\begin{center}
\includegraphics[scale=1.0]{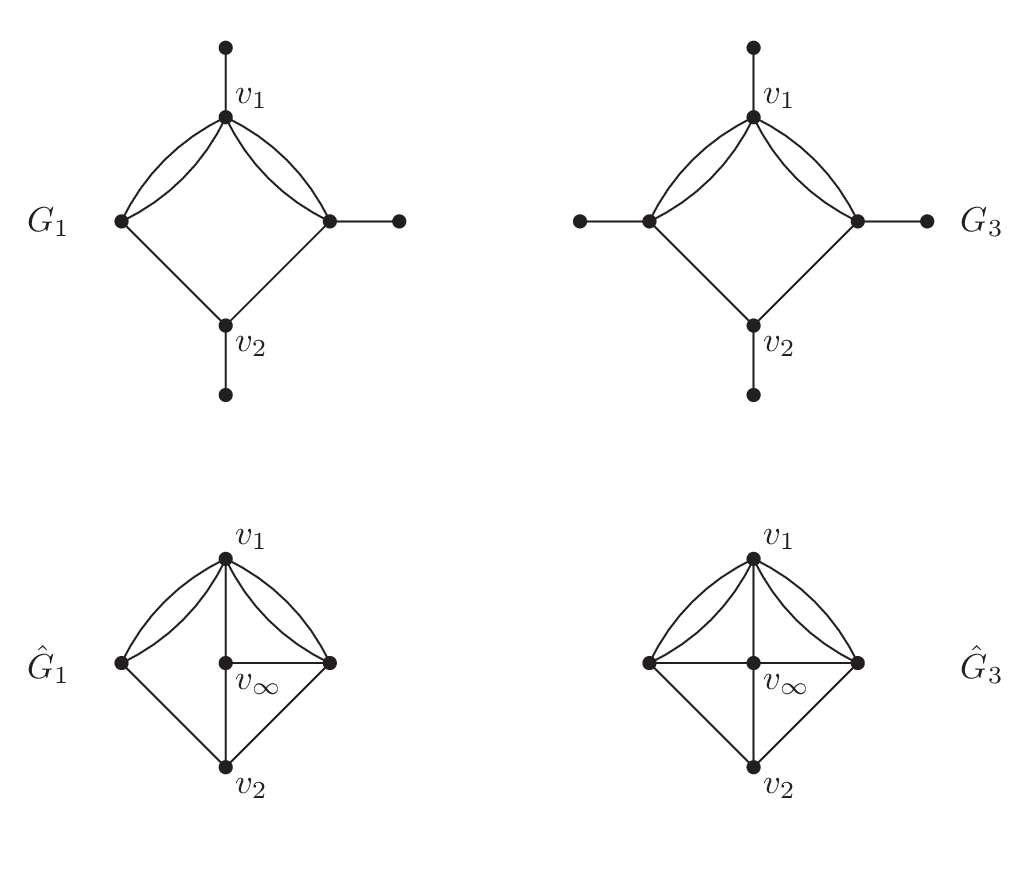}
\end{center}
\caption{\label{appendix_solutions::fig_twisting_solution} 
The upper part shows the graphs $G_1$ and $G_3$ with two labelled vertices.
The lower part shows the graphs $\hat{G}_1$ and $\hat{G}_3$.}
\end{figure}
show
\bq
 \mathcal{U}\left(G_1\right) \; = \; \mathcal{U}\left(G_2\right),
 & &
 \mathcal{U}\left(G_3\right) \; = \; \mathcal{U}\left(G_4\right).
\eq
Consider fig.~\ref{appendix_solutions::fig_twisting_solution}.
We obtain $G_2$ from $G_1$ by twisting at the vertices $v_1$ and $v_2$ of $G_1$.
Similarly, we obtain $G_4$ from $G_3$ by twisting at the vertices $v_1$ and $v_2$ of $G_3$.
Hence $\mathcal{U}(G_1)=\mathcal{U}(G_2)$ and $\mathcal{U}(G_3)=\mathcal{U}(G_4)$.

For the second graph polynomial $\mathcal{F}$ we have to consider the graphs $\hat{G}_1$, $\hat{G}_2$, $\hat{G}_3$ and $\hat{G}_4$.
The graph $\hat{G}_1$ is shown in the lower left part of fig.~\ref{appendix_solutions::fig_twisting_solution}.
By twisting at the vertices $v_1$ and $v_2$ of $\hat{G}_1$ we obtain $\hat{G}_2$ and 
the relation $\mathcal{F}(G_1)=\mathcal{F}(G_2)$ follows.
The graph $\hat{G}_3$ is shown in the lower right part of fig.~\ref{appendix_solutions::fig_twisting_solution}.
For the twisting operation we have to split the graph into two disjoint pieces by cleaving at exactly two vertices.
In order to obtain $\hat{G}_4$ from $\hat{G}_3$ we would have to cleave at the three vertices $v_1$, $v_2$ and $v_\infty$.
But the twisting operation requires cleaving at exactly two vertices.
Hence we cannot obtain $\hat{G}_4$ from $\hat{G}_3$ by the operations of vertex identifications, vertex cleavings and twisting.
A short calculation shows that for generic external momenta $\mathcal{F}(G_3) \neq \mathcal{F}(G_4)$.
}
\es
\\
\\
%
%
\bs
{\it \refstepcounter{exercise}
{\bf Exercise \theexercise}: 
Derive eq.~(\ref{chapter_qft:gluon_propagator}) from eq.~(\ref{chapter_qft:gluon_prop_inverse_x_space}) 
and eq.~(\ref{chapter_qft:gluon_prop_Fourier_trafo}).
\\
\\
{\bf Solution}: 
With
\bq
 P^{\mu\nu\;ab}(x) & = & \partial_\rho \partial^\rho g^{\mu\nu} \delta^{ab}
                  - \left( 1 - \frac{1}{\xi} \right) \partial^\mu \partial^\nu \delta^{ab}
\eq
and
\bq
 \left( P^{-1} \right)_{\mu\nu}^{ab}(x) & = & 
  \int \frac{d^D q}{(2 \pi)^D} e^{-i q \cdot x} \left( \tilde{P}^{-1} \right)_{\mu\nu}^{ab}(q).
\eq
we have
\bq
 P^{\mu\sigma\;ac}(x) \left( P^{-1} \right)_{\sigma\nu}^{cb}(x-y) 
 & = & 
  \int \frac{d^D q}{(2 \pi)^D} e^{-i q \cdot \left(x-y\right)}
  q^2 \left[ - g^{\mu\sigma} 
                  + \left( 1 - \frac{1}{\xi} \right) \frac{q^\mu q^\sigma}{q^2} 
 \right] \delta^{ac}
 \; \left( \tilde{P}^{-1} \right)_{\sigma\nu}^{cb}(q).
 \nonumber
\eq
This should be equal to
\bq
 g^\mu_{\;\;\nu} \delta^{ab} \delta^D(x-y)
 & = &
  \int \frac{d^D q}{(2 \pi)^D} e^{-i q \cdot \left(x-y\right)} g^\mu_{\;\;\nu} \delta^{ab}.
\eq
We have for
\bq
 M^{\mu\nu} = - g^{\mu\nu} + \left( 1 - \frac{1}{\xi} \right) \frac{q^\mu q^\nu}{q^2}
 & \mbox{and} &
 N_{\mu\nu} = - g_{\mu\nu} + \left( 1 -\xi \right) \frac{q_\mu q_\nu}{q^2}
\eq
the relation
\bq
 M^{\mu \sigma} N_{\sigma \nu} & = & g^\mu_{\;\;\nu}.
\eq
Therefore
\bq
 \begin{picture}(100,20)(0,5)
 \Gluon(20,10)(70,10){-5}{5}
 \Text(15,12)[r]{\footnotesize $\mu, a$}
 \Text(75,12)[l]{\footnotesize $\nu, b$}
\end{picture} 
 & = & 
  \frac{i}{q^2} \left( - g_{\mu\nu} + \left( 1 -\xi \right) \frac{q_\mu q_\nu}{q^2} \right) \delta^{ab}.
\eq
}
\es
\\
\\
\bs
{\it \refstepcounter{exercise}
{\bf Exercise \theexercise}: 
Compute the four-gluon amplitude ${\mathcal A}_4^{(0)}$ from the four diagrams shown in fig.~\ref{chapter_qft:fig1}.
Assume that all momenta are outgoing.
Derive the Mandelstam relation
\bq 
 s + t + u & = & 0.
\eq
{\bf Solution}: Let us start with the Mandelstam relation.
Using momentum conservation and the on-shell relations we have
\bq
 0 & = &
 p_4^2
 \;\; = \;\; 
 \left( p_1 + p_2 + p_3 \right)^2
 \;\; = \;\; 
 2 p_1 \cdot p_2 + 2 p_2 \cdot p_3 + 2 p_1 \cdot p_3
 \nonumber \\
 & = &
 \left( p_1 + p_2 \right)^2 + \left( p_2 + p_3 \right)^2 + \left( p_1 + p_3 \right)^2
 \;\; = \;\; 
 s + t + u.
\eq
We then turn to the computation of the amplitude.
Let us first examine the colour factors.
The first diagrams has a colour factor
$C_s=(i f^{a_1 a_2 b}) (i f^{b a_3 a_4})$.
The second diagram we may equally well draw with legs $1$ and $4$ exchanged.
The colour factor is then given by $C_t=(i f^{a_2 a_3 b}) (i f^{b a_1 a_4})$.
(If we read off the colour factor directly from diagram $2$, we find 
$(i f^{a_2 a_3 b}) (i f^{b a_4 a_1}) = -(i f^{a_2 a_3 b}) (i f^{b a_1 a_4})$.
The minus sign cancels with another minus sign from the kinematic part.)
The third diagram has the colour factor $C_u=(i f^{a_3 a_1 b}) (i f^{b a_2 a_4})$.
The fourth diagram with the four-gluon vertex gives three terms, one contributing to each colour structure.
We may therefore write the amplitude as
\bq
 {\mathcal A}_4^{(0)}
 & = &
 i g^2
 \left[
  \frac{(i f^{a_1 a_2 b}) (i f^{b a_3 a_4}) \; N_s}{s}
  +
  \frac{(i f^{a_2 a_3 b}) (i f^{b a_1 a_4}) \; N_t}{t}
  +
  \frac{(i f^{a_3 a_1 b}) (i f^{b a_2 a_4}) \; N_u}{u}
 \right].
\eq
$N_s$ is given by
\bq
 N_s 
 & = &
 \left\{
 \left[ g^{\mu_1\mu_2} \left( p_1^{\nu} - p_2^{\nu} \right)
         +g^{\mu_2\nu} \left( p_2^{\mu_1} - p_{34}^{\mu_1} \right)
         +g^{\nu\mu_1} \left( p_{34}^{\mu_2} - p_1^{\mu_2} \right)
   \right]
 g_{\nu\rho}
 \left[ g^{\mu_3\mu_4} \left( p_3^{\rho} - p_4^{\rho} \right)
         +g^{\mu_4\rho} \left( p_4^{\mu_3} - p_{12}^{\mu_3} \right)
 \right. \right.
 \nonumber \\
 & & \left. \left.
         +g^{\rho\mu_3} \left( p_{12}^{\mu_4} - p_3^{\mu_4} \right)
   \right]
 +
 2 p_1 \cdot p_2 \left( g^{\mu_1\mu_3} g^{\mu_2\mu_4} - g^{\mu_2\mu_3} g^{\mu_1\mu_4} \right)
 \right\}
 \eps_1^{\mu_1} \eps_2^{\mu_2} \eps_3^{\mu_3} \eps_4^{\mu_4},
\eq
where we used the notation $p_{ij} = p_i + p_j$.
Using momentum conservation and $p_j \cdot \eps_j=0$ we may simplify this expression to
\bq
\label{appendix_solutions:to_be_contracted}
 N_s 
 & = &
 \left\{
 \left[ g^{\mu_1\mu_2} \left( p_1^{\nu} - p_2^{\nu} \right)
         + 2 g^{\mu_2\nu} p_2^{\mu_1} 
         - 2 g^{\nu\mu_1} p_1^{\mu_2}
   \right]
 g_{\nu\rho}
 \left[ g^{\mu_3\mu_4} \left( p_3^{\rho} - p_4^{\rho} \right)
         + 2 g^{\mu_4\rho} p_4^{\mu_3}
         - 2 g^{\rho\mu_3} p_3^{\mu_4} 
   \right]
 \right.
 \nonumber \\
 & & \left.
 +
 2 p_1 \cdot p_2 \left( g^{\mu_1\mu_3} g^{\mu_2\mu_4} - g^{\mu_2\mu_3} g^{\mu_1\mu_4} \right)
 \right\}
 \eps_1^{\mu_1} \eps_2^{\mu_2} \eps_3^{\mu_3} \eps_4^{\mu_4}.
\eq
The contraction of indices for long expressions is best done with the help
of a computer algebra program.
Here is a short {\tt FORM} program, which performs the contractions in eq.~(\ref{appendix_solutions:to_be_contracted}):
{\footnotesize
\begin{verbatim}
* Example program for FORM

V p1,p2,p3,p4, e1,e2,e3,e4;
I mu1,mu2,mu3,mu4,nu,rho;

L  Ns = ((d_(mu1,mu2)*(p1(nu)-p2(nu)) + 2*d_(mu2,nu)*p2(mu1) - 2*d_(nu,mu1)*p1(mu2))*
         d_(nu,rho)*
         (d_(mu3,mu4)*(p3(rho)-p4(rho)) + 2*d_(mu4,rho)*p4(mu3) - 2*d_(rho,mu3)*p3(mu4))
         +2*p1(nu)*p2(nu)*(d_(mu1,mu3)*d_(mu2,mu4)-d_(mu2,mu3)*d_(mu1,mu4)))*
        e1(mu1)*e2(mu2)*e3(mu3)*e4(mu4);

print;

.end
\end{verbatim}
}
\noindent 
The same can be done in C++, using the {\tt GiNaC} library:
{\footnotesize
\begin{verbatim}
// Example in C++ with GiNaC

#include <iostream>
#include <string>
#include <sstream>
#include <ginac/ginac.h>
using namespace std;
using namespace GiNaC;

string itos(int arg)
{
  ostringstream buffer;
  buffer << arg; 
  return buffer.str(); 
}

int main()
{
  varidx mu1(symbol("mu1"),4), mu2(symbol("mu2"),4),
         mu3(symbol("mu3"),4), mu4(symbol("mu4"),4), 
         nu(symbol("nu"),4), rho(symbol("rho"),4);

  symbol p1("p1"), p2("p2"), p3("p3"), p4("p4"),
         e1("e1"), e2("e2"), e3("e3"), e4("e4");

  vector<ex> p_vec = { p1, p2, p3, p4 };
  vector<ex> e_vec = { e1, e2, e3, e4 };

  scalar_products sp;
  for (int i=0; i<4; i++)
    {
      for (int j=i+1; j<4; j++)
        {
          sp.add(p_vec[i],p_vec[j],symbol( string("p")+itos(i+1)+string("p")+itos(j+1) ));
          sp.add(e_vec[i],e_vec[j],symbol( string("e")+itos(i+1)+string("e")+itos(j+1) ));
          sp.add(p_vec[i],e_vec[j],symbol( string("p")+itos(i+1)+string("e")+itos(j+1) ));
          sp.add(e_vec[i],p_vec[j],symbol( string("p")+itos(j+1)+string("e")+itos(i+1) ));
        }
    }

  ex Ns = ((lorentz_g(mu1,mu2)*(indexed(p1,nu)-indexed(p2,nu)) 
	    + 2*lorentz_g(mu2,nu)*indexed(p2,mu1) - 2*lorentz_g(nu,mu1)*indexed(p1,mu2))
	   * lorentz_g(nu.toggle_variance(),rho.toggle_variance())
	   * (lorentz_g(mu3,mu4)*(indexed(p3,rho)-indexed(p4,rho))
	      + 2*lorentz_g(mu4,rho)*indexed(p4,mu3) - 2*lorentz_g(rho,mu3)*indexed(p3,mu4))
           + 2*indexed(p1,nu)*indexed(p2,nu.toggle_variance())
	   *(lorentz_g(mu1,mu3)*lorentz_g(mu2,mu4) - lorentz_g(mu2,mu3)*lorentz_g(mu1,mu4)))
    *indexed(e1,mu1.toggle_variance())*indexed(e2,mu2.toggle_variance())
    *indexed(e3,mu3.toggle_variance())*indexed(e4,mu4.toggle_variance());

  Ns = Ns.expand();
  Ns = Ns.simplify_indexed(sp);

  cout << Ns << endl;
}
\end{verbatim}
}
\noindent
After a few additional simplifications one finds
\bq
 N_s & = &
 4 \left( p_1 \cdot \eps_2 \right) \left( p_3 \cdot \eps_4 \right) \left( \eps_1 \cdot \eps_3 \right)
 - 4 \left( p_1 \cdot \eps_2 \right) \left( p_4 \cdot \eps_3 \right) \left( \eps_1 \cdot \eps_4 \right)
 + 4 \left( p_2 \cdot \eps_1 \right) \left( p_4 \cdot \eps_3 \right) \left( \eps_2 \cdot \eps_4 \right)
 \nonumber \\
 & &
 - 4 \left( p_2 \cdot \eps_1 \right) \left( p_3 \cdot \eps_4 \right) \left( \eps_2 \cdot \eps_3 \right)
 + 4 \left[ \left( p_1 \cdot \eps_3 \right) \left( p_2 \cdot \eps_4 \right) 
          - \left( p_1 \cdot \eps_4 \right) \left( p_2 \cdot \eps_3 \right) \right] \left( \eps_1 \cdot \eps_2 \right)
 \nonumber \\
 & &
 + 4 \left[ \left( p_3 \cdot \eps_1 \right) \left( p_4 \cdot \eps_2 \right) 
          - \left( p_3 \cdot \eps_2 \right) \left( p_4 \cdot \eps_1 \right) \right] \left( \eps_3 \cdot \eps_4 \right)
 + 2 \left( p_1 \cdot p_2 \right) \left( \eps_1 \cdot \eps_3 \right) \left( \eps_2 \cdot \eps_4 \right)
 \nonumber \\
 & &
 - 2 \left( p_1 \cdot p_2 \right) \left( \eps_1 \cdot \eps_4 \right) \left( \eps_2 \cdot \eps_3 \right)
 - 2 \left( p_2 \cdot p_3 - p_1 \cdot p_3 \right) \left( \eps_1 \cdot \eps_2 \right) \left( \eps_3 \cdot \eps_4 \right).
\eq
The numerator $N_t$ is obtained from the numerator $N_s$ by the substitution $(1,2,3) \rightarrow (2,3,1)$,
the numerator $N_u$ is obtained from the numerator $N_s$ by the substitution $(1,2,3) \rightarrow (3,1,2)$.

Let us add the following remark:
The colour factors satisfy (obviously) the Jacobi identity
\bq
 C_s + C_t + C_u & = & 0.
\eq
It is an easy exercise to check, that the numerators $N_s$, $N_t$ and $N_u$,
as determined above, 
satisfy the Jacobi-like identity
\bq
 N_s + N_t + N_u & = & 0.
\eq
}
\es
\\
\\
\bs
{\it \refstepcounter{exercise}
{\bf Exercise \theexercise}: 
Let $n \in {\mathbb N}$. Show that the action of $({\bf j}^+)^n$ on the integrand  
of the Schwinger parameter representation is given by
\bq
 \left( {\bf j}^+ \right)^n I_{\nu_1 \dots \nu_j \dots \nu_{\ninternal}}\left(D\right)  & = &
 \frac{e^{\loopnumber \eps \Eulerconstant}}{\prod\limits_{k=1}^{\ninternal}\Gamma(\nu_k)}
 \;
 \int\limits_{\alpha_k \ge 0}  d^{\ninternal}\alpha \;
 \left( \prod\limits_{k=1}^{\ninternal} \alpha_k^{\nu_k-1} \right)
 \frac{\alpha_j^n}{{\mathcal U}^{\frac{D}{2}}}
 e^{- \frac{{\mathcal F}}{{\mathcal U}}}.
\eq
{\bf Solution}: 
By definition the operator $({\bf j}^+)^n$ acts on $I_{\nu_1 \dots \nu_j \dots \nu_{\ninternal}}(D)$ as
\bq
 \left( {\bf j}^+ \right)^n I_{\nu_1 \dots \nu_j \dots \nu_{\ninternal}}\left(D\right)  & = &
 \nu_j \left( \nu_j+1\right) \cdot \dots \left( \nu_j+n-1 \right) \cdot 
 I_{\nu_1 \dots \left(\nu_j+n\right) \dots \nu_{\ninternal}}\left(D\right)
 \nonumber \\
 & = &
 \frac{\Gamma\left(\nu_j+n\right)}{\Gamma\left(\nu_j\right)}
 I_{\nu_1 \dots \left(\nu_j+n\right) \dots \nu_{\ninternal}}\left(D\right).
\eq
For $I_{\nu_1 \dots (\nu_j+n) \dots \nu_{\ninternal}}(D)$ we use the Schwinger parameter representation
\bq
I_{\nu_1 \dots \left(\nu_j+n\right) \dots \nu_{\ninternal}}\left(D\right)
 & = &
 \frac{e^{\loopnumber \eps \Eulerconstant}}{\Gamma(\nu_j+n)\prod\limits_{\substack{k=1 \\ k \neq j}}^{\ninternal}\Gamma(\nu_k)}
 \;
 \int\limits_{\alpha_k \ge 0}  d^{\ninternal}\alpha \;
 \left( \prod\limits_{k=1}^{\ninternal} \alpha_k^{\nu_k-1} \right)
 \frac{\alpha_j^n}{{\mathcal U}^{\frac{D}{2}}}
 e^{- \frac{{\mathcal F}}{{\mathcal U}}}.
\eq
Thus
\bq
 \left( {\bf j}^+ \right)^n I_{\nu_1 \dots \nu_j \dots \nu_{\ninternal}}\left(D\right)  
 & = &
 \frac{\Gamma\left(\nu_j+n\right)}{\Gamma\left(\nu_j\right)}
 I_{\nu_1 \dots \left(\nu_j+n\right) \dots \nu_{\ninternal}}\left(D\right)
 \nonumber \\
 & = &
 \frac{e^{\loopnumber \eps \Eulerconstant}}{\prod\limits_{k=1}^{\ninternal}\Gamma(\nu_k)}
 \;
 \int\limits_{\alpha_k \ge 0}  d^{\ninternal}\alpha \;
 \left( \prod\limits_{k=1}^{\ninternal} \alpha_k^{\nu_k-1} \right)
 \frac{\alpha_j^n}{{\mathcal U}^{\frac{D}{2}}}
 e^{- \frac{{\mathcal F}}{{\mathcal U}}},
\eq
as claimed.
}
\es
\\
\\
\bs
{\it \refstepcounter{exercise}
{\bf Exercise \theexercise}: 
Work out the corresponding formula for
\bq
\int \frac{d^{D }k}{i\pi^{D /2}} k^{\mu_1} k^{\mu_2} k^{\mu_3} k^{\mu_4} k^{\mu_5} k^{\mu_6} f(k^2).
\eq
{\bf Solution}: 
The result must be proportional to a 
symmetric tensor $T^{\mu_1\mu_2 \mu_3 \mu_4 \mu_5 \mu_6}$ build from the metric tensor.
Let's first construct this tensor.
It has $15$ terms. This can be seen as follows:
Start from index $\mu_1$: There are five possibilities how this index can be paired with another index $\mu_j$
into $g^{\mu_1 \mu_j}$: Any choice $j \in \{2,3,4,5,6\}$ is allowed.
The remaining four indices must form a symmetric tensor of rank $4$, 
such a rank $4$ tensor has three terms (compare with eq.~(\ref{chapter_qft:symmetric_integration})).
Thus
\bq
 T^{\mu_1\mu_2 \mu_3 \mu_4 \mu_5 \mu_6}
 & = &
   g^{\mu_1 \mu_2} g^{\mu_3 \mu_4} g^{\mu_5 \mu_6}
 + g^{\mu_1 \mu_2} g^{\mu_3 \mu_5} g^{\mu_4 \mu_6}
 + g^{\mu_1 \mu_2} g^{\mu_3 \mu_6} g^{\mu_4 \mu_5}
 \nonumber \\
 & &
 + g^{\mu_1 \mu_3} g^{\mu_2 \mu_4} g^{\mu_5 \mu_6}
 + g^{\mu_1 \mu_3} g^{\mu_2 \mu_5} g^{\mu_4 \mu_6}
 + g^{\mu_1 \mu_3} g^{\mu_2 \mu_6} g^{\mu_4 \mu_5}
 \nonumber \\
 & &
 + g^{\mu_1 \mu_4} g^{\mu_3 \mu_2} g^{\mu_5 \mu_6}
 + g^{\mu_1 \mu_4} g^{\mu_3 \mu_5} g^{\mu_2 \mu_6}
 + g^{\mu_1 \mu_4} g^{\mu_3 \mu_6} g^{\mu_2 \mu_5}
 \nonumber \\
 & &
 + g^{\mu_1 \mu_5} g^{\mu_3 \mu_4} g^{\mu_2 \mu_6}
 + g^{\mu_1 \mu_5} g^{\mu_3 \mu_2} g^{\mu_4 \mu_6}
 + g^{\mu_1 \mu_5} g^{\mu_3 \mu_6} g^{\mu_4 \mu_2}
 \nonumber \\
 & &
 + g^{\mu_1 \mu_6} g^{\mu_3 \mu_4} g^{\mu_5 \mu_2}
 + g^{\mu_1 \mu_6} g^{\mu_3 \mu_5} g^{\mu_4 \mu_2}
 + g^{\mu_1 \mu_6} g^{\mu_3 \mu_2} g^{\mu_4 \mu_5}.
\eq
Thus we have the ansatz
\bq
 \int \frac{d^{D }k}{i\pi^{D /2}} k^{\mu_1} k^{\mu_2} k^{\mu_3} k^{\mu_4} k^{\mu_5} k^{\mu_6} f(k^2)
 & = &
 T^{\mu_1\mu_2 \mu_3 \mu_4 \mu_5 \mu_6}
 \int \frac{d^{D }k}{i\pi^{D /2}} g(k^2) f(k^2)
\eq
for some unknown function $g(k^2)$.
We contract both sides with $g_{\mu_1 \mu_2} g_{\mu_3 \mu_4} g_{\mu_5 \mu_6}$.
On the left-hand side we obtain
\bq
 g_{\mu_1 \mu_2} g_{\mu_3 \mu_4} g_{\mu_5 \mu_6} k^{\mu_1} k^{\mu_2} k^{\mu_3} k^{\mu_4} k^{\mu_5} k^{\mu_6}
 & = &
 \left( k^2 \right)^3,
\eq
on the right-hand side we obtain
\bq
 g_{\mu_1 \mu_2} g_{\mu_3 \mu_4} g_{\mu_5 \mu_6} T^{\mu_1\mu_2 \mu_3 \mu_4 \mu_5 \mu_6}
 & = &
 D \left(D+2\right) \left(D+4\right),
\eq
hence
\bq
 \int \frac{d^{D }k}{i\pi^{D /2}} k^{\mu_1} k^{\mu_2} k^{\mu_3} k^{\mu_4} k^{\mu_5} k^{\mu_6} f(k^2)
 & = &
 - \frac{T^{\mu_1\mu_2 \mu_3 \mu_4 \mu_5 \mu_6}}{D \left(D+2\right) \left(D+4\right)}
 \int \frac{d^{D }k}{i\pi^{D /2}} \left(-k^2\right)^3 f(k^2).
\eq
}
\es
\\
\\
\bs
{\it \refstepcounter{exercise}
{\bf Exercise \theexercise}: 
Prove eqs.~(\ref{chapter_qft:Dirac_trace_rule_1})-(\ref{chapter_qft:Dirac_trace_rule_3}).
\\
\\
{\bf Solution}: 
We start with
\bq
 \mathrm{Tr}\left(\gamma_{(4)}^\mu \gamma_{(4)}^\nu\right) 
 & = & 
 4 g_{(4)}^{\mu \nu}.
\eq
From the cyclic property of trace and the 
anti-commutation relation of eq.~(\ref{chapter_qft:Dirac_anticommutation}) we have
\bq
 2 \mathrm{Tr}\left(\gamma_{(4)}^\mu \gamma_{(4)}^\nu\right) 
 & = & 
 \mathrm{Tr}\left(\gamma_{(4)}^\mu \gamma_{(4)}^\nu\right) + \mathrm{Tr}\left(\gamma_{(4)}^\nu \gamma_{(4)}^\mu\right)
 \; = \; 
 2 g_{(4)}^{\mu \nu} \mathrm{Tr} \; {\bf 1}
 \; = \; 
 8 g_{(4)}^{\mu \nu}.
\eq
In order to prove
\bq
 \mathrm{Tr}\left(\gamma_{(4)}^{\mu_1} \gamma_{(4)}^{\mu_2} ... \gamma_{(4)}^{\mu_{2 n}}\right)
 & = & 
 \sum\limits_{j=2}^{2n} \left(-1\right)^j
 g_{(4)}^{\mu_1 \mu_j} \mathrm{Tr}\left(\gamma_{(4)}^{\mu_2} ... \gamma_{(4)}^{\mu_{j-1}} \gamma_{(4)}^{\mu_{j+1}} ... \gamma_{(4)}^{\mu_{2 n}}\right)
\eq
we anti-commute the first Dirac matrix $\gamma_{(4)}^{\mu_1}$ from the first place to the last place, using
the anti-commutation relation of eq.~(\ref{chapter_qft:Dirac_anticommutation}).
This yields
\bq
 \mathrm{Tr}\left(\gamma_{(4)}^{\mu_1} \gamma_{(4)}^{\mu_2} ... \gamma_{(4)}^{\mu_{2 n}}\right)
 & = & 
 2 \sum\limits_{j=2}^{2n} \left(-1\right)^j
 g_{(4)}^{\mu_1 \mu_j} \mathrm{Tr}\left(\gamma_{(4)}^{\mu_2} ... \gamma_{(4)}^{\mu_{j-1}} \gamma_{(4)}^{\mu_{j+1}} ... \gamma_{(4)}^{\mu_{2 n}}\right)
 +
 \mathrm{Tr}\left(\gamma_{(4)}^{\mu_2} ... \gamma_{(4)}^{\mu_{2 n}} \gamma_{(4)}^{\mu_1} \right).
\eq
From the cyclicity of the trace we have
\bq
 \mathrm{Tr}\left(\gamma_{(4)}^{\mu_2} ... \gamma_{(4)}^{\mu_{2 n}} \gamma_{(4)}^{\mu_1} \right)
 & = &
 \mathrm{Tr}\left(\gamma_{(4)}^{\mu_1} \gamma_{(4)}^{\mu_2} ... \gamma_{(4)}^{\mu_{2 n}}\right)
\eq
and the result follows.
In order to show
that the trace of an odd number of Dirac matrices vanishes
\bq
 \mathrm{Tr}\left(\gamma_{(4)}^{\mu_1} \gamma_{(4)}^{\mu_2} ... \gamma_{(4)}^{\mu_{2 n-1}}\right) 
 & = & 0
\eq
we use $\gamma_5^2={\bf 1}$, the anti-commutation relations of $\gamma_5$ (we anti-commute one $\gamma_5$ from the second place to the last place) and the cyclicity of the trace:
\bq
 \mathrm{Tr}\left(\gamma_{(4)}^{\mu_1} \gamma_{(4)}^{\mu_2} ... \gamma_{(4)}^{\mu_{2 n-1}}\right) 
 & = &
 \mathrm{Tr}\left(\gamma_5 \gamma_5 \gamma_{(4)}^{\mu_1} \gamma_{(4)}^{\mu_2} ... \gamma_{(4)}^{\mu_{2 n-1}}\right) 
 \; = \; 
 - \mathrm{Tr}\left(\gamma_5 \gamma_{(4)}^{\mu_1} \gamma_{(4)}^{\mu_2} ... \gamma_{(4)}^{\mu_{2 n-1}} \gamma_5 \right) 
 \nonumber \\
 & = &
 - \mathrm{Tr}\left(\gamma_5 \gamma_5 \gamma_{(4)}^{\mu_1} \gamma_{(4)}^{\mu_2} ... \gamma_{(4)}^{\mu_{2 n-1}} \right) 
 \; = \;
 - \mathrm{Tr}\left(\gamma_{(4)}^{\mu_1} \gamma_{(4)}^{\mu_2} ... \gamma_{(4)}^{\mu_{2 n-1}}\right).
\eq
Thus
\bq
 2 \mathrm{Tr}\left(\gamma_{(4)}^{\mu_1} \gamma_{(4)}^{\mu_2} ... \gamma_{(4)}^{\mu_{2 n-1}}\right) 
 & = &
 0.
\eq
For traces involving $\gamma_5$ 
\bq
 \mathrm{Tr}\left(\gamma_5\right) 
 & = & 0,
 \nonumber \\
 \mathrm{Tr}\left(\gamma_{(4)}^{\mu_1} \gamma_{(4)}^{\mu_2} \gamma_5\right) 
 & = & 0,
 \nonumber \\
 \mathrm{Tr}\left(\gamma_{(4)}^{\mu_1} \gamma_{(4)}^{\mu_2} \gamma_{(4)}^{\mu_3} \gamma_{(4)}^{\mu_4} \gamma_5\right) 
 & = & 4 i \eps^{\mu_1 \mu_2 \mu_3 \mu_4}
\eq
we insert the definition of $\gamma_5$
\bq
 \gamma_5 & = & 
 \frac{i}{24} \eps_{\nu_1 \nu_2 \nu_3 \nu_4} \gamma_{(4)}^{\nu_1} \gamma_{(4)}^{\nu_2} \gamma_{(4)}^{\nu_3} \gamma_{(4)}^{\nu_4}.
\eq
We then have to evaluate traces with an even number of Dirac matrices, which we can do with the rule
proven above.
In the first two cases we have to evaluate a trace over four, respectively six Dirac matrices.
Each term of the result necessarily contains a factor $g^{\nu_i \nu_j}$, which vanishes when contracted into
$\eps_{\nu_1 \nu_2 \nu_3 \nu_4}$.
In the third case we have to evaluate a trace over eight Dirac matrices.
Here, terms of the form
\bq
 g^{\mu_1 \nu_{\pi(1)}} g^{\mu_2 \nu_{\pi(2)}} g^{\mu_3 \nu_{\pi(3)}} g^{\mu_4 \nu_{\pi(4)}}
\eq
survive, where $\pi$ is a permutation of $(1,2,3,4)$.
We may either work out the result by brute force, or -- more elegantly -- first establish that the
final result must be proportional to $\eps^{\mu_1 \mu_2 \mu_3 \mu_4}$, as any symmetric part vanishes:
\bq
 \mathrm{Tr}\left(\gamma_{(4)}^{\mu_2} \gamma_{(4)}^{\mu_1} \gamma_{(4)}^{\mu_3} \gamma_{(4)}^{\mu_4} \gamma_5\right) 
 & = &
 - \mathrm{Tr}\left(\gamma_{(4)}^{\mu_1} \gamma_{(4)}^{\mu_2} \gamma_{(4)}^{\mu_3} \gamma_{(4)}^{\mu_4} \gamma_5\right) 
 +
 2 g_{(4)}^{\mu_1 \mu_2} \mathrm{Tr}\left(\gamma_{(4)}^{\mu_3} \gamma_{(4)}^{\mu_4} \gamma_5\right) 
 \nonumber \\
 & = &
 - \mathrm{Tr}\left(\gamma_{(4)}^{\mu_1} \gamma_{(4)}^{\mu_2} \gamma_{(4)}^{\mu_3} \gamma_{(4)}^{\mu_4} \gamma_5\right).
\eq
Thus
\bq
 \mathrm{Tr}\left(\gamma_{(4)}^{\mu_1} \gamma_{(4)}^{\mu_2} \gamma_{(4)}^{\mu_3} \gamma_{(4)}^{\mu_4} \gamma_5\right) 
 & = & c \eps^{\mu_1 \mu_2 \mu_3 \mu_4}
\eq
for some constant $c$. We then obtain the constant by contracting with $i \eps_{\mu_1 \mu_2 \mu_3 \mu_4}/24$.
On the left-hand side we find
\bq
 \frac{i}{24} \eps_{\mu_1 \mu_2 \mu_3 \mu_4}
 \mathrm{Tr}\left(\gamma_{(4)}^{\mu_1} \gamma_{(4)}^{\mu_2} \gamma_{(4)}^{\mu_3} \gamma_{(4)}^{\mu_4} \gamma_5\right) 
 & = &
 \mathrm{Tr}\left(\gamma_5 \gamma_5\right) 
 \; = \;
 \mathrm{Tr}\left({\bf 1}\right) 
 \; = \;
 4.
\eq
On the right-hand side we obtain (using $\eps_{\mu_1 \mu_2 \mu_3 \mu_4} \eps^{\mu_1 \mu_2 \mu_3 \mu_4}=-24$)
\bq
 \frac{i}{24} \eps_{\mu_1 \mu_2 \mu_3 \mu_4}
 \cdot
 c \eps^{\mu_1 \mu_2 \mu_3 \mu_4}
 & = &
 -i c,
\eq
and hence $c=4i$.
}
\es
\\
\\
\bs
{\it \refstepcounter{exercise}
{\bf Exercise \theexercise}: 
Show that with the definitions and conventions as above the rules for the traces of Dirac matrices carry over 
to $D$ dimensions.
In detail, show:
\begin{enumerate}
\item Traces of an even number of Dirac matrices are evaluated with the rules
\bq
 \mathrm{Tr}\left(\gamma_{(D)}^\mu \gamma_{(D)}^\nu\right) 
 & = & 
 4 g_{(D)}^{\mu \nu},
 \nonumber \\
 \mathrm{Tr}\left(\gamma_{(D)}^{\mu_1} \gamma_{(D)}^{\mu_2} ... \gamma_{(D)}^{\mu_{2 n}}\right)
 & = & 
 \sum\limits_{j=2}^{2n} \left(-1\right)^j
 g_{(D)}^{\mu_1 \mu_j} \mathrm{Tr}\left(\gamma_{(D)}^{\mu_2} ... \gamma_{(D)}^{\mu_{j-1}} \gamma_{(D)}^{\mu_{j+1}} ... \gamma_{(D)}^{\mu_{2 n}}\right).
\eq
\item Traces of an odd number of Dirac matrices vanish:
\bq
 \mathrm{Tr}\left(\gamma_{(D)}^{\mu_1} \gamma_{(D)}^{\mu_2} ... \gamma_{(D)}^{\mu_{2 n-1}}\right) 
 & = & 0
\eq
\item For traces involving $\gamma_5$ we have
\bq
 \mathrm{Tr}\left(\gamma_5\right) 
 & = & 0,
 \nonumber \\
 \mathrm{Tr}\left(\gamma_{(D)}^{\mu} \gamma_{(D)}^{\nu} \gamma_5\right) 
 & = & 0,
 \nonumber \\
 \mathrm{Tr}\left(\gamma_{(D)}^{\mu} \gamma_{(D)}^{\nu} \gamma_{(D)}^{\rho} \gamma_{(D)}^{\sigma} \gamma_5\right) 
 & = & 
 \left\{ \begin{array}{ll}
 4 i \eps^{\mu \nu \rho \sigma}, & \mu,\nu,\rho,\sigma \in \{0,1,2,3\}, \\
 0, & \mbox{otherwise}. \\
 \end{array} \right.
\eq
\end{enumerate}
{\bf Solution}: 
The proof of point 1 follows exactly the proof in four space-time dimensions.
Note that we assume the normalisation as in eq.~(\ref{chapter_qft:Dirac_normalisation})
\bq
 \mathrm{Tr}\left( {\bf 1} \right)
 & = & 4,
\eq
otherwise there would be small modifications.

In order to prove that the trace of an odd number of Dirac matrices vanishes, we proceed by induction.
For $n=1$ we have
\bq
 D \mathrm{Tr}\left(\gamma_{(D)}^{\mu}\right) 
 & = &
 \mathrm{Tr}\left(\gamma^{(D)}_{\nu} \gamma_{(D)}^{\nu} \gamma_{(D)}^{\mu}\right) 
 \; = \;
 2 g_{(D)}^{\mu\nu} \mathrm{Tr}\left(\gamma^{(D)}_{\nu} \right) 
 -
 \mathrm{Tr}\left(\gamma^{(D)}_{\nu} \gamma_{(D)}^{\mu} \gamma_{(D)}^{\nu} \right) 
 \nonumber \\
 & = &
 \left(2-D\right) \mathrm{Tr}\left(\gamma_{(D)}^{\mu} \right) 
\eq
and hence
\bq
 2 \left(D-1\right) \mathrm{Tr}\left(\gamma_{(D)}^{\mu}\right) 
 & = &
 0.
\eq
As this has to hold for any $D$, we conclude
\bq
 \mathrm{Tr}\left(\gamma_{(D)}^{\mu}\right) 
 & = &
 0.
\eq
Let us now assume that a trace over $(2n-3)$ Dirac matrices vanishes.
We consider
\bq
\lefteqn{
 D \mathrm{Tr}\left(\gamma_{(D)}^{\mu_1} \gamma_{(D)}^{\mu_2} ... \gamma_{(D)}^{\mu_{2 n-1}}\right) 
 =  
 \mathrm{Tr}\left(\gamma^{(D)}_{\nu} \gamma_{(D)}^{\nu} \gamma_{(D)}^{\mu_1} \gamma_{(D)}^{\mu_2} ... \gamma_{(D)}^{\mu_{2 n-1}}\right) 
 }
 \nonumber \\
 & = &
 2 \sum\limits_{j=1}^{2n-1} \left(-1\right)^{j-1}
 \mathrm{Tr}\left(\gamma_{(D)}^{\mu_j} \gamma_{(D)}^{\mu_1} ... \gamma_{(D)}^{\mu_{j-1}} \gamma_{(D)}^{\mu_{j+1}} ... \gamma_{(D)}^{\mu_{2 n-1}}\right) 
 - D \mathrm{Tr}\left(\gamma_{(D)}^{\mu_1} \gamma_{(D)}^{\mu_2} ... \gamma_{(D)}^{\mu_{2 n-1}}\right) 
 \nonumber \\
 & = &
 \left[2 \left(2n-1\right)- D \right] \mathrm{Tr}\left(\gamma_{(D)}^{\mu_1} \gamma_{(D)}^{\mu_2} ... \gamma_{(D)}^{\mu_{2 n-1}}\right),
\eq
and hence
\bq
 2 \left( D - 2n + 1\right)\mathrm{Tr}\left(\gamma_{(D)}^{\mu_1} \gamma_{(D)}^{\mu_2} ... \gamma_{(D)}^{\mu_{2 n-1}}\right) 
 & = & 0,
\eq
from which the claim follows.

The proof of point 3 is to a large extent identical to the proof for four space-time dimensions.
It remains to show that
\bq
 \mathrm{Tr}\left(\gamma_{(D)}^{\mu} \gamma_{(D)}^{\nu} \gamma_{(D)}^{\rho} \gamma_{(D)}^{\sigma} \gamma_5\right) 
 & = & 0,
\eq
if at least one index is not an element of $\{0,1,2,3\}$.
Assume $\mu \notin \{0,1,2,3\}$. We insert the definition of $\gamma_5$
\bq
 \gamma_5 & = & i \gamma_{(D)}^0 \gamma_{(D)}^1 \gamma_{(D)}^2 \gamma_{(D)}^3
\eq
and evaluate a trace over eight Dirac matrices.
Each term which does nor vanish for other reasons, will contain a factor
\bq
 g^{\mu \tau},
\eq
with $\mu \notin \{0,1,2,3\}$ and $\tau \in \{0,1,2,3\}$.
As the metric tensor is diagonal, any non-diagonal element vanishes.
}
\es
\\
\\
\bs
{\it \refstepcounter{exercise}
{\bf Exercise \theexercise}: 
Derive eq.~(\ref{chapter_qft:trace_anomaly}).
\\
\\
{\bf Solution}: 
In order to show
\bq
 \mathrm{Tr}\; {\slashed q}_1 \gamma_\beta {\slashed q}_0 \gamma_\alpha {\slashed q}_2 {\slashed k}_{(-2\eps)} \gamma_5 
 & = & 
 k_{(-2\eps)}^2 \cdot 4 i \eps_{\alpha\lambda\beta\kappa} p_1^\lambda p_2^\kappa + \dots,
 \nonumber \\
 \mathrm{Tr}\; {\slashed q}_2 \gamma_\alpha {\slashed q}_0 \gamma_\beta {\slashed q}_1 {\slashed k}_{(-2\eps)} \gamma_5
 & = & 
 k_{(-2\eps)}^2 \cdot 4 i \eps_{\alpha\lambda\beta\kappa} p_1^\lambda p_2^\kappa + \dots,
\eq
we decompose $q_0$, $q_1$ and $q_2$ into a four-dimension part and a $(-2\eps)$-dimensional part:
\bq
 q_0 \; = \; k_{(4)} + k_{(-2\eps)},
 \;\;\;\;\;\;
 q_1 \; = \; k_{(4)} + k_{(-2\eps)} - p_2,
 \;\;\;\;\;\;
 q_2 \; = \; k_{(4)} + k_{(-2\eps)} + p_1.
\eq
The external momenta $p_1$ and $p_2$ are four-dimensional.
Inside the traces we already have one ${\slashed k}_{(-2\eps)}$, we need to pick up from the substitution of $q_0, q_1, q_2$
exactly one other $(-2\eps)$-dimensional part.
In all other cases the traces vanish.
Thus
\bq
\lefteqn{
 \mathrm{Tr}\; {\slashed q}_1 \gamma_\beta {\slashed q}_0 \gamma_\alpha {\slashed q}_2 {\slashed k}_{(-2\eps)} \gamma_5 
 =  
 \mathrm{Tr}\; {\slashed k}_{(-2\eps)} \gamma_\beta {\slashed k}_{(4)} \gamma_\alpha \left({\slashed k}_{(4)} + {\slashed p}_1 \right) {\slashed k}_{(-2\eps)} \gamma_5 
 } & & \\
 & &
 +
 \mathrm{Tr}\; \left({\slashed k}_{(4)} - {\slashed p}_2 \right) \gamma_\beta {\slashed k}_{(-2\eps)} \gamma_\alpha \left({\slashed k}_{(4)} + {\slashed p}_1 \right) {\slashed k}_{(-2\eps)} \gamma_5 
 +
 \mathrm{Tr}\; \left({\slashed k}_{(4)} - {\slashed p}_2 \right) \gamma_\beta {\slashed k}_{(4)} \gamma_\alpha {\slashed k}_{(-2\eps)} {\slashed k}_{(-2\eps)} \gamma_5.
 \nonumber 
\eq
We permute the ${\slashed k}_{(-2\eps)}$ next to each other and use eq.~(\ref{chapter_qft:Dirac_trace_rule_3_D_dim}):
\bq
\label{appendix_solutions:eval_trace_1}
\lefteqn{
 \mathrm{Tr}\; {\slashed q}_1 \gamma_\beta {\slashed q}_0 \gamma_\alpha {\slashed q}_2 {\slashed k}_{(-2\eps)} \gamma_5 
 = } & &
 \nonumber \\
 & &
 \left( k_{(-2\eps)}^2 \right)
 \cdot 
 4 i \eps_{\alpha \lambda \beta \kappa}
\left[
 k_{(4)}^\kappa \left( k_{(4)}^\lambda + p_1^\lambda \right)  
 -
 \left( k_{(4)}^\kappa - p_2^\kappa \right) \left( k_{(4)}^\lambda + p_1^\lambda \right)
 +
 \left( k_{(4)}^\kappa - p_2^\kappa \right) k_{(4)}^\lambda
 \right].
\eq
Here we used the fact, that traces like
\bq
 \mathrm{Tr}\; \left({\slashed k}_{(4)} - {\slashed p}_2 \right) \gamma_\beta \left({\slashed k}_{(4)} + {\slashed p}_1 \right) {\slashed k}_{(-2\eps)} \gamma_5 
 & = & 0
\eq
vanish.
In eq.~(\ref{appendix_solutions:eval_trace_1})
terms quadratic in $k_{(4)}$ vanish due to the contraction with the totally anti-symmetric tensor,
terms linear in $k_{(4)}$ vanish after integration.
Thus
\bq
 \mathrm{Tr}\; {\slashed q}_1 \gamma_\beta {\slashed q}_0 \gamma_\alpha {\slashed q}_2 {\slashed k}_{(-2\eps)} \gamma_5 
 & = &
 \left( k_{(-2\eps)}^2 \right)
 \cdot 
 4 i \eps_{\alpha \lambda \beta \kappa}
 p_1^\lambda p_2^\kappa + \dots.
\eq
The derivation of
\bq
 \mathrm{Tr}\; {\slashed q}_2 \gamma_\alpha {\slashed q}_0 \gamma_\beta {\slashed q}_1 {\slashed k}_{(-2\eps)} \gamma_5
 & = & 
 k_{(-2\eps)}^2 \cdot 4 i \eps_{\alpha\lambda\beta\kappa} p_1^\lambda p_2^\kappa + \dots
\eq
follows along the same lines.
}
\es
\\
\\
\bs
{\it \refstepcounter{exercise}
{\bf Exercise \theexercise}: 
Show
\bq
 \int \frac{d^Dk}{(2 \pi)^D i} \frac{k_{(-2\eps)}^2}{k_0^2 k_1^2 k_2^2} 
 & = & 
 - \frac{1}{2} \frac{1}{(4 \pi)^2}
 + {\mathcal O}\left(\eps\right).
\eq
{\bf Solution}: 
From eq.~(\ref{chapter_basics:k2_eps_numerator}) it follows that
\bq
 \int \frac{d^Dk}{(2 \pi)^D i} \frac{k_{(-2\eps)}^2}{k_0^2 k_1^2 k_2^2} 
 & = & 
 4 \pi \eps
 \int \frac{d^{D+2}k}{(2 \pi)^{D+2} i} \frac{1}{k_0^2 k_1^2 k_2^2}.
\eq
The Feynman parameter representation of the integral in $D+2=6-2\eps$ space-time dimensions
reads
\bq
 \int \frac{d^{D+2}k}{(2 \pi)^{D+2} i} \frac{1}{k_0^2 k_1^2 k_2^2}
 & = &
 - \frac{\Gamma\left(\eps\right)}{\left(4\pi\right)^{3-\eps}} 
 \left(\mu^2\right)^{-\eps}
 \int\limits_{a_j \ge 0} d^3a \; \delta\left(1-a_0-a_1-a_2\right) \;  
 \frac{1}{\left[ {\mathcal F}\left(a\right) \right]^{\eps}}
\eq
with
\bq
 {\mathcal F}
 & = & 
 a_0 a_2 \left(\frac{-p_1^2}{\mu^2} \right)
 + a_0 a_1 \left(\frac{-p_2^2}{\mu^2} \right)
 + a_1 a_2 \left(\frac{-\left(p_1+p_2\right)^2}{\mu^2} \right).
\eq
We only need the pole part. The prefactor $\Gamma(\eps)$ delivers a pole in $\eps$.
The Feynman parameter integral is finite and we may therefore set $\eps=0$ in the exponent of ${\mathcal F}$.
Then the Feynman parameter integral becomes trivial:
\bq
 \int\limits_{a_j \ge 0} d^3a \; \delta\left(1-a_0-a_1-a_2\right) \;  
 \frac{1}{\left[ {\mathcal F}\left(a\right) \right]^{\eps}}
 & = &
 \int\limits_{a_j \ge 0} d^3a \; \delta\left(1-a_0-a_1-a_2\right) 
 + {\mathcal O}\left(\eps\right)
 \; = \;
 \frac{1}{2} + {\mathcal O}\left(\eps\right).
 \;\;\;\;\;\;\;\;
\eq
Thus
\bq
 \int \frac{d^{D+2}k}{(2 \pi)^{D+2} i} \frac{1}{k_0^2 k_1^2 k_2^2}
 & = &
 - \frac{1}{2\eps}
 \frac{1}{\left(4\pi\right)^{3}} 
 + {\mathcal O}\left(\eps^0\right)
\eq
and hence we obtain in $D=4-2\eps$ space-time dimensions
\bq
 \int \frac{d^Dk}{(2 \pi)^D i} \frac{k_{(-2\eps)}^2}{k_0^2 k_1^2 k_2^2} 
 & = & 
 - \frac{1}{2} \frac{1}{(4 \pi)^2}
 + {\mathcal O}\left(\eps\right).
\eq
}
\es
\\
\\
%
%
\bs
{\it \refstepcounter{exercise}
{\bf Exercise \theexercise}: 
Reduce the tensor integral
\bq
 A^{\mu_1 \mu_2 \mu_3 \mu_4}(m) 
 & = &
 e^{\eps \Eulerconstant} \mu^{2\eps}
 \int \frac{d^{D}k}{i \pi^{D/2}}
        \frac{k^{\mu_1} k^{\mu_2} k^{\mu_3} k^{\mu_4}}{(-k^{2}+m^{2})}
\eq
to $A_0(m)$.
\\
\\
{\bf Solution}: 
The integral does not depend on any external momenta, so it must be proportional to a symmetric tensor build from 
$g^{\mu \nu}$:
\bq
 A^{\mu_1 \mu_2 \mu_3 \mu_4}(m) 
 & = &
  \left( g^{\mu_1\mu_2} g^{\mu_3\mu_4} + g^{\mu_1\mu_3} g^{\mu_2\mu_4} + g^{\mu_1\mu_4} g^{\mu_2\mu_3} \right) 
 A_4(m).
\eq
Contraction with $g_{\mu_1\mu_2} g_{\mu_3\mu_4}$ yields
\bq
 A_4(m)
 & = &
 \frac{1}{D\left(D+2\right)}
 e^{\eps \Eulerconstant} \mu^{2\eps}
 \int \frac{d^{D}k}{i \pi^{D/2}}
        \frac{\left(k^2\right)^2}{(-k^{2}+m^{2})}.
\eq
We further have
\bq
 \int \frac{d^{D}k}{i \pi^{D/2}}
        \frac{\left(k^2\right)^2}{(-k^{2}+m^{2})}
 & = &
 \left(m^2\right)^2
 \int \frac{d^{D}k}{i \pi^{D/2}}
        \frac{1}{(-k^{2}+m^{2})}
 - m^2 \int \frac{d^{D}k}{i \pi^{D/2}}
 + \int \frac{d^{D}k}{i \pi^{D/2}} (-k^{2})
 \nonumber \\
 & = &
 \left(m^2\right)^2
 \int \frac{d^{D}k}{i \pi^{D/2}}
        \frac{1}{(-k^{2}+m^{2})},
\eq
since the scaleless integrals vanish for $D \neq 0,-2$ (see eq.~(\ref{chapter_basics:basic_eq_negative_dimensions})).
Thus
\bq
 A^{\mu_1 \mu_2 \mu_3 \mu_4}(m) 
 & = &
 \frac{\left(m^2\right)^2}{D\left(D+2\right)}
  \left( g^{\mu_1\mu_2} g^{\mu_3\mu_4} + g^{\mu_1\mu_3} g^{\mu_2\mu_4} + g^{\mu_1\mu_4} g^{\mu_2\mu_3} \right) 
 A_0(m). 
\eq
}
\es
\\
\\
\bs
{\it \refstepcounter{exercise}
{\bf Exercise \theexercise}: 
Reduce 
\bq
 g_{\mu_1 \mu_2} g_{\mu_3 \mu_4} C^{\mu_1 \mu_2 \mu_3 \mu_4}(p_{1},p_{2},0,0,0)
 & = & 
 - g_{\mu_1 \mu_2} g_{\mu_3 \mu_4} 
 e^{\eps \Eulerconstant} \mu^{2\eps}
 \int \frac{d^{D}k}{i \pi^{D/2}}
        \frac{k^{\mu_1} k^{\mu_2} k^{\mu_3} k^{\mu_4}}{k^{2}
                 (k-p_{1})^{2}
                 (k-p_{1}-p_{2})^{2}}
 \;\;\;\;\;\;\;\;\;
\eq
to scalar integrals.
\\
\\
{\bf Solution}: 
We have
\bq
 g_{\mu_1 \mu_2} g_{\mu_3 \mu_4} C^{\mu_1 \mu_2 \mu_3 \mu_4}(p_{1},p_{2},0,0,0)
 & = & 
 - e^{\eps \Eulerconstant} \mu^{2\eps}
 \int \frac{d^{D}k}{i \pi^{D/2}}
        \frac{\left(k^2\right)^2}{k^{2}
                 (k-p_{1})^{2}
                 (k-p_{1}-p_{2})^{2}}.
\eq
There are two powers of $k^2$ in the numerator. One factor of $k^2$ in the numerator cancels the factor
of $k^2$ in the denominator:
\bq
\label{appendix_solutions:reduced_two_point_function}
 g_{\mu_1 \mu_2} g_{\mu_3 \mu_4} C^{\mu_1 \mu_2 \mu_3 \mu_4}(p_{1},p_{2},0,0,0)
 & = & 
 - e^{\eps \Eulerconstant} \mu^{2\eps}
 \int \frac{d^{D}k}{i \pi^{D/2}}
        \frac{k^2}{
                 (k-p_{1})^{2}
                 (k-p_{1}-p_{2})^{2}}.
\eq
However, there is no factor $k^2$ left in the denominator to cancel the one in the numerator.
The purpose of this exercise is to show how to handle this case.
We realise that the integral in eq.~(\ref{appendix_solutions:reduced_two_point_function})
is no longer a three-point function, but just a two-point function.
With $k'=k-p_1$ we have
\bq
 g_{\mu_1 \mu_2} g_{\mu_3 \mu_4} C^{\mu_1 \mu_2 \mu_3 \mu_4}(p_{1},p_{2},0,0,0)
 & = & 
 - e^{\eps \Eulerconstant} \mu^{2\eps}
 \int \frac{d^{D}k'}{i \pi^{D/2}}
        \frac{\left(k'+p_1\right)^2}{
                 k'{}^{2}
                 (k'-p_{2})^{2}}
 \\
 & = &
 - g_{\mu_1\mu_2} B^{\mu_1\mu_2}(p_2,0,0)
 - 2 p_{1,\mu_1} B^{\mu_1}(p_2,0,0)
 - p_1^2 B_0(p_2,0,0).
 \nonumber
\eq
With
\bq
 g_{\mu_1\mu_2} B^{\mu_1\mu_2}(p_2,0,0) & = & 0
\eq
and 
\bq
 B^{\mu_1}(p_2,0,0)
 & = &
 \frac{1}{2} p_2^{\mu_1} B_0(p_2,0,0)
\eq
we finally obtain
\bq
 g_{\mu_1 \mu_2} g_{\mu_3 \mu_4} C^{\mu_1 \mu_2 \mu_3 \mu_4}(p_{1},p_{2},0,0,0)
 & = & 
 - p_1 \cdot \left( p_1 + p_2 \right) B_0(p_2,0,0).
 \nonumber
\eq
}
\es
\\
\\
\bs
{\it \refstepcounter{exercise}
{\bf Exercise \theexercise}: 
Determine the constants $\alpha_1$ and $\alpha_2$ in eq.~(\ref{chapter_one_loop_def_l_1_l_2}) from the requirement that $l_1$ and $l_2$ are light-like, 
i.e. $l_1^2=l_2^2=0$.
Distinguish the cases
\begin{description}
\item{(i)} $p_i$ and $p_j$ are light-like.
\item{(ii)} $p_i$ is light-like, $p_j$ is not.
\item{(iii)} both $p_i$ and $p_j$ are not light-like.
\end{description}
\noindent
{\bf Solution}:
We start with the case (i): If $p_i^2=p_j^2=0$ there is not much to be done:
We set $\alpha_1=\alpha_2=0$ and
\bq
 l_1 \; = \; p_i, & & l_2 \; = \; p_j
\eq
is the desired solution.
Let now consider the case (ii): We assume $p_i^2=0$ and $p_j^2 \neq 0$.
(The case $p_i^2 \neq 0$, $p_j^2=0$ is similar and obtained by $p_i \leftrightarrow p_j$.)
With
\bq
 \alpha_1 \; = \; 0,
 & &
\alpha_2 \; = \; \frac{p_j^2}{2p_i p_j}
\eq
we have
\bq
l_1 = p_i, 
& &
l_2 = -\alpha_2 p_i + p_j.
\eq
We verify that $l_2$ is light-like:
\bq
 l_2^2 & =& p_j^2 - 2 \alpha_2 p_i \cdot p_j \; = \; 0.
\eq
Let us now turn to case (iii): 
We assume that both $p_i$ and $p_j$ are not light-like,
i.e. $p_i^2 \neq 0$ and $p_j^2 \neq 0$.
For 
$2p_ip_j>0$ we set 
\bq
\alpha_1 = \frac{2p_ip_j-\sqrt{\Delta}}{2p_j^2},
 & &
\alpha_2 = \frac{2p_ip_j-\sqrt{\Delta}}{2p_i^2}.
\eq
For $2p_ip_j<0$ we set
\bq
\alpha_1 = \frac{2p_ip_j+\sqrt{\Delta}}{2p_j^2},
 & &
\alpha_2 = \frac{2p_ip_j+\sqrt{\Delta}}{2p_i^2}.
\eq
Here,
\bq
\Delta & = & \left( 2p_ip_j \right)^2 - 4p_i^2 p_j^2.
\eq
The signs are chosen in such away that the light-like limit
$p_i^2 \rightarrow 0$ (or $p_j^2 \rightarrow 0$) is approached smoothly.
Note that $l_1$, $l_2$ are real for $\Delta>0$.
For $\Delta<0$, $l_1$ and $l_2$ acquire imaginary parts.
}
\es
\\
\\
\bs
{\it \refstepcounter{exercise}
{\bf Exercise \theexercise}: 
The method above does not apply to a tensor two-point function, as there is only one linear independent external
momentum.
However, the tensor two-point functions is easily reduced with standard methods to the scalar two-point function.
In this exercise you are asked to work this out for the massless tensor two-point function.
The most general massless tensor two-point function is given by
\bq
I_2^{\mu_1 ... \mu_r,s} & = &
 e^{\eps \Eulerconstant} \mu^{2\eps} \int \frac{d^Dk}{i \pi^{\frac{D}{2}}}
  \left(-k_{(-2\eps)}^2\right)^s \frac{k^{\mu_1} ... k^{\mu_r}}{k^2 (k-p)^2}.
\eq
Reduce this tensor integral to a scalar integral.
\\
\\
{\bf Solution}: We first use Feynman parametrisation and obtain
\bq
I_2^{\mu_1 ... \mu_r,s} 
 & = &
 e^{\eps \Eulerconstant} \mu^{2\eps} \int\limits_0^1 da \int \frac{d^Dk}{i \pi^{\frac{D}{2}}}
   \left(-k_{(-2\eps)}^2\right)^s
   \left( k+a p \right)^{\mu_1} ... \left( k+a p \right)^{\mu_r} 
   \left[ -k^2 + a(1-a)\left(-p^2\right) \right]^{-2}.
 \nonumber
\eq
Expanding $\left( k+a p \right)^{\mu_1} ... \left( k+a p \right)^{\mu_r}$ yields terms
of the form
\bq
a^{r-2t} k^{\mu_{\sigma(1)}} ... k^{\mu_{\sigma(2t)}} p^{\mu_{\sigma(2t+1)}} ... p^{\mu_{\sigma(r)}}.
\eq
Note that terms with an odd number of $k^{\mu}$'s vanish after integration (see eq.~(\ref{chapter_qft:odd_power_loop_momenta})).
For terms with an even number of $k^{\mu}$'s let us recall eq.~(\ref{chapter_qft:symmetric_integration})
and let us generalise the formulae given there to arbitrary even tensor rank.
We find 
\bq
\int \frac{d^Dk}{i \pi^{\frac{D}{2}}} k^{\mu_1} ... k^{\mu_{2w}} f(k^2)
 & = & 
  2^{-w} \frac{\Gamma\left(\frac{D}{2}\right)}{\Gamma\left(\frac{D}{2}+w\right)}
  \left( g^{\mu_1 \mu_2} ... g^{\mu_{2w-1} \mu_{2w}} + \mbox{permutations}
                         \right)
  \int \frac{d^Dk}{i \pi^{\frac{D}{2}}} \left( k^2 \right)^w f(k^2).
 \nonumber
\eq
The fully symmetric tensor structure 
\bq
 S^{\mu_1 .... \mu_{2w}} & = & g^{\mu_1 \mu_2} ... g^{\mu_{2w-1} \mu_{2w}} + \mbox{permutations}
\eq
has $(2w-1)!!=(2w-1)(2w-3)...1$ terms.
We obtain in the absence of powers of $k_{(-2\eps)}^2$
\bq
\label{appendix_solutions:scal_two_point_no_eps}
\lefteqn{
 e^{\eps \Eulerconstant} \mu^{2\eps} \int\limits_0^1 da \; a^{r-2t}
   \int \frac{d^Dk}{i \pi^{\frac{D}{2}}}
   k^{\mu_1} ... k^{\mu_{2t}}
   \left[ -k^2 + a(1-a)\left(-p^2\right) \right]^{-2}
 = } & &\nonumber \\
& = &
 \left(-\frac{p^2}{2} \right)^t S^{\mu_1 .... \mu_{2t}}
 \frac{\Gamma(1+r-t-\eps) \Gamma(2-2\eps)}
      {\Gamma(1-\eps) \Gamma(2+r-2\eps)} I_2
 \nonumber \\
& = &
 \left( - \frac{p^2}{2} \right)^t S^{\mu_1 .... \mu_{2t}}
 \frac{(r-t)!}{(r+1)!} \left\{ 1 + \eps \left[ 2 S_1(r+1) - S_1(r-t) - 2 \right] + {\cal O}(\eps^2) \right\} I_2,
\eq
where $S_1(n)$ is a harmonic sum
\bq
 S_1(n) & = & \sum\limits_{j=1}^n \frac{1}{j},
\eq
and $I_2$ is the scalar two-point function:
\bq
I_2 & = & 
 e^{\eps \Eulerconstant} \left( \frac{-p^2}{\mu^2} \right)^{-2\eps}
  \frac{\Gamma(-\eps) \Gamma(1-\eps)^2}{\Gamma(2-2\eps)}
 = 
 \frac{1}{\eps} + 2 - \ln\left( \frac{-p^2}{\mu^2} \right) + {\cal O}(\eps).
\eq
Since $I_2$ starts at $1/\eps$ we can neglect ${\cal O}(\eps^2)$ terms in eq.~(\ref{appendix_solutions:scal_two_point_no_eps}).
If powers of $k_{(-2\eps)}^2$ are present, we obtain if all indices are contracted into four-dimensional quantities
\bq
\lefteqn{
 e^{\eps \Eulerconstant} \mu^{2\eps} \int\limits_0^1 da \; a^{r-2t}
   \int \frac{d^Dk}{i \pi^{\frac{D}{2}}}
   \left(-k_{(-2\eps)}^2\right)^s
   k^{\mu_1} ... k^{\mu_{2t}}
   \left[ -k^2 + a(1-a)\left(-p^2\right) \right]^{-2}
 = } & &\nonumber \\
& = &
 - \eps \left( p^2 \right)^s \left( - \frac{p^2}{2} \right)^t 
  S^{\mu_1 .... \mu_{2t}}
 \frac{(s-1)! (r+s-t)!}{(r+2s+1)!} I_2 + {\cal O}(\eps)
 \nonumber \\
 & &
 + \mbox{terms, which vanish when contracted into $4$-dimensional quantities}.
\eq
}
\es
\\
\\
\bs
{\it \refstepcounter{exercise}
{\bf Exercise \theexercise}: 
Consider a one-loop four-point function with external momenta $p_1, p_2, p_3, p_4$ and 
$p_1^2=p_2^2=p_3^2=p_4^2=0$. The external momenta satisfy momentum conservation $p_1+p_2+p_3+p_4=0$.
For $j \in \{1,2,3,4\}$ set $q_j=k-p_j^{\mathrm{sum}}$.
Solve the equations for the quadruple cut
\bq
 q_{1}^2 = q_{2}^2 = q_{3}^2 = q_{4}^2 = 0.
\eq
Hint: Start from an ansatz
\bq
\label{appendix_solutions:ansatz_quadruple_cut}
 k_\mu & = & c \left\langle a- | \gamma_\mu | b- \right\rangle,
\eq
with $c \in {\mathbb C}$ and $a, b$ light-like.
\\
\\
{\bf Solution}: 
We have $q_4=k-p_1-p_2-p_3-p_4=k$ due to momentum conservation.
The ansatz of eq.~(\ref{appendix_solutions:ansatz_quadruple_cut}) automatically satisfies the equation $q_4^2=0$:
\bq
 q_4^2
 \; = \;
 k^2
 \; = \;
 c^2 \left\langle a- | \gamma_\mu | b- \right\rangle \left\langle a- | \gamma^\mu | b- \right\rangle
 \; = \;
 2 c^2 \left\langle a- | a+ \right\rangle \left[b+ | b- \right]
 \; = \; 0,
\eq
as $\langle a- | a+ \rangle = 0$ and $[b+ | b- ]=0$.
With $k^2=0$ (and $p_1^2=p_2^2=p_3^2=p_4^2=0$) the other three equations which we have to satisfy reduce to
\bq
 2 k \cdot p_1 & = & 0,
 \nonumber \\
 2 k \cdot p_2 & = & 2 p_1 \cdot p_2 ,
 \nonumber \\
 2 k \cdot p_4 & = & 0
\eq
Here we used $p_4=-p_1-p_2-p_3$.
Let's consider the first equation $2 k \cdot p_1 = 0$. If we choose $a=p_1$ (or $b=p_1$) this equation is trivially satisfied:
\bq
 c \left\langle p_1- | \gamma_\mu | b- \right\rangle p_1^\mu
 \; = \;
 c \left\langle p_1- | p_1+ \right\rangle \left[ p_1+ | b- \right] 
 \; = \; 0.
\eq
If we choose $a=p_1$ we may then use the freedom to choose $b$ to satisfy the third equation $2 k \cdot p_4=0$.
This will lead to the choice $b=p_4$.
We now have one remaining parameter $c$ left in our ansatz
\bq
 k_\mu & = & c \left\langle p_1- | \gamma_\mu | p_4- \right\rangle,
\eq
which we use to satisfy the second equation $2 k \cdot p_2 = 2 p_1 \cdot p_2$.
This yields
\bq
 c & = & \frac{\left[p_2 p_1 \right]}{\left[p_2 p_4 \right]}.
\eq
Of course, we could have chosen as well $a=p_4$ in the beginning. This will give us the second solution.
In summary we obtain the two solutions
\bq
 k_\mu^+ \; = \;
 \frac{\left[p_2 p_1 \right]}{\left[p_2 p_4 \right]} \left\langle p_1- | \gamma_\mu | p_4- \right\rangle,
 & &
 k_\mu^- \; = \;
 \frac{\left\langle p_1 p_2 \right\rangle}{\left\langle p_4 p_2 \right\rangle} \left\langle p_4- | \gamma_\mu | p_1- \right\rangle.
\eq
}
\es
\\
\\
\bs
{\it \refstepcounter{exercise}
{\bf Exercise \theexercise}: 
Consider the one-loop eight-point amplitude in massless $\phi^4$ theory.
Verify eq.~(\ref{chapter_one_loop:box_coefficient}) for the box coefficient.
\\
\\
{\bf Solution}: 
We consider the box coefficient, where the external momenta are distributed as in fig.~\ref{chapter_solutions:fig_box_coefficient}.
\begin{figure}
\begin{center}
\includegraphics[scale=1.0]{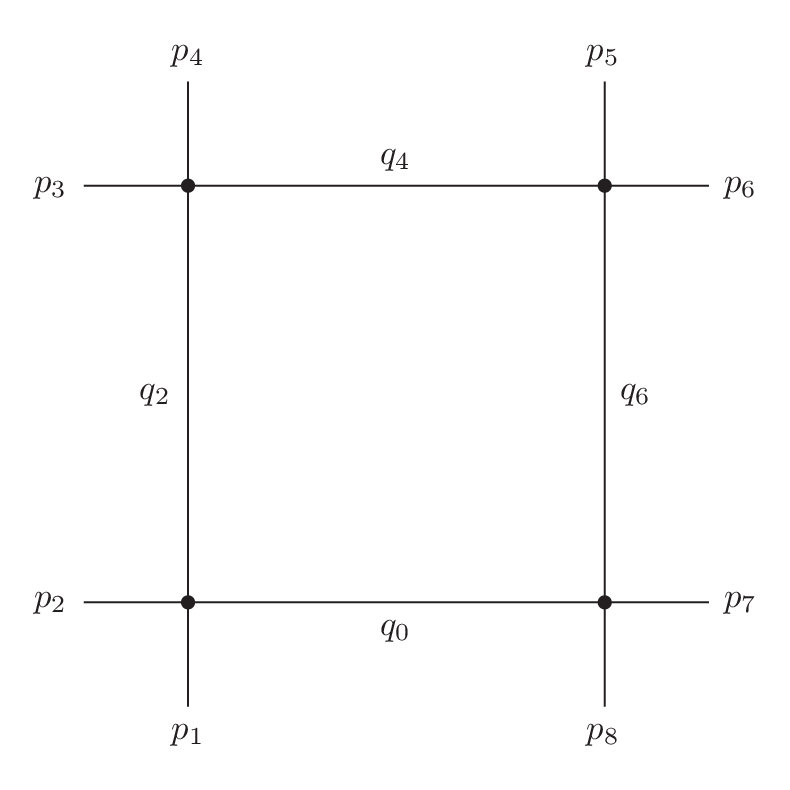}
\end{center}
\caption{
A diagram in $\phi^4$ theory.
}
\label{chapter_solutions:fig_box_coefficient}
\end{figure}
We first perform the calculation with standard Feynman rules. There is only one Feynman diagram contributing to the
coefficient $c^{(0)}_{0 2 4 6}$. This is the diagram shown in fig.~\ref{chapter_solutions:fig_box_coefficient}.
The contribution of this Feynman diagram to the one-loop amplitude is given by
(including a factor $(4\pi)^{-\eps} e^{\eps \Eulerconstant}$ corresponding to the $\overline{\mathrm{MS}}$-scheme)
\bq
 \left. i {\mathcal A}_{8}^{(1)} \right|_{\mathrm{Box}}
 & = & 
 \left( i \lambda \right)^4
 \left(4\pi \right)^{-\eps}
 e^{\eps \Eulerconstant} \mu^{2\eps} 
 \int \frac{d^Dk}{\left(2\pi\right)^D}
 \frac{i}{q_0^2}
 \frac{i}{q_2^2}
 \frac{i}{q_4^2}
 \frac{i}{q_6^2}
 \; = \;
 \frac{i \lambda^4}{\left(4\pi\right)^{2}} I_4^{(0 2 4 6)}.
\eq
The coefficient of the box integral is therefore given by
\bq
 c^{(0)}_{0 2 4 6} & = &
 \frac{\lambda^4}{\left(4\pi\right)^{2}}.
\eq
Let us now consider eq.~(\ref{chapter_one_loop:box_coefficient}).
There are four tree amplitudes. 
We have 
$I_1=\{1,2\}$,
$I_2=\{3,4\}$,
$I_3=\{5,6\}$ and
$I_4=\{7,8\}$.
In this simple example each tree amplitude equals
\bq
 i {\mathcal A}_{|I_1|+2}^{(0)}
 \; = \; 
 i {\mathcal A}_{|I_2|+2}^{(0)}
 \; = \; 
 i {\mathcal A}_{|I_3|+2}^{(0)}
 \; = \; 
 i {\mathcal A}_{|I_4|+2}^{(0)}
 \; = \; 
 i \lambda^4,
\eq
and is independent of the external momenta. Thus we do not need to know the concrete solutions of the
on-shell conditions
\bq
 q_0^2 
 \; = \;
 q_2^2 
 \; = \;
 q_4^2 
 \; = \;
 q_6^2 
 \; = \;
 0,
\eq
it suffices to know that there are two solutions.
We consider a scalar theory, hence there are no spins.
Eq.~(\ref{chapter_one_loop:box_coefficient}) gives us then
\bq
 c^{(0)}_{0 2 4 6}
 & = &
 \frac{1}{2}
 \frac{1}{\left(4\pi\right)^2}
 \sum\limits_{\sigma=\pm}
 {\mathcal A}_{|I_1|+2}^{(0)}
 {\mathcal A}_{|I_2|+2}^{(0)}
 {\mathcal A}_{|I_3|+2}^{(0)}
 {\mathcal A}_{|I_4|+2}^{(0)}
 \; = \;
 \frac{\lambda^4}{\left(4\pi\right)^{2}},
\eq
which agrees with our previous calculation.
}
\es
\\
\\
%
%
\bs
{\it \refstepcounter{exercise}
{\bf Exercise \theexercise}: 
Show that eq.~(\ref{chapter_iterated_integrals:basic_ibp_relation}) holds for $q=k$.
\\
Hint: Consider the scaling relation eq.~(\ref{chapter_basics:scaling_integral_measure}).
\\
\\
{\bf Solution}: 
The scaling relation reads
\bq
 \int \frac{d^Dk}{i \pi^{\frac{D}{2}}} f\left(k\right) 
 & = & 
 \left(1+\lambda\right)^D \int \frac{d^Dk}{i \pi^{\frac{D}{2}}} f\left( k+\lambda k\right).
\eq
The right-hand side has to be independent of $\lambda$.
This implies in particular that the ${\mathcal O}(\lambda)$-term has to vanish.
There are now two contributions to the ${\mathcal O}(\lambda)$-term.
On the one hand we have
\bq
 \left(1+\lambda\right)^D & = & 1 + D \lambda + {\mathcal O}\left(\lambda^2\right),
\eq
on the other hand we have
\bq
 f\left( k+\lambda k\right)
 & = &
 f\left(k\right) + \lambda k^\mu \frac{\partial}{\partial k^\mu} f\left(k\right) + {\mathcal O}\left(\lambda^2\right).
\eq
Note that
\bq
 k^\mu \frac{\partial}{\partial k^\mu} f\left(k\right)
 & = &
 \frac{\partial}{\partial k^\mu} \left[ k^\mu \cdot f\left(k\right) \right] - D f\left(k\right) 
\eq
and in summary the ${\mathcal O}(\lambda)$-term is given by
\bq
 \int \frac{d^Dk}{i \pi^{\frac{D}{2}}} \frac{\partial}{\partial k^\mu} \left[ k^\mu \cdot f\left(k\right) \right].
\eq
This proves eq.~(\ref{chapter_iterated_integrals:basic_ibp_relation}) for $q=k$.
}
\es
\\
\\
\bs
{\it \refstepcounter{exercise}
{\bf Exercise \theexercise}: 
Repeat the derivation with $q_{\mathrm{IBP}}=k$ and show
\bq
 \left( D - \nu_1 - 2 \nu_2 \right) I_{\nu_1 \nu_2}
 - \nu_1 I_{(\nu_1+1) (\nu_2-1)} + \nu_1 \left(2+x\right) I_{(\nu_1+1) \nu_2} 
 + 2 \nu_2 I_{\nu_1 (\nu_2+1)} 
 & = & 0.
\eq
{\bf Solution}: 
We have
\bq
 0
 & = &
 e^{\eps \Eulerconstant} \left(m^2\right)^{\nu_{12}-\frac{D}{2}}
 \int \frac{d^Dk}{i \pi^{\frac{D}{2}}} 
 \frac{\partial}{\partial k^\mu} 
 \frac{k^\mu}{\left(-q_1^2+m^2\right)^{\nu_1} \left(-q_2^2+m^2\right)^{\nu_2}}
 \nonumber \\
 & = &
 e^{\eps \Eulerconstant} \left(m^2\right)^{\nu_{12}-\frac{D}{2}}
 \int \frac{d^Dk}{i \pi^{\frac{D}{2}}} 
 \left[ 
 \frac{\nu_1 \left(q_1^2+q_2^2-p^2\right)}{\left(-q_1^2+m^2\right)^{\nu_1+1} \left(-q_2^2+m^2\right)^{\nu_2}}
 +
 \frac{2 \nu_2 q_2^2}{\left(-q_1^2+m^2\right)^{\nu_1} \left(-q_2^2+m^2\right)^{\nu_2+1}}
 \right. \nonumber \\
 & & \left.
 +
 \frac{D}{\left(-q_1^2+m^2\right)^{\nu_1} \left(-q_2^2+m^2\right)^{\nu_2}}
 \right]
 \nonumber \\
 & = &
 \nu_1 \left[ - I_{\nu_1 \nu_2} - I_{(\nu_1+1) (\nu_2-1)} + \left(2+x\right) I_{(\nu_1+1) \nu_2} \right]
 + 2 \nu_2 \left[ - I_{\nu_1 \nu_2} + I_{\nu_1 (\nu_2+1)} \right] 
 + D I_{\nu_1 \nu_2}.
\eq
Thus
\bq
 \left( D - \nu_1 - 2 \nu_2 \right) I_{\nu_1 \nu_2}
 - \nu_1 I_{(\nu_1+1) (\nu_2-1)} + \nu_1 \left(2+x\right) I_{(\nu_1+1) \nu_2} 
 + 2 \nu_2 I_{\nu_1 (\nu_2+1)} 
 & = & 0.
\eq
}
\es
\\
\\
\bs
{\it \refstepcounter{exercise}
{\bf Exercise \theexercise}: 
Derive the integration-by-parts identity for the integral
\bq
 I_{0 \nu_2}
 & = &
 e^{\eps \Eulerconstant} \left(m^2\right)^{\nu_{2}-\frac{D}{2}}
 \int \frac{d^Dk}{i \pi^{\frac{D}{2}}} 
 \frac{1}{\left(-k^2+m^2\right)^{\nu_2}}.
\eq
Verify the identity with the explicit result from eq.~(\ref{chapter_basics:result_tadpole}).
\\
\\
{\bf Solution}: The tadpole integral does not depend on any external momentum, hence the only 
integration-by-parts identity is
\bq
 0
 & = &
 e^{\eps \Eulerconstant} \left(m^2\right)^{\nu_{2}-\frac{D}{2}}
 \int \frac{d^Dk}{i \pi^{\frac{D}{2}}} 
 \frac{\partial}{\partial k^\mu} 
 \frac{k^\mu}{\left(-k^2+m^2\right)^{\nu_2}}
 \nonumber \\
 & = &
 e^{\eps \Eulerconstant} \left(m^2\right)^{\nu_{2}-\frac{D}{2}}
 \int \frac{d^Dk}{i \pi^{\frac{D}{2}}} 
 \left[
 \frac{2 \nu_2 k^2}{\left(-k^2+m^2\right)^{\nu_2+1}}
 +
 \frac{D}{\left(-k^2+m^2\right)^{\nu_2}}
 \right]
 \nonumber \\
 & = &
 \left(D- 2 \nu_2\right) I_{0 \nu_2} + 2 \nu_2 I_{0 (\nu_2+1)}.
\eq
From eq.~(\ref{chapter_basics:result_tadpole}) we have
\bq
 I_{0 \nu_2}
 & = &
 \frac{e^{\eps \Eulerconstant} \Gamma\left(\nu_2-\frac{D}{2}\right)}{\Gamma\left(\nu_2\right)}.
\eq
Substituting this into the integration-by-parts identity and using $\Gamma(x+1)=x\Gamma(x)$
we obtain
\bq
 \left(D- 2 \nu_2\right) I_{0 \nu_2} + 2 \nu_2 I_{0 (\nu_2+1)}
 & = &
 \left(D- 2 \nu_2\right) \frac{e^{\eps \Eulerconstant} \Gamma\left(\nu_2-\frac{D}{2}\right)}{\Gamma\left(\nu_2\right)} 
 + 2 \nu_2 \frac{e^{\eps \Eulerconstant} \Gamma\left(\nu_2+1-\frac{D}{2}\right)}{\Gamma\left(\nu_2+1\right)}
 \nonumber \\
 & = &
 \left(D- 2 \nu_2\right) \frac{e^{\eps \Eulerconstant} \Gamma\left(\nu_2-\frac{D}{2}\right)}{\Gamma\left(\nu_2\right)} 
 + 2 \left(\nu_2-\frac{D}{2}\right) \frac{e^{\eps \Eulerconstant} \Gamma\left(\nu_2-\frac{D}{2}\right)}{\Gamma\left(\nu_2\right)}
 \nonumber \\
 & = & 0.
\eq
}
\es
\\
\\
\bs
{\it \refstepcounter{exercise}
{\bf Exercise \theexercise}: 
Consider the double-box graph $G$ shown in fig.~\ref{chapter_basics:fig_doublebox} and the auxiliary graph $\tilde{G}$ with nine
propagators shown in fig.~\ref{chapter_basics:fig_doublebox_auxiliary}.
This exercise is about the family of Feynman integrals
\bq 
 I_{\nu_1 \nu_2 \nu_3 \nu_4 \nu_5 \nu_6 \nu_7 \nu_8 \nu_9}
\eq 
with $\nu_8, \nu_9 \le 0$. 
Use the notation of the momenta as in fig.~\ref{chapter_basics:fig_doublebox_auxiliary}.
Assume that all external momenta are light-like ($p_1^2=p_2^2=p_3^2=p_4^2=0$) and that all internal propagators are massless. 
Use one of the public available computer programs {\tt Kira}, {\tt Reduze} or {\tt Fire} to reduce the Feynman integral
\bq 
 I_{1 1 1 1 1 1 1 (-1) (-1)}
\eq 
to master integrals.
For the choice of master integrals you may use the default ordering criteria of the chosen computer program. 
\\
\\
{\bf Solution}:
We are considering the integral
\bq
 I_{\nu_1 \nu_2 \nu_3 \nu_4 \nu_5 \nu_6 \nu_7 \nu_8 \nu_9}
 & = &
 e^{2 \eps \Eulerconstant} \left(\mu^2\right)^{\nu-D}
 \int \frac{d^Dk_1}{i \pi^{\frac{D}{2}}} \frac{d^Dk_2}{i \pi^{\frac{D}{2}}} 
 \prod\limits_{j=1}^{9} \frac{1}{\left(-q_j^2\right)^{\nu_j}}
\eq
with
\begin{align}
 q_1 & = k_1-p_1, & q_2 & = k_1 - p_1 - p_2, & q_3 & = k_1, 
 \nonumber \\
 q_4 & = k_1+k_2, & q_5 & = k_2+p_1+p_2, & q_6 & = k_2,
 \nonumber \\
 q_7 & = k_2+p_1+p_2+p_3, & q_8 & = k_1-p_1-p_3, & q_9 & = k_2+p_1+p_3.
\end{align}
The aim of this exercise is to get acquainted with one of the integration-by-parts reduction programs
{\tt Kira}, {\tt Reduze} or {\tt Fire}.
We show how each of the three programs can be applied to the problem at hand.
The actual syntax may differ for different versions of the same program.
The solutions shown below refer to 
{\tt Kira} version 2.0, {\tt Reduze} version 2.4 and {\tt Fire} version 6.4.2.
In addition one should consult the manuals of these programs.
Please note that {\tt Kira}  and {\tt Reduze} use the convention
\bq
 \frac{1}{q_j^2-m_j^2}
 & \mbox{instead of} &
 \frac{1}{-q_j^2+m_j^2}
\eq
for propagators. This implies a minus sign for every integral where $\nu$ is odd between the {\tt Kira}/{\tt Reduze} notation
and the notation used in this book.
The CPU timings refer to a standard laptop with $2.6 \; \mathrm{GHz}$.

We start with {\tt Kira}.
We prepare the following files
{\footnotesize
\begin{verbatim}
 job.yaml
 myreduction.in
 config/integralfamilies.yaml
 config/kinematics.yaml
\end{verbatim}
}
\noindent
The files \verb|job.yaml| and \verb|myreduction.in| reside in a directory.
This directory also contains a subdirectory \verb|config|.
The \verb|config|-subdirectory contains the files \verb|integralfamilies.yaml| and \verb|kinematics.yaml|.

The file {\tt job.yaml} is the main file and specifies what should be done:
{\footnotesize
\begin{verbatim}
jobs:
  - reduce_sectors:
      reduce:
         - {topologies: [doublebox], sectors: [127], r: 8, s: 2}
      select_integrals:
        select_mandatory_recursively:
          - {topologies: [doublebox], sectors: [127], r: 8, s: 2}
  - kira2form:
     target:
     - [doublebox,myreductions.in]
     reconstruct_mass: true
\end{verbatim}
}
\noindent
The file {\tt myreduction.in} contains the integrals, which should be reduced. In our case it only 
contains a single line
{\footnotesize
\begin{verbatim}
doublebox[1,1,1,1,1,1,1,-1,-1]
\end{verbatim}
}
\noindent
We need to give the information on the family of Feynman integrals we are interested in.
This is done in the file {\tt integralfamilies.yaml}:
{\footnotesize
\begin{verbatim}
integralfamilies:
  - name: "doublebox"
    loop_momenta: [k1, k2]
    top_level_sectors: [127]
    propagators:
      - [ "k1-p1", 0 ]
      - [ "k1-p1-p2", 0 ]
      - [ "k1", 0 ]
      - [ "k1+k2", 0 ]
      - [ "k2+p1+p2", 0 ]
      - [ "k2", 0 ]
      - [ "k2+p1+p2+p3", 0 ]
      - [ "k1-p1-p3", 0 ]
      - [ "k2+p1+p3", 0 ]
\end{verbatim}
}
\noindent
Finally, we need to specify the kinematics. This is done in 
the file {\tt kinematics.yaml}:
{\footnotesize
\begin{verbatim}
kinematics : 
  incoming_momenta: []
  outgoing_momenta: [p1, p2, p3, p4]
  momentum_conservation: [p4,-p1-p2-p3]
  kinematic_invariants:
    - [s,  2]
    - [t,  2]
  scalarproduct_rules:
    - [[p1,p1],  0]
    - [[p2,p2],  0]
    - [[p3,p3],  0]
    - [[p1+p2,p1+p2],  s]
    - [[p2+p3,p2+p3],  t]
    - [[p1+p3,p1+p3],  -s-t]
  symbol_to_replace_by_one: s
\end{verbatim}
}
\noindent
With these preparations we may now run {\tt Kira}
with the command
{\footnotesize
\begin{verbatim}
 kira job.yaml
\end{verbatim}
}
\noindent
This will produce a file \verb|kira_myreductions.in.inc| in the directory {\tt results/doublebox}
with the content
{\footnotesize
\begin{verbatim}
id doublebox(1,1,1,1,1,1,1,-1,-1) = 
 + doublebox(1,1,1,1,1,1,1,-1,0)*((-4*t-3*s)*den(2))
 + doublebox(1,1,1,1,1,1,1,0,0)*(-t^2+(-s)*t)
 + doublebox(1,0,1,1,1,0,1,0,0)*((-18*t^2+(-27*s)*t-9*s^2)*den((2*s)*t))
 + doublebox(1,1,1,1,0,0,1,0,0)*(((12*d-36)*t+(9*d-27)*s)*den((d-4)*t))
 + doublebox(1,0,0,1,0,0,1,0,0)*(((162*d^3-1458*d^2+4356*d-4320)*t+(99*d^3-891*d^2+2662*d
 -2640)*s)*den(((2*d^3-24*d^2+96*d-128)*s)*t^2))
 + doublebox(1,0,0,1,1,1,0,0,0)*(((12*d^2-76*d+120)*t^2+((66*d^2-418*d+660)*s)*t+(27*d^2
 -171*d+270)*s^2)*den(((2*d^2-16*d+32)*s^2)*t))
 + doublebox(0,1,1,0,1,1,0,0,0)*(((4*d^2-16*d+12)*t+(4*d^2-16*d+12)*s)*den((d^2-8*d+16)
 *s^2))
 + doublebox(0,0,1,1,1,0,0,0,0)*(((-144*d^2+864*d-1280)*t^2+((54*d^3-630*d^2+2316*d-2720)
 *s)*t+(81*d^3-729*d^2+2178*d-2160)*s^2)*den(((2*d^3-24*d^2+96*d-128)*s^3)*t))
;
\end{verbatim}
}
\noindent
This gives the desired reduction in {\tt FORM} notation.
{\tt Kira} uses by default a ISP-basis.
There are eight master integrals.
The total run time on a standard laptop is about $30 \; \mathrm{s}$.

Let us now turn to {\tt Reduze}.
Using {\tt Reduze} we prepare the following files:
{\footnotesize
\begin{verbatim}
 job.yaml
 myreduction.in
 config/integralfamilies.yaml
 config/kinematics.yaml
 config/global.yaml
\end{verbatim}
}
\noindent
The syntax of {\tt Reduze} is very similar to the syntax of {\tt Kira}.
The file {\tt job.yaml} reads now
{\footnotesize
\begin{verbatim}
jobs:
  - setup_sector_mappings: {}
  - reduce_sectors:
      conditional: true
      sector_selection:
        select_recursively: [ [doublebox, 127] ]
      identities:
        ibp:
          - { r: [t, 8], s: [0, 2] }
        lorentz:
          - { r: [t, 8], s: [0, 2] }
        sector_symmetries:
          - { r: [t, 8], s: [0, 2] }
  - select_reductions:
      input_file: "myreductions.in"
      output_file: "myreductions.tmp.1"
  - reduce_files:
      equation_files:
          - "myreductions.tmp.1"
      output_file: "myreductions.tmp.2"
  - export:
      input_file: "myreductions.tmp.2"
      output_file: "myreductions.sol"
      output_format: "maple"
\end{verbatim}
}
\noindent
The file {\tt myreduction.in} contains again a list of the integrals to be reduced, for the case at hand
it is given by
{\footnotesize
\begin{verbatim}
{
INT["doublebox",{1,1,1,1,1,1,1,-1,-1}]
}
\end{verbatim}
}
\noindent
The file {\tt integralfamilies.yaml} specifies the family of Feynman integrals under consideration:
{\footnotesize
\begin{verbatim}
integralfamilies:
  - name: "doublebox"
    loop_momenta: [k1, k2]
    propagators:
      - [ "k1-p1", 0 ]
      - [ "k1-p1-p2", 0 ]
      - [ "k1", 0 ]
      - [ "k1+k2", 0 ]
      - [ "k2+p1+p2", 0 ]
      - [ "k2", 0 ]
      - [ "k2+p1+p2+p3", 0 ]
      - [ "k1-p1-p3", 0 ]
      - [ "k2+p1+p3", 0 ]
    permutation_symmetries: []
\end{verbatim}
}
\noindent
The file {\tt kinematics.yaml} is identical to the corresponding file for {\tt Kira}:
{\footnotesize
\begin{verbatim}
kinematics : 
  incoming_momenta: []
  outgoing_momenta: [p1,p2,p3,p4]
  momentum_conservation: [p4,-p1-p2-p3]
  kinematic_invariants:
    - [s,  2]
    - [t,  2]
  scalarproduct_rules:
    - [[p1,p1],  0]
    - [[p2,p2],  0]
    - [[p3,p3],  0]
    - [[p1+p2,p1+p2],  s]
    - [[p2+p3,p2+p3],  t]
    - [[p1+p3,p1+p3],  -s-t]
  symbol_to_replace_by_one: s
\end{verbatim}
}
\noindent
In addition, there is a file {\tt global.yaml} containing
{\footnotesize
\begin{verbatim}
global_symbols:
  space_time_dimension: d

paths:
  fermat: /usr/local/fermat/ferl64/fer64
\end{verbatim}
}
\noindent
The last line gives the absolute path to the {\tt Fermat}-executable and should 
be modified accordingly.
Running
{\footnotesize
\begin{verbatim}
 reduze job.yaml
\end{verbatim}
}
\noindent
will produce a file {\tt myreductions.sol}
{\footnotesize
\begin{verbatim}
myreductions := [

INT("doublebox",7,127,7,2,[1,1,1,1,1,1,1,-1,-1]) = 
  INT("doublebox",7,127,7,1,[1,1,1,1,1,1,1,-1,0]) * 
      (-2*t-3/2*s) +
  INT("doublebox",7,127,7,0,[1,1,1,1,1,1,1,0,0]) * 
      (-t^2-t*s) +
  INT("doublebox",5,93,5,0,[1,0,1,1,1,0,1,0,0]) * 
      (-9/2*(2*t^2+3*t*s+s^2)*t^(-1)*s^(-1)) +
  INT("doublebox",5,79,5,0,[1,1,1,1,0,0,1,0,0]) * 
      (-3*(4*t-t*d)^(-1)*(3*d*s-12*t+4*t*d-9*s)) +
  INT("doublebox",4,57,4,0,[1,0,0,1,1,1,0,0,0]) * 
      (1/2*(12*t^2*d^2+66*t*d^2*s+120*t^2-171*d*s^2-76*t^2*d+660*t*s+27*d^2*s^2+270*s^2
      -418*t*d*s)*(16*t-8*t*d+t*d^2)^(-1)*s^(-2)) +
  INT("doublebox",4,54,4,0,[0,1,1,0,1,1,0,0,0]) * 
      (-4*(16+d^2-8*d)^(-1)*(4*d*s-3*t+4*t*d-d^2*s-3*s-t*d^2)*s^(-2)) +
  INT("doubleboxx123",3,28,3,0,[0,0,1,1,1,0,0,0,0]) * 
      (-1/2*(2662*d*s-4320*t+4356*t*d+99*d^3*s-891*d^2*s+162*t*d^3-2640*s-1458*t*d^2)
      *(12*t^2*d^2+64*t^2-t^2*d^3-48*t^2*d)^(-1)*s^(-1)) +
  INT("doublebox",3,28,3,0,[0,0,1,1,1,0,0,0,0]) * 
      (1/2*(144*t^2*d^2+630*t*d^2*s+1280*t^2-54*t*d^3*s-2178*d*s^2-864*t^2*d+2720*t*s
      +729*d^2*s^2+2160*s^2-81*d^3*s^2-2316*t*d*s)*(64*t-48*t*d-t*d^3+12*t*d^2)^(-1)
      *s^(-3))
];
\end{verbatim}
}
\noindent
This gives the reduction in {\tt Maple} format.
{\tt Reduze} uses by default a ISP-basis.
The running time on a standard laptop is about $1270 \; \mathrm{s}$.

Let us now turn to {\tt Fire}.
In order to get the same number of master integrals we use it in combination with {\tt Litered} \cite{Lee:2012cn,Lee:2013mka}.
The program {\tt Litered} provides symmetry relations to {\tt Fire} and ensures that we end up in the example under consideration
with eight master integrals as above.
We prepare the following files:
{\footnotesize
\begin{verbatim}
 prepare1.m
 prepare2.m
 prepare3.m
 readout.m
 work/data.m
 work/myreductions.m
 work/doublebox.config
\end{verbatim}
}
\noindent
The file {\tt data.m} defines the family of Feynman integrals 
{\footnotesize
\begin{verbatim}
Internal = {k1, k2};
External = {p1, p2, p3};
Propagators = { -(k1-p1)^2, -(k1-p1-p2)^2, -k1^2, -(k1+k2)^2, -(k2+p1+p2)^2, -k2^2, 
-(k2+p1+p2+p3)^2, -(k1-p1-p3)^2, -(k2+p1+p3)^2 };
Replacements = { p1^2 -> 0, p2^2 -> 0, p3^2 -> 0, p1 p2 -> s/2, p2 p3 -> t/2, 
p1 p3 -> (-s-t)/2 };
\end{verbatim}
}
\noindent
The file {\tt myreductions.m} contains the list of Feynman integrals, which we would like to reduce. For the case at hand
{\footnotesize
\begin{verbatim}
{
 {1,{1,1,1,1,1,1,1,-1,-1}}
}
\end{verbatim}
}
\noindent
{\tt Fire} runs partly within {\tt Mathematica} and partly in {\tt C++}.
First, three preparation steps are done within {\tt Mathematica}, specified by the files
{\tt prepare1.m}, {\tt prepare2.m} and {\tt prepare3.m}.
The file {\tt prepare1.m} reads 
{\footnotesize
\begin{verbatim}
Get["FIRE6.m"];
Get["work/data.m"];
PrepareIBP[];
Prepare[AutoDetectRestrictions -> True];
SaveStart["work/doublebox"];
\end{verbatim}
}
\noindent
Issuing in {\tt Mathematica} the command
{\footnotesize
\begin{verbatim}
 Get["prepare1.m"];
\end{verbatim}
}
\noindent
will generate the file
{\footnotesize
\begin{verbatim}
 work/doublebox.start
\end{verbatim}
}
\noindent
The file {\tt prepare2.m} reads 
{\footnotesize
\begin{verbatim}
SetDirectory["extra/LiteRed/Setup/"];
Get["LiteRed.m"];
SetDirectory["../../../"];
Get["FIRE6.m"];
Get["work/data.m"];
CreateNewBasis[doublebox, Directory -> "work/litered"];
GenerateIBP[doublebox];
AnalyzeSectors[doublebox, {__,0,0}];
FindSymmetries[doublebox, EMs->True];
DiskSave[doublebox];
\end{verbatim}
}
\noindent
Issuing in {\tt Mathematica} the command
{\footnotesize
\begin{verbatim}
 Get["prepare2.m"];
\end{verbatim}
}
\noindent
will generate a subdirectory
{\footnotesize
\begin{verbatim}
 work/litered
\end{verbatim}
}
\noindent
containing several files.
The file {\tt prepare3.m} reads 
{\footnotesize
\begin{verbatim}
Get["FIRE6.m"];
LoadStart["work/doublebox"];
TransformRules["work/litered", "work/doublebox.lbases", 1];
SaveSBases["work/doublebox"]; 
\end{verbatim}
}
\noindent
Issuing in {\tt Mathematica} the command
{\footnotesize
\begin{verbatim}
 Get["prepare3.m"];
\end{verbatim}
}
\noindent
will generate the two files
{\footnotesize
\begin{verbatim}
 work/doublebox.lbases
 work/doublebox.sbases
\end{verbatim}
}
\noindent
After these preparation step the {\tt C++} program can be called.
We need a configuration file {\tt doublebox.config} containing
{\footnotesize
\begin{verbatim}
#threads           1
#variables         d, s, t
#start
#folder            work/
#problem           1 doublebox.sbases
#lbases            doublebox.lbases
#integrals         myreductions.m
#output            myreductions.tables
\end{verbatim}
}
\noindent
Running
{\footnotesize
\begin{verbatim}
 bin/FIRE6 -c work/doublebox
\end{verbatim}
}
\noindent
will generate the file
{\footnotesize
\begin{verbatim}
 work/myreductions.tables
\end{verbatim}
}
\noindent
The readout is again done within {\tt Mathematica}.
The file {\tt readout.m} reads
{\footnotesize
\begin{verbatim}
Get["FIRE6.m"];
LoadStart["work/doublebox", 1];
Burn[];
LoadTables["work/myreductions.tables"];
res = F[1, {1,1,1,1,1,1,1,-1,-1}];
Save["work/myreductions.out",res];
\end{verbatim}
}
\noindent
Issuing in Mathematica the command
{\footnotesize
\begin{verbatim}
 Get["readout.m"];
\end{verbatim}
}
\noindent
will generate the file \verb|myreductions.out| with
{\footnotesize
\begin{verbatim}
res = ((-10 + 3*d)*(-8 + 3*d)*(-1350*s^2 + 990*d*s^2 - 234*d^2*s^2 + 
        18*d^3*s^2 - 238*s*t - 56*d*s*t + 65*d^2*s*t - 9*d^3*s*t + 1616*t^2 - 
        1448*d*t^2 + 404*d^2*t^2 - 36*d^3*t^2)*
       G[1, {0, 0, 1, 1, 1, 0, 0, 0, 0}])/(2*(-5 + d)^2*(-4 + d)^3*s^3*t) - 
     (2*(-3 + d)*(46*s - 33*d*s + 5*d^2*s + 58*t - 40*d*t + 6*d^2*t)*
       G[1, {0, 1, 1, 0, 1, 1, 0, 0, 0}])/((-5 + d)*(-4 + d)^2*s^2) - 
     ((-3 + d)*(-10 + 3*d)*(-5760*s^2 + 2772*d*s^2 - 414*d^2*s^2 + 
        18*d^3*s^2 - 21744*s*t + 12352*d*s*t - 2326*d^2*s*t + 145*d^3*s*t - 
        13152*t^2 + 8368*d*t^2 - 1768*d^2*t^2 + 124*d^3*t^2)*
       G[1, {0, 1, 1, 1, 0, 0, 1, 0, 0}])/(8*(-6 + d)*(-5 + d)^2*(-4 + d)^2*
       s^2*t) + (5*(-3 + d)*(-10 + 3*d)*(-8 + 3*d)*(2*s + 3*t)*
       G[1, {1, 0, 0, 1, 0, 0, 1, 0, 0}])/((-4 + d)^3*s*t^2) + 
     (3*(-3 + d)*(64 - 18*d + d^2)*(3*s + 4*t)*
       G[1, {1, 0, 0, 1, 1, 1, 1, 0, 0}])/(2*(-6 + d)*(-5 + d)*(-4 + d)*t) - 
     (3*(s + t)*(3*s + 5*t)*G[1, {1, 0, 1, 1, 1, 0, 1, 0, 0}])/(s*t) + 
     ((-42*s^2 + 9*d*s^2 - 64*s*t + 14*d*s*t - 16*t^2 + 4*d*t^2)*
       G[1, {1, 1, 1, 1, 1, 1, 1, 0, 0}])/(4*(-4 + d)) - 
     ((-6 + d)*s*t*(3*s + 4*t)*G[1, {1, 1, 1, 1, 1, 1, 2, 0, 0}])/
      (4*(-5 + d)*(-4 + d))
\end{verbatim}
}
\noindent
{\tt Fire} uses by default a dot-basis.
The running time for the individual parts sum up to about $30 \; \mathrm{s}$ on a standard laptop.
}
\es
\\
\\
\bs
{\it \refstepcounter{exercise}
{\bf Exercise \theexercise}: 
Consider the example of the one-loop two-point function with equal internal masses, discussed 
below eq.~(\ref{chapter_iterated_integrals:example_equal_mass_bubble}).
Let 
\bq
 \vec{I} & = & \left( \begin{array}{c} I_{10}\left(D,x\right) \\ I_{11}\left(D,x\right) \\ \end{array} \right)
\eq
be a basis in $D$ space-time dimensions and 
\bq
 \vec{I}' & = & \left( \begin{array}{c} I_{10}\left(D+2,x\right) \\ I_{11}\left(D+2,x\right) \\ \end{array} \right)
\eq
be a basis in $(D+2)$ space-time dimensions.
Work out the $2 \times 2$-matrices $S$ and $S^{-1}$.
\\
\\
{\bf Solution}: 
For the one-loop two-point function the graph polynomial ${\mathcal U}$ is given by
\bq
 {\mathcal U}\left(\alpha_1,\alpha_2\right)
 & = &
 \alpha_1 + \alpha_2.
\eq
Hence
\bq
 I_{10}\left(D,x\right)
 & = &
 {\mathcal U}\left( {\bf 1}^+, {\bf 2}^+\right)
 I_{10}\left(D,x\right)
 \; = \;
 \left( {\bf 1}^+ + {\bf 2}^+\right) I_{10}\left(D+2,x\right)
 \nonumber \\
 & = &
 I_{20}\left(D+2,x\right).
\eq
Note that
\bq
 {\bf 2}^+ I_{10}\left(D+2,x\right)
 & = &
 0 \cdot I_{11}\left(D+2,x\right)
 \; = \; 0,
\eq
which shows that an index, which is zero, cannot be raised.
For $I_{10}$ we could also have used alternatively the graph polynomial for the one-loop one-point function, which is simpler and
given by  ${\mathcal U}(\alpha_1) = \alpha_1$, yielding the same result.

For the second master integral we have
\bq
 I_{11}\left(D,x\right)
 & = &
 {\mathcal U}\left( {\bf 1}^+, {\bf 2}^+\right)
 I_{11}\left(D,x\right)
 \; = \;
 \left( {\bf 1}^+ + {\bf 2}^+\right) I_{11}\left(D+2,x\right)
 \nonumber \\
 & = &
 I_{21}\left(D+2,x\right) + I_{12}\left(D+2,x\right)
 \; = \;
 2 I_{21}\left(D+2,x\right).
\eq
Here we used the fact that for the equal-mass two-point function we have the symmetry $I_{\nu_1 \nu_2} = I_{\nu_2 \nu_1}$.

In the next step we reduce each of the integrals in $I_{20}(D+2,x)$ and $I_{21}(D+2,x)$ to a linear combination of the master integrals
$I_{10}(D+2,x)$ and $I_{11}(D+2,x)$.
From eq.~(\ref{chapter_iterated_integrals:ibp_equations_tadpole}) we obtain
\bq
 I_{20}\left(D+2,x\right)
 & = &
 \left(1-\frac{\left(D+2\right)}{2} \right) I_{10}\left(D+2,x\right)
 \; = \;
 - \frac{D}{2} I_{10}\left(D+2,x\right).
\eq
Please note that we use eq.~(\ref{chapter_iterated_integrals:ibp_equations_tadpole}) with $D$ substituted by $(D+2)$.
From eq.~(\ref{chapter_iterated_integrals:ibp_equations}) we obtain
\bq
\lefteqn{
 x \left(4+x\right) I_{21}\left(D+2,x\right)
 = }
 \nonumber \\
 & = &
 \left[ 3 - \left(D+2\right) \right] x I_{11}\left(D+2,x\right)
 + \left(2+x\right) I_{20}\left(D+2,x\right)
 - 2 I_{02}\left(D+2,x\right)
 \nonumber \\
 & = &
 -\left(D-1\right) x I_{11}\left(D+2,x\right)
 + x I_{20}\left(D+2,x\right)
 \nonumber \\
 & = &
 - \left(D-1\right) x I_{11}\left(D+2,x\right)
 - \frac{D}{2} x I_{10}\left(D+2,x\right).
\eq
Putting everything together we have
\bq
 \left( \begin{array}{c} I_{10}\left(D,x\right) \\ I_{11}\left(D,x\right) \\ \end{array} \right)
 & = &
 \left( \begin{array}{cc} 
 - \frac{D}{2} & 0 \\
 - \frac{D}{4+x} & - \frac{2\left(D-1\right)}{4+x} \\
 \end{array} \right)
 \left( \begin{array}{c} I_{10}\left(D+2,x\right) \\ I_{11}\left(D+2,x\right) \\ \end{array} \right),
\eq
and therefore
\bq
 S \; = \;
 \left( \begin{array}{cc} 
 - \frac{D}{2} & 0 \\
 - \frac{D}{4+x} & - \frac{2\left(D-1\right)}{4+x} \\
 \end{array} \right),
 & &
 S^{-1} \; = \;
 \left( \begin{array}{cc} 
 - \frac{2}{D} & 0 \\
 \frac{1}{D-1} & - \frac{4+x}{2\left(D-1\right)} \\
 \end{array} \right).
\eq
}
\es
\\
\\
\bs
{\it \refstepcounter{exercise}
{\bf Exercise \theexercise}: 
Show that
\bq
 \sum\limits_{j=1}^{\NB+1}
 x_j \frac{\partial}{\partial x_j} 
 I_{\nu_1 \dots \nu_{\ninternal}}
 & = &
 \left( \frac{\loopnumber D}{2} - \nu \right)
 \cdot
 I_{\nu_1 \dots \nu_{\ninternal}}.
\eq
Hint: Consider the Feynman parameter representation.
\\
\\
{\bf Solution}: 
We apply the differential operator to the Feynman parameter representation
\bq
 I_{\nu_1 \dots \nu_{\ninternal}}
 & = &
 \frac{e^{\loopnumber \eps \Eulerconstant}\Gamma\left(\nu-\frac{\loopnumber D}{2}\right)}{\prod\limits_{k=1}^{\ninternal}\Gamma(\nu_k)}
 \int\limits_{a_k \ge 0} d^{\ninternal}a \; \delta\left(1-\sum\limits_{k=1}^{\ninternal} a_k \right) \; 
 \left( \prod\limits_{k=1}^{\ninternal} a_k^{\nu_k-1} \right)
 \frac{\left[ {\mathcal U}\left(a\right) \right]^{\nu-\frac{\left(\loopnumber+1\right) D}{2}}}{\left[ {\mathcal F}\left(a\right) \right]^{\nu-\frac{\loopnumber D}{2}}}.
\eq
We obtain
\bq
\lefteqn{
 \sum\limits_{j=1}^{\NB+1}
 x_j \frac{\partial}{\partial x_j} 
 I_{\nu_1 \dots \nu_{\ninternal}}
 = } & &
 \nonumber \\
 & = & 
 \left( \frac{\loopnumber D}{2} - \nu \right)
 \sum\limits_{j=1}^{\NB+1}
 \frac{e^{\loopnumber \eps \Eulerconstant}\Gamma\left(\nu-\frac{\loopnumber D}{2}\right)}{\prod\limits_{k=1}^{\ninternal}\Gamma(\nu_k)}
 \int\limits_{a_k \ge 0} d^{\ninternal}a \; \delta\left(1-\sum\limits_{k=1}^{\ninternal} a_k \right) \; 
 \left( \prod\limits_{k=1}^{\ninternal} a_k^{\nu_k-1} \right)
 \frac{\left[ {\mathcal U}\left(a\right) \right]^{\nu-\frac{\left(\loopnumber+1\right) D}{2}} x_j \cdot {\mathcal F}_{x_j}'\left(a\right) }{\left[ {\mathcal F}\left(a\right) \right]^{\nu-\frac{\loopnumber D}{2}+1}}
 \nonumber \\
 & = &
 \left( \frac{\loopnumber D}{2} - \nu \right)
 \frac{e^{\loopnumber \eps \Eulerconstant}\Gamma\left(\nu-\frac{\loopnumber D}{2}\right)}{\prod\limits_{k=1}^{\ninternal}\Gamma(\nu_k)}
 \int\limits_{a_k \ge 0} d^{\ninternal}a \; \delta\left(1-\sum\limits_{k=1}^{\ninternal} a_k \right) \; 
 \left( \prod\limits_{k=1}^{\ninternal} a_k^{\nu_k-1} \right)
 \frac{\left[ {\mathcal U}\left(a\right) \right]^{\nu-\frac{\left(\loopnumber+1\right) D}{2}}}{\left[ {\mathcal F}\left(a\right) \right]^{\nu-\frac{\loopnumber D}{2}}}
 \nonumber \\
 & = &
 \left( \frac{\loopnumber D}{2} - \nu \right)
 I_{\nu_1 \dots \nu_{\ninternal}}.
\eq
}
\es
\\
\\
\bs
{\it \refstepcounter{exercise}
{\bf Exercise \theexercise}: 
The steps from eq.~(\ref{chapter_iterated_integrals:example_partial_derivative_1}) 
to eq.~(\ref{chapter_iterated_integrals:example_partial_derivative_2}) can still be carried out by hand. Fill in the missing details.
\\
\\
{\bf Solution}: 
From eq.~(\ref{chapter_iterated_integrals:example_partial_derivative_1}) we have
\bq
 \frac{\partial}{\partial x} I_{1 0}\left(D,x\right)
 & = &
 0,
 \nonumber \\
 \frac{\partial}{\partial x} I_{1 1}\left(D,x\right)
 & = &
 - I_{2 2}\left(D+2,x\right).
\eq
We have to express $I_{2 2}(D+2,x)$ as a linear combination of $I_{1 0}(D,x)$ and $I_{1 1}(D,x)$.
We first express $I_{2 2}(D+2,x)$ as a linear combination of $I_{1 0}(D+2,x)$ and $I_{1 1}(D+2,x)$.
Using eq.~(\ref{chapter_iterated_integrals:ibp_equations}) with $(D+2)$ we obtain
\bq
 I_{2 2}\left(D+2,x\right)
 & = &
 \frac{4+\left(4-D\right)x}{x\left(4+x\right)} I_{2 1}\left(D+2,x\right)
 - \frac{4}{x\left(4+x\right)} I_{3 0}\left(D+2,x\right).
\eq
Here we used the symmetry $I_{\nu_2 \nu_1}(D+2,x)=I_{\nu_1 \nu_2}(D+2,x)$.
Using again eq.~(\ref{chapter_iterated_integrals:ibp_equations}), this time for $I_{2 1}(D+2,x)$ we obtain
\bq
 I_{2 1}\left(D+2,x\right)
 & = &
 \frac{1-D}{4+x} I_{1 1}\left(D+2,x\right)
 + \frac{1}{4+x} I_{2 0}\left(D+2,x\right).
\eq
From eq.~(\ref{chapter_iterated_integrals:ibp_equations_tadpole}) we have
\bq
 I_{2 0}\left(D+2,x\right)
 & = &
 - \frac{D}{2} I_{1 0}\left(D+2,x\right),
 \nonumber \\
 I_{3 0}\left(D+2,x\right)
 & = &
 \frac{D\left(D-2\right)}{8} I_{1 0}\left(D+2,x\right),
\eq
and hence
\bq
 I_{2 2}\left(D+2,x\right)
 & = &
 \frac{\left(1-D\right) \left[4+\left(4-D\right)x\right]}{x\left(4+x\right)^2} I_{1 1}\left(D+2,x\right)
    - \frac{D \left[2\left(D-1\right)+x\right]}{x\left(4+x\right)^2} I_{1 0}\left(D+2,x\right).
\eq
From exercise~\ref{chapter_iterated_integrals:exercise_dimensional_shift} we have
\bq
 I_{1 0}\left(D+2,x\right)
 & = &
 - \frac{2}{D} I_{1 0}\left(D,x\right),
 \nonumber \\
 I_{1 1}\left(D+2,x\right)
 & = &
 \frac{1}{D-1} I_{1 0}\left(D,x\right) - \frac{4+x}{2\left(D-1\right)} I_{1 1}\left(D,x\right),
\eq
and therefore
\bq
 I_{2 2}\left(D+2,x\right)
 & = &
 \frac{D-2}{x\left(4+x\right)} I_{1 0}\left(D,x\right)
 +
 \frac{4+\left(4-D\right)x}{2 x\left(4+x\right)} I_{1 1}\left(D,x\right).
\eq
}
\es
\\
\\
\bs
{\it \refstepcounter{exercise}
{\bf Exercise \theexercise}: 
This example depends on two kinematic variables $x_1$ and $x_2$, hence the integrability condition is non-trivial.
Check explicitly the integrability condition
\bq
 d A + A \wedge A & = & 0.
\eq
{\bf Solution}: 
With $A = A_{x_1} dx_1 + A_{x_2} dx_2$ we have
\bq
 d A & = & \left( \partial_{x_1} A_{x_2} - \partial_{x_2} A_{x_1} \right) dx_1 \wedge dx_2,
 \nonumber \\
 A \wedge A & = & \left( A_{x_1} A_{x_2} - A_{x_2} A_{x_1} \right) dx_1 \wedge dx_2
 \; = \; \left[ A_{x_1}, A_{x_2} \right] dx_1 \wedge dx_2.
\eq
The integrability condition translates to
\bq
 \partial_{x_1} A_{x_2} - \partial_{x_2} A_{x_1}  + \left[ A_{x_1}, A_{x_2} \right] & = & 0.
\eq
With the explicit expressions for $A_{x_1}$ and $A_{x_2}$ given in eq.~(\ref{chapter_iterated_integrals:A_box}) one verifies
that this equation holds.
We have
\bq
 d A
 & = &
 - A \wedge A
 \\
 & = &
 \left( \begin{array}{cccc} 
 0 & 0 & 0 & 0 \\
 0 & 0 & 0 & 0 \\
 0 & 0 & 0 & 0 \\
  \frac{2\left(D-3\right)\left(x_1-x_2\right)}{x_1\left(1-x_1\right)x_2\left(1-x_2\right)\left(1-x_1-x_2\right)} 
  & 
  \frac{2\left(D-3\right)\left(1-2x_1\right)}{x_1^2\left(1-x_1\right)x_2\left(1-x_1-x_2\right)} 
  & 
  - \frac{2\left(D-3\right)\left(1-2x_2\right)}{x_1x_2^2\left(1-x_2\right)\left(1-x_1-x_2\right)} 
  & 0 \\
 \end{array} \right)
 dx_1 \wedge dx_2.
 \nonumber
\eq
}
\es
\\
\\
\bs
{\it \refstepcounter{exercise}
{\bf Exercise \theexercise}: 
Let $X={\mathbb C}^2$ and 
\bq
 Y & = & \left\{ x \in X | x_1+x_2 \; = \; 0 \right\}.
\eq
Compute
\bq
 \mathrm{Res}_Y\left( \frac{x_1 x_2^2 dx_1 \wedge dx_2}{x_1+x_2} \right).
\eq
{\bf Solution}: 
We have
\bq
 \frac{x_1 x_2^2 dx_1 \wedge dx_2}{x_1+x_2}
 & = &
 \frac{d\left(x_1+x_2\right)}{x_1+x_2} \wedge \left( x_1 x_2^2 dx_2 \right),
\eq
hence
\bq
 \mathrm{Res}_Y\left( \frac{x_1 x_2^2 dx_1 \wedge dx_2}{x_1+x_2} \right)
 & = &
 \left. x_1 x_2^2 dx_2 \right|_Y
 \; = \;
 \left. - x_2^3 dx_2 \right|_Y.
\eq
Note that alternatively we could have used
\bq
 \frac{x_1 x_2^2 dx_1 \wedge dx_2}{x_1+x_2}
 & = &
 \frac{d\left(x_1+x_2\right)}{x_1+x_2} \wedge \left( - x_1 x_2^2 dx_1 \right)
\eq
and 
\bq
 \mathrm{Res}_Y\left( \frac{x_1 x_2^2 dx_1 \wedge dx_2}{x_1+x_2} \right)
 & = &
 \left. - x_1 x_2^2 dx_1 \right|_Y
 \; = \;
 \left. - x_1^3 dx_1 \right|_Y.
\eq
On $Y$ we have $x_2=-x_1$ and $dx_2=-dx_1$. Therefore the two forms agree on $Y$. 
}
\es
\\
\\
\bs
{\it \refstepcounter{exercise}
{\bf Exercise \theexercise}: 
Let 
\bq
 \omega & = & 3 dx_1 + (5+x_1) dx_2 + x_3 dx_3
\eq
and
\bq
 \gamma & : & [0,1] \rightarrow {\mathbb C}^3,
 \;\;\;\;\;\;
 \gamma\left(\lambda\right)
 = \left( \begin{array}{c}
 \lambda \\ \lambda^2 \\ 1 + \lambda \\ \end{array} \right).
\eq
Compute
\bq
 \int\limits_\gamma \omega.
\eq
{\bf Solution}: 
We have
\bq
 \int\limits_\gamma \omega 
 & = & 
 \int\limits_0^1 d\lambda \;
 \left[ 3 \frac{d}{d\lambda} \lambda + (5+\lambda) \frac{d}{d\lambda} \lambda^2 + (1+\lambda) \frac{d}{d\lambda} (1+\lambda) \right]
 \nonumber \\
 & = &
 \int\limits_0^1 d\lambda \;
 \left[ 3 + 2 \lambda (5+\lambda) + (1+\lambda) \right]
 =
 \int\limits_0^1 d\lambda \;
 \left[ 4 + 11 \lambda + 2 \lambda^2 \right]
 \nonumber \\
 & = & 4 + \frac{11}{2} + \frac{2}{3} = \frac{61}{6}.
\eq
}
\es
\\
\\
\bs
{\it \refstepcounter{exercise}
{\bf Exercise \theexercise}: 
Prove eq.~(\ref{chapter_iterated_integrals:combined_path}) for the case $r=2$, i.e. show
\bq
 I_{\gamma_2 \circ \gamma_1}\left(\omega_1,\omega_2;\lambda\right)
 & = &
 I_{\gamma_1}\left(\omega_1,\omega_2;\lambda\right)
 +
 I_{\gamma_2}\left(\omega_1;\lambda\right)
 I_{\gamma_1}\left(\omega_2;\lambda\right)
 +
 I_{\gamma_2}\left(\omega_1,\omega_2;\lambda\right).
\eq
{\bf Solution}: 
Without loss of generality we take $a=0$ and $b=\lambda=1$, i.e. we consider
\bq
 \gamma_1 & : & \left[0,1\right] \; \rightarrow X,
 \nonumber \\
 \gamma_2 & : & \left[0,1\right] \; \rightarrow X,
 \nonumber \\
 \gamma_2 \circ \gamma_1 & : & \left[0,1\right] \; \rightarrow X.
\eq
We decompose the full integration region $0 \le \lambda_1 \le 1$, $0 \le \lambda_2 \le \lambda_1$
of $I_{\gamma_2 \circ \gamma_1}(\omega_1,\omega_2;1)$
\begin{figure}
\begin{center}
\includegraphics[scale=1.0]{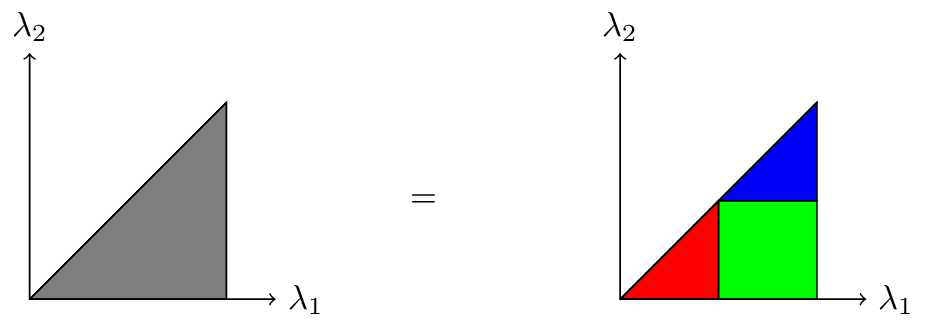}
\end{center}
\caption{
The iterated integral of depth $2$ for the combined path $\gamma_2\circ \gamma_1$:
The full integration region (shown in grey on the left) is decomposed into three regions (shown in red, green and blue on the right). 
}
\label{chapter_solutions:fig_path_composition}
\end{figure}
into three regions as shown in fig.~\ref{chapter_solutions:fig_path_composition}.
The integration over the region 
$0 \le \lambda_1 \le 1/2$, $0 \le \lambda_2 \le \lambda_1$ (red region)
gives $I_{\gamma_1}(\omega_1,\omega_2;1)$,
the integration over the region 
$1/2 \le \lambda_1 \le 1$, $0 \le \lambda_2 \le 1/2$ (green region)
gives
$I_{\gamma_2}(\omega_1;1) I_{\gamma_1}(\omega_2;1)$
and the integration over the region
$1/2 \le \lambda_1 \le 1$, $1/2 \le \lambda_2 \le \lambda_1$ (blue region)
gives
$I_{\gamma_2}(\omega_1,\omega_2;1)$.
}
\es
\\
\\
\bs
{\it \refstepcounter{exercise}
{\bf Exercise \theexercise}: 
Prove eq.~(\ref{chapter_iterated_integrals:reversed_path}).
\\
\\
{\bf Solution}: 
Without loss of generality we take $[a,b]=[0,1]$.
For the path $\gamma : [0,1] \rightarrow X$ 
and a differential one-form $\omega_j$ we write for the pull-back
\bq
 f_j\left(\lambda\right) d\lambda 
 & = & 
 \gamma^\ast \omega_j.
\eq
For the reversed path $\gamma^{-1} : [0,1] \rightarrow X$ 
we write for the pull-back of $\omega_j$
\bq
 f^{\mathrm{rev}}_j\left(\lambda'\right) d\lambda' 
 & = & 
 \left(\gamma^{-1}\right)^\ast \omega_j.
\eq
We have
\bq
 f^{\mathrm{rev}}_j\left(\lambda'\right)
 & = &
 - f_j\left(1-\lambda'\right).
\eq
We have to consider
\bq
 I_{\gamma^{-1}}\left(\omega_1,\dots,\omega_r;1\right)
 & = &
 \int\limits_0^1 d\lambda_1' f_1^{\mathrm{rev}}\left(\lambda_1'\right)
 \dots
 \int\limits_0^{\lambda_{r-1}'} d\lambda_r' f_r^{\mathrm{rev}}\left(\lambda_r'\right).
\eq
With $\lambda_j=1-\lambda_j'$ we have for the integrand
\bq
 f_1^{\mathrm{rev}}\left(\lambda_1'\right) \dots f_r^{\mathrm{rev}}\left(\lambda_r'\right)
& = &
 \left(-1\right)^r
 f_1\left(1-\lambda_1'\right) \dots f_r\left(1-\lambda_r'\right)
 \nonumber \\
 & = &
 \left(-1\right)^r
 f_1\left(\lambda_1\right) \dots f_r\left(\lambda_r\right).
\eq
The integration domain
\bq
 0 \; \le \; \lambda_r' \; \le \; \dots \; \le \lambda_1' \; \le \; 1
\eq
transforms into
\bq
 0 \; \le \; \lambda_1 \; \le \; \dots \; \le \lambda_r \; \le \; 1.
\eq
Thus
\bq
 I_{\gamma^{-1}}\left(\omega_1,\dots,\omega_r;1\right)
 & = &
 \left(-1\right)^r
 \int\limits_0^1 d\lambda_r f_r\left(\lambda_r\right)
 \dots
 \int\limits_0^{\lambda_1} d\lambda_1 f_1\left(\lambda_1\right)
 \nonumber \\
 & = &
 \left(-1\right)^r
 I_{\gamma}\left(\omega_r,\dots,\omega_1;1\right).
\eq
}
\es
\\
\\
\bs
{\it \refstepcounter{exercise}
{\bf Exercise \theexercise}: 
Show that $I_4'$ is given at the kinematic point $(x_1,x_2)=(1,1)$ by
\bq
 I_4'
 & = &
 e^{\eps \Eulerconstant} 
 \frac{\Gamma\left(1+\eps\right)\Gamma\left(1-\eps\right)^2}{\Gamma\left(1-2\eps\right)}
 \left( 1 - \sum\limits_{k=2}^\infty \zeta_k \eps^k \right).
\eq
Hint: Use the trick from exercise~\ref{chapter_basics:exercise_oneloopbox_Feynman_parameter_trick} 
and the Mellin-Barnes technique.
\\
\\
{\bf Solution}: 
From exercise~\ref{chapter_basics:exercise_oneloopbox_Feynman_parameter_trick} we know that we may write $I_4'$
with generic $x_1$ and $x_2$ as
\bq
 I_4'
 & = &
 \frac{\eps^2 e^{\eps \Eulerconstant} \Gamma\left(2+\eps\right) \Gamma\left(-\eps\right)^2}{2 \Gamma\left(-2\eps\right)}
 x_1 x_2 
 \int\limits_0^1 da
 \int\limits_0^1 db
 \left[ a b x_1 + \bar{a} \bar{b} x_2 + \bar{a} b \right]^{-2-\eps},
\eq
with $\bar{a}=1-a$ and $\bar{b}=1-b$.
Specialising to $x_1=x_2=1$ we have
\bq
 I_4'
 & = &
 \frac{\eps^2 e^{\eps \Eulerconstant} \Gamma\left(2+\eps\right) \Gamma\left(-\eps\right)^2}{2 \Gamma\left(-2\eps\right)}
 \int\limits_0^1 da
 \int\limits_0^1 db
 \left[ 1 - a + a b \right]^{-2-\eps}.
\eq
We now use the Mellin-Barnes technique and split $1-a+ab$ into $(1-a)$ and $(ab)$:
\bq
 I_4'
 & = &
 \frac{\eps^2 e^{\eps \Eulerconstant} \Gamma\left(-\eps\right)^2}{2 \Gamma\left(-2\eps\right)}
 \frac{1}{2\pi i}
 \int d\sigma \;
 \Gamma\left(-\sigma\right) \Gamma\left(\sigma+2+\eps\right)
 \int\limits_0^1 da
 \int\limits_0^1 db \;
 a^\sigma b^\sigma \left(1-a\right)^{-\sigma-2-\eps}.
\eq
The integrals over $a$ and $b$ are now easily done, yielding
\bq
 I_4'
 & = &
 \frac{\eps^2 e^{\eps \Eulerconstant} \Gamma\left(-\eps\right)}{2 \Gamma\left(-2\eps\right)}
 \frac{1}{2\pi i}
 \int d\sigma \;
 \frac{\Gamma\left(-\sigma\right) \Gamma\left(-\sigma-1-\eps\right) \Gamma\left(\sigma+1\right)^2 \Gamma\left(\sigma+2+\eps\right)}{\Gamma\left(\sigma+2\right)}.
\eq
We close the contour to the right and sum up the residues from $\Gamma(-\sigma)$ and $\Gamma(-\sigma-1-\eps)$.
This yields
\bq
 I_4'
 & = &
 \frac{\eps^2 e^{\eps \Eulerconstant} \Gamma\left(-\eps\right)}{2 \Gamma\left(-2\eps\right)}
 \left\{
 \sum\limits_{n=0}^\infty \frac{\left(-1\right)^n}{n!} 
 \frac{\Gamma\left(-n-1-\eps\right) \Gamma\left(n+1\right)^2 \Gamma\left(n+2+\eps\right)}{\Gamma\left(n+2\right)}
 \right. \nonumber \\
 & & \left.
 +
 \sum\limits_{n=0}^\infty \frac{\left(-1\right)^n}{n!} 
 \frac{\Gamma\left(-n+1+\eps\right) \Gamma\left(n-\eps\right)^2 \Gamma\left(n+1\right)}{\Gamma\left(n+1-\eps\right)}
 \right\}.
\eq
With the help of
\bq
 \left(-1\right)^n \Gamma\left(-n-1-\eps\right) \Gamma\left(n+2+\eps\right)
 & = & 
 \Gamma\left(-1-\eps\right) \Gamma\left(2+\eps\right),
 \nonumber \\
 \left(-1\right)^n \Gamma\left(-n+1+\eps\right) \Gamma\left(n-\eps\right)
 & = & 
 \Gamma\left(1+\eps\right) \Gamma\left(-\eps\right),
\eq
and
\bq
 \Gamma\left(-1-\eps\right) \Gamma\left(2+\eps\right),
 & = & 
 - \Gamma\left(1+\eps\right) \Gamma\left(-\eps\right)
\eq
this simplifies to
\bq
 I_4'
 & = &
 \frac{e^{\eps \Eulerconstant} \Gamma\left(1-\eps\right)^2 \Gamma\left(1+\eps\right)}{\Gamma\left(1-2\eps\right)}
 \left[ 1 - \eps \sum\limits_{n=1}^\infty \left( \frac{1}{n}-\frac{1}{n-\eps} \right) \right].
\eq
Finally, expanding the geometric series
\bq
 \frac{1}{n-\eps}
 & = & 
 \frac{1}{n} \sum\limits_{k=0}^\infty \left( \frac{\eps}{n} \right)^k
\eq
we obtain
\bq
 I_4'
 & = &
 \frac{e^{\eps \Eulerconstant} \Gamma\left(1-\eps\right)^2 \Gamma\left(1+\eps\right)}{\Gamma\left(1-2\eps\right)}
 \left[ 1 - \sum\limits_{k=2}^\infty \eps^k \sum\limits_{n=1}^\infty \frac{1}{n^k} \right]
 \nonumber \\
 & = &
 \frac{e^{\eps \Eulerconstant} \Gamma\left(1-\eps\right)^2 \Gamma\left(1+\eps\right)}{\Gamma\left(1-2\eps\right)}
 \left[ 1 - \sum\limits_{k=2}^\infty \zeta_k \eps^k\right].
\eq
}
\es
\\
\\
\bs
{\it \refstepcounter{exercise}
{\bf Exercise \theexercise}: 
Show the equivalence of the ${\mathcal O}(\eps^2)$-term of $I_4'$ 
between eq.~(\ref{chapter_iterated_integrals:result_box_v1}) and eq.~(\ref{chapter_iterated_integrals:result_box_v2}).
\\
\\
{\bf Solution}: 
We have to show that the expressions
\bq
\label{appendix_solutions:Box_weight_2_expression_v1}
          G\left(0,0;x_1\right) + G\left(0,0;x_2\right)
        - G\left(1,0;x_1\right) - G\left(1,0;x_2\right)
        + G\left(0;x_1\right) G\left(0;x_2\right)
        + \frac{1}{2} \zeta_2
\eq
and
\bq
\label{appendix_solutions:Box_weight_2_expression_v2}
 & &
          - \mathrm{Li}_2\left(x_1\right)
          - \mathrm{Li}_2\left(x_2\right)
          + \frac{1}{2} \ln^2\left(x_1\right) 
          + \frac{1}{2} \ln^2\left(x_2\right) 
          + \ln\left(x_1\right) \ln\left(x_2\right) 
          - \ln\left(x_1\right) \ln\left(1-x_1\right)
          - \ln\left(x_2\right) \ln\left(1-x_2\right)
 \nonumber \\
 & &
        + \frac{1}{2} \zeta_2
\eq
agree.
We have
\bq
 G\left(0;x\right) & = & \ln\left(x\right),
 \nonumber \\
 G\left(0,0;x\right) & = & \frac{1}{2} \ln^2\left(x\right)
\eq
and
\bq
 G\left(1,0;x\right)
 & = &
 \int\limits_0^x dt \frac{\ln t}{t-1} 
 \; = \; 
 \int\limits_1^{1-x} dt \frac{\ln\left(1-t\right)}{t} 
 \; = \; 
 \int\limits_0^{1-x} dt \frac{\ln\left(1-t\right)}{t} 
 -
 \int\limits_0^{1} dt \frac{\ln\left(1-t\right)}{t} 
 \nonumber \\
 & = &
 \zeta_2 - \mathrm{Li}_2\left(1-x\right)
\eq
From eq.~(\ref{chapter_one_loop:dilog_identities}) we have
\bq
\mathrm{Li}_{2}(x) & = & - \mathrm{Li}_{2}(1-x) + \frac{1}{6} \pi^{2} - \ln(x) \ln(1-x),
\eq
and therefore
\bq
 G\left(1,0;x\right)
 & = &
 \mathrm{Li}_{2}(x)
 + \ln(x) \ln(1-x).
\eq
Plugging all expressions into eq.~(\ref{appendix_solutions:Box_weight_2_expression_v1}) shows
the equivalence of eq.~(\ref{appendix_solutions:Box_weight_2_expression_v1}) with
eq.~(\ref{appendix_solutions:Box_weight_2_expression_v2}).
}
\es
\\
\\
\bs
{\it \refstepcounter{exercise}
{\bf Exercise \theexercise}: 
Derive eq.~(\ref{chapter_iterated_integrals:relation_pull_backs}) from eq.~(\ref{chapter_iterated_integrals:relation_sections}).
\\
\\
{\bf Solution}: 
Let $\gamma : [a,b] \rightarrow M$ be a curve in $M$ with $\gamma(0)=x$.
A tangent vector at $x$ is given by
\bq
 X & = & \left. \frac{d}{dt} \gamma(t) \right|_{t=0}
\eq
In order to keep the notation simple we will suppress maps between a manifold and an appropriate
coordinate chart.
We have to show
\bq
 A_2\left(X\right)
 & = &
 \left( U A_1 U^{-1} + U d U^{-1} \right)\left(X\right),
\eq
where $A_1$ and $A_2$ are defined by
\bq
 A_1 \;\; = \;\; \sigma_1^\ast \omega,
 & &
 A_2 \;\; = \;\; \sigma_2^\ast \omega,
\eq
and the sections $\sigma_1$ and $\sigma_2$ are related by
\bq
 \sigma_2 & = & \sigma_1 U^{-1}.
\eq
Let us choose a local trivialisation $(x,g)$ of $P(M,G)$ and work out $\sigma_{2 \ast} X$.
With $U_0=U(\gamma(0))$ we have
\bq
 \sigma_{2 \ast} X 
 & = &
 \sigma_{2 \ast} \left( \left. \frac{d}{dt} \gamma(t) \right|_{t=0} \right)
 \;\; = \;\; 
 \left. \frac{d}{dt} \left( \gamma(t), \sigma_2\left(\gamma(t) \right) \right) \right|_{t=0}
 \;\; = \;\; 
 \left. \frac{d}{dt} \left( \gamma(t), \sigma_1\left(\gamma(t) \right) U\left(\gamma(t) \right)^{-1} \right) \right|_{t=0}
 \nonumber \\
 & = &
 \left( X, \left. \frac{d}{dt} \sigma_1\left(\gamma(t) \right) \right|_{t=0} U_0^{-1}
        + \sigma_1\left(\gamma(0) \right) \left. \frac{d}{dt} U\left(\gamma(t) \right)^{-1} \right|_{t=0} \right)
 \nonumber \\
 & = &
 R_{U_0^{-1} \ast} \left( \sigma_{1 \ast} X \right)
 + \left( X, \sigma_2\left(\gamma(0) \right) U_0 \left. \frac{d}{dt} U\left(\gamma(t) \right)^{-1} \right|_{t=0} \right).
\eq
We then have, using (CF1) and (CF2),
\bq
 A_2\left(X\right)
 & = &
 \omega\left( \sigma_{2 \ast} X \right)
 \;\; = \;\;
 \omega_{(x,\sigma_2(\gamma(0)))}\left( R_{U_0^{-1} \ast} \left( \sigma_{1 \ast} X \right) \right)
 + \omega_{(x,\sigma_2(\gamma(0)))}\left( U_0 \left. \frac{d}{dt} U\left(\gamma(t) \right)^{-1} \right|_{t=0} \right)
 \nonumber \\
 & = &
 U_0 \left( \omega_{(x,\sigma_1(\gamma(0)))}\left( \sigma_{1 \ast} X \right) \right) U_0^{-1}
 + \left( U_0 d U^{-1} \right)\left(X\right)
 \nonumber \\
 & = & 
 \left( U A_1 U^{-1} + U d U^{-1} \right)\left(X\right).
\eq
}
\es
\\
\\
\bs
{\it \refstepcounter{exercise}
{\bf Exercise \theexercise}: 
Work out the maximal cut of the double box integral $I_{111111100}$ shown 
in fig.~\ref{chapter_iterated_integrals:maximal_cut_graph}.
Use the notation as in example 2 in section~\ref{chapter_iterated_integrals:deriving_the_dgl}.
To work out the maximal cut it is simpler to use the loop-by-loop approach as discussed in 
section~\ref{chapter_basics:sect_Baikov_representation}.
\\
\\
{\bf Solution}: 
We label the internal edges as in fig.~\ref{chapter_basics:fig_doublebox_auxiliary}.
We first consider the loop with edges $e_1$, $e_2$, $e_3$ and $e_4$ and then the second loop with edges 
$e_5$, $e_6$, $e_7$ and an edge with momentum $k_2+p_1$.
The latter edge is introduced by integrating out the first loop with loop momentum $k_1$.
The loop-by-loop approach has the advantage that we only need eight Baikov variables $z_1$-$z_7$ and $z_9$, the variable
$z_8$ is absent.
The Baikov representation reads
\bq
 I_{111111100}
 & = &
 \frac{e^{2 \eps \Eulerconstant} \left(\mu^2\right)^{7-D}}{64 \pi^3 \Gamma\left(\frac{D-3}{2}\right)^2}
 \left[G\left(p_1,p_2,p_3\right)\right]^{\frac{4-D}{2}}
 \\
 & &
 \int\limits_{\mathcal C} d^{8}z \;
 \left[G\left(p_1,p_2,k_2\right)\right]^{\frac{4-D}{2}}
 \left[G\left(k_1,p_1,p_2,k_2\right)\right]^{\frac{D-5}{2}}
 \left[G\left(k_2,p_1,p_2,p_3\right)\right]^{\frac{D-5}{2}}
 \frac{1}{z_1 z_2 z_3 z_4 z_5 z_6 z_7}.
 \nonumber 
\eq
We have
\bq
 G\left(p_1,p_2,p_3\right)
 & = & 
 \frac{1}{4} s t \left(s+t\right),
 \nonumber \\
 \left. G\left(p_1,p_2,k_2\right) \right|_{z_1=z_2=z_3=z_4=z_5=z_6=z_7=0}
 & = &
 \frac{1}{4} s \left(s+t-z_9\right) \left(t-z_9\right),
 \nonumber \\
 \left. G\left(k_1,p_1,p_2,k_2\right) \right|_{z_1=z_2=z_3=z_4=z_5=z_6=z_7=0}
 & = &
 \frac{1}{16} s^2 \left(t-z_9\right)^2,
 \nonumber \\
 \left. G\left(k_2,p_1,p_2,p_3\right) \right|_{z_1=z_2=z_3=z_4=z_5=z_6=z_7=0}
 & = &
 \frac{1}{16} s^2 z_9^2.
\eq
Thus 
\bq
\lefteqn{
 \mathrm{MaxCut} \; I_{111111100}
 = } & &
 \\
 & &
 \left(2\pi i\right)^7
 \frac{e^{2 \eps \Eulerconstant} \left(\mu^2\right)^{7-D} 2^{6-2D}}{\pi^3 \Gamma\left(\frac{D-3}{2}\right)^2}
 s^{D-6} t^{\frac{4-D}{2}} \left(s+t\right)^{\frac{4-D}{2}}
 \int\limits_{\mathcal C'} dz_9 \;
 z_9^{D-5} 
 \left(t-z_9\right)^{\frac{D-6}{2}}
 \left(s+t-z_9\right)^{\frac{4-D}{2}}.
 \nonumber
\eq
The integration region ${\mathcal C}'$ is obtained from the conditions
\bq
 & &
 \left. \frac{G\left(k_2,p_1,p_2,p_3\right)}{G\left(p_1,p_2,p_3\right)} \right|_{z_1=z_2=z_3=z_4=z_5=z_6=z_7=0}
 \; = \; 
 \frac{1}{4} \frac{s z_9^2}{t\left(s+t\right)}
 \; > \; 0,
 \nonumber \\
 & &
 \left. \frac{G\left(k_1,p_1,p_2,k_2\right)}{G\left(p_1,p_2,k_2\right)} \right|_{z_1=z_2=z_3=z_4=z_5=z_6=z_7=0}
 \; = \; 
 \frac{1}{4} \frac{s \left(t-z_9\right)}{\left(s+t-z_9\right)}
 \; > \; 0.
\eq
Let's assume $t>0$ and $s<-t$. Then
\bq
 {\mathcal C}' & = & \left] -\infty, s+t \right] \cup \left[ t, \infty \right[,
\eq
and
\bq
\lefteqn{
 \int\limits_{-\infty}^{s+t} dz_9 \;
 z_9^{D-5} 
 \left(t-z_9\right)^{\frac{D-6}{2}}
 \left(s+t-z_9\right)^{\frac{4-D}{2}}
 = } & &
 \nonumber \\
 & &
 \frac{\Gamma\left(\frac{6-D}{2}\right)\Gamma\left(5-D\right)}{\Gamma\left(\frac{16-3D}{2}\right)}
 \left(s+t\right)^{D-5}
 {}_2F_1\left(5-D,\frac{6-D}{2},\frac{16-3D}{2};\frac{t}{s+t}\right),
 \nonumber \\
\lefteqn{
 \int\limits_{t}^{\infty} dz_9 \;
 z_9^{D-5} 
 \left(t-z_9\right)^{\frac{D-6}{2}}
 \left(s+t-z_9\right)^{\frac{4-D}{2}}
 = } & & 
 \nonumber \\
 & &
 -
 \frac{\Gamma\left(\frac{D-4}{2}\right)\Gamma\left(\frac{5-D}{2}\right)}{\sqrt{\pi}}
 2^{4-D} t^{D-5}
 {}_2F_1\left(5-D,\frac{D-4}{2},\frac{6-D}{2};\frac{s+t}{t}\right).
\eq
Combining the results we obtain
\bq
 \mathrm{MaxCut} \; I_{111111100}\left(4-2\eps\right)
 & = &
 \left(2\pi i\right)^7
 \frac{\left(\mu^2\right)^{3}}{4 \pi^4 s^2 t \eps}
 + {\mathcal O}\left(\eps^0\right).
\eq
}
\es
\\
\\
\bs
{\it \refstepcounter{exercise}
{\bf Exercise \theexercise}: 
Show that the Landau equations imply ${\mathcal F}=0$.
\\
\\
{\bf Solution}: 
The graph polynomial ${\mathcal F}$ is homogeneous of degree $(\loopnumber+1)$
in the Feynman parameters $a_j$.
Hence
\bq
 \sum\limits_{j=1}^{\ninternal} a_j \frac{\partial}{\partial a_j} {\mathcal F}
 & = & 
 \left(\loopnumber+1\right) {\mathcal F}.
\eq
The Landau equations imply that the left-hand side vanishes: We either have 
$a_j=0$ (note that $\partial{\mathcal F}/\partial a_j$ is again a polynomial
in the Feynman parameters) or $\partial{\mathcal F}/\partial a_j=0$.
Since $(\loopnumber+1)\neq 0$ it follows that ${\mathcal F}=0$.
}
\es
\\
\\
\bs
{\it \refstepcounter{exercise}
{\bf Exercise \theexercise}: 
Work out the Landau discriminant for the double box graph
discussed in exercise~\ref{chapter_iterated_integrals:exercise_doublebox_ibp}.
\\
\\
{\bf Solution}: 
The graph polynomial ${\mathcal F}$ reads
\bq
 {\mathcal F} & = & \left[ a_2 a_3 \left( a_4+a_5+a_6+a_7 \right)
                        + a_5 a_6 \left( a_1+a_2+a_3+a_4 \right)
                        + a_2 a_4 a_6 + a_3 a_4 a_5 \right] x
      + a_1 a_4 a_7.
\eq
The Landau equations for the leading Landau singularity read
\bq
 a_4 a_7 + a_5 a_6 x & = & 0,
 \nonumber \\
 \left[ a_3 \left(a_4+a_5+a_6+a_7\right) + \left(a_4+a_5\right) a_6 \right] x & = & 0,
 \nonumber \\
 \left[ a_2 \left(a_4+a_5+a_6+a_7\right) + \left(a_4+a_6\right) a_5 \right] x & = & 0,
 \nonumber \\
 a_1 a_7 + \left(a_2+a_5\right) \left(a_3+a_6\right) x & = & 0,
 \nonumber \\
 \left[ a_6 \left(a_1+a_2+a_3+a_4\right) + \left(a_2+a_4\right) a_3 \right] x & = & 0,
 \nonumber \\
 \left[ a_5 \left(a_1+a_2+a_3+a_4\right) + \left(a_3+a_4\right) a_2 \right] x & = & 0,
 \nonumber \\
 a_1 a_4 + a_2 a_3 x & = & 0.
\eq
We then solve these equation for $(a_1,\dots,a_7,x)$.
For the leading Landau singularity we are interested in solutions with $a_j \neq 0$. We find that only for $x=-1$ we have such a solution and hence
\bq
 D_{\mathrm{Landau}} & = &
 \left\{ -1 \right\}.
\eq 
}
\es
\\
\bs
{\it \refstepcounter{exercise}
{\bf Exercise \theexercise}: 
Show that for $\varphi$ as in eq.~(\ref{chapter_iterated_integrals:representative_left}) 
and $\omega$ as in eq.~(\ref{chapter_iterated_integrals:def_omega})
the differential $n$-form $\varphi$ is closed with respect to $\nabla_\omega$:
\bq
 \nabla_\omega \varphi & = & 0.
\eq
{\bf Solution}: 
Let $1 \le j \le n$. We have
\bq
 \nabla_\omega & = & 
 \sum\limits_{j=1}^n  \left( \frac{\partial}{\partial z_j} + \omega_j \right) dz_j
 +
 \sum\limits_{j=1}^n  \left( \frac{\partial}{\partial \bar{z}_j} \right) d\bar{z}_j.
\eq
Hence
\bq
 \left( \frac{\partial}{\partial z_j} \frac{q}{p_1^{n_1} \dots p_m^{n_m}} + \omega_j \right) \; dz_j \wedge dz_n \wedge \dots \wedge dz_1 & = & 0,
 \nonumber \\
 \left( \frac{\partial}{\partial \bar{z}_j} \frac{q}{p_1^{n_1} \dots p_m^{n_m}} \right) \; d\bar{z}_j \wedge dz_n \wedge \dots \wedge dz_1 & = & 0.
\eq
In the first line, the wedge product contains $dz_j \wedge dz_j$, while in the second line 
the derivative in the bracket vanishes.
}
\es
\\
\\
\bs
{\it \refstepcounter{exercise}
{\bf Exercise \theexercise}: 
Proof eq.~(\ref{chapter_iterated_integrals:intersection_beta_fct}) for the special case $n_1=n_2=n_3=n_4=0$.
\\
\\
{\bf Solution}:
We have to show that for
\bq
 \omega
 & = &
 \gamma_1 \frac{dz}{z} - \gamma_2 \frac{dz}{1-z}
\eq
and 
\bq
 \varphi_L \; = \; \frac{dz}{z \left(1-z\right)},
 & &
 \varphi_R \; = \; \frac{dz}{z \left(1-z\right)},
\eq
we have
\bq
 \left\langle \varphi_L \right. \left| \varphi_R \right\rangle_\omega
 & = &
 \frac{\left(\gamma_1+\gamma_2\right)}{\gamma_1 \gamma_2}
 \; = \;
 \frac{1}{\gamma_1} + \frac{1}{\gamma_2}.
\eq
For the case at hand, $\psi_{L,1}$ is given by
\bq
 \psi_{L,1}
 & = &
 \frac{1}{\gamma_1} + {\mathcal O}\left(z\right),
\eq
and $\psi_{L,2}$ is given by
\bq
 \psi_{L,1}
 & = &
 - \frac{1}{\gamma_2} + {\mathcal O}\left(z-1\right).
\eq
We further have
\bq
 \mathrm{Res}_{D_1}\left(\psi_{L,1} \varphi_R\right)
 & = & 
 \mathrm{Res}_{\{0\}}\left( \frac{dz}{\gamma_1 z \left(1-z\right)} \right)
 \; = \; \frac{1}{\gamma_1},
 \nonumber \\
 \mathrm{Res}_{D_2}\left(\psi_{L,2} \varphi_R\right)
 & = & 
 \mathrm{Res}_{\{1\}}\left( - \frac{dz}{\gamma_2 z \left(1-z\right)} \right)
 \; = \; \frac{1}{\gamma_2}.
\eq
Therefore
\bq
 \left\langle \varphi_L \right. \left| \varphi_R \right\rangle_\omega
 & = &
 \frac{1}{\gamma_1} + \frac{1}{\gamma_2}.
\eq
}
\es
\\
\\
\bs
{\it \refstepcounter{exercise}
{\bf Exercise \theexercise}: 
Consider the monomials
\bq
 p_1 \; = \; x_1^2 x_2 x_3,
 & &
 p_2 \; = \; x_1 x_2^3.
\eq
Order the two monomials with respect to the degree lexicographic order
and the degree reverse lexicographic order (assuming $x_1>x_2>x_3$).
\\
\\
{\bf Solution}: 
Both polynomials have total degree four.
Let us write
\bq
 p_1 
 \; = \;
 x_1^2 x_2 x_3
 \; = \; 
 x_1^{m_1} x_2^{m_2} x_3^{m_3}
 & \Rightarrow &
 \left(m_1,m_2,m_3\right) \; = \; \left(2,1,1\right),
 \nonumber \\
 p_2
 \; = \;
 x_1 x_2^3
 \; = \; 
 x_1^{m_1'} x_2^{m_2'} x_3^{m_3'}
 & \Rightarrow &
 \left(m_1',m_2',m_3'\right) \; = \; \left(1,3,0\right).
\eq
We have
\bq
 \left(m_1-m_1',m_2-m_2',m_3-m_3'\right)
 & = &
 \left(1,-2,1\right).
\eq
For the degree lexicographic order we have
\bq
 p_1 & >_{\mathrm{deglex}} & p_2,
\eq
as $m_1-m_1'=1>0$, while for the degree reverse lexicographic order
we have
\bq
 p_1 & <_{\mathrm{degrevlex}} & p_2,
\eq
as $m_3'-m_3=-1<0$.
}
\es
\\
\\
\bs
{\it \refstepcounter{exercise}
{\bf Exercise \theexercise}: 
Assume $C_{\lambda+1}=C_{\lambda-1}=0$. Show that this implies
\bq
 \dim H_{\lambda+1}\left(M\right) \; = \; 0,
 \;\;\;\;\;\;
 \dim H_{\lambda}\left(M\right) \; = \; C_\lambda,
 \;\;\;\;\;\;
 \dim H_{\lambda-1}\left(M\right) \; = \; 0.
\eq
{\bf Solution}: 
Let us denote $b_k= \dim H_k(M)$.
We write down the Morse inequalities for $(\lambda+1)$, $\lambda$, $(\lambda-1)$ and $(\lambda-2)$, using $C_{\lambda+1}=C_{\lambda-1}=0$.
Multiplying in addition the first and third equation by $(-1)$ we obtain 
\bq
 \sum\limits_{k=0}^{\lambda+1} \left(-1\right)^{\lambda-k} b_k
 & \ge &
 C_{\lambda} + \sum\limits_{k=0}^{\lambda-2} \left(-1\right)^{\lambda-k} C_k,
 \nonumber \\
 \sum\limits_{k=0}^\lambda \left(-1\right)^{\lambda-k} b_k
 & \le &
 C_{\lambda}
 +
 \sum\limits_{k=0}^{\lambda-2} \left(-1\right)^{\lambda-k} C_k,
 \nonumber \\
 \sum\limits_{k=0}^{\lambda-1} \left(-1\right)^{\lambda-k} b_k
 & \ge &
 \sum\limits_{k=0}^{\lambda-2} \left(-1\right)^{\lambda-k} C_k,
 \nonumber \\
 \sum\limits_{k=0}^{\lambda-2} \left(-1\right)^{\lambda-k} b_k
 & \le &
 \sum\limits_{k=0}^{\lambda-2} \left(-1\right)^{\lambda-k} C_k.
\eq
From the first two inequalities we obtain
\bq
 \sum\limits_{k=0}^\lambda \left(-1\right)^{\lambda-k} b_k
 & \le &
 \sum\limits_{k=0}^{\lambda+1} \left(-1\right)^{\lambda-k} b_k
\eq
and therefore
\bq
 0 & \le & - b_{\lambda+1}.
\eq
As $b_{\lambda+1}$ cannot be negative, it follows that $b_{\lambda+1} = \dim H_{\lambda+1}(M) = 0$.
In a similar way we obtain from the third and fourth equation $b_{\lambda-1} = \dim H_{\lambda-1}(M) = 0$.

Having shown $b_{\lambda+1}=b_{\lambda-1}=0$, the first two equations simplify to
\bq
 b_\lambda + \sum\limits_{k=0}^{\lambda-2} \left(-1\right)^{\lambda-k} b_k
 & \ge &
 C_{\lambda} + \sum\limits_{k=0}^{\lambda-2} \left(-1\right)^{\lambda-k} C_k,
 \nonumber \\
 b_\lambda
 +
 \sum\limits_{k=0}^{\lambda-2} \left(-1\right)^{\lambda-k} b_k
 & \le &
 C_{\lambda}
 +
 \sum\limits_{k=0}^{\lambda-2} \left(-1\right)^{\lambda-k} C_k,
\eq
and this implies
\bq
 b_\lambda
 +
 \sum\limits_{k=0}^{\lambda-2} \left(-1\right)^{\lambda-k} b_k
 & = &
 C_{\lambda}
 +
 \sum\limits_{k=0}^{\lambda-2} \left(-1\right)^{\lambda-k} C_k,
\eq
In a similar way one obtains from the third and the fourth equation
\bq
 \sum\limits_{k=0}^{\lambda-2} \left(-1\right)^{\lambda-k} b_k
 & = &
 \sum\limits_{k=0}^{\lambda-2} \left(-1\right)^{\lambda-k} C_k.
\eq
Hence $b_\lambda=C_\lambda=\dim H_{\lambda}(M)$.
}
\es
\\
\\
\bs
{\it \refstepcounter{exercise}
{\bf Exercise \theexercise}: 
Derive eq.~(\ref{chapter_iterated_integrals:relation_Euler_characteristic_Morse}) from eq.~(\ref{chapter_iterated_integrals:Morse_inequalities}).
\\
\\
{\bf Solution}: 
The proof is similar to the previous exercise.
Let us denote $b_k= \dim H_k(M)$.
With $\dim M = n$ we have trivially $b_{n+1} = \dim H_{n+1}(M) = 0$ and $C_{n+1}=0$.
We write down the Morse inequalities for $(n+1)$ and $n$, using $b_{n+1}=C_{n+1}=0$.
Multiplying in addition the first equation by $(-1)$ we obtain 
\bq
 \sum\limits_{k=0}^{n} \left(-1\right)^{n-k} b_k
 & \ge &
 \sum\limits_{k=0}^{n} \left(-1\right)^{n-k} C_k,
 \nonumber \\
 \sum\limits_{k=0}^n \left(-1\right)^{n-k} b_k
 & \le &
 \sum\limits_{k=0}^{n} \left(-1\right)^{n-k} C_k.
\eq
Hence
\bq
 \sum\limits_{k=0}^n \left(-1\right)^{n-k} b_k
 & = &
 \sum\limits_{k=0}^{n} \left(-1\right)^{n-k} C_k.
\eq
and
\bq
 \chi\left(M\right)
 & = &
 \sum\limits_{k=0}^n \left(-1\right)^{k} b_k
 \; = \;
 \sum\limits_{k=0}^{n} \left(-1\right)^{k} C_k.
\eq
}
\es
\\
%
%
\bs
{\it \refstepcounter{exercise}
{\bf Exercise \theexercise}: 
Let $\Nmaster=1$, $\NB=2$ and
\bq
 A & = & 
 d \ln\left(\frac{x_1}{x_1+x_2}\right).
\eq
Show that $B_\lambda$, defined as in eq.~(\ref{chapter_transformations:diff_eq_lambda}), equals zero.
\\
\\
{\bf Solution}: 
We have
\bq
 A
 & = &
 A_{x_1} dx_1 + A_{x_2} dx_2
 \; = \;
 \left( \frac{1}{x_1} - \frac{1}{x_1+x_2} \right) dx_1
 - \frac{1}{x_1+x_2} dx_2.
\eq
For $\alpha=[\alpha_1:1]$ we have
\bq
 B_\lambda
 & = & 
 \alpha_1 \left( \frac{1}{\alpha_1 \lambda} - \frac{1}{\alpha_1\lambda+\lambda} \right) 
 - \frac{1}{\alpha_1\lambda+\lambda} 
 \; = \; 0.
\eq
}
\es
\\
\\
\bs
{\it \refstepcounter{exercise}
{\bf Exercise \theexercise}: 
Prove the two relations in eq.~(\ref{chapter_transformations:conversion_Euler_operator}).
\\
\\
{\bf Solution}: 
We start with
\bq
 \frac{d^k}{dx^k} & = & x^{-k} \prod\limits_{j=0}^{k-1} \left(\theta - j\right).
\eq
We prove this relation by induction.
For $k=1$ the right-hand side equals
\bq
 x^{-1} \prod\limits_{j=0}^{0} \left(\theta - j\right)
 & = &
 x^{-1} \theta \; = \; \frac{d}{dx}.
\eq
Let us now assume that the relation is correct for $(k-1)$.
We have
\bq
 \frac{d^k}{dx^k}
 & = &
 \frac{d}{dx} \frac{d^{k-1}}{dx^{k-1}}
 \; = \; 
 \frac{1}{x} \theta \left[ x^{-k+1} \prod\limits_{j=0}^{k-2} \left(\theta - j\right) \right].
\eq
We further have the operator relation
\bq
 \theta x^{-k+1}
 & = & 
 x^{-k+1} \left( \theta -k+1\right)
\eq
and therefore
\bq
 \frac{d^k}{dx^k}
 & = &
 x^{-k} \left( \theta -k+1\right) \prod\limits_{j=0}^{k-2} \left(\theta - j\right)
 \; = \;
 x^{-k} \prod\limits_{j=0}^{k-1} \left(\theta - j\right).
\eq
Let us now look at
\bq
 \theta^k & = & \sum\limits_{j=1}^{k} S\left(k,j\right) x^j \frac{d^j}{dx^j}.
\eq
For $k=1$ the right-hand side equals
\bq
 \sum\limits_{j=1}^{1} S\left(1,j\right) x^j \frac{d^j}{dx^j}
 & = &
 S\left(1,1\right) x \frac{d}{dx}
 \; = \; \theta,
\eq
where we used $S(1,1)=1$.
Let us now assume that the relation is correct for $(k-1)$.
We have
\bq
 \theta^k
 & = &
 \theta \theta^{k-1}
 \; = \;
 x \frac{d}{dx} \left[ \sum\limits_{j=1}^{k-1} S\left(k-1,j\right) x^j \frac{d^j}{dx^j} \right]
 \nonumber \\
 & = &
 \sum\limits_{j=1}^{k-1} S\left(k-1,j\right) \left[ j x^j \frac{d^j}{dx^j} + x^{j+1} \frac{d^{j+1}}{dx^{j+1}} \right]
 \nonumber \\
 & = &
 \sum\limits_{j=1}^{k} \left[ j S\left(k-1,j\right) + S\left(k-1,j-1\right) \right] x^j \frac{d^j}{dx^j}.
\eq
In the last line we used $S(k-1,0)=S(k-1,k)=0$.
The Stirling numbers of the second kind satisfy the recurrence relation
\bq
 S\left(k,j\right) 
 & = &
 j S\left(k-1,j\right) + S\left(k-1,j-1\right).
\eq
With the help of this relation the claim follows:
\bq
 \theta^k
 & = &
 \sum\limits_{j=1}^{k} S\left(k,j\right) x^j \frac{d^j}{dx^j}.
\eq
}
\es
\\
\bs
{\it \refstepcounter{exercise}
{\bf Exercise \theexercise}: 
Rewrite
\bq
 L_2 & = &
 x \left(x+1\right)\left(x+9\right) \frac{d^2}{dx^2}
 + \left(3x^2+20x+9\right) \frac{d}{dx}
 + x + 3
\eq
in Euler operators.
(This is the differential operator of eq.~(\ref{chapter_transformations:L2_sunrise}) multiplied with $x(x+1)(x+9)$).
\\
\\
{\bf Solution}: 
With
\bq
 \frac{d^2}{dx^2} \; = \; \frac{1}{x^2} \theta \left(\theta-1\right),
 & &
 \frac{d}{dx} \; = \; \frac{1}{x} \theta
\eq
we first obtain
\bq
 L_2
 & = &
 \frac{\left(x+1\right)\left(x+9\right)}{x} \theta^2 + 2 \left(x+5\right) \theta + x + 3.
\eq
Multiplication with $x$ gives
\bq
 \tilde{L}_2
 & = &
 \left(x+1\right)\left(x+9\right) \theta^2 + 2 x \left(x+5\right) \theta + x \left(x + 3\right).
\eq
}
\es
\\
\\
\bs
{\it \refstepcounter{exercise}
{\bf Exercise \theexercise}: 
Consider
\bq
 \tilde{L} & = & \left(\theta-\alpha\right)^\lambda.
\eq
Show that the solution space is spanned by
\bq
\label{appendix_solutions:example_Frobenius_multiple_roots}
 x^\alpha, \; x^\alpha \ln\left(x\right), \; \dots, \; 
 \frac{x^\alpha \ln^{\lambda-1}\left(x\right)}{\left(\lambda-1\right)!}.
\eq
\\
\\
{\bf Solution}: 
Set
\bq
 f_j\left(x\right)
 & = &
 \frac{1}{j!} x^\alpha \ln^{j}\left(x\right).
\eq
We have
\bq
 \left(\theta - \alpha \right) f_0\left(x\right)
 & = &
 \left(x \frac{d}{dx} - \alpha \right) x^\alpha
 \; = \; 0.
\eq
For $j>0$ we have
\bq
 \left(\theta - \alpha \right) f_j\left(x\right)
 & = &
 f_{j-1}\left(x\right),
\eq
hence all functions in eq.~(\ref{appendix_solutions:example_Frobenius_multiple_roots}) are annihilated by $(\theta-\alpha)^\lambda$.
}
\es
\\
\\
\bs
{\it \refstepcounter{exercise}
{\bf Exercise \theexercise}: 
Consider the differential operators
\bq
 \tilde{L}_a & = & \left(\theta-1\right) \left(\theta-x\right),
 \nonumber \\
 \tilde{L}_b & = & \left(\theta-x\right) \left(\theta-1\right).
\eq
Construct for both operators two independent solutions around $x_0=0$.
\\
\\
{\bf Solution}: 
The indicial equation reads in both cases
\bq
 \left(\alpha-1\right) \alpha & = & 0,
\eq
hence the indicials are in both cases $\alpha_1=0$ and $\alpha_2=1$.

Let's first consider the case $\tilde{L}_a$. 
We have
\bq
 \tilde{L}_a & = & \left(\theta-1\right) \left(\theta-x\right)
 \; = \; x^2 \frac{d^2}{dx^2} - x^2 \frac{d}{dx}
\eq
and it is clear that $f_{a,1}(x)=1$ is a solution.
$f_{a,1}$ corresponds to the indicial $\alpha_1=0$. We may view $f_{a,1}$ as a power series which terminates after the first term.
We construct the second solution (corresponding to the indicial $\alpha_2=1$) from the ansatz
\bq
 x + \sum\limits_{j=2}^\infty c_{2,j} x^j.
\eq
One finds
\bq
 f_{a,2}\left(x\right) & = & 
 \sum\limits_{j=1}^\infty \frac{x^j}{j!}
 \; = \; e^x -1.
\eq
Let us now look at $\tilde{L}_b$.
We have
\bq
 \tilde{L}_a & = & \left(\theta-x\right) \left(\theta-1\right)
 \; = \; x^2 \frac{d^2}{dx^2} - x^2 \frac{d}{dx} + x.
\eq
One solution is again trivial: One easily checks that
\bq
 f_{b,2}\left(x\right) & = & x
\eq
is a solution.
The purpose of this exercise is to show how the solution $f_{b,1}(x)$, which starts at order $x^0$, is extended 
to higher orders in $x$.
We start from the ansatz
\bq
 f_{b,1}\left(x\right)
 & = &
 1 + \sum\limits_{j=1}^\infty \left[ c_{1,j,0} \ln\left(x\right) + c_{1,j,1} \right] x^j.
\eq
Inserting this ansatz into the differential equation one finds 
\bq
 0 & = &
 x + \sum\limits_{j=1}^\infty
 \left\{
 \left[ j \left(j-1\right) c_{1,j,1} + \left(2j-1\right) c_{1,j,0} \right] x^j
 -
 \left[ \left(j-1\right) c_{1,j,1} + c_{1,j,0} \right] x^{j+1}
 \right. \nonumber \\
 & & \left.
 + j \left(j-1\right) c_{1,j,0} x^j \ln\left(x\right)
 - \left(j-1\right) c_{1,j,0} x^{j+1} \ln\left(x\right)
 \right\}.
\eq
The coefficients of $x^j$ and $x^j \ln(x)$ have to vanish separately.
From the term $x^1$ and the logarithmic terms we obtain
\bq
 c_{1,1,0} \; = \; -1,
 & &
 c_{1,j,0} \; = \; 0
 \;\;\;\;\;\; \mbox{for} \; j \ge 2.
\eq
The differential equation does not constrain $c_{1,1,1}$, as this term corresponds to the second independent solution
$f_{b,2}(x)$. Therefore we may set $c_{1,1,1}=0$. For the higher terms one finds $c_{1,2,1}=-1/2$ and the recursion formula
\bq
 c_{1,j,1} & = & \frac{\left(j-2\right)}{j\left(j-1\right)} c_{1,j-1,1},
 \;\;\;\;\;\; j \; \ge \; 3.
\eq
Thus
\bq
 f_{b,1}\left(x\right)
 & = &
 1 - x \ln\left(x\right)
 - \frac{1}{2} x^2
 - \frac{1}{12} x^3
 - \frac{1}{72} x^4
 - \frac{1}{480} x^5
 + {\mathcal O}\left(x^6\right).
\eq
}
\es
\\
\bs
{\it \refstepcounter{exercise}
{\bf Exercise \theexercise}: 
Show the equivalence of eq.~(\ref{chapter_transformations:solution_path_ordered_exponential})
with eq.~(\ref{chapter_transformations:solution_infinte_series}).
\\
\\
{\bf Solution}: 
We have to show
\bq
 {\mathcal P}
 \exp\left( - \int\limits_0^x dx_1 A\left(x_1\right) \right)
 & = &
 1
 - \int\limits_0^x dx_1 A\left(x_1\right)
 + \int\limits_0^x dx_1 A\left(x_1\right)
   \int\limits_0^{x_1} dx_2 A\left(x_2\right)
 \nonumber \\
 & & 
 - \int\limits_0^x dx_1 A\left(x_1\right)
   \int\limits_0^{x_1} dx_2 A\left(x_2\right)
   \int\limits_0^{x_2} dx_3 A\left(x_3\right)
 + \dots
\eq
We start with the left-hand side and expand the exponential function:
\bq
\label{appendix_solutions:expansion_path_ordered_exponential}
 {\mathcal P}
 \exp\left( - \int\limits_0^x dx_1 A\left(x_1\right) \right)
 & = &
 {\mathcal P}
 \sum\limits_{n=0}^\infty \frac{1}{n!} \left[ - \int\limits_0^x dx_1 A\left(x_1\right) \right]^n.
\eq
The $n$-th term in this sum has $n$ integrations. Let's label the integration variables $x_1, \dots, x_n$.
The integration region is an $n$-dimensional cube
\bq
 \left[ 0, x\right]^n.
\eq
We divide the $n$-dimensional cube into $n!$ simplices defined by
\bq
 x \; \ge \; x_{\sigma_1} \; > \; s_{\sigma_2} \; > \; \dots \; > \; x_{\sigma_n} \; \ge \; 0.
\eq
Each simplex is uniquely specified by a permutation $\sigma \in S_n$.
For each simplex, the path ordering operator gives
\bq
 {\mathcal P}\left( A\left(x_1\right) A\left(x_2\right) \dots A\left(x_n\right) \right)
 & = &
 A\left(x_{\sigma_1}\right) A\left(x_{\sigma_2}\right) \dots A\left(x_{\sigma_n}\right).
\eq
By a relabelling of the integration variables we see that each of the $n!$ simplices gives exactly the same contribution,
cancelling the $1/n!$ factor in eq.~(\ref{appendix_solutions:expansion_path_ordered_exponential}).
Thus we obtain
\bq
 {\mathcal P}
 \exp\left( - \int\limits_0^x dx_1 A\left(x_1\right) \right)
 & = &
 \sum\limits_{n=0}^\infty \left(-1\right)^n
 \int\limits_0^x dx_1 A\left(x_1\right)
 \int\limits_0^{x_1} dx_2 A\left(x_2\right)
 \dots
 \int\limits_0^{x_{n-1}} dx_n A\left(x_n\right).
\eq
}
\es
\\
\\
\bs
{\it \refstepcounter{exercise}
{\bf Exercise \theexercise}: 
Prove eq.~(\ref{chapter_transformations:Magnus_diagonal_nilpotent}).
\\
\\
{\bf Solution}: 
We have to show
\bq
 \frac{d}{dx} 
 \left( e^{\Omega\left[D_x\right]\left(x\right)} \; e^{\Omega\left[N_x'\right]\left(x\right)} \right)
 & = &
 - 
 \left( D_x\left(x\right) + N_x\left(x\right) \right) 
  e^{\Omega\left[D_x\right]\left(x\right)} \;
 e^{\Omega\left[N_x'\right]\left(x\right)}.
\eq
In general, the Magnus series $\Omega[A_x](x)$ satisfies
\bq
 \frac{d}{dx} 
 \left( e^{\Omega\left[A_x\right]\left(x\right)} \right)
 & = &
 - A_x\left(x\right) e^{\Omega\left[A_x\right]\left(x\right)}.
\eq
}
Therefore
\bq
 \frac{d}{dx} 
 \left( e^{\Omega\left[D_x\right]\left(x\right)} \; e^{\Omega\left[N_x'\right]\left(x\right)} \right)
 & = &
 - D_x\left(x\right) e^{\Omega\left[D_x\right]\left(x\right)} \; e^{\Omega\left[N_x'\right]\left(x\right)}
 - e^{\Omega\left[D_x\right]\left(x\right)} \; N_x'\left(x\right) e^{\Omega\left[N_x'\right]\left(x\right)}.
\eq
We have
\bq
 - e^{\Omega\left[D_x\right]\left(x\right)} \; N_x'\left(x\right) e^{\Omega\left[N_x'\right]\left(x\right)}
 & = & 
 - e^{\Omega\left[D_x\right]\left(x\right)} e^{-\Omega\left[D_x\right]\left(x\right)} N_x\left(x\right) e^{\Omega\left[D_x\right]\left(x\right)} e^{\Omega\left[N_x'\right]\left(x\right)}
 \nonumber \\
 & = &
 - N_x\left(x\right) e^{\Omega\left[D_x\right]\left(x\right)} e^{\Omega\left[N_x'\right]\left(x\right)}
\eq
and therefore
\bq
 - D_x\left(x\right) e^{\Omega\left[D_x\right]\left(x\right)} \; e^{\Omega\left[N_x'\right]\left(x\right)}
 - e^{\Omega\left[D_x\right]\left(x\right)} \; N_x'\left(x\right) e^{\Omega\left[N_x'\right]\left(x\right)}
 & = &
 - 
 \left( D_x\left(x\right) + N_x\left(x\right) \right) 
  e^{\Omega\left[D_x\right]\left(x\right)} \;
 e^{\Omega\left[N_x'\right]\left(x\right)}.
 \;\;\;\;\;\;\;\;\;
\eq
\es
\\
\\
\bs
{\it \refstepcounter{exercise}
{\bf Exercise \theexercise}: 
Show that the transformation in eq.~(\ref{chapter_transformations:Magnus_trafo_A0}) 
transforms the differential equation eq.~(\ref{chapter_transformations:Magnus_linear_dgl}) into
eq.~(\ref{chapter_transformations:Magnus_trafo_A0_eps_form}).
\\
\\
{\bf Solution}: 
We have to show that 
\bq
 \vec{I}' & = & U \vec{I},
 \;\;\;\;\;\;\;\;\;
 U \; = \; e^{-\Omega[A_x^{(0)}]\left(x\right)}
\eq
transforms the differential equation
\bq
 \left( \frac{d}{dx} + A_x^{(0)} + \eps A_x^{(1)} \right) \vec{I} & = & 0
\eq
into
\bq
 \left( \frac{d}{dx} + A_x' \right) \vec{I}' & = & 0,
\eq
with $A_x'$ given by
\bq
 A_x'
 & = &
 \eps U A_x^{(1)} U^{-1}.
\eq
$A_x'$ is given by
\bq
 A_x'
 & = &
 U A_x U^{-1} + U \frac{d}{dx} U^{-1}.
\eq
We have $U^{-1}= e^{\Omega[A_x^{(0)}]\left(x\right)}$ and
\bq
 \frac{d}{dx} U^{-1}
 & = &
 - A_x^{(0)} e^{\Omega[A_x^{(0)}]\left(x\right)}
 \; = \;
 - A_x^{(0)} U^{-1}.
\eq
Thus
\bq
 A_x'
 & = &
 U \left( A_x^{(0)} + \eps A_x^{(1)} \right) U^{-1} - U A_x^{(0)} U^{-1}
 \; = \;
 \eps U A_x^{(1)} U^{-1}.
\eq
}
\es
\\
\\
\bs
{\it \refstepcounter{exercise}
{\bf Exercise \theexercise}: 
Assume that $A_x$ is in Fuchsian form (i.e. of the form as in eq.~(\ref{chapter_transformations:moser_A_x_Fuchsian_form})).
Show that the matrix residue at $x=\infty$ is given by
\bq
 M_{\infty,1}\left(\eps\right)
 & = &
 -
 \sum\limits_{x_i \in S'} 
 M_{x_i,1}\left(\eps\right).
\eq 
{\bf Solution}: 
We map the point $x=\infty$ to $x'=0$ with the transformation $x'=1/x$.
We have
 \bq
 dx & = & - \frac{dx'}{x'{}^2}
\eq
and with $x_i'=1/x_i$
\bq
 A_x dx
 & = &
 \sum\limits_{x_i \in S'} 
 M_{x_i,1}\left(\eps\right)
 \frac{1}{\left(x-x_i\right)}
 dx
 \; = \;
 - \sum\limits_{x_i \in S'} 
 M_{x_i,1}\left(\eps\right)
 \frac{x_i'}{x' \left(x_i'-x'\right)}
 dx'
 \nonumber \\
 & = &
 - \sum\limits_{x_i \in S'} 
 M_{x_i,1}\left(\eps\right)
 \frac{1}{x'}
 dx'
 + \dots,
\eq
where the dots stand for terms regular at $x'=0$. Thus
\bq
 M_{\infty,1}\left(\eps\right)
 & = &
 -
 \sum\limits_{x_i \in S'} 
 M_{x_i,1}\left(\eps\right).
\eq 
}
\es
\\
\\
\bs
{\it \refstepcounter{exercise}
{\bf Exercise \theexercise}: 
Show that $P$ and ${\bf 1}-P$ are projectors, i.e.
\bq
 P^2 \; = \; P,
 & &
 \left({\bf 1}-P\right)^2 \; = \; {\bf 1}-P.
\eq
Show further
\bq
 \left[ \left({\bf 1}-P\right) + \frac{x-x_2}{x-x_1} P \right]
 \left[ \left({\bf 1}-P\right) + \frac{x-x_1}{x-x_2} P \right]
 & = &
 {\bf 1}.
\eq
{\bf Solution}: 
We have
\bq
 P^2
 \; = \;
 \frac{\vec{v}_{R,x_1} \; \vec{v}_{L,x_2}^T}{\left( \vec{v}_{L,x_2}^T \cdot \vec{v}_{R,x_1} \right)}
 \cdot
 \frac{\vec{v}_{R,x_1} \; \vec{v}_{L,x_2}^T}{\left( \vec{v}_{L,x_2}^T \cdot \vec{v}_{R,x_1} \right)}
 \; = \;
 \frac{\vec{v}_{R,x_1} \; \left( \vec{v}_{L,x_2}^T \cdot \vec{v}_{R,x_1} \right) \; \vec{v}_{L,x_2}^T}{\left( \vec{v}_{L,x_2}^T \cdot \vec{v}_{R,x_1} \right)^2}
 \; = \;
 \frac{\vec{v}_{R,x_1} \; \vec{v}_{L,x_2}^T}{\left( \vec{v}_{L,x_2}^T \cdot \vec{v}_{R,x_1} \right)}
 \; = \; P.
\eq
Further
\bq
 \left({\bf 1}-P\right)^2
 \; = \;
 {\bf 1} - 2 P + P^2
 \; = \;
 {\bf 1} - 2 P + P
 \; = \;
 {\bf 1} - P.
\eq
With
\bq
 P \left({\bf 1}-P\right)
 \; = \; 
 \left({\bf 1}-P\right) P
 \; = \; 0
\eq
we also have
\bq
\lefteqn{
 \left[ \left({\bf 1}-P\right) + \frac{x-x_2}{x-x_1} P \right]
 \left[ \left({\bf 1}-P\right) + \frac{x-x_1}{x-x_2} P \right]
 = } & &
 \nonumber \\
 & = &
 \left({\bf 1}-P\right)^2
 + \frac{x-x_2}{x-x_1} P \left({\bf 1}-P\right)
 + \frac{x-x_1}{x-x_2} \left({\bf 1}-P\right) P
 + P^2
 \; = \;
 \left({\bf 1}-P\right)
 + P
 \; = \; 
 {\bf 1}.
\eq
}
\es
\\
\\
\bs
{\it \refstepcounter{exercise}
{\bf Exercise \theexercise}: 
Consider the square roots
\bq
 r_1 \; = \; \sqrt{x\left(4+x\right)}
 & \mbox{and} &
 r_2 \; = \; \sqrt{x\left(36+x\right)}.
\eq
Find a transformation, which simultaneously rationalises $r_1$ and $r_2$.
\\
\\
{\bf Solution}: 
We rationalise the two roots sequentially. We start with the root $r_1$.
We already know that the transformation
\bq
 x & = & \frac{\left(1-x'\right)^2}{x'}
\eq
rationalises $r_1$.
The root $r_2$ expressed in terms of $x'$ reads
\bq
 r_2 & = & \frac{1-x'}{x'} \sqrt{1+34x'+x'{}^2}
\eq
We have 
\bq
 1+34x'+x'{}^2
 & = &
 \left( x' + 17 - 12 \sqrt{2} \right)\left( x' + 17 + 12 \sqrt{2} \right)
\eq
and eq.~(\ref{chapter_transformations:square_root_quadratic_polynomial_rationalisation_II}) gives us
\bq
 x' & = &
 - \frac{6\sqrt{2}}{x''}\left( 1 + \frac{17}{12}\sqrt{2} x'' + x''{}^2 \right).
\eq
The variable $x''$ simultaneously rationalises $r_1$ and $r_2$:
\bq
 r_1
 & = &
 12 \sqrt{2} \frac{\left(x''+\sqrt{2}\right)\left(2x''+\sqrt{2}\right)\left(3x''{}^2+4\sqrt{2}x''+3\right)}{x''\left(3x''+2\sqrt{2}\right)\left(4x''+3\sqrt{2}\right)},
 \nonumber \\
 r_2
 & = &
 36 \sqrt{2} \frac{\left(x''+\sqrt{2}\right)\left(2x''+\sqrt{2}\right)\left(1-x''{}^2\right)}{x''\left(3x''+2\sqrt{2}\right)\left(4x''+3\sqrt{2}\right)}.
\eq
The appearance of the root $\sqrt{2}$ is unaesthetic, but not a principal problem.
It stems from the fact that in using eq.~(\ref{chapter_transformations:square_root_quadratic_polynomial_rationalisation_II})
we have implicitly chosen the non-rational point $(r,x')=(0,-17-12\sqrt{2})$ on the hypersurface
\bq
 f(r,x') & = &
 r^2 - x'{}^2 - 34 x' - 1.
\eq
This hypersurface has rational points, for example $(r,x')=(6,1)$.
Repeating the exercise one finds for example that the transformation
\bq
 x & = & \frac{\left(\tilde{x}^2-9\right)^2}{\tilde{x}\left(\tilde{x}+1\right)\left(\tilde{x}+9\right)}
\eq
rationalises both $r_1$ and $r_2$
\bq
 r_1
 \; = \;
 \frac{\left(\tilde{x}^2-9\right)\left(\tilde{x}^2+2\tilde{x}+9\right)}{\tilde{x}\left(\tilde{x}+1\right)\left(\tilde{x}+9\right)},
 & &
 r_2
 \; = \;
 \frac{\left(\tilde{x}^2-9\right)\left(\tilde{x}^2+18\tilde{x}+9\right)}{\tilde{x}\left(\tilde{x}+1\right)\left(\tilde{x}+9\right)},
\eq
and only contains rational coefficients.
}
\es
\\
\\
%
%
\bs
{\it \refstepcounter{exercise}
{\bf Exercise \theexercise}: 
Prove eq.~(\ref{chapter_multiple_polylogarithms:conversion_Li_to_G}).
\\
\\
{\bf Solution}: 
We have to show
\bq
 \mathrm{Li}_{m_1 \dots m_k}(x_1,\dots,x_k)
 & = & (-1)^k 
 G_{m_1 \dots m_k}\left( \frac{1}{x_1}, \frac{1}{x_1 x_2}, \dots, \frac{1}{x_1...x_k};1 \right),
\eq
where we may assume that 
\bq
\left| x_1 x_2 \dots x_j \right| \le 1 & & \mbox{for all} \; j \in \{1,\dots,k\} \;\; \mbox{and} \;\; (m_1,x_1) \neq (1,1).
\eq
Set $r=m_1+\dots+m_k$.
The integral representation has depth $r$.
Let us introduce some notation to facilitate the proof:
We set
\bq
b_j & = & \frac{1}{x_1 x_2 ... x_j}.
\eq
and introduce the following notation for iterated integrals
\bq
 \int\limits_0^y \frac{dt}{t-z_1} \circ ... \circ \frac{dt}{t-z_r} 
 & = & 
 \int\limits_0^y \frac{dt_1}{t_1-z_1} \int\limits_0^{t_1} \frac{dt_2}{t_2-z_2} \dots \int\limits_0^{t_{r-1}} \frac{dt_r}{t_r-z_r},
\eq
together with the short hand notation
\bq
 \int\limits_0^y \left( \frac{dt}{t} \circ \right)^{m} \frac{dt}{t-z}
 & = & 
 \int\limits_0^y
 \underbrace{\frac{dt}{t} \circ ... \frac{dt}{t}}_{m \;\mathrm{times}} \circ \frac{dt}{t-z}.
\eq
The integral representation $G_{m_1 \dots m_k}( b_1, \dots, b_k; 1)$
reads then
\bq
 (-1)^k 
 G_{m_1 \dots m_k}\left( b_1, \dots, b_k; 1\right)
 & = &
 (-1)^k \int\limits_0^1 
 \left( \frac{dt}{t} \circ \right)^{m_1-1} \frac{dt}{t-b_1} 
 \circ ... \circ
 \left( \frac{dt}{t} \circ \right)^{m_k-1} \frac{dt}{t-b_k}.
\eq
For all integration variables we have $|t_j| \le 1$ (with $j \in \{1,\dots,r\}$).
We have $k$ terms of the form $1/(t_j-b_j)$. As $|t_j/b_j| = |x_1 x_2 \dots x_j t_j| \le 1$ we expand the geometric series
(the case $x_1 x_2 \dots x_j t_j=1$ is handled by first replacing the outermost integration limit $1$ by $y<1$ and taking the limit $y\rightarrow 1$ in the end.
With $y<1$ we have $|t_j| < 1$, which is sufficient for the convergence of the geometric series. 
The limit $y\rightarrow 1$ may be taken for $(m_1,x_1) \neq (1,1)$):
\bq
 \frac{1}{t_j-b_j}
 & = &
 - \frac{1}{b_j} \sum\limits_{i=0}^{\infty} \left( \frac{t_j}{b_j} \right)^{i}
 \; = \;
 - \sum\limits_{i=1}^{\infty} \frac{t_j^{i-1}}{b_j^i}.
\eq
Integrating term-by-term gives
\bq
 (-1)^k 
 G_{m_1 \dots m_k}\left( b_1, \dots, b_k; 1\right)
 & = &
 \sum\limits_{i_1=1}^{\infty} \dots \sum\limits_{i_k=1}^{\infty}
 \frac{1}{\left(i_1+\dots+i_k\right)^{m_1}}
 \frac{1}{b_1^{i_1}}
 \dots
 \frac{1}{\left(i_{k-1}+i_k\right)^{m_{k-1}}}
 \frac{1}{b_{k-1}^{i_{k-1}}}
 \frac{1}{i_k^{m_k}}
 \frac{1}{b_k^{i_k}}
 \nonumber \\
 & = &
 \sum\limits_{i_1=1}^{\infty} \dots \sum\limits_{i_k=1}^{\infty}
 \frac{x_1^{i_1}}{\left(i_1+\dots+i_k\right)^{m_1}}
 \dots
 \frac{\left(x_1 \dots x_{k-1}\right)^{i_{k-1}}}{\left(i_{k-1}+i_k\right)^{m_{k-1}}}
 \frac{\left(x_1 \dots x_k\right)^{i_k}}{i_k^{m_k}}
 \nonumber \\
 & = &
 \sum\limits_{i_1=1}^{\infty} \dots \sum\limits_{i_k=1}^{\infty}
 \frac{x_1^{i_1+\dots+i_k}}{\left(i_1+\dots+i_k\right)^{m_1}}
 \dots
 \frac{x_{k-1}^{i_{k-1}+i_k}}{\left(i_{k-1}+i_k\right)^{m_{k-1}}}
 \frac{x_k^{i_k}}{i_k^{m_k}}.
\eq
Changing the summation indices according to $n_1=i_1+\dots+i_k$, $n_2=i_2+\dots+i_k$, $\dots$, $n_{k-1}=i_{k-1}+i_k$ and $n_k=i_k$ yields
\bq
 (-1)^k 
 G_{m_1 \dots m_k}\left( b_1, \dots, b_k; 1\right)
 & = &
 \sum\limits_{n_1=1}^{\infty} 
 \sum\limits_{n_2=1}^{n_1-1} 
 \dots 
 \sum\limits_{n_{k-1}=1}^{n_{k-2}-1}
 \sum\limits_{n_k=1}^{n_{k-1}-1}
 \frac{x_1^{n_1}}{n_1^{m_1}}
 \frac{x_2^{n_2}}{n_2^{m_2}}
 \dots
 \frac{x_{k-1}^{n_{k-1}}}{n_{k-1}^{m_{k-1}}}
 \frac{x_k^{n_k}}{n_k^{m_k}}
 \nonumber \\
 & = &
 \mathrm{Li}_{m_1 \dots m_k}(x_1,\dots,x_k).
\eq
}
\es
\\
\\
\bs
{\it \refstepcounter{exercise}
{\bf Exercise \theexercise}: 
Consider the alphabet $A=\{l_1,l_2\}$ with $l_1<l_2$. Write down all Lyndon words of depth $\le 3$.
\\
\\
{\bf Solution}: 
At depth $1$ we have the words $l_1$ and $l_2$. Both of them are Lyndon words.
At depth $2$ we have to consider the words 
$l_1 l_1$, $l_1 l_2$, $l_2 l_1$ and $l_2 l_2$.
Out of these only
\bq
 l_1 l_2
\eq
is a Lyndon word. For example $w=l_1 l_1$ may be written as $w=u v$, $u=l_1$, $v=l_1$ and $v<w$.
At depth $3$ we have to consider the words 
$l_1 l_1 l_1$, $l_1 l_1 l_2$, $l_1 l_2 l_1$, $l_1 l_2 l_2$,
$l_2 l_1 l_1$, $l_2 l_1 l_2$, $l_2 l_2 l_1$ and $l_2 l_2 l_2$.
Out of these the Lyndon words are
\bq
 l_1 l_1 l_2,
 & &
 l_1 l_2 l_2.
\eq
}
\es
\\
\\
\bs
{\it \refstepcounter{exercise}
{\bf Exercise \theexercise}: 
Express the product
\bq
 G_2\left(z;y\right) \cdot G_3\left(z;y\right)
\eq
as a linear combination of multiple polylogarithms.
\\
\\
{\bf Solution}: 
We have
\bq
 G_2\left(z;y\right) \; = \; G\left(0,z;y\right),
 & &
 G_3\left(z;y\right) \; = \; G\left(0,0,z;y\right).
\eq
We first work out the shuffle product for
$G\left(z_1,z_2;y\right) \cdot G\left(z_3,z_4,z_5;y\right)$
and set $z_1=z_3=z_4=0$ and $z_2=z_5=z$ in the end.
We have
\bq
\lefteqn{
 G\left(z_1,z_2;y\right) \cdot G\left(z_3,z_4,z_5;y\right)
 = } & &
 \nonumber \\
 & &
 G\left(z_1,z_2,z_3,z_4,z_5;y\right)
 + G\left(z_1,z_3,z_2,z_4,z_5;y\right)
 + G\left(z_1,z_3,z_4,z_2,z_5;y\right)
 + G\left(z_1,z_3,z_4,z_5,z_2;y\right)
 \nonumber \\
 & &
 + G\left(z_3,z_1,z_2,z_4,z_5;y\right)
 + G\left(z_3,z_1,z_4,z_2,z_5;y\right)
 + G\left(z_3,z_1,z_4,z_5,z_2;y\right)
 + G\left(z_3,z_4,z_1,z_2,z_5;y\right)
 \nonumber \\
 & &
 + G\left(z_3,z_4,z_1,z_5,z_2;y\right)
 + G\left(z_3,z_4,z_5,z_1,z_2;y\right).
\eq
Setting $z_1=z_3=z_4=0$ and $z_2=z_5=z$ yields
\bq
 G\left(0,z;y\right) \cdot G\left(0,0,z;y\right)
 & = &
 G\left(0,z,0,0,z;y\right)
 + 3 G\left(0,0,z,0,z;y\right)
 + 6 G\left(0,0,0,z,z;y\right),
\eq
or
\bq
 G_2\left(z;y\right) \cdot G_3\left(z;y\right)
 & = &
 G_{2 3}\left(z,z;y\right)
 + 3 G_{3 2}\left(z,z;y\right)
 + 6 G_{4 1}\left(z,z;y\right).
\eq
}
\es
\\
\\
\bs
{\it \refstepcounter{exercise}
{\bf Exercise \theexercise}: 
Work out the quasi-shuffle product
\bq
 \mathrm{Li}_{m_1 m_2}(x_1,x_2) \cdot \mathrm{Li}_{m_3 m_4}(x_3,x_4).
\eq
\\
\\
{\bf Solution}: 
We obtain
\bq
\lefteqn{
 \mathrm{Li}_{m_1 m_2}(x_1,x_2) \cdot \mathrm{Li}_{m_3 m_4}(x_3,x_4).
 = } & &
 \nonumber \\
 & &
 \mathrm{Li}_{m_1 m_2 m_3 m_4}(x_1,x_2,x_3,x_4)
 + \mathrm{Li}_{m_1 m_3 m_2 m_4}(x_1,x_3,x_2,x_4)
 + \mathrm{Li}_{m_1 m_3 m_4 m_2}(x_1,x_3,x_4,x_2)
 \nonumber \\
 & &
 + \mathrm{Li}_{m_3 m_1 m_2 m_4}(x_3,x_1,x_2,x_4)
 + \mathrm{Li}_{m_3 m_1 m_4 m_2}(x_3,x_1,x_4,x_2)
 + \mathrm{Li}_{m_3 m_4 m_1 m_2}(x_3,x_4,x_1,x_2)
 \nonumber \\
 & &
 + \mathrm{Li}_{\left(m_1+m_3\right) m_2 m_4}(x_1 \cdot x_3,x_2,x_4)
 + \mathrm{Li}_{\left(m_1+m_3\right) m_4 m_2}(x_1 \cdot x_3,x_4,x_2)
 + \mathrm{Li}_{m_1 \left(m_2+m_3\right) m_4}(x_1,x_2 \cdot x_3,x_4)
 \nonumber \\
 & &
 + \mathrm{Li}_{m_3 \left(m_1+m_4\right) m_2}(x_3,x_1 \cdot x_4,x_2)
 + \mathrm{Li}_{m_1 m_3 \left(m_2+m_4\right)}(x_1,x_3,x_2 \cdot x_4)
 + \mathrm{Li}_{m_3 m_1 \left(m_2+m_4\right)}(x_3,x_1,x_2 \cdot x_4)
 \nonumber \\
 & &
 + \mathrm{Li}_{\left(m_1+m_3\right) \left(m_2+m_4\right)}(x_1 \cdot x_3,x_2 \cdot x_4).
\eq
}
\es
\\
\\
\bs
{\it \refstepcounter{exercise}
{\bf Exercise \theexercise}: 
Use the (regularised) double-shuffle relations to show
\bq
 \zeta_2^2 & = & \frac{5}{2} \zeta_4.
\eq
{\bf Solution}: 
As in eq.~(\ref{chapter_multiple_polylogarithms:regularised_zeta_1}) we set
\bq
 L & = & - \ln \lambda \; = \; \mathrm{Li}_1\left(1-\lambda\right) \; = \; - G\left(1;1-\lambda\right).
\eq
In the quasi-shuffle algebra we have
\bq
 L \cdot \zeta_3
 & = &
 L \cdot \mathrm{Li}_{3}\left(1\right)
 \; = \; 
 \mathrm{Li}_{1 3}\left(1-\lambda,1\right)
 +
 \mathrm{Li}_{3 1}\left(1,1-\lambda\right)
 +
 \mathrm{Li}_{4}\left(1-\lambda\right).
\eq
In the shuffle algebra we consider
\bq
 - L \cdot G\left(0,0,1;1-\lambda\right)
 & = &
 G\left(1;1-\lambda\right) \cdot G\left(0,0,1;1-\lambda\right)
 \\
 & = &
 G\left(1,0,0,1;1-\lambda\right)
 +
 G\left(0,1,0,1;1-\lambda\right)
 +
 2 G\left(0,0,1,1;1-\lambda\right).
 \nonumber
\eq
In the $\mathrm{Li}$-notation this is equivalent to
\bq
 L \cdot \mathrm{Li}_{3}\left(1-\lambda\right)
 & = &
 \mathrm{Li}_{1 3}\left(1-\lambda,1\right)
 +
 \mathrm{Li}_{2 2}\left(1-\lambda,1\right)
 +
 2 \mathrm{Li}_{3 1}\left(1-\lambda,1\right).
\eq
We have
$\mathrm{Li}_{3}\left(1-\lambda\right) - \zeta_3 = {\mathcal O}\left(\lambda\right)$
and subtracting the quasi-shuffle relation from the shuffle relation
\bq
 \zeta_{2 2} + \zeta_{3 1} - \zeta_4 & = & 0
\eq
follows. From eq.~(\ref{chapter_multiple_polylogarithms:example_zeta_31}) 
we know already that $\zeta_{3 1} = \frac{1}{4} \zeta_4$, and therefore
\bq
 \zeta_{2 2} & = & \frac{3}{4} \zeta_4.
\eq
Substituting this result into eq.~(\ref{chapter_multiple_polylogarithms:example_zeta_31_quasi_shuffle})
the sought-after relation
\bq
 \zeta_2^2 & = & \frac{5}{2} \zeta_4.
\eq
follows.
}
\es
\\
\\
\bs
{\it \refstepcounter{exercise}
{\bf Exercise \theexercise}: 
Let
\bq
 f_0\left(x\right) \; = \; \frac{1}{r!} \ln^r\left(x\right),
 & &
 f_1\left(x\right) \; = \; \frac{\left(-1\right)^r}{r!} \ln^r\left(1-x\right).
\eq
Determine
\bq
 {\mathcal M}_0 f_0\left(x\right),
 \;\;\;\;\;\;
 {\mathcal M}_0 f_1\left(x\right),
 \;\;\;\;\;\;
 {\mathcal M}_1 f_0\left(x\right),
 \;\;\;\;\;\;
 {\mathcal M}_1 f_1\left(x\right).
\eq
{\bf Solution}: 
Let us first discuss the case $r=1$:
\bq
 h_0\left(x\right) \; = \; \ln\left(x\right),
 & &
 h_1\left(x\right) \; = \; - \ln\left(1-x\right).
\eq
We already know that (recall that $\ln(x)$ is regular at $x=1$ and $\ln(1-x)$ is regular at $x=0$)
\begin{align}
 {\mathcal M}_0 h_0\left(x\right) & = h_0\left(x\right) + 2 \pi i,
 &
 {\mathcal M}_0 h_1\left(x\right) & = h_1\left(x\right),
 \nonumber \\
 {\mathcal M}_1 h_0\left(x\right) & = h_0\left(x\right),
 &
 {\mathcal M}_1 h_1\left(x\right) & = h_1\left(x\right) - 2 \pi i.
\end{align}
We then have
\bq
 {\mathcal M}_0 f_0\left(x\right) & = &
 \frac{1}{r!} \left[ {\mathcal M}_0 h_0\left(x\right) \right]^r
 \; = \; 
 \frac{1}{r!} \left[ \ln\left(x\right) + 2 \pi i\right]^r,
 \nonumber \\
 {\mathcal M}_0 f_1\left(x\right) & = &
 \frac{1}{r!} \left[ {\mathcal M}_0 h_1\left(x\right) \right]^r
 \; = \; 
 f_1\left(x\right),
 \nonumber \\
 {\mathcal M}_1 f_0\left(x\right) & = &
 \frac{1}{r!} \left[ {\mathcal M}_1 h_0\left(x\right) \right]^r
 \; = \; 
 f_0\left(x\right),
 \nonumber \\
 {\mathcal M}_1 f_1\left(x\right) & = &
 \frac{1}{r!} \left[ {\mathcal M}_1 h_1\left(x\right) \right]^r
 \; = \; 
 \frac{1}{r!} \left[ -\ln\left(1-x\right) - 2 \pi i\right]^r.
\eq
}
\es
\\
\\
\bs
{\it \refstepcounter{exercise}
{\bf Exercise \theexercise}: 
Compute the monodromy of $G(1,1;y)$ around $y=1$.
\\
\\
{\bf Solution}: The simple solution is as follows: As $G(1,1;y) = \frac{1}{2} \ln^2(1-y)$
it follows from exercise~\ref{chapter_multiple_polylogarithms:exercise_monodromy_log} that
\bq
 {\mathcal M}_1 G\left(1,1;y\right) 
 & = &
 \frac{1}{2} \left[ -\ln\left(1-y\right) - 2 \pi i\right]^2
 \nonumber \\
 & = &
 G\left(1,1;y\right) + 2 \pi i G\left(1;y\right) + \frac{1}{2} \left(2 \pi i\right)^2.
\eq
The purpose of this exercise is to compute the monodromy with the help of eq.~(\ref{chapter_multiple_polylogarithms:monodromy_Glog}).
Of course, we should obtain the same result.
With $G(1;y) = \ln(1-y)$ and ${\mathcal M}_1 G\left(1;y\right) = G\left(1;y\right) + 2 \pi i$ 
we obtain from eq.~(\ref{chapter_multiple_polylogarithms:monodromy_Glog})
and eq.~(\ref{chapter_multiple_polylogarithms:monodromy_Glog_z_equals_z1})
\bq
 {\mathcal M}_1 G\left(1,1;y\right)
 & = & 
 G\left(1,1;y\right)
 +
 \lim\limits_{\eps \rightarrow 0}
 \left\{
 \oint \frac{dy'}{y'-1} G\left(1;y'\right)
 + 
 \int\limits_{1+\eps}^y \frac{dy'}{y'-1} 
  \left[ {\mathcal M}_z G\left(1;y'\right) - G\left(1;y'\right) \right]
 \right\}
 \nonumber \\
 & = &
 G\left(1,1;y\right)
 +
 2 \pi i
 \lim\limits_{\eps \rightarrow 0}
 \left\{
 \int\limits_0^1 dt G\left(1;1+\eps e^{2\pi i t}\right)
 + 
 \int\limits_{1+\eps}^y \frac{dy'}{y'-1} 
 \right\}.
\eq
For the first integral we have
\bq
 \int\limits_0^1 dt G\left(1;1+\eps e^{2\pi i t}\right)
 & = &
 \int\limits_0^1 dt \ln\left(-\eps e^{2\pi i t} \right)
 \; = \;
 \ln\left(-\eps\right)
 + 2 \pi i \int\limits_0^1 t \; dt
 \; = \;
 \ln\left(-\eps\right)
 + \frac{1}{2} \left( 2 \pi i \right),
 \;\;\;
\eq
the second integral gives
\bq
 \int\limits_{1+\eps}^y \frac{dy'}{y'-1} 
 & = &
 \ln\left(1-y\right) - \ln\left(-\eps\right).
\eq
In total we obtain 
\bq
 {\mathcal M}_1 G\left(1,1;y\right) 
 & = &
 G\left(1,1;y\right) + 2 \pi i G\left(1;y\right) + \frac{1}{2} \left(2 \pi i\right)^2,
\eq
in agreement with our previous result.
}
\es
\\
\\
%
%
\bs
{\it \refstepcounter{exercise}
{\bf Exercise \theexercise}: 
Prove eq.~(\ref{chapter_multiple_polylogarithms:Li_zero_index})
from chapter~\ref{chapter_multiple_polylogarithms}.
\\
\\
{\bf Solution}:
We may write the sum representation of
$\mathrm{Li}_{m_1 \dots 0 \dots m_k}(x_1,\dots,x_i,\dots,x_k)$
as
\bq
 \mathrm{Li}_{m_1 \dots 0 \dots m_k}(x_1,\dots,x_i,\dots,x_k)
 = 
 \sum\limits_{n_1=1}^{\infty} 
 \frac{x_1^{n_1}}{n_1^{m_1}}
 \;
 \dots
 \;
 \sum\limits_{n_{i-1}=1}^{n_{i-2}-1}
 \frac{x_{i-1}^{n_{i-1}}}{n_{i-1}^{m_{i-1}}}
 \;
 Z_{0,m_{i+1},\dots,m_k}(x_i,x_{i+1},\dots,x_k;n_{i-1}-1).
\eq
Consider now
\bq
 Z_{0}(x_i;n_{i-1}-1)
 Z_{m_{i+1},\dots,m_k}(x_{i+1},\dots,x_k;n_{i-1}-1).
\eq
On the one hand we have
\bq
 Z_{0}(x_i;n_{i-1}-1)
 & = & 
 \frac{x_i}{1-x_i} - \frac{x_i^{n_{i-1}}}{1-x_i}
 \; = \;
 \mathrm{Li}_{0}(x_i)
 - \mathrm{Li}_{0}(x_i) x_i^{n_{i-1}-1},
\eq
on the other hand we may use the quasi-shuffle product for the $Z$-sums:
\bq
\lefteqn{
 Z_{0}(x_i;n_{i-1}-1)
 Z_{m_{i+1},\dots,m_k}(x_{i+1},\dots,x_k;n_{i-1}-1)
 = } & &
 \nonumber \\
 & &
 \sum\limits_{j=i}^k
 Z_{m_{i+1} \dots m_j 0 m_{j+1} \dots m_k}(x_{i+1},\dots,x_j,x_i,x_{j+1},\dots,x_k;n_{i-1}-1)
 \nonumber \\
 & &
 + \sum\limits_{j=i+1}^k
 Z_{m_{i+1} \dots m_j \dots m_k}(x_{i+1},\dots,x_i \cdot x_j,\dots,x_k;n_{i-1}-1)
\eq
Combining these two equations 
and noting that
\bq
 \frac{1}{x_i} \mathrm{Li}_{0}(x_i)
 & = & 
 1 + \mathrm{Li}_{0}(x_i)
\eq
proves eq.~(\ref{chapter_multiple_polylogarithms:Li_zero_index}).
}
\es
\\
\\
\bs
{\it \refstepcounter{exercise}
{\bf Exercise \theexercise}: 
Consider $I_{111}$ from eq.~(\ref{chapter_nested_sums:example_1_transcendental_function})
with $\mu^2=-p_3^2$ and $x=p_1^2/p_3^2$ in $D=4-2\eps$ space-time dimensions:
\bq
I_{111} & = &
 e^{\eps\Eulerconstant}
 \frac{\Gamma(-\eps)\Gamma(1-\eps)}{\Gamma(1-2\eps)}
 \sum\limits_{n=0}^\infty
 \frac{\Gamma(n+1+\eps)}
      {\Gamma(n+2)}
 \left(1-x\right)^{n}.
\eq
Expand the sum in $\eps$ and give the first two terms of the $\eps$-expansion for the full expression.
\\
\\
{\bf Solution}:
With the substitution $n \rightarrow n+1$ we have
\bq
I_{111} & = &
 e^{\eps\Eulerconstant}
 \frac{\Gamma(-\eps)\Gamma(1-\eps)}{\Gamma(1-2\eps)}
 \sum\limits_{n=1}^\infty
 \frac{\Gamma(n+\eps)}
      {\Gamma(n+1)}
 \left(1-x\right)^{n-1}.
\eq
Expanding $\Gamma(n+\eps)$ according to eq.~(\ref{chapter_nested_sums:expansiongamma}) 
one obtains:
\bq
I_{111}
 & = & 
 e^{\eps\Eulerconstant}
  \frac{\Gamma(-\eps)\Gamma(1-\eps)\Gamma(1+\eps)}{\Gamma(1-2\eps)}
 \frac{1}{1-x}
 \sum\limits_{n=1}^\infty
 \eps^{n-1}
 H_{\underbrace{1 \dots 1}_{n}}(1-x).
\eq
In this special case all harmonic polylogarithms can be expressed in terms of powers of the standard logarithm:
\bq
H_{\underbrace{1 \dots 1}_{n}}(1-x) & = & 
 \frac{(-1)^n}{n!} \left( \ln x \right)^n.
\eq
Therefore
\bq
\label{appendix_solutions:expansion_one_loop_triangle}
I_{111}
 & = & 
 - e^{\eps\Eulerconstant}
  \frac{\Gamma(1-\eps)\Gamma(1-\eps)\Gamma(1+\eps)}{\eps^2 \Gamma(1-2\eps)}
 \frac{1}{1-x}
 \sum\limits_{n=1}^\infty
 \eps^{n}
 \frac{(-1)^n}{n!} \left( \ln x \right)^n.
\eq
The expansion of the prefactor is
\bq
 e^{\eps\Eulerconstant}
  \frac{\Gamma(1-\eps)\Gamma(1-\eps)\Gamma(1+\eps)}{\eps^2 \Gamma(1-2\eps)}
 & = &
 \frac{1}{\eps^2} \left( 1 - \frac{1}{2} \zeta_2 \eps^2 \right) + {\mathcal O}\left(\eps\right).
\eq
Thus
\bq
\label{appendix_solutions:expansion_one_loop_triangle_final_result}
I_{111}
 & = & 
 \frac{1}{1-x} 
 \left[ \frac{1}{\eps} \ln x - \frac{1}{2} (\ln x)^2 \right]
 + {\mathcal O}\left(\eps\right).
\eq
The result for this example is particular simple and one recovers from eq.~(\ref{appendix_solutions:expansion_one_loop_triangle})
the well-known
all-order result
\bq
 I_{111}
 & = &
 e^{\eps\Eulerconstant}
  \frac{\Gamma(1-\eps)^2\Gamma(1+\eps)}{\Gamma(1-2\eps)}
 \frac{1}{\eps^2}
 \frac{1-x^{-\eps}}{1-x},
\eq
which (for this simple example)
can also be obtained by direct integration. 
If we expand this result in $\eps$ we recover eq.~(\ref{appendix_solutions:expansion_one_loop_triangle_final_result}).
}
\es
\\
\\
\bs
{\it \refstepcounter{exercise}
{\bf Exercise \theexercise}: 
Consider the $k$-dimensional standard simplex in ${\mathbb R}^{k+1}$. This is the polytope with vertices
given by the $(k+1)$ standard unit vectors $e_j \in {\mathbb R}^{k+1}$.
Show that the standard simplex has Euclidean volume $1/k!$ and therefore the normalised volume $1$.
\\
\\
{\bf Solution}:
The Euclidean volume of the standard $k$-dimensional simplex $\Delta$ is given by the integral
\bq
 \mathrm{vol}\left(\Delta\right)
 & = &
 \int\limits_{\alpha_j \ge 0} d^{k+1}\alpha \; \delta\left(1-\sum\limits_{j=1}^{k+1} \alpha_j \right).
\eq
From eq.~(\ref{chapter_basics:multi_beta_fct}) we have
\bq
 \mathrm{vol}\left(\Delta\right)
 & = & 
 \frac{1}{\Gamma(k+1)}
 \; = \; 
 \frac{1}{k!},
\eq
and hence
\bq
 \mathrm{vol}_0\left(\Delta\right)
 & = & 
 k! \; \mathrm{vol}\left(\Delta\right)
 \; = \; 1.
\eq
}
\es
\\
\\
\bs
{\it \refstepcounter{exercise}
{\bf Exercise \theexercise}: 
Show that the left picture of fig.~\ref{chapter_nested_sums:fig_example_regular_triangulation}
defines a regular triangulation.
\\
\\
{\bf Solution}:
We label the points as in fig.~\ref{appendix_solutions:fig_example_regular_triangulation_labelling}.
\begin{figure}
\begin{center}
\includegraphics[scale=1.0]{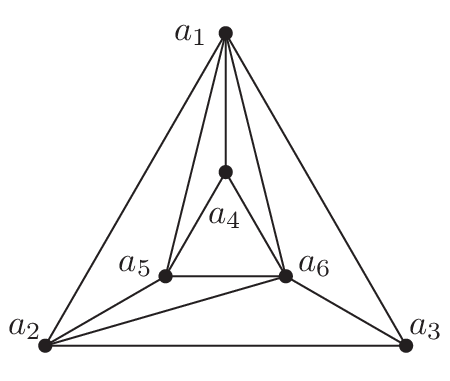}
\hspace*{10mm}
\includegraphics[scale=1.0]{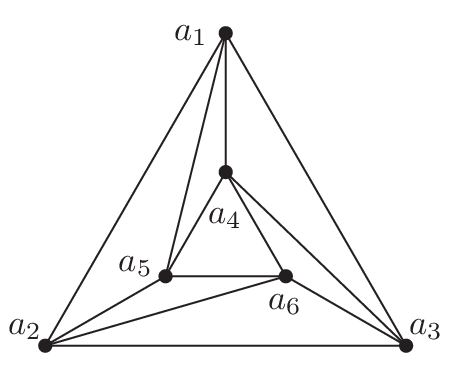}
\end{center}
\caption{
The labelling of the points for the regular triangulation of the polytope $P$ (left) and the
non-regular triangulation of the polytope $P$ (right).
}
\label{appendix_solutions:fig_example_regular_triangulation_labelling}
\end{figure}
In order to show that the triangulation is regular, we have to give a height vector.
The height vector
\bq
 h & = & \left( 1,2,3,0,0,0 \right)^T
\eq
does the job. This is most easily seen by drawing $(\tilde{a}_1,\tilde{a}_2,\tilde{a}_3,\tilde{a}_4,\tilde{a}_5,\tilde{a}_6) \in {\mathbb R}^4$.
As $(a_1,a_2,a_3,a_4,a_5,a_6) \in {\mathbb R}^3$ lie in a plane with normal vector $n=(1,1,1)$, it is sufficient to draw the points  
$(\tilde{a}_1,\tilde{a}_2,\tilde{a}_3,\tilde{a}_4,\tilde{a}_5,\tilde{a}_6)$ in ${\mathbb R}^3$, with two coordinates for the plane and one
coordinate the height.
One easily sees that the lift is convex.
The lift is convex for $h_4=h_5=h_6=0$ and
\bq
 0 \; < \; h_1 \; < \; h_2 \; < \; h_3.
\eq 
The chosen height vector fulfils this condition.
}
\es
\\
\\
\bs
{\it \refstepcounter{exercise}
{\bf Exercise \theexercise}: 
Show that the right picture of fig.~\ref{chapter_nested_sums:fig_example_regular_triangulation}
defines a non-regular triangulation.
\\
\\
{\bf Solution}:
We label the points as in fig.~\ref{appendix_solutions:fig_example_regular_triangulation_labelling}.
Let us look at the points $a_1$, $a_2$, $a_4$ and $a_5$.
These points satisfy
\bq
\label{appendix_solutions:eq_non_regular_triangulation}
 a_1 - a_2 - 4 a_4 + 4 a_5 & = & 0.
\eq
This is easily verified by plugging in the defining coordinates.
Now consider the simplex $\sigma_{125}$, i.e. the triangle with vertices $a_1$, $a_2$ and $a_5$.
Let $r_{125} \in {\mathbb R}^3$ be the vector such that
\bq
 r_{125} \cdot a_j & = & h_j
 \;\;\;\;\;\; j \; \in \; \left\{ 1,2,5 \right\},
 \nonumber \\
 r_{125} \cdot a_j & < & h_j
 \;\;\;\;\;\; j \; \in \; \left\{ 3,4,6 \right\}.
\eq 
Contracting eq.~(\ref{appendix_solutions:eq_non_regular_triangulation}) with $r_{125}$ we obtain
\bq
 0 & = & h_1 - h_2 - 4 r_{125} \cdot r_4 + 4 h_5 \; > \; h_1 - h_2 - 4 h_4 + 4 h_5
\eq
and hence 
\bq
\label{appendix_solutions:eq1_non_regular_triangulation}
 h_1 - h_2 - 4 h_4 + 4 h_5 & < & 0.
\eq
Repeating the argumentation for the points $a_2$, $a_3$, $a_5$, $a_6$ and the triangle $\sigma_{236}$ we obtain
\bq
\label{appendix_solutions:eq2_non_regular_triangulation}
 h_2 - h_3 - 4 h_5 + 4 h_6 & < & 0.
\eq
Repeating once more the argumentation for the points $a_1$, $a_3$, $a_4$, $a_6$ and the triangle $\sigma_{134}$ we obtain
\bq
\label{appendix_solutions:eq3_non_regular_triangulation}
 h_3 - h_1 - 4 h_6 + 4 h_4 & < & 0.
\eq
Adding up eq.~(\ref{appendix_solutions:eq1_non_regular_triangulation}), 
eq.~(\ref{appendix_solutions:eq2_non_regular_triangulation}) and
eq.~(\ref{appendix_solutions:eq3_non_regular_triangulation}) we obtain
\bq
 0 & < & 0,
\eq
this is a contradiction, hence a height vector does not exist and the triangulation is non-regular.
}
\es
\\
\\
\bs
{\it \refstepcounter{exercise}
{\bf Exercise \theexercise}: 
In this exercise we are going to prove theorem~\ref{chapter_nested_sums:theorem_A_hypergeometric}.
Let $G(z,x')=G(z_1,\dots,z_{\ninternal},x_1',\dots,x_n')$ be a generalised Lee-Pomeransky polynomial 
such that the associated $(\ninternal+1)\times n$-matrix ${\mathcal A}$ 
satisfies eq.~(\ref{chapter_nested_sums:conditions_A}).
Consider the integral
\bq
 I
 & = &
 C
 \int\limits_{z_j \ge 0}  d^{\ninternal}z \;
 \left( \prod\limits_{j=1}^{\ninternal} z_j^{\nu_j-1} \right)
 \left[G\left(z,x'\right)\right]^{-\frac{D}{2}}.
\eq
Show that $I$ satisfies the differential equations in eq.~(\ref{chapter_nested_sums:GKZ_diff_eq_set_1}) and eq.~(\ref{chapter_nested_sums:GKZ_diff_eq_set_2}) with
$c=(-D/2,-\nu_1,\dots,-\nu_{\ninternal})^T$.
\\
\\
{\bf Solution}:
The entries $a_{ij}$ of ${\mathcal A}$ define $G(z,x')$:
\bq
 G\left(z,x'\right)
 & = &
  \sum\limits_{j=1}^n x_j' \; z_1^{a_{1 j}} \dots z_{\ninternal}^{a_{\ninternal j}}.
\eq
We have
\bq
 \frac{\partial}{\partial x_j'} G\left(z,x'\right)
 & = &
 z_1^{a_{1 j}} \dots z_{\ninternal}^{a_{\ninternal j}}.
\eq
We have to verify the set of equations in eq.~(\ref{chapter_nested_sums:GKZ_diff_eq_set_1}) and eq.~(\ref{chapter_nested_sums:GKZ_diff_eq_set_2}).
These equations are equivalent to eq.~(\ref{chapter_nested_sums:GKZ_diff_eq_set_1_v2}) and eq.~(\ref{chapter_nested_sums:GKZ_diff_eq_set_2_v2}).
We start with eq.~(\ref{chapter_nested_sums:GKZ_diff_eq_set_1_v2}).
Let $u,v \in {\mathbb N}_0^n$ with ${\mathcal A} u = {\mathcal A} v$ and set
\bq
 \left|u\right| \; = \; \sum\limits_{j=1}^{n} u_j,
 & &
 \left|v\right| \; = \; \sum\limits_{j=1}^{n} v_j.
\eq
Then
\bq
 \partial^u I 
 & = &
 \frac{\Gamma\left(1-\frac{D}{2}\right)}{\Gamma\left(1-\frac{D}{2}-|u|\right)}
 C
 \int\limits_{z_j \ge 0}  d^{\ninternal}z \;
 \left( \prod\limits_{j=1}^{\ninternal} z_j^{\nu_j+a_{jk}u_k-1} \right)
 \left[G\left(z,x'\right)\right]^{-\frac{D}{2}-|u|},
 \nonumber \\
 \partial^v I 
 & = &
 \frac{\Gamma\left(1-\frac{D}{2}\right)}{\Gamma\left(1-\frac{D}{2}-|v|\right)}
 C
 \int\limits_{z_j \ge 0}  d^{\ninternal}z \;
 \left( \prod\limits_{j=1}^{\ninternal} z_j^{\nu_j+a_{jk}v_k-1} \right)
 \left[G\left(z,x'\right)\right]^{-\frac{D}{2}-|v|}.
\eq
From ${\mathcal A} u = {\mathcal A} v$ we have $a_{jk}u_k=a_{jk}v_k$ (a sum over $k$ is implied).
The first row of ${\mathcal A}$ is $(1,\dots,1)$. This implies $|u|=|v|$.
Hence
\bq
 \partial^u I 
 & = &
 \partial^v I.
\eq
Let us now turn to eq.~(\ref{chapter_nested_sums:GKZ_diff_eq_set_2_v2}).
We start with the first row of ${\mathcal A}$ with the entries $a_{01}=\dots=a_{0n}=1$:
We have
\bq
 \sum\limits_{j=1}^n x_j' \frac{\partial}{\partial x_j'} I
 & = &
 - \frac{D}{2}
 C
 \int\limits_{z_j \ge 0}  d^{\ninternal}z \;
 \left( \prod\limits_{j=1}^{\ninternal} z_j^{\nu_j-1} \right)
 \left[G\left(z,x'\right)\right]^{-\frac{D}{2}}
 \; = \;
 -\frac{D}{2} I.
\eq
We then consider the other rows ($1 \le i \le \ninternal$):
We first have 
\bq
 \sum\limits_{j=1}^n a_{ij} x_j' \frac{\partial}{\partial x_j'} G\left(z,x'\right)
 & = &
  \sum\limits_{j=1}^n a_{ij} x_j' \; z_1^{a_{1 j}} \dots z_{\ninternal}^{a_{\ninternal j}}
 \; = \; 
 z_i \frac{\partial}{\partial z_i} G\left(z,x'\right).
\eq
Hence
\bq
 \sum\limits_{j=1}^n a_{ij} x_j' \frac{\partial}{\partial x_j'} I
 & = &
 C
 \int\limits_{z_j \ge 0}  d^{\ninternal}z \;
 \left( \prod\limits_{j=1}^{\ninternal} z_j^{\nu_j-1} \right)
 z_i \frac{\partial}{\partial z_i}
 \left[G\left(z,x'\right)\right]^{-\frac{D}{2}}
 \nonumber \\
 & = &
 - \nu_i C
 \int\limits_{z_j \ge 0}  d^{\ninternal}z \;
 \left( \prod\limits_{j=1}^{\ninternal} z_j^{\nu_j-1} \right)
 \left[G\left(z,x'\right)\right]^{-\frac{D}{2}}
 \nonumber \\
 & =&
 - \nu_i I,
\eq
where we used partial integration.
Therefore we have with $c=(-D/2,-\nu_1,\dots,-\nu_{\ninternal})^T$
\bq
 \left( {\mathcal A} {\bm \theta} - c \right) I & = & 0.
\eq
}
\es
\\
\\
%
%
\bs
{\it \refstepcounter{exercise}
{\bf Exercise \theexercise}: 
Show that for a Feynman integral as 
in eq.~(\ref{chapter_sector_decomposition:Feynman_parameter_representation}) 
we have in any primary sector $c=0$ and therefore the additional factor is absent.
\\
\\
{\bf Solution}:
For the Feynman integral in the Feynman parameter representation of eq.~(\ref{chapter_sector_decomposition:Feynman_parameter_representation})
we have
\bq
 a_i + \eps b_i & = & \nu_i - 1,
 \;\;\;\;\;\;
 1 \; \le\; i \; \le \; \ninternal
\eq
and $r=2$. We take $P_1 = {\mathcal U}$ and $P_2={\mathcal F}$.
We have
\bq
 c_1 + \eps d_1 \; = \; \nu- \frac{\left(l+1\right)D}{2},
 & &
 c_2 + \eps d_2 \; = \; \frac{l D}{2} - \nu.
\eq
We further know that ${\mathcal U}$ is homogeneous of degree $h_1=\loopnumber$ and ${\mathcal F}$ is homogeneous 
of degree $h_2=(\loopnumber+1)$.
Thus
\bq
 c & = &
 -\ninternal - \sum\limits_{i=1}^{\ninternal} \left( a_i+\eps b_i \right) - \sum\limits_{j=1}^2 h_j \left( c_j+\eps d_j\right)
 \nonumber \\
 & = &
 -\ninternal - \left( \nu-\ninternal \right) 
 - \loopnumber \left(\nu- \frac{\left(l+1\right)D}{2}\right) 
 - \left(\loopnumber+1\right) \left( \frac{l D}{2} - \nu \right)
 \nonumber \\
 & = & 0.
\eq
}
\es
\\
\\
%
%
\bs
{\it \refstepcounter{exercise}
{\bf Exercise \theexercise}: 
Show that the convolution product is associative:
\bq
 \left( \varphi_1 \ast \varphi_2 \right) \ast \varphi_3
 & = &
 \varphi_1 \ast \left( \varphi_2 \ast \varphi_3 \right).
\eq
{\bf Solution}: 
Let $a \in C$. We have
\bq
 \left( \left( \varphi_1 \ast \varphi_2 \right) \ast \varphi_3 \right)\left(a\right)
 & = &
 \cdot \left( \left( \varphi_1 \ast \varphi_2 \right) \otimes \varphi_3 \right) \Delta\left(a\right)
 \; = \;
 \cdot \left( \cdot \left( \varphi_1 \otimes \varphi_2 \right) \Delta \otimes \varphi_3 \right) \Delta\left(a\right)
 \nonumber \\
 & = &
 \cdot \left( \cdot \otimes \mathrm{id} \right) \left( \varphi_1 \otimes \varphi_2 \otimes \varphi_3 \right) \left( \Delta  \otimes \mathrm{id} \right) \Delta\left(a\right)
 \nonumber \\
 & = &
 \cdot \left( \mathrm{id} \otimes \cdot \right) \left( \varphi_1 \otimes \varphi_2 \otimes \varphi_3 \right) \left( \mathrm{id} \otimes \Delta  \right) \Delta\left(a\right)
 \nonumber \\
 & = &
 \cdot \left( \varphi_1 \otimes \cdot \left( \varphi_2 \otimes \varphi_3 \right) \Delta  \right) \Delta\left(a\right)
 \; = \;
 \cdot \left( \varphi_1 \otimes \left( \varphi_2 \ast \varphi_3 \right) \right) \Delta\left(a\right)
 \nonumber \\
 & = &
 \left( \varphi_1 \ast \left( \varphi_2 \ast \varphi_3 \right) \right)\left(a\right).
\eq
}
\es
\\
\\
\bs
{\it \refstepcounter{exercise}
{\bf Exercise \theexercise}: 
Show that $1_{\mathrm{Hom}} = e \bar{e} \in \mathrm{Hom}(C,A)$ is a neutral element for the convolution product, i.e.
\bq
 \varphi \ast 1_{\mathrm{Hom}} 
 \; = \;
 1_{\mathrm{Hom}} \ast \varphi
 & = & 
 \varphi.
\eq
{\bf Solution}: 
Let $a \in C$ and write $\Delta(a)=a^{(1)} \otimes a^{(2)}$, using Sweedler's notation.
We have
\bq
 \left(\varphi \ast 1_{\mathrm{Hom}}\right)\left(a\right)
 & = & 
 \cdot \; \left(\varphi \otimes e \bar{e} \right) \Delta\left(a\right)
 \; = \;
 \cdot \; \left(\varphi \otimes e \bar{e} \right) \left( a^{(1)} \otimes a^{(2)} \right).
\eq
Using the axiom of the counit in eq.~(\ref{chapter_hopf:axiom_counit}) this equals
\bq
 \left(\varphi \ast 1_{\mathrm{Hom}}\right)\left(a\right)
 & = & 
 \cdot \; \left(\varphi \otimes e \right) \left( a \otimes 1 \right),
\eq
where $1$ denotes the unit in $R$.
We obtain
\bq
 \left(\varphi \ast 1_{\mathrm{Hom}}\right)\left(a\right)
 & = & 
 \cdot \; \left(\varphi\left(a\right) \otimes e\left(1\right) \right)
 \; = \; 
 \varphi\left(a\right).
\eq
The proof for $1_{\mathrm{Hom}} \ast \varphi = \varphi$ is similar.
}
\es
\\
\\
\bs
{\it \refstepcounter{exercise}
{\bf Exercise \theexercise}: 
Show that $\varphi^{-1} = \varphi S$ is an inverse element to $\varphi \in \mathrm{AlgHom}(H,A)$.
\\
\\
{\bf Solution}:
We have to show that
\bq
 \varphi \ast \varphi^{-1}
 \; = \;
 \varphi^{-1} \ast \varphi
 & = &
 1_{\mathrm{AlgHom}}.
\eq
We have
\bq
 \left( \varphi \ast \varphi^{-1} \right)\left(a\right)
 & = &
 \cdot \; \left( \varphi \otimes \left( \varphi S \right) \right) \Delta\left(a\right)
 \; = \; 
 \cdot \; \left(\varphi \otimes \varphi \right) \left( \mathrm{id} \otimes S\right) \Delta\left(a\right)
 \nonumber \\
 & = &
 \varphi \cdot \left( \mathrm{id} \otimes S\right) \Delta\left(a\right).
\eq
In the last step we used the fact that $\varphi$ is an algebra homomorphism.
Using the axiom of the antipode eq.~(\ref{chapter_hopf:antipode_def1}) we have
\bq
 \left( \varphi \ast \varphi^{-1} \right)\left(a\right)
 & = &
 \varphi e_H \bar{e}\left(a\right)
 \; = \;
 e_A \bar{e}\left(a\right)
 \; = \; 
 1_{\mathrm{AlgHom}}\left(a\right),
\eq
since $\varphi(e_H)=e_A$.
The proof of $\varphi^{-1} \ast \varphi = 1_{\mathrm{AlgHom}}$ is similar.
}
\es
\\
\\
\bs
{\it \refstepcounter{exercise}
{\bf Exercise \theexercise}: 
Which rooted trees are primitive elements in the Hopf algebra of rooted trees?
\\
\\
{\bf Solution}:
For a primitive element $t_{\mathrm{prim}}$ the coproduct has to be 
\bq
 \Delta\left(t_{\mathrm{prim}}\right) & = &
  t_{\mathrm{prim}} \otimes e + e \otimes t_{\mathrm{prim}}.
\eq
For a rooted tree $t$, the coproduct is in general
\bq
 \Delta(t) 
 & = & 
 t \otimes e + e \otimes t + \sum\limits_{\mathrm{adm. cuts} \; C \mathrm{of} \; t} P^C(t) \otimes R^C(t),
\eq
thus for a primitive rooted tree the sum over all admissible cuts has to be absent.
This is the case if the rooted tree consists only of a single vertex, the root.
}
\es
\\
\\
\bs
{\it \refstepcounter{exercise}
{\bf Exercise \theexercise}: 
Show that the map $R$ in eq.~(\ref{chapter_hopf:MSbar}) fulfils the Rota-Baxter equation~(\ref{chapter_hopf:rotabaxter}).
\\
\\
{\bf Solution}:
Let 
\bq
 a_1 \; = \; \sum\limits_{j=-L_1}^\infty b_j \eps^j \; \in \; A,
 & &
 a_2 \; = \; \sum\limits_{k=-L_2}^\infty c_k \eps^k \; \in \; A.
\eq
Then
\bq
\lefteqn{
 R\left( a_1 a_2 \right) + R\left( a_1 \right) R\left( a_2 \right) 
 - R\left( a_1 R\left( a_2 \right) \right) - R\left( R\left( a_1 \right) a_2 \right)
 = } & & \nonumber \\
 & = &
 \sum\limits_{j+k<0} b_j c_k \eps^{j+k}
 +
 \sum\limits_{j<0,k<0} b_j c_k \eps^{j+k}
 -
 \sum\limits_{j+k<0,k<0} b_j c_k \eps^{j+k}
 -
 \sum\limits_{j+k<0,j<0} b_j c_k \eps^{j+k}
 \nonumber \\
 & = & 0.
\eq
}
\es
\\
\\
\bs
{\it \refstepcounter{exercise}
{\bf Exercise \theexercise}: 
Resolve the operator overloading: In eq.~(\ref{chapter_hopf:condition_coaction})
the symbols ``$\cdot$'', $\bar{e}$ and $\Delta$ appear in various places.
Determine for each occurrence to which operation they correspond.
\\
\\
{\bf Solution}:
We start with the equation
\bq
 \cdot \left( \bar{e} \otimes \mathrm{id} \right) \Delta\left(v\right) & = & v.
\eq
We read the left-hand side right-to-left.
$\Delta(v)$ with $v \in M$ corresponds to the map $\Delta : M \rightarrow C \otimes M$ of eq.~(\ref{chapter_hopf:def_coaction}).
Let us write
\bq
\label{appendix_solutions:sweedler_coaction}
 \Delta\left(v\right) & = & a^{(1)} \otimes v^{(2)},
 \;\;\;\;\;\;\;\;\;\;\;\;
 a^{(1)} \; \in \; C, 
 \;\;\;\;\;\;
 v, v^{(2)} \; \in \; M.
\eq
Then
\bq
 \left( \bar{e} \otimes \mathrm{id} \right) \Delta\left(v\right)
 & = &
 \bar{e}\left(a^{(1)}\right) \otimes v^{(2)},
\eq
and $\bar{e} : C \rightarrow R$ denotes the counit in $C$ (there is no other counit).
Finally, as $\bar{e}(a) \in R$, the final multiplication ``$\cdot$'' is the scalar multiplication of the $R$-module:
$\cdot : R \times M \rightarrow M$.

Let us now turn to
\bq
 \left( \Delta \otimes \mathrm{id} \right) \Delta\left(v\right)
 & = &
 \left( \mathrm{id} \otimes \Delta \right) \Delta\left(v\right).
\eq
Again, we read both sides right-to-left. The first operation $\Delta(v)$ is the operation
$\Delta : M \rightarrow C \otimes M$ of eq.~(\ref{chapter_hopf:def_coaction}).
With eq.~(\ref{appendix_solutions:sweedler_coaction}) our original equation becomes
\bq
 \Delta\left(a^{(1)}\right) \otimes v^{(2)}
 & = &
 a^{(1)} \otimes \Delta\left(v^{(2)}\right).
\eq
On the left-hand side $\Delta(a^{(1)})$ refers to the comultiplication in $C$, i.e. $\Delta : C \rightarrow C \otimes C$.
On the right-hand side $\Delta(v^{(2)})$ refers again to the operation
$\Delta : M \rightarrow C \otimes M$ of eq.~(\ref{chapter_hopf:def_coaction}).
}
\es
\\
\\
\bs
{\it \refstepcounter{exercise}
{\bf Exercise \theexercise}: 
Work out $\Delta(\ln^{\mathfrak m}(x) )$. Note that $\ln^{\mathfrak m}(x) = I^{\mathfrak m}(1;0;x)$.
\\
\\
{\bf Solution}:
From eq.~(\ref{chapter_hopf:I_coaction}) we have
\bq
 \Delta\left( I^{\mathfrak m}(1;0;x) \right)
 & = &
 I^{\mathfrak{d}\mathfrak{R}}(1;0;x) \otimes 1
 + 1 \otimes I^{\mathfrak m}(1;0;x),
\eq
and therefore
\bq
\label{appendix_solutions:coaction_logarithm}
 \Delta\left( \ln^{\mathfrak m}(x) \right)
 & = &
 \ln^{\mathfrak{d}\mathfrak{R}}(x) \otimes 1 + 1 \otimes \ln^{\mathfrak m}(x).
\eq
}
\es
\\
\\
\bs
{\it \refstepcounter{exercise}
{\bf Exercise \theexercise}: 
Consider 
\bq
 I\left(0;x,x;y\right)
 & = & 
 G\left(x,x;y\right) 
 \; = \;
 G_{1 1}\left(1,1;\frac{y}{x}\right)
 \; = \;
 \mathrm{Li}_{1 1}\left(\frac{y}{x},1\right)
 \; = \;
 H_{1 1}\left(\frac{y}{x}\right).
\eq
With the techniques of chapter~\ref{chapter_multiple_polylogarithms} it is not too difficult to show that the derivatives
with respect to $x$ and $y$ are
\bq
 \frac{\partial}{\partial x} I\left(0;x,x;y\right)
 & = &
 \frac{y}{x\left(x-y\right)} \ln\left(\frac{x-y}{x}\right),
 \nonumber \\
 \frac{\partial}{\partial y} I\left(0;x,x;y\right)
 & = &
 \frac{1}{y-x} \ln\left(\frac{x-y}{x}\right).
\eq
Re-compute the derivatives using eq.~(\ref{chapter_hopf:coaction_derivative_all}).
\\
\\
{\bf Solution}:
We first compute the coaction
\bq
 \Delta\left( I^{\mathfrak m}\left(0;x,x;y\right) \right)
 & = &
 1 \otimes I^{\mathfrak m}\left(0;x,x;y\right)
 + \left( I^{\mathfrak{d}\mathfrak{R}}\left(0;x;x\right) + I^{\mathfrak{d}\mathfrak{R}}\left(x;x;y\right) \right) \otimes I^{\mathfrak m}\left(0;x;y\right)
 + I^{\mathfrak{d}\mathfrak{R}}\left(0;x,x;y\right) \otimes 1,
 \nonumber \\
\eq
and therefore
\bq
 \Delta_{1,1}\left( I^{\mathfrak m}\left(0;x,x;y\right) \right)
 & = &
 \left( I^{\mathfrak{d}\mathfrak{R}}\left(0;x;x\right) + I^{\mathfrak{d}\mathfrak{R}}\left(x;x;y\right) \right) \otimes I^{\mathfrak m}\left(0;x;y\right)
 \nonumber \\
 & = &
 \left( -\ln^{\mathfrak{d}\mathfrak{R}}\left(-x\right) + \ln^{\mathfrak{d}\mathfrak{R}}\left(y-x\right) \right) \otimes \ln^{\mathfrak m}\left(\frac{x-y}{x}\right).
\eq
We then have
\bq
 \frac{\partial}{\partial x} I^{\mathfrak m}\left(0;x,x;y\right)
 & = &
 \cdot \left( \frac{\partial}{\partial x} \otimes 1 \right) \Delta_{1,1}\left(I^{\mathfrak m}\left(0;x,x;y\right)\right)
 \nonumber \\
 & = &
 \cdot \left( \frac{\partial}{\partial x} \otimes 1 \right) \left[ \left( -\ln^{\mathfrak{d}\mathfrak{R}}\left(-x\right) + \ln^{\mathfrak{d}\mathfrak{R}}\left(y-x\right) \right) \otimes \ln^{\mathfrak m}\left(\frac{x-y}{x}\right) \right]
 \nonumber \\
 & = &
 \cdot \left[ \left( -\frac{1}{x} - \frac{1}{y-x} \right) \otimes \ln^{\mathfrak m}\left(\frac{x-y}{x}\right) \right]
 \; = \;
 \frac{y}{x\left(x-y\right)} \ln^{\mathfrak m}\left(\frac{x-y}{x}\right).
\eq
In a similar way we obtain
\bq
 \frac{\partial}{\partial y} I^{\mathfrak m}\left(0;x,x;y\right)
 & = &
 \cdot \left( \frac{\partial}{\partial y} \otimes 1 \right) \Delta_{1,1}\left(I^{\mathfrak m}\left(0;x,x;y\right)\right)
 \nonumber \\
 & = &
 \cdot \left( \frac{\partial}{\partial x} \otimes 1 \right) \left[ \left( -\ln^{\mathfrak{d}\mathfrak{R}}\left(-x\right) + \ln^{\mathfrak{d}\mathfrak{R}}\left(y-x\right) \right) \otimes \ln^{\mathfrak m}\left(\frac{x-y}{x}\right) \right]
 \nonumber \\
 & = &
 \frac{1}{y-x} \ln^{\mathfrak m}\left(\frac{x-y}{x}\right).
\eq
}
\es
\\
\\
\bs
{\it \refstepcounter{exercise}
{\bf Exercise \theexercise}: 
Work out the symbols
\bq
 S\left( -\ln\left(x\right) \right) 
 & \mbox{and} &
 S\left( \ln\left(-x\right) \right).
\eq
\\
\\
{\bf Solution}:
We start with $S( -\ln(x) )$. We have
\bq
 S\left( -\ln\left(x\right) \right) 
 & = &
 -  S\left( \ln\left(x\right) \right) 
 \; = \; - \left(x\right).
\eq
On the other hand, we have for $S( \ln(-x) )$
\bq
 S\left( \ln\left(-x\right) \right) 
 & = &
 S\left( \ln\left(x\right) \right) 
 \; = \; \left(x\right).
\eq
}
\es
\\
\\
\bs
{\it \refstepcounter{exercise}
{\bf Exercise \theexercise}: 
Fill in the details for the derivation of $\mathrm{sv}^{\mathfrak m}(\mathrm{Li}^{\mathfrak m}_1(x))$
and $\mathrm{sv}^{\mathfrak m}(\mathrm{Li}^{\mathfrak m}_2(x))$.
\\
\\
{\bf Solution}:
We start with $\mathrm{Li}^{\mathfrak m}_1(x)$. From eq.~(\ref{chapter_hopf:I_coproduct}) we have
\bq
 \Delta\left(\mathrm{Li}^{\mathfrak{d}\mathfrak{R}}_1(x)\right)
 & = &
 \mathrm{Li}^{\mathfrak{d}\mathfrak{R}}_1(x) \otimes 1
 + 1 \otimes \mathrm{Li}^{\mathfrak{d}\mathfrak{R}}_{1}(x).
\eq
Eq.~(\ref{chapter_hopf:I_antipode}) gives us
\bq
 S\left(\mathrm{Li}^{\mathfrak{d}\mathfrak{R}}_1(x)\right)
 & = &
 - \mathrm{Li}^{\mathfrak{d}\mathfrak{R}}_1(x).
\eq
We have
\bq
 \Delta^{\mathfrak m}\left(\mathrm{Li}^{\mathfrak{d}\mathfrak{R}}_1(x)\right)
 & = &
 \mathrm{Li}^{\mathfrak m}_1(x) \otimes 1
 + 1 \otimes \mathrm{Li}^{\mathfrak m}_{1}(x),
\eq
and
\bq
 \Sigma\left(1\right) & = & 1,
 \nonumber \\
 \Sigma\left(\mathrm{Li}^{\mathfrak m}_{1}(x)\right) & = & \mathrm{Li}^{\mathfrak m}_{1}(x).
\eq
Thus
\bq
 \mathrm{sv}^{\mathfrak m}\left(\mathrm{Li}^{\mathfrak m}_1(x)\right)
 & = &
 \mathrm{period}\left(\mathrm{Li}^{\mathfrak m}_{1}(x)\right)
 + \mathrm{period}\left(F_\infty\mathrm{Li}^{\mathfrak m}_{1}(x)\right)
 \; = \; 
 \mathrm{Li}_{1}\left(x\right) + \mathrm{Li}_{1}\left(\overline{x}\right).
\eq
Let us now turn to $\mathrm{Li}^{\mathfrak m}_2(x)$. 
From eq.~(\ref{chapter_hopf:I_coproduct}) we have
\bq
 \Delta\left(\mathrm{Li}^{\mathfrak{d}\mathfrak{R}}_2(x)\right)
 & = &
 \mathrm{Li}^{\mathfrak{d}\mathfrak{R}}_2(x) \otimes 1
 + 1 \otimes \mathrm{Li}^{\mathfrak{d}\mathfrak{R}}_{2}(x)
 + \ln^{\mathfrak{d}\mathfrak{R}}(x) \otimes \mathrm{Li}^{\mathfrak{d}\mathfrak{R}}_{1}(x).
\eq
Eq.~(\ref{chapter_hopf:I_antipode}) gives us
\bq
 S\left(\mathrm{Li}^{\mathfrak{d}\mathfrak{R}}_2(x)\right)
 & = &
 - \mathrm{Li}^{\mathfrak{d}\mathfrak{R}}_2(x)
 + \ln^{\mathfrak{d}\mathfrak{R}}(x) \mathrm{Li}^{\mathfrak{d}\mathfrak{R}}_1(x).
\eq
We work out 
\bq
\lefteqn{
 \left( \mathrm{id} \otimes F_\infty \Sigma \right) \Delta^{\mathfrak m}\left(\mathrm{Li}^{\mathfrak{d}\mathfrak{R}}_2(x)\right)
 = } & & \nonumber \\
 & = &
 \left( \mathrm{id} \otimes F_\infty \Sigma \right)
  \left[ \mathrm{Li}^{\mathfrak m}_2(x) \otimes 1
         + 1 \otimes \mathrm{Li}^{\mathfrak m}_{2}(x)
         + \ln^{\mathfrak m}(x) \otimes \mathrm{Li}^{\mathfrak m}_{1}(x)
 \right]
 \nonumber \\
 & = &
 \mathrm{Li}^{\mathfrak m}_2(x) \otimes 1
 + 1 \otimes F_\infty \left( - \mathrm{Li}^{\mathfrak m}_2(x) + \ln^{\mathfrak m}(x) \mathrm{Li}^{\mathfrak m}_1(x) \right)
 + \ln^{\mathfrak m}(x) \otimes \left( F_\infty \mathrm{Li}^{\mathfrak m}_{1}(x) \right),
\eq
and therefore
\bq
 \mathrm{sv}^{\mathfrak m}\left(\mathrm{Li}^{\mathfrak m}_2(x)\right)
 & = &
 \mathrm{Li}_2\left(x\right)
 - \mathrm{Li}_2\left(\overline{x}\right) + \ln\left(\overline{x}\right) \mathrm{Li}_1\left(\overline{x}\right) 
 + \ln\left(x\right) \cdot \mathrm{Li}_{1}\left(\overline{x}\right)
 \nonumber \\
 & = &
 \mathrm{Li}_2\left(x\right)
 - \mathrm{Li}_2\left(\overline{x}\right) 
 + \ln\left(\left|x\right|^2\right) \cdot \mathrm{Li}_{1}\left(\overline{x}\right).
\eq
}
\es
\\
\\
\bs
{\it \refstepcounter{exercise}
{\bf Exercise \theexercise}: 
Show that eq.~(\ref{chapter_hopf:result_bubble_v1}) and eq.~(\ref{chapter_hopf:result_bubble_v2}) agree in a neighbourhood of $x=0$.
\\
\\
{\bf Solution}:
We have to show that
\bq
 I_{2,a}'{}^{(2)}\left(x\right)
 & = & 
 2 \left[ G\left(0,0;x'\right) - 2 G\left(-1,0;x'\right) - \zeta_2 \right],
 \nonumber \\
 I_{2,b}'{}^{(2)}\left(x\right)
 & = & 
 - 4 \; \mathrm{Li}_2\left( y_1 \right) + 2 \ln^2\left(y_2\right) - \ln^2\left(f_1\right) + 2 \zeta_2
\eq
agree in a neighbourhood of $x=0$. 
$I_{2,a}'{}^{(2)}(x)$ and $I_{2,b}'{}^{(2)}(x)$ are functions of a single variable $x$.
Two functions are identical, if their derivatives are identical and the two functions have the same value at a single point.

We first check the value at $x=0$ (corresponding to $x'=1$ and $y_1=y_2=1/2$):
\bq
 I_{2,a}'{}^{(2)}\left(0\right)
 \; = \; 
 0,
 & & 
 I_{2,b}'{}^{(2)}\left(0\right)
 \; = \; 
 0.
\eq
In the second step we check the derivatives. We set $r=\sqrt{x(4+x)}$. Carrying out the derivatives and collecting terms we first obtain
\bq
 \frac{d}{dx} I_{2,a}'{}^{(2)}\left(x\right)
 & = & 
 2 \frac{x-r}{r\left(4+x-r\right)} \ln\left(\frac{2+x-r}{2}\right),
 \nonumber \\
 \frac{d}{dx} I_{2,b}'{}^{(2)}\left(x\right)
 & = & 
 - \frac{2}{r} \ln\left(\frac{y_2}{1-y_1}\right) - \frac{2}{4+x} \ln\left(\left(1-y_1\right)y_2\left(4+x\right)\right).
\eq
We then simplify these expressions. We first make the denominator of $dI_{2,a}'{}^{(2)}/dx$ rational. We multiply the numerator and the denominator
with $(4+x+r)$. Noting that
\bq
 \left(4+x-r\right) \left(4+x+r\right) & = & 4 \left(4+x\right),
 \nonumber \\
 \left(x-r\right) \left(4+x+r\right) & = & - 4 r
\eq
we obtain
\bq
 \frac{d}{dx} I_{2,a}'{}^{(2)}\left(x\right)
 & = & 
 - \frac{2}{\left(4+x\right)} \ln\left(\frac{2+x-r}{2}\right).
\eq
In order to simplify $dI_{2,b}'{}^{(2)}/dx$ we first notice that
\bq
 y_2 & = & 1 - y_1.
\eq
Thus
\bq
 \frac{d}{dx} I_{2,b}'{}^{(2)}\left(x\right)
 \; = \;
 - \frac{2}{4+x} \ln\left(y_2^2\left(4+x\right)\right)
 \; = \;
 - \frac{2}{4+x} \ln\left(\frac{2}{2+x+r}\right)
 \; = \;
 - \frac{2}{4+x} \ln\left(\frac{2+x-r}{2}\right).
 \;\;\;
\eq
}
\es
\\
\bs
{\it \refstepcounter{exercise}
{\bf Exercise \theexercise}: 
Let $f_1, f_2, g_1, g_2$ be algebraic functions of the kinematic variables $x$.
Determine the symbols of
\bq
 \mathrm{Li}_{2 1}\left(f_1, f_2\right)
 & \mbox{and} &
 G_{2 1}\left(g_1,g_2;1\right).
\eq
Assume then $g_1=1/f_1$ and $g_2=1/(f_1 f_2)$. Show that in this case the two symbols agree.

From the two symbols deduce the constraints on the arguments $f_1, f_2$ of $\mathrm{Li}_{2 1}(f_1, f_2)$
and on the arguments $g_1, g_2$ of $G_{2 1}(g_1,g_2;1)$.
\\
\\
{\bf Solution}:
We start with $G_{2 1}(g_1,g_2;1)$. According to eq.~(\ref{chapter_multiple_polylogarithms:differential_Glog})
we have
\bq
 d G_{2 1}\left(g_1,g_2;1\right)
 \; = \;
 d G\left(0,g_1,g_2;1\right)
 & = &
 - G\left(g_1,g_2;1\right) d\ln\left(g_1\right) 
 +
 G\left(0,g_2;1\right) d\ln\left(\frac{g_1}{g_2-g_1}\right)
 \nonumber \\
 & &
 +
 G\left(0,g_1;1\right) d\ln\left(\frac{g_2-g_1}{g_2}\right),
\eq
and therefore
\bq
\lefteqn{
 S\left( G_{2 1}\left(g_1,g_2;1\right) \right)
 = } & & \\
 & &
 - \left( g_1 \otimes S\left( G\left(g_1,g_2;1\right) \right) \right)
 - \left( \frac{\left(g_2-g_1\right)}{g_1} \otimes S\left( G\left(0,g_2;1\right) \right) \right)
 + \left( \frac{\left(g_2-g_1\right)}{g_2} \otimes S\left( G\left(0,g_1;1\right) \right) \right).
 \nonumber
\eq
The symbols of the functions of weight two are
\bq
 S\left( G\left(g_1,g_2;1\right) \right)
 & = & 
 \left( \frac{\left(1-g_1\right)}{\left(g_2-g_1\right)} \otimes \frac{\left(1-g_2\right)}{g_2} \right)
 + \left( \frac{\left(g_2-g_1\right)}{g_2} \otimes \frac{\left(1-g_1\right)}{g_1} \right),
 \nonumber \\
 S\left( G\left(0,g_2;1\right) \right)
 & = & 
 - \left( g_2 \otimes \frac{\left(1-g_2\right)}{g_2} \right),
 \nonumber \\
 S\left( G\left(0,g_1;1\right) \right)
 & = & 
 - \left( g_1 \otimes \frac{\left(1-g_1\right)}{g_1} \right).
\eq
Putting everything together we obtain
\bq
\label{appendix_solutions:symbol_G_21}
 S\left( G_{2 1}\left(g_1,g_2;1\right) \right)
 & = &
 \left( g_1 \otimes \frac{\left(g_2-g_1\right)}{\left(1-g_1\right)} \otimes \frac{\left(1-g_2\right)}{g_2} \right)
 + \left( \frac{\left(g_2-g_1\right)}{g_1} \otimes g_2 \otimes \frac{\left(1-g_2\right)}{g_2} \right)
 \nonumber \\
 & &
 - \left( g_1 \otimes \frac{\left(g_2-g_1\right)}{g_2} \otimes \frac{\left(1-g_1\right)}{g_1} \right)
 - \left( \frac{\left(g_2-g_1\right)}{g_2} \otimes g_1 \otimes \frac{\left(1-g_1\right)}{g_1} \right)
\eq
Let us now turn to $\mathrm{Li}_{2 1}(f_1, f_2)$.
According to eq.~(\ref{chapter_multiple_polylogarithms:dLi}) we have
\bq
 d \mathrm{Li}_{2 1}\left(f_1, f_2\right)
 & = &
 \mathrm{Li}_{1 1}\left(f_1, f_2\right) d\ln\left(f_1\right)
 +
 \mathrm{Li}_{2 0}\left(f_1, f_2\right) d\ln\left(f_2\right).
\eq
We may rewrite $\mathrm{Li}_{20}$ with the help of eq.~(\ref{chapter_multiple_polylogarithms:Li_zero_index})
as
\bq
 \mathrm{Li}_{2 0}\left(f_1, f_2\right)
 & =&
 \mathrm{Li}_{0}\left(f_2\right) \mathrm{Li}_{2}\left(f_1\right)
 - \mathrm{Li}_{2}\left(f_1 f_2\right)
 - \mathrm{Li}_{0}\left(f_2\right) \mathrm{Li}_{2}\left(f_1 f_2\right).
\eq
This generates terms with $\mathrm{Li}_0$. We haven't defined the symbol of a weight zero function.
Let's work out the prescription: We consider $\mathrm{Li}_1(x)$.
We know its symbol from eq.~(\ref{chapter_hopf:def_symbol_recursively}):
\bq
 S\left(\mathrm{Li}_1\left(x\right)\right)
 & = &
 S\left(-\ln\left(1-x\right)\right) \; = \; - \left(1-x\right).
\eq
On the other hand, eq.~(\ref{chapter_multiple_polylogarithms:dLi}) gives us
\bq
 d \mathrm{Li}_{1}\left(x\right)
 & = &
 \mathrm{Li}_{0}\left(x\right) d\ln\left(x\right).
\eq
Since 
\bq
 \mathrm{Li}_{0}\left(x\right) & = & \frac{x}{1-x}
\eq
we have
\bq
 \mathrm{Li}_{0}\left(x\right) d\ln\left(x\right)
 & = & 
 \frac{x}{1-x} \cdot \frac{dx}{x}
 \; = \;
 \frac{dx}{1-x} 
 \; = \;
 - d\ln\left(1-x\right).
\eq
Hence we have
\bq
 d \mathrm{Li}_{1}\left(x\right)
 & = &
 - d\ln\left(1-x\right),
 \nonumber \\
 S\left(\mathrm{Li}_{1}\left(x\right)\right)
 & = & - \left(1-x\right).
\eq
Thus we combine any $\mathrm{Li}_0$ function with the accompanying dlog-form.
This gives
\bq
 S\left(\mathrm{Li}_{2 1}\left(f_1, f_2\right)\right)
 & = &
 \left( f_1 \otimes S\left(\mathrm{Li}_{1 1}\left(f_1, f_2\right)\right)\right)
 - \left( f_2 \otimes S\left(\mathrm{Li}_{2}\left(f_1 f_2\right)\right)\right)
 \nonumber \\
 & &
 + \left( \left(1-f_2\right) \otimes S\left(\mathrm{Li}_{2}\left(f_1 f_2\right)-\mathrm{Li}_{2}\left(f_1\right)\right)\right).
\eq
With
\bq
 S\left(\mathrm{Li}_{1 1}\left(f_1, f_2\right)\right)
 & = &
 \left( \left(1-f_2\right) \otimes \left(1-f_1\right) \right)
 + \left( \frac{\left(1-f_1\right) f_2}{\left(1-f_2\right)} \otimes \left(1-f_1f_2\right) \right)
\eq
we arrive at
\bq
\label{appendix_solutions:symbol_Li_21}
 S\left(\mathrm{Li}_{2 1}\left(f_1, f_2\right)\right)
 & = &
 \left( f_1 \otimes \left(1-f_2\right) \otimes \left(1-f_1\right) \right)
 + \left( \left(1-f_2\right) \otimes f_1 \otimes \left(1-f_1\right) \right)
 \nonumber \\
 & &
 + \left( f_1 \otimes \frac{\left(1-f_1\right) f_2}{\left(1-f_2\right)} \otimes \left(1-f_1f_2\right) \right)
 + \left( \frac{f_2}{1-f_2} \otimes f_1 f_2 \otimes \left(1-f_1f_2\right) \right).
\eq
Now let us substitute $g_1=1/f_1$ and $g_2=1/(f_1f_2)$ in eq.~(\ref{appendix_solutions:symbol_G_21}):
\bq
 S\left( G_{2 1}\left(\frac{1}{f_1},\frac{1}{f_1 f_2};1\right) \right)
 & = &
 \left( f_1 \otimes \frac{\left(1-f_1\right) f_2}{\left(1-f_2\right)} \otimes \left(1-f_1f_2\right) \right)
 - \left( \frac{1-f_2}{f_2} \otimes f_1 f_2 \otimes \left(1-f_1f_2\right) \right)
 \nonumber \\
 & &
 + \left( f_1 \otimes \left(1-f_2\right) \otimes \left(1-f_1\right) \right)
 + \left( \left(1-f_2\right) \otimes f_1 \otimes \left(1-f_1\right) \right).
\eq
This agrees with eq.~(\ref{appendix_solutions:symbol_Li_21}).

Let's assume that $f_1$ and $f_2$ are power products of the letters of the alphabet.
From the symbol of $\mathrm{Li}_{2 1}(f_1, f_2)$ we deduce that then
\bq
 1-f_1,
 \;\;\;
 1-f_2,
 \;\;\;
 1-f_1f_2
\eq
should also be power products of the letters of the alphabet.

Assuming that $g_1$ and $g_2$ are power products of the letters of the alphabet,
we deduce from the symbol of $G_{2 1}(g_1,g_2;1)$ that
\bq
 1-g_1,
 \;\;\;
 1-g_2,
 \;\;\;
 g_2-g_1
\eq
should also be power products of the letters of the alphabet.
}
\es
\\
\\
%
%
\bs
{\it \refstepcounter{exercise}
{\bf Exercise \theexercise}: 
Show that the mutation of the matrix $B$ at a fixed vertex $v_k$ is an involution, i.e. mutating twice at the same vertex
returns the original matrix $B$.
\\
\\
{\bf Solution}:
We mutate the matrix $B$ at the vertex $v_k$ twice. We denote the matrix after the first mutation by $B'$, the
one after the second mutation by $B''$.
For $i=k$ or $j=k$ we have
\bq
 b_{ij}'' & = & - b_{ij}' \; = \; b_{ij}.
\eq
For $i \neq k$ and $j \neq k$ we have
\bq
 b_{ij}'' & = &
 b_{i j}' + \mathrm{sign}\left(b_{i k}'\right) \cdot \max\left(0, b_{i k}' b_{k j}'\right)
 \nonumber \\
 & = &
 b_{i j} + \mathrm{sign}\left(b_{i k}\right) \cdot \max\left(0, b_{i k} b_{k j}\right) + \mathrm{sign}\left(-b_{i k}\right) \cdot \max\left(0, b_{i k} b_{k j}\right)
 \; = \; b_{ij}.
\eq
}
\es
\\
\bs
{\it \refstepcounter{exercise}
{\bf Exercise \theexercise}: 
Derive the transformation in eq.~(\ref{chapter_cluster:mutation_X}) 
from eq.~(\ref{chapter_cluster:def_X_coordinates}), eq.~(\ref{chapter_cluster:mutation_A}) and eq.~(\ref{chapter_cluster:mutation_B}).
\\
\\
{\bf Solution}:
We start with the case $j=k$: We have
\bq
 x_k' & = &
 \prod\limits_{i} \left( a_i' \right)^{b_{i k}'}
 \; = \; 
 \prod\limits_{i} \left( a_i \right)^{-b_{i k}}
 \; = \;
 \frac{1}{x_k}.
\eq
The case $j \neq k$ requires more work:
\bq
 x_j'
 & = &
 \prod\limits_{i} \left( a_i' \right)^{b_{i j}'}
 \; = \;
 \left( a_k' \right)^{b_{k j}'} \prod\limits_{i \neq k} \left( a_i' \right)^{b_{i j}'}
 \nonumber \\
 & = &
 a_k^{b_{k j}} \left( \prod\limits_{i \; | \; b_{i k} > 0} a_i^{b_{i k}} + \prod\limits_{i \; | \; b_{i k} < 0} a_i^{-b_{i k}}  \right)^{-b_{k j}} 
 \prod\limits_{i \neq k} a_i^{b_{i j} + \mathrm{sign}\left(b_{i k}\right) \cdot \max\left(0, b_{i k} b_{k j}\right)}
 \nonumber \\
 & = &
 x_j \left( \prod\limits_{i \; | \; b_{i k} > 0} a_i^{b_{i k}} + \prod\limits_{i \; | \; b_{i k} < 0} a_i^{-b_{i k}}  \right)^{-b_{k j}} 
 \prod\limits_{i \neq k} a_i^{\mathrm{sign}\left(b_{i j}\right) \cdot \max\left(0, b_{i j} b_{k j}\right)}
 \nonumber \\
 & = &
 x_j \left( \prod\limits_{i} a_i^{b_{i k}} + 1 \right)^{-b_{k j}} 
 \left( \prod\limits_{i \; | \; b_{i k} < 0} a_i^{b_{i k} b_{k j}} \right)
 \left( \prod\limits_{i \neq k} a_i^{\mathrm{sign}\left(b_{i k}\right) \cdot \max\left(0, b_{i k} b_{k j}\right)} \right)
 \nonumber \\
 & = &
 x_j \left( 1 + x_k \right)^{-b_{k j}} 
 \left( \prod\limits_{i \; | \; b_{i k} < 0} a_i^{b_{i k} b_{k j}} \right)
 \left( \prod\limits_{i \neq k} a_i^{\mathrm{sign}\left(b_{i k}\right) \cdot \max\left(0, b_{i k} b_{k j}\right)} \right).
\eq
We now distinguish the cases $b_{k j}>0$ and $b_{k j}<0$.
For $b_{k j}>0$ we have
\bq
 \left( \prod\limits_{i \; | \; b_{i k} < 0} a_i^{b_{i k} b_{k j}} \right)
 \left( \prod\limits_{i \neq k} a_i^{\mathrm{sign}\left(b_{i k}\right) \cdot \max\left(0, b_{i k} b_{k j}\right)} \right)
 & = & 
 \prod\limits_{i} a_i^{b_{i k} b_{k j}}
 \; = \; 
 x_k^{b_{k j}},
\eq
while for $b_{k j}<0$ we have
\bq
 \left( \prod\limits_{i \; | \; b_{i k} < 0} a_i^{b_{i k} b_{k j}} \right)
 \left( \prod\limits_{i \neq k} a_i^{\mathrm{sign}\left(b_{i k}\right) \cdot \max\left(0, b_{i k} b_{k j}\right)} \right)
 & = & 
 1.
\eq
We therefore obtain
\bq
 x_j' & = &
 \left\{
 \begin{array}{ll}
 x_j \left( \frac{x_k}{1+x_k} \right)^{b_{k j}} \; = \; x_j \left(1 + \frac{1}{x_k} \right)^{-b_{k j}}, & b_{k j}>0, \\
 x_j \left(1+x_k\right)^{-b_{k j}}, & b_{k j}<0. \\
 \end{array}
 \right.
\eq
Combining the two cases into one formula we find
\bq
 x_j ' & = &
 x_j \left( 1 + x_k^{-\mathrm{sign}\left(b_{k j}\right)} \right)^{-b_{k j}}.
\eq
Note that we never used the anti-symmetry of $b_{i j}$.
}
\es
\\
\\
\bs
{\it \refstepcounter{exercise}
{\bf Exercise \theexercise}: 
Determine the cluster $A$-variables for the ice quiver $Q'$ of fig.~\ref{chapter_cluster:fig_example_ice_quiver}
in terms of the cluster variables of the ice quiver $Q$.
\\
\\
{\bf Solution}:
We have
\bq
 a_1' & = & \frac{a_3 a_6 + a_2 a_7}{a_1},
 \nonumber \\
 a_2' & = & a_2.
\eq
The frozen $A$-variables are not changed.
}
\es
\\
\\
\bs
{\it \refstepcounter{exercise}
{\bf Exercise \theexercise}: 
Mutate the ice quiver $Q'$ of fig.~\ref{chapter_cluster:fig_example_ice_quiver} at the vertex $v_2$ to obtain
an ice quiver $Q''$.
Determine the cluster $A$-variables for the ice quiver $Q''$ 
in terms of the cluster variables of the ice quiver $Q$.
\\
\\
{\bf Solution}:
The mutated quiver $Q''$ is shown in fig.~\ref{appendix_solutions:fig_example_double_mutated_ice_quiver}.
\begin{figure}
\begin{center}
\includegraphics[scale=0.8]{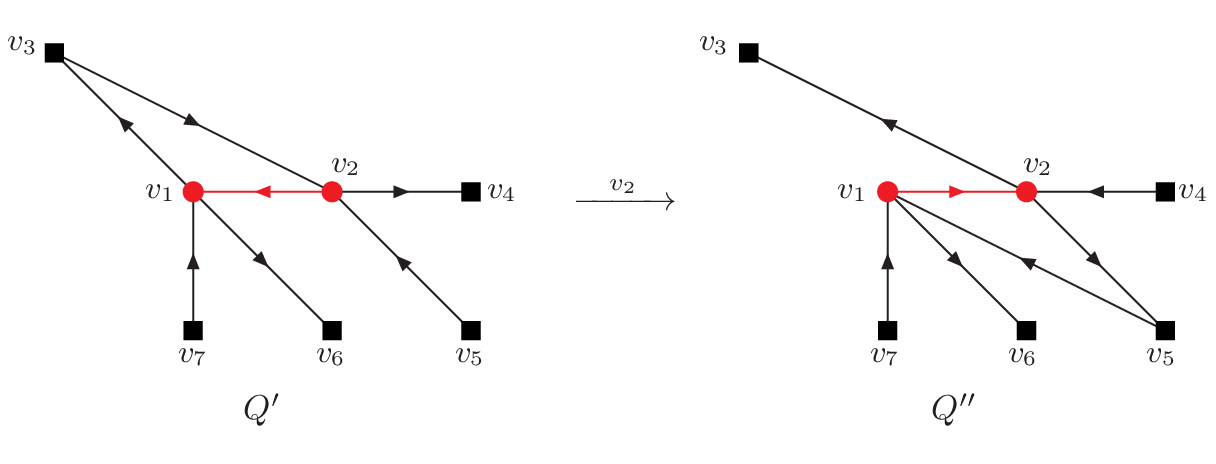}
\caption{\label{appendix_solutions:fig_example_double_mutated_ice_quiver}
The mutation of the ice quiver $Q'$ at the vertex $v_2$ yields the ice quiver $Q''$.
}
\end{center}
\end{figure}
For the cluster $A$-variables we have
\bq
 a_1'' & = & a_1' \; = \; \frac{a_3 a_6 + a_2 a_7}{a_1},
 \nonumber \\
 a_2'' & = & \frac{a_3'a_5'+a_1'a_4'}{a_2'}
 \, = \; \frac{a_1 a_3 a_5 + a_2 a_4 a_7 + a_3 a_4 a_6}{a_1 a_2}
\eq
}
\es
\\
\\
\bs
{\it \refstepcounter{exercise}
{\bf Exercise \theexercise}: 
The $B_2$-cluster algebra:
Determine the cluster variables from the initial seed
\bq
 B \; = \; 
 \left(\begin{array}{rr}
  0 & -1 \\ 
  2 & 0 \\
 \end{array} \right),
 & &
 a \; = \; \left(a_1,a_2\right).
\eq
\\
\\
{\bf Solution}:
We start from the seed $(B,a)$.
As two mutations on the same vertex will give us back the original seed, we alternate the vertices where we perform mutations.
The exchange matrix $B$ changes sign under each mutation.
We start with a mutation at $v_1$. This yields
\bq
 \left(a_1',a_2'\right)
 & = &
 \left( \frac{1+a_2^2}{a_1}, a_2 \right).
\eq
We set $a_3 = (1+a_2^2)/a_1$. 
We then mutate at vertex $v_2$.
This yields
\bq
 \left(a_1'',a_2''\right)
 & = &
 \left( \frac{1+a_2^2}{a_1}, \frac{1+a_1+a_2^2}{a_1a_2} \right).
\eq
We set $a_4=(1+a_1+a_2^2)/(a_1a_2)$.
Continuing in this way we obtain
\bq
 a_{n+1} & = & 
 \left\{ \begin{array}{ll}
 \frac{1+a_n^2}{a_{n-1}}, & \mbox{if $n$ is even}, \\
 \frac{1+a_n}{a_{n-1}}, & \mbox{if $n$ is odd}. \\
 \end{array}
 \right.
\eq
This will give a sequence with period $6$:
\bq
 a_{n+6} & = & a_n.
\eq
The first six terms are the cluster variables:
\bq
 a_1,
 \;\;\;
 a_2,
 \;\;\;
 \frac{1+a_2^2}{a_1},
 \;\;\;
 \frac{1+a_1+a_2^2}{a_1a_2},
 \;\;\;
 \frac{1+2a_1+a_1^2+a_2^2}{a_1a_2^2},
 \;\;\;
 \frac{1+a_1}{a_2}.
\eq
}
\es
\\
\\
%
%
\bs
{\it \refstepcounter{exercise}
{\bf Exercise \theexercise}: 
Consider the elliptic curve $y^2=4x^3-g_2x-g_3$ . Show that
\bq
 dz & = & \frac{dx}{y},
\eq
where $y=\sqrt{4x^3-g_2x-g_3}$. This shows that $dx/y$ is a holomorphic differential. 
\\
\\
{\bf Solution}:
The variables $z$ and $x$ are related by eq.~(\ref{chapter_elliptics:relation_x_to_z}):
\bq
 z 
 & = &
 \int\limits_\infty^x \frac{dt}{\sqrt{4t^3-g_2t-g_3}}.
\eq
We therefore have
\bq
 dz & = & \frac{dx}{\sqrt{4x^3-g_2x-g_3}} \; = \; \frac{dx}{y}.
\eq
}
\es
\\
\\
\bs
{\it \refstepcounter{exercise}
{\bf Exercise \theexercise}: 
Determine two independent periods for the elliptic curve defined by
a quartic polynomial:
\bq
\label{appendix_solutions:general_quartic_elliptic_curve}
 y^2 & = & \left(x-x_1\right) \left(x-x_2\right) \left(x-x_3\right) \left(x-x_4\right).
\eq
{\bf Solution}:
We would like to express $y$ as the square root of the right hand side of eq.~(\ref{appendix_solutions:general_quartic_elliptic_curve}).
We denote by $[x_i,x_j]$ the line segment from $x_i$ to $x_j$ in the complex plane.
We may express $y$ as a single-valued and continuous function on ${\mathbb C} \backslash ([x_l,x_i] \cup [x_j,x_k])$ through
\bq
 y & = &
 \pm
 \left(x_i-x_l\right) \left(x_k-x_j\right)
 \sqrt{\frac{x-x_l}{x_i-x_l}}
 \sqrt{\frac{x-x_i}{x_i-x_l}}
 \sqrt{\frac{x-x_j}{x_k-x_j}}
 \sqrt{\frac{x-x_k}{x_k-x_j}}.
\eq
For a given choice $(i,j,k,l)$ of branch cuts $[x_l,x_i]$ and $[x_j,x_k]$ 
the transformation
\bq
 T\left(x\right) & = & \frac{\left(x_k-x_l\right)}{\left(x_k-x_i\right)} \frac{\left(x-x_i\right)}{\left(x-x_l\right)}
\eq
maps the points $x_i$, $x_k$, $x_l$ to $0$, $1$, $\infty$, respectively.
The point $x_j$ is then mapped to
\bq
 \lambda & = &
 \frac{\left(x_k-x_l\right)}{\left(x_k-x_i\right)} \frac{\left(x_j-x_i\right)}{\left(x_j-x_l\right)}.
\eq
We denote the cross-ratio by
\bq
 \left[ j,k | i,l \right]
 & = &
 \frac{\left(x_k-x_l\right)}{\left(x_k-x_i\right)} \frac{\left(x_j-x_i\right)}{\left(x_j-x_l\right)}.
\eq
The cross-ratios satisfy
\bq
 \left[ i,j | k,l \right] & = & \left[ k,l | i,j \right],
 \nonumber \\
 \left[ i,j | k,l \right] & = & \left[ j,i | l,k \right],
 \nonumber \\
 \left[ i,j | k,l \right] & = & \left[ i,j | l,k \right]^{-1},
 \nonumber \\
 \left[ i,j | k,l \right] + \left[ i,k | j,l \right] & = & 1.
\eq
Let $\delta_1$ and $\delta_2$ be two independent cycles as shown in fig.~\ref{appendix_solutions:fig_choice_periods}.
\begin{figure}
\begin{center}
\includegraphics[align=c,scale=1.0]{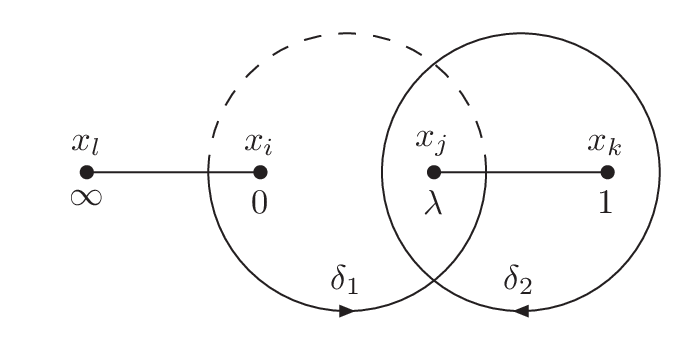}
\end{center}
\caption{
Branch cuts and cycles for the computation of the periods of an elliptic curve.
}
\label{appendix_solutions:fig_choice_periods}
\end{figure}
The cycles $\delta_1$ and $\delta_2$ have intersection number $+1$.
We define the periods by
\bq
 \psi_1
 & = &
 2 \int\limits_{x_i}^{x_j} \frac{dx}{y}
 \; = \;
 \frac{4}{\sqrt{\left(x_j-x_l\right)\left(x_k-x_i\right)}}
 K\left(\sqrt{\frac{\left(x_j-x_i\right)\left(x_k-x_l\right)}{\left(x_j-x_l\right)\left(x_k-x_i\right)}}\right),
 \nonumber \\
 \psi_2
 & = &
 2 \int\limits_{x_k}^{x_j} \frac{dx}{y}
 \; = \;
 \frac{4i}{\sqrt{\left(x_j-x_l\right)\left(x_k-x_i\right)}}
 K\left(\sqrt{\frac{\left(x_i-x_l\right)\left(x_k-x_j\right)}{\left(x_j-x_l\right)\left(x_k-x_i\right)}}\right).
\eq
For $\psi_2$ the square root $y$ is evaluated in the complex $x$-plane below the cut.
Let us now discuss the possibilities of choosing $x_i,x_j,x_k,x_l$: 
Due to the symmetries $[ i,j | k,l ] = [ k,l | i,j ]$ and
$[ i,j | k,l ] = [ j,i | l,k ]$ we may fix $x_l=x_4$.
This leaves six possibilities for $\lambda$.
These are
\bq
 \left(i,j,k,l\right) = \left(2,3,1,4\right):
 & & 
 \left[ 3,1 | 2,4 \right] \; = \; \lambda
 \; = \; 
 \frac{\left(x_1-x_4\right)}{\left(x_1-x_2\right)} \frac{\left(x_3-x_2\right)}{\left(x_3-x_4\right)},
 \nonumber \\
 \left(i,j,k,l\right) = \left(2,1,3,4\right):
 & & 
 \left[ 1,3 | 2,4 \right] \; = \; \frac{1}{\lambda},
 \nonumber \\
 \left(i,j,k,l\right) = \left(3,1,2,4\right):
 & & 
 \left[ 1,2 | 3,4 \right] \; = \; \frac{\lambda-1}{\lambda},
 \nonumber \\
 \left(i,j,k,l\right) = \left(3,2,1,4\right):
 & & 
 \left[ 2,1 | 3,4 \right] \; = \; \frac{\lambda}{\lambda-1},
 \nonumber \\
 \left(i,j,k,l\right) = \left(1,2,3,4\right):
 & & 
 \left[ 2,3 | 1,4 \right] \; = \; \frac{1}{1-\lambda},
 \nonumber \\
 \left(i,j,k,l\right) = \left(1,3,2,4\right):
 & & 
 \left[ 3,2 | 1,4 \right] \; = \; 1-\lambda.
\eq
It is worth noting that we 
have three possibilities for $\lambda (1-\lambda)$. These are
\bq
 \left[ 3,1 | 2,4 \right] \; \cdot \; \left[ 3,2 | 1,4 \right] 
 & = & \lambda \left(1-\lambda\right),
 \nonumber \\
 \left[ 1,2 | 3,4 \right] \; \cdot \; \left[ 1,3 | 2,4 \right] 
 & = & -\frac{\left(1-\lambda\right)}{\lambda^2},
 \nonumber \\
 \left[ 2,3 | 1,4 \right] \; \cdot \; \left[ 2,1 | 3,4 \right] 
 & = & - \frac{\lambda}{\left(1-\lambda\right)^2}.
\eq
}
\es
\\
\\
\bs
{\it \refstepcounter{exercise}
{\bf Exercise \theexercise}: 
Express the modulus squared $k^2$ and the complementary modulus squared $k'{}^2$ as a quotient of eta
functions.
\\
\\
{\bf Solution}:
From eq.~(\ref{chapter_elliptics:Jacobi_theta_relation_1}) and eq.~(\ref{chapter_elliptics:Jacobi_theta_relation_2}) we have
\bq
 k^2
 & = &
 \frac{\theta_2^4\left(0,q\right)}{\theta_3^4\left(0,q\right)}
 \; = \;
 16 
 \frac{\eta\left(\frac{\tau}{2}\right)^8 \eta\left(2\tau\right)^{16}}{\eta\left(\tau\right)^{24}}
\eq
and
\bq
 k'{}^2
 & = &
 \frac{\theta_4^4\left(0,q\right)}{\theta_3^4\left(0,q\right)}
 \; = \;
 \frac{\eta\left(\frac{\tau}{2}\right)^{16} \eta\left(2\tau\right)^8}{\eta\left(\tau\right)^{24}}.
\eq
}
\es
\\
\\
\bs
{\it \refstepcounter{exercise}
{\bf Exercise \theexercise}: 
Show that
\bq
 \left(f \slashoperator{\gamma_1}{k}\right) \slashoperator{\gamma_2}{k}
 & = &
 f \slashoperator{\left(\gamma_1 \gamma_2\right)}{k}.
\eq
{\bf Solution}:
Let
\bq
 \gamma_1 \; = \; \left(\begin{array}{cc} a_1 & b_1 \\ c_1 & d_1 \\ \end{array} \right),
 \;\;\;
 \gamma_2 \; = \; \left(\begin{array}{cc} a_2 & b_2 \\ c_2 & d_2 \\ \end{array} \right),
 \;\;\;
 \gamma_{12} \; = \; \left(\begin{array}{cc} a_{12} & b_{12} \\ c_{12} & d_{12} \\ \end{array} \right)
\eq
with $\gamma_{12} = \gamma_1 \cdot \gamma_2$.
From matrix multiplication we have
\bq
 c_{12} \; = \; c_1 a_2 + d_1 c_2,
 \;\;\;
 d_{12} \; = \; c_1 b_2 + d_1 d_2.
\eq
On the one hand we have
\bq
 \left(f \slashoperator{\gamma_1}{k}\right) \slashoperator{\gamma_2}{k}
 & = & 
 \left( \left(c_1\tau+d_1\right)^{-k} f\left(\gamma_1\left(\tau\right)\right) \right) \slashoperator{\gamma_2}{k}
 \nonumber \\
 & = &
 \left(c_2\tau+d_2\right)^{-k} \left(c_1\gamma_2\left(\tau\right)+d_1\right)^{-k} f\left(\gamma_1\left(\gamma_2\left(\tau\right)\right)\right) 
 \nonumber \\
 & = &
 \left(\left(c_1 a_2+d_1 c_2\right)\tau+c_1b_2+d_1d_2\right)^{-k} f\left(\gamma_1\left(\gamma_2\left(\tau\right)\right)\right) 
 \nonumber \\
 & = &
 \left( c_{12} \tau + d_{12} \right)^{-k} f\left(\gamma_{12}\left(\tau\right)\right).
\eq
On the other hand we have
\bq
 f \slashoperator{\left(\gamma_1 \gamma_2\right)}{k}
 & = & 
 f \slashoperator{\gamma_{12}}{k}
 \; = \;
 \left( c_{12} \tau + d_{12} \right)^{-k} \cdot f\left(\gamma_{12}\left(\tau\right)\right).
\eq
}
\es
\\
\\
\bs
{\it \refstepcounter{exercise}
{\bf Exercise \theexercise}: 
Let $f \in \mathcal{M}_k(\Gamma(N))$ and $\gamma \in \mathrm{SL}_2(\mathbb{Z}) \backslash \Gamma(N)$.
Show that $f\slashoperator{\gamma}{k} \in \mathcal{M}_k(\Gamma(N))$.
\\
\\
{\bf Solution}:
Let $\gamma_1 \in \Gamma(N)$.
We have to show
\bq
 \left( f\slashoperator{\gamma}{k}\right) \slashoperator{\gamma_1}{k}
 & = & 
 f\slashoperator{\gamma}{k}.
\eq
As $\Gamma(N)$ is a normal subgroup of $\mathrm{SL}_2(\mathbb{Z})$ there exist for $\gamma \in \mathrm{SL}_2(\mathbb{Z})$ and 
$\gamma_1 \in \Gamma(N)$ a $\gamma_2 \in \Gamma(N)$ such that
\bq
 \gamma \gamma_1
 & = &
 \gamma_2 \gamma.
\eq
As $\gamma_2 \in \Gamma(N)$ we have
\bq
 f\slashoperator{\gamma_2}{k} & = & f.
\eq
We have with
$(f \slashoperator{\gamma_1}{k}) \slashoperator{\gamma_2}{k} = f \slashoperator{\left(\gamma_1 \gamma_2\right)}{k}$
\bq
 \left( f\slashoperator{\gamma}{k}\right) \slashoperator{\gamma_1}{k}
 & = & 
 \left( f\slashoperator{\left(\gamma \gamma_1\right)}{k}\right) 
 \; = \;
 \left( f\slashoperator{\left(\gamma_2 \gamma\right)}{k}\right) 
 \; = \;
 \left( f\slashoperator{\gamma_2}{k}\right) \slashoperator{\gamma}{k}
 \; = \;
 f\slashoperator{\gamma}{k}.
\eq
}
\es
\\
\\
\bs
{\it \refstepcounter{exercise}
{\bf Exercise \theexercise}: 
Let $\chi$ be a Dirichlet character with modulus $N$ and
$f \in \mathcal{M}_k(N,\chi)$.
Let further $\gamma_1, \gamma_2 \in \Gamma_0(N)$ and set $\gamma_{12} = \gamma_1 \gamma_2$.
Show that
\bq
 f\left(\gamma_1\left(\gamma_2\left(\tau\right)\right)\right)
 & = &
 f\left(\gamma_{12}\left(\tau\right)\right).
\eq
{\bf Solution}:
Let
\bq
 \gamma_1 \; = \; \left(\begin{array}{cc} a_1 & b_1 \\ c_1 & d_1 \\ \end{array} \right),
 \;\;\;
 \gamma_2 \; = \; \left(\begin{array}{cc} a_2 & b_2 \\ c_2 & d_2 \\ \end{array} \right),
 \;\;\;
 \gamma_{12} \; = \; \left(\begin{array}{cc} a_{12} & b_{12} \\ c_{12} & d_{12} \\ \end{array} \right).
\eq
Since $\gamma_{12} = \gamma_1 \cdot \gamma_2$ we have
\bq
 c_{12} \; = \; c_1 a_2 + d_1 c_2,
 \;\;\;
 d_{12} \; = \; c_1 b_2 + d_1 d_2.
\eq
Let us set $\tau'=\gamma_2(\tau)$.
We have
\bq
 f\left(\gamma_1\left(\gamma_2\left(\tau\right)\right)\right)
 & = &
 \chi(d_{1}) \left(c_{1}\tau'+d_{1}\right)^k f\left(\tau'\right)
 \; = \;
 \chi(d_{1}) \chi(d_{2}) \left(c_{1}\tau'+d_{1}\right)^k \left(c_{2}\tau+d_{2}\right)^k f\left(\tau\right)
 \nonumber \\
 & = &
 \chi(d_{1}) \chi(d_{2}) \left(c_{12}\tau+d_{12}\right)^k f\left(\tau\right).
\eq
On the other hand
\bq
 f\left(\gamma_{12}\left(\tau\right)\right)
 & = &
 \chi(d_{12}) \left(c_{12}\tau+d_{12}\right)^k f\left(\tau\right).
\eq
Thus we see that $f(\gamma_{12}(\tau)) = f(\gamma_1(\gamma_2(\tau)))$ requires $\chi(d_{12})=\chi(d_{1}) \chi(d_{2})$.
Since $\gamma_1 \in \Gamma_0(N)$ we have $c_1 = 0 \mod N$ and therefore
\bq
 \chi\left(d_{12}\right)
 & = &
 \chi\left(c_1 b_2 + d_1 d_2\right)
 \; = \; 
 \chi\left(d_1 d_2\right)
 \; = \; 
 \chi\left(d_1\right) \chi\left(d_2\right).
\eq
}
\es
\\
\\
\bs
{\it \refstepcounter{exercise}
{\bf Exercise \theexercise}: 
Consider
\bq
 f\left(\tau\right) & = & 
 e_2\left(\tau\right) - 2 e_2\left(2\tau\right)
\eq
and work out the transformation properties under $\gamma \in \Gamma_0(2)$.
\\
\\
{\bf Solution}:
Let 
\bq
 \gamma & = &
 \left( \begin{array}{cc} a & b \\ c & d \\ \end{array} \right)
 \; = \; \Gamma_0(2).
\eq
This implies that $c$ is even and hence
\bq
 \left( \begin{array}{cc} a & 2b \\ \frac{c}{2} & d \\ \end{array} \right)
 & \in &
 \mathrm{SL}_2({\mathbb Z}).
\eq
We have
\bq 
 \tau' \;= \; \frac{a\tau+b}{c\tau+d}
 & \mbox{and} &
 2 \tau' \;= \; \frac{a\left(2\tau\right)+2b}{\frac{c}{2}\left(2\tau\right)+d}.
\eq
Hence
\bq
 f\left(\tau'\right) & = & 
 e_2\left(\tau'\right) - 2 e_2\left(2\tau'\right)
 \nonumber \\
 & = &
 \left(c\tau+d\right)^2 e_2\left(\tau\right)
 - 2 \pi i c \left(c\tau+d\right)
 -2 \left[ 
  \left(\frac{c}{2}\left(2\tau\right)+d\right)^2 e_2\left(2\tau\right)
 - 2 \pi i \frac{c}{2} \left(\frac{c}{2}\left(2\tau\right)+d\right)
 \right]
 \nonumber \\
 & = &
 \left(c\tau+d\right)^2 \left[ e_2\left(\tau\right) - 2 e_2\left(2\tau\right) \right]
 \; = \; 
 \left(c\tau+d\right)^2 f\left(\tau\right).
\eq
This shows that $f(\tau)$ is a modular form of weight $2$ for $\Gamma_0(2)$.
}
\es
\\
\\
\bs
{\it \refstepcounter{exercise}
{\bf Exercise \theexercise}: 
Show eq.~(\ref{chapter_elliptics:def_Eisenstein_h_k_N_r_s}).
\\
\\
{\bf Solution}:
Let us introduce the short-hand notation
\bq
 \sum\limits_{n_1 \in {\mathbb Z}} f\left(n_1\right)
 & = &
 \lim\limits_{N_1\rightarrow \infty} \sum\limits_{n_1=-N_1}^{N_1}
 f\left(n_1\right).
\eq
Then we may write Eisenstein's summation prescription as
\bq
 \sideset{}{_e}\sum\limits_{(n_1,n_2) \in {\mathbb Z}^2} f\left(z+n_1 + n_2\tau \right)
 & = &
 \lim\limits_{N_2\rightarrow \infty} \sum\limits_{n_2=-N_2}^{N_2}
 \left(
 \lim\limits_{N_1\rightarrow \infty} \sum\limits_{n_1=-N_1}^{N_1}
 f\left(z + n_1 + n_2 \tau \right)
 \right)
 \nonumber \\
 & = &
 \sum\limits_{n_2 \in {\mathbb Z}} 
 \left( \sum\limits_{n_1 \in {\mathbb Z}} f\left(z + n_1 + n_2 \tau \right) \right).
\eq
Let's now turn to the problem at hand. We split the outer $n_2$-summation into $n_2=0$ and $n_2 \neq 0$
and obtain
\bq
 \sideset{}{_e}\sum\limits_{(n_1,n_2) \in {\mathbb Z}^2\backslash (0,0)} 
 \frac{e^{\frac{2\pi i}{N} \left(n_1s-n_2r\right)}}{\left(n_1+n_2\tau\right)^k}
 & = &
 \mathrm{Li}_k\left(e^{2\pi i\frac{s}{N}}\right)
 +
 \left(-1\right)^k
 \mathrm{Li}_k\left(e^{-2\pi i\frac{s}{N}}\right)
 \nonumber \\
 & &
 +
 \sum\limits_{n_2=1}^{\infty}
 \sum\limits_{n_1 \in {\mathbb Z}}
 \frac{e^{\frac{2\pi i}{N} \left(n_1s-n_2r\right)}+\left(-1\right)^k e^{-\frac{2\pi i}{N} \left(n_1s-n_2r\right)}}{\left(n_1+n_2\tau\right)^k}.
\eq
The essential trick is the following identity
\bq
\label{appendix_solutions_eq_cotangens_k_eq_1}
 \sum\limits_{n_1 \in {\mathbb Z}} \frac{1}{n_1+\tau}
 & = & 
 \pi \cot\left(\pi \tau\right)
 \; = \;
 - 2 \pi i \left[ \frac{1}{2} + \sum\limits_{n=1}^\infty \bar{q}^n \right],
 \;\;\;\;\;\;\;\;\;
 \bar{q} \; = \; e^{2\pi i \tau}.
\eq
Taking $(k-1)$-times the derivative with respect to $\tau$ gives for $k \ge 2$
\bq
\label{appendix_solutions_eq_cotangens_higher_k}
 \sum\limits_{n_1 \in {\mathbb Z}} \frac{1}{\left(n_1+\tau\right)^k}
 & = & 
 \frac{\left(-2\pi i\right)^k}{\left(k-1\right)!}
 \sum\limits_{n=1}^\infty n^{k-1} \bar{q}^n.
\eq
Let us first consider the case $k \ge 2$.
We apply eq.~(\ref{appendix_solutions_eq_cotangens_higher_k}) to
\bq
 \sum\limits_{n_1 \in {\mathbb Z}} \frac{\left(e^{2\pi i \frac{s}{N}}\right)^{n_1}}{\left(n_1+n_2\tau\right)^k}.
\eq
For $k \ge 2$ the sum is absolutely convergent and we may reorder the terms.
For $N \in {\mathbb N}$, $s \in {\mathbb N}_0$  the numerator is periodic with period $N$.
We therefore have
\bq
\label{appendix_solutions:eq_Eisenstein_reordering}
 \sum\limits_{n_1 \in {\mathbb Z}} \frac{\left(e^{2\pi i \frac{s}{N}}\right)^{n_1}}{\left(n_1+n_2\tau\right)^k}
 & = &
 \sum\limits_{c_1=0}^{N-1} \sum\limits_{n_1' \in {\mathbb Z}}
 \frac{\left(e^{2\pi i \frac{s}{N}}\right)^{n_1'N+c_1}}{\left(n_1'N+c_1+n_2\tau\right)^k}
 \nonumber \\
 & = &
 \frac{1}{N^k}
 \sum\limits_{c_1=0}^{N-1} e^{2\pi i \frac{s c_1}{N}}
 \sum\limits_{n_1' \in {\mathbb Z}}
 \frac{1}{\left(n_1'+\frac{c_1}{N}+\frac{n_2\tau}{N}\right)^k}
 \nonumber \\
 & = &
 \frac{\left(-2 \pi i\right)^k}{\left(k-1\right)! N^k}
 \sum\limits_{c_1=0}^{N-1} 
 \sum\limits_{d=1}^\infty d^{k-1} e^{2\pi i\left(\frac{s c_1}{N}+\frac{c_1 d}{N}+ \frac{n_2 d}{N}\tau\right)}.
\eq
Thus
\bq
\label{appendix_solutions:eq_Eisenstein_triple_sum}
\lefteqn{
 \frac{1}{2} \frac{\left(k-1\right)!}{\left(2\pi i\right)^k}
 \sideset{}{_e}\sum\limits_{(n_1,n_2) \in {\mathbb Z}^2\backslash (0,0)} 
 \frac{e^{\frac{2\pi i}{N} \left(n_1s-n_2r\right)}}{\left(n_1+n_2\tau\right)^k}
 = 
 \frac{1}{2} \frac{\left(k-1\right)!}{\left(2\pi i\right)^k}
 \left( \mathrm{Li}_k\left(e^{2\pi i\frac{s}{N}}\right) + \left(-1\right)^k \mathrm{Li}_k\left(e^{-2\pi i\frac{s}{N}}\right) \right)
 } & &
 \nonumber \\
 & &
 +
 \frac{1}{2} 
 \frac{\left(-1\right)^k}{N^k}
 \sum\limits_{n_2=1}^\infty
 \sum\limits_{c_1=0}^{N-1} 
 \sum\limits_{d=1}^\infty d^{k-1} 
 \left[ 
   e^{2\pi i\left(\frac{s c_1}{N}-\frac{n_2 r}{N}+\frac{c_1 d}{N}+ \frac{n_2 d}{N}\tau\right)}
   + 
   \left(-1\right)^k e^{2\pi i\left(-\frac{s c_1}{N}+\frac{n_2 r}{N}+\frac{c_1 d}{N}+ \frac{n_2 d}{N}\tau\right)}
 \right].
\eq
The first term on the right-hand side yields with eq.~(\ref{chapter_iterated_integrals:Glaisher_to_Bernoulli})
\bq
 a_0 & = &
 \frac{1}{2} \frac{\left(k-1\right)!}{\left(2\pi i\right)^k}
 \left( \mathrm{Li}_k\left(e^{2\pi i\frac{s}{N}}\right) + \left(-1\right)^k \mathrm{Li}_k\left(e^{-2\pi i\frac{s}{N}}\right) \right)
 \; = \;
 - \frac{1}{2k} B_k\left(\frac{s}{N}\right).
\eq
The second term on the right-hand side of eq.~(\ref{appendix_solutions:eq_Eisenstein_triple_sum}) we may rearrange as follows:
\bq
\label{appendix_solutions:eq_Eisenstein_reordering_second_term}
\lefteqn{
 \frac{1}{2} 
 \frac{\left(-1\right)^k}{N^k}
 \sum\limits_{n_2=1}^\infty
 \sum\limits_{c_1=0}^{N-1} 
 \sum\limits_{d=1}^\infty
 d^{k-1} 
 \left[ 
   e^{2\pi i\left(\frac{s c_1}{N}-\frac{n_2 r}{N}+\frac{c_1 d}{N}+ \frac{n_2 d}{N}\tau\right)}
   + 
   \left(-1\right)^k e^{2\pi i\left(-\frac{s c_1}{N}+\frac{n_2 r}{N}+\frac{c_1 d}{N}+ \frac{n_2 d}{N}\tau\right)}
 \right]
 } & & \nonumber \\
 & = &
 \frac{\left(-1\right)^k}{2 N^k}
 \sum\limits_{n=1}^\infty
 \sum\limits_{c_1=0}^{N-1} 
 \sum\limits_{d | n}
 d^{k-1} 
 \left[
   e^{2\pi i\left(\frac{s c_1}{N}-\frac{n r}{d N}+\frac{c_1 d}{N}\right)}
   +
   \left(-1\right)^k e^{2\pi i\left(-\frac{s c_1}{N}+\frac{n r}{d N}+\frac{c_1 d}{N}\right)}
 \right]
 \bar{q}^n_N
 \nonumber \\
 & = &
 \frac{1}{2 N^k}
 \sum\limits_{n=1}^\infty
 \sum\limits_{c_1=0}^{N-1} 
 \sum\limits_{d | n}
 d^{k-1} 
 \left[
   e^{\frac{2\pi i}{N}\left(r \frac{n}{d}-\left(s -d\right)c_1\right)}
   +
   \left(-1\right)^k e^{-\frac{2\pi i}{N}\left(r \frac{n}{d}-\left(s + d \right) c_1\right)}
 \right]
 \bar{q}^n_N.
\eq
For $k=1$ we have to be more careful about absolute convergence and reordering of terms.
The easiest approach is to consider eq.~(\ref{appendix_solutions:eq_Eisenstein_reordering}) for $k=2$ and to
integrate in $\tau'=n_2 \tau$. This yields
\bq
 \sum\limits_{n_1 \in {\mathbb Z}} \frac{\left(e^{2\pi i \frac{s}{N}}\right)^{n_1}}{n_1+n_2\tau}
 & = &
 \tilde{C}
 - \frac{2 \pi i}{N}
 \sum\limits_{c_1=0}^{N-1} 
 \sum\limits_{d=1}^\infty e^{2\pi i\left(\frac{s c_1}{N}+\frac{c_1 d}{N}+ \frac{n_2 d}{N}\tau\right)},
\eq
with some unknown constant $\tilde{C}$.
We then repeat the steps as in eq.~(\ref{appendix_solutions:eq_Eisenstein_triple_sum}) 
and eq.~(\ref{appendix_solutions:eq_Eisenstein_reordering_second_term}) and obtain
\bq
 \frac{1}{2} \frac{1}{2\pi i}
 \sideset{}{_e}\sum\limits_{(n_1,n_2) \in {\mathbb Z}^2\backslash (0,0)} 
 \frac{e^{\frac{2\pi i}{N} \left(n_1s-n_2r\right)}}{n_1+n_2\tau}
 = 
 a_0
 +
 \frac{1}{2 N}
 \sum\limits_{n=1}^\infty
 \sum\limits_{c_1=0}^{N-1} 
 \sum\limits_{d | n}
 \left[
   e^{\frac{2\pi i}{N}\left(r \frac{n}{d}-\left(s -d\right)c_1\right)}
   - e^{-\frac{2\pi i}{N}\left(r \frac{n}{d}-\left(s + d \right) c_1\right)}
 \right]
 \bar{q}^n_N,
\eq
with another unknown constant $a_0$. We determine $a_0$ by evaluating both sides at $\tau=i\infty$.
On the right-hand side only $a_0$ survives.
On the left-hand side we consider the three cases
(i) $s=r=0 \bmod N$,
(ii) $s=0 \bmod N$, $r \neq 0 \bmod N$ and
(iii) $s \neq 0 \bmod N$.
We start with case (i):
We have
\bq 
 \lim\limits_{\tau \rightarrow i \infty}
 \frac{1}{2} \frac{1}{2\pi i}
 \sideset{}{_e}\sum\limits_{(n_1,n_2) \in {\mathbb Z}^2\backslash (0,0)} 
 \frac{1}{n_1+n_2\tau}
 & = &
 \lim\limits_{\tau \rightarrow i \infty}
 \frac{1}{2} \frac{1}{2\pi i}
 e_1\left(\tau\right) 
 \; = \; 0,
\eq
and therefore $a_0=0$.
In the case (ii) we have
\bq 
 \lim\limits_{\tau \rightarrow i \infty}
 \frac{1}{2} \frac{1}{2\pi i}
 \sideset{}{_e}\sum\limits_{(n_1,n_2) \in {\mathbb Z}^2\backslash (0,0)} 
 \frac{e^{-2\pi i\frac{r}{N} n_2}}{n_1+n_2\tau}
 & = &
 \lim\limits_{\tau \rightarrow i \infty}
 \frac{1}{2} \frac{1}{2\pi i}
 \sum\limits_{n_2=1}^\infty
 \left[ 
  e^{-2\pi i \frac{r}{N} n_2} \sum\limits_{n_1 \in {\mathbb Z}} \frac{1}{n_1+n_2\tau}
  +
  e^{2\pi i \frac{r}{N} n_2} \sum\limits_{n_1 \in {\mathbb Z}} \frac{1}{n_1-n_2\tau}
 \right]
 \nonumber \\
 & = &
 \lim\limits_{\tau \rightarrow i \infty}
 \frac{1}{4 i}
 \sum\limits_{n_2=1}^\infty
 \left[ 
  e^{-2\pi i \frac{r}{N} n_2} - e^{2\pi i\frac{r}{N} n_2} 
 \right]
  \cot\left(\pi \tau n_2 \right) 
 \nonumber \\
 & = &
 - \frac{1}{4}
 \sum\limits_{n_2=1}^\infty
 \left[ 
  e^{-2\pi i \frac{r}{N} n_2} - e^{2\pi i\frac{r}{N} n_2} 
 \right]
 \; = \;
 \frac{i}{2}
 \sum\limits_{n_2=1}^\infty
 \sin\left( 2 \pi \frac{r}{N} n_2 \right)
 \nonumber \\
 & = &
 \frac{i}{4}
 \cot\left( \pi \frac{r}{N} \right),
\eq
and therefore $a_0=\frac{i}{4} \cot(\frac{r}{N}\pi)$.
In the case (iii) one first shows that
\bq
 \lim\limits_{\tau \rightarrow i \infty}
 \sum\limits_{n_1 \in {\mathbb Z}} 
 \frac{e^{2\pi i \frac{s}{N} n_1}}{n_1+n_2\tau}
 & = & 0.
\eq
Then
\bq
 \lim\limits_{\tau \rightarrow i \infty}
 \frac{1}{2} \frac{1}{2\pi i}
 \sideset{}{_e}\sum\limits_{(n_1,n_2) \in {\mathbb Z}^2\backslash (0,0)} 
 \frac{e^{\frac{2\pi i}{N} \left(n_1s-n_2r\right)}}{n_1+n_2\tau}
 & = &
 \frac{1}{4\pi i}
 \sum\limits_{n_1=1}^\infty
 \frac{e^{2\pi i\frac{s}{N} n_1}-e^{-2\pi i\frac{s}{N} n_1}}{n_1}
 \nonumber \\
 & = &
 \frac{1}{4\pi i}
 \left[
  \mathrm{Li}_1\left(e^{2\pi i\frac{s}{N}}\right) - \mathrm{Li}_1\left(e^{-2\pi i\frac{s}{N}}\right)
 \right]
 \nonumber \\
 & = &
 \frac{1}{2\pi}
 \mathrm{Gl}_1\left(2\pi \frac{s}{N}\right)
 \; = \; 
 \frac{1}{4} - \frac{s}{2N},
\eq
and therefore $a_0=\frac{1}{4}-\frac{s}{2N}$.
}
\es
\\
\\
\bs
{\it \refstepcounter{exercise}
{\bf Exercise \theexercise}: 
Prove eq.~(\ref{chapter_elliptics:trafo_Eisenstein_h_2}) for the case $k \ge 3$.
\\
\\
{\bf Solution}:
Let
\bq
 \tau' & = & \frac{a \tau +b}{c\tau +d}.
\eq
We set
\bq
 \gamma \; = \; \left( \begin{array}{cc} a & b \\ c & d \\ \end{array} \right),
 & &
 \gamma^{-1} \; = \; \left( \begin{array}{rr} d & -b \\ -c & a \\ \end{array} \right).
\eq
In this exercise we only consider the case $k \ge 3$.
In this case the sums are absolutely convergent and we may drop 
the Eisenstein summation prescription.
We consider
\bq
 2 \frac{\left(2\pi i\right)^k}{\left(k-1\right)!}
 \left(h_{k,N,r,s}\slashoperator{\gamma}{k}\right)\left(\tau\right)
 & = &
 \left(c\tau+d\right)^{-k} 
 \sum\limits_{(n_1,n_2) \in {\mathbb Z}^2\backslash (0,0)} 
 \frac{e^{\frac{2\pi i}{N} \left(n_1s-n_2r\right)}}{\left(n_1+n_2\tau'\right)^k}.
\eq
We have
\bq
 n_1+n_2\tau'
 & = &
 \frac{1}{c\tau+d} 
 \left(n_2,n_1\right) 
 \left( \begin{array}{cc} a & b \\ c & d \\ \end{array} \right)
 \left( \begin{array}{c} \tau \\ 1 \\ \end{array} \right).
\eq
Thus
\bq
 \left(c\tau+d\right)^{-k} 
 \sum\limits_{(n_1,n_2) \in {\mathbb Z}^2\backslash (0,0)} 
 \frac{e^{\frac{2\pi i}{N} \left(n_1s-n_2r\right)}}{\left(n_1+n_2\tau'\right)^k}
 & = &
 \sum\limits_{(n_1,n_2) \in {\mathbb Z}^2\backslash (0,0)} 
 \frac{e^{\frac{2\pi i}{N} \left(n_1s-n_2r\right)}}
      { \left[ \left(n_2,n_1\right) \left( \begin{array}{cc} a & b \\ c & d \\ \end{array} \right)
                                    \left( \begin{array}{c} \tau \\ 1 \\ \end{array} \right) 
        \right]^k}.
\eq
For $k \ge 3$ the sum is absolutely convergent and we may sum over the individual terms in a different order.
We set
\bq
 \left(n_2',n_1'\right) 
 \; = \;
 \left(n_2,n_1\right) 
 \left( \begin{array}{cc} a & b \\ c & d \\ \end{array} \right),
 & &
 \left(n_2,n_1\right) 
 \; = \;
 \left(n_2',n_1'\right) 
 \left( \begin{array}{rr} d & -b \\ -c & a \\ \end{array} \right)
\eq
and sum over $(n_1',n_2')$.
Then
\bq
 \left(c\tau+d\right)^{-k} 
 \sum\limits_{(n_1,n_2) \in {\mathbb Z}^2\backslash (0,0)} 
 \frac{e^{\frac{2\pi i}{N} \left(n_1s-n_2r\right)}}{\left(n_1+n_2\tau'\right)^k}
 & = &
 \sum\limits_{(n_1',n_2') \in {\mathbb Z}^2\backslash (0,0)} 
 \frac{e^{\frac{2\pi i}{N} \left[n_1'\left(as+cr\right)-n_2'\left(bs+dr\right)\right]}}
      { \left( n_1' + n_2' \tau \right)^k}.
\eq
Thus we have shown for $k \ge 3$
\bq
 \left(h_{k,N,r,s}\slashoperator{\gamma}{k}\right)\left(\tau\right)
 & = &
 h_{k,N,\left(rd+sb\right) \bmod N,\left(rc+sa\right) \bmod N}\left(\tau\right).
\eq
}
\es
\\
\\
%
%
\bs
{\it \refstepcounter{exercise}
{\bf Exercise \theexercise}: 
Show that a relative boundary is a relative cycle.
\\
\\
{\bf Solution}:
Elements in the relative chain group $C_k(B,A)$ are equivalence classes in $C_k(B)$.
If $c_k \in C_k(B)$ we denote the corresponding equivalence class by $[c_k]=c_k+A$.

Now, let us consider a boundary $[b_k] \in B_k(B,A)$. By definition, there exists a 
$c_{k+1} \in C_{k+1}(B,A)$ such that
\bq
\label{appendix_solutions:relative_boundary}
 b_k - \partial c_{k+1} & \in & A.
\eq
We have to show that $[b_k]$ is a relative cycle, i.e. $\partial b_k \in A$.
As $A$ is a subcomplex, we have for any $a_k \in A$ that $\partial a_k \in A$.
From eq.~(\ref{appendix_solutions:relative_boundary}) we have then
\bq
 \partial \left( b_k - \partial c_{k+1} \right) 
 \; = \;
 \partial b_k - \partial \partial c_{k+1}
 & \in & A.
\eq
Since $\partial \partial c_{k+1} = 0$ we have $\partial b_k \in A$.
}
\es
\\
\\
\bs
{\it \refstepcounter{exercise}
{\bf Exercise \theexercise}: 
We now have two bases of $H^1_{\mathrm{dR}}(E)$: on the one hand $(\omega_1,\omega_2)$, on the other hand $(dz,d\bar{z})$. We already know $\omega_1=dz$. Work out the full relation between the two bases.
\\
\\
{\bf Solution}: We first relate $(\omega_1,\omega_2)$ to $(\gamma_1^\ast,\gamma_2^\ast)$:
We make the ansatz $\omega_i = c_{i,1} \gamma_1^\ast + c_{i,2} \gamma_2^\ast$.
Integrating over $\gamma_j$ gives
\bq
 \left\langle \omega_i, \gamma_j \right\rangle
 & = & 
 c_{i,1} \left\langle \gamma_1^\ast, \gamma_j \right\rangle + c_{i,2} \left\langle \gamma_2^\ast, \gamma_j \right\rangle.
\eq
Using eq.~(\ref{chapter_motives:def_duality_Betti}) we find
\bq
 \left(\begin{array}{c}
  \omega_1 \\
  \omega_2 \\
 \end{array} \right)
 & = & 
 \left(\begin{array}{cc}
  \psi_1 & \psi_2 \\
  \phi_1 & \phi_2 \\
 \end{array} \right)
 \left(\begin{array}{c}
  \gamma_1^\ast \\
  \gamma_2^\ast \\
 \end{array} \right).
\eq
As $dz=\omega_1$ we have
\bq
 \left(\begin{array}{c}
  dz \\
  d\bar{z} \\
 \end{array} \right)
 & = & 
 \left(\begin{array}{cc}
  \psi_1 & \psi_2 \\
  \overline{\psi}_1 & \overline{\psi}_2 \\
 \end{array} \right)
 \left(\begin{array}{c}
  \gamma_1^\ast \\
  \gamma_2^\ast \\
 \end{array} \right).
\eq
The inverse of the period matrix is
\bq
 \left(\begin{array}{cc}
  \psi_1 & \psi_2 \\
  \phi_1 & \phi_2 \\
 \end{array} \right)^{-1}
 & = & 
 \frac{1}{2\pi i}
 \left(\begin{array}{rr}
  \phi_2 & -\psi_2 \\
  -\phi_1 & \psi_1 \\
 \end{array} \right)
\eq
and we obtain
\bq
 \left(\begin{array}{c}
  dz \\
  d\bar{z} \\
 \end{array} \right)
 & = & 
 \left(\begin{array}{cc}
  1 & 0\\
  \frac{1}{2\pi i}\left(\overline{\psi}_1\phi_2 - \overline{\psi}_2\phi_1\right)  & -\frac{1}{2\pi i}\left(\overline{\psi}_1\psi_2 - \overline{\psi}_2\psi_1 \right) \\
 \end{array} \right)
 \left(\begin{array}{c}
  \omega_1 \\
  \omega_2 \\
 \end{array} \right).
\eq
}
\es
\\
\\
\bs
{\it \refstepcounter{exercise}
{\bf Exercise \theexercise}: 
Work out all $V^{p,q}$ and show that $V_{\mathbb Q}$ is mixed Tate.
\\
\\
{\bf Solution}:
The four conditions
\bq
 W_0 V_{\mathbb Q} = V_{\mathbb Q}, 
 \;\;\;\;\;\;
 W_{-2n-1} V_{\mathbb Q} = 0, 
 \;\;\;\;\;\;
 F^1 V_{\mathbb C} = 0,
 \;\;\;\;\;\;
 \overline{F^1 V_{\mathbb C}} = 0
\eq
allow only a finite number of $V^{p,q}$'s to be non-zero, namely the ones with
\bq
 p \; \le \; 0,
 \;\;\;\;\;\;
 q \; \le \; 0,
 \;\;\;\;\;\;
 p + q \; \ge \; -2n.
\eq
For $n=2$ these are shown in fig.~\ref{chapter_solutions:fig_mixed_hodge_structure}.
\begin{figure}
\begin{center}
\includegraphics[scale=0.65]{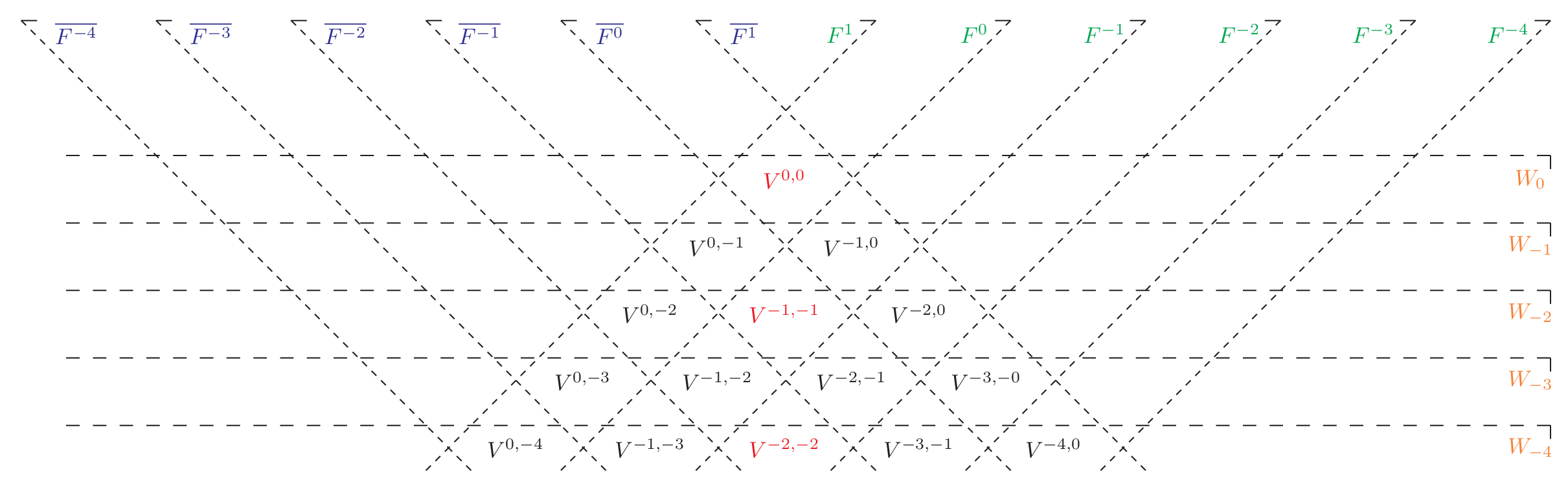}
\end{center}
\caption{
For $n=2$ the four conditions $W_0 V_{\mathbb Q} = V_{\mathbb Q}$, $W_{-2n-1} V_{\mathbb Q} = 0$, 
$F^1 V_{\mathbb C} = 0$ and $\overline{F^1 V_{\mathbb C}} = 0$ allow only the shown $V^{p,q}$ (shown in red and black) to be non-zero.
In the text we then show that only the $V^{p,q}$ shown in red are non-zero. 
}
\label{chapter_solutions:fig_mixed_hodge_structure}
\end{figure}
We have over ${\mathbb C}$
\bq
 \mathrm{Gr}_{0}^W V_{\mathbb C}
 & = & 
 W_0 V_{\mathbb C} / W_{-1} V_{\mathbb C}
 = 
 \left\langle v_0, v_1, v_2, \dots, v_n  \right\rangle / \left\langle v_1, v_2, \dots, v_n  \right\rangle
 = 
 \left\langle e_0, v_1, v_2, \dots, v_n  \right\rangle / \left\langle v_1, v_2, \dots, v_n  \right\rangle
 \nonumber \\
 & \cong &
 \left\langle e_0 \right\rangle,
 \nonumber \\
 \mathrm{Gr}_{-1}^W V_{\mathbb C}
 & = & 
 W_{-1} V_{\mathbb C} / W_{-2} V_{\mathbb C}
 = 
 \left\langle v_1, v_2, \dots, v_n  \right\rangle / \left\langle v_1, v_2, \dots, v_n  \right\rangle
 \nonumber \\
 & \cong &
 0,
 \nonumber \\
 \mathrm{Gr}_{-2}^W V_{\mathbb C}
 & = & 
 W_{-2} V_{\mathbb C} / W_{-3} V_{\mathbb C}
 = 
 \left\langle v_1, v_2, \dots, v_n  \right\rangle / \left\langle v_2, \dots, v_n  \right\rangle
 = 
 \left\langle 2 \pi i e_1, v_2, \dots, v_n  \right\rangle / \left\langle v_2, \dots, v_n  \right\rangle
 \nonumber \\
 & \cong &
 \left\langle 2 \pi i e_1 \right\rangle,
\eq
and more generally for $0 \le j \le n$
\bq
 \mathrm{Gr}_{-2j}^W V_{\mathbb C}
 \; \cong \;  
 \left\langle \left( 2\pi i \right)^j e_j \right\rangle,
 & &
 \mathrm{Gr}_{-2j-1}^W V_{\mathbb C}
 \; \cong \;
 0. 
\eq
Let's look at the even weights $(-2j)$:
As the $e_j$'s are independent we have
\bq
 (2\pi i)^j e_j 
 & \notin &
 F^{j-1} V_{\mathbb C} \; = \; \langle e_0, e_{-1}, \dots, e_{-j+1} \rangle
\eq 
and therefore $V^{p,q}=0$ for $p>q$ and $p+q=-2j$.
From $V^{q,p}=\overline{V^{p,q}}$ it follows then that also $V^{p,q}=0$ for $p<q$ and $p+q=-2j$.
This shows that at weight $(-2j)$ only $V^{-j,-j}$ can be non-zero. This proves that 
$V_{\mathbb Q}$ is mixed Tate and we have
\bq
 V^{-j,-j} & \cong & \left\langle \left( 2\pi i \right)^j e_j \right\rangle.
\eq
}
\es
\\
\bs
{\it \refstepcounter{exercise}
{\bf Exercise \theexercise}: 
Derive eq.~(\ref{chapter_motives:example_Griffiths}).
\\
\\
{\bf Solution}:
The starting point is $\nabla v_j = 0$ for $0 \le j \le n$ together with the relation of the $v_j$'s to the $e_{-j}$'s
given in eq.~(\ref{chapter_motives:example_polylog_v_e}).
We prove eq.~(\ref{chapter_motives:example_Griffiths}) recursively, starting with $e_{-n}$.
As $v_n = (2\pi i)^n e_{-n}$, the equation $\nabla v_n=0$ implies
\bq
 \nabla e_{-n} & = & 0.
\eq
We further have
\bq
 v_{n-1} & = & \left(2\pi i\right)^{n-1} \left[ e_{-n+1} + \ln(x) e_{-n} \right]
\eq
and $\nabla v_{n-1}=0$ implies
\bq
 0 & = & \nabla \left[ e_{-n+1} + \ln(x) e_{-n} \right]
 \; = \;
 \nabla e_{-n+1} + \frac{dx}{x} e_{-n},
\eq
and therefore 
\bq
 \nabla e_{-n+1} & = & - \frac{dx}{x} e_{-n}.
\eq
Let now assume that
\bq
\label{chapter_solutions:Griffiths_recursion}
 \nabla e_{-j} & = & - \frac{dx}{x} e_{-j-1}
\eq
for $j \in \{n,n-1,\dots,k+1\}$ and $k>0$. We show that eq.~(\ref{chapter_solutions:Griffiths_recursion}) holds also for $j=k$:
From $\nabla v_k=0$ we have
\bq
 0 & = &
 \nabla \left[ \sum\limits_{j=k}^{n} \frac{\ln^{j-k}(x)}{(j-k)!} e_{-j}\right] 
 \; = \;
 \nabla e_{-k}
 +
 \sum\limits_{j=k+1}^{n} \frac{\ln^{j-k-1}(x) dx}{(j-k-1)! x} e_{-j}
 -
 \sum\limits_{j=k+1}^{n-1} \frac{\ln^{j-k}(x) dx}{(j-k)! x} e_{-j-1}
 \nonumber \\
 & = &
 \nabla e_{-k}
 +
 \frac{dx}{x} e_{-j}.
\eq
In a similar way one derives from $\nabla v_0=0$
\bq
 0
 & = &
 \nabla \left[ e_0 - \sum\limits_{j=1}^n \mathrm{Li}_j(x) e_{-j} \right]
 \; = \;
 \nabla e_0 
 - \sum\limits_{j=1}^n \frac{\mathrm{Li}_{j-1}(x) dx}{x} e_{-j}
 + \sum\limits_{j=1}^{n-1} \frac{\mathrm{Li}_j(x) dx}{x} e_{-j-1}
 \nonumber \\
 & = &
 \nabla e_0 
 - \frac{\mathrm{Li}_{0}(x) dx}{x} e_{-1}
 \; = \;
 \nabla e_0 
 + \frac{dx}{x-1} e_{-1}.
\eq
}
\es
\\
\bs
{\it \refstepcounter{exercise}
{\bf Exercise \theexercise}: 
Consider $X={\mathbb C} \backslash \{0\}$ and $Y=\emptyset$.
Take $\omega=dx/x$ as a basis of $H^1_{\mathrm{alg \; dR}}(X)$
and let $\gamma$ be a small counter-clockwise circle around $x=0$.
$\gamma$ is a basis of $H_1^{\mathrm{B}}(X)$. Denote by $\gamma^\ast$ the dual basis of $H^1_{\mathrm{B}}(X)$.
Work out
\bq
 F_\infty\left(\gamma^\ast\right).
\eq
{\bf Solution}:
Let's first work out the comparison isomorphism
\bq
 \mathrm{comparison}
 & : & 
 H^1_{\mathrm{alg \; dR}}\left(X\right) \otimes {\mathbb C}
 \rightarrow
 H^1_{\mathrm{B}}\left(X,{\mathbb Q}\right) \otimes {\mathbb C}
\eq
The period is
\bq
 p & = & \int\limits_\gamma \omega \; = \; 2\pi i.
\eq
We then have with eq.~(\ref{chapter_motives:def_comparison}) 
\bq
 \mathrm{comparison}\left(\omega\right) & = & 2 \pi i \gamma^\ast.
\eq
Let us now consider $\omega \otimes 1 \in H^1_{\mathrm{alg \; dR}}\left(X\right) \otimes {\mathbb C}$.
On the one hand we have
\bq
 \mathrm{comparison}\left(\mathrm{conj}_{\mathrm{alg \; dR}}\left( \omega \otimes 1 \right)\right)
 & = & 
 \mathrm{comparison}\left( \omega \otimes 1 \right)
 \nonumber \\
 & = & 
 \gamma^\ast \otimes \left(2\pi i\right).
\eq
On the other hand we have
\bq
 \left(F_\infty \otimes \mathrm{id}\right) \; \mathrm{conj}_{\mathrm{B}}\left( \mathrm{comparison}\left( \omega \otimes 1 \right)\right)
 & = &
 \left(F_\infty \otimes \mathrm{id}\right) \; \mathrm{conj}_{\mathrm{B}}\left( \gamma^\ast \otimes \left(2 \pi i\right) \right)
 \nonumber \\
 & = &
 \left(F_\infty \otimes \mathrm{id}\right) \; \left( \gamma^\ast \otimes \left(-2 \pi i\right) \right)
 \nonumber \\
 & = &
 - F_\infty\left(\gamma^\ast\right) \otimes \left(2 \pi i\right).
\eq
This should be equal to $\gamma^\ast \otimes \left(2\pi i\right)$ and therefore
\bq
 F_\infty\left(\gamma^\ast\right) & = & - \gamma^\ast.
\eq
The dual of $(-\gamma^\ast)$ is $(-\gamma)$, a small circle in the clockwise direction around $x=0$.
}
\es
\\
\\
\bs
{\it \refstepcounter{exercise}
{\bf Exercise \theexercise}: 
Let ${\mathbb F}$ be a sub-field of ${\mathbb C}$. Show that $\mathrm{GL}_n({\mathbb F})$
(the group of $(n \times n)$-matrices with entries from ${\mathbb F}$ and non-zero determinant)
can be defined by a polynomial equation.
\\
\\
{\bf Solution}:
An element $g \in \mathrm{GL}_n({\mathbb F})$ is given by $n^2$ elements $z_{ij} \in {\mathbb F}$
\bq
 g & = &
 \left( \begin{array}{cccc}
  z_{11} & z_{12} & \dots & z_{1n} \\
  \vdots & \vdots & & \vdots \\
  z_{n1} & z_{n2} & \dots & z_{nn} \\
 \end{array} \right),
\eq
such that
\bq
 \det g & \neq & 0.
\eq
The determinant $\det g$ is a polynomial in the $z_{ij}$'s.
In order to convert this inequality to a polynomial equation, we introduce another variable $z_0 \in {\mathbb F}$.
$\mathrm{GL}_n({\mathbb F})$ is then isomorph to the set of points in ${\mathbb F}^{n^2+1}$ satisfying
the polynomial equation
\bq 
 z_0 \det g - 1 & = & 0.
\eq
}
\es
\\
\bs
{\it \refstepcounter{exercise}
{\bf Exercise \theexercise}: 
Consider the one-loop two-point function with equal internal masses
\bq
 I_{11}\left(2,x\right)
 & = &
 \int \frac{d^2k}{i \pi}
 \frac{m^2}{\left(-q_1^2+m_1^2\right) \left(-q_2^2+m_2^2\right)},
 \;\;\;\;\;\;
 x \; = \; - \frac{p^2}{m^2}.
\eq
Derive with the methods of this section the differential equation for $I_{11}(2,x)$ with respect to the kinematic variable $x$.
\\
\\
{\bf Solution}:
The Feynman parametrisation reads
\bq
 I_{11}\left(2,x\right)
 & = &
 \int\limits_{\Delta} 
 \frac{\omega}{{\mathcal F}},
\eq
with
\bq
 {\mathcal F} \; = \;
 a_1 a_2 x + \left( a_1 + a_2 \right)^2,
 \;\;\;
 \omega \; = \; 
 a_1 da_2
 - a_2 da_1,
 \;\;\;
 \Delta \; = \; 
 {\mathbb R} {\mathbb P}^{1}_{\ge 0}.
\eq
We set 
\bq
 \varphi & = & \frac{\omega}{{\mathcal F}}.
\eq
We first look for a differential equation of the form
\bq
\label{appendix_solutions:basic_picard}
 L^{(r)} \varphi & = & d \beta,
\eq
where
\bq
\label{appendix_solutions:picard_fuchs_operator}
 L^{(r)} & = & \sum\limits_{j=0}^r R_j\left(x\right) \frac{d^j}{dx^j},
 \;\;\;\;\;\;
 R_r\left(x\right) \; = \; 1.
\eq
is a Picard-Fuchs operator of order $r$.
For the example of the one-loop two-point function
$\beta$ is a $0$-form, depending on the Feynman parameters $a_i$. 
The differential $d$ is with respect to the Feynman parameters $a_i$.
For $\beta$ we make the following ansatz
\bq
\label{appendix_solutions:ansatz_beta}
 \beta & = & 
 \frac{1}{{\mathcal F}^{r}}
 \left( a_{1} q_{2} - a_{2} q_{1} \right),
\eq
where the $q_i$ are homogeneous polynomials of degree $(2r - 1)$ in the variables $a_i$.
The ansatz is based on the fact, that the singularities of the integrand $\varphi$ 
are given by powers of the graph polynomial ${\mathcal F}$.
Acting with $L^{(r)}$ on the integrand $\varphi$ 
will only increase the power of ${\mathcal F}$ in the denominator by $r$, but will not introduce singularities on new algebraic varieties.
Each polynomial $q_i$ contains only a finite number of monomials in the Feynman parameters 
with a priori unknown coefficients.
We therefore take these coefficients and the variables $R_0$, ..., $R_{r-1}$ as the set of our unknown variables.
Plugging the ansatz~(\ref{appendix_solutions:ansatz_beta}) in eq.~(\ref{appendix_solutions:basic_picard}) gives a linear system of equations for the unknown variables.
This system may or may not have a solution.
In order to find the differential equation of minimal order we start at $r=1$ and try to solve the linear system of equations.
If no solution is found, we increase $r$ by one and repeat the exercise, until a solution is found.
This is then the solution of minimal order $r$.

For the case of the one-loop two-point function we obtain a solution for $r=1$.
Thus we obtain the Picard-Fuchs operator
\bq
 L^{(1)} & = & \frac{d}{dx} + \frac{x+2}{x(x+4)}
\eq
and a possible solution for $\beta$ is given by
\bq
 \beta & = & \frac{1}{\mathcal F} \frac{1}{x(x+4)} \left[ (x+2) a_1 a_2 + 2 a_2^2 \right].
\eq
The boundary of $\Delta$ is given by the two points $[1:0]$ and $[0:1]$. 
The integration of the inhomogeneous term yields
\bq
\int\limits_{\partial \Delta} \beta & = & \frac{2}{x(x+4)}.
\eq
Putting everything together, we obtain the differential equation
\bq
 \left[ x(x+4) \frac{d}{dx} +x+2 \right] I_{11}\left(2,x\right) & = & 2.
\eq
This agrees with eq.~(\ref{chapter_iterated_integrals:example_partial_derivative_2}) if we use the result
for the tadpole integral
\bq
 I_{10}\left(2-2\eps,x\right) & = & \frac{1}{\eps} + {\mathcal O}\left(\eps^0\right).
\eq
}
\es
\\
%
%
\bs
{\it \refstepcounter{exercise}
{\bf Exercise \theexercise}: 
Derive eq.~(\ref{chapter_numerics:Li2Bernoulli}).
\\
\\
{\bf Solution}:
We start from the integral representation, substitute $t=1-e^{-\tilde{z}}$, expand and integrate term-by-term:
\bq
 \mathrm{Li}_{2}(x)
 & = &
 - \int\limits_{0}^{x} \frac{dt}{t} \ln\left(1-t\right)
 \; = \; 
 \int\limits_{0}^{z} d\tilde{z} \frac{\tilde{z}}{e^{\tilde{z}}-1} 
 \; = \;
 \int\limits_{0}^{z} d\tilde{z} \sum\limits_{j=0}^\infty \frac{B_j}{j!} \tilde{z}^j
 \; = \;
 \sum\limits_{j=0}^\infty \frac{B_j}{\left(j+1\right)!} z^{j+1}.
\eq
}
\es
\\
\\
%
%
\bs
{\it \refstepcounter{exercise}
{\bf Exercise \theexercise}: 
Rewrite the differential one-form 
\bq
 \omega & = &
 - \frac{\left(1-x_2\right)^2 dx_1}{x_1 \left[\left(1-x_2\right)^2-x_1x_2\right]}
 - \frac{x_1\left(1+x_2\right)dx_2}{\left(1-x_2\right) \left[\left(1-x_2\right)^2-x_1x_2\right]}
\eq
as a dlog-form.
\\
\\
{\bf Solution}:
We first look at the polynomials which appear in the denominator of $\omega$: 
There are three distinct polynomials
\bq
 p_1 \; = \; x_1,
 \;\;\;\;\;\;
 p_2 \; = \; x_2-1,
 \;\;\;\;\;\;
 p_3 \; = \; \left(1-x_2\right)^2-x_1x_2.
\eq
We then make the ansatz
\bq
 \omega^{\mathrm{ansatz}} & = & d\ln\left( p_1^{n_1} p_2^{n_2} p_3^{n_3} \right),
 \;\;\;\;\;\;\;\;\;
 n_1, n_2, n_3 \; \in \; {\mathbb Z}.
\eq
Let's now study $\omega-\omega^{\mathrm{ansatz}}$. We partial fraction the terms proportional to $dx_1$ with respect to $x_1$,
and the terms proportional to $dx_2$ with respect to $x_2$.
This yields
\bq
 \omega-\omega^{\mathrm{ansatz}}
 & = &
 \left[ - \frac{1+n_1}{x_1} - \frac{x_2\left(1-n_3\right)}{\left(1-x_2\right)^2-x_1x_2} \right] dx_1
 +
 \left[ - \frac{2+n_2}{x_2-1} - \frac{\left(1-n_3\right)\left(2+x_1-2x_2\right)}{\left(1-x_2\right)^2-x_1x_2} \right] dx_2.
 \nonumber \\
\eq
This gives the linear system of equations
\bq
 1+n_1 & = & 0,
 \nonumber \\
 2+n_2 & = & 0,
 \nonumber \\
 1-n_3 & = & 0,
\eq
whose unique solution is $(n_1,n_2,n_3)=(-1,-2,1)$.
Hence
\bq
 \omega & = & d\ln p_3 - d\ln p_1 - 2 d\ln p_2.
\eq
}
\es
\\
\\
\bs
{\it \refstepcounter{exercise}
{\bf Exercise \theexercise}: 
Let
\bq
 \tilde{f}_1 & = & \lambda\left(x_2,x_3,1\right) + 8x_3 - x_1 \left(x_2-x_3\right) - r_4 r_5,
 \nonumber \\
 \tilde{f}_2 & = & \lambda\left(x_2,x_3,1\right) + 8x_2 + x_1 \left(x_2-x_3\right) - r_4 r_5,
 \nonumber \\
 \tilde{f}_3 & = & \lambda\left(x_2,x_3,1\right) - x_1 \left(x_2-x_3\right) + r_4 r_5,
 \nonumber \\
 \tilde{f}_4 & = & \lambda\left(x_2,x_3,1\right) + x_1 \left(x_2-x_3\right) + r_4 r_5,
\eq
Show that
\bq
 \tilde{\omega} & = & d\ln\left(\frac{\tilde{f}_1\tilde{f}_2}{\tilde{f}_3\tilde{f}_4}\right)
\eq
has a pole along $x_1=0$, while
\bq
 \omega & = & d\ln\left(x_1^2 \frac{\tilde{f}_1\tilde{f}_2}{\tilde{f}_3\tilde{f}_4}\right)
\eq
does not.
\\
\\
{\bf Solution}:
There are several ways to show this:
We may compute the residue of $\tilde{\omega}$ and $\omega$ along $Y=\{x_1=0\}$.
One finds
\bq
 \mathrm{Res}_Y\left(\tilde{\omega}\right) \; = \; -2,
 & &
 \mathrm{Res}_Y\left(\omega\right) \; = \; 0.
\eq
Alternatively we may compute the limits $x_1\rightarrow 0$ of $\tilde{f}_1$-$\tilde{f}_4$:
\bq
 \tilde{f}_1 \tilde{f}_2 & = &
 4 \left[1-\left(x_2-x_3\right)^2\right]^2
 + {\mathcal O}\left(x_1\right),
 \nonumber \\
 \tilde{f}_3 \tilde{f}_4 & = &
 4 x_1^2 x_2 x_3 \left[\frac{1-\left(x_2-x_3\right)^2}{\lambda\left(x_2,x_3,1\right)}\right]^2
 + {\mathcal O}\left(x_1^3\right).
\eq
The simple pole of $\tilde{\omega}$ along $x_1=0$ originates from $\tilde{f}_3 \tilde{f}_4$, which vanishes as $x_1^2$ in the limit $x_1\rightarrow 0$.
}
\es
\\
\\
\bs
{\it \refstepcounter{exercise}
{\bf Exercise \theexercise}: 
Rewrite the differential one-form 
\bq
 \omega & = &
 - \frac{\left(2-2x_2-x_1x_2\right)r_1 dx_1}{x_1\left(4+x_1\right) \left[\left(1-x_2\right)^2-x_1x_2\right]}
 - \frac{r_1 dx_2}{\left[\left(1-x_2\right)^2-x_1x_2\right]}
\eq
where $r_1=\sqrt{x_1(4+x_1)}$
as a dlog-form.
\\
\\
{\bf Solution}:
We use the rationalisation
\bq
 x_1 \; = \; \frac{\left(1-x_1'\right)^2}{x_1'},
 & &
 x_1' \; = \; \frac{1}{2} \left( 2+x_1-r_1 \right).
\eq
We have
\bq
 dx_1 & = & - \frac{\left(1-x_1'^2\right)}{x_1'^2} dx_1'
\eq
and
\bq
 \omega & = &
 \left[ \frac{1}{x_1'} + \frac{1}{x_1'-x_2} + \frac{x_2}{1-x_1' x_2} \right] dx_1'
 +
 \left[ \frac{1}{x_2-x_1'} + \frac{x_1'}{1-x_1' x_2} \right] dx_2.
\eq
We are now back to the rational case and we find
\bq
 \omega & = &
 d \ln\left(\frac{x_1'\left(x_1'-x_2\right)}{\left(1-x_1'x_2\right)}\right).
\eq
We substitute back from $x_1'$ to $x_1$ and obtain
\bq
 \omega & = &
 d\ln\left( \frac{2+x_1-2x_2-r_1}{2+x_1-2x_2+r_1} \right)
 \; = \;
 2 d\ln\left(2+x_1-2x_2-r_1\right) - d\ln\left(\left(1-x_2\right)^2-x_1x_2\right).
\eq
}
\es
\\
\bs
{\it \refstepcounter{exercise}
{\bf Exercise \theexercise}: 
Let
\bq
 g & = & \frac{2+x_1-r_1}{2+x_1+r_1}.
\eq
Express $g$ and $(1-g)$ as a power product in the letters of the alphabet 
defined by eq.~(\ref{chapter_final_project:rational_alphabet}), eq.~(\ref{chapter_final_project:algebraic_alphabet}) 
and the constant $f_0=2$.
\\
\\
{\bf Solution}:
We start with $g$: We multiply the numerator and the denominator with $(2+x_1+r_1)$:
\bq
 g & = & \frac{2+x_1-r_1}{2+x_1+r_1}
 \; = \;
 \frac{\left(2+x_1-r_1\right)^2}{\left(2+x_1+r_1\right)\left(2+x_1-r_1\right)}.
\eq
The third binomial formula gives
\bq
 \left(2+x_1+r_1\right)\left(2+x_1-r_1\right) & =& \left(2+x_1\right)^2 -r_1^2
 \; = \; 
 \left(4+4x_1+x_1^2\right) - x_1\left(4+x_1\right) \; = \; 4.
\eq
Hence
\bq
 g & = & \frac{\left(2+x_1-r_1\right)^2}{4}
 \; = \;
 f_0^{-2} f_{10}.
\eq
Let us now turn to $(1-g)$. We have
\bq
 1-g & = &
 1-\frac{2+x_1-r_1}{2+x_1+r_1}
 \; = \;
 \frac{2r_1}{2+x_1+r_1}
 \; = \; 
 \frac{2r_1\left(2+x_1-r_1\right)}{4}
\eq
From eq.~(\ref{chapter_final_project:roots_in_alphabet}) we have $r_1^2=f_1f_2$ and therefore
\bq
 1-g & = & f_0^{-1}f_1f_2f_{10}.
\eq
Thus $g$ is an allowed argument of $\mathrm{Li}_n$.
}
\es
\\
\\
\bs
{\it \refstepcounter{exercise}
{\bf Exercise \theexercise}: 
The master integral $J_{15}$ starts at order ${\mathcal O}(\eps^2)$.
The weight two term of $J_{15}$ is given in terms of iterated integrals 
by
\bq
 J_{15}^{(2)}
 & = &
 2 i I_\gamma\left(2\omega_{15}-\omega_{3}-\omega_{5},\omega_{5}-\omega_{3};1\right)
 + 2 J_7^{(2)}(0,1,1),
\eq
where $J_7^{(2)}(0,1,1)$ denotes the boundary value of eq.~(\ref{chapter_final_project:boundary_J7_J8}):
\bq
 J_7^{(2)}(0,1,1) & = & \frac{3}{2i} \overline{H}_2\left(e^{\frac{2\pi i}{3}}\right)
 \; = \; 
 \frac{3}{2i} \left[ \mathrm{Li}_2\left(e^{\frac{2\pi i}{3}}\right) - \mathrm{Li}_2\left(e^{-\frac{2\pi i}{3}}\right) \right].
\eq
Express $J_{15}^{(2)}$ in terms of multiple polylogarithms.
\\
\\
{\bf Solution}:
We have to convert the iterated integral $I_\gamma(2\omega_{15}-\omega_{3}-\omega_{5},\omega_{5}-\omega_{3};1)$
to multiple polylogarithms.
We have
\bq
 2\omega_{15}-\omega_{3}-\omega_{5}
 \; = \;
 d\ln\left(\frac{f_{15}^2}{f_{3}f_{5}}\right),
 & &
 \omega_{5}-\omega_{3}
 \; = \; 
 d\ln\left(\frac{f_{5}}{f_{3}}\right)
\eq
and
\bq
 \frac{f_{15}^2}{f_{3}f_{5}}
 \; = \;
 -2
 \frac{\left(2x_2-x_3+ir_3\right)}{x_2},
 & & 
 \frac{f_{5}}{f_{3}}
 \; = \; 
 \frac{x_3}{x_2}.
\eq
This involves only the square root $r_3$, which we may rationalise.
With the rationalisation of eq.~(\ref{chapter_final_project:rationalisation_r2_r3}) we obtain
\bq
 2\omega_{15}-\omega_{3}-\omega_{5}
 & = &
 d\ln\left(x_2'+i\right)-d\ln\left(x_2'-i\right),
 \nonumber \\
 \omega_{5}-\omega_{3}
 & = & 
 - d\ln\left(x_2'+i\right)- d\ln\left(x_2'-i\right).
\eq
We are in the lucky situation that after the change of variables $(x_2,x_3) \rightarrow (x_2',x_3')$
the integrand depends only on $x_2'$ but not on $x_3'$.
The inverse transformation is given by
\bq
 x_2' \; = \; \sqrt{\frac{4x_2-x_3}{x_3}}.
\eq
The boundary point $(x_2,x_3)=(1,1)$ corresponds to $x_2'=\sqrt{3}$.
Thus we have to integrate in $x_2'$-space from $\sqrt{3}$ to the final value $x_2'$.
We obtain
\bq
 I_\gamma\left(2\omega_{15}-\omega_{3}-\omega_{5},\omega_{5}-\omega_{3};1\right)
 & = &
 \int\limits_{\sqrt{3}}^{x_2'} dt_1 \left( \frac{1}{t_1-i} - \frac{1}{t_1+i} \right)
 \int\limits_{\sqrt{3}}^{t_1} dt_2 \left( \frac{1}{t_2-i} + \frac{1}{t_2+i} \right).
\eq
This is not quite yet the standard definition of multiple polylogarithms, the lower integration boundary equals $\sqrt{3}$, not $0$.
However this is easily adjusted, for example
\bq
 \int\limits_{\sqrt{3}}^{t_1} \frac{dt_2}{t_2-i}
 & = &
 \int\limits_{0}^{t_1} \frac{dt_2}{t_2-i}
 -
 \int\limits_{0}^{\sqrt{3}} \frac{dt_2}{t_2-i}
\eq
and we obtain
\bq
\lefteqn{
 I_\gamma\left(2\omega_{15}-\omega_{3}-\omega_{5},\omega_{5}-\omega_{3};1\right)
 = } & &
 \nonumber \\
 & &
 G\left(i,i;x_2'\right)
 -G\left(-i,i;x_2'\right)
 +G\left(i,-i;x_2'\right)
 -G\left(-i,-i;x_2'\right)
 \nonumber \\
 & &
 -
 \left[ G\left(i;\sqrt{3}\right) + G\left(-i;\sqrt{3}\right) \right]
 \left[ G\left(i;x_2'\right) - G\left(-i;x_2'\right) \right]
 \nonumber \\
 & &
 +G\left(i,i;\sqrt{3}\right)
 +G\left(-i,i;\sqrt{3}\right)
 -G\left(i,-i;\sqrt{3}\right)
 -G\left(-i,-i;\sqrt{3}\right).
\eq
We could stop here, as we managed to express everything in terms of multiple polylogarithms.
However it is instructive to consider also the bootstrap approach. 
As a benefit, this will allow us to simplify the expression above.
The symbol of $I_\gamma(2\omega_{15}-\omega_{3}-\omega_{5},\omega_{5}-\omega_{3};1)$ is
\bq
 S\left(I_\gamma\left(2\omega_{15}-\omega_{3}-\omega_{5},\omega_{5}-\omega_{3};1\right)\right)
 & = & \frac{f_{15}^2}{f_3 f_5} \otimes \frac{f_5}{f_3}.
\eq
We expect $1 \pm i x_2'$ to be possible arguments of the $\mathrm{Li}_n$-functions.
Let us check if they are allowed.
We extend the alphabet by the constants $f_{-1}=-1$ and $f_0=2$.
We have
\bq
 i x_2' \; = \; f_{-1}^\frac{1}{2} f_{7}^{\frac{1}{2}} f_{5}^{-\frac{1}{2}},
 \;\;\;\;\;\;
 1-ix_2'\; = \; f_5^{-1} f_{15},
 \;\;\;\;\;\;
 1+ix_2'\; = \; f_0^2 f_3 f_{15}^{-1}.
\eq
In a similar way one checks that also $(1+ix_2')/(1-ix_2')$ and $(1-ix_2')/(1+ix_2')$ are admissible arguments 
of the $\mathrm{Li}_n$-functions.
We calculate a few symbols:
\bq
 S\left[ 2 \mathrm{Li}_2\left(1-ix_2'\right)-2 \mathrm{Li}_2\left(1+ix_2'\right)\right]
 & = &
 \left( \frac{f_{15}}{f_{3} f_{5}} \otimes \frac{f_{5}}{f_{7}} \right),
 \nonumber \\
 S\left[ \mathrm{Li}_2\left(\frac{1+ix_2'}{1-ix_2'}\right) - \mathrm{Li}_2\left(\frac{1-ix_2'}{1+ix_2'}\right)\right]
 & = &
 \left( \frac{f_{15}}{f_{3} f_{5}} \otimes \frac{f_{7}}{f_{3}} \right).
\eq
We see that the sum of the two terms matches the symbol.
We then check the derivative with respect to $x_2'$. This requires us to add the term
\bq
 i \pi \ln\left(x_2'{}^2+1\right)
\eq
to our ansatz.
Finally we check the value at a specific point.
This will instruct us that we should add the constant
\bq
 i C_7^{(2)} 
\eq
to our ansatz for $I_\gamma(2\omega_{15}-\omega_{3}-\omega_{5},\omega_{5}-\omega_{3};1)$.
Putting everything together
we arrive at
\bq
 J_{15}^{(2)}
 = 
 2i \left[
  2 \mathrm{Li}_2\left(1-ix_2'\right)-2 \mathrm{Li}_2\left(1+ix_2'\right)
  + \mathrm{Li}_2\left(\frac{1+ix_2'}{1-ix_2'}\right) - \mathrm{Li}_2\left(\frac{1-ix_2'}{1+ix_2'}\right)
  + i\pi \ln\left(x_2'{}^2+1\right)
 \right].
\eq
}
\es
\\
%
%
\bs
{\it \refstepcounter{exercise}
{\bf Exercise \theexercise}: 
Prove eq.~(\ref{appendix_trancendental:dgl_horn_function}).
\\
\\
{\bf Solution}:
$P_j({\bf \theta})$ acts on a monomial ${\bf x}^{{\bf i}} = x_1^{i_1} \cdot \dots \cdot x_n^{i_n}$ as
\bq
 P_j\left({\bf \theta}\right) {\bf x}^{{\bf i}}
 & = &
 P_j\left({\bf i}\right) {\bf x}^{{\bf i}}.
\eq
Therefore 
\bq
 \left(1+\theta_j\right) P_j\left({\bf \theta}\right) H\left({\bf x}\right)
 & = &
 \sum\limits_{{\bf i} \in {\mathbb N}_0^n} C_{{\bf i}} \; \left(1+\theta_j\right) P_j\left({\bf \theta}\right) {\bf x}^{\bf i}
 \; = \; 
 \sum\limits_{{\bf i} \in {\mathbb N}_0^n} C_{{\bf i}} \; \left(1+i_j\right) P_j\left({\bf i}\right) {\bf x}^{\bf i}.
\eq
Let's now study
\bq
 \left(1+\theta_j\right)
 Q_j\left({\bf \theta}\right) \frac{1}{x_j} 
 H\left({\bf x}\right)
 & = &
 \sum\limits_{{\bf i} \in {\mathbb N}_0^n} C_{{\bf i}} \; \left(1+\theta_j\right) Q_j\left({\bf \theta}\right) {\bf x}^{{\bf i}-e_j}
 \; = \;
 \sum\limits_{{\bf i} \in {\mathbb N}_0^n} C_{{\bf i}} \; i_j \; Q_j\left({\bf i}-e_j\right) {\bf x}^{{\bf i}-e_j}.
\eq
Due to the factor $i_j$ all terms with $i_j=0$ vanish and the sum over $i_j$ starts at $i_j=1$.
The substitution $i_j \rightarrow i_j-1$ transforms the summation range back to ${\mathbb N}_0$.
We thus have
\bq
 \left(1+\theta_j\right)
 Q_j\left({\bf \theta}\right) \frac{1}{x_j} 
 H\left({\bf x}\right)
 & = &
 \sum\limits_{{\bf i} \in {\mathbb N}_0^n, i_j\ge1} C_{{\bf i}} \; i_j \; Q_j\left({\bf i}-e_j\right) {\bf x}^{{\bf i}-e_j}
 \; = \;
 \sum\limits_{{\bf i} \in {\mathbb N}_0^n} C_{{\bf i}+e_j} \; \left(1+i_j\right) \; Q_j\left({\bf i}\right) {\bf x}^{{\bf i}}.
 \;\;\;
\eq
Hence
\bq
 \left(1+\theta_j\right)
 \left[ 
  Q_j\left({\bf \theta}\right) \frac{1}{x_j} 
  - P_j\left({\bf \theta}\right) 
 \right] H\left({\bf x}\right)
 & = &
 \sum\limits_{{\bf i} \in {\mathbb N}_0^n} 
  \left(1+i_j\right) \left[ C_{{\bf i}+e_j} \; Q_j\left({\bf i}\right) - C_{{\bf i}} \; P_j\left({\bf \theta}\right) \right] {\bf x}^{{\bf i}}.
\eq
From eq.~(\ref{appendix_trancendental:horn_ratio}) we have
\bq
 C_{{\bf i}+e_j} \; Q_j\left({\bf i}\right) - C_{{\bf i}} \; P_j\left({\bf \theta}\right)
 & = &
 0
\eq
and therefore
\bq
 \left(1+\theta_j\right)
 \left[ 
  Q_j\left({\bf \theta}\right) \frac{1}{x_j} 
  - P_j\left({\bf \theta}\right) 
 \right] H\left({\bf x}\right)
 & = &
 0.
\eq
}
\es
\\
\\
%
%
\bs
{\it \refstepcounter{exercise}
{\bf Exercise \theexercise}: 
Show that $[X, X]=0$ implies the anti-symmetry of the Lie bracket $[X,Y]=-[Y,X]$.
Show further that also the converse is true, provided $\mathrm{char}\; {\mathbb F} \neq 2$.
Explain, why the argument does not work for $\mathrm{char}\; {\mathbb F} = 2$.
\\
\\
{\bf Solution}:
\bq
 0 
 \;\; = \;\; 
 \left[ X+Y, X+Y \right]
 \;\; = \;\; 
 \left[ X, Y \right] + \left[ Y, X \right]
\eq
and therefore $[X,Y]=-[Y,X]$.
Now let us consider the other direction. Assuming $[X,Y]=-[Y,X]$ we have for $X=Y$
the relation $[X,X]=-[X,X]$ or equivalently
\bq
\label{appendix_solutions:char_F_relation}
 2 \left[ X,X \right] & = & 0.
\eq
For $\mathrm{char}\; {\mathbb F} \neq 2$ it follows that $[X,X]=0$.
For $\mathrm{char}\; {\mathbb F} = 2$ we have $2=0 \mod 2$
and eq.~(\ref{appendix_solutions:char_F_relation}) does not give any constraint on $[X,X]$.
}
\es
\\
\\
\bs
{\it \refstepcounter{exercise}
{\bf Exercise \theexercise}: 
Derive eq.~(\ref{appendix_lie_algebra:secular_equation}) from eq.~(\ref{appendix_lie_algebra:def_root}).
\\
\\
{\bf Solution}:
Substituting 
\bq
 A \; = \; \sum\limits_{a=1}^n c_a T^a,
 & & 
 X \; = \; \sum\limits_{a=1}^n x_a T^a
\eq
into eq.~(\ref{appendix_lie_algebra:def_root}) we obtain
\bq
 i c_a x_b f^{abc} T^c & = & \rho x_c T^c.
\eq
This is equivalent to
\bq
 \left( c_a x_b i f^{abc} - \rho x_c \right) & = & 0
\eq
and 
\bq
 \left( c_a i f^{abc} - \rho \delta^{bc} \right) x_b & = & 0.
\eq
}
\es
\\
\\
\bs
{\it \refstepcounter{exercise}
{\bf Exercise \theexercise}: 
Consider the Lie algebra ${\mathfrak s}{\mathfrak u}(2)$:
Start from the generators
\bq
 & &
 I^1 =
 \frac{1}{2} 
 \left( \begin{array}{rr}
 0 & 1 \\
 1 & 0 \\
 \end{array} \right),
\;\;\;
 I^2 =
 \frac{1}{2} 
 \left( \begin{array}{rr}
 0 & -i \\
 i & 0 \\
 \end{array} \right),
\;\;\;
 I^3 =
 \frac{1}{2} 
 \left( \begin{array}{rr}
 1 & 0 \\
 0 & -1 \\
 \end{array} \right).
\eq
These generators are proportional to the Pauli matrices and normalised as
\bq
 \mathrm{Tr}\left( I^a I^b \right) & = & \frac{1}{2} \delta^{ab}.
\eq
The commutators are given by 
\bq
 \left[ I^a, I^b \right] & = & i \eps^{abc} I^c,
\eq
where $\eps^{abc}$ denotes the totally antisymmetric tensor.
Start from $A=I^3$. Determine for this choice the roots, the Cartan standard form and the root vectors.
\\
\\
{\bf Solution}:
A root satisfies the equation 
\bq
 \left[ I^3, X \right] & = & \rho X.
\eq
The secular equation reads
\bq
 \det\left( i \eps^{3bc} - \rho \delta^{bc} \right) & = & 0.
\eq
Working this out yields
\bq
 \left|
 \begin{array}{ccc}
 -\rho & i & 0 \\
 -i & -\rho & 0 \\
 0 & 0 & -\rho \\
 \end{array}
 \right|
 & = & 0,
 \nonumber \\
 -\rho^3 +\rho & = & 0,
 \nonumber \\
 \rho \left( \rho^2 - 1 \right) & = & 0.
\eq
Therefore the roots are $0,\pm 1$.
We have
\begin{alignat}{4}
 \rho & = 0: & \left[ I^3, X \right] & = 0 & & \Rightarrow & X & = I^3 = H_1,
 \nonumber \\
 \rho & = 1: & \left[ I^3, X \right] & = X & & \Rightarrow & X & = \frac{1}{\sqrt{2}} \left( I^1 + i I^2 \right) = E_1,
 \nonumber \\
 \rho & = -1: \;\;\; & \left[ I^3, X \right] & = -X \;\;\; & & \Rightarrow \;\;\; & X & = \frac{1}{\sqrt{2}} \left( I^1 - i I^2 \right) = E_{-1}.
\end{alignat}
Thus we obtain the Cartan standard form of ${\mathfrak s}{\mathfrak u}(2)$:
\bq
 & &
 H_1 =
 \frac{1}{2} 
 \left( \begin{array}{rr}
 1 & 0 \\
 0 & -1 \\
 \end{array} \right),
\;\;\;
 E_1 = 
 \frac{1}{\sqrt{2}}
 \left( \begin{array}{rr}
 0 & 1 \\
 0 & 0 \\
 \end{array} \right),
\;\;\;
 E_{-1} =
 \frac{1}{\sqrt{2}} 
 \left( \begin{array}{rr}
 0 & 0\\
 1 & 0 \\
 \end{array} \right).
\eq
The roots are
\bq
 \left[ H, E_1 \right] = E_1,
 & &
 \left[ H, E_{-1} \right] = -E_1.
\eq
The Lie algebra ${\mathfrak s}{\mathfrak u}(2)$ has rank $1$ (there is one generator denoted by $H$)
and the root vectors are one-dimensional. They are given by
\bq
 \vec{\alpha}\left(E_1\right) \; = \; \left(1\right),
 & &
 \vec{\alpha}\left(E_{-1}\right) \; = \; \left(-1\right).
\eq
}
\es
\\
\\
%
%
\bs
{\it \refstepcounter{exercise}
{\bf Exercise \theexercise}: 
Let $(z_2,\dots,z_{n-2})$ be simplicial coordinates and $(x_2,\dots,x_{n-2})$ the corresponding cubical coordinates.
Show that
\bq
 \mathrm{Li}_{m_{n-2} \dots m_3 m_2}\left(x_{n-2},\dots,x_3,x_2\right)
 & = &
 \left(-1\right)^{n-3}
 G_{m_{n-2} \dots m_3 m_2}\left(z_{n-2},\dots,z_3,z_2;1\right)
\eq
\\
\\
{\bf Solution}:
Let's first work out the relation between the simplicial coordinates $z_j$ and the cubical coordinates $x_j$.
From eq.~(\ref{appendix_moduli_space:eq_z_coordinates}) and eq.~(\ref{appendix_moduli_space:eq_x_coordinates}) we have
\bq
 z_i & = &
 \frac{1}{\prod\limits_{j=i}^{n-2} x_j},
 \;\;\;\;\;\;\;\;\;\;\;\;
 i \in \{2,\dots,n-2\}.
\eq
Hence, we have to show
\bq
 \mathrm{Li}_{m_{n-2} \dots m_3 m_2}\left(x_{n-2},\dots,x_3,x_2\right)
 & = &
 \left(-1\right)^{n-3}
 G_{m_{n-2} \dots m_3 m_2}\left(\frac{1}{x_{n-2}},\dots,\frac{1}{x_3 \dots x_{n-2}},\frac{1}{x_2 \dots x_{n-2}};1\right)
 \nonumber \\
\eq
However, this follows immediately from eq.~(\ref{chapter_multiple_polylogarithms:conversion_Li_to_G}).
}
\es
\\
\\
\bs
{\it \refstepcounter{exercise}
{\bf Exercise \theexercise}: 
Let
\bq
 \gamma \; = \; \left(\begin{array}{rr} -1 & 0 \\ 0 & -1 \\ \end{array}\right),
 & &
 \vec{n} \; = \; \left( 0, 0 \right).
\eq
Work out $z'$ and $\tau'$.
\\
\\
{\bf Solution}:
Let start with $\tau'$:
With $a=-1, b=0, c=0, d=-1$ we have
\bq
 \tau' & = & \frac{a\tau +b}{c\tau +d} \; = \; \frac{-\tau +0}{-1} \; = \; \tau.
\eq
Let's now work out $z'$. We have in addition $n_1=n_2=0$ and therefore
\bq
 z'
 & = & 
 \frac{z+n_2 \tau + n_1}{c \tau + d}
 \; = \;    
 \frac{z}{-1}
 \; = \; -z.
\eq
Thus we see that the variable $z$ changes sign.
}
\es
\\
\\
%
%
\bs
{\it \refstepcounter{exercise}
{\bf Exercise \theexercise}: 
Let $R$ be a commutative ring with $1$ and $P$ a prime ideal. Set $S_P = R \backslash P$.
Show that $S_P$ is closed under multiplication.
\\
\\
{\bf Solution}:
Let $a \in S_P$ and $b \in S_P$.
Assume that $a \cdot b \notin S_P$. This means
\bq
 a \cdot b & \in & P.
\eq
However, $P$ is assumed to be a prime ideal. This implies if $a \cdot b \in P$ then either $a \in P$ or $b \in P$.
This contradicts our assumption that $a \notin P$ and $b \notin P$. Hence $a \cdot b \in S_P$ and we have shown
that $S_P$ is closed under multiplication.
}
\es
\\
\\
\bs
{\it \refstepcounter{exercise}
{\bf Exercise \theexercise}: 
Consider the commutative ring ${\mathbb Z}$ and the prime ideal $P=\lideal 5 \rideal$.
Define $S_P = {\mathbb Z}\backslash \lideal 5 \rideal$.
Describe the quotient ring $R_{S_P}$ and its maximal ideal $P_{S_P}$.
\\
\\
{\bf Solution}: The prime ideal $P=\lideal 5 \rideal$ consists of all integer numbers (zero included), which are divisible by $5$.
The set $S_P$ is then the set of all integer numbers, which are not divisible by $5$.
We have $0 \notin S_P$.
The ring ${\mathbb Z}$ is an integral domain, hence the condition
\bq
 s \left(s_2 r_1 - s_1 r_2 \right) & = & 0, 
 \;\;\;\;\;\;
 s, s_1, s_2 \; \in \; S_P,
 \;\;\;
 r_1, r_2 \; \in \; {\mathbb Z},
\eq
implies that either $s=0$ or $s_2 r_1 - s_1 r_2=0$. As $0 \notin S$ it follows that $s_2 r_1 - s_1 r_2=0$.
Hence $R_{S_P}$ is the set
\bq
 R_{S_P} & = &
 \left\{ \, \frac{p}{q} \, | \, p \in {\mathbb Z}, \; q \in {\mathbb N}, \; \mathrm{gcd}(p,q)=1, \; 5 \nmid q \, \right\} \; \subset \; {\mathbb Q}.
\eq
In plain text: $R_{S_P}$ is the set of rational numbers $p/q$ where $q$ is not divisible by $5$
(and in order to have a unique representative we require $\mathrm{gcd}(p,q)=1$ and $q>0$).

Now let's look at the non-invertible elements in the ring $R_{S_P}$:
These are all elements, where $p$ in the numerator is divisible by $5$: If we would invert these
elements in ${\mathbb Q}$, we would get a prime factor $5$ in the denominator, however these inverses are not 
in $R_{S_P}$. Let's define
\bq
 P_{S_P}
 & = &
 \left\{ \, \frac{p}{q} \, | \, p \in {\mathbb Z}, \; q \in {\mathbb N}, \; \mathrm{gcd}(p,q)=1, \; 5| p, \; 5 \nmid q \, \right\}.
\eq
One easily shows that $P_{S_P}$ is an ideal. From theorem~\ref{appendix_algebraic_geometry:theorem_local_ring}
we conclude that $P_{S_P}$ is the only maximal ideal in $R_{S_P}$.
It is easy to see that $P_{S_P}$ is generated by $5=\frac{5}{1}$ in $R_{S_P}$:
\bq
 P_{S_P}
 & = &
 \left\lideal 5 \right\rideal.
\eq 
This motivates the abuse of notation $P=P_{S_P}$. Note that
\bq
 \lideal 5 \rideal \subset {\mathbb Z} & : & \lideal 5 \rideal \; = \; \left\{ \; 5 r \; | \; r \in {\mathbb Z} \, \right\},
 \nonumber \\
 \lideal 5 \rideal \subset R_{S_P} & : & \lideal 5 \rideal \; = \; \left\{ \; 5 r \; | \; r \in R_{S_P} \, \right\}.
\eq
}
\es
\\
%
%
\bs
{\it \refstepcounter{exercise}
{\bf Exercise \theexercise}: 
Let $r$ be defined by eq.~(\ref{appendix_finite_fields:chinese_remainder_theorem_r}). Determine
\bq
 r \bmod n_i.
\eq
{\bf Solution}:
We have with $s_i \tilde{n}_i + t_i n_i = 1$
\bq
 r \bmod n_i
 & = &
 \left( \sum\limits_{j=1}^k r_j s_j \tilde{n}_j \bmod n \right) \bmod n_i
 \; = \; 
 \sum\limits_{j=1}^k r_j s_j \tilde{n}_j \bmod n_i
 \nonumber \\
 & = &
 r_i s_i \tilde{n}_i \bmod n_i
 \; = \; 
 r_i \left(1-t_i n_i\right) \bmod n_i
 \; = \;
 r_i \bmod n_i
 \nonumber \\
 & = &
 r_i.
\eq
}
\es